\documentclass[10pt]{article}

\usepackage{amsmath}
\usepackage{array}
\usepackage{appendix}
\usepackage{graphicx}
\usepackage{amsfonts}
\usepackage{amssymb}
\usepackage{mathrsfs}
\usepackage{yfonts}
\usepackage{euscript}
\usepackage{upgreek}
\usepackage{slantsc}
\usepackage{calligra}
\usepackage[T1]{fontenc}
\usepackage{epsf}
\usepackage{latexsym}

\usepackage{tipa}

\def\be{\begin{equation}}
\def\ee{\end{equation}}
\def\beq{\begin{equation}}
\def\eeq{\end{equation}}
\def\bea{\begin{eqnarray}}
\def\eea{\end{eqnarray}}

\def\ni{\noindent}
\def\foo{\footnote}
\def\bu{\noindent$\bullet$}

\def\hat{\widehat}
\def\tilde{\widetilde}
\def\!{\hspace{-1.6667em}}

\def\mA{\mbox{A}}   
\def\mB{\mbox{B}}  
\def\mC{\mbox{C}}   
\def\mD{\mbox{D}}
\def\mE{\mbox{E}}
\def\mF{\mbox{F}}
\def\mG{\mbox{G}}
\def\mH{\mbox{H}} 
\def\mI{\mbox{I}}
\def\mJ{\mbox{J}}  \def\nJ{\mbox{J}}
\def\mK{\mbox{K}}
\def\mL{\mbox{L}}
\def\mM{\mbox{M}}
\def\mN{\mbox{N}} 

\def\mP{\mbox{P}}
\def\mQ{\mbox{Q}}
\def\mR{\mbox{R}}
\def\mS{\mbox{S}}
\def\mT{\mbox{T}} 
\def\mU{\mbox{U}}
\def\mV{\mbox{V}}
\def\mW{\mbox{W}}
\def\mX{\mbox{X}}
\def\mY{\mbox{Y}}
\def\mZ{\mbox{Z}} 
\def\ma{\mbox{a}}
\def\mb{\mbox{b}}
\def\mc{\mbox{c}} 
\def\md{\mbox{d}} 
\def\me{\mbox{e}}

\def\mg{\mbox{g}}
\def\mh{\mbox{h}}
\def\mi{\mbox{i}}
\def\mj{\mbox{j}}  \def\nj{\mbox{j}} 
\def\mk{\mbox{k}}
\def\ml{\mbox{l}}  \def\nl{\mbox{l}} 
\def\mm{\mbox{m}}  \def\nm{\mbox{m}} 
\def\mn{\mbox{n}}  \def\nn{\mbox{n}} 
\def\mo{\mbox{o}}
\def\mp{\mbox{p}} 

\def\mr{\mbox{r}}
\def\ms{\mbox{s}}
\def\mt{\mbox{t}}
\def\muu{\mbox{u}}
 
\def\mw{\mbox{w}}

\def\my{\mbox{y}}

\def\nl{\mbox{{l}}}
\def\nm{\mbox{{m}}}
\def\nn{\mbox{{j}}}
\def\nr{\mbox{{r}}}
\def\nuu{\mbox{{u}}}

\def\nM{\mbox{{M}}}
\def\nP{\mbox{{P}}}

\def\ttA{\mbox{\tt A}}
\def\ttB{\mbox{\tt B}}
\def\ttC{\mbox{\tt C}}
\def\ttD{\mbox{\tt D}}
\def\ttE{\mbox{\tt E}}
\def\ttF{\mbox{\tt F}}
\def\ttG{\mbox{\tt G}}
\def\ttH{\mbox{\tt H}}
\def\ttI{\mbox{\tt I}}
\def\ttJ{\mbox{\tt J}}
\def\ttK{\mbox{\tt K}}
\def\ttL{\mbox{\tt L}}
\def\ttN{\mbox{\tt N}}

\def\ttP{\mbox{\tt P}}
\def\ttQ{\mbox{\tt Q}}
\def\ttS{\mbox{\tt S}}
\def\ttT{\mbox{\tt T}}
\def\ttU{\mbox{\tt U}}
\def\ttV{\mbox{\tt V}}
\def\ttW{\mbox{\tt W}}
\def\ttg{\mbox{\tt g}}
\def\tta{\mbox{\tt a}}
\def\ttb{\mbox{\tt b}}
\def\ttc{\mbox{\tt c}}

\def\tte{\mbox{\tt e}}
\def\ttf{\mbox{\tt f}}                          %
\def\ttg{\mbox{\tt g}}                          

\def\ttp{\mbox{\tt p}}
\def\ttq{\mbox{\tt q}}
\def\ttr{\mbox{\tt r}}
\def\tts{\mbox{\tt s}}
\def\ft{\mbox{\tt t}} 
\def\ttu{\mbox{\tt u}}
\def\ttv{\mbox{\tt v}}

\def\chij{\upchi_{\mbox{\scriptsize j}}}         
\def\urho{{\underline{\rho}}}
\def\uupbeta{{\underline{\upbeta}}}
\def\suupbeta{\mbox{\scriptsize$\underline{\upbeta}$}}

\def\uupkappa{{\underline{\upkappa}}}
\def\uupi{{\underline{\underline{\pi}}}}

\def\Bigalpha{\mbox{\Large $\alpha$}}               
\def\Bigbeta{\mbox{\Large $\alpha$}}                
\def\normalbeta{\mbox{\normalsize $\beta$}}         
\def\normalalpha{\mbox{\normalsize $\beta$}}        
\def\Bigupalpha{\mbox{\Large $\alpha$}}             
\def\Biggamma{\mbox{\Large $\gamma$}}               
\def\Bigkappa{\mbox{\Large $\kappa$}}

\def\BigupRho{\mbox{\Large $\rho$}}                 
\def\Bigdelta{\mbox{\Large $\delta$}}               
\def\Bigvarepsilon{\mbox{\Large $\varepsilon$}}
\def\Bigeta{\mbox{\Large $\eta$}}
\def\lepsilon{\mbox{\Large $\epsilon$}}

\def\brho{\mbox{\boldmath$\rho$}}          
\def\bnu{\mbox{\boldmath$\nu$}}            
\def\bdelta{\mbox{\boldmath$\delta$}}  

\def\uupbeta{\mbox{\underline{$\upbeta$}}}

\def\bsigma{\mbox{\boldmath$\sigma$}}
\def\bupSigma{\mbox{\boldmath$\Sigma$}}      
\def\sbSigma{\mbox{\scriptsize\boldmath$\Sigma$}} 
\def\bGamma{\mbox{\boldmath$\Gamma$}}
\def\btheta{\mbox{\boldmath$\theta$}}
\def\bTheta{\mbox{\boldmath$\theta$}}
\def\buppi{\mbox{\boldmath$\uppi$}}

\def\fa{\mbox{\sffamily a}}

\def\fc{\mbox{\sffamily c}}

\def\ff{\mbox{\sffamily f}}
\def\fg{\mbox{\sffamily g}}
\def\fh{\mbox{\sffamily h}}

\def\fl{\mbox{\sffamily l}}
\def\fm{\mbox{\sffamily m}}
\def\fn{\mbox{\sffamily n}}

\def\fp{\mbox{\sffamily p}}

\def\fs{\mbox{\sffamily s}}
\def\ft{\mbox{\sffamily t}}

\def\fw{\mbox{\sffamily w}}

\def\fA{\mbox{\sffamily A}}
\def\fB{\mbox{\sffamily B}}
\def\fC{\mbox{\sffamily C}}
\def\fD{\mbox{\sffamily D}}
\def\fE{\mbox{\sffamily E}}
\def\fF{\mbox{\sffamily F}}
\def\fG{\mbox{\sffamily G}}
\def\fH{\mbox{\sffamily H}}
\def\fI{\mbox{\sffamily I}}
\def\fJ{\mbox{\sffamily J}}

\def\fL{\mbox{\sffamily L}}
\def\fM{\mbox{\sffamily M}}
\def\fN{\mbox{\sffamily N}}

\def\fP{\mbox{\sffamily P}}
\def\fQ{\mbox{\sffamily Q}}

\def\fS{\mbox{\sffamily S}}
\def\fT{\mbox{\sffamily T}}
\def\fU{\mbox{\sffamily U}}
\def\fV{\mbox{\sffamily V}}
\def\fW{\mbox{\sffamily W}}

\def\fZ{\mbox{\sffamily Z}}

\def\cE{{\mathscr E}}
\def\cF{{\mathscr F}}
\def\cG{{\mathscr G}}

\def\cL{{\mathscr L}}

\def\cO{{\mathscr O}}
\def\cP{{\mathscr P}}

\def\cS{{\mathscr S}}
\def\cT{{\mathscr T}}
\def\cU{{\mathscr U}}
\def\cV{{\mathscr V}}
\def\cW{{\mathscr W}}

\def\cs{\mbox{\scriptsize$\cS$}}

\def\uA{\underline{A}}
\def\uB{\underline{B}}

\def\uF{\underline{\mbox{F}}}

\def\uR{\underline{R}}

\def\up{\underline{p}} 
\def\uq{\underline{q}}

\def\ux{\underline{{x}}} 
\def\uy{\underline{{y}}}

\def\urr{\underline{r} }
\def\uPP{\underline{P}}

\def\upp{\underline{p}}
\def\uqq{\underline{q}}

\def\R{\underline{R}}
\def\P{\underline{P}}
\def\r{\underline{r}}

\def\p{\underline{p}}
\def\q{\underline{q}}

\def\bh{\underline{\underline{\mbox{h}}}  }            
\def\sbh{\underline{\underline{\mbox{\scriptsize h}}}  }     

\def\suF{\underline{\mbox{\scriptsize F}}}

\def\barp{\bar{\tt p}}
\def\barq{\bar{\tt q}}
\def\barr{\bar{\tt r}}
\def\ip{i^{\prime}}
\def\jp{j^{\prime}}
\def\kp{k^{\prime}}
\def\lp{l^{\prime}}
\def\ipp{i^{\prime\prime}}
\def\jpp{j^{\prime\prime}}
\def\kpp{k^{\prime\prime}}
\def\lpp{l^{\prime\prime}}

\def\bn{\mbox{\bf n}}
\def\bU{\mbox{\bf U}}
\def\bp{\mbox{\bf p}}
\def\bA{\mbox{\bf A}}

\def\bP{\mbox{\bf P}}
\def\bM{\mbox{\bf M}}
\def\bQ{\mbox{\bf Q}}
\def\bS{\mbox{\bf S}}

\def\bg{\mbox{{\bf g}}}
\def\bM{\mbox{{\bf M}}}
\def\bM{\mbox{{\bf M}}}

\def\bh{\mbox{{\bf h}}}

\def\bn{\mbox{{\bf n}}}
\def\bq{\mbox{{\bf q}}}
\def\br{\mbox{{\bf r}}}

\def\bx{\mbox{{\bf x}}}
\def\by{\mbox{{\bf y}}}

\def\bV{\mbox{\bf V}}
\def\bd{\mbox{\bf d}}

\def\bfQ{\mbox{{\bf \sffamily Q}}}
\def\bfP{\mbox{{\bf \sffamily P}}}
\def\bfM{\mbox{{\bf \sffamily M}}}

\def\bfQ{\mbox{\boldmath {\sffamily Q}}}

\def\scA{\mbox{\scriptsize ${\cal A}$}}

\def\scC{\mbox{\scriptsize ${\cal C}$}}          
\def\scD{\mbox{\scriptsize ${\cal D}$}}          
\def\scE{\mbox{\scriptsize ${\cal E}$}}          

\def\scH{\mbox{\scriptsize ${\cal H}$}}          
\def\scI{\mbox{\scriptsize ${\cal I}$}}

\def\scL{\mbox{\scriptsize ${\cal L}$}}          
\def\scM{\mbox{\scriptsize ${\cal M}$}}          
\def\scN{\mbox{\scriptsize ${\cal N}$}}

\def\scP{\mbox{\scriptsize ${\cal P}$}}
\def\scQ{\mbox{\scriptsize ${\cal Q}$}}          

\def\scU{\mbox{\scriptsize ${\cal U}$}}          

\def\mathfrakV{\mbox{\LARGE $\mathfrak{v}$}}
\def\bigr{\mbox{\boldmath$\mathfrak{R}$}}

\def\Fr{\mbox{\Large $\mathfrak{r}$}}

\def\FP{\mbox{\Large $\mathfrak{p}$}}
\def\FrQ{\mbox{\Large $\mathfrak{q}$}}
\def\FrB{\mbox{\boldmath$\mathfrak{B}$}}
\def\FrU{\mbox{\boldmath$\mathfrak{U}$}}
\def\FrP{\mbox{\boldmath$\mathfrak{P}$}}
\def\FrH{\mbox{\boldmath$\mathfrak{H}$}}
\def\FrV{\mbox{\boldmath$\mathfrak{V}$}}
\def\sFrH{\mbox{\boldmath\scriptsize$\mathfrak{H}$}}
\def\FrCC{\mbox{$\mathfrak{C}$}}
\def\FrC{\mbox{\Large $\mathfrak{c}$}}
\def\sFrC{\mbox{\normalsize $\mathfrak{c}$}}
\def\FrX{\mbox{\Large $\mathfrak{x}$}}
\def\sFrX{\mbox{\normalsize $\mathfrak{x}$}}
\def\FrD{\mbox{\boldmath$\mathfrak{D}$}}
\def\FrT{\mbox{\boldmath$\mathfrak{T}$}}                        
\def\sFrT{\mbox{\boldmath\scriptsize$\mathfrak{T}$}}
\def\sFrG{\mbox{\boldmath\scriptsize$\mathfrak{G}$}}
\def\sFrM{\mbox{\boldmath\scriptsize$\mathfrak{M}$}}
\def\FrF{\mbox{$\mathfrak{F}$}}                                 
\def\FrL{\mbox{$\mathfrak{L}$}}                                 
\def\FrM{\mbox{\Large $\mathfrak{m}$}}                          
\def\FrMgen{\mbox{\boldmath$\mathfrak{M}$}}                              
\def\sFrT{\mbox{\large $\mathfrak{T}$}}
\def\tFrT{\mbox{\normalsize $\mathfrak{T}$}}
\def\sFQ{\mbox{$\mathfrak{q}$}}
\def\sFR{\mbox{$\mathfrak{r}$}}
\def\sFG{\mbox{$\mathfrak{g}$}}
\def\sFS{\mbox{$\mathfrak{s}$}}
\def\sFP{\mbox{$\mathfrak{p}$}}
\def\FrR{\mbox{\Large $\mathfrak{r}$}}
\def\bFrR{\mbox{$\mathfrak{R}$}}
\def\bsFrR{\mbox{\scriptsize$\mathfrak{R}$}}
\def\sbFrR{\mbox{\scriptsize$\mathfrak{R}$}}
\def\tFrRR{\mbox{\tiny$\mathfrak{R}$}}
\def\FrS{\mbox{\Large $\mathfrak{s}$}}
\def\sFrP{\mbox{\normalsize $\mathfrak{p}$}}
\def\FrG{\mbox{\Large $\mathfrak{g}$}}
\def\capFrG{\mbox{\boldmath$\mathfrak{G}$}}                    
\def\capFrC{\mbox{\boldmath$\mathfrak{C}$}}                    
\def\scapFrG{\mbox{\boldmath\scriptsize$\mathfrak{G}$}}        
\def\scapFrC{\mbox{\boldmath\scriptsize$\mathfrak{C}$}}        

\def\FrO{\mbox{\Large $\mathfrak{o}$}}
   
\def\FA{\mbox{\Large $\mathfrak{a}$}}

\def\FC{\mbox{\Large $\mathfrak{c}$}}
\def\FO{\mbox{\Large $\mathfrak{o}$}}

\def\stS{\mbox{\large\tt s}}                        
\def\FS{\mbox{\LARGE\tt s}}                         
\def\lm{\mbox{\Large $m$}}                          
\def\le{\mbox{\Large $e$}}                          
\def\lH{\mbox{\Large $H$}}                          
\def\lme{\mbox{\Large e}}

\def\sa{\mbox{\scriptsize a}}
\def\sb{\mbox{\scriptsize b}}
\def\scc{\mbox{\scriptsize c}}
\def\sd{\mbox{\scriptsize d}}
\def\se{\mbox{\scriptsize e}}
\def\sf{\mbox{\scriptsize f}}
\def\sg{\mbox{\scriptsize g}} 
\def\sh{\mbox{\scriptsize h}} 
\def\si{\mbox{\scriptsize i}}
\def\sj{\mbox{\scriptsize j}} 
\def\sk{\mbox{\scriptsize k}}
\def\sll{\mbox{\scriptsize l}}  
\def\sm{\mbox{\scriptsize m}}
\def\sn{\mbox{\scriptsize n}} 
\def\so{\mbox{\scriptsize o}} 
\def\sp{\mbox{\scriptsize p}}
\def\sq{\mbox{\scriptsize q}}
\def\sr{\mbox{\scriptsize r}}
\def\sss{\mbox{\scriptsize s}}  
\def\st{\mbox{\scriptsize t}}
\def\su{\mbox{\scriptsize u}}
\def\sv{\mbox{\scriptsize v}}
\def\sw{\mbox{\scriptsize w}}
\def\sx{\mbox{\scriptsize x}}
\def\sy{\mbox{\scriptsize y}} 

\def\sA{\mbox{\scriptsize A}} 
\def\sB{\mbox{\scriptsize B}}
\def\sC{\mbox{\scriptsize C}}
\def\sD{\mbox{\scriptsize D}}
\def\sE{\mbox{\scriptsize E}}
\def\sF{\mbox{\scriptsize F}}
\def\sG{\mbox{\scriptsize G}}
\def\sH{\mbox{\scriptsize H}}
\def\sI{\mbox{\scriptsize I}}
\def\sJ{\mbox{\scriptsize J}}
\def\sK{\mbox{\scriptsize K}}
\def\sL{\mbox{\scriptsize L}} 
\def\sM{\mbox{\scriptsize M}} 
\def\sN{\mbox{\scriptsize N}} 
\def\sO{\mbox{\scriptsize O}}
\def\sP{\mbox{\scriptsize P}} 
\def\sQ{\mbox{\scriptsize Q}} 
\def\sR{\mbox{\scriptsize R}}
\def\sS{\mbox{\scriptsize S}}
\def\sT{\mbox{\scriptsize T}}
\def\sU{\mbox{\scriptsize U}}
\def\sV{\mbox{\scriptsize V}}
\def\sW{\mbox{\scriptsize W}}
\def\sX{\mbox{\scriptsize X}} 
\def\sY{\mbox{\scriptsize Y}} 
\def\sZ{\mbox{\scriptsize Z}} 

\def\iB{\mbox{\scriptsize$B$}}   
\def\iD{\mbox{\scriptsize$D$}}   
\def\iK{\mbox{\scriptsize$K$}}   
\def\iC{\mbox{\scriptsize$B$}}   
\def\tiK{\mbox{\tiny$K$}}        

\def\sfa{\mbox{\sffamily{\scriptsize a}}}
\def\sfb{\mbox{\sffamily{\scriptsize b}}}
\def\sfc{\mbox{\sffamily{\scriptsize c}}}
\def\sfd{\mbox{\sffamily{\scriptsize d}}}
\def\sfe{\mbox{\sffamily{\scriptsize e}}}
\def\sff{\mbox{\sffamily{\scriptsize f}}}
\def\sfg{\mbox{\sffamily{\scriptsize g}}}

\def\sfl{\mbox{\sffamily{\scriptsize l}}}

\def\sfp{\mbox{\sffamily{\scriptsize p}}}
\def\sfq{\mbox{\sffamily{\scriptsize q}}}

\def\sfs{\mbox{\sffamily{\scriptsize s}}}
\def\sft{\mbox{\sffamily{\scriptsize t}}}

\def\sfA{\mbox{\sffamily{\scriptsize A}}}
\def\sfB{\mbox{\sffamily{\scriptsize B}}}
\def\sfC{\mbox{\sffamily{\scriptsize C}}}
\def\sfD{\mbox{\sffamily{\scriptsize D}}}
\def\sfE{\mbox{\sffamily{\scriptsize E}}}
\def\sfF{\mbox{\sffamily{\scriptsize F}}}

\def\sfI{\mbox{\sffamily{\scriptsize I}}}
\def\sfJ{\mbox{\sffamily{\scriptsize J}}}
\def\sfK{\mbox{\sffamily{\scriptsize K}}}

\def\sfO{\mbox{\sffamily{\scriptsize O}}}

\def\sfQ{\mbox{\sffamily{\scriptsize Q}}}

\def\sfS{\mbox{\sffamily{\scriptsize S}}}
\def\sfT{\mbox{\sffamily{\scriptsize T}}}
\def\sfU{\mbox{\sffamily{\scriptsize U}}}
\def\sfV{\mbox{\sffamily{\scriptsize V}}}
\def\sfW{\mbox{\sffamily{\scriptsize W}}}   
\def\sfX{\mbox{\sffamily{\scriptsize X}}}
\def\sfY{\mbox{\sffamily{\scriptsize Y}}}
\def\sfZ{\mbox{\sffamily{\scriptsize Z}}}

\def\stta{\mbox{\scriptsize ${\tta}$}}
\def\sttB{\mbox{\scriptsize ${\tta}$}}
\def\sttg{\mbox{\scriptsize ${\ttg}$}}
\def\sttp{\mbox{\scriptsize ${\ttp}$}}
\def\sttq{\mbox{\scriptsize ${\ttq}$}}
\def\sttr{\mbox{\scriptsize ${\ttr}$}}

\def\sbm{\mbox{{\bf \scriptsize m}}}
\def\sbh{\mbox{{\bf \scriptsize h}}}

\def\sbd{\mbox{{\bf \scriptsize d}}}

\def\sbM{\mbox{{\bf \scriptsize M}}}
\def\sbN{\mbox{{\bf \scriptsize N}}}

\def\sbS{\mbox{{\bf \scriptsize S}}}

\def\sbU{\mbox{{\bf \scriptsize U}}}

\def\sbfM{\mbox{\bf \scriptsize\sffamily M}}

\def\sbttM{\mbox{\scriptsize\boldmath{\tt M}}}

\def\sttc{\mbox{\scriptsize $\bar{\ttc}$}}

\def\stts{\mbox{\scriptsize $\bar{\tts}$}}

\def\tip{\tilde{\mbox{\scriptsize$\ttp$}}}
\def\tiq{\tilde{\mbox{\scriptsize$\ttq$}}}
\def\tir{\tilde{\mbox{\scriptsize$\ttr$}}}
\def\tis{\tilde{\mbox{\scriptsize$\tts$}}}

\def\usF{\underline{\mbox{\scriptsize F}}}

\def\ta{\mbox{\tiny a}}
\def\tb{\mbox{\tiny b}}
\def\tc{\mbox{\tiny c}}
\def\td{\mbox{\tiny d}}
\def\te{\mbox{\tiny e}}
\def\tf{\mbox{\tiny f}}

\def\th{\mbox{\tiny h}}
\def\ti{\mbox{\tiny i}}

\def\tk{\mbox{\tiny k}}
\def\tl{\mbox{\tiny l}}

\def\tm{\mbox{\tiny m}}
\def\tn{\mbox{\tiny n}}
\def\to{\mbox{\tiny o}}
\def\tp{\mbox{\tiny p}}

\def\tr{\mbox{\tiny r}}
\def\ts{\mbox{\tiny s}}
\def\ttt{\mbox{\tiny t}}   
\def\tu{\mbox{\tiny u}}

\def\ty{\mbox{\tiny y}}

\def\tA{\mbox{\tiny A}}
\def\tB{\mbox{\tiny B}}
\def\tC{\mbox{\tiny C}}
\def\tD{\mbox{\tiny D}}
\def\tE{\mbox{\tiny E}}
\def\tF{\mbox{\tiny F}}
\def\tG{\mbox{\tiny G}}
\def\tH{\mbox{\tiny H}}

\def\tJ{\mbox{\tiny J}}
\def\tK{\mbox{\tiny K}}

\def\tM{\mbox{\tiny M}}
\def\tN{\mbox{\tiny N}}
\def\tO{\mbox{\tiny O}}
\def\tP{\mbox{\tiny P}}

\def\tR{\mbox{\tiny R}}
\def\tS{\mbox{\tiny S}}
\def\tT{\mbox{\tiny T}}

\def\tV{\mbox{\tiny V}}

\def\tY{\mbox{\tiny Y}}

\def\tfA{\mbox{\sffamily{\tiny A}}}
\def\tfB{\mbox{\sffamily{\tiny B}}}

\def\tfZ{\mbox{\sffamily{\tiny Z}}}

\def\tcH{\mbox{\tiny ${\cal H}$}} 
\def\tcM{\mbox{\tiny ${\cal M}$}} 
\def\tcD{\mbox{\tiny ${\cal D}$}} 
\def\tcL{\mbox{\tiny ${\cal L}$}} 

\def\biiP{\mbox{\boldmath$P$}}
\def\biP{\mbox{\scriptsize\boldmath$P$}}
\def\biK{\mbox{\scriptsize\boldmath$K$}}

\def\Q{\overline{\ttQ}}

\def\bttM{\mbox{\boldmath \ttM}}
\def\ttH{\mbox{\tt{H}}}
\def\ttM{\mbox{\tt{M}}}
\def\sttM{\mbox{\scriptsize\tt{M}}}
\def\bari{\bar{p}}
\def\ttP{\mbox{\tt P}}
\def\B{\underline{B}}

\def\ft{\mbox{\sffamily t}}
\def\lft{\mbox{\Large \sffamily t}}
\def\lt{\mbox{\Large $t$}}
\def\lmt{\mbox{\Large t}}
\def\sft{\mbox{\scriptsize\tt t}}

\def\K{Kucha\v{r} }

\def\NSI{Na\"{\i}ve Schr\"{o}dinger Interpretation }
\def\CPI{Conditional Probabilities Interpretation }
\def\NSII{Na\"{\i}ve Schr\"{o}dinger Interpretation}
\def\CPII{Conditional Probabilities Interpretation}

\def\pa{\partial}
\def\d{\textrm{d}}

\def\ordial{\bd\hspace{-0.088in}\pa}          
\def\ordional{\bdelta\hspace{-0.08in}\bd}     
\def\partional{\bdelta\hspace{-0.08in}\pa}    
\def\sordial{{\sbd\hspace{-0.07in}\mbox{\scriptsize$\pa$}}}       
\def\spartional{\bdelta\hspace{-0.063in}\pa}

\def\Circ{\mbox{\Large$\circ$}}               
\def\Last{\mbox{\Large$\ast$}}                
\def\5Star{\mbox{\Large$\star$}}              
\def\Ast{\mbox{\Large$\ast$}}                 %
\def\Rec{\mbox{\textcircled{$\star$}}}        

\def\Diamond{\mbox{\Large$\diamond$}}          
\def\Heart{\mbox{$\heartsuit$}}            
\def\Spade{\mbox{$\spadesuit$}}           
\def\Club{\mbox{$\clubsuit$}}             

\def\SStar{\mbox{\normalsize$\ast$}}

\def\cr{\mbox{\scriptsize{\bf $\mbox{ } \times \mbox{ }$}}}

\def\suma{\sum\mbox{}_{\mbox{}_{\mbox{\scriptsize $i$}}}}
\def\sumi3{\sum\mbox{}_{\mbox{}_{\mbox{\scriptsize $i$=1}}}^3}
\def\sumin{\sum\mbox{}_{\mbox{}_{\mbox{\scriptsize $i$=1}}}^{n}}
\def\sumiN{\sum\mbox{}_{\mbox{}_{\mbox{\scriptsize $I$=1}}}^{N}}
\def\sumIN{\sum\mbox{}_{\mbox{}_{\mbox{\scriptsize $I$=1}}}^{N}}
\def\sumj3{\sum\mbox{}_{\mbox{}_{\mbox{\scriptsize $j$=1}}}^3}
\def\sumk3{\sum\mbox{}_{\mbox{}_{\mbox{\scriptsize $k$=1}}}^3}

\def\prodbp{\prod\mbox{}_{\mbox{}_{\mbox{\scriptsize $\sfA$ = 1}}}}

\begin{document}

\begin{titlepage}

\normalfont

\vspace{.7in}
\begin{center}
\Large{\bf THE PROBLEM OF TIME AND QUANTUM COSMOLOGY}

\Large{\bf IN THE RELATIONAL PARTICLE MECHANICS ARENA}\normalsize 

\vspace{.1in}

\normalsize

\vspace{.4in}

{\large \bf Edward Anderson}$^1$

\vspace{.2in}

\large {\em APC AstroParticule et Cosmologie, Universit\'{e} Paris Diderot CNRS/IN2P3, CEA/Irfu, 
Observatoire de Paris, Sorbonne Paris Cit\'{e}, 10 rue Alice Domon et L\'{e}onie Duquet, 75205 Paris Cedex 13, France,} \normalsize

\vspace{.2in}

\large and {\em DAMTP, Centre for Mathematical Sciences, Wilberforce Road, Cambridge CB3 OWA.  } \normalsize


\end{center}

\begin{abstract}

\mbox{ }

This article contains a local solution to the notorious Problem of Time in Quantum Gravity at the conceptual level and which is actually realizable for the relational triangle.
The Problem of Time is that `time' in GR and `time' in ordinary quantum theory are mutually incompatible notions, 
which is problematic in trying to put these two theories together to form a theory of Quantum Gravity. 
Four frontiers to this resolution in full GR are identified, alongside three further directions not yet conquered even for the relational triangle.

This article is also the definitive review on relational particle models originally due to Barbour (2003: dynamics of pure shape) and Barbour and Bertotti (1982: 
dynamics of shape and scale).
These are exhibited as useful toy models of background independence, which I argue to be the `other half' of GR to relativistic gravitation, as well as the originator of the 
Problem of Time itself.  
Barbour's work and my localized extension of it are shown to be the classical precursor of the background independence that then manifests itself 
at the quantum level as the full-blown Problem of Time.  
In fact 7/8ths of the Isham--Kucha\v{r} Problem of Time facets are already present in classical GR; even classical mechanics in relational particle mechanics 
formulation exhibits 5/8ths of these!
In addition to Isham, Kucha\v{r} and Barbour, the other principal authors whose works are drawn upon in building this Problem of Time approach are Kendall 
(relational models only: pure-shape configuration spaces), Dirac, Teitelboim and Halliwell (Problem of Time resolving components).  
The recommended scheme is a combination of the Machian semiclassical approach, histories theory and records theory.

\end{abstract}

%
%

\vspace{2.5in}

\noindent This piece is far from necessarily to be read linearly, it can instead be treated as a reference book, 
in particular using keyword searches to find all mentions for each topic of interest to the reader.  
Use the Preface or the Index to identify the particular topics covered.  
As well as `Quantum Gravity' and `Semiclassical Quantum Cosmology', these in good part interdisciplinary works are of interest in 
Theoretical Mechanics, Shape Geometry and Shape Statistics, Molecular Physics/Theoretical Chemistry and the Physics and Philosophy of Time, Space and Relationalism.  
A number of conceptually interesting innovations in the Principles of Dynamics are supplied, as befit Physics cast in a relational language.  

\noindent As regards searching the File for which Problems of Time manifested by whichever strategy of interest, these are tagged by `X with Y' for X the problem and Y the strategy. 
See Fig \ref{Evol-Fac} for the evolution from the familiar names of the Problem of Time Facets to the corrected names actually used in this Article, 
and Figs \ref{Temporary-Web-Cl} and \ref{Temporary-Web-QM} for lists of the strategies considered in the Article.

\mbox{ }

\noindent It is my intent to update this Preprint every year or two in the manner of a `Living Review'.  

\mbox{ } 

\noindent ea212@cam.ac.uk

\end{titlepage}

\noindent\Large\bf Acknowledgements \normalfont\normalsize

\mbox{ }

Foremost, I thank my Wife, Sophie Reed and Amelia Booth.  
Without your support, I could have never written this.

\noindent For, over the years, discussions, references and help with my career, I thank 
Professor Enrique Alvarez,
Dr Ioannis Alevizos, 
Dr Julio Arce, 
Professors Julian Barbour, 
Harvey Brown, 
Martin Bojowald, 
Jeremy Butterfield,
and Louis Crane, 
Mr Alexis de Saint-Ours,  
Dr Andreas Doering, 
Professor Fay Dowker, 
Mr Chris Duston, 
Professor George Ellis, 
Dr Cecilia Flori,
Dr Brendan Foster, 
Ms Anne Franzen,  
Professor Belen Gavela, 
Dr Marc Geiller, 
Professor Gary Gibbons, 
Dr Henrique Gomes, 
Dr Alexei Grinbaum,  
Dr Sean Gryb, 
Professor Jonathan Halliwell, 
Professor Petr Ho\v{r}ava, 
Mr Pui Ip, 
Professor Chris Isham,
Dr Smaragda Kessari, 
Professor Claus Kiefer, 
Ms Sophie Kneller, 
Dr Tim Koslowski, 
Professor Karel Kucha\v{r},
Professor Marc Lachi\`{e}ze--Rey, 
Professor H. Le,  
Mr Matteo Lostaglio,
Professors Jorma Louko, 
Malcolm MacCallum 
and Lionel Mason,
Dr Flavio Mercati, 
Professor Niall \'{O} Murchadha,
Dr Jonathan Oppenheim,  
Professors Daniele Oriti,
Don Page,
Renaud Parentani,
Jorge Pullin  
and Carlo Rovelli, 
Dr Ntina Savvidou,  
Mr David Schroeren,  
Mr Eduardo Serna,
Professors Christopher Small, 
Rafael Sorkin, 
Gerard t'Hooft and 
Reza Tavakol,
Dr Jurgen Theiss,  
Professor Paolo Vargas-Moniz, 
Dr James Yearsley 
and 
Professor Jimmy York.  
I also thank the organizers of the ``Second Conference on Time and Matter", at Bled, Slovenia, of the ``Space and Time 
100 Years after Minkowski" Conference at Bad Honnef, Germany, the Perimeter Institute, Queen Mary University of London, Imperial College 
London, The Spinoza Institute in Utrecht, and the ``Do We Need a Physics of Passage?" Conference in Cape Town for invitations to speak.
I thank in anticipation all those of you, and others, who go on to provide further lucid discussion/examples/critique/references 
about topics contained in this Article now that it is publically available.

I thank Peterhouse for funding me in 2004--2008 bar in 2005, during which I thank the Killam Foundation for funding me at the University of Alberta.      
Some of the work in the 2008--2010 academic years was supported by travel money via Grant Number RFP2-08-05 from The Foundational Questions Institute (fqXi.org).  
I thank Universidad Autonoma de Madrid for funding me in 2010--2011.  
I thank the fqXi for subsequently funding me via Grant FQXi-RFP3-1101.

I also thank my friends 
Adam,
Alex,
Alicia, 
Amy,
Anne, 
Anya,  
Becky,
Beth, 
Bryony,
Cecilia,  
Charlotte, 
Coryan, 
Duke,
Ed, 
Emma,   
Emilie, 
Emily,
Eve, 
Gue,
Gabrielle, 
Hannah, 
Hettie, 
Jaya,
Jenny,  
Katie,
Liz,
Luke,  
Lynnette, 
Maria,
Marwah,
Rosie,
Sally,   
Sharon, 
Simeon, 
Sophie 
and Xander 
for keeping my spirits up.

I finally wish to pay my respects to Profs John Archibald Wheeler and David Kendall, who passed away during the years I have been working on this problem.  
Your life's works have been of great value and inspiration in my own studies, as this Article makes clear.  

\mbox{ } 

\noindent{\bf \Large Preface: this Article as a Quantum Gravity program} 

\mbox{ }

\noindent This Article represents a new such program.  
Here, Quantum Gravity is viewed as a piecing together General Relativity (GR) and ordinary Quantum Theory. 
How to do this is far from simple or clear since these two individually successful branches of Physics 
rest on what at least appear to be distinct and incompatible foundations. 
GR does not only have an aspect as a relativistic theory of gravity that renders Newtonian gravity compatible with relativity, 
but also a second aspect as a freeing from absolute/background structure.
Following on from this double nature, I argue that ``Quantum Gravity" is a misnomer since only making reference to the former.    
I.e. a truer name would then be something like "Quantum theory of the relativistic gravitation--background independence gestalt", 
which I abbreviate by QG with the `G' standing for `gestalt' rather than just for `gravity'.\footnote{As to other names, `Quantum GR' will not do in this 
role since it implies the specific Einstein field equations, whereas Quantum Gestalt remains open-minded as to {\sl which} relativistic theory of gravitation is 
involved; this distinction parallels that between `Quantum GR' and `Quantum Gravity'.  
Quantum Gestalt is also in contradistinction to `Background Independent Quantum Gravity' since the latter carries connotations of the background 
independent and gravitational parts being {\sl separate}, whereas in the former they are held to be {\sl joint}.} 
Now, much as one can model Quantum Gravity without Background Independence (at least for some purposes), 
one can likewise model Quantum Background Independence without Relativistic Gravitation; 
from the Quantum Gestalt perspective, these two ventures are complementary, 
suggesting that one may learn considerably from the latter as well as from the former.
Thus the present Article's quantum relational particle mechanics (RPM) = `quantum Background Independence without SR' (Special Relativity) models are  
GR-like as regards this second aspect. 
I.e. a valid, albeit hitherto less considered alternative to other partial piecings together of Theoretical Physics programs such as the quantum relativistic theory not of gravity 
(alias QFT: Quantum Field Theory) \cite{qftcorr, Weinberg}, or quantum nonrelativistic theory of Newtonian Gravity without Background Independence (see e.g. \cite{QNG2, QNG1}).  
Moreover, there is value in multiple partial perspectives when the total perspective is intractably hard; one is to use of multiple toy models (see below). 
The present Article is then a first review of how the partial features of QG 
that are present in RPM's are of comparable value to the full problem as the partial features of, say, GR, QFT and Quantum Newtonian Gravity are.  
Finally I strongly argue in this Article that Background Independence implies a Problem of Time (POT) [as opposed to smuggling in Background-Dependent features 
as POT resolutions, as suggested in certain fairly recent programs].  

\mbox{ }

\noindent I make the following choices (partly based on the above arguments).

\noindent a) I choose a background-independent/canonical approach \cite{DeWitt67, Kuchar92, I93} that pays careful attention to GR 
features rather than a `QM first' covariant or perturbative strings approach \cite{DeWitt2, GSW1, GSW2, Berkovits}.  

\noindent b) I choose the simpler non-supersymmetric case (see \cite{Arelsusy} for further supersymmetric considerations).  
Whilst I do not necessarily choose the dimension of space $d$ in the gravitational theory to be 3, I want the theory to have the usual kind of degrees of 
freedom ($d = 2$ has no local such), so I do require $d \geq 3$ gravitation.  
3-$d$ space is then the simplest case that has this nontriviality, and appears to correspond to reality, on which grounds it remains preferable. 
I regard the capacity to extend such a scheme to $d \geq 4$ as an occasionally relevant bonus rather than something to specifically aim for or generalize to).  

\noindent c) I choose an approach that rather more closely parallels Geometrodynamics than the Ashtekar Variables/Loop Quantum Gravity (LQG) approach 
(which I term Nododynamics -- knot dynamics -- as per the argument in Sec \ref{Nodos}). 
I make this choice in part because 

\noindent i) Geometrodynamics has particularly solid conceptual foundations. 

\noindent ii) This Article does, among other things, consider the possibility of making do without canonical transformations and phase space enlargements, 
particularly in programs intended to involve quantization; on the other hand, Ashtekar variables based approaches rely on these.    

\noindent iii) The POT, which is the principal interest in this Article, has been studied in greater bredth and depth in this setting 
(in part due to it having been around for longer). 

\noindent Moreover, the Leibniz--Mach--Barbour (LMB) meaning of `relational' and the study of Background Independence in Physics as per the present Article 
will also be of interest in Nododynamics, as well as being relevant toward a broad understanding 
of some of the features which M-theory (but not perturbative string theory) would be expected to have.  
Additionally, this Article's principal theme of the POT and an understanding of its various possible resolutions is 
an important one for both of these theories due to their background independence.  
Finally, the above principal connections with other QG programs by no means exhausts the totality of such connections made in this Article.  

\noindent d) I proceed via considering a toy model for Geometrodynamics (it is less close to Nododynamics; 
other theories such as electromagnetism, Yang--Mills theory \cite{PullinGambini}, the Husain--\K model \cite{HusKu} 
and BF theory \cite{PullinGambini, BF, Smolin08}, serve as toy models in that case).  
I have particular interest in midisuperspace-like aspects (i.e. features beyond minisuperspace, but within simpler models than actual 
midisuperspace: structure formation and nontrivial linear constraints), and in certain approaches to the POT (see below). 
\noindent As well as minisuperspace and inhomogeneous perturbations thereabout \cite{HallHaw}, other toy models of comparable complexity to 
those in the present Article are Carlip's work on 2 + 1 gravity \cite{Carlip}, the parametrized particle \cite{Kuchar92}, parametrized field 
theories \cite{Kuchar91, Kuchar92}, strong gravity \cite{I76, Pilatilit12,  Pilatilit3, San}, 
the Montesinos--Rovelli--Thiemann model \cite{MRT} and the bosonic string as a toy model of Geometrodynamics \cite{KuTorre, Kuchar92, KK02}.

\noindent e)  This Article's principal toy model is relational particle mechanics (RPM); this builds on how it has been shown \cite{RWR, Phan} 
that GR is relational in an LMB-like sense (see \cite{BB82, B94I, Phan} and the present Article), 
which GR itself has also been shown to obey \cite{RWR, Phan}.    
I.e. RPM's are background-independent in a similar (but simpler) way to how GR is.  

\noindent f) Within the possible RPM's, I go for the simplest, so as to maximize tractability.  
Thus I choose 

\noindent 1) to consider plain over mirror-image-identified configurations (though that possibly sins 
by toy-modelling ordinary rather than affine Geometrodynamics \cite{Klauder} and Secs \ref{ComRel-KinQ}, \ref{Affine3}). 

\noindent 2) To study distinguishable particles over (partly) indistinguishable ones (though that 
probably sins against being as Leibnizian as possible as per Secs \ref{Intro} and \ref{Examples}).  

\noindent 3) I need scaled models as regards realistic Quantum Cosmology, semiclassicality and internal time 
(for some other purposes, pure-shape -- i.e. scale-free -- models are simpler, and these occur as subproblems within scaled models).  

\noindent 4) Each of 1-$d$, 2-$d$ and 3-$d$ models have an increasing number of midisuperspace-like features. 
2-$d$ is the first scaled model to have a nontrivial linear constraint.  
Namely, the zero angular momentum constraint on the whole model universe. 
However, 1-$d$ already has a notion of localization. 
[These two notions are interlinked  {\sl in GR}, where they are both related to spatial derivatives being nontrivial, but these notions are separate for RPM's.]  
However, for 2-$d$ the associated quotiented-out group (analogue of the spatial diffeomorphisms in GR) is $SO(2)$, which is rather simple (e.g. Abelian, trivial group-orbit structure). 
Considering 3-$d$'s $SO(3)$ thus contributes a bit more. 
[However, it still in no way emulates a number of the more specific features of the diffeomorphisms, which lie entirely outside the scope of this Article's toy models.]  
Finally, I note that 3-$d$ is considerably less tractable than 2-$d$, so, for detailed considerations, I restrict myself to at most 2-$d$ in the present Article. 
To be entirely clear, I consider 2-$d$ RPM as a toy model of 3-$d$ Geometrodynamics because this analogy itself is very insensitive to the toy model's $d$ provided that it is $> 1$.  

\noindent 5) RPM's also have much freedom in the form of their potential.  
No potential is in many ways the simplest possibility, but it lacks in realisticness and in normalizability.  
Multiple HO-like potentials are also simple to study (far more exactly tractable than more complicated cases, and now with nice 
boundedness properties on the QM states), though such potentials are also known to be atypical in their simpleness and properties. 
I consider it an improvement on the modelling to use the potential freedom so as to better match up with Quantum Cosmology -- to have a slow heavy dynamics of RPM's scale. 
[This is the moment of inertia of the system, or, more usually, its square root.]
It is in direct parallel with GR (semi)classical cosmological scale dynamics. 
This Cosmology--Mechanics analogy determines which potentials more complicated than HO's to consider.  
Another qualification of `simple' is lacking in change of relative shape momenta (angular and/or relative distance momenta, see Sec \ref{Dyn1}). 
However, this is disappointing from the relational and emergence of dynamics perspectives, so I do what I can to get round this simpleness.  

\noindent 6) Zero angular momentum has the good fortune of simplifying mechanics from a fibre bundles/topological defects perspective.
One gets very simple mathematics, particularly for 1-$d$ models and 3 particles in 2-$d$ for the above plain and distinguishable choices ($\mathbb{S}^{N - 2}$ and $\mathbb{S}^{2}$ 
mathematics that is very close to textbook QM of one particle in 3-$d$ and straightforward generalizations of these).  

\noindent This simpleness is a triumph, because then one can do and check many POT calculations 
(which would {\sl not} make sense if done for atoms or molecules, say, due to the absolutist underpinnings in these latter cases). 
Thus the current Article concerns taking very old and well-known mathematics and giving it a new interpretation by 
which it makes sense in a model Quantum Cosmology context to which many POT schemes and a number of other Quantum Gravity subtleties are relevant.\foo{Moreover, 
4 particles in 2-$d$ is a tougher problem with less standard mathematics -- it is $\mathbb{CP}^2$, whose 
isometry group is $SU$(3) rather than some $SO(p)$ group as occurs for $N$ particles in 1-$d$ or 3 particles in 2-$d$ and which requires somewhat harder mathematics.  
4 particles in 2-$d$ is currently the frontier of my research, with only a few starting notes in the present Article; see \cite{QuadI, QuadII, QuadIII} for further developments. 
Passing from 3 to 4 particles in 2-$d$ is comparable in increase of complexity to passing from diagonal to non-diagonal Bianchi IX minisuperspace.
The extension to $N$  particles in 2-$d$ is only slightly harder (unlike in 3-$d$, where increasing particle number very quickly leads to serious mathematical difficulties).}
%
And it probes a distinct set of features from minisuperspace, having the abovementioned midisuperspace-like attributes while not having fully 
GR-inherited potentials or kinetic term indefiniteness (as well as not having counterparts of some field theoretic and/or diffeomorphism-specific features).  
This is rather in the spirit of the POT reviews of Kucha\v{r} \cite{Kuchar92} and Isham \cite{I93}; 
the present Article could be  viewed as a framework to improve on those by exploiting the newly opened-up RPM toy model arena.  

\mbox{ } 

\noindent My specific RPM models permit dilational hidden time, emergent semiclassical time, timeless approaches 
(Na\"{\i}ve Schr\"{o}dinger Interpretation, Conditional Probabilities Interpretation, Records Theory), Histories Theory and some observables/beables 

\noindent -based approaches.  
Excluded approaches include `Riem time' (supplied by GR's indefinite kinetic metric) and various diffeomorphism-related approaches. 
I also see there to be sizeable difficulties with hidden and matter time approaches using RPM and relational considerations.  
The Thin Sandwich Conjecture and its Best Matching generalization to arbitrary theories via RPM's are of particular significance in this approach, 
and lead to a particular POO/B resolution.  
I consider strategization about each POT facet rather than just the Frozen Formalism Problem. 
Of the approaches I favour and RPM's are capable of investigating, RPM's were largely selected 
due to their capacity for investigation of timeless approaches, and constructed to be compatible with the Semiclassical Approach.  

\noindent One further goal then is to piece together Records Theory, the Semiclassical Approach and Histories Theory.  
I provide an arena for systematic study of the Semiclassical Approach's usually omitted terms: back-reaction, diabatic terms, averaged terms and terms that bring in higher derivatives.
I also provide a wider range of dilational hidden times.  
I further develop more on Records Theory, including notions of distance and some considerations of notions of information.  
I consider the histories--records--Semiclassical Approach combination along the lines of Halliwell \cite{H99, H03, H09} for RPM's.  
This remains work in progress.  
This Article also provides a substantial treatment of the nature of relationalism, which different authors 
(in particular  Barbour, Rovelli, Crane and I) have taken to mean rather different things.
A main part of the program developed in this Article are the ties between relationalism and an improved understanding of the POT: 
classical relationalism is demonstrated here to be the same as the classical precursor of the POT, and the quantum-level POT itself is seen as ordering of 
configurational relationalism, Temporal Relationalism, quantization and observables/beables finding steps. 
I provide a number of new particularities of the variational principles of dynamics for the specifically-relational and thus POT-geared setting. 


As regards other foundational issues in Quantum Cosmology, RPM's additionally exhibit a number of aspects of the operator-ordering problem, 
including new midisuperspace ones and a connection between operator-ordering and the absolute versus relational debate. 
I also give useful qualitative treatment of uniform states and structure formation.   


Classical and quantum antecedents of this Article's program are as follows.  

\noindent 1) Barbour's relationalism \cite{BB82, B94I}. 

\noindent 2) Hojman--Kucha\v{r}--Teitelboim's \cite{T73, Tei73b, HKT} work on geometrodynamics.

\noindent 3) My previous program with Barbour, Foster, Kelleher and O'Murchadha \cite{RWR, AB, ABFO, ABFKO, Phan, Lan2}.

\noindent 4) That program's other descendants (see Sec 2 \cite{Kuchar92, I93} for details) by such as Gryb, Mercati, Koslowski, Gomes and Budd.  
%

\noindent 5) Isham and \K's POT reviews; I demonstrate in the present Article that Barbour's work and mine provides and resolves 
a classical precursor for the POT, as well as making quite a number of subsequent suggestions for use at the quantum level.  

\noindent 6) Halliwell's approach to Quantum Cosmology \cite{HallHaw, Halliwell87, H99, H03, H09, H11}.

\mbox{ }
 
\noindent{\bf Prerequisites for the present Article}. 
Principles of Dynamics                           as in Lanczos                         \cite{Lanczos}
                                           and also in Dirac                           \cite{Dirac},
GR                                               as in Wald                            \cite{Wald}, 
QM                                               as in e.g. Landau and Lifschitz       \cite{LLQM} 
                                           and also in Isham                           \cite{Ibook} 
and elementary Geometry and Geometrical Methods  as in e.g. Nakahara                   \cite{Nakahara}.  
A strong familiarity with 
real linear Methods of Mathematical Physics is also desirable,                               
                                                 as in e.g. Courant and Hilbert        \cite{CH}.  

\mbox{ }

\noindent {\bf Questions}. The text contains around 100 of these.  These have begun to lead to a number of papers and projects. 
Some of v1's questions have already been answered in the current v3; these are no longer listed as questions.
Stars indicate which questions are expected to be harder to answer.
7 particular ones are pointed out as the present program's main frontiers.

\vspace{10in}

\tableofcontents

\vspace{10in}

\section{Introduction} \label{Intro}

The present Article is a study of the Problem of Time (POT) and other foundational and conceptual issues in `Quantum Gravity' and Quantum Cosmology.  
Parts II and IV of the present Article review the POT, affording a rather more detailed (and separated out into classical and quantum parts) overview than the recent publications 
\cite{APOT}, \cite{APOT2}, as well as containing my own recommended POT program (some small portions of which are in \cite{ARel2, ACos2, AHall}).  
As explained in the Preface, the present Article exposits Quantum Gestalt (QG) to be a more suitable renaming of, and conceptual content for, `Quantum Gravity' that 
places comparable weighting on coming up with a quantum version of GR's aspect as a freeing of Physics from background structure as well as of GR's aspect as a relativistic 
theory of gravitation.

The present Article's study them makes substantial use of a simple concrete examples of Quantum Background Independence: relational particle models (RPM's), as well as of associated 
relational considerations (not just restricted to these models); Parts I and III are the definitive Reviews of these models.
The word `relational' is used very differently by Barbour (and various collaborators) and I \cite{BB82, B94I, RWR, ABFKO, Phan, FORD} 
on the one hand, and by Crane \cite{Crane93, Crane} and Rovelli \cite{Rov96, Rovellibook}  
(with subsequent use in Loop Quantum Gravity (`LQG') \cite{Montes, Rovellibook, Thiemann, GPPT, BojoBook, Rfqxi, Bojo1}) on the other hand.  
This Article mostly concerns the former and my own improvement and extension of that, though it does contrast, and to some extent compose, this with the latter.  
Moreover, the position on relationalism developed here (an abridged version of which has been submitted for publication in \cite{ARel}) mostly lies {\sl between} Barbour's \cite{B94I, 
Bfqxi} and Rovelli's \cite{Rovellibook, Rfqxi} extremes regarding the fine detail of the Machian notion of time as being abstracted from change.  
A central tenet of this quartet of Theses is that Barbour and I's work on relationalism constitutes the classical precursor to many quantum POT issues; it is by this that it 
makes sense to consider classical and quantum Theses on relational models concurrently with classical and quantum Theses on the POT: these are very richly, novelly and insightfully 
interconnected.  
Unlike the POT, Relationalism is a subject in which there have been various recent reviews by other authors. 
Thus, to contrast with the present Article, I comment that Pooley's review \cite{Pooley12} is a 3-way comparison of which three Barbour's alone features in the present Article  
(I cover the approaches that enter Quantum Gravity programs, whereas Pooley goes for the past two decades' philosophizing).  
Barbour's article \cite{B11} and Gryb's Thesis \cite{GrybTh} largely only cover Barbour's approach [see also the Third (2012) Edition of Kiefer's book \cite{Kieferbook}].

The RPM's considered in the present Article are as follows.  

\mbox{ }

\noindent 1) {\bf Scaled RPM} alias Euclidean RPM (abbreviated to ERPM) alias Barbour--Bertotti 1982 theory \cite{BB82} (after its first proponents and in
distinction to their 1977 theory \cite{BB77}) is a mechanics in which only relative times, relative angles and relative separations are meaningful. 
In the Physics literature, it was further studied or featured in reviews in \cite{B86, BS, Rovelli, Smolin, Kuchar92, B94I, Buckets, 
EOT, Gergely, GergelyMcKain, Paris, 06I, Kieferbook, TriCl, 08I, MGM, Cones, ScaleQM, 08III, SemiclIII, GrybTh, AHall, ARel, APOT2, ACos2}. 
It has additionally been substantially discussed in the Philosophy of Physics literature (see footnote \ref{Spinne} for references).  

\mbox{ } 

\noindent 2) {\bf Pure-shape RPM} alias Similarity RPM (abbreviated to SRPM) alias Barbour 2003 theory \cite{B03} is a mechanics in which only relative
times, relative angles and ratios of relative separations are meaningful.  
In the Physics literature, it has been further studied in \cite{Piombino, Paris, 06II, TriCl, FORD, 08I, 08II, AF, +Tri, QShape, QSub, B11, QuadI, QuadII, PhasePaper, 
LostaglioMA, QuadIII}.  

\mbox{ } 

\noindent
RPM's occur within the following relational program.\foo{The historical background for this relational   
program 	 is the absolute or relational motion debate.  
[I use `relational' here rather than `relative' here to avoid confusion with Einstein's relativity, which 
also features prominently.    
Philosophers of Physics have also taken to using `substantival' instead of `absolute'.]
This debate has been around since the inception of Newtonian Mechanics.

In outline, Newton's \cite{Newton} traditional formulation of mechanics defined `true' motion to be 
relative to an absolute space that is all-pervasive, infinite, invisible and can not be acted upon.  
Newton also considered motion to occur {\sl in} time.  
His notion of time was additionally an absolute one.  
I.e., external, continuous, and uniformly flowing as `needed to transform kinematic geometry into physical dynamics' \cite{DOD}. 
(In fact, much of this conception of time was not new to Newton, c.f. Barrow \cite{Barrow}, Gassendi \cite{Gassendi} and even Ptolemy \cite{Ptolemy}.)
The virtue of Newton's scheme is working well in practise, nature immediately vindicating its practical consequences in detail.

However, Newton's formulation of mechanics has been argued to be bankrupt from a philosophical perspective. 
Absolute space does not comply with Leibniz's famous {\bf identity of indiscernibles} \cite{L}, by which e.g. our universe and a 
copy in which all material objects are collectively displaced by a fixed distance surely share all observable properties and thus are one and the same.  
Also, Leibniz already considered time not to be a separate external entity with respect to which things change.  
Bishop Berkeley \cite{Berkeley} and Mach \cite{M} (see also e.g. \cite{DOD, Buckets} for further discussion) added to these arguments, 
e.g. Einstein attributed holding absolute space to be a non-entity on account of its not being actable upon to Mach.    
The alternative is for mechanics to be {\bf relational}: spatial properties are to be entirely about the relations between 
material objects, and Mach's notion of `time to be abstracted from change'. 
Historically, there was a lack in alternative theories with such features, though the comparatively recent RPM's of Barbour--Bertotti (1982) 
and Barbour (2003) have made up for this deficiency.

Furthermore, Leibniz, Bishop Berkeley and Mach's arguments are philosophically compelling enough that they ought to apply to not just Mechanics but to Physics as a whole.
As argued in Secs \ref{Intro-Cl} and \ref{Rel-Gdyn}, GR is also relational, and indeed relationalism is one foundation from which GR 
{\sl can} be derived, for all that this was not the historical route to GR.  
For, while Einstein was interested in `Machian issues', the way in which he viewed these does not coincide with the Barbour-relational view 
\cite{WheelerGRT, DOD, RWR}, nor did Einstein's historical route to GR \cite{Einstein1, Einstein2} constitute a {\sl direct} implementation of Machian ideas.

See e.g. \cite{B86, Buckets, Barbourphil3, B99, EOT, Hofer, Pooley012, PooleyBr, Saunders12, ButterBar, Earman, Ryn, Sklar, Smolin08, Kon, Thebault1, Thebault2, Pooley12} 
and Sec \ref{AORM-Concl} as regards arguments for (and against) the value of Barbour and Bertotti's 1982 theory toward the absolute or relative motion debate.   
Relationalism would appear to have a good case for use in whole-universe situations, both from a 
philosophical perspective and because GR and at least some further GR-like theories do implement it.   
\label{Spinne}} 

\mbox{ }

{         \begin{figure}[ht]
\centering
\includegraphics[width=0.65\textwidth]{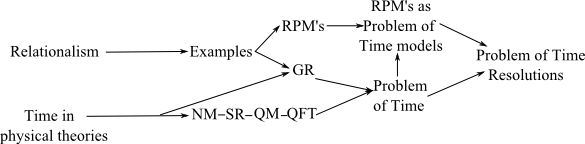}
\caption[Text der im Bilderverzeichnis auftaucht]{        \footnotesize{We first concentrate on outlining the nature of this relational program (Secs \ref{Underlay} to \ref{Dir-CR}). 
Then we outline RPM's in a bit more detail in Sec \ref{Intro-RPM-Ex}, further motivating them by showing how GR is a relational theory too in Sec \ref{GR-Is-Relational}. 
We then outline Parts I and III on classical and Quantum RPM's in Secs \ref{Intro-Cl} and \ref{Intro-QM}.
We then need to start a new thread on the nature of time, background independence and the POT in Secs \ref{Time} and \ref{POTiQG}.
[This is the first manifestation of the frontpage's warning that this Article is nonlinear...]
This done, we can outline parts II and IV -- classical and quantum POT exemplified chiefly by RPM's. 
We end the Introduction by listing other fundamental questions in Quantum Cosmology touched upon by this Article in Sec \ref{Q-Cos}.  } }
\label{Shape-of-Sec-1} 
\end{figure}          }

\subsection{Plausible underlaying postulates for this Article's Relational Program}\label{Underlay}

Relationalism 0) {\bf Relational Physics is to be solely about the relations between tangible entities}.

\mbox{ }

\noindent This is taken to be 's universalization to all Physics of footnote \ref{Spinne}'s statement of relationalism for mechanics.

\mbox{ } 

\noindent I use `{\bf tangible entities}' rather than `material objects' to make it clear that this is 
open to fields and `force mediators' as well as `matter building blocks', as befits modern Physics.
The key properties of tangible entities are as follows. 

\mbox{ }

\noindent Relationalism 1) [Einstein attributed this to Mach] They {\bf act testably and are actable upon}. 

\mbox{ }

\noindent Relationalism 2) [Leibniz] Any such which are {\bf indiscernible are held to be identical}. 

\mbox{ } 

%
\noindent Note 1) That which is not acting testably or actable upon is held to be a {\sl physical} non-entity. 
[This is still held to be a type of thing as regards being able to {\sl philosophize} about it or {\sl mathematically represent} it;  absolute space is an obvious archetypal example.] 

%
\noindent Note 2) As regards Relationalism 2), in Physics, relationalism posits that physical indiscernibility {\sl must} trump multiplicity of mathematical representation; 
this multiplicity still exists mathematically, but the mathematics corresponding to the {\sl true} physics in question is the equivalence class spanning that multiplicity.
See e.g. Butterfield and Caulton \cite{BC11} for a range of attitudes to the identity of indiscernibles and alternatives to it.  
One would then only wish to attribute physical significance to calculations of tangible entities which succeed in being independent 
of the choice of representative of the equivalence class; the archetype of this is Gauge Theory.  
%
%
An important lesson from this   is that a set of part-tangible and part-non entities is often altogether more straightforward to represent mathematically.

%
\noindent Note 3) The word `testable' also requires caution, as to whether it means now or ever. 
\noindent [Better data can split two hitherto indiscernible models.  
\noindent But `ever' could well be regarded as containing part-non-entities such as the infinite future or the attainability of experimental/observational perfection.
On the other hand, some uses of `ever' are based on principle, e.g. the indiscernibility of the universe from 
itself under a uniform translation does not depend on how accurately we know the universe (unless this somehow 
provided evidence for a new paradigm of Physics in which absolute position was a tangible entity).
I merely state, rather than resolve these dilemmas.]

%
\noindent Note 4) In this Article's main sense of relational, one treats instantaneous configurations and time separately as befits the great 
conceptual heterogeneity between them.


%
\noindent Note 5) I do not take this SSec's principles too seriously; it is the standard use of configuration space, and Relationalism 4--6) and 8), 
perhaps supplemented by Relationalism 3) and 7), that are the actually mathematically-implemented starting point of the Leibniz--Mach--Barbour relational 
approach (and variants).  
As such, Relationalism 0--2) for now being vague or loose as compared to Relationalism 3--8) is not a problem for the work actually done in this Article, 
but only for this present SSec, which represents a tentative under{\sl laying} rather than a rigorous foundational under{\sl lying}.

\subsection{Configuration space. And the portmanteau notation} \label{Config}

As regards the instantaneous configuration entities of a given theory, which I denote by ${\fQ}^{\sfC}$, a key further concept is the 

\noindent{\sl configuration space} $\FrQ$, i.e. the space of all the possible instantaneous configurations 
(each of the points in configuration space represents one state of the instantaneous configuration, i.e. a set of one value per ${\fQ}^{\sfC}$).  

\mbox{ }

\noindent Example 1) Particle positions are configurations; see \cite{Lanczos} for a clear exposition of the configuration space notion for these.

\noindent Example 2) The more reduced inter-particle relative separations or relative cluster separations (see Sec \ref{Examples}) are configurations.

\noindent Example 3) Sec \ref{Q-Geom} covers even more reduced RPM configurations.  

\noindent Example 4) The values at each point of continuous extended objects, whether fields or geometrical objects that have values everywhere in space, which include both the 3-metrics 
(or the more reduced 3-geometries) of GR (see Sec \ref{Examples}) and objects that have their own separate {\it notion of space (NOS)} of extent such as strings or membranes. 

\mbox{ }

\noindent I denote the general NOS by $\bupSigma_p$ for $p \leq d$ the dimension of space, and I coordinatize it by $x^{\mu_p}$. 
However, in this Article all specific examples of this are for $p = d$ (field theories), in which case I drop the $p$ subscripts, or for $p = 0$, in which case $\bupSigma_0$ is just a point 
corresponding to the particle position, which point ceases to possess any internal coordinatization. 
\noindent I denote integration over whatever NOS is appropriate by $\int_{\bupSigma_p}\d \bupSigma_p$. 
I take this to collapse to a multiplicative 1 for particle mechanics and other finite theories.  

\mbox{ } 

\noindent I denote finite theory cases' instantaneous configurations by $Q^{\sfC}$.
For field theory cases, I denote the instantaneous configurations by $\mQ^{\sfC}(\ux)$. 
Thus the sans serif object $\fQ^{\sfC}$ is a portmanteau of the upright space-dependent case for field theories and the slanty case for finite theories.   
That is the default notation in this Article as regards typeface, straight and slanty symbols.
{\sl Such portmanteaux embody in shorthand many of the analogies that are in place in toy-modelling an infinite field theory via a finite toy model}.
In particular, they allow for a single expression to encompass both the geometrodynamical formulation of GR and its toy-modelling 
by the RPM's that are this Article's principal toy models. 
This Article uses quite a number of fonts in similarly systematic ways.  
If lost at any stage, consult Appendix A.1.  

\mbox{ } 

\noindent Note 1) The field theory part is {\sl curved}, so later parts of this analogy are more subtle than simply jointly treating particles and fields, 
for all that the latter is another plausible use for a distinct portmanteau.

\noindent Note 2) The $\fC$ in use above is a suitable multi-index, over particle and/or continuous extended object 
species.\footnote{I use $\cP\lfloor\fQ^{\tfA}\rfloor$ as the portmanteau of $P(Q^{\tfA})$ for finite
theories and $\mP(x^{\mu_p}; \mQ^{\tfA}]$ for infinite theories, where round brackets indicate function dependence, square brackets indicate 
functional dependence and ( ; ] is mixed function-functional dependence, with the former prior to the semi-colon and the latter after it. 
I use braces for actual brackets.  
I then use a special font (Large typeface) for such portmanteaux that come integrated over their corresponding notion of space (the action $\mbox{\tt S}$ is a such); 
these are therefore pure functionals in terms of what they depend on, but differ in the portmanteau way in terms of what features on the computational side of the equation.}

\noindent N.B. configuration space is generally not just a set, being, rather, further equipped with topological space, metric space and geometrical structure.   
A broad example of the last of these is Finsler metric geometry $\FrQ = \langle\FrS, F \rangle$ for 
$\FrS$ a topological manifold and $F$ the metric function.\footnote{I generally use
$\langle \mbox{ }  \rangle$ to denote `space of'.
I use lower-case Greek indices for spatial indices 1 to $d$.  
I use bold font as shorthand for configuration space indices and underline for spatial indices.}
%
(Semi-)Riemannian geometry is then a common subcase of this which is usually written in terms of the metric 
${\fM}_{\sfA\sfB}$ itself: $\FrQ = \langle\FrS, {\mbox{\boldmath${M}$}}\rangle$. 
Some of the relational program's examples are in fact infinite-dimensional generalizations of such geometries. 
(So $F \rightarrow$ F or to $\fF$ to cover both.   
But I still refer to such by the usual finite-dimensional geometries' names, i.e. I elevate those names to be finite and field-theoretic portmanteaux of the usual finite notions). 

\mbox{ } 

\noindent `Relationalism 3)' [Barbour] This concerns the possibility that the {\bf configurations are the primary entities}.  
I.e. the configuration space $\FrQ$ has a primary role in the conceptualization of Physics.  

\mbox{ }

%
\noindent In fact,   
Relationalism 3) is not technically a relationalist postulate but a {\bf minimalist} one [`Minimalism i)']; such go quite often hand in hand with actual relational postulates.  
This does not necessarily mean that relationalists discard structures, but rather that they are prepared to consider the outcome of doing so.  
Some relational ideas are a subset of minimalism, e.g. doubting and discarding absolute space and absolute time.  
See also Secs \ref{Cl-Soln} and \ref{Conclusion} as regards whether scale is to be discarded from Physics, 
%
%
Secs \ref{Examples} and \ref{+temJBB} as regards relationalists' questioning of Minkowski's view on spacetime and its subsequent primacy over how GR is conceived of.
[I.e. arguing that the dynamical tradition favours the primality of space, and that this, despite amounting to supposing less structure, 
nevertheless turns out to serve as an alternative foundation for GR.]  

\mbox{ }

\noindent There are in fact various strengths of this.  
For now, one can consider just $\fQ^{\sfC}$ but also consider these alongside a notion of change $\ordial \fQ^{\sfC}$, corresponding to paths in configuration space.
We shall see later that upgrading this to a history or including momenta offer further variety away from the strictest possible $\fQ^{\sfC}$ minimality.  

\mbox{ } 

\noindent [The finite--field theoretic portmanteau implies not one but three notions of portmanteau derivative. 
First of all, I use $\delta$     for functional      derivatives; then I use 
      $\ordial$    for {\it ordial}    derivatives (ordinary, partial derivative portmanteau), 
      $\ordional$  for {\it ordional}  derivatives (ordinary, functional derivative portmanteau), and  
	  $\partional$ for {\it partional} derivatives (partial, functional derivative portmanteau).     
It is convenient at this point to explain further notation: I hang `cov', `abs' suffixes on these for covariant and absolute derivatives, using $\mD$ for $\pa$ and ${\cal D}$ 
for $\delta$ in these contexts, and I use $\triangle$ for Laplacians (suitably suffixed to say of which type), $\mathbb{D}$ for measures and oversized $\mbox{\large $\delta$}$ 
for variations.]

\subsection{The aim of building relational actions and then using the Principles of Dynamics}\label{Build-Actions}

Relationalism 4) [Barbour] One aims to build up {\bf relational actions} from natural compound objects derived from the $\fQ^{\sfC}$.  

\mbox{ } 

%
\noindent Note 1) In making use of actions, we are drawing upon the Principles of Dynamics.  
Many of the steps below follow from this as natural consequences of having an action, rather than of relationalism per se. 
But the particular actions used embody relationalism, and the some of the subsequent Principles of Dynamics workings assist in enforcing that. 
This valuing and making specific use of the Principles of Dynamics differs from Mach's own thinking; it is due to Barbour and concerns 
positing that relational actions are the most fundamental starting-point for (classical) whole-universe physics.\footnote{Barbour tells me that \K called 
this additional part of Barbour's thinking a `very Cartesian', meaning that Barbour puts a high premium on clear/manifest concepts.
That is relevant background for many of my relational reformulations of the standard Principles of Dynamics in Secs \ref{Intro} and \ref{Examples}.}
 
%
\noindent Note 2) This is a mathematical {\it structural composition} postulate; it is relational only in that the primary objects themselves are to be taken to be relational.   

%
\noindent Note 3) The compound objects readily include NOS derivatives (whenever the NOS is nontrivial) and contracted objects.  
From what the reader knows about actions, one might expect velocities to be required too, and these involve an incipient notion of time, 
so we need to discuss that next before making progress with the construction of actions. 
Moreover, the manner of that progress may well be unexpected to the reader.  
In particular, actions conventionally involve velocities.

\subsection{Temporal Relationalism and actions implementing it}\label{TR}

\subsubsection{Temporal Relationalism}\label{TR-Itself}

Relationalism 5) [Leibniz] A physical theory is {\bf temporally relational} if there is {\bf no meaningful 
primary notion of time for the system as a whole} (e.g. the universe) \cite{BB82, RWR, FORD}.  

\mbox{ }

\noindent The very cleanest implementation of this is by using actions that are 

\mbox{ }

\noindent i) {\bf manifestly parametrization irrelevant (MPI)}, and

\mbox{ } 

\noindent ii) {\bf free of extraneous time-related variables}. 
[E.g. one is not to involve external absolute Newtonian time, or the dynamical formulation of GR's `lapse' variable (explained in Sec \ref{Sec-ADM}).]

\subsubsection{A preliminary, more conventional view on constructing temporally relational actions}\label{MPI-TR_Actions}

Perhaps the reader is objecting that the preceding SSec does not help with obtaining a notion of velocity since it denies rather than 
provides a primary notion of time for the system as a whole.  
A simple way of getting round this involves how MPI actions are trivially equivalent to actions that are 

\mbox{ }  

\noindent i$^{\prime}$) {\bf manifestly reparametrization-invariant (MRI)}.  

\mbox{ }  

\noindent 
[For MRI, any choice of (monotonic) parameter will do, so that choice is empty, and any such choice is equivalent to making no choice by not parametrizing i.e. using the MPI action.] 
Next, MRI {\sl has} a parameter $\lambda$, so that, for all that it is a non-entity, one can interpret it as a {\it label time} non-entity, and write down 
\beq
\mbox{velocity} := \ordial \mbox{(configuration variable)}/\ordial (\mbox{label time}) \mbox{ i.e. } \mbox{ } \ordial\fQ^{\sfA}/\d\lambda \mbox{ } .
\eeq 
One can then straightforwardly build the kinetic term 
$
\cT := ||\Circ\fQ||_{\sbfM}\mbox{}^2/2 = \fM_{\sfA\sfB}\Circ\fQ^{\sfA}\Circ\fQ^{\sfB}/2
$.
I am assuming for now that this takes the most physically standard form that is homogeneous quadratic in the velocities; I lift this assumption in Sec 1.4.5.
%
%
${\fM}_{\sfA\sfB}$ is the configuration space metric portmanteau of ${M}_{\sfA\sfB}({Q}^{\sfC})$ for finite theories and 
${\mM}_{\sfA\sfB}({\mQ}^{\sfC}(\ux))$ for infinite theories (this is assuming {\it ultralocality}, 
i.e. no derivative dependence, collapsing the object from a functional to a mere function, 
which mathematical simplicity holds over the entirety of the standardly accepted fundamental theories of Physics).
Its determinant is $\fM$ and its inverse is $\fN^{\sfA\sfB}$.  
$||\mbox{ }||_{\sbfM}\mbox{}$ denotes the corresponding ${\bfM}$-`norm'.
$\Circ := \ordial/\ordial\lambda$
This is intended as type of `dot symbol'; I often use large preceding derivative symbols rather than overhead derivative symbols 
for graphical ease of accommodation of the subtleties of Configurational Relationalism \ref{Indir-CR}
by hanging suffixes on the large-symbol versions.

One can then form the Lagrangian portmanteau, $\cL\lfloor \dot{\fQ}^{\sfA}, \fQ^{\sfA} \rfloor$ of the Lagrangian $L(Q^{\sfA}, \dot{Q}^{\sfA})$ for finite theories and the 
Lagrangian density $\fL(\ux, \dot{\mQ}\mbox{}^{\sfA}(\ux); \mQ^{\sfA}(\ux)]$ that is taken to be part-ultralocal (in the velocities) for field theories.
Then integrating over $\lambda$ and the NOS, one obtains the relational action.
Compliance with MRI does make the Lagrangian portmanteau in question {\it look} somewhat unusual, i.e. not be of difference-type form 
$\cL = \cT - \cV$, but rather of {\it product form} $\cL = 2\sqrt{\cT\cW}$.  
For now we take on trust that this ${\cW} = {\cE} + {\cU}$ for ${\cU} = -{\cV}$, where $\cV\lfloor\fQ^{\sfC}\rfloor$ is the potential energy portmanteau: 
potential $V(Q^{\sfA})$ for finite theories and potential density ${\mV}(\ux; \mQ^{\sfA}(\ux)]$ for field theories.  
Thus it is assumed to be 

\noindent A) independent of the velocities [this is another mathematical simplicity that happens to be in accord with the standardly-accepted Fundamental Physics]. 

\noindent B) time-independent [which makes good sense for fundamental whole-universe setting as opposed to the setting for dissipative subsystems 
or approximate modelling e.g. systems involving friction].
This time-independence of the potential, alongside time-independence of the kinetic metric is part of what is covered by the 
`no extraneous time-like objects' part of the given mathematical implementation of Temporal Relationalism.


\noindent $\cE$ is the energy-like portmanteau: total energy $E$ for finite theories and, for field theories, some kind of total energy density (this includes a Jacobian factor).  
[$\cU$ and $\cW$ are, for now, just to be considered as the notational tidyings defined above; we will justify this trust in Sec \ref{J-EL}.]
The MRI relational action is thus 
\be
\FS^{\sM\sR\sI}_{\sJ} = \int\d\lambda\int_{\sbSigma_p}\d\bupSigma_p\cL^{\sM\sR\sI}_{\sJ} = 
2\int\d\lambda\int_{\bupSigma_p}\d\bupSigma_p\sqrt{{\cT}{\cW}} \mbox{ } ,    
\label{action}
\ee
where the J stands for `Jacobi' since the finite case of this is {\bf Jacobi's Principle} \cite{Lanczos}.
%

\subsubsection{Relational dethroning of velocity as a primary notion}\label{Dethrone-Vel}

A more manifestly relational manoeuvre, however, is to acknowledge that if there is no meaningful notion of time, then conventional physicist's definition of velocity, 
\be
\mbox{velocity := $\ordial$(configuration variable)/$\ordial$(some notion of time)} \mbox{ } , \label{Flutt}
\ee
{\sl reads that} there is consequently no meaningful primary notion of velocity! 
Subsequently, nothing defined in terms of velocities such as kinetic terms, Lagrangians or canonical momenta, is guaranteed to meaningfully exist either.  
What {\sl does} have tangible physical content, however, are the
\beq
\mbox{$\ordial$(configuration variable)} \mbox{ } \mbox{ i.e. the} \mbox{ } \ordial\fQ^{\sfA}
\eeq
themselves, so that velocities are replaced by differentials of the configuration variables.  
I.e., changes {\sl in time} are meaningless in Relational Physics, the tangible content resides, rather, among the changes of one configuration variable with respect to another, 
\beq
\mbox{$\ordial$(configuration variable 1)/$\ordial$(configuration variable 2) } 
\mbox{ } \mbox{ i.e. } \mbox{ } \ordial\fQ_1/\ordial\fQ_2 \mbox{ } . \label{erina}
\eeq

\subsubsection{The subsequent fully relational view on constructing temporally relational actions}\label{MPI-TR-Actions}

As a follow-up of Relational Physics' dethroning of velocity, the usual homogeneous quadratic kinetic term 
is supplanted by $\ordial\cs^{2}$/2; given that I am taking the latter's paradigm to be primary, it is better that I recast this as the 
dikinetic term being supplanted by $\ordial\cs^2$ -- the square of the arc element that corresponds to the kinetic metric, 
\beq
\mbox{(kinetic arc element)$^2$} := \ordial \cs^2 := ||\ordial \bfQ ||_{\sbfM}\mbox{}^2 = \fM_{\sfA\sfB}\ordial\fQ^{\sfA}\ordial\fQ^{\sfB} \mbox{ } .  
\label{KAE}
\eeq
%
The action is, in complete generality, to be homogeneous linear in the d(configuration variable) so as to comply with MPI/MRI;   
in the homogeneous quadratic case this amounts to taking the square root of (\ref{KAE}).
Then the expression under the action's integral in the MPI action is of the form $\ordial\widetilde{\cs}$, where the tilde indicates 
a rescaling by some weight that is homogeneous of degree zero in the $\ordial\fQ$. 
[This is usually taken to be independent of the weights so as to match the conventional approach to Fundamental Physics having 
velocity-independent potential.]
I use 
$
\mbox{(physical arc element)} := \ordial\widetilde{\cs} := \sqrt{2\cW}\ordial \cs = \sqrt{2\cW}||\ordial \bfQ ||_{\sbfM}  
$
and then the action is 
\beq
\FS^{\sM\sP\sI}_{\sJ} := \int\int_{\sbSigma_p}\d {\bupSigma_p}\ordial\widetilde{\cs} \mbox{ } . 
\label{GeneralAction}
\eeq
\mbox{ } \mbox{ } Such assembly of an action may be subject to some limitations, whether from implementing physical or philosophical 
principles (see e.g. Sec \ref{TR+CR}), or from purely mathematical simplicity postulates such as `no derivatives higher than 
first (or, in a few special cases, including the important example of GR, second)'.
\noindent The above particular action coming from the homogeneous quadratic physics only assumption is of {\it Jacobi type}.  
In parallel to the Lagrangian portmanteau, $\ordial\cs$ and $\ordial\widetilde{\cs}$ are in fact portmanteaux of arclength elements in the finite 
case and nontrivially Riemannian elements (no signature restriction implied) in the field-theoretic case.

\noindent Note 1) The MPI formulation does not have a primary concept of Lagrangian -- it has the arc element instead.  
In other words, it is a `{\bf geodesic principle}' (see Appendix \ref{Examples}.B for more detail). 

\noindent Note 2) The MRI or MPI action of the above form for homogeneous quadratic $\cT$ or $\ordial\cs^2$ physics is the Jacobi-type 
action principle; in the finite case of mechanics, it is the Jacobi action principle itself. 
Whilst probably more commonly encountered in the MRI form, though I note that Jacobi himself did use the MPI geodesic principle form.
It is also worth noting that the Jacobi-type action more commonly arises in the literature (see e.g. \cite{Hsiang1, Lanczos}) not from the desire to be 
relational but rather from the desire to provide the natural mechanics for a given geometry.

\subsubsection{More general temporally relational actions}\label{+General-TR-Actions}

\beq
\FS^{\sM\sP\sI}_{\sJ\sS} = \int\int_{\sbSigma_p}\d \bupSigma_p \ordial\widetilde{\cs} = \int\int_{\sbSigma_p}\d \bupSigma_p \ordial\lambda \cL  = \FS^{\sM\sR\sI}_{\sJ\sS}
\mbox{ } 
\eeq
complies with Temporal Relationalism, and has mutual compatibility, by 
$
\cL\left\lfloor {\ordial\fQ}/{\ordial \lambda}, \fQ \right\rfloor\ordial\lambda  = 
 \{{\ordial \widetilde{\cs}}/{\ordial \lambda}\}\ordial\lambda = {\ordial \widetilde{\cs}} = 
$

\noindent 
$
\{{\ordial \cs}/{\ordial \mu}\}\ordial\mu = \cL\left\lfloor {\ordial\fQ}{\ordial \mu}, \fQ,\right\rfloor\ordial \mu 
$.  
[This uses Euler's Theorem for homogeneous functions, and takes $\lambda$ and $\mu$ to be any two 
monotonically-related parameters; also, by this equivalence, I henceforth drop MPI and MRI labels.]   
The `square root of homogeneous quadratic' Jacobi-type case is then an obvious subcase of this.
Moreover, the above generalization includes passage from Riemannian geometry (no signature connotation intended) to 
Finsler geometry (no nondegeneracy connotation intended; $\cL$ is cast in the role of the portmanteau generalization of the metric function, $\cF$).
It is still a geodesic principle, just for a more complicated notion of geometry.  
Synge's work starting with \cite{Synge} generalized this to the more general geometry above \cite{Lanczos}.  
The geometry in question enters by playing the role of configuration space geometry.
In their honour, I denote the geometrically-natural construction of a mechanics given a metric geometry by the map $\fJ\fS: \langle{\FrS }, 
\fF\rangle \longrightarrow \FS^{\sM\sP\sI}_{\sJ\sS}$.

\subsubsection{Temporally relational formulation of the notion of conjugate momentum}\label{TR-Mom}

The standard definition of the conjugate momentum to the portmanteau $\fQ^{\sfA}$ is
$
\fP_{\sfA} := \partional\cL/\partional\dot{\fQ}\mbox{}^{\sfA} \mbox{ } .
$
Then computing this from the MRI form of the Jacobi action gives
$
\fP_{\sfA} = \sqrt{\cW/\cT}\fM_{\sfA\sfB}\Circ\fQ^{\sfA} \mbox{ } .  
$
However, manifest relationalism has been argued to have no place for velocities or Lagrangians, so either the above definition of momentum, 
or the conceptual role played by momentum, need to be reformulated. 
In this case, it turns out that the notion of conjugate momentum itself survives, by there being an equivalent manifestly-relational form 
for the defining formula, 
$
\fP_{\sfA} := \partional\ordial\widetilde{\cs}/\partional\ordial\fQ^{\sfA} \mbox{ } .  
$
Then computing this gives
$
\fP_{\sfA} = \sqrt{2\cW}\fM_{\sfA\sfB}\ordial\fQ^{\sfB}/|| \ordial\bfQ||_{\sbfM} \mbox{ } , 
$
which is manifestly within the scope of notion (\ref{erina}).

\subsubsection{Outline of the Dirac method}\label{Dirac-Method} 

Next, one inspects to see if there are any inter-relations among these momenta due to the form of the action: ({\it primary constraints} \cite{Dirac}).  
Then variation of the action with respect to the base objects provides some evolution equations and perhaps some {\it secondary constraints} \cite{Dirac}. 
One then finds the evolution equations.
Finally, one checks whether even more constraints may arise by the requirement that the evolution equations propagate the constraints; 
this is best done at the level of Poisson brackets (or Dirac brackets that take their place if required by the need to eliminate so-called 
{\it second-class} constraints \cite{Dirac}, though I take these to be Poisson bracket-like entities of a more reduced classical theory).   
See e.g. \cite{HenTei} for a more modern perspective on this.

\subsubsection{Quadratic constraint: momenta as `direction cosines'}\label{Quad-Con}

The MRI implementation of Temporal Relationalism leads to constraints via the following general argument of Dirac \cite{Dirac} 
(as straightforwardly modified by me to concern differentials rather than velocities, and MPI instead of MRI actions).    
Since MPI actions are homogeneous of degree 1 in the differentials, the $k$ momenta arising from these are homogeneous of degree 0 in the differentials.  
Hence the momenta are functions of at most $k - 1$ independent ratios of differentials. 
Thus the momenta must have at least 1 relation between them (which is by definition a primary constraint).  

\noindent Moreover, for the above Jacobi-type action, there is precisely one such constraint (per relevant NOS point). 
This is due to 

\noindent A) the square-root form of the action, by which the momenta are much like direction cosines and `their squaring to 1' property 
by which they are not all independent.  
Here, instead, these square (using the $\bfM$ matrix) to $2\cW$.
Thus, the kinetic term in terms of momenta (a legitimately relational object since the momenta are), which is half of the above square, 
is equal to $\cW$.

\noindent B) in cases with nontrivial space of extent, due to the particular {\it local} ordering of having the square root (and sum over 
$\fA$, $\fB$ implicit in the kinetic arc element or kinetic term) {\sl inside} the integral-over-space-of-extent sign as opposed to outside it.  
All-in-all, one has an energy-type constraint which is quadratic and not linear in the momenta, 
\beq
\scQ\scU\scA\scD := {\fN}^{\sfA\sfB}{\fP}_{\sfA}{\fP}_{\sfB}/2 + \cW = 0 \mbox{ } .   
\label{GEnCo}
\eeq
%

\subsubsection{Evolution equations}\label{Evol-Eq}

Again, the usual form of the evolution equations for $\fQ^{\sfA}$ makes reference to Lagrangians, times and velocities:
\beq
\frac{\ordial}{\ordial\lambda}\left\{\frac{\partional \cL}{\partional \dot{\fQ}\mbox{}^{\sfA}}\right\}  = \frac{\partional \cL}{\partional\fQ^{\sfA}} \mbox{ } .  
\eeq
There is however a manifestly relational version of this available,  
\beq
\ordial \{\partional\ordial\widetilde{\cs}/\partional \ordial\fQ^{\sfA}\}  = \partional\d\widetilde{\cs}/\partional\fQ^{\sfA} \mbox{ } .  
\eeq
Then, computationally,   
\beq
\frac{\sqrt{2\cW}}{||\ordial \bfQ||_{\sbfM}}  \ordial 
\left\{
\frac{\sqrt{2\cW}}{||\ordial \bfQ||_{\sbfM}}  \ordial \fQ^{\sfA}
\right\} 
+ \Gamma^{\sfA}\mbox{}_{\sfB\sfC} 
\frac{\sqrt{2\cW}}{||\ordial \bfQ||_{\sbfM}} \ordial \fQ^{\sfB}
\frac{\sqrt{2\cW}}{||\ordial \bfQ||_{\sbfM}} \ordial \fQ^{\sfC} = 
\frac{\partional\cW}{\partional \fQ_{\sfA}} \mbox{ } ,  
\label{Evol}
\eeq
where $\Gamma^{\sfA}\mbox{}_{\sfB\sfC}$ are the configuration space Christoffel symbols.

\subsubsection{Emergent Jacobi time}\label{JET}

\noindent Leibnizian timelessness for the universe as a whole is to be resolved for everyday subsystem physics as follows. 

\mbox{ }  

\noindent Relationalism 6) [Mach's Time Principle] {\bf Time is to be abstracted from change}: 
\beq
\ft_{\sM\sa\scc\sh} \mbox{ } \mbox{ is of the form } \mbox{ } \cG\lfloor \fQ^{\sfA}, \ordial\fQ^{\sfA}\rfloor \mbox{ } .
\eeq
\noindent {\sl What} change \cite{ARel, ARel2}?  
Two diametrically opposite views on this are as follows.

\mbox{ }

\noindent Relationalism 7-AMR) [Aristotle--Mach--Rovelli]\footnote{This is along the lines that Rovelli \cite{Rovellibook} advocates; 
I point out its similarity to `{\it tot tempora quot motus}' (so many times as there are motions, which is the Scholastics' view of 
Aristotle, see e.g. p 54 of \cite{Jammerteneity}.}  
{\bf Time is to be abstracted from ANY change}. 

\mbox{ } 

\noindent Relationalism 7-LMB) [Leibniz--Mach--Barbour]  {\bf THE MOST PERFECT time is to be abstracted from 
ALL change},\footnote{This is Barbour's position, based on Leibnizian whole-universe considerations.}
\beq
\ft_{\sL\sM\sB} = \cG\lfloor\fQ^{\sfA}, \mbox{ all } \ordial\fQ^{\sfA}\rfloor \mbox{ } .
\eeq

\mbox{ }  

\noindent For reasons further detailed in Secs \ref{Time} and \ref{Cl-POT}, I adopt an `LMB-CA' position that is intermediate between these two (and rather closer to Barbour's).  
The C in the name stands for `Clemence', the American Astronomer whose 1950's work \cite{Clemence} was also a substantial inspiration to Barbour, 
albeit I favour Clemence's view even further (at the expense of part of the Leibnizian whole-universe considerations).  

\mbox{ }  

\noindent A particular implementation of LMB in principle (and, as argued in Secs \ref{Cl-POT} and \ref{+temJBB}, LMB-CA in practise) follows from how 
both the conjugate momenta and the evolution equations simplify if one uses
\beq
\Last := 
\sqrt{\frac{\cW}{\cT}} \Circ := 
\frac{1}{\fN}\Circ := 
\frac{1}{\dot{\fI}}\Circ =  
\frac{\ordial}{\ordial\fI} 
\mbox{ } .
\label{emtime}
\eeq
\mbox{ } \mbox{ } This amounts to finding a $\ft^{\se\sm(\sJ)}$ (`emergent Jacobi time')
that is, ab initio, a portmanteau of $t^{\se\sm(\sJ)}[Q^{\sfA}, \ordial Q^{\sfB}]$ and $\mt^{\se\sm(\sJ)}(\ux; \mQ^{\sfA}, \delta\mQ^{\sfA}]$.  
N.B. that this is, following Mach, a highly dependent variable rather than the independent variable role that time is usually taken to play.
This is because this `usually' presumes one knows what time is beforehand, whereas the present move involves operationally establishing that (see Secs \ref{Cl-POT} and \ref{+temJBB} 
for examples and discussion).  
Moreover, upon finding a satisfactory such for one's error tolerance, one can pass to the more usual picture in which it is taken to be an independent variable, 
i.e. $Q^{\sfA} = Q^{\sfA}(t^{\se\sm(\sJ)})$ or $\mQ^{\sfA} = \mQ^{\sfA}(\ux, \mt^{\se\sm(\sJ)}(\ux)]$. 
[The latter is not the most independent possible choice in the field-theoretic case, but is rather the local proper time or `bubble time' type choice of formulation.]
Now the emergent Jacobi time formula itself belongs to the former conceptualization, being given by 
\beq
\ft^{\se\sm(\sJ)} - \ft^{\se\sm(\sJ)}(0) = \int||\ordial\bfQ||_{\sbfM}/\sqrt{2\cW} \mbox{ } .\label{clemerg}
\eeq
This found, in the latter conceptualization, 
\beq
\Last := \ordial/\ordial\ft \mbox{ } 
\eeq
when acting on such as $\fQ^{\sfC}$, or
\beq
\Last := \partional/\partional\ft \mbox{ } 
\eeq
when acting on objects that are functionally dependent on $\fQ^{\sfC}$.  

\mbox{ } 

\noindent Note 1) The emergent Jacobi time, and modifications of this (see Sec \ref{TR+CR}, have long appeared in Barbour's work (e.g. \cite{BB82, B94I}). 
Though I have also seen it in earlier Russian literature \cite{Ak} for mechanics, and as Christodoulou's Chronos Principle 
\cite{Christodoulou1} for the GR case, so I leave it as an open question who its first proponent was.  

\noindent Note 2) (\ref{clemerg}) is a rearrangement of the energy-type equation (\ref{GEnCo}) by use of the momentum--velocity relation, 
taking the square root and integrating so as to make $\ft^{\se\sm(\sJ)}$ the subject.

\noindent Note 3) Detailed use of the alteration in the assumed function--functional dependence between time-finding and time-using is new to the 
present Article and \cite{ACos2}.  

\noindent Note 4) As I argue in \cite{ARel2} and Sec \ref{Cl-POT}, (\ref{clemerg}) in practise implements 

\mbox{ } 

\noindent Relationalism 7-LMB-CA) {\bf time is to be abstracted from a STLRC} (sufficient totality of locally significant change).  

\mbox{ }  

\noindent In (\ref{emtime}), the first equality is the MRI computational formula, the second relates the generalized lapse $\fN$ to this, and the third recasts 
generalized lapse as a truer {\it velocity of the instant} notion $\dot{\fI}$ \cite{FEPI}. 
The fourth recasts this as the even truer {\it ordial of the instant} notion $\ordial \fI$, by which the instant $\fI$ itself 
is identified with the emergent time $\ft^{\se\sm(\sJ)}$ that labels that instant. 
The above generalized lapse notion $\fN$ is a portmanteau of $N$ for finite theories and $\mN(\ux)$ for field theories. 
Similarly, the above generalized instant notion $\fI$ is a portmanteau of $I$ for finite theories and $\mI(\ux)$ for field theories. 

The momenta in terms of $\Last$ are then just 
\beq
\fP_{\sfA} = \fM_{\sfA\sfB}\Last\fQ^{\sfB} \mbox{ } , 
\eeq
while the evolution equations are 
\beq
{D}_{\sa\sb\sss}\mbox{}^2 {\fQ}^{\sfA} = \Last\Last\fQ^{\sfA} + 
\Gamma^{\sfA}\mbox{}_{\sfB\sfC}\Last\fQ^{\sfB}\Last\fQ^{\sfC} = \partional\cW/\partional\fQ_{\sfA} 
\label{parag} 
\mbox{ } .
\eeq 
N.B. that this is but a {\bf parageodesic equation} with respect to the kinetic metric (meaning it has a forcing term arising from the $\cW$-factor), 
but is a true geodesic equation with respect to the physical (tilded) metric [see Appendix \ref{Examples}.B].

One can supplant one of the evolution equations (finite case) or one per space point (infinite case) 
with the Lagrangian form of the quadratic `energy-type' constraint, 
\beq
{\fM}_{\sfA\sfB}\Last{\fQ}^{\sfA}\Last{\fQ}^{\sfB}/2 + \cW = 0 \mbox{ }  .  
\label{ENERGY}
\eeq
This supplanting is useful in practical calculations. 
As first remarked upon in the subcase of mechanics \cite{B94I}, the above reveals $\ft^{\se\sm(\sJ)}$ to be a recovery on relational premises of the same quantity that is more usually 
assumed to be the absolute external Newtonian $t^{\sN\se\sw\st\so\sn}$; there is a conceptually-similar recovery in the general case 
which covers a further number of well-known notions of time (see Sec \ref{Examples}).   

\noindent Another form for (\ref{parag}) is 
$
\Last \fP_{\sfA} = \partional\cW/\partional\fQ^{\sfA}.  
$
\noindent As $\ft^{\se\sm(\sJ)}$  has been cast as an expression solely in terms of the $\fQ^{\sfA}$ and $\ordial \fQ^{\sfA}$, 
\beq
\mbox{$\ordial$(configuration variable/d(emergent Jacobi time)} = \ordial\fQ^{\sfA}/\ordial\ft^{\se\sm(\sJ)} 
\eeq
{\sl do} have tangible physical content, through being a type of (\ref{Flutt}) and not of (\ref{erina}).  
Thus in the relational approach the above equations entirely make sense from a temporally relational perspective, unlike in the absolute 
Newtonian counterparts of them that have the same mathematical form but pin an absolute time interpretation on the times present.    
The emergent time is  {\sl provided by} the system.  
For now, it furthermore gives the appearance of being provided by the whole of the subsystem.
These last two sentences fit Mach's own conception of time.   
Also, the usually-assumed notion of time as an independent variable is un-Leibnizian and un-Machian.  
However, though it is to be overall abstracted from motion, once this is done it {\sl is} a convenient choice for (emergent) independent variable.  

\mbox{ } 

\noindent  We shall from now on work with MPI and MRI forms held to be interchangeable, with preference for writing MPI ones. 
We will likewise present specific equations in terms of $\Last$ in the geodesic/Lagrangian picture (the geodesic picture is in terms of configuration variables and their 
differentials in parallel to how the Lagrangian picture is in terms of configuration variables and their velocities).

\subsubsection{Inter-relating Jacobi product actions and Euler--Lagrange actions}\label{J-EL}

The much more well-known difference alias Euler--Lagrange-type actions are, rather, of the form  
\be
\FS_{\sE\sL} = \int\d \ft\int_{\sbSigma_p}\d \bupSigma_p \cL = \int\d \ft\int_{\sbSigma_p}\d \bupSigma_p\{{\cT}_{\sft} - {\cV}\} \mbox{ } . 
\label{Lagaction}
\ee
Here ${\cT}_{\sft}$ is the kinetic energy formulated in terms of $\ordial/\ordial\ft$ derivatives for $\ft$ conventionally an 
external notion of time (absolute time $t$ for mechanics, coordinate time $\mt(\ux)$ for GR, which is a label time because GR is {\sl already}-parametrized). 
See e.g. \cite{Lanczos} or  Sec \ref{Examples} for how (\ref{action}) $\Rightarrow$ (\ref{Lagaction}) in the case of mechanics by passage to the Routhian, 
and  \cite{SemiclI} or Sec \ref{Examples} for (\ref{Lagaction}) $\Rightarrow$ (\ref{action}) by the emergence of some lapse-like quantity.\foo{Such 
mathematics conventionally appears in the mechanics literature under the name of the {\it parametrization procedure}.  
Namely, one may adjoin the space of the original notion of time's time variable to the configuration space $\mbox{\large $\mathfrak{q}$}$ $\longrightarrow$ 
$\mbox{\large $\mathfrak{q}$} \times \sFrT$ by rewriting one's action in terms of a label-time parameter $\lambda \in \sFrT$ (see e.g. \cite{Lanczos}).  
However, in the relational context in which Barbour and I work, one rather at this stage adopts 1) above.}
%
Various advantages for product-type actions over difference-type actions stem from this as regards 
consideration of whole-universe Fundamental Physics, which is the setting for Quantum Cosmology.  
The above justifies the identification of $\cW$ as the combination of well-known physical entities $\cE - \cV$.

\subsection{Configurational Relationalism and its indirect implementation}\label{Indir-CR}

\noindent Relationalism 8) [my extension and part-reformulation of Barbour]. 
The action is to be {\bf configurationally relational} as regards a group of transformations if a $\FrG$-transformed 
world-configuration is indiscernible from one that is not.  
I.e., one is to build into one's theory that a certain group of transformations $\FrG$ acting upon the theory's configuration space $\FrQ$ are 
to be irrelevant, i.e. physically meaningless \cite{BB82, RWR, Lan, Phan, FORD, FEPI, Cones}, transformations.  
More specific detail of the $\FrQ$--$\FrG$ pairing is postponed to Sec \ref{Patti}, i.e. until after we have benefited from various specific examples.  

\mbox{ }

\noindent One way to implement this is to use not `bare' configurations and their composites as above, but rather their 

\noindent arbitrary-$\FrG$-frame-corrected counterparts.\foo{I adopt
the {\it passive}, as opposed to the {\it active}, point of view of transformations. 
That is the opposite of how Barbour conceives Configurational Relationalism (see e.g. spatial relationalism 
in \cite{BB82} and Sec \ref{Examples}), which places the primary importance on the constituent objects of the universe. 
Moreover, these two a priori distinct conceptualizations turn out to be mathematically equivalent representations 
\cite{AMP}, at least in the context that is the main subject of this Article.
Thus, within this context, I use ``one way" and ``the only way" to encompass both the passive {\sl and} the active.}
%
This is the only known way that is sufficiently widespread for a relational program to underlie the 
whole of the classical Fundamental Physics status quo of GR coupled to the Standard Model (see Sec \ref{RWR}).
The corrections are with respect to auxiliary variables $\fg^{\sfZ}$ [a portmanteau of the finite case $g^{\sfZ}$ and the position-dependent 
case $\mg^{\sfZ}(\ux)$ of field theory] that are paired with the infinitesimal generators of $\FrG$.

\noindent Originally the $\fg^{\sfZ}$ employed  were in the form of multiplier coordinate corrections to the velocities,  
\beq
\dot{\fQ}^{\sfA} - \sum\mbox{}_{\mbox{}_{\mbox{\scriptsize $\fZ$}}}
                               \stackrel{\rightarrow}{\FrG_{\sttg^{\tfZ}}} {\fQ}^{\sfA} \mbox{ } .  
\eeq
Here $\stackrel{\longrightarrow}{\FrG}$ denotes the infinitesimal group action.
However, this formulation is not consistent with manifest Temporal Relationalism since the multiplier corrections spoil this. 
Nor is it relational insofar as it is formulated in terms of meaningless label-time velocities.
Nor is it an arbitrary-$\FrG$ frame approach.  
The next SSec amends these things, and then the rest of the present SSec applies just as well within this amendment.

The above may also seem to be taking a step in the wrong direction as regards $\FrG$ being irrelevant in passing 
from the already-frame-redundant $\FrQ$ to the principal fibre bundle\foo{This 
being a bona fide bundle may require excision of certain of the degenerate configurations. 
Recall also that a fibre bundle being principal means that the fibres and the structure group coincide \cite{Nakahara}.}
%
over $\FrQ$, $P(\FrQ, \FrG)$.     
For, from the perspective of just counting degrees of freedom, this can be locally regarded as the product space $\FrQ \times \FrG$, 
which has more degrees of freedom, as opposed to passing to the quotient space $\FrQ/\FrG$, which has less.  
However, we shall also resolve this contention below.

Moreover, the essential line of thought of this SSec is the {\sl only} known approach to Configurational Relationalism 
that is general enough to cover the Einstein--Standard Model presentation of Physics.\foo{For 
mechanics in 1- and 2-$d$ \cite{FORD}, I have found that working directly on the least redundant/most relational configuration space provides workable relational theories. 
Moreover, these coincide with the restriction to those dimensions of Barbour's theories as formulated in arbitrary $\sFG$-frame form and then reduced (Sec \ref{Q-Geom}).} 

\mbox{ } 

\noindent This SSec's notion of Configurational Relationalism is my \cite{Lan} portmanteau that spans both of the following.

\mbox{ }

\noindent 1) {\bf Spatial Relationalism}, as in the traditional mechanics setting and the (geometro)dynamical formulation of GR. 
and which was Barbour's notion prior to this extension.  

\mbox{ }  

\noindent 2) {\bf Internal Relationalism}, as in electromagnetism, Yang--Mills Theory and the associated scalar and fermion gauge theories.  

\mbox{ }

\noindent This wide range of cases is afforded by correspondingly wide varieties of $\FrQ$ and $\FrG$. 
(Although adopting a $\FrQ$ may carry connotations of there being some underlying NOS with which that is compatible.  
Also, given a $\FrQ$, there are consistency limitations on what variety of $\FrG$ can then have -- see Sec \ref{Q-G-Comp}).
For the scaled version of RPM, $\FrG$ is the Euclidean group of translations and rotations, Eucl($d$), while for pure-shape version 
it is the similarity group of translations, rotations and dilations, Sim($d$).

This extension and partial reformulation of Barbour's spatial relationalism shows that it and the conventional 
notion of Gauge Theory (usually internal but also applicable to spatial concepts) bear a very close relation.  
Gauge Theory is indeed Leibnizian in this sense and indeed an example of how such can usually only be implemented indirectly by mathematics 
of a mixed set of tangible entities and non-entities (respectively true dynamical degrees of freedom and gauge degrees of freedom).
\noindent Configurational relationalism is generalized in Sec \ref{Cl-Str}.  

\mbox{ }

\noindent Finally, I here propose to extend the meaning of Configurational Relationalism to have a second clause: {\bf `no extraneous space structures'}.
This matches Temporal Relationalism's second clause of no extraneous time variables, and also serves to exclude the Nambu--Goto string 
action (which, without this extra criterion, can indeed be included under the previous literature's formulation of relationalism).
For such reasons \cite{bigciteA, bigciteB, bigcite2, bigcite3, Carlip01, bigcite5, bigcite6, bigcite7, bigcite8, I93, Kieferbook, Dirac}, theories in fixed-
\noindent background  contexts are `less relational' (or only relational in a weaker sense).  
It is, rather, with background-independent M-theory (or at least some limiting classical action for this) that the most purely relational program would be expected to make contact. 
\noindent N.B. How generally this clause should be imposed requires further discussion in Sec \ref{RPM-vs-Strings}.

\subsection{Combining Temporal and Configurational Relationalism}\label{TR+CR}

Typically,
the potential term ${\cV}$ is manifestly a good $\FrG$-scalar but the kinetic term ${\cT}$ has correction terms due to `$\FrG$-transformations and differentials not commuting'.   
Consider for now theories whose configuration space carry kinetic arc elements that are homogeneous quadratic and whose associated metric 
has at most dependence on the ${\fQ}^{\sfA}$ -- i.e. a (semi)Riemannian as  opposed to Finslerian or even more general geometry.  
MRI is then attained by supplanting the previous $\cT$ by 
\be
{\cT} := ||\Circ_{\sttg}{\bfQ}  ||_{\mbox{\scriptsize\boldmath${\sbfM}$}}\mbox{}^2/2
\mbox{ } \mbox{ for } \mbox{ } \Circ_{\sttg}{\fQ}^{\sfA} := \Circ{\fQ}^{\sfA} - 
\sum\mbox{}_{\mbox{}_{\mbox{}_{\sfZ}}}\stackrel{\rightarrow}{\FrG}_{\dot{\sttg}^{\tfZ}}{\fQ}^{\sfA} \mbox{ } . 
\label{Taction}
\ee  
The auxiliary variable in use here is now interpreted as the velocity corresponding to a cyclic coordinate; doing so requires 

\noindent taking into account the variational subtleties laid out in Appendix \ref{Examples}.A.
I term the ensuing variational principle 

\noindent {\fJ\fB\fB}[$P(\langle \FrS$, $\fF \rangle$, $\FrG$)], the BB standing for Barbour--Bertotti.  
[As a map, {\fJ\fB\fB} = {\fJ\fS} $\circ \mbox{ } \FrG$-bundle, for 

\noindent $\FrG$-bundle: $\langle \FrS, \fF \rangle \times \FrG \longrightarrow 
P(\langle \FrS$, $\fF \rangle$, $\FrG$).]
On relational grounds, however, I prefer the equivalent MPI form, the argumentation behind which never involves the label-time 
non-entity or the subsequently-dethroned notion of velocity.

Here, instead of the above, I supplant the previous configurational-relationally unsatisfactory arclength (density) $\d \cs$ by 
the $\FrG$-corrected arclength
\beq
\ordial \cs_{\sJ\sB\sB} = ||\ordial_{\sttg}\bfQ||_{\sbfM}  \mbox{ } \mbox{ for } \mbox{ } \ordial_{\sttg}{\fQ}^{\sfA} := \ordial{\fQ}^{\sfA} - 
\sum\mbox{}_{\mbox{}_{\mbox{}_{\sfZ}}}\stackrel{\rightarrow}{\FrG}_{\sordial{\sttg}^{\tfZ}}{\fQ}^{\sfA} \mbox{ } .   
\eeq 
The auxiliary variable in use here is now interpreted as the differential corresponding to a cyclic coordinate; this   
has its own close parallel to the abovementioned variational subtleties, also laid out in Appendix \ref{Examples}.A.

One can then repeat the preceding SSec's treatment of conjugate momenta, quadratic constraint, evolution equations and the beginning of the treatment of emergent time 
`by placing $\fg$ suffices' on the $\d$, $\fI$, $\ft^{\se\sm}$ and $\Last$ (and, if one ever makes use of them, the $\Circ = \ordial/\ordial\lambda$ and $\fN$; 
I also find it helpful in considering the formulae inter-relating these to place a $\fg$ suffix on the $\cT$ as a reminder of the $\fg$-dependence residing within it).
\noindent $\ft^{\se\sm(\sJ\sB\sB)}_{\sttg}$ is no longer a relationally-satisfactory notion of time due to manifest configurational non-relationalism. 
As such, I term this a $\FrG$-dependent proto-time.
I also name it, and its eventual $\FrG$-independent successor, after JBB rather than just after Jacobi, to reflect the upgrade to a configurational-relationally nontrivial context.

\subsubsection{Linear constraints: enforcers of Configurational Relationalism}\label{Lin-Con}

Variation with respect to each $\FrG$-auxiliary produces one independent secondary constraint.  
In this Article's principal examples, each of these uses up {\sl two} degrees of freedom.
(That this is the case is testable for in a standard manner \cite{Dirac}, and amounts to these constraints being first- and not second-class.)
Thus one ends up on the quotient space $\FrQ/\FrG$ of equivalence classes of $\FrQ$ under $\FrG$ motions.  
This resolves the above-mentioned controversy, confirming the arbitrary $\FrG$-frame method both to indeed 
implement Configurational Relationalism and to be an indirect implementation thereof.  
I denote these linear constraints by $\scL\scI\scN_{\sfZ}$: 
\beq
0 = \partional \cL/ \partional \dot{\fc}^{\sfZ} \mbox{ (or } \mbox{ } \partional\ordial\widetilde{\cs}/\partional\ordial\fc^{\sfZ}) := \scL\scI\scN_{\sfZ} = 
\fP_{\sfA} \, \mbox{\Large $\delta$} \ordial\fc^{\sfZ} / \mbox{\Large $\delta$} \{\stackrel{\rightarrow}{\FrG_{\sordial\sttc}}\fQ^{\sfA}\} \mbox{ } ,   
\label{LinZ}
\eeq
where the last form manifestly demonstrates these being purely linear in the momenta.
The variational procedure behind obtaining these is justified in Sec \ref{Aux}.  

\mbox{ }

\noindent One then applies the Dirac procedure to be sure that the constraints breed no further constraints (see Sec \ref{Patti} for what the consequences are if they do).

\subsubsection{Outline of `Best Matching' as a procedure}\label{Best-Match}

Barbour's central notion of `best matching' \cite{BB82, B03} (which has played a part in almost all of his articles and research seminars) amounts to 

\noindent Best Matching 1) construct an arbitrary $\FrG$-frame corrected action.  

\noindent Best Matching 2) Vary with respect to the $\FrG$-auxiliary $\fg^{\sfZ}$ to obtain the linear constraint $\scL\scI\scN_{\sfZ} = 0$.

\noindent Best Matching 3) Solve the Lagrangian form of $\scL\scI\scN_{\sfZ} = 0$ for $\fg^{\sfZ}$. 
(This is often an impasse.) 

\noindent Best Matching 4) Substitute this back in the action to obtain a new action.  

\noindent Best Matching 5) Elevate this new action to be one's primary starting point.  

\mbox{ }  
  
\noindent Note 1) Best Matching 2) to 4) can be viewed as a minimization (or, at least, as an extremization).
One is searching for a minimizer to establish the least incongruence between adjacent physical configurations.
See Sec \ref{MIS-Demise} for more.  

\noindent Note 2) Best Matching 2) to 5) can be viewed as a configuration space reduction procedure (Fig \ref{Red}).  
%
{            \begin{figure}[ht]
\centering
\includegraphics[width=0.4\textwidth]{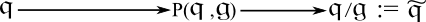}
\caption[Text der im Bilderverzeichnis auftaucht]{        \footnotesize{One passes from the configuration space $\FrQ$ to the principal 
bundle by adjoining $\FrG$ generators and then to the corresponding quotient space if one succeeds in performing reduction.} }
\label{Red} \end{figure}          }

\subsubsection{Emergent Jacobi--Barbour--Bertotti time} \label{temJBB}

As a distinct application of Best Matching 3), substitute its solution into the  $\FrG$-dependent proto-time $\ft^{\se\sm(\sJ\sB\sB)}_{\sttg}$
to render it configurational-relationally acceptable. 
[Albeit this is at the cost of being technically implicit and subject to the threats of nonuniqueness and of the extremization not being solvable in practise.]
I.e. the JBB emergent time is
\beq
\ft^{\se\sm(\sJ\sB\sB)} - \ft^{\se\sm(\sJ\sB\sB)}(0) = \stackrel{\mbox{\scriptsize extremum $\fg \mbox{ } \in \mbox{ }$} \sFG}
                                                               {\mbox{\scriptsize of $\stS^{\tr\te\tl\ta\ft\ti\to\tn\ta\tl}$}}                                                              
\left(                                                              
                                                               \int||\ordial_{\sg}\bfQ||_{\sbfM}/\sqrt{\cW} 
\right)  \mbox{ } . 
\label{Kronos}
\eeq
N.B. that this expression does {\sl not} contain a spatial integral: the field-theoretic $\mt^{\se\sm(\sJ\sB\sB)}$ is {\sl local}.

\subsection{Direct implementation of Configurational Relationalism}\label{Dir-CR}

Configurational relationalism can also be implemented, at least in lower-$d$ RPM's, by directly constructing one's formulation to be $\FrG$-invariant \cite{FORD, Cones}. 
One can view this as working directly on the relational configuration space.
There is now no need for any arbitrary $\FrG$-frame variables, nor then do any linear constraints arise nor are these to be used as a basis for reduction to pass to a new action. 
Here, instead, one's action is already directly $\FrG$-invariant i.e. on $\FrQ/\FrG =: \widetilde{\FrQ}$ by construction; this is reflected by the more complicated form taken by the 
kinetic arc element/kinetic term.  
This approach \cite{FORD, Cones} builds on work of Kendall \cite{Kendall} on the spaces of shapes and the `cones' over \cite{Cones} 
over these (a notion explained in Sec \ref{Q-Geom}) so as to include scale, by then using the Jacobi--Synge approach to build a natural mechanics from a metric geometry.  
I.e. the actions $\fJ\fS$(shape space) and $\fJ\fS(\mC$(shape space)) = $\fJ\fS$(relationalspace).   
%

\mbox{ } 

\noindent This {\bf direct relationalspace construction} (Sec \ref{DRIII}) can be viewed as a gestalt of both the geometrization of mechanics 
that motivated Jacobi and Synge themselves and of the establishment of a temporally relational theory as above.

\mbox{ } 

\noindent Note 1) Sec \ref{Q-Geom} establishes that the direct relationalspace\footnote{I use {\it relationalspace} 
as the portmanteau of relational space for scaled RPM and shape space for pure-shape RPM, i.e. each case's non-redundant configuration space.}
implementation coincides with the configuration space reduction procedure `Best Matching 2-5)' in the case of 1- and 2-$d$ RPM's for which both are explicitly calculable.  
{\it This provides a second foundation for RPM's that is independent of, but the output from which is coincident with, Barbour's work}  \cite{FORD, Cones}.
I.e., it coincides with what arises from Barbour's formulation upon performing reduction (see also Sec \ref{DRIII}).
By this, emergent JBB time is itself a J time for a more reduced configuration space geometry (though by this stage I consider BB to deserve 
sufficient credit, so I use `emergent JBB time' whenever Configurational Relationalism has been taken into account, no matter by which means).  

\noindent Note 2) This formulation possesses emergent time as per (\ref{emtime}), quadratic constraint (\ref{GEnCo}), no linear constraints, 
evolution equations in form (\ref{Evol}) and energy constraint (\ref{ENERGY}) [the last two now have no $\fg$ subscripts].

\subsection{Outline of scaled and pure-shape Relational Particle Mechanics (RPM) examples}\label{Intro-RPM-Ex} 

In both cases, one has $N$ particles in dimension $d$ is $\FrQ = \{\mathbb{R}^{d}\}^{N} = \mathbb{R}^{dN}$.  
One then implements Temporal Relationalism by building a MRI/MPI Jacobi-type \cite{Lanczos} action.
For scaled RPM \cite{BB82, B86, BS, Rovelli, Smolin, Kuchar92, B94I, Buckets, LB, EOT, Gergely, GergelyMcKain, RWR, Lan, Kieferbook, Paris, 06I, 06II, TriCl, FORD, 08I, Cones, ScaleQM, 
08III, SemiclIII}, let the group of irrelevant motions $\FrG$ is the $d$-dimensional Euclidean group of translations and rotations, explaining this theory's alias being `Euclidean RPM' 
(ERPM). 
Thus only relative times, relative angles and relative separations are meaningful. 
E.g. for 3 particles in dimension $d$ $> 1$, scaled RPM is a dynamics of the (scaled) triangle that the 3 particles form.    
This Configurational Relationalism is implemented indirectly (Sec \ref{ERPM}) by introducing auxiliary variables that represent arbitrary Eucl($d$)-frame corrections.   
The resulting action can be written as\foo{I use 
upper-case Latin indices for particle labels 1 to $N$. 
I use $\underline{q}^I$ for particle labels, with corresponding masses $m_I$ and conjugate momenta $\underline{p}_I$.
The configuration space metric $m_{I\mu J\nu} = m_I\delta_{IJ}\delta_{\mu\nu}$      with inverse 
                               $n^{I\mu J\nu} =    \delta_{IJ}\delta_{\mu\nu}/m_I$. 
For scaled RPM, I use $T$, $W$, $U$, $V$, $E$, and sans serif script versions thereof for pure-shape RPM's counterparts that have different physical dimensions. 
Thus the {\it pseudo-energy} $\ttE$ has dimensions of energy $\times \,\, \mI$  and $\ttV$, $\ttU$ and $\ttW$ likewise, while $\ttT$ has dimensions of energy/$\mI$.}
\beq
\FS^{\sE\sR\sP\sM}_{\sJ\sB\sB} = 2\int\d\lambda\sqrt{T^{\sE\sR\sP\sM}_{\sJ\sB\sB}W} = \sqrt{2}\int\sqrt{W}\d s^{\sE\sR\sP\sM}_{\sJ\sB\sB} 
\mbox{ } 
\eeq
\beq
\mbox{ for } \mbox{ } T^{\sE\sR\sP\sM}_{\sJ\sB\sB}    = ||\Circ_{\underline{A}, \underline{B}}\mbox{\boldmath$q$}||_{\mbox{\scriptsize\boldmath$m$}}\mbox{}^2/2 \mbox{ }
\mbox{ or } \mbox{ } \d s^{\sE\sR\sP\sM}_{\sJ\sB\sB} = ||\d_{\underline{A}, \underline{B}}\mbox{\boldmath$q$}||_{\mbox{\scriptsize\boldmath$m$}}
\label{ERPMaction}
\eeq
\beq
\mbox{ and }         \Circ_{{\underline{A}}, {\underline{B}}}\uqq^{I}: = \dot{\uqq}^{I} - \dot{\uA} - \dot{\uB} \cr \uqq^{I} \mbox{ } 
\mbox{ or } \mbox{ } \d_{{\underline{A}}, {\underline{B}}}\uqq^{I}: = \d{\uqq}^{I} - \d{\uA} - \d{\uB} \cr \uqq^{I} \mbox{ } . 
\eeq 
For pure-shape RPM, instead, now the group of irrelevant motions $\FrG$ is the $d$-dimensional similarity group of translations, 
rotations and dilatations \cite{B03, Paris, 06II, TriCl, FORD, 08I, 08II} explaining this theory's alias as Similarity RPM (SRPM).
Thus it is a theory only of relative times, relative angles and ratios of relative separations, to which I refer to as {\bf pure shape} (taken to exclude size).
The action for it is 
\beq
\FS^{\sS\sR\sP\sM}_{\sJ-\sA} = 2\int\d\lambda\sqrt{{\ttT^{\sS\sR\sP\sM}_{\sJ-\sA}}{W}}   
= \sqrt{2}\int\sqrt{W}\d \tts^{\sS\sR\sP\sM}_{\sJ-\sA} \mbox{ } .
\eeq
\beq
\mbox{ for } \mbox{ } \ttT^{\sS\sR\sP\sM}    = ||\Circ_{\underline{A}, \underline{B}, C}\mbox{\boldmath$q$}||_{\mbox{\scriptsize\boldmath$m$}}\mbox{}^2/2\mI \mbox{ }
 \mbox{ or } \mbox{ } \d \tts^{\sS\sR\sP\sM\, 2} = ||\d_{\underline{A}, \underline{B}, C}\mbox{\boldmath$q$}||_{\mbox{\scriptsize\boldmath$m$}}\mbox{}^2/\mI 
\label{SRPMaction}
\eeq
\beq
\mbox{ and } \Circ_{{\underline{A}}, {\underline{B}}, C}\uqq^{I}: = \dot{\uqq}^{I} - \dot{\uA} - \dot{\uB} \cr \uqq^{I} + + \dot{C}\uq^I 
\mbox{ } \mbox{ or } \mbox{ } \d_{{\underline{A}}, {\underline{B}}, C}\uqq^{I}: = \d{\uqq}^{I} - \d{\uA} - \d{\uB} \cr \uqq^{I} + \d C\uq^I 
\mbox{ } . 
\eeq 
Furthermore, in 1- or 2-$d$, RPM can be cast in direct relationalspace/reduced form  \cite{FORD, Cones} (the two are proven equivalent in Sec \ref{Q-Geom} and jointly denoted by `r').  
Here, e.g. for ERPM, 
\beq
\FS^{\sE\sR\sP\sM}_{\sr} = 2\int\d\lambda\sqrt{T^{\sE\sR\sP\sM}_{\sr}W} = \sqrt{2}\int\sqrt{W}\d s^{\sE\sR\sP\sM}_{\sr} 
\eeq
\beq
\mbox{ for } \mbox{ } T^{\sE\sR\sP\sM}_{\sr} = ||\Circ\mbox{\boldmath$Q$}||_{\mbox{\scriptsize \boldmath$M$}(\mbox{\scriptsize\boldmath$Q$})}\mbox{}^2/2 \mbox{ } , \mbox{ or } \mbox{ } 
\d\tts_{\sr}^{\sE\sR\sP\sM} = ||\d\mbox{\boldmath$Q$}||_{\mbox{\scriptsize\boldmath$M$}(\mbox{\scriptsize\boldmath$Q$})}
\eeq
Here, $\mbox{\boldmath$M$}_{\sfA\sfB}(\mbox{\boldmath$Q$})$ are in-general-curved configuration space metrics, which are established in Sec \ref{Q-Geom} to be e.g. the usual metric on 
$\mathbb{S}^{N - 2}$ for $N$-particle 1-$d$ SRPM, and the Fubini--Study metric on $\mathbb{CP}^{N - 2}$ for $N$ particle 2-$d$ SRPM.

The primary quadratic energy constraint is, for the indirectly-formulated versions of scaled and pure-shape RPM respectively, the `energy constraints'
\beq
\scE := n^{I\mu J\nu}p_{I\mu}p_{J\nu}/2 + V = E \mbox{ } \mbox{ } , \mbox{ } \mbox{ } 
\scE := \mI \, n^{I\mu J\nu}P_{i\mu}P_{j\nu}/2 + \ttV = \ttE \mbox{ }. \mbox{ }
\label{EnCo}
\eeq
The quadratic constraint is still an energy constraint of form (\ref{GEnCo}) built from inverses of the above-mentioned metrics.

For RPM's, the zero total momentum, zero total angular momentum and zero total dilational momentum constraints 
\beq
\underline{\scP} := \sumiN\mbox{ } \upp_I = 0 
\mbox{ } \mbox{ } , \mbox{ } \mbox{ }
\underline{\scL} := \sumiN\mbox{ } \uqq^I \cr \upp_I = 0 
\mbox{ } \mbox{ } , \mbox{ } \mbox{ } 
\scD := \sumiN\mbox{ } \uqq^I \cdot \upp_I = 0 
\label{LD}
\eeq
follow from variation with respect to $\underline{A}^{\mu}$, $\underline{B}^{\mu}$ and $C$ (so that only the former two constraints occur in scaled RPM).

The emergent time is given by (e.g. for ERPM)
\beq
t^{\se\sm(\sJ\sB\sB)} - t^{\se\sm(\sJ\sB\sB)}_0 = 
\stackrel{\mbox{\scriptsize extremum $\underline{A}, \underline{B} \mbox{ } \in \mbox{ } \mbox{\scriptsize Eucl($d$)}$}}
         {\mbox{\scriptsize of $\stS^{\tE\tR\tP\tM}_{\tJ\tB\tB}$}}
\int \sqrt{T/W}\d\lambda = \int ||\d_{\underline{A}, \underline{B}} \mbox{\boldmath$q$}||_{\mbox{\scriptsize\boldmath$m$}}/\sqrt{2W} \mbox{ } .
\eeq

\subsection{Principal motivation: GR is a relational theory}\label{GR-Is-Relational}

Further motivation\foo{As well as this and the aforementioned relevance to the absolute versus 
relative motion debate, yet further motivation is furbished by how Configurational Relationalism is closely related \cite{Lan, FEPI, Gryb3} to gauge theory.  
See also Sec \ref{QM-Intro} for some QM motivation.} 
for the relational scheme are the following arguments that the geometrodynamical and conformogeometrodynamical 
formulations of GR are also relational \cite{RWR, Phan, FEPI, AB, San, OM02, OM03, Van, Lan, Than, Phan, Lan2}.    
There are furthermore results about {\sl not just casting, but deriving}, GR from less than the usual number of assumptions \cite{RWR, Phan, Lan2}.

\subsubsection{The usual spacetime formulation of GR}\label{GR-Spacetime}

The usual covariant spacetime tensor presentation of the Einstein field equations follows from the Einstein--Hilbert action
\beq
\FS^{\sG\sR}_{\sE\sH} = \int_{\sFrM} \d^4x\sqrt{|\mg|} \, \mbox{${\cal R}\mi\mc$}(X^{\Gamma}; \mg_{\Gamma\Lambda}] \mbox{ }  
\label{EinHilb}  
\eeq
for ${\FrM}$ the spacetime topological manifold and $\bg$ the indefinite spacetime 4-metric that obeys the Einstein field equations (EFE's),  
with determinant $\mg$ and Ricci scalar ${\cal R}$ic.  
This came about due to Einstein reconceptualizing the nature of space and time, in good part due to the compellingness of Mach's arguments. 
However, there is considerable confusion as to how GR is and is not Machian (see Sec \ref{Time} and Sec \ref{BI-Rel-POT}); the next two SSSecs' 
materials provide a strong resolution of this.

\subsubsection{GR as Geometrodynamics}\label{GR-Gdyn}
%
{            \begin{figure}[ht]
\centering
\includegraphics[width=0.28\textwidth]{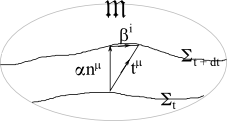}
\caption[Text der im Bilderverzeichnis auftaucht]{        \footnotesize{Arnowitt--Deser--Misner 3 + 1 split of $({\FrM}, \mg_{\Gamma\Delta})$. 
One sees that the lapse (`time elapsed') and shift (an example of `point identification map' \cite{Stewart} between adjacent spatial hypersurfaces) form a 
strutting.   }  }
\label{Fig-1-First}\end{figure}          }

While the role of spacetime in GR has often been touted, one should not forget that GR also admits a 
dynamical interpretation in terms of evolving spatial 3-geometries, i.e. Geometrodynamics.  
In this approach to GR, one has to preliminarily choose a residual NOS in the sense of a spatial hypersurface of fixed topology $\bupSigma$. 
This, I take to be a compact without boundary one for simplicity, and 3-$d$ as this suffices to match current observations.  
Under the ADM split of the spacetime metric\foo{I use capital Greek letters as spacetime indices.  
%
The spatial topology $\bupSigma$ is taken to be compact without boundary. 
$\sh_{\mu\nu}$ is a spatial 3-metric thereupon [within the ADM perspective, it is the {\it induced metric} upon this by the spacetime metric 
$\sg_{\Gamma\Delta}$, with determinant $\mh$, covariant derivative $\sD_{\mu}$, Ricci scalar Ric($\ux; \bh$] and conjugate momentum $\uppi^{\mu\nu}$.  
$\upalpha$ is the lapse function and $\upbeta^{\mu}$ is the shift vector.
$\pounds_{\suupbeta}$ is the Lie derivative with respect to $\upbeta^{\mu}$.  
$t$ is GR coordinate time.  
$\Lambda$ is the cosmological constant.
In dynamical language, DeWitt's 2-index to 1-index map \cite{DeWitt67} (which also inverts upstairs and downstairs) recasts $\mh_{\mu\nu}$ as $\mh^A$ 
with the capital Latin indices in this context running from 1 to 6.  
Thus $\mh$ is a configuration space object (bold shorthand) as well as a spatial 2-tensor (double-underlined shorthand).  
Then $\pi^{\mu\nu}$ maps to $\pi_A$.  
The GR configuration space metric is then $\sM_{AB} = {\sM}^{\mu\nu\rho\sigma} := \{\sh^{\mu\rho}\mh^{\nu\sigma} - \sh^{\mu\nu}\mh^{\rho\sigma}\}$, 
i.e. the undensitized inverse DeWitt supermetric \cite{DeWitt67} with determinant ${\sM}$ and inverse $\sN^{AB} = {\sN}_{\mu\nu\rho\sigma}$ that is 
itself the undensitized DeWitt supermetric, $\sh_{\mu\rho} \sh_{\nu\sigma} - \sh_{\mu\nu}\sh_{\rho\sigma}/2$.   
The densitized versions are $\sqrt{\sh}$ times the former and $1/\sqrt{\sh}$ times the latter.}
\beq
\mg_{\Gamma\Delta} =
\left(
\stackrel{    \mbox{$ \upbeta_{\mu}\upbeta^{\mu} - \upalpha^2$}    }{ \mbox{ }  \mbox{ }  \upbeta_{\delta}    }
\stackrel{    \mbox{$\upbeta_{\gamma}$}    }{  \mbox{ } \mbox{ }  \mh_{\gamma\delta}    }
\right)
\mbox{ }   .   
\eeq

\noindent The {\it extrinsic curvature} of a hypersurface is the rate of change of the normal $\mn^{\Gamma}$ along that hypersurface and 
thus of the bending of that hypersurface relative to its ambient space,
\beq
\mK_{\Gamma\Delta} = \mh_{\Gamma}\mbox{}^{\Lambda}\nabla_{\Lambda}\mn_{\Delta} = \pounds_{\sn}\mh_{\Gamma\Delta}/2 \mbox{ } .  
\eeq
Or (since it is a hypersurface object so it can be thought of as a spatial tensor as well as a spacetime one: $\mK_{\Gamma\Lambda}\mn^{\Lambda} = 0$, just  
like $\mh_{\Gamma\Lambda}\mn^{\Lambda} = 0$),  
\beq
\mK_{\mu\nu} =  \{\dot{\mh}_{\mu\nu} - \pounds_{\suupbeta}\mh_{\mu\nu}\}/{2\upalpha} = \{\dot{\mh}_{\mu\nu} - 2\mD_{(\mu}\upbeta_{\nu)}\}/{2\upalpha}  \mbox{ } .
\eeq

With respect to a foliation by such spatial hypersurfaces, the Einstein--Hilbert action then splits into the ADM--Lagrangian action  (here presented in 
terms of the inverse DeWitt metric),
\beq
\FS^{\sG\sR}_{\sA\sD\sM-\sL\sD} = 
\int_{\sbSigma}\d\mt\int\d^3x\sqrt{\mh}\,\upalpha 
\left\{
\mT^{\sG\sR}_{\sA\sD\sM-\sL\sD}/\upalpha^2 +   \mbox{${\cal R}\mi\mc$}(\ux; \bh]   - 2\Lambda
\right\} \mbox{ } , \mbox{ } \mbox{ for }
\label{ADM-L}
\mT^{\sG\sR}_{\sA\sD\sM-\sL} = ||\dot{\bh}     - \pounds_{\suupbeta}\bh||^2_{\mbox{\scriptsize ${\bM}$}}/4 \mbox{ } .  
\eeq
As well as extrinsic curvature being an important characterizer of hypersurfaces, it is relevant due to its close connection to the GR momenta, 
\beq
\uppi^{\mu\nu} := {\delta \mL_{\sG\sR}}/{\delta \dot{\mh}_{\mu\nu}} = - \sqrt{\mh}\{ \mK^{\mu\nu} - \mK \mh^{\mu\nu} \}
\label{Gdyn-momenta}
\eeq
for 
\beq
\mK := \mK_{\mu\nu}\mh^{\mu\nu} \mbox{ the {\sl constant mean curvature} } .
\label{CMC-Def}
\eeq

This action encodes the GR Hamiltonian constraint 
\beq
\scH := \mN_{\mu\nu\rho\sigma}\uppi^{\mu\nu}\uppi^{\rho\sigma} - \sqrt{\mh}\{\mbox{${\cal R}\mi\mc$}(\ux; \bh] - 2\Lambda\} = 0
\label{Hamm}
\eeq
from variation with respect to the lapse $\upalpha$ and the momentum constraint 
\beq
\scM_{\mu} := - 2\mD_{\nu}{\uppi^{\nu}}_{\mu} = 0  
\label{Momm}
\eeq
from variation with respect to $\upbeta^{\mu}$.
\noindent The GR momentum constraint is straightforwardly interpretable in terms of the physicality residing not in the choice of coordinatization/point-identification 
but rather solely in terms of the remaining 3-geometry of space information that is also contained in the 3-metric.  
This is how GR is, more closely, a dynamics of 3-geometries in this sense: {\bf Geometrodynamics}. 
However, interpreting the GR Hamiltonian constraint is tougher; it leads to the Frozen Formalism Facet of the POT (Secs \ref{POTiQG}, \ref{QM-POT}).

\subsubsection{GR in relational form}\label{GR-Relational}

The Barbour-type indirect formulation of RPM's (\ref{GeneralAction},\ref{Taction}) then makes the relational parallels particularly clear. 
In the geometrodynamical counterpart of this, $\FrQ$ = Riem($\bupSigma$) -- the space of Riemannian 3-metrics on some 
spatial manifold of fixed topology $\bupSigma$ (taken to be compact without boundary for simpleness).  
The group of irrelevant motions $\FrG$ is Diff($\bupSigma$), i.e., the diffeomorphisms on $\bupSigma$.  
This Configurational Relationalism is implemented indirectly by introducing auxiliary variables that represent arbitrary-Diff($\bupSigma$) corrections. 
Temporal Relationalism is implemented by building a MRI/MPI action \cite{RWR, Lan} [this is similar to the Baierlein--Sharp--Wheeler action 
(\cite{BSW}, see also Sec \ref{Gdyn-Interp}) but now properly combining the temporal and Configurational Relationalisms].    
The relational formulation of Geometrodynamics is valuable in providing guidance in yet further investigations of alternative conceptual foundations for GR \cite{BB82, B94I, RWR, Phan}, 
and as regards addressing the POT in QG (see Sec \ref{POTiQG} and Parts II and IV).\foo{$\dot{\mF}^{\mu}$ 
is the velocity of the frame; in the manifestly relational formulation of GR, this cyclic velocity , or the cyclic partial differential $\pa\mF^{\mu}$, 
plays the role more usually played by the shift Lagrange multiplier coordinate.}
\beq
\FS_{\sG\sR}^{\sr\se\sll\sa\st\si\so\sn\sa\sll} = 
2\int\d\lambda\int_{\sbSigma}\d^{3}x\sqrt{\mh}\sqrt{    \mT_{\sG\sR}^{\sr\se\sll\sa\st\si\so\sn\sa\sll}  \{\mbox{${\cal R}\mi\mc$}(\ux; \bh] - 2\Lambda\}  } = 
\sqrt{2}\int\d\lambda\int_{\sbSigma}\d^{3}x\sqrt{\mh}\sqrt{\mbox{${\cal R}\mi\mc$}(\ux; \bh] - 2\Lambda} \, \d\ms_{\sG\sR}^{\sr\se\sll\sa\st\si\so\sn\sa\sll}
\eeq
\beq
\mbox{for } \mbox{ } 
\mT_{\sG\sR}^{\sr\se\sll\sa\st\si\so\sn\sa\sll}   := ||\Circ_{\usF}\bh||_{\sbM}\mbox{}^2/4 
\mbox{ }  \mbox{ or } \mbox{ }
\d\ms_{\sG\sR}^{\sr\se\sll\sa\st\si\so\sn\sa\sll} := ||\pa_{\usF}\bh||_{\sbM}/2 
\mbox{ and } \mbox{ } 
\Circ_{\suF}\mh_{\mu\nu} := \dot{\mh}_{\mu\nu} - \pounds_{\dot{\usF}}\mh_{\mu\nu}   
\mbox{ }  \mbox{ or } \mbox{ }
\pa_{\suF}\mh_{\mu\nu} := \pa{\mh}_{\mu\nu} - \pounds_{\pa{\usF}}\mh_{\mu\nu} 
\label{GRaction} \mbox{ } .  
\eeq
In this case, $\FrQ$ is the space Riem($\bupSigma$) of Riemannian 3-metrics on a fixed spatial topology $\bupSigma$, and $\FrG$ is the corresponding 3-diffeomorphism group, Diff($\bupSigma$).  
So this action is a JBB$[P(\langle \mbox{Riem}(\bupSigma), \mbox{ } \bM\rangle, \mbox{ } \mbox{Diff}(\bupSigma))]$.

Now, MPI/MRI gives \cite{Dirac} GR's Hamiltonian constraint (\ref{Hamm}), which parallels (\ref{GEnCo}).
Also, variation with respect to the auxiliary Diff($\bupSigma$)-variables $\mF^{\mu}$ gives the GR momentum constraint (\ref{Momm}).

The zero total dilational momentum constraint (\ref{LD} iii) closely parallels GR's well-known maximal slicing condition \cite{Lich44}, 

\noindent
\beq
\uppi := \mh_{\mu\nu}\uppi^{\mu\nu} = 0 \mbox{ } . 
\label{maxsl}
\eeq
Additionally, much like generalizing maximal slicing to constant mean curvature (CMC) slicing \cite{York72}\footnote{Clearly  
(\ref{Gdyn-momenta}) and (\ref{CMC-Def}) imply that (\ref{maxsl}) is equivalent to $\mK = 0$ and (\ref{CMCsl}) to $\mK =$ constant, hence the CMC name.}
turns on a `York time' variable \cite{York72, York73, Kuchar81, Kuchar92, I93} 
\beq 
\mt^{\sY\so\sr\sk} := \mbox{$\frac{2}{3}$}\mh_{\mu\nu}\uppi^{\mu\nu}/\sqrt{\mh} = c(\lambda \mbox{ coordinate label time alone, 
i.e. a spatial hypersurface constant}) 
\mbox{ } ,
\label{CMCsl}
\eeq 
one can think of the passage from pure-shape RPM to scaled RPM as involving an extra `Euler time' variable \cite{Paris, 06II, SemiclI, Cones}       
\beq
t^{\sE\su\sll\se\sr} := \suma \uq^i \cdot \up_i \mbox{ } .
\label{TEuler}
\eeq 
%
%
This is all underlied for both GR and RPM's by scale--shape splits, the role of scale being played by 
$\sqrt{\mI}$ or $\mI$ for RPM's and by such as the scalefactor $a$ or $\sqrt{\mh}$ in GR.  
In both cases it is then tempting to use the singled-out scale as a time variable but this runs into monotonicity problems. 
These are avoided by using as times the quantities conjugate to (a function of) the scale, such as $\mt^{\sY\so\sr\sk}$ or $t^{\sE\su\sll\se\sr}$.  
Moreover, GR in the conformogeometrical formulation on CMC slices \cite{York72, York73} can be reformulated \cite{ABFKO} as a relational theory in which $\FrG$ consists both of 
Diff($\bupSigma$) and a certain group of conformal transformations (see Sec \ref{CS+V}).

GR has an analogue of emergent JBB time,
\beq
\mt^{\se\sm(\sJ\sB\sB)}(\ux) - \mt^{\se\sm(\sJ\sB\sB)}_0(\ux) = 
\stackrel{    \mbox{\scriptsize extremum} \mbox{ } \sF^{\mu} \mbox{ }  \in  \mbox{ }  \mbox{\scriptsize Diff}(\sbSigma)        }   
         {    \mbox{\scriptsize of} \mbox{ } \stS^{\mbox{\tiny relational}}_{\tG\tR}    }  
\left. \int ||\pa_{\suF}\bh||_{\sbM} \right/   \sqrt{     \mbox{${\cal R}\mi\mc$}(\ux; \bh] - 2\Lambda    }  \mbox{ } .
\label{GRemt}
\eeq
This represents the same quantity as the usual spacetime-assumed formulation of GR's {\it proper time}, and, in the (predominantly) homogeneous cosmology setting, the cosmic time. 
We shall also see how (Sec \ref{Semicl}) an approximation to it also coincides with GR's own version of emergent semiclassical (alias WKB) time.

See Sec \ref{POTiQG}, \ref{Analogy-Diff} and the Conclusion for further GR--RPM analogies.
The resemblance between RPM's and GR is to a comparable but different extent to the resemblance between 
GR and the more habitually studied minisuperspace models \cite{mini, Misref, Magic, minimetric2, HH83, Wiltshire, eath, Bojowald}.
The main useful modelling points in this regard are as follows.   

\noindent 1) that RPM's linear zero total angular momentum constraint $\scL_{\mu}$ is a nontrivial analogue of GR's linear momentum constraint $\scM_{\mu}$. 
(This is a structure which minisuperspace only possesses in a trivial sense and one which is important 
as regards the details of a large number of applications, including aspects of the POT. 

\noindent 2) Also, RPM's (unlike minisuperspace) have notions of locality in space and thus of 
clustering/structure formation; this is useful, among other things, for cosmological modelling.

\mbox{ }  

\noindent Note 1) In GR, 1) and 2) are tightly related as both concern the nontriviality of the spatial derivative operator.  
However, for RPM's, the nontriviality of angular momenta and the notion of structure/inhomogeneity/particle clumping are unrelated.  
Thus, even in the simpler case of 1-$d$ models, RPM's have nontrivial notions of structure formation/inhomogeneity/localization

\noindent /correlations between localized quantities.  

\noindent Note 2) Minisuperspace is, however, closer to GR in having more specific and GR-inherited potentials and indefinite kinetic terms. 
Thus it and RPM's are to some extent complementary in their similarities to GR, and thus in the ways in which they are useful as toy models of GR.\foo{Midisuperspace 
(see e.g. \cite{midi, midi21, midi22, Kuchar94}) unites all these desirable features but is unfortunately then calculationally too hard for many of the aspects of the POT.}  

\subsection{Classical study of RPM's: outline of Part I}\label{Intro-Cl}

I build up a reasonable set of concrete RPM models.
A first key step in understanding these is Sec \ref{Examples}'s use of relative Jacobi coordinates.    
Sec \ref{Examples} also reviews various ways of setting up RPM's; its detail is mostly expected to be of interest to people who have worked with RPM's.    
A second key step (Sec \ref{Q-Geom}) is in restricting attention for now to 1-$d$ and 2-$d$ models; for $N$ particles, I term these, respectively, 
{\it N-stop metroland} and {\it N-a-gonland}. 
(The first two nontrivial $N$-a-gonlands I furthermore refer to as {\it triangleland} and {\it quadrilateralland}). 
This is a key step because their configuration spaces are mathematically highly tractable \cite{Kendall84, Kendall, FORD}:  
in the pure-shape case $\mathbb{S}^{N - 2}$ spheres for 1-$d$ and $\mathbb{CP}^{N - 2}$ complex projective spaces for 2-$d$. 
The third key step is that the scaled case's configuration spaces are the cones over these (\cite{Cones}, Sec \ref{Q-Geom}) which include 
C$(\mathbb{S}^{N - 2}) = \mathbb{R}^{N - 1}$ 
The above configuration spaces are all for the choice of plain rather than mirror-image-identified shapes.
%
%
I argue for plain shapes, at least to start off with, since these giving simpler mathematics and thus are conducive to a wider range of toy 
model calculations being completeable, comparable and combineable, which is precisely what the study of the POT needs!
While pure-shape models are more straightforward than models with scale for a number of purposes, scaled models shall also be required as 
regards reasonably realistic semiclassical quantum cosmologies. 
Also, as a consequence of the cone structure, pure-shape problems occur as subproblems in models with scale \cite{BGS03, FORD, 08I}, 
so studying these first also makes sense even from this semiclassical quantum cosmological perspective).    
N.B. that the interesting theoretical parallels between GR and RPM's are unaffected by the choice of plain shapes and of low-$d$ RPM's.

The smallest nontrivial relational examples are scaled 3-stop metroland, pure-shape 4-stop metroland and pure-shape triangleland.
N.B. that while both involve spheres, 4-stop metroland has a simpler physical realization of this than triangleland \cite{Cones}.  
Thus study of 4-stop metroland will generally precede that of triangleland in this Article. 
Also note that many 3- and 4-stop metroland results readily extend to the $N$-stop case.  
Further noteworthy features of the models are that 4-stop metroland has disjoint nontrivial subsystems and hierarchies of relationally 
nontrivial subsystems, while triangleland can simultaneously possess scale and nontrivial linear constraints.  
Quadrilateralland has the useful property of simultaneously possessing all of these features, as well as being more geometrically 
typical for an $N$-a-gonland than triangleland (at the cost of this mathematics being somewhat harder than triangleland's.
Quadrilateralland is one of the principal frontiers in the first edition of this Article. (see \cite{QuadI, QuadII, QuadIII}).  
Other possible extensions are to $N$-a-gonland (3-$d$ models rapidly become too hard to handle with increasing $N$  -- see Sec \ref{Q-Geom} for more on this),  
mirror-image-identified counterparts and/or counterparts with (partly) indistinguishable particles.

Since Classical Dynamics and QM both benefit considerably from study of the underlying configuration space at the topological and geometrical level, I provide this for pure-shape RPM 
and then scaled RPM in Sec \ref{Q-Geom},  doing so in great enough generality to anchor mirror-image-identified and indistinguishable particle cases.   
Since pure-shape $N$-stop metroland and $N$-a-gonland have standard and tractable geometries, one can work directly on these 
spaces/reduce down to them, as well as subsequently having available numerous useful coordinate systems and Methods of Mathematical Physics.  
For triangleland,  $\mathbb{CP}^{1}  = \mathbb{S}^{2}$, so one has `twice as many techniques'.
Thus I focus on triangleland in particular -- this meets many of the nontrivialities required by POT strategies.  
Scaled RPM has as configuration spaces the `cones' corresponding to each of the above pure-shape RPM configuration spaces.    
This makes clear the {\it scale--shape} representation of scaled RPM, which plays a big role in the present Article, input through  
pure-shape RPM also occuring as a subproblem within the scale--shape split of scaled RPM. 
I give a description of shape quantities, including the Cartesian versus Dragt \cite{Dragt} correspondence which distinguishes 
between the ways the 4-stop metroland and triangleland 2-spheres are realized; this amount to how one geometrically interprets  suitable variables.  
I also present the useful technique of tessellation of the mass-weighted shape space sphere by its physical interpretation.

Some indirect formulations of relational theories can be reduced to direct ones (SSec \ref{Red-App}), or independently conceived of 
in direct terms (SSec \ref{DRIII}) -- the relationalspace approach.  
For scaled RPM, see \cite{LB,Gergely, 06I, TriCl, FORD}.  
It was also done in 2-$d$ for pure-shape RPM \cite{06II, TriCl, FORD}.  
What one then gets in these examples coincides with what one gets from direct formulation (as I demonstrate in Sec \ref{Dyn1}).  
I also provide comparison with the relational-absolute split of Newtonian Mechanics.

I provide a number of less usual and new variational principles of dynamics moves and objects as are appropriate for the relational treatment:  
free-end NOS variation, more relational parallels of the total Hamiltonian, the Dirac procedure and of phase space. 
This includes also a `rigged' notion of phase space that might suffice for approaches in which $\FrQ$ is primary.

In Sec \ref{Dyn1}, I consider the dynamical equations following from the actions of RPM's in relationalspace/reduced form. 
This includes useful analogies with more commonly encountered physical systems: rotors, central problems, the Kepler--Coulomb problem.
I also provide physical interpretation for RPM's momenta, isometries and conserved quantities (the middle of these are mathematically $SO(p)$ and $SU(p)$ for 1- and 2-$d$ models, 
but are realized by quantities which are more general in interpretation in physical space than angular momenta: they have a `shape momenta' interpretation).
These are conserved for certain sufficiently shape-independent in loose parallel to the centrality condition in ordinary mechanics implying conservation of angular momentum.  
%
%
I also discuss gauge-invariant quantities for RPM's and outline how indirectly-formulated RPM's avoid the absence of configuration space monopole problems.

In Sec \ref{Cl-Soln}, I consider power-law potential terms.  
I then consider potentials chosen via analogy with Cosmology \cite{Cones}.  
Here, the heavy slow dynamics of $\sqrt{\mI}$ or $\mI$  parallels that of the GR scalefactor $a$. 
However, this is now coupled to a simpler light fast finite dynamics of pure shape [finite and tied to well-studied configuration space geometry 
such as $\mathbb{S}^{N - 2}$ or $\mathbb{CP}^{N - 2}$] rather than the rather more complicated light fast infinite dynamics of small GR inhomogeneities.  
Thus also RPM's make for a useful qualitative model of the quantum-mechanical origin of structure formation.

I then provide simple exact and approximate solutions for all of these, many of are familiar from elsewhere in Physics. 
I furthermore interpret them in RPM terms (which is new, and overall amounts to using established 
knowledge to understand RPM models which then have significant further POT and quantum cosmological properties).  
I provide qualitative descriptions of solutions, and mathematical forms of some solutions 
or approximate solutions (mostly in parallel with either ordinary mechanics work or GR Cosmology work). 
These give a reasonable understanding of RPM dynamics.  
I also include Appendix \ref{Cl-Soln}.C as regards whether the observed universe can be directly modelled with RPM's.  

\noindent Finally, in Sec \ref{Cl-Str} I give an account of the further levels of classical structure required for this paper's POT applications. 
These are notions of distance and localization in each of space and configuration space, localization, sub-statespace, union of 
statespaces, grainings, information, correlation and propositions. 
I illustrate the variety of structures by giving a detailed account of notions of distance on configuration space (`between two shapes').  
I give a $\FrG$-act, $\FrG$-all generalization of the Best-Matching indirect implementation of Configurational Relationalism.
Grainings and information carry SM connotations.
My treatment of correlations draws from Statistics as well as from Physics.  
Propositions are simple at the classical level (for the most part), forming the usual Boolean logic.

\subsection{Quantum study of RPM's: outline of Part III}\label{Intro-QM}

This Article's approach to quantization is laid out in Sec \ref{QM-Intro} in parallel with Quantum Cosmology/Quantum Geometrodynamics.  
While it is true that various infinite-dimensional, diffeomorphism related and indefinite configuration 
space metric signature related aspects are not captured by RPM's, however, the following major issues {\sl are} present for RPM's.  

\noindent 1) Kinematical quantization, including ambiguities paralleling i) affine versus ordinary Geometrodynamics and ii) absolute versus relational choices.  

\noindent 2) Time-independent Schr\"{o}dinger equations (TISE's), and thus the frozen-formalism aspect of the Problem of 
Time (see next SSec and Part III for more on this -- the principal focus of the present Article.) 

\noindent 3) Accompanying linear quantum constraints (unless already reduced out, and indeed reduced versus Dirac approach issues \cite{Smolin, 06I, 08II, 08III}).   

\noindent 4) Operator-ordering issues \cite{Banal, Cones}, including some novel connections between these and 
absolute versus relational motion issues, Dirac versus reduced quantization issues, and inner product issues. 


\noindent 5) -- 9) C.f. also the Quantum Cosmology issues in SSec \ref{Q-Cos}.

\mbox{ } 

\noindent I then consider specific examples already familiar from Part I, that go for the simplest possible mathematics 
to maximize tractability and checkability of subsequent POT manoeuvres, while remaining meaningful as whole-universe models.  
A further filter on models is that which of the potentials arising from the Cosmology--Mechanics analogy  
analogy have the further useful features of being more analytically tractable well-behaved at the quantum level.   
As $N$-stop metroland has no Dirac-reduced difference or operator-ordering ambiguity, I start in Sec \ref{Q-3-Stop} with scaled 3-stop metroland.   
This gives simple Bessel/Laguerre mathematics which I then reinterpret in terms of the less usual relational physics.
I use this to point out and sometimes resolve various foundational issues: closed universe effects lessening the 
spectrum unless interpreted in a multiverse sense (by energy-interlocking) and non-badness of its semiclassicality.
4-stop metroland and triangleland both have spherical harmonics mathematics, and with potentials included, also have mathematics that is well-known from Molecular Physics.  
I use asymptotic as well as exact and perturbative methods of solution.  
I consider some scaled RPM's in Sec \ref{Q-ERPM}.  
Scaled 4- and $N$- stop metroland also give simple Bessel/Laguerre mathematics, while 
scaled triangleland can be tackled with well-known spherical and parabolic coordinate systems, giving spherical harmonics, Bessel and Laguerre mathematics.

This further establishes the useful presence of simple mathematics.  
This Article carefully exports the usefulnesses and limitations of  Molecular Physics methods and concepts to Quantum Cosmology in Sec \ref{RPM-for-QC}.  
In Sec \ref{QM-Str}, I provide the QM counterpart of Sec \ref{Cl-Str}'s levels of structure, 
with particular use of quantum information and nontrivial theory of propositions (Heyting Algebras, Topos Theory).

\subsection{The Nature of Time} \label{Time}

The other main theme (c.f. Fig 1) of this Article is the Problem of Time (POT) in QG.  

\mbox{ } 

\noindent Part II is on the POT at the classical level, which results from Background Independence/Relationalism and 
to some extent closed-universe considerations also; it is already inherent in GR, and, as this Article argues, in certain formulations of Classical Mechanics too.

\noindent Part IV is on the quantum POT, which comes about because the `time' of GR and the `time' of ordinary Quantum Theory are mutually incompatible notions.
This incompatibility leads to a number of problems with trying to replace these two branches of Physics with a 
single framework in situations in which the premises of both apply, such as in black holes and in the very early universe.  
There are then yet more analogies between RPM's and GR (see \cite{Battelle, DeWitt67, Kuchar81, PW83, Kuchar91, Kuchar92, B94I, B94II, Kuchar99, EOT, Kieferbook, 06II, Smolin08, 08II, 
AF, +Tri, Cones, ScaleQM, 08III, SemiclIII} and Part IV of this Article itself)  at the level of various strategies toward the resolution of the POT \cite{Kuchar92, I93}.

\subsubsection{Time in Philosophy and Physics; desirable properties for a candidate time to indeed be a time}\label{tfns}

It is unfortunately often tacitly assumed in the literature that once one allots the word `time' to a variable 
or parameter, then this quantity will then satisfactorily serve as a such. 
I hold that, rather, such an allocation is of a {\bf candidate time}, and only detailed study of it will reveal whether it is a successful candidate.  
One needs to check a fairly extensive property list before one can be satisfied that it is a) indeed a time or a clock and 
b) that it is useful for accurate work in comparing theory and experiment.  
Of course, philosophers have had conflicting views about what time is since the beginning of civilization; Heraclitus and Paramenides were 
already at odds over whether the world contains a `flow of time', whilst this and the next SSec tackles Saint Augustin's \cite{Augustin} question: 
 
\mbox{ } 

{\it ``What then is time? If no one asks me, I know what it is. If I wish to explain it to him who asks, I do not know."}   
 
\mbox{ }   
  
\noindent Finally, what is a sufficient set of properties for a candidate time in modern Physics is theory-dependent: 
each grouping Newtonian Mechanics and Quantum Mechanics, SR  and QFT, and GR, has a different notion of time.   
Consequently what time is to mean in QG is heavily disputed \cite{I93, APOT2}...

\mbox{ }  


\noindent Time-1) {\bf Time as an ordering} \cite{Leibniz2, Jammerteneity}.  
Simultaneity enters here, i.e. a notion of present that separates future and past notions, which are perceived differently (e.g. one remembers the past but not the future).  
The structure of the present {\it instant} is the whole-universe configuration that includes one's theory's notion of space 
as well as whatever internal configuration spaces along the lines of gauge theory.

\noindent Time-2) {\bf Causation} \cite{Leibniz2, Jammerteneity}, i.e. that one phenomenon brings about another at a later value of `the time'.    

\noindent Time-3) {\bf Temporal logic}: this extends more basic (atemporal) logic with ``and then" and ``at time $t_1$" constructs.

\noindent Time-4) In (e.g. Newtonian) dynamics, one encounters the idea of {\bf change in time} (so that time is a container): 
a parameter of choice with respect to which change is manifest.   
Newtonian absolute time is a such (and external to the system itself and continuous).  

\noindent One state of a system ``{\bf becoming}" another state of a system is another phrasing of this dynamical facet of time.

\noindent {\bf Duration}  is defined to be magnitude of time, the container property that parallels `extent' in the case of space.  
See e.g. \cite{Poincare98, Bergson, Bachelard, Whitrow} for emphasis on the importance of this property. 

\mbox{ }

\noindent Note 1) The first two sentences in Time-4) should be contrasted with Mach's [Relationalism 7))] viewpoint that {\sl time is abstracted from change}.  
This is considerably developed in Sec \ref{Cl-POT}.  
Such a secondary time would, moreover, be expected to have many of the other properties listed in this SSec.  

\noindent Note 2) There is an {\bf arrow of time} notion present in ordering, causation and temporal logic. (see Sec \ref{AOT} for more).    
This distinction is observed in practise and yet is not manifest in fundamental dynamical equations (which are time-reversal invariant or CPT invariant \cite{qftcorr}).  
In such cases, the arrow is built in by hand at the level of the solutions selected (and the correlations in direction of various aspects of the arrow remain mysterious). 

\mbox{ }

\noindent Time-5) Mathematically, time is often taken to be modelled by the real line or an interval of this or a discrete approximation of this (`atoms of time').    
Though time can easily be {\sl position-dependent} in field theories: $\mt(\ux)$ in place of $t$.  
Some notions of time are more complicated (e.g. parallel times and branching times in Fig \ref{TimeTop} or many-fingered times in GR in the next SSec).   

{            \begin{figure}[ht]
\centering
\includegraphics[width=0.7\textwidth]{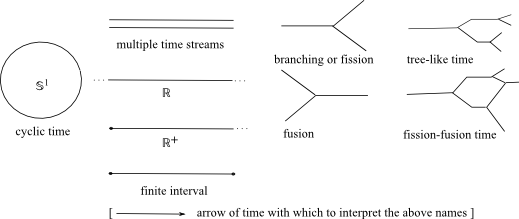}
\caption[Text der im Bilderverzeichnis auftaucht]{        \footnotesize{Different topologies proposed for single-function time.  
As compared to space, these are rather restricted by time's dimension being one.  
The options then are the real line, the half-real line (e.g. from big bang to heat death), the finite interval (e.g. big bang 
to big crunch), the cycle (as in Hindu philosophy or the ekpyrotic universe), multiple time streams and then the following non-Hausdorff options.  
{\sl Branching} (or {\sl fission}) is solely to the future, whilst the time-reversal of that is termed {\sl fusion}; if 
multiple time streams are possible, then one or both of these features might also be present, tying time streams together.}   }
\label{TimeTop} 
\end{figure}  } 

\noindent Time-6) Time is also usually taken to be {\bf monotonic} (rather than direction-reversing) 
This makes sense in the context of time having further ordering and causation properties.    
It is also a part of the arrow of time property, in that there is a direction involved (the further part being that the various directions are then correlated).  
Apart from this SSec, see e.g. \cite{UW89, Hartle96, Zeh09} for emphasis on many other properties of time. 
   
\mbox{ }

\noindent Time-7) There is to be freedom in prescribing a timefunction as to the {\bf choice of calendar year zero} or {\bf choice of when to start the stop-watch} 
and of {\bf year-length} or {\bf tick-duration}. 
I.e.,  
\beq
\mbox{if $\ft$ is a timefunction, so is $A + B\,\ft$ for $A$, $B$ constants } . 
\eeq
Notation: I use $\lft$ (`calendar-year-zero-adjusted $\ft$') for the combination $\ft - \ft (0)$ that is ubiquitous in this Article.

\mbox{ }

\noindent Time-8) A good time function is {\bf globally valid} \cite{Kuchar92, anhHaj, Hartle96} both over time (antagonist to the half-finite and finite interval times unless there 
is a good physical reason for this) and over space (made necessary by field theories and, to a greater extent generic curved space/GR).   
%

\noindent Time-9) It also makes sense for a time function to be {\bf operationally meaningful} (computable from observable quantities -- tangible and practically accessible).

\subsubsection{Time in the main physical theories}\label{Time-in-Physics}

\noindent {\bf Newtonian Absolute Time} is the basis of classical (pre-relativistic) Physics and of ordinary QM.
This is a {\sl fixed} background parameter that enters ordinary QM in the following ways.

\noindent Time-QM 1) It is represented in QM by an {\sl anti}-hermitian operator (unlike the representation of other quantities). 

\noindent Time-QM2) The time--energy uncertainty relation 
$\Delta t^{\sN\se\sw\st\so\sn}\Delta E \geq \hbar/2$ is also given an entirely different meaning to that of the other uncertainty relations.  
[I subsequently use $\hbar = c = G = 1$ units, except where required by semiclassical considerations.]

\noindent Time-QM3) There is unitary evolution in time, i.e. that probabilities always sum to one.
This evolution is in accord with the theory's time-dependent wave equation (TDWE, for now a time-dependent Schr\"{o}dinger equation: TDSE).  
The {\it scalar product} on the Hilbert space of states leads to conserved probability currents.

\noindent Time-QM4) Moreover, there is a second dynamical process: collapse of the wavefunction that is held to occur in ordinary QM 
despite its not being described by the evolution equation of the theory.  

\mbox{ }

\noindent SR brings in the further notions of 

\noindent Time-SR1) a proper time corresponding to each observer, and of 

\noindent Time-SR2) time as another coordinate on (for now, flat) spacetime, as opposed to the external absolute time of Newtonian Physics.  

\noindent Whilst in SR it is often said that space and time can also be regarded as fused into the spacetime of Minkowski, SR breaks only isolation of 
space and time, not their distinction \cite{Broad}, with Whitrow \cite{Whitrow} and Barbour \cite{B11} also being dismissive of Minkowski's \cite{Minkowski}  
considering the individual notions of time and space to be  ``{\it doomed to fade away}".

\noindent Time-SR3) Moreover, time in SR is also external and absolute in the sense of it having its own presupposed set of privileged inertial frames 
(the quantum theory can be made independent of a choice of frame if it carries a unitary representation of the Poincar\'e group).    
In that sense in SR all one has done is trade one kind of absolute time for another, so the passage from Newtonian Mechanics to SR and 
thus from ordinary QM to relativistic QFT does {\sl not} greatly affect the role of time.  
(See \cite{Kiefer09bis} for another recent account of this position.)
The new privileged structures are underlied by SR's Minkowski spacetime's possessing suitable Killing vectors (in fact a maximal number of Killing vectors).

\noindent The changes are, rather, in

\noindent Time-SR4) The notion of simultaneity (and how to set up the simultaneity convention) changes in passing from Newtonian Mechanics to SR \cite{Jammerteneity}: 
the demise of universal slices of simultaneity and the rise of the physical significance of null cones. 

\noindent Time-QM3$^{\prime}$) Also the type of TDWE is now e.g. Klein--Gordon or Dirac 
(and these then require a QFT interpretation, including a distinct kind of inner product).  

\mbox{ }  

It is only GR that frees one from absolute time, and so that the SR to GR conceptual leap is in many ways bigger than the one from Newtonian Mechanics to SR.  

\noindent Time-GR1) In GR matter now exerts an influence on the form of space and time, by which it is in general curved rather than flat as Minkowski spacetime is.   
The conventional {\bf GR spacetime formulation} is in terms of $({\FrM}, \mg_{\Gamma\Delta})$. 
[I usually restrict to the vacuum case for simplicity, but extending this Article's study to include the ordinary matter fields is unproblematic.] 

\noindent Time-GR2) Time is a {\sl general} spacetime coordinate in GR, clashing with ordinary QM's having held time to be a unique and sui generis extraneous quantity.   

\noindent Time-GR3) GR's generic solutions have no Killing vectors (and so no timelike Killing vector pointing out a privileged time), 
and GR is ultimately taken to be about generic solutions.
One has to pass from having privileged frame classes to dealing with the diffeomorphisms of each of space, spacetime and split space-time.
This means that much of the structure of ordinary QM simply ceases to have an analogue.  

\noindent Time-GR4) In GR, time has the ordering property, whilst causality reigns where simultaneity reigned in Newtonian Mechanics as extension 
of the situation in SR except now matter and gravitation influence the larger-scale causal properties.  

\noindent Time-GR5) GR additionally has a time non-orientability notion \cite{Wald}, and a closed timelike curve notion (both of these 
are usually held to be undesirable features for a physical solution to possess).  

\noindent Time-GR6) While in some senses GR has a Machian character through the influence in Time-GR1) and the absence of globally preferred frames, 
Einstein's inception of GR did not concretely build up on Mach's ideas \cite{WheelerGRT, DOD}, so whether the theory actually implements these has been a source of quite some argument. 
%
%
However, this Introduction's passing to Geometrodynamics, and then the relational formulation thereof, irons this out.  

\noindent Time-GR7) Time as a general choice of coordinate is embodied geometrically in how GR spacetime is sliced into a 
{\bf sequence of} (or {\bf foliation by}) {\bf spacelike hypersurfaces} corresponding to constant values of that time.  
There is one timefunction per choice of foliation; thus time in GR is said to be `{\bf many-fingered}' (see Secs \ref{Examples} and \ref{Cl-POT} for further detail).  
This corresponds to such as the geometrodynamical formulation of GR; for dynamical formulations of GR, one assumes {\bf global hyperbolicity}, 
which amounts to determinability of GR evolution from GR initial data and excludes e.g. the properties listed under Time-GR5).

\noindent In general, if there is more than one plausible conceptual approach providing a timefunction, one is then interested in whether the various timefunctions are aligned.  

\noindent In the GR setting, each is for a given sequence of slices, each of which is experienced by a family of observers moving in a particular way.  

\noindent That space and time can be thought of as distinct and yet spacetime also carries insights accounts for why 
a number of POT facets are already present in classical GR (see Secs 3 and \ref{Facets}).\footnote{This is furtherly so
given that the former can be formulated relationally in the sense of LMB.
9) also accounts for some parts of the classical POT facets.}  


\noindent Time-GR8)  Directly building up on some of Mach's ideas happens to lead one to (a portion of) GR, arriving to it in its 
(geometro)dynamical form \cite{Buckets, DOD, RWR, Phan}, so that Einstein's GR does, in any case, happen to contain this philosophically desirable kernel.

\mbox{ }  

\noindent Time-QG1) Indeed, are {\sl spacetime} or any of its subaspects meaningful in QG, and how do the other aspects emerge in the classical limit?  
Is (spatial or spacetime) classical {\sl geometry} or any of its subaspects meaningful in QG, and how do the other aspects emerge in the classical limit. 
Is there something resembling the classical notion of {\sl causality} in QG, and if so, which 
aspects of classical causality are retained as fundamental, and how do the other aspects emerge in the classical limit?

\mbox{ }

\noindent Time-Closed-1) External time is furthermore incompatible with describing truly {\sl closed} systems, the ultimate of which is closed-universe Quantum Cosmology. 
Here, Page and Wootters \cite{PW83} have given convincing arguments that such a system's only physical states are {\sl eigenstates} of the 
Hamiltonian operator, whose time evolution is essentially trivial. 
\noindent How QM should be {\sl interpreted} for the universe as a whole is then a recurring theme (see e.g. Secs \ref{Q-3-Stop} ... \ref{RPM-for-QC} and \ref{QM-POT-Strat}).  

\mbox{ }

\noindent Also, the idea of events happening at a single time plays a crucial role in the technical and conceptual foundations of Quantum Theory, as follows.  

\mbox{ }

\noindent Time-QM5) {\sl Measurements} made at a particular time are a fundamental ingredient of the conventional Copenhagen 
interpretation (which is anchored on the existence of a privileged time).  
Then in particular \cite{I93}, an {\bf observable} is a quantity whose value can be measured at a `given time'.  
On the other hand, a `history' has no direct physical meaning except in so far as it refers to the outcome of a sequence of time-ordered measurements.

\noindent Time-QM6) In constructing a Hilbert space for a quantum theory, one is to select a complete set of observables that are required to 
{\sl commute} at a fixed value of time -- i.e. obey {\bf equal-time commutation relations}.
\noindent Again, this notion straightforwardly extends to the SR case by considering not one Newtonian time but the set of relativistic inertial frames; 
then one's QM can be made independent of frame choice via it carrying a unitary representation of the Poincar\'e group. 
For a relativistic QFT, the above equal-time commutation relations point is closely related to requiring {\it microcausality} \cite{I93}: 
\beq
[\widehat{\phi}(X^{\Gamma}),\widehat{\phi}(Y^{\Gamma})]    = 0     
\label{microcausal}
\eeq
for relativistic quantum field operators $\hat{\psi}$ and all spacelike-separated spacetime points $X^{\Gamma}$ and $Y^{\Gamma}$.  

\noindent Time-QG2) However, observables and some notion of time-equal commutation relations pose significant difficulties in the context of GR.

\subsubsection{Extra issues in using the word `clock'}\label{Clocks}

\noindent Here are some properties that it may well be desirable for any object/subsystem claimed to be a `clock' to possess (I do not claim this list to be complete).  

\noindent Clock-1) Clocks are usually taken to count occurrences that are considered to be regular in one's notion of time. 
One finds out which occurrences are regular by comparison between candidate clocks, and by extent of these candidate clocks' 
predictive power in the study of further dynamical systems rather than just of themselves.   
Clocks are often\footnote{Some (very poor) clocks can be elsewise, e.g. based on repeatable processes 
that need re-setting -- the hourglass and the water-clock.} 
taken to involve periodical processes.

\noindent Clock-2) We wish to use a notion of time in terms of which the motion is simplest, and then desires clocks that read off such a time.     
I.e. clocks can be argued to be as convenient conventions \cite{NS}; in antiquity, uniform rotation was argued to be the best standard {\sl because it is easiest to count} \cite{Aristo}, 
and the `circular motion of the heavens' provided an excellent example of a such for that epoch (see Sec \ref{Cl-POT}'s account of sidereal time for more).  

\noindent Clock-3) Clocks should actually read the purported timefunction rather than whatever they please; how this comes about 
(material versus spacetime property alignment) is not entirely clear, due to some standard Physics uses of {\sl postulation} rather than of explanation.  
If this fails, one has a bad clock or a timefunction that is at best secondary in practise (if no clock can be found that reads it).

\noindent Clock-4) Clocks should actually be physically constructible within the physically-relevant regime (and this is 
difficult to envisage in early universe regimes, parts of black hole spacetimes and in the extremely small).  

\noindent Clock-5) Clocks should be accurate enough to deal with the physics at hand.  

\mbox{ }  

\noindent Note 1) Once one has accurate clocks,\footnote{Portable mechanical clocks suitable for such purposes passed from theoretical 
design to practical sufficiently accurate engineering via Galileo, Huygens, Gemma and Harrison \cite{Jammerteneity}.
[The heavens will not serve for this purpose, since at a given moment of time they will look different from different points 
on the surface of the earth: the problem of keeping track of time-and-position at sea.]  
See \cite{Thomson1884, Einstein1905, CW} for further significant contributions to the theory of simultaneity.}
one can consider nonlocal simultaneity conventions from a practical perspective \cite{Jammerteneity}.  
Thus clocks come before, from an operational perspective, position-dependent timefunctions.  
That what it reads is synchronized with other such of relevance \cite{Jammerteneity}.  
Indeed, then, in an operational sense, clocks come before timefunctions (as opposed to mere mathematical imagining of foliations of a 
spacetime without thought as to how to populate them with suitable observers and clocks).    

\noindent Note 2) A wider problem comes from the conceptual distinctions between timefunctions, clouds of clocks (observables of a particular kind) and clouds of observers.  
Can the use of each, and the relationship between each, be carefully justified?   
See the Conclusion for more detailed treatment of this.    

\noindent Note 3) Barbour follows Leibniz (in the particular sense explained e.g. around p.41 of \cite{Whitrow}) in suggesting 
that the only perfect clock is the whole universe (at the classical level).
This should be contrasted with Newton's position that the universe {\sl contains} clocks.  
I add that it is entirely impractical to use the whole universe as a clock, since it takes considerable effort to monitor 
and one only has very limited knowledge of many of its constituent parts. 
It also looks paradoxical to regard any subsystem as imperfect, since in getting to the stage of including everything in the universe so as to `perfect' one's clock, 
one has transcended to physics beset by the Frozen Formalism Problem that is the most well-known facet of the POT.  
See also Secs \ref{Cl-POT}, \ref{QM-Clock} for further comparison of LMB time and Newtonian time.

\mbox{ } 

\noindent These things exposited, I can now turn to the POT.

\subsection{The Problem of Time (POT) in QG (Introduction to Parts II and IV)} \label{POTiQG}

\subsubsection{Background Independence, Relationalism and the POT}\label{BI-Rel-POT}

Our attitude (as per the Preface) is that Background Independence is philosophically-desirable and a lesson to be taken on board from GR, 
quite separately from any detail of the form of its field equations that relativize gravitation.
`Background Independence' is to taken here in the sense of freedom from absolute structures, i.e. of adopting a suitable relationalism, 
which is the above-developed (with some extensions in Secs 2, 4, 6 and 9).  
Moreover, relationalism already implies many POT facets at the classical level, and the POT at the quantum level can be taken to be the quantum analogue of classical relationalism.
Thus adopting background independence/relationalism entails the notorious POT as a direct consequence that is to be faced rather than simply avoided by starting to deny background 
independence/relationalism on mere grounds of practical convenience (assuming absolute structures leads to understood conceptualization and more standard and 
tractable calculations).

\mbox{ } 

\noindent Background independence criteria are held to be an important feature in e.g. Geometrodynamics and LQG (which is better emphasized by renaming this as Nododynamics, 
as per the Preface and Sec \ref{Rel-AV}),  and the relational reformulability of each of these ensures contact despite Einstein's original indirectness.  


\noindent On the other hand, perturbative covariant quantization involves treating the metric as a small perturbation about a fixed background metric.  
This is neither background-independent nor successful on its own terms, at least in the many steps of this program that have been completed, due to 
nonrenormalizability, or alternatively due to non-unitarity in some higher-curvature cases \cite{Stelle}.
%
%
Perturbative String Theory is another background-dependent approach.  
This amounts to taking time to be a fixed-background SR-like notion rather than a GR-like one in this formulation's deepest level.
Here, GR's field equations are emergent, so that problems associated with them are not held to be fundamental but should be referred further along to 
background spacetime metric structure that the strings move in.  
Moreover, technical issues then drive one to seek nonperturbative background-independent strategies and then POT issues resurface among the various 
possibilities for the background-independent nature of M-Theory. 

\noindent While Nododynamics/LQG has a greater degree of background independence than perturbative string theory -- it is independent of background {\sl metric} structure -- 
the argument for Background Independence can be repeated at the level of background {\sl topological structure}, for which one is presently very largely 
unprepared (see Appendix \ref{Cl-Str}.A and \ref{TopCha} for more).

\subsubsection{The Frozen Formalism Problem}\label{FFP-Intro}

\noindent From the relational perspective, the {\bf classical Frozen Formalism Problem} is there ab initio for the whole universe by Leibniz's Temporal Relationalism 5).
This is then implemented by MRI alongside no extraneous timelike variables, and the reparametrization-invariance implies a Hamiltonian constraint $\scH$ that is quadratic 
but not linear in the momenta, which feature and consequence are shared by $\scE$.

\noindent The {\bf Quantum Frozen Formalism Problem} then comes from elevating $\scH$ to a quantum equation produces a stationary i.e timeless or frozen wave equation -- 
the  Wheeler-DeWitt equation 
\beq
\hat\scH\Psi = 0
\label{WDE} \mbox{ } ,   
\eeq
in place of ordinary QM's time-dependent one, 
\beq
i\pa\Psi/\pa t = \hat{H}\Psi \mbox{ } .   
\eeq 
Here, $H$ is a Hamiltonian, $t$ is absolute Newtonian time and $\Psi$ is the wavefunction of the universe. 
In the hypothetical GR counterpart, one would use $\delta$ in place of $\pa$, so it is, more generally a partional derivative, $\partional$.
The GR case of (\ref{WDE}) is, in more detail, the Wheeler--DeWitt equation (WDE) 
\beq
\hat\scH\Psi := - \hbar^2\frac{1}{\sqrt{\mM}}  \frac{\delta}{\delta \mh^{\mu\nu}}
\left\{
\sqrt{\mM}\mN^{\mu\nu\rho\sigma}  \frac{\delta\Psi}{\delta \mh^{\rho\sigma}}
\right\} 
- \xi \,\mbox{${\cal R}\mi\mc$}_{\mbox{\scriptsize\boldmath$M$}}(\ux; \bh]\mbox{'}\,\Psi - \sqrt{\mh}\mbox{${\cal R}\mi\mc$}(\ux; \bh]\Psi + 2\sqrt{\mh}\Lambda\Psi + \hat\scH^{\mbox{\scriptsize matter}}\Psi = 0  \label{WDE2} \mbox{ } ,     
\eeq
where `\mbox{ }' implies in general various well-definedness and operator-ordering issues (see Sec \ref{QM-Intro} for a summary). 
Correspondingly, e.g. for indirectly formulated ERPM's in r-formulation, 
\beq
\hat\scE\Psi := - N^{\sfA\sfB}\frac{\pa}{\pa Q^{\sfA}}\frac{\pa}{\pa Q^{\sfB}}\Psi + V(Q^{\sfC})\Psi =  E\Psi \mbox{ } .  
\label{RPMWDE}
\eeq

\subsubsection{Outline of strategies for the Frozen Formalism Problem}\label{FFP-Strat}

\noindent Some of the strategies toward resolving the POT are as follows (all are presented here in the simpler case of RPM's). 
These reflect the longstanding philosophical fork between `time is fundamental' and `time should be eliminated from one's conceptualization of the world'.  
Approaches of the second sort are to reduce questions about `being at a time' and `becoming' to, merely, questions of `being'.  

A finer classification \cite{Kuchar92,I93,APOT} is into Tempus Ante Quantum (time before quantum), Tempus Post Quantum (after), Tempus Nihil Est (timelessness) 
\cite{Kuchar92, I93} and Non Tempus sed Historia (not time but history) \cite{APOT}. 
[My separation out of the last of these from \K and Isham's timeless approaches is due to its departure from conventional Physics' 
dynamics and quantization of {\sl configurations} and conjugates.]      
This Article is organized into classical parts of the POT in Part II and quantum parts of the POT in Part IV.  

\mbox{ }

\noindent Type 0 `Barbour' Tempus Ante Quantum) Take $\ft^{\se\sm(\sJ\sB\sB)}$ as a resolution of the classical Frozen Formalism Problem (Secs 7, 9); however, this fails to unfreeze at the QM level.

\noindent Type 1 Tempus Ante Quantum) Perhaps one is to find a hidden time at the classical level (\cite{Kuchar92}, Sec \ref{TAQ}) 
by performing a canonical transformation under which the quadratic constraint is sent to 
\beq
\fp_{t^{\th\ti\td\td\te\tn}} + \fH_{\st\sr\su\se} = 0 \mbox{ } .
\label{Casta}
\eeq 
Here, $\fp_{t^{\th\ti\td\td\te\tn}}$ is the momentum conjugate to some new coordinate $\ft^{\sh\si\sd\sd\se\sn}$ that is a candidate timefunction. 
$\fH_{\st\sr\su\se}$ is each such scheme's {\it true Hamiltonian candidate}.
(\ref{Casta}) is then promoted to a hidden-TDSE 
\beq
i\partional\Psi/\partional{\ft_{\sh\si\sd\sd\se\sn}} = \hat{\fH}_{\st\sr\su\se}\Psi \mbox{ } .
\eeq
The already-mentioned York time is a candidate internal time for GR, an RPM analogue for which is the Euler time.  
[While JBB time is indeed emergent whilst still at the classical level, it does not provide a linear momentum dependence and so does not serve the above purpose. 
It is however aligned with the timestandard that is recovered in B) below.]
The parabolic form (\ref{Casta}) can also be achieved by not rearranging GR but instead appending `reference fluid' matter 
[see Secs \ref{QM-POT} and \ref{TAQ}, which both conceptually and mathematically generalize the way that A) is presented above.]   

\mbox{ }  

\noindent Tempus Post Quantum 1) Perhaps (Secs \ref{TPQ} and \ref{Semicl}) one has slow, heavy `$\fh$'  variables that provide an approximate 

\noindent timestandard with respect to which the other fast, light `l' degrees of freedom evolve \cite{HallHaw, Kuchar92, Kieferbook}.  
In the Halliwell--Hawking \cite{HallHaw} scheme for GR Quantum Cosmology, $\fh$ is scale (and homogeneous matter modes) and $\fl$ are small inhomogeneities.  
The Semiclassical Approach involves making

\noindent i) the Born--Oppenheimer \cite{BO27} ansatz 
\beq
\Psi(\fh, \fl) = \psi(\fh)|\chi(\fh, \fl)\rangle
\label{BO}
\eeq 
and the WKB ansatz 
\beq
\psi(\fh) = \mbox{exp}(i\,S(\fh)/\hbar) \mbox{ } 
\label{WKB}
\eeq 
(in each case making a number of associated approximations).  

\noindent ii) One forms the $\fh$-equation, which is 
\beq
\langle\chi| \widehat{\scQ\scU\scA\scD} \, \Psi = 0
\eeq 
for RPM's.  
Then, under a number of simplifications, this yields a Hamilton--Jacobi equation\foo{For simplicity, I 
present this in the case of 1 $\fh$ degree of freedom and with no linear constraints; more generally it involves $\spartional$ derivatives contracted into inverse kinetic metric 
$\fN$ and an accompanying linear constraint, see Sec \ref{Semicl} for these in detail.} 
\beq
\{\pa S/\pa h\}^2 = 2\{E - V(h)\} \mbox{ }  
\eeq
where $V(h)$ is the $h$-part of the potential. 
One way of solving this is for an approximate emergent semiclassical time $t^{\se\sm(\sW\sK\sB)}(h)$. 

\noindent iii) One then forms the $\fl$-equation 
\beq
\{1 - |\chi\rangle\langle\chi|\}\widehat{\scQ\scU\scA\scD}\,\Psi = 0 \mbox{ } . 
\eeq 
This fluctuation equation can be recast (modulo further approximations) into an emergent-WKB-TDSE for the $l$-degrees of freedom, the mechanics/RPM form of which is\footnote{More 
generally, this involves the partional derivative and a $\scL\scI\scN$ correction, see Sec \ref{Semicl} for further details.}
\beq
i\hbar\pa|\chi\rangle/\pa t^{\se\sm(\sW\sK\sB)}  = \widehat\scE_{l}|\chi\rangle \mbox{ }
\label{TDSE2}
\eeq
the emergent-time-dependent left-hand side arising from the cross-term $\pa_{h}|\chi\rangle\pa_{h}\psi$. 
$\widehat\scE_{l}$ is the remaining surviving piece of $\widehat\scE$, acting as a Hamiltonian for the $\fl$-subsystem.  

\mbox{ } 

\noindent N.B. that the working leading to such a TDWE ceases to work in the absence of making the WKB ansatz and 
approximation, which, additionally, in the quantum-cosmological context, is not known to be a particularly strongly supported ansatz and approximation to make.    
This is crucial for this Article since propping this up requires considering one or two further POT strategies from the classical level upwards.  
[B) and conceptually-related schemes are further discussed in Secs \ref{TPQ} and \ref{Semicl}.]
$t^{\se\sm(\sW\sK\sB)}$ aligns with $t^{\se\sm(\sJ\sB\sB)}$ at least to first approximation [whose properties are substantially further covered in Secs \ref{Cl-POT} and \ref{+temJBB}, 
and which also ends up playing a role in C) and D) as explained in Secs 20 and 22.

\mbox{ }

\noindent Tempus Nihil Est 1) A number of approaches take timelessness at face value (Secs \ref{TNE}, \ref{Cl-Nihil}, \ref{QM-Nihil-Intro}, \ref{QM-Nihil}). 
One considers only questions about the universe `being', rather than `becoming', a certain way.  
This can cause at least some practical limitations, but nevertheless can address at least {\sl some} questions of interest. 

\noindent i) A first example is the {\bf Na\"{\i}ve Schr\"{o}dinger Interpretation}. 
(This is due to Hawking and Page \cite{HP86,HP88}, though its name itself was coined by Unruh and Wald \cite{UW89}.)  
This concerns the `being' probabilities for universe properties such as: what is the probability that the universe is large? 
Flat? 
Isotropic? 
Homogeneous?   
One obtains these via consideration of the probability that the universe belongs to region $\FrR$ of the configuration space that corresponds 
to a quantification of a particular such property, 
\beq
\mbox{Prob}(R) \propto \int_{\bsFrR}|\Psi|^2\mathbb{D}\fQ \mbox{ } , 
\eeq 
for $\mathbb{D}\fQ$ the configuration space volume element.
This approach is termed `na\"{\i}ve' due to it not using any further features of the constraint equations.  
It was considered for RPM's in \cite{AF, +Tri, ScaleQM, 08III}.

\noindent ii) The Page--Wootters {\bf Conditional Probabilities Interpretation} \cite{PW83} goes further by addressing conditioned questions of `being'. 
The conditional probability of finding $B$ in the range $b$, given that $A$ lies in $a$, and to allot it the value
\beq
\mbox{Prob}(B\in b | A\in a; \BigupRho) = \frac{\mbox{tr}
\big(
\mP^B_{b}\,\mP^A_{a}\,\BigupRho\,\mP^A_a
\big)                            }{
\mbox{tr}
\big(
\mP^A_a\,\BigupRho
\big)
} \mbox{ } ,    
\label{Pr:BArhoProto}
\eeq
where $\BigupRho$ is a density matrix for the state of the system and the $\mP^A_a$ denote projectors.  
Examples of such questions for are `what is the probability that the universe is flat given that it is isotropic'?  
Or, in the present context, e.g. `what is the probability that the triangular model universe is nearly isosceles given that it is nearly collinear?'  

\noindent iii) {\bf Records Theory} \cite{PW83, GMH, B94II, EOT, H99, Records, Records2} involves localized subconfigurations of a single instant of time.  
It concerns issues such as whether these contain useable information, are correlated to each other, and whether a semblance of dynamics or history arises from this.  
This requires 

\noindent 1) notions of localization in space and in configuration space. 

\noindent 2) Notions of information and correlation.  

\noindent This Article has further novel analysis of Records Theory in Secs (\ref{Cl-POT-Strat}, \ref{Cl-Nihil}, \ref{QM-POT-Strat}, \ref{QM-Nihil}).  

\mbox{ }

\noindent Non Tempus sed Historia 1) Perhaps instead it is the histories that are primary ({\it Histories Theory} \cite{GMH, Hartle}, Secs \ref{Cl-Hist}, \ref{QM-Hist}).    

\mbox{ } 

\noindent Combining the semiclassical, records and histories approaches (for all of which RPM's are well-suited) is a particularly interesting prospect \cite{H03}, 
along the following lines (see \cite{GMH, H99, H09} and Sec \ref{QM-Combo} for more).   
There is a Records Theory within Histories Theory \cite{GMH, H99}.  
Histories decohereing is one possible way of obtaining a semiclassical regime in the first place, i.e. 
finding an underlying reason for the crucial WKB assumption without which the Semiclassical Approach does not work.    
What the records are will answer the also-elusive question of which degrees of freedom decohere which others in Quantum Cosmology.

\mbox{ }  

\noindent Rovelli-type Tempus Nihil Est. Distinct timeless approaches involve {\it evolving constants of the motion} (`Heisenberg' rather than `Schr\"{o}dinger' style QM), 
or {\it partial observables} \cite{Rovellibook} (see Secs \ref{Cl-Pers}, \ref{Cl-POB}, \ref{QM-SubS}, \ref{QM-Beables}, \ref{Rov-vs-Other} for more on these). 
These are used in Nododynamics/LQG's {\it master constraint program} \cite{Thiemann}, and can also be studied in the RPM arena.   

\mbox{ }

\noindent Some approaches to the POT that do {\sl not} have an RPM analogue include the `Riem time' approach
(which requires indefinite configuration spaces) and ones involving finer details about the diffeomorphisms.

\subsubsection{That the POT has further facets: the `Ice Dragon'}\label{Facets}

Over the past decade, it has become more common to argue or imply that the POT {\sl is} the Frozen Formalism Problem.  
However, a more long-standing point of view \cite{Kuchar92, I93} that the POT contains a number of further facets; I argue in favour of this in this Article.  
It should first be stated that the problems encountered in trying to quantize gravity largely interfere with each other rather than standing as independent obstacles.
Kucha\v{r} \cite{Kuchar93} presented this as a `many gates' problem, in which one attempting to enter the 
gates in sequence finds that they are no longer inside some of the gates they had previously entered. 
(The object being described is presumably some kind of enchanted castle, or, at least, a topologically nontrivial one).  
Additionally, the various of these problems that are deemed to be facets of the POT do bear conceptual and technical relations that makes it likely to be advantageous to treat them as 
parts of a coherent package rather than disassembling them into a mere list of problems to be addressed piecemeal.
For, these facets arise from a joint cause, i.e. the mismatch of the notions of time in GR and Quantum Theory.

{            \begin{figure}[ht]
\centering
\includegraphics[width=0.9\textwidth]{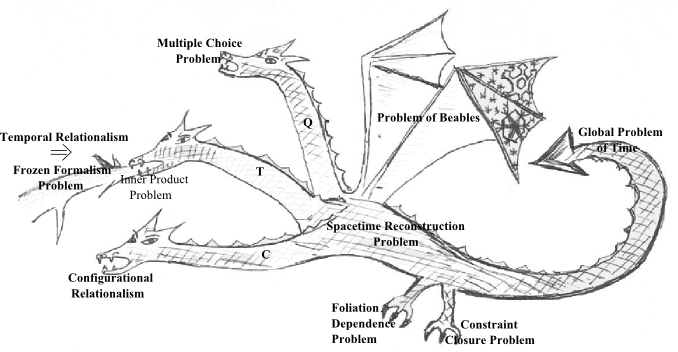}
\caption[Text der im Bilderverzeichnis auftaucht]{        \footnotesize{The Ice Dragon as a mythical mnemonic for the multi-facetedness of the POT: 
multiple problems underlied by one common cause, the conceptual incompatibility between what one calls `time' in each of GR and QM.
I hold this mnemonic to be {\sl particularly} useful given the growing proportion of theoretical physicists who do not know even this much about the time-related foundational 
incompatibilities between GR and QM, for all that these theories are the key cornerstones upon which all of the rest of the field is founded upon.} }
\label{IceDrac}\end{figure}            }
%
As such, I put forward Fig \ref{IceDrac}'s mythological mnemonic for the multi-facetedness of the POT.  
One notices that the physical equations get frozen.  
One reaction is to set about trying to unfreeze them.
However, another perspective is that an Ice Dragon is on the loose, a beast \cite{GRRM} that not only possesses the `freezing breath' of 
the Frozen Formalism (and ensuing `teeth' of the Inner Product Problem that one usually encounters next right behind the source of 
the frozenness) but which is the joint cause of other observed devastations, through coming simultaneously equipped with the following further facets.  

\mbox{ } 

\noindent I now clean up the presentation of the facets (further developed in Secs \ref{Cl-POT}, \ref{QM-POT}).   
1), 2), 3) are the heads-and-necks {\sc t} (Temporal Relationalism), {\sc c} (Configurational Relationalism) and {\sc q} (quantization) respectively

\noindent As well as the frozen breath of the Frozen Formalism Problem, the {\sc t}-head has teeth: the Inner Product Problem.

\noindent The {\sc c}-Head traditionally consists of the Sandwich Problem, that I argue generalizes to a Best Matching Problem and thus, procedural ordering independently 
to Configurational Relationalism itself.

\noindent The {\sc q}-head has the teeth of the Multiple Choice Problem. 

\noindent The limbs of the Ice Dragon are the 

\noindent 4) Foliation Dependence Problem, and 

\noindent 5) Constraint Closure Problem.  

\noindent 6) The `wings' of the Ice Dragon depict the Problem of Beables,

\noindent 7) the `thunderbolt tail', the Global POT,  and  

\noindent 8) the `scaled armour' of the Spacetime Reconstruction Problem.  
This is the last facet {\sl to encounter in attempts at practical resolutions}, but not the last in the build-up of the {\sl conceptual understanding} of the problem; 
this accounts for the conceptual-ordered numbering in the next SSSec being distinct from the present one's strategic-ordered numbering.  

\noindent Nor do the attributes of the Ice Dragon take an entirely fixed form.  
E.g. in path-integral rather than canonical approaches, one has to reckon with a Measure Problem instead of some of the difficulties of interpreting a frozen wave equation 
(see Sec \ref{QM-POT-Strat} for more).  

\mbox{ } 

\noindent The joint cause, or unity, of the Ice Dragon indeed stems from the common origin of these 
facets from the conceptual disparity between the GR and ordinary Quantum Theory notions of time.  
Now, setting about unfreezing the physics is likely more straightforward than defeating an Ice Dragon.
Some might even claim to be defeating the Ice Dragon by just unfreezing the physics, by taking its 
rather feared `POT' name and rebranding it to mean just its `Frozen Formalism Facet'.  
However, this amounts to leaving oneself open to the `claws', `tail', `scales' and `wings' by choosing not to take any advantage of the presence of a joint cause -- the mismatch of the 
notion of time in GR and Quantum Theory -- as a warning that these, too, would also be expected to be present as obstacles needing overcoming before one's 
QG program has genuine long-term viability. 
[Though it may take long for such deficiencies in QG programs to be demonstrated, through requiring detailed analysis of many difficult calculations.]   
Finally, there are the problems of other beasts: there are other multi-faceted incompatibilities between GR and ordinary 
Quantum Theory each anchored on its own common cause (e.g. the differences in role played by observers, see  Sec \ref{ObsApp}).

\mbox{  }

\noindent Thus there are seven further facets to examine, alongside various of the inter-relations between them.  
The Frozen Formalism Problem itself has the following addendum (`teeth').

\mbox{ }  

\noindent 
{\bf Hilbert Space} alias {\bf Inner Product Problem}, i.e. how to turn the space of solutions of the frozen equation in question into a Hilbert space.
See Sec \ref{TPQ} for why this is a problem (i.e. why ordinary QM inspired guesses for this will not do for GR).
It is a time problem due to the ties between inner products, conservation and unitary evolution.

\subsubsection{Outline of the other facets}\label{Other-Facets}

2) {\bf Configurational Relationalism} itself is the procedural ordering independent generalization of the {\bf Best Matching Problem} theory 
generalization of the usually-listed {\bf Thin Sandwich Problem} \cite{WheelerGRT, TSC1, TSC2, Kuchar92, I93}. 
The linear constraints often involved entwine with many other aspects of the POT, particularly in the cases in which these are 
momentum constraints that correspond to 3-diffeomorphisms or similar. 

\mbox{ }

\noindent{\bf Dirac observables} \cite{DiracObs} are quantities that commute\footnote{This means Poisson-brackets-commute at the classical level,
and commutator-commute at the quantum level.}
with all of a given theory's constraints.  
 
\noindent {\bf \K observables} \cite{Kuchar93, Kuchar99, BF08} are quantities that commute with all of a given theory's linear constraints.
These are also known as {\bf gauge-invariant quantities}, under the understanding that quadratic constraints are not part of this 
conception of gauge theory (and any second-class constraints have been previously removed).  

\noindent Rovelli's {\bf partial observables} \cite{Rov91a, Rov91b, Rov91c, 0111, Rovelli02, Rovellibook, Rfqxi} do not require commutation with any constraints.
These contain unphysical information but one is to consider correlations between pairs of them that are physical.  

\noindent{\bf Beables} are Bell's name for these \cite{Speakable, Bell}, which is more appropriate fro the whole-universe (quantum) 
cosmological context, through carrying no connotations of external observing, but rather simply of being.
These are e.g. more appropriate to talk about in a universe in which decoherence, rather than observing, occurs.

\mbox{ } 

\noindent 3) {\bf The Problem of Beables} (usually stated as `of observables') is then that it is hard to construct a set of these, particularly for gravitational theory.  
Clearly \K observables are more straightforward to construct than Dirac ones, whilst partial ones are trivial
(though opinions differ as to whether partial observables or even \K observables are sufficient to resolve the Problem, see Secs \ref{Cl-POT} and \ref{QM-POT}). 

\mbox{ } 

\noindent 4) {\bf Constraint Closure Problem}. 
This refers to closure of the brackets suitable to the level of one's formalism.
It concerns closure under evolution. 
At the quantum level for field theories it has previously been called the {\bf Functional Evolution Problem} \cite{Kuchar92, I93}.
However, `functional' here refers to the functional derivatives $\delta$ taken in the process, which is specific to a field theory. 
Thus to generalize this quantum notion to a finite--field theory portmanteau, one has to take into account the finite theory's corresponding derivatives are partial ones $\pa$, 
and so one is to term it {\bf Partional Evolution Problem} after the partional derivative portmanteau, $\partional$.  
This quantum-level problem is then, technically, a subset of the {\bf possibility of anomalies} \cite{Dirac, ModernAnomalies1, ModernAnomalies2}.  
Anomalies involve breakdown of the immediate closure of the algebraic structure formed by the constraints which occurs at the classical level due to the appearance of obstruction terms.  
Only a subset of these are involved, since not all anomalies bear any relation to time and frame issues pertinent to the POT.   
In particular, non-closure here is a way in which the Foliation Dependence Problem can manifest itself, through the non-closure becoming entwined with details of the foliation.

\mbox{ } 

\noindent 5) {\bf Foliation Dependence Problem}: for all that one would like a quantization of GR to retain classical GR's nice property of 
refoliation invariance (see Fig \ref{Teit} and Sec \ref{Trident}), at the quantum level there ceases to be an established way of guaranteeing this property.    
That this is obviously a time problem follows from each foliation by spacelike hypersurfaces being orthogonal to a GR timefunction: 
each slice corresponds to an instant of time for a cloud of observers distributed over the slice.
Each foliation corresponds to the cloud of observers moving in a particular way.   

\mbox{ } 

\noindent 6) {\bf Spacetime (Reconstruction or Replacement) Problem}.  
Internal space or time coordinates to be used in the conventional classical spacetime context need to be scalar field functions on the spacetime 4-manifold.    
In particular, these do not have any foliation dependence.
However, the canonical approach to GR uses functionals of the canonical variables, and which there is no a priori reason for such to be scalar fields of this type.  
Thus one is faced with either finding functionals with this property, or coming up with a new means of reducing to the standard spacetime meaning at the classical level.   
There are further issues involving properties of spacetime being problematical at the quantum level.  
Quantum Theory implies fluctuations are unavoidable, but now that this amounts to fluctuations of 
3-geometry, these are moreover too numerous to fit within a single spacetime (see e.g. \cite{Battelle}).  
Thus (something like) the superspace picture (considering the set of possible 3-geometries) might be expected to take over from the spacetime picture at the quantum level.  
It is then not clear what becomes of causality (or of locality, if one believes that the quantum replacement for spacetime is `foamy' \cite{Battelle}).  
There is also an issue of recovering continuity in suitable limits in approaches that treat space or spacetime as discrete at the most fundamental level (\cite{APOT2} lists 
some suitable references for this).  

\mbox{ }

\noindent 7) {\bf Multiple Choice Problem} \cite{Kuchar92}.  
This is the purely quantum-mechanical problem that different choices of time variable may give inequivalent quantum theories.  
\noindent Foliation Dependence is one of the ways in which the Multiple Choice Problem can manifest itself.
Moreover, the Multiple Choice Problem is known to occur even in some finite toy models, so that foliation issues are not the only source of the Multiple Choice Problem.  
For instance, another way the Multiple Choice Problem can manifest itself is as a subcase of how making different 
choices of sets of variables to quantize may give inequivalent quantum theories, as follows from e.g. the Groenewold--van Hove phenomenon (See Sec \ref{MCP}).

\mbox{ } 

\noindent 8) {\bf Global POT}.  
This is a problem whose origin is more basic and already present at the classical level, as an obstruction to making globally-valid constructs.
These include global gauge choices being obstructed in parallel with the well-known Gribov effect of Yang--Mills theory, except that in the present 
situation, gauge choices involve choices of times and frames, thus making it a part of the POT.
Its subfacets are, in one sense,
\noindent i) that timefunctions may not be globally defined in space.
\noindent ii) That timefunctions may not be globally defined in time itself.  
As we shall see, the classical version is readily understandable in terms of meshing conditions for coordinates but 
the quantum version is far less clear: can one mesh together unitary evolutions, and, if so, how?  

\mbox{ } 

\noindent Common combinations of these are as follows (in the previous SSSec's order).

\noindent 1, 2, 3, 4, 5, 6 concerns {\bf a local resolution} of the POT, which is a well-defined sub-problem both conceptually and technically.  
The `local' avoids the Global POT and the `a' avoids the Multiple Choice Problem.

\noindent 1, 2, 7: the order in which to challenge the three necks of Temporal Relationalism, Configurational Relationalism and the quantum sometimes with 3 
(the `wings' of the Problem of Beables) ordered in too, is the main basis of the current Article's classification of POT strategies 
(see Sec \ref{Cl-POT-Strat}), and of its main `{\it Magic Sword}' combination of Strategies (Sec \ref{Conclusion}). 
(This only very limitedly touches upon 7: the neck but not the Multiple Choice Problem `teeth').

\noindent 4, 5, 6: the limbs and scales are also significant combinations, corresponding to the innermost defenses of the 
Ice Dragon that are indeed usually faced last in attempting to defeat it.  
The `{\it Trident}' (Sec \ref{Trident}) combination of strategies for these in GR is largely beyond the scope of the present Article.  

\noindent The grouping 2, 4, 5, 7 that I'd previously stressed by depicting them as the 4 legs in my older pictures \cite{v1} 
of the Ice Dragon remain a useful grouping of four due to interconnections.
They are now depicted as 2 limbs and 2 heads (without dangerous breath), so it still makes sense for these four to coordinate as an intermediate-range of defenses.

\noindent The facets rendered harder by diff-specific modelling are 2, 3, 4, 5, 6 [and probably 7 and 8 also, but those remain open questions].

\subsection{This Article's principal POT strategies}

\noindent Our main thrust is Semiclassical, which recovers the approximate $t^{\se\sm(\sJ\sB\sB)}$ and then a slightly unaligned first approximation 
(quantum change is not classical change, so the times abstracted are to be different).  
We present this as a semiclassical quantum relational scheme.  
It has further significance as a qualitative model of Halliwell--Hawking Quantum Cosmology.

\noindent This Article's semiclassical approach is furthermore formulated as a Machian scheme.  

\noindent However, I also explain why the semiclassical approach is unsatisfactory even within its own ballpark.
We are to advance from that using a Histories--Records--Semiclassical Combination, so we first present quantum Records Theory and quantum Histories Theory for RPM's. 
As the histories--records--semiclassical combination currently in use is a regions-based implementation of propositions, the \NSI is also a relevant warm-up as well 
as a further simple interpretation of the material in Part III.  

\noindent We then give the histories--records--semiclassical scheme for the triangle, along with an account of what remains unsettled (Zeno, regions, extent of applicability 
of Halliwell, this still being a semiclassical scheme).  

\noindent The Conclusion lists the main results, alongside commentaries on: comparisons between strategies, combinations of strategies, the case for scale, 
what other foundational inconsistencies apart from time there are between GR and QM.

\subsection{Analogues of RPM's in various other aspects of Quantum Cosmology.}\label{Q-Cos} 

To motivate Quantum Cosmology itself, via e.g. inflation, it may then contribute to the understanding 
and prediction of cosmic microwave background fluctuations and the origin of galaxies \cite{Inf, Inf2, Inf3, HallHaw}.
Inflation is currently serving reasonably well \cite{WMAP, WMAP2} at providing an explanation for these.  
This remains an `observationally active area', with the Planck experiment \cite{Planck} about to be performed.
The underlying Quantum Cosmology, however, remains conceptually troubled  (see e.g. \cite{HT, H03}). 
It is via the above contact with testable assertions that hitherto philosophical contentions about QM 
(in particular as applied to closed systems such as the whole universe) may enter into mainstream Physics.     
Suggestions as to how one might approach such conceptual issues in Quantum Cosmology include \cite{JLBell, GMH, Hartle, +Hartle1, IGUS, H99, HD, 
HT, H03, B94II, EOT, HP86, HP88, UW89, PW83, Kuchar92, I93, Kuchar99, Page1, Page2, Kieferbook, +closed1, +closed2,ToposI, ToposTalk, ToposRev}.

Further foundational issues in Quantum Cosmology that may be addressable at least qualitatively 
using RPM toy models are as follows (expanded upon in Sec \ref{RPM-for-QC}).  

\noindent 5) Does structure formation in the universe have a quantum-mechanical origin?  
In GR, this requires midisuperspace or at least inhomogeneous perturbations about minisuperspace, and these are hard to study.   
(Scaled RPM's are a tightly analogous, simpler version of Halliwell and Hawking's \cite{HallHaw} model for this; moreover scalefree RPM's such
as this Article's occurs as a subproblem within scaled RPM's, corresponding to the light fast modes/inhomogeneities.)

\noindent 6) There are also a number of difficulties associated with closed-system Physics and observables.

\noindent 7) What are the meaning and form of, the wavefunction of the universe? 
(E.g. whether a uniform state is to play an important role; in classical and quantum Cosmology, these are held to be conceptually important notions).

\noindent 8) Quantum Cosmology has robustness issues, as regards whether ignoring certain degrees of freedom compromises the outcome of calculations \cite{KR89}.  

\noindent 9) Like other branches of Physics, Quantum Cosmology has Arrow of Time issues, and, moreover, 
may have something to say as regards the origin of various arrows of time \cite{HH83, EOT, H03, Rovellibook}.  
However, I feel that this largely lies outside the scope of the current Article.

\subsection{Some means of judging formulations}\label{Judge}  

I judge by the following criteria (I do not claim this list to furnish a complete judgement). 

\mbox{ } 

\noindent Criterion 1) In all aspects {\sl convincing at the conceptual level}.  
Moreover, one has to take care since this can shift as `one examines in more detail' \cite{Kuchar92, I93}. 

\noindent Criterion 2) It {\sl affords insights that the original formalism did not}. 

\noindent Criterion 3) It {\sl affords smoother interpolation between theories usually formulated more heterogeneously}. 

\noindent Criterion 4) It {\sl suggests new extensions/alternative theories} that are now natural but which were not apparent from the original formulation.

\noindent Criterion 5) (of theories, not formulations).  The same theory arises along multiple different conceptually-well-founded routes.  

\mbox{ }

\noindent Note 1) We shall see that GR is of this nature in Sec \ref{Many-Routes}, and question the extent to which Supergravity and M-Theory are of this nature in Sec \ref{RPM-Susy}.  

\noindent Note 2) There is a trade-off between Criterion 4) in theory selection and {\bf universality} 
-- a type of formulation that holds for any theory (at least within a broad class, e.g. based continuum geometry).

\noindent Note 3) Is Criterion 5) a manifestation of deepness or, (as Malcolm MacCallum suggested to me) of a lack of imagination?  
Point that these theories turn out to be able to handle things other than those they were {\sl designed} to handle.  

\noindent Note 4) Criteria 1) and 5) tend to clash.

\mbox{ }  

\noindent Also, due to the specific nature of the research in this Article, I judge in particular by [what are, strictly, 2 subcases of Criterion 1)]:

\noindent \bu that it can be formulated within the relational program exposited in the present Article [especially Mach's Time Principle which fulfils criteria 1), 2), 3)].  

\noindent \bu In those formulations that do involve a timefunction, that it fulfils the criteria by which candidate timefunctions are approved as actual timefunctions 
as per Sec \ref{tfns}.

\vspace{10in}

\noindent{\huge\bf PART I: CLASSICAL THEORY OF RPM's}\normalsize

\section{Examples of relational theories} \label{Examples}

\subsection{Setting up Relational Particle Mechanics (RPM's)}\label{Setup-RPM}

The first task of the relational approach is to provide an unreduced configuration space $\FrQ$ and a group $\FrG$ of transformations that are held to be physically irrelevant.
If one is then to proceed by the indirect means of considering arbitrary $\FrG$-frame corrected objects, 
then by the nature of this correcting procedure, $\FrG$ should consist entirely of continuous transformations.  
One can only sometimes proceed directly instead (see Sec \ref{Dyn1} for examples).
I denote configuration space dimension by $k$.

\subsubsection{First choices of a $\FrQ$ for RPM's} \label{Q-for-RPM}

1) In the indirect approach, one's incipient notion of space (NOS) is {\it Absolute space} $\FA(d)$ of dimension $d$.   
For usual studies of particle mechanics, this is held to be $\mathbb{R}^{d}$ (equipped with the standard inner product 
$(\mbox{ } , \mbox{ })$ with corresponding norm $|| \mbox{ } ||$ built from the $d$-dimensional identity matrix).  
Then an incipient {\it configuration space} $\FrQ(N, d)$ for particle mechanics is ``$N$ labelled possibly superposed 
material points in $\mathbb{R}^{d}$", i.e. $\mathbb{R}^{N d}$, with coordinates $q^{I\mu}$.        
The inner product $(\mbox{ }, \mbox{ })_{\mbox{\scriptsize\boldmath$m$}}$ with corresponding norm $||\mbox{ }||_{\mbox{\scriptsize\boldmath$m$}}$ is defined on this.  

\noindent 2) Because the indirect implementation of Configurational Relationalism below encodes continuous $\FrG$ below is held to be continuous, I furthermore need to consider here, 
at the preliminary level of choosing $\FrQ$ itself, whether physical irrelevance of the non-continuous reflection operation is to be an option or, indeed, obligatory.
I.e. should $\FrQ$ be a space of {\it plain} configurations $\FrQ(N, d)$ or of {\it mirror-image-identified} configurations $\FrO\FrQ(N, d) = \FrQ(N, d)/\mathbb{Z}_2$ 
[the O stands for `orientation-identified', meaning that the clockwise and anticlockwise versions are identified, O being a more pronounceable prefix than M].  
Mirror-image-identified configurations are always at least a mathematical option at this stage by treating all shapes 
and their mirror images as one and the same: $\mathbb{R}^{N d}/\mathbb{Z}_2$.  
However, one shall see below that in some cases the model has symmetry enough that declaring overall rotations themselves to be irrelevant already includes such an identification. 
In this case one can not meaningfully opt out of including the reflection, so this fork degenerates to a single prong.  

\noindent 3) In this Article, I just consider the case of plain distinguishable particles bar brief mention in this and the next Sec, which serves as a base for a wider range of theories.  
Distinguishable particles covers the case of generic masses or, via some further unstated mysterious classical labels, the case of equal masses. 
For indistinguishable particles, consider quotients by the likewise-discrete permutation group on $N$ objects, 
$S_{N}$ or the even permutation group $A_N$ so as to continue to exclude the extra reflection generator present in $S_N$.  
(Alternatively, one could use the $P < N$ versions of these if only a subset are distinguishable, or a product of such whose 
$P$'s sum up to $\leq N$ if the particles bunch up into internally indistinguishable but mutually distinguishable).
This would be more Leibnizian (by identifying even more indiscernibles), as well as, more specifically, concordant with the
nonexistence of meaningless distinguishing labels at the quantum level. 
(Though it is fine for particles to be distinguishable by differing in a {\sl physical} property such as mass or 
charge or spin, which then actively enters the physical equations, and spin only arises at the quantum level.) 
%
%

\noindent 4) Some applications require further excision from the configuration space of of degenerate configurations (e.g. collinearities or some collisions).  

\mbox{ }

\noindent dim($\FrQ(N, d)$) = $N d$.  
Note throughout this Sec that quotienting out discrete transformations such as reflections or permutations does not affect the 
configuration space dimension (these amount to taking a same-dimensional portion).

\subsubsection{Choice of $\FrG$ for RPM's} \label{G-for-RPM}

Various possibilities for the continuous group of physically-irrelevant transformations $\FrG$ are as follows.  
$\mbox{Tr}(d)$ are the $d$-dimensional {\it translations}. 
Tr($d$) is the noncompact commutative group, $(\mathbb{R}^{d}; +)$.  
dim(Tr($d$)) = $d$.  
$\mbox{Rot}(d)$ are the $d$-dimensional {\it rotations}.
dim(Rot($d$)) = $d$\{$d$ -- 1\}/2; this being 0 in 1-$d$ corresponds to the obvious triviality of continuous rotations in 1-$d$.  
Rot($d$) is the special orthogonal group $SO(d)$ of $d \times d$ matrices.  
$\mbox{Dil}(d)$ are the $d$-dimensional {\it dilations}, alias {\it homotheties}.  
Dil($d$) is the noncompact commutative group $(\mathbb{R}^+; \cdot)$ independently of $d$, so I denote this by Dil from now on.  
dim(Dil) = 1.

The {\it Euclidean group} of $d$-dimensional continuous isometries is\foo{Here, \textcircled{S}  denotes  
semidirect product, see e.g. \cite{I84}.  
Strictly speaking, Eucl($d$), Sim($d$) and Nonrot($d$) are the `proper' versions of these groups via not being taken to include the discontinuous reflections.}
$\mbox{Eucl}(d) =$ Tr($d$) $\mbox{\textcircled{S}}$ Rot($d$).   
dim(Eucl($d$)) = $d\{d + 1\}/2$.    
The {\it proper linear group} of $d \times d$ matrices is Pl($d$) = Rot($d$) $\times$ Dil.
dim(Sl($d$)) = $d\{d - 1\}/2 + 1$.
The $d$-dimensional `non-rotation isometry' group Nonrot($d$) = Tr($d$) $\times$ Dil.   
dim(Nonrot($d$)) = $d$ + 1.    
Finally, the {\it similarity group} of $d$-dimensional continuous isometries and homotheties is $\mbox{Sim}(d)$ = Tr($d$) $\mbox{\textcircled{S}}$ Rot($d$) $\times$ Dil. 
dim(Sim($d$)) = $d\{d + 1\}/2 + 1$.

Then if one chooses $\FrQ$ as in the preceding section alongside $\FrG$ = Eucl($d$), one has scaled RPM (Sec 2.2), or, alongside 

\noindent$\FrG$ = Sim($d$), one has pure-shape RPM (Sec \ref{SRPM}), or, alongside $\FrG$ = Nonrot($d$), one has non-rotational RPM (Sec \ref{NonRot}).

\subsubsection{Further choice of a $\FrQ$ for RPM's}\label{+Q-for-RPM}

One can also consider the centre of mass motion of the RPM universe to be ab initio meaningless.  

\noindent Then one's incipient configuration space is {\it relative space} $\Fr(N, d) = \mathbb{R}^{n d}$ for $n := N - 1$  or {\it Orelative space} $\FO\Fr(N, d) = 
\mathbb{R}^{n d}/\mathbb{Z}_2$
This is equivalent to using $\FrQ(N, d)$ and quotienting out the translations to render absolute position meaningless, 
but this is devoid of mathematical structure or any extra analogy of GR, so using $\Fr(N, d)$ makes for a clearer presentation. 
[Both are presented below because it is not immediately straightforward to see that mechanics on $\Fr(N, d)$ can be cast in extremely close parallel to mechanics on $\FrQ(N, d)$.  
For, making this manifest requires passing to relative Jacobi coordinates and noting how all the relevant mechanical formulae then look the same (Sec \ref{Rel-Jac}), which point was 
missed in the RPM literature prior to my involvement and yet turned out to be the crucial first key in unlocking the detailed understanding of these models.] 
In this second scheme, then, one is to use Rot($d$), Pl($d$) and Dil to form scaled, pure-shape and non-rotational RPM's respectively.

\subsubsection{A deeper relational consideration: which $\FrQ(N, d)$ are actually discernible}\label{Discerning}

It is next necessary to point out a probably non-obvious flaw in the way this Sec has hitherto been presenting its mathematical structure. 
Namely, that $N$ and $d$ have been introduced as if they were on an equal footing as independent free parameters.  
That does not however comply with relational thinking.
For, particle number $N$ is, undeniably, a material property. 
However, given $N$ particles as the entirety of the contents of one's relational model universe, there only exist $n = N - 1$ 
independent relative separation vectors, and therefore only the following worlds are discernible.  
\beq
\mbox{dimension $d$ = 1 to $n$ with or without mirror image identification} \mbox{ } .  
\label{Wyman}
\eeq
However, if one tries to use dimension $d > n$, one finds that this is {\sl indiscernible} form the $d = n$ mirror image 
identified world, and thus {\sl identical} to it as per Leibniz.  
%
%
Thus in RPM's, dimension $d$ is {\sl not} a free parameter; {\sl the different values dimension $d$ can meaningfully 
take are contingent on the particle number, $N$.}

Furthermore, this perspective is a very clear way of both anticipating the following next SSec's Note 1)'s subtlety and the ways 
in which Appendix \ref{Q-Geom}.E's 3 particles in 3-$d$ model is distinct only being at the cost of it being less relational.  

A further issue is that some values of the relationalspace (i.e. entirely physical configuration space) dimension $k$ are 

\mbox{ } 

\noindent i) utterly trivial (0 degrees of freedom).  

\mbox{ } 

\noindent ii) {\bf Relationally trivial} (1 degree of freedom).  
This is trivial because the relational program concerns expressing one material change in terms of another material 
change rather than in terms of an meaningless arbitrarily-reparametrizable label-time, so $k >1$ is required.

\noindent It is however an interesting issue whether relational triviality continues to hold at the quantum level (see Sec \ref{Leela}). 
%

\mbox{ }

\noindent By these criteria, further of the (\ref{Wyman}) are knocked out as per the next SSSec.

\subsubsection{Outline of the subsequent quotient spaces $\FrQ/\FrG$} \label{Quot}

If absolute orientation (in the rotational sense) is also to have no meaning, then one is left on a configuration space 

\noindent {\bf relational space} $\bigr(N, d) = \FrQ(N, d)/\mbox{Eucl($d$)}$ or $\FrR(N, d)/\mbox{Rot($d$)}$
Or, in the case of O-shapes, on {\it Orelational space} $\FrO\bigr(N, d)$ 

\noindent $= \FrO\FrQ(N, d)/\mbox{Eucl}(d)$ or $\FO\Fr(N, d)/\mbox{Rot($d$)}$.  
These have dimension $k = nd$ -- $d\{d - 1\}/2$ = d\{2$n$ + 1 -- $d$\}/2, i.e. $N - 1$ in 2-$d$, $2N - 3$ in 2-$d$ and $3N - 6$ in 3-$d$.  

If, instead, absolute scale is also to have no meaning, then one is left on a configuration space termed {\it preshape space} by Kendall 
\cite{Kendall}, $\FP(N, d) = \FrQ(N, d)/\mbox{Nonrot}(d)$ or $\Fr(N, d)/\mbox{Dil}$. 
Or, one is left on {\it Opreshape space} 

\noindent$\FrO\FP(N, d) = \FrO\FrQ(N, d)/\mbox{Nonrot}(d)$ or $\FO\FrR(N, d)/$Dil.  
These have dimension $k = nd$ -- 1.

If both absolute orientation and absolute scale are to have no meaning, then one is left on what Kendall \cite{Kendall} termed 

\noindent{\bf shape space}, $\FrS(N, d) = \FrQ(N, d)/\mbox{Sim}(d)$ or 
Or, one is left on  {\it Oshape space} $\FrO\FrS(N, d) = \FrO\FrQ(N, d)/\mbox{Sim}(d)$.  
These have dimension $k = d$\{2$n$ + 1 -- $d$\}/2 -- 1, i.e. $N - 2$ in 2-$d$, $2N - 4$ in 2-$d$ and $3N - 7$ in 3-$d$.
For later use in shortening equations, I denote these particular k's by 
\beq
k(N, d) := \mbox{dim}(\FrS(N, d)) = \mbox{dim}(\mathbb{R}^{Nd}/\mbox{Sim}(d)) = Nd - \{d\{d + 1\}/2 + 1\} = nd - 1 - d\{d - 1\}/2 \mbox{ } .
\label{qNd}
\eeq
N.B. that shape spaces have no place for a maximal collision.
One can regard both preshape space and relational space as intermediaries in reaching shape space.
We will really consider the above spaces as augmented to be normed spaces, metric spaces, topological spaces, and, where possible, Riemannian geometries (see Sec \ref{Q-Geom}).    
Finally note that  
\beq
\FP(N, 1) = \FrS(N, 1)
\eeq
and likewise for all O and indistinguishable-particle cases. 
This is simply because there are no rotations in 1-$d$.  

\mbox{ }  

\noindent Note 1) the above is all within the confines of $d \leq n$ that suffices to span the discernible RPM worlds.
Were one to use $d > n$, it could then only be Rot($n$) and not Rot($d$) which acts on the physical configurations.
Consequently Eucl($n$) and not Eucl($d$), and Sim($n$) and not Sim($d$) would be used in these cases.
Without this observation, the formulae for the degrees of freedom indeed manifestly fail! 
This action by less than Rot($d$) itself has a counterpart within the absolutist setting, most notably as regards the well-known number of degrees of freedom count in linear molecules
\cite{AtF}.  

\noindent Note 2) Additionally picking off the utterly trivial and relationally trivial cases, the full set of discernible and nontrivial RPM's are as in Fig \ref{Allow}.  
The small but nontrivial $k$'s among these are the most likely candidates for tractable models.  
As we shall see, most applications require $d \geq 2$, some require $d \geq 3$ and some require $k \geq 4$.  

{            \begin{figure}[ht]
\centering
\includegraphics[width=1.0\textwidth]{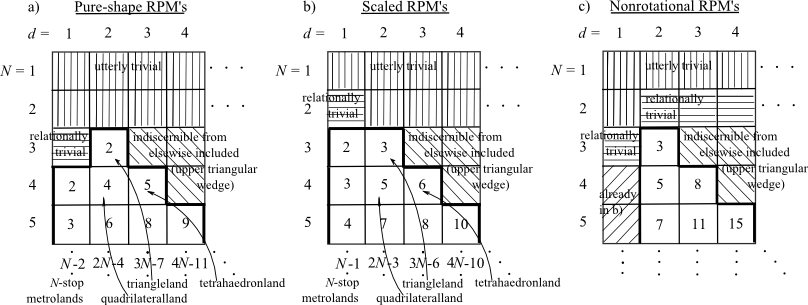}
\caption[Text der im Bilderverzeichnis auftaucht]{        \footnotesize{ The full set of discernible and nontrivial a) pure-shape 
RPM's/shape spaces, b) scaled RPM's/relational spaces and c) non-rotational RPM's/preshape spaces.  
The numbers indicated are the configuration spaces' dimension, $k$.  
} }
\label{Allow}\end{figure}            }

\noindent I term 1-$d$ RPM's {\boldmath$N$}{\bf -stop metroland}s since they look like an underground train line. 
I term 2-$d$ RPM's {\boldmath$N$}{\bf -a-gonland}s since each point in them is a planar $N$-sided polygon. 
The mathematically highly special $N$ = 3 is {\bf triangleland}, and the first mathematically-generic $N = 4$ case is {\bf quadrilateralland}.   
Finally, I term the 3-$d$ such {\it N-cornerland}s as each point in them is a generally nonplanar solid with $N$ corners. 
The mathematically highly special $N$ = 4 case I refer to as {\it tetrahaedronland}.
3-cornerland is even more mathematically special, but it it also indiscernible from Otriangleland.  

\mbox{ }

\noindent N.B. for more than 3 particles, the configurations are really {\bf constellations}: sets of points, which, 
as the material point-particles, are the relationally-primary content of the theory, rather than what shapes can be made by joining up the dots. 
For, beyond triangles, the latter becomes ambiguous; while certain coordinate systems and particle labels may imply 
certain orders of `joining up the dots', these are practitioner-dependent/particle label dependent rather than intrinsic.    
This furthermore means that, beyond triangles, some features of laminar/solid shapes are less relevant than others through depending on more than the actual physical entity which is 
the constellation; this is then reflected in terms of what questions I choose to ask of such shapes.
E.g. whether a shape has edges that cross over is a labelling or laminar/solid shape construction dependent property 
and thus less interesting than configurations with coincident particles.  
This is an early indication that one will need to think carefully about how to separate out properties of/questions about shapes according to different levels of structure.
This paragraph clearly explains why I use the $N$-cornerland name aside from in the special case of tetrahaedronland, 
for which there is both a common word and the coincidence that the number of haedra (faces) equals the number of particles.

\subsubsection{Poincar\'{e}'s Principle}\label{Poincare}

This concerns how Physics ought to have complete data sets, usually stated as {\it ``a point and a direction in configuration space"}.  
Whilst Newtonian theory is of this form in its absolute setting, in the shape space setting it requires five extra pieces of data 
(no matter how large the particle number is), and this is considered by Barbour to be a sign of a defective theory \cite{B11}.
On the other hand, RPM's on relationalspace and shape space do not exhibit this `defect'; this has some ties to how 
zero angular momentum Newtonian Mechanics is a considerably structurally simpler theory as outlined in Sec \ref{Dyn1}.
I also note that Wheeler also ends his discussion in \cite{Battelle} with a very similar statement for the case of Geometrodynamics:  
{\it ``Give a point in superspace and give a fully developed direction at this point in superspace.  
Then (hypothesis!) this information is sufficient, together with Einstein's equations, uniquely to determine the entire 4-geometry".}
I note for more general use that the key feature of this `defect' concerns a small fixed number of extra data 
rather than the totality of the data having to come in point--direction pairs, which are manifestly second-order-bosonic. 
See Appendix \ref{Examples}.C for the extension of this to fermions.

\subsection{Scaled RPM's} \label{ERPM}

In scaled RPM \cite{BB82, B86, B94I, EOT}, only relative times, relative angles and relative separations are meaningful.    
E.g. for $N = 3$ in $d > 1$, scaled RPM is a dynamics of the triangle formed by the 3 particles.    
Scaled RPM was originally \cite{BB82} conceived for $\FrQ = \FrQ(N, d) = \mathbb{R}^{N d}$ and $\FrG$ = Eucl($d$).
One can use $\FrO\FrQ(N, d) =  \mathbb{R}^{Nd}/\mathbb{Z}_2$ instead in order to accommodate reflections.  
In each case `best-matching' involves a different extremization, but is carried out by the same method [Fig \ref{Std-Ori-Shuffle}].  

{            \begin{figure}[ht]
\centering
\includegraphics[width=0.6\textwidth]{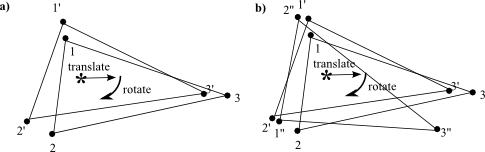}
\caption[Text der im Bilderverzeichnis auftaucht]{        \footnotesize{a) Barbour demonstrates the concept of best matching in this way 
(with wooden triangles in many of his seminars).  
For the plain shapes, one keeps one configuration (illustrated here by a triangle, unprimed) fixed, 
while shuffling the other (primed) around using translations and rotations so as to seek out how to minimize the incongruence between the two.   
The star is the centre of mass of the primed triangle.  

b) In the case of the mirror-image-identified shapes, one proceeds likewise, except that one shuffles the second triangle {\sl and} its mirror image (double-primed) around.} }
\label{Std-Ori-Shuffle}\end{figure}            }

\subsubsection{Scaled RPM in terms of particle positions}\label{06I-BB82} \label{q-ERPM}

Let $\FrQ$ be the na\"{\i}ve $Nd$-dimensional configuration space in $d$ dimensions for $N$ particles, 
The classical particle mechanics notion of Configurational Relationalism is implemented by passing to a suitable notion of arbitrary frame.
This is achieved by the introduction of a translational auxiliary d-vector $\underline{A}$ and whichever rotational auxiliary corresponds to $d$. 
There is none for $d$ = 1, a scalar ${B}$ for $d$ = 2 or a 3-vector $\underline{B}$ for $d$ = 3. 
All these cases are encoded by
\be
\q^{I} - \underline{A} -  \underline{B} \cr \q^{I} \mbox{ } .  \label{abcorr}
\ee
if one allows for $\underline{B}$ = (0, 0, $B$) in 2-$d$ and $\underline{B} = 0$ in 1-$d$.

My procedure for constructing actions is that they are to be built as best as possible out of objects that transform well under $\lambda$-dependent $\FrG$-transformations.    
I.e. $\d/\d\lambda$ acts (and MRI ensures this is $\lambda$-independent so equivalently $\d$ acts in the MPI form).  
For scaled RPM,  the restriction 
\be
V = V(\{\q^{I} -\q^{J}\} \cdot \{\q^{K} -\q^{L}\}  \mbox{ alone}) \mbox{ } , 
\label{Vrel}
\ee
guarantees that the auxiliary corrections straightforwardly cancel each other out within.    
In examples encountered in practise, these are usually of the form $V(||\q^{I} - \q^{J}|| \mbox{ alone})$.
The situation with the kinetic arc element is more complicated in this sense, due to $\d$ not being a tensorial operation under $\d$-dependent Eucl($d$) transformations.
(This looks more familiar if $\lambda$'s  are inserted: ${\d}/{\d\lambda}$ should rather be seen as the Lie derivative 
$\pounds_{{\d}/{\d\lambda}}$ in a particular frame, as per e.g. \cite{Stewart}.)
This leads to
\be
\d s^{\sE\sR\sP\sM}_{\sJ\sB\sB} = 
||\d_{\underline{A}, \underline{B}}\mbox{\boldmath$q$}||_{\mbox{\scriptsize\boldmath$m$}} \mbox{ } \mbox{ for } \mbox{ } 
\d_{\underline{A},\underline{B}}\uq^{I} := \d{\uq}^{I} - \d{\underline{A}} - \d{\underline{B}} \cr \uq^{I} \mbox{ } .    
\label{T}
\ee
Finally, the proposed action (a variant of the JBB action) is  
\be
\FS_{\sJ\sB\sB}^{\sE\sR\sP\sM} = \sqrt{2}\int \sqrt{W} \d s^{\sE\sR\sP\sM}_{\sJ\sB\sB} 
\label{Jac}
\ee
with (\ref{Vrel}) and (\ref{T}) substituted into it.

Then the momenta conjugate to the $q_{\mu A}$ are 
\be
p_{\mu I} = m_{I\mu J\nu}\Last_{\underline{A}, \underline{B}}q^{\nu I} \mbox{ } ,    
\mbox{ }  
\mbox{ for } \mbox{ } \Last_{\underline{A}, \underline{B}} := {\d}/{\d t^{\se\sm(\sJ\sB\sB)}_{\underline{A}, \underline{B}}} := 
\sqrt{2W}\d/||\d_{\underline{A}, \underline{B}}\mbox{\boldmath$q$}||_{\mbox{\scriptsize\boldmath$m$}} \mbox{ } .    
\label{Glim}
\eeq
Then by virtue of the MPI and particular square-root form of the Lagrangian, the momenta obey a primary constraint,  
\be
\scE := ||\mbox{\boldmath$p$}||_{\mbox{\scriptsize\boldmath$n$}}^2/2 + V = E \mbox{ } . 
\label{calE}
\ee
N.B. that this is quadratic and not linear in the momenta and that it is physically interpreted as an `energy constraint'.   
%

Then variation with respect to $\underline{A}$ gives $\sum_{I = 1}^{N}\p_I = \underline{C}$, constant, which is 0 at the free end-point (FEP) and thus $\underline{C} = 0$ everywhere.  
Thus, one obtains
\be
\underline{\scP} := \sum\mbox{}_{\mbox{}_{\mbox{\scriptsize I = 1}}}^{N}\upp_{I} = 0 \mbox{ } \mbox{ } \mbox{ (zero total momentum constraint) } .  
\label{ZM}
\ee
Next, variation with respect to $\underline{B}$ gives $\sum_{I = 1}^{N}\q^I \cr \p_I = \underline{D}$, constant, which is 0 at the FEP and thus $\underline{D} = 0$ everywhere.  
Thus 
\be
\underline{\scL} := \sum\mbox{}_{\mbox{}_{\mbox{\scriptsize I = 1}}}^{N} \underline{q}^{I} \cr 
\underline{p}_{I} = 0 \mbox{ } \mbox{ } \mbox{ (zero total angular momentum constraint) }
\label{ZAM} \mbox{ }.
\ee
[Or whichever portion of this that is relevant in the corresponding dimension, i.e.  in 1-$d$ there is no $\underline\scL$ constraint at all, and 
in 2-$d$ $\underline\scL$ has just one component that is nontrivially zero: $\scL = \sum_{I = 1}^{N} \{q^{I1}p_{I2} - q^{I2}p_{I1}\} = 0$.]
 
\mbox{ } 

\noindent Note 1) These constraints are linear in the momenta: examples of $\scL\scI\scN_{\sfZ}$.  
 
\noindent Note 2) $\underline{\scL}$ is interpretable as the centre of mass motion for the dynamics of the whole universe being irrelevant rather than physical.  
All the tangible physics is in the remaining relative vectors between particles.     

\noindent Note 3) $\scE$ is interpretable as the absolute angles being 
irrelevant rather than physical, so that it is in relative angles and relative separations that the physics resides.  

\mbox{ } 

\noindent The evolution equations are
\beq
\Last_{\underline{A}, \underline{B}} p_{I\mu} = - \pa V/\pa q^{I\mu} \mbox{ } .
\eeq
The constraints then obey the Poisson bracket algebra \cite{Gergely} whose nonzero brackets are\foo{1 
and 2-$d$ versions are included in the usual sense. } 
\be
\{{\scP}_{\mu}, {\scL}_{\nu} \} = {\epsilon_{\mu\nu}}^{\gamma}{\scP}_{\gamma} 
\mbox{ } , \mbox{ }
\{{\scL}_{\mu}, {\scL}_{\nu}\} = {\epsilon_{\mu\nu}}^{\gamma}{\scL}_{\gamma} 
\mbox{ } . \mbox{ }
\label{first6}
\ee
As this is closed, there are no further constraints.  
These Poisson brackets are just the usual statement that momentum is a vector under rotations and the usual algebra of rotations. 

\mbox{ } 

Inverting (\ref{Glim}) and applying the Configurational Relationalism implementing extremization,
\be
\mbox{\Large$t$}^{\se\sm(\sJ\sB\sB)} = \stackrel{\mbox{\scriptsize extremum $A, B$ of Eucl($d$)}}
                                                            {\mbox{\scriptsize of $\stS_{\mbox{\tiny JBB}}^{\mbox{\tiny ERPM}}$}} 
\left(
\int||\d_{\underline{{A}}, \underline{{B}}} \mbox{\boldmath$q$}||_{\mbox{\scriptsize\boldmath$m$}}/\sqrt{2W}
\right)  
\mbox{ } .
\label{JBB}
\ee
\noindent Note 4) As well as incorporating the freedom of choice of time-origin/`{calendar year zero}', one 
One is also free to choose time-unit/`{\bf tick-duration}' as per Appendix \ref{Examples}.B.3.  

\noindent Note 5) Moreover, for the emergent time to be uniquely defined rather than a $\FrG$-dependent proto-time, 
the procedure has to imply some means of freeing itself from dependence on $\d{\underline{A}}$ and $\d{\underline{B}}$.  
The most obvious such is  -- illustrated for the subcase of extremization with respect to $\underline{B}$ of the above time-functional 
-- followed by solving the subsequent equation for $\d\underline{B}$,
\beq
{W}^{-1/2}             {\pa ||\d_{\underline{A},\underline{B}}\mbox{\boldmath$q$}||_{\mbox{\scriptsize\boldmath$m$}}}/
                       {\pa   \d{\underline{B}}} = 0 \mbox{ } ,
\label{Siicut}
\eeq
and substituting this back into the integral.  
There is a problem however in general with this choice of extremization, as explained in Fig \ref{TEET} and backed up by the below GR example.

The alternative is using the extremization of the action; for RPM's this produces
\beq
\sqrt{W}{\pa ||\d_{\underline{A},\underline{B}}\mbox{\boldmath$q$}||_{\mbox{\scriptsize\boldmath$m$}}}/{\pa\d\underline{B}} = 0 \mbox{ } ,
\label{Diixit}
\eeq
and this is equivalent to (\ref{Siicut}) [at least away from zeros of $W$, for which the relational action has problems anyway, c.f. Sec 
\ref{POZ}].  
This subtlety becomes very necessary in the GR case however, because here the difference between $1/\sqrt{\mW^{\sG\sR}}$ and 
$\sqrt{\mW^{\sG\sR}}$ becomes entangled within the derivative operators that arise `by parts' in the spatial integration, so that the two extremizations produce different equations.
%
{            \begin{figure}[ht]
\centering
\includegraphics[width=0.3\textwidth]{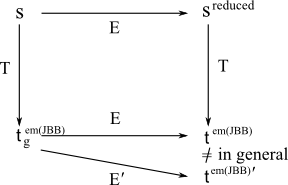}
\caption[Text der im Bilderverzeichnis auftaucht]{        \footnotesize{ I denote the map from an action to the corresponding 
emergent (proto)time by T.
I denote the map consisting of substituting in the g-extremum of the action by $\mE$, and the map consisting of substituting in the g-extremum of the action by $\mE^{\prime}$
Then E and T naturally commute: TE = ET, but in general TE$^{\prime} \neq$ ET. 
This is why I use the g-extremum of S in order to free t of g-dependence.} }
\label{TEET}\end{figure}            }

\noindent See Appendix \ref{Examples}.A as regards RPM's in Hamiltonian form  and more relationally meaningful generalizations thereof.

\subsubsection{Relative Lagrangian coordinates} \label{Lag}

C.f. Fig \ref{Fig-A} for the $N = 3$ in 2-$d$ example of this.
N.B. this notion immediately extends to arbitrary $N$ and $d$.  

{            \begin{figure}[ht]
\centering
\includegraphics[width=0.6\textwidth]{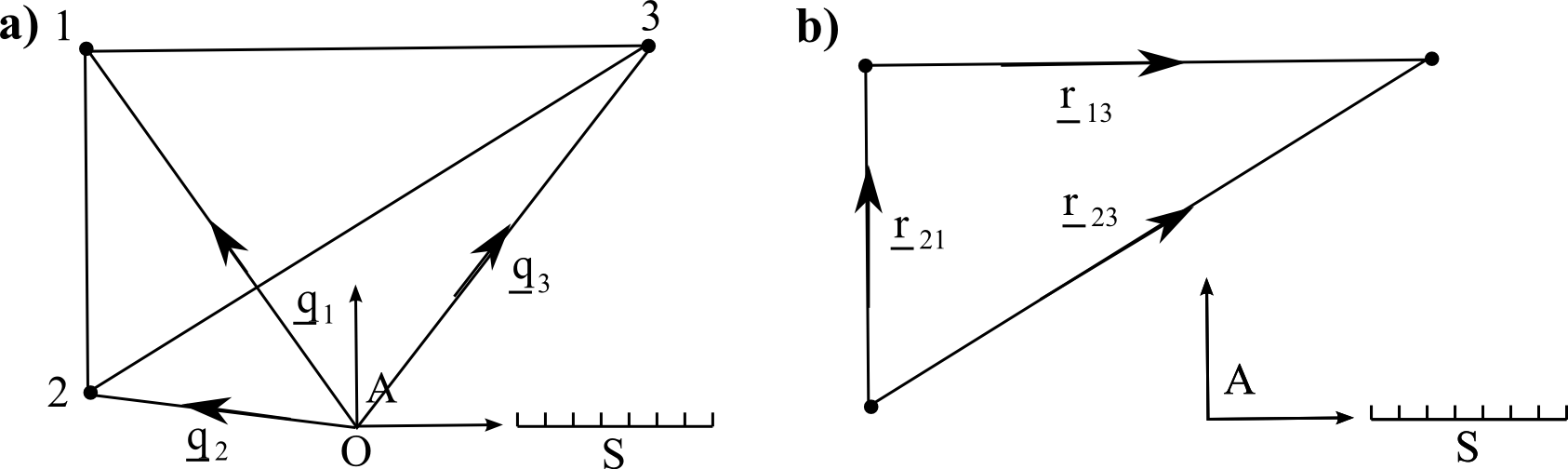}
\caption[Text der im Bilderverzeichnis auftaucht]{        \footnotesize{Coordinate systems for 3 particles. 

\noindent a) and b) Absolute particle position coordinates ($\q_1$, $\q_2$, $\q_3$) in 1- and 2-$d$.  
These are defined with respect to, where they exist, fixed axes A, a fixed origin O and a fixed scale S.  

\noindent c) and d) Relative particle position coordinates $\br = \{\r^{IJ}, I > J\}$.
Their relation to the $\q^I$ are obvious: $\r^{IJ} = \q^J - \q^{I}$.  
For 3 particles, any 2 of these form a basis. 
No fixed origin enters their definition, but they are in no way freed from fixed coordinate axes A or scale S.} }
\label{Fig-A}\end{figure}            }

\subsubsection{Scaled RPM in relative Lagrangian coordinates}\label{Lag-Action}  

One renders absolute position irrelevant, by passing to any sort of relative coordinates, leaving one on relative space = $\FrR(n, d) = \mathbb{R}^{n d}$.  
The most obvious such are the relative Lagrange coordinates.  
I use that the $d$ = 3 working contains everything under the provisos (in Sec \ref{Rel-Jac}), and then comment on each individual case $d$ = 1, 2, 3 as these differ significantly.    
I begin with (\ref{Jac}) and eliminate $\d{\uA}$ from the Lagrangian form of $\underline\scP = 0$  
\be
\d{\uA} = {M}^{-1}\sum\mbox{}_{\mbox{}_{\mbox{\scriptsize I = 1}}}^N m_I\{{\d{\q}}^{I} - \d\B \cr {\q}^{I}\}
\ee 
[the Lagrangian counterpart of the Hamiltonian expression (\ref{ZAM})]. 
This results in the (semi-)eliminated `Jacobi--Lagrange--Gergely' \cite{Gergely} action\foo{Here,
the total mass $M := \sum_{I = 1}^Nm_I$.}
\be
\FS^{\sE\sR\sP\sM}_{\sJ-\sL\sG} = \sqrt{2} \int \sqrt{E - V(\urr^{IJ}\cdot\urr^{KL} \mbox{ alone})}
\d s^{\sE\sR\sP\sM}_{\sJ-\sL\sG}  
\mbox{ } \mbox{ for } \mbox{ }
\d s^{\sE\sR\sP\sM\, 2}_{\sJ-\sL\sG} = 
\sum\sum\mbox{}_{\mbox{}_{\mbox{\scriptsize I $<$ J}}} \frac{m_Im_J}{M}||\d_{\underline{B}}{\urr}^{IJ}||^2 
\mbox{ } . \label{semi}
\ee
\noindent Note 1) What happens to Barbour's best matching demonstration in relative coordinates? 
Given two triangles, instead of leaving one's vertices fixed and shuffling the other's around by rigid 
translations and rotations, one rather 1) lines up the two centres of mass once and for all.  
2) One rigidly rotates the relative separation vectors of the second triangle so as to try to minimize the incongruence between the two [Fig \ref{Rel-Shuffle}]. 

{            \begin{figure}[ht]
\centering
\includegraphics[width=0.25\textwidth]{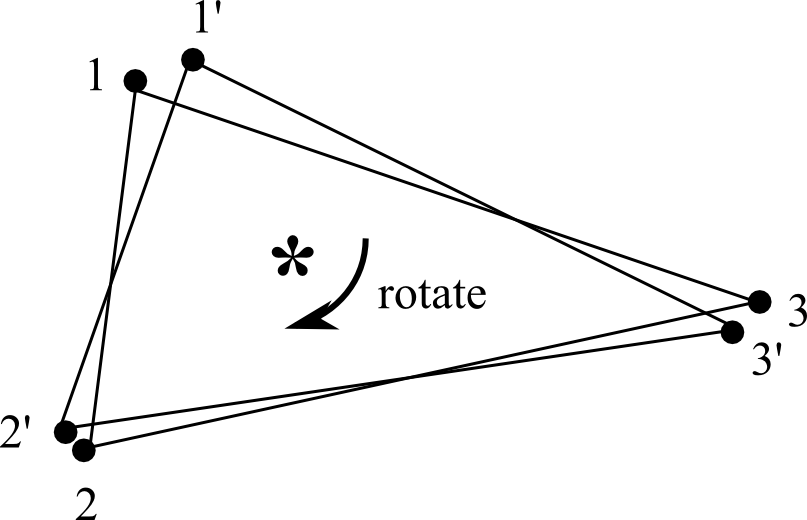}
\caption[Text der im Bilderverzeichnis auftaucht]{        \footnotesize{The star is now the coincidence of the centre of mass of each system. 
Shuffling now just involves rotations.} }
\label{Rel-Shuffle}\end{figure}            }

\noindent Note 2) As written, these expressions depend on all of the $\urr^{IJ}$'s rather than on an $n$-member basis. 
I first redress this by noting that 
\be
\r^{IJ}, I = 1 \mbox{ to } n - 1, J = I + 1 \mbox { form a basis for the relative separation coordinates, }
\label{basis}
\ee
and recasting the previous section's symmetric but redundant `Lagrangian' expressions asymmetrically 
yet nonredundantly in terms of these alone [e.g using $\urr_{13} = \urr_{12} + \urr_{13}$].   

\noindent Note 3) However, once written in terms of a basis subset of the relative Lagrange variables, 
the kinetic term /arc element is non-diagonal (more widely, the kinetic metric within is not).  
This hinders progress (e.g. it is more straightforward to spot classical and quantum separability for some simple potentials with diagonal coordinates).    
I will get round this by changing to relative Jacobi coordinates in Sec \ref{Rel-Jac}.

\noindent Note 4) On the other hand, using relative Lagrange coordinates enables contrast of scaled RPM 
with the BB77 theory for which these are the original and most natural variables.   
This is a distinct, non-Newtonian, experimentally refuted and often rediscovered\foo{It 
was obtained by Reissner (and subsequently rediscovered by Schr\"{o}dinger and by Barbour--Bertotti \cite{BB77, Buckets}). 
This theory was however not widely known, and is incompatible with somewhat more modern mass-anisotropy experiments (e.g. \cite{HughesGrav, Drever}).}   
theory formulated directly in terms of the $r^{IJ}$ and without auxiliaries referring to absolute space.  
This theory has a kinetic arc element that is exceptionally simple from the Lagrangian perspective,  
\be
\d s^2 = \frac{\delta_{\mu\nu}\delta_{IK}\delta_{JL}}{|r^{PQ}|}\d{r}^{\mu IJ}\d{r}^{\nu IK} 
\mbox{ } ;
\label{BB77T}
\ee  
this may explain why this theory has been rediscovered quite a few times.  
The question then arises whether scaled and pure-shape RPM's can {\sl also} be cast as a presumably more complicated direct formulation of this kind. 
If this is the case, not even indirect or unphysical reference to absolute space is required.  
This is contained in Sec \ref{DRIII} in the 1- and 2-$d$ cases.  
See also Appendix \ref{Q-Geom}.B for further discussion of the BB77 theory.

\subsubsection{Relative Jacobi coordinates}\label{Rel-Jac}

Relative Jacobi coordinates are sets of $n$ inter-particle (cluster) separations chosen such that the kinetic term/arc element  
is diagonal; they are very well-known in basic Celestial Mechanics \cite{Marchal} and Molecular Physics \cite{LR97}.\foo{The more common  
Jacobi coordinate vectors themselves consist of the relative Jacobi coordinate vectors plus one extra centre of mass absolute Jacobi coordinate vector.
I use lower-case Latin indices for a basis of relative separation labels 1 to $n$.} 
%
Denote the mass matrix alias kinetic metric in the new basis by  $\mu_{ij\mu\nu} = \mu_i\delta_{ij}\delta_{\mu\nu}$, where $\mu_i$ are the Jacobi masses (see below).  
This has determinant $\mu$ and inverse $\nu^{i\mu j\nu}$.
I also use $\uPP_i$ for the momentum conjugate to $\uR^i$, and $\mI$ for the moment of inertia, $\sum_{i = 1}^n\mu_i|\uR^i|^2$.  
$\rho = \sqrt{I}$ is termed the {\it hyperradius} in the Molecular Physics literature; its use apparently also dates back to Jacobi \cite{ACG86}. 
In the present Article, I usually choose to call this by the more descriptive name {\bf configuration space radius}.   
It was used in QM at least as far back as the 50's by Fock \cite{Fock} and by Morse and Feschbach \cite{MFII}; see also e.g. \cite{Smith62, Mont96, Hsiang1}.
I is used e.g. in \cite{Dragt, Iwai87, LR97} though these are for $\theta$ running over half of the range most often used in the present 
article (but there is an analogous use of I in the case involving the whole range too).

A further useful tidying, particularly convenient at the post-variational level of solving classical and quantum equations, are the  
mass-scaled relative Jacobi coordinates  $\underline{\rho}^i := \sqrt{\mu_i}\uR^i$.  
I denote the conjugates of these by $p_i$.   
The norms of these are then the partial moments of inertia $\mI^i = \mu_i|{\uR^i}|^2$ 
Finally `normalized' mass-scaled relative Jacobi coordinates are obtained by dividing the preceding by $\sqrt{\mI} = \rho$: $\underline{\mn}^i = 
\underline{\rho^i}/\rho$, and `normalized' partial moments of inertia $\mN^i = \mI^i/\mI$.  
Then $\sum_i \mn^{i\,2} = \sum_i \mN^i =  1$.  
All of these variables are as primary and as natural as the $\underline{R}^i$.   
I use $\underline{\mp}_i$ for the conjugates of the $\underline{\mn}^i$.

The first specific nontrivial case for RPM's has $N$ = 3 and so $n$ = 2 relative Jacobi coordinates.  
We take one interparticle separation vector as the first relative Jacobi vector.  
There are 3 ways of choosing the two particles involved, and an additional $2^2 = 4$ ways to orient the vectors. 
The second Jacobi vector is from the centre of mass of the first two particles to the third particle; there is no residual freedom at this stage.  
The total number of ways is thus 3$!$ = 6 = number of permutations on 3 objects, while the 3 choices without 
ascribing an orientation amount to 3 possible different clusterings for 3 particles (i.e. 2 + 1 partitions into clusters).  

\mbox{ } 

\noindent Note 1) The two Jacobi vectors thus chosen amount to `placing a {\bf Jacobi tree}' \cite{ACG86} on the constellation of points.  
The name `tree' is meant in the graph-theoretic sense of `connected, with no cycles'; more precisely, these are irreducible trees.  

\noindent Note 2) The above counting arguments and the form of the tree are both dependent on $N$ but not on $d$.  
Moreover, for $N$ particles, the tree has $n$ branches corresponding to the $n$ relative Jacobi vectors.  

\noindent Note 3) For this simplest case, there is a unique shape of Jacobi tree, which for $d \geq 2$ is a T-shape [Fig \ref{Fig1}d)] 
into which I include the 1-$d$ shape [Fig \ref{Fig1}a)] by terming it a `squashed T'  (an inclusion and nomenclature I extend to all other types of tree).   

\noindent Note 4) Each of the T's is then labelled according to its clustering structure: I use $\{\ma, \mb\mc\}$ read left-to-right in 1-$d$ and 
anticlockwise in $\geq$ 2 $d$, which I abbreviate by (\ma).\footnote{My general cluster and clustering notation if as follows.  
I use \{a...c\} to denote a cluster composed of particles a, ... c, ordered left to right in 1-$d$ and anticlockwise in 2-$d$; I take these to be 
distinct from their right to left and clockwise counterparts i.e. I consider plain configurations.  
I insert commas and brackets to indicate a clustering, i.e. a partition into clusters.
These notations also cover collisions, in which constituent clusters collapse to a point.}

\noindent Note 5) It is the constellation and not the tree, cluster-label or shape made by joining the dots that is relational; 
moreover, we shall see that the choice of tree and of cluster label can have particular {\sl perspectival} meaning as regards the regime of study or the propositions being addressed.
For now I give as a simple example that if the three particles are the Sun, Earth and Moon, then there is particular significance to having the apex be the Sun 
(as the Earth and Moon base pair are far more localized) or the Moon (as it is considerably lighter and thus amenable to treatment as a perturbation).  

\mbox{ } 

\noindent For $N$ = 4, not only are the relative Jacobi coordinates nonunique as regards permutations of the particle labels, but also that 
there are now two different shapes of Jacobi tree: Jacobi H-coordinates [2 + 2 split: Fig \ref{Fig1}e)] and Jacobi K-coordinates [(2 + 1) + 1 split: Fig \ref{Fig1}f)]
For the H-coordinates, 1) pick two particles and take the first relative Jacobi vector to be the particle separation vector of these.
2) Take the second relative Jacobi vector to be the particle separation vector of the other two.  
3) The third relative Jacobi vector is then the separation vector of the centres of masses of these two clusters. 
This H-construction can be chosen in 3 ways (number of choices of the 2 + 2 split), 
and then there are 8 ways of ascribing the orientations of the 3 vectors (one overall orientation and 2 internal ones). 
Its labelling is \{ab,cd\}, for which I use the shorthand (Hb)
For the K-coordinates, step 1) is the same, but then take the second relative Jacobi vector to be between the centre of mass of the first two particles and the third particle.  
Then the third relative Jacobi vector to be between the centre of mass of this triple cluster and the final particle.  
This K-construction can be chosen in 4 ways (number of choices of the 3 + 1 split) and then there are 3 
ways to decide how to coordinatize the triple cluster, and then 8 possible ways to orient the 3 vectors.  
Its labelling is \{a, bcd\} for which I use the shorthand (Ka), or, more precisely picking the clustering within 
the triple subcluster, \{a, \{b, cd\}, for which I use the shorthand (Kab).  

{           \begin{figure}[ht]
\centering
\includegraphics[width=0.77\textwidth]{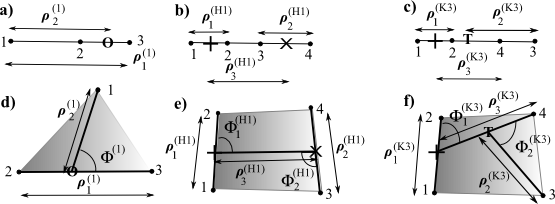}
\caption[Text der im Bilderverzeichnis auftaucht]{  \footnotesize{O, +, $\times$ and T denote COM(23), 
COM(12), COM(34) and COM(124) respectively, where COM(ab) is the centre of mass of particles a and b.
a) For 3 particles in 1-$d$, one particular choice of mass-weighted relative Jacobi coordinates are as indicated. 
For 4 particles in 1-$d$, b) and c) give, respectively, a particular choice of mass-weighted relative Jacobi H-coordinates and of Jacobi K-coordinates.
d) For 3 particles in 2-$d$, one particular choice of mass-weighted relative Jacobi coordinates are as indicated. 
I furthermore define $\Phi^{(\sa)}$ as the `Swiss army knife' angle 
$\mbox{arccos}\big( \brho_1^{(\sa)} \cdot \brho_3^{(\sa)} / \rho_1^{(\sa)} \rho_3^{(\sa)} \big)$.
For 4 particles in 2-$d$ e) and f) give, respectively, a particular choice of mass-weighted relative Jacobi 
H-coordinates [with $\Phi_1^{(\sH\sb)}$ and $\Phi_2^{(\sH\sb)}$ as the `Swiss army knife' angles 
$\mbox{arccos}\big(\brho_1^{(\sH\sb)}\cdot\brho_3^{(\sH\sb)}/\rho_1^{(\sH\sb)}\rho_3^{(\sH\sb)}\big)$ 
and 
$\mbox{arccos}\big(\brho_2^{(\sH\sb)}\cdot\brho_3^{(\sH\sb)}/\rho_2^{(\sH\sb)}\rho_3^{(\sH\sb)}\big)$ respectively] and of 
K-coordinates [with $\Phi_1^{(\sK\sa)}$ and $\Phi_2^{(\sK\sa)}$ as the `Swiss army knife' angles  
$\mbox{arccos}\big(\brho_1^{(\sK\sa)}\cdot\brho_3^{(\sK\sa)}/\rho_1^{(\sK\sa)}\rho_3^{(\sK\sa)} \big)$ 
and  
arccos$\big( \brho_2^{(\sK\sa)}\cdot\brho_3^{(\sK\sa)}/\rho_2^{(\sK\sa)}\rho_3^{(\sK\sa)} \big)$   respectively].  
These are all relative angles: unlike the $\rho's$, they {\sl do not} make reference to axes A or scale S.}        } 
\label{Fig1} \end{figure}         } 
%
\noindent Note 6) For 4 particles in 2-$d$, `joining the dots' is no longer unique and which way they are joined is not particularly meaningful.  
Thus quadrilateralland is in truth rather less about quadrilaterals than triangleland is about triangles.  
That is relevant as regards understanding which physical propositions make particular sense for quadrilateralland. 

\noindent Note 7) H-coordinates are particularly suited to the study of two pairs of binary stars/diatomics/H atoms,  
whilst K-coordinates are suited to a binary/diatomic/H atom alongside two single bodies.  

\mbox{ } 

\noindent The specific forms of the vectors and associated masses are needed for various further applications in this paper, 
as well as serving as useful examples to ensure the student readers understand the form of the Jacobi construction.
For 3 particles, in the (1) cluster [and suppressing the cluster suffixes],  
\beq
\R_1 = \q_3 - \q_2 \mbox{ } \mbox{ and } \mbox{ } \R_2 = \q_1 - \frac{m_2\q_2 + m_3\q_3}{m_2 + m_3} 
\mbox{ } .
\eeq
The inversion of this (i.e. expressing the $r$'s in terms of the $R$'s) is 
\beq
\urr_{12} =  - \uR_2 - \frac{m_3}{m_2 + m_3}\uR_1 \mbox{ } , \mbox{ } \urr_{13} =  - \uR_2 + \frac{m_2}{m_2 + m_3}\uR_1 
\mbox{ } \mbox{ and } \mbox{ } \urr_{23} = \uR_1 \mbox{ } .  
\eeq
The corresponding Jacobi interparticle (cluster) reduced masses $\mu_i$ that feature in the diagonal kinetic term are then  
\beq
\mu_1 = \frac{m_2m_3}{m_2 + m_3} \mbox{ } \mbox{ and } 
\mbox{ } \mu_2 = \frac{m_1\{m_2 + m_3\}}{m_1 + m_2 + m_3} \mbox{ } . 
\eeq
For the case of equal masses that I commonly use in this Article, the above equations reduce to 
\beq
\R_1 = \q_3 - \q_2 \mbox{ } \mbox{ and } \mbox{ } \R_2 = \q_1 - \{\q_2 + \q_3\}/2 \mbox{ } .
\eeq
The corresponding Jacobi interparticle (cluster) reduced masses $\mu_i$ that feature in the diagonal kinetic term are then 
\beq
\mu_1 = 1/2 \mbox{ } \mbox{ and } \mbox{ } \mu_2 = 2/3 \mbox{ } . 
\eeq

\noindent For the (H2) coordinates, 
\beq
\R_1 = \q_2 - \q_1 \mbox{ } , \mbox{ } \R_2 = \q_4 - \q_3 \mbox{ } , \mbox{ } 
\R_3 = \frac{m_3\q_3 + m_4\q_4}{m_3 + m_4} - \frac{m_1\q_1 + m_2\q_2}{m_1 + m_2} \mbox{ } , 
\eeq
with corresponding cluster masses 
\beq
\mu_1 = \frac{m_1 m_2}{m_1 + m_2} \mbox{ } , \mbox{ } \mu_2 = \frac{m_3m_4}{m_3 + m_4} 
\mbox{ and } \mu_3 = \frac{\{m_1 + m_2\}\{m_3 + m_4\}}{m_1 + m_2 + m_3 + m_4} \mbox{ } .  
\eeq
For the case of equal masses that I commonly use in this Article, the above equations reduce to 
\beq
\R_1 = \q_2 - \q_1 \mbox{ } , \mbox{ } \R_2 = \q_4 - \q_3 \mbox{ } , \mbox{ } 
\R_3 = \{\q_3 + \q_4\}/2 - \{\q_1 + \q_2\}/2
\eeq
with corresponding cluster masses
\beq
\mu_1 = 1/2 \mbox{ } , \mbox{ } \mu_2 = 1/2 \mbox{ and } \mu_3 = 1 \mbox{ } .  
\eeq
Finally, indeed as announced, $T^{\sE\sR\sP\sM}$ or $\d s^{\sE\sR\sP\sM}$ are indeed diagonal in these coordinates.
Kinetic arc element $\d s = {||\d\mbox{\boldmath$q$}||_{\mbox{\scriptsize\boldmath$m$}}}$ maps to $||\d\mbox{\boldmath$R$}||_{\mbox{\scriptsize\boldmath$\mu$}} = ||\d{\brho}||$.
Thus it looks look very similar to those in particle position coordinates, just involving $n$ instead of $N$ objects.
(This amounts to rendering centre of mass motion irrelevant in passing to a relative coordinate description and then the  
$\underline{\rho}^i$ are a set of Cartesian coordinates for the relative space.) 
Thus there is a neat {\it `Jacobi map'} $q^{I\mu} \longrightarrow \mR^{i\mu}$  which will clearly be useful in transcribing the action, 
Hamiltonian and constraints of RPM's from particle position expressions to very similar relative Jacobi coordinate expressions.  
This and the structural uninterestingness of eliminating the translations makes Jacobi coordinates a tidier `starting point'. 
As further substance for this analogy, the total moment of inertia: $\mI^{\st\so\st} =  
{||\mbox{\boldmath$q$}||_{\sbm}}^2$ maps to the total barycentric moment of inertia  $\mI = ||\mbox{\boldmath$R$}||_{\mbox{\scriptsize\boldmath$\mu$}}\mbox{}^2 = ||\brho||^2$ 
The `Euler quantity' $\scD = ( \mbox{\boldmath$q$} , \mbox{\boldmath$p$})$ maps likewise to its obvious Jacobi coordinate counterpart $(\mbox{\boldmath$R$}, \mbox{\boldmath$P$}) = (\brho, {\mbox{\boldmath $p$}})$ 
and likewise the angular momentum: $\underline\scL = \sumiN \q^I \cr \p_I$ to $\sumin \R^i \cr \P_i = \sumin\underline{\rho}^i \cr \underline{p}_i$.     
Thus the relative Jacobi coordinates not only diagonalize the kinetic arc element but also have many further properties which are particularly well-suited to the study of RPM's.\foo{I 
also use $\mI^i, \ttD_i, \underline{L}_i$ for the partial counterparts of $\mI$, $\tcD$ and $\underline{\tcL}$.} 

\subsubsection{Scaled RPM action in relative Jacobi coordinates} \label{ERPM-Rel-Jac}

As relative Jacobi coordinates are inter-particle (cluster) separations, I take $\FrQ = \FrR(N, d)$ = $\mathbb{R}^{\sn d}$, and $\FrG$ = Rot($d$).  
Then\footnote{Mass-weighting relative Jacobi coordinates does not change the $B$ and $C$ auxiliaries thanks to 
homogeneous linearity in $\R^i$, $\dot{\R}^j$.}  
\be
\FS^{\sE\sR\sP\sM}_{\sJ-\sJ\sA} = 
\sqrt{2}\int\sqrt{ E - V(\underline{\rho}^j\cdot\underline{\rho}^k \mbox{ alone)}}\d s^{\sE\sR\sP\sM}_{\sJ-\sJ\sA}      
\mbox{ } ,
\label{wasT}
\ee
\be
\mbox{ for } \mbox{ }  
\d s^{\sE\sR\sP\sM}_{\sJ-\sJ\sA} = ||\d_{{\underline{B}}}\brho|| \mbox{ } , \mbox{ } \mbox{ } 
\d_{\underline{B}}\underline{\rho}^{i} := \d\underline{\rho}^i - \d{\underline{B}} \cr \underline{\rho}^{i} \mbox{ } .  
\ee
Then the conjugate momenta are  
 \be
p_{\mu i} = \delta_{\mu\nu}\delta_{ij}\Last_{\underline{B}}\rho^{\nu i} \mbox{ } ,
\mbox{ } 
\mbox{ where }
\mbox{ } 
\Last_{\underline{B}} := {\d }/{\d t^{\se\sm(\sJ\sB\sB)}_{\underline{B}}} := \sqrt{2W}\d/||\d_{\underline{B}}\brho|| \mbox{  }  .  
\eeq 
The surviving constraints are, as a primary constraint, 
\be
\scE := ||\mbox{\boldmath $p$}||^2/2 + V(\brho) = E \mbox{ } ,
\label{HamT}
\ee 
and, as a secondary constraint 
\be
\underline\scL =  \sumin \underline{\rho}^{i} \cr \underline{p}_{i} =  0 \mbox{ }  
\label{DamT}
\ee 
from FEP variation with respect to $\underline{B}$.  
The evolution equations are
\beq
\Last_{\underline{B}} p_{i\mu} = - \pa V/\pa\rho^{i\mu} \mbox{ } .
\eeq  
By applying the Jacobi map, the Poisson brackets are as before other than no longer involving any $\scP_{\mu}$.  

The emergent JBB time is now given by 
\beq
\lt^{\se\sm(\sJ\sB\sB)} = 
\stackrel{\mbox{\scriptsize extremum $\underline{B} \mbox{ } \in \mbox{ } \mbox{Rot(d)}$}}
         {\mbox{\scriptsize of $\stS^{\tE\tR\tP\tM}_{\tJ-\tJ\tA}$}}                                                              
\left(                                                              
\int     ||\d_{\underline{B}}\mbox{\boldmath$\rho$}||/\sqrt{2W(\mbox{\boldmath$\rho$})}\} 
\right)
\mbox{ } . 
\label{Anima}
\eeq

\noindent Note that Barbour's best matching demonstration in terms of relative Jacobi coordinates is presentationally important due 
to touting Jacobi coordinates as an alternative starting point.
This is clearly just a recoordinatization of Fig \ref{Rel-Shuffle}.

\subsection{Pure-shape RPM's}\label{SRPM}

In pure-shape RPM \cite{B03, 06II, TriCl}, only relative times, relative angles and ratios of relative separations are meaningful.    
I.e., it is a dynamics of shape excluding size: a {\sl dynamics of pure shape}.  
E.g. 3 particles in dimension $d > 1$, pure-shape RPM is the dynamics of the shape of the triangle that the 3 particles form.      
$\FrQ = \FrQ(N, d) = \mathbb{R}^{Nd}$ as before. 
$\FrG$ is now  Sim($d$).

\subsubsection{Pure-shape RPM in particle position coordinates: Barbour's form}\label{SRPM-B-Form} 

[This does not lie within the present Article's general format, but can be reformulated as such, c.f. the next SSSec.]
Take a MPI action 
\beq
\FS^{\sS\sR\sP\sM}_{\sJ-\sB} = \sqrt{2}\int\sqrt{E - V}\d s^{\sS\sR\sP\sM}_{\sJ-\sB}      
\label{Gine}
\eeq
\beq
\mbox{ with } \mbox{ } \d s^{\sS\sR\sP\sM}_{\sJ-\sB} = ||\d_{{\underline{A}}, {\underline{B}}, {c}}\mbox{\boldmath$q$}||_{\mbox{\scriptsize\boldmath$m$}}/2
\mbox{ for }  \mbox{ }
\d_{{\underline{A}}, {\underline{B}}, {c}}\q^A = 
c\{\q^{A} - \underline{\dot{\mA}} - \underline{\dot{B}} \cr \q^{A} + \{\dot{c}/c\}\q^{A}\} \mbox{ } .   
\eeq
[I also note that $\mI$ in $c\q^I$-coordinates is $c^2\mI$ in $\q^I$-coordinates.
Moreover, it is also in Barbour's style to not yet make any impositions on the form that the potential is to take.]

The conjugate momenta are then
\beq
\mp_{I\mu} = m_I c \delta_{\mu\nu}\delta_{IJ}\Ast_{{\underline{A}}, {\underline{B}}, {c}} \q^{\nu J} 
\mbox{ }
\mbox{ where }
\mbox{ } 
\Last_{\underline{A}, \underline{B}, c} := {\d }/{\d t^{\se\sm(\sJ\sB\sB)}_{\underline{A}, \underline{B}, c}} 
                                        := \sqrt{2W}\d/||\d_{\underline{A}, \underline{B}, c}\mbox{\boldmath$q$}||_{\mbox{\scriptsize\boldmath$m$}}  \mbox{  }  
\eeq 
Then, as a primary constraint,   
\beq
\scE := ||\mbox{\boldmath$p$}||_{\mbox{\scriptsize\boldmath$n$}}\mbox{}^2/2 + V(\mbox{\boldmath$q$}) = E \mbox{ } .  
\eeq
The secondary constraints (\ref{ZM}, \ref{ZAM}) also hold.
Barbour varies with respect to $c$ an a somewhat unusual manner, which he viewed as a toy model of some 
of the more complicated situation of some of the conformogeometrodynamical approaches in Sec \ref{Rel-CGdyn}. 
However, I pointed out \cite{TriCl} that this model can be recast in a simpler form for which the analogy is absent (see the next SSSec).  
In any case, the outcome is the constraint  
\beq
\scD = \sum\mbox{}_{\mbox{}_{\mbox{\scriptsize $i = 1$}}}^{N}\q^i\cdot\p_i = 0 \mbox{ } .
\label{ZDM}
\eeq
Note 1) This is also linear in the momenta and interpreted as zero total dilational momentum constraint.  

\noindent Note 2) Barbour then subsequently finds that for the constraints to propagate, $W = E - V$ must be homogeneous of degree --2.
This can be seen to follow from the Poisson bracket 
\beq
\{{\scD}, {\scE}\} = 2{\scE} + \big\{ 2 W + \sum\mbox{}_{\mbox{}_{\mbox{\scriptsize$A$ = 1}}}^{N}\q^{A}\cdot{\pa W}/{\pa \q^{ A}} \big\}
\mbox{ }  
\ee
and subsequent lack of propagation unless the right hand side is killed off strongly.  

\noindent Note 3) The preceding result invalidates actual energy $E$ from occurring in this theory 
[$E =0$ in (\ref{Gine}) due to elsewise $E$ not being homogeneous of degree --2].   
By this, alongside the conformogeometrodynamical counterpart of this \cite{BO}, 
Barbour hoped to use to do away with the analogous (and often problematic) notion of cosmological constant.  

\noindent Note 4) This attitude of Barbour's toward the potential is less restrictive in its assumptions but also renders the treatment less systematic. 
To date the less restrictive aspect has not given us anything more, while the more systematic scheme is clearer and somewhat faster to do calculations with.  
The more systematic approach is also based on a more overarching principle that treats kinetic and potential inputs on the same footing as per Sec \ref{Intro}. 
This ought to be deeper and clearer than just having a rule for correcting velocities/differentials.    

\noindent Note 5) In pure-shape RPM, the moment of inertia $\mI$ is a conserved quantity. 
[This follows from (\ref{ZDM})].  

\noindent Note 6)  The association of potentials that are homogeneous of degree -- 2 and of conserved $\mI$ with scale invariance are well-known 
elsewhere in the literature (see e.g.  \cite{Saari2, Mont96, ChMon, Conf, Conf2, Conf3, Conf4, Conf5}).
What was new to Barbour's program was the combination of these two elements so as to strike a close analogy with Newtonian Physics by 
using powers of the post-variationally constant $\mI$ in the potential as some of the powers which have to add up to --2. 
See Sec \ref{Cl-Soln} for examples of this.  
By this means the homogeneity condition on the potential is argued to be not overly restrictive.  
This is through it permitting e.g. some level of capacity to mimic of standard potentials such as linear combinations of distinct-power-law potentials, over extensive regimes.

\subsubsection{Pure-shape RPM in particle position coordinates: my form}\label{SRPM-my-form} 

Some additional differences from Barbour's original formulation in my subsequent good Sim($d$) object procedure 
alias arbitrary Sim($d$)-frame case of arbitrary $\FrG$-frame method are as follows.

\noindent 1) I employ a different parageodesic principle splitting conformal transformation (PPSCT) representation (explained in Appendix \ref{Examples}.B): 
the one in which the scale invariance is most natural, through each of $\ttT$, $\ttW = \ttE - \ttV$ being homogeneous of degree 0 rather than homogeneous of 
degree +2 and --2 respectively. 
This also later turns out to be more natural as regards the geometry of the reduced configuration space.
Both of the above criteria are clear geometrical naturalities (the former as regards the conformal-geometric meaning of scale invariance itself, 
while the latter is in terms of obtaining recognizable and natural Riemannian geometry on the reduced configuration space. 
Barbour's representation, on the other hand, is the mechanically-natural one. 
In particular, in it the physical dimensions of the various quantities are the usual ones, while in my  
geometrically natural representation the kinetic term has units of energy/I while $\ttE$ and $\ttV$ have units of energy $\times \mbox{ } \mI$.
[However, moving between these and my for of pure-shape RPM's quantities is particularly straightforward as the conversion factor, $\mI$, is but a constant in pure-shape RPM].
It is then clear that while $E$ is banished, a new constant $\ttE$ (`{\it pseudo-energy}) of units energy/I occupies its place, playing a very similar role whereby one does not 
succeed in banishing cosmological constant type notions by scale-relational thinking, rather just replacing them or transmuting them.  

\noindent 2) The arbitrary Sim($d$)-frame approach takes  $\ttW = \ttE - \ttV$ to be homogeneous of degree 0 as 
part of the starting point rather than a deduction, for it is such that is a good Sim($d$) object.  

\noindent 3) I also use a new auxiliary $C$, which is related to \cite{B03}'s $c$ by $C$ = ln$\,c$. 
This simplifies the working, which then follows through without any unusual procedures in the variation whatsoever.  
As such, it largely does away with Barbour's suggestion that pure-shape RPM is a good model of some of the 
complications found in conformal gravity and the \{CS + Vol\}($\bupSigma$) formulation of GR. 
On the other hand, Barbour's formulation has useful comparison with the scale--shape split of scaled RPM (c.f. Sec \ref{SS-Split}) 
and has mechanical/dimensionally natural quantities coincident with scaled RPM's.

Thus my formulation is as follows.  
The $A$th particle's position $\underline{q}^{A}$ is replaced by an arbitrary-frame position
 \beq
\underline{q}^{A} - \underline{A} - \underline{B} \cr \underline{q}^{A} + C\underline{q}^{A} \mbox{ } .  
\eeq
Then the good Sim($d$) object for the potential factor is
$
\fW = \fW\big(\mbox{ratios of} \{\underline{q}^{A} - \underline{q}^{B}\}\cdot\{\underline{q}^K - \underline{q}^L\} \mbox{ alone }\big) \mbox{ } .
$
This is free of auxiliary variables by cancellation though the overall homogeneity of degree 0.   
The nontrivial part of the good Sim($d$) objects construction is, once again, the building of the kinetic term.  
This is again more straightforward to present in the MRI picture along the lines in \cite{Stewart}: ${\pa q^{A\mu}}/{\pa\lambda}$ 
is not in itself a tensorial operation under $\lambda$-dependent Sim($d$) transformations.  
It should rather be seen as the Lie derivative $\pounds_{{\pa \sq^{A\mu}}/{\pa\lambda}}$ in a particular frame. 
This transforms to the Lie derivative with respect to `$\d\q^{A\mu}/\d\lambda$ corrected 
additively by generators of the translations, rotations and scalings', which gives the combination 
\beq
\d_{\underline{A}, \underline{B}, C}\q^{A} = \d\q^{A} -\d{\underline{A}} - \d{\underline{B}} \cr \q^{A} + \d{C}\q^{A} \mbox{ } .
\eeq
Then the kinetic arc element is  
\be
\d\tts^{\sS\sR\sP\sM\, 2}_{\sJ-\sA} = ||\d_{\underline{A}, \underline{B}, C} \mbox{\boldmath$q$} ||_{\mbox{\scriptsize\boldmath$m$}}\mbox{}^2/\mI  
\mbox{ }  .  
\label{18b}
\ee
The Jacobi-type action for pure-shape RPM is then  
\be
\FS^{\sS\sR\sP\sM}_{\sJ-\sA} = \sqrt{2}\int \sqrt{\ttE - \ttV\big(\mbox{ ratios of } 
\{\underline{q}^I - \underline{q}^J\}\cdot\{\underline{q}^K - \underline{q}^L\} \mbox{ alone }\big)}\d \tts^{\sS\sR\sP\sM}_{\sJ-\sA} 
\label{SRPMJac} 
\mbox{ } .
\ee
%
%
\noindent In this formulation, Barbour's triangles demonstration can now be done by using one wooden triangle and an overhead projection image of the compared triangle.

The conjugate momenta are then
\be
\p_{A} = \delta_{AB}m_{A}\{\5Star_{\underline{A},\underline{B},C}{\q}^{B}\}/\mI
\mbox{ } ,  
\mbox{ where } \mbox{ } 
\5Star_{\underline{A}, \underline{B}, C} := \d/{\d t^{\se\sm(\sJ\sB\sB)}_{\underline{A}, \underline{B}, C}} = 
\sqrt{2\ttW}\d/||\d_{\underline{A}, \underline{B}, C}\mbox{\boldmath$q$}||_{\mbox{\scriptsize\boldmath$m$}} \mbox{ } . 
\label{mom4}
\eeq 
[I explain the difference between $\5Star$ and $\Last$ in App \ref{Examples}.B.]
These obey, as a primary constraint,
\beq
\scE := \mI ||\mbox{\boldmath$p$}||_{\mbox{\scriptsize\boldmath$n$}}\mbox{}^2/2 + \ttV(\mbox{\boldmath$q$}) = \ttE \mbox{ } ,  
\label{Dilen}
\eeq
which is again quadratic and not linear in the momenta. 
Again, the momenta additionally also obey as secondary constraints  (\ref{ZM}), (\ref{ZAM}), 
and now also by standard FEP variation of $C$, a zero dilational momentum constraint  (\ref{ZDM}); there are various reasons to name the left-hand-side term the `Euler quantity'.

The evolution equations are  
\beq
\5Star_{\underline{A},\underline{B},C} \up_I = -{\pa\ttV}/{\pa\uq^I} \mbox{ } . 
\eeq
The scaled RPM Poisson brackets retain the same form in pure-shape RPM, while the new nonzero bracket is 
\be
\{{\scD}, {\scP}_{\mu}\} = {\scP}_{\mu} 
\label{other4} \mbox{ } .
\eeq
I note that $\underline\scL$ and $\scD$ commute, i.e. the rotations and dilatations do not interfere with each other.  

\noindent The emergent JBB time is now 
\beq
\lt^{\se\sm(\sJ\sB\sB)} = 
\stackrel{\mbox{\scriptsize extremum $\underline{A}, \underline{B}, C \mbox{ } \in \mbox{ } \mbox{Sim($d$)}$}}
         {\mbox{\scriptsize of $\stS^{\tS\tR\tP\tM}_{\tJ-\tA}$}}                                                              
\left(                                                              
\int||\d_{\underline{A}, \underline{B}, C}\mbox{\boldmath$q$}||_{\mbox{\scriptsize\boldmath$m$}}/\sqrt{2\ttW(\mbox{\boldmath$q$}}
\right)   \mbox{ } .
\ee

\subsubsection{Pure-shape RPM in relative Jacobi coordinates}\label{R-SRPM}

Here $\FrQ = \Fr(N, d)$ and $\FrG$ = Sim($d$).
Then $\d_{\underline{B},C}\underline{\rho}^{i} := \d\underline{\rho}^{i} - \d \underline{B}\cr\underline{\rho}^{i} + \d C\underline{\rho}^{i}$.
The kinetic arc element built from this is 
\beq
\d\tts^{\sS\sR\sP\sM\, 2}_{\sJ-\sJ\sA}  = 
||\d_{\underline{B}, C}\brho||\mbox{}^2/\mI  
\mbox{ } .   
\eeq 
$\ttV = \ttV(\mbox{ratios of } \underline{\rho}^i\cdot\underline{\rho}^j \mbox{ alone})$.  
The relational action is then 
\beq
\FS^{\sS\sR\sP\sM}_{\sJ-\sJ\sA} = \sqrt{2}\int\sqrt{\ttE - \ttV(\mbox{ratios of } \underline{\rho}^i\cdot\underline{\rho}^j 
\mbox{ alone}})\d\tts^{\sS\sR\sP\sM}_{\sJ-\sJ\sA} \mbox{ } . 
\label{SRPMAc}
\eeq
\mbox{ } \mbox{ } The conjugate momenta are then 
\beq
p_{i\mu}= \delta^{IJ}\delta^{\mu\nu}\5Star_{{\underline{B}},{C}}\rho_{J\nu}/\mI 
\mbox{ } ,
\mbox{ for } \mbox{ } 
\5Star_{\underline{B}, C} := \d/\d t^{\se\sm(\sJ\sB\sB)}_{\underline{B}, C} = \sqrt{2\ttW}\d/||\d_{\underline{B}, C}\brho||\mbox{ }  .
\eeq
These obey as a primary constraint  
\beq
\mI||\mbox{\boldmath$p$}||^2/2 + \ttV = \ttE
\mbox{ } .
\label{Quaad}
\eeq
There is also a secondary zero angular momentum constraint (\ref{DamT}) and a secondary zero dilational momentum constraint,
\beq
\scD := \sumin \underline{\rho}^{i}\cdot\underline{\mbox{$p$}}_{i} = 0 \mbox{ } . 
\label{ZDM2}
\eeq
The evolution equations are 
\beq
\5Star_{{\underline B}, C}p_{\mu i} = -{\pa\ttV}/{\pa\rho^{\mu i}} \mbox{ } . 
\eeq
Exploiting the Jacobi map, the Poisson brackets are as before but with no $\scP_{\mu}$ involved anymore.  
The emergent JBB time is 
\beq
\lt^{\se\sm(\sJ\sB\sB)} = \stackrel{\mbox{\scriptsize extremum $\underline{B}, C \mbox{ } \in \mbox{ } \mbox{Pl($d$)}$}}
         {\mbox{\scriptsize of $\stS^{\tS\tR\tP\tM}_{\tJ-\tJ\tA}$}}                                                              
\left(                                                              
                                                               \int||\d_{\underline{B}, C}\brho||/\sqrt{2\ttW(\brho)}
\right)  \mbox{ } .
\label{Gus}
\ee

\subsection{Nonrotational RPM}\label{NonRot}

I give this one straight off in my preferred conceptualization of relative Jacobi coordinates and geometrically natural objects.  
One needs $d > 1$ (i.e. rotational nontriviality) for this to be distinct from SRPM.
This theory is a more tractable arena for investigation of implications of scale invariance than Barbour's theory itself, in particular for $d \geq 3$.   
Here,
\be
\d\tts^{\sN\sR\sP\sM}_{\sJ-\sJ\sA}\mbox{ }^{2} = ||\d_{C}{\brho}||\mbox{}^2/\mI  \mbox{ and } \mbox{ } 
\d_{C}\rho = \d\rho_i + \d C \rho^i \mbox{ }
\ee
alongside $\ttV = \ttV$(a function of ratios of the $\rho^i$ components alone).    
Note that there is now no `dot product' rotational restriction!
Then the action is 
\beq
\FS^{\sN\sR\sP\sM}_{\sJ-\sJ\sA} = \sqrt{2}\int\sqrt{\ttE - \ttV\mbox{( a function of ratios of the $\rho^i$ components alone)}} \d\tts^{\sS\sR\sP\sM}_{\sJ-\sJ\sA} \mbox{ } . 
\label{NRPMAc}
\eeq
\mbox{ } \mbox{ } The conjugate momenta are
\be
p_{i\mu} = \delta_{ij}\delta_{\mu\nu}\{\5Star{\rho}^{j\nu} + \5Star{C}\rho^{j\nu}\}/\mI 
\mbox{ } 
\mbox{ for } \mbox{ } 
\5Star_{C} := \d/\d t^{\se\sm(\sJ\sB\sB)}_{C} := \sqrt{2\ttW}\,\d/||\d_{C}\brho||\mbox{ }  .
\eeq
These momenta obey energy and dilational momentum constraints as above: (\ref{Quaad}--\ref{ZDM2}).

The evolution equations are
\beq
\5Star_{C}p_{\mu i} = -{\pa\ttV}/{\pa\rho^{\mu i}} \mbox{ } . 
\eeq
The emergent time is 
\beq
\lt^{\se\sm(\sJ\sB\sB)}  = \stackrel{\mbox{\scriptsize extremum $C \mbox{ } \in \mbox{ } \mbox{Dil}$}}
         {\mbox{\scriptsize of $\stS^{\tN\tR\tP\tM}_{\tJ-\tJ\tA}$}}                                                              
\left(                                                              
                                                               \int||\d_{C}\brho||/\sqrt{2\ttW(\brho)}
\right)   \mbox{ } .
\ee
\noindent Note: nonrotational RPM suffices to investigate theoretical changes due to considering length ratios and angles alone (which is a theoretically desirable situation to 
investigate since only ratios of lengths are in practise measured, even if absolute angles still remain in the model).
Also, this theory might have some value toward determining whether scale-invariant physics can reproduce Astrophysics and Cosmology 
(to the same kind of extent that Newtonian Cosmology can reproduce actual Cosmology, see Appendix \ref{Cl-Soln}.C).

\subsection{Scale--scalefree split of scaled RPM}\label{SS-Split}

This is a useful further way of looking at scaled RPM (and is a working original to this Article, which may be viewed as a `missing link' between various of the preceding workings.
I begin working in q's because it is valuable to compare this with Barbour's formulation of SRPM.  
For the mass-normalized coordinates $u^{I\mu} = \sqrt{m_{I}}q^{I\mu}$, apply the homothetic ansatz 
\beq 
u^{I\mu} = \sigma s^{I\mu} 
\mbox{ } \mbox{ such that }   
\scC := \sum\mbox{}_{\mbox{}_{\mbox{\scriptsize I = 1}}}^N\underline{s}_{I}\underline{s}^{I} - 1 = 0 \mbox{ } .   
\label{quu}
\eeq 
Here, $s^{I\mu}$ is the {\it scalefree part} (which in the present context means shape part plus absolute-position and absolute-angle part).
Then one has the usual relational action 
\beq
\FS^{\sE\sR\sP\sM}_{\sJ-\sS\sS\sS} = \sqrt{2}\int\sqrt{W}\d s^{\sE\sR\sP\sM}_{\sJ-\sS\sS\sS} 
\mbox{ } \mbox{  with } \mbox{ } 
\d s^{\sE\sR\sP\sM}_{\sJ-\sS\sS\sS} = ||\sigma\,\d_{\underline{A},\underline{B} }\mbox{\boldmath$s$} + \d_{\sigma}\mbox{\boldmath$s$}||_{\mbox{\scriptsize\boldmath$m$}} \mbox{ } .
\eeq
[One could write this as $\d s = ||\d_{\underline{A},\underline{B},\sigma}||_{\mbox{\scriptsize\boldmath$m$}}$ with this symbol taking the same meaning as in Barbour's formulation 
of pure-shape RPM, however the key difference is that $\sigma$ is {\sl not} now an auxiliary, so the FEP condition does {\sl not} now apply to it.  
The cross-term is also zero, by the constraint (\ref{quu}) implying $\sum_{I = 1}^N\underline{s}^I\dot{\underline{s}}_I = 0$, 
by which the kinetic term cleanly splits into shape and scale blocks, 
\beq
\d s^{\sE\sR\sP\sM\, 2}_{\sJ-\sS\sS\sS} =  \d{\sigma}^2 + \sigma^2||\d_{\underline{A},\underline{B} }\mbox{\boldmath$s$}||_{\mbox{\scriptsize\boldmath$m$}}^2 \mbox{ } .]
\eeq

Then the conjugate momenta are 
\beq
P_{\sigma} = \Last \sigma \mbox{ }  \mbox{ and } \mbox{ } 
p^{s}_{I\mu} =  \sigma^2\Last_{\underline{A},\underline{B}} m_Is_{I\mu} \mbox{ } .  
\eeq
These obey the constraint 
\beq
\sigma\, P_{\sigma} = p^{s}_{I\mu}s^{I\mu}/m_I
\label{predil}
\eeq
either as a primary constraint from the first form of the action or as an enforcer of (\ref{quu}), 
the primary constraint  
\beq
\scE = ||\mbox{\boldmath$p$}^{s}||_{\mbox{\scriptsize\boldmath$n$}}\mbox{}^2/2\sigma^2 + \ttV = \ttE \label{splitHam}
\eeq
and the secondary constraints
\beq
\underline\scP := \sumIN \underline{p}_{I} = \sumIN \{\sigma \, \d\underline{s}_{I} + \d{\sigma} \, \underline{s}_{I} \} = 0  
\mbox{ } , \mbox{ } \mbox{ } 
\underline\scL := \sumIN \underline{s}^{I} \cr \underline{p}^{s}_{I} = 0 \mbox{ } .
\label{PDsplit}  
\eeq
Thus the zero total angular momentum constraint is identified as being a scalefree constraint. 
Another difference between the unreduced scaled RPM's scale--scalefree spilt and pure-shape RPM is that the Hamiltonian 
now has a multiplier (or cyclic velocity if one applies the almost-Dirac procedure, see Appendix \ref{Examples}.A) that will enter Hamilton's equations.  
In the Barbour-style approach, (\ref{predil}) with $q$ for $\sigma$ is replaced by (including FEP variation where now appropriate) 
\beq
P_{\sigma} = 0 \mbox{ } ,  \mbox{ and so }  \mbox{ } 0 = \sumin \q^i \cdot \p_i = \scD \mbox{ }   
\eeq
follows. 
This makes pure-shape RPM and the scale--scalefree split of scaled RPM quite different at the uneliminated level, with Barbour's formulation of pure-shape RPM being closer 
than my own one.

One could alternatively work at the level of the Jacobi coordinates with a scale--non-scale split 
\beq
\rho^{i\mu} = \Sigma \, S^{i\mu} \mbox{ } \mbox{ such that } \mbox{ }  
\scC^{\prime} := \sumin \underline{S}_i \cdot \underline{S}^i - 1 = 0 \mbox{ } 
\label{puu}
\eeq 
(for scale $\Sigma$), which works out exactly as above except that there is now no momentum constraint (that being automatically incorporated).  
In this context, scalefree means shape plus absolute angular information.

\mbox{ } 

\noindent Note 1: by the definition of moment of inertia and (\ref{quu}) or (\ref{puu}), the scale $\Sigma$ 
here is the square root of moment of inertia $\rho = \sqrt{\mI}$, which is the dimensionally natural quantity for it to be.  

\noindent Note 2) If one uses $\ttT$, $\ttV$, $\ttE$ instead, then the $\sigma^2$ factor in (\ref{splitHam})'s 
kinetic term gets cancelled out, but reappears in the total `pseudoenergy' and the potential.  
This is the usual PPSCT with $\Omega^2 = \mI$ (see Appendix \ref{Examples}.B) and permits alignment with the pure-shape RPM convention.  

\noindent Note 3) Pure-shape RPM and scale--scalefree split of scaled RPM will be shown to be much more closely linked in the reduced = direct relationalspace formulation; 
in this context, the scalefree part is indeed the shape.   
Also, pure-shape RPM can then be seen as a $\sigma$-nondynamical version of this, in which (\ref{quu}) is now unnecessary and the $P_{\sigma} = 0$ form of (\ref{predil}) 
arises in a new way.  
I.e. as a secondary constraint from variation with respect to pure-shape RPM's new $C$-auxiliary.  
However, the way in which pure-shape RPM is the shape portion of scaled RPM comes out more cleanly in the direct formulation of Sec \ref{Q-Geom}.

\subsection{Relational formulation of GR-as-Geometrodynamics}\label{Rel-Gdyn}

\subsubsection{Many routes to relativity}\label{Many-Routes}

General relativity (GR) is often thought of in the spacetime formulation of  Einstein's original conceptualization and derivation 
\cite{Ein15, Ein16} (supplemented by Vermeil--Weyl--Cartan--Lovelock mathematical simplicity theorems \cite{Vermeil17, Weyl21, Cartan22, Lovelock71}).  
The corresponding action principle is Einstein--Hilbert's (\ref{EinHilb}).  
However, GR can be arrived at along many different routes\footnote{That is the perspective  
of the well-known physicist John Archibald Wheeler \cite{Battelle, MTW}, who, among other things, was a co-founder of Geometrodynamics.  
It is a prime example of Judging Criterion 5).
This perspective bears some similarities to Poincar\'{e}'s conventionalism as subsequently championed by Reichenbach \cite{Reichenbach50} and Gr\"{u}nbaum \cite{Grunbaum}.  
On the other hand, Kant had long previously argued in denial the diversity of possible scientific points of view. 
This is an example of Judging Criteria 1) and 5) clashing.
I here comment that physically equivalent formulations can still be {\sl conceptually} different (whether in terms of which 
philosophy they implement  or via which mathematical axioms are needed for such results to exist).  
As such, two routes turning out to be (partly) equivalent formulations of a single theory constitutes an interesting {\sl theorem}, whether rigorous or heuristic. 
This is how I view recovery of Newtonian mechanics from BB82 and also how Sec 3 arrives at the same structures as Sec 2.}

\noindent Route 1) is the above.

\noindent Route 2) is Cartan's route \cite{Cartan25}.

\noindent Routes 3) and 4) are the 2-way passage between the spacetime and Arnowitt--Deser--Misner (ADM) 
\cite{ADM} split space-time formulations in which GR is a dynamics of 3-geometries (see below).  

\noindent Route 5) is Sakharov's \cite{VSak}, in which GR is the elasticity of space due to Particle Physics. 

\noindent Route 6) is Fierz--Pauli's \cite{VFP}, in which GR emerges from consideration of a spin-2 field on Minkowski spacetime.  


\noindent Note 1) There are a number of other (mostly later) programs that could well be considered as routes. 
See e.g. the rest of this and the next SSecs.  
One could also view in this way various programs involving `beins' or spinors \cite{Stewart, Wald, PenRind} (including Ashtekar variables based approaches \cite{Ashtekar, Thiemann}).  

\noindent Note 2) Additionally, various (partial or total) unifications lead to aspects of GR (such as the closed relativistic string necessarily containing a spin-2 mode \cite{GSW1}).  

\mbox{ }

\noindent The analogies between RPM's and GR are via GR as Geometrodynamics and Confomogeometrodynamics, so I concentrate on these in this SSec and the next respectively.

\subsubsection{Further detail of ADM's form of Geometrodynamics} \label{Sec-ADM}

In dynamical/canonical approaches, the spacetime manifold $\FrM$ is 3 + 1 split (3 spatial dimensions and 1 time dimension) into $\bupSigma \times \mathbb{R}$;  
this assumes that the spatial topology $\bupSigma$ does not change with coordinate time $t \in \mathbb{R}$ (or some subinterval thereof). 
This furthermore assumes that GR is globally hyperbolic. 
Some considerations focus around a single $\bupSigma$ embedded into GR spacetime (or without that assumption, usually leading to its deduction, c.f. Sec \ref{RWR}).   
Moreover, some further considerations require a {\it foliation} ${\FrF} = \{{\FrL}_{\sfA}\}_{\sfA \in A}$, which is a decomposition of an $m$-dimensional manifold ${\FrMgen}$ 
into a disjoint union of connected $p$-dimensional subsets (the {\it leaves} $\FrL_{\sfA}$ of the foliation) such that the following holds. 
$m \in \FrMgen$ has a neighbourhood $\FrU$ in which coordinates ($x^1, ..., x^m$): $\FrU \rightarrow \mathbb{R}^m$ such that for each leaf 
$\FrL_{\sfA}$ the components of $\FrU \bigcap \FrL_{\sfA}$ are described by $x^{p + 1}$ to $x^m$ constant [see Fig 1a)].
The codimension of the foliation is then $c = m - p$. 
For 3 + 1 GR, $\FrMgen = \FrM$, so $m = 4$, the leaves $\FrL_{\sfA}$ are 3-space hypersurfaces $\bupSigma_{\sfA}$ so $p = 3$ and thus $c = 1$: the time dimension.  

{            \begin{figure}[ht]
\centering
\includegraphics[width=0.4\textwidth]{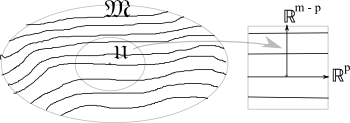}
\caption[Text der im Bilderverzeichnis auftaucht]{        \footnotesize{Picture supporting the text's definition of foliation.}  }
\label{Fig-1-Second}\end{figure}          }

\noindent Given 2 neighbouring hypersurfaces, one can \{3 + 1\}-decompose the spacetime metric \cite{ADM} $\mg_{\Gamma\Delta}$ [see Fig \ref{Fig-1-Second}] into (\ref{ADM-split}).  
The relation between the foliation vector $n^{\mu}$ and the above \{3 + 1\}-split quantities 
is $\mt^{\Gamma} = \upalpha \mn^{\Gamma} + \upbeta^{\Gamma}$ and so $\mn^{\Gamma} = \upalpha^{-1}[1, - \uupbeta]$.

Using the spacetime metric split and extrinsic curvature of Sec \ref{GR-Gdyn} alongside various geometrical identities, the Einstein--Hilbert action 
(\ref{EinHilb}) can be rearranged to the ADM action
\be
\FS^{\sG\sR}_{\sA\sD\sM} = 
\int\d\mt\int_{\sbSigma}\d^3x\sqrt{\mh}\,\upalpha\,
\{\mK_{\mu\nu}\mK^{\mu\nu} - \mK^2 + \mbox{${\cal R}\mi\mc$}(\ux; \bh] - 2\Lambda\} \mbox{ } .  
\ee
The corresponding manifestly-Lagrangian form of this action is then 
\beq
\FS^{\sG\sR}_{\sA\sD\sM-\sL} = 
\int\d\mt\int_{\sbSigma}\d^3x\sqrt{\mh}\,\upalpha 
\left\{
\mT^{\sG\sR}_{\sA\sD\sM-\sL}/\upalpha^2 +   \mbox{${\cal R}\mi\mc$}(\ux; \bh]   - 2\Lambda
\right\} \mbox{ } , 
\label{ADM}
\eeq
\beq
\mT^{\sG\sR}_{\sA\sD\sM-\sL} = \{\mh^{\mu\rho}\mh^{\nu\sigma} - \mh^{\mu\nu}\mh^{\rho\sigma}\}
\{\dot{\mh}_{\mu\nu}     - \pounds_{\suupbeta}\mh_{\mu\nu}\}
\{\dot{\mh}_{\rho\sigma} - \pounds_{\suupbeta}\mh_{\rho\sigma}\}/4 \mbox{ } .  
\eeq
Note: $\pounds_{\suupbeta}h_{\mu\nu}  = 2\mD_{(\mu}\upbeta_{\nu)} = \{\mbox{\scriptsize$\mathbb{K}$}\upbeta\}_{\mu\nu}$ (the {\it Killing form}).  
Here the first equality is computational, whilst the second is another geometrical identification (however, the Lie derivative notion is a 
highly minimalist geometrical notion \cite{Stewart} as well as most lucidly depicting the infinitesimal action of diffeomorphisms).

From this action, or from the split of the spacetime EFE's (here given in vacuo and in terms of each of $\mK_{\mu\nu}$ and of $\uppi^{\mu\nu}$), the constraints are 
\beq
\mbox{Hamiltonian constraint, } \scH := 2{\mG}^{(4)}_{\perp\perp} = \mK^2 - \mK_{\mu\nu} \mK^{\mu\nu}  + \mbox{${\cal R}\mi\mc$}(\ux; \bh]  = 
\mN_{\mu\nu\rho\sigma}\uppi^{\mu\nu}\pi^{\rho\sigma}/\sqrt{\mh} - \sqrt{\mh}\mbox{${\cal R}\mi\mc$}(\ux; \bh]  = 0  \mbox{ } ,   
\eeq
\be
\mbox{momentum constraint, } \scM_{\mu} :=  2{\mG}^{(4)}_{\mu\perp} = 2\{\mD_{\nu}{\mK^{\nu}}_{\mu} - \mD_{\mu }\mK\}  = -2\mD_{\nu}\uppi^{\nu}\mbox{}_{\mu} = 0 \mbox{ } 
\label{Mom}
\eeq
The $\mK_{\mu\nu}$-forms of these serve to identify these as contractions of the Gauss--Codazzi equations for the embedding of spatial 3-slice into spacetime \cite{Wald} 
(a higher-$d$ indefinite-signature generalization of Gauss' Theorema Egregium).

The GR `constraint algebra' is the Dirac algebroid \cite{Dirac} 
\be 
\{    \underline{\scM}(\uupbeta^{\prime}),  \underline{\scM}(\uupbeta)    \} = \pounds_{\suupbeta} \underline{\scM}(\uupbeta^{\prime}) 
\mbox{ } , 
\label{Mom-Mom} 
\ee 
\be 
\{    \scH(\upalpha),  \underline{\scM}(\uupbeta)\} = \pounds_{\suupbeta}\scH(\upalpha) 
\mbox{ } , 
\label{Ham-Mom} 
\ee 
\be 
\{\scH(\upalpha), \scH(\upalpha^{\prime})\} =  \underline{\scM}(\uupkappa) 
\mbox{ }\mbox{ for  } \mbox{ } 
\upkappa^{\mu} := \mh\mh^{\mu\nu}\{\upalpha\pa_{\nu} \upalpha^{\prime} - \upalpha^{\prime}\pa_{\nu}\upalpha\} 
\mbox{ } ,  
\label{Ham-Ham} 
\ee 
Note that the third bracket, possesses {\sl structure functions}, so that this is far more complicated than a Lie algebra: a {\it Lie algebroid} \cite{BojoBook} 
(it is usually called `Dirac algebra', though `Dirac algebroid' is both more mathematically correct and not confusable with the Dirac algebra of fermionic theory). 
(\ref{Mom-Mom}) is the closure of the Lie algebra of 3-diffeomorphisms on the 3-surface, whilst (\ref{Ham-Mom}) just means that $\scH$ 
is a scalar density; neither of these have any dynamical content.
It is the third bracket then that has dynamical content.

The Dirac algebroid has remarkable properties at the classical level, as exposited in Fig \ref{Teit}b) and Sec \ref{Cl-POT-Strat}.  
Yet that the Dirac algebroid of the classical GR constraints is not a Lie algebra does limit many a quantization approach \cite{I84}, Sec \ref{QM-Intro}.

\subsubsection{Some interpretations of Geometrodynamics}\label{Gdyn-Interp}

Geometrodynamics has been interpreted in a number of ways, in great part envisaged by Wheeler.  

\noindent Interpretation 1) His original attempt (from around 1960) as an interpretation, widely regarded as failed \cite{Geometrodynamica}, was to 
consider vacuum GR in these dynamical terms as a Theory of Everything.  
In later interpretations, matter source terms have been considered a necessity; fortunately, these are straightforward to handle.  

\noindent Interpretation 2) The most standard interpretation, however, is the Introduction's superspace one (1967 \cite{DeWitt67, Battelle}) 
as underlying the ADM action (which work itself was consolidated by 1962 \cite{ADM}).  
Then $\mT^{\sG\sR}_{\sA\sD\sM-\sL}$ can be rewritten in the Lagrangian--DeWitt form  
\beq
\mT^{\sG\sR}_{\sA\sD\sM-\sL-\sD} = ||\pa{\bh}/\pa\mt - \pounds_{\suF}\bh||_{\mbox{\scriptsize\boldmath${\sbM}$}}\mbox{}^2/4 \mbox{ } ;
\eeq
this is the form used in the Introduction.
Note also that this is in terms of the undensitized GR kinetic metric.

Then via Sec \ref{GR-Gdyn}'s interpretation of $\scM_{\mu}$, GR is a Geometrodynamics) on the quotient configuration space superspace($\bupSigma$).   
This description is less (rather than completely non-)redundant, the partial redundancy being due to the Hamiltonian constraint not yet being addressed.
Interpreting the Hamiltonian constraint is more problematic.  
In this interpretation, instead of any kind of unification, the attitude is that Standard Model matter can be `added on' (very straightforwardly 
for bosons \cite{ADM, Teitelboim, IN77b} and via passing to triad formalism for Dirac fields \cite{BIY85}.  

\mbox{ } 

\noindent Interpretation 3) In the {\it thin sandwich approach}, Baierlein, Sharp and Wheeler \cite{SharpTh, BSW} regarded $\mh_{\mu\nu}$ and $\dot{\mh}_{\mu\nu}$ as knowns.
This solved for the spacetime `filling' in between, in analogy with the QM set-up of transition amplitudes between states at two different times \cite{WheelerGRT}.    
[It is the `thin' limit of taking the bounding `slices of bread': the hypersurfaces $\mh_{\mu\nu}^{(1)}$ and $\mh_{\mu\nu}^{(2)}$ as knowns.]  
In more detail, 

\noindent BSW 0) They take as primary a conventional difference-type action for Geometrodynamics.             
  
\noindent BSW 1) They vary this with respect to the ADM lapse  $\upalpha$.
  
\noindent BSW 2) They solve resulting equation for $\upalpha$.   
  
\noindent BSW 3) They substitute this back in the ADM action to obtain a new action: the BSW action, 
\beq
\FS^{\sG\sR}_{\sB\sS\sW} = 2\int\d \mt \int_{\sbSigma}\d^3x\sqrt{\mh}\sqrt{\mT^{\sG\sR}_{\sA\sD\sM-\sL-\sD}\{\mbox{${\cal R}\mi\mc$}(\ux; \bh] - 2\Lambda\}} \mbox{ } .
\eeq

\noindent BSW 4) They then take this action as one's primary starting point.

\noindent BSW 5) They vary with respect to the shift $\upbeta^{\mu}$ to obtain the GR linear momentum constraint $\scM_{\mu} = 0$.

\noindent BSW 6) They posit to solve the Lagrangian form of $\scM_{\mu} = 0$ for the shift $\upbeta^{\mu}$.

\noindent BSW 7) They posit to then substitute this into the computational formula for the lapse $\upalpha$.   

\noindent BSW 8) Finally, they substitute everything into the formula for the extrinsic curvature to construct the region of spacetime contiguous to the original slice datum. 

\mbox{ } 

\noindent Note 1) Unfortunately the p.d.e. involved in this attempted elimination is the so-called {\it reduced thin sandwich equation} \cite{BF}. 
The {\bf Thin Sandwich Problem} consists of obstructions at this stage (e.g. to existence or uniqueness, see Sec \ref{TSC} for more), and this is furthermore 
one of the facets of the POT as per the Introduction. 
Counterexamples to its solubility have been found \cite{TSC1, counterTS2}, while good behaviour {\sl in a restricted sense} took many years to establish \cite{TSC2}.  
See Secs \ref{TSC}, \ref{+Sand} for more.  

\noindent Note 2) BSW 0--3) together constitute a {\it multiplier elimination}.  

\noindent Note 3) This approach has mostly only been considered for phenomenological matter \cite{WheelerGRT, BF, FodorTh}, though \cite{TSCG} considered it for Einstein--Maxwell theory.

\subsubsection{Relational reformulation of GR-as-Geometrodyamics}\label{GR-Actual-Relational}

Interpretation 4) The Barbour-type indirect formulation of RPM's (\ref{GeneralAction},\ref{Taction}) makes these parallels particularly clear. 
In the 

\noindent geometrodynamical counterpart of this, $\FrQ$ = Riem($\bupSigma$) -- the space of Riemannian 3-metrics on some 
spatial manifold of fixed topology $\bupSigma$ (taken for simplicity to be compact without boundary).  
This choice of 3-metric objects is a simple one (one geometrical object, rather than more than one), moreover one that is 

\noindent A) useful for Physics given its significance in terms of lengths.

\noindent B) Mathematically rich -- one then gets a connection for free -- the metric connection -- as well as the notions of curvature and of Hodge star.

\noindent Nevertheless, one is entitled to view this choice as a possibly inessential simplicity postulate.  

\noindent See Sec \ref{Top-Rel} for discussion of the relational undesirability of the underlying residual NOS in the sense of a fixed spatial topology $\bupSigma$. 

\noindent Next, Diff($\bupSigma$) the group of 3-diffeomorphisms on $\bupSigma$, is a suitable $\FrG$, i.e. the 
physically irrelevant transformations between the indiscernible 3-metrics that correspond to the same 3-geometries.  
\noindent The quotient Riem($\bupSigma$)/Diff($\bupSigma$) is called superspace($\bupSigma$) [see Sec \ref{GRCtr} for more].  
The arbitrary Diff($\bupSigma$)-frame expressions are most easily discussed in the form $\Circ_{\suF} \mh_{\mu\nu} := 
\dot{\mh}_{\mu\nu} - {\pounds}_{\dot{\suF}}\mh_{\mu\nu}$ rather than `bare' $\dot{\mh}_{\mu\nu}$.
From these base objects, one can construct the frame-corrected objects of the spatial metric geometry.  
As these transform well under the 3-diffeomorphisms of the spatial 3-metric geometry, the Diff$(\bupSigma)$ corrections are manifest only as corrections to the metric velocities.
\noindent Here, one uses the cyclic velocity of the grid, $\dot{\mF}^{\mu}$ (the type of frame auxiliary relevant to Geometrodynamics) 
instead of the Lagrange multiplier shift $\upbeta^{\mu}$.  

\noindent Then $\pounds_{\dot{\suF}}$ is the Lie derivative with respect to $\mF^{\mu}$, 
$\Circ_{\suF} := \Circ{\mbox{ }} - \pounds_{\dot{\suF}}$ is the geometrodynamical `best-matching' derivative.  
This is the same mathematical quantity as the hypersurface derivative $\delta_{\suupbeta}$, albeit now conceived of from within the foundational assumption of space alone rather than 
spacetime, and using a cyclic velocity of the grid rather than a shift multiplier at the level of the Principles of Dynamics.  
The MPI counterpart is $\pa_{\suF} := \pa - \pounds_{\pa\suF}$.

{            \begin{figure}[ht]
\centering
\includegraphics[width=0.6\textwidth]{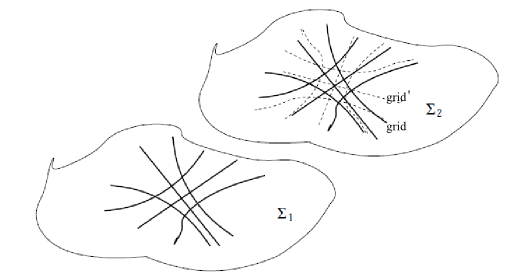}
\caption[Text der im Bilderverzeichnis auftaucht]{        \footnotesize{ Barbour's demonstration for 3-metrics: 
one now shuffles grids on $\bupSigma_2$ so as to minimize incongruence with $\bupSigma_1$'s grid.}   }
\label{GridBM} 
\end{figure}  } 

The action that one builds so as to implement Temporal Relationalism is \cite{RWR, Lan, ABFO, Phan} 
\beq
\FS_{\sB\sF\sO-\sA}^{\sG\sR} = 2\int\d\lambda\int_{\sbSigma}\d^3{x}\sqrt{\mh}\sqrt{\mT^{\sG\sR}_{\sB\sF\sO-\sA}\{\mbox{${\cal R}\mi\mc$}(\ux; \bh] - 2\Lambda\}} = 
\sqrt{2}\int\d\lambda\int_{\sbSigma}\d^3{x}\sqrt{\mh}\sqrt{\mbox{${\cal R}\mi\mc$}(\ux; \bh] - 2\Lambda} \, \pa\ms^{\sG\sR}_{\sB\sF\sO-\sA} \mbox{ } ,
\label{BFOA}
\eeq
\beq
\mbox{ where } \mT_{\sB\sF\sO-\sA}^{\sG\sR} = ||\Circ_{\suF}\mh_{\mu\nu}||_{\sbM}\mbox{}^2/4
\mbox{ }  \mbox{ or } \mbox{ } 
\pa\ms_{\sB\sF\sO-\sA}^{\sG\sR}\mbox{}^2 = ||\pa_{\suF}\mh_{\mu\nu}||_{\sbM}\mbox{}^2/2 \mbox{ } .
\label{BFO-action}
\eeq 
In this case, the action is a JBB$[P(\langle \mbox{Riem}(\bupSigma), \mbox{ } \mbox{\boldmath${\bM}$}\rangle, \mbox{ } \mbox{Diff}(\bupSigma))]$.   
The relational formulation of Geometrodynamics is valuable in providing guidance in yet further investigations of alternative conceptual foundations 
for GR \cite{BB82, B94I, RWR, Phan, FEPI}, and as regards addressing the POT in QG (see Sec \label{1.7} and Part III).  


\noindent The BFO-A action bears many similarities to the better-known BSW one, but supercedes it in attaining manifest Temporal Relationalism.    

\mbox{ } 

\noindent Interpretation 5) One can also obtain \cite{FEPI} the BFO--A action from another action \cite{FEPI}.
This is via 

\noindent 1) using the instant-frame version of the 3 + 1 ADM split of the spacetime metric,
\beq
\mg_{\Gamma\Delta} =
\left(
\stackrel{    \mbox{$\dot{\mF}_{\mu}\dot{\mF}^{\mu} - \dot{\upgamma}^2$}    }{ \mbox{ }  \mbox{ }  \dot{\mF}_{\delta}    }
\stackrel{    \mbox{$\dot{\mF}_{\gamma}$}    }{  \mbox{ } \mbox{ }  \mh_{\gamma\delta}    }
\right)
\mbox{ }    . 
\label{ADM-split} 
\eeq
See Appendix \ref{Examples}.A.5 for the interpretation of  $\dot{\upgamma}$ and explanation of its similarities and differences with $\upalpha$, $\dot{\mI}$ and $\mN$.  
 
\noindent 2) Using the corresponding `instant--frame' formulation of the extrinsic curvature, 
\beq
\mK_{\mu\nu} = -\{1/2\,\dot{\mI}\}\{\dot{\mh}_{\mu\nu} - \pounds_{\dot{\suF}}\mh_{\mu\nu}\} 
\mbox{ } , 
\eeq
to obtain, from the Einstein--Hilbert action,  
\be
\FS^{\sG\sR}_{\sA} = \int\d\mt \int_{\sbSigma}\d^3x\sqrt{\mh}\,\dot{\mI}\{\mK_{\mu\nu}\mK^{\mu\nu} - \mK^2 + \mbox{${\cal R}\mi\mc$}(\ux; \bh] - 2\Lambda\} 
\mbox{ } .  
\label{A-Ac}
\ee
Then the corresponding manifestly-Lagrangian form is 
\beq 
\FS^{\sG\sR}_{\sL-\sA} = \int\d \mt \int_{\sbSigma}\d^3x\sqrt{\mh}\,\dot{\mI}
\{||\Circ_{{\suF}}\bh||_{\mbox{\scriptsize\boldmath${\bM}$}}\mbox{}^2/4\,\dot{\mI}^2 + \mbox{${\cal R}\mi\mc$}(\ux; \bh] - 2\Lambda\}
\mbox{ } .
\label{AAAAAc}
\eeq
\noindent 3) The BFO-A action now follows from the A action by using passage to the Routhian to eliminate $\dot{\mI}$ (an even 
closer parallel of Jacobi's deparametrization than BSW's lapse multiplier elimination procedure).  

\mbox{ } 

\noindent Of these actions, $\FS^{\sG\sR}_{\sL-\sB\sF\sO-\sA}$ is the one of most interest to the relationalist, so I perform the variation for it below.  
The conjugate momenta are 
\beq
\uppi^{\mu\nu} =  \sqrt{\mh}{\mM}^{\mu\nu\rho\sigma}\Last_{\suF}\mh_{\rho\sigma} 
\label{tumvel}
\mbox{ } \mbox{ where } \mbox{ } 
\Last_{\suF} := {\delta }/{\delta \mt^{\se\sm(\sJ\sB\sB)}_{\suF}} := \sqrt{2\{\mbox{${\cal R}\mi\mc$}(\ux; \bh] - 2\Lambda\}}\delta/||\delta_{\suF}\bh||_{\sbM} \mbox{ } .  
\eeq 
(\ref{GRaction}) being MRI, there must likewise be at least one primary constraint, which is 
in this case the GR Hamiltonian constraint (\ref{Hamm}) to which the momenta contribute quadratically but not linearly. 
Also, variation with respect to $\mF^{\mu}$ yields as a secondary constraint the GR momentum constraint (\ref{Momm}), which is linear in the gravitational momenta.

\noindent As for standard ADM, the momentum constraint is interpretable geometrically as GR not being just a dynamics of 3-metrics 
(`metrodynamics') but furthermore that moving the spatial points around with 3-diffeomorphisms does not affect the physical content of the theory.    
The remaining information in the metric concerns the `underlying geometrical shape'.  
This is how GR is, more closely, a dynamics of 3-geometries in this sense (a geometrodynamics \cite{Battelle, DeWitt67} on the 
quotient configuration space superspace($\bupSigma$) = Riem($\bupSigma$)/Diff($\bupSigma$). 

\noindent Variation with respect to $\mh_{\mu\nu}$ provides the usual GR evolution equations (modulo functions of the constraints), 
which very straightforwardly propagate the above constraints via the Bianchi identities. 
\noindent The constraint algebroid closes as per (\ref{Mom-Mom}, \ref{Ham-Mom}, \ref{Ham-Ham}).

In the relational approach, GR has an analogue of emergent JBB time 
\beq
\lmt^{\se\sm(\sJ\sB\sB)}(\ux) = 
\stackrel{    \mbox{\scriptsize extremum} \mbox{ } \sF^{\mu} \mbox{ }  \in  \mbox{ }  \mbox{\scriptsize Diff}(\bupSigma)        }   
         {    \mbox{\scriptsize of} \mbox{ } \stS^{\mbox{\tiny relational}}_{\tG\tR}    }  
\left. \int ||\pa_{\suF}\bh||_{\sbM} \right/   \sqrt{     \mbox{${\cal R}\mi\mc$}(\ux; \bh] - 2\Lambda    }  \mbox{ } .
\label{GRemt2}
\eeq
This represents the same quantity as the usual spacetime-assumed formulation of GR's {\it proper time}, and, in the (predominantly) homogeneous cosmology setting, the {\it cosmic time}. 
It also coincides to lowest order with GR's own version of {\it emergent semiclassical} (alias {\it WKB}) {\it time} (explained in Secs \ref{QM-POT-Strat} and \ref{Semicl}).

The relational approach extends to standard model matter (again straightforwardly for bosonic fields \cite{RWR, AB, Van, Than}   
but requiring the generalization of Appendix \ref{Examples}.C for Dirac fields (\cite{Van}, Sec \ref{AFermi}).

\subsubsection{Best Matching versus Baierlein--Sharp--Wheeler Sandwich and Christodoulou Chronos Principle}\label{BM-BSW-Chronos}

The Multiplier elimination BSW 1)--3) can be reformulated (Appendix \ref{Examples}.A.1) as a cyclic velocity elimination, i.e. a passage to 
the Routhian \cite{Lanczos, Goldstein} directly equivalent to the Introduction's Euler--Lagrange action to Jacobi action. 

\noindent BSW 4--6) directly parallel Best Matching 1--3) modulo the upgrade from multiplier shift to cyclic velocity/differential of the instant. 
Thus in the Geometrodynamics case, Best Matching is hampered in practise by the Thin Sandwich Problem.  

\noindent The Thin Sandwich Approach is both generalized, and placed on relational terms, by the Best Matching Procedure   
to 
\beq
\mbox{solving $\scL\scI\scN_{\sfZ} = 0$ for whichever form of $\fg^{\sfZ}$ auxiliaries it contains} \mbox{ } ;    
\label{BMMMM}
\eeq
if this is obstructed, that constitutes the Best Matching Problem.  
Moreover, `sandwich' is clearly a geometrodynamical, or more generally Diff(3)- or foliability-specific name, whereas `best matching' is a more universal one, 
so in fact I choose the name {\bf Best Matching Problem} for the general theoretical problem (\ref{BMMMM}) of which the Thin Sandwich Problem is but the particular 
case corresponding to theories with some kind of NOS Diff-invariance.  

\noindent BSW did not consider the culmination move Best Matching 4) of this which (merely formally) produces an action on superspace.  

\noindent BSW 7) is, however, a primitive version of the $\ft^{\se\sm(\sJ\sB\sB)}$ computation (via the $\fN$ to $\dot{\fI}$ to 
$\d\fI = \d  \ft^{\se\sm(\sJ\sB\sB)}$ and hence $\ft^{\se\sm(\sJ\sB\sB)}$ progression).  
There is a presupposed (BSW) versus emergent (relational approach) difference in the status of the computed object too.

\noindent As far as I am aware, Christodoulou \cite{Christodoulou1, Christodoulou2} was the first person to cast BSW 7) as an emergent time 
[amounting to the differential form of (\ref{Kronos}), which he termed the `{\it chronos principle}'].  
Moreover, like Wheeler, he did not base this on relational first principles. 
That said, Christodoulou was certainly aware in doing so of some of what Barbour and I have held to be the Leibniz--Mach position on 
relational time: ``{\it They contain the statement that time is not a separate physical entity in which the changing of the physical system takes place.  
It is the measure of the changing of the physical system itself that is time}".  
He deduced the generalization of (\ref{BMMMM}) to include gauge fields minimally-coupled to GR involving shift 
and the Yang--Mills generalizations of the electric potential \cite{Christodoulou3}. 
I credit Alevizos for pointing out to me this connection between Barbour's and Christodoulou's work.  

\noindent BSW 8) is  subsequent step concerning dynamical evolution which for Geometrodynamics has nontrivial geometrical content.  

\noindent I have thus clearly laid out how the technical content of Barbour's Best Matching notion resides within a very immediate generalization of the original BSW paper.  
Wheeler did not, however, have the relational significance that underpins  this procedure, or the cyclic auxiliary coordinate formulation of the next SSec.

\subsubsection{Some more interpretations of Geometrodynamics}\label{Gdyn-+Interps}

\noindent Wheeler also asked \cite{Battelle} why $\scH$ takes the form it does and whether this 
could follow from first principles (`7th route' to GR) rather than from mere rearrangement of the Einstein equations.  
To date, there are the following two different answers which tighten wide classes of ans\"{a}tze down to the GR form (see \cite{Phan} for a comparison of these two 
).  

\mbox{ } 

\noindent Interpretation 6) [Hojman--Kucha\v{r}--Teitelboim] (HKT) \cite{HKT} is from {\it deformation algebroid} first principles.
Whilst the action of $\underline{\scM}$ involves but shifting points around within a given hypersurface, the action of $\scH$ involves deforming theat hypersurface itself.

\noindent N.B. This approach assumes embeddability into spacetime.  
%
{            \begin{figure}[ht]
\centering
\includegraphics[width=0.8\textwidth]{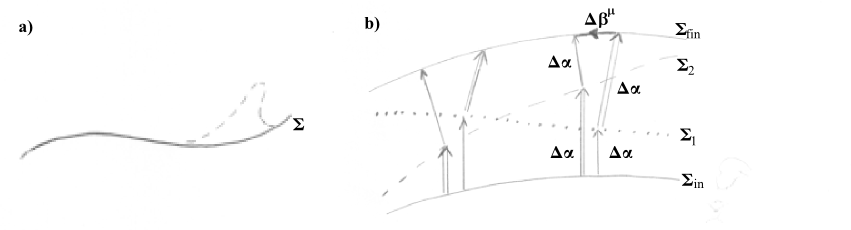}
\caption[Text der im Bilderverzeichnis auftaucht]{        \footnotesize{a) 
As already covered, ${\scM}_{\mu}$ generates stretches {\sl within} $\bupSigma$.
On the other hand, $\scH$ acts on spatial hypersurface $\bupSigma$ by deforming it into another hypersurface (dotted line).  

b) Illustration of Teitelboim's Refoliation-Invariance construct that holds at the classical level 
thanks to the form of the Dirac algebroid.  
One then considers going between $\bupSigma_{\si\sn}$ and $\bupSigma_{\sf\si\sn}$ via 2 different orderings of deformations, 
the first proceeding via intermediate hypersurface $\bupSigma_{1}$ and the second via intermediate hypersurface $\bupSigma_{2}$.  
Then one {\sl only} needs the stretch associated with $\Delta\beta^{\mu}$ to compensate for the non-commutativity of the two deformations involved.
Note that the diagram is itself a graphical formulation of the $\scH$, $\scH$ non-commutation.}  }
\label{Teit}\end{figure}          }
%
\noindent Interpretation 6b) Is HKT's work redone based on the A-, rather than ADM-, action.  
Here the Dirac Algebra has $\upalpha \longrightarrow \dot{\upgamma}$ and $\underline{\upbeta} \longrightarrow \underline{\dot{\mF}}$ smearings, and the Teitelboim path independence 
diagram has $\dot{\mF}^{\mu}\d x_{\mu}$ in place of $\upbeta^{\mu}\d x_{\mu}$.  

\noindent Interpretation 6c) is HKT's work redone based on the BFO--A relational action.  
Here the Dirac Algebra has $\upalpha \longrightarrow \dot{\mI}$ smearing, with $\underline{\dot{\mF}}$ retaining its various roles.  
This corresponds to adopting relationalist principles but also {\sl still} assuming spacetime structure.  

\noindent Interpretation 7) [Barbour--Foster--O'Murchadha and I] \cite{RWR, AB, San, OM02, OM03, Van, Than, Lan, Phan, Lan2} 
the `{\it relativity without relativity}' approach, which derives the relational form of GR from less structure assumed (including not presupposing spacetime) as the next SSSec outlines.

\subsubsection{Relativity without Relativity}\label{RWR}

Furthermore, adopting the relational first principles, {\sl without} assumption of additional features derived in ADM's approach, leads to the 
{\sl recovery} of the BFO-A action of GR as one of very few consistent choices within a large class of such actions. 
E.g. one does not need to assume the GR form of the kinetic metric or of the potential. 
One then has a family of trial actions $\fJ\fB\fB[P(\langle \mbox{Riem}(\bupSigma), \mbox{ } \bM^W\rangle, \mbox{ } \mbox{Diff}(\bupSigma))]$ for $\bM^{W}$ 
the general ultralocal supermetric ${\mM}^{\st\sr\si\sa\sll}_{\sq\su\sa\sd}\mbox{}^{\mu\nu\rho\sigma} := 1/Y\{\mh^{\mu\rho}\mh^{\nu\sigma} - W \mh^{\mu\nu}\mh^{\rho\sigma}\}$  
and a potential ansatz such as $\mW^{\st\sr\si\sa\sll} = A + B\,\mbox{${\cal R}\mi\mc$}(\ux; \bh]$ for $A$, $B$ constant,   
\beq
\FS^{\sR\sW\sR}_{\st\sr\si\sa\sll} = \sqrt{2}\int\int_{\sbSigma}\d^3x\sqrt{h}\sqrt{\mW^{\st\sr\si\sa\sll}}\pa\ms^{\st\sr\si\sa\sll}_{\sq\su\sa\sd}  \mbox{ } .  
\eeq
Then relational postulates alongside a few simplicities already give this since the Dirac procedure \cite{Dirac} prevents most other choices of potential term 
$\d \ms^{\st\sr\si\sa\sll}$ from working \cite{RWR, San, Than, Lan, Phan, OM02, OM03} alongside fixing $W$ to take the DeWitt value 1 of GR as one of very few possibilities.
More concretely \cite{Phan, Lan2},
\beq
H_{\sg\se\sn\se\sr\sa\sll} = \frac{1}{Y}\{\mh^{\mu\rho}\mh^{\nu\sigma} -  W \mh^{\mu\nu}\mh^{\rho\sigma}\}\mK_{\mu\nu}\mK_{\rho\sigma} - A - B \, \mbox{${\cal R}\mi\mc$}(\ux; \bh] = 0
\eeq
Then one obtains the constraint-propagation obstruction term
\beq
\dot{H}_{\sg\se\sn\se\sr\sa\sll} \approx \frac{2}{\dot{\mI}} Y A \{W - 1\} \mD_{\mu}\{ {\dot{\mI}}^2 \mD^{\mu} \uppi\} \mbox{ } .  
\eeq
Then considering each factor present in the obstruction term in turn, 
$Y = 0$ is the Galilean option ($c$ infinite) \cite{Lan, Than}.
$A = 0$ is the opposite Carollian option ($c$ zero); this corresponds to strong gravity, which is a reasonable model in the vicinity of curvature singularities \cite{San}.
$W = 1$ is the GR option: $c$ finite and the supermetric coefficient fixed to take the DeWitt value that characterizes GR as a dynamical theory \cite{RWR}.
The last factor \cite{RWR} allows for conformal options, via its $\uppi = 0$ and $\uppi/\sqrt{\mh} =$ constant subcases \cite{ABFO, ABFKO}, Sec \ref{Rel-CGdyn}.


\noindent Note 1) The last factor covers privileged foliation theories as well as ones that are reslicing invariant and thus formulations of (a restriction of) GR.

\noindent Note 2) It is curious for the three types of local relativity to come accompanied by a privileged foliation (or hidden time or frozen) option.  
The obstruction term's factors thus represent a joint packaging not seen before in Theoretical Physics; this could lend credibility to privileged foliations 
and/or York time as arising as alternatives to local special, Galilean and Carrollian relativities.
This satisfies all of Criteria 1) to 4).  

\noindent Note 3) the third factor includes ++++ as well as --+++ signature GR via the two possible signs of $B$.  

\noindent Note 4) $A = 0$ was found \cite{San} not to require $W = 1$ (the usual strong gravity limit of GR); $W \neq 1$ are the strong gravity limits of scalar--tensor theories. 

\noindent Note 5) Talking about local relativities only makes sense once one is considering multiple matter fields. 
Usually one assumes that these have to obey the same local relativity.
In the present approach, one {\sl derives this as a consistency condition}, in the sense that simple matter fields' propagation modes are found to be forced to share the null cone 
(or its degenerate plane and line cases corresponding to $c = \infty$ and $c = 0$ respectively) of the gravitation, and thus all end up sharing null cone with each other.   

\noindent Note 6) A word of caution: the word `simple' here is crucial; it means `minimally coupled'.  
Without this, the matter and gravitation are more interlinked and are both needed to establish the propagation modes.

\noindent Note 7) A further word of caution: whilst the earlier papers on this claimed to {\sl uniquely pick out} not only locally-SR physics null cones but also gauge theory and even 
the Equivalence Principle, these results did not stand up robustly to \cite{Lan, Than, Lan2} using knowledge of the split-spacetime formulations of more canonically complicated matter 
theories.  
Other simplicities involve neglecting the possibility of metric--matter cross-terms and other terms that one knows appear in the canonical 
formulation of more general fields, as well as the restriction on allowed orders of derivatives that is much more ubiquitous in Theoretical Physics.  
By such means, some more general/complicated possibilities can be included among the relationally-formulable theories:

\noindent 1) Brans--Dicke theory \cite{RWR}, 

\noindent 2) Proca theory \cite{Lan}, and 

\noindent 3) Local-SR-cone-violating and Equivalence Principle violating vector--tensor theories \cite{Lan2} (a subset of Einstein--Aether theories \cite{AET}).

\mbox{ }  

\noindent Thus, indeed, one can only have recovery of the SR null cone if the matter fields are kept simple.  
The equivalence principle derivation claim was particularly unravelled due to \cite{RWR} having been found to tacitly 
assume what had been known to be a `geometrodynamical' form of equivalence principle since Hojman--Kuchar--Teitelboim's much earlier paper \cite{HKT}.  

\mbox{ }

\noindent {\bf Question 1} A further gap, not yet addressed, is that in RWR the possibility of second-class constraints is not properly 
catered for in any uniqueness by exhaustion proof so far given.  
[I am studying this with Flavio Mercati.]

\subsubsection{Minisuperspace}\label{MSS}

These are GR models that are spatially homogeneous; this simplifies working with them but at the expense of having any notion of structure (formation), 
linear constraints or diffeomorphisms.
${\mM}^{\mu\nu\rho\sigma}(\mh_{\gamma\delta}(x^{\omega}))$ collapses to an ordinary $6 \times 6$ matrix ${M}_{AB}$ or 
further in the diagonal case (a $3 \times 3$ matrix $M_{\sfA\sfB}$) -- the `minisupermetric'.  
Since there are no diffeomorphisms to implement, the BFO-A to BSW action distinction dissolves here, 
likewise the pair of actions (\ref{ADM}, \ref{AAAAAc}) and this case requires no Best Matching and thus has an explicitly-evaluated t$^{\se\sm(\sJ\sB\sB)}$ too,
\beq
\lt^{\se\sm(\sJ\sB\sB)} = \left.\int \d \ms_{\mbox{\scriptsize\boldmath$M$}}\right/ 
\sqrt{Ric(\mbox{\boldmath$h$}) - 2\Lambda} \mbox{ } .  
\eeq
That gives it very much the same status as for Jacobi Mechanics, except that one has kept a remnant of GR's kinetic indefiniteness and of GR's specific potential.  

\noindent See Secs \ref{Cl-Jac-POT} and \ref{Q-Jac-POT} for outlines of which POT facets these manifest at the classical and quantum levels respectively.

\subsection{Relational formulations of Conformogeometrodynamics}\label{Rel-CGdyn}

In the geometrodynamical formulation, GR's true dynamical degrees of freedom are Wheeler's ``2/3 of superspace", 
which his subsequent postdoc Jimmy York worked out the nature of \cite{York71, York72, York73, York74}.     
Conformal superspace [CS($\bupSigma$)] is the space of all conformal 3-geometries on a fixed spatial topology, i.e. the result of quotienting Riem($\bupSigma$) by the semidirect product of 
Diff($\bupSigma$) and Conf($\bupSigma$), the conformal transformations, gives the right count for this whilst being geometrically natural.  
However do bear in mind that the 2/3 of Superspace($\bupSigma$) that CS($\bupSigma$) picks out might, however, 
not be directly related with the 2/3 of Superspace($\bupSigma$) picked out by the Hamiltonian constraint itself 
-- let us call this True($\bupSigma$) whilst acknowledging its purely formal character.
This may cause me to doubt a subset of this Sec due to Criterion 1).

\subsubsection{Lichnerowicz's and York's Conformogeometrodynamics}\label{CGdyn}

Traditionally, Conformogeometrodynamics is viewed as a convenient decoupling leading to substantial mathematical and numerical tractability.  
Lichnerowicz's work in this direction \cite{Lich44} is based on maximal slicing (\ref{maxsl}), which York  \cite{York72, York73} 
generalized to CMC slicing (\ref{CMCsl}) const $= \mK$ (the trace of the extrinsic curvature, which is also equal to $-\uppi/2\sqrt{\mh}$).  
This is an important generalization as regards spatially-compact spacetimes, for which maximal slicing cannot be propagated (and important as regards numerical GR too 
\cite{BS03, Gour}).   
We will subsequently see that the CMC is proportional to a hidden internal time (York time).

This can be viewed as a departure from the Thin-Sandwich program. 
Both have by now much POT strategization built upon them.
Conformogeometrodynamics has developed much more as regards classical constraint solving. 
A particular adaptation of this is presented below.

The constraints are decoupled in this formulation [meeting Criterion 2)] because the momentum constraint happens to be conformally covariant so 
one can solve it irrespective of the subsequent solution of the conformally-transformed Hamiltonian constraint for the physical scale.  
In the maximal slicing case this last equation takes the form 
\beq
8 \triangle_{\sbh}\upphi = \mbox{${\cal R}\mi\mc$}(\ux; \bh]\upphi - {\uppi_{\mu\nu}\uppi^{\mu\nu}}/{\sqrt{\mh}}\upphi^{7}  \mbox{ } ,
\label{Lich}
\eeq
which is known as the {\it Lichnerowicz equation}.
In the CMC case, this generalizes to the {\it Lichnerowicz--York equation},
\beq
8\triangle_{\sbh}\upphi = \mbox{${\cal R}\mi\mc$}(\ux; \bh]\upphi - {\uppi_{\mu\nu}\uppi^{\mu\nu}}/{\sqrt{\mh}}\upphi^{7} + {\uppi^2}/{6\sqrt{\mh}}\upphi^5 \mbox{ } .   
\label{LYE} 
\eeq
\noindent Note 1) The above are complicated nonlinear equations, though they do have the benevolent feature of being quasilinear; this 
allows for substantial existence and uniqueness theorems for them \cite{CBY80, I95} and numerical relativity work \cite{BS03} albeit very much not for exact solutions work.
In this respect, this 1972 approach of York's \cite{York72} to the GR IVP (initial-value problem) is far better off than Wheeler's earlier thin sandwich conceptualization of the IVP. 

\noindent Note 2) For minisuperspace, the  Lichnerowicz--York equation becomes a merely algebraic polynomial equation 
\beq
\mbox{${\cal R}\mi\mc$}(\ux; \bh]\upphi - {\uppi_{\mu\nu}\uppi^{\mu\nu}}/{\sqrt{\mh}}\upphi^{7} + {\uppi^2}/{6\sqrt{\mh}}\upphi^5 = 0
\mbox{ } ,    
\label{LYE-MSS} 
\eeq
which does have at least some exactly-solvable cases. 

\noindent Note 3) Maximal slicing is maintained if the lapse solves the maximal lapse-fixing equation (LFE) 
\beq
\triangle_{\sbh} \upalpha = \upalpha\,\mbox{${\cal R}\mi\mc$}(\ux; \bh] 
\eeq
albeit this is frozen for compact spatial topology. 
CMC slicing is maintained if the lapse solves the CMC LFE 
\beq
2\{\upalpha\,\mbox{${\cal R}\mi\mc$}(\ux; \bh] - \triangle_{\sbh} \upalpha\} + \upalpha \, \uppi^2/2h = \{\uppi/\sqrt{\mh}\}\,\dot{\mbox{}} \mbox{ } .  
\label{CMCLFE}
\eeq

\noindent Note 4) The conformogeometrodynamical approach straightforwardly extends to standard model matter fields for certain particular scalings of these \cite{IOY, IN77}.  

\mbox{ } 

\noindent The next three SSSecs consider some possible foundations for the conformogeometrodynamical approach; for now, these all represent LMB-relational work.  

\mbox{ }

\noindent {\bf Question 2$^*$}  Is there any conformogeometrodynamical parallel of the HKT route to Geometrodynamics?

\subsubsection{Conformal Relationalism: the frozen theory}\label{CGdyn-Frozen}

Consider $\FrQ$ = Riem($\bupSigma$) and $\FrG$ = Diff($\bupSigma$) $\mbox{\textcircled{S}}$ 
Conf($\bupSigma$) where Conf($\bupSigma$) is the group of conformal transformations on $\bupSigma$.    
The configuration spaces had been previously studied e.g. in \cite{York72, York74, FM, FM96}.  
Then one has somewhat more of a problem than usual in getting `good' Conf($\bupSigma$)-objects.
Then conformal superspace CS($\bupSigma$) = Riem($\bupSigma$)/Diff($\bupSigma$)\mbox{\textcircled{S}} Conf($\bupSigma$). 
One needs a $\upphi$ such that the metric and it form a simple, internally-conformally-invariant pair
\beq
\upphi \longrightarrow \upomega^{-1}\upphi \mbox{ } , \mbox{ } \mh_{\mu\nu} \longrightarrow \upomega^4\mh_{\mu\nu} 
\eeq
so $\mh_{\mu\nu}\upphi^4$ is Conf($\bupSigma$)-invariant and it is out of this that the action is to be built.  
Writing this is a {\sl local} scale--shape split (since $\upphi$ depends on position).  
Then such an action is 
\beq
\FS^{\sG\sR}_{\sp\sr\so\st\so-\sA\sB\sF\sO} = \int\d\int_{\sbSigma}\d^3x\sqrt{\mh}
\upphi^6\sqrt{  \upphi^{-4}\{\mbox{${\cal R}\mi\mc$}(\ux; \bh] -8\triangle_{\sbh}\upphi/\upphi\}  }\d \ms^{\sG\sR }_{\sp\sr\so\st\so-\sA\sB\sF\sO}
\label{GRconfaction}
\eeq
where\foo{For simplicity, I 
present this with no $\Lambda$ term; see \cite{ABFO} for inclusion of that.  
Also, they themselves considered this using two separate multipliers instead of a single more general 
auxiliary whose velocity also features in the action and has to be free end hypersurface varied. 
The way this is presented here is that of \cite{ABFKO}, whose variation is justified in Appendix \ref{Examples}.A.3.}

\noindent
$$
\pa\ms^{\sG\sR\, 2}_{\sB\sO} = \{  \upphi^{-4}\mh^{\mu\rho}\upphi^{-4}\mh^{\nu\sigma} - \upphi^{-4}\mh^{\mu\nu}\upphi^{-4}\mh^{\rho\sigma}\}
\Circ_{{\suF}}\{\upphi^4\mh_{\mu\nu}\}\Circ{{\suF}}\{\upphi^4\mh_{\rho\sigma}\} = 
$$
\beq
\{\mh^{\mu\rho}\mh^{\nu\sigma} - \mh^{\mu\nu}\mh^{\rho\sigma}\}
\{\pa_{{\suF}} \mh_{\mu\nu} +     4 \mh_{\mu\nu}    \pa\upphi/\upphi\}
\{\pa_{{\suF}} \mh_{\rho\sigma} + 4 \mh_{\rho\sigma}\pa\upphi/\upphi\} \mbox{ } .
\eeq
(Note that the combination $\{\mbox{${\cal R}\mi\mc$}(\ux; \bh] - 8\triangle_{\sbh}\}\upphi$ is conformally invariant.)

This action then gives as a primary constraint the Lichnerowicz equation (\ref{Lich}), $\scM_{\mu}$ as a secondary constraint from 
$\mF^{\mu}$-variation, and the maximal slicing condition (\ref{maxsl}) from a part of the FESH $\upphi$-variation.
However the other part of this last variation entails frozenness for the CWB $\bupSigma$'s of interest. 
[This parallels the shortcoming with Lichnerowicz's work that York overcame, though I now recast this as a frozen 
VOTIFE -- i.e. a fixing equation for the velocity of the instant $\dot{\mI}$  -- rather than for the lapse $\mN$.] 
Ultimately this entailment is as a frozen `POTIFE'.

\subsubsection{Conformal Relationalism: the alternative theory}\label{CG}

Barbour and \'{O} Murchadha \cite{BO} and ABFO \cite{ABFO} got round this frozenness by considering a new action 
obtained by dividing the above one by $\mbox{Vol}^{2/3}$ (a manoeuvre reminiscent of the Yamabe Conjecture \cite{Yamabe, Schoen} familiar from the GR IVP literature).  
This amounts to again fully using $\FrG$ = Diff($\bupSigma$) \textcircled{S} Conf($\bupSigma$). 
The subsequent variational principle is 
\beq
\FS_{\sA\sB\sF\sO} = \FS_{\sB\sF\sO-\sA}[\upphi^4 \mh_{\mu\nu}]/\mbox{Vol}^{2/3} = 
\int \int_{\sbSigma}\d^3x\sqrt{\mh}\upphi^4
\sqrt{        \{\mbox{${\cal R}\mi\mc$}(\ux; \bh] - 8\{\triangle_{\sbh}\upphi\}/  \upphi   \}                         }
            ||\pa{\bh} + 4\pa{\upphi}\bh/\upphi ||_{\sbM}   /         \mbox{Vol}^{2/3} \mbox{ } .       
\label{Action-ABFO}
\eeq
Note however that this is an alternative theory and not GR; moreover it is questionable as an alternative theory \cite{ABFO, Lan, Than}) due to there being no means by which it can 
explain observational cosmology; I include this variational principle here only for comparison with the next SSSec and for an entirely different application in Sec \ref{Dist}.    
ABFKO \cite{ABFKO} got round the frozenness in a different way (see the next SSSec).

\subsubsection{Conformal Relationalism: the recovery of York's formulation}\label{CS+V}

Instead by using $\FrG$ = Diff($\bupSigma$) \textcircled{S} VPConf($\bupSigma$), where VPConf($\bupSigma$) are the volume-preserving conformal transformations on $\bupSigma$ as implemented by using 
\beq
{\upphi}_{\sV\sP} := \upphi\left/\left\{\int_{\sbSigma}\d^3x\sqrt{\mh}\upphi^6\right\}^{1/6}\right. \mbox{ } .  
\eeq
These are associated with the constant mean curvature condition (\ref{CMCsl}).
Then \{CS + V\}($\bupSigma$) = Riem($\bupSigma$)/Diff($\bupSigma$)  \textcircled{S} VPConf($\bupSigma$) \cite{ABFKO} (which previously featured in e.g. \cite{York72}.  

\mbox{ }

\noindent {\bf Question 3}.   \{CS + V\}($\bupSigma$) has not to my knowledge yet been studied in detail from a geometrical perspective. 

\mbox{ }  

\noindent [The CRiem($\bupSigma$) = Riem($\bupSigma$)/Conf($\bupSigma$) analogue of preshape space has been 
studied geometrically in e.g. \cite{DeWitt67, FM, FM96}, in the last of which it is termed `pointwise conformal superspace'].

The action for this formulation is
\beq\FS_{\sA\sB\sF\sK\sO} = \FS_{\sB\sF\sO-\sA}[\upphi_{\sV\sP}^4 \mh_{\mu\nu}] = 
\int \int_{\sbSigma}\d^3x\sqrt{\mh}\upphi_{\sV\sP}^4
\sqrt{        \{\mbox{${\cal R}\mi\mc$}(\ux; \bh] - 8\{\triangle_{\sbh}\upphi_{\sV\sP}\}/  \upphi_{\sV\sP}   \}                         }
            ||\pa{\bh} + 4\pa{\upphi_{\sV\sP}}\bh/\upphi_{\sV\sP} ||_{\sbM}   \mbox{ } ,       
\label{Action-ABFKO}
\eeq
which is a JBB$(\langle \mbox{Riem}(\bupSigma), \bM \rangle, \mbox{Diff}(\bupSigma) \mbox{\textcircled{S}} \mbox{VPConf}(\bupSigma))$.

In this scheme the primary constraint (now denoted by $\scH_{\hat{\sV\sP}}$) becomes the Lichnerowicz--York equation (\ref{LYE}) 
and a more complicated realization of FESH $\upphi$-variation (that rests on the Lemma in Appendix \ref{Examples}.A.3) now gives the CMC condition 
(\ref{CMCsl}) alongside a LFE (or, fully relationally speaking, a DOTIFE) that successfully maintains this.

The emergent JBB time is now 
\beq
\lmt^{\se\sm(\sJ\sB\sB)} =  \stackrel{\mbox{\scriptsize extremum $\underline{\mF}, \upphi \in$ Diff($\bupSigma$) \textcircled{S} VPConf($\bupSigma$)}}
                                               {\mbox{\scriptsize of $\stS^{\tA\tB\tF\tK\tO}$}}
\left.\int\pa \ms^{\sA\sB\sF\sK\sO}(\ux; \bh, \uF,\upphi]\right/\sqrt{2\mW(\ux; \bh, \upphi]}  \mbox{ } .
\eeq

\mbox{ }

\noindent{\bf Question 4}. There are hints that there is {\sl not} a `relativity without relativity' result for Conformogeometrodynamics \cite{Lan2}. 
Can this be verified by working within a fully \{CS + Vol\}($\bupSigma$) picture with general trial actions?

\subsubsection{Conformal Relationalism: the Linking Theory}\label{275}

\noindent The `linking theory' shape dynamics program by Gryb, Koslowski, Gomes, Mercati and Budd \cite{BONew, Kos1, Kos2, Gomes, Kos3, Kos4, Gom2, GK12} 
is somewhat related to this, though it is as yet not clear to me whether this continues to be anchored to the relational first principles.
Certainly, their use of a Hamiltonian formulation is cleaner; however see Appendix \ref{Dyn1}.B about their non-minimalism. 
There have been some attempts ( see e.g. \cite{Kos1, GrybTh, GK12}) to relate the linking theory to some of the AdS--CFT literature.

\subsubsection{Conformal Relationalism: Doubly General Relativity}\label{276}

A `doubly general relativity' shape dynamics program by Koslowski and Gomes has also now appeared \cite{GomKos12}, which purports to place rods and clocks on an equal footing.
[However, I caution that there are considerable practical and material inequivalences between these, c.f. Secs \ref{Clocks} and \ref{ObsApp}.]

\subsubsection{Conformal Relationalism: Cartan alias hamster-ball geometry}\label{277}

Gryb and Mercati \cite{Hamster} have considered GR shape dynamics as arising in this way (a manifold probed by rolling a homogeneous space around on it).
This approach simplifies considerably for the 2 + 1 GR toy models.

\subsubsection{Discussion}

\noindent{\bf Question 5}. The addition of matter to the ABFKO scheme has not to date been extensively studied. 

\mbox{ }

\noindent (So far, no significant hindrances have been found \cite{MacSweeney}. 
Other at least partially-relevant results are the addition of matter to the ABFO scheme \cite{ABFO} and in the below `shape dynamics' program \cite{Gomes, Gom2}. 
These have the common feature of respecting the same scalings as in \cite{ABFO}.)  

\mbox{ } 

\noindent{\bf Question 6$^*$}  Are the three approaches of Secs \ref{275}-7 to be taken as another case of `many routes' to the same physics [Criterion 1)]?  
Or are there concrete reasons [Criterion 1)] to favour one of these three approaches over the others?  
To what extent are the three approaches physically equivalent?

\mbox{ }

\noindent The present Article is for now quite clearly about a {\sl different} path from \cite{ABFO, ABFKO} and the preceding 3 SSSec's papers 
toward conceptual quantum gravity considerations; it is not yet clear whether the two shall overlap or cross-fertilize.  
It is also not clear whether the linking theory and AdS--CFT can be technically meshed together due to being of too great a conceptual heterogeneity.

\subsection{RPM's and GR: 118 analogies and 49 differences}\label{Analogy-Diff}

These analogies are my main motivation for studying RPM's.  
Take the below to be a summary of those encountered so far, a few new ones that are appropriate at the present position of development in this Article. 
On the other hand, I also refer forward to other Secs for many further ones, due to it being best for the Article to lay out some more context before these appear.

\subsubsection{Analogies and Differences at the level of GR's most standard features}\label{AD-GR-Std}

\noindent Difference 1) RPM's are not explicitly theories of gravitation.
E.g. they are not rigidly restricted to having Newtonian-gravitation-like potentials or a GR cosmology-like dynamical structure.

\noindent Analogy 1) RPM's are, however, explicitly theories of Background Independence, which is a feature of GR 
(and theories beyond GR) emphasized by Einstein, in Nododynamics/LQG and in the build-up to M-Theory, and which I am arguing to be 
conceptually important enough that GR is a Gestalt Theory of Gravity and Background Independence.  

\noindent Thus RPM's are valuable in isolating this aspect so as to see what comes of its study in the absence of some of the other complicating features. 

\noindent Analogy 2) RPM's are appropriate as toy models for closed system/whole universe physics.  

\noindent Difference 2) RPM's do not possess any elements of SR. 

\noindent This is less debilitating than might na\"{\i}vely be expected, due to the NM to SR passage just involving trading one set of absolute structures for another.  

\noindent Difference 3) RPM's do not possess a nontrivially GR-like notion of spacetime.  

\noindent However, there is a case that all Physics prior to SR was based on dynamics, and this current has continued to run in all 
physical theories since and might ring truer, pace Minkowski (c.f. Sec \ref{tfns}).

\noindent RPM's do moreover possess a notion of {\bf strutting} (lapse and point identification map). 
But the RPM space-time is not furtherly a recoordinatizing and reslicing invariant geometrical structure like GR spacetime is [see after Differences 8-11) for why].  

\noindent Analogy 3) RPM's are a dynamical formulation parallelling in particular GR's geometrodynamical formulation.  

\noindent Analogy 4) RPM's and GR are both relational theories in this Article's LMB(-CA) sense.    

\noindent Difference 4) RPM's are finite rather than infinite/field-theoretic.  

\noindent This is in part a difference with benefits due to it rendering RPM's more tractable than GR, 
e.g. no regularization issues and less equation well-definedness issues (see Sec \ref{Wave-Eqs}).

\noindent On the other hand, it does preclude some interesting aspects; as a first example for now is that it precludes analogies of the 
manifestly field-theoretic fifth and sixth routes to GR.  

\noindent Finally, I comment that in some cases the presumption that field theory is required for a number of further issues to 
manifest themselves is in fact mythical (see Sec \ref{FacetsDet}).  

\noindent Difference 5) RPM's do not make contact with the inherently nonlinear nature of GR.  

\noindent This is not unexpected, since the nonlinearity signifies that gravity gravitates and is thus a feature of the Gravity side 
of the Gestalt rather than the Background Independence part which RPM's model.  

\noindent Analogy 5) Analogies 1), 2) and 3) render RPM's useful as qualitative models for classical and quantum Cosmology.

\noindent Analogy 6)  RPM's are furthermore useful thus through having an analogous notion of clumping/inhomogeneity/structure and thus of structure formation. 

\noindent Difference 6) There is no nontrivial analogue of which I am aware of black holes in RPM's. 
[I.e. of the principal arena other than cosmology in which QM and GR can be simultaneously relevant.]

\noindent Analogies 3), 4), 5) make it clear Criterion 3) of smoother interpolation between mechanics and GR applies to the relational formulation.

\subsubsection{Analogies at the level of configuration spaces}\label{Analogy-Q}

\noindent Analogy 7) In GR-as-Geometrodynamics, the incipient straightforward configuration space role of $\FrR$($N$, $d$)'s 
is played by the space Riem($\bupSigma$) of Riemannian 3-metrics on a fixed spatial topology $\bupSigma$. 

\noindent Analogy 8) Both RPM's and GR are constrained dynamical systems. 

\noindent Analogy 9) Both can be cast to obey Configurational Relationalism implemented indirectly by arbitrary $\FrG-$frame corrections. 

\noindent Analogy 10) The role of Rot($N$, $d)$'s as group of irrelevant transformations $\FrG$ is played by the 3-diffeomorphism group, Diff($\bupSigma$).  
This analogy goes further through both groups being nonabelian and having nontrivial orbit structure 
(though, as that requires $d > 2$, this further aspects are beyond the scope of this Article's specific examples).  

\noindent Moreover, I am not aware of any consequences specifically of nonabelianness of $\FrG$ for the POT.  
N.B. this is not one of the groups that plays a global role in \cite{I84}'s account. 

\noindent Analogy 11) The role of Dil as a further contribution to the group of irrelevant transformations is played in a 
certain sense by the conformal transformations Conf($\bupSigma$) and in a certain sense by VPConf($\bupSigma$) \cite{ABFKO}.  
This analogy goes further in the sense that these are all scales, though Dil is global. 
(Here this is global in the sense of pertaining to the whole system rather than to any particular cluster therein.)
Conf($\bupSigma$) is local and VPConf($\bupSigma$) is `local excluding one global degree of freedom' (the global volume).  

\noindent Analogy 12) Relational space = $\FrR(N, d)$/Rot($d$) corresponds to superspace($\bupSigma$) = 
Riem($\bupSigma$)/Diff($\bupSigma$) \cite{DeWitt67, DeWitt70, Fischer70}.  

\noindent Analogy 13) Preshape space is the analogue of 
``CRiem($\bupSigma$)" \cite{08I} alias ``pointwise version of CS($\bupSigma$)" \cite{FM96}.  

\noindent Analogy 14) Shape space = $\FrR(N, d)$/Rot $\times$ Dil for analogous to conformal superspace \cite{York74, ABFO}.   

\noindent CS($\bupSigma$) = Riem($\bupSigma$)/Diff($\bupSigma$) $\times$ Conf($\bupSigma$) \cite{York72, York73, Yorktime1, York74, ABFO, FM96}.

\noindent Analogy 15) Both RPM's and GR admit scale--shape splits.  
See Sec \ref{PS} for details about RPM and GR scale variables, and below and Secs \ref{Riem}, 
\ref{RiemS}, \ref{Shape-Quant}, \ref{GRCtr} for details about the shape variables in each case.
 
\noindent Analogy 16) \{CS + V\}($\bupSigma$) is analogous also to relationalspace (now in scale-shape split form, see Sec \ref{Q-Geom} for more on this).

\subsubsection{Analogies between RPM's and GR at the level of actions}\label{Analogy-Action}  

\noindent Analogy 17) Both can be cast to obey Temporal Relationalism implemented by MRI/MPI in each case.

\noindent Analogy 18) There is a relational action (\cite{RWR, Phan}, \ref{BFOA}) for GR-as-Geometrodynamics that is the 
direct counterpart of the indirectly-formulated scaled RPM action in particular.  

\noindent Analogy 19) There is also a relational \{CS + Vol\}($\bupSigma$) action (\cite{ABFKO}, \ref{Action-ABFKO}) that has further parallels 
with scale--shape split form of scaled RPM action. 

\noindent Partial Analogy 20) Also, the conformal gravity action has further parallels with pure-shape RPM; 
homogeneity requirements are incorporated in both via powers of some dimensionful quantity.  
Namely, $\mI$ in pure-shape RPM and the volume of the universe Vol in conformal gravity \cite{ABFO}. 
This is not GR, but this action is argued to have an application to GR in Sec \ref{Cl-Str}.

\noindent Analogy 21) Reading off (\ref{action}) and (\ref{GRaction}), energy $E$ and cosmological constant 
$\Lambda$ (up to factor of $- 2$) play an analogous role at the level of the relational actions. 
Thus, to some extent $E$ is a toy model for the mysteries surrounding $\Lambda$ (c.f. discussion in Sec \ref{SRPM}, but also contrast Analogy 51).

\noindent The way in which that the physical equations follow from the relational action for 
GR-as-Geometrodynamics and from the indirectly formulated RPM actions then have many parallels.

\noindent Analogy 22) Dirac's argument from MRI/MPI and the square root form of the action gives each a primary constraint quadratic in the momenta. 

\noindent Analogy 23) The ensuing quadratic constraints -- the Hamiltonian constraint (\ref{Hamm}) and the RPM `energy constraint' (\ref{EnCo}i) -- are analogous.   
This is then the basis of a number of further RPM-GR analogies.   
N.B. this is the outcome of demanding Temporal Relationalism.  

\noindent Analogy 24) Variation with respect to the auxiliary $\FrG$-variables gives each relational theory constraints linear in the momenta.

\noindent Analogy 25) Forms of these: for GR, the momentum constraint $\scM_{\mu} := - 2\mD_{\nu}{\uppi^{\nu}}_{\mu} = 0$ 
from variation with respect to $\mF^{\mu}$, and, for RPM's, the zero total angular momentum constraint 
$\underline\scL := \sumin\mbox{ } \underline{\rho}^i \cr \underline{p}_i = 0$.   

\noindent Analogy 26) pure-shape RPM's zero total dilational momentum constraint, $\scD := \sumin\mbox{ } 
\underline{\rho}^i \cdot \underline{p}_i = 0$ closely parallels the well-known GR maximal slicing condition \cite{Lich44, York72, York73}, $\mh_{\mu\nu}\uppi^{\mu\nu} = 0$.  

\noindent Analogy 27) The best matching originally conceived of for RPM's is essentially a portion of BSW's approach to GR.  

\noindent Analogy 28) Both subsequent workings involve their own version of an emergent JBB time, each aligned with a number of other well-known notions of time.  

\noindent Analogy 29) `Dilational' conjugates to scale quantities are e.g. $\scD$ itself (now taken to be just an object -- the  
{\it Euler quantity} \cite{06II} -- rather than being a constraint equal to zero) that is conjugate to $\mbox{ln}\,\rho$. 
And e.g. the York quantity \cite{Yorktime1} $Y = \frac{2}{3}\mh_{\mu\nu}\uppi^{\mu\nu}/\sqrt{\mh}$ (which is a rescaling and reweighting 
of the object involved in the maximal slicing condition likewise regarded as a quantity rather than being equal to zero by the slicing condition) that is conjugate to $\sqrt{\mh}$. 
The York quantity is 4/3 times the mean curvature $\mK$; it indexes constant mean curvature (CMC) slices of (regions of) spacetime.  
This additional geometrical interpretation would appear to have no counterpart in RPM's.

\subsubsection{Some differences between RPM's and Geometrodynamics}\label{RPM-Gdyn-Diff}

\noindent Helpful Difference 7) In GR, nontrivial structure and nontrivial linear constraint are tightly related as both concern the nontriviality of the spatial derivative operator.  
In particular, minisuperspace has neither.
However, for RPM's, the nontriviality of angular momenta and the notion of structure/inhomogeneity/particle clumping are unrelated.  
Thus, even in the simpler case of 1-$d$ models, RPM's have nontrivial notions of structure formation/inhomogeneity/localization/correlations between localized quantities.
This isolates one of these 2 midisuperspace aspects in the scaled 1-$d$ models, whilst having both in the pure-shape and 2-$d$ models.  

\mbox{ } 

\noindent Next, the rotations cannot emulate a number of further features that are more specifically of the 3-diffeomorphisms themselves. 

\mbox{ } 

\noindent Difference 8) The constraint algebras for RPM's (\ref{first6}, \ref{other4}) do not have  $\scL\scI\scN$-as-$\scQ\scU\scA\scD$-integrability -- 
unlike in the Dirac algebroid of Geometrodynamics.
Thus, for GR but not for RPM's, if one has only $\scH$, one discovers $\scH$ \cite{Dirac,T73,OM02,San}.  
It follows from this that in RPM's temporal and Configurational Relationalism are separate issues, but 
in GR they have to be taken together as a package, reflecting extra `togetherness' of the GR notion of spacetime.  

\noindent Difference 9) The presence of the $\mh^{\mu\nu}$ factor in the right-hand-side of the Poisson bracket of two Hamiltonian 
constraints for diffeomorphisms causes these to substantially differ from rotations in the following ways (paraphrasing \cite{T73, Tei73b}).  
One may speak of rotations without saying `whether it is a black cube, a green cube or a yellow cat' that is being acted upon by the rotation.  
Furthermore, the bracket of rotations is well-defined in terms of the original rotations and nothing else.
However in Geometrodynamics, in order to evaluate the bracket of two normal deformations, one requires the original deformations 
{\sl and} the spatial hypersurface with metric $\mh_{\mu\nu}$ that they were acting upon. 

\noindent Difference 10) Also, RPM's also have nothing like the embeddability/hypersurface deformation 
interpretation of the Dirac algebroid (which is very diffeomorphism-specific). 
Consequently, RPM's have nothing like the HKT \cite{HKT} first-principles route to Geometrodynamics (see also the next SSSec).  

\noindent Difference 11) RPM's have nothing like the GR spacetime 4-diffeomorphisms. 

\noindent One can now see why RPM space-time lacks the properties of GR spacetime that go beyond the strutting structure. 
This is because the RPM does not involve $\FrG$ = Diff($\bupSigma$) and its associated momentum constraint, or the Hamiltonian constraint, or the subsequent Dirac/deformation constraint 
algebroid or the spacetime diffeomorphisms, and it is these that give GR the extra recoordinatization of space, reslicing and unrestricted recoordinatization of spacetime properties 
that make it more than just a privileged slicing evolving with respect to a single finger's worth of time.

\noindent Difference 12) There is also no RPM counterpart of the `relativity without relativity' \cite{RWR, Phan, Lan2} first-principles route to Geometrodynamics.  

\noindent Analogy 30) One can however view the scale--shape split approach to scaled RPM as paralleling the first-principles 
route to Conformogeometrodynamics (`8th route to relativity') in \cite{ABFKO}, as per Analogies 26) and 29).   

\noindent Difference 13) However, this RPM model has no analogue of the VOTIFE for RPM's, 
which limits their usefulness as toy models as regards some of the as yet little-understood features of the scheme.  

\mbox{ }

\noindent Further  discussion of configuration spaces, RPM's whose scale parts mimic the cosmological scale dynamics, QM, and both 
facets of and strategies for the POT each bring out a number of further analogies and differences.
Finally, I present a figure in the Conclusion that gathers these together into a flowchart of dependency

\subsubsection{Many routes to Newtonian Mechanics}\label{Many-Routes-NM}

Newtonian Mechanics itself can also be considered as lying at the end of many routes [Criterion 1)].  

\noindent Criterion 5) Moreover there are analogies between some of these routes and some of those to GR. 

\mbox{ } 

\noindent NM Route 0) Newton's own route is sui generis rather than pairable with any GR route.  

\noindent NM Route 1) Cartan's route to Newtonian mechanics bears close parallels to Einstein's traditional route to GR.    

\noindent NM  Route 2) The variational work of Euler and Lagrange bears parallels with the Einstein--Hilbert action and, 
even more closely, with the ADM--Lagrangian action.   

\mbox{ } 

\noindent Note 1) Various of the above differences then apply, while scaled RPM parallels BFO-A's route 7B) to GR.  
Despite the above resemblances, I argue that there is {\sl not} a close parallel between the BB82 route to Newtonian Mechanics and the {\sl tightening} 
aspects of the relativity without relativity (RWR) and HKT deformation algebroid routes to GR.  
For, that tightening arises from the restrictiveness of the emergent or assumed  $\scL\scI\scN$-as-$\scQ\scU\scA\scD$-integrability. 
of GR, while the RPM's lack in this [i.e. Difference 12)].  

\noindent Note 2) Also, for scaled RPM, the variant of RWR that \'{O} Murchadha and I investigated \cite{OM02, San} (see also Sec \ref{Q-G-Comp}) fails here. 
This is because it relies on linear constraints first appearing via being integrabilities of $\scH$, but the  
analogue $\scE$  of this does not have $\scL\scI\scN$-as-$\scQ\scU\scA\scD$-integrability.    
  
\noindent Note 3) In scaled RPM, the standard RWR technique of using consistency under constraint propagation leads to the 
result that if one sets $V$ arbitrary, then one is led to $V(|\urr^{ab}| \mbox{ alone})$.  
In fact, that's how the original BB82 paper proceeds.   
In pure-shape RPM, the RWR technique furthermore gives the homogeneity property of $\ttV$.  
No more than this can be gleaned.  
Done the HKT way [regarding in each case a particular algebra, now Eucl($d$) and Sim($d$) respectively, as 
fundamental and demanding that constraint ans\"{a}tze close in precisely that way], for the BB82 formulation I find that 
\be
\sum\mbox{}_{\mbox{}_{\sfA = 0}}^{\infty} f(|\urr^{IJ}|)^{I_1...I_{\sfA}} p_{I_1} ... p_{I_{\sfA}}
\label{HKT1}
\ee
survives, while in pure-shape RPM furthermore  
$
f^{I_1...I_{\sfA}} = l^{I_1...I_{\sfA}}(\urr^{IJ}) k(\urr^{IJ})
$
for $l^{I_1...I_{\sfA}}$ homogeneous functions of degree {\sffamily A} and $k$ a homogeneous function of degree $- 2$.
These represent considerable generalizations of scaled and pure-shape RPM, corresponding to even more complicated MRI actions 
than the fermionic-including extension of Appendix \ref{Examples}.C.   

\noindent Note 4) One can view the `Newtonian Mechanics for the island universe subsystems within RPM' account 
of Appendix \ref{Cl-Soln}.C as a relational route to Newtonian Mechanics.

\subsection{Further compatibility restrictions between $\FrG$ and $\FrQ$}\label{Patti}\label{Q-G-Comp}

Now we have covered some examples, we can return to this...

\noindent Relationalism 9) [E.A.] 
\noindent A) {\bf Nontriviality} $\FrG$ cannot be too big [if dim($\FrG$) $\geq$ dim($\FrQ$) -- 1 
then the relational procedure will yield a theory with not enough degrees of freedom to be relational, or even an inconsistent theory]. 

\noindent B) Further {\bf structural compatibility} is required.   

\mbox{ } 

\noindent Example 1) If one uses such as Sim($d_1$) or Eucl($d_1$) with $\FrQ(N, d_2)$, it is usually for $d_1 = d_2$ (or at least $d_1 \leq d_2$).  

\noindent Example 2) If one uses Diff($\bupSigma_1$) to match to Riem($\bupSigma_2$), it is usually for $\bupSigma_1 = \bupSigma_2$, or at least for the 
two to be obviously related (e.g. taking out Diff($\mathbb{S}^1$) from Riem($\mathbb{S}^1 \times \mathbb{S}^1$)

\mbox{ }  

\noindent C) More concretely, $\FrG$ {\bf has to have a (preferably natural and faithful) group action on} $\FrQ$.

\mbox{ }

\noindent Relationalism 10) [Barbour and E.A.] In looking to do fundamental modelling, one may well have a strong taste for {\bf lack of extraneity}, 
e.g. always taking out all rotations rather than leaving a preferred axis.  

\noindent I) Aut($\FA$) is then a very obvious designate for this, though some subgroup of Aut($\FA$) might also be desirable, 
and there is also the issue of automorphisms of up to which level of structure.  
Then Aut($\FA$) and its subgroups definitely comply with A) and stand a good chance of suitably satisfying criteria B) and C).

\noindent II) One may also wish to eliminate as many extraneous properties as possible [Barbour's view], 
though under some circumstances there may be strong practical reasons to not eliminate some [E.A.'s counter-view]. 

\mbox{ }

\noindent Example 1) One may wish to retain the scale, thus using Eucl($d$) rather than Sim($d$) for RPM's, or and superspace($\bupSigma$) 
or \{CS + V\}($\bupSigma$) instead of CS($\bupSigma$) for Geometrodynamics, so as to have a viable cosmological theory and possibly for reasons 
of time provision [see Secs \ref{+temJBB}, \ref{Semicl}].  
Scale is possibly undesirable from the relational perspective via being a single heterogeneous addendum to the shapes \cite{Piombino, BONew}.
The cone structure makes scale especially extraneous in RPM's -- it has the same sort of dynamics irrespective of what the shapes are.
It may then seem quite conceptually undesirable for scale to play so prominent a roles in Cosmology. 
However, 

\noindent 1) there are currently no credible alternatives to this as regards explaining Cosmology \cite{Than}.  

\noindent 2) This heterogeneity can be useful by alignment with choices that would be harder to make without it, renders scale more presentable from a conceptual perspective.
This is the position I took in the upcoming `Conformal Nature of the Universe' Conference at the Perimeter Institute.  
See Sec \ref{ScaleDisc} for how various approaches fare in this regard. 

\noindent Example 2) Quotienting out Diff($\bupSigma$) and Conf($\bupSigma$) does not interfere with the existence of spinors, 
but for further/other choices one has a lack of such guarantees \cite{PenRind}, likely compromising one's ability to model spin-1/2 fermion matter. 
Moreover, it is well-known that some choices of $\bupSigma$ by themselves preclude the existence of fermions 
(see e.g. \cite{Nakahara}), and that orientability of $\bupSigma$ is also often desired.
(But supersymmetric extensions are quite clearly also compatible in this sense: Sec \ref{RPM-Susy} and \cite{Arelsusy}.)
%

\mbox{ }

\noindent In more detail, one has a background NOS ${\FA}$, which has properties deemed redundant, all or some of which can be modelled as such. 
There is then some level of structure $\langle {\FA}, \mbox{Prop}_{\sm\so\sd\se\sll\sll\se\sd} \rangle$ within the $\langle {\FA}, 
\mbox{Prop}_{\sT\so\st\sa\sll}\rangle = \langle {\FA}, \langle\mbox{Prop}_{\sm\so\sd\se\sll\sll\se\sd}, \mbox{Prop}_{\su\sn\sm\so\sd\se\sll\sll\se\sd} \rangle\rangle$.  
One then builds the configuration space of actually mathematically manipulable part-tangible part-intangible entities to lean on the structure of 
${\FA}$, so I denote it in this Sec by $\FrQ({\FA})$. 
Then finally one passes to 
\beq
\FrQ({\FA})/\mbox{Aut}(\langle {\FA}, \mbox{Prop}_{\sm\so\sd\se\sll\sll\se\sd} \rangle) \mbox{ } .
\eeq 
BB82 has a means of modelling continuous transformations at the level of Riemannian geometry.
As per Sec \ref{Setup-RPM}, one can also conceive of ${\FA}$ in such a way that discrete isometries can be considered as well as continuous ones.  
If one fails to be able to model $\mbox{Prop}_{\sT\so\st\sa\sll}$, then the automorphisms in question will not be freeing one of all the levels of structure/aspects of $\FA$. 
This is clearly the case for superspace($\bupSigma$): quotienting out Diff($\bupSigma$) from Riem($\bupSigma$) 
clearly does not free one from what choice of $\bupSigma$ one made in setting up Riem($\bupSigma$).  
This is because $\bupSigma$ pertains to the level of topological structure, and rendering the diffeomorphisms irrelevant is only a freeing from background {\sl metric} structure.

The relationalist who takes minimalism to an extreme would then aim to remove all sources of indiscernibility and actors that cannot be acted upon.
Taking out Rot(2) but not Rot(3) in a 3-$d$ world would make little sense.
But if a type of transformation is held to play a physical role, it is kept.
See the Conclusion as regards the status of scale, and Sec \ref{Cl-Str} as regards whether spatial topology in GR is discernible 
and can be acted upon; another example is how dimension is not primary for RPM's (Sec \ref{Examples} explains how 3-particle RPM's cannot distinguish between dimension 2 or higher).  

\mbox{ } 

\noindent The objective of relationalism is thus 
\beq
\FrQ(\FA)/\mbox{Aut}(\langle \FA, \mbox{Prop}_{\mbox{\scriptsize held to be } \su\sn\sp\sh\sy\si\scc\sa\sll}\rangle), 
\mbox{ or possibly even } \mbox{ }  
\FrQ(\FA)/\mbox{Aut}(\langle\FA, \mbox{Prop}_{\sT\so\st\sa\sll} \rangle)   \mbox{ } .  
\eeq
\noindent Relationalism 11) [BFO--A].
Note 1) $\FrQ$ (the entity taken to have some tangible physical content) has the {\bf a posteriori right to reject} \cite{RWR, Lan, Phan, Lan2} 
a proposed $\FrG$ by triviality or inconsistency that go beyond 1), arising instead via, for the moment, the Dirac procedure yielding further constraints.

\mbox{ } 

\noindent Example 1) If one attempts to use $\FrG =$ id with Riem($\bupSigma$) and a BSW-like trial action (like \'{O} Murchadha \cite{OM02} or I \cite{San}), 
one finds that enlarging $\FrG$ to Diff($\bupSigma$) is {\sl enforced} by the $\scM_{\mu}$ arising as an integrability of $\scH$.  
This also exemplifies how not all subgroups of Aut are guaranteed to serve.  

\mbox{ }

\noindent Note 2) Additionally, a given $\FrQ$, $\FrG$ pair still constitutes a substantial ambiguity as to the form of the action principle.
This can sometimes be truncated with simplicity postulates like that the action is not to contain 
higher than first (or occasionally second) derivatives, but ambiguities beyond that still often remain.   

\mbox{ }  

\noindent Example 2) The potential is completely free in scaled RPM and free up to homogeneity in pure-shape RPM. 

\noindent Example 3) One can have GR or Euclidean GR or strong gravity for the Riem($\bupSigma$), Diff($\bupSigma$) pair 
with action restricted to contain at most first label-time derivatives (featuring quadratically) and at most second spatial derivatives.  

\mbox{ } 

\noindent Note 3) This scheme retains considerable freedom.  
Relationalism does {\sl not} exert a highly unique control over the form that Theoretical Physics is to take [though Relationalism 10) and 11) can sometimes help with this].  

\mbox{ }  

\noindent Relationalism 11) Quotienting out the $\FrG$ is to {\bf remove all the structure in the} $\FrQ$ {\bf that is actually extraneous}. 
However, precisely what structures/features are extraneous in the case of the universe we live in is not so clear.  
[E.g., again, whether to discard scale.]  
One passes to 
\beq
\FrQ({\FA})/\mbox{Aut}(\langle {\FA}, \mbox{Prop} \rangle) \mbox{ } 
\eeq 
for $\mbox{Prop}$ the set of properties of structures up to a certain mathematical level (e.g. metric geometry or topology). 
One might ideally wish for relationalism to remove {\sl all} levels of structure of $\FA$.  
\noindent $\FrG$ = id is too trivial for most of Theoretical Physics, whilst a number of further GR features 
come from $\FrG$ = Diff or some generalization (Diff $\times$ Conf, the SuperDiff of Supergravity) or possibly some slight weakening, such as by 
attitude to the large diffeomorphisms \cite{Giu} or restriction to the transverse diffeomorphisms \cite{Alvarez}.

\subsection{Relationalism in other theoretical settings}\label{ROTS}

\subsubsection{Relationalism and the Ashtekar variables formulation of GR}\label{Rel-AV}

I first present some basic nuts and bolts for this approach.
\noindent Pass from $\mh_{\mu\nu}$, $\uppi^{\mu\nu}$ to a $SU(2)$ connection ${\mathbb{A}_{\mu}}^{\mbox{\scriptsize\tt AB\normalfont\normalsize}}$ 
and its conjugate momentum ${\mathbb{E}^{\mu}}_{\mbox{\scriptsize\tt AB\normalfont\normalsize}}$ 
[which is related to the 3-metric by $\mh_{\mu\nu} = - \mbox{tr}(\mathbb{E}_{\mu}\mathbb{E}_{\nu})$].\foo{The 
capital typewriter indices denote the Ashtekar variable use of spinorial $SU(2)$ indices.  
tr denotes the trace over these. $\mathbb{D}_{\mu}$ is here the $SU(2)$ covariant derivative as defined in the first equality of (\ref{ashgauss}). 
$|[\mbox{ }, \mbox{ } ]|$ denotes the classical Yang--Mills-type commutator.}  
%
The split space-time action for this is, schematically 
\beq
\FS \propto \int\d\lambda \int_{\sbSigma}\d^3x \sqrt{\mh} \, \upalpha \, \{         \mathbb{F}^{\mu\nu\mbox{\scriptsize\tt AB\normalfont\normalsize}}
                                                                 \mathbb{F}_{\mu\nu\mbox{\scriptsize\tt AB\normalfont\normalsize}} - 
\mathbb{E}^{\mu\mbox{\scriptsize\tt AB\normalfont\normalsize}}   \mathbb{E}_{\mu \mbox{\scriptsize\tt AB\normalfont\normalsize}}\} \mbox{ } .  
\eeq
The subsequent constraints are
\be
\mathbb{D}_{\mu}  
{\mathbb{E}^{\mu}}_{\mbox{\scriptsize\tt AB\normalfont\normalsize}} 
\equiv \pa_{\mu}
{\mathbb{E}^{\mu}}_{\mbox{\scriptsize\tt AB\normalfont\normalsize}}  +  
|[\mathbb{E}_{\mu}, \mathbb{E}^{\mu}]|_{\mbox{\scriptsize\tt AB\normalfont\normalsize}} = 0 
\label{ashgauss} 
\mbox{ } , 
\ee
\be
\mbox{tr}(\mathbb{E}^{\mu} \mathbb{F}_{\mu\nu}) = 0 
\label{ashmom} 
\mbox{ } , 
\ee
\be
\mbox{tr}(\mathbb{E}^{\mu}\mathbb{E}^{\nu}\mathbb{F}_{\mu\nu}) = 0 
\label{ashham} 
\mbox{ } .
\ee
(\ref{ashgauss}) arises because one is using a first-order formalism; in this particular case, it is an $SU(2)$ (Yang--Mills--)Gauss constraint.  
(\ref{ashmom}) and (\ref{ashham}) are the polynomial forms taken by the GR momentum and Hamiltonian constraints respectively, where 
$\mathbb{F}^{\mbox{\scriptsize\tt AB\normalfont\normalsize}}_{ij} := 2
\pa_{[i}\mathbb{A}^{\mbox{\scriptsize\tt AB\normalfont\normalsize}}_{j]} + 
|[\mathbb{A}_{i}, \mathbb{A}_{j}]|^{\mbox{\scriptsize\tt AB\normalfont\normalsize}}$ 
is the Yang--Mills field strength corresponding to ${\mathbb{A}_{i}}^{\mbox{\scriptsize\tt AB\normalfont\normalsize}}$. 
One can see that (\ref{ashmom}) is indeed associated with momentum since it is the condition for a vanishing (Yang--Mills--)Poynting vector.  
Again, the Hamiltonian constraint (\ref{ashham}) lacks such a clear-cut interpretation.  
Finally, the Ashtekar variables formulation given above is of complex GR.  
But then one requires troublesome \cite{Kuchar93} {\it reality conditions} in order to recover real GR.  
Nowadays so as to avoid reality conditions, one usually prefers to work not with Ashtekar's original complex variables 
but with Barbero's real variables \cite{Barbero} that now depend on an Immirzi parameter, $\gamma$. 
LQG is then a particular QM scheme for this \cite{Rovellibook, Thiemann}.
Ease of manipulation in Ashtekar variables leads to Criterion 2 being met.

Now, proponents of Ashtekar variables often use the word `relational' of their approach. 
However, this has not hitherto been systematically used in that program in the full LMB or LMB-CA sense; 
instead they use this word with less detail and/or to carry (a subset of) the perspectival meanings exposited in Sec \ref{Cl-Str}.  
Nevertheless, I state here that the Ashtekar variables scheme can be formulated in the full LMB or LMB-CA relational scheme.
This is modulo a small glitch at the start of such an investigation, which nevertheless turns out to be entirely surmountable.  
Namely, that the pure gravity case of the Ashtekar variables approach is not castable in temporally relational form at the level of the classical Lagrangian.  
From the point of view in which spacetime is presupposed \cite{Van}, in which one sets about elimination of the lapse to form the temporally 
relational action, this glitch is due to the following.  
The lapse-uneliminated action is purely linear in the lapse.
This is due to the well-known `pure-$\mT$' character of Ashtekar's canonical action for pure GR, in contrast to the more usual `$\mT - \mV$' form of the geometrodynamical action.
Thus the variation with respect to the lapse produces an equation independent of the lapse. 
Thus the lapse cannot now be eliminated from its own variational equation. 
However, it is clear that addition of matter fields breaks this pure linearity in the lapse, so 
this unusual accident that blocks the elimination of the lapse from its own variational equation ceases to occur and a temporally relational action can be obtained.  
Thus that the Barbour/LMB-(CA) relational literature's appears  to be fixated with GR in geometrodynamical form as regards which examples it uses 
is in fact down to a matter of taste/of each paper's applications, rather than being in any way being indicatory of LMB-relationalism only 
applying to the older Geometrodynamics and not the newer Ashtekar variables work that underlies LQG.
It is then surely of foundational importance for Ashtekar variables programs for them to be revealed to be relational in this additional well-motivated and historically and 
philosophically well-founded way in addition to the various other meanings that they have hitherto been pinning on the word `relational' in the Ashtekar variables/LQG program.

Configurational relationalism for Ashtekar variables formulations has  $\FrG = SU(2) \times$ Diff(3).
Then variation with respect to the additional $SU(2)$-auxiliaries produces the $SU(2)$ Yang--Mills--Gauss constraint that is 
ubiquitous in the Ashtekar variables literature, whilst variation with respect to the Diff(3)-auxiliaries produces the Ashtekar variables form of the GR momentum constraint.   
The Hamiltonian constraint arises as a primary constraint, and the usual form for a JBB time emerges.    
An Ashtekar variables formulation of a type of GR shape dynamics has recently appeared \cite{BST12}.  

\noindent The LMB(-CA) relational action for the canonical formulation of GR in Ashtekar Variables is
\beq
\FS^{\sG\sR}_{\sA\sV-\sr\se\sll} = 2\int\d\lambda \int_{\sbSigma}\d\bupSigma\sqrt{\{\mT^{\sG\sR}_{\sA\sV} + \mT^{\sm\sa\st\st\se\sr}\}\mV^{\sm\sa\st\st\se\sr}} = 
                        2\int          \int_{\sbSigma}\d\bupSigma\sqrt{\{\pa \ms^{\sG\sR\,2}_{\sA\sV} + \pa \ms^{\sm\sa\st\st\se\sr\,2}\}\mV^{\sm\sa\st\st\se\sr}} \mbox{ } .  
\eeq
The Ashtekar Variables form of the GR Hamiltonian constraint then follows as a primary constraint. 
The Ashtekar Variables form of the momentum constraint (\ref{ashmom}) and this formulation's further Yang--Mills--Gauss constraint (\ref{ashgauss}) follow as per usual, 
except that one is now to proceed via FENOS variation with respect to frame and Yang--Mills versions of the `action per unit charge'\footnote{Or `magnetic flux 
due to change in shape of surface over time' integrated over time.} 
auxiliary variable replacement for the (generalized) electric potential multiplier, whose form is given for electromagnetism in \cite{FEPI} in order for the MRI/MPI 
of the action to remain respected.
The Ashtekar Variables form of GR's emergent Jacobi--Barbour--Bertotti time is 
\beq
\lmt^{\sJ\sB\sB}_{\sA\sV-\sr\se\sll} = \stackrel{\mbox{\scriptsize extremum $\suF, \Psi \in$ Diff($\bupSigma$) \textcircled{S} $SU(2)$}}
                             {\mbox{\scriptsize of $\stS^{\tG\tR}_{\tA\tV}$}}
\left.
\int \sqrt{\pa s^{\sG\sR\,2}_{\sA\sV} + \pa \ms^{\sm\sa\st\st\se\sr\,2}}
\right/
\sqrt{2\mV^{\sm\sa\st\st\se\sr}}    \mbox{ } .  
\eeq
\noindent See \cite{Ashtekar, Thiemann} and the POT book for further discussion of this approach's algebroid of constraints, 
which is very similar in its properties to the geometrodynamical one.

\mbox{ }

\noindent{\bf Question 7$^*$} Do analogous `relativity without relativity' results hold in the Ashtekar variables context?  

\mbox{ }  

\noindent The present Article's main interest however is in RPM's, for which there are particularly many parallels 
with the geometrodynamical and conformogeometrodynamical formulations of GR.  
I note that these are fine for conceptual consideration of the POT (and are indeed the arena for which such has been most highly developed \cite{Kuchar92, I93}).  

\mbox{ }

\noindent Analogy 31) There is an analogy between loops and preshapes [i.e. the configurations prior to the main part of the reduction -- Dil($d$) but not Rot($d$) 
taken out for RPM's or $SU(2)$ but not Diff($\bupSigma$) taken out for LQG].

\noindent Analogy 32) There is a loose conceptual analogy between pure-shape RPM's shapes or scaled RPM's scale-and-shapes ({\sl classical} resolutions of the linear constraints) 
and Nododynamics/LQG's knot states \cite{PullinGambini} ({\sl quantum} resolutions of the linear constraints).   
It is not hard to pass to a quantum resolution of the RPM linear constraints to render this analogy tighter: the 
space of wavefunctions that depend solely on the pure shapes or on the scale-and-shapes.  
One might then view the moves in passing to knot equivalence classes of loops as analogous to the transformations used in \cite{FORD, Cones} 
unveil the shape space/relational space variables.  
However the transformations in Sec \ref{Examples} and 3 for RPM's are purely configuration space manoeuvres, whereas passing to Ashtekar variables is itself a canonical transformation.  
Thus RPM's fit better the philosophy of the central importance of the configuration space (as opposed to phase space or of any `polarizations' of it that aren't physically configurations).

\mbox{ } 

\noindent Most uses of the word `relational' in Nododynamics (i.e. Ashtekar variables/LQG) are termed `perspectival' in the present Article, 
which additionally explains a number of other senses in which the word `relational' and pieces together various of these uses.    
Thus e.g. GR in general and the Nododynamical formulation do have manifest many other  senses of `relational' outside of the more usually-ascribed perspectival ones.  

\mbox{ } 

\noindent {\sl LQC} is for now LQG's minisuperspace analogue (trivial Configurational Relationalism); it usually includes scalar matter. 
%

\subsubsection{Question 8$^{*}$) is there a supersymmetric extension of RPM and does it toy-model Supergravity?} \label{RPM-Susy}

Supersymmetric particle mechanics exists (see e.g. \cite{Freund}).
Presumably then supersymmetric RPM also exists.
Here the indirect formulation's incipient NOS $\FA = \mathbb{R}^d$ is replaced by a Grassmann space.
What are then the sequence of configuration spaces? 
The appropriate $\FrG$'s? 
The isometry groups of the reduced configuration spaces? (See Sec \ref{Q-Geom} for these for the usual RPM's.)
Does this toy-model the canonical Geometrodynamics form of Supergravity to any significant extent 
(e.g. do this Article's Analogies and Differences carry over; and do supersymmetric RPM's and Supergravity theories share more?)

\mbox{ } 

\noindent Note 1) This circumvents Sec \ref{Q-G-Comp} lack of spinors impasse. 

\noindent Note 2) Compared to theories preceding it, supersymmetry is conceptually strange in how the product of two supersymmetry operations produces a {\sl spatial} transformation.

\noindent Note 3) Also, the supersymmetry constraint in Supergravity is the `square root' of the Hamiltonian constraint\cite{SqrtTeitelboim}, 
along the same sort of lines that the Dirac operator is the `square root' of the Klein--Gordon operator.

\noindent Note 4)  Notes 1) and 2) indicate that there is some chance that the supersymmetry transformation is one of: not conceptually sound, or conceptually deeper than,   
other transformations, and the relational perspective could be one that has more to say about this matter.  
Thus, {\sl the relational program might be able to provide its own prediction} as to whether to expect supersymmetry in nature. 
Conversely, insisting on supersymmetry might expose the current conceptualization of Relationalism to be insufficient.  
See Secs \ref{Q-G-Comp} and \cite{Arelsusy} for more.  
See also Secs \ref{Trident} and \ref{App-B-1}.

\subsubsection{Relationalism versus String Theory}\label{RPM-vs-Strings}

Aside from this Sec's earlier use of the preclusion of perturbative strings to sharpen up criteria 
for relationalism, there are some conceptual parallels between passing from studying point particles to 
studying strings and passing from point particles to studying relational quantities only. 
These are comparable from a conceptual perspective; whether each of these is rich, tractable and has anything to say about QG  
is another matter.  
The latter is moreover more conservative -- the relational quantities come from careful thought about the 
original problem rather than replacing it  with a distinct problem as in String Theory.  
This gives less new structure leading perhaps to less mathematical richness, but it it would appear to be a safer bet as regards the ``{\it hypotheses non fingo}" 
tradition of Physics. 

\noindent Exporting the Configurational Relationalism/Best Matching ideas to other parts of Physics could conceivably be of interest 
even if  not accompanied by the further demands of no background spatial structures or of Temporal Relationalism.    
E.g. one could consider what form these ideas take for target space theories of which strings on a given background.    
Finally, one would expect nonperturbative M-Theory to have {\sl both} of these features (relationalism {\sl and} extended objects).

\mbox{ }

\noindent{\bf Question 9$^{**}$}  Provided that supersymmetry can be accommodated into (a suitable generalization of) Relationalism, 
I conjecture that relational notions along the lines of those in this Article are necessary for any truly GR-embracing notion of background independence for M-Theory\foo{To avoid  
confusion, I note here that some other uses of the term `background (in)dependence' in the String Theory literature have a different meaning, 
namely concerning the effect of choice of vacuum on string perturbations.} 
Since background-independence amounts to a POT, it would then be exceedingly likely for M-Theory to have one of those too.  
Thus study of the POT is likely to be a valuable investment from the perspective of developing and understanding M-Theory.  
Establish or refute these statements.

\mbox{ } 

\noindent Geometrodynamics (very easily generalizable to arbitrary-$d$) could be seen as a first toy model of spatially 10-$d$ M-Theory.  
There may be better RPM's and Geometrodynamics models for this particular setting, e.g. spatially 10-$d$ Supergravity (which is a low-energy/`semiclassical' limit of M-Theory). 

\mbox{ }  

\noindent The situation with extended objects that occupy/furnish a multiplicity of notions of space is, 
in addition to supersymmetry, an  interesting frontier in which to further explore the relational postulates.

\begin{subappendices}
\subsection{Auxiliary variables: multipliers versus cyclic velocities/ordials}\label{Aux}

\subsubsection{FENOS variation renders these representations equivalent}\label{FENOS}

Consider for the moment that the auxiliary $\fg^{\sfZ}$ variables take their usual guise as multipliers in 
the generalized sense (coordinates whose velocities do not occur in the action). 
E.g. one conventionally views in this way the electric potential $\Phi(\ux)$, and the shift 
$\upbeta^{\mu}(\ux)$ and lapse $\upalpha(\ux)$ of GR split with respect to spatial hypersurfaces. 
This is the way that translation and rotation correction variables were viewed in the original BB82 paper \cite{BB82}.        
Then variation with respect to the $\fg^{\sfZ}$ produces multiplier equations, 
\beq 
{\partional\cL}/{\partional \fg^{\sfZ}} = 0 \mbox{ } ,  \mbox{ } \mbox{ or } \mbox{ } {\partional\ordial\widetilde{\fs}}/{\partional \fg^{\sfZ}} = 0
\label{multipliereq}
\eeq
which are the constraints that one is expecting if one is indeed to obtain thus a gauge theory with gauge group $\FrG$.
These constraints then use up both the auxiliary variables' degrees of freedom and an equal number of degrees of freedom from the $\FrQ$. 
So indeed one is left with a theory in which the dynamical variables pertain to the quotient space $\FrQ/\FrG$. 
(I.e. variables among the original $\FrQ$ which are entirely unaffected by which choice of $\FrG$-frame is made).

However, this multiplier viewpoint is not the only viable interpretation.  
One can also use an interpretation as velocities associated with an auxiliary cyclic coordinate.  
This is possible because of the entirely unphysical nature of auxiliary variables.
As it is physically meaningless for these to take a fixed value anywhere, in particular, they must take arbitrary rather than fixed values at the `end points' of the variation.    
More precisely, the appropriate type of variation is {\bf free end notion of space (FENOS)} variation (alias 
{\it variation with natural boundary conditions}) \cite{CH, Fox, BrMa, Lanczos}.  
This is in particular a portmanteau of {\it free end point (FEP)} variation for finite theories, and 
{\it free end spatial hypersurface (FESH) variation} for field theories. 
This argument can be viewed as non-entities not acting or as the various end-NOS values producing indiscernible and thus identical physics.

Next, note that FENOS variation involves more freedom than standard fixed end variation, i.e. it involves a space 
of varied curves of notions of space (which are just varied curves in the FEP subcase) that is larger.
It is by this means that it is tolerable for it to impose more conditions than the more usual fixed-end variation does.
Then consider configurations $\{\fQ^{\sfA}, \fg^{\sfZ}\}$ for $\fg^{\sfZ}$ entirely arbitrary through being unphysical.  
With this in mind, I start from scratch, with a more general Lagrangian portmanteau or arclength parameter 
in which both the auxiliaries $\fg^{\sfZ}$ and their velocities $\dot{\fg}^{\sfZ}$ or ordials $\ordial{\fQ}^{\sfA}$ in general occur: 
$\cL = \cL\lfloor\dot{\fQ}^{\sfA}, \dot{\fg}^{\sfZ}, \fQ^{\sfA}, \fg^{\sfZ}\rfloor$ or 
$\ordial\widetilde{\fs}\lfloor\ordial{\fQ}^{\sfA}, \ordial{\fg}^{\sfZ}, \fQ^{\sfA}, \fg^{\sfZ}\rfloor$.

Then, because $\fg^{\sfZ}$ is auxiliary, variation with respect to $\fg^{\sfZ}$ is free, so $\Bigdelta \fg^{\sfZ}|_{\mbox{\scriptsize end-NOS}}$ and are not controllable. 
Thus one obtains 3 conditions per variation, 
\beq
{\partional\cL}/{\partional \fg^{\sfZ}} = \dot{\fp}_{\sfZ} \mbox{ , } \mbox{ } \mbox{ or } \mbox{ } 
{\partional\ordial\widetilde{\fs}}/{\partional \fg^{\sfZ}} = \ordial{\fp}_{\sfZ} \mbox{ } , \mbox{ } \mbox{ each alongside } \mbox{ }  
\left.
\fp_{\sfZ}
\right|_{\mbox{\scriptsize end-NOS}} = 0 \mbox{ } .  
\label{correct}
\eeq
\mbox{ } \mbox{ } If the auxiliaries $\fg^{\sfZ}$ are multipliers $\fm^{\sfZ}$, (\ref{correct}) reduces to
\beq 
\fp_{\sfZ} = 0 \mbox{ } , \mbox{ } {\partional\cL}/{\partional \fm^{\sfZ}} = 0
\eeq
and redundant equations.
I.e. the end-NOS terms automatically vanish in this case by applying the multiplier equation to the first factor of each.  
This is the case regardless of whether the multiplier is not auxiliary and thus standardly varied, or
auxiliary and thus FENOS varied, as this difference translates to whether or not the cofactors of the above zero factors are themselves zero or not.  
Thus the FENOS subtlety in no way affects the outcome in the multiplier coordinate case.

If the auxiliaries $\fg^{\sfZ}$ are cyclic coordinates $\fc^{\sfZ}$, the above reduces to 
\beq 
\left.
\fp_{\sfZ}
\right|_{\se\sn\sd-\sN\sO\sS} = 0 \mbox{ } 
\label{ckill}
\eeq
and
\beq 
\dot{\fp}_{\sfZ} = 0 \mbox{ }  \mbox{ or equivalently } \mbox{ } \ordial{\fp}_{\sfZ} = 0 \mbox{ }
\label{hex}
\eeq
which implies that 
\beq 
\fp_{\sfZ} = C(x^{\mu_p}) \mbox{ , invariant along the curve of notions of space} \mbox{ } . 
\eeq 
Then $C(x^{\mu_p})$ is identified as 0 at either of the two end NOS's (\ref{ckill}), and, being invariant along the curve of notions of space, is therefore zero everywhere.  
Thus (\ref{hex}) and the definition of momentum give 
\beq 
{\partional \cL}/{\partional \dot{\fc}^{\sfZ}} := \fp_{\sfZ} \mbox{ }  \mbox{ or equivalently } \mbox{ } {\partional \ordial\widetilde{\fs}}/{\partional \ordial{\fc}^{\sfZ}}
 = 0  \mbox{ } .  
\eeq
So, after all, one does get equations that are equivalent to the multiplier equations (\ref{multipliereq}) at the classical level. 
Albeit, there are conceptual and foundational reasons to favour the latter -- now that fits in with Temporal Relationalism as laid out in the Introduction.  
For previous discussion of this case in the literature, see \cite{B03, ABFO, Lan, ABFKO, FEPI}.

\subsubsection{Multiplier elimination is then equivalent to passage to the Routhian}\label{ME-Routh}

We also need to establish that the a priori distinct procedures of cyclic velocity elimination (known as passage to the Routhian \cite{Lanczos}) and multiplier elimination are equivalent.  
Consider $\cL\lfloor\fQ^{\sfA}, \dot{\fQ}^{\sfA}, \fg^{\sfZ}\rfloor$ or $\d\widetilde{\cs}\lfloor\fQ^{\sfA}, \dot{\fQ}^{\sfA}, \fg^{\sfZ}\rfloor$.     
If these are taken to be multiplier coordinates $\fm^{\sfZ}$, then variation yields $0 = {\partional\cL}/{\partional \fm^{\sfZ}}$ or 
$\partional\ordial\widetilde{\fs}/\partional\fm^{\sfZ}$.    
If this is soluble for $\fm^{\sfZ}$, one can replace it by $\fm^{\sfZ} = \fm^{\sfZ}(\fQ^{\sfA}, \dot{\fQ}^{\sfA})$, and then substitute that into $\cL$.  
If the $\fg^{\sfZ}$ are taken to be velocities corresponding to cyclic coordinates $\fc^{\sfZ}$, then FENOS variation yields 
\beq
{C}^{\sfZ} = {\partional\cL}/{\partional \dot{\fc}^{\sfZ}} \mbox{ or } {\partional\ordial\fs}/{\partional \ordial{\fc}^{\sfZ}}   = \fp^{\sfZ} =  0 \mbox{ } . 
\label{sku}
\eeq

\noindent
This is soluble for $\dot{\fc}^{\sfZ}$ or $\d\fc^{\sfZ}$ iff the above is soluble for $\fm^{\sfZ}$.  
However one now requires passage to the Routhian portmanteau $\mbox{\sffamily R}$ in eliminating $\dot{\fc}^{\sfZ}$ from the action: 
$\int\d\lambda\int_{\sbSigma_p}\d \bupSigma_p\mbox{\sffamily R} = 
\int_{\bupSigma_p}\d \sbSigma_p\{\cL - \dot{\fc}^{\sfZ}\fp_{\sfZ}\}$ or 
$\int\int_{\sbSigma_p}\d \sbSigma_p\d\mbox{\sffamily R} = 
\int_{\sbSigma_p}\d \sbSigma_p\{\ordial\widetilde{\fs} - \ordial{\fc}^{\sfZ}\fp_{\sfZ}\}$  
But the last term is zero, by (\ref{sku}), so passage to the Routhian in the cyclic velocity interpretation is equivalent to multiplier elimination in the multiplier coordinate interpretation.

\subsubsection{The AK Lemma}\label{AK-Lemma}

{\bf Lemma 1} [A and Kelleher]. This was used in the variational procedure for the ABFO and ABFO actions \cite{ABFO, ABFKO}.   
As a third and more general case of the discussion in Appendix A.1, if an auxiliary and its velocity are simultaneously present in an action, then FENOS variation is equivalent 
to varying with respect to the auxiliary and its velocity as if they were separate multiplier coordinates.  
By this, e.g. \cite{ABFO} and the earlier 2-multiplier-coordinates formulation of \cite{BO} are equivalent. 
The advantage of using a single auxiliary whose velocity is simultaneously present is, once again, manifest 
Temporal Relationalism, now for more complicated theories for which a cyclic auxiliary velocity does not 
suffice, due to the now the potential also requiring a best-matching correction. 
This is because e.g. the Ricci 3-scalar is not a conformal tensor, so one requires instead the potential combination in actions (\ref{BFOA}, \ref{Action-ABFKO}).

\subsubsection{(Ordial-)almost-phase spaces, -Hamiltonians and -Dirac procedures}\label{Almost}

{            \begin{figure}[ht]
\centering
\includegraphics[width=0.6\textwidth]{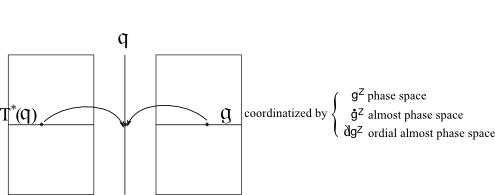}
\caption[Text der im Bilderverzeichnis auftaucht]{        \footnotesize{Phase space, almost phase space APhase and ordial almost phase 
space OAPhase are types of `bi-bundle' as indicated. }        }
\label{DAPS}\end{figure}           }

\noindent
Phase  has a                     Hamiltonian   $\fH(\fQ^{\sfA}, \fP{\sfA}, \fm^{{\sfZ}})$, 
APhase has an                  A-Hamiltonian   $\fA(\fQ^{\sfA}, \fP{\sfA}, \dot{\fc}^{{\sfZ}})$, and 
OAPhase has an                OA-Hamiltonian   $\ordial\fA(\fQ^{\sfA}, \fP{\sfA}, \ordial\fc^{{\sfZ}})$.  
The A stands for `almost', referring to it almost being Phase, except that the unphysical auxiliary variables' velocities (rather than their 
momenta) remain present; thus its physical content is equivalent to that of Phase.  
The OA stands for `ordial almost' i.e. the parametrization-irrelevant form in which the unphysical auxiliary variables' ordials
$\ordial\ttg$ are present rather than their velocities $\dot{\ttg}$.
Clearly if one has no $\FrG$ the first two reduce to the same, whilst the second and third are MRI and MPI and thus interchangeable by (de)parametrization.

The Dirac procedure involves appending constraints by multipliers in passing to the total Hamiltonian; this is in line with the above $\FrG$-extension of phase space. 
However, if one wishes to remain temporally relational by MRI, one would append constraints by cyclic velocities in passing to a total A-Hamiltonian, which I term the A-Dirac procedure.
And, if one wishes to remain temporally relational by MPI, one would append constraints by cyclic ordials in passing to a total OA-Hamiltonian, which I term the OA-Dirac procedure.

Likewise, one has now not Hamilton's equations
\beq
{\partional\fH}/{\partional \fP_{\sfA}} =  \dot{\fQ}^{\sfA} \mbox{ } , \mbox{ } 
{\partional\fH}/{\partional \fQ^{\sfA}} = -\dot{\fP}_{\sfA} \mbox{ } , \mbox{ augmented by } 
{\partional\fH}/{\partional \fm^{\sfZ}} =  0 \mbox{ } ,
\eeq
but rather A-Hamilton's equations 
\beq
{\partional\fA}/{\partional \fP_{\sfA}} =  \dot{\fQ}^{\sfA} \mbox{ } , \mbox{ } 
{\partional\fA}/{\partional \fQ^{\sfA}} = -\dot{\fP}_{\sfA} \mbox{ } , \mbox{ augmented by }
{\partional\fA}/{\partional \dot{\fc}^{\sfZ}} =  0 \mbox{ } ,
\eeq
or the manifestly relational OA-Hamilton's equations
\beq
{\partional\ordial\fA}/{\partional \fP_{\sfA}} =  \ordial{\fQ}^{\sfA} \mbox{ } , \mbox{ } 
{\partional\ordial\fA}/{\partional \fQ^{\sfA}} = -\ordial{\fP}_{\sfA} \mbox{ } , \mbox{ augmented by }
{\partional\ordial\fA}/{\partional \ordial {\fc}^{\sfZ}} =  0 \mbox{ } .
\eeq
\noindent Example 1) The ERPM total A-Hamiltonian and OA-Hamiltonian are
\beq 
{\fA}^{\sR\sP\sM}_{\sT\so\st\sa\sll} = \Circ I \,\scH + 
                                       \Circ\underline{A}\cdot\underline{\scP} + 
                                       \Circ\underline{B}\cdot\underline{\scL} 
\mbox{ } \mbox{ } , \mbox{ } \mbox{ }
\d{\fA}^{\sR\sP\sM}_{\sT\so\st\sa\sll} = \d I \, \scH + 
                                    \d\underline{A}\cdot\underline{\scP} + 
                                    \d\underline{B}\cdot\underline{\scL} 
\mbox{ } . 
\eeq
\noindent Example 2) In the case of Geometrodynamics, the well-known form of the (total) Hamiltonian shared between the ADM and BSW approaches is \cite{Dirac,ADM}     
\be
\mbox{\sffamily{H}}_{\sT\so\st\sa\sll} = \int\d\lambda\int_{\sbSigma} \d^3x\sqrt{\mh}\{\upalpha\scH + \upbeta^{\mu}\scM_{\mu}\} 
\mbox{ } .
\ee
On the other hand, the A and BFO-A approaches share instead the total A-Hamiltonian and partial-differential almost Hamiltonian are given by 
\be
\mbox{\sffamily{A}}_{\sT\so\st\sa\sll} = 
\int\d\lambda\int_{\sbSigma} \d^3x\sqrt{\mh}\{\dot{\mI}\,\scH + \dot{\mF}^{\mu}\scM_{\mu}\}  
\mbox{ } , \mbox{ } \mbox{ or } \mbox{ }
\d\mbox{\sffamily{A}}_{\sT\so\st\sa\sll} = 
\int \int_{\sbSigma} \d^3x_{\bupSigma}\sqrt{\mh}\{\d\mI\,\scH + \d{\mF}^{\mu}\scM_{\mu}\}  
\mbox{ } .  
\ee
\noindent {\bf The Hamiltonian collapse problem}.  
This Article's relational perspective at the classical level suffers from how, despite the different appearance from the usual of its 
Lagrangians, both its Lagrangians and the usual approach to Physics' lead to post-variationally equivalent (O(A))-Hamiltonians.  
Hamiltonians being crucial in many ways for the passage to quantum theory, the classical advances in the 
understanding of relationalism have a large tendency to {\sl not} directly provide any insights as to how to think relationally at the quantum level.

\mbox{ }  

\noindent Note: the {\sl physical} part of these `(O(A))'s' remains unchanged.  
One is driven to this subtle difference in representation of the unphysical parts by the wish to work with manifestly relational formulations of the Principles of Dynamics.

\subsubsection{One further family of forms for the action}\label{+Forms-Action}

Starting off with a Legendre transformation to the (O(A))-Hamiltonian variables, 
\beq
\FS =                    \int_{\sbSigma} \d\bupSigma \int \{\dot{\fQ}^{\sfA}\fP_{\sfA} - \fA_{\sT\so\st\sa\sll}\}\d\lambda  
     =                    \int_{\sbSigma} \d\bupSigma \int \{\dot{\fQ}^{\sfA}\fP_{\sfA} - \dot{\fI}\,{\scQ\scU\scA\scD} - \dot{\fc}^{\sfZ}
     \scL\scI\scN_{\sfZ}\}\d\lambda
      \stackrel{\mbox{$= \mbox{\LARGE $\int$}_{\sbSigma} \d\bupSigma \mbox{\LARGE $\int$} 
                                          \{\ordial \fQ^{\sfA}\fP_{\sfA} - \ordial\fI\,{\scQ\scU\scA\scD} - \ordial\fc^{\sfZ}\,\scL\scI\scN_{\sfZ}\} $ \hspace{0.4in}}  }
                   { = \int_{\sbSigma} \d\bupSigma \int \{  \Last \fQ^{\sfA}\fP_{\sfA} - {\scQ\scU\scA\scD}  - \Last \fc^{\sfZ}\scL\scI\scN_{\sfZ}\}\d\ft^{\se\sm(\sJ\sB\sB)}}
\mbox{ } ,
\label{Aac-1}
\eeq
where the second form is MRI thanks to using the A-Hamiltonian, the upstairs form is MPI, and 
the downstairs form is obtained by setting the label time to being the emergent JBB time. 
The downstairs form (or the same as the downstairs form but with different interpretations 
pinned on the time-variable) is commonly found in the literature; the above establishes equivalence between this and MRI forms.

As they will be needed in Part III's path integral type constructions, and a number of the expressions within 
are well-known, I now specialize the above equation to the GR and ERPM cases. 
\beq
\FS_{\sG\sR} = 
\int_{\sbSigma} \d^3x \int \{\dot{\mh}_{\mu\nu}\uppi^{\mu\nu} - \mA_{\sT\so\st\sa\sll}\}\d\mt
      =                     \int_{\sbSigma} \d^3x \int \{\dot{\mh}_{\mu\nu}\uppi^{\mu\nu} - \dot{\mI}\,\scH - \dot{\mF}^{\mu}\scM_{\mu}\}\d\mt
      \stackrel{\mbox{$= \mbox{\LARGE $\int$}_{\sbSigma} \d^3x \mbox{\LARGE $\int$} 
                                          \{\pa  {\mh}_{\mu\nu}\uppi^{\mu\nu} - \pa  {\mI}\,\scH - \pa  {\mF}^{\mu}\scM_{\mu}\} $ \hspace{0.4 in}}}
                   { = \int_{\sbSigma} \d^3x \int \{   \Last{\mh}_{\mu\nu}\uppi^{\mu\nu} - \scH          -  \Last{\mF}^{\mu}\scM_{\mu}\}\d  \mt^{\se\sm(\sJ\sB\sB)}}
\mbox{ } ,
\label{Aac-2}
\eeq
for  which the second form's multiplier counterpart is the commonest (ADM) form in the literature, and the 
downstairs form coincides mathematically with the proper time formulation.  
\beq
\FS_{\sE\sR\sP\sM} =                      \int \{\dot{\underline{\rho}}^{i}\cdot\underline{p}_{i} - A_{\sT\so\st\sa\sll}\}\d\lambda  
                =                      \int \{\dot{\underline{\rho}}^{i}\cdot\underline{p}_{i} - \dot{I}\,\scE -   \underline{\dot{B}}\cdot\underline{\scL} 
                \mbox{ }            \}\d\lambda
      \stackrel{  \mbox{$ =   \mbox{\LARGE $\int$} 
                                          \{\d  \underline{\rho}^{i}\underline{p}_{i} - \d I\,\scE - 
                                          \d  \underline{B}\cdot\underline{\scL}
                                           \} $}\hspace{0.6in} }
                   { =  \int \{   \Last\underline{\rho}^{i}\underline{p}_{i} - \scE          
                     -  \Last \underline{B}\cdot\underline {\scL} \} \d t^{\se\sm(\sJ\sB\sB)}}
\mbox{ } ,
\label{Aac-3}
\eeq
for which the downstairs form reinterprets and extends the usual mechanics form for this action.  
In the specific case of scaled $N$-stop metroland (needed for Sec \ref{PaB-RPM}), (\ref{Aac-3}) becomes 
\beq
\FS_{\sN-\sss\st\so\sp\, \sE\sR\sP\sM} =                      \int \{\dot{\rho}^{i}p_{i} - A_{\sT\so\st\sa\sll}\}\d\lambda  
    =                      \int \{\dot{\rho}^{i}p_{i} - \dot{I}\,\scE \}\d\lambda
      \stackrel{\mbox{$ =  \mbox {\LARGE $\int$} 
                                          \{\d\rho^{i}p_{i} - \d  {I}\,\scE  \} $}}
                   { =  \int \{   \Last\rho^{i}p_{i} - \scE \}\d t^{\se\sm(\sJ\sB\sB)}}
\mbox{ } .
\label{Aac-4}
\eeq
In the specific case of the indirect presentation of the scaled triangle (needed for Sec \ref{PaB-RPM}), this becomes
\beq
\FS_{\triangle -\sE\sR\sP\sM} =                      \int \{\dot{\underline{\rho}}^{i}\cdot\underline{p}_{i} - 
                            A_{\sT\so\st\sa\sll}\}\d\lambda  
    =                      \int \{\dot{\underline{\rho}}^{i}\cdot\underline{p}_{i} - \dot{I}\,\scE - \dot{B}\,{\scL} \}\d\lambda
      \stackrel{\mbox{$ =  \mbox {\LARGE $\int$} 
                                          \{\d  \underline{\rho}^{i}\cdot\underline{p}_{i} - \d I\,\scE - \d  B\,{\scL}\} $}}
                   { =  \int \{   \Last\underline{\rho}^{i}\cdot\underline{p}_{i} - \scE -  \Last B{\scL} \}\d t^{\se\sm(\sJ\sB\sB)}}
\mbox{ } .
\label{Aac-5}
\eeq

\subsubsection{On $\mN$, $\dot{\mI}$, $\upalpha$ and $\dot{\upgamma}$}\label{NIAG}

These are four symbols for mathematically the same quantity but with each resting on different foundations.
The last two expressions come from presupposing spacetime and performing on it the ADM split and my parallel split (\ref{emtime}) respectively.    
These are called the lapse and the velocity of the instant (VOTI).  
Next, $\mN$ and $\dot{\mI}$ come from relational approaches: the non-manifestly temporally relational BSW formulation and the manifestly temporally relational BFO-A one. 
These are, respectively, the {\it emergent lapse} and the {\it emergent VOTI}; moreover, these last two quantities come with a 
common computational formula that is not present for the other two (unless one performs manoeuvres that transcend to the BSW or BFO-A action).

The labeller of the instants is identifiable by comparison with the standard ADM formulation and its interpretation as being 
the {\it proper time} of the constituent elements (taken to include an idealization of observers where relevant) on the instants in question.  
This makes good physical sense. 
$\upgamma = \tau$ is particularly identifiable thus, with $\dot{\upgamma}$ then being the derivative of proper time with respect 
to coordinate time, whereas $\dot{\mI}$ is then to be viewed as the emergent version of this derivative.  
The lapse has long been known to be of this nature,
\beq
\upalpha(\ux, \mbox{coordinate time}) := \pa \mbox{(proper time)}/\pa 
\mbox{(coordinate time)} \mbox{ } , 
\eeq
the $\upalpha$ to $\dot{\upgamma}$ difference reflecting rather a difference prior to variation: 
to use the lapse itself as an auxiliary variable or to cast the instant that sits within the definition of the lapse in that auxiliary role. 
This was not considered prior to the relational program probably because of the need for subtlety in the variational procedure in 
the second case represented a large increase in subtlety from the first case.  
On the other hand, whether to use $\dot{\mI}$ or $\mN$ is but a difference in naming this combination that is widespread throughout the results of the variational procedure.  
The $\dot{\mI}$ name is then the more consistent one given the relational approach's obligation to use $\dot{\mF}^{\mu}$ rather than 
the manifest temporal-relationalism-breaking $\upbeta^{\mu}$, and also to reflect the homogeneous factor of $1/\d\lambda$ within the emergent object.

In parallel, for RPM's $I$ is identifiable as Newtonian time on an emergent footing, with $\dot{I}$ then being the derivative of this by label time.
This fits in that GR is already-MRI so its coordinate time has label status. 
The general case has 
\beq
\dot{\fI} = \ordial \mbox{(instant-labelling time)}/\ordial\mbox{(meaningless label parameter time)}
\eeq
The instant-labelling time has special status by being a property of the instants rather than being just a mere parametrization of the `time-line' itself.  
However, it is not unique in that sense, e.g. scale time is a property of instants, so there is more to it than that.
Proper time is very observer-associated, though more broadly it is actual of fictitious material blobs that it is attached to. 
It is interesting that Newtonian and proper times have a common emergent origin from the relational perspective.  
%

\noindent In the MPI relational presentation, one uses the OOTI $\d\fI$ (ordial of the instant) in place of the VOTI $\dot{\fI}$.
This can be identified as an emergent version of the differential of the proper time or of the Newtonian time in the GR and mechanics contexts respectively.  

\mbox{ } 

\noindent $t^{\se\sm(\sJ\sB\sB)}$ itself {\sl starts} as the most dependent variable of them all -- $t^{\se\sm(\sJ\sB\sB)}[Q, \ordial Q]$; 
only when found to sufficient accuracy is it cast as the most convenient independent variable to use, i.e. $t^{\se\sm(\sJ\sB\sB)}$ such that $Q = Q(t^{\se\sm(\sJ\sB\sB)}) =: 
t^{\se\sm(\sJ\sB\sB)}_{\si\sn\sd\se\sp}$.  
$\mt^{\se\sm(\sJ\sB\sB)}$ is to some extent like that too, except that its final form is taken to be a local function of $\ux$.  
Then if we wish to identify these with the conventional formulations of Newtonian and proper times respectively, we should note that this identification pertains to the {\sl final} forms.  
There is a certain amount of freedom as regards where to break the exact equality in the sequence `$\ft^{\se\sm(\sJ\sB\sB)} = \fI = \ft$ for $\ft$ the Newtonian--proper time portmanteau.
$t^{\se\sm(\sJ\sB\sB)}_{\si\sn\sd\se\sp} = t^{\sN\se\sw\st\so\sn)}$ but $\ft^{se\sm(\sJ\sB\sB)}[\Q, \ordial Q] \neq t^{\sN\se\sw\st\so\sn)}$ {\sl until} a sufficiently good approximation 
is found to be used in that role once no longer regarded as a highly dependent variable.
\noindent I then choose to interpret $\fI$ as the {\sl instant-labelling} role that is conceptually distinct from (but indeed ends up being `dual to') the Newtonian--proper time 
portmanteau. 
[This duality is of a hypersurface-primality versus timefunction primality nature, much as between the 3 + 1 `ADM' and 1 + 3 `threading' \cite{HE} formulations of GR.]  
We consider this role to played secondarily to the $\fQ$ and $\ordial\fQ$ pertaining to the hypersurface in question, so that instant-labelling also has prior and posterior forms.   
Thus, in summary, `$\ft^{\se\sm(\sJ\sB\sB)} = \fI$ priorly and posteriorly, but `$\ft^{\se\sm(\sJ\sB\sB)} = \fI = \ft$ only holds all the way through as a posterior equality. 
Thus the $\ft^{\se\sm(\sJ\sB\sB)} = \fI$ identity is entire, the two symbols used solely relating to the aforementioned duality.
On the other hand, $\fI = \ft$ is also a double-naming due to this duality, but only refers to the posterior half of $\fI$'s meaning.  
 
\vspace{10in}

\subsection{Parageodesic principle split conformal transformations (PPSCT's)}\label{Banal}

\subsubsection{Jacobi's Principle: geodesic principle versus parageodesic principle}\label{Parageodesic}

The physical quantity is $\ordial\widetilde{\fs}$.
In terms of this, Jacobi's principle is a {\bf geodesic principle} for some geometry (a Riemannian one in Jacobi's own setting, though 
readily extendible to arbitrary-signature Riemannian geometry, and to, following Synge, Finslerian geometry; I additionally mean each of these in a finite--infinite portmanteau sense).  
In the (arbitrary-signature) Riemannian case, the equations of motion are indeed here of geodesic form, 
\beq
\widetilde{\ordional}_{\sa\sb\sss}\mbox{}^2 {\fQ}^{\sfA} = \widetilde{\Last}\widetilde{\Last}\fQ^{\sfA} + 
\widetilde{\Gamma}^{\sfA}\mbox{}_{\sfB\sfC}\widetilde{\Last}\fQ^{\sfB}\widetilde{\Last}\fQ^{\sfC} = 0 \mbox{ } .
\eeq
However,

\noindent Difficulty 1) this is entirely on a case-by-case basis: one requires a different geometrization for each system.  

\noindent Difficulty 2) Specific models come in families. 
Restricting for now to the (arbitrary-signature) Riemannian case, all the possible potential factors $\fW$'s for a given 
configuration space's material contents; the configuration space may well then have a natural $\fW$-independent kinetic line-element $\d\fs$ such that 
\beq
\ordial\widetilde{\cs} = \fw \, \ordial\cs
\eeq 
for weight function $\fw = \sqrt{2\fW}$. 

\noindent It would then often be desirable to use this geometrization.  
In terms of such a geometrically-natural $\ordial\cs$, Jacobi's Principle is a {\bf parageodesic principle}, the equations of motion now being (\ref{parag}).  
However, geometrical naturality (and, even more widely) convenience, need not be uniquely defined.  
Thus if one splits $\ordial\widetilde{\cs}$, how one splits it is nonunique.  
\beq
\ordial\widetilde{\cs} = \{\fw/\Omega\} \{\Omega\ordial\cs\}
\eeq
will also do.  
I term this a {\bf parageodesic principle splitting conformal transformation (PPSCT)}, the first factor being an ordinary conformal transformation of 
the metric arc element, and the second factor being the corresponding {\it compensatory transformation} of the potential factor.
This occurring for GR and RPM's constitutes Analogy 33).

\subsubsection{PPSCT-tensorialities of the objects of the Principles of Dynamics}\label{PPSCT-Tensor}

1) Squaring and assigning the transformation to the (arbitrary-signature) Riemannian kinetic metric within the $\d \fs^2$, the transformation is  
\beq
{\fM}_{\sfA\sfB} \longrightarrow \overline{\fM}_{\sfA\sfB} = \Omega^2 {\fM}_{\sfA\sfB} \mbox{ } \mbox{ } , \mbox{ } \mbox{ } 
{\fW} \longrightarrow \overline{\fW} = {\fW}/\Omega^2 \mbox{ }  .  
\label{Warh}
\eeq
\noindent 2) I adopt the (for metric geometry, very usual) convention that the basic transformation involves 2 powers of the conformal factor $\Omega$. 
Then the kinetic metric is a PPSCT-vector and the potential factor is a PPSCT-covector (and so, overall, the product-type parageodesic principle action is PPSCT-invariant).  
Thus the inverse of the configuration space metric scales as 
\noindent
\beq
\fN_{\sfA\sfB} \rightarrow \overline{\fN}_{\sfA\sfB} = \Omega^{-2}\fN_{\sfA\sfB}
\label{PPSCTinv}
\eeq 
and the square-root of the determinant of the configuration space metric to scale as 
\beq
\sqrt{|\fM|} \rightarrow \sqrt{|\overline{\fM}|} = \Omega^r \sqrt{|\fM|} \mbox{ } .  
\label{PPSCTdet}
\eeq

\noindent 3) Moreover, $\Last$ is then a PPSCT-covector [immediately from its formula (\ref{emtime})]; 
this means it is in fact highly nonuniquely defined, at least at this stage in the argument.\footnote{To not confuse `$\ft^{\te\tm(\tJ\tB\tB)}$ 
as present in the previous literature'
\cite{B94I, SemiclI} and the PPSCT-covector discovered in \cite{Banal} and explained in this Appendix, I denote the latter by $\vec{\ft}$.
One can also think of this as the emergent lapse, VOTI and DOTI scaling as PPSCT-vectors.} 
%
I.e.,
\beq
\ordial \ft^{\se\sm(\sJ\sB\sB)} \longrightarrow \ordial \overline{\ft}^{\se\sm(\sJ\sB\sB)} = \Omega^2\d \ft^{\se\sm(\sJ\sB\sB)} \mbox{ } .  
\label{7iv)}
\eeq
\noindent 4) Clearly from the invariance of the action, performing such a transformation should not (and does not) affect the 
physical content of  one's classical equations of motion.

\noindent 5) If one then considers the emergent timefunction's PPSCT-vectoriality to carry over to the difference-type action formulations' 
timefunction (including to the Newtonian/proper time {\sl interpretation} usually pinned upon these timefunctions therein), 
then one has unravelled a more complicated manifestation of the conformal invariance for difference-type actions too.
$$
\overline{\FS} = 
\int\int_{\sbSigma} \d\bupSigma\{\overline{\fT}_{\sft} - 
                           \overline{\fV}\}\ordial\overline{\vec{\ft}} 
= \int\int_{\sbSigma}\d\bupSigma
\left\{
\overline{\fM}_{\sfA\sfB}\overline{\Last}_{\sg}{\fQ}^{\sfA}\overline{\Last}_{\sg}{\fQ}^{\sfB}/2 
- \overline{\fV}
\right\}
\ordional\overline{\vec{\ft}} = 
\int\int_{\sbSigma}\d\bupSigma
\{\Omega^2{\fM}_{\sfA\sfB}\Omega^{-2}\Last_{\sg}{\fQ}^{\sfA}\Omega^{-2}\Last_{\sg}{\fQ}^{\sfB}/2 
- \Omega^{-2}{\fV}\}\Omega^2\ordial \vec{\ft}
$$
\beq
= \int\int_{\sbSigma}\d\bupSigma\{{\fM}_{\sfA\sfB}\Last_{\sg}{\fQ}^{\sfA}\Last_{\sg} {\fQ}^{\sfB}/2  - {\fV}\}\ordial \vec{\ft} = 
\int\int_{\sbSigma} \d\bupSigma\{{\fT}_{\sft} - {\fV}\}\d{\vec{\ft}}= \FS \mbox{ } .  
\label{Whela}
\eeq
I.e., one has a {\it 3-part conformal transformation} 
$({\bfM}, {\fW}, \Last) \longrightarrow (\overline{\bfM}, \overline{\fW}, \overline{\Last}) = (\Omega^2{\bfM}, \Omega^{-2}{\fW}, \Omega^{-2}\Last)$.  
[Note that in the case of the relational formulation, the PPSCT of $\bfM$, ${\fW}$ directly implies the 3-part conformal transformation, 
so that these are not then distinct entities, but the 3-part conformal transformation {\sl is} a distinct entity if one {\sl does not} presuppose relational foundations.]  
Working through how the scaling of ${\bfM}$, ${\fW}$ and the timefunction $\ft$ conspire to cancel out at the level of the classical equations of motion reveals interesting 
connections (Sec \ref{+temJBB}) between the simplifying effects of using the emergent timefunction on the equations of motion and those of the rather better-known affine 
parametrization \cite{Wald, Stewart}.

\noindent 6) The evolution equations of motion following from a product-type action (\ref{Taction} in \ref{action}) 
will clearly be invariant under the PPSCT, as the action that they follow from is.  

\noindent 7) The conjugate momenta are PPSCT-invariant: 
\beq
\overline{\fP}_{\sfB} = \overline{\fM}_{\sfA\sfB}\overline{\Last}_{\sg}\fQ^{\sfA} = {\fM}_{\sfA\sfB}{\Last_{\fg}}\fQ^{\sfA} = \fP_{\sfB} 
\mbox{ } .  
\label{20}
\eeq
Thus PPSCT concerns, a fortiori, configuration space rather than phase space.
 
\noindent 8) As $\fN^{\sfA\sfB}$ is the inverse of $\fM_{\sfA\sfB}$, it scales as a PPSCT-covector,  
\beq
\fN^{\sfA\sfB} \longrightarrow \overline{\fN}^{\sfA\sfB} = \Omega^{-2}\fN^{\sfA\sfB} \mbox{ } .  
\label{22}
\eeq
Then (\ref{Warh}ii), (\ref{20}) and (\ref{22}), $\scQ\scU\scA\scD$ is a PPSCT-covector.

\noindent 9) In cases with nontrivial Configurational Relationalism (for which the above arc elements and stars, and thus 
momentum--emergent-time-velocity relations and evolution equations, are $\FrG$-corrected), there are also linear constraints from variation with respect to $\FrG$-auxiliaries.  
Then by the pure linearity in the momenta of these as manifested in (\ref{LinZ}) and by 6), these are conformally invariant.

\noindent 10) The total Hamiltonian is PPSCT-invariant by $\fN$ and $\scQ\scU\scA\scD$ vector--covector cancellation 
and neither $\fm^{\sfZ}$ nor $\scL\scI\scN_{\sfZ}$ scaling in the first place.  
The same shape of argument immediately apply also to the total (O(A))-Hamiltonian.  

\noindent 11) Finally, the Liouville operator $\fP_{\sfA}\dot{\fQ}^{\sfA}$ is PPSCT-invariant, so that the last SSSec of the previous Appendix's 
form for the action is also indeed PPSCT-invariant.

\subsubsection{Further applications}\label{+PPSCT} 

\noindent 1) The constant conformal factor case gives the tick-duration freedom: $\ft^{\se\sm}$ to $k^2\ft^{\se\sm} =  \ft^{\se\sm}_k$, so
\beq 
\ft^{\se\sm}_k - \ft^{\se\sm}_k(0) =: \lft^{\se\sm}_{k} = k^2\int \ordial\fs/\sqrt{2\fW} 
\eeq
for $\ft^{\se\sm}_{\sk}(0) := k^2\ft^{\se\sm}(0)$.  

\noindent 2) The evolution that follow have two simplifications for or particular choices of parameter that are generally different (the two coincide if the potential is constant).   

\noindent Simplification A) use of emergent time. 

\noindent Simplification B) use of geodesic rather than parageodesic form, the former corresponding to `the dynamical curve 
being an affinely-parametrized geodesic on configuration space'. 
In this case I denote the timefunction by $\ft^{\sa\sf\sf-\sg\se\so} = \ft^{\sg\se\so}$, the latter 
name coming from the shape-space geometrical naturality of this time.

\noindent 3) There is a link to the affine transformations of geodesic equations exposited in Sec \ref{+temJBB}.  

\mbox{ } 

\noindent Note: considering 3) and 4) together amounts to following Misner's Hamiltonian treatment \cite{Magic}, now taking it further in tracing the variational principle origin of his conformal transformation of the Hamiltonian constraint back to the relational product-type Jacobi parageodesic principle.

\subsubsection{Further notation for this Article's PPSCT representations}\label{PPSCT-Notation}

\noindent For scaled RPM, the PPSCT representations are centred about the mechanical representation $\Omega = 1$ for $M_{\sfA\sfB}$, $E - V$; 
the corresponding time is $t_{\sm\se\scc\sh}$, usually abbreviated to $t$.  
The scaled triangleland flat representation has $\Omega^2 = 4I$ and objects in this representation are denoted by checks.
Occasionally I also use $\Omega^2 = I$, which I denote by bars.

For pure-shape RPM, I choose to centre the PPSCT representations about the geometrically-natural representation, 
so that $\Omega = 1$ for $\fM_{\sfA\sfB}$ and $\fE - \fV$; the corresponding emergent time is $\ft_{\sg\se\so}$, usually abbreviated to $\ft$.   
I then need to use the five-pointed star $\5Star$ for its derivative to avoid confusion with ERPM's six-pointed star * which is distinct 
due to the difference in PPSCT representation-centring conventions.
In SRPM, the mechanical representation (`mech' subscript) then has $\Omega = \sqrt{I}$.
Also, the flat representation of pure-shape 4-stop metroland has  $\Omega = \{1 + {\cal R}^2\}/2$ and objects in this representation are then denoted by bars.   
The pure-shape triangleland flat representation has  $\Omega = 1 + {\cal R}^2$ and objects in this representation are also denoted by bars.
The pure-shape triangleland breved representation has $\Omega = 2$.

\subsubsection{Problems with (para)geodesic principles}\label{POZ}

A principal hope/application of configuration space studies is representing motions by geodesics on configuration space. 

\mbox{ }  

\noindent {\bf Problem of edges/singularities of configuration space}
Geodesics can however run into the these, (e.g. the $N$ = 3, $d$ = 1 straight (half-) line geodesics that go into the triple collision). 
Thus it is interesting to investigate the nature of these edges and singularities (c.f. Sec \ref{Q-Geom}).
It is then helpful that the study of singularities for particle mechanics is well-developed \cite{Diacu}, 
particularly for the classical $1/|\r^{IJ}|$ potentials though also for similar potentials such as $1/|\r^{IJ}|^l$, $l > 0$.    
These can include curvature singularities on configuration space, c.f. Sec \ref{Obgon} and \ref{Obgon2}.  
While studying this at the level of the relationalspaces is less common, at least some such studies have been done \cite{Montgomery2, Gergely, GergelyMcKain}.    
Ideally, one would like to know how typical it is for the motion to hit such a boundary and what happens to the motion after hitting the boundary.  
One possibility is that boundaries are singular, a simple subcase of this being when the boundaries 
represent curvature singularities of the configuration space.  
E.g. the $N$ = 3, $d$ = 2 or $3$ relational space metric blows up at the triple collision \cite{GergelyMcKain}, 
while the shape metric is better behaved in being of constant positive curvature \cite{Montgomery2}. 
Also, the $N$ = 4, $d$ = 1 shape metric is of finite curvature.  


\noindent{\bf Problem of zeros, infinities and nonsmoothnesses (POZIN)}.  
Additionally note that one requires not the `bare' metrics ${\fM}_{\sfA\sfB}$ but the physical 
$\widetilde{\fM}_{\sfA\sfB} = \fW {\fM}_{\sfA\sfB}$ for each ${\fW}$ in order to encode motions as geodesics.  
Now, clearly, performing such a conformal transformation generally alters geodesics and curvature.   
Additionally, this transformation cannot necessarily be performed on extended regions since it requires 
a smooth nonzero conformal factor while $\fW$ in general has zeros or unbounded/rough behaviour.  
This limits such use of Jacobi-type actions as geodesics.
In the case of Newtonian gravitational potentials, this is a condition of having to stay within the so-called Hill's regions (see e.g. \cite{Marchal}). 
Here, such zeros are  always `halting points' in the sense that $p_{i} = 0$ there by  
$0 = W = T(p_i)$ (conservation of energy and positive-definiteness of the mechanical $T$). 

\mbox{ } 

\noindent Analogy 34) Both GR and RPM's possess this POZIN.  

\noindent Difference 14) Moreover, the nature of such zeros is different in GR. 
This is due to RPM definiteness versus GR indefiniteness of the kinetic term [Difference 17)], which is relevant by zeros of the 
potential corresponding to zeros of the kinetic term via the quadratic constraint.
Definiteness causes a barrier which indefiniteness causes merely spurious zeros \cite{Magic} since the 
kinetic term now need not get stuck at zero momentum, but, rather, the motion may continue through the 
zero on the Superspace null cone, which is made up of perfectly reasonable Kasner universes.   

\mbox{ }  

\noindent Note 1) To illustrate that the presence of zeros in the potential term is an important occurrence in GR, note 
that Bianchi IX spacetimes have an infinity of such zeros as one approaches the cosmological singularity.  
Furthermore, these spacetimes are important due to the BKL conjecture.  
The above argument was put forward in \cite{TB93} to argue against the validity of the use of the 
`Jacobi principle' to characterize chaos in GR \cite{VSydlowski}.  

\noindent Note 2) more generally, 
\beq
0 < \Omega(\fQ^{\sfA}) < \infty \mbox{ } , \mbox{ }  \mbox{ }  \Omega(\fQ^{\sfA}) \mbox{ smooth }
\eeq
is needed throughout the region of physical interest for the two-way conformal transformation in question to be defined.

\subsection{Accommodating Fermions into Relationalism}\label{AFermi}

These can be accommodated via lifting the homogeneous quadraticity restriction \cite{Lanczos, Van} (which is not per se relational).
In particular this allows for mechanics with linear `gyroscopic' terms \cite{Lanczos}, as well as moving charges, spin-1/2 fermions 
and the usual theories coupling these to gauge fields, scalars and GR. 
The form of action here is of Randers type \cite{Randers, ARel2} (a subcase of Finsler geometry, modulo degeneracy issues)  
\beq
\FS^{\sR\sW\sR}_{\st\sr\si\sa\sll-\sA} = \int\int_{\sbSigma}\d\bupSigma \big\{\sqrt{2}\sqrt{\mW^{\st\sr\si\sa\sll}}\ordial \ms^{\st\sr\si\sa\sll}_{\sq\su\sa\sd} + 
\ordial\ms^{\st\sr\si\sa\sll}_{\sll\si\sn} \big\} \mbox{ } .
\label{62}
\eeq
$\ordial s^{\st\sr\si\sa\sll}_{\sll\si\sn}$ does not then contribute to the quadratic constraint or to the emergent JBB time (thus this is not entirely universal, 
though the potential associated with such fields does sit inside the square root and thus itself does contribute to these things; see also the scheme in Appendix \ref{+temJBB}.A.
$\ordial \ms^{\st\sr\si\sa\sll}_{\sll\si\sn}$ does however in general contribute to the linear constraints.

\noindent Note 1) This action is clearly MPI and thus implements Temporal Relationalism.    

\noindent Note 2) This clearly generalizes to a much wider range of combinations of roots and sums.

\noindent Note 3) This also renders clear how to extend Christodoulou's Chronos Principle working to include fermions.  

\noindent Note 4) Finally, Poincar\'{e}'s Principle now has to concern one position and direction per bosonic degree of freedom alongside just one position per fermionic degree 
of freedom.

\mbox{ } 

\noindent I further discuss these variants on relationalism involving boson--fermion distinctions (and the supersymmetric case) in \cite{Arelsusy}.  

\end{subappendices}

\vspace{10in}

\section{Geometry of relationalspaces and RPM's thereupon}\label{Q-Geom}

Understanding the configuration space $\FrQ$ is very important for, firstly, classical dynamics and quantum mechanics in general.
Secondly this Article's Relationalism 3) postulate is that $\FrQ$ is primary (though I {\sl discuss whether} to adhere to this postulate rather than always taking it for granted).
Thirdly, $\FrQ$ is important as regards a number of POT approaches (the timeless case in particular very much concerns the study of configurations and configuration spaces).

{            \begin{figure}[ht]
\centering
\includegraphics[width=0.43\textwidth]{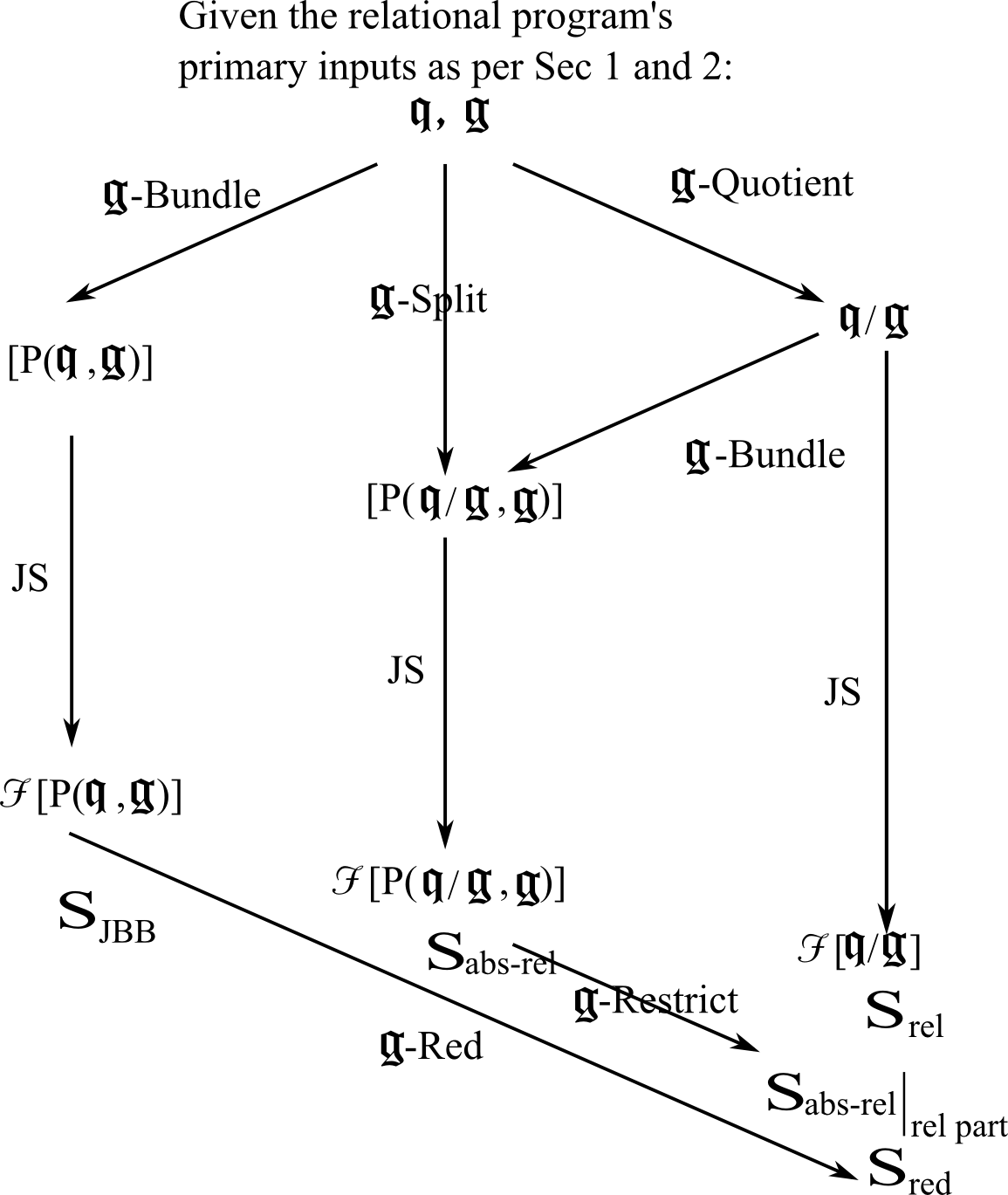}
\caption[Text der im Bilderverzeichnis auftaucht]{        \footnotesize{A, B, C.I) and D can be identified as paths of moves on this diagram.  
The ${\cal F}$'s are just the spaces the actions themselves live on.  A priori, it is not clear whether the cluster of three actions in the bottom-right corner will coincide.}        }
\label{ABCD}\end{figure}            }

\noindent In this chapter, I approach $\FrQ$ for RPM's from first principles.  
I do so due to the availability of Kendall's study \cite{Kendall84, Kendall} (along with his book's coauthors Barden, Carne and Le and his former student Small, who wrote an earlier book on this subject \cite{Small}). 
For, despite this study not being done in the context of mechanics, turns out to be the right mathematics for the configuration space study of pure-shape RPM.  
Furthermore, the cone construction gives the scale--shape split scaled RPM configuration spaces from pure-shape RPM's configuration spaces (shape spaces).
Then I term constructing the natural mechanics action $\FS^{\sr\se\sll\sa\st\si\so\sn\sa\sll\sss\sp\sa\scc\se}$ associated with this by the Jacobi--Synge parageodesic principle

\mbox{ } 

\noindent A) the {\bf relationalspace approach}.
I identify this as a map $\FrG$-Rel: $(\FrQ, \FrG) \longrightarrow {\cal F}(\FrQ/\FrG)$, with, via Fig \ref{ABCD},
\beq
\mbox{$\FrG$-Rel} = \mbox{JS} \circ \mbox{$\FrG$-Quotient} \mbox{ } .
\eeq

\noindent This is a priori disjoint from the work so far in this Article, i.e.    

\mbox{ }

\noindent B) the {\bf indirect approach} that constitutes Sec 2, which I identify via Fig \ref{ABCD} as the map 
\beq
\mbox{(Best Matching with respect to $\FrG$)} := \mbox{$\FrG$-BM} = 
\mbox{$\FrG$-Red} \circ \mbox{$\FrG$-JBB} := \mbox{$\FrG$-Red} \circ \mbox{JS} \circ \mbox{$\FrG$-Bundle} \mbox{ } ,  
\eeq
which is, a priori, another map $\FrG$-{Rel}:$(\FrQ, \FrG) \longrightarrow {\cal F}(\FrQ/\FrG)$ and eventually leads to an $\FS_{\sr\se\sd}$ 
(the best matching has not been performed yet in this Article other than in the very trivial case of 1-$d$ scaled RPM).  

\noindent Thus one will for the moment need to take it on trust that this investigation will indeed join up 
with Sec \ref{Examples}'s, representing therefore a second, independent foundation (a `second route' to RPM's in Wheeler's parlance).   
I demonstrate in Secs \ref{DRIII} to \ref{RelRedNM} that the two match up, at the level of   
 
\mbox{ }  
 
\noindent C.I) {\bf configuration space reduction}, which is in this context one and the same as the best matching procedure 
and the first 5/6ths of the relational-mechanical counterpart of BSW.  

\mbox{ }

\noindent I note that reduction can also be done at other levels: 

\mbox{ } 

\noindent C.II) (O(A))Phase space reduction, and 

\noindent C.III) Reduction at the level of the QM equations (e.g. in Sec \ref{QM-Intro}).  

\noindent [There is also 

\noindent C.IV) reduction {\sl after} quantization alias the {\it Dirac quantization scheme} which is at the level of the QM solutions themselves.]

\mbox{ }

\noindent Note also the additional scheme

\mbox{ }

\noindent D) rearranging Newtonian Mechanics by {\bf passing to absolute--relational split generalized coordinates}, so as to look at the relational portion sitting within this.

\mbox{ }  

\noindent Note 1) [A generalization of D)] in Fig \ref{ABCD} can be viewed as part of a third map 
$\FrG$-split $\circ$ JS $\circ \,\, \FrG$-Restrict: $(\FrQ, \FrG)  \longrightarrow{\cal F}(\FrQ/\FrG)$.  
However, I only make use of the first two steps of D) in the present Article.  

\noindent Note 2) These are useful through D) sharing kinematical structure with A) to C), and this kinematical structure then being well-known in D) as part and parcel of D) 
being an elsewhere considerably studied subject!    
In other words, D) is a source of coordinate systems that continue to be useful in setting up A) and in the reduction C) of B).  
E.g. Jacobi coordinates (Sec \ref{Rel-Jac}), spherical-type and inhomogeneous coordinates (Sec \ref{Riem}), parabolic-type coordinates, Dragt coordinates \cite{Dragt}), 
and democratic invariants \cite{Zick, LR95} (all in \ref{Shape-Quant}), and no doubt further such in the study of 3-$d$ cases beyond the scope of this Article.
For D) see e.g. Iwai \cite{Iwai87}, Montgomery \cite{Mont96, Mont98}, Hsiang \cite{Hsiang1}, and in work reviewed by Littlejohn and Reinsch \cite{LR97}.  
The Molecular Physics--RPM connection was first envisaged by Gergely, though he tapped this connection to a far smaller extent than my own research program has done.  
The lack of awareness of $N$-body Physics among Theoretical Physicists is a somewhat worrisome trend, e.g. basic and Theoretical Physics textbook QM is often misguiding 
as regards it being configuration space rather than space that is central (by overuse of 1-particle examples at the expense of $N$-body problem examples). 
The present Article looks to partly reverse this trend, both by making use of Molecular Physics techniques and analogies and 
by showing that $N$-body Physics has its own versions of background-dependence versus background-independence and   
nontrivially exhibit many of Theoretical Physics's conceptual problems as regards time and closed universes. 
 
\noindent Note 3) 4 particles in 2-$d$ is {\sl not} habitually covered in the Molecular Physics literature, however I provide contact with 
distinct pre-existing sources of useful coordinate systems that are readily applicable to this problem in Sec \ref{Shape-Quant} and \cite{QuadI}.  

\noindent Note 4) The absolute--relational split action $\FS_{\sa\sb\sss-\sr\se\sll}$ is useful in Sec \ref{QM-Intro} in pointing out 
quantum discrepancies between the particular mathematical forms of absolute and relational QM.  

\noindent Note 5) This Sec's success with the reduction approach for 1- and 2-$d$ RPM's amounts to the following.  

\mbox{ }

\noindent Difference 15) The Thin Sandwich/Best Matching Problem is solved for 1- and 2-$d$ RPM's, unlike in GR.  

\noindent Analogy 35) If one wishes to continue to have such an obstruction, however, 3-$d$ RPM's (for which the entirely 
physically reasonable collinerities have singular inertia tensors in 3-$d$, obstructing the elimination of the rotational auxiliary) remain impassible here.  

\mbox{ }

\noindent Note 6) This Sec's scheme A) has {\sl no} counterpart of Barbour's wooden triangles demonstration: in it, one is held to already have the extremizing sequence of shapes.  

\noindent Note 7) Butterfield \cite{BuBe1} uses `relational' for my C.I) (configuration space reduction) and `reductive' for my C.II) (phase space reduction);  
Belot \cite{BuBe2, BuBe3} also compares these two approaches.  
However, I consider these both to be reductions in different formalisms and coincident for many purposes.   

\noindent Note 8) I will subsequently find that (Sec \ref{QM-Intro})  for this Article's RPM examples, all of A to C look to be coincident, 
bar C.IV, which can differ due to interference of operator-ordering issues. 
I also find that D) (the relational--absolute split of Newtonian Mechanics) can likewise be distinct.  
A remaining question is: are these classically distinct as well as quantum-mechanically distinct?

\noindent Note 9) The indirect formulation A) retains the virtue of being the closest 
analogue to the geometrodynamical formulation of GR alongside all of its uses and developments in Quantum Cosmology and the POT.

\subsection{Preliminary definitions}\label{Wiiii}

I use the following additional index types. 
${\cal A}$ for preshape space coordinates, $\tta$ for shape space coordinates, with $\bar{\tta}$ and $\tilde{\tta}$ for radial and angular such in 2-$d$ respectively.

\noindent I denote by Quot$_{N, d}$ the nontrivial {\it quotient map}: $\FrS(N, d) \longrightarrow \FP(N, d)$.  
I usually abbreviate this to Quot.  

\mbox{ }

\noindent{\bf Lemma 2 (real representations)} The configuration space $\FrQ(\mN, d)$ can be represented by $\bq = \{{\q}^I\}$, the particle position coordinates.   
The relative configuration space $\FrR(N, d)$ can be represented by $\br = \{ \r^i\}$, a basis set of relative coordinates.  
Examples of such are 1) a basis subset of the relative particle positions $\r^{IJ} = \q^{I} - \q^{J}$. 
2) the relative Jacobi coordinates $\R^i$.  
Preshape space $\FP(N, d)$ can be represented by $\bar{\br} = \{ \bar{\r}^{\cal A}\}$ obtained from the preceding by normalization.
N.B. that preshape space is all that one needs for nonrotational RPM.

\subsection{Topological structure of (pre)shape space}\label{Top}

At the topological level, $\FA(d) = \mathbb{R}^{d}$, $\FrQ(N, d) = \mathbb{R}^{Nd}$ and $\FrR(N, d) = \mathbb{R}^{nd}$ are all simple and well-known.
Next, it is natural to ask what topology the various reduced configuration spaces have.  
Figs \ref{3-1} and \ref{Top-S} set about a first-principles investigation of this for simple cases (along the lines of \cite{06II}).

{            \begin{figure}[ht]
\centering
\includegraphics[width=0.25\textwidth]{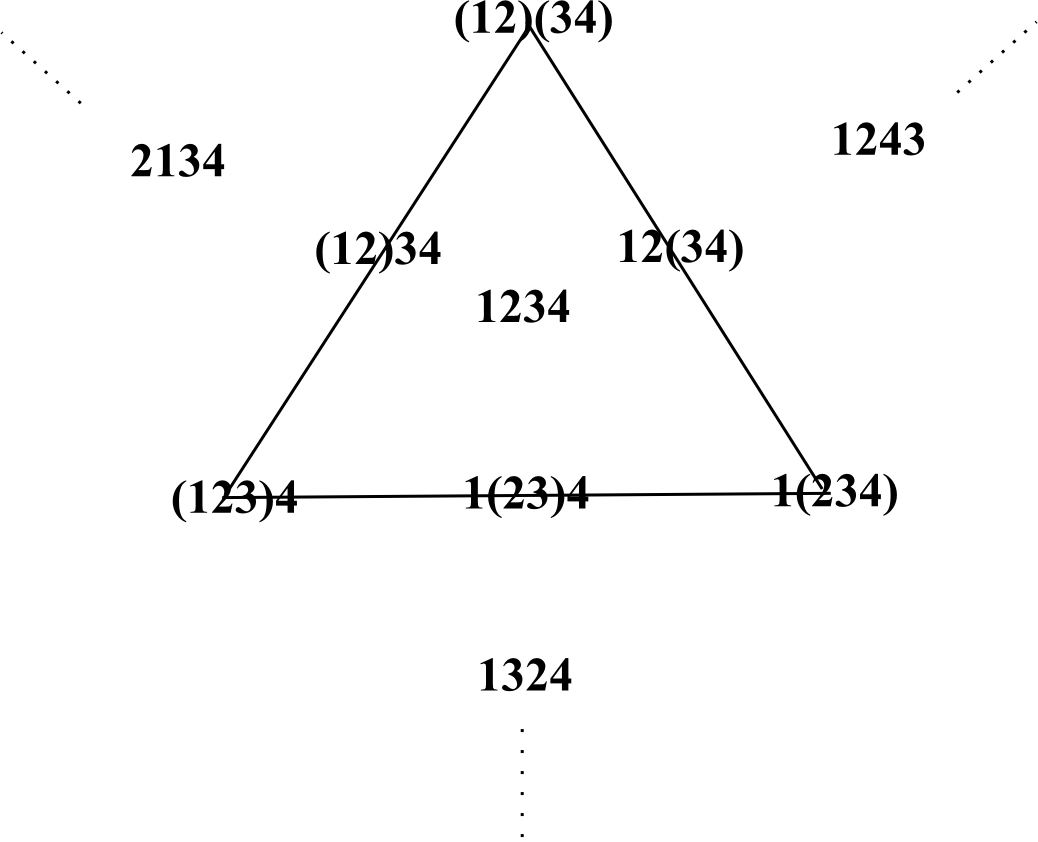}
\caption[Text der im Bilderverzeichnis auftaucht]{        \footnotesize{ 
A sketch of the construction for the example of 4-stop metroland, regarding the particles 1, 2, 3, 4 as 
distinguishable and the order 1234 as distinct from 4321.
Starting in the order 1234, one has a 2-$d$ region of shape space bounded by three double-collision line segments, and simultaneous double and triple collision points.  
Next, deduce which regions lie adjacent to this by continuous deformation into permutations one swap away from the preceding 2-$d$ region.  
Keep on going until all permutations of the labels have been covered and thus the whole configuration space has been covered 
(permit no identically-labelled region to occur more than once, by use of identification as necessary.}        }
\label{3-1}\end{figure}            }

{            \begin{figure}[ht]
\centering
\includegraphics[width=0.8\textwidth]{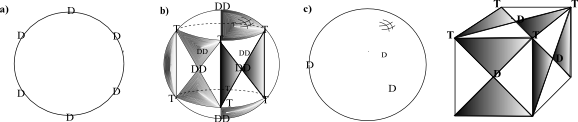}
\caption[Text der im Bilderverzeichnis auftaucht]{        \footnotesize{Applying Figure \ref{3-1}'s 
procedure to various cases results in the following.  

\noindent a) $\FrS(3, 1) = \mathbb{S}^1$.  
The six D's are double collisions, and they cut up the circle into 6 arcs.  

\noindent b) $\FrS(4, 1) = \mathbb{S}^2$.  
The 8 T's are triple collisions and the 6 DD's are double--double collisions. 
Each DD is attached to 4 T's, and each T to 3 T's and 3 DD's, in each case by single double collision lines.  
There are also 36 line segments that are double collisions (D). 
These tessellate the sphere into 24 identical disjoint regions.

\noindent c) $\FrS(3, 2)  = \mathbb{S}^2$. 
There are three D points, so, if these points are excised, this is the `pair of pants' \cite{ArchRat, Montgomery2}.  
Note that these 3 points are the half/mirror-image-identified  version of a) [i.e. Fig \ref{Fig-3-3}a)].  

\noindent d) N.B. quadrilateralland is harder to visualise than the previous subsections' shape spaces, due to greater dimensionality 
as well as greater geometrical complexity and a larger hierarchy of special regions of the various possible codimensions. 
The topologically-defined decorations on $\mathbb{CP}^2$ are 6 pairs of spheres (each of which is a triangleland) which share 4 T points and 3 DD points. 
As these are 2-$d$, how they fit together is likewise visualizable. 
They are arrayed as the corners and face centres of the identified cube. 
Thus each shaded triangle in fact represents a pair of spheres. 
Removing this set is like forming the connected pants rather than 4-stop metroland's slice-up into 24 disjoint regions, since the codimension is 2 and not 1.
There is also a net of 4-stop metrolands with one being a D point on an $\mathbb{RP}^2$ mirror image
identified by rotation in 2-$d$; note the similarity between these and the half/mirror-image-identified part of b) [i.e. Fig\ref{Fig-3-3}b)].\normalsize }        }
\label{Top-S}\end{figure}            }

\noindent I subsequently found that systematized investigations into the question of shape space topology had 
already been done in 1- and 2-$d$ by Kendall and co-workers \cite{Kendall, Kendall80, Kendall84}. 
Whilst these investigations did not consider such as the configuration space of a mechanics, I bridged the gap via the Jacobi--Synge procedure 
(\cite{FORD}, Sec \ref{DRIII}--\ref{RelRedNM}).  
Kendall's own interests in these spaces have been pure-mathematical geometry and statistical applications. 
(Given a set of standing stones, or quasars, say, how many approximate alignments are 
needed therein for further explanations beyond mere chance to be necessary? See Sec 24 for more.)  
Kendall also worked on the 3-$d$ case, but found that to be much harder \cite{Kendall, Kendall43A, Kendall43B, Kendall43C}.  
In contrast, 1- and 2-$d$ are be straightforward, leading to their immediate exploitability via RPM's as whole-universe models being toy models for a wide range of POT strategies.
Kendall's work proceeds via the simpler preliminary treatment of preshape space, to which I next turn.

\mbox{ } 

\noindent{\bf Lemma 3 (`Preshape space is always simple')}. $\FP(N, d) = \mathbb{S}^{n d - 1}$ homeomorphically.  

\noindent\underline{Proof} Immediate from the definition of $\FP(N, d)$.

\mbox{ }

\noindent{\bf Corollary 1} $\FrS(N, 1) = \FP(N, 1) = \mathbb{S}^{n - 1} = \mathbb{S}^{N - 2}$.

\mbox{ }

\noindent{\bf Lemma 4  (Beltrami coordinates)}.
Straightforwardly, $\mathbb{S}^{nd - 1}$ can be coordinatized by $b^{\cal A} = \rho^{a\upalpha}/\rho^{11}$ for $a,\mu \neq 1,1$.
 
\mbox{ } 
 
\noindent Note: these are clearly `projective' rather than spherical coordinates.
 
\mbox{ } 

\noindent{\bf Lemma 5 (complex representations).} 
i) In 2-$d$, relative configurations can be represented by $n$ complex numbers $\mbox{boldmath$z$}^{a}$ -- the {\it homogeneous coordinates}.  

\noindent ii) Assume that not all of these coordinates are simultaneously 0 (i.e. exclude the {\it maximal collision}).  
Then 2-$d$ shapes can be represented by $N$ -- 1 independent complex ratios, the {\it inhomogeneous coordinates} ${Z}^{{\mbox{\scriptsize \tt a}}}$.

\noindent 
[This is straightforwardly established by dividing the complex numbers in i) by a particular $z^{a}$ and ignoring the 1 among the new string of complex numbers.]

\noindent I denote the corresponding polar forms by $z^{a} = \rho^{a}\mbox{exp}({i\phi^{a}})$ and $Z^{{\stta}} = {\cal R}^{\bar{\stta}}\mbox{exp}({i\Phi^{\widetilde{\stta}}})$.  
These are independent ratios of the $z^{a}$, and so, physically their magnitudes ${\cal R}^{\bar{\stta}}$ are ratios of 
magnitudes of Jacobi vectors, and their arguments $\Phi^{\widetilde{\stta}}$ are now angles between Jacobi vectors, which are entirely relational quantities.   
Also, I use $|{Z}|_{\scc}\mbox{}^2 := \sum_{\stta}|{Z}^{{\stta}}|^2$, $( \mbox{ } \cdot \mbox{ } )_{\scc}$ 
for the corresponding inner product, overline to denote complex conjugate and $|\mbox{ }|$ to denote complex modulus.

\subsubsection{The simpler shape spaces as topological surfaces}\label{SS-Top}

\noindent{\bf `Partial Periodic Table' Theorem [1)]}. At the topological level, the discernible shape spaces include 3 simple series [Fig \ref{Discern}]: 
\noindent i) $\FrS(N, 1) = \mathbb{S}^{n - 1} = \mathbb{S}^{N - 2}$,

\noindent ii) $\FrS(N, 2) {=} \mathbb{CP}^{N - 2}$ and

\noindent iii) $\FrS(d + 1, d) {=} \mathbb{S}^{\{d - 1\}\{d + 2\}/2}$.

\noindent\underline{Proof}
\noindent i) Corollary 1). 
ii) Begin with Lemma 5.ii)'s standard coordinatization of $\mathbb{CP}^{N - 2}$. 
Then quotient exp(${i\omega}$), i.e. $U(1)$, i.e. the $SO$(2) rotations:  
in the form $\mathbb{CP}^{n - 1} = \mathbb{S}^{2 n - 1}/U(1)$, this is a well-known result of Hopf 
[which generalizes the most basic Hopf fibration, $\mathbb{CP}^1 = \mathbb{S}^2 = \mathbb{S}^3/U(1)$].
\noindent iii) is Casson's Theorem (proven e.g. on p20-22 of \cite{Kendall}).

{            \begin{figure}[ht]
\centering
\includegraphics[width=0.45\textwidth]{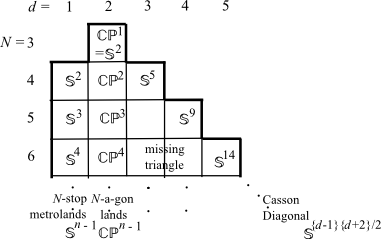}
\caption[Text der im Bilderverzeichnis auftaucht]{        \footnotesize{`Partial Periodic Table' of the discernible shape spaces 
at the topological level. 
This and its metric counterpart (Fig \ref{PerPer}) are the second key to unlocking RPM's.}        } \label{Discern}\end{figure}            }

\noindent Note 1) there is indeed agreement on the overlap of i) and iii) by the well-known homeomorphic equivalence of $\mathbb{CP}^1$ and $\mathbb{S}^2$.  
Thus $\FrS(3, 2) = \mathbb{S}^2$ homeomorphically.

\noindent Note 2) There are further shape spaces for $N \leq d$ but these are all indiscernible from each other and from $\FO\FrS(d + 1, d)$, so I treat them in Sec \ref{Top2}.

\noindent Note 3)  i), ii) and iii) are series with particular simple patterns, at least at the topological level (Fig \ref{Discern}); 
this is the first half of the second key observation toward understanding simple concrete examples of RPM's.  

\noindent Note 4) Once one also takes into account only the discernible shape spaces as per Sec \ref{Setup-RPM}, this is all the shape spaces bar the 
missing triangular wedge  in Fig \ref{Discern}. 
The missing triangular wedge's spaces are likely new in this context rather than known from elsewhere in mathematics, even at the topological level.  
See Chapters 2-5 of \cite{Kendall} for a partial characterization of the new spaces at the topological level.\footnote{Also see e.g. 
\cite{Kendall, Kendall43A, Kendall43B, Kendall43C} for 4 particles in 3-$d$.
} 

\noindent Note 5) $\mathbb{CP}^{n - 1}$ involves $n$ lines, whilst $n$ lines can be used to form whichever Jacobi tree for an $N$-a-gon.
This lucidly explains why $\mathbb{CP}^{n - 1}$ is naturally representable as the space of all $N$-a-gons.

\subsubsection{Some detailed topological properties of (pre)shape space}
\label{Det}

Use this as a topological  characterization, parts of which have further physical significance.  
In the 3-$d$ case, this topological characterization is as good a description of what the topology {\sl is} for the spaces in question \cite{Kendall}.  

\noindent Now that $\FP(N, d)$, $\FrS(N, 1)$ and $\FrS(N, 2)$ have been identified as spheres and complex projective spaces, the following classically and quantum-mechanically useful 
topological information about paths on, and obstructions in, these configuration spaces becomes available.    
$\FP(N, d) = \mathbb{S}^{nd - 1}$, $\FrS(N, 1) = \mathbb{S}^{n - 1}$ and $\FrS(N, 2) = 
\mathbb{CP}^{n - 1}$ are compact without boundary and Hausdorff.  
%

\noindent Following from \cite{Hatcher}, the homotopy groups (useful for the classification of classical paths) 
of $\FP(N, d)$ exhibit the simple pattern  
$\pi_{p}(\FP(N, d)) = \pi_{p}(\mathbb{S}^{n d - 1}) = 
\mbox{\Huge \{}
\stackrel{    \mbox{$\mathbb{Z}$}    }{    0    }
\stackrel{    \mbox{ } \mbox{ $p$ = $nd$ -- 1 $> 1$}    }{  \mbox{ } \mbox{ $p$ $ < $ $nd$ -- 1} } 
\mbox{ } .$ 

\noindent The first few homotopy groups in the remaining wedge are \cite{Toda}

\noindent
\begin{tabbing}
                 \hspace{1.8in}                 \= 
                 \hspace{0.2in}                 \=
$\pi_3$          \hspace{0.4in}                 \=
$\pi_4$          \hspace{0.4in}                 \=
$\pi_5$          \hspace{0.4in}                 \=
$\pi_6$          \hspace{0.4in}                 \=   
$\pi_7$          \hspace{0.4in}                 \=
$\pi_8$          \hspace{0.4in}                 \=
$\pi_9$          \hspace{0.4in}                 \=   
$\pi_{10}$       \hspace{0.4in}                 \=
$\pi_{11}$       \hspace{0.4in}                 \=     \\ 
\FP(4, 1) = \FrS(3, 2) = \> $\mathbb{S}^{2}$     \>
$\mathbb{Z}$                                    \>
$\mathbb{Z}_2$                                  \>
$\mathbb{Z}_2$                                  \>
$\mathbb{Z}_{12}$                               \>
$\mathbb{Z}_{2}$                                \>
$\mathbb{Z}_{2}$                                \>
$\mathbb{Z}_{3}$                                \>
$\mathbb{Z}_{15}$                               \>
$\mathbb{Z}_{2}$                                \>   \\
\FP(5, 1) = \FP(3, 2) = \> $\mathbb{S}^{3}$     \>
                                                \>
$\mathbb{Z}_2$                                  \>
$\mathbb{Z}_2$                                  \>
$\mathbb{Z}_{12}$                               \>   
$\mathbb{Z}_{2}$                                \>
$\mathbb{Z}_{2}$                                \>
$\mathbb{Z}_{3}$                                \>
$\mathbb{Z}_{15}$                               \>
$\mathbb{Z}_{2}$                                \>   \\
\FP(6, 1) = \> $\mathbb{S}^{4}$                 \>
                                                \>
                                                \>
$\mathbb{Z}_2$                                  \>
$\mathbb{Z}_2$                                  \>
$\mathbb{Z} \times \mathbb{Z}_{12}$             \>
$\mathbb{Z}_{2} \times \mathbb{Z}_2$            \> 
$\mathbb{Z}_{2} \times \mathbb{Z}_2$            \>
$\mathbb{Z}_{24} \times \mathbb{Z}_3$           \>
$\mathbb{Z}_{15}$                               \>    \\
\FP(7, 1) = \FP(4, 2) = \FrS(4, 3) \> $\mathbb{S}^{5}$     \>
                                                \>
                                                \>
                                                \>
$\mathbb{Z}_2$                                  \>
$\mathbb{Z}_{2}$                                \>
$\mathbb{Z}_{24}$                               \>
$\mathbb{Z}_{2}$                                \>
$\mathbb{Z}_{2}$                                \>
$\mathbb{Z}_{2}$                                \>    \\
\FP(8, 1) = \> $\mathbb{S}^{6}$                 \>
                                                \>
                                                \>
                                                \>
                                                \>
$\mathbb{Z}_2$                                  \>
$\mathbb{Z}_{2}$                                \>
$\mathbb{Z}_{24}$                               \>
0                                               \>
$\mathbb{Z}$                                    \>    \\
\FP(9, 1) = \FP(5, 2) = \> $\mathbb{S}^{7}$     \>
                                                \>
                                                \>
                                                \>
                                                \>
                                                \>
$\mathbb{Z}_2$                                  \>
$\mathbb{Z}_{2}$                                \>
$\mathbb{Z}_{24}$                               \>
0                                               \>    \\
\end{tabbing} 
By $\FrS(N, 1) = \FP(N, 1)$, the corresponding results for the $N$-stop metrolands $\FrS(N, 1)$ are readily obtained from the above by setting $d$ = 1.  
The pattern is    
$\pi_{p}(\FP(N, 1)) = \pi_{p}(\mathbb{S}^{N - 1}) = 
\mbox{\Huge \{}
\stackrel{    \mbox{$\mathbb{Z}$}    }{    0    }
\stackrel{    \mbox{ } \mbox{ $p$ = $n$ -- 1 $> 1$}    }{  \mbox{ } \mbox{ $p$ $ < $ $n$ -- 1} } $, 
while the remaining cases' table above includes the other $\FrS(N, 1)$ results.  
Likewise, the Casson diagonal is included by using $\{d - 1\}\{ d + 2\}/2$ in place of $nd$.  

\mbox{ }

\noindent From \cite{Hatcher}, it follows that the homology and cohomology groups of preshape space are  

\noindent $H_{p}(\FP(N, d)) = H_{p}(\mathbb{S}^{n d - 1}) = 
\mbox{\Huge \{}
\stackrel{    \mbox{$\mathbb{Z}$}    }{    0    }  
\stackrel{    \mbox{ } \mbox{ $p$ $= 0$ or $nd$ -- 1}    }{   \mbox{ } \mbox{ otherwise}    } \mbox{\Huge \}} = 
H^{p}(\mathbb{S}^{nd - 1}) = H^{p}(\FP(N, d)).$
Thus, for $N$-stop metroland,  

\noindent
$H_{p}(\FrS(N, 1)) = 
\mbox{\Huge \{}
\stackrel{    \mbox{$\mathbb{Z}$}    }{    0    }  
\stackrel{    \mbox{ } \mbox{ $p$$ = 0$ or $n$ -- 1}    }{   \mbox{ } \mbox{ otherwise}    } 
\mbox{\Huge \}} 
= H^{p}(\FrS(N, 1))$. 

The first and second Stiefel--Whitney classes are trivial for all spheres \cite{Nakahara}.  
The main use of this is that it implies that $\FP(N, d)$ are all orientable and admit a nontrivial spin structure, and likewise for $\FrS(N, 1)$. 
This is required e.g. for there to exist a fermionic theory thereupon.

\mbox{ }

\noindent From \cite{Whitehead}, it follows that the homotopy groups $\pi_{p}(\FrS(N, 2)) = 
\pi_{p}(\mathbb{CP}^{N - 2}) = \mbox{\Huge \{} \stackrel{\mbox{$\mathbb{Z}$}}{\pi_{p}(\mathbb{S}^{2N - 3})} \stackrel{\mbox{$p$ = 2}}{\mbox{ otherwise}}$.

\noindent From \cite{Hatcher}, it follows that the homology and cohomology groups are 

\noindent $H_{p}(\FrS(N, 2)) = H_{p}(\mathbb{CP}^{n - 1}) = 
\mbox{\Huge \{}
\stackrel{   \mbox{$\mathbb{Z}$}    }{    0    }  
\stackrel{   \mbox{ } \mbox{ $p$ even up to $2\{n - 2\}$}    }{  \mbox{ } \mbox{ }  \mbox{ otherwise}    } 
\mbox{\Huge \}}
= H^{p}(\FrS(N, 2)) = H^{p}(\mathbb{CP}^{n - 1})$.

The first Stiefel--Whitney classes are trivial for all complex projective spaces \cite{Nakahara}, which implies that all the $\FrS(N, 2)$ are orientable.  
The second Stiefel--Whitney classes are trivial for $N - 2$ an odd integer. 
Thus $\FrS(N, 2)$ for $N$ odd admit a nontrivial spin structure.
On the other hand, they are nontrivial [equal to the generator of $H^2(\mathbb{CP}^{n - 1}, \mathbb{Z})]$ for $N - 2$ an even integer. 
Thus nontrivial spin structures do not exist for $\FrS(N, 2)$ with $N$ even due to topological obstruction.
However, generalized `$\mbox{spin}^c$' structures do exist \cite{Pope} in these cases.

\subsubsection{The simpler shape spaces as metric spaces}\label{Met}

The $\mathbb{R}^{d}$ inner product serves to have a notion of `localized in space', which survives in some form for all the configuration spaces considered.  
This is useful for discussing observable configurations.

\noindent One also has have a notion of localized in configuration space -- i.e. of which configurations look 
alike (this is important as in Physics one does not know precisely what configuration one has; see Sec \ref{Cl-Str} for more).  
For this one has available the possibly-weighted $\mathbb{R}^{N d}$ norm  $||\mbox{ }||$ as per Sec \ref{Q-for-RPM}.
Likewise, one has the possibly-weighted $\mathbb{R}^{n d}$ norm  $||\mbox{ }||$ and the corresponding inner product for $\FrR(N, d)$.
[If the Jacobi coordinates themselves are mass-weighted, these norms are unweighted i.e. associated with the unit $n d \times n d$ matrix.]   
One could use the $\mathbb{R}^{nd}$ norm in $\FP(N, d)$ too, as a chordal distance, but there 
are also intrinsic distances that could be used thereupon [based on angular separations in 
$\mathbb{S}^{n d - 1}$ that are available once one has a notion of intrinsic metric]; see below. 

\mbox{ } 

\noindent{\bf Structure A}  $ \langle \FrQ(N, d), ||\mbox{ }||\rangle$ and $ \langle \FrR(N, d), ||\mbox{ }||\rangle$ are appropriate as metric spaces to work with.  
Then $\langle \FP(N, d), \mbox{(Chordal Dist)} \rangle$ is inherited as a metric space.  
This then turns out to be topologically equivalent (see p 13 of \cite{Kendall}) to the metric space $\langle \FrQ(N, d), \mbox{(Great Circle Dist)}\rangle$, which is a geodesic distance.
Furthermore note that $\mbox{(Chordal Dist)}$ and $\mbox{(Great Circle Dist)}$ are 
related by $\mbox{(Chordal Dist)} = \mbox{sin}(\mbox{Great Circle Dist})$ (p 205 of \cite{Kendall}).  
Then $\mbox{(Great Circle Dist)}(\bar{A}, \bar{B}) = \mbox{arccos}(\bar{A}\bar{B}).$  
This carries over to shape space: one has the metric space $ \langle \FrS(N, d), D\rangle$, for the quotient metric
\be 
\mbox{Dist}(Q(\bar{A}, \bar{B})) = \mbox{ } \stackrel{\mbox{\scriptsize min}}{\mbox{\scriptsize $T \in SO(d)$}} 
\mbox{(Great Circle Dist)}(\bar{A}, T\bar{B}) = \mbox{ } \stackrel{\mbox{\scriptsize min}}{\mbox{\scriptsize $T \in SO(d)$}} 
\mbox{arccos}(\bar{A} , T\bar{B}) \mbox{ } .  
\label{KenMin} 
\ee

%
\noindent Appropriate topological spaces to work with are 1) $ \langle \FP(N, d); \tau_{\sFP}\rangle$ for $\tau_{\sFP}$ the set of open sets (obeying topological space axioms) 
determined by $\mbox{(Great Circle Dist)}$ or $\mbox{(Great Circle Dist)}$ 
2) $ \langle \FrS(N, d); \tau_{\sFS}\rangle$ e.g. obtained from the preceding as the quotient topology corresponding to the map Quot. 
Equivalently, it can be obtained as the set of open sets determined by $\mbox{(Great Circle Dist)}$ .  
[For later use, $\mbox{(Great Circle Dist)}$ generalizes to $\mbox{(Geodesic Dist)}$.]

\subsection{Topological structure of O(pre)shape space}\label{Top2}

For all that one may wish to focus on the plain shapes, the O-shapes do provide some contrast, and characterizing their greater difficulty is of interest.  
Also, some of the Orelationalspaces feature as meaningful subspaces within plain relationalspaces, so 
at least a moderate understanding of the O-case in the study of the plain case is needed.

Note that at the topological level $\FrO\FA(\d) = \mathbb{R}^{d}/\mathbb{Z}_2 = \mathbb{R}^d_+$, 
                                   $\FrO\FrQ(N, d) = \mathbb{R}^{N d}/\mathbb{Z}_2 = \mathbb{R}^{N d}_+$ and  
                                   $\FrO\FrR(N, d) = \mathbb{R}^{n d}/\mathbb{Z}_2 = \mathbb{R}^{n d}_+$.  
What about the various reduced configuration spaces? 
My choice here is to study the plain shape case due to its greater tractability. 
However, in the present set-up chapter, I do provide some non-laborious aspects of the O case's working 
for contrast and possibly future treatment of the O case in parallel with the rest of the present Article.  
Note that these spaces are going to be orbifolds at the level of differential geometry (see SSec \ref{Top3}).  
Some first-principles construction figures (along the lines of \cite{06II}) for O(pre)shape spaces are then as follows.

{            \begin{figure}[ht]
\centering
\includegraphics[width=0.6\textwidth]{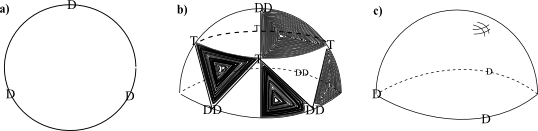}
\caption[Text der im Bilderverzeichnis auftaucht]{        \footnotesize{Repeating the procedure of Fig \ref{Top-S}   
in the O case leads to the following.  

a) $\FrO\FrS$($3$, $1$) can be represented as a semicircle and as a circle in double angles.
In each case there are 3 D's (double collisions) so that the semicircle or circle is cut into 3 arcs.  
One can view this as the equator of c), or that of $\FrS(3, 2)$.
 
\noindent
b) $\FrO\FrS$($4$, $1$) = $\mathbb{S}^2/\mathbb{Z}_2$ acting inversively = $\mathbb{RP}^2$ \cite{06II}.  
For this, there are 12 spherical triangle regions of shape space, bounded by 18 D (double collision) line segments, and 3 DD (double-double) and 4 T (triple) collision points.  
This also plays a role as the collinear configurations of $\FrS(4, 2)$.

\noindent
c) $\FrO\FrS$($3$, $2$) = $\mathbb{S}^2/\mathbb{Z}_2$ acting reflexively.                        ]  
This is a hemisphere whose equator edge is punctuated by 3 D points that split it up into 3 arcs. 

\noindent 
d) The excised regions from $\mathbb{CP}^2/\mathbb{Z}_2^{\scc\so\sn\sj}$ are 6 $\mathbb{RP}^2$ Otrianglelands. that share  
4 T points and 3 DD points, i.e. Fig \ref{Top-S} d).  \normalsize
}        }
\label{Fig-3-3}\end{figure}            }

\subsubsection{The simpler Oshape spaces as topological surfaces}\label{TopSur2}

\noindent{\bf Corollary 3) [of `Partial Periodic Table' Theorem 1)]} In each case homeomorphically, 

\noindent i)  $\FrO\FrS(N, 1) {=} \mathbb{RP}^{N - 2}$.

\noindent ii) $\FrO\FrS(N, 2) {=} \mathbb{CP}^{N - 2}/\mathbb{Z}_2^{\scc\so\sn\sj}$.

\noindent iii) $\FrO\FrS(A, B) {=} \mathbb{S}^{\{B - 1\}\{B + 2\}/2}/\mathbb{Z}_2^{\scc\so\sn\sj}$ 
for $A \leq B + 1$ (and these are identified).   
%

\noindent\underline{Proof} For the last part, quotienting out twice is the same as quotienting out once. $\Box$

\mbox{ } 

\noindent Note 1) The superscript `conj' denotes complex conjugate action at the level of the configuration space.  
N.B. that this is different from inversive action, as is clearest from the simplest case: $\FrO\FrS(3,\mbox{ }2) = \mbox{hemisphere with edge} \neq \mathbb{RP}^2$. 
Demonstrating it is a reflection rather than an inversion follows from the transformation mapping a labelled shape to its mirror image being 
$z^{a} = \rho^{a}\mbox{exp}(i\phi^{a}) \rightarrow \rho^a\mbox{exp}(i\{\pi - \phi^a\})$, which is a reflection about the y-axis rather than an inversion. 

\noindent Note 2)  A notion of {\it weighted projective space}, that encapsulates the spaces of 
interest to us as particular weightings among many other possibilities, is present in the literature.   
These other possibilities correspond to the other possible ways in which the $\mathbb{Z}_2$ can act upon such a space.  
Unfortunately the examples of such that I have seen in the literature do not appear to coincide with 
the spaces of interest in this Article, on account of having different weightings. 
Thus these would only serve as a rough guide to what deviations from $\mathbb{CP}^{N - 2}$ one 
might expect upon applying some $\mathbb{Z}_2$ action, and as such I relegate detailed discussion of this literature to Appendix \ref{Q-Geom}.A  

{            \begin{figure}[ht]
\centering
\includegraphics[width=0.45\textwidth]{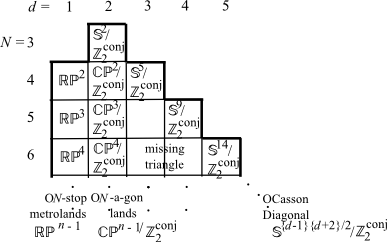}
\caption[Text der im Bilderverzeichnis auftaucht]{        \footnotesize{The `Partial Periodic Table' of discernible 
Oshape spaces at the topological level. }        } \label{DiscernOSS}\end{figure}            }


\noindent {\bf Corollary 4) (of Kuiper's Theorem)}.  
From establishing that the following are diffeomorphic by Kuiper's Theorem (see Sec \ref{Kuiper} or \cite{Kuiper, Kuroki}),
and diffeomorphic $\Rightarrow$ homeomorphic, $\FrO\FrS$(4,2) = $\mathbb{CP}^2/\mathbb{Z}_2^{\scc\so\sn\sj} = \mathbb{S}^4$ at the topological level.

\subsubsection{Some detailed topological properties of O(pre)shape spaces}\label{Det2}

\noindent We have identified $\FrO\FP(N, d)$ and $\FrO\FrS(N, 1)$ and $\FrO\FrS(N, 2)$ as 
$\mathbb{RP}^{\sk}$ and $\mathbb{CP}^{\sk}/\mathbb{Z}_2^{\scc\so\sn\sj}$ manifolds.  
This permits the following classically and quantum-mechanically useful topological information about paths on and obstructions thereupon.    
1) $\FrO\FP(N, d) = \mathbb{RP}^{n d - 1}$, $\FrO\FrS(N, 1) = \FrO\FP(N, 1)$ and $\FrO\FrS(N, 2) = 
\mathbb{CP}^{n - 1}/\mathbb{Z}_2^{\scc\so\sn\sj}$ are compact without boundary.   

\noindent 2) Hausdorffness is not in general inherited in forming a quotient space, by the following basic counterexample.  
%
%
All open intervals in $\mathbb{R}$ contain rationals and irrationals. 
Thus $\mathbb{Q}$, $\mathbb{R}/\mathbb{Q}$ is not open in $\mathbb{R}$.  
This can be viewed as a 2-class equivalence relation that is the quotient of a Hausdorff space which has 2 points (the classes) and no open singleton, and so is not itself Hausdorff.  
Thus more specific results are required. 
\noindent $\mathbb{RP}^{n}$ is straightforwardly Hausdorff \cite{Lee}. 
So is $\mathbb{CP}^{2}/\mathbb{Z}_2^{\scc\so\sn\sj}$, via it being homeomorphic with $\mathbb{S}^4$.  

\mbox{ }  

\noindent Topological groups for Opreshapespace are as follows.
\noindent The homotopy groups corresponding to these spaces are  
$\pi_0(\FrO\FP(N, d)) = \pi_0(\mathbb{RP}^{n d - 1}) = 0$ 
for, trivially examining loops on the identified $k$-spheres, there is nothing for these to wrap around, so they are all contractible.
$\pi_1(\FrO\FP(N, d)) = \pi_1(\mathbb{RP}^{n d - 1}) = \mathbb{Z}_2$ for $nd \geq$ 3 (pp. 79-80 of \cite{Spanier}). 
$\pi_{p}(\FrO\FP(N, d)) = \pi_{p}(\mathbb{RP}^{nd - 1}) = \pi_1(\mathbb{S}^{nd - 1})$ 
for $nd$ $\geq$ 3 by pp. 35-36 of \cite{GreenbergHarper}, whereupon one can read them off (\ref{Det}).  
By 
\beq
\FrO\FrS(N, 1) = \FrO\FP(N, 1) \mbox{ } , 
\label{1-$d$}
\eeq
the corresponding results for the $N$-stop metrolands $\FrS(N, 1)$ are readily obtained from the above by setting $d$ = 1: 

\noindent $\pi_0(\FrO\FrS(N, 1)) = 0$,  
$\pi_1(\FrO\FrS(N, 1)) = \mathbb{Z}_2$ for $n \geq$ 3 and  
$\pi_p(\FrO\FP(N, 1)) = \pi_p(\mathbb{RP}^{n - 1}) = \pi_1(\mathbb{S}^{n - 1})$.   

\mbox{ } 

\noindent The homology groups are $H_{p}(\FrO\FP(N, d)) = H_{p}(\mathbb{RP}^{n d - 1}) = 
\mbox{\Huge \{}
\stackrel{\mbox{$\mathbb{Z} \mbox{ }  \mbox{ } p = 0$}}
         {\stackrel{\mbox{$\mathbb{Z}_2 \mbox{ }  \mbox{ } \mp \mbox{ odd , } 0 <  p < n d - 1$}}
                   {0 \mbox{ } \mbox{ otherwise .}}}$
Also, by (\ref{1-$d$}) and the corresponding results for the O$N$-stop metrolands $\FO\FrS(N, 1)$, 
$H_{p}(\FrO\FrS(N, 1)) = H_{p}(\mathbb{RP}^{n - 1}) =   
\mbox{\Huge \{}
\stackrel{\mbox{$\mathbb{Z} \mbox{ }  \mbox{ } p = 0$}}
         {\stackrel{\mbox{$\mathbb{Z}_2 \mbox{ }  \mbox{ } p \mbox{ odd , } 0 < p < n - 1$}}
                   {0 \mbox{ } \mbox{ otherwise .}}}$

\noindent The cohomology groups are $H^{p}(\FrO\FP(N, d)) = H^{p}(\mathbb{RP}^{n d - 1}) = 
\mbox{\Huge \{}
\stackrel{\mbox{$\mathbb{Z} \mbox{ }  \mbox{ } p = 0 \mbox{ or $p$} \mbox{ odd and } 0 < p < n d$}}
         {\stackrel{\mbox{$\mathbb{Z}_2 \mbox{ }  \mbox{ } p \mbox{ even , } 1 < p < n d $}}
                   {0 \mbox{ } \mbox{ otherwise .}}}$
By (\ref{1-$d$}), the corresponding results for the $N$-stop metrolands $\FrS(N, 1)$ are then 
$H^{p}(\FrO\FrS(N, 1)) =  
\mbox{\Huge \{}
\stackrel{\mbox{$\mathbb{Z} \mbox{ }  \mbox{ } p = 0 \mbox{ or } p \mbox{ odd and } 0 < p < 
n - 1 $ }}{\stackrel{\mbox{$\mathbb{Z}_2 \mbox{ }  \mbox{ } p \mbox{ even , } 0 < p < n$}}
                   {0 \mbox{ } \mbox{ otherwise .}}}$

The first Stiefel--Whitney class of $\mathbb{RP}^k$ is trivial for $k$ odd and nontrivial for $k$ even.  
Hence, setting $k$ = $n d$ -- 1, for at least one of $n$ and $d$ even, the first Stiefel--Whitney class is trivial and so $\FO\FP(N, d)$ is orientable.  
Conversely, if both $n$ and $d$ are odd, the first Stiefel--Whitney class is nontrivial and so $\FO\FP(N, d)$  is non-orientable.  
In particular 1) for $d$ = 1 and $N$ odd one has an orientable Opreshape space, so odd-stop metroland is also orientable.  
2) For $d$ = 1 and $N$ even, one has a non-orientable Opreshape space, so even-stop metroland is also nonorientable. 
3) For $d$ = 2 all Opreshape spaces are orientable.  


The $\FrO\FrS(3, 2)$ hemisphere is also topologically standard.  
For Otriangleland, $\pi_1 = \mathbb{Z}_2$, and the rest of their homotopy groups coincide with those of $\mathbb{S}^2$, as per Sec \ref{Det}.


\mbox{ }  

\noindent Next, if 2 spaces are diffeomorphic, they must they be homeomorphic. 
That $\mathbb{CP}^2/\mathbb{Z}_2^{\scc\so\sn\sj}$ is homeotopic to  $\mathbb{S}^4$ implies the following.  
$\pi_{p}( \mathbb{CP}^2/\mathbb{Z}_2^{\scc\so\sn\sj}  ) = \pi_{p}(\mathbb{S}^4) = $ 0 for 
p $<$ 4, $\mathbb{Z}$ for $p$ = 4 and subsequently the third line of the table in Sec \ref{Det}.  

\noindent
$H_{p}(\mathbb{CP}^2/\mathbb{Z}_2^{\scc\so\sn\sj}) =  H_{p}(\mathbb{S}^4 = \mbox{\Huge \{} \stackrel{\mbox{$\mathbb{Z} \, \,$  $p$ = 0 or 4}}{0 \mbox{ otherwise}} \mbox{\Huge \}}  
 =  H^{p}(\mathbb{S}^4) = H^{p}(\mathbb{CP}^2/\mathbb{Z}_2^{\scc\so\sn\sj})$  
The first and second Stiefel-Whitney classes for this are trivial, so this space is orientable and with nontrivial spin structure.  
I have not explored beyong this. 
A next modest step would be to heed what differences occur between $\mathbb{CP}^{n - 1}$ and the weighted projective spaces in Appendix \ref{Q-Geom}.A).

\subsubsection{Oshape spaces as metric spaces}\label{Met2}

In general, the quotient of a metric space only has a pseudometric on it, i.e. a structure Pseu such that 
\beq
\mbox{Pseu}([x],[y]) = 0 \not{\hspace{-0.05in}\Rightarrow} [x] = [y] \mbox{ } .
\label{pseudo}
\eeq 
For M compact, $M/\,\widetilde{\mbox{ }}$ has the quotient topology.  
One then needs to check if $(\mathbb{S}^2, \mbox{distance})/\mathbb{Z}_2$ is a metric space with respect 
to the corresponding notion of distance, and this is tied to $\mathbb{RP}^2$'s equivalents of chords and great circles.

Geodesic distance between 2 points $x$, $y$ in $\mathbb{RP}^{n}$ is 
\beq
\mbox{(Geodesic Dist)}_{\mathbb{RP}^n} = {\mbox{min}}(\mbox{Geodesic Dist}_{\mathbb{S}^n}(x, y), 
\mbox{Geodesic Dist}_{\mathbb{S}^n}(x, -y))
\eeq
which gives a suitable definition of metric on $\mathbb{RP}^{n}$ for the current level of structure.   
Also, the map $\mathbb{S}^{n} \longrightarrow \mathbb{RP}^{n}$ sends great circles to what will be 
$\mathbb{RP}^{n}$ geodesics at the level of Riemannian geometry.  
Also note (paralleling Lemma 5): there is an analogue of homogeneous coordinates on 
$\FrO\FrS(N, 2) = \mathbb{CP}^{n - 1}/\mathbb{Z}_2^{\scc\so\sn\sj}$, just with different coordinate ranges.  
For the triangleland case, define $\underline{y}^{\scc\so\sn\sj} = (r\mbox{cos}\theta, -r\mbox{sin}\theta$) corresponding to $\underline{y} = (r\mbox{cos}\theta, r\mbox{sin}\theta$). 
Note that this is not just a reflection, but also a conjugation; it is the latter that extends to arbitrary particle number case.
Then a geodesic distance is 
\beq
\mbox{(Geodesic Dist)}_{\mathbb{RP}^n} = {\mbox{min}}\,{\mbox{\scriptsize$\mathbb{S}^n$}}(\mbox{(Geodesic Dist)}_{\mathbb{S}^n}(x, y), 
\mbox{(Geodesic Dist)}_{\mathbb{S}^n}(x, y^{\scc\so\sn\sj}))
\label{geodist}
\eeq

\subsubsection{The case of (partly) indistinguishable particles}\label{PIndist}

One could likewise set out to repeat this Sec's study so far in this setting.  
This involves quotienting out larger groups and thus produces portions of Fig \ref{Top-S}'s configuration spaces, as in Fig \ref{Indis-Top}. 

{            \begin{figure}[ht]
\centering
\includegraphics[width=0.45\textwidth]{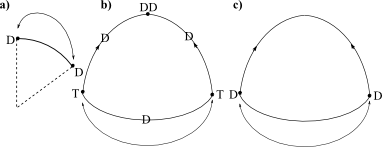}
\caption[Text der im Bilderverzeichnis auftaucht]{        \footnotesize{
Fully indistinguishable particle (Leibniz space) cases of a) 3-stop metroland b) 4-stop metroland: $\mathbb{S}^{2}/S_4$ 
with the $S_4$ acting as one handedness of the cube group.   
and c) triangleland, all at the topological level:  $\mathbb{S}^2/S_3$ with the $S_3$ acting as $\mathbb{D}_3$. 
In each case, the A- instead of S-version is obtained by doubling the picture: to a 120 degree wedge in the first case, 
and to a full lune by adding a lower spherical triangle in the other two cases.  
}        } \label{Indis-Top}\end{figure}            }

\noindent I use the prefix A for $A_N$ quotients and the prefix S for $S_N$ quotients just as I previously used the prefix O for $\mathbb{Z}_2^{\sr\se\sf}$ quotients. 
Thus I define {\it ARelative space} $\FA\FrR(N, d)$,  {\it APreshape space} $\FA\FP(N, d)$, 
              {\it ARelational space} $\FA\bigr(N, d)$, {\it AShape space} $\FA\FrS(N, d)$,    
              {\it SRelative space} $\FrS\FrR(N, d)$,  {\it SPreshape space} $\FrS\FP(N, d)$, 
              {\it SRelational space}  $\FrS\bigr(N, d)$ and SShape space, which, both being hard to 
pronounce and the most very Leibnizian of the possible configuration spaces for mechanics with equal particle masses, 
I rechristen as {\it Leibniz space} Leib($N, d) := \FrS(N, d)/S_N$  

\mbox{ }

\noindent{\bf Question 10}. What is known about $\mathbb{S}^p/\FrG$ and $\mathbb{CP}^p/\FrG$ where $\FrG$ is (a subgroup of) a permutation group? 

\mbox{ }

\noindent [Do these occur elsewhere in Mathematics or Physics?  
In the first case, yes: they are a subcase of the {\it spherical primes} \cite{Giu1}, 
which occur also in the study of the large-scale shape of space study \cite{MLR} and in dynamical systems \cite{MacCor}.
In the second case, I have not seen these `complex projective primes' elsewhere, other than for $\FrG = \mathbb{Z}_2$, as per Appendix \ref{WPS}).]  

\mbox{ } 

\noindent {\bf Question 11} Investigate whether further sorts of indistiguishability as pointed out by e.g. Butterfield and Caulton \cite{BC11} can be manifested by RPM's.    
Furthermore, following the correspondence in Fig \ref{Fig2also}, can these be manifested in GR?

\subsection{Topological structure of (plain, O, A or S)relational spaces}\label{Top3}

\subsubsection{Relational space as the cone over shape space}\label{RSACOSS}

The third key step in understanding RPM's is that relational space $\bigr(N, d)$ can be viewed as the {\bf cone}\footnote{\noindent Early papers  
involving such a notion of cone are Lema\^{\i}tre \cite{Lemaitre1}, Deprit--Delie \cite{DD1}, and also implicitly in the 3-body Celestial Mechanics work of Moeckel \cite{MoeckelQ} 
and of Saari \cite{Saari} (see also \cite{Hsiang1, Mont98, Montgomery2}). 
(See also 3-body Celestial Mechanics work such as \cite{MoeckelQ, Saari,Montgomery2, Hsiang1, Mont98}.)}
over shape space, C$(\FrS(N, d))$.
At the topological level, for C($\FrX$) to be a cone over some topological manifold $\FrX$, 
\beq
\mbox{C(\FrX) = \FrX $\times$ [0, $\infty$)/\mbox{ }$\widetilde{\mbox{ }}$} \mbox{ } , 
\eeq
where $\widetilde{\mbox{ }}$ means that all points of the form \{p $\in$ \FrX, 0 $\in [0, \infty)$\} are 
`squashed' or identified to a single point termed the {\it cone point}, and denoted by 0.

Cones are `examples' of {\it orbifolds} (making them so can involve defining more structure than one would have otherwise).  
At the level of differential geometry, the additional extra structure for a real or complex orbifold \cite{GSW2, Orbi1, Orbi2} involves 
charts to quotients $\mathbb{R}^k/\FrG$ and $\mathbb{C}^k/\FrG$. 
This parallels (and generalizes) how real and complex manifolds are defined in terms of charts to $\mathbb{R}^k$ and $\mathbb{C}^k$.  

\mbox{ } 

\noindent{\bf Question 12$^*$}  For the relational program, I would like to push the above definition of cone far as possible toward 
cases in which $\FrX$ is a stratified manifold (collection of manifolds of in general different dimensionality glued together in a 
particular way [See Sec \ref{GRCtr}'s references for more detail] or orbifold 
\beq 
\mbox{C(\FrX) for \FrX = $\FrM/\FrG$} \mbox{ } , 
\eeq 
i.e. the cone over some topological manifold $\FrMgen$ quotiented by a group $\FrG$ (which may be continuous, discrete or a mixture of both).
E.g. C($\FrS(N, d)$) = C($\mathbb{R}^{n d}/SO(d) \times$ Dil), including C$(\FrS(N, 1)) = $
C$(\mathbb{S}^{n - 1})$ and C$(\FrS(N, 2)) = C(\mathbb{CP}^{n - 1})$, C($\FrO\FrS(N,\d)$) = 
C($\mathbb{R}^{n d}/SO(d) \times$ Dil $\times \mathbb{Z}_2$), includes both  C($\FrO\FrS(N, 1)$) = 
C($\mathbb{S}^{\sn - 1}/\mathbb{Z}_2$) and C($\FrO\FrS(N, 2)$) = C($\mathbb{CP}^{n - 1}/\mathbb{Z}_2$).  
E.g. further examples involving bigger discrete groups in the case of (partially) indistinguishable particles.  
Does the definition of cone indeed survive the weakening from `$\FrX$ a manifold' to `$\FrX$ an orbifold'? 
Which theorems survive such a weakening? 

\mbox{ } 

\noindent{\bf Partial Answer.} Witten and Atiyah \cite{W1, W3} consider cones over weighted projective spaces (Sec \ref{WPS}), and they can, at 
least in some cases, possess orbifold singularities. 
More general such are also developed by Emmerich and R\"{o}mer \cite{ER90}.

\subsubsection{Many of this Article's specific examples' cones are straightforward}\label{Easy-Cones}

These being cones and these cones being mathematically straightforward is the third key to unlocking RPM's. 
Moreover, via the scale--shape split, the above pure-shape RPM study is useful again within scaled RPM's.  

C($\FrS(n, 1)$) = C($\mathbb{S}^{\sn - 1}$) = $\mathbb{R}^{n}$ using just elementary results (c.f. e.g. \cite{Kendall, Rotman}).   
C($\FrS(3, 2)$) = C($\mathbb{CP}^{1}$) = C($\mathbb{S}^{2}$) = $\mathbb{R}^{3}$ (see e.g. \cite{Hsiang1}).
C($\FrO\FrS(n, 1)$) = C($\mathbb{S}^{\sn - 1}/\mathbb{Z}_2$) = $\mathbb{R}^{n}_+$, the half-space i.e. half of the possible generalized deficit 
angle.  

\noindent C($\FrO\FrS$(3, 2)) = C($\mathbb{S}^2/\mathbb{Z}_2^{\scc\so\sn\sj}$) = $\mathbb{R}^3_+$.   
These last two, however, have edge issues. 
In 3-stop metroland, the identification coincides with the gluing in constructing the `everyday cone'.  
Oriented 4-stop metroland is $\mathbb{RP}^2$.    
On the other hand, Otriangleland is {\sl not} $\mathbb{RP}^2$ but rather the sphere with reflective 
rather than inversive $\mathbb{Z}_2$ symmetry about the equator; one can think of this space loosely as a `half-onion'.
 
\noindent In the case of plain shapes, one has $\FrR(N, \d) = \mC(\mathbb{S}^{nd - 1})$, $\bigr(N, 1) = 
\mC(\mathbb{S}^{n - 1})$ and $\bigr(N, 2) = \mC(\mathbb{CP}^{n - 1})$. 
$\bigr(3, 2) = \mC(\mathbb{CP}^1) = \mC({\mathbb{S}^2})$ is then a special case of this studied in e.g. \cite{Hsiang1, Mont98}.  
It is said to be better-behaved than its O-counterpart in that collinearities can be included (see above reference for description of the desingularization)).  
In the O case, the spaces one has are $\FrO\FrR(N, \d) = \mC(\mathbb{RP}^{nd - 1})$, 
$\FrO\bigr(N, 1) = \mC(\mathbb{RP}^{n - 1})$ and $\bigr(N, 2) = \mC(\mathbb{CP}^{n - 1}/\mathbb{Z}_2)$.
$\FrO\bigr(3, 2) = \mC(\mathbb{CP}^1/\mathbb{Z}_2) = \mC(\mathbb{S}^2/\mathbb{Z}_2) = \mC(\mathbb{RP}^2)$ is then a special case of this.

\mbox{ }

\noindent Note 1) Atiyah and Witten have studied C$(\mathbb{CP}^3)$, directly relevant to the present Article's relational program.  

\noindent Note 2) By the scale--shape split, shape space is both the entirety of the reduced configuration space for the 
pure-shape theory and the shape part of the scale--shape split of the corresponding scaled theory.

\noindent Note 3) Because of pathologies at the origin for certain purposes, relational space is in some ways a less advantageous intermediate to study than preshape space.

{            \begin{figure}[ht]
\centering
\includegraphics[width=0.42\textwidth]{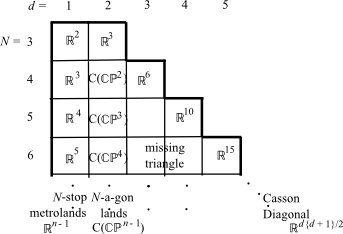}
\caption[Text der im Bilderverzeichnis auftaucht]{        \footnotesize{`Partial periodic table' of the discernible relational spaces at the topological level. 
}        } \label{DiscernR}\end{figure}            }

\subsubsection{More detailed topological properties of this Article's relational space cones}\label{Det-Top-Cones}

All the RPM shape spaces $\FrX$ over which I take cones in this Article are compact and Hausdorff.
Coning in general sends compact spaces to noncompact ones, due to the scale itself being noncompact. 
This Article's relational space cones are all Hausdorff.
They are also path-connected \cite{Mont98}.

Cones have a tendency to be straightforward from the topological point of view.  
Certainly the main particular examples of this Article, which reduce to $\mathbb{R}^{n}$ (that plain 
triangleand is homeomorphic to $\mathbb{R}^{3}$: \cite{Hsiang1}), and $\mathbb{R}^{n}_+$, are topologically straightforward.  
As additional results, cones are 1) contractible (pp. 23-24 of \cite{Rotman}), and so has the same homotopy 
type as the point, and  2) acyclic, by which they have no nontrivial homology groups (pp. 43-46 of \cite{Munkres}). 
The situation with cohomologies is also straightforwardly related to that of the constituent shape space in the role of $\FrX$ \cite{KW06}.

That C$(\mathbb{S}^{n - 1}) = \mathbb{R}^n$ finishes off the topological study of ${\bigr}(N, 1)$.  
The further results that C$(\mathbb{CP}^1)$ = C$(\mathbb{S}^2) = 
\mathbb{R}^3$ and \noindent C$(\mathbb{CP}^1/\mathbb{Z}_2^{\scc\so\sn\sj})$ = C$(\mathbb{S}^2/
\mathbb{Z}_2^{\scc\so\sn\sj}) = \mathbb{R}^3_+$,  and, since $\mathbb{CP}^2/\mathbb{Z}_2 = \mathbb{S}^4$, 
$\mC(\mathbb{CP}^2/\mathbb{Z}_2) = \mC(\mathbb{S}^4) = \mathbb{R}^5$  finish off some more cases.  

\mbox{ }

\noindent{\bf Question 13}. What are the other $\mC(\mathbb{CP}^{N - 2})$, C($\mathbb{RP}^{N - 2}$) and 
$\mC(\mathbb{CP}^{N - 2}/\mathbb{Z}_2^{\scc\so\sn\sj})$ from a detailed topological perspective?

\subsubsection{Relational spaces at the level of metric spaces}\label{Met3}

\noindent{\bf Question 14}.  Does this case possess an analogue of chordal distance?  
Of great circle distance?  
These questions are meant to be applied in cases where the relational space is not just $\mathbb{R}^{n}$.  
Can one combine an a $\mathbb{CP}^{n - 1}$ distance with a radial distance to make a well-defined notion of distance?  

\mbox{ }  

\noindent A crude suggestion for now is to use the following extension of the pure-shape metric
\beq
\mbox{Dist$_{\sFR(N,d)}$($\rho_1$, $s_1^{\sfa}$; $\rho_2$, $s_2{\sfa}$) = $(\rho_1 - \rho_2)^2$ + 
Dist$_{\sFS(N, d)}$($s_1^{\sfa}$, $s_2^{\sfa}$)} \mbox{ } ,
\eeq
which does indeed straightforwardly satisfy the distance axioms. 
It is not necessarily clear if this Dist$_{\sFR(N,d)}$ is the most natural measure however, e.g. due to its heterogeneity.
Given the Riemannian structure and the considerations of Sec \ref{Cl-Str}, more satisfactory notions of distance for these spaces are forthcoming.

\subsection{The collision set and singular potentials}\label{Sing-Pot}

The above results about cones being topologically straightforward do not, however, cover the following cases.

\noindent 1) Some applications do require the cone point to be excised, i.e. {\it punctured cones}.    
%
%
As an example of how this changes topological properties, as regards homotopies, this effectively amounts to a 
return to the shape space by means of a retract, and thus now possesses the same nontrivial homotopies as for that.  

\noindent 2) The binary collisions define distinguished submanifolds of codimension d in $\FrQ$. 
For $N$ particles, there are obviously $\mbox{\LARGE(} \stackrel{N}{\mbox{\scriptsize 2}}\mbox{\LARGE)}$ such. 
Let us label them $\FrCC_{IJ}$ for coincident masses $m_I$ and $m_J$.  
Then the collision set $\FrCC := \bigcup_{I, J}$ $\FrCC_{IJ}$.  
Then the collision-free configurations are $\FrQ^* = \FrQ \backslash$ \FrCC.

The study of singular potentials can require excision of some points.  
If these potentials involve negative powers of all the $|\mr^{IJ}|$, what one must excise is the collision set $\FrCC$
(the set of configurations that include collisions, whether non-maximal or maximal).
While the HO-type potentials I study the most are not of this form, some of the further cosmologically-motivated potentials in Sec (\ref{Scram}) are.

To demonstrate that this wider class of excisions is indeed capable of altering -- and substantially complicating -- the topology, I give the following example. 
The topologically trivial $\FrQ(N, d) = \mathbb{R}^{n d}$, upon excision, picks up the nontrivial homotopy group 
\beq
\pi_1(\FrQ(N, d) \backslash \mbox{\FrCC}) = \{\mbox{coloured braid group}\} \mbox{ } , 
\eeq
the colouring *notation* here being a rephrasing of the notion of particle distinguishability.\foo{That these are 
isomorphic is clear given that the particles are assumed distinguishable and the orbits can wind around each 
of the binary collisions in whatever order but not intersect with them.
This matches the definition of a coloured braid if each distinguishable particle is taken to represent a colour.)  
In fact, the mathematical structure nowadays known as the coloured braid group was first discovered 
by Hurwitz in 1891 \cite{Hurwitz} in this very mechanics context.   
Thus, the discovery of this mathematical structure preceded Artin's realization in 1925 \cite{Artin} of its 
braid interpretation; these two works were first interconnected in \cite{FN62}.
See e.g. \cite{Birman} for a review and updates on the theory of the braid group.
See \cite{Mont98} for further study of this in the case of 3 particles in 2-$d$, and \cite{T90, AG} for further applications.}
%
Additional work in dynamics involves   
\beq
\mbox{projective coloured braid group} = \mN(\FrB) := \FrB/\mZ(\FrB) \mbox{ } , 
\eeq
where Z($\FrG$) and N($\FrG$) denote the centre and normal subgroup of a group $\FrG$, so as to consider the 
case in which each path and hence each braid is to return to its starting point.
For $N$ = 3, the configuration space is homotopic to $\mathbb{S}^2$ with 3 points (the D-points) excised and then the $\pi_1$ of this is the `free group on two letters'.   
Braid group applications to configuration space are also mentioned in \cite{T90, AG}.   

\mbox{ }

\noindent{\bf Question 15$^{*}$} What are the other homotopy groups for these collision-excised configuration spaces?

\subsection{Riemannian geometry of shape spaces}\label{Riem}

\subsubsection{Preshape space and the shape space of $N$-stop metroland}\label{RiemN}

{\bf Lemma 6 (`Preshape space is metrically simple')}. $\FP(N, d)$ is isometric to the Riemannian geometry $\langle \mathbb{S}^{n d - 1}, 
\mbox{\boldmath$\ttM$}^{\sss\sp\sh\se}(1/2)\rangle$. [I.e. the standard spherical metric on the sphere of radius 1/2.]

\noindent\underline{Proof} $\FP(N, d)$ is described by $\sum_{\sfA \, = \, 1}^{n d}\bar{r}_{\sfA}^2$ = 
constant (normalization condition, which is clearly the $\mathbb{S}^{n d - 1}$  sphere embedded in the usual way in $\mathbb{R}^{n d}$.
[Here $\sfA$ is a multi-index for i$\mu$.] $\Box$  

\mbox{ } 

\noindent{\bf Lemma 7} Straightforwardly, in Lemma 4's Beltrami coordinates, the standard metric on the \{$nd$--1\}-sphere has arc element 
\be
\d \tts^2 = \ttM_{{\cal I}{\cal J}}\d{b}^{\cal I}\d{b}^{\cal J}  = 
\{     \{  1 + ||\mbox{\boldmath$b$}||^2  \}||\d\mbox{\boldmath$b$}||^2 - 
(\mbox{\boldmath$b$}, \d\mbox{\boldmath$b$})^2 \}/\{1 + ||\mbox{\boldmath$b$}||^2 \}^2 \mbox{ } .  
\label{Belt}
\ee
\noindent [N.B. these coordinates bring out the parallel with $\mathbb{CP}^k$, and occur in the reduction procedure in Sec 
\ref{Red-App}) but are not themselves that useful to work with.]


\noindent{\bf Lemma 8} On $\FP(N, d)$ the geodesics are great circles.  

\noindent\underline{Proof} By Lemma 6), $\FP(N, d)$ is isometric to  $\mathbb{S}^{nd - 1}$.  
Then it is well-known that the geodesics of spheres are great circles. $\Box$  

\mbox{ } 

\noindent {\bf Lemma 9} $\FrS(N, 1) = \FP(N, 1)$, $\bigr(N, 1) = \FrR(N, 1)$ (both homeomorphically and isometrically).

\mbox{ } 

\noindent{\bf Corollary 5)} $\FrS(N, 1) = \FP(N, 1) {=} \mathbb{S}^{n - 1}$ homeomorphically.  

\noindent\underline{Proof} $\FrS(N, 1) = \FP(N, 1)$ by the rotations being trivial in 1-$d$.  
Then use Lemma 3):  homeomorphically, $\FP(N, 1) {=} \mathbb{S}^{nd - 1} = \mathbb{S}^{n - 1}$. $\Box$  

\mbox{ } 

\noindent{\bf Corollary 6)} $\FrS(N, 1)$ is isometrically $\FP(N, 1)$ which is isometrically $\langle \mathbb{S}^{N - 2}, \bttM^{\sss\sp\sh\se}(1/2)\rangle$.  

\noindent\underline{Proof} $\FrS(N, 1) = \FP(N, 1)$ by rotations being trivial in 1-$d$.  
Then use Lemma 7: $\FP(N, 1)$ is isometrically $\mathbb{S}^{n d - 1} := \mathbb{S}^{N - 2}$. $\Box$

\mbox{ }

\noindent{\bf Corollary 7)} The metric on 1-$d$ shape space is the appropriate subcase of (\ref{Belt}), corresponding to the arc element 
\beq
\d \tts^{\sr\se\sll\sa\st\si\so\sn\sa\sll\sss\sp\sa\scc\se \, 2}_{N-\sss\st\so\sp\,\sS\sR\sP\sM}  =         
\left.            
\big\{\big\{1 + ||\mbox{\boldmath${\cal R}$}||^2\big\}||\d \mbox{\boldmath${\cal R}$}||^2 - 
(\mbox{\boldmath${\cal R}$} \cdot \d{\mbox{\boldmath${\cal R}$}})^2\big\}
\right/   
\{1 + ||\mbox{\boldmath${\cal R}$}||^2\}^2    \mbox{ } ;                  
\eeq
alternatively, this is 
\beq
\d \tts^{\sr\se\sll\sa\st\si\so\sn\sa\sll\sss\sp\sa\scc\se \, 2}_{N-\sss\st\so\sp\,\sS\sR\sP\sM}  = 
\{||\mbox{\boldmath${\rho}$}||^2||\d\mbox{\boldmath${\rho}$}||^2 - (\mbox{\boldmath${\rho}$} \cdot \d \mbox{\boldmath$\rho$})^2\}/||\brho||^2
\eeq
(the $\brho$ are homogeneous coordinates to the ${\cal R}^{{\stta}}$ being inhomogeneous coordinates; the former provides an in-space interpretation 
of what are mathematically the Beltrami coordinates in the configuration space for this problem).

\mbox{ }

Recast in the much more common ultraspherical coordinates by the transformation 
\beq
\theta_{\stta} = \mbox{arctan}
\left(
{    \sqrt{\sum\mbox{}_{\mbox{}_{\mbox{\scriptsize $\sfA$ = 1}}}^{\stta}{\cal R}_{\sfA}^2}    }/
{    {\cal R}_{\stta + 1}    }
\right) \mbox{ } .  
\label{Ultra}
\eeq
which diagonalizes the metric, by which these coordinates are often also much more useful.  
The corresponding Riemannian arc element is  
\beq
\d \tts^{\sr\se\sll\sa\st\si\so\sn\sa\sll\sss\sp\sa\scc\se \, 2}_{N-\sss\st\so\sp\,\sS\sR\sP\sM} = 
||\d\mbox{\boldmath$\theta$}||_{\mbox{\boldmath$\sttM$}_{\ts\tp\th\te}}\mbox{}^2 =
\sum\mbox{}_{\mbox{}_{\mbox{\scriptsize $\tta = 1$}}}^{n - 1}
\prod\mbox{}_{\mbox{}_{\mbox{\scriptsize $\hat{\sttp} = 1$}}}^{\stta - 1}
\mbox{sin}^2\theta_{\hat{\mbox{$\sttp$}} } \, \d{\theta}_{\stta}^2 \mbox{ } 
\label{Tween}
\eeq
(\ref{Ultra}), where $\prod_{i = 1}^{0}$ terms are defined to be 1. 

\noindent As subexamples, for 4-stop metroland, in either H or K-coordinates, the coordinate transformation is, in terms of $\rho^i$,  
\beq
\theta = \mbox{arctan}\left(\sqrt{\rho_1\mbox{}^2 + \rho_2\mbox{}^2}/\rho_3\right) 
\mbox{ } , \mbox{ } 
\phi = \mbox{arctan}\left({\rho_2}/{\rho_1}\right) \mbox{ } , 
\label{Var}
\eeq
or, inversely, the even more familiar form  
\beq 
\rho_1 = \rho \,\mbox{sin}\,\theta\,\mbox{cos}\,\phi \mbox{ } , \mbox{ } 
\rho_2 = \rho \,\mbox{sin}\,\theta\,\mbox{sin}\,\phi \mbox{ } , \mbox{ } 
\rho_3 = \rho \,\mbox{cos}\,\theta               \mbox{ } . \mbox{ } 
\eeq
The coordinate ranges are $0 < \theta < \pi$ and $0 \leq \phi < 2\pi$, so these are geometrically the 
standard azimuthal and polar spherical angles on the unit shape space sphere $\mathbb{S}^2$.

\subsubsection{Explicit geometrical objects for $N$-stop metroland}\label{ObjN}

For $\FrS(N, 1) = \mathbb{S}^{n - 1}$ and in ultraspherical coordinates \{$\theta^{\sa}$\}, the inverse metric is 
\beq
\ttN^{\sttp\sttq} = \delta^{\sttp\sttq}\prod\mbox{}_{\mbox{}_{\mbox{\scriptsize $\sfA = 1$}}}^{\sttp - 1}
\mbox{sin}^{-2}\theta_{\sfA}\d{\theta}_{\sttp}^2 \mbox{ } .  
\eeq
and the square root of the determinant is 
\beq
\sqrt{\ttM} = \prod \mbox{}_{\mbox{}_{\mbox{\scriptsize $\sttr = 1$}}}^{n - 1}
\prod \mbox{}_{\mbox{}_{\mbox{\scriptsize $\sfA = 1$}}}^{\sttr} \mbox{sin}\,\theta_{\sfA} \mbox{ } . 
\eeq 
The nonzero Christoffel symbols in these coordinates are 
\be
{\Gamma^{\sttp}}_{\sttq\sttq} = -\mbox{sin}\,\theta_{\sttp}\,\mbox{cos}\,\theta_{\sttp}
\,\prod\mbox{}_{\mbox{}_{\sfA = \mbox{\scriptsize 1}, \sfA \neq \sa}}^{\sttr - 1}\mbox{sin}^2\theta_{\sttq} \mbox{ } , \mbox{ }  
\Gamma^{\sttp}_{\sttq\sttp} = \mbox{cos}\,\theta_{\sttq}/\mbox{sin}\,\theta_{\sttq}  \mbox{ } .
\ee
The Ricci tensor is 
\be
Ric_{\sbttM}\mbox{}_{\sttp\sttq} = \{\mn d - 2\}{\ttM}_{\sttp\sttq}
\ee
(so $\mathbb{S}^{n d - 1}$ is Einstein) and hence have constant Ricci scalar curvature
\be
Ric_{\sbttM} = \{n d - 1\}\{n d - 2\} \mbox{ } .  
\label{whine}
\ee 
Finally, these spaces are all conformally flat, as an easy consequence of their being maximally symmetric. 
These results obviously immediately extend to $\FP(N, d) = \mathbb{S}^{n d - 1}$.

$\FrS(N, 1)$ and $\FP(N, d)$ are real manifolds; $\mathbb{S}^{2} = \mathbb{CP}^1$ alone among them is also a complex manifold. 
The Euler and Pontrjagin classes of these as real manifolds and the Chern classes and characters of 
$\mathbb{S}^2$ as a complex manifolds are readily computable. 
(These are defined in e.g. \cite{Nakahara} and are important as obstructions to quantization, and as regards issues concerning instantons and magnetic charges).   
We need some Chern classes for the discussion of some global issues in quantization.  
$c_1(\mathbb{S}^2) = 2$ and 
%
%
$c_2(\mathbb{S}^2) = 0$.

\subsubsection{Isotropy groups and orbits for $d > 1$ shape spaces}\label{+SS} 

Associated with the nontrivial quotient map Quot are the orbits Orb($\bar{X}$) = Quot$^{-1}$(Quot$(\bar{X}$)) 
$= \langle T\bar{X} | T \in SO(d) \rangle$ and the stabilizers Stab($\bar{X}$) = $\langle T \in SO(d) | T\bar{X} = \bar{X} \rangle$.  
These can furthermore be thought of as fibres and isotropy groups.  
The orbits or fibres are, for $\bar{X}$ of rank $e$, Quot$(\bar{X}) =\left\{ \stackrel{\mbox{$SO$($d$), $e$ $\geq d$ + 1}}{\mbox{Stie($d$, $e$), $e$ 
$< d$ + 1}}\right.$ for Stie($d$, $e$) = $SO$($d$)/$SO$($d$ -- $e$) the Stiefel manifolds \cite{Hatcher} of orthonormal $e$ frames in $\mathbb{R}^{d}$. 
[Note that Stie(3,1) = $SO$(3)/$SO$(2) = $\mathbb{S}^2$: spherical orbits.]

$N$-a-gonlands are homogeneous spaces.
This follows from $\mathbb{CP}^{n} \mbox{ } \widetilde{=} \mbox{ } SU(n + 1)/\{SU(n) \times U(1)\}$, which is a 
composite of Hopf's $\mathbb{CP}^{n - 1} = \mathbb{S}^{2 n - 1}/U(1)$  and $SU(n)/SU(n - 1) = \mathbb{S}^{2 n - 1}$  (see e.g. p 219 of \cite{W82b}).

\mbox{ }

\noindent Let ${\FrD}_{e}(N, d)$ be the subset of $\FP(N, d)$ corresponding to rank $\leq$ $e$ and let ${\FrD}_{e}^{\scc}(N, d)$ be its complement.  
I will often drop the $(N, d)$ from the notation.
Note that restricting attention to the discernible shape spaces on or above the Casson diagonal, ${\FrD}_{d - 2}$ is the set with nontrivial isotropy groups.  
\noindent Quot$({\FrD}_{d - 2})$ is the {\it singularity set} of $\FrS(N, d)$, while Quot$({\FrD}_{d - 2}^{\scc})$ is the {\it nonsingular part} of $\FrS(N, d)$.  

\mbox{ } 

\noindent In 1-$d$ case, note that, from the triviality of the rotations involved, nothing needs to be induced from the sphere, nor is any minimization required.  
Additionally, the singularity set $\FrD$ is empty in this case.  

\mbox{ }

\noindent{\bf Lemma 10} On Quot$({\FrD}_{d - 2}^{\scc})$ there is a unique Riemannian metric compatible 
with the differential structure and with respect to which Q is particularly well behaved is inherited from $\FP(N, d)$.

\noindent\underline{Proof} 1) by Riemannian submersion \cite{Neill12}.  

\noindent\underline{Proof} 2) Alternatively, from first principles according to the steps below up to and including Structure B. $\Box$

\mbox{ } 

\noindent Note that the geodesic joining $\bar{X}$ and $\bar{Y}$ takes the form 
\be
\Gamma_{\bar{Y}}(\tts) = \bar{X}\,\mbox{cos}\,\tts + \bar{Y}\,\mbox{sin}\,\tts
\label{GeoForm}
\ee 
parametrized by geodesic distance, for $0 \leq \tts \leq \pi$.
The tangent vector to the geodesic is ${\d \Gamma_{\bar{Y}}(\tts)}/{\d \tts}|_{\stts = 0} = \bar{Y}$.  

\mbox{ } 

\noindent Next, define the {\it exponential map} by $T_{\bar{X}}(\FP(N, d)) \longrightarrow \FP(N, d)$ $\bar{Y} \longmapsto \Gamma_{{\bar{Y}}/{||Y||}}(||Y||)$.
It restricts to a diffeomorphism of $\langle \bar{Y} \in T_{\bar{X}}(\FP(N, d))  | \mbox{ } 
||\bar{Y}|| < \pi \rangle$ onto $\langle \FP(N, d)/\mbox{antipode of } \bar{X} \rangle$.  

\mbox{ }

\noindent
$\hat{\Gamma}_{A}$ is a curve in $SO(d)$ starting from the unit matrix $\mathbb{I}$ so that the tangent vector at 
$\tts = 0$, ${\d\hat{\Gamma}_{A}}/{\d \tts}|_{\stts = 0} = A$ is tangent to $SO(d)$ at $\mathbb{I}$. 
$\mbox{exp}(\tts A)$ is in $SO(d)$ iff $\mbox{exp}(\tts A)\mbox{exp}(\tts A)^{\sT} = \mathbb{I}$ iff 
exp$(\tts\{A + A^{\sT}\}) = \mathbb{I}$  iff $A + A^{\sT} = 0$. 
So any skew-symmetric matrix $\bar{A}$ represents a vector tangent to $SO(d)$ at $\mathbb{I}$.  
As the space of $d \times d$ skew symmetric matrices has dimension 

\noindent $d$\{$d$ -- 1\}/2 = dim($SO$($d$)), that is the entire tangent space to $SO$($d$) at $\mathbb{I}$.  
As exp$(\tts\bar{A})$ lies in $SO$($d$) whenever $A^{\sT} = -A$, $\hat{\gamma}_A(\tts) = \mbox{exp}(\tts A) \bar{X}$ lies in the fibre orbit through $\bar{X}$.  

The subspace of tangent vectors ${\d\hat{\gamma}_A(\tts)}/{\d \tts}|_{\stts = 0} = A\bar{X}$ to such curves at 
$\bar{X}$ is the {\it vertical} tangent subspace at $\bar{X}$, $\FrV_{\bar{X}} = \langle A\bar{X}| A^{\sT} = A\rangle$. 
Its orthogonal complement $\FrH_{\bar{X}}$ i.e. such that $T_{\bar{X}}(\FP(N, d)) = 
\FrV_{\bar{X}}(N, d) \oplus \FrH_{\bar{X}}(N, d)$ is the {\it horizontal} tangent subspace at $\bar{X}$.

Note that for Quot$(\bar{X})$ nonsingular, i.e. $\bar{X} \not{\in\hspace{0.075in}} {\FrD}_{d - 2}$, 
$\FrV_{\bar{X}}(N, d)$ is isomorphic to $SO(d)$ at $\mathbb{I}$. 
On the other hand, at a singular point $A$ is tangent to the isotropy subgroup, so $\FrV_{\bar{X}}(N, d)$ is isomorphic to Stie($d$, $e$) at $\mathbb{I}$.

\mbox{ } 

\noindent{\bf Proposition 1} 
\noindent i) If a geodesic in $\FP(N, d)$ starts out in a horizontal direction, then its tangent 
vectors remain horizontal along it.    

\noindent ii) Distance-parametrization and horizontality are preserved under $SO(d)$. 

\noindent\underline{Proof} i) by definition, the geodesic $\Gamma_{\bar{Y}}(\tts)$ is horizontal at $\tts = 0$ iff $\bar{X}\bar{Y}^{\sT} = \bar{Y}\bar{X}^{\sT}$.  
Then $\forall \mbox{ } \tts$, 
$$
\Gamma_{\bar{Y}}\left\{\frac{\d \Gamma_{\bar Y}}{\d \tts}\right\} = \frac{\d \Gamma_{\bar{Y}}}{\d \tts}\Gamma_{\bar{Y}}(\tts)^{\sT}
$$ 
via the explicit formula (\ref{GeoForm}) for the geodesics and trivial algebra.  
Thus each tangent vector ${\d \Gamma_{\bar{Y}}}/{\d \tts}$ complies with the definition of horizontal at $\Gamma_{\bar{Y}}(\tts)$.  

\noindent ii) $T\Gamma_{\bar{Y}}(\tts) = T\bar{X}\,\mbox{cos}\,\tts + T\bar{Y}\,\mbox{sin}\,\tts$ is a distance-preserving geodesic by (\ref{GeoForm}).

\noindent $\mbox{tr}(T\bar{X}\{T\bar{Y}\}^{\sT}) = \mbox{tr}(T\bar{X}\bar{Y}^{\sT}T^{\sT}) = \mbox{tr}(T^{\sT}T\bar{X}\bar{Y}^{\sT}) = 
\mbox{tr}(\bar{X}\bar{Y}^{\sT})$ by the cyclic identity and $T$ orthogonal.

\noindent Next, $T\bar{X}\{T\bar{Y}\}^{\sT} = T\bar{X}\bar{Y}^{\sT}T^{\sT} = T\bar{Y}\bar{X}^{\sT}T^{\sT} = T\bar{Y}\{T\bar{X}\}^{\sT}$ 
(with the third step using horizontality of the untransformed geodesic), which reads overall that the transformed geodesic is then also horizontal. $\Box$

\mbox{ } 

\noindent{\bf (Riemannian) Structure B} i) Thus the exp function restricted to $\FrH_{\bar{X}}$, exp$|_{\sFrH_{\bar{X}}}$, maps 
$\langle$ vectors of length $< \pi$ $\rangle$ onto the submanifold $\FrH_{\bar{X}}$ of $\FrS(N, d)$ 
defined by $\FrH_{\bar{X}} =  \langle \tts \in \FrS(N, d) |$ all tangent vectors at $\bar{X}$ are horizontal $\rangle$.  

\noindent ii) The tangent spaces to the fibre and to $\FrH_{\bar{X}}$ are clearly perpendicular.  
Thus there is a neighbourhood $\FrU_{\bar{X}}$ such that $\forall \mbox{ } Y \in \FrU_{\bar{X}}$ the tangent spaces to the fibre and to $\FrH_{\bar{X}}$ remain transverse.  
Thus the fibre at Y meets $\FrU_{\bar{X}}$ only at Y. 
Thus, one has established that given X outside ${\FrD}_{d - 2}$, through each point $T\bar{X}$ of the fibre at $\bar{X}$ 
there is a submanifold $\FrU_{\bar{X}}$ traced out by local horizontal geodesics through $T\bar{X}$ with the following properties.   

\noindent a) $\FrU_{T_1\bar{X}}$ and $\FrU_{T_2\bar{X}}$ are disjoint if $T_1 \neq T_2$.

\noindent b) Each submanifold $\FrU_{T\bar{X}}$ is mapped by the quotient mapping Q bijectively and thus  
homeomorphically with respect to the quotient topology onto a neighbourhood of Quot$(\bar{X}) \in \FrS(N, d)$.  

\noindent c) The action of each $S \in$ $SO(d)$ restricts to a diffeomorphism of $\FrU_{T\bar{X}}$ that also preserves the Riemannian metric. 
I.e., it maps geodesics to geodesics of the same length and its derivative maps horizontal tangent vectors at $T\bar{X}$ to horizontal tangent vectors of the same length at $ST\bar{X}$.  
Thus one can use $\FrU_{\bar{X}}$, Quot$|_{\FrU_{\bar{X}}}$ to determine a differential structure on the nonsingular part of shape space: Quot$({\FrD}_{d - 2})$.  
This is since, for any other choice ($\FrU_{T\bar{X}}$, Quot$|_{\FrU_{T\bar{X}}}$), the composition 
(Quot$|_{\FrU_{T\bar{X}}})^{-1} \circ$ Quot$|_{\FrU_{\bar{X}}}$ is just the diffeomorphism $T|_{\FrU_{\bar{X}}}$.  

\noindent iii) The above ensures independence of which point on the fibre is used. 
Thus one has a Riemannian metric on the nonsingular part of shape space.  

\mbox{ } 

\noindent Note that this metric is naturally induced from $\FP(N, d) = \mathbb{S}^{n d  - 1}$.   
It has been defined such that 

\noindent Quot$: \FrH_{\bar{X}}(N, d) \longrightarrow \FP(N, d) \longrightarrow T_{\mbox{\scriptsize Quot}(\bar{X})}(\FrS(N, d))$, which is a Riemannian submersion 
(the second arrow is an isometric map).  

\mbox{ } 

%
\noindent A {\it geodesic} in $\FrS(N, d)$ is the image of any horizontal geodesic in $\FP(N, d)$.  
Note that this permits geodesics to pass between strata.
Thus geodesics can be extended beyond the nonsingular part of the space, and this serves to extend the 
above Riemannian structure (see \cite{Kendall} for these results, which not used in the present Article).
%

\mbox{ } 

\noindent{\bf Proposition 2} The geodesics of (\ref{GeoForm}), Lemma 6 (or the associated Riemannian metric of Structure B) provide the same metric distance $D$ as Structure A.  

\noindent\underline{Proof} By definition, the geodesic between two shapes Quot$(\bar{X})$ and Quot$(\bar{Y})$ 
is the image of a horizontal geodesic $\Gamma$ from $\bar{X}$ to some point $T\bar{Y}$ in fibre Quot$(\bar{Y})$. 
Since $\Gamma$ meets the fibres orthogonally at these points [$\Gamma$ being horizontal and the fibres 
being vertical], so the induced distance that follows from the geodesics/associated Riemannian metric is indeed [c.f. (\ref{KenMin})]
$$
\stackrel{\mbox{\scriptsize min}}{\mbox{\scriptsize $T \in SO(d)$}} 
\left( 
\mbox{Dist}(\bar{X}, T\bar{Y})
\right) = 
\mbox{arccos} 
\left(
\stackrel{\mbox{\scriptsize max}}{\mbox{\scriptsize $T \in SO(d)$}} \mbox{tr}(\bar{X}^{\sT} T \bar{Y}) 
\right) 
:= \mbox{Dist}(\mbox{Quot}(\bar{X}), \mbox{Quot}(\bar{Y})) \mbox{ } \mbox{ } . \mbox{ } \Box 
$$
%
\noindent Note that what one has constructed thus is a Riemannian structure on Quot$({\FrD}_{d - 2}^{\scc})$.
In general, one would have to worry about the geometry on ${\FrD}_{d - 2}$ -- the name `singularity set' does indeed carry curvature singularity connotations.  
But this Article circumvents that by considering only 1- and 2-$d$, for which the singularity set is empty.    
Thus for these cases, what one has constructed above is a Riemannian structure everywhere on shape space 
(and one can then show computationally that there are no curvature singularities within these shape spaces).

\subsubsection{Natural metric on the shape space of $N$-a-gonland}\label{Riemgon}

In 2-$d$ rotations are simpler than in higher-$d$ while being nontrivial.  
It is this that lies behind Lemma 5's straightforward complex representations for $\FP(N, 2)$ and $\FrS(N, 2)$.  
The latter representation, $\langle {Z}^{\mbox{\scriptsize \tt a}}, \mbox{ } \{\mbox{\tt a} = 1 
\mbox{ to } n - 1 \rangle$ has two manifest symmetries: $\mathbb{Z}_2$ conjugation and $U(N - 1)/U(1)$ permutations of coordinates.  
Each of these commutes with the $SO(2)$ action.

Other simplifications in 2-$d$ are that $S(N, 2) {=} \mathbb{CP}^{n - 1}$ topologically [ii) of the `Partial Periodic Table' Theorem 1)]. 
Also, the minimization in the quotient metric (in the metric space sense) of Structure A may be carried out explicitly in 2-$d$ by use of the complex representation as follows.   

\mbox{ } 

\noindent{\bf Proposition 3} $\mbox{cos}\big(\mbox{Dist}(\mbox{Quot}(\mbox{\boldmath$z$}), \mbox{Quot}(\mbox{\boldmath$w$}))\big) = 
{|(\mbox{\boldmath$z$} \cdot \mbox{\boldmath$w$})_{\tC}|}/{||\mbox{\boldmath$z$}||_{\tC}||\mbox{\boldmath$w$}||_{\tC}}$. 

\noindent\underline{Proof} By the definition of Dist, the left-hand side is 
$
\stackrel{\mbox{\scriptsize max}}{\mbox{\scriptsize $\omega \in [0, 2\pi)$}}\mbox{tr}(|\mbox{\boldmath$z$}|^{-1}|\mbox{\boldmath$w$}|^{-1}\mbox{e}^{i\omega}\mbox{\boldmath$z$}\mbox{\boldmath$w$}^{\sT}) 
\mbox{ } ,
$
the numerator of which contains 

\noindent $
\stackrel{\left(\mbox{Re($\mbox{\boldmath$z$}$)} \mbox{ } \mbox{Im($\mbox{\boldmath$z$}$)}\right)}{\mbox{ }}
\mbox{\Huge $($}
\stackrel{\mbox{ }  \mbox{cos\,$\omega$} \mbox{ } \mbox{sin\,$\omega$}}
         {\mbox{--} \mbox{sin\,$\omega$} \mbox{ } \mbox{cos\,$\omega$}}
\mbox{\Huge $)$}
\mbox{\Huge $($}
\stackrel{\mbox{Re($\mbox{\boldmath$w$}$)}}{\mbox{Im($\mbox{\boldmath$w$}$)}}
\mbox{\Huge $)$}
= A\,\mbox{cos}\,\omega + B\,\mbox{sin}\,\omega \mbox{ } ,
$
where $A = \mbox{Re}(\mbox{\boldmath$w$} \cdot \mbox{\boldmath$z$})_{\sC}$ and $B = \mbox{Im}(\mbox{\boldmath$w$} \cdot \mbox{\boldmath$z$})_{\sC}$.  
The maximum condition which follows from this is then tan$\,\omega = B/A$, for which the maximum value is $\sqrt{A^2 +  B^2}/||\mbox{\boldmath$z$}||_{\sC}||\mbox{\boldmath$w$}||_{\sC}$. $\Box$    

\mbox{ } 

\noindent{\bf Kendall's Theorem [2]} The corresponding Riemannian arc element is 
\be
\d \, \mbox{Dist}^2 = 
\left.
\{||\mbox{\boldmath$z$}||_{\sC}^2 ||\d\mbox{\boldmath$z$}||_{\sC}^2 - |(\mbox{\boldmath$z$} \cdot \d \mbox{\boldmath$z$})_{\sC}|^2\}
\right/
{||\mbox{\boldmath$z$}||_{\sC}^4} = 
\left.
\{\{1 + ||\mbox{\boldmath$Z$}||_{\sC}^2\} ||\d\mbox{\boldmath$Z$}||_{\sC}^2 -  |\mbox{\boldmath$Z$} \cdot \d \mbox{\boldmath$Z$})_{\sC}|^2\}\right/ 
{\{1 + ||\mbox{\boldmath$Z$}||_{\sC}^2\}^2} =: \d\tts^{\sr\se\sll\sa\st\si\so\sn\sa\sll\sss\sp\sa\scc\se}_{N-\sa-\sg\so\sn\,\sS\sR\sP\sM}\mbox{}^2 \mbox{ } .  
\label{Easter}
\ee
in terms of homogeneous and inhomogeneous coordinates respectively.

\noindent\underline{Proof}  Consider $\mbox{\boldmath$w$} = \mbox{\boldmath$z$} + \delta \mbox{\boldmath$z$}$.  
Then 
\be
\delta \mbox{Dist}^2 = \mbox{sin}^2\delta \mbox{Dist} + O(\delta \mbox{Dist}^4) = 
1 - \mbox{cos}^2 \mbox{Dist}(\mbox{Quot}(\mbox{\boldmath$z$}), \mbox{Quot}(\mbox{\boldmath$z$} + \delta \mbox{\boldmath$z$}) + O(\delta \mbox{Dist}^4) 
\ee
then use the Proposition 3, linearity, the binomial expansion and take the limit as $\delta \mbox{\boldmath$z$} \longrightarrow 0$ to get the first form.  
Then divide top and bottom by $||{\mbox{\boldmath$z$}}||_{\sC}^4$ and use the definition of $\mbox{\boldmath$Z$}^{\mbox{\scriptsize \tt a}}$ to get the second form. $\Box$ 


\noindent N.B. this arc element (which indeed is Riemannian, its positive-definiteness following from the Schwarz inequality) 
is the classical Fubini--Study \cite{FS, FS2} arc element on $\mathbb{CP}^{N - 2}$.\footnote{As regards appearance of these results in the $N$-body dynamics literture, 
that the Fubini--Study metric occurs for these was known to Iwai in 1987 \cite{Iwai87}, whilst Montgomery \cite{Mont96, Mont98, ArchRat} states    
that this result was almost certainly known prior to that too.
Indeed, I found that Kendall cites this result starting with a brief mention in 1977 \cite{1977} and built on this in his Whitehead Lecture 
given to the London Mathematical Society in 1980 and in his 1984 Article \cite{Kendall84}.
Montgomery's work and a large amount of literature from Molecular Physics \cite{LR95} involves an absolute--relative split of Newtonian Mechanics, 
while, as argued in the present Article, Kendall's work turns out to be more highly desirable as regards a whole-universe relational program.}  
%
This is the natural arc element thereupon, such that its constant curvature is 4.\foo{Note that 
Fubini-Study metrics of this form are available for any $N$ for ratios and relative angles paired together as complex coordinates on $\mathbb{CP}^{N - 2}$. 
This is promising as regards the arbitrary-$N$ 2-$d$ case, as $\mathbb{CP}^{N - 2}$ is the reduced 
configuration space for scale, translation and rotation free shapes in 2-$d$ \cite{Kendall}.} 

Thus the following has been proven.  

\mbox{ }

\noindent{\bf Corollary 8} $S(N, 2) = \langle \mathbb{CP}^{N - 2}; \bttM^{\sF\sS} \rangle$ isometrically (with curvature constant 4).   

\mbox{ } 

\noindent{\bf Corollary 9} $\FrS(3, 2) = \langle \mathbb{S}^2; \bttM^{\sss\sp\sh\se} \mbox{ of radius 1/2 }\rangle$ isometrically. 

\noindent\underline{Proof} Now $|| \mbox{ } ||_{\sC} = | \mbox{ } |$, so two terms cancel in the second form of (\ref{Easter}), leaving
\be
\d\tts^{\sr\se\sll\sa\st\si\so\sn\sa\sll\sss\sp\sa\scc\se \, 2}_{\triangle-\sS\sR\sP\sM} = {\d{Z}\d\overline{{Z}}}/{\{1 + {Z}^2\}^2} = 
\{\d {\cal R}^2 + {\cal R}^2\d\Phi^2\}/{\{1 + {\cal R}^2\}^2} = \{\d\Theta^2  + \mbox{sin}^2\Theta \d\Phi^2\}/4
\label{Pessach}
\ee
by using the polar form for the complex numbers and the coordinate transformation ${\cal R} = \mbox{tan}\frac{\Theta}{2}$. $\Box$

\mbox{ } 

One can also use
\beq
\d  \tts^{\sr\se\sll\sa\st\si\so\sn\sa\sll\sss\sp\sa\scc\se \, 2}_{\triangle-\sS\sR\sP\sM} = \dot{\cal R}^2 + {\cal R}^2\dot{\Phi}^2  
\label{flatbrod}
\eeq
by performing a PPSCT with conformal factor $\{1 + {\cal R}^2\}^2$.  
This is geometrically trivial, while the other above forms are both geometrically natural and mechanically natural 
(equivalent to $\ttE$ appearing as an eigenvalue free of weight function).
Also using the conformal factor $\{1 + {\cal R}^2\}$ is the conformally-natural choice.  

\mbox{ }

\noindent Note 1) In 2-$d$, the $SO(d)$ action is free (i.e. if $g x = x$, then $g$ = id). 
Thus the stabilizers $\{g \in \FrG \mbox{ } | \mbox{ } g x = x\}$ are all trivial. 
Thus, by the Orbit--Stabilizer Theorem, everything lies on the one orbit. 
I.e. there is no stratification/no singularity set. 
There are then no complications in considering geodesics.  
Also, the natural metric on shape space is everywhere-defined and everywhere of finite curvature.  

\noindent Note 2) In contrast, in 3-$d$, for which, for collinear configurations, it is $SO(2)$ rather than $SO(3)$ that is relevant.  
This gives a number of further reasons why the 3-$d$ case is harder.  
 
\noindent Note 3) See Sec \ref{CPST} for geometrical/mechanical interpretations of ${\cal R}$, $\Theta$ and $\Phi$. 
Useful coordinates for pure-shape triangleland are then the azimuthal angle $\Theta = 2\,\mbox{arctan}\,{\cal R}$ and the `Swiss army knife' angle between the 2 Jacobi vectors, 

\noindent $\Phi = \mbox{arccos}(\underline{\rho}_1\cdot\underline{\rho}_2/|\rho_1||\rho_2|)$ (c.f. Fig \ref{Fig1}).

\noindent Note 4) The $\Theta$'s for $N$-stop metrolands and the $Z = \{{\cal R}, {\Theta}\}$'s for $N$-a-gonland 
are objects that have position/direction data as per the statement of Poincar\'e's Principle.

\subsubsection{Explicit geometrical objects for pure-shape $N$-a-gonland}\label{Obgon}

Using the multipolar form ${Z}^{\stta} = {\cal R}^{\bar{\stta}}\mbox{exp}(i\Theta^{\widetilde{\stta}})$,       
the configuration space metric can be written in two blocks: $\ttM_{\bar{\sttp}\tiq} = 0$, 
\be
{\ttM}_{\bar{\sttp}\bar{\sttq}} =   \delta_{\bar{\sttp}\bar{\sttq}}/\{1 + ||\mbox{\boldmath${\cal R}$}||^2\} - 
{\cal R}_{\bar{\sttp}}{\cal R}_{\bar{\sttq}}/\{1 + ||\mbox{\boldmath${\cal R}$}||^2\}^2 
\mbox{ } ,
\label{B1}
\ee 
\be
{\ttM}_{\tip\tiq} = \{\delta_{\tip\tiq}/ - {\cal R}_{\tip}{\cal R}_{\tiq}/\{1 + ||\mbox{\boldmath${\cal R}$}||^2\}\}  
{\cal R}_{\tip}{\cal R}_{\tiq}/\{1 + ||\mbox{\boldmath${\cal R}$}||^2\}
\mbox{   (no sum) .}
\ee 
\mbox{ } \mbox{ } Then the inverse metric is ${\cal N}^{\bar{\sttp}\tiq} = 0$,   
\be
{\ttN}^{\bar{\sttp}\bar{\sttq}} = \{ 1 + ||\mbox{\boldmath${\cal R}$}||^2 \}\{ \delta^{\bar{\sttp}\bar{\sttq}} + 
{\cal R}^{\bar{\sttp}}{\cal R}^{\bar{\sttq}} \} 
\mbox{ } ,  \mbox{ } \mbox{ }
{\ttN}^{\tip\tiq} = \{1  + ||\mbox{\boldmath${\cal R}$}||^2 \}\{ \delta^{\tip\tiq}/{\cal R}^2_{\tip} + 1|^{\tip\tiq} \}  
\mbox{ (no sum) ,}
\ee
for $1|^{\tip\tiq}$ the matrix whose entries are all 1.  
Then the only nonzero first partial derivatives of the metric are (no sum)
\be
{\ttM}_{\bar{\sttp}\bar{\sttq},\bar{\sttr}} = 
\{{4{\cal R}_{\bar{\sttp}}{\cal R}_{\bar{\sttq}}{\cal R}_{\bar{\sttr}}}/{1 + ||\mbox{\boldmath${\cal R}$}||^2} - 
\{2{\cal R}_{\bar{\sttr}}\delta_{\bar{\sttp}\bar{\sttq}} + {\cal R}_{\bar{\sttq}}\delta_{\bar{\sttp}\bar{\sttr}} + 
{\cal R}_{\sttp}\delta_{\sttq\sttr} \}\}/{\{1 + ||\mbox{\boldmath${\cal R}$}||^2\}^2} \mbox{ } , 
\ee
\be
{\ttM}_{\tip\tiq,\bar{\sttr}} = \frac{2{\cal R}_{\tip}{\cal R}_{\tiq}}{\{1 + ||\mbox{\boldmath${\cal R}$}||^2\}^2}
\left\{
\frac{2{\cal R}_{\tip}{\cal R}_{\tiq}{\cal R}_{\bar{\sttr}}}{1 + ||\mbox{\boldmath${\cal R}$}||^2} - 
\{{\cal R}_{\bar{\sttr}}\delta_{\tip\tiq} + {\cal R}_{\tiq}\delta_{\tip\barr} + 
{\cal R}_{\tip}\delta_{\tiq\bar{\sttr}} \}
\right\} + 
\frac{\delta_{\tip\tiq}\{{\cal R}_{\tip}\delta_{\tiq\bar{\sttr}} + {\cal R}_{\tiq}{\delta}_{\tip\bar{\sttr}}
\}}{1 + ||\mbox{\boldmath${\cal R}$}||^2} \mbox{ } .  
\ee
The only nonzero Christoffel symbols are  
\be
{\Gamma^{\bar{\sttp}}}_{\bar{\sttq}\bar{\sttr}} = - \{  {\cal R}_{\bar{\sttr}}{\delta^{\bar{\sttp}}}_{\bar{\sttq}} + 
{\cal R}_{\bar{\sttq}}{\delta^{\bar{\sttp}}}_{\bar{\sttr}}  \}/\{1 + ||\mbox{\boldmath${\cal R}$}||^2\} 
\mbox{ } ,
\ee
\be
{\Gamma^{\bar{\sttp}}}_{\tiq\tir} = \delta_{\tiq\tir}{\delta_{\tir}}^{\bar{\sttp}}{\cal R}_{\tiq} - 
{\cal R}_{\tiq}{\cal R}_{\tir}
\{  {\cal R}_{\tir}{\delta^{\barp}}_{\tiq} + 
{\cal R}_{\tiq}{\delta^{\barp}}_{\tir}  \}/\{1 + ||\mbox{\boldmath${\cal R}$}||^2\} \mbox{ } ,
\ee 
\be
{\Gamma^{\tip}}_{\tiq \bar{\sttr}} = \left\{\frac{\delta^{\tip\tis}}{{\cal R}^2_{\tip}} + 1|^{\tip\tis}\right\}
\left\{\frac{{\cal R}_{\tis}{\cal R}_{\tiq}}{\{1 + ||\mbox{\boldmath${\cal R}$}||^2\}^2}
\left\{
\frac{2{\cal R}_{\tis}{\cal R}_{\tiq}{\cal R}_{\bar{\sttr}}}{1 + ||\mbox{\boldmath${\cal R}$}||^2} - 
\{{\cal R}_{\bar{\sttr}}\delta_{\tiq\tis} + {\cal R}_{\tis}\delta_{\tiq \bar{\sttr}} + 
{\cal R}_{\tiq}\delta_{\tis \bar{\sttr}} \}\right\} + 
\frac{\delta_{\tis\tiq}\{{\cal R}_{\tis}\delta_{\tiq \tir} + {\cal R}_{\tiq}{\delta}_{\tis \bar{\sttr}}\}}
{1 + ||\mbox{\boldmath${\cal R}$}||^2}\right\} \mbox{ } .  
\ee
\mbox{ } \mbox{ } These spaces have Ricci tensor 
\be
Ric_{\mbox{\scriptsize$\ttp\ttq$}}({\bttM}) = 2 \, n \, {\ttM}_{\mbox{\scriptsize$\ttp\ttq$}}
\label{isEinstein}
\ee 
so $\mathbb{CP}^{n - 1}$ is Einstein (hence its appearance in the literature on GR instantons \cite{GiPo,PageInst}), and thus these are also spaces of constant Ricci scalar curvature 
\beq
Ric({\bttM}) = 4\,n\,\{n - 1\} \mbox{ } .  
\label{RicCP}
\eeq
However, for $N > 3$, they have nonzero Weyl tensor (as checked by Maple \cite{Maple}) and so are not conformally flat; 
their Weyl tensor, moreover, takes a very simple form \cite{W82a}.  
None of the abovementioned curvatures, or curvature scalars constructed from them and the metric, blow up for finite ${\cal R}_{\barp}$.

The second Chern class for quadrilateralland is $c_2(\mathbb{CP}^2) = 1$ \cite{Iwai92}.

\subsubsection{Isometry groups for $N$-stop metroland}\label{Isometries}

$N$-stop metroland's shape spaces $\FrS(N, 1)$ are $\mathbb{S}^{n - 1}$, which are the positively-curved maximally symmetric spaces. 
Thus they have the maximum possible isometry group dimension. 
For configuration space dimension $q$, the maximal isometry group has dimension $q\{q + 1\}/2$), so here $n\{n - 1\}/2$.  
$\mbox{Isom}(\FrS(N, 1)) = \mbox{Isom}(\mathbb{S}^{n - 1}) = PSO(n)$ (the $n$-dimensional projective special orthogonal group) = $SO(n)$.  
These results obviously immediately extend to $\FP(N, d) = \mathbb{S}^{nd - 1}$.
Also, $\mbox{Isom}(\FrO\FrS(N, 1)) = \mbox{Isom}(\mathbb{RP}^{n - 1}) =$ the projective orthogonal group $PO(n) = O(n)/\mZ(O(n))$, which is $SO(n)$ again.

\subsubsection{Isometry groups for pure-shape $N$-a-gonlands}\label{Isom2}

Isom($\FrS(N, 2)$) = Isom($\mathbb{CP}^{n - 1}$) = the projective unitary group $PSU(n)$ = $SU(n)$/Z($SU(n)$) = $SU(n)/\mathbb{Z}_{n}$.
Thus these are fairly symmetrical but not maximally symmetric for $N > 1$.
For, there are $n^2 - 1$ Killing vectors associated with $SU(n)$, while the maximum number of Killing vectors for a configuration space of dimension 2\{$n$ -- 1\} is 

\noindent\{$n$ -- 1\}\{2$n$ -- 1\} = 2$n^2$ -- 3$n$ + 1.  
One can in fact characterize $\mathbb{CP}^{n - 1}$ as the spaces of constant sectional curvature, 
which are in a sense the next most symmetric spaces after the maximally symmetric spaces of constant curvature.

E.g. for quadrilateralland, $\mathbb{CP}^2$ has the 8 Killing vectors \cite{GiPo} associated with $SU$(3), rather than the maximal 10.  
This case's representation theory is particularly well-known due to its appearing in the Particle Physics' approximate flavour symmetry and exact colour symmetry.
For pentagonland, $\mathbb{CP}^3$ has 15 Killing vectors associated with $SU$(4) rather than the maximum possible of 21. 
For hexagonland, there are 24 compared to the maximum possible of 36.

One way of studying the Killing Vectors for $\mathbb{CP}^{n - 1}$ is that these can be obtained by 
projecting down the Hopf fibration of $\mathbb{CP}^{n - 1}$ by $\mathbb{S}^1$ with base manifold  $\mathbb{S}^{2n - 1}$ \cite{W82b}.  

{            \begin{figure}[ht]
\centering
\includegraphics[width=0.4\textwidth]{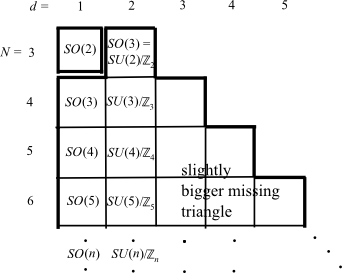}
\caption[Text der im Bilderverzeichnis auftaucht] {\footnotesize{The `partial Periodic Table of Periodic Tables' illustrating the two series 
of known isometry groups that coincide with the well-known families of Lie groups $SO(p)$ and $SU(p)$. 
Taking out $\mathbb{Z}_n$ amounts to rendering a connected component, a step which, of course, makes no difference infinitesimally. 
In any case, there is a good (and relational!) reason for colour modelling of the strong force to involve this connected component also (namely 
that the colour labels red, green and blue are frivolous, in the sense that the theory is invariant under permuting these names).  
Thus these isometry groups and the subsequent representation theory are widely known both in Mathematics and in Physics.
The Casson diagonal is, as far as I know, only a simple series at the topological level.}        }
\label{PerPer}\end{figure}            }

Finally, I note that $\mathbb{CP}^k/\mathbb{Z}_2$ has in general a different number of Killing vectors from 
$\mathbb{CP}^k$, at least under some of some the weighted projective space group actions (see SSec \ref{WPS}).  
A toy model (see e.g. Sec \ref{Isom3}) for this is how identifications can cause loss of some Killing vectors from $\mathbb{R}^2$. 
This is due to these becoming discontinuous at the identification points, so that the identified topology does not globally support such Killing vectors.     
The triangleland case's number of Killing vectors is unaffected by whether one identifies mirror images (that case keeps all 3 of the $SU(2)$ Killing vectors).  
I have not yet considered any specific $\mathbb{CP}^k/\mathbb{Z}_2^{\scc\so\sn\sj}$ beyond that.

\subsection{Riemannian geometry of O(pre)shapespace (mostly open questions)}\label{RiemS}

\subsubsection{Oriented $N$-stop metroland: $\mathbb{RP}^{n}$}\label{RiemSN}

Metric geometry is local, so this is as in the previous section as regards metric, inverse, determinant, Christoffel symbols and curvature tensors.\footnote{As 
analogy elsewhere within Mechanics, $\mathbb{RP}^2$ has also been used as a simple model of the falling cat \cite{Mont93}. 
This consists of 2 jointed rods, for which one can see how the number of degrees of freedom matches the present problem's  2 Jacobi vectors.  
However, this analogy is at most partial since the cat model is physically/geometrically distinct in space.} 

\subsubsection{Question 16) what is the Riemannian geometry of $N$-a-gonland's $\mathbb{CP}^{n - 1}/\mathbb{Z}_2^{\scc\so\sn\sj}$ ?}\label{RiemSgon}

As a partial answer, recollect that the Corollary [4)] (to Kuiper's Theorem) of Sec \ref{Kuiper}) is that $\mathbb{CP}^2/\mathbb{Z}_2$ is diffeomorphic to $\mathbb{S}^4$.  
The conj action is a means of representing reflection. 
Moreover, the metric here is not the standard spherical one. 
For, locally the metric is as for $\mathbb{CP}^2$: Fubini--Study (8 local Killing vectors) whilst the 4-sphere's spherical metric has 10.

\subsubsection{Questions outside scope of this Article on higher-$d$ Oshape spaces}\label{NSS}

{\bf Question 17$^*$)} What is the mathematics of pure-shape `N$d$-simplexland' for $d \geq 3$? 

\mbox{ } 

\noindent It is considerably harder \cite{Kendall} than for $d \leq 2$ though at least pure-shape RPM is free of maximal collisions. 
Though at least pure-shape RPM is free of collisions that are {\sl maximal} (i.e. between all the particles at once). 
For increasing $N$, even the topology quickly cease to be standard or even well-studied, while the singular set begins to play a prominent and obstructory role. 
\cite{Kendall} does however describe and provide references for a partial study of (easier subcases of) these shape spaces.

\subsection{Riemannian geometry of relational space}\label{Riem2}

Relational space $\bigr(N, d)$ can then be envisaged in scale--shape split form as the 
{\it cone} $C(\FrS(N, d))$ over the corresponding shape space $\FrS(N, d)$ \cite{Cones}.

\subsubsection{Parametrizing scale}\label{PS}

In GR, one uses e.g. the scalefactor $a$ such that $\mh_{\mu\nu} = \ma^2\muu_{\mu\nu}$ for det$(\mu_{\mu\nu})$ = 1; 
then the alternative $\sqrt{\mh} = \sqrt{\mbox{det}\, (\mh_{\mu\nu})} = \ma^{3}$, which is local volume.  
The alternative Misner variable is, in the homogeneous $\ma = a$ setting, $\Omega := \mbox{ln}(\sqrt{\mh}) = 3\mbox{ln}(a)$  
This is sometimes called `Misner time' in the literature; see Sec \ref{TAQ} for how this and other scale variables fare as candidate times.

\mbox{ }

\noindent Analogy 36), the RPM counterparts of these are, respectively, $\rho$ or $\mI$ as analogues of scalefactor. 
The above shape analysis can be uplifted to above to scaled RPM, alongside a further size variable, by the cone construction/scale--shape split.   
This further variable can be taken to be the configuration space radius $\rho = \sqrt{\mI}$ for $N$-stop metroland, or $\mI$ for triangleland.
I term $\rho^{\sd\si\sm({\cal R}(N, d))} = \rho^{nd - d\{d - 1\}/2}$ {\it pseudo-volume scales} due to their parallels with GR's local volume scale.  
Finally, I term quantities proportional to ln$\,\rho$ I term {\it pseudo-Misner scales} due to their parallel with GR's Misner variable.

\subsubsection{Shape--scale split as a cone at the level of metric geometry}\label{Cone2}

At the level of Riemannian geometry, a {\bf cone} C($\langle \FrMgen, \mbox{ } \bg \rangle$) over a $m$-dimensional Riemannian space 
$\langle \FrMgen, \mbox{ } \bg \rangle$ is an \{$m$ + 1\}-dimensional metric space with 1) topological properties as in the preceding SSec. 
2) Now, it additionally possesses also an arc element  
\beq
\d s^2 = \d \, \sigma^2 + \sigma^2\d \tts^2 \mbox{ } ,
\eeq
where $\d s^2$ the arc element of $\FrMgen$ itself and $\sigma$ is a suitable `radial variable' running over [0, $\infty$), which represents the distance from the cone point.  
This metric is smooth everywhere except (in general) at the troublesome cone point.   
The above arc element is the second half of the third key point, and has the shape--scale split application, by which 
shape space geometry, dynamics, QM is re-applicable as a subproblem of the scaled case.  
Some comments on this are as follows.  

\mbox{ }

\noindent \bu The everyday-life cone C($\mathbb{S}^1$) can indeed be viewed as a simple example of these constructs, using $\FrMgen$ = (a piece of) $\mathbb{S}^1$.  

\noindent \bu At the Riemannian level, there is a notion of distance and hence (for sufficiently 
nontrivial dimension) of angle, so that one can talk in terms of deficit angle.  
C($\mathbb{S}^1$) itself has no deficit angle, while using a p-radian piece entails a deficit angle of $2\pi - p$.  
The presence of deficit angle, in turn, gives issues about `conical singularities', e.g. \cite{KS91}.   

\noindent \bu the previous two points (generalized to `deficit solid angle') straightforwardly generalize to $\mathbb{S}^k$. 

\mbox{ }  

\noindent \bu {\bf Lemma 11}. The simplest few cases of RPM's involve very straightforward cones.  
C($\FrS$(N, 1)) = C($\mathbb{S}^{n - 1}$) = $\mathbb{R}^{n}$ using just elementary results (c.f. e.g. 
\cite{Kendall, Rotman}).   
C($\FrS$(3, 2)) = C($\mathbb{CP}^{1}$) = C($\mathbb{S}^{2}$) of radius 1/2 = $\mathbb{R}^{n}$ (see e.g. 
\cite{Hsiang1, PP87}) up to a conformal factor at the metric level, which can be `passed' to the 
potential, c.f. Sec \ref{Dyn1}.  

\noindent \bu C($\FrO\FrS$(n, 1)) = C($\mathbb{S}^{n - 1}/\mathbb{Z}_2$) = $\mathbb{R}^{n}_+$, the half-space i.e. 
half of the possible generalized deficit angle.  
C($\FrO\FrS$(3, 2)) = C($\mathbb{S}^2/\mathbb{Z}_2$) = $\mathbb{R}^3_+$.   
These last two, however, have edge issues. 3-stop metroland: the identification coincides with the gluing in constructing the `everyday cone'.  
Oriented 4-stop metroland is $\mathbb{RP}^2$.    
Finally, triangleland might include the boundary or might not, depending on the application, and should not be glued as antipodal points in the z = 0 plane do not match.

\subsubsection{Scaled $N$-stop metroland as a cone}\label{ConeN}

The plain $N$-stop metroland case is straightforward as C($\mathbb{S}^{n - 1}$) is $\mathbb{R}^{n}$  (c.f. \cite{PP87}, and elsewhere in this Sec).
The O case can be viewed as having a deficit angle of $\pi$ for 3 particles and a deficit solid angle of 2$\pi$ for 4 particles. 
This involves coordinate ranges rather than form of arc element, so below arc elements serve for both.    
[Oriented cases do have further issues as regards whether conical singularities show up.  
These could lead to having to excise the cone point.]

\noindent $\mC(\mathbb{CP}^{k})$ is mentioned  by Montgomery \cite{Mont98}, who also considers the corresponding collision set.

\noindent Minus the cone point, plain triangleland is diffeomorphic to $\mathbb{R}^3$ \cite{Hsiang1}.  
Is the version I consider in \cite{08I}.  
If one keeps the $1/4\mI$ factor, the geometry is curved, with a curvature singularity at 0, and, 
obviously, conformally flat where the conformal transformation is defined (everywhere bar 0).  
In any case the $E/\mI$ `potential term' in  \cite{08I, 08III} is singular there too.  
Clearly 
\beq
\d s^{\sr\se\sll\sa\st\si\so\sn\sa\sll\sss\sp\sa\scc\se \, 2}_{N-\sss\st\so\sp\,\sE\sR\sP\sM}    = \d\rho^2 + \rho^2\d \tts^2_{\sss\sp\sh\se} \mbox{ } .  
\label{Uber}
\eeq

\subsubsection{Scaled $N$-a-gonland as a cone}\label{ConeTri}

For $N$-a-gonland, 
\beq
\d s^{\sr\se\sll\sa\st\si\so\sn\sa\sll\sss\sp\sa\scc\se}_{N-\sa-\sg\so\sn\,\sE\sR\sP\sM}\mbox{}^2 = \d\rho^2 + \rho^2\d \tts^2_{\sF\sS} \mbox{ } .
\label{Pentecost}
\eeq
Then e.g. for scaled triangleland, as the shape space is the sphere of radius 1/2,
\beq
\d s^{\sr\se\sll\sa\st\si\so\sn\sa\sll\sss\sp\sa\scc\se \, 2}_{\triangle\,\sE\sR\sP\sM} = \d\rho^2 + \{\rho^2/4\}\{\d\Theta^2 + \mbox{sin}^2\Theta\d\Phi^2\}
\mbox{ } , 
\eeq
which inconvenience in coordinate ranges can be overcome\footnote{See 
\cite{J83} for a distinct way of doing this.  
Also, from now on I upgrade `configuration space radius' to 
mean the portmanteau of `usually $\rho$ but $\rho^2 = \sI$ in the case of the spherical presentation of triangleland.  
By this subtlety, I am not in the end taking `configuration space radius' to be a entirely synonymous to `hyperradius'. 
One use for the spherical presentation of triangleland  is use that it allows the shape part to be studied 
in $\mathbb{S}^2$ terms which more closely parallel the Halliwell--Hawking \cite{HallHaw} analysis of GR inhomogeneities over $\mathbb{S}^3$.}
by using, instead, $\mI$ as the configuration space radius variable, 
\beq
\d s^{\sr\se\sll\sa\st\si\so\sn\sa\sll\sss\sp\sa\scc\se \, 2}_{\triangle\,\sE\sR\sP\sM} = \{1/4\mI\}\{\d \mI^2 + \mI^2\{\d\Theta^2 + \mbox{sin}^2\Theta\d\Phi^2\}\} \mbox{ } .
\label{Michaelmas}
\eeq 
This renders the metric manifestly conformally flat, the flat metric itself being 
\beq
\d \check{s}^{\sr\se\sll\sa\st\si\so\sn\sa\sll\sss\sp\sa\scc\se \, 2}_{\sf\sll\sa\st \, \triangle-\sE\sR\sP\sM} = 
\d \mI^2 + \mI^2\{\d\Theta^2 + \mbox{sin}^2\Theta\d\Phi^2\}
\label{Eostre}
\eeq
(spherical polar coordinates with $\mI$ as radial variable). 
This using of $\mI$ as radial variable: the start of significant differences between the triangleland and 4-stop metroland spheres and half-spheres.
In the O case, one can also use a double angle variable running over the usual range of angles e.g. \cite{Smith62, LMAC98}.  

\mbox{ }
 
\noindent {\bf Question 18} Figure out the metric-level counterpart of Question 13.

\mbox{ }

\noindent Minus the cone point, plain triangleland is diffeomorphic to $\mathbb{R}^3$ \cite{Hsiang1}.  
Is the version I consider in \cite{08I}.  
If one keeps the $1/4\,\mI$ factor, the geometry is curved, with a curvature singularity at 0, and, 
obviously, conformally flat where the conformal transformation is defined (i.e. elsewere than 0).  
In any case the $E/\mI$ `potential term' in  \cite{08I, 08III} is singular there too.

\subsubsection{Aside: comments on e.g. `The End of Time' \cite{EOT}'s pyramidal presentation of scaled triangleland}\label{Pyramid}

I mention this due to it appearing in discussions in Sec \ref{Wedge}.  
Parametrizing triangleland by the three sides gives the pyramidal presentation of the relational configuration space in \cite{B94II, EOT} (also mentioned in \cite{LR97}).
Note that this is limited in its usefulness compared to the spherical presentation, because it lacks naturality at the metric level.

\subsubsection{Question 19) what is the geometry of scaled O$N$-a-gonland?}\label{ConeC}

I.e. of the cone C($\mathbb{CP}^{n - 1}/\mathbb{Z}_2^{\scc\so\sn\sj}$).
Of possible relevance to this study, Witten, Acharya and Joyce \cite{W1, W2, Acharya, Joyce} (see Sec \ref{WPS}) consider cones over both $\mathbb{CP}^3$ and a $\mathbb{WCP}^3$.

\subsubsection{Explicit geometrical objects for scaled $N$-a-gonland}\label{Obgon2}

Using the multipolar form; then the configuration space metric can be written in three blocks (with all other components zero): 

\noindent
\beq
{M}_{00} = 1 \mbox{ } , \mbox{ } \mbox{ }
{M}_{\barp\barq} = \rho^2\{\{1 + ||{\cal R}||^2\}^{-1}\delta_{\barp\barq} - 
\{1 + ||{\cal R}||^2\}^{-2}{\cal R}_{\barp}{\cal R}_{\barq}\}
\mbox{ } ,
\label{B12}
\eeq 
\beq
{M}_{\tip\tiq} = \rho^2\{\{\{1 + ||{\cal R}||^2\}^{-1}\delta_{\tip\tiq} - \{1 + ||{\cal R}||^2\}{\cal R}_{\tip}{\cal R}_{\tiq}\} {\cal R}_{\tip}{\cal R}_{\tiq}\} \mbox{   (no sum) ,}
\eeq 
\beq
{M}_{\barp\tiq} = 0 \mbox{ } .  
\label{B2}
\ee
Then the inverse metric's nonzero components are 
\beq
{N}^{00} = 1 \mbox{ } , \mbox{ } \mbox{ } 
{N}^{\barp\barq} = \{1/\rho^2\}\{\{1 + ||{\cal R}||^2\}\{\delta^{\barp\barq} + 
{\cal R}^{\barp}{\cal R}^{\barq}\}\} \mbox{ } ,  
\eeq
\be
{N}^{\tip\tiq} = \{1/\rho^2\}\{\{1 + ||{\cal R}||^2\}\{\delta^{\tip\tiq}/{\cal R}^2_{\tip} + 1^{\tip\tiq}\}\}  
\mbox{ (no sum) .}
\ee
Then the only nonzero first partial derivatives of the metric are (no sum) 
\be
{M}_{\barp\barq,\barr} = \frac{1}{\{1 + ||{\cal R}||^2\}^2}
\left\{
\frac{4{\cal R}_{\barp}{\cal R}_{\barq}{\cal R}_{\barr}}{1 + ||{\cal R}||^2} - 
\{2{\cal R}_{\barr}\delta_{\barp\barq} + {\cal R}_{\barq}\delta_{\barp\barr} + 
{\cal R}_{\barp}\delta_{\barq\barr} \}
\right\} \mbox{ } , 
\ee
\be
{M}_{\tip\tiq,\barr} = \frac{2{\cal R}_{\tip}{\cal R}_{\tiq}}{\{1 + ||{\cal R}||^2\}^2}
\left\{
\frac{2{\cal R}_{\tip}{\cal R}_{\tiq}{\cal R}_{r}}{1 + ||{\cal R}||^2} - 
\{{\cal R}_{\barr}\delta_{\tip\tiq} + {\cal R}_{\tiq}\delta_{\tip\barr} + 
{\cal R}_{\tip}\delta_{\tiq\barr} \}
\right\} + 
\frac{\delta_{\tip\tiq}\{{\cal R}_{\tip}\delta_{\tiq\barr} + {\cal R}_{\tiq}{\delta}_{\tip\barr}\}}{1 + ||{\cal R}||^2} \mbox{ } .  
\ee
The only nonzero Christoffel symbols are (no sum except over $\tilde{s}$) the same as for $\mathbb{CP}^{n}$ with the following extra nonzero ones: 
\beq
\Gamma^{\rho}\mbox{}_{\barp\barq} = {M}_{\barp\barq}/\rho \mbox{ } , \mbox{ } 
\Gamma^{\barp}\mbox{}_{\barp\rho} (\mbox{no sum}) = 1/\rho \mbox{ } .
\eeq
These spaces have Ricci tensor 
\be
Ric_{\mbox{\scriptsize$\ttp\ttq$}}(\mbox{\boldmath$M$}) = 3{M}_{\mbox{\scriptsize$\ttp\ttq$}}/\rho^2
\label{isEinstein2}
\ee 
(run from 1 to 2\{$n$ - 1\}) with all other components zero, and 
\beq
Ric(\mbox{\boldmath$M$}) = 6\,\{n - 1\}/\rho^2  \mbox{ } ,   
\label{So}
\eeq
which is {\sl not} constant (as reported in \cite{06II} for the triangleland case).   
Additionally, for $N > 3$, they have nonzero Weyl tensor (as checked by Maple \cite{Maple}) and so are not conformally flat.

Then the Ricci scalar blows up as $\rho \longrightarrow 0$, so this is a curvature singularity.  
The Ricci tensor and Weyl tensor's components are finite (remember ${M}_{\barp\barq}$ already has an $\rho^2$ factor).

\mbox{ } 

\noindent As regards $\mC(\mathbb{RP}^{k})$ and $\mC(\mathbb{CP}^{k}/\mathbb{Z}_2^{\scc\so\sn\sj})$ geometries, these are locally as before.
However, they have scope for new singularities due to the identifications...
The latter is now the coning of a metrically nonstandard 4-sphere. 
It is also not clear how one associates a $\mathbb{R}^{5}$ with a 6-$d$ system's objects (three 2-$d$ relative Jacobi vectors). 
This is like triangleland's $\mathbb{R}^3$ for a 4-$d$ system's objects, but now there is no longer a Hopf map to provide the answer? 
[The 4-$d$ case trivially contains $\mathbb{S}^3$, which Hopf-maps to $\mathbb{S}^2$ which is trivially contained in 3-$d$. 
Now, however the 6-$d$ case is $\mathbb{S}^5$ which does not Hopf-map to $\mathbb{S}^4$.  
It is $\mathbb{S}^7$ which does.  
That gives a small chance to 8-$d$ of absolute space trivially containing $\mathbb{S}^7$, which Hopf-maps to $\mathbb{S}^4$ which sits trivially in $\mathbb{R}^5$.  
In any case, there are only 3 Hopf maps of this kind, so one would run out eventually. 
However, the Hopf map admits a generalization via its $\mathbb{S}^3 \longrightarrow \mathbb{CP}^1$ aspect to $\mathbb{S}^{2\sn + 1} \longrightarrow \mathbb{CP}^{n}$.]  
%
%

\subsubsection{Isometry groups for scaled $N$-stop metrolands}\label{Isom3}

$\mbox{Isom}(\bigr(N, 1)) = \mbox{Isom}(\mathbb{R}^{n}) = \mbox{Eucl}(n)$, which is the flat case of maximally symmetric space, and thus possesses 

\noindent $n\{n + 1\}/2$ independent Killing vectors.  
Next, $\mbox{Isom}(\FrO\bigr(N, 1)) = \mbox{Isom}(\mathbb{R}^{n}/\mathbb{Z}_2)$.  
For $n$ = 3, this gives $SO(3)$ $\times \,\,\, \mathbb{R}^2$ 

\noindent($\pa/\pa z$ being discontinuous).
This result also extends easily to arbitrary $n$.

\subsubsection{Isometry groups for scaled $N$-a-gonlands}\label{++Isom}

The $N$-stop metroland case is somewhat of a fluke in terms of computibility.  
$N$-a-gonland requires more general considerations.

\mbox{ } 

\noindent {\bf Lemma 12} Coning i) preserves Killing vectors and ii) gives a new conformal Killing vector.  
\noindent (The proof is straightforward).  

\mbox{ } 

\noindent{\bf Question 20$^*$} Does coning $\mathbb{CP}^{n - 1}$ and $\mathbb{CP}^{n - 1}/\mathbb{Z}_2^{\scc\so\sn\sj}$  produce any further Killing vectors?
Conformal Killing vectors?  

\mbox{ }

\noindent We do control the 3-particle case of course, through it being conformal to the flat metric.  
We also know that C($\mathbb{CP}^2/\mathbb{Z}_2^{\scc\so\sn\sj}$) is diffeomorphic to $\mathbb{R}^5$ as a Corollary of Kuiper's Theorem. 
However the metric on this is not the standard flat one, as as its Ricci scalar is proportional to $1/\rho^2$.  
It is not conformally flat either, since its Weyl tensor does not vanish.

\subsection{Edges and singularities of configuration space}\label{Edge}

Nontrivial group orbit structure is in general relevant here.  
Also, one sometimes needs to excise due to singular potentials, though detail of this is diverted to Sec \ref{Cl-Soln}.

\subsection{Some more useful coordinatizations of specific RPM's configuration spaces}\label{Shape-Quant}

\subsubsection{$\mathbb{S}^k$ within $\mathbb{R}^{k + 1}$}

Spherical coordinates [($\alpha, \chi$) for the most common case of the 2-sphere, see App \ref{DynSphe}] are good for solving many 
aspects of the dynamics of the sphere no matter what its physical interpretation is to be.  
Working at the level of a general joint treatment, one can recast the k-sphere in terms of $k + 1$ 
variables $u^{\Gamma}$ such that $\sum_{\Gamma = 1}^k u^{\Gamma\,2} = 1$.  
These describe a Euclidean \{$k + 1$\}-space that surrounds the sphere; 
it is often convenient to use $u_x$, $u_y$ and $u_z$ for the components of $u^{\Gamma}$ in the case of the 2-sphere in 3-$d$ Euclidean space.   
Then the $u^{\Gamma}$ are related to the $\alpha$ and $\chi$ through being the components of the corresponding unit Cartesian vector in spherical polar coordinates:
\beq
u_x = \mbox{sin}\,\alpha\,\mbox{cos}\,\chi\, \mbox{ } , \mbox{ } \mbox{ }
u_y = \mbox{sin}\,\alpha\,\mbox{sin}\,\chi\, \mbox{ } , \mbox{ } \mbox{ }  
u_z = \mbox{cos}\,\alpha \mbox{ } . 
\eeq 
For $\mathbb{S}^2$ in `actual space', the $\mathbb{R}^3$ {\sl is} `actual space' with the physical radius $r$ in the role of the radial coordinate.

\subsubsection{Coordinates for $N$-stop metroland} \label{NM-Coords}

For 4-stop metroland, I make subsequent use of 
\beq
{\cal R} = \mbox{tan$\frac{\theta}{2}$} = \sqrt{\{1 - \mn_z\}/\{1 + \mn_z\}} \mbox{ } ,
\label{Rdef}
\eeq
which is a radial stereographic coordinate corresponding to the tangent plane at the `North' DD-pole.
As regards surrounding coordinates for the pure-shape $N$-stop metroland configuration (or coordinates for the 
scaled $N$-stop metroland configuration space), here, the $\rho^i$ serve straightforwardly as Cartesian coordinates.  
These are subject to the on-hypersphere condition 
\beq
\sumin \mI^i = 
\sumin \rho^{i\,2} = \rho^2 = \mI \mbox{ }  \mbox{ } \mbox{ (constant) } , \mbox{ } 
\mbox{ or } \mbox{ } 
\sumin \mN^i = \sumin \mn^{i\,2} = 1 \mbox{ } .  
\label{New19}
\eeq
The $\mn^i$ are then the components of the unit Cartesian vector [e.g. $\mn^i = 
(\mbox{sin}\,\theta\,\mbox{cos}\,\phi, \mbox{sin}\,\theta\,\mbox{sin}\,\phi, \mbox{cos}\,\theta)$ in spherical polar coordinates in the 4-stop case].  
I term passing from $\ux$ in $n$-$d$ space to $\rho^{i}$ in $N$-stop metroland configuration space the {\it Cartesian correspondence}.

\subsubsection{$N$-stop metrolands relationally interpreted}\label{SQNS}

There are pure-shape versions and scaled versions of these, for which my notation is lower case for the former and upper-case for the latter.  
The former are related to the latter by multiplication in the obvious way by the applicable size quantity $\rho :=$ Size or $\mI :=$ SIZE.

Scaled $N$-stop metroland's Cartesian components are all cluster-dependent ratio quantities.  
These concern how large a given cluster is relative to the whole model or how well-separated the two 
clusters are in the latter case that has enough particles to build up such a quantity.  
For 3-stop metroland \cite{AF}, $\mn_1^{(\sH \sb)}$ is a measure of how large the universe is relative to cluster bc.
$\mn_2^{(\sH \sb)}$ is a measure of how large the universe is relative to the separation between cluster bc and particle a.  
For 4-stop metroland \cite{AF}, in Jacobi H-coordinates or K-coordinates, my convention is to position the $\rho^i$ as in (\ref{Var}). 
In the H-coordinates case, these are, respectively a measure of the size of the universe's contents relative to the size of the whole 
model universe, and a measure of inhomogeneity among the contents of the universe (whether one of the constituent clusters is larger than the other one.)  
Also, $\mn_3^{(\sH \sb)}$ := Relsize(1b, cd) is a measure of how large the universe is relative to the separation between clusters 1b and cd. 
This being large means physically that clusters \{1b\} and \{cd\} each cover but a small portion of the model universe, and corresponds geometrically to the polar caps. 
$\mn_1^{(\sH \sb)}$ := RelSize(1b) is a measure of how large the universe is relative to cluster 1b. 
This being small means physically that cluster \{1b\} is but a speck in the firmament, and corresponds geometrically to a belt around the `Bangladeshi' meridian. 
This being large means physically that cluster \{1b\} engulfs the rest of the model universe, and corresponds 
geometrically to an antipodal pair of caps around each of the intersections of the equator and the Greenwich meridian.
$\mn_2^{(\sH \sb)}$ is a measure of how large the universe is relative to cluster cd.  
This being small corresponds to clusters \{1b\} and \{cd\} being merged, and corresponds geometrically to the equatorial belt.

In the K-coordinates case, these are now respectively a measure the sizes of the \{12\} and \{T3\} clusters relative to the whole model universe 
and of the sizes of the \{12\} and \{T3\} clusters relative to each other.  
\noindent 
$\mn_1^{(\sK \sa)}$ = RelSize(ab) is a measure of how large the universe is relative to cluster ab.  
$\mn_2^{(\sK \sa)}$ = RelSize(ab, c)   is a measure of how large the universe is relative to the separation between subcluster ab and particle c. 
$\mn_3^{(\sK \sa)}$ = RelSize(abc, d) is a measure of how large the universe is relative to the separation between cluster abc and particle d. 
Note that these have a more symmetric meaning in H coordinates than in K coordinates or for 3 particles.

\subsubsection{Stereographic and Dragt coordinate systems for triangleland}\label{CPST}

\noindent A sometimes convenient coordinatization is in terms of ${\cal R} = \mbox{tan}\frac{\Theta}{2}$.  
This is geometrically the stereographic coordinate in the tangent plane to the North Pole, and is physically another 
coordinatization of the non-angle ratio that is the tallness of the triangle. 
Two further useful coordinatizations for this are as in Fig \ref{Fig-3tri}.  

{            \begin{figure}[ht]
\centering
\includegraphics[width=0.4\textwidth]{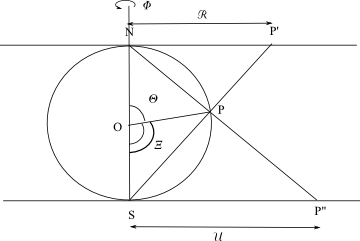}
\caption[Text der im Bilderverzeichnis auftaucht]{        \footnotesize{
Interrelation between the following useful coordinates ${\cal R}$ and ${\cal U}  = 1/{\cal R}$, 
and its supplementary angle $\Xi = \pi - \Theta$.
I term the underlying operation between each of these pairs the {\it duality map}; in $\rho_i$ or $\mI_i$ variables, it takes the form of interchanging the 1-indices and the 2-indices.
N is the North Pole, S is the South Pole, O is the centre.  Sometimes N and S are E and $\bar{\mE}$, and sometimes they are D and M 
(see Sec \ref{TessiTri} for this Article's axis systems for triangleland).  }        }
\label{Fig-3tri}\end{figure}            }

\noindent Like for 4-stop metroland, the shape space sphere for triangleland can be taken to sit in embedding relational space.  
However, for triangleland, one no longer has the above straightforward Cartesian map.  
Pure-shape triangleland's $\underline{\rho}_1$ and $\underline{\rho}_2$ are not related to the Cartesian 
coordinates of the surrounding relational space in the above familiar Cartesian way.  
Rather, they are related in the less straightforward extended Hopf map/Dragt correspondence way. 
For the particular case of triangleland, this is the fourth key step in understanding the Physics, 
and again is a kinematical manoeuvre with precedent in the Theoretical Molecular Physics literature.    

That corresponds to having to use not $\rho$ but $\mI$ as radial variable in triangleland.  
(This is due to triangleland's shape sphere arises from $\mathbb{CP}^1$, which gives its natural radius 
an unusual factor of 1/2, which is absorbed by the coordinate transformation to radial variable $\mI$ 
This makes triangleland quite unlike 4-stop metroland or actual space as regards the physical meaning of its $u^{\Gamma}$.) 
I.e. the triangleland $u^{\Gamma}$ are Dragt coordinates \cite{Dragt} are related to the 
configuration space's coordinates, rather, by the mathematics of the $\mathbb{S}^3 \longrightarrow \mathbb{S}^2$ Hopf-type map'\footnote{See 
\cite{Gronwall,Smith62} for earlier literature and e.g. \cite{Aqui93, ML99,Iwai87,PP87,LR97} for some applications of these coordinates.
Note that in the plain case, Hopf applies. 
Iwai \cite{Iwai87} then wrote down the O-counterpart of this.
Also note that Dragt's own work is for 3-cornerland's $\mathbb{R}^3_+$; outside of this context, I call the analogous coordinates ``of Dragt type".  
There also being some slight differences here between 2-$d$ and 3-$d$, plain and O triangles, and the scaled and pure-shape cases, I refer to the coordinates used as of `Dragt-type'.
(Most e.g. Dragt \cite{Dragt}, Iwai \cite{Iwai87}, Littlejohn and Reinsch \cite{LR97} have the O-case's 
range for angles, though e.g. Hsiang \cite{Hsiang1} and I \cite{08I, +Tri} do not [which is natural from the intrinsically 2-$d$ perspective].).}
I then term substituting $\ux$ by $Dra^{\Gamma}$, or, alternatively, substituting the $u^{\mu}$ 
of space for: $dra^{\Gamma}$ the {\it Dragt correspondence}.  
Dropping (a) labels, 
\beq
dra_x = \mbox{sin}\,\Theta\,\mbox{cos}\,\Phi = 
2 \mn_1 \mn_2 \ ,\mbox{cos}\,\Phi = 
2 \{\underline{\mn}_1 \cr \underline{\mn}_2\}_3 
\mbox{ } ,
\label{dragt1}
\eeq
\beq
dra_y = \mbox{sin}\,\Theta\,\mbox{sin}\,\Phi = 
2 \mn_1\mn_2\,\mbox{sin}\,\Phi = 2 
\underline{\mn}_1 \cdot \underline{\mn}_2
\mbox{ } ,
\label{dragt2}
\eeq
\beq
dra_z = \mbox{cos}\,\Theta = \mN_2 - \mN_1 
\mbox{ } .   
\label{dragt3}
\eeq
The 3 component in the first of these indicates the component in the fictitious third dimension of this cross-product.
In these depending on squared quantities, one can see consequences of $\rho^2 = \mI$ and not $\rho$ being the radial coordinate.
So one goes from ($\alpha, \chi$) to $u^{\Gamma}$, $\Gamma$ = 1 to 3, and then, for 4-stop metroland, to 
$\mn^i_{(\sH \sb)}$ via the Cartesian correspondence, and, for triangleland to $dra$ via the Dragt correspondence.  
The metric is then 
\beq
\d s_{\triangle\,\sE\sR\sP\sM}^{\sr\se\sll\sa\st\si\so\sn\sa\sll\sss\sp\sa\scc\se}\mbox{}^{\,2} = 
\sum \mbox{}_{\mbox{}_{\mbox{\scriptsize $\Gamma = 1$}}}^3\d{Dra}^{\Gamma\, 2} \mbox{ } .  
\label{Minerva}
\eeq
\noindent Note: the usual Dragt coordinates are related to ours by $Dra^{\Gamma} = \mI\,dra^{\Gamma}$; 
moreover in the literature it is usually the O-case's half-space for which these are presented.

\subsubsection{My geometrical interpretation of Dragt-type coordinates}\label{triSQNS}

I drop (a) labels.  
Now, $dra_z$ is the `ellipticity' 
\be 
ellip = \mN_2 - \mN_1
\ee 
of the two `normalized' partial moments of inertia involved in the (a)-clustering. 
This (and $\Theta$ itself) is a function of a pure ratio of relative separations rather than of relative angle.  
It is closely related to sharp = $\mN_2$, a sharpness quantity -- how sharp the triangle is 
with respect to the (a)-clustering, and flat = $\mN_1$, which is likewise a flatness quantity.     
Moreover sharp and flat are but linear functions of $ellip$:\footnote{Barbour calls sharp configurations `needles',  
whilst Kendall refers collectively to considerably sharp and flat configurations as `splinters'.}   
\beq
sharp = \{1 + ellip\}/2 \mbox{ } \mbox{ and } \mbox{ } 
flat = \{1 -  ellip\}/2.  
\label{libellula}
\eeq
The rearrangement $ellip$ = 1 -- 2\,$flat$ is furthermore useful interpretationally as $flat$ = (length of the \{bc\} base) per unit $\mI$, 
which is an obviously primary quantity within the mass-weighted triangle.  
$\Theta$ can also be considered as a ratio variable.  
That $ellip$ is cos$\,\Theta$ subsequently plays an important role in this Article.

$\Phi$ is the relative angle `rightness variable' right corresponding to each clustering.
Triangles with $\Phi$ = $\pi/2$ and $3\pi/2$ are (a)-isosceles, which is as (a)-right as possible, while 0 and $\pi$ are collinear -- which is as unright as possible.  
Thus, dra$_x$ and dra$_y$ provide mixed ratio and relative angle information. 
The ratio information for both of these of these is a $2\mn_1\mn_2 = \sqrt{1 - ellip^2}$ factor.  
On the other hand, the relative angle information is in the cos$\,\Phi$ and sin$\,\Phi$ factors.
The relation cos$\,\Phi$ = $ellip$/$\sqrt{1 - \{4 \times area\}^2}$ is useful in the below analysis.

One can view $dra_x$ as a measure of `anisoscelesness' [i.e. departure from (a)'s notion of isoscelesness, c.f. anisotropy as a departure from isotropy in GR Cosmology), 
so I denote it by $aniso$ [with an (a)-label].
It is specifically a measure of anisoscelesness in that aniso(a) $\times I$ per unit base length in mass-weighted space is the $\ml_1 - \ml_2$ indicated in Fig 3. 
I.e., it is the amount by which the perpendicular to the base fails to bisect it (which it would do if the triangle were isosceles).

{            \begin{figure}[ht]
\centering
\includegraphics[width=0.23\textwidth]{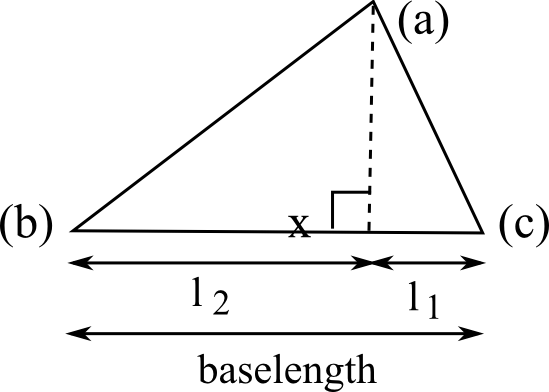}
\caption[Text der im Bilderverzeichnis auftaucht]{        \footnotesize{ Definitions of $l_1$ and $l_2$ in interpretation of this 
article's notion of anisoscelesness.]                          }        }
\label{Fig-3tri-2}\end{figure}            }

\noindent One can likewise view dra$^{(\sa)}_y$ as a measure of noncollinearity.  
%
%
Moreover this is actually clustering-independent/a democracy invariant \cite{Zick, ACG86, LR95}. 
It is furthermore equal to $4 \mbox{ } \times$ $area$ (the area of the triangle per unit $I$ in mass-weighted space).  
In contrast, pure-shape 4-stop metroland has no democratic invariant.

\beq
\mbox{$physical area$} = \mI\sqrt{\frac{m_1 + m_2 + m_3}{m_1m_2m_3}}\frac{\mbox{4 $\times$ area}}{4}  = \frac{\mI\sqrt{3}}{m} area\mbox{ }  
\label{area}
\eeq
(with the second equality holding for the equal-mass case) and collinearity retains its precise meaning (and area-minimizing property) in the arbitrary-mass case.  
However, the meanings of some of the quantities in this Subsection change.  
Equilateral becomes, in all clusterings, to equal-magnitude mass-weighted Jacobi vectors (i.e. 
equal moments of inertia) which are perpendicular to each other; it remains area-maximizing.  
(a)-isosceles retains this perpendicularity with respect to the (a) clustering, so that anisoscelesness generalizes to unrightness in this sense.   
The definition of ellipticity is unchanged since that already involved moments of inertia.

Finally, $Ellip$(a) $:= \mI$\, $ellip$(a), $Aniso$(a) $:= \mI$\, $aniso$(a), and $Area$ $:= \mI$\, $area$.

\subsubsection{Democratic invariants}

Moving between different clusterings involves linear transformations $u^{\Gamma} = 
{D^{\Gamma}}_{\Lambda}u^{\Lambda}$, termed `democracy transformations' in e.g. \cite{LR95}.  
The present Article's notion of `clustering invariant' thus coincides with the Theoretical Molecular Physics literature's notion of `democracy invariant'.  

\noindent Note: these are $d$-independent transformations but the number and nature of democracy invariants is dimension-dependent.  

\noindent The $area$ Dragt coordinate is a democratic invariant demo(3) and is useable as a measure of uniformity \cite{QShape}, 
its modulus running from maximal value at the most uniform configuration (the equilateral triangle) to minimal value for the collinear configurations.

\noindent Note that this is new to triangleland: $N$-stop metroland had no shape variables of this nature (the configuration space radius is always of this nature.)  
See \cite{QShape} for derivation of the demo(4) counterpart of this for quadrilateralland.

\subsubsection{Parabolic-type coordinates for triangleland}

One sometimes also swaps the dra$_{2}$ for the scale variable $\mI$ in the non-normalized version 
of the coordinates to obtain the \{$\mI$, $aniso$, $ellip$\} system (see the next SSec for further interpretation).  
Then a simple linear recombination of this is \{$\mI_1$, $\mI_2$, aniso\}, i.e. the 
two partial moments of inertia and the dot product of the two Jacobi vectors.  
This is in turn closely related \cite{08I} to the parabolic coordinates on the flat $\mathbb{R}^3$ conformal to the triangleland relational space, which are $\{\mI_1, \mI_2, \Phi\}$.  
The arc element is then 
\beq
\d s^2 = \frac{\d \mI_1^2}{4\mI_1} + \frac{\d \mI_2^2}{4\mI_2} + \frac{\mI_1\mI_2}{\mI}\d\Phi^2 = \frac{1}{\mI}
\left\{
\frac{\mI}{4}
\left\{ 
\frac{\d \mI_1^2}{\mI_1} + \frac{\d \mI_2^2}{\mI_2}
\right\} 
+ \mI_1\mI_2\d\Phi^2 
\right\} 
\mbox{ } .  
\eeq
i.e. conformal to the (confocal) parabolic coordinates.  
This is a useful identification to make as regards solving and understanding the classical dynamics and quantum mechanics of the system.  

\noindent N.B. the existence of coordinate systems of this kind is irrespective of whether one is making the plain or O choice.  
I note also the following inversions from the definitions of the Dragt coordinates,
\beq
Flat  := \mI_1 = \{\mI + Dra_z\}/2 = \{\mI + Ellip\}/2 = \mI\{1 + \mbox{cos}\,\Theta\}/2
\mbox{ } , \mbox{ } \mbox{ } 
Sharp := \mI_2 = \{\mI - Dra_z\}/2 = \{\mI - Ellip\}/2 = \mI\{1 - \mbox{cos}\,\Theta\}/2 \mbox{ } . 
\eeq
These invert to 
\beq
\mI = \mI_1 + \mI_2 \mbox{ } , \mbox{ } \Theta = \mbox{arccos}
\left(
\{\mI_2 - \mI_1\}/\{\mI_1 + \mI_2\}   
\right) \mbox{ } .
\eeq
A physical interpretation for these is that $\mI_1$, $\mI_2$ are the partial moments of inertia of the base and the median, with $\Phi$ the `Swiss army knife' angle between these. 
They are clearly a sort of subsystem-split coordinates and thus useful in applications concerning subsystems/perspectival ideas (see Sec \ref{Cl-Str}).  

{\begin{figure}[ht]
\centering
\includegraphics[width=0.8\textwidth]{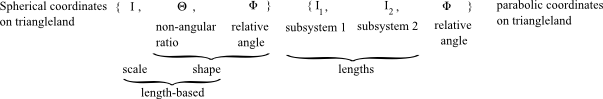}
\caption[Text der im Bilderverzeichnis auftaucht]{\footnotesize{The shape, scale, length ratio, relative 
angle and subsystem statuses of the spherical and parabolic-type coordinate systems.}} 
\label{Fig-2}\end{figure} } 

\noindent For triangleland, spherical, Dragt and parabolic coordinates are keys 4, 5 and 6 to unlocking the mathematics of the model.

\subsubsection{Useful coordinate systems for quadrilateralland. I. Shape and Kuiper coordinates} 

This is a first departure from standard kinematics, which would here, rather follow its 3-$d$ prejudice and consider 
tetrahaedronland.  
Nevertheless, quadrilateralland possesses $\mathbb{CP}^2$ mathematics and coordinate systems for this have been 
worked out in detail for a number of other applications of this space to  Theoretical Physics.

\noindent The Dragt correspondence does not straightforwardly carry though to larger 2- and 3-$d$ models; 
the situation in my 2-$d$ case of interest is rather different from that in the more usually studied 3-$d$ case.  
Even for 4 particles, there are multiple ways of building analogues of the Dragt coordinates in 3-$d$. 
For 2-$d$, see \cite{QShape}.

Further useful coordinate systems for quadrilateralland are as follows (see also Sec \ref{GiPo}). 

\mbox{ }   

\noindent 1) \{$s^{\sfA}$\} are a redundant set of six shape coordinates (see \cite{QShape} for their explicit forms), 
which, according to the construction in \cite{LR95, QShape}, are a quadrilateralland analogue of triangleland's Dragt coordinates.   

\mbox{ }

\noindent 2) The coordinate system \{$\mI$, $s^{\sfa}$\} turns out to be more useful in this case \cite{QShape}.  
These are the counterpart of triangleland's \{I, $aniso$, $ellip$\} albeit they are now redundant.  

\mbox{ } 

\noindent 3) Kuiper's coordinates are then a simple linear combination of 2) (mixing the $\mI$, $s_1$ and $s_5$ coordinates).   
These consist of all the possible inner products between pairs of Jacobi vectors, i.e. 3 magnitudes of Jacobi vectors per unit MOI 
$\mN^e$, alongside 3  $\underline{\mn}_f\cdot\underline{\mn}_g$ ($e \neq f$), which are very closely related to the three relative angles.  
As such, they are, firstly, very much an extension of the parabolic coordinates for the conformally-related flat $\mathbb{R}^3$ of triangleland \cite{08I}.  
Secondly, in the quadrilateralland setting, they are a clean split into 3 pure relative angles (of which any 2 are independent 
and interpretable as the anioscelesnesses of the coarse-graining triangles) and 3 magnitudes (supporting 2 independent non-angular ratios).
I therefore denote this coordinate system by \{$\mN^e$, $aniso$($e$)\}.  
Thirdly, whilst they clearly contain 2 redundancies, they are fully democratic in relation to the constituent 
Jacobi vectors and coarse-graining triangles made from pairs of them. 
This coordinate system is a close analogue of the parabolic coordinates for triangleland (c.f. Fig \ref{Fig-2} and Fig \ref{Fig-2GP}), albeit now redundant.  
The very closest parabolic coordinate analogue involves the three partial moments of inertia alongside the three relative angles themselves.  

\noindent Note: the above can be viewed as quadrilateralland counterparts of the Dragt coordinates, the \{I, $aniso$, $ellip$\} 
coordinate system and the parabolic coordinate system (i.e. key steps 4 and 5).  
%
{            \begin{figure}[ht]
\centering
\includegraphics[width=0.95  \textwidth]{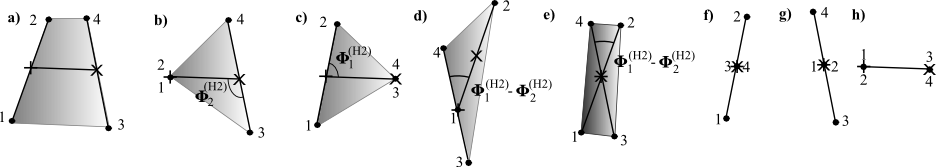}
\caption[Text der im Bilderverzeichnis auftaucht]{        \footnotesize{Figure of the coarse-graining 
triangles.  
a) For Jacobi H-coordinates coordinates for a given quadrilateral, collapsing each of $\rho_1^{(\tH 2)}$, 
$\rho_2^{(\tH 2)}$ and $\rho_3^{(\tH 2)}$ in turn gives the coarse-graining triangles b), c) and d) \cite{QShape}.  
[One can use d) to interpret the associated anisoscelesness, but the true physical situation is that of e) i.e. a rhombus.]
Collapsing pairs of these, one obtains f), g) and h).}         }
\label{CTQH}\end{figure}          }
%
{            \begin{figure}[ht]
\centering
\includegraphics[width=0.8  \textwidth]{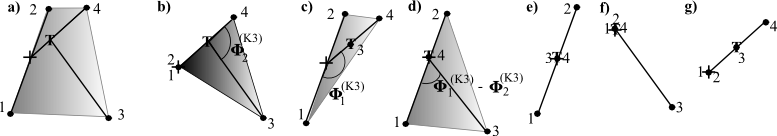}
\caption[Text der im Bilderverzeichnis auftaucht]{        \footnotesize{Figure of the coarse-graining 
triangles in Jacobi K-coordinates a): 
collapsing each of $\rho_1^{(\tK 3)}$, $\rho_2^{(\tK 3)}$ and $\rho_3^{(\tK 3)}$ in turn gives 
the coarse-graining triangles b), c) and d).  Collapsing pairs of these at once gives e), f) and g).  }         }
\label{FigX2}\end{figure}          }

\noindent In Kuiper coordinates, each of the three on-$\mathbb{S}^2$ conditions for whichever of H or K coordinates involves losing one magnitude and two inner products.  
Thus the survivors are two magnitudes and one inner product.

\subsubsection{The split of quadrilateralland into hemi-$\mathbb{CP}^2$'s of oriented quadrilaterals}\label{Kuiper}

{\bf Definition}: The {\it Veronese surface} Ver \cite{Harris} is the space of conics through a point (parallel to how 
a projective space is a set of lines through a point).  
It is the embedding $\mathbb{CP}^2 \longrightarrow \mathbb{CP}^5$ Ver: [$x$, $y$, $z$] $\longrightarrow$ [$x^2$, $y^2$, $z^2$, 
$yz$, $xz$, $xy$] where $x$, $y$, $z$ are homogeneous coordinates \cite{Harris} (and can be taken to formalize the statement that 5 points determine a conic). 

\mbox{ }

\noindent {\bf Kuiper's Theorem} i) The map  

\beq
\eta: \stackrel{\mbox{$\mathbb{CP}^2$}}{ (z_1, z_2, z_3) }                  
\mbox{ } \mbox{ } \mbox{ }
      \stackrel{\mbox{$\longrightarrow$}}{   \stackrel{}{\mbox{$\longmapsto$}}   }  
\mbox{ } \mbox{ } \mbox{ }
      \stackrel{\mbox{$\mathbb{E}^5$}}{(|z_1|^2, |z_2|^2,|z_3|^2, 
\{z_2\bar{z}_3 + z_3\bar{z}_2\}/2, \{z_3\bar{z}_1 + z_1\bar{z}_3\}/2, \{z_1\bar{z}_2 + z_2\bar{z}_1\}/2)}
\eeq  
induces a piecewise smooth embedding of $\mathbb{CP}^2/\mathbb{Z}_2^{\scc\so\sn\sj}$  onto the 
boundary of the convex hull of Ver in $\mathbb{E}^5$, which moreover has the right properties to be the usual smooth 4-sphere \cite{Kuiper}. 

\mbox{ } 

\noindent Restricting $\{{z}_1, {z}_2, {z}_3\}$ to the real line corresponds in the quadrilateralland interpretation 
to considering the collinear configurations, which constitute a $\mathbb{RP}^2$ space as per Fig \ref{Top-S}.    
Moreover the above embedding sends this onto Ver itself.  
Proving this proceeds via establishing that, as well as the on-$\mathbb{S}^5$ condition $\mN_1 + \mN_2 + \mN_3 = 1$, a second restriction holds, 
which in the quadrilateralland interpretation, reads $\sum_e aniso(e)^2\mN^e - 4\mN_1\mN_2\mN_3 + aniso(1)\,aniso(2)\, aniso(3) = 0$. 
(Knowledge of this restriction should also be useful in kinematical quantization \cite{I84}, and it is clearer in the \{$\mN^e$, $aniso$($e$)\} 
system, which is both the quadrilateralland interpretation of Kuiper's redundant coordinates and a simple linear recombination of the  

\noindent \{$\mI$, $s^{\gamma}$\} coordinates obtained in \cite{QShape}, than in these other coordinates themselves.)  

\mbox{ }

\noindent {\bf Another form for Kuiper's Theorem} \cite{Kuiper}.
Moreover, $\mathbb{CP}^2$ itself is topologically a double covering of $\mathbb{S}^4$ branched along the $\mathbb{RP}^2$ 
of collinearities which itself embeds onto $\mathbb{E}^5$ to give Ver. 

\mbox{ } 

\noindent Here, branching is meant in the sense familiar from the theory of Riemann surfaces \cite{Riemann}. 
Moreover,  the $\mathbb{RP}^2$ itself embeds {\sl non-smoothly} into Ver.

\mbox{ }

\noindent
Then the quadrilateralland interpretation of these results is in direct analogy with the plain shapes case of triangleland consisting of two hemispheres of opposite orientation bounded 
by an equator circle of collinearity, the mirror-image-identified case then consisting of one half plus this collinear edge.  
Thus plain quadrilateralland's distinction between clockwise- and anticlockwise-oriented figures is strongly anchored 
to this geometrical split, with the collinear configurations lying at the boundary of this split.

\subsubsection{Useful coordinate systems for quadrilateralland. II. Gibbons--Pope type coordinates.}\label{GiPo}

\noindent Useful intrinsic coordinates (which extend the spherical coordinates on the triangleland and 4-stop metroland shape spheres) are the Gibbons--Pope type coordinates 
\{$\chi$, $\beta$, $\phi$, $\psi$\} \cite{GiPo, Pope} are also useful for the study of quadrilateralland.  
These are, in some senses, a generalization of triangleland's $\Theta$ and $\Phi$. 
Their ranges are $0 \leq \chi \leq \pi/2$, 
                 $0 \leq \beta \leq \pi$, 
                 $0 \leq \phi \leq 2\pi$ (a reasonable range redefinition since it is the third relative angle), and
                 $0 \leq \psi \leq 4\pi$.
These are related to the bipolar form of the Fubini--Study coordinates by
\beq
\psi^{\prime} = - \{\Phi_1 + \Phi_2\} \mbox{ } , \mbox{ } \mbox{ } \phi^{\prime} = \Phi_2 -\Phi_1, 
\eeq
with then $\psi = - \psi^{\prime}$ (measured in the opposite direction to match Gibbons--Pope's convention) 
and $\phi$ is taken to cover the coordinate range $0$ to $2\pi$, which is comeasurate with it itself being the third relative angle between the Jacobi vectors involved.
Also,
\beq
\beta = 2\,\mbox{arctan}\,({\cal R}_2/{\cal R}_1)  \mbox{ } , \mbox{ } \mbox{ } \chi = \mbox{arctan}\big(\sqrt{{\cal R}_1\mbox{}^2 + 
{\cal R}_2\mbox{}^2}\big) \mbox{ } .  
\eeq
However now in each of the conventions I use for H and K coordinates, a different interpretation is to be attached to these last two formulae in terms of the $\rho^e$.
For H-coordinates in my convention, 
\beq
\beta = 2\,\mbox{arctan}\,(\rho_2/\rho_1)  \mbox{ } , \mbox{ } \mbox{ } \chi = \mbox{arctan}\big(\sqrt{\rho_1\mbox{}^2 + \rho_2\mbox{}^2}/\rho_3\big) \mbox{ } ,
\eeq
whilst for K-coordinates in my convention, 
\beq
\beta = 2\,\mbox{arctan}\,(\rho_1/\rho_3)  \mbox{ } , \mbox{ } \mbox{ } \chi = \mbox{arctan}\big(\sqrt{\rho_1\mbox{}^2 + \rho_3\mbox{}^2}/\rho_2\big) \mbox{ } .
\eeq
Note 1) By their ranges, $\beta$ and $\phi$ parallel azimuthal and polar coordinates on the sphere.
[In fact, $\beta$, $\phi$ and $\psi$ take the form of Euler angles on $SU(2)$, with the remaining coordinate $\chi$ playing the role of a compactified radius. 
Thus Gibbons--Pope coordinates are a particular example of $SU(2)$-adapted coordinates, for the $SU(2)$ in question 
being a subgroup of the natural $SU(3)$, by which such coordinates are useful for a number of applications. 

\noindent Note 2) In the quadrilateralland interpretation, $\beta$ has the same mathematical form as triangleland's azimuthal coordinate $\Theta$.
Additionally, $\chi$ parallels 4-stop metroland's azimuthal coordinate $\theta$, except that it is over half of the range of that, reflecting 
that the collinear 1234 and 4321 orientations have to be the same due to the existence of the second dimension via which one is rotateable into  the other.  

\noindent Note 3) These coordinates represent the simplest available choice of block structure (a weakening of diagonalization, which is 
not here possible: a 2 by 2 block and two diagonal entries), with the Gibbons--Pope choice of radii further simplifying the formulae within each of these blocks. 
In this way, Gibbons--Pope coordinates are analogous to spherical coordinates, 
Fubini--Study coordinates themselves being more like stereographic coordinates (projective, not maximally block-simplified).  

{\begin{figure}[ht]
\centering
\includegraphics[width=0.99\textwidth]{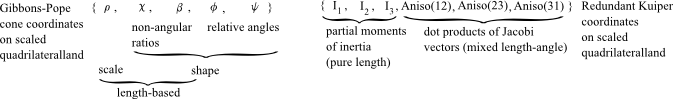}
\caption[Text der im Bilderverzeichnis auftaucht]{\footnotesize{The shape, scale, length ratio, relative 
angle and subsystem statuses of the Gibbons--Pope type and Kuiper coordinate systems.}} 
\label{Fig-2GP}\end{figure} } 
 
The Gibbons--Pope type coordinates have the following quadrilateralland interpretations.  
In an H-clustering, the $\phi$ is the difference of the relative angles [see Fig \ref{Fig1}e)], 
so the associated momentum represents a counter-rotation of the two constituent subsystems ($\times$ relative to \{12\} and + relative to \{34\}).       
The $\psi$ is minus the sum of the relative angles, so the associated momentum represents a co-rotation of these two 
constituent subsystems (with counter-rotation in $\times$ relative to $+$ so as to preserve the overall zero angular momentum condition).  
The $\beta$ is a measure of contents inhomogeneity of the model universe: the ratio of the sizes of the 2 constituent subclusters.  
Finally, the $\chi$ is a measure of the selected subsystems' (`posts') sizes relative to that of the whole-universe model (`crossbar').  
These last two are conjugate to quantities that involve relative distance momenta in addition to relative angular momenta.

On the other hand, in a K-clustering, the $\phi$ is the difference of the two relative angles [see Fig 2f)], but by the 
directions these are measured in, the associated momentum now represents a co-rotation of the two constituent subsystems 
(which are now \{12\} relative to 4 and \{+4\} relative to 3, and with  counter-rotation in T relative to $+$ so as to preserve the overall zero angular momentum condition).   
The $\psi$ is minus the sum of the two relative angles, so the associated momentum represents a counter-rotation of these two constituent subsystems. 
The $\beta$ is now a comparer between the sizes of the \{12\} subcluster and the separation between the non-triple cluster particle 3 and T.
Finally, the $\chi$ is a comparer between the sizes of the above two contents of the universe (\{12\} (blade width) and \{T3\}) (handle length) 
on the one hand, and the separation between them on the other hand (\{4+\} (blade length). 
These last two are again conjugate to quantities that involve relative distance momenta in addition to relative angular momenta.

\noindent Note how in both H and K cases, the Gibbons--Pope type coordinates under the quadrilateralland interpretation involve a split 
into two pure relative angles and two pure non-angular ratios of magnitudes.

In Gibbons--Pope type coordinates, the Fubini--Study metric then takes the form  
\beq
\d \tts^2 = \d\chi^2 + 
\mbox{sin}^2\chi\big\{\d \beta^2 + \mbox{cos}^2\chi\{\d\phi^2 + \d\psi^2 + 
2\,\mbox{cos}\,\beta\, \d\phi\,\d\psi\} + \mbox{sin}^2\chi\,\mbox{sin}^2\beta \,\d\phi^2 
\big\}/4 \mbox{ } .
\label{GP-FS}
\eeq

\subsubsection{$N$-a-gonland generalization of the quadrilateralland coordinates used so far}\label{Quier}

\noindent Extending the Dragt/parabolic/shape/Kuiper type of redundant coordinates is itself straightforward, though it is not clear the extent 
to which the resulting coordinates will retain usefulness for the study of each $\mathbb{CP}^{N - 2}$.  
Certainly the number of Kuiper-type coordinates (based on inner products of pairs of Jacobi vectors, 
of which there are $n\{n - 1\}/2$) further grows away from 2$\{n - 1\}$ = dim($\mathbb{CP}^{n - 1}$) as $N$ gets larger.
The present Article also finds a surrounding space of just {\sl one} dimension more for $\mathbb{CP}^2$.  
The $N$-a-gonland significance of two half-spaces of different orientation separated by an orientationless manifold of 
collinearities gives reason for double covers to the $N - 2 \geq 3$ $\mathbb{CP}^{N - 2}$ spaces to exist for all $N$.  
However, there is no known guarantee that these will involve geometrical entities as simple as or tractable as 
quadrilateralland's $\mathbb{S}^4$ for the half-spaces, or of the Veronese surface Ver as the place of branching.  
However, one does have the simple argument of Sec \ref{Quier} that the manifold of collinearities within $N$-a-gonland's 
$\mathbb{CP}^{N - 2}$ is $\mathbb{RP}^{N - 2}$, so at least that is a known and geometrically-simple result 
for the structure of the general $N$-a-gonland.
Whether the intrinsic Gibbons--Pope type coordinates can be extended to $N$-a-gonland in a way that maintains their 
usefulness in characterizing conserved quantities \cite{QuadII} and via separating the free-potential TISE, remains to be seen.

\subsection{Tessellation by the physical/shape-theoretical interpretation}\label{Tessi}

The configuration space splits up into a number of physically/shape-theoretically significant regions. 
I restrict myself to the equal-mass case in this Article\footnote{This applies in mass-weighted space. 
This is a further stipulation since equal masses does not imply equal Jacobi masses (c.f. \ref{Rel-Jac}), so if one 
unweights the mass-weighted relative Jacobi coordinates, there is somewhat of a distortion of the tessellation.  
This distortion does preserve the perpendicularity of the axes, but stretches each by a different amount 
so that angles interior to each quadrant/octant are indeed distorted (e.g. the 120 degree angle in the next SSSec changes by a few percent under this distortion). 
However, I principally work with mass-weighted variables in this Article (so my notion of shape is, strictly, of mass-weighted shape). 
Then the below tessellations are exact.}  
and this case definitely confers a large amount of symmetry in the allocation of physical significance to these regions, by splitting it up into a number of {\sl equal} regions.
Such a split is known as a {\bf tessellation} \cite{Magnus}, i.e. a tiling of the configuration space with equal tiles that completely cover it.
Tessellations are most well-known for flat space, but have also been studied for e.g. the sphere (see e.g. Magnus' book \cite{Magnus}),  
which is significant to us as the shape space of both 4-stop metroland and triangleland.  
This method is indeed particularly useful for low configuration space dimension \cite{AF, +Tri, Cones, ScaleQM, 08III}, for which pictorial 
representation of the whole configuration space is straightforward.   
Faces, edges and vertices therein are physically significant (and not in all cases physically equivalent), so that one really has a {\bf labelled tessellation}.

One is then to interpret classical trajectories as paths upon these, and classical potentials and 
quantum-mechanical probability density functions as height functions over these.
Thus this SSec's figures will often feature as an `interpretational back-cloth' in subsequent Secs.  
This parallels Kendall's `spherical blackboard' technique for the visualization of shape statistics of triangles in e.g. \cite{Kendall89}; see also Fig \ref{Blackboard}.    
See \cite{Tessi} for a distinct application of tessellations in Theoretical Molecular Physics.  

\mbox{ }  

\noindent Note 1) that some such features were already meaningful at the topological level, and so have already been covered in Secs \ref{Top}, \ref{Top2} and \ref{Top3}.  
This SSec considers a number of further metrically-significant notions, as well as providing an overall picture of the physical/shape-theoretical interpretation.  

\noindent Note 2) At the topological level, one does have a notion of coincidence and hence of collision. 
However, further concepts such as equilateral, isosceles and the various mergers of subsystems require additional metric structure to be in place.   

\noindent Note 3) The tessellations in question are underlied by the $S_{N} \times \mathbb{Z}_2$ symmetry group of particle label 
permutations and orientation allocation (of order $2N!$). 
There is still an issue of how these act geometrically, which I treat case by case below.

\subsubsection{3-stop metroland tessellation}\label{Tessi3Stop}

In the plain case \cite{06I}, the shape space for this is the circle with 6 regularly-spaced double collision (D) points on it. 
[C.f. the rim of Fig \ref{Top-S}a) but now these are metrically-meaningfully symmetrically placed].  
One can furthermore fine-tile with merger points (M).\footnote{C.f. the use of the terms `fine graining' and `coarse graining' in Statistical Mechanics (SM) and Histories Theory 
(Sec \ref{Cl-Hist}, \ref{QM-Hist}).  
I likewise call the opposite notions of these `coarse-tiling' and `coarse-labelling'.}

Here, the $S_3 \times \mathbb{Z}_2$ symmetry group acts as\footnote{ 
$\mathbb{D}_p$ is the dihaedral group of order 2$p$.}
$\mathbb{D}_6 \mbox{ } \widetilde{=} \mbox{ } \mathbb{D}_3 \times \mathbb{Z}_2$. 
[Ignoring the labelling, the `clock-face' fine-tiling has the larger symmetry group $\mathbb{D}_{12}$, but this is broken down to 
the preceding group by the M and D labels (which are indeed physically distinct as well as labelled distinctly).]

Next, by forming diametrically-opposite pairs, the 6 D's pick out 3 preferred D-axes, and the 6 M's likewise pick out 3 further preferred M-axes, each perpendicular to a D-axis.      
Each orthogonal (D, M) axis system corresponds to one of the 3 permutations of Jacobi coordinates.  
The polar angle $\varphi^{(\sa)}$ about each of these axes is then natural for the study of the clustering corresponding to that choice of Jacobi coordinates, \{a, bc\}.

The O-counterpart [Fig \ref{Fig-3-3}b)] has 3 D's and 3 M's and the symmetry group $\mathbb{D}_3 \mbox{ } \widetilde{=} \mbox{ } S_3$. 
This can also be represented using double angles as the rim of a whole pie [Fig \ref{Fig-3-3}c)].    
Now each D is opposite to an M, so these D--M pairs pick out 3 preferred axes.  
Forming a Cartesian system now requires a second axis that has no particular extra physical significance; 
I thus term the points on this second axis {\it spurious}, and denote them with an S [Fig \ref{Fig-3-3}d)).

\mbox{ }  

{            \begin{figure}[ht]
\centering
\includegraphics[width=0.82\textwidth]{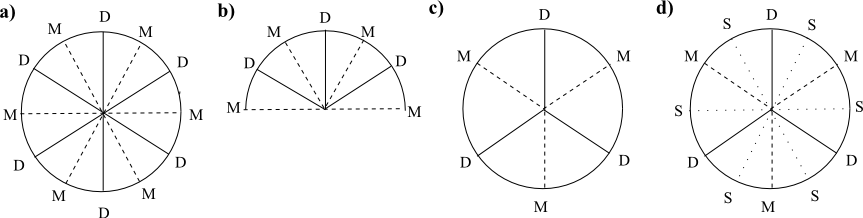}
\caption[Text der im Bilderverzeichnis auftaucht]{        \footnotesize{A sketch of the tessellation of 
3-stop metroland's relational spaces (the corresponding shape spaces are the rims of these).}        }
\label{Cones-Fig2}    
\end{figure}  }

\noindent The relational spaces for the scaled 3-stop metroland theories are then the cones over these decorated shape spaces.  
I.e., they are (the infinite extension of) 1) Fig\ref{Cones-Fig2}a)'s pie of 12 slices with 6 D half-lines and 6 M half-lines in the plain case. 
2) likewise Fig \ref{Cones-Fig2}b)'s half-pie of 6 slices [or Fig \ref{Cones-Fig2}c)'s pie of 3 slices] with 3 D half-lines and 3 M half-lines in the O case.   
All of these half-lines emanate from the triple collision at the cone point, 0.

\subsubsection{4-stop metroland tessellation}\label{Tessi4Stop}

There are 8 triple collision (T) points and 6 double-double (DD) collision points.   
Each DD is attached to 4 T's, and each T to 3 T's and 3 DD's.
This forms a tessellation with 24 identical spherical isosceles triangle faces, 36 single double collision (D) edges and 14 vertices.

{            \begin{figure}[ht]
\centering
\includegraphics[width=0.3\textwidth]{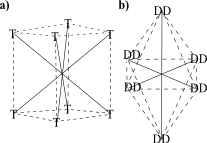}
\caption[Text der im Bilderverzeichnis auftaucht]{        \footnotesize{
a) The octahaedron of DD-vertices and the non-perpendicular diagonals between them.  
b) The cube of T-vertices and the set of perpendicular axes between them.                  }        }
\label{Fig2AFa}\end{figure}            }

The T's and DD's form respectively the vertices of a cube [Fig \ref{Fig2AFa}a)] and the octahaedron dual to it [Fig \ref{Fig2AFa}b)]. 
Thus the $S_4 \times \mathbb{Z}_2$ group acts as the symmetry group of the cube/octahaedron. 
[This order-48 group is indeed isomorphic to this as is clear by the action of the permutations on the four diagonals between opposite vertices.]\footnote{Cubes 
and octahaedra are also involved in quadrilateralland, e.g. in the form of the collision set, c.f. Fig \ref{Fig-3-3}d).}  
%
See p 72-75 of \cite{Magnus} for further mathematical discussion of this tessellation and SSSec \ref{Sing-Pot}, \cite{Tessi} for other occurrences of it in mechanics.
In this case the labelling does not break the symmetry group of the unlabelled tessellation.

The T's and DD's form 7 antipodal pairs, thus picking out 7 preferred axes.  
The 3 axes corresponding to antipodal DD pairs are related to the 3 permutations of Jacobi H-coordinates, which are thus adapted to `seeing' DD collisions. 
Likewise, the 4 axes corresponding to antipodal T  pairs are related to the 4 permutations of Jacobi K-coordinates.
Being perpendicular, (DD, DD, DD) is a suitable Cartesian axis system. 
Extending each T to such an axis system requires more work below.

Then the spherical polar coordinates about each of these axes -- a \big($\phi_{(\sH\sb)}$, 
$\theta_{(\sH\sb)}$\big) or a \big($\phi_{(\sK\sa)}$, $\theta_{(\sK\sa)}$\big) coordinate system -- are natural for the study of the corresponding H or K structure.  
I.e. each choice of H or K has a different natural spherical polar coordinate chart.  
Any two of these natural charts suffice to form an atlas for the sphere (each goes bad solely at its poles, where its axial angle ceases to be defined).  
To look very close to a pole, one can `cartesianize' e.g. after projecting the relevant hemisphere onto the equatorial disc.

Note the inclusion of six copies of 3-stop metroland (with each's 6 D's labelled as DD,T,T,DD,T,T).  
These correspond to the 6 ways of fusing 2 of the 4 particles to produce a 3-particle problem.  
In the O case (Fig \ref{Fig2AFa} d), one has this picture with antipodal identification. 
Or, correspondingly, the (without loss of generality) Northern hemisphere portion (so the numbers of 
faces and vertices down by a factor of 2, and the corresponding group is now $S_4$).
(O)relational spaces are then the solid cones made from each corresponding shape space decorated by its tessellation.  
In making this a cone, points go to radial half-lines, arcs go to 2-$d$ sectors and faces go to 3-$d$ sectors of solid angle. 
All emanate from the maximal (quadruple) collision at the cone point, 0.  

{            \begin{figure}[ht]
\centering
\includegraphics[width=0.25\textwidth]{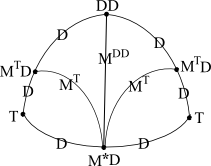}
\caption[Text der im Bilderverzeichnis auftaucht]{        \footnotesize{There are M-points for 4-stop metroland also. 
This figure illustrates the notions of merger within a single triangular face.
 }        }
\label{41Merger}
\end{figure}            }

\noindent{\bf Fine-tessellation by notions of merger}.   
Since this further complicates the picture, I present it at the end as a fine-tiling and fine-labelling 
of one of the preceding diagram's tiles (Fig \ref{41Merger}).  
For 4-stop metroland, there are 2 kinds of great circles of mergers, and these furthermore contain 2 kinds of special points that are both a merger and a double collision. 
In fact, there are 2 kinds of merger: $\nM^{\sT}$ with one particle at the centre of mass of the other 3 
and $\nM^{\sD\sD}$ with the centres of mass of two pairs of particles coinciding.  
I denote the special points on these lines by appending a D; the diagram is consistent since 
$\nM^{\sD\sD}\mD$ can also be viewed as two different $\nM^{\sT}\mD$'s, so I give these triple confluence points a new notation $\mM^*\mD$. 
Then there are in total 50 vertices (now also 12 $\nM^{\sD\sD}\mD$ as these subdivide each edge of the cube and 24 $\nM^{\sT}\mD$ as there are 4 per face of the cube) and 144 edges. 
(Each old D is cut into 2 new D's and there are also 48 $\nM^{\sT}$ edges -- 8 per face of the cube -- and 24 $\nM^{\sD\sD}$ edges -- 4 per face of the cube).  
It has 96 faces (by each old tile being subdivided into 4).  
These tiles are not all labelled the same, however.  
The fundamental unit is one pair of such tiles, i.e. a T-DD-M$^*$D triangle. 
There are then 48 of these, and considering them amounts to not using the $\nM^{\sT}$D as vertices or the M$^{\sT}$ as edges, in which case there remain 26 vertices and 72 edges.  
These are also the types of merger present in quadrilateralland.
As can be seen from how the new vertices democratically decorate the cube, the above two fine-tilings with labellings continue to respect the symmetry group of the cube.   
These merger points define a number of further axes, allowing for the T-axes to be extended to form a Cartesian system of the form (T, $\mM^*\mD$, $\mM^*\mD$).

\subsubsection{Triangleland tessellation: Kendall's Spherical Blackboard}\label{TessiTri}
%
{\begin{figure}[ht]
\centering
\includegraphics[width=0.9\textwidth]{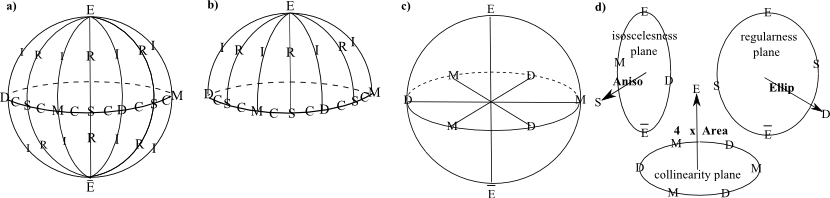}
\caption[Text der im Bilderverzeichnis auftaucht]{\footnotesize{ a) Tessellation of triangleland.  
\noindent b) Tessellation of Otriangleland.
\noindent c) Axes through the equilateral triangle poles (E, $\bar{\mE}$) and through pairings of double 
collisions (D) and mergers (M). 
\noindent d) The relation between triangleland's basis of Cartesian vectors and the planes perpendicular 
to each, clearly illustrating that 4 $\times$ $area$ is a departure from the plane of collinearity, 
that anisoscelesness is indeed a departure from the plane of isoscelesness, and that ellipticity is a departure from the plane of regularity.  } } \label{Fig-Noob}\end{figure}     }  
%
{            \begin{figure}[ht]
\centering
\includegraphics[width=1.0\textwidth]{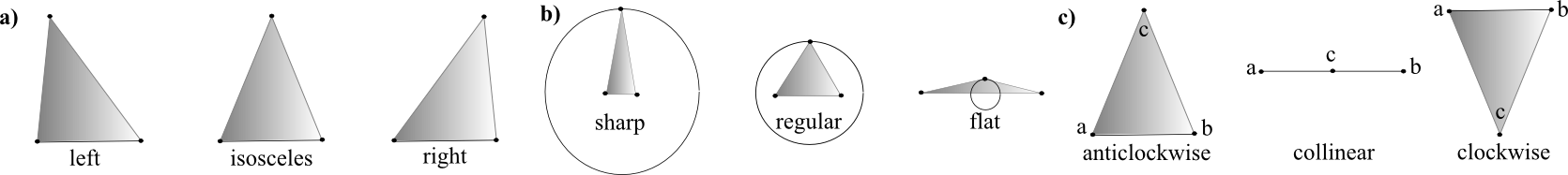}
\caption[Text der im Bilderverzeichnis auftaucht]{        \footnotesize{The most physically meaningful 
great circles on the triangleland shape correspond to the isosceles, regular and collinear triangles.
These respectively divide the shape sphere into hemispheres of right/left triangles, sharp/flat 
triangles, and anticlockwise/clockwise triangles.} }
\label{Fig6} \end{figure}          }
%
Beyond the previously encountered topologically-meaningful points of double collision (D), 
triangleland's shape space's further notions of distance and angle at the level of metric geometry furnish the following [Fig \ref{Fig-3tri-2})]. 

\noindent 1) a notion of collinearity C that comprise the equator (split into 3 equal arcs by the D's, so that these lie at $2\pi/3$ to each other).
N.B. that this equator of collinearity is the same as O3-stop metroland, the rotation in the second dimension forcing the O-choice.
This equator of collinearity is orientationless and separates the 2 hemispheres of distinct orientation.   
These correspond to clockwise- and anticlockwise labelled triangles [Fig \ref{Fig-Noob} a, d)].  

\noindent 2) A notion of equilaterality E (and its labelled-triangle mirror image $\bar{\mE}$.) which occur at the poles.  

\noindent 3) The half-meridians I that join the D's to the E's correspond to the 3 distinct notions of isosceles triangles that a labelled triangle possesses (3 choices of base). 
Each of these is associated with a particular clustering (the one that picks out the particular base with respect to which one is to have mirror symmetry about).
On the other hand, the notions of collinearity and equilaterality are clearly clustering-independent. 
These are the plane of zero area and points of extremal area per given $\mI$, i.e. the E-direction perpendicular to the collinearity plane is proportional to the $area$ vector.

However, it is not just half, but the whole of, the abovementioned great circles that correspond to isosceles triangles. 
There are planes of zero anisoscelesness, the perpendiculars to each of which are proportional to the corresponding $aniso$(a) vector.      
That a) only captured half of each isoscelesness great circle I also attracts attention to the other  
intersection points of this and the collinearity equator C (which are antipodal to the D's).  
These are mergers, M: configurations such that one particle is at the centre of mass of the other 
two.  
I depict this fine-tiling of the preceding paragraph in Fig \ref{Fig-Noob} b, e).
Finally, the meridian of isoscelesness separates hemispheres of $left$(a) [left-slanting of (a)'s notion 
of isoscelesness] and $right$(a) [right-slanting of (a)'s notion of isoscelesness].

The meridians perpendicular to the I's also have lucid physical meaning.
These are the {\it regular configurations}, R (given by $\mI_1^{(\sa)} = \mI_2^{(\sa)}$, i.e. median = 
base in mass-weighted length), of which, indeed, a labelled triangle also possesses three different permutations.  
These are planes of zero ellipticity, the perpendiculars to each of which are proportional to the corresponding $ellip$(a) vector.
They separate two hemispheres of clustering-dependent notions, sharp(a) (median $>$ base, i.e. `sharp' 
triangles) from flat(a) (median $<$ base, i.e. `flat' triangles) [Fig \ref{Fig-Noob} c, f)].
I then know of no further interesting feature possessed by the intersection points of the equator 
of collinearity C and the meridians of regularness R, so, again, I use the letter S (for `spurious') for these.

Thus there are 12 vertices on the equator forming an hour-marked clock-face labelled D,S,M,S,D,S,M,S,D,S,M,S 
(with the hours fully distinguished by further clustering-labels), and there are 14 vertices in total.
There are then 12 edges of collinearity, 12 of isoscelesness and 12 of regularness, and thus 36 edges in total. 
This corresponds to a tessellation by (Fig \ref{Fig-3tri-2}c) 24 equal isosceles spherical triangles 
in the `austroboreal zodiac' shape (the split of the night sky into the 12 signs of the Zodiac and then further partitioned into boreal and austral skies). 
The symmetry group corresponding to this unlabelled tessellation is $\mathbb{D}_{12} \times \mathbb{Z}_2$, of order 48.\footnote{Note that these are all  
 members of the family of dihaedral tessellations studied on p 71-72 of \cite{Magnus}. 
Moreover, $\mathbb{D}_3 \times \mathbb{Z}_2$ is trigonal bipyramidal to the 4-stop metroland's 
$S_4\times \mathbb{Z}_2$ that is tetragonal bipyramidal and regular and hence octahaedral and so dual to cubic.
Also note the contrast between the extra labelling breaking the tessellation group here with how the 
4-stop metroland's fine labelling preserves that case's `octahaedron--cube' symmetry group.}  
%
However, the labels partly break down this group to $\mathbb{D}_3 \times \mathbb{Z}_2$, which is indeed isomorphic to the expected $S_3 \times \mathbb{Z}_2$ of order 12.  
The 48-triangle tessellation's constituent triangles are labelled either D,I,E,R,S,C or M,I,E,R,S,C rather than all being labelled the same.  
Thus the `fundamental region' if some or all labels are included is any tile of the tessellation with 
12 faces, 8 vertices (2 E's, 3 D's and 3 M's) and  18 edges (6 C's, 6 I's and 6 R's). 
%
{            \begin{figure}[ht]
\centering
\includegraphics[width=0.5\textwidth]{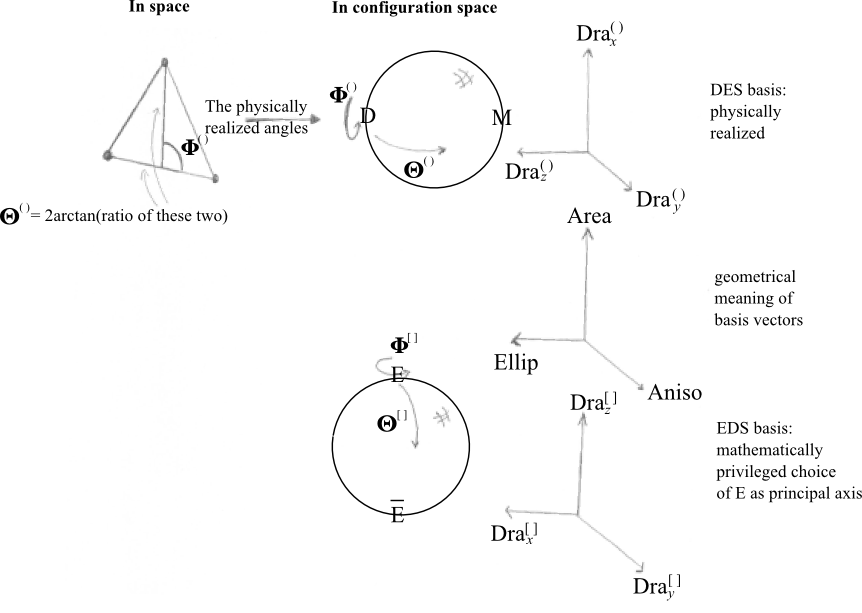}
\caption[Text der im Bilderverzeichnis auftaucht]{        \footnotesize{A useful summary: relation between the physical angles and the spherical coordinates in the DES and EDS bases.} }
\label{DES-and-EDS} \end{figure}          }
%
{            \begin{figure}[ht]
\centering
\includegraphics[width=0.28\textwidth]{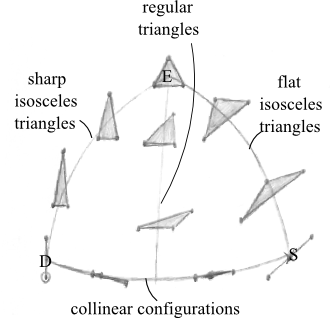}
\caption[Text der im Bilderverzeichnis auftaucht]{        \footnotesize{Kendall's Spherical Blackboard marked with with constituent triangles.   
This is the fundamental region: 1/3 of a hemisphere. It is the space of all unlabelled triangles.  }        }
\label{Blackboard}\end{figure}            }

\noindent Next, note that the D and M vertices pick out 3 particularly distinguished axes at $\pi/3$ to each other in the collinearity plane. 
Each of these corresponds to a different choice of particle permutations in building the 3-particle Jacobi coordinates.  
Each such axis corresponds to an $ellip$(a).   
The axes perpendicular to each of these correspond to the 3 notions of $aniso$(a).  
I denote spherical polar coordinates about each of these as principal axis by \{$\Theta_{(\sa)}$, 
$\Phi_{(\sa)}$\}, where $\Theta_{(\sa)}$ is the azimuthal spherical angle as measured from each D. 
Also, I denote spherical polar coordinates with E as North pole and a as the second axis by $\{I, \Theta_{[\sa]}, \Phi_{[\sa]}\}$.
I.e. a (DM, $\mE\overline{\mE}$, SS) axis system or (D, E, S) for short for a particular choice of orientation for the axes.

One can take E$\bar{\mE}$ as principal axis by interchanging the roles of the second and third coordinate axes.   
I.e. the ($\mE\overline{\mE}$, DM, SS) axis system or likewise (E, D, S) for short.  
I denote the subsequent quantities by square brackets [a] for a the axis system's choice of second axis. 
The principal axis, however, now measures 4 $\times$ $area$ independently of any clustering considerations.  
On the other hand, each $\Phi_{[\sa]}$ is a regularness and isoscelesness quantity.  

The O-counterpart (Fig \ref{Fig-3tri}) is the preceding with antipodal identification. 
Thus it consists of 12 equal isosceles forming the signs of the zodiac but now just for the boreal sky.  
There is now but one equilateral configuration E. 
The other 12 vertices are as for the O case.  
There are then the obvious subset of 24 edges of the preceding, with the same interpretation.  
The symmetry group for the labelled tessellation is then 
The corresponding physical group for this tessellation additionally labelled in this way is $\mathbb{D}_3 \widetilde{=} \mbox{ } S_3$ (order 6). 
This corresponds to 3 labelling permutations but now without additionally ascribing an overall orientation.

Triangleland's relational space is the corresponding cone over the above tessellation-decorated shape space.
E, D, M, S become half-lines, C, I, R become sectors of 2-$d$ angle, and faces become sectors of solid 
angle, all emanating from the triple collision at the cone point, 0.

\subsubsection{Quadrilateralland tessellation with some dimensions suppressed: my $\mathbb{CP}^2$ chopping board}

Beyond the topological distinctions between types of configuration in Sec \ref{Top}, at the metric level, collinearities become meaningful. 
These form $\mathbb{RP}^2$ in the distinguishable particle cases and $\mathbb{RP}^2/A_4$ in the indistinguishable particle cases.
Here is a demonstration that $\mathbb{RP}^{N - 2}$ plays this role within the general $N$-a-gonland $\mathbb{CP}^{N - 2}$ case.
Collinear configurations involve the relative angle coordinates being 0 or multiples of $\pi$, by which the complex projective space definition collapses to the 
real projective space definition (for which the ${\cal R}$'s provide the in-space interpretation of the Beltrami $b$'s). 
Moreover, abcd... can be rotated via the second dimension into ...dcba (and that there 
is no further such identification) and so the real submanifold of collinear configurations is $\mathbb{RP}^{N - 2}$.  
The above spaces of collinearity are in each case like an equator in each case, e.g. separating two  $\mathbb{S}^4$'s in
$\mathbb{E}^5$ in the first case (see below for why the two halves are, topologically, $\mathbb{S}^4$ and for further issues of the geometry involved).

Finally, there are also 6, 3, 2 and 1 distinguishable labelled squares in each case.
This motivates a new action on quadrilateralland, in which e.g. the left-most particle is preserved and the other 3 are permuted or just evenly-permuted.

\noindent Note: it is also easy to write conditions in these coordinates for rectangles, kites, trapezia... but
these are less meaningful 1) from a mathematical perspective (e.g. they are not topologically defined).   
2) From a physical perspective (the square is additionally a configuration for which various notions of uniformity are maximal). 
However, squares are not the only notion of maximal uniformity by the demo(4) quantifier described in \cite{QShape, QSub}.  
%
{            \begin{figure}[ht]
\centering
\includegraphics[width=0.55\textwidth]{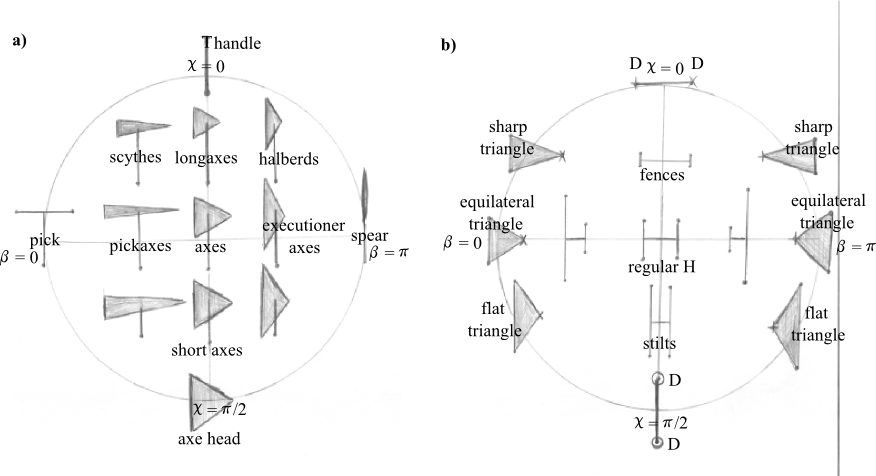}
\caption[Text der im Bilderverzeichnis auftaucht]{        \footnotesize{a) My `$\mathbb{CP}^2$ chopping board' counterpart of Kendall's spherical blackboard, 
displayed in pure-relative-ratio coordinates (i.e. with relative angle coordinates $\phi$ and $\psi$ suppressed; it is dr	wn in the $^\psi = \pi/2 = \phi$ case). 
This is for the Jacobi-K tree/triple clustering, depicted as an `axe' whose blade is the coordinate-privileged 3-cluster and whose handle 
runs from the COM T of that to the final particle.  

\noindent b)  The corresponding space of `fences' for the Jacobi-H case.  

\noindent N.B. how Figs \ref{CTQH} and \ref{FigX2} supply the limiting cases.   }        }
\label{EndGame}\end{figure}            }

\subsubsection{The inclusion of trianglelands and 4-stop metroland within quadrilateralland}\label{trincl}

In Gibbons--Pope type coordinates based on Jacobi H-coordinates, when $\rho_3 = 0$, $\chi = \pi/2$ and the metric reduces to 
\beq
\d \tts^2 = \{1/2\}^2\{\d \beta^2 + \mbox{sin}^2\beta\,\d\phi^2\}
\eeq
i.e. a sphere of radius 1/2, which corresponds to the conformally-untransformed $\mathbb{CP}^1$.  
When $\rho_2$ = 0, $\beta = 0$ and the metric reduces to 
\beq
\d \tts^2 = \{1/2\}^2\{\d \Theta_1^2 + \mbox{sin}^2\Theta_1\d\Phi_1^2\}
\eeq
for $\Theta_1 = 2\chi$ having the correct coordinate range for an azimuthal angle.  
Finally, when $\rho_1$ = 0, $\beta = 0$, the metric reduces to 
\beq
\d \tts^2 = \{1/2\}^2\{\d \Theta_2^2 + \mbox{sin}^2\Theta_2\d\Phi_2^2\}
\eeq
for $\Theta_2 = 2\chi$ again having the correct coordinate range for an azimuthal angle.   
The first two of these spheres are a triangleland shape sphere included within quadrilateralland as per Sec \ref{trincl} 
The first is for \{12\}, 4 and 3 as the particles.  
The second is for 1, 2 and 34 as the particles. 
The third of these spheres corresponds, rather, to a {\bf merger}, of + and $\times$, i.e. a merger of type 
$\mM^{\sD\sD}$  -- the space of parallelograms labelled as in Fig \ref{CTQH}e). 
[See Sec \ref{Tessi} for detailed definition and classification of types of merger.]

In Gibbons--Pope type coordinates based on the Jacobi K, when $\rho_3 = 0$, $\chi = \pi/2$ and the metric reduces to 
\beq
\d \tts^2 = \{1/2\}^2\{\d \beta^2 + \mbox{sin}^2\beta\,\d\phi^2\}
\eeq
i.e. a sphere of radius 1/2, which corresponds to the conformally-untransformed $\mathbb{CP}^1$.  
When $\rho_2$ = 0, $\beta = 0$ and the metric reduces to 
\beq
\d \tts^2 = \{1/2\}^2\{\d \Theta_1^2 + \mbox{sin}^2\Theta_1\d\Phi_1^2\}
\eeq
for $\Theta_1 = 2\chi$ again having the correct coordinate range for an azimuthal angle.  
Finally, when $\rho_1$ = 0, $\beta = 0$ and the metric reduces to 
\beq
\d \tts^2 = \{1/2\}^2\{\d \Theta_2^2 + \mbox{sin}^2\Theta_2\d\Phi_2^2\}
\eeq
for $\Theta_2 = 2\chi$ yet again having the correct coordinate range for an azimuthal angle.   
The second of these is a triangleland shape space sphere with 12, 4 and 3 as the particles. 
The first and third correspond rather to mergers, of +, T and 4 in the first case, and of T and 3 in the second case. 
%
{            \begin{figure}[ht]
\centering
\includegraphics[width=0.80\textwidth]{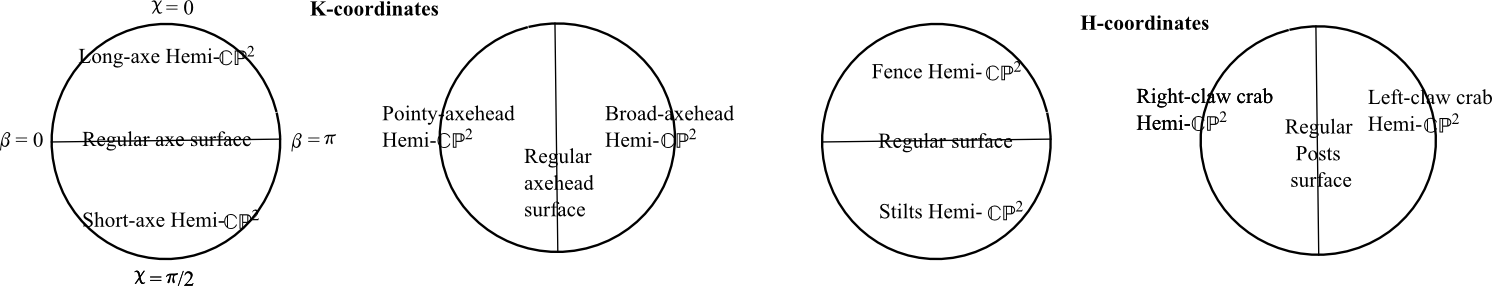}
\caption[Text der im Bilderverzeichnis auftaucht]{        \footnotesize{Meanings of the hemi-$\mathbb{CP}^2$'s for quadrilateralland.}        }
\label{Sec-3-extra}\end{figure}            }
%
\subsubsection{Configuration spaces for the indistinguishable particle cases at the metric level} 
%
{            \begin{figure}[ht]
\centering
\includegraphics[width=0.43\textwidth]{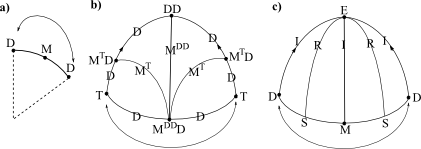}
\caption[Text der im Bilderverzeichnis auftaucht]{        \footnotesize{Combining the topological 
figure for these with the newly-gained metric information, one arrives at the indicated 
configurations spaces for fully indistinguishable particle versions of cases of a) 3-stop metroland. b) 4-stop metroland. c) triangleland.         }        }
\label{Indis-Met}\end{figure}            }

\subsection{GR counterpart: configuration space study}\label{GRCtr}

At the level of dimension-counting, GR configuration spaces are obviously all infinite-dimensional.

\mbox{ } 

\noindent Moreover [Smale]\footnote{Wheeler \cite{Battelle} credits Smale with all points marked with his name here.}, superspace's points are not of equal dimension.  
In particular, points corresponding to metrics with maximal and non-maximal symmetry differ in dimension.
For, different 3-metrics $\mh_{\mu\nu}$ have isometry groups Isom($\mh_{\mu\nu}$) of different dimension:
$\mbox{dim}(\mbox{Isom}(\delta_{\mu\nu})) = \mbox{dim(Eucl)} = 6$, $\mbox{dim}(\mbox{Isom}(\mbox{generic } \mh_{\mu\nu})) = 0$.  
I note that this argument extends to CS($\bupSigma$) and \{CS + Vol\}($\bupSigma$) using conformal isometries and VP-conformal isometries.  

\noindent Topological studies of superspace are in \cite{Fischer70, Fischer86, Giu1, Giu2}.
These are rather more involved than the above topological computations for the simpler RPM's, 
and are arrived at by largely different methods such as Cerf's Theorem \cite{Cerf} and in relation to the Hatcher Conjecture \cite{Hatcher-Conj}.  

\noindent Moncrief's work \cite{MonCS} (see also the review \cite{Carlip}) shows that the 2 + 1 GR version of CS is much more tractable than the 3 + 1 case ; 
in this case the geometry of {\sl Teichm\"{u}ller spaces} ensues. 
%

\subsubsection{Edges and singularities in GR configuration spaces}\label{GREdge}

By the preceding non-constant dimensionality point, Superspace is not a manifold.    
This is meant in the sense that  that one has to represent a whole class of points in superspace by a single point for geometries with less than maximal symmetry.
Fischer \cite{Fischer70} (see also DeWitt's accompanying commentary \cite{DeWitt70}) then demonstrated that superspace is a stratified manifold.  
He tabulated the allowed isometry groups on various different spatial topologies.  
The orbit map $\Pi:$ Riem $\longrightarrow$ Superspace identifies all isometric metrics into a single class.  
(Diffeomorphisms which leave $\mh_{\mu\nu}$ fixed) = stabilizers = isotropy group = isometry group Isom($\langle \bupSigma$, $\mh_{\mu\nu} \rangle$; see \cite{Fischer70} for the third equality.)  
The Lie algebra corresponding to this is isomorphic to that of the Killing vector fields of $\langle \bupSigma$, $\mh_{\mu\nu} \rangle$. 

\mbox{ } 

\noindent{\bf Superspace Decomposition Theorem} (\cite{Fischer70} p 325):  The decomposition of Superspace($\bupSigma$) into 
orbits is a countable partially-ordered ${\cal C}\mt\ms^{\infty}$-Fr\'{e}chet manifold partition. 
(A Fr\'{e}chet manifold is a differential structure in which the charts are Fr\'{e}chet spaces, i.e. topological spaces that are metrizable and complete \cite{AMP}. 
This generalizes the notion of Banach manifolds, for which the charts are Banach spaces \cite{AMP}.)    

\mbox{ } 

\noindent{\bf Superspace Stratification Theorem} (\cite{Fischer70} p 326): The manifold partition of superspace is an inverted stratification indexed by symmetry type.  

\mbox{ } 

\noindent{\bf Superspace Stratum Theorem} (\cite{Fischer70} p 344): This is another theorem that involves a large table of cases. 

\mbox{ } 

\noindent The gist of this (i.e. presence of stratification rather than the specific tabulations or theorems concerning it) 
carries over to CS($\bupSigma$) (and I have no reason to believe \{CS + Vol\}($\bupSigma$) is substantially different in this respect).  
Fischer and Moncrief \cite{FM96} provide some of the counterparts of this detailed work for CS($\bupSigma$).  

\mbox{ } 

\noindent Analogy 37) Relative spaces and shape spaces for RPM's for which $d \geq 3$ is a meaningful (discernible) concept have nontrivial orbit structure/stratification, 
as do the superspace($\bupSigma$), CS($\bupSigma$) and \{CS + Vol\}($\bupSigma$) of GR.  

\noindent This is a tidier, more whole-universe-model relevant version of the $N$-body problem analogy already made by DeWitt in the original literature \cite{DeWitt70}.  

\mbox{ }  

\noindent {\bf Question 21}. I leave the extent to which these harder RPM's have counterparts of Fischer's detailed results to the reader.  

\mbox{ } 

\noindent Difference 16) For the $d \leq 2$ cases explicitly covered in this Article, however, there is only nontrivial orbit structure/stratification 
(other than possibly due to having to delete certain regions due to potential singularness there).  

\mbox{ }

\noindent DeWitt and Fischer \cite{DeWitt67, DeWitt70, Fischer70} also reported the presence of singular boundaries in superspace.  
DeWitt showed that the natural geometry inherited on Superspace is geodesically-incomplete.  
Fischer \cite{Fischer70} showed that the barriers in question correspond to the stratification.

\mbox{ } 

\noindent Analogy 38) Both for GR and RPM's, many of the configuration spaces have physically-significant bad points. 
\noindent E.g. $a = 0$ is the Big Bang and $\mI = 0$ is the maximal collision, which are furthermore analogous through being related to scale variables.
Furthermore, in both cases stratification causes problems with continuations of dynamical trajectories.  

\mbox{ } 

\noindent {\bf Question 22$^{**}$} How satisfactorily can {\sl all} GR singularities be treated from a configuration space perspective? 

\mbox{ } 

\ni Leutwyler and Wheeler \cite{Battelle} appear to have been the first to ask about initial/boundary conditions on superspace.  
DeWitt suggested \cite{DeWitt70} that when one reaches the edge of one of the constituent manifolds 
(where the next stratum starts), one could reflect the path in Superspace that represents the evolution of the 3-geometry.  
C.f. also on p 352 of \cite{Fischer70}, with examples in \cite{Misref}.  
Fischer subsequently alternatively proposed to deal with the extensions of these motions by working instead with 
a nonsingular extended space which no longer has the stratified manifold's problems with differential equations becoming questionable on edges between strata.    
He explicitly built a such in \cite{Fischer86} based on the theory of fibre bundles.

\subsubsection{Some further GR--RPM analogies at the level of configuration space differential geometry}\label{GRAnal}

The GR configuration space supermetric that features in the GR action and whose inverse, the DeWitt supermetric, 
features in the Hamiltonian constraint, are defined on Riem($\bupSigma$), i.e. (\ref{Hamm}).   

\mbox{ } 

\noindent Analogy 39) RPM's relationalspaces and GR's Riem both have curved configuration space metrics.  

\noindent Difference 17) RPM's have positive-definite kinetic arc elements, whilst GR's is indefinite.  

\noindent We shall see that this helps a number of applications and hinders a number of other ones.

\noindent Analogy 40) Additionally, CS($\bupSigma$) is also like shape space in being a space of shapes!   
Though now with `shape' being taken to have a different, more complicated meaning than in the study of RPM's: that of conformal 3-geometries.
Each notion of shape is then a good candidate for true degrees of freedom of the theory in question.

\mbox{ } 

\noindent Analogy 41) {\bf Shape-definiteness alignment Lemma 13} 
This is based firstly on DeWitt's observation \cite{DeWitt67} that the part of the GR configuration 
space metric that causes it to be indefinite is the overall scale part of the metric. 
In particular, DeWitt established that CRiem is positive-definite, and then CS($\bupSigma$) must also be since it is contained within CRiem.
The analogy part of this is with shape space also being positive-definite.
The rest of the Lemma is that in the homogeneous/inhomogeneous split, some shape parts -- the anisotropies -- are 
present within the homogeneous part of the split, as are homogeneous matter modes.    
In this way, this split differs in GR from the scale--shape split.
Another common split is into `scale part and homogeneous mode of scalar field matter' and the remainder.
Finally, one can nest these splits hierarchically.  
One can conclude the following.

\mbox{ }  

\noindent The split-out parts other than the scale can be taken to be positive-definite.
In particular, subconfiguration spaces built out of (the GR notion of conformal-geometric) shapes and conventional matter fields have positive-definite kinetic metrics.  
I will be putting this fact to various good uses in Part III.  

\mbox{ } 

\noindent Possible Analogy 42) Relational space containing a better-behaved shape space might give hope that CS($\bupSigma$) is better-behaved than superspace.  
Whether CRiem($\bupSigma$) is better-behaved than Riem($\bupSigma$) is a more accessible pre-test of this speculation; DeWitt already found some evidence for this \cite{DeWitt67}.  

\mbox{ } 

\noindent N.B. the geometrical nature of superspace and conformal superspace is extremly complicated, placing limitations on what insights one can get from their study.

\mbox{ }  

\noindent As regards putting a metric on superspace itself, investigating whether the DeWitt supermetric has a vertical--horizontal split is relevant \cite{Superspace4}.    
Infinite dimensionality means that transversality alone is insufficient to make a direct sum split of this kind.  
Also one needs to check that the summands are topologically closed subspaces, which typically involves 
regularity properties of elliptic operators, and which is necessary for the projection maps to be continuous.  
Because of the nature of the metric at a point corresponding to a geometry with isometries, the tangent space does not exist. 
By a metric at such a point, what is meant is the induced metric in the horizontal subspaces of the covering points.  

\mbox{ }  

\noindent Difference 18) The GR counterpart of relational space being the cone over shape space in RPM's is the configuration space \{CS + Vol\}($\bupSigma$) = Riem($\bupSigma$)/Diff($\bupSigma$) 
$\times$ VPConf($\bupSigma$), where the Vol stands for global volume. 
Now clearly the 2 roles played by relational space are played by obviously different (e.g. dimensionally different) spaces superspace and \{CS + Vol\}($\bupSigma$).  
Additionally, I have shown that \{CS + Vol\}($\bupSigma$) is not the cone over CS($\bupSigma$).

\mbox{ }

\noindent {\bf Question 23}. Is \{CRiem + Vol\}$(\bupSigma)$ a cone over CRiem$(\bupSigma)$?  

\mbox{ } 

\noindent \{CS + Vol\}($\bupSigma$) is the closest thing known to a `space of true dynamical degrees of freedom for GR' \cite{York72, York73, ABFKO}.
%

\mbox{ } 

\noindent Analogy 43) Each of relational space and superspace \cite{Kuchar81} has a conformal Killing vector associated with scale.

\subsubsection{Shape variables in GR}\label{GRSVA}

\noindent Analogy 44) Scale in scaled RPM's is a direct counterpart of scale (and homogeneous matter modes in cases in which these are present) as slow, heavy h degrees of freedom. 
Pure shape is a direct counterpart of anisotropies, inhomogeneities and corresponding matter modes as fast, light l degrees of freedom. 
(In GR, anisotropy could be an alternative l or part of a bigger h or part of a 3-tier hierarchy: homogeneous isotropic h degrees of freedom, anisotropic `middling heaviness and 
middling slowness' `m' degrees of freedom and inhomogeneous l degrees of freedom).  

\mbox{ } 

\noindent In GR, shape variables include the $\beta_{\pm}$ anisotropy parameters in diagonal Bianchi IX, and the conformal 3-geometries in the York method.  
Considering the (more) isotropic subcases largely does not constitute a tessellation.  
This is through lacking in permutation symmetries.  
However, pictures concerning special lower-$d$ regions (e.g. Taub in mixmaster or isotropic solutions) are used at the classical level, e.g. \cite{WainwrightEllis}.)
I do not know if these have seen much use at the quantum level yet.  

\mbox{ } 

\noindent Studying inhomogeneities along these lines, beyond \cite{HallHaw}'s perturbative treatment, is less clear-cut.

\mbox{ } 

\noindent Analogy 45) The tessellation by geometrical/physical interpretation method has an also-useable counterpart in 
minisuperspace examples as regards reading off the meaning of the QM wavefunction or of of classical dynamical trajectories \cite{WainwrightEllis}.

\subsubsection{Large diffeomorphisms, quantum superspace etc.}

Large diffeomorphisms LDiff($\bupSigma$) are those diffeomorphisms which cannot be connected to the identity, i.e. 
those which are generated by more than just infinitesimal diffeomorphism generators.   

\mbox{ } 

\noindent Analogy 46) RPM analogues of LDiff($\bupSigma$) are $\mathbb{Z}_2^{\sr\se\sf}$, $A_N$ and their combinations $S_N$.  

\noindent Then Riem($\bupSigma$)/LDiff($\bupSigma$)   is analogous to ORelative   space, Arelative   space or Srelative   space. 

\noindent QSuperspace($\bupSigma$) = Superspace($\bupSigma$)/LDiff($\bupSigma$) is analogous to ORelational space, Arelational space or Srelational space.

\noindent QCRiem($\bupSigma$) = CRiem($\bupSigma$)/LDiff($\bupSigma$)           is analogous to OPreshape   space, APreshape   space or SPreshape   space 
QCS($\bupSigma$) =    CS($\bupSigma$)/LDiff($\bupSigma$)                        is analogous to OShape      space, AShape      space or SShape      space 
= Leibniz space, and 
QCS($\bupSigma$) + Vol = 

\noindent \{CS + Vol\}($\bupSigma$)/LDiff($\bupSigma$)     is analogous to the cone presentations of 
                                                             is analogous to ORelational space, Arelational space or Srelational space.  

\mbox{ } 

\noindent{\bf Question 24$^*$} Does the simpler study of RPM's shed any light on the necessity (or otherwise) and other implications of studying 
`quantum' rather than `full' configuration spaces?

\subsubsection{Summary of this Article's configuration spaces and the relations and analogies between them}\label{Clough}
%
{            \begin{figure}[ht]
\centering
\includegraphics[width=0.92\textwidth]{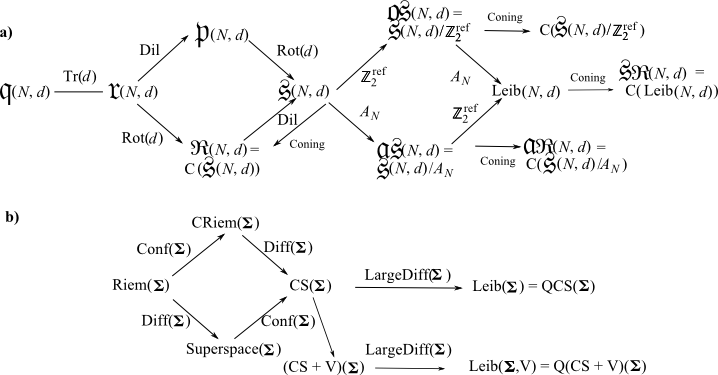}
\caption[Text der im Bilderverzeichnis auftaucht]{        \footnotesize{Summary. a) Specific sequence of 
configuration spaces of this paper. 
b) Corresponding sequence of configuration spaces for GR.  
 } }
\label{Fig2also} \end{figure}          }

{            \begin{figure}[ht]
\centering
\includegraphics[width=0.67\textwidth]{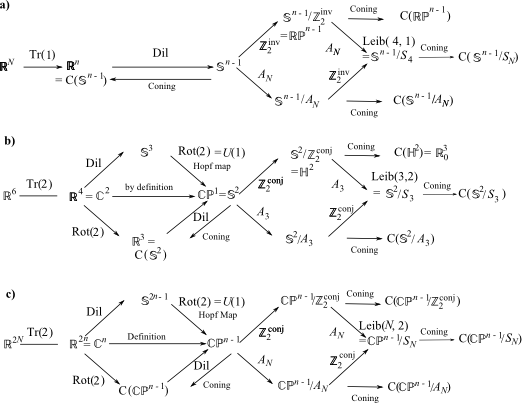}
\caption[Text der im Bilderverzeichnis auftaucht]{        \footnotesize{The sequence of configuration spaces for 
a) $N$-stop metroland,
b) the exceptional case of triangleland ($\mathbb{H}^2$ is the hemisphere with edge and $\mathbb{R}^3_0$ the half-space with edge), and
c) $N$-a-gonland.} }
\label{Fig-Uni} \end{figure}          }

\subsection{Scheme A): direct relationalspace implementation of Configurational Relationalism}\label{DRIII}

Given a Riemannian geometry $\langle\FrMgen$, $\mbox{\boldmath$M$}\rangle$ [standard finite-$d$ Riemannian suffices in the present context], 
the natural mechanics in the sense of Jacobi and of Synge \cite{Lanczos} associated with it, $\ttJ\ttS[\langle\FrMgen, \mbox{\boldmath$M$}\rangle$], is  
\beq
\FS^{\sr\se\sll\sa\st\si\so\sn\sa\sll\sss\sp\sa\scc\se} = \sqrt{2}\int\sqrt{W}\d\tts^{\sr\se\sll\sa\st\si\so\sn\sa\sll\sss\sp\sa\scc\se} 
\mbox{ } .
\eeq
This is explicitly directly constructible for the following of Kendall's shape space geometries and the cones thereover. 
The shape space of $N$-stop metroland is $\langle\mathbb{S}^{n - 1}$, $\mbox{\boldmath$\ttM$}_{\sss\sp\sh\se}$ $\rangle$, so the associated pure-shape RPM is
\beq
\FS^{\sr\se\sll\sa\st\si\so\sn\sa\sll\sss\sp\sa\scc\se}_{N-\sss\st\so\sp\,\sS\sR\sP\sM} =  
\sqrt{2}\int \d\lambda\sqrt{\ttW}\,\d\tts^{\sr\se\sll\sa\st\si\so\sn\sa\sll\sss\sp\sa\scc\se}_{N-\sss\st\so\sp\,\sS\sR\sP\sM}
\label{genac}
\eeq
for $\d\tts_{N-\sss\st\so\sp\,\sS\sR\sP\sM}^{\sr\se\sll\sa\st\si\so\sn\sa\sll\sss\sp\sa\scc\se}\mbox{}^2$ as given by (\ref{Tween})   
in terms of ultraspherical coordinates.

The shape space of $N$-a-gonland is $\langle\mathbb{CP}^{n - 1}$, $\mbox{\boldmath$\ttM$}_{\sF\sS}$$\rangle$, so the associated pure-shape RPM is   
\beq
\FS^{\sr\se\sll\sa\st\si\so\sn\sa\sll\sss\sp\sa\scc\se}_{N-\sa-\sg\so\sn\,\sS\sR\sP\sM} =  
\sqrt{2}\int \d\lambda\sqrt{\ttW}\,\d\tts^{\sr\se\sll\sa\st\si\so\sn\sa\sll\sss\sp\sa\scc\se}_{N-\sa-\sg\so\sn\,\sS\sR\sP\sM} 
\label{genac2}
\eeq
for $\d\tts_{N-\sa-\sg\so\sn\,\sS\sR\sP\sM}^{\sr\se\sll\sa\st\si\so\sn\sa\sll\sss\sp\sa\scc\se} \mbox{}^2$ given by (\ref{Easter}) in inhomogeneous coordinates.   
In the case of triangleland, this simplifies to (\ref{Pessach}) in polar coordinate form.

The relational space of $N$-stop metroland is C($\mathbb{S}^{n - 1}$, $\mbox{\boldmath$M$}_{\sss\sp\sh\se}$) = 
$\langle\mathbb{R}^{n}$, $\mbox{\boldmath$M$}_{\sf\sll\sa\st}\rangle$, so the associated scaled RPM is
\beq
\FS^{\sr\se\sll\sa\st\si\so\sn\sa\sll\sss\sp\sa\scc\se}_{N-\sss\st\so\sp\,\sE\sR\sP\sM} =  
\sqrt{2}\int \sqrt{W}\d s^{\sr\se\sll\sa\st\si\so\sn\sa\sll\sss\sp\sa\scc\se}_{N-\sss\st\so\sp\,\sE\sR\sP\sM}
\label{genac4}
\eeq
for $\d s_{N-\sss\st\so\sp\,\sE\sR\sP\sM}^{\sr\se\sll\sa\st\si\so\sn\sa\sll\sss\sp\sa\scc\se}\mbox{}^2$  given by (\ref{Uber}) 
in the scale--shape split's ultraspherical polar coordinates.

The relational space of $N$-a-gonland is C($\langle\mathbb{CP}^{n - 1}, \mbox{\boldmath$M$}_{\sF\sS}\rangle) = 
\langle\mC(\mathbb{CP})^{n - 1}, \mbox{\boldmath$M$}_{\sC(\sF\sS)}\rangle$, so the associated scaled RPM is 
\beq
\FS^{\sr\se\sll\sa\st\si\so\sn\sa\sll\sss\sp\sa\scc\se}_{N-\sa-\sg\so\sn\,\sE\sR\sP\sM} =  
\sqrt{2}\int \sqrt{W} \d s^{\sr\se\sll\sa\st\si\so\sn\sa\sll\sss\sp\sa\scc\se}_{N-\sa-\sg\so\sn\,\sE\sR\sP\sM}
\label{genac5}
\eeq
with $\d s_{N-\sa-\sg\so\sn\,\sE\sR\sP\sM}^{\sr\se\sll\sa\st\si\so\sn\sa\sll\sss\sp\sa\scc\se}\mbox{}^2$ given by (\ref{Pentecost})
in $\rho$ alongside inhomogeneous coordinates ${Z}^{{\sttr}}$. 
In the case of triangleland, this simplifies to (\ref{Michaelmas}) in polar coordinate form.  
This triangleland case can also be cast in terms of $\langle\mathbb{S}^2$, $\mbox{\boldmath$M$}_{\sss\sp\sh\se}$(radius 1/2)$\rangle$, so that e.g. 
\beq
\FS_{\triangle-\sE\sR\sP\sM}^{\sr\se\sll\sa\st\si\so\sn\sa\sll\sss\sp\sa\scc\se} =  
\sqrt{2}\int \sqrt{\check{W}}\d \check{s}_{\triangle-\sE\sR\sP\sM}^{\sr\se\sll\sa\st\si\so\sn\sa\sll\sss\sp\sa\scc\se}
\mbox{ } ,
\label{genac2b}
\eeq
with $\d\check{s}_{\triangle-\sE\sR\sP\sM}^{\sr\se\sll\sa\st\si\so\sn\sa\sll\sss\sp\sa\scc\se}\mbox{}^{\,2}$ given by (\ref{Eostre})
(away from $\mI = 0$ in which place this conformal transformation is invalid) and with $\check{W} = W/4\,\mI$
Cartesianizing that, one ends up using Dragt coordinate form (\ref{Minerva}).

\subsection{Scheme C.I): reduction approach}\label{Red-App}

\noindent In the indirect approach of Sec \ref{Examples} using mass-weighted Jacobi coordinates, $\underline\scL$ and $\scD$ give, in Lagrangian form, 
\beq
\underline{\scL} = \sumin\urho^{i}\cr\{{\urho}^{i\star} - {\underline{B}^{\5Star}} \cr \urho^{i} + {C}^{\star}\urho^{i}\} = 0 \mbox{ } , \mbox{ } 
\scD = \sumin\urho^{i}\cdot\{{\urho}^{i\star} - \dot{\underline{B}} \cr \urho^{i} + {C}^{\star}\urho^{i}\} = 0 
\mbox{ } .
\label{Lag-form}
\eeq
%
Now the third term of the first equation and the second term of the second equation are 0 by symmetry-antisymmetry, so eliminating $\d{\underline{B}}$ from the first and $\d{C}$ from the second in no way interfere with each other. 
The first equation then gives, $\sum_{i = 1}^{n}{\urho}\cr\{\urho\cr\d{\underline{B}}\} = -\sum_{i = 1}^n\urho^i\cr\d{\urho}^i$. 
This is recastable, at least formally, as $\d{\underline{B}} = -\underline{\underline{\mI}}^{-1}\d\underline{L}$ for $\underline{L}$ the stationary frame version of $\underline\scL$ 
and $\underline{\underline{\mI}}$ the barycentric inertia tensor,  
\beq
{\mI}_{\mu\nu} = \sumin \left\{ |\rho^i|^2\delta_{\mu\nu} - \delta_{ij}\rho^i_{\mu}\rho^i_{\nu} \right\} \mbox{ }  .  
\eeq
This is realizable in 2-$d$ away from the cone-point $\mI = 0$. 
However, it has further singularities in 3-$d$ on the collinear configurations. 
These are due to these having a zero principal moment factor for collinearities in 
the 3-$d$ case, and these not being cases one has any particular desire to exclude on physical grounds.  
The second equation gives $\d{C} = -\mI^{-1}\d\scD$ (realizable away from $\mI = 0$, which is never in any case included in pure-shape RPM).

Then 
\beq
\d\tts_{\sS\sR\sP\sM}^{\sr\se\sd}\mbox{}^2 = \{\mI\{\d s^2 - \d A^2\} - \d\scD^2\}/\mI^2 \mbox{ } 
\label{Centauri}
\eeq 
for $\d A^2 = \d\underline{L}\,\underline{\underline{\mI}}\,\d\underline{L}$ (twice the rotational kinetic arc element).   
%
%
So, for 1-$d$, as $\underline{L} = 0$, $\d A = 0$ and the expressions for $I$, $\d \tts^2$ and $\d\scD$ give this to be the ultraspherical kinetic arc element in Beltrami coordinates, 
\beq
\d\tts_{N-\sss\st\so\sp\,\sS\sR\sP\sM}^{\sr\se\sd}\mbox{}^2 = \{||\mbox{\boldmath${\rho}$}||^2||\d{\mbox{\boldmath${\rho}$}}||^2 - 
(\mbox{\boldmath${\rho}$} \cdot\d{\mbox{\boldmath${\rho}$}})^2\}/||\mbox{\boldmath${\rho}$}||^4 \mbox{ } ,  
\eeq
which suffices to identify  $\d\tts_{N-\sss\st\so\sp\,\sS\sR\sP\sM}^{\sr\se\sd}\mbox{}^2$ to be the same as  
$\d \tts_{N-\sss\st\so\sp\,\sS\sR\sP\sM}^{\sr\se\sll\sa\st\si\so\sn\sa\sll\sp\sa\scc\se}\mbox{}^2$.  
%
%
For 2-$d$, $A$ has another form by the inertia tensor collapsing to just a scalar, the definition of $\underline{L}$ and the Kronecker Delta Theorem, 
$\d A^2 = \mI^{-1}\sum_i\sum_j\{(\urho^{i} \cdot \urho^j)(\d{\urho}^{i} \cdot \d{\urho}^j) - 
                         (\urho^{i} \cdot \d{\urho}^j)(\urho^{j} \cdot \d{\urho}^i)\}$. 
Then using multipolar Jacobi coordinates, 
\beq
\d^2A = \mI^{-1}\sum\mbox{}_{\mbox{}_{\mbox{\scriptsize i = 1}}}^{n} \sum\mbox{}_{\mbox{}_{\mbox{\scriptsize j = 1}}}^{n}
          \rho^{i\,2}\rho^{j\, 2}\d{\theta}^i\d{\theta}^j \mbox{ } \mbox{ and } \mbox{ }
\scD\mbox{}^2 =     \sum\mbox{}_{\mbox{}_{\mbox{\scriptsize i = 1}}}^{n} \sum\mbox{}_{\mbox{}_{\mbox{\scriptsize j = 1}}}^{n}
          \rho^i\d{\rho}^i\rho^j\d{\rho}^j \mbox{ } , 
\eeq
so 
\beq
\mI\, \d^2A + \d\scD^2 = |(\bar{\mbox{\boldmath$z$}}\cdot\d\mbox{\boldmath$z$})_{\scc}|^2
\eeq
in complex notation.  
Then as also $\mI = ||\mbox{\boldmath${\rho}$}||^2 = ||\mbox{\boldmath$z$}||_{\scc}^2$ and $\d s = ||\d{\mbox{\boldmath${\rho}$}}|| = ||\d{\mbox{\boldmath$z$}}||_{\scc}$, one obtains 
the kinetic arc element to be the Fubini--Study one in inhomogeneous coordinates, thus identifying  
$d\tts_{N-\sa-\sg\so\sn\,\sS\sR\sP\sM}^{\sr\se\sd}$ to be the same as   
$\d\tts_{N-\sa-\sg\so\sn\,\sS\sR\sP\sM}^{\sr\se\sll\sa\st\si\so\sn\sa\sll\sss\sp\sa\scc\se}$.
The triangleland case is then simpler and rearrangeable as per the preceding SSec.

Next, since there is now no dilational constraint, and in the mechanical rather than geometrical PPSCT representation, 
\beq
\d s_{\sE\sR\sP\sM}^{\sr\se\sd}\mbox{}^2 = \d s^2 - d^2A = 
\{\d\scD/\sqrt{I}\}^2 + I\d\tts_{\sS\sR\sP\sM}^{\sr\se\sd\su\scc\se\sd}\mbox{ }^2  \mbox{ } .
\eeq
[The second equality is by (\ref{Centauri}).]     
Then $\d\scD = (\mbox{\boldmath${\rho}$} \cdot \d{\mbox{\boldmath${\rho}$}}) = \d{\mI}/2 = \rho\d{\rho}$, so 
$\d\scD/\sqrt{\mI} = \d{\rho}$, and so 
\beq
\d s_{\sE\sR\sP\sM}^{\sr\se\sd}\mbox{}^2 = \d{\rho}^2 + \rho^2\d\tts_{\sS\sR\sP\sM}^{\sr\se\sd}\mbox{}^2 \mbox{ } , 
\eeq
i.e. twice the reduced scaled RPM kinetic term is the one whose metric is the cone over the metric corresponding to twice the reduced pure-shape RPM kinetic term.  
Note furthermore that this derivation is independent of the spatial dimension.
Next, in the cases for which $\d\tts_{\sS\sR\sP\sM}^{\sr\se\sd}$ has been specifically derived, $\d s_{\sE\sR\sP\sM}^{\sr\se\sd}$ can then also immediately be derived.  
Namely, in 1-$d$, $\d\tts_{N-\sss\st\so\sp\,\sE\sR\sP\sM}^{\sr\se\sd}\mbox{}$ = $\d\tts_{N-\sss\st\so\sp\,\sE\sR\sP\sM}^{\sr\se\sll\sa\st\si\so\sn\sa\sll\sss\sp\sa\scc\se}\mbox{}$.  
Of course, this case could have been established more trivially, since in this case there are no constraints to eliminate, leaving this working being but a change to the 
scale--shape-split-abiding ultraspherical coordinates.
However, the 2-$d$ case is less trivial, amounting to $\d\tts_{N-\sa-\sg\so\sn\,\sS\sR\sP\sM}^{\sr\se\sd}\mbox{}$ being identified to 
be the same as the $\d\tts_{N-\sa-\sg\so\sn\,\sS\sR\sP\sM}^{\sr\se\sll\sa\st\si\so\sn\sa\sll\sss\sp\sa\scc\se}\mbox{}$. 
The simpler triangleland example is furthermore rearrangeable as per the preceding SSec.

\subsection{Comparison of schemes A), B), C), and D) at the classical level}\label{RelRedNM}

Thus I have obtained the following theorem (which holds for the O-case as well, due to this just changing coordinate ranges, but these play no part in  the above proof).   
This exemplifies the many-routes Criterion 5).

\mbox{ } 

\noindent{\bf Direct = Best-Matched Theorem [3)]}: For 1- and 2-$d$ RPM's, the direct relationalspace implementation of Configurational Relationalism A) 
is equivalent to the `indirect implementation B) followed by reduction C.I)' (the whole of which package is termed `Best Matching').  


\noindent This can be expressed, in each case for $d = 1, 2$, as a connection of research programs,
\beq
\mbox{JS(Kendall) =    Red(Barbour 03)} \mbox{ } ,
\eeq
\beq
\mbox{JS(C(Kendall)) = Red(Barbour--Bertotti 82) }\mbox{ } .
\eeq
\noindent As maps,  
\beq
\mbox{JS} \circ \mbox{$\FrQ(N,d)$-Quotient} = \mbox{Sim($d$)-Red} \circ \mbox{JS} \circ \mbox{Sim($d$)-Bundle} \mbox{ } ,
\eeq
\beq
\mbox{JS} \circ \mbox{$\FrQ(N,d)$-Quotient} = \mbox{Eucl($d$)-Red} \circ \mbox{JS} \circ \mbox{Eucl($d$)-Bundle} \mbox{ } .  
\eeq
\noindent And, as actual formulae, 
\beq
\mbox{JS}(\FS(N, d)) = \mbox{ } \stackrel{\mbox{\scriptsize extremum}}{\mbox{\scriptsize $\underline{A}, \underline{B}, C \in$ Sim($d$)}} 
\mbox{JBB}(\FrQ(N, d), \mbox{Sim}(d)) \mbox{ }  \mbox{ } ,
\eeq
\beq
\mbox{JS}(\bigr(N, d)) = \mbox{ } \stackrel{\mbox{\scriptsize extremum}}{\mbox{\scriptsize $\underline{A}, \underline{B} \in$ Eucl($d$)}} 
\mbox{JBB}(\FrQ(N, d), \mbox{Eucl}(d))  \mbox{ }  \mbox{ } .  
\eeq
The O versions of these statements also apply throughout.  
Thus indeed Kendall already knew the answer to `what is the non-redundant configuration space for RPM's?', but, since he never 
crossed paths with Barbour, it was left to me to make this connection between these two research programs, with the consequence of greatly strengthening the RPM program.  

\mbox{ } 

\noindent Note 1) To celebrate this result, I henceforth jointly abbreviate the demonstratedly equivalent `relationalspace' and `reduced' labels by `r'.  

\noindent Note 2) I later comment on the generalization of this `Direct = Best-Matched' Theorem in its map form 
\beq
\mbox{JS} \circ \mbox{$\FrQ$-Quotient} = \mbox{$\FrG$-Red} \circ \mbox{JS} \circ \mbox{$\FrG$-Bundle} \mbox{ } , 
\eeq
and  actual formula-based version 
\beq
\mbox{JS}(\FrQ/\FrG) = \mbox{ } \stackrel{\mbox{\scriptsize extremum}}{\mbox{\scriptsize $\ttg \in \sFG$}} \mbox{JBB}(P(\FrQ, \FrG))  \mbox{ } .  
\eeq
I will sometimes speculate that this result holds more widely than in the case of this Article's specific models for which I proved it above.
I will term this (and its further generalization in Sec \ref{Cl-Str}) the {\bf Direct = Best-Matched conjecture}.

\noindent Note 3) NRPM has relationalspace = reduced for all dimensions, and is always explicitly evaluable 
(no Best Matching Problem and thus always a computable r-action and $\ft^{\se\sm(\sJ\sB\sB)}$.  

\noindent Note 4) The coincidence of this result was by no means guaranteed.  
E.g. Bookstein (see the account in \cite{Kendall}) has considered a different metric geometry on the space of shapes; if I had studied that one instead, I would 
have found a non-coincidence at this point. 
I did not study this one because it gave material significance to the plane figures themselves (a kind of non-total crushability), whereas 
relational mechanics considers the constellation of points to be the primary entity.

\subsection{Reduction in GR: Thin Sandwich Problem}\label{TSC}

The analogues of (one or both of) eq's (\ref{Lag-form}) is now the so-called {\it reduced thin sandwich equation} 
(this use of `reduced' referring to the absence of lapse/VOTI, by which one is not treating the Hamiltonian constraint as a fourth simultaneous

\noindent equation). 
The relational version of this equation (not that this change makes any mathematical difference, but due to the context of 
arising within the relational setting and being prominent in some approaches rooted in relationalism) is as follows.
\beq
\scM_{\mu} \propto \mD_{\nu}\left\{ {\sqrt{\frac{2\Lambda - \mbox{${\cal R}\mi\mc$}(\ux; \bh]}
                                {\{\mh^{\gamma\epsilon}\mh^{\delta\upsilon} - \mh^{\gamma\delta}\mh^{\epsilon\upsilon}\}
                                 \{\dot{\mh}_{\gamma\delta} - 2\mD_{(\gamma}\dot{\mF}_{\delta)}\}
                                 \{\dot{\mh}_{\epsilon\upsilon} - 2\mD_{(\epsilon}\dot{\mF}_{\upsilon)}\}  }}} 
                                 \{\mh^{\nu\rho} \delta^{\sigma}_{\mu} - \delta^{\nu}_{\mu}\mh^{\rho\sigma}   \}
                                 \{\dot{\mh}_{\rho\sigma} - 2\mD_{(\rho}\dot{\mF}_{\sigma)}\}\right\} = 0 \mbox{ } .  
\eeq
Note 1) This is proportional to momentum flux $\Pi_{\mu}$ in the case of phenomenomological matter \cite{WheelerGRT}, which rests on the action given in Appendix \ref{Examples}.C.  

\noindent Note 2) It is significant for the complexity of the problem that this is a p.d.e, rather than algebraic, like eq's (\ref{Lag-form}).  

\noindent Note 3) The star/square root being trapped inside the derivative gives a fairly complicated p.d.e.

\noindent Note 4) The thin sandwich exhibits the POZ via the use of the form for N or $\dot{\mI}$ or $\mt^{\se\sm(\sJ\sB\sB)}$ in its formulation and of the condition 
$2\Bigvarepsilon$ - Ric$(\ux; \bh] > 0$ (for $\Bigvarepsilon$ the energy density) 
in the subsequent Belasco--Ohanian \cite{TSC1} uniqueness Theorem, and Bartnik--Fodor \cite{TSC2} existence and uniqueness Theorem that holds under less restrictive conditions.
Bartnik--Fodor also require localization so that the linearized approximation is reasonable throughout the neighbourhood of 
the data in question, and for this neighbourhood to contain no metrics with Killing vectors.

\noindent Note 5) The partialness of the theorems is not just due to lack of proof; there are counterexamples. 

\noindent i) Thick sandwiches are already not well-posed for mechanics \cite{FodorTh}: a simple periodic example exhibit both nonexistence and nonuniqueness.  
The electromagnetic counterpart of this drives one to consider the thin-sandwich limit.  

\mbox{ } 

\noindent Analogy 47) sharpened up is between having to excise the collinear configurations and having to excise the metrics with Killing vectors in Thin Sandwich proofs. 

\mbox{ } 

\noindent Then a {\sl local} solution to BM reduction exists for local (i.e. away from the collinearities.  
One can at least do some Physics for motions away from the collinearities.  
Such localizing (and its GR metrics-with-Killing-vectors excluding counterpart) however has less clear a status in the quantum theory, which have a number of globally-dependent features.

\subsection{Detailed justification for renaming Loop Quantum Gravity as Nododynamics}\label{Nodos}

Following on from the Analogies given in \ref{Rel-AV}, one source of good names for physical programs that purport to be relational and thus concern tangible entities, is to name 
the theories after the tangible entities rather than after convenient but meaningless parts of the formalism that happen to be in use in the mathematical study of that theory.  
By this criterion, `Shape Theory'/`Shape Dynamics', `Geometrodynamics' and `Conformogeometrodynamics' are very good relational names, whereas `Loop Quantum Gravity' 
leaves quite a lot to be desired.  
For, in this program the $SU(2)$ gauge group is put in by hand and it is only the subsequent removal of its significance that brings loops into the program's formalism, 
but the lion's share of the work in {\sl any} canonical quantum GR program resides, rather, in removing Diff($\Sigma$) and in interpreting the Hamiltonian constraint. 
Even without the `put in by hand' issue, the parallel naming for RPM's would be `sphere theory' or `sphere dynamics' from this being the form the preshape space takes before the 
lion's share of the work is done in removing Rot($d$) (and, in the dynamics case, interpreting the energy-type constraint), and this is rather clearly {\sl not} enlightening 
naming because the preshape space spheres do {\sl not} play a deep conceptual role in shape theory.
As such, I would suggest names along the lines of `Quantum Nododynamics' or `Knot Quantum Gestalt'.
This is also particularly appropriate via the Preface's Gestalt conceptualization of GR being, at least at the level of metric structure, 
very much emphasized in the `LQG' program itself, just not in the LQG program's {\sl name}.
Thiemann's alternative name `Quantum Spin Dynamics (QSD)' \cite{Thiemann} itself is somewhat confuseable with the Ising model and its ilk, 
and also the spin connotations once again come from the $SU(2)$ gauge group rather than the even more important diffeomorphisms.

\begin{subappendices}
\subsection{Weighted projective spaces}\label{WPS}

\noindent These are valuable as examples of departures from how $\mathbb{CP}^p$ behaves, albeit not exactly the spaces of direct interest in this Article.  
A simple example of weighted projective space is $\mathbb{WCP}^2_{2, 1, 1}$. 
[The bracketed numbers are the weights: here identify $(z_1, z_2, z_3)$ with $(\lambda^2z_1, \lambda z_2, \lambda z_3)$.]   
This occurs as a moduli space in field theory and string theory, in which context it is studied geometrically e.g. by Auzzi et al \cite{ASY}. 
E.g. this has only an $SU$(2)'s worth of Killing vectors and a $\mathbb{Z}_2$ orbifold singularity.  
Witten, Acharya and Atiyah \cite{W1, W2, W3}, Acharya \cite{Acharya} and Joyce \cite{Joyce} consider a $\mathbb{WCP}^3$ in the context of M-Theory (it is a 6-$d$ compact space).  
Collinucci \cite{Colli} mentions a $\mathbb{WCP}^4$.
Auzzi et al. \cite{ASY} also mention $\mathbb{WCP}^5_{2,2,2,1,1,1}$, 
Witten \cite{W4} considers a $\mathbb{WCP}^p$ (including its cohomology) and Eto et al. \cite{Eto} state that some such spaces are singular.  
Witten and Atiyah \cite{W1} also consider C($\mathbb{WCP}^3_{q,q,1,1}$) for $q \in \mathbb{N}$.

\subsection{Parallels to Barbour--Bertotti 1977 Theory from deeper layers of structure}\label{ParaBB77}

Note 1) The BB77 ansatz is exposed as not being general enough by perusal of the relative Lagrangian coordinate forms of 
ERPM and SRPM breaking its mathematical simplicities whilst remaining just as relational. 

\noindent Note 2) BB77 no longer looks as simple as it did in relative Lagrangian coordinates once one passes to relative Jacobi coordinates. 
Thus some of that simpleness was coordinate-dependent.
Moreover, in e.g. Jacobi coordinates and in Dragt coordinates for the triangleland case, one can re-apply the BB77 mathematical 
simplicities to obtain other simple-looking theories in terms of each of these. 
(Again, these only retain that simple form in that particular coordinate system, and the three of them are indeed not inter-convertible into 
each other and so constitute three separate theories.)  
E.g. clusterwise analogue of BB77's kinetic arc element is 
\beq
\d s^2 = \frac{\mu_i\delta_{\mu\nu}\delta_{ij}}{||\uR^i||}\d{\mR}^{\mu i}\d{\mR}^{\nu j}
\eeq
which does not reduce to BB77 as the $\R^i$ become linear combinations of $\urr^{IJ}$'s inside the mod signs. 
On the other hand, the Dragt analogue of BB77 for a triangle universe has kinetic arc element  
\beq
\d s^2 = \frac{\delta_{\Gamma\Lambda}}{||\underline{Dra}||}\d{Dra}^{\Gamma}\d{Dra}^{\Lambda} 
\mbox{ } .
\eeq

\noindent Note 3) On the other hand, the argumentation leading to BB82 and B03 is unaffected by changes of coordinate system and of amount of reducedness.
This furnishes a further theoretical reason to favour these over BB77 (and its above two counterparts).

\subsection{Geometry, conformal geometry and geometry of just relative angles}\label{DisRatAng}

Riemannian geometry has all of distance, relative distance and angle, whilst conformal geometry has just the last two of these notions.   
[I use $g_{\sfA\sfB}$ for a general Riemannian metric, the unit-determinant factor of which is the conformal metric $u_{\sfA\sfB}$.] 
$v^{\sfA}$ and $w^{\sfA}$ are vectors associated with the space in question.
\beq
\mbox{distance} = ||v|| = \sqrt{g_{\sfA\sfB}v^{\sfA}v^{\sfB}} \mbox{ } , 
\eeq
\beq
\mbox{relative distance} = \frac{||v||}{||w||} = 
\sqrt{    \frac{g_{\sfA\sfB}v^{\sfA}v^{\sfB}}
               {g_{\sfC\sfD}w^{\sfC}w^{\sfD}}  } =
\sqrt{    \frac{u_{\sfA\sfB}v^{\sfA}v^{\sfB}}
               {u_{\sfC\sfD}w^{\sfC}w^{\sfD}}  } 
\mbox{ } ,
\eeq
\beq
\mbox{angle} = \mbox{arccos}
\left(\frac{(v, w)}{||u||| |v||}\right) = 
\mbox{arccos}\left(\frac{g_{\sfA\sfB}v^{\sfA}w^{\sfB}}{\sqrt{ g_{\sfC\sfD}v^{\sfC}v^{\sfD} }\sqrt{ g_{\sfE\sfF}w^{\sfE}w^{\sfF} } }\right)  = 
\mbox{arccos}\left(\frac{u_{\sfA\sfB}v^{\sfA}w^{\sfB}}{\sqrt{ u_{\sfC\sfD}v^{\sfC}v^{\sfD} }\sqrt{ u_{\sfE\sfF}w^{\sfE}w^{\sfF} } }\right) 
\mbox{ } .
\eeq
I let these definitions apply both to space and to configuration space, and be possibly modulo involving mass-weighted quantities.
$\rho$ is a special case of distance.

\mbox{ }  

\noindent Note: Barbour has at times argued in terms of just relative angles, as a general possibility and in connection with the specific example 
of the celestial sphere in astronomy, most recently in \cite{Barbour-Essay-2012}, but this is an erroneous over-simplification. 
In pure-shape theories, one can have, and often separately need, a notion of ratios of relative distances as well as of angles.  
As a finer detail, the split into relative angle information and information in ratios of relative separations is not irreducible, as is clear from how any of 2 angles, 
the ratio of 2 sides and the angle between them or whichever 2 ratios of 2 sides all characterize triangles modulo scale (i.e. the shape alias similarity class of the triangle).
The celestial sphere modelled as above this way entails adopting a fixed vantage point and using the corresponding projective model as one's picture of the world [Fig \ref{Celeste}a)]; 
this is {\sl not} what one usually models in RPM's, as the Figure below makes clear by unveiling this as a further RPM different from the usually-studied ones. 

{            \begin{figure}[ht]
\centering
\includegraphics[width=0.97\textwidth]{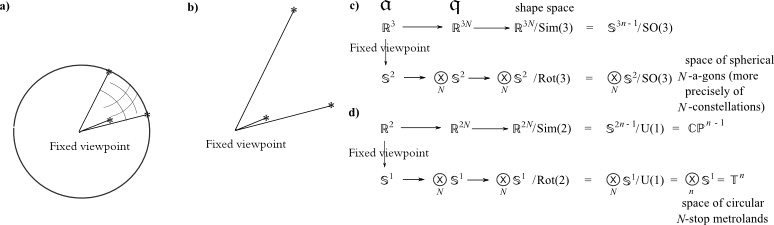}
\caption[Text der im Bilderverzeichnis auftaucht]{        \footnotesize{The geometrical difference between a) the projective single vantage point celestial sphere model and 
b) the usual model that also allows for relative distances.
c) If one adopts $\mathbb{S}^2$ rather than $\mathbb{R}^3$ as one's incipient notion of space $\FA$, one obtains the indicated model.
I also include a more tractable toy model d).
These constitute a new type of RPM's within the general scheme laid out in Secs \ref{Intro} to \ref{Q-Geom}.
More generally, albeit less motivatedly, one can consider the counterpart of RPM's formulated indirectly on whichever other spatial Riemannian geometry.} }
\label{Celeste} \end{figure}          }

\subsection{Further r-formulation of the scaled triangleland action}

This is the form obtained by Legendre transformation to the (O(A))-Hamiltonian variables (in parallel to Sec \ref{+Forms-Action}).  
This case is most cleanly presented in Dragt coordinates, and is needed for Sec \ref{PaB-RPM}.
\beq
\FS^{\sr}_{\triangle-\sE\sR\sP\sM} =              \int \{\dot{Dra}^{\Gamma}\Pi^{Dra}_{\Gamma} - 
                         \fA(Dra^{\Gamma}, \Pi^{Dra}_{\Gamma}, \dot{I}\})\d\lambda  
  =                      \int \{\dot{Dra}^{\Gamma}\Pi^{Dra}_{\Gamma} - \dot{I}\,\scE \}\d\lambda
                         \stackrel{  \mbox{$ =  \mbox {\LARGE $\int$} 
                                          \{\d  Dra^{\Gamma}\Pi^{Dra}_{\Gamma} - \d  {I}\,\scE  \} $} \hspace{0.4in} }
                   { =  \int \{   \check{*}Dra^{\Gamma}\Pi^{Dra}_{\Gamma} - \scE\}\d \check{t}^{\se\sm(\sJ\sB\sB)}}
\mbox{ } .
\label{Aac-6}
\eeq

\subsection{The O3-cornerland fallacy}\label{03}

This model exists, but is less relational as per Sec \ref{Setup-RPM}.
\noindent O3-Cornerland has three moment of inertia components while 
\noindent (O)Triangleland other has just one.  
The relational argument is that, since only two dimensions are meaningful, there is no sense in considering such a secondary object that has 
components that come about via supposing a third dimension.  
This failure to be totally relational lies outside of Butterfield and Caulton's \cite{BC11} classification, suggesting one way for that classification to be expanded upon.  

\mbox{ } 

\noindent This Difference then enters into the non-invertibility of larger than relationally maximal inertia tensors.  

\noindent Difference I) This causes the version presupposing 3-$d$ to at best be locally reducible (local in configuration space -- 
in regions that do not intersect with the collinearity set).  

\noindent Difference II) Presupposition of 3-$d$ also occurs in Gergely \cite{Gergely} and Gergely--McKain \cite{GergelyMcKain}'s work in the 3-particle case.  
Their work picked out this case as being qualitatively different, as follows.  
In contrast with the present Article, they obtained the metric on the space of orbits of group of translations and rotations rigged by the 
generators (this sense of `rigged' is advocated more widely in \cite{Schouten}).  
They then compute the Riemann tensors of the 3$N$ space and the 3$N$ -- 6 space and their interrelation by a (higher codimension generalization of the) Gauss equation; 
this comes out in terms of sums of products of pairs of vorticity tensors (these having a complicated form, see the original paper for their definition).
They then contract to form the Ricci scalar, expressing it in terms of the principal moments of inertia and the number of particles:
\beq
R = 6\{N - 2\}\sum_{\mu}/{\mI_{\mu}} - \{{3}/{2\mI_1\mI_2\mI_3}\}\sum_{\mu}\mI_{\mu}\mbox{}^2
\eeq
This permits one to figure out whether the collinear configurations can be included.  
For, without loss of generality $\mI_1 = 0$, $\mI_2 = \mI_3 = \mI$, 
\beq
R =  {6\{2N - 5\}}/{\mI} + {6\{N - 3\}}/{\mI_1}
\eeq 
which is infinite hence revealing a curvature singularity, unless $N$ = 3.  
They show that for O3-cornerland the metric is conformally flat (a result in fact already noted by Iwai \cite{Iwai87}).  
They also demonstrate that 3-cornerland the metric has a conformal symmetry, enabling computation of 
extrinsic curvature of ellipsoids orthogonal to the conformal Killing vector.  
They furthermore consider a prescription for allowing geodesics to reach and leave the boundaries.  

\mbox{ }

\noindent{\bf Question 25} The above study should have an interesting, distinct and fully relational 2-$d$ counterpart, 
and also a pure-shape RPM counterpart in both 2-$d$ and 3-$d$.  
I encourage some PhD student with a keen interest in differential geometry to perform these calculations in parallel to Gergely's and Gergely--McKain's work.  
Inclusion of coincident configurations in such a study is claimed to be done in \cite{BDGMS90}. 
However, this is controversial insofar as \cite{T90} asserts not knowing how to give such a space topological or differential structure. 

\mbox{ }  

\noindent Difference III) In 3-cornerland, the collinear configurations entail passing to a smaller isotropy group (i.e. to 
belonging to a distinct orbit), causing a jump that usually leads to the collinear configurations being excluded.  
This does not happen for triangleland, as in 2-$d$ the collinear configurations are no different in symmetry.
Such an exclusion then requires consideration of a boundary condition.  

\noindent Difference IV) Triangleland and 3-cornerland differ in the mathematical severity of binary collisions.    
For triangleland, one ends up considering $\mathbb{S}^2$ itself \cite{Kendall}, while in 3-cornerland  
one ends up considering $\mathbb{S}^2\backslash$\{3 D-points\}, i.e. the `pair of pants' \cite{Montgomery2}. 


\noindent Difference V) Secs \ref{RPM-for-QC} and \ref{OOP-2} contain a further O3-Cornerland--Triangleland Difference at the quantum level.  

\end{subappendices}

\vspace{11in}

\section{Dynamics: variation of reduced actions}\label{Dyn1}

The next two sections explore RPM's at the classical level, which is an important prequel to Parts II and III's studies of time and QM.     
For a dynamical system, conserved quantities correspond to isometries of the kinetic metric that are also 
respected by the potential; the isometries for RPM's were given in Secs \ref{Isometries}-7 and \ref{Isom2}-9.

\subsection{Dynamical equations for pure-shape RPM}\label{Dyn-Pure-Shape}

\subsubsection{4-stop metroland}\label{PSNM-Dyn}

One can immediately obtain the momenta, constraints and equations of motion from  Sec \ref{DynSphe} under 
$\alpha \longrightarrow \theta$, $\chi \longrightarrow \phi$, $\sfS_{\sfA} \longrightarrow \sfD_{\Gamma}$ [these are $SO(3)$ objects]. 
If $\ttV$ is $\phi$-independent, $\sfD = \sfD_3$ is conserved (which I term the {\it special case}). 
If $\ttV$ is also $\theta$-independent and thus constant, all 3 $\sfD_{\Gamma}$ are conserved (which I term the {\it very special case}).
N.B. special is a clustering-dependent property but very special is not (having no preferred axis is a basis-independent property).

This is in clear analogy with well-known rotor and planar problems in ordinary mechanics. 
A first analogy involves  
\beq
\mbox{$\theta$ and $\phi$ in place of their spatial counterparts $\theta_{\sss\sp}$, $\phi_{\sss\sp}$ } \mbox{ } ,
 \mbox{( moment of inertia of the rotor ), } \mI_{\sr\so\st}  \longrightarrow 1 \mbox{ } \mbox{ } \mbox{ } \mbox{ } .  
\label{ana4} 
\eeq
A second analogy involves transforming $\theta$ to the radial stereographic coordinate ${\cal R} = \mbox{tan}\frac{\theta}{2}$ and passing to the `barred' PPSCT representation.   
Then the kinetic term becomes that for the flat plane in polar coordinates and one can read off the analogies ${\cal R}$ to $r$, $\phi$ to $\phi_{\sss\sp}$,  
$
\mbox{(test particle mass) } m \longrightarrow 1.
\label{an3}
$
These analogies will be furthermore fruitful in analyzing 4-stop metroland's equations of motion and conserved quantities in the next 2 subsections, as well as when further 
specifics about the potential are brought in (see Sec \ref{Scram}).

Finally, $\widetilde{\ttV}_{\se\sf\sf} := \widetilde{\ttV} + {\sfD^2}/\mbox{sin}^2\theta  - \widetilde{\ttE}$ is the potential quantity that is significant for motion in time. 
On the other hand, combining (\ref{MPW}) and (\ref{CF}), $\widetilde{\ttU}_{\so\sr\sb} := - \mbox{sin}^4\theta \, \widetilde{\ttV}_{\se\sf\sf}$ 
is the potential quantity that is significant as regards featuring in the integrals for the shapes of the classical orbits.  
The whole-universe aspect of RPM modelling means that it is desirable to study these with $\ttE$ fixed and $\sfD$ a free parameter, which is somewhat unusual.

\subsubsection{$N$-stop metroland generalization in ultraspherical coordinates}

In straightforward extension, the conjugate momenta are 
\beq  
P_{\sttq} = 
\left\{
\prodbp^{\sttq - 1}\mbox{sin}^2\theta_{\sttp}
\right\}
{\theta}_{\sttq}^{\star} \mbox{ } .  
\eeq
These momenta obey as a primary constraint the quadratic `energy equation'
\beq
\scE := \frac{1}{2}\sum\mbox{}_{\mbox{}_{\mbox{\sttr = 1}}}^{n - 1}
{P_{\sttq}^2}/{\prodbp^{\sttq - 1}\mbox{sin}^2\theta_{\sttp}} + \ttV(\theta_{\sttp}) 
= \ttE \mbox{ } ,
\label{Belc}
\eeq
the middle expression of which also serves as the Hamiltonian.
Also, the $d = 1$ case of the expression (\ref{Gus}) for $\lt^{\se\sm(\sJ\sB\sB)}$ can be explicitly evaluated to be 
\beq
\lt^{\se\sm(\sJ\sB\sB)}_{\sN-\sss\st\so\sp} = \int \d\tts_{\mathbb{S}^{n - 1}}(\theta_{\sttp})/\sqrt{2\{\ttE - \ttV(\theta_{\sttp})\}} \mbox{ } , 
\eeq
for the line element given by (\ref{Tween}).
The evolution equations are 
\beq
\left\{
\prodbp^{\sttq - 1} \mbox{sin}^2\theta_{\sttp} \theta_{\sttq}^{\star}\right\}^{\star} - 
\sum\mbox{}_{\mbox{}_{\mbox{\scriptsize $\sttr = \sttq + 1$}}}^{n - 1} 
\left\{
\prod\mbox{}_{\mbox{}_{\mbox{\scriptsize $\sttp = 1, \sttp \neq \sttq$}}}^{\sttr - 1}
\mbox{sin}^2\theta_{\sttp}\mbox{sin}\theta_{\sttq}\mbox{cos}\theta_{\sttq}
\right\} 
\theta_{\sttr}^{\star\,2} = - {\pa \ttV}/{\pa\theta_{\sttq}}
\mbox{ } ,  
\eeq
one of which can be supplanted by the Lagrangian form of (\ref{Belc}), 
\beq
\sum\mbox{}_{\mbox{}_{\mbox{\sttr = 1}}}^{n - 1}
      \prodbp^{\sttr - 1}\mbox{sin}^2\theta_{\sfA}{\theta}_{\sttr}^{\star\,2}/{2} + \ttV(\theta_{\sttr}) = \ttE \mbox{ } .  
\eeq
If $\ttV$ is $\theta_1$-independent, there is one $SO(2)$ conserved quantity $\sfD$.  
If $\ttV$ is $\theta_1$ and $\theta_2$ independent, there are three $SO(3)$ conserved quantities.  
This `very$^p$ special' tower continues through to $\ttV$ being independent of all the $\theta_{\sttr}$ i.e. constant 
for which there are $n\{n - 1\}$ $SO(n)$ quantities correspond to the full underlying isometry group.

\subsubsection{Triangleland in ($\Theta, \Phi$) spherical coordinates}\label{Full-Red-Var}

One can immediately obtain the momenta, constraints and equations of motion from  App \ref{DynSphe} under 
$\alpha \longrightarrow \Theta$, $\chi \longrightarrow \Phi$, $\sfS_{\sfA} \longrightarrow \sfS_{\Gamma}$ [these are $SO(3)$ objects]. 
Also note that, to make analogy with ordinary mechanics on the sphere,
\beq
1 \mbox{ } \longleftrightarrow \mbox{ } m = \mbox{ (test particle mass) }  
\mbox{ } .  
\label{PaMa}
\eeq
If $\ttV$ is $\phi$-independent, $\sfJ = \sfS_3$ is conserved (which, again, I term the {\it special case}). 
If $\ttV$ is also $\theta$-independent and thus constant, all 3 $\sfS_{\Gamma}$ are conserved (which, again, I term the {\it very special case}).

\subsubsection{Triangleland in (${\cal R}, \Phi$) stereographic coordinates}

The above analogy furthermore points to the well-known $u = 1/r$ substitution of ordinary mechanics being paralleled by the 
${\cal U} = 1/{\cal R}$ one (Fig \ref{Fig-3tri}), which turns out to be exceedingly useful in my study below.  
The momenta are 
\beq
P_{{\cal R}} = {\cal R}^{\overline{\star}} \mbox{ } \mbox{ } , \mbox{ } \mbox{ }  
P_{{\Phi}} = {\cal R}^2{\Phi}^{\overline{\star}} \mbox{ } .
\eeq
These momenta obey as a primary constraint the quadratic `energy equation'
\beq
\overline\scE = \frac{1}{2}
\left\{
P_{{\cal R}}^2 + \frac{P_{\Phi}^2}{{\cal R}^2}
\right\}  
+ \overline{\ttV}({\cal R}, \Phi) = \overline{\ttE}({\cal R}) \mbox{ } , 
\label{grawp}
\eeq
the middle expression of which also serves as the Hamiltonian.  
The evolution equations are 
\beq
{\cal R}^{\overline{\star}\overline{\star}} - {\cal R}\Phi^{\overline{\star}\,2} = {\pa \{\overline{\ttE} - \overline{\ttV}\}}/{\pa {\cal R}} 
\mbox{ } \mbox{ } , \mbox{ } \mbox{ } 
\{{\cal R}^2\Phi^{\overline{\star}}\}^{\overline{\star}} =- {\pa \overline{\ttV}}/{\pa \Phi} \mbox{ } ,  
\label{PhiEL}
\eeq
one of which can be supplanted by the Lagrangian form of (\ref{grawp}),
\beq
{\cal R}^{\overline{*} 2}/2 + {\cal R}^2\Phi^{\overline{*} 2}/2 + \overline{\ttV}({\cal R}, \Phi) = \overline{\ttE}({\cal R})
\label{OI} \mbox{ } .
\eeq
\mbox{ } \mbox{ } If $\widetilde{\ttV}$ is $\Phi$-independent,
\beq
\sfJ = {\cal R}^2 \Phi^{\overline{*}} = P_{\Phi} \mbox{ } , \mbox{ } \mbox{ }  
\label{MIA}
\eeq
and the energy equation becomes 
\beq
\{\d {\cal R}/\d \overline{t}\}^2/2 + \sfJ^2/2{\cal R}^2 + \widetilde{\ttV} = \widetilde{\ttE}     \mbox{ } ,  
\eeq
or
\beq
P_{\cal R}^2/2 + \sfJ^2/2{\cal R}^2 + \overline{\ttV} = \overline{\ttE}  \mbox{ } .   
\eeq
Finally, $\overline{\ttV}_{\se\sf\sf} := \overline{\ttV} + {\sfJ^2}/{{\cal R}^2} - \overline{\ttE}$ is the potential quantity that is significant for motion in time.  
Also, combining (\ref{MIA}) and (\ref{OI}), $\overline{\ttU}_{\so\sr\sb} := - {\cal R}^4\overline{\ttV}_{\se\sf\sf}$ is the potential quantity 
that is significant as regards featuring in the integrals for the shapes of the classical orbits.  
Translating into spherical coordinates and the breved PPSCT presentation, the corresponding quantities are 
$\breve{\ttV}_{\se\sf\sf} = \breve{\ttV} + {\sfJ^2}/{\mbox{sin}^2\Theta} - \breve{\ttE}$ and $\breve{\ttU}_{\so\sr\sb} = - \mbox{sin}^4{\Theta} \, \breve{\ttV}_{\se\sf\sf}$.

\subsubsection{$N$-a-gonland extension in inhomogeneous coordinates}\label{Ngon-Dyn}

For pure-shape $N$-a-gonland, the momenta are ($Z$ is clearly not an index)
\beq
P^{Z}_{\sfa} = \ttM^{\sF\sS}_{\sfa\sfb}\d\overline{Z}^{\sfb}/\d t^{\se\sm(\sJ\sB\sB)} \mbox{ } 
\eeq
and the energy constraint is
\beq
\scE :=  \ttN^{\sfa\sfb}_{\sF\sS}P^{Z}_{\sfa}P^{Z}_{\sfa} - \ttV(\mbox{\boldmath$Z$}, \overline{\mbox{\boldmath$Z$}}) \mbox{ } . 
\eeq
Also, the $d = 2$ case of the expression (\ref{Gus}) for $\lt^{\se\sm(\sJ\sB\sB)}$ can be explicitly evaluated to be 
\beq
\lt^{\se\sm(\sJ\sB\sB)}_{\sN-\sa-\sg\so\sn} = \int       \d\tts_{\mathbb{CP}^{n - 1}-\sF\sS}(\mbox{\boldmath$Z$})    /
                                                         \sqrt{2\{\ttE - \ttV(\mbox{\boldmath$Z$})\}}                      
\mbox{ } , 
\label{Vent}
\eeq
for the line element given by (\ref{Easter}), or, in polar form $Z_{\sttp} = {\cal R}_{\sttp}\mbox{exp}(i\Phi_{\widehat{\sttp}})$, piece by piece in Sec \ref{Obgon}.
The triangleland subcase of this simplifies to 
\beq
\lt^{\se\sm(\sJ\sB\sB)}_{\triangle} = \sqrt{2}\int\d\tts_{\mathbb{S}^{2}}(\Theta, \Phi)/\sqrt{\ttE - \ttV(\Theta, \Phi)} 
\mbox{ } ,
\eeq
with reference to the last form of (\ref{Pessach}) for the line element.
We do not need to provide the equations of motion in this formulation, because most applications use the next version.

In the multi-polars version, the conjugate momenta are 
\beq
P^{{\cal R}}_{\barp} = 
\left\{
\frac{\delta_{\barp\barq}}{1 + ||{\cal R}||^2}   - 
\frac{{\cal R}_{\barp}{\cal R}_{\barq}}{\{1 + ||{\cal R}||^2\}^2}
\right\}
{\cal R}_{\barq}^{{\star}}
\mbox{ } \mbox{ } , \mbox{ } \mbox{ }  
P^{\Theta}_{\widetilde{\sp}} = 
\left\{
\frac{\delta_{\overline{\sp}\overline{\sq}}}{1 + ||{\cal R}||^2} - 
\frac{{\cal R}_{\overline{\sp}}{\cal R}_{\overline{\sq}}}{\{1 + ||{\cal R}||^2\}^2}
\right\}
{\cal R}_{\overline{\sp}}{\cal R}_{\overline{\sq}}\Theta_{\widetilde{\sp}}^{{\star}} \mbox{ } .  
\eeq
These momenta obey as a primary constraint the quadratic `energy equation'
\beq
\scE := \frac{1}{2\{1 + ||{\cal R}||^2\}}
\left\{
\{\delta^{\barp\barq} + {\cal R}^{\barp}{\cal R}^{\barq}\}  P^{{\cal R}}_{\barp}  P^{{\cal R}}_{\barq} + 
\left\{
\frac{\delta^{\overline{\sp}\overline{\sq}}}{{\cal R}_{\barp}\mbox{}^2} + 1|^{\overline{\sp}\overline{\sq}}
\right\}
P^{\Theta}_{\widetilde{\sp}} 
P^{\Theta}_{\widetilde{\sq}}
\right\}
+ \ttV(\mbox{\boldmath${\cal R}$}, \mbox{\boldmath${\Theta}$}) = \ttE
\mbox{ } ,
\eeq
the middle expression of which also serves as the Hamiltonian.
The corresponding evolution equations are (for $Q^{\sfA} := \{{\cal R}^{\barp},{\Theta}^{\tilde{\sttp}}\}$)  
\beq
D^2_{\sa\sb\sss} Q^{\sfA}/D t^{\se\sm(\sJ\sB\sB)\,2} = {Q}^{\sfA\star\star} + \Gamma^{\sfA}\mbox{}_{\sfB\sfC} {Q}^{\sfB\star} {Q}^{\sfC\star} = \pa^{\sfA}{\ttW} \mbox{ } , 
\label{ELE!!}
\eeq
with the geometrical objects of Sec \ref{Obgon} substituted in.

\subsubsection{Quadrilateralland in Gibbons--Pope-type coordinates}\label{PSQuad-Dyn}

These are particularly useful coordinates as regards dynamics on $\mathbb{CP}^2$ and its quadrilateralland interpretation.
The conjugate momenta are  
\beq
P_{\psi}  = \mbox{sin}^2\chi\mbox{cos}^2\chi\{\psi^{\star} + \mbox{cos}\,\beta\,\phi^{\star}\}/4 \mbox{ } , \mbox{ } \mbox{ } 
P_{\phi}  = \mbox{sin}^2\chi\{\mbox{cos}^2\chi\{\phi^{\star} + \mbox{cos}\,\beta\,\psi^{\star}\} + \mbox{sin}^2\chi\,\mbox{sin}^2\beta\, \phi^{\star}\}/4 
\mbox{ } ,  \mbox{ } \mbox{ } 
P_{\beta}  =  \mbox{sin}^2\chi\,\beta^{\star}/4 \mbox{ } , \mbox{ } \mbox{ }  
P_{\chi}  = \chi^{\star} \mbox{ } .  
\eeq
These momenta obey as a primary constraint the quadratic `energy equation'
\beq
\scE = \frac{P_{\chi}^2}{2} + \frac{2}{\mbox{sin}^2\chi}\left\{P_{\beta}^2 + \frac{1}{\mbox{sin}^2\beta}\{P_{\phi}^2  \mbox{ } +   \mbox{ } 
P_{\psi}^2  \mbox{ } -  \mbox{ } 2\, P_{\phi}P_{\psi} \mbox{cos}\,\beta \}\right\} + \frac{2}{\mbox{cos}^2\chi}P_{\psi}^2  + \ttV = \ttE \mbox{ } ,
\label{hior}
\eeq
the middle expression of which also serves as the Hamiltonian.
Also, (\ref{Vent}) now further simplifies to 
\beq
\lt^{\se\sm(\sJ\sB\sB)}_{\Box} = \int       \d\tts_{\mathbb{CP}^{2}-\sF\sS}(\chi,\beta,\phi,\psi)    /
                                              \sqrt{2\{\ttE - \ttV(\chi,\beta,\phi,\psi)\}}                     
											  \mbox{ } , 
\eeq
for form (\ref{GP-FS}) of the line element.
The evolution equations are 
\beq
\{\mbox{sin}^2\chi\,\mbox{cos}^2\chi\{\psi^{\star} + \mbox{cos}\,\beta\,\phi^{\star}\}/4\}^{\star} = - {\pa \ttV}/{\pa \psi}
\mbox{ } ,
\eeq
\beq
\{\mbox{sin}^2\chi\{\mbox{cos}^2\chi\{\phi^{\star} + \mbox{cos}\,\beta\,\psi^{\star}\} +  \mbox{sin}^2\chi\,\mbox{sin}^2\beta\,\phi^{\star}\}/4\}^{\star} 
= - {\pa \ttV}/{\pa \phi}
\mbox{ } ,
\eeq
\beq
\{\mbox{sin}^2\chi\,\beta^{\star}/4\}^{\star} = 
\mbox{sin}^2\chi\,\mbox{sin}\,\beta\{\mbox{sin}^2\chi\mbox{cos}\,\beta\,\phi^{\star} - \mbox{cos}^2\chi\,\psi^{\star}\}\phi^{\star}/4  - {\pa \ttV}/{\pa \beta}
\mbox{ } ,
\eeq
\beq
\chi^{{\star}{\star}} = \mbox{sin}\,\chi\,\mbox{cos}\,\chi \{\beta^{{\star}\,2} + \mbox{cos}\,2\chi\,\{\phi^{{\star}\,2} + 
\psi^{{\star}\,2} + 2\,\phi^{\star}\,\psi^{\star}\mbox{cos}\,\beta\} + 2\mbox{sin}^2\chi\,\mbox{sin}^2\beta\,\phi^{{\star}\,2} \}/4  - {\pa \ttV}/{\pa \chi}
\mbox{ } ,
\eeq
one of which can be supplanted by the Lagrangian form of (\ref{hior}),
\beq
{\chi^{{\star} \, 2}}/{2} + {\mbox{sin}^2\chi}
\{\beta^{{\star} \, 2} + \mbox{cos}^2\chi\{\phi^{{\star}\,2} + \psi^{{\star}\,2} + 2\,\phi^{\star}\psi^{\star}\mbox{cos}\,\beta\} + 
\mbox{sin}^2\chi\,\mbox{sin}^2\beta\,\phi^{{\star}\, 2}\}/8 + \ttV = \ttE
\mbox{ } .
\eeq

\subsection{Dynamical equations for scaled RPM}\label{3Stop-EOM}

\subsubsection{3-stop metroland in Cartesian coordinates}\label{EOM-3StopCart}

This is the $k = 2$, $X^{\sfA} \longrightarrow \rho^i$ case of App \ref{Land}.

\subsubsection{3-stop metroland in ($\rho, \varphi$) polar coordinates}\label{EOM-3StopPol}

This is the $R, \chi \longrightarrow \rho, \varphi$ and $\sfS \longrightarrow \sfD$ case of App \ref{Hope}.

\subsubsection{4-stop metroland in Cartesian coordinates}\label{EOM-4StopCart}

This is the $k = 3$, $X^{\sfA} \longrightarrow \rho^i$ case of App \ref{Land}.

\subsubsection{4-stop metroland in ($\rho, \theta, \phi$) spherical polar coordinates}\label{EOM-4Stop}

This is the $R, \alpha, \chi \longrightarrow \rho, \theta, \phi$ and $\sfS_{\sfA} \longrightarrow \sfD_{\Gamma}$ case of App \ref{Glory}.

\subsubsection{$N$-stop metroland extension}\label{EOM-NStop}

In Cartesian coordinates, this is the  $k = n$ counterpart of the above.
On the other hand, in ultraspherical polar coordinates, the conjugate momenta are 
\beq
P_{\sttq} = \rho^2\left\{
\prodbp^{\sttq - 1}\mbox{sin}^2\theta_{\sttp}
\right\}
{\theta}_{\sttq}^* \mbox{ } , \mbox{ }  \mbox{ }  P_{\rho} = \rho^* \mbox{ } .  
\eeq
These momenta obey as a primary constraint the quadratic `energy equation'
\beq
\scE := \frac{P_{\rho}^2}{2} + \frac{1}{2\rho^2}\sum\mbox{}_{\mbox{}_{\mbox{\scriptsize $\sttr = 1$}}}^{n - 1}
{P_{\sttq}^2}/{\prodbp^{\sttq - 1}\mbox{sin}^2\theta_{\sfA}} + V(\theta_{\sttp}) = E \mbox{ } ,
\label{Rapac}
\eeq
the middle expression of which also serves as the Hamiltonian.
Also, the $d = 1$ case of the expression (\ref{Anima}) for $\lt^{\se\sm(\sJ\sB\sB)}$ collapses to 
\beq
\lt^{\se\sm(\sJ\sB\sB)}_{\sN-\sss\st\so\sp} = \int \sqrt{\d\rho^2 + \rho^2 \d\tts_{\mathbb{S}^{n - 1}}(\theta_{\sttp})\mbox{}^2}/\sqrt{2\{E - V(\theta_{\sttp})\}} \mbox{ } , 
\eeq
for line element given by (\ref{Uber}).
The evolution equations are now  
\beq
\left\{
\rho^2
\prodbp^{\sttq - 1} \mbox{sin}^2\theta_{\sttp} \theta_{\sttq}^*
\right\}^* 
- \rho^2 
\left\{
\sum\mbox{}_{\mbox{}_{\mbox{\scriptsize $\sttr = \sttq + 1$}}}^{n - 1} 
\left\{
\prod\mbox{}_{\mbox{}_{\mbox{\scriptsize $\fA = 1, \fA \neq \sttq$}}}^{\sttr - 1}\mbox{sin}^2\theta_{\sttp}
\right\}
\right\} 
\mbox{sin}\theta_{\sttq}\mbox{cos}\theta_{\sttq}\theta_{\sttr}^{*2} = - \frac{\pa V}{\pa\theta_{\sttq}}
\mbox{ } ,    
\eeq
one of which can be supplanted by the Lagrangian form of (\ref{Rapac}),
\beq
{\rho}^{*\,2}/2 + \rho^2\sum\mbox{}_{\mbox{}_{\mbox{\scriptsize $\sttr = 1$}}}^{n - 1}
      \prod\mbox{}_{\sttp = 1}^{\sttr - 1}\mbox{sin}^2\theta_{\sttp}{\theta}_{\sttr}^{*2}/{2} + V(\theta_{\sttr}) = E \mbox{ } .  
\eeq

\subsubsection{Triangleland in Cartesian (i.e. Dragt) coordinates}

This is the $k = 3$, $X^{\sfA}, P_{\sfA} \longrightarrow Dra^{\Gamma}, P^{Dra}_{\Gamma}$ case of App \ref{Land}; 
I reproduce the energy constraint for this since I need it for future reference:  
\beq
\sum\mbox{}_{\mbox{}_{\mbox{\scriptsize $\Gamma$}}} Dra^{\Gamma * \,2} + V(Dra^{\Lambda}) = E \mbox{ } .  
\label{EnDragta}
\eeq
I also comment that the Dragt correspondence $\underline{\rho}^i \longrightarrow Dra^{\Gamma}$ 
does preserve the form of a number of objects, though it is not quite as nice as the Jacobi map in this way:  
$|\sum_{\Gamma = 1}^3Dra^{\Gamma}|^2 = \mI^2 = |\sum_{i = 1}^2\mI^i|^2$, $|\sum_{\Gamma = 1}^3P^{Dra}_{\Gamma}|^2 = 2T = |\sum_{i = 1}^2p_i|^2$  and 
%
%
$\sum_{\Gamma = 1}^3\underline{Dra}^{\Gamma} \cdot \underline{P}^{Dra}_{\Gamma} = 
2 \, \scD = 2\sum_{i = 1}^2 \underline{\rho}^i \cdot \underline{p}_{i}$. 
However, $ \sum_{\Gamma = 1}^3\underline{Dra}^{\Gamma} \cr \underline{P}^{Dra}_{\Gamma}$ is nothing like $\sum_{i = 1}^2 \underline{\rho}^i \cr \underline{p}_i$.

\subsubsection{Triangleland in ($\mI, \Theta, \Phi$) spherical polar coordinates.}\label{+Tri-Dyn}

This is the $R, \alpha, \chi \longrightarrow \mI, \Theta, \Phi$, with $\{E - V\}/4\mI$ in place of $E - V$ (giving an extra $\pa \{E/4\mI\}/\pa\mI = - E/4\mI^2$ 
in the I-evolution equation), $\Last \longrightarrow \check{\Last}$ and $\sfS_{\sfA} \longrightarrow \sfD_{\Gamma}$ case of App \ref{Glory}.

Also, the $d = 2$, $N = 3$ case of the expression (\ref{Anima}) for $\lt^{\se\sm(\sJ\sB\sB)}$ can be explicitly evaluated to be 
\beq
\lt^{\se\sm(\sJ\sB\sB)}_{\triangle} = \sqrt{2}\int \sqrt{\d \mI^2 + \mI^2 \d\tts_{\mathbb{S}^{2}}(\Theta, \Phi)\mbox{}^2}/
                                                     \sqrt{\{E - V(\Theta, \Phi)\}/\mI} \mbox{ } ,
\eeq
with reference to (\ref{Eostre}) for the line element.

\subsubsection{Triangleland in ($\mI_1, \mI_2, \Phi)$ parabolic-type coordinates}\label{Para-Dyn-2}

These subsystem-split coordinates are useful in the context of a $\Phi$-independent potential energy in the special and very special cases, for which the conjugate momenta are
\beq
P_{i} = {\mI_{i}^{\overline{*}}}/4{\mI_i} \mbox{ } , \mbox{ } 
P_{\Phi} = {\mI_1\mI_2\Phi^{\overline{*}}}/{\mI} = \sfJ \mbox{ } , \mbox{ constant , from the $\Phi$  
evolution equation being } 
\{ {\mI_1\mI_2}\Phi^{\overline{*}}/\mI\}^{\overline{*}} = -{\pa V}/{\pa \Phi} \mbox{ } .   
\label{posiassur}
\eeq
These momenta obey as a primary constraint the quadratic `energy equation'
\beq
\overline\scE := 2{\mI_1P_1^2} + 2{\mI_2P_2^2} + 
\frac{\sfJ^2}{2}\left\{\frac{1}{\mI_1} + \frac{1}{\mI_2}\right\} + \frac{K_1\mI_1 + K_2\mI_2}{8} = 
\frac{E}{4} \mbox{ } ,  
\eeq
the middle expression of which also serves as the Hamiltonian.

If $V$ is independent of $\Phi$, then $\Phi$ is a cyclic coordinate and the $\Phi$ evolution 
equation simplifies considerably: ${\mI_1\mI_2}\Phi^{*}/\mI = \sfJ, \mbox{constant}$, so 
\beq
\Phi^{*} = \sfJ\{{1}/{\mI_1} + {1}/{\mI_2}\} 
\mbox{ } .  
\label{angleq}
\eeq
This can be used to remove $\Phi^{\star}$ from the other equations of motion, 
\beq
\rho_i^{**} - {\sfJ^2}/{\rho_i^3} = - {\pa V}/{\pa\rho_i} \mbox{ } , 
i = 1 , 2 \mbox{ } .  
\eeq

\subsubsection{General scale--shape split momenta}

For $T = \{\dot{\sigma}^2 + \sigma^2\ttT_{\sS}\}/2$, the momenta are
\beq
p_{\sigma} = \sigma^* \mbox{ } , \mbox{ } \mbox{ } p^{\sS}_{\sfa} = \sigma^2 {\ttM}_{\sfa\sfb}\mS^{\sfb *} \mbox{ } . 
\eeq
These momenta obey as a primary constraint the quadratic `energy equation'
\beq
\scE = \{P_{\sigma}^2 + {\ttN}^{\sfa\sfb}P^{\sS}_{\sfa}P^{\sS}_{\sfb}/\sigma^2\}/2 + V  = E \mbox{ } ,
\label{Niudan}
\eeq
the middle expression of which also serves as the Hamiltonian.  
The evolution equations are 
\beq
\sigma^{**}  = \sigma\ttM_{\sfa\sfb}\mS^{\sfa}\mS^{\sfb} - \pa V/\pa\sigma \mbox{ } , 
\mbox{ } \mbox{ } \{ \sigma^2\mS^{\sfb *} \}^* + \Gamma^{\sfb}_{\sfa\sfc}
\mS^{\sfa *}\mS^{\sfc *} = -{\ttN}^{\sfa\sfb}\pa V/\pa\mS^{\sfb} \mbox{ } ,
\eeq 
(into which one can substitute Sec \ref{Q-Geom}'s Christoffel symbol terms), one of which can be supplanted by the Lagrangian form of (\ref{Niudan}),
\beq
\{\sigma^{*\,2} + \sigma^2{\ttM}_{\sfa\sfb}\mS^{\sfa *}\mS^{\sfb *}\}/2 + V = E \mbox{ } .  
\eeq

\vspace{10in}

\subsection{Physical interpretation of RPM momenta}\label{Pisuke}

The conjugates of the three kinds of quantity in App \ref{DisRatAng} could be termed distance or dilational momenta 
(one case of which is scale, and mathematically, radial), angular or rotational momenta and relative distance or relative dilational momenta. 
I choose to use the first option in each case.  
I use dilation, rather of the $\ttD_i$ (partial dilations).

\subsubsection{3- and 4-stop metroland momenta}

For 3-stop metroland in polar coordinates, the momenta are [dropping (a) labels and using $\mp_i$ to mean the conjugate of $\mn^i$],
\beq
\sfD := p_{\varphi} = \mn_1\mp_2 - \mn_2\mp_1 = \ttD_2\mn_1/\mn_2 - \ttD_1\mn_2/\mn_1 
\eeq
for $\ttD_i$ the {\it partial dilations}.
The second form of this is manifestly a shape-weighted {\it relative dilational quantity} corresponding to a 
particular exchange of dilational momentum between the \{bc\} and \{a\} clusters.
It is indeed conceptually clear that the conjugate to the non-angular length ratio $\varphi^{(\sfa)}$ will be a relative distance momentum.

For 4-stop metroland in spherical coordinates, the momenta are [dropping (Hb) or (Ka) labels] 
\beq
\ttD_{\phi}   := p_{\phi} = \mn_1\mp_2 - \mn_2\mp_1  = \ttD_2\mn_1/\mn_2 - \ttD_1\mn_2/\mn_1 \mbox{ } ,   
\label{WillBeA3}
\eeq
i.e. a a weighted relative dilational quantity corresponding to a particular exchange of dilational momentum between the \{ab\} and \{cd\} 
clusters in the H case or the \{bc\} and \{Ta\} clusters in the K case, and
\beq
\ttD_{\theta} := p_{\theta} \mbox{ } . 
\eeq
A formula for the latter can readily be deduced from Sec \ref{Sprintina}.

\subsubsection{Triangleland momenta}

For triangleland in spherical coordinates, the momenta are [dropping (a) labels] 
\beq
\sfJ =: p_{\Phi} = dra_1P^{dra}_2 -  dra_1P^{dra}_2  \mbox{ } , 
\eeq
which, given that $\Phi$ is a relative angle in space, is a relative angular momentum in space (hence the notation $\sfJ$), whilst 
\beq
\ttD_{\triangle} := p_{\Theta}
\eeq
is indeed a relative dilational quantity since $\Theta$ is a function of a relative distance (length ratio).
A formula for the latter can readily be deduced from Sec \ref{Sprague}.

As regards $\sfJ$'s interpretation as a relative angular momentum \cite{08I},  
\beq
\sfJ = \mI_1\mI_2\Phi^{*}/\mI = \mI_1\mI_2\{\theta_2^* - \theta_1^*\}/\mI = \{\mI_1L_2 - \mI_2L_1\}/\mI = 
L_2 = - L_1 = \{L_2 - L_1\}/2 \mbox{ } 
\label{MaxZ}
\eeq
(the fourth equality uses the zero angular momentum constraint).   
Thus it is the angular momentum of one of the two constituent subsystems, minus the angular momentum of the other, or half of the difference between the two subsystems' angular momenta.

Franzen and I \cite{AF} termed the overall set of angular momenta and relative distance momenta (and mixtures of the two) {\it rational momenta} 
since they correspond to the general-ratio generalization of the angle-ratio's angular momenta.
Rational momenta was previously called {\it generalized angular momenta} by Smith \cite{Smith60}; we jettisoned that name since it is not conceptually descriptive.
Moreover, Serna and I found it to be conceptually cleaner to introduce this notion at the level of the momenta (as followed in this SSSec)
rather than at the level of the isometries/conserved quantities.
That makes it clear that these quantities' `true name' \cite{WheelerInt, Kvothe} ought to be {\bf shape momenta}, 
since what are mathematically ratio variables can also be seen to be dimensionless shape variables, and the quantity in question is the momentum conjugate to such a quantity.   
Serna and I celebrate this by passing from the previous notation ${\cal R}$ for `rational' to $\sfS$ for `shape'.

This `true naming' becomes clear in moving, away from the previous idea of interpreting in physical space the $SO(n)$ mathematics of 
the first few RPM models studied, to the following line of thought.  

 \noindent 
1) scale--shape splits are well defined, and then there are corresponding splits into conjugate scale momenta and shape momenta.  

\noindent
2) The shape momenta correspond to dimensionless variables, i.e. ratios (or functions of ratios), accounting for why the previously encountered objects were termed rational momenta. 

\noindent 
3) Then in some cases, shape momentum mathematics coincides with (arbitrary-dimensional) angular momentum mathematics, and also 
some shapes/ratios happen to be physically angles in space, so the interpretation in space {\sl indeed is} as angular momentum.  
  
\noindent 
4) But in other cases, shapes can correspond physically to ratios other than those that go into angles in space, e.g. ratios of two lengths 
(then one's momentum is a pure relative distance momentum) or a mixture of angle and non-angle in space ratios (in which case one has a general shape momentum).
Moreover, there is no a priori association between shape momenta and $SO(n)$ groups; this {\sl happens} to be the case for the first few examples encountered 
($N$-stop metroland, triangleland) but ceases to be the situation for quadrilateralland (and $N$-a-gonlands beyond that).

\subsubsection{Quadrilateralland momenta}

The Gibbons--Pope-type coordinates for quadrilateralland extend the above triangleland spherical polar coordinates 
in constituting a clean split into pure non-angle ratios and pure angle ratios (two of each).
Thus their conjugates are again cleanly-split pure relative angular momenta and relative distance momenta as their conjugates (two of each): 
\beq
\sfJ_{\psi}  := p_{\psi}  \mbox{ } , \mbox{ } \mbox{ }
\sfJ_{\phi}  := p_{\phi}  \mbox{ } , \mbox{ } \mbox{ }
\ttD_{\beta} := p_{\beta} \mbox{ } , \mbox{ } \mbox{ } 
\ttD_{\chi}  := p_{\chi}  \mbox{ } .  
\eeq
The corresponding Hamiltonian is then 
\beq
\scE = \frac{\ttD_{\chi}^2}{2} + \frac{2}{\mbox{sin}^2\chi}\left\{\ttD_{\beta}^2 + \frac{1}{\mbox{sin}^2\beta}
\{\sfJ_{\phi}^2 + \sfJ_{\psi}^2 - 2\,\sfJ_{\phi}\sfJ_{\psi}\mbox{cos}\,\beta \}\right\} + \frac{2}{\mbox{cos}^2\chi}
\sfJ_{\psi}^2  + \ttV \mbox{ } . \hspace{0.1in} 
\eeq
See \cite{QuadII} for how various combinations of $\scD_{\beta}$, $\sfJ_{\phi}$ and $\sfJ_{\psi}$ can be constants of the motion for various types of potential 
($N$-a-gonlands have a more intricate version of the very$^p$ special tower) and for further study of quadrilateralland.

\subsubsection{Scale momentum interpretation} 

In terms of the overall dilation object, 
\beq
P_{\rho} = \scD/\rho \mbox{ } , \mbox{ } \mbox{ } 
P_{\sI} = \scD/2\mI  \mbox{ } .
\eeq

\subsection{Physical interpretation of RPM's relationalspace isometries/conserved quantities}\label{Cons}

\subsubsection{$N$-stop metroland momenta}\label{Sprintina}

The pure-shape case of 3-stop metroland is relationally trivial as per Sec \ref{Discerning}, but it is part of dynamically nontrivial scaled 3-stop metroland problem.  
Here the generator of Isom$(\FrS(3, 1))$ = Isom$(\mathbb{S}^1) = U(1) = SO(2)$ is just the above-described $\sfD$.   
This is mathematically the `component out of the plane' of  `angular momentum', albeit in {\sl configuration space}, 
there clearly being no meaningful physical concept of angular momentum in 1-$d$ space itself.  

\mbox{ }

For 4-stop metroland, the three generators of Isom($\FrS(4, 1)$) = Isom($\mathbb{S}^2$) = $SO(3)$ are 
\beq
\sfD_i = {{\epsilon_{ij}}^k}\mn^j\mp_{k} \mbox{ } .
\label{DDD}
\eeq 
(\ref{DDD}) are mathematically the three components of `angular momentum' albeit again in configuration space.  
Their physical interpretation (for the moment in the setting of H-coordinates) in space is an immediate extension of that of the already-encountered 3-component of this object 
(\ref{WillBeA3}):
\beq
\sfD_i = \scD_k\mn^j/\mn^k - \scD_j\mn^k/\mn^j \mbox{ } .  
\eeq
Moreover, this example's interpretation relies, somewhat innocuously, on the three conserved quantities $\sfD_i$ corresponding to three mutually perpendicular directions 
(the three DD axes picked out by using H-coordinates), as is brought out more clearly by the next example.

For 4-stop metroland in K-coordinates one has the above formulae again [dropping (Ka) labels instead of (Hb) ones]. 
They are clearly still all relative distance momenta, albeit corresponding to a different set of ratios.  
Then e.g. $\sfD_3$ is a (weighted) {\it relative dilational quantity} corresponding to a particular exchange of dilational momentum between the \{12\} and \{T3\} clusters. 
Here, one needs to use an axis system containing only one T-axis, e.g. a \{$\mT, \mM^*\mD, \mM^*\mD$\} axis system.

Finally, for 4-stop metroland the {\it total shape momentum} counterpart of the total angular momentum is 

\noindent $\sfT\sfO\sfT = \sum_{i = 1}^{3}\sfD_i\mbox{}^2 = \sfD_{\theta}\mbox{}^2 + \mbox{sin}^{-2}\theta\,\sfD_{\phi}\mbox{}^2 = 2\ttT$. 
[For 3-stop metroland, this is just $\sfT\sfO\sfT = \sfD^2$.]  

\mbox{ }

The above pattern repeats itself, giving,  for $N$-stop metroland, $n$ -- 1 hyperspherical coordinates interpretable as a sequence of ratios of relative inter-particle cluster 
separations, shape space isometry group $SO(n)$ and a set of $n(n - 1)/2$ isometry generators which are, mathematically, components of `angular momentum' in configuration space.

\subsubsection{Triangleland momenta}\label{Sprague}

Here, the three Isom $(\mathbb{S}^2) = \mbox{Isom}(\FrS(3, 2)) = SO(3) = SU(2)/\mathbb{Z}_2$ generators are given by 
\beq
\sfS_i = {{\epsilon_{ij}}^k}dra^j P^{dra}_{k} \mbox{ } ,
\label{SSS}
\eeq
which are mathematically the three components of `angular momentum' albeit yet again in configuration space rather than in space.  
Now on this occasion, there is a notion of relative angular momentum in space.
There are even three natural such, one per clustering: $\sfJ$  
Are these the three components of $\sfS_{\Gamma}$?
No! 
The three are coplanar and at 120 degrees to each other, so only can only pick one of these for any given orthogonal coordinate basis, 
much as in the above K-coordinate example. 
The other components point in an E and an S direction (c.f. Fig \ref{Fig-Noob}).
E and D are then the two main useful choices of principal axes (I later study QM with respect to both of these bases), furnishing the \{E, D, S\} and \{D, E, S\}.   
Moreover the component pointing in the D direction has the form of a  pure relative angular momenta, 
$\sfS_3 = \sfJ$ of the \{23\} subsystem relative to the 1 subsystem.   
The other two $\sfS_{\Gamma}$'s are mixed dilational and angular momenta with shape-valued coefficient [dropping (a) labels]: 
\beq
\mbox{sin}\,\Phi\, \sfD_{\triangle} + \mbox{cos}\,\Phi\,\mbox{cot}\,\Theta\,\sfJ   \mbox{ } \mbox{ and } 
\mbox{ } \mbox{ }
-\mbox{cos}\,\Phi\, \sfD_{\triangle} + \mbox{sin}\,\Phi\,\mbox{cot}\,\Theta\,\sfJ   \mbox{ } .
\eeq 
Finally, for triangleland, the {\it total shape momentum} counterpart of the total angular momentum is 

\noindent $\sfT\sfO\sfT = \sum_{\Gamma = 1}^3\sfS_{\Gamma}\mbox{}^2 = \sfJ^2 + \mbox{sin}^{-2}\Theta\,\sfD_{\triangle}\mbox{}^2 = 2\ttT$.

\subsubsection{Quadrilateralland momenta} 

\noindent Isom$(\FrS(4,2))$ = Isom($\mathbb{CP}^2$) = $SU$(3) [up to a usually not important quotienting by $\mathbb{Z}_3$] giving the same representation theory and mathematical form of 
conserved quantities as in the idealized flavour $SU$(3) or the colour $SU$(3) of Particle Physics [these {\sl also} have this quotienting].   
Our use of 1, 2, 3, +, and -- is the standard one of $SU(2)$ mathematics.
$SU(3)$ contains three overlapping such ladders (in fact three overlapping $SU(2) \times U(1)$'s, with the 
$SU(2)$'s being isospins $I_+$, $I_-$, $I_3$, $V_+$, $V_-$, $V_3$ and $U_+$, $U_-$, $U_3$ and the 
$U(1)$'s being hypercharges $Y$, $Y_V$ and $Y_U$).  
The usual set of independent such objects, $I_3$, $I_+$, $I_-$, $V_+$, $V_-$, $U_+$, $U_-$ 
and ${Y}$, are then  represented by the Gell-Mann $\lambda$-matrices up to proportion.
For quadrilateralland itself, I use the calligraphic-font analogues of these; Serna and I \cite{QuadII} relate some of these quantities 
to the Gibbons--Pope momenta that pick out one of the $SU(2) \times U(1)$'s 
\beq
\sfY = 2p_{\psi} = 2\sfJ_{\psi} \mbox{ } \mbox{ } , \mbox{ }  \mbox{ } \sfI_3 = {p}_{\phi} = \sfJ_{\phi} \mbox{ } .
\eeq
In terms of the quadrilateralland-significant inhomogeneous bipolar coordinates, these are
\beq
\sfY = - 2\{ p_{\Phi_1} + p_{\Phi_2}  \} \mbox{ } , \mbox{ }  \mbox{ }
\sfI_3 = p_{\Phi_2} - p_{\Phi_1} \mbox{ } .  
\eeq
In H coordinates, the momentum associated with the $\phi$ coordinate represents a counter-rotation of the two constituent subsystems (`posts') 
($\times$ relative to \{12\} and + relative to \{34\}).       
The momentum associated with the $\psi$ coordinate represents a co-rotation of the two posts 
(with counter-rotation of the cross-bar so as to preserve the overall zero angular momentum condition).
In K-coordinates, the momentum associated with the $\phi$ coordinate now represents a co-rotation of the blade-face and the handle
with counter-rotation of the blade-width so as to preserve the overall zero angular momentum condition.   
The momentum associated with the $\psi$ now represents the corresponding counter-rotation.  
Also, 
\beq
\sfI_1 = - \mbox{sin}\,\phi\, p_{\beta} + \frac{\mbox{cos}\,\phi}{\mbox{sin}\,\beta}\{p_{\psi} - \mbox{cos}\,\beta\,p_{\phi}\} = 
- \mbox{sin}\,\phi\, \ttD_{\beta} + \frac{\mbox{cos}\,\phi}{\mbox{sin}\,\beta}\{\sfJ_{\psi} - \mbox{cos}\,\beta\,\sfJ_{\phi}\}                                                       \mbox{ } , \mbox{ } 
\eeq
\beq
\sfI_2 = \mbox{cos}\,\phi\,p_{\beta} + \frac{\mbox{sin}\,\phi}{\mbox{sin}\,\beta}\{p_{\psi} - \mbox{cos}\,\beta\,{p}_{\phi}\} =
\mbox{cos}\,\phi\,\ttD_{\beta} + \frac{\mbox{sin}\,\phi}{\mbox{sin}\,\beta}\{\sfJ_{\psi} - \mbox{cos}\,\beta\,\sfJ_{\phi}\} 
\mbox{ } .  
\eeq 
Finally, 
\beq
\sfI_{\sT\so\st} := \sfI^2 = p_{\beta}\mbox{}^2 + \frac{1}{\mbox{sin}^2\beta}
\{p_{\phi}\mbox{}^2 - 2\mbox{cos}\,\beta\, p_{\psi}{p}_{\phi} + p_{\psi}\mbox{}^2\} = 
\ttD_{\beta}\mbox{}^2 + \frac{1}{\mbox{sin}^2\beta}\{\sfJ_{\phi}\mbox{}^2 - 2\mbox{cos}\,\beta\,\sfJ_{\phi}\sfJ_{\psi} + 
\sfJ_{\psi}\mbox{}^2\} 
\mbox{ } .  
\eeq
Thus, whether for H's or for K's there is also a pair of coordinates $\beta$ and $\chi$: additionally dependent on only one corresponding  ratio of relative separations, 
i.e. the $\sfI_1$ and $\sfI_2$ depend on $\beta$ alone rather than on $\chi$.  
These are conjugate to quantities that involve relative distance momenta in addition to relative angular momenta.

The other expressions ($\sfU_{\pm}, \sfV_{\pm}$) come out much more messily and less insightfully in these particular ${\sfI}$-adapted Gibbons--Pope type coordinates.
Of course, $\sfU$ and $\sfV$ adapted Gibbons--Pope type coordinates exist as well, via omitting in each case a different choice of Jacobi vector.

\subsubsection{Scaled counterparts}

Isom$({\cal R}(N, 1))$ = Eucl($n$) covering the preceding and the translations, and Isom$({\cal R}(3, 2))$ = Eucl(3) likewise.

\subsubsection{Internal `arrow' degrees of freedom}\label{Rat-Mom}

As the word `spin' itself has rotational and hence angular momentum connotations, Franzen and I called its generalization to 
RPM's that do not necessarily possess any notion angular momentum `arrow', ${\cal A}$.  
Adding arrows to the present model could be nontrivial in the sense that the 1-$d$ arrows need not just obey separate `tensored-on' occupation rules. 
This is via addition of arrow and relative distance momentum quantities.
By this means, spatially 1-$d$ models can have arrow-relative distance momentum interactions that parallel the 
spin-orbital angular momentum couplings that occur in higher spatial dimensions.

\subsection{Gauge-invariant quantities for RPM's}\label{Gau}

Whilst these are to be the theory's classical \K beables as well (Sec \ref{Cl-POT-Strat}), for now they are presented but 
in terms of internal practicalities of the mechanical theory itself.  
These are \iB's that solve  
\beq
\{{\scL}_{\mu}, \iB\} = 0 \mbox{ } ,                     
\label{OLi0}
\eeq
and, also, in the pure-shape case, 
\beq
\{\scD, \iB\}       =  0 \mbox{ }                    
\label{OD0}  
\eeq
in pure-shape case, and just the first of these in the scaled case.  
This holds for \cite{AHall, QuadII} functionals of the shape variables elucidated in Sec \ref{Q-Geom} alongside their conjugate momenta derived in the present Sec.

In particular, in the pure-shape case, these are, formally, functionals ${\cal F}$[$\mS^{\sfA}, P^{\sS}_{\sfA}$].   
They have to be ${\cal C}\mt\ms^1$ or smoother for the defining Poisson brackets to be meaningful.  
Cases for which the independent-function arguments are known specifically are as follows. 
For $N$-stop metroland, these are ${\cal F}$[$\btheta$, $\mbox{\boldmath$P$}^{\theta}$], and for $N$-a-gonland, these are 
${\cal F}$[$\mbox{\boldmath$Z$}, \mbox{\boldmath$P$}^{Z}$] = 
${\cal F}$[$\mbox{\boldmath${\cal R}$}, \bTheta, \mbox{\boldmath$P$}^{\cal R}, \mbox{\boldmath$P$}^{\Theta}$];   
for triangleland \cite{AHall} and quadrilateralland \cite{QuadII} 
${\cal F}$[$\Theta,\Phi, P_{\Theta}, P_{\Phi}$] and 
${\cal F}$[$\chi,\beta,\phi,\psi,P_{\chi},P_{\beta},P_{\phi},P_{\psi}$] are probably the most natural formulations respectively.

Also, in the scaled case, these are, formally, functionals ${\cal F}$[$\bS, \sigma, \mbox{\boldmath$P$}^{\sbS}, p_{\sigma}$]. 
Cases for which the independent-function arguments are known specifically are as follows. 
For $N$-stop metroland, these are ${\cal F}$[$\btheta$, $\rho$, $\mbox{\boldmath$P$}^{\theta}, p_{\rho}$], 
and for $N$-a-gonland,  these are ${\cal F}$[$\mbox{\boldmath$Z$}, \rho, \mbox{\boldmath$P$}^{Z}, p_{\rho}$] = 
${\cal F}$[$\mbox{\boldmath${\cal R}$}, \btheta, \rho, \mbox{\boldmath$P$}^{\cal R}, \mbox{\boldmath$P$}^{\theta}, P_{\rho}$];  
for triangleland, ${\cal F}$[$\Theta, \Phi, \mI, P_{\Theta}, P_{\Phi}, P_{\sI}$] 
is probably the most natural formulation, though one also has 
${\cal F}$[$Dra^{\Gamma}, P^{Dra}_{\Gamma}$] and the subsystem-specific ${\cal F}$[$\mI_1, \mI_2, \Phi, P_1, P_2, P_{\Phi}$].   
For quadrilateralland, the most natural formulation is ${\cal F}$[$\chi,\beta,\phi,\psi,\rho, P_{\chi},P_{\beta},P_{\phi},P_{\psi},P_{\rho}$].
This is the end of this Article's systematic coverage of quadrilateralland structure.  
Parallels of this Article's subsequent classical and quantum solution work is in \cite{QuadI, QuadII, QuadIII, QuadIV}.

\subsection{Monopole issues at the classical level}\label{Gau-Mono}

Guichardet \cite{Guichardet}, Iwai \cite{Iwai87}, and Shapere and Wilczek \cite{SW89} realized that certain gauge fields play an important role in reductions/reduced dynamics.

Monopole issues are known to affect classical, and in particular, quantum-mechanical, study of a system.
E.g. consider a charged particle in 3-$d$. 
[Here, Cartesian coordinates are $\underline{x} = (x^1, x^2, x^3)$ and ($r$, $\theta_{\sss\sp}$, 
$\phi_{\sss\sp}$) in spherical polars, with corresponding mechanical momentum $\underline{p}$. 
$m$ is the particle's mass and $e$ is its charge.] 
Let this be in the presence of a Dirac monopole \cite{DirMon, Dirac48, YW} of monopole strength $g$, corresponding to field strength 
\beq
\mF_{\nu\gamma} = \epsilon_{\mu\nu\gamma}g{\ux}/{r^3} \mbox{ } .  \label{FiSt}
\eeq
If one looks for a vector potential $\underline{A}$ corresponding to this in just the one chart, one 
finds that it is singular somewhere -- the monopole has a Dirac string emanating from it in some direction or other.  
However, there is no physical content in the direction in which it emanates, and one can avoid having 
such strings by using more than one chart (each chart's choice of Dirac string lying outside the chart). 
One such choice is to have an 
N-chart $\{\theta_{\sss\sp}, \phi_{\sss\sp} \mbox{ } | \mbox{ } 
0 \leq \theta_{\sss\sp} \leq \pi/2 + \epsilon\}$, and an 
S-chart $\{\theta_{\sss\sp}, \phi_{\sss\sp} \mbox{ } | \mbox{ } 
\pi/2 - \epsilon \leq \theta_{\sss\sp} \leq \pi\}$, with the vector potential in each of these being given by 
\beq
\underline{\mA}^{\sN} \cdot \d\underline{x} =  g\{x_1 \d x_2 - x_2 \d x_1\}/{r\{r + x_3\}} 
\mbox{ } , \mbox{ } \mbox{ } 
\underline{\mA}^{\sS} \cdot \d\underline{x} =  g\{x^1 \d x_2 - x_2 \d x_1\}/{r\{r - x_3\}} 
\mbox{ } .
\eeq 
Then the classical Hamiltonian for the charged particle is built from the canonical momentum combination $\underline{p} - e\underline{\mA}$ :
\beq
H = \{\underline{p} - e\underline{\mA}\}^2/2m + V(\underline{x}) \mbox{ } ,
\eeq
with $\underline{A}$ taking the above monopole form, i.e. $\underline{\mA}^{\sN}$ in the N-chart and 
                                                          $\underline{\mA}^{\sS}$ in the S-chart.   

\mbox{ }
														  
\noindent N.B. however how this working collapses in the case of an uncharged particle: the monopole is then not 
`felt' so one has the mathematically-usual particle Hamiltonian (see Sec \ref{No-Poles} for further QM consequences).

\mbox{ }

Now, monopole issues involving the triple collision are somewhat well-known to occur in 3-body problem configuration spaces.   
In the case of RPM's, is the preceding an indication of somewhat unusual mathematics arising analogously to in the above Dirac monopole considerations?

I begin by considering this for plain scaled triangleland in the Newtonian context of this sitting inside absolute space.
Firstly, the connection involved in this case is Guichardet's \cite{Guichardet} rotational connection (c.f. Appendix \ref{Dyn1}.A.1).  
This is indeed a nontrivial connection as it has a nontrivial field strength (\ref{nontri}).  
In this case, passing to e.g. $Dra^{\Gamma}$ coordinates gives precisely the Dirac monopole 
(\ref{FiSt}) under the correspondence $Dra^{\Gamma}$ for $x^{\mu}$, clustering (1)'s D in the role of N and its M in that of S.

Next, the Hamiltonian is (for relational space coordinates $Q^{\sfA}$ with corresponding mechanical momenta $P_{\sfA}$ with metric $g_{\sfA\sfB}$ on configuration space)  
\beq
H = \underline{L}\,\underline{\underline{\mI}}^{-1}\,\underline{L} + g^{\sfA\sfB}\{{P}_{\sfA} - \underline{L}\cdot\underline{\mA}_{\sfA}\}
      \{{P}_{\sfB} - \underline{L}\cdot\underline{\mA}_{\sfB}\}/2 + V({Q}^{\sfC}) \mbox{ } . 
\label{bigham}
\eeq
From here, monopole effects would spread into the TISE and its solutions.  

\mbox{ }

However the key point is that the case of interest in RPM's is the one with {\sl zero total angular momentum} $\underline{L} = \underline\scL = 0$. 
For this, analogously to the special case of an uncharged particle in a Dirac monopole field, the Hamiltonian (\ref{bigham}) collapses to a much simpler form, 
\beq
H = g^{\sfA\sfB}{P}_{\sfA}{P}_{\sfB}/2 + V({Q}^{\sfC})
\eeq
corresponding to `not feeling' the monopole.  
Thus monopole effects do not enter the TISE and its solutions in this way. 
(Nor does a distinction between mechanical and canonical momentum change the form of the relative rational momentum operator in this case of zero total angular momentum).
Thus the associated $SO$(3) mathematics is standard. 
And, in the case of a `central' potential energy $V = V(\mI \mbox{ alone})$, the angular part of the TISE does indeed give the spherical harmonics. 
(Determining this requires additional working due to operator ordering issues, which is provided in Sec \ref{QM-Intro}).   
Thus the pure-shape RPM study of these in \cite{08II, +Tri} turns out to be reuseable in the study of the scaled triangle as the shape part of the scale--shape split.

Suppose that one chooses the case of O-shapes instead. 
(These are much more common in the literature due to the bias that the real world is 3-$d$.)  
Then the monopole is not now of Dirac-type in 3-$d$ space but rather Iwai's monopole \cite{Iwai87} in 3-$d$ half-space.   
The mathematics remains similar to that of the Dirac monopole in any given chart and gauge. 
(Though now other choices of string are more convenient \cite{Iwai87, LR97} and the flux is halved due to involving half as much solid angle as before).  
In particular, the Hamiltonian remains of the form (\ref{bigham}).

For more than three particles, the above simplifying effect of being in a $\scL = 0$ theory.  
This is a useful result e.g. toward developing the quadrilateralland model.  
However, I have not as yet considered in these cases whether the $g_{\sfA\sfB}$ coincides with the 
metric obtained from first principles/from reduction of BB82-type formulations.  
  
We straightforwardly eliminated the translations early on in this Article's treatment. 
I should next therefore reassure the reader that, if these had been left alone, they would not have given rise to a further significant connection.
This is because, as a further manifestation of the mathematical simpleness of the translations, their 
analogue of the Guichardet connection form has trivial field strength (Appendix \ref{Dyn1}.A.2).  
Thus one can be sure that for scaled $N$-stop metroland (involving at most translations), there are not any monopole effects to worry about.    
This protects my treatment of $N$-stop metroland as `ordinary Euclidean space Physics', among the 
coordinate systems for which the $\mathbb{S}^{n - 1}$ polar coordinates have a particularly lucid scale-shape interpretation.

Next, in the case of pure-shape theories, can the dilations cause analogous monopole effects?   
I establish in Appendix \ref{Dyn1}.A.2 that this is not the case because the dilational analogue of the Guichardet connection form also has trivial field strength.   
Thus combining this result and the above rotational results, pure-shape theories carry no vestiges of the excluded maximal collision through its action as a monopole. 
(Without establishing this, one could have feared e.g. that the shape spaces could contain a bad point of gauge-dependent position.  
Such could have arisen from the corresponding relational space's gauge and chart choice's Dirac string casting a `shadow' on the shape space at 
the point intersection in relational space between the Dirac string and the surrounding shape space, c.f. \cite{Dirac48}).
[Finally, translations, rotations and dilations do not interfere with each other if treated together in this sense.]

\begin{subappendices} 
 
\subsection{`Guichardet connection' for various transformation groups}\label{Guich}

\subsubsection{The Guichardet connection for rotations} \label{Guich-1} 

\noindent Working in mass-weighted Jacobi coordinates, 
\beq
\underline{\scL} = 
\sum\mbox{}_{\mbox{}_{\mbox{\scriptsize $i = 1$}}}^{n}\underline{\rho}^i \cr \underline{p}_i = 
\sum\mbox{}_{\mbox{}_{\mbox{\scriptsize $i = 1$}}}^{n}\underline{\rho}^i \cr 
\{\dot{\underline{\rho}}^i + \underline{\dot{B}} \cr \underline{\rho}^i\} = 
\sum\mbox{}_{\mbox{}_{\mbox{\scriptsize $i = 1$}}}^{n}\underline{\rho}^i \cr 
\big\{
\dot{Q}^{\sfA}{\pa\underline{\rho}^i}/{\pa {Q}^{\sfA}}
 + \underline{\dot{B}} \cr \underline{\rho}^i
\big\} 
\eeq
(for $\underline{p}_i$ the momentum conjugate to $\underline{\rho}^i$), then let 
\beq
\underline{\scL} = \underline{\underline{\mI}}\,\{\underline{\dot{B}} + 
\underline{\mA}_{\sfA}\dot{Q}^{\sfA}\}
\eeq
for $\underline{\underline{\mI}}$ the inertia tensor.  
The last term in this defines the Guichardet-type \cite{Guichardet} gauge potential 
$\underline{\mA}_{\sfA} = \underline{\underline{\mI}}^{-1}\underline{a}_{\sfA}$ for 
$\underline{a}_{\sfA} = \sum_{i = 1}^{n}\underline{\rho}^{i} \cr 
{\pa \underline{\rho}^{i}}/{\pa {Q}^{\sfA}}$.  
For vanishing angular momentum, ${\dot{\underline{\mB}}} = - \underline{\mA}_{\sfA}\dot{Q}^{\sfA}$ i.e. the mapping between change of shape and corresponding infinitesimal rotation.  
In 2-$d$ and for plain scaled triangleland, use 1) ${Q}^{\sfA}$ =($\rho_1$, $\rho_2$, $\Phi$) coordinates (closely related to parabolic coordinates \cite{08I, 08III}). 
2) Use what is referred to in the literature as the `xxy gauge' \cite{LR97}: $\underline{\rho}_1 = \rho_1(1, 0)$ and $\underline{\rho}_2 = \rho_2(\mbox{cos}\,\Phi, \mbox{sin}\,\Phi)$.  
Then the nonzero component of $\mA_{\sfA}$ is 
\beq
{\mA}_{\Phi} = sharp \mbox{ } , 
\eeq
which, in passing to Dragt-type coordinates, gives 
\beq
\underline{\mA}_{\sfA}\d {Q}^{\sfA} = \{
     { dra_1\d \{dra_2\} - dra_2\d \{dra_1\} }\}/
     { 2\{1 - dra_3\}    } \mbox{ } .  
\eeq
This is in direct correspondence with Wu--Yang's \cite{YW} $\mA^{\sS}_{\mu}$ (S for `South') for the Dirac monopole if one applies the Dragt correspondence. 
It sets the monopole strength $g$ to be 1/2.  
The current triangleland context has several further lucid interpretative points.
Firstly, in configuration space, its `South' is the clustering in question's merger point, M.
Secondly, in space itself, the base of the triangle is aligned with the absolute x-axis, so I rename this gauge the {\it `base = x' gauge}.  
Thirdly, mathematically it is most simply and clearly presented as the gauge-fixing ${\cal F}_{\sM}:= \theta_1 = 0$.

Likewise, if one inverts the roles of $\underline{\rho}_1$ and $\underline{\rho}_2$, the nonzero component of $\mA_{\Gamma}$ in the resulting chart and gauge is 
\beq
\mA_{\Phi} = flat \mbox{ } , 
\eeq
which, in passing to Dragt coordinates, gives 
\beq
\mA_{\Gamma}\d {Q}^{\Gamma} = \mbox{$\frac{1}{2}$}
\{   {dra_1\d dra_2 - dra_2\d dra_1}\}/  
     {1\{1 + dra_3\}    } \mbox{ } .  
\eeq
This is in direct correspondence with Wu--Yang's $\mA^{\sN}_{\mu}$ (N for `North') for the Dirac monopole if one applies the Dragt correspondence.  
It likewise sets the monopole strength $g$ to be 1/2.  
In the current triangleland context, then, its `North' in configuration space is the clustering in question's double collision point, D.  
In space itself, it is the median of the triangle that is now aligned with the absolute x-axis, so I rename this gauge the {\it `median = x' gauge}. 
Finally, mathematically it is most simply and clearly presented as the gauge-fixing ${\cal F}_{\sD}:= \theta_2 = 0$.

Then this D-chart and M-chart provide a full stringless description of this relational space monopole, just as the N-chart and S-chart do for the usual Dirac monopole in space.  
Given the precise nature of the correspondence between these, it is clear that the field strength is 
\beq
\mF_{\Gamma\Lambda} =  \epsilon_{\Sigma\Gamma\Lambda}\,{dra^{\Sigma}}/{\mI^2} \mbox{ } .  
\label{nontri}
\eeq
\mbox{ } \mbox{ }  For scaled Otriangleland, these workings still hold except that 1) the $dra_3 = \mbox{4 $\times$ area} < 0$ 
half-plane has ceased to be part of the configuration space. 
2) Other charts and gauges which position the Dirac string elsewhere are now more convenient. 
See e.g. \cite{Iwai87, LR97}.  
In this case one has an {\sl Iwai monopole} on $\mathbb{R}^3_+$ (the Iwai string is restricted to the one hemisphere).

\subsubsection{Translations and dilations give but trivial analogues} \label{TrDil} 

\noindent The below results hold for all particle numbers and spatial dimensions.  

\noindent In the case of translations, the coordinates are mass-weighted particle positions, 
$\underline{X}^I = \sqrt{m_I}\underline{q}^{I}$ rather than mass-weighted Jacobi coordinates.  
\beq
\scP_{\mu} = \sum\mbox{}_{\mbox{}_{\mbox{\scriptsize $I = 1$}}}^{N}\underline{p}_I = 
\sum\mbox{}_{\mbox{}_{\mbox{\scriptsize $I = 1$}}}^{N}m_I\{\underline{q}_{I} + \underline{\dot{\mA}}\} = 
M \sum\mbox{}_{\mbox{}_{\mbox{\scriptsize $I = 1$}}}^{N} \big\{ \underline{\dot{\mA}} +  
M^{-1}\sum\mbox{}_{\mbox{}_{\mbox{\scriptsize $I = 1$}}}^{N} 
\sqrt{m_I}\dot{Q}^{\sfA}{\pa\underline{X}^I}/{\pa {Q}^{\sfA}} \big\}  \mbox{ } .  
\eeq
Here, the total mass is the analogue of inertia tensor, so $\underline{\mA}^{\st\sr\sa\sn\sss}_{\sfA} = 
\underline{a}^{\st\sr\sa\sn\sss}_{\sfA} = \pa_{\sfA}\{\sum_{I = 1}^{\sN}\sqrt{m_I} \underline{X}^{I}/M \}$.  
As this is of gradient form $\pa_{\sfA}\eta$, the corresponding translational field strength $\mF_{\sfA\sfB} = 2\pa_{[\sfA}\pa_{\sfB]}\eta = 0$ by symmetry--antisymmetry.  
Thus this connection is flat and so geometrically trivial.

In the case of dilations, taking the coordinates to be mass-weighted Jacobi coordinates, 
\beq
\sfD = \sum\mbox{}_{\mbox{}_{\mbox{\scriptsize $i = 1$}}}^{n}\underline{\rho}_i \cdot \underline{p}_i = 
\sum\mbox{}_{\mbox{}_{\mbox{\scriptsize $i = 1$}}}^{n}\underline{\rho}_i\cdot\{\underline{\rho}_i + 
{\dot{C}}\underline{\rho}_i\} = 
\sum\mbox{}_{\mbox{}_{\mbox{\scriptsize $i = 1$}}}^{n}\underline{\rho}_i \cdot 
\big\{ \dot{Q}^{\sfA}{\pa\underline{\rho}_i}/{\pa {Q}^{\sfA}} + {\dot{C}}\underline{\rho}_i \big\} \mbox{ } , 
\eeq
so
\beq
\sfD =  \mI\{\dot{C} +  {\mA}_{\sfA}\dot{Q}^{\sfA}\} \mbox{ } .  
\eeq
Here the scalar moment of inertia $\mI = \sum_{i = 1}^{n}\{\rho^{i\,2}$ is the analogue of the inertia 
tensor, and ${\mA}^{\sd\si\sll}_{\sfA} = I^{-1}a^{\sd\si\sll}_{\sfA}$, $a^{\sd\si\sll}_{\sfA} = 
\sum_{i = 1}^{n}\underline{\rho}^i\cdot{\pa \underline{\rho}^i}/{\pa {Q}^{\sfA}}$.  
Then for vanishing dilation, $\dot{C} = - \mA^{\sd\si\sll}_{\sfA}\dot{Q}^{\sfA}$ so it is a mapping between change of shape and corresponding infinitesimal size change.  
But again this can be cast in gradient form: $\mA^{\sd\si\sll}_{\sfA}=\pa_{\sfA}\{ \mbox{ln}(\rho)\}$, 
so the corresponding field strength is also zero and so this connection is also flat/geometrically trivial.  

\noindent Finally, composition of translational, rotational and dilational corrections is additive, so outcomes for each of these things do not affect each other.   
(To consider combinations involving the translations, note that the above presentations for mass-weighted 
relative Jacobi coordinates for rotations and dilations continue to hold identically under $i$ to $I$, $n$ to $N$ = $n$ + 1 and $\rho^i$ to $X^{I}$. 
I.e., under the trivial position to relative Jacobi coordinates map, see e.g. \cite{06I, 06II}.)

\subsubsection{Electromagnetism and GR conterparts}\label{EM-GR}

{\bf Question 26} The Guichardet connection \cite{Guichardet} and configuration space monopole issues of RPM's have at least formal counterparts for GR and electromagnetism.
To what extent are these amenable to study?

\subsection{Further layers of structure}

\subsubsection{Phase space and rigged phase space}\label{App-B-1}

I consider passing from conceiving $\FrQ$ to be primary to suggesting that $\fQ$ and $\fP$ are operationally distinct; I refer to the ensuing state space as {\bf RigPhase}.  
From the more usual perspective, this is a privileged {\it polarization} \cite{Woodhouse, Ashtekar} of Phase (phase space) 
that maintains the distinction between the physical $\fP$'s and $\fQ$'s.  
On the other hand, a distinct and far more common viewpoint is that phase space
\beq
\mbox{Phase} = (T^*(\FrQ), \{ \mbox{ }  , \mbox{ } \}) 
\eeq
itself is central. 
Here, $T^*(\FrQ)$ is the cotangent bundle (for simple bosonic theories) of momenta over the configurations and 

\noindent\{ \mbox{ }  , \mbox{ }  \} is the Poisson bracket, so that this is a Poisson algebra. 
N.B. the status of RigPhase in this program is as an {\sl option} that is more structurally minimalist.  

\mbox{ } 

\noindent 
RigPhase Motivation 1) As argued in Sec \ref{Q-Geom},  considering more minimalist possibilities is very much part of relational thinking.    
One such avenue then would be to investigate what happens if one weakens the amount of structure assumed in the study of the `associated spaces' of the principles of dynamics. 
There are very strong reasons why $\FrQ$ (or something with an equivalent amount of physical information in it) is indispensable, 
but some parts of the conventional structure of phase space are more questionable.  
Much as one can envisage preferred-foliation counterparts of GR spacetime, one can also construe of preferred-polarization versions of phase space.
Whereas relational perspectives usually view preferred foliations of spacetime as less physical, I contend that, at the level of the associated abstract spaces 
of the Principles of Dynamics, the relationalist's ontological hierarchy as regards preferred versions of these spaces is the reverse of that for spacetime itself.
My argument then is that the operational distinction between configuration and impact measurements can be encoded by using a preferred  
polarization centred about the {\sl physical} $\FrQ$ (as opposed to any other nontrivially canonically-related {\sl mathematical} $\FrQ$).  
\noindent As a more general point, one should not necessarily postulate that physical entities that are operationally distinct 
{\sl have} to be mixable just because transformations that mix them can be mathematically defined.  
One should not confuse `is often useful to transform' with giving unwarrantedly unconditional physical significance to the transformations in question.

It then makes sense to question the `associated spaces' of Physics. 
There are very strong reasons why $\FrQ$ (or something with an equivalent amount of physical information in it) 
is indispensable, but the conventional structure of phase space may be questionable.  
Much as one can envisage preferred-foliation counterparts of GR spacetime, one can also construe of preferred-polarization versions of phase space.  
My argument then is that one can encode the operational distinction between configuration and impact measurements by using a preferred-polarization centred about the physical $\FrQ$.  

\mbox{ }

\noindent I argue the practical implementation of RigPhase to be as follows. 
I hold that the physicality of $\FrQ$ is {\sl not} the centre of this argument; that is, rather,  operational distinguishability between the 
$\fP$'s on the one hand and the $\fQ$'s on the other.   
If one includes fermions, one can operationally tell apart fermionic quantities that manage to be positions and momenta at once from quantities 
that solely manifest themselves as configurations and quantities that solely manifest themselves as momenta.  
Thus the scheme extends to cases where phase spaces are not just simple tangent bundles over configuration spaces, via operational distinguishability now being tripartite \cite{Arelsusy}.
Furthermore, were supersymmetry to exist in nature it could well impose limitations on applicability of Relationalism 3) and its RigPhase implementation. 

\noindent RigPhase Motivation 2) One might doubt canonical transformations due to their clash with the common notion of `adding in a potential'.
If this is to be considered as a structurally minor change, as is often done, then one has a problem due to the asymmetry between the 
complication caused by adding a $Q^4$ potential, say, as compared to the major structural upheaval of adding a $P^4$ kinetic term to one's Hamiltonian.  
Of course, this can be dealt with by accepting that canonical transformations imply that `adding a potential' is as great a structural upheaval 
(e.g. the above two additions to an HO are identical to each other under canonical transformation). 
This translates to changing the habitual simplicity requirements on Hamiltonians and Lagrangians due to their canonical disparity, 
and that doubtlessly cuts down on widely applicable theorems that depended on such canonically-disparate simplicity requirements. 
One might then hold all such results to have been empty anyway, but one might attempt to keep them by allotting particular 
significance to the Physics in its presentation centred about $\FrQ$.  
By the above, it is also clear that canonical transformations are at odds with statements of simplicity such as `at most quadratic in the momenta'.  
Yet further motivations for weakening Phase space at the QM level are given in Sec \ref{QM-Intro}, along with some theoretical consequences.


\noindent As continuation of MRI and MPI formulations, APhase, OAPhase, ARigPhase and OARigPhase options are relevant alternatives at this point .  
These then have defined on them (O(A))-Hamiltonians; this amounts to a fullest distinction at the level 
of `total' \cite{Dirac} objects: 
$\fH_{\sT\so\st\sa\sll} = \fN\,\scQ\scU\scA\scD + \fg_{\sfZ}\scL\scI\scN^{\sfZ}$, versus 
$\fA_{\sT\so\st\sa\sll} = \dot{\fI}\,\scQ\scU\scA\scD + \dot{\fg}_{\sfZ}\scL\scI\scN^{\sfZ}$ and
$\ordial\fA_{\sT\so\st\sa\sll} = \ordial\fI\,\scQ\scU\scA\scD + \ordial \fg_{\sfZ}\scL\scI\scN^{\sfZ}$.   
\noindent Upon performing classical reduction -- taking out $\scL\scI\scN$ to pass to a tilded $\widetilde{\FrQ}$ with just a $\widetilde{\fH}$ on it, as per Fig \ref{RigPhase} 
-- there is no longer any ((O)A)-Hamiltonian distinction here, though there does remain a distinction in smeared objects: $\fH_{\sss\sm\se\sa\sr} = 
\fN \, \scQ\scU\scA\scD$ versus $A_{\sss\sm\se\sa\sr} = \dot{\fI} \, \scQ\scU\scA\scD$ and $\ordial \fA_{\sss\sm\se\sa\sr} = \ordial\fI  \, \scQ\scU\scA\scD$. 

\mbox{ }  

\noindent I also mention a set-back  as regards the distinctiveness of the LMB relational approach (Hamiltonian Collapse problem). 
In passing to the Hamiltonian formalism prior to quantization, a lot of LMB-relational versus nonrelational formalism differences are ironed out.  
Now both have sum-form total ((O)A)-Hamiltonians whereas the actions were product-form versus difference-form; moreover the objects of interest 
are Hamiltonians rather than the smeared objects that exhibit the `(O(A))' trichotomy.    
However, as we shall see in Parts III and IV, the LMB-relational {\sl ideas}, if not precisely the same implementations of these that occur at 
the classical level, can be recycled at the quantum level.

\subsubsection{Further discussion of $\FrQ$ Primality}

Firstly I only wish to consider {\sl direct} measurements rather than inferences 
(which could e.g. involve inferring distances from momentum-type measurements connected to photon momenta as in sight or spectral lines). 
One can dissociate from this problem by thinking in terms of simple `blind feeling out' measurements on a scale small enough that one
 can reach out to feel (for all that there will often be strong practical limitations on this).   
The virtue of this approach is that one's configuration measurements for particle mechanics are all {\sl geometrical}: 
local angles, adjoining rulers to the separations between objects; the blind Geometer's work is free from the impact connotations that enter his sighted colleague's work. 
[The latter's methodology will however often have greater {\sl practicality}, e.g. if both worked as surveyors...]

Next, in GR in geometrodynamical form, Riemannian 3-geometry measurements of space are again cleanly separated out from 
gravitational momentum/extrinsic curvature.  
Note how in the mechanics--Geometrodynamics sequence of physical theories, one can pass from a simple notion of relational 
geometry measurements being operationally primary to a more complicated one (argued for e.g. in \cite{Christodoulou1, Christodoulou3}. 
A theme that may then support part of the argument for the primality of configurational measurements is that geometrical measurements 
are primary;\footnote{Moreover, this is {\sl not} to be because of geometry constituting a layer of structure prior to the account of the 
physical objects themselves, since the context in which configurations are being considered as primary is a relational one and one which is 
specifically to encompass GR, for which there is indeed no prior fixed notion of Riemannian geometry.}
this exemplifies Criterion 3).

From this perspective, it is somewhat incongruous for triad information to be considered as momentum-like within the 
Ashtekar variables type approaches given that this is another form of space-geometric information.    
It would likewise be somewhat incongruous if the Ashtekar $\underline{\mathbb{A}}$ were to be only an indirectly inferible quantity.\footnote{Here one 
usually argues (see e.g. \cite{PullinGambini} instead for Criterion 3) in terms of the progression from Electromagnetism to Yang--Mills Theory to GR in Ashtekar variables 
form as theories based on notions of connection.  
As regards the current Appendix's theme of operational distinction and most primary types of measurements, firstly I note that in classical 
electromagnetism it is $\underline{\mB}$ that is measured (rather than $\underline{\mA}$), and that this is operationally distinct from  measuring $\underline{\mE}$. 
Secondly, mentioning Yang--Mills Theory brings one face to face with problems with centring one's thinking in terms of classical measurements; 
of course, QM offers much wider challenges than this the idea of operational primality of measurements of configurations.   
[This idea may only reflect a simple top-down thinking, c.f. Sec \ref{QM-Intro}.]
As regards measuring Ashtekar's $\underline{\mathbb{A}}$, one usually argues that one can make do with measuring curvature quantities and eventually 
inferring $\underline{\mathbb{A}}$ from them.

Moreover, the solidity of the above sequence as a first-principles scheme partly rests on the notion of `connection' receiving a conceptually 
clear treatment (to be more precise, the corresponding notion of holonomy is particularly important for LQG).  
For Ashtekar variables, this is itself a geometrical perspective, for a yet more complicated notion of geometry than GR's usual 
(semi)Riemannian geometry. 
[Both Ashtekar's $\underline{\mathbb{E}}$ and $\underline{\mathbb{A}}$ can be construed of as geometrical objects.  
However, it is the $\underline{\mathbb{E}}$ whose sense of geometry (triad variable) ties more directly to the more straightforwardly realized spatial Riemannian geometry.  
The $\underline{\mathbb{B}}$-field does have more purely geometrical connotations by its being the curvature associated with the connection, 
but the physical realization of this geometry is more subtle than that of mechanics, 3-metric geometries or their triad counterpart.]

Moreover, the Ashtekar variables case does run into some Criterion 1)-type contentions (e.g. \cite{Samuel, MarcMarc}) as regards which features are appropriate for a 
connection serving this purpose, though Thiemann \cite{Thiemann} terms the first of these references `aesthetic').  
On the other hand, the solidity of this scheme as obtained from rearrangement of the usual geometrodynamical conceptualization of canonical 
GR rests on whether it is appropriate to extend the classical phase space to include degenerate configurations, as well as on canonical transformations.}  
%
On the other hand, the foundations of Geometrodynamics are {\sl not} incongruous in this way, though the above-linked footnote does place 
limitations on, and give alternatives to, this particular sense of incongruence fostered by this Sec's tentative expansion of the point 
of view that $\FrQ$ is primary.  
\noindent The debate about whether lengths, impacts, times, frequencies... are the most primary things to measure is indeed an old and 
unsettled one (see also Sec \ref{ObsApp}).  
%
%
%
%
I wish to make a somewhat different point too: that, regardless of primality, one can tell apart whether 
what one is measuring is instantaneous-configuration information or impact information.  

\mbox{ } 

\noindent If canonical transformations are held in doubt, this affects internal time, Histories Theory, Ashtekar variables and the recent 
linking theory approach; the second and fourth of these make {\sl more} than the usual amount of use of canonical transformations.
This Appendix suggests the possibility of developing Physics in the {\sl opposite} direction.

\subsubsection{Categorization extension in anticipation of quantization?}

One possible motivation for this extension is that quantization can be formally understood as a (bad) functor, linking the classical to the quantum-mechanical.
I argue against this in a number of ways in Sec \ref{QM-Intro}.  

\mbox{ } 

\noindent{\bf Categories} \textgoth{C} = (\textgoth{O}, \textgoth{M}) consist of objects \textgoth{O} and {\it morphisms} \textgoth{M} (the maps between the objects, 
\textgoth{M}: \textgoth{O} $\longrightarrow$ \textgoth{O}, obeying the axioms of domain and codomain assignment, identity relations, associativity 
relations and book-keeping relations, see e.g. \cite{Lawvere, LawRose, MacLane} for details.)

\mbox{ } 

\noindent {\bf Functors} are then maps \textgoth{\Large F}: \textgoth{C}$_1$ $\longrightarrow$ \textgoth{C}$_2$ that obey various further 
axioms concerning domain, codomain, identity and action on composite morphisms.  

\mbox{ }

\noindent It is then clear that so far we have only been studying objects [Relationalism 3), Appendix \ref{Dyn1}.B], but we should have also been studying morphisms.
In particular the morphisms corresponding to $\FrQ$ are the so-called {\it point transformations} (p. 15 of \cite{Lanczos}) {\bf Point}: $\FrQ \longrightarrow \FrQ$.  

\noindent Note 1) For 1- and 2-$d$ RPM's, this makes $\mbox{Point}(\FrQ/\FrG)$ the coordinate transformations of known manifolds as listed in Sec \ref{Q-Geom},  
so this situation is well under control.

\noindent Note 2) $\FrG$-quotient = $\FrG$-constrain = $\FrG$-reduce: ($\FrQ, \mbox{Point}) \longrightarrow ({\FrQ}/\FrG, {\mbox{Point}(\FrQ/\FrG)}$).

\noindent The morphisms of phase space are  the canonical transformations Can: Phase $\longrightarrow$ Phase.  

\mbox{ } 

\noindent The RigPhase perspective {\sl considers} resisting the suggestion that what mathematically preserves the Poisson bracket 
should be the associated morphisms due to the previous SSec's argument of configuration variables and momenta being operationally distinguishable. 
Thus it is questionable to take bracket preserving morphisms that do not respect this physical insight.  
An alternative would be a rigged version RigPhase of the phase space that preserves this distinction.
Here, the $\FrQ$-first interpretation of this is that the $\fQ$'s are fundamental and each $\fP$ then follows by conjugation and only transforms on $\FrQ$ are primarily meaningful 
(If the $\fQ$'s change, then the $\fP$'s follow suit by being the new conjugates, without any extra freedom in doing so.) 
Thus the morphisms of this are just Point again, the conjugate momenta being held to follow whatever the coordinates are rather than having their own morphisms 
(this approach's intent {\sl is} to put configurations first, so one should not be surprised at these and not the momenta having the primary transformation properties).  
See Fig \ref{RigPhase} for the ensuing options.  

{            \begin{figure}[ht]
\centering
\includegraphics[width=0.95\textwidth]{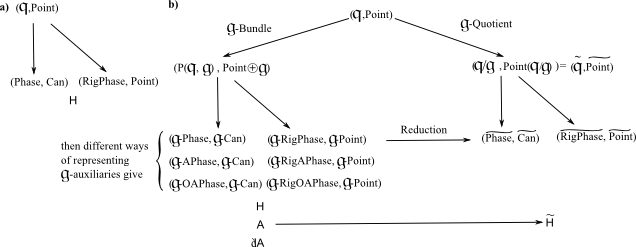}
\caption[Text der im Bilderverzeichnis auftaucht]{        \footnotesize{The options without and with the involvement of a $\FrG$. 
[Once a $\FrG$ is involved, manifestly temporally relational APhase and OAPhase alternatives arise. } }
\label{RigPhase} \end{figure}          }

\noindent As regards further involvement of categories, we shall see in Sec\ref{Cl-Str} and Sec \ref{QM-Str} that Perspecting and Propositioning, which are themselves at 
least partly inter-related, lead toward Topos Theory, a Topos being a category armed with some extra structures that make it behave much 
like the category of sets, $\underline{\mbox{\textgoth Sets}}$ (see Sec \ref{QM-Str} for more details).  

\end{subappendices}

\vspace{10in}

\section{Classical RPM solutions}\label{Cl-Soln}

\subsection{Some particular potentials for RPM's}\label{Scram}

Dynamics involves all of the kinetic metric, the potential and the value of the energy (often its sign in particular).
Thus this section considers which specific potentials to use for RPM's.   
A priori, pure-shape RPM has a lot of potential freedom (any potential homogeneous of degree 0 is valid), while scaled RPM can have any potential at all.  
Thus I consider scaled RPM first.  
The no/constant potential case is often the easiest case, but tends to exhibit unbounded/non-normalizable behaviour,\footnote{`Often' and `tends' 
are needed here due to generally curved nature of configuration spaces and use of nontrivial PPSCT representations.}  
so I tend to go beyond this case. 
(Moreover, the free problem does serve as as an approximation is applicable to scattering problems, and, elsewise, being a thing to solve perturbatively about.)
I look at HO-type potentials and then at particular cosmologically-inspired potentials.

\subsubsection{Power-law potentials}\label{PLP}

Many potentials in Physics are proportional to some power of the separation between two particles, 
$k_{IJ}||\underline{q}^I - \underline{q}^J||^{\sn}$, or are linear combinations of these.  
\noindent In the 3-particle case, I use $k_1$ as shorthand for $k_{23}$ etc.  
I also replace some of the $k$'s with specific labelling letters in the below examples.   

\mbox{ } 

\noindent For scaled RPM then one can just consider the above power-law potentials (and sums of these).    
Then, in terms of relative Jacobi coordinates, power-law potentials take on forms built out of $||\brho^i||$ or, more generally, $\underline{\brho}^i\cdot \underline{\brho}^j$.

\subsubsection{Special potentials}\label{Special}

\noindent In pure-shape RPM, power-law potentials and their $\{\q^i - \q^j\}\cdot\{\q^j - \q^l\}$ generalization need to occur multiplied by suitable powers of $\mI$ 
(i.e. $\mI^{-\sn/2}$) in order to satisfy the homogeneity of degree zero consistency condition exposited in Sec \ref{Examples}.  
This is not too much of a problem in that powers of $\mI$ are post-variationally constant in pure-shape RPM.
Thus overall, whilst power-law potentials (other than the constant) are disallowed in pure-shape RPM, they can nevertheless be {\sl mimicked} by forming dimensionless ratios with $\mI$.  

\noindent Examples of {\sl separable} power laws for e.g. triangleland are then  
\beq
\ttV_{(\sn, \sm)} \propto  \mn_1^{\sn - \sm}\mn_2^{\sm} \propto \{{\cal R}/{\sqrt{ 1 + {\cal R}^2}}\}^{\sn - \sm}\{{1}/{\sqrt{ 1 + {\cal R}^2}}\}^{\sm} 
\propto \mbox{sin}^{\sn - \sm}\mbox{$\frac{\Theta}{2}$}\mbox{cos}^{\sm}\mbox{$\frac{\Theta}{2}$}
\mbox{ }\mbox{ or }\mbox{ }
V_{(\sn, \sm)} \propto {{\cal R}^{\sn - \sm}}/{\{1 + {\cal R}^2\}^{{\sn}/{2} + 2}} \mbox{ } \mbox{ } .  
\eeq
The most physically relevant subcase of this are the power-law-mimicking potentials $\nm = 0$.    
These correspond to potential contributions solely between particles 2, 3.  
The case $\nm = \mn$ are potential contributions solely between particle 1 and the centre of mass of particles 2, 3 (which is less widely physically meaningful). 
It also turns out that Sec \ref{CPST}'s duality map sends $\ttV_{(\sn,\sm)}$ to $\ttV_{(\sn,\sm - \sn)}$.

\mbox{ } 

\noindent It turns out that `special case' can be concretely characterized by the number of conserved quantities (as per Sec \ref{Dyn1}). 
The other side of this coin however is that special cases are too much of a simplification for some purposes, since these conserved quantities 
signify restrictions to the physics such as no angular momentum or dilational momentum {\sl exchange between subsystems}.

\subsubsection{HO-like potentials: motivation and limitations}\label{Arg-for-HO}

In pure-shape RPM, I term these `HO-like' because they come divided by I as per above.   
In scaled RPM, however, one has entirely standard HO potentials.  
\noindent HO-like potentials are motivated by 

\noindent 1) HO potentials being ubiquitous in Theoretical Physics. 

\noindent 2) Being among the most analytically tractable potentials.  

\noindent 3) Being quantum-mechanically well behaved/bounded.

\noindent 4) The scaled-RPM version of these giving relevant terms from the perspective of analogue models of Cosmology (see \cite{Cones} or Sec \ref{MechCos}). 

\noindent Caveat 1) Such models are however atypical in their simpleness.

\noindent Caveat 2) Nor do they by themselves parallel the dominant scale dynamics of commonly-used cosmological models.

\subsection{HO(-like) potentials for RPM's}\label{HO-An}

\subsubsection{Analogies with ordinary Physics}\label{HO-An-1}

Some analogies for 4-stop metroland are as follows. 
The potential $V_0\{1 - \mbox{cos}\,2\theta_{\sss\sp}\}$ [c.f. form 3 of (\ref{39})] occurs in modelling the rotation of a linear molecule in a crystal \cite{P,Stern,PW,WS}.  
Here, the further aspects of the analogy are that the axis and rotor in question are provided by the linear molecule itself, `energy' $\longleftrightarrow$ energy up to a constant, 
\beq
K_1/2 \longleftrightarrow 2V_0 \mbox{ up to the same constant difference as in the energy analogy } ,
\eeq
\beq
B \longleftrightarrow -2V_0 \mbox{ } .  
\eeq
The potential $-\Bigupalpha_{||}{\lme}^2\,\mbox{cos}^2\theta$, for $\Bigupalpha_{||}$ the polarizability along the axis occurs in the study 
\cite{C,TS} of e.g. the $CO_2$ molecule in a background electric field $\lme$ is involved in the theoretical underpinning for Raman spectroscopy.     
The analogy here is (\ref{ana4}), 
\beq
B \longleftrightarrow - {\Bigupalpha_{||}}{\le}^2/2 \mbox{ } . 
\eeq

\noindent Some useful mathematical analogies for pure-shape triangleland with multiple HO potentials are as follows.   
\beq
\mbox{ very special HO } \longleftrightarrow \mbox{ linear rigid rotor } ,
\eeq
\beq
\mbox{ special HO } \longleftrightarrow 
\mbox{linear rigid rotor in a background homogeneous electric field in the axial (`z')-direction } , 
\eeq
\beq
\mbox{ general HO } \longleftrightarrow 
\mbox{ linear rigid rotor in a background homogeneous electric field in an arbitrary direction }. 
\eeq
In particular, this classical problem has 
\beq
T_{\sr\so\st\so\sr} = \mI_{\sr\so\st\so\sr}\{\dot{\theta}_{\sss\sp}^2 + \mbox{sin}^2\theta_{\sss\sp}\dot{\phi}_{\sss\sp}^2\}/{2} 
\mbox{ } , \mbox{ } V_{\sr\so\st\so\sr} = - \lm\,\le\,\mbox{cos}\theta \mbox{ } ,
\eeq
where $\mI_{\sr\so\st\so\sr}$ is the single nontrivial value of the moment of inertia of the linear rigid rotor, 
and ${\lm}$ is the dipole moment component in that direction.  
Thus the correspondence is (\ref{PaMa}), 
\beq
\mbox{(energy)/4 -- (sum of mass-weighted Jacobi--Hooke coefficients)/16 } 
\mbox{ } = \mbox{ }  
{\ttE}/{4} - A 
\mbox{ } = \mbox{ } 
\overline{\ttE} - A \mbox{ } \longleftrightarrow \mbox{ } E \mbox{ = (energy) } ,
\eeq
\beq
\mbox{ (difference of mass-weighted Jacobi--Hooke coefficients) = } \mbox{ } 
B \mbox{ } \longleftrightarrow \mbox{ } - {\lm\,\le} \mbox{ } . \mbox{ } 
\eeq
These all being well-studied at the quantum level \cite{TS, Messiah, Hecht}, this identification is of considerable value in solving the relational problem in hand.  

\mbox{ } 

Next, note that ($N$, 1) scaled RPM's are analogous to ordinary multi-HO's.
Also, the very special scaled RPM HO PPSCT-maps to the Kepler problem with 
\beq
\mbox{ (radius) } =  r \mbox{ } \longleftrightarrow \mbox{ } \mI \mbox{ (total moment of inertia) } , 
\mbox{ } 
\eeq
\beq
\mbox{ (test mass) } =  m \mbox{ } \longleftrightarrow \mbox{ } 1 \mbox{ } , \mbox{ } 
\eeq
\beq
\mbox{ (angular momentum) }   =       L \mbox{ } \longleftrightarrow \mbox{ } \sfJ \mbox{ (relative angular momentum) } , \mbox{ } 
\eeq
\beq
\mbox{ (total energy) } = E \mbox{ } \longleftrightarrow \mbox{ } - A = 
\mbox{ -- (sum of mass-weighted Jacobi--Hooke coefficients)/16 }     \mbox{ } ,  
\eeq
\beq
\mbox{and } \mbox{ } 
\mbox{ (Newton's gravitational constant)(massive mass)(test mass) } = GMm \mbox{ } \longleftrightarrow 
\mbox{ } \overline{E} \mbox{ (total energy)/4 }  
\eeq
[or to the 1-electron atom Coulomb problem with the last analogy replaced by 
\beq
\mbox{(nuclear charge)(test charge of electron)/4$\pi$(permettivity of free space) } = (Ze)e/4\pi\epsilon_0 \mbox{ } \longleftrightarrow \mbox{ } \overline{E}  \mbox{ (total energy)/4] } 
\mbox{ } .
\eeq
Note 1) The positivity of the Hooke's coefficients translates to the requirement that the gravitational or atomic energy be negative, i.e. to bound states.  
Also, the positivity of $\ttE$ required for classical consistency corresponds to attractive problems like 
the Kepler problem or the atomic problem being picked out, as opposed to repulsive Coulomb problems.

\noindent Note 2) The special case corresponds to the same `background electric field'  that the rotor was subjected to above.  
Moreover, this is proportional to cos$\,\Theta$, which is analogous to cos$\,\theta_{\sss\sp}$, which is in the axial (`$Z$') direction. 
But it is {\sl not} the well-known mathematics of the axial (`$z$') direction {\sl Stark effect} for the atom, which involves, rather, $r\, \mbox{cos}\,\theta_{\sss\sp}$.  
Nevertheless, it is both closely related to the rotor and to the mathematics of the atom in parabolic coordinates (see e.g. \cite{LLQM, Hecht}).  
The general case then corresponds to the `electric field' pointing in an arbitrary direction.

\subsubsection{3-stop metroland HO's}\label{HO-Pots}

The 3-HO-like potential for pure-shape 3-stop metroland is $\ttV = h_{23}||\q_2 - \q_3||^2/2\mI +$ cycles for 
$h_{IJ}$ the Hooke's coefficients for the springs between the $I$th and the $J$th particles.  
This potential can be re-expressed as 
\beq
\ttV = K_1 \mn_1^2/2 + K_2 \mn_2^2/2 + L \mn_1 \mn_2  
  =  A + B\,\mbox{cos}\,2\varphi + C\mbox{sin}\,2\varphi  
  =  A + B\,\mY_{2c} + C\,\mY_{2s} \mbox{ } ,  
\eeq
\beq
\mbox{ from }  \mbox{ } K_1 = \big\{ h_{23} + \{h_{13}m_2^2 + h_{12}m_3^2\}/\{m_2 + m_3\}^2\big\}/\mu_1  
, \mbox{ }
K_2 = \{h_{12} + h_{13}\}/\mu_2 
, \mbox{ }
L = 2\{h_{13}m_2 - h_{12}m_3\}/\{m_2 + m_3\}\sqrt{\mu_1\mu_2} 
\mbox{ } , \mbox{ }
\eeq
\beq
A = \{K_1 + K_2\}/2 \mbox{ } , \mbox{ } \mbox{ } B = \{K_2 - K_1\}/2 
\mbox{ } , \mbox{ } \mbox{ } C = L/2 \mbox{ } .
\eeq
I also use $\omega_{\sttr} = \sqrt{K_{\sttr}}$, which are frequencies, and likewise in terms of subsequent K's below.   
This has a special case, for $C = 0$ corresponding to $m_2h_{13} = m_3h_{12}$:
\beq
\ttV = K_1\mn_1^2/2 + K_2\mn_2^2/2 = A + B\, \mbox{cos} \, 2\varphi  \mbox{ } .  
\eeq
Its physical meaning is that the resultant force of the second and third `springs' points along the line joining the centre of mass of 23 to the position of particle 1. 
There is also a very special case, for $B = 0 = C$, corresponding to $K_1 = K_2$, which amounts to $m_1h_{23} = m_2h_{13} = m_3h_{12}$, 
high-symmetry situation the various potential contributions balance out to produce the constant potential 
\beq
\ttV = A \mbox{ } .  
\eeq

\subsubsection{4-stop metroland HO's}\label{HO-4Stop}

Here, the HO-type potential is in general a linear combination of 6 inter-particle springs divided by I.
This can be re-expressed as
$$
\ttV   = \sum\mbox{}_{a = 1}^3\{\ttK_{a}\mn^{a\, 2}/2 + \ttL_{a}\mn^{b}\mn^{c}\}
       = \ttA + \ttB\,\mbox{cos}\,2\theta + \ttC\,\mbox{sin}^2\,\theta\,\mbox{cos}\,2\phi + 
\ttF\,\mbox{sin}^2\theta\,\mbox{sin}\,2\phi + \ttG\,\mbox{sin}\,2\theta\,\mbox{cos}\,\phi + 
\ttH\,\mbox{sin}\,2\theta\,\mbox{sin}\,\phi
$$
\beq
= a + b\,\mY_{2,0}(\theta) + c\,\mY_{2, 2\scc}(\theta, \phi) + f\,\mY_{2,2\sss}(\theta, \phi) + g\,\mY_{2,1\scc}(\theta, \phi) + h\,\mY_{2,1\sss}(\theta, \phi)  
\label{39}
\eeq  
\beq
\mbox{ for } \mbox{ } \ttA = \frac{1}{4}\left\{\ttK_3 + \frac{\ttK_1 + \ttK_2}{2}\right\} \mbox{ } , \mbox{ } 
\ttB = \frac{1}{4}\left\{\ttK_3 - \frac{\ttK_1 + \ttK_2}{2}\right\} \mbox{ } , \mbox{ } \ttC = \frac{\ttK_1 - \ttK_2}{4} \mbox{ } .  
\eeq
E.g. the \{12,34\} H-cluster is simplified ($\ttE$, $\ttF$, $\ttG$ = 0) by considering 12, 34 and +$\times$ springs only.  
One can alternatively remove $\ttE$, $\ttF$, $\ttG$ by diagonalization \cite{AF}, though this case does rotate the physical interpretation so that directions picked out by the potential 
no longer coincide with kinematically picked out directions of a 3 DD axis system or a \{T, M$^*$D, M$^*$D\} axis system.   
The $\mY$'s are spherical harmonics (their c and s subscripts standing for cosine and sine $\phi$-parts) 
and the precise form of the constants $a$, $b$, $c$ is not required for this Article.  
One can imagine whichever of these problems' potentials as a superposition of familiar `orbital shaped' lumps. 
Though such a superposition will of course in general alter the number, size and position of 
peaks and valleys according to what coefficients each harmonic contribution has.

This has a special case, for $\ttC = 0$ corresponding to $\ttK_1 = \ttK_2$, i.e. that cluster 1 and cluster 2 
has the same `constitution': the same Jacobi--Hooke coefficient per Jacobi cluster mass.  
This is a kind of `homogeneity requirement' on the `structure formation' in the cosmological analogy,
\beq
\ttV = \ttA + \ttB\, \mbox{cos}\,2\theta  \mbox{ } .  
\eeq
There is also a very special case, for $\ttB = 0 = \ttC$, corresponding to $\ttK_1 = \ttK_2 = \ttK_3$, for which 
high-symmetry situation the various potential contributions balance out to produce the constant, 
\beq
\ttV = \ttA \mbox{ } .  
\eeq
\mbox{ } \mbox{ }  Additionally the $\ttB << \ttA$ perturbative regime about the very special case signifies $\ttK_1 + \ttK_2 << \ttK_3$ 
so the inter-cluster spring is a lot stronger than the intra-cluster springs. 
This is in some ways is analogous to scalefactor dominance over inhomogeneous dynamics in cosmology.  
On the other hand, the $\ttC << \ttA$ regime corresponds to either or both of the conditions 
$\ttK_1 + \ttK_2 << \ttK_3$, $\ttK_1 \approx \ttK_2$ the latter of which signifies high {\it contents homogeneity in the material coefficients}. 
(This means that the particle clusters that make up the model universe are, among themselves, of similar constitution.) 
The multiplicity of forms of writing the potential above is useful to bear in mind in searching for 
mathematical analogues for the present problem in e.g. the Molecular Physics literature (c.f. Sec \ref{41Mol}).

\subsubsection{$N$-stop metroland HO's}\label{HO-NStop}

This extension is 
\beq
\ttV = \sum\mbox{}_{p = 1}^{n}\ttK_{p}\mn_{p}^2(\theta_{\barr})/2 + 
\sum\sum\mbox{}_{p > q}\ttL_{pq}{\mn}_{q}(\theta_{\barr}){\mn}_{q}(\theta_{\barr}) \mbox{ } , 
\eeq
can similarly be recast in terms of ultraspherical harmonics and admits a hierarchy of (very)$^k$ special subcases whose significance extends the previous example's.

\subsubsection{Triangleland HO's}\label{HO-Tri}

The HO-type potential can here be re-expressed as 
\beq
\ttV = K_1\mn_1\mbox{}^2/2 + K_2\mn_2\mbox{}^2/2 + L \underline{\mn}^1 \cdot \underline{\mn}^2 = 
       A + B\,\mbox{cos}\,\Theta + C\,\mbox{sin}\,\Theta\,\mbox{cos}\,\Phi
\eeq 
with the $A$, $B$, $C$ each being 1/4 of what they were for the 3-stop metroland example.
The occurrence and significance of special and very special cases then follows suit (note now that one requires the special case in order to have separability).  

\mbox{ }  

\noindent Note 1) Contrast with the 4-stop metroland counterpart is also interesting at this point -- here $\mY_{0,0}$ and just two of the first-order spherical harmonics arose. 
(This is because the also-quadratic $|\brho_1 \cr \brho_2|_3$ is not a piece of the most general multi-harmonic oscillator potential between 3 particles in 2-$d$).  

\noindent Note 2) The quadrilateralland counterpart of this is given in \cite{QuadII}.

\subsubsection{Qualitative analysis for pure-shape 4-stop metroland HO}\label{Desc-Qual}

This is important in simplifying both this Sec's solution study and Part II's QM counterpart.  
4-stop metroland's special potential ($\ttC$ = 0) either having wells at both poles or an equatorial bulge that is shallower at the poles depending on the sign of $\ttB$: 
here both of the near-polar regimes are simultaneously realisable. 
%
{            \begin{figure}[ht]
\centering
\includegraphics[width=0.8\textwidth]{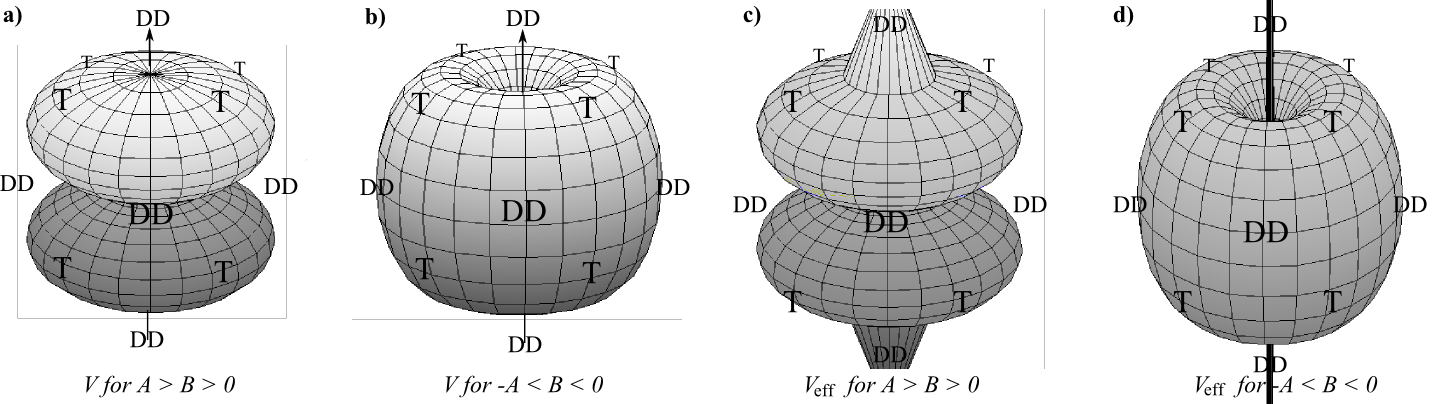}
\caption[Text der im Bilderverzeichnis auftaucht]{        \footnotesize{    Sketches of $\ttV$ over the 
sphere for the mechanically significant cases a) the `peanut' $A > B > 0$ and b) the `tyre' 
$-A < B < 0$.  
The first has barriers at the poles and a well around the equator, while the second has wells 
at the pole and a barrier around the equator.  
In each case, considering $\ttV_{\se\sf\sf}$ for $\sfD \neq 0$ adds a spike at each pole. 
(This now 
means that for $\sfD \neq 0$ both islands cannot simultaneously collapse to their generally 
distinct centre of mass points).
Finally, note that this potential is axisymmetric and has reflection symmetry about its equator so its symmetry 
group is $\mathbb{D}_{\infty} \times \mathbb{Z}_2$ ($\mathbb{D}$ denotes dihaedral).  
If this is aligned with a DD axis of the physical interpretation, the overall problem retains a 
$\mathbb{D}_4 \times \mathbb{Z}_2$ symmetry group, of order 16.  
    }        }
\label{AF-Fig3}\end{figure}           }

\noindent Next, I consider small and large regimes for the special case.  
More precisely, these are near-North Pole and near-South Pole regimes in $\theta$ but become large and small regimes in terms of ${\cal R} = \mbox{tan$\frac{\theta}{2}$}$.  
For this (including changing to the tilded PPSCT representation), and using `shifted energy' 
$\ttE^{\prime} := \ttE - \ttA - \ttB$
\beq
\overline{\ttW} := \overline{\ttE} - \overline{\ttV} = 4\ttE^{\prime}/{\{1 + {\cal R}^2\}^2} + {32\,\ttB\,{\cal R}^2}/{\{1 + {\cal R}^2\}^4} \mbox{ } . 
\label{r}
\eeq
Then the near-North Pole regime (${\cal R} << 1$) maps to the problem with flat polar kinetic term and  
\beq
\ttW = 4\ttE^{\prime} + 8\{4\,\ttB - \ttE^{\prime}\}{\cal R}^2 
\label{oc}
\eeq
up to $O({\cal R}^4)$.  
This has the mathematics of a 2-$d$ isotropic harmonic oscillator,  
\beq
\ttW = \lepsilon - \Omega^2{\cal R}^2/2 \mbox{ } ,
\label{qus} 
\eeq
provided that the `classical frequency' (for us with units of $\mI$/time) $\Omega \mbox{ } < 0$ 
(else it would be a constant potential problem or an upside-down harmonic oscillator problem), 
alongside $\lepsilon > 0$ to stand a chance of then meeting classical energy requirements.  
Writing $\lepsilon$ and $\Omega^2$ out by comparing the previous two equations, these inequalities signify that $2\ttE > \ttK_3$ and $2\ttE > \ttK_3 + 2\{\ttK_3 - \ttK_1\}$. 
The latter is more stringent if $\ttK_3   >  \ttK_1$    (`stronger inter-cluster binding') 
          and less stringent if $\ttK_3 \leq \ttK_1$ (`weaker inter-cluster binding').
One can also deduce from the first of these and $\ttK_3 \geq 0$ (spring) that $\ttE > 0$.

Next, note that the near-South Pole regime (${\cal R} << 1$) maps to the problem with flat polar kinetic term and  
\beq
\ttW = 4\ttE^{\prime}/{{\cal R}^4} + {8\{4\ttB - \ttE^{\prime}\}}/{{\cal R}^6} 
\label{w}
\eeq
up to  $O({1}/{\cal R}^8)$. 
Moreover, ${\cal U} = 1/{\cal R}$ maps the large case's (\ref{r}) to the small case's (\ref{w}), so this is also an isotropic harmonic oscillator -- in $({\cal U}$, $\phi$) 
coordinates and with the same $\lepsilon$ and $\Omega$ as above.  
This is an exact `large--small' or `antipodal' duality.  
It halves the required solving to understand $\theta$ $\approx 0$ and $\approx \pi$. 

\mbox{ }

\noindent Note: Another lesson learnt during \cite{08I, 08II, AF} is that study of {\sl second} large/small approximations is considerably more profitable than that of first ones.

\subsubsection{Qualitative analysis for pure-shape triangleland HO}

{            \begin{figure}[ht]
\centering
\includegraphics[width=0.9\textwidth]{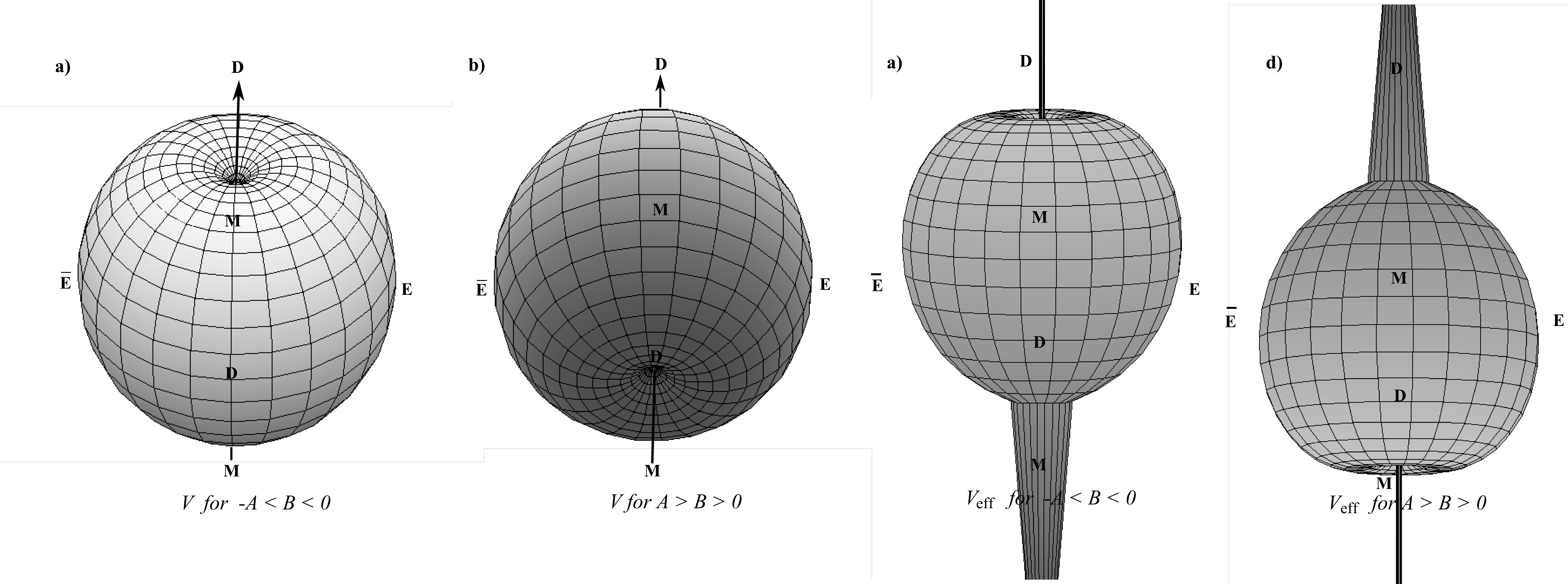}
\caption[Text der im Bilderverzeichnis auftaucht]{        \footnotesize{
The heart-shaped potential for triangleland.    
The first has a well about the North `D' pole and a barrier about the South `M' pole, and the second has vice versa. 
In each case, considering $\ttV_{\se\sf\sf}$ for $\sfS \neq 0$ adds a spike at each pole. 
(This means that for $\sfS \neq 0$, the corresponding cluster cannot collapse to a point or have the third particle lie on its centre of mass.) 
Finally, note that this potential is axisymmetric so its symmetry group is $\mathbb{D}_{\infty}$.   
If this is aligned with a D-M axis of the physical interpretation as depicted, the overall problem retains a $\mathbb{D}_3$ symmetry group, of order 6.  }        }
\label{Heart}
\end{figure}  }
  
\noindent For triangleland, the regime in which the (1)-clustering's notion of special applies is $|B| \leq A$.  
In spherical polars, this corresponds to a heart or spheroidal shaped potential each with one end bulkier than the other. 
Thus one gets a well centred on either on North pole (D point) or the South pole (M point), depending on the sign of $B$ (i.e. which of $K_1$ and $K_2$ is larger).  
[The $\sfJ \neq 0$ case adds narrow infinite skewers to this that pass through the poles.]
$K_1 > K_2$ has its well centred on the North pole, i.e. sharp(1) triangles. 
The physics here is that the inter-cluster spring binding \{23\} to the `external particle' 1 is stronger than \{23\}'s intra-cluster spring. 
$K_1 < K_2$ has its well centred on the South pole, i.e. flat(1) triangles. 
The physics here is that the is intra-cluster spring of \{23\} is more tightly binding than the inter-cluster spring between \{23\} and 1. 
Thus only one of the near-polar regimes is actualized at once for a given problem, as, while small motions about the thick end of the potential are also admissible. 
These are unstable to escaping by rolling to where the potential is thinner.

In triangleland, the heart/spheroidal potential, even if inclined at an angle to the D-M axis (general case), continues to pick out a `small regime' near its thin end.  
The physical meaning of this region does, however, vary with the angle.  
The potential confining $\Theta_{(\gamma)}$ to be small means that $ellip(\gamma) = \mbox{cos}\,\gamma$ in this region, so that any value of sharpness or flatness can now be picked out 
(if it is not immediately clear to the Reader, consult the next SSec for the definition of $\Theta_{(\gamma)}$).  
E.g. $\gamma = \pi/2$ (pure $C$ term) picks out regular triangles (neither sharp nor flat).

Also note that the triangleland potential now furthermore breaks the tessellation group. 
For, the heart/spheroidal potential has symmetry group $\mathbb{D}_{\infty}$ and involves an axis perpendicular to the E$\bar{\mE}$ axis. 
Thus the overall problem retains just a $\mathbb{Z}_2$ reflection symmetry about the plane of collinearity.

Working in spherical coordinates, set $0 = {\pa{\ttV}}/{\pa\Theta} = - B\,\mbox{sin}\,\Theta + 
C\mbox{cos}\,\Theta\,\mbox{cos}\,\Phi$, $0 = {\pa{\ttV}}/{\pa\Phi} = -C\,\mbox{sin}\,\Theta\,\mbox{sin}\,\Phi$ to find the critical points.
These are at $(\Theta, \Phi) = (\mbox{arctan}(C/B), 0)$,  $(-\mbox{arctan}(C/B), \pi)$ which are 
antipodal (see Fig \ref{Fig4-too}); in fact the potential is axisymmetric about the axis these lie on. 
The critical points are, respectively, a maximum and a minimum.  
[The very special case $B = C = 0$ is also critical, for all angles -- this case ceases to have a preferred axis.] 

{            \begin{figure}[ht]
\centering
\includegraphics[width=0.3\textwidth]{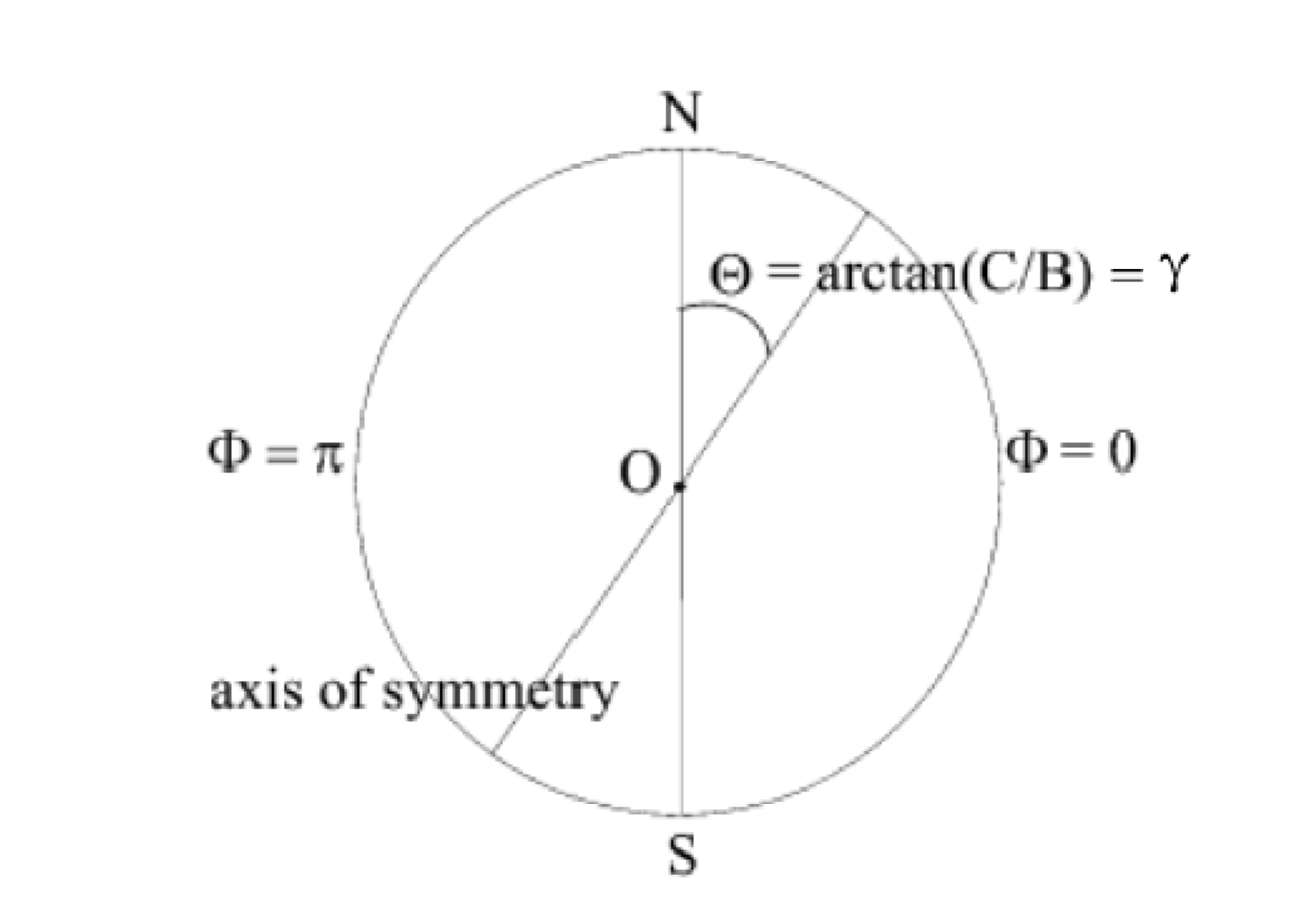}
\caption[Text der im Bilderverzeichnis auftaucht]{        \footnotesize{    The preferred axis. 
The North pole N is physically a D and the South pole S an M.  
}        }
\label{Fig4-too}
\end{figure}            }

A move useful for the study of the potential is to pass to (${\cal R}$, $\Phi$) coordinates; then the special potential is of form 
\beq
2\overline{\ttV} = \{K_1{\cal R}^2 + K_2\}/\{1 + {\cal R}^2\}^3 \mbox{ } .
\eeq
Now, ${\cal R}$ small corresponds to $\Theta$ small, for which 
\beq
\breve{\ttU} + \breve{\ttE} = \{\breve{\ttE} - A - B\} - C\Theta\,\mbox{cos}\,\Phi  + B\Theta^2/2 + O(\Theta^3) \mbox{ } , \mbox{or}   
\eeq
\beq
2\{\breve{\ttU} + \breve{\ttE}\} = 2\ttE - K_2 - L\,{\cal R}\,\mbox{cos}\,\Phi - \{4\fE + K_1 - 3K_2\}{\cal R}^2 + O({\cal R}^3) := 
u_0 - L\,{\cal R}\,\mbox{cos}\Phi  - u_2 {\cal R}^2+ O({\cal R}^3)
\mbox{ } .  
\label{assmall}
\eeq
Thus the leading term is a constant, unless $u_0 = 2\ttE - K_2 \mbox{ }(\mbox{ } \propto \overline{\ttE} - A - B \mbox{ }) = 0$.  
In this case, it is linear in $\Theta$ or ${\cal R}$ and with a cos$\Phi$ factor, unless also 
$L \mbox{ }(\mbox{ } \propto C \mbox{ }) = 0$ (which is also the condition for the `special' case). 
In this case, it is quadratic in $\Theta$ or ${\cal R}$, unless $B = 0$. 
(Given previous conditions, this is equivalent to $u_2 = 4\ttE + K_1 - 3K_2 = 0$). 
This means that one is in the $K_2 = 2\ttE$ subcase of the `very special' case, for which $\ttU + \ttE$ has no terms at all.

${\cal R}$ large corresponds to the supplementary angle $\Xi := \pi - \Theta$ being small, so 
\beq
\overline{\ttU} + \overline{\ttE} = \{\overline{\ttE} - A - B\} + C\Xi\,\mbox{cos}\,\Phi + B\Xi^2/2 + O(\Xi^3) \mbox{ } , \mbox{ or }
\eeq
\beq
2\{\widetilde{\ttU} + \widetilde{\ttE}\} = \{2\ttE - K_1\}{\cal R}^{-4} - L{\cal R}^{-5}\mbox{cos}\,\Phi - \{4\fE + K_2 - 3K_1\}{\cal R}^{-6} + O({\cal R}^{-7}) := 
{\mw_0}{\cal R}^{-4} - L{\cal R}^{-5}\mbox{cos}\,\Phi - w_2{\cal R}^{-6} + O({\cal R}^{-7}) \mbox{ } .  
\label{aslar}
\eeq
Thus the leading term goes as a constant in $\Xi$ or as ${\cal R}^{-4}$, unless $w_0 = 2\fE - K_1 \mbox{ }(\mbox{ }\propto \overline{\ttE} - A + B \mbox{ }) = 0$. 
In this case, it goes linearly in $\Xi$ or as ${\cal R}^{-5}$ in each case also with a cos$\,\Phi$ factor. 
This holds unless also $L \mbox{ }(\mbox{ } \propto C \mbox{ }) = 0$ (`special' case), in which case it goes quadratically in $\Xi$ or as ${\cal R}^{-6}$, unless $B = 0$. 
(Given previous conditions, this is equivalent to $w_2 = 4\ttE + K_2 - 3K_1 = 0$).  
This means that one is in the $K_1 = 2\ttE$ subcase of the `very special' case, for which $\ttE + \ttU$ has no terms at all.)

Finally note that the large and small asymptotics are dual to each other. 
(The difference of 4 powers is accounted for by how the kinetic energy scales under the duality map).  
So that {\sl one need only analyse the parameter space for one of the two regimes and then obtain everything about the other regime by simple transcription}.
I.e., for triangleland, there is a duality under size inversion (swapping round what is large and what is small, or, alternatively, the antipodal map). 
Now, only the very special case is self-dual, the special case requiring additionally for $K_1$ and $K_2$ to be swapped around or alternatively the sign of B to be reversed.

\subsubsection{Normal coordinates for triangleland HO}\label{Tri-Norm}

A rotation sending the general case to the special case in new coordinates is as follows. 
One can avoid having a $C$-term by using normal modes/adapted bases \cite{08I}, again at the cost of making the physical interpretation more complicated. 
One can get to these by inserting such a rotation into the preceding solution process.  
I denote the new coordinates with $N$-subscripts ($N$ for normal).   
This is done by inserting $N$-subscripts and, at the end of the calculation, rotating back from normal 
Jacobi coordinates to a more complicated form in terms of the original Jacobi coordinates.

What rotation angle is required? 
From the matrix equation $\underline{\uR}  (\gamma, dra_y) \underline{dra} = \underline{dra}_{N}$ (rotation through angle $\gamma$ about the $dra_y$-axis),
\beq
\mbox{\fontsize{1.4cm}{1.4cm}\selectfont (}
\stackrel{    \stackrel{  \mbox{\scriptsize cos$\gamma$} \mbox{ } \mbox{\scriptsize 0} \mbox{ } \mbox{\scriptsize --sin$\gamma$} }
                       {  \mbox{\scriptsize 0} \mbox{ }  \mbox{ } \mbox{\scriptsize 1} \mbox{ }  \mbox{ } \mbox{\scriptsize 0}  }    }
                       {  \mbox{\scriptsize sin$\gamma$} \mbox{ } \mbox{\scriptsize 0} \mbox{ } \mbox{\scriptsize cos$\gamma$}  }
\mbox{\fontsize{1.4cm}{1.4cm}\selectfont )}
\mbox{\fontsize{1.4cm}{1.4cm}\selectfont (}
\stackrel{\stackrel{\mbox{\scriptsize $C$}}{\mbox{\scriptsize 0}}}{\mbox{\scriptsize $B$}}
\mbox{\fontsize{1.4cm}{1.4cm}\selectfont )}
=
\mbox{\fontsize{1.1cm}{1.1cm}\selectfont (}
\stackrel{    \stackrel{  \mbox{\scriptsize 0}  }{  \mbox{\scriptsize 0}  }    }{    \mbox{\scriptsize $B_{N}$}    } 
\mbox{\fontsize{1.1cm}{1.1cm}\selectfont )} \mbox{ }  
\eeq
it is 
\beq
\gamma = \mbox{arctan}(C/B) \mbox{ } . 
\eeq
This sends $B\,\mbox{cos}\,\Theta + C\,\mbox{sin}\,\Theta\,\mbox{cos}\,\Phi$ to $B_{N}\,\mbox{cos}\,\Theta_{N}$.

In addition to the axis systems described in Sec \ref{TessiTri}, one can also take E$\bar{\mE}$ as principal axis 
while placing the 2-axis at a general angle $\gamma$ within the collinearity plane (measured without loss of generality from the (1)-axis.    
I denote these bases by $[\gamma]$ superscripts (the axis systems in Sec \ref{TessiTri} are the $[\gamma = 0]$ case of these.   
These bases serve to align the axis system with the general HO problem's symmetry axis (potential-adapted bases).  
For general $B$ and $C$ the harmonic oscillator-like potential acts as a well centred on some point of the equator of collinearity.    
Clearly the azimuthal angle is unaffected by passing to this set-up, so cos$\,\Theta_{[\gamma]}$ is also 4 $\times$ $area$, while $\Phi_{[\gamma]} = \Phi_{[1]} - \gamma$.  
[This is measured without loss of generality from the (1)-axis.]  
Thus $\mbox{cos}\,\Phi_{[\gamma]}$ and $\mbox{sin}\,\Phi_{[\gamma]}$ are obtainable by 2-angle formulae.
Finally, one could take the $\gamma$-axis as principal axis and E$\bar{\mE}$ as second axis. 
I denote these basis by $(\gamma)$; this is in various ways even better-adapted to the general problem of Sec \ref{Tri-Norm}.  
Unfortunately, formulae for $\Theta_{(\gamma)}$ and $\Phi_{(\gamma)}$ are somewhat more complicated; they can be derived e.g. by spherical trigonometry similar to that in \cite{08I}.  
Note that $[\gamma]$ and $(\gamma)$ include the preceding three bases under the identifications $\gamma = 0 \mapsto 1$,  $\gamma = 2\pi/3 \mapsto 2$ and $\gamma = 4\pi/3 \mapsto 3$.

Despite the new axis being confined to the plane of collinearity, it can nevertheless still pass through/near a range of qualitatively different points.  
Thus $C \neq 0$ is an even richer dynamics than $B \neq 0$ $C = 0$ was found to be in \cite{08I}. 
Whether this richness translates to interestingness from the perspective of dynamical systems remains to be seen.
(This did turn out to some extent to be the case \cite{WainwrightEllis} in the minisuperspace counterpart.)

Also, 
\beq 
\mbox{sin}\,\gamma = {C}/{\sqrt{B^2 + C^2}} \mbox{ } , \mbox{ } 
\mbox{cos}\,\gamma = {B}/{\sqrt{B^2 + C^2}} \mbox{ } , 
B_{N} = \sqrt{B^2 + C^2} \mbox{ } .     
\eeq
\beq
\mbox{cos}\,\Theta_{N} = \{\mbox{cos}\,\gamma\,\mbox{cos}\,\Theta + \mbox{sin}\,\gamma\,\mbox{sin}\,\Theta\,\mbox{cos}\,\Phi\}
\mbox{ }
\mbox{ and } 
\mbox{ }
\Phi_{N} = \mbox{arctan}
\left(
{\mbox{sin}\,\Theta\,\mbox{sin}\,\Phi}/\{\mbox{cos}\,\gamma\,\mbox{sin}\,\Theta\,\mbox{cos}\,\Phi - 
\mbox{sin}\,\gamma\,\mbox{cos}\,\Theta\}
\right) \mbox{ } .  
\label{Rex}
\eeq

The HO-like potential is now
\beq
\{K_1^{N}R_{1_{N}}^2 + K_2^{N}R_{2_{N}}^2  \}/{8\mI} = A_{N} + B_{N}\,\mbox{cos}\,\Theta_{N} 
\mbox{ } .
\eeq
It is also useful to note for later use the following coefficient interconversions: 
\beq
A_{N} = A \mbox{ } , \mbox{ } B_{N} = \sqrt{B^2 + C^2} \mbox{ } ,
\eeq
\beq
K_1^{N} = 8\big\{A - \sqrt{B^2 + C^2}\big\}
\mbox{ } , \mbox{ }
K_2^{N} = 8\big\{A + \sqrt{B^2 + C^2}\big\} 
\mbox{ } , 
\label{Lex}
\eeq
\beq
K_1^{N} = K_1 + K_2 - \sqrt{\{K_1 - K_2\}^2 + 4Q^2} 
\mbox{ } , \mbox{ } 
K_2^{N} = K_1 + K_2 + \sqrt{\{K_1 - K_2\}^2 + 4Q^2} \mbox{ } .  
\eeq
\mbox{ } \mbox{ } From the spherical perspective, the normal coordinates solution has the same form as the special solution in the original coordinates.  
However, now one is to project onto the general tangent plane rather than the tangent plane at the North Pole.  
One is interpreting the general stereographic coordinate to now be the ratio of the square roots of the barycentric partial moments of inertia.  
This permits use of tessellation to interpret the `shifted' pattern in $N$-variables.

\subsection{Mechanics--Cosmology analogy as chooser of potentials}\label{MechCos}

As regards (sums of) power-law potentials; these are motivated by being are common in mechanics 
and by their mapping to the commonly-studied terms in the Friedmann equation of Cosmology also.  
E.g. for scaled 3-stop metroland,  
\beq
V = C_1|q_2 - q_3|^{\sn} + C_2|q_3 - q_1|^{\sn} + C_3|q_1 - q_2|^{\sn}
\label{char}
\eeq
becomes 
\beq
V  =  \rho^{\sn}\{ C_1|\mbox{cos}\,\varphi|^{\sn} + 
                   C_2|\mbox{cos}\,\varphi + \sqrt{3}\mbox{sin}\,\varphi|^{\sn} + 
                   C_3|\mbox{cos}\,\varphi - \sqrt{3}\mbox{sin}\,\varphi|^{\sn}\}  \mbox{ } .  
\label{3stoppot}
\eeq
On the other hand, for scaled triangleland, (\ref{char}) becomes
\beq
V  =  
\rho^{\sn}\{C_1\big|\mbox{sin}\mbox{$\frac{\Theta}{2}$}\big|^{\sn} + 
            C_2\big|\mbox{cos}\mbox{$\frac{\Theta}{2}$} - \mbox{$\frac{1}{2}$}\mbox{sin}\mbox{$\frac{\Theta}{2}$}\big|^{\sn} + 
            C_3\big|\mbox{cos}\mbox{$\frac{\Theta}{2}$} + \mbox{$\frac{1}{2}$}\mbox{sin}\mbox{$\frac{\Theta}{2}$}\big|^{\sn} \}
\mbox{ } .  
\eeq
More widely, these analogies with cosmology require that any shape factors present being slowly-varying so that one 
can carry out the following scale-dominates-shape approximation holding at least in some region of interest.

\subsubsection{The `scale dominates shape' family of approximations}\label{SSA}

I originally considered an action-level scale-dominates shape approximation \cite{Cones}, most clearly formulated as 
\beq
||\d_{\underline{B}}\bS||_{\sbttM}/\d\{ \mbox{ln} \rho\}  << 1 \mbox{ } .  
\eeq
However, further consideration (Sec \ref{+temJBB}) reveals that this assumption is better justified if first made at the level 
of the equations of motion/forces, so that it involves  
\beq
\rho \, \mS^{*\,2}/\rho^{**} = \epsilon_{\sss\sd\sss 1} << 1 
\label{SSA1}
\eeq
(which is encompassed under `S is fast' compared to $\rho$), whilst $\pa V_{\rho}/\pa \mS^{\sfa}$ does not in any case contribute to the shape evolution equations).
%
%
Also, 
\beq
|\pa J_{\rho \sS}/\pa {\mS}|/|\pa V_{\sS}/\pa \mS| = \epsilon_{\sss\sd\sss 2} <<  1 \mbox{ } .  
\label{SSA2}
\eeq
[Without the `scale dominates shape' approximation, one cannot separate out the heavy (here scale) part so that it can provide the approximate 
timefunction with respect to which the light (here shape) part's  dynamics runs.
Moreover, N.B. that isotropic cosmology itself similarly suppresses small anisotropies and inhomogeneities, 
so that exact solutions of this are really approximate solutions for more realistic universes too.]
For the spherical presentation of triangleland, the barred, $\rho \longrightarrow \mI$ counterpart of this holds.  

\mbox{ }

\noindent Analogy 48) GR counterparts of the scale-dominates-shape analogy are the leading-order neglect of scalar field terms in 
e.g. \cite{GibGri}, of anisotopy in e.g. \cite{Amsterdamski} and of inhomogeneity in e.g. \cite{HallHaw}.  
The inhomogeneities case of this -- the most important for Cosmology -- is well-known to be of the order of 1 part in $10^5$.

\mbox{ }

\noindent Note: isotropic cosmology itself similarly suppresses small anisotropies and inhomogeneities, 
so that exact solutions thereof are really approximate solutions for more realistic universes too.  
Take the barred, $\rho \longrightarrow \mI$ counterpart of all this for the $\mathbb{S}^2$ presentation of triangleland.
Throughout, these analogies are subject to any shape factors present being slowly-varying so that one 
can carry out the approximation (\ref{SSA1}--\ref{SSA2}), at least in some region of interest.

\subsubsection{Ordinary 1-particle Mechanics--Cosmology analogy}\label{Cos-Mech}

Next, I use the analogy between Mechanics and Cosmology to broaden the range of potentials under 
consideration and pinpoint ones which parallel classical and Quantum Cosmology well. 
This would seem to be making more profitable use of the potential freedom than previous mere use of simplicity.

Isotropic cosmology (in $c = 1$ units) has the Friedmann equation
\beq
\lH^2 := \left\{\frac{{a}^*}{a}\right\}^2 = - \frac{k}{a^2} + \frac{8\pi G\epsilon}{3} + \frac{\Lambda}{3}  
= -\frac{k}{a^2} + \frac{2GM_{\sd\su\sss\st}}{a^3} + \frac{2GM_{\sr\sa\sd}}{a^4} + \frac{\Lambda}{3}
\mbox{ } ,    
\eeq
the second equality coming after use of energy--momentum conservation and assuming non-interacting matter components. 
Here, $a$ is the scalefactor of the universe, $\Last = \d/\d t^{\scc\so\sss\sm\si\scc}$ (aligned with $\d/\d t^{\sN\se\sw\st\so\sn}$ here). 
$\lH$ is the Hubble quantity. 
$k$ is the spatial curvature which is without loss of generality normalizable to 1, 0 or --1. 
$G$ is the gravitational constant. 
$\epsilon$ is matter energy density.
$\Lambda$ is the cosmological constant.  
$M$ is the mass of that matter type that is enclosed up to the radius $a(t)$.

A fairly common analogy is then between this and 
\beq
\left\{\frac{{r}^*}{r}\right\}^2 = \frac{2E}{r^2} + \frac{K_{\mbox{\scriptsize Newton}}}{r^3} + 
\frac{K_{\mbox{\scriptsize Conformal}}}{r^4} + \frac{K_{\mbox{\scriptsize Hooke}}}{3}
\mbox{ }    \label{EnRel}
\eeq
i.e. the unit-mass ordinary mechanics energy equation)$/r^2$.  
Here and elsewhere in this Article the various $K$'s are constant coefficients, and 
$\Last$ is here $\d/\d t^{\sN\se\sw\st\so\sn}$.  
A particularly well-known subcase of this is that 1-$d$ mechanics with a $1/r$ Newtonian Gravity type 
potential is analogous to isotropic GR cosmology of dust. 
This extends to an analogy between the Newtonian dynamics of a large dust cloud and the GR isotropic 
dust cosmology \cite{Milne,Mc1,Mc2,BGS03}. 
(Here, shape is least approximately negligible through its overall averaging 
out to approximately separated out shape and cosmology-like scale problems.)  
For \cite{Cones} and the present Article's purposes, enough of these parallels (\cite{Harrison67, MTW, Pee, Rindler}) survive 
the introduction of a pressure term in the cosmological part, and the using of an $N$-stop metroland RPM in place of ordinary Mechanics. 
%

In considering a large number of particles, another way in which shape could be at least approximately negligible arises \cite{BGS03}. 
This is through its overall averaging out to produce a highly radial problem (in 
a factorization into a cosmology-like scale problem and a shape problem).  
In the case of dust in 3-$d$, the many Newtonian gravity potential terms average out to produce the 
effective dust, and one's equations are split into 1) the standard dust cosmological scale equation. 
2) The also well-known central configuration problem for shape.\footnote{This being well-known is, however, in a 
somewhat different context, i.e. the Celestial Mechanics literature considers the few-particle case 
(see e.g. \cite{Moeckel} for an introduction and further references).  
On the other hand, the late universe cosmological interest is in the large-particle limit \cite{BGS03}.}
%
The Newton--Hooke problem, amounting to cosmology with dust and cosmological constant, has also been 
studied in a somewhat similar context (see e.g. \cite{GP03}).  
It is an interesting question to me whether the averaging out to produce a radial equation and a shape 
equation occurs for other power-law potentials and their superpositions.  
Furthermore, are any of the resulting shape problems of an already mathematically-known form?

\subsubsection{$N$-stop metroland--Cosmology analogy}\label{Cos-NStop}

Analogy 49) The $N$-stop metroland--Cosmology analogy itself (all of the  rest of this SSSec above 
applies also to the $\mathbb{CP}^{N - 2}$ presentation of $N$-a-gonland) is between the above Friedmann equation and (RPM energy relation)/$\rho^2$,
\beq
\left\{\frac{\dot{\rho}}{\rho}\right\}^2 =  
                                             \frac{2E}{\rho^2} 
                                             - \frac{\sfT\sfO\sfT}{\rho^4} 
                                             - \frac{2V(\rho, \mbox{S}^{u})}{\rho^2} = 
\frac{2E}{\rho^2} + \frac{2K}{\rho^3} + \frac{2R - \sfT\sfO\sfT}{\rho^4} - 2A \mbox{ } , \mbox{ } \mbox{ } 
\eeq 
with (\ref{EnRel})'s $\Last$ taking this Article's usual meaning of $\d/\d t^{\se\sm(\sJ\sB\sB)}$.  
In fact, one needs to make this identification at the level of the Lagrangian, for, if not, the momentum form of the equations 
would entail a different (and not physically insightful) correspondence.

Then,  

\mbox{ } 

\noindent Analogy 50) the spatial curvature term $k$ becomes --2 times the energy $E$.   

\noindent Analogy 51) The cosmological constant term $\Lambda/3$ becomes --2 times the net $A$ from the (upside down) HO potentials.

\noindent Analogy 52) The dust term $2GM/a^3$'s coefficient $2GM$ becomes --2 times the net coefficient $K$ from the 
Newtonian Gravity potential terms. 
N.B. the $K$ of interest has a particular sign shared by dust and Newtonian Gravity (and the attractive subcase of Coulomb).   

\noindent Analogy 53) The radiation term coefficient $2GM/a^4$'s coefficient $2GM$ becomes $- \sfT\sfO\sfT         
+ 2R$ for $2R$ the coefficient of the $V_{(0)}$ contribution from the $1/|r^{IJ}|^2$ terms.  
It corresponds to the conformally invariant potential term, which is quite well-studied in Classical and Quantum Mechanics. 

\noindent Difference 19) Note that $\sfT\sfO\sfT$ itself is of the wrong sign to match up with the ordinary radiation term of cosmology.  
In the GR cosmology context, `wrong sign' radiation fluid means that it still has $p = \epsilon/3$ 
equation of state (for $p$ the pressure), but its density $\epsilon$ is negative.  
Thus it violates all energy conditions, making it unphysical in a straightforward GR cosmology context, and also having the effect of Singularity Theorem evasion by `bouncing'.  
From the ordinary mechanics perspective, however, this is just the well-known repulsion of the centrifugal barrier preventing collapse to zero size.

\mbox{ } 

\noindent Note 1)  This difference in sign is underlied by an important limitation in the Mechanics--Cosmology analogy. 
For, in mechanics the kinetic energy is positive-definite, while in GR the kinetic energy is indefinite, 
with the scale part contributing negatively.  
This does not affect most of the analogy, due to the following.  
The energy and potential coefficients can be considered to come with the opposite sign. 
On the other hand, the relative sign of the shape and scale kinetic terms cannot be 
changed thus, and it is from this that what is cosmologically the `wrong sign' arises. 

\noindent Note 2) Two distinct attitudes to wrong-sign term are as follows.  
Firstly, one can suppress this $\sfT\sfO\sfT$ term by taking toy models in which this term. 
(The significance of this is the relative distance momentum of two constituent subsystems is zero or 
small, or swamped by `right-sign' $1/|r^{IJ}|^2$ contributions to the potential.) 
Secondly, it should also be said that more exotic geometrically complicated scenarios such as brane 
cosmology can possess `dark radiation' including of the `wrong sign' \cite{DarkRad}. 
(Indeed, this can possess what appears to be energy condition violation from the 4-$d$ spacetime 
perspective due to projections of higher-dimensional objects \cite{AT05}).  
Thus such a term is not necessarily unphysical.    

\mbox{ } 

\noindent Difference 20) The analogy does not however have a metric interpretation or a meaningful 
interpretation in terms of an energy density $\epsilon$, it is after all just a particle mechanics model.

\subsubsection{Spherical presentation of triangleland--Cosmology analogy}\label{Cos-Tri}

\noindent Analogy 49$^{\prime}$) It is between the above Friedmann equation and (\ref{EnRel})/$\mI^2$,
\beq
\left\{\frac{{\mI}^{\star}}{\mI}\right\}^2 =  \frac{2E}{\mI^3} - \frac{\sfT\sfO\sfT}{\mI^4} 
                                                     - \frac{2V(\mI, \mS^{u})}{\mI^3} = 
\frac{-2A}{\mI^2} + \frac{2E}{\mI^3} + \frac{2R - \sfT\sfO\sfT}{\mI^4} - 2S \mbox{ } . 
\label{plix}
\eeq 
Thus now 

\mbox{ } 

\noindent Analogy 50$^{\prime}$) the spatial curvature term $k$ becomes the net HO terms' $2A$.   

\noindent Analogy 51$^{\prime}$) The cosmological constant term $\Lambda/3$ becomes -- 2 times the surviving lead 
term from the $|r^{IJ}|^6$ potentials' coefficient $S$.  

\noindent Analogy 52$^{\prime}$)  The dust term $2GM/a^3$'s coefficient $2GM$ becomes $2E$ so to have an analogy with a physical 
kind of dust it is $E > 0$ that is of interest.

\noindent Analogy 53$^{\prime}$) just coincides with analogy 53) again.

\mbox{ }  

\noindent In the spherical triangleland analogy, the Newtonian $1/|r^{IJ}|$ type potentials that one might 
consider to be mechanically desirable to include produce $1/\mI^{7/2}$ terms. 
These are analogous to an effective fluid with equation of state $P = \epsilon/6$ i.e. an interpolation `halfway between' radiation fluid and dust.

\subsubsection{Stability of the approximation and sketches of unapproximated potentials}\label{Stab-ClSol}

HO/$\Lambda < 0$ problems  include as a subcase finite-minimum wells about the poles of the 4-stop metroland configuration space sphere ($C$ = 0 case) \cite{08I, AF, +Tri}. 
Adding a $\sfT\sfO\sfT$ effective potential term adds spokes to the wells. 
These models are exactly soluble.
Positive power potentials are still finite-minimum wells, but cease to be exactly soluble.    
For Newtonian Gravity/dust models (or negative power potentials more generally), near the corresponding 
lines of double collision, the potential has abysses or infinite peaks.  
Thus the small-shape approximation is definitely not valid there, and so some assumptions behind the Semiclassical Approach fail in the region around these lines.  
Thus for negative powers of relative separations the heavy approximation only makes sense in certain wedges of angle. 
There is then the possibility that dynamics set up to originally run in such regions falls out from them: a stability analysis 
is required to ascertain whether semiclassicality is stable. 
Fig \ref{Fig--5} illustrates this for single and triple potential terms.  
This can be interpreted as a tension between the procedure used in Semiclassical Quantum Cosmology and the example of {\bf trying to approximate a 3-body problem by a 2-body one}.

{            \begin{figure}[ht]
\centering
\includegraphics[width=0.5\textwidth]{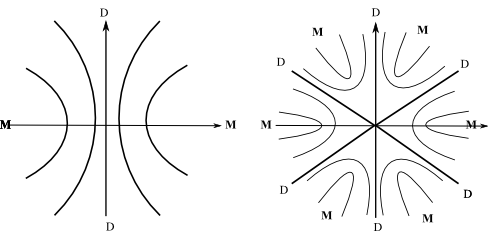}
\caption[Text der im Bilderverzeichnis auftaucht]{        \footnotesize{Contours for single and triple 
negative power potentials. 
These have abysses along the corresponding double collision lines and high ground in between these (for negative-power 
coefficients such as for the attractive Newtonian Gravity potential.  
}        }
\label{Fig--5}
\end{figure}  }
%
Moreover, at least in the simple minisuperspace quantum cosmologies I have looked at so far, it is
potential powers that are common, and these then do not possess the abysses and instabilities of the $1/|r^{IJ}|$ potentials.) 
This furnishes is one further reason why I mostly concentrate on positive power models in the current Article.
E.g. the Cosmology--spherical presentation of triangleland analogy cause this Article's cosmologically-inspired set of potentials 
to all be such positive powers.

\subsection{Classical solutions for pure-shape 4-stop metroland}\label{4Stop-ClSol}

Classical study of solution includes finding the shapes of the orbits (which is a relational and timeless concern) 
as well as consideration of traversal times (which Franzen and I \cite{AF} treat partly qualitatively and partly in analogy with cosmology).

\subsubsection{Classical solutions for $\sfD = 0$} \label{D=0-ClSol}

Here, $0 = \sfD = \mbox{sin}^2\theta\,\phi^{\star}$. 
Thus, either $\sin\theta = 0$, so the motion is stuck on a pole (the DD's corresponding to the 
H-coordinates with respect to which $\theta$ is defined, taken to be \{12,34\} and \{34, 12\}), or, 
$\phi$ is constant, so that the motion is stuck on a meridian.  
This situation corresponds to losing a dilational momentum exchange freedom, which renders this SSSec 's solutions of somewhat limited interest.  
How far one goes in the $\theta$ direction, and whether the motion turns around, is determined by 
considering the other equation of motion (see below for examples).

Some significant examples of such meridians include 1) the one in the direction of the 1-axis that corresponds to cluster \{34\} always 
being collapsed while cluster \{12\} varies in size.  
As well as the two DD's, this passes through the \{134\} and \{234\} T's.   
2)The \{12\} $\longleftrightarrow$ \{34\} of this going in the direction of the 2-axis.  
3, 4) Going in the direction of a $\mn_y = \pm \mn_x$ lines, that correspond to the clusters always being 
of the same size (contents homogeneity) but with that size varying from zero (at the \{12,34\} and 
\{34,12\} DD collisions) to maximal. 
[Here, the two clusters are superposed into the \{13,24\} or \{14,23\} DD collisions].  
[This SSSec's results, and those of SSSec 
\ref{J=0-ClSol}, clearly hold for any special potential rather than just a multi-HO-type one. 
There being no `centrifugal' barrier term, it has the simpler mathematics of motion in a 1-$d$ 
potential.]

\subsubsection{Classical solution in the very special case: the geodesic solution}\label{Geo-ClSol}

For $\sfD \neq 0$ in general, the very special case is solved by the geodesics on the shape space 
sphere, 
\beq
\mbox{cos}(\phi - \phi_0) = \kappa\,\mbox{cot}\,\theta \mbox{ } 
\eeq  
for $\kappa = \sfD/\sqrt{2\{\ttE - \ttA\} - \sfD^2}$ and $\phi_0$ constants.  
Then in terms of the $\mn^i$ (alias RelSize's), 
\beq
\kappa \, \mn_z = \mn_x\,\mbox{cos}\,\phi_0 + \mn_y\,\mbox{sin}\,\phi_0 \mbox{ } ,
\label{plane}
\eeq
i.e. restriction to a plane through the origin, with arbitrary normal (cos$\,\phi_0$, sin$\,\phi_0$, $-\kappa$).      
Since also $\sumi3 \mn^{i\, 2} = 1$, one is restricted to the intersection of the sphere and the arbitrary plane through its centre, 
i.e. another well-known presentation of the great circles as circles within $\mathbb{R}^3$.

The disc in the equatorial plane is useful in considering the mechanics of the problem with clusters 
\{12\} and \{34\} picked out by the choice of Jacobi H-coordinates.  
Eliminating $\mn_z$ projects an ellipse onto this disc,
\beq
\kappa^2 = \{\kappa^2 + \mbox{cos}^2\phi_0\}{\mn_x}^2 + 2\,\mbox{cos}\,\phi_0\,\mbox{sin}\,\phi_0\mn_x\mn_y 
+ \{\kappa^2 + \mbox{sin}^2\phi_0\}{\mn_y}^2 \mbox{ } ,  
\eeq
centred on the origin with its principal axes in general not aligned with the coordinates.  
E.g. for $\phi_0 = 0$, the ellipse is
\beq
1 =  \left\{{\mbox{RelSize(1,2)}}/{\{1 + \kappa^{-2}\}^{-1/2}}\right\}^2 + {\mbox{RelSize(3,4)}}^2 
\mbox{ } ,
\eeq
which has major axis in the RelSize(3,4) = $\mn_y$ direction and minor axis in the RelSize(1,2) = $\mn_x$ 
direction, while the value of RelSize(12,34) = $\mn_z$ around the actual curve can then be read off 
(\ref{plane}) to be $\mn_z = \mn_x/\kappa$.  
With reference to the tessellation of 4-stop metroland, as $\sfD \longrightarrow \infty$, the 
dynamical trajectory is the equator, corresponding to maximally-merged configurations including four DD 
 collisions.  
For $\sfD$ small, the motion approximately goes up and down a meridian. 
E.g. this forms a basic unit of a narrow cycle from the polar DD to slightly around the T on the 
Greenwich meridian (reflections of) which is repeated various times to form the whole trajectory. 
[The actual limiting on-axis motion $\sfD = 0$ gives back the meridians considered in the preceding 
section; the motion now goes round and round the entirety of whichever of these meridians.]


Other $\phi_0$ straightforwardly correspond to rotated ellipses.  
However the mechanical meaning of these differs. 
E.g. about $\pi/2$ clusters \{12\} and \{34\} are interchanged, while about $\pm \pi/4$ also has 
distinct sharp physical significance.  
Throughout, note the periodicity of the motion (already clear in the spherical model as the great 
circles are closed curves).  
The tessellation lines are great circles, projecting to the disc rim, the axes, the lines at $\pi/4$ to 
the axes and ellipses with principal directions aligned with the preceding.  
C.f. the great circles in Figs \ref{41Merger} and \ref{Fig-Noob}.

\subsubsection{Approximate solutions for the special case of HO-like problem}\label{Spec-ClSol}

In the first approximation for stereographic radius ${\cal R}$ small\footnote{Not 
all dynamical orbits enter such a regime -- sometimes the quadrature's integral goes complex before the 
small ${\cal R}$ regime is attained (${\cal R}$ large `classically forbidden'), for $\widetilde{\fU}
({\cal R}) {\cal R}^2 - \sfD^2 < 0$.}
gives, by integrating the ${\cal R}$ quadrature, the orbits 
\beq
\pm\sqrt{2\nuu}\,\mbox{sec}(\phi - \bar{\phi}) = {\cal R} = \mbox{tan$\frac{\theta}{2}$} = \mn_1/\mn_2 
\mbox{ }  .  
\label{exactsoln}
\eeq
I cast the answer in terms of 
straightforward relational variables, and using the $\sfD$-absorbing constant 
$f_0 = 2{\lepsilon}/\sfD^2$, $f_2 = \Omega^2/\sfD^2$ and $g = \sqrt{f_0\mbox{}^2 - f_2}$.  
However, the ${\cal R}$ form of the orbits are parallel straight lines (vertical for $\phi_0 = 0, \pi$).    
Such are known not to be very good approximands in that they totally neglect the non-constant part of 
the potential (i.e. deflections due to forces) and thus are precisely rectilinear motions.  
This argument for the use of the second approximation pervades this Article.

In the second approximation,  
\beq
\sqrt{\ttf_0 + \ttg\,\mbox{cos}(2\{\phi - \phi_0\})} = {1}/{{\cal R}} = \mbox{cot$\frac{\theta}{2}$} 
\mbox{ } .  
\label{ellipse}
\eeq
In terms of ${\cal R}$, this is straightforwardly rearrangeable into quite a standard form (e.g 
\cite{Moulton, Whittaker})
\beq
{\cal R}^2 = 1\left/ \big\{\ttu_0 + \sqrt{\ttu_0 - \ttv}\, \mbox{cos}(2\{\Phi - \Phi_0\})\big\}  \right. \mbox{ } ,
\eeq
the case-by-case analysis of which is provided in Fig \ref{Fig-4AF}. 
Casting into a well-known form (e.g. \cite{Moulton}):
\beq
{{\cal R}_{}}^2 = 1/\{\Biggamma + \Bigeta\,\mbox{cos}^2(\phi_{} - \phi_{0})\} 
\mbox{ } ,   
\eeq  
where $\Biggamma$ and $\Bigeta$ are also constants.  
These describe closed curves around the DD collision the potential is centred about.
The cases of interest are then ellipses (including the bounding case of circles but excluding the other bounded case of pairs of straight lines, as in \label{FIG4-AF}).
[N.B. this is isotropic HO mathematics rather than Kepler problem mathematics, so that the ellipses are centred on the origin rather than having the origin at one focus.]  

{            \begin{figure}[ht]
\centering
\includegraphics[width=0.75\textwidth]{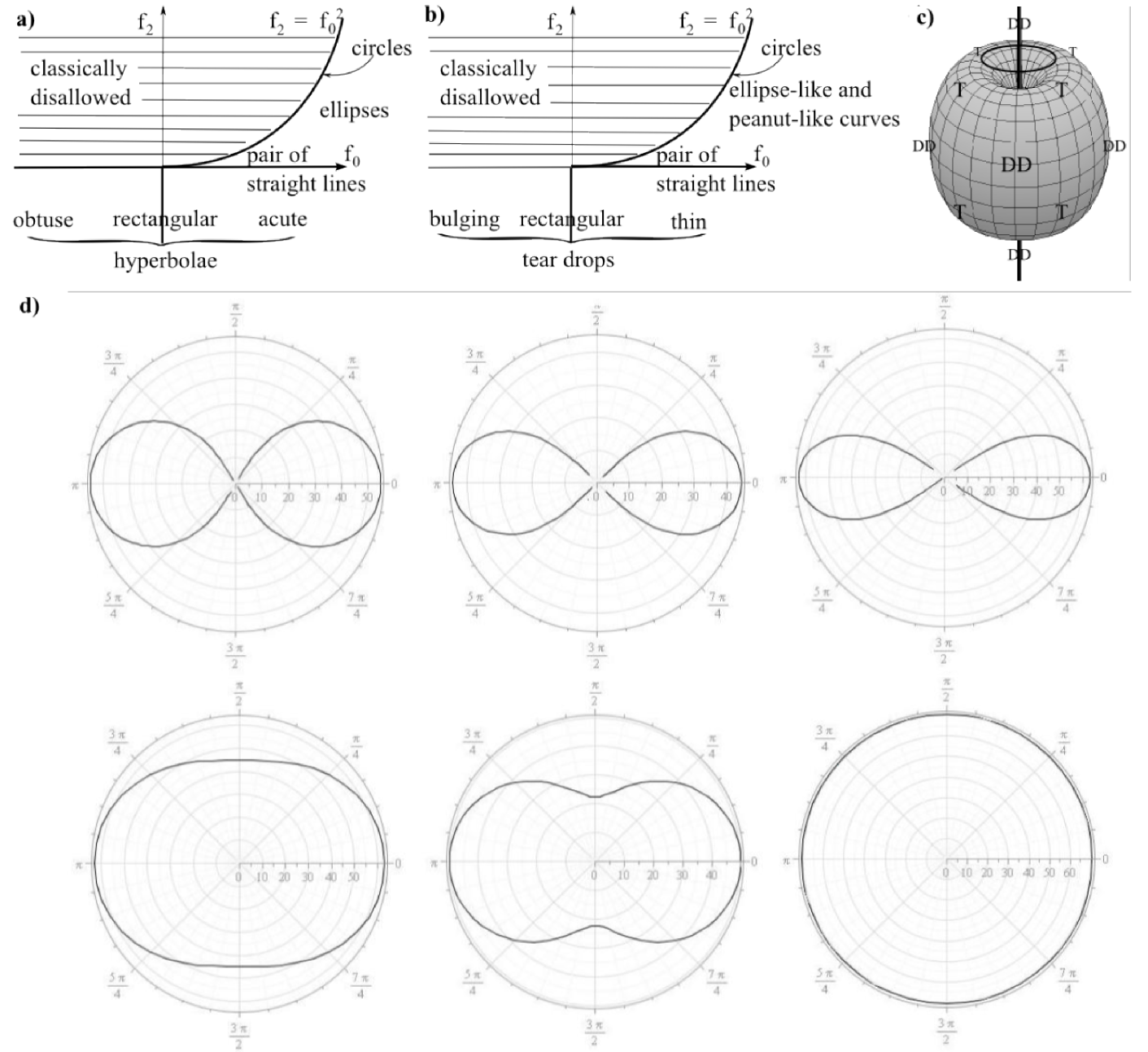}
\caption[Text der im Bilderverzeichnis auftaucht]{        \footnotesize{    a) For the small 
approximation in (${\cal R}$, $\phi$) coordinates, the desired parameter space is the indicated wedge populated by ellipses. 
On the bounding parabola, I get circles, coinciding with the edge of the $\mn_x$, $\mn_y$ disc for $\ttf_0 = 1 = \ttf_2$ and becoming smaller in either direction. 
This is standard isotropic HO mathematics.

\noindent b) The large approximation in (${\cal R}$, $\phi$) coordinates involves somewhat more unusual curves, and yet same breakdown of parameter space as for a).
This is explained by it taking the same form as a) in (${\cal U}$, $\phi$) coordinates by the size duality/antipodal map.

\noindent c) The relevant small behaviour are ellipses (or pieces of these or other of the curves that then exit the region where the approximation is appropriate).  
I illustrate how this is qualitatively clear by providing the form of the potential as a backdrop.  
As regards RPM interpretation, these are curves for which DD is small.

\noindent d) I provide the forms of the second large approximation's curves, since these often illustrate 
the large behaviour of pure-shape RPM solutions (thus ignore inner parts of curves where this does not apply well).     
The scale of the figure is too small to distinguish between all the DD's and T's (bar the South Pole DD, which is at infinity), 
these all lying piled up on/very near the origin in the figures.  
}        }
\label{Fig-4AF}
\end{figure}            }

\noindent N.B. there is now a third inequality on ${\lepsilon}$, $\Omega$:   
\beq
4{\lepsilon}^2 \geq \sfD^2\Omega^2 \mbox{ } , 
\eeq
that replaces ${\lepsilon} > 0$ as it is more stringent.
Thus in terms of $\ttE$ and the $K_i$ I get the allowed wedge of parameter space to be
\beq
2\{2\ttE - \ttK_3\}^2/\sfD^2 \geq \{2\ttE - \ttK_3\} + 2\{\ttK_1 - \ttK_3\} > 0  \mbox{ } . 
\eeq
Saturation of this corresponds to circular trajectories.   
For such circles to exist, the discriminant gives the condition 

\noindent $\{\sfD/4\}^2 \geq K_3 - K_1$, so that the relative dilational quantity is bounded from below by the amount by which the inter-cluster spring dominates.

Then by elementary trigonometry (\ref{Rdef}), (\ref{ellipse}) become   
\beq
\sqrt{        \ttf_0 + \ttg \frac{    \{{\mn_x}^2 - {\mn_y}^2\}\,\mbox{cos}\,2\phi_0 + 
                                 2 \mn_x\mn_y \,\mbox{sin}\,2\phi_0    }
                           {    {\mn_x}^2 + {\mn_y}^2                 }         } = 
\sqrt{\frac{1 + \mn_z}{1 - \mn_z}} \mbox{ } .
\eeq
So, solving for $\mn_z$ and applying the on-sphere condition for $\phi_0 = 0$, say, gives 
\beq
\sqrt{    1 - {\mn_x}^2 - {\mn_y}^2    } = \mn_z = 
\frac{    \{\ttf_0 + \ttg - 1\}{\mn_x}^2 + \{\ttf_0 - \ttg - 1\}{\mn_y}^2    }
     {    \{\ttf_0 + \ttg + 1\}{\mn_x}^2 + \{\ttf_0 - \ttg + 1\}{\mn_y}^2    } \mbox{ } .  
\eeq
Then one can write down a curve in terms of two independent variables such as RelSize(1,2) and RelSize(3,4). 
Either RelSize(1,2) + RelSize(3,4) = 0 (so both are 0 because they are positive and so both clusters have collapsed) or
\beq
\big\{\{\ttf_0 + \ttg + 1\}\mbox{RelSize(1,2)}^2 + \{\ttf_0 - \ttg + 1\}\mbox{RelSize(3,4)}^2\big\}\mbox{}^2 = 
4\{\{\ttf_0 + \ttg\}\mbox{RelSize(1,2)}^2 + \{\ttf_0 - \ttg\}\mbox{RelSize}(3,4)^2\} \mbox{ } .
\eeq
\mbox{ } \mbox{ } The large regime then has
\beq
\sqrt{\ttf_0 + \ttg\,\mbox{cos}(2\{\phi - \phi_0\})} = {1}/{\cal U} = {\cal R} = \mbox{tan$\frac{\theta}{2}$} 
\mbox{ } ,
\eeq
which is, for the cases of interest, an ellipse-like or peanut-like curve in the $\{{\cal R}, \Phi\}$ plane \cite{08I}.  
Applying the interpretation in terms of $\mn^i$ or RelSize variables, the {\sl same} answer as for the small regime arises again.    
This conclusion just reflects the potential imposed possessing the large--small duality/antipodal 
symmetry, which physically translates to shapes and their mirror images behaving in the same fashion.

\subsubsection{Discussion of  more general HO-like problems}\label{+Springs-ClSol}

The $\ttL_i$ terms (or, equivalently, $\ttF$, $\ttG$ and $\ttH$ terms) can be dropped in the sense that one can 
pass to normal coordinates for which the symmetric matrix of Jacobi--Hooke coefficients has been 
diagonalized. 
Note that, unlike in the below triangleland counterpart, this does not send one to the special case -- 
the $\ttC$-term survives and so requires addressing separately (e.g. perturbatively).  
The elimination thus of $\ttF$, $\ttG$ and $\ttH$ terms is also subject to the mechanical interpretation of the 
normal-coordinate problem being more difficult algebraically than for $\ttF$ = $\ttG$ = $\ttH$ = 0.  
Thus it is conceivable that one might prefer to retain this simpler interpretation and treat $\ttF$, $\ttG$ and $\ttH$ perturbatively.

\subsection{Pure-shape triangleland}\label{Tri-ClSol}

\subsubsection{Classical solutions for $\sfJ = 0$}\label{J=0-ClSol}

This follows \cite{+Tri}, which tidies up \cite{TriCl, 08I}. 
$0 = \sfJ = \mbox{sin}^2\Theta\,\Phi^{\bar{*}}$.
Thus, either $\sin\Theta = 0$, so one is stuck on a pole (the D or M corresponding to the clustering 
with respect to which $\Theta$ is defined, taken to be \{1, 23\}),or, $\Phi$ is constant, so that one is 
stuck on one of the corresponding meridians.  
These correspond to losing an angular momentum exchange freedom, which renders this subsection's 
solutions to be of somewhat limited interest.  
Again, how far one goes in the $\Theta$ direction, and whether the motion turns around, is determined 
by considering the other equation of motion for $\Theta$ (in analogy to the above examples).

These meridians include 1) the equator of collinearity of Fig \ref{Fig-3tri-2} that goes through all 3 D's and all 3 M's. 
2) The one in the direction of the $area$ axis, that goes though the equilateral configurations E, $\bar{\mE}$ as well as the \{1,23\} cluster's M and D.

\subsubsection{Exact solution for the `very special case' of HO-like problem}\label{VSpec-ClSol}

For $\sfJ \neq 0$ in general, the very special case is solved by the geodesics on the shape space,  
\beq
\Bigkappa^{-1}\mbox{cos}(\Phi - \Phi_{\scc}) = 2\,\mbox{cot}\,\Theta  =  \{1 - {\cal R}^2\}/2{\cal R} = 
\{\mn_2^2 - \mn_1^2\}/\mn_1\mn_2  \mbox{ }   
\label{squik}
\eeq  
(following through \cite{08I} with variable transformations so as to cast it in terms of 
straightforward relational variables, and using $\left.\sfJ\Bigkappa^{-1} = 2\sqrt{2\{\overline{\ttE} - A\} - 
\sfJ^2} = \sqrt{2\ttE - K_2 - 4\,\sfJ^2} = \sqrt{u_0 - 4\,\sfJ^2}\right)$.  
$\Bigkappa$ and $\Phi_0$ are constants.
Further note that the ${\cal R}$ form above can be rearranged to the equation of a generally off-centre 
circle in the stereographic plane.  
Also, in terms of shape quantities, this solution can be written as 
\beq
aniso\,\mbox{cos}\,\Phi_0 + \mbox{4 $\times$ $area$}\,\mbox{sin}\,\Phi_0 = 2\Bigkappa\, ellip
\eeq
with the cos$\,\Phi_0$ = 0 or sin$\,\Phi_0$ = 0 cases obviously being simpler.

Of particular significance among these are the various great circles visible in Fig \ref{Fig-Noob}'s tessellation of triangleland. 
As well as the abovementioned equator of collinearity, these now include A) all three of the meridians of 
isoscelesness that go through a particular cluster's D and M as well as the two equilateral configurations, E and $\bar{\mE}$. 
B) All three of the meridians of regularity that go through the two E's and are constituted of 
configurations with $\mI_1 = \mI_2$ that cycle round all possible values of the relative angle $\Phi$.

\subsubsection{Small relative scale asymptotic behaviour}\label{Small-As}

This is the $\theta \longrightarrow \Theta$, $\phi \longrightarrow \Phi$, $\sfD \longrightarrow \sfS$ of Sec \ref{Spec-ClSol}.
The motion is now trapped in a well about a D-point as per Fig \ref{Heart}.  
Thus the first approximation is 
\beq
\mbox{sec}(\Phi - \Phi_{\scc}) =  \mn_1/\mn_2 \mbox{ } ,  
\eeq
but, likewise, this is not a sufficiently good approximation for almost any purposes).
Then, for the second approximation,  
\beq
\pm 1/\sqrt{u_0 + v \, \mbox{cos}(2\{\Phi - {\Phi}_{\scc}\})} = {\cal R} = \mbox{tan$\frac{\Theta}{2}$} = \mn_1/\mn_2 \mbox{ }  
\label{Sexactsoln}
\eeq
(following through with variable transformations so as to cast it in terms of straightforward 
relational variables, and using the $\sfJ$-absorbing constant $u_2 = u_2/\sfJ^2$) and $v = \sqrt{u_0\mbox{}^2 - u_2}$. 
Thus they are ellipses centred about the equilateral configuration at the pole, E.

In terms of shape quantities, this is
\beq
ellip \pm \sqrt{4 - 3\, ellip^2} = \pm \sqrt{u + v
\left\{ 
\frac{ aniso^2 - 4 \times area^2}{4}\mbox{cos}(2{\Phi}_{\scc}) + 
\frac{ aniso   \, \,  4 \times area}{2}\mbox{sin}(2{\Phi}_{\scc})
\right\}} \mbox{ } .  
\eeq
In the special case, now there is a well trapping paths around e.g. the (1)-clustering's D or M -- a 
(1)-sharp triangle region or a (1)-flat triangle region.  
Then [dropping (1) labels] for $\sfJ \neq 0$, these have $\Phi$ go round and round, so all regular 
and all isosceles configurations are periodically attained for small trajectories.   
In the small approximation, to second order, one gets, as the case most relevant to subsequent QM work, 
the isotropic harmonic oscillator in (${\cal R}, \Phi$) variables.  
This is solved by ellipses centred about the origin \cite{08I}.
Now one only has duality up to $K_1 \longleftrightarrow K_2$, meaning that the coefficients are different in each case.

\subsubsection{Large relative scale asymptotic behaviour} \label{Big-As}

For analogous notions of first and second {\sl large} approximations, now the quadrature in ${\cal U} = 
1/{\cal R}$ takes the same form as the ${\cal R}$-quadrature in the above workings with 
\beq
u_0 \longrightarrow w_0 \mbox{ } , \mbox{ } v \longrightarrow x
\eeq
($w_0$, $w_2$ and $x$ below are then the obvious analogues of $u_0$, $u_2$ and $v$).  
Hence the solutions are dual to those of Sec \ref{Small-As}.  
Thus all of Sec \ref{Small-As}'s results apply again under the duality substitutions. 
[The subsequently-induced language changes are as follows. 
`small' $\longrightarrow$ `large'. 
``2, 1, 3 collinearity with particle 1 at the centre of mass of particles 2, 3" $\longrightarrow$ 
``collision between particles 2, 3, which is also interpretable as particle 1 escaping to infinity".]

The first approximation is then 
\beq
\pm\sqrt{2w}\,\mbox{cos}(\Phi - \Phi_0) = {\cal R} = \mbox{tan$\frac{\Theta}{2}$} = \mn_1/\mn_2 \mbox{ } .
\eeq
In the (${\cal R}, \Phi$) plane and for $\Phi_0 = 0$, this takes the form of a family of circles of radius 
$\sqrt{w/2}$ and centre $(\sqrt{w/2}, 0)$, so that they are all tangent to the vertical axis through the origin \cite{TriCl}.  
2) The second approximation gives 
\beq
\pm\sqrt{w + \sqrt{w^2 - x}\, \mbox{cos}(2\{\Phi - \bar{\Phi}\})} = {\cal R} = 
\mbox{tan$\frac{\Theta}{2}$} = \mn_1/\mn_2 \mbox{ } .  
\eeq
In the ${\cal R}, \Phi$ plane and for $\Phi_0 = 0$, this takes the forms in Fig \ref{Fig-4AF}d) in the parameter regions of Fig \ref{Fig-4AF}b).

\noindent Note: for general $\widetilde{\ttV}_{(\sn, 0)}$ with constant of proportionality $\Lambda_{\sn}$, one gets exactly the same 
large-asymptotics analysis as here, with $u_0 = 2\ttE - \Lambda_{\sn}$, $u_2 = 4\ttE - \{4 + \mn\}\Lambda_{\sn}$.  
Thus generally this duality to the isotropic HO of the universal large-scale asymptotics of pure-shape 
triangleland is a useful and important result for this Machian mechanics. 
This usefulness is due to the classical and quantum mechanics of isotropic HO's being rather 
well-studied and thus a ready source of classical and quantum methods, results, and insights.

\noindent The second large approximation's solutions are likewise [by the duality] ellipses but in the corresponding (${\cal U}, \Phi)$-chart. 
These ellipses map to more unusual closed curves (given in \cite{08I}) in the (${\cal R}, \Phi)$-chart, 
and correspond to a sequence of very flat triangles.

\subsection{Scaled 3-stop metroland with free and HO potentials}\label{S3Stop-ClSol}

Regardless of the scale-dependence of the potential, $\sfD = 0$ orbit shape is always a radial half-line.  
Some of these are physically/shape-theoretically special by being along an M or D direction.  
The general such line corresponds to $\rho_1$ and $\rho_2$ being in a fixed proportion, 
so that the shape is fixed and one has a dynamics of pure scale.  

\noindent $E \leq 0, A \geq 0$  is classically impossible due to producing a negative radicand.

\subsubsection{Traversal times for free and (upside-down) isotropic HO}

\noindent{\boldmath$\sfD = 0$}

\mbox{ } 

\noindent 1) $E = 0 = A$    is solved by $\rho$ = constant.  

\noindent 2) $E = 0, A < 0$ is solved by $\lt^{\se\sm(\sJ\sB\sB)} = \{1/\sqrt{-2A}\}$ln$\,\rho$, which inverts to $\rho = \mbox{exp}(\sqrt{-2A}\,\lt^{\se\sm(\sJ\sB\sB)})$.

\noindent 3) $E > 0, A = 0$ is solved by $\lt^{\se\sm(\sJ\sB\sB)} = \rho/\sqrt{2E}$, which inverts to $\rho = \sqrt{2E}\,\lt^{\se\sm(\sJ\sB\sB)}$.

\noindent 4) $E > 0, A > 0$ is solved by $\lt^{\se\sm(\sJ\sB\sB)} = \{1/\sqrt{2A}\} \mbox{arcsin}(\sqrt{A/E}\,\rho)$,  which inverts to 
$\rho = \sqrt{E/A}\,  \mbox{sin}(\sqrt{2A}\,\lt^{\se\sm(\sJ\sB\sB)})$.

\noindent 5) $E > 0, A < 0$ is solved by $\lt^{\se\sm(\sJ\sB\sB)} = \{1/\sqrt{-2A}\}\mbox{arsinh}(\sqrt{-A/E}\,\rho)$, 
which inverts to $\rho = \sqrt{E/-A}\, \mbox{sinh}(\sqrt{2A}\,\lt^{\se\sm(\sJ\sB\sB)})$.

\noindent 6) $E < 0, A < 0$ is solved by $\lt^{\se\sm(\sJ\sB\sB)} = \{1/\sqrt{-2A}\}\mbox{arcosh}(\sqrt{A/E}\,\rho)$ , 
which inverts to $\rho = \sqrt{E/A}\,  \mbox{cosh}(\sqrt{-2A}\,\lt^{\se\sm(\sJ\sB\sB)})$.

\mbox{ }

\noindent{\boldmath$\sfD \neq 0$}

\mbox{ } 

\noindent 1) $E = 0 = A$ is now classically impossible by producing a negative radicand.

\noindent 2) $E = 0, A < 0$ is solved by $\lt^{\se\sm(\sJ\sB\sB)} = \{1/2\sqrt{-2A}\} \mbox{arcosh}(\{\sqrt{-2A}/|\sfD|\}\rho^2)$, 

\noindent                        which inverts to $\rho = \sqrt{|\sfD|/\sqrt{-2A}}\mbox{cosh}(2\sqrt{-2A}\,\lt^{\se\sm(\sJ\sB\sB)})$.
 
\noindent 3) $E > 0, A = 0$ is solved by $\lt^{\se\sm(\sJ\sB\sB)} = \sqrt{\rho^2 - \sfD^2/2E}$, 

\noindent which inverts to $\rho = \sqrt{\lt^2 + \sfD^2/2E}$.

\noindent 4) $E > 0, A > 0$ is solved by $\lt^{\se\sm(\sJ\sB\sB)} = \{{1}/{2\sqrt{2A}}\}\mbox{arcsin}\left(\{2A\rho^2 - E\}/{\sqrt{E^2 - 2A\,\sfD^2}    }\right)$, 

\noindent                        which inverts to $\rho = \sqrt{\{\sqrt{E^2 - 2A\sfD^2}  \mbox{sin}(2\sqrt{2A}\,\lt^{\se\sm(\sJ\sB\sB)}) + E\}/2A}$.

\noindent 5) $E > 0, A < 0$ is solved by $\lt^{\se\sm(\sJ\sB\sB)} = \{1/2\sqrt{-2A}\}\mbox{arcosh}\left(\{2A\rho^2 - E\}/\sqrt{E^2 - 2A\,\sfD^2}\right)$, 
                        which inverts to 
						
\noindent $\rho = \sqrt{\{\sqrt{E^2 - 2A\,\sfD^2}\mbox{cosh}(2\sqrt{-2A}\,\lt^{\se\sm(\sJ\sB\sB)}) + E\}/2A}$. 

\noindent 6) $E < 0, A < 0$ is as for 5).

\noindent 4), 5), 6) are subjected to 
\beq 
E^2 > 2A\,\sfD^2 
\label{Eexisto}
\eeq 
in order to exist.

\subsubsection{Shapes of the orbits for free and (upside-down) isotropic HO cases}\label{llaut}

\noindent{\boldmath$\sfD$}{\bf = 0} forces all orbits to lie upon the radial half-line.  

\mbox{ }

\noindent{\boldmath$\sfD$}{\bf $\neq 0$}

\mbox{ } 

\noindent 2) $E = 0, A < 0$ is solved by $\rho = \sqrt{    |\sfD|    /    \sqrt{-2A}    }/\mbox{sin}(\sqrt{-2A}\{\{\varphi - \varphi_{\scc}\}/\sfD^2)$; 
this blows up as befits an upside-down HO.  

\noindent Those parts of 3), 4), 5), 6)  that exist due to obeying (\ref{Eexisto}) are solved by 
$\rho = |\sfD|/\sqrt{\sqrt{E^2 - 2A\,\sfD^2}   \mbox{sin}(2\{\varphi - \varphi_{\scc}\}) + E}$ 
with as its free subcase the simpler form $\rho = |\sfD|/\sqrt{E(1 + \mbox{sin}(2\{\varphi - \varphi_{\scc}\})}$.  

\mbox{ } 

\noindent Note 1) This simpler form can be beaten into the conventional `sec' form of the general line not through the origin that meets basic physical intuition, 
via a redefined constant turning the sine into a cosine, and then use of 1 + cos$\,2\kappa$ = 2\,cos$^2\,\kappa$ and taking the --1/2th power.

\noindent Note 2) The former expression remains finite for HO's ($A > 0$ from $E^2 - 2A\,\sfD < E$) but blows up for upside-down HO's (opposite signs).

\subsubsection{Special HO solved in Cartesians}

In the special HO case, 
\beq
\lt^{\se\sm(\sJ\sB\sB)} = {\sqrt{K_{\barr}}}^{-1}\mbox{arcsin}
\left(
\sqrt{{K_{\barr}}/{2E_{\barr}}\rho_{\barr}}
\right) \mbox{ } ,
\eeq
and the shapes of the orbits are 
\beq
{K_1}^{-1}\mbox{arcsin}
\left(
\sqrt{{K_{1}}/{2E_{1}}\rho_{1}}
\right)
= {K_2}^{-1}\mbox{arcsin}
\left(
\sqrt{{K_{2}}/{2E_{2}}\rho_{2}}
\right) + \mbox{const ,}
\eeq
the constant arising from the two subsystem `calendar year zero's' not being the same in general.

\subsection{Exact solutions for scaled triangleland}\label{STri-ClSol-1}

The free problem in Dragt coordinates gives
\beq
Dra^{\Gamma} = A^{\Gamma} t^{\se\sm(\sJ\sB\sB)} + B^{\Gamma}
\eeq
for $A^{\Gamma}, B^{\Gamma}$ constants. 
Then
\beq
\{Dra_1 - B_1\}/A_1 = \{Dra_2 - B_2\}/A_2 = \{Dra_3 - B_3\}/A_3 \mbox{ } .
\eeq
This method is of limited use upon the introduction of potential terms however, since the usual potentials tend to 
become complicated nonseparable combinations of the Dragt coordinates.

\subsubsection{Simple exact solutions for triangleland with $\Phi$-free potentials}
\label{STri-ClSol}

Each of the free-free, attractive Newton--Coulomb-free, HO-free and the  
aforementioned special multiple HO setting problems separate into single-variable problems.

In the case of zero relative angular momentum, the motion is linear (and indeed equivalent to the 1-$d$ problem at the classical level.
The free-free problem's solution is 
\beq
\rho_2 = \mbox{const}\,\rho_1 + \mbox{Const} \mbox{ } , \mbox{ } \Phi \mbox{ fixed .}
\eeq
The HO-free problem's solution is
\beq
\rho_1 = \sqrt{\frac{E_1}{K_1}}\mbox{sinh}
\left(
\sqrt{ \frac{K_1}{E_2 + K_2} \sqrt{\mu_2} \{\rho_2 - \mbox{ const}\}    }
\right) 
\mbox{ } , \mbox{ } \Phi \mbox{ fixed .}
\eeq

\subsubsection{Very special case}\label{VSpecial-SClSol}

Here,
\beq
{  \mI^{\check{*}2}  }/{  2  } + {  \sfJ^2  }/{  2\mI^2\mbox{sin}^2{\Theta}_0  } + A = {E}/{\mI}
\eeq
[usually one would set $\Theta_0 = \pi/2$ without loss of generality, however the present physical 
interpretation has the value of $\Theta_0$ be meaningful, as 
$\Theta_0 = 2\mbox{arctan}(\rho_1/\rho_2)$].  
Thus the solutions are conic sections  
\beq
\mI = {l}/\{1 + \mbox{$e$ cos}(\Phi - \Phi_0)\}
\eeq
where the semi-latus rectum and the eccentricity are given by 
\beq
l = {\sfJ^2}/{\breve{E}\,\mbox{sin}^2\Theta_0}  
\mbox{ } , \mbox{ } 
e = \sqrt{1 - {2A\,\sfJ^2}/{\breve{E}\mbox{}^2\mbox{sin}^2\Theta_0}} \mbox{ } .  
\eeq
So, in terms of straightforward relational variables,  
\beq
\mI_1 + \mI_2 = {l}/\{1 + \mbox{$e$ cos}(\Phi - \Phi_0)\} \mbox{ } . 
\eeq

In moment of inertia--relative angle space, for $2A = \{\breve{E}\,\mbox{sin}\,\Theta_0/\sfJ\}^2$ one has circles. 
For $0 < 2A < \{\breve{E}\,\mbox{sin}\,\Theta_0/\sfJ\}^2$ one has ellipses. 
For $A = 0$ one has parabolae (corresponding to the case with no springs).  
The hyperbolic solutions ($A < 0$) are not physically relevant here because this could only be attained with negative Hooke's coefficient springs.  
The circle's radius is $\mI = l = \breve{E}/2A$ while for the ellipses $\mI$ is bounded to lie between $\sfJ/\sqrt{2A}\,\mbox{sin}\,\Theta_0$ and $\breve{E}/2A$.  
The smallest $\mI$ attained in the parabolic case is $\sfJ^2/2\breve{E}\,\mbox{sin}^2\Theta_0$.  
The period of motion for the circular and elliptic cases is $\pi\breve{E}/\sqrt{2A^3}$.

As regards the individual subsystems, combining the fixed plane equation and the $\mI = \mI(\Phi)$ relation,
\beq
\mI_1 = {   l\,\mbox{sin}^2\mbox{$\frac{\Theta}{2}$}    }/
           \{   1 + \mbox{$e$ cos}(\Phi - \Phi_0)    \} 
\mbox{ } \mbox{ } , \mbox{ } \mbox{ } 
\mI_2 = {   l\, \mbox{cos}^2\mbox{$\frac{\Theta}{2}$}    }/ 
           \{   1 + \mbox{$e$ cos}(\Phi - \Phi_0)    \} 
\eeq
so each of these behave individually similarly to the total $\mI$.  
In the $\Theta_0 = \pi/2$ plane, they are both equal (and so equal to I/2).  
The circle and ellipse cases have $\mI_1$ and $\mI_2$ as closed bounded curves which sit inside the curve that $\mI$ traces.  
The parabolic case has $\mI_1$, $\mI_2$ curves to the `inside' of the parabola that $\mI$ traces.

\subsubsection{Special case}\label{Special-SClSol}

This is solved by 
\beq
\overline{\lt}^{\se\sm(\sJ\sB\sB)} = \{2/\sqrt{K_i}\}\mbox{arccos}
\left(
\{\mI_iK_i - E_i\}\left/{\sqrt{E_i^2 - K_i\sfJ^2}} 
\right.\right)
\eeq
(in agreement with \cite{TriCl}, once differences in convention are taken into account).  
Thus, synchronizing, one part of the equation for the orbits is 
\beq
\sqrt{K_2}\mbox{arccos}
\left(
\{\mI_1K_1 - E_1\}\left/{\sqrt{E_1^2 - K_1\sfJ^2}} 
\right.\right) = 
\sqrt{K_1}\mbox{arccos}
\left(
\{\mI_2^2K_2 - E_2\}\left/{\sqrt{E_2^2 - K_2\sfJ^2}} 
\right.\right) \mbox{ } .  
\eeq
[One can see how the arccosines cancel in the very special case... 
Then $E_1 = E_2 = E/2$ gives  ${\rho_1} = {\rho_2}$ i.e. $\Theta = 2\,\mbox{arctan} \left(\rho_1/\rho_2\right) = 2$, $\mbox{arctan}(1) = {\pi}/{2}$, 
so they are confined to the plane perpendicular to the chosen Z-axis.]  
Then the $\Phi$ evolution equation implies
$$
\Phi - \Phi_{\scc} = \sfJ\int\d\overline{\lt}
\left\{
\frac{1}{\mI_1} + \frac{1}{\mI_2}
\right\} = 
\frac{\sfJ}{2}\sum\mbox{}_{\mbox{}_{\mbox{\scriptsize i = 1}}}^{2}\sqrt{K_i}\int\frac{\d\tau_i}{F_i\mbox{cos}\,\tau_i + E_i} = 
\sum\mbox{}_{\mbox{}_{\mbox{\scriptsize i = 1}}}^2 \mbox{arctan}
\left(
\sqrt{\frac{\{E_i - F_i\}\{F_i - G_i\}}{\{E_i + F_i\}\{F_i + G_i\}}}
\right)
$$  
(for $\tau_i = 2\sqrt{K_i}\lt$ and $F_i := \sqrt{E_i^2 - K_iJ^2}$, $G_i = K_i\mI_i - E_i$, ), which simplifies to 
\beq
\Phi - \Phi_{\scc} = \sum\mbox{}_{\mbox{}_{\mbox{\scriptsize i = 1}}}^{2}\mbox{arctan}
\left(
\sqrt{\left.\left\{\left\{\sqrt{E_i^2 - K_i\sfJ^2} - E_i\right\}\mI_i  + \sfJ^2\right\}\right/
            \left\{\left\{\sqrt{E_i^2 - K_i\sfJ^2} + E_i\right\}\mI_i  - \sfJ^2\right\}   }
\right)
\eeq
in the straightforward relational variables.

\subsubsection{Separate working required for single HO case}\label{SingleHO-ClSol}

For $K_1 = K_2 = 0$, the trajectories are given by, after some manipulation, 
\beq
\sqrt{    {E_2}/{E_1}    }\mI_2 = \mI_1 = \mbox{sec}
(E_1\{\Phi - \Phi_0\}/E)/{\sqrt{2E_1}} \mbox{ } ,  
\eeq
which is obviously the expected straight line in the absence of forces.

For $K_2 = 0$, $K_1 \neq 0$, the trajectories are given by, in straightforward relational variables,  
\beq
\{\mI_1 K_1  - E_1\}/\sqrt{E_1^2 - K_1 \sfJ^2} = 
\mbox{cos}
\left(
\sqrt{{2K_1}/{E_2}}\sqrt{2E_2\mI_2 - \sfJ^2}
\right)
\label{melenes}
\eeq 
and
\beq
\Phi - \Phi_0 = \mbox{arctan}
\left(
\sqrt{       \left.\left\{\left\{\sqrt{E_1^2 - K_1\sfJ^2} - E_1 \right\}\mI_1  + \sfJ^2 \right\}\right/
             \left\{\left\{\sqrt{E_1^2 - K_1\sfJ^2} + E_1 \right\}\mI_1  - \sfJ^2  \right\}        }
\right)
+ \mbox{arctan}
\left(
\sqrt{        \frac{2E_2\mI_2\mbox{}^2}{\sfJ^2}  - 1   }
\right) \mbox{ } .  
\label{hiperboliques}
\eeq

\subsubsection{A brief interpretation of the previous two SSSecs' examples}\label{Interp-ClSol}

In Sec \ref{SingleHO-ClSol}'s example, $\rho_2$ (or $\mI_2$) makes a good time-standard as the absolute space 
intuition of it `moving in a straight line' survives well enough to confer monotonicity.   
It is convenient then to rewrite (\ref{melenes}, \ref{hiperboliques}) as a curve in parametric form 
with $\mI_2$ playing the role of parameter. 
 
\noindent In Sec \ref{Special-SClSol}'s example, $\rho_1$ and $\rho_2$ oscillate boundedly, so neither of 
these (or $\mI_1$ or $\mI_2$) is globally a good clock parameter from the point of view of monotonicity.
There is again some scope for variation in relative angle $\Phi$, including `sporadic' amplitude  variations.

\subsubsection{General HO counterpart for triangleland}\label{SNormal-ClSol}

The working of Sec \ref{Tri-Norm} holds again (now capitalizing $Dra$: $Dra_{N}$) but radii are 
unaffected by rotations and so cancel out giving the same rotation and $\Theta$, $\Phi$ to 
$\Theta_{N}$, $\Phi_{N}$ coordinate change as before.

The evolution equations and energy integral here are, respectively and after discovering the 
conserved quantity $J$ and eliminating it, as for the special case but with $N$-subscripts appended.  
The momenta, Hamiltonian and the energy 
constraint are as before with $N$-subscripts appended to the various quantities.  
One can then rotate the above two exact solutions for the special case to obtain solutions to the 
general case, but these are too lengthy to include in this Article.

\subsection{Types of behaviour of cosmologically-inspired (approximately) classical solutions}\label{TOB}

{            \begin{figure}[ht]
\centering
\includegraphics[width=0.9\textwidth]{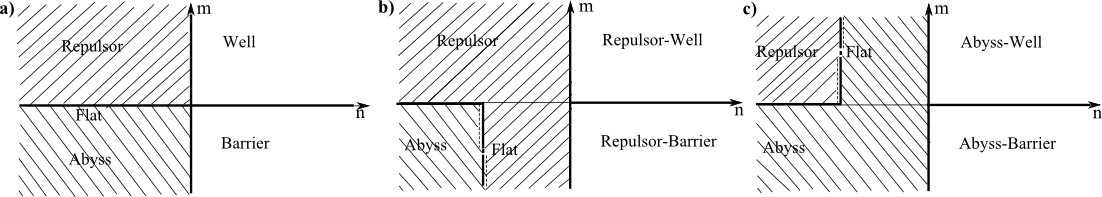}
\caption[Text der im Bilderverzeichnis auftaucht]{        \footnotesize{a) 5 classical behaviours for 
$V = \mm\rho^{\sn}$ potentials (these could just as well be scaled triangleland $\mm \mI^{\sn}$ potentials).  
Solid lines denote included edges and dashed lines denote excluded edges.
In the top-right quadrant, a representative is the HO.  
In the bottom-right quadrant, a representative is the upside-down HO.  
In the bottom-left quadrant, a representative `abyss' is the bound l = 0 analogue of the hydrogen atom 
model.
In the top-left quadrant, a representative `repulsor' is the L = 0 analogue of the electron--electron model.
The fifth region is the axes, for which the potential is constant.  

\noindent b) Next consider the also commonly occuring (under the 
$\rho \longrightarrow r$, Net = $\sfT\sfO\sfT - 2R \longrightarrow L_{\sT\so\st}$  analogy) case of 
$V = \mm\rho^{\sn}$ + Net/$\rho^2$, i.e. with a Net $ > 0$ `centrifugal barrier' added.
This sends the HO to the repulsor HO, the upside-down HO to the repulsor upside-down HO.  
Also now the repulsor behaviour takes over the axes and a strip plus partial boundary of the bottom-left 
quadrant, pushing out the `abyss' behaviour to $\mn < -2$ and the lower part of $\mn = 2$.  
The only case exhibiting the flat behaviour now is the critical value of $\mm$ for $\mn = - 2$.  

\noindent Note: one of the representatives of repulsor, however, (L $\neq$ 0 bound Hydrogen analogue) has a well next to the repulsor.

\noindent c) Finally, consider the Net $ < 0$ case c), which is not mechanically 
standard but is cosmologically standard (radiation term).  
This sends the HO to the abyss--HO, the upside-down HO to the abyss--upside-down-HO.  
Also now it is the abyss behaviour that takes over the axes and a strip plus partial boundary of the 
top-left quadrant, pushing out the `repulsor' behaviour to $\mn < -2$ and the upper part of $\mn = 2$.   }        }
\label{Fig-3}
\end{figure}  }

As there are many, one needs a more efficient method of qualitatively understanding these than the previous SSecs on HO-type potentials. 
\noindent The shape of the potentials can be classified into the regions of Fig \ref{Fig-3}. 
The corresponding qualitative classical behaviours of these are in Fig 4; therein I make use of   
Robertson's kind of \cite{Robertson33}'s notation for types of solution, also used in e.g. \cite{Harrison67, Rindler}. 
This involves using 
$O$ for oscillatory models of the types 
$O_1$ ($0 \leq \rho \leq \rho_{\sm\sa\sx}$), 
$O_2$ ($\rho_{\sm\si\sn} \leq \rho \leq \rho_{\sm\sa\sx}$) and  
$O_3$ ($\rho_{\sm\si\sn} \leq \rho \leq \infty$).
$M$ for monotonic solutions of type 
$M_1$ ($0 \leq \rho \leq \infty$), and 
$M_2$ ($\rho_{\sm\si\sn} \leq \rho \leq \infty$).
$S$ for static solutions of types 
$S_1$ unstable, 
$S_2$ stable and 
$S_3$ stable for all $\rho$.
$A$ for solutions asymptotic to static solutions at a finite $\rho_{\sA}$, of types 
$A_1$ coming in from 0 and 
$A_2$ coming in from $\infty$.

{            \begin{figure}[ht]
\centering
\includegraphics[width=0.8\textwidth]{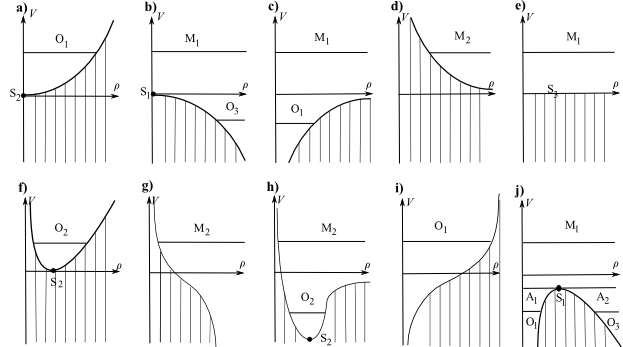}
\caption[Text der im Bilderverzeichnis auftaucht]{        \footnotesize{ Qualitative classical 
behaviour and identification of corresponding cosmological solutions.
Shaded regions are classically forbidden.  
\noindent a) The HO model corresponds to a Milne in AdS cosmology (I vii) in Appendix \ref{Cl-Soln}.A); the potential is nonsingular at 0.  

\noindent b) For zero and positive energy, the upside-down HO model corresponds to the ($k = 0$, ordinary) de Sitter and $k = -1$ de Sitter sinh cosmologies,  
with instability causes separation to keep on increasing and faster, which motion corresponds to a significant expansion phase in cosmology. 
For negative energy, it can be taken to bounce ($O_3$) at a finite $\rho_{\tm\ti\tn}$.  
This corresponds to the $k = -1$ de Sitter cosh solution.

\noindent c) This behavior covers pure dust and corresponds to Newtonian Gravity mechanics models and pure radiation 
corresponding to conformal potential mechanics models.
The flat and negatively curved (open in cosmology) models therein involve expanding forever ($M_2$).  
On the other hand, the positively curved  (closed in cosmology) models involve recollapse ($O_3$) from and to a point at which the potential is singular.  

\noindent d) This behaviour covers pure wrong-sign radiation, corresponding to mechanics with centrifugal term and/or positive conformal potential.   

\noindent e) This constant-potential case involves expansion forever ($M_2$) or everywhere static stable behaviour.

\noindent f) This is wrong-sign radiation (centrifugal term and/or positive conformal potential) alongside negative cosmological constant  (alias HO).    

\noindent g) This is wrong-sign radiation corresponding to conformal potential alongside positive cosmological constant/HO.    

\noindent h) This is the L $\neq$ 0 Hydrogen analogue, corresponding to Newton-conformal potential problem and has oscillatory (`bound') and monotonic (`ionized') regions.  

\noindent i) This is radiation (sufficiently negative conformal potential) alongside negative cosmological constant/HO and 
involves recollapse from and to a point at which the potential is singular. 

\noindent j) This is radiation (sufficiently negative conformal potential) alongside positive 
cosmological constant/upside-down HO, covering many different possible behaviours.   

\noindent [Models with all of dust, radiation and cosmological constant \cite{Harrison67,Vajk,CG82,DS86} exhibit behaviours 
f), g), i) and j), with scope for turning points in the intermediate region.]  
}        }
\label{Fig4-three}
\end{figure}  }

Some solutions corresponding to QM models studied in this Article are as follows.  
$\rho =  \mbox{sin}(\sqrt{2A}t^{\se\sm(\sJ\sB\sB)})/\sqrt{2A}$ is analogous to the Milne in anti de Sitter 
solution.
$\rho =  \mbox{cosh}(\sqrt{-2A}t^{\se\sm(\sJ\sB\sB)})/\sqrt{-2A}$ is analogous to the positively-curved de Sitter model. 
[Both are models with just analogues of $k$, $\Lambda$ of various signs.]
$\rho = \{-9K/2\}^{1/3}t^{\se\sm(\sJ\sB\sB)\,2/3}$ is analogous to the flat dust model, with well known 
cycloid and hyperbolic counterpart solutions in the positively and negatively-curved cases.  
$\rho = \{4\{2R - \sfT\sfO\sfT\}\}^{1/4}t^{\se\sm(\sJ\sB\sB) \, 1/2}$ is analogous to 
the flat radiation model, with well-known curved counterparts (one analogous to the Tolman model). 
The present problem also requires cosmologically less familiar wrong-sign radiation models familiar 
(but these are familiar in ordinary mechanics as solutions with central term), such as $\rho = \sqrt{t^{\se\sm(\sJ\sB\sB) \, 2} + \sfT\sfO\sfT}$.

\subsection{Scaled RPMs' approximate scale solutions (semiclassically relevant)}\label{Approx-ClSol}

I build these by the mathematical analogy with mostly Appendix \ref{Cl-Soln}.A and occasionally Appendix \ref{Cl-Soln}.B.

\subsubsection{$N$-stop metroland analogue cosmology approximate scale equation solutions}\label{NStop-CosSol}

I argue that $E-$normalizing suffices for the current qualitative treatment. 
I also point out that this work also applies to the $\mathbb{CP}^{N - 1}$ presentation of $N$-a-gonland.  

\noindent I) Models with energy and (upside-down) HO potentials' scale contribution are as follows.     

\noindent i)   $E = 0$, $\ttA = 0$ is $\rho$ constant: a static model universe.  

\noindent ii)  $E = 1/2$, $\ttA = 0$ is $\rho = t^{\se\sm(\sJ\sB\sB)}$.

\noindent iii) $E = -1/2$ and $\ttA \geq 0$ is impossible.  

\noindent iv)  $E = 0$ and    $\ttA < 0$ is $\rho = \mbox{exp}(\sqrt{-2 \ttA}t^{\se\sm(\sJ\sB\sB)})$.

\noindent v)   $E = 1/2$ and  $\ttA < 0$ is $\rho = \{1/\sqrt{-2 \ttA}\}\mbox{sinh}(\sqrt{-2\ttA}t^{\se\sm(\sJ\sB\sB)})$.  

\noindent vi)  $E = -1/2$ and $\ttA < 0$ is $\rho = \{1/\sqrt{-2 \ttA}\}\mbox{cosh}(\sqrt{-2\ttA}t^{\se\sm(\sJ\sB\sB)})$. 

\noindent vii) $E = 1/2$ and  $\ttA > 0$ is $\rho = \{1/\sqrt{2 \ttA}\}\mbox{sin}(\sqrt{2 \ttA}t^{\se\sm(\sJ\sB\sB)})$.


\noindent II) Models with energy and Newtonian potentials' scale contribution are as follows.   

\noindent i)   $E = -1/2$ is a cycloid solution.  

\noindent ii)  $E = 0$ is $\rho = \{-9K/2\}^{1/3}t^{\se\sm(\sJ\sB\sB)\,2/3}$.  

\noindent iii) $E = 1/2$ is a hyperbolic analogue of the cycloid.


\noindent III) Models with energy and scale-invariant potential terms have the following approximate heavy-scale solutions 
(the $R$ is but the constant lead term of an expansion in the shape variables).  
For $2R - \sfT\sfO\sfT > 0$ (corresponding to right-sign radiation in Cosmology), 

\noindent i)   $E = -1/2$: $\rho = \sqrt{2R - \sfT\sfO\sfT}\{1 - \{1 - t^{\se\sm(\sJ\sB\sB)}/\sqrt{2R - \sfT\sfO\sfT}\}^2\}^{1/2}$.

\noindent ii)  $E = 0$: $\rho = \{4\{2R - \sfT\sfO\sfT\}\}^{1/4}t^{\se\sm(\sJ\sB\sB)) \, 1/2}$.

\noindent iii) $E = 1/2$: $\rho = \sqrt{2R - \sfT\sfO\sfT}\{\{1 + t^{\se\sm(\sJ\sB\sB)}/\sqrt{2R - \sfT\sfO\sfT}\}^2 - 1\}^{1/2}$

\noindent For $2R - \sfT\sfO\sfT < 0$, including $\sfT\sfO\sfT \neq 0$, $R = 0$, and corresponding to wrong-sign radiation 
in Cosmology,  so that the direct analogy is with the mechanics of Appendix \ref{Cl-Soln}.B,

\noindent iv) $E = -1/2$ is impossible, 

\noindent v)  $E = 0$ is also impossible, and 

\noindent vi) $E = 1/2$ gives $\rho = \sqrt{t^{\se\sm(\sJ\sB\sB) \, 2} + \sfT\sfO\sfT - 2R}$.  


\noindent V) Some further approximate (the $K$ is but the constant lead term of an expansion in the shape variables) heavy-scale solutions are as follows.  

\noindent i) $E = 0$, with upside-down HO $\ttA < 0$ and Newtonian potential terms, solved by $\rho = \{\{K/2 \ttA\}\{\mbox{cosh}(3\sqrt{-2 \ttA}t^{\se\sm(\sJ\sB\sB)}) - 1\}\}^{1/3}$.  

\noindent ii)  $E = 0$, with HO  $\ttA > 0$ and Newtonian potential terms, solved by $\rho = \{\{-K/2 \ttA\}\{1 - \mbox{cos}(3\sqrt{2 \ttA}t^{\se\sm(\sJ\sB\sB)})\}\}^{1/3}$.  
 
\mbox{ } 
 
The forms with the $E$'s not scaled out of the following out of the above models are subsequently needed in this Article. 

\noindent For model I v), $t^{\se\sm(\sJ\sB\sB)} = \{1/\sqrt{-2 \ttA}\} \mbox{arsinh}(\sqrt{- \ttA/E}\rho)$ + const.  
This model's full name is `positive energy upside-down HO RPM analogous to negative-curvature dS' (de Sitter).  

\noindent For model I vi), $t^{\se\sm(\sJ\sB\sB)} = \{1/\sqrt{-2 \ttA}\} \mbox{arcosh}(\sqrt{\ttA/E}\rho)$ + const.
This model's full name is `negative energy upside-down HO RPM analogous to positive-curvature dS'.  

\noindent For model I vii), $t^{\se\sm(\sJ\sB\sB)} = \{1/\sqrt{2 \ttA}\}  \mbox{arcsin}(\sqrt{\ttA/E}\rho)$ + const.  
This model's full name is `HO RPM analogous to Milne-in-AdS'. (HO necessarily implies positive energy; all the positives and negatives in this SSSec are strict).    

\noindent For model II ii), $t^{\se\sm(\sJ\sB\sB)} = \sqrt{-2K/9}\rho^{3/2}$ + const
This model's full name is `zero-energy Newtonian potential RPM analogous to zero-curvature dust cosmology'.

\noindent I note that all of the above approximate heavy-scale timefunctions are monotonic apart from 
model I vii), which nevertheless has a reasonably long era of monotonicity as regards modelling early-universe Quantum Cosmology.  
Model I vii) has periods proportional to $\sqrt{E/ \ttA}$ for $N$-stop metroland, which plays a role proportional to that of $1/\sqrt{\Lambda}$ in GR Cosmology.  
Finally, for model I vi) there is a nonzero minimum size.

\subsubsection{Triangleland analogue cosmology approximate scale equation solutions}\label{Tri-CosSol}

\noindent  I) Models with (upside down) HO and $|r^{IJ}|^6$ potential terms are as follows 
(these are just approximate heavy solutions in cases with $S \neq 0$).

\noindent i)   $A = 0$, $S = 0$ is $\mI$ constant: static universe.  

\ noindent ii)  $A = -1/2$, $S = 0$ is $\mI = t^{\se\sm(\sJ\sB\sB)}$. 

\noindent iii) $A = 1/2$ and $S \leq 0$ is impossible.  

\noindent iv)  $A = 0$ and $S > 0$ is $\mI = \mbox{exp}(\sqrt{2S}t^{\se\sm(\sJ\sB\sB)})$.

\noindent v)   $A = -1/2$ and $S > 0$ is $\mI = \{1/\sqrt{2S}\}\mbox{sinh}(\sqrt{2S}t^{\se\sm(\sJ\sB\sB)})$.  

\noindent vi)  $A = 1/2$ and $S > 0$ is $\mI = \{1/\sqrt{2S}\}\mbox{cosh}(\sqrt{2S}t^{\se\sm(\sJ\sB\sB)})$. 

\noindent vii) $A = -1/2$ and $S < 0$ is $\mI = \{1/\sqrt{-2S}\}\mbox{sin}(\sqrt{-2S}t^{\se\sm(\sJ\sB\sB)})$.  


\noindent II) Models with (upside down) HO and energy have the following solutions.  

\noindent
i)   $A = 1/2$ is a cycloid.  

\noindent ii)  $A = 0$ $\mI = \{9E/2\}^{1/3}t^{\se\sm(\sJ\sB\sB) \, 2/3}$.  

\noindent iii) $A = - 1/2$ is the hyperbolic analogue of the cycloid. 


\noindent III) Models with conformally invariant potential and (upside down) HO include the following solutions. 
For $2R - \sfT\sfO\sfT > 0$ (corresponding to right-sign radiation in Cosmology), 

\noindent i) $A = 1/2$: $\mI = \sqrt{2R - \sfT\sfO\sfT}
\{1 - \{1 - t^{\se\sm(\sJ\sB\sB)}/\sqrt{2R - \sfT\sfO\sfT}\}^2\}^{1/2}$. 

\noindent ii) $A = 0$: $\mI = \{4\{2R - \sfT\sfO\sfT\}   \}^{1/4}t^{\se\sm(\sJ\sB\sB) \, 1/2}$.

\noindent iii) $A = - 1/2$: $\mI = \sqrt{2R - \sfT\sfO\sfT}
\{\{1 + t^{\se\sm(\sJ\sB\sB)}/\sqrt{2R - \sfT\sfO\sfT}\}^2 - 1\}^{1/2}$.  

\noindent For $2R - \sfT\sfO\sfT < 0$, including $\sfT\sfO\sfT \neq 0$, $R = 0$, and corresponding to wrong-sign radiation in Cosmology, 

\noindent iv) $A = 1/2$ is impossible, 

\noindent v)  $A = 0$ is also impossible, and 

\noindent vi) $A = -1/2$ gives $\mI = \sqrt{t^{\se\sm(\sJ\sB\sB) \, 2} +    \sfT\sfO\sfT - 2R}$.  


\noindent IV) The model with Newtonian potentials and $E = 0$ has the approximate heavy solution 
$\mI = \{2E\}^{13/21} \{4t^{\se\sm(\sJ\sB\sB)}/3\}^{4/7}$ which parallels the flat cosmology with $P = \rho/6$.    


\noindent V) Some further approximate heavy-scale solutions are as follows.  

\noindent i) $A = 0$, $S > 0$ and E-term of suitable sign: $\mI = \{\{E/2S\}\{\mbox{cosh}(3\sqrt{2S}t^{\se\sm(\sJ\sB\sB)}) - 1\}\}^{1/3}$. 

\noindent ii) $A = 0$, $S < 0$ and E-term of suitable sign: $\mI = \{\{-E/2S\}\{1 - \mbox{cos}(3\sqrt{-2S}t^{\se\sm(\sJ\sB\sB)})\}\}^{1/3}$. 

\mbox{ } 
 
\noindent With the $A$'s not scaled out, the particular model among these which are used later on in the present Article is as follows 
(presented in the subsequently-used inverted form for $t^{\se\sm(\sJ\sB\sB)}$).

\noindent A slight generalization of model ${\cal A}$ii) with $\bar{t}^{\se\sm(\sJ\sB\sB)} = \sqrt{-2A}\,\mI$ + const. 
This model's full name is `zero-energy upside-down HO spherical triangleland analogous to negative-curvature vacuum cosmology'.

\subsubsection{The preceding in terms of $t^{\sg\se\so}$}

$t^{\sg\se\so} = \int \d t^{\se\sm(\sJ\sB\sB)}/\rho^2$ is in a particular sense more geometrically natural for shape space 
(the $\rho^2$ factor is that from $\d \fs^2$ versus $\rho^2 \d\fs^2$) and will later be found to be particularly useful.  
In all of the cases considered (both here and more extensively in \cite{ScaleQM, 08III}) this is analytically possible, and invertible so that 
one can make the scale a function of $t^{\sg\se\so\sm}$ analytically, by composition of two analytical inversions.  
For model I v)   $t^{\sg\se\so} = - \sqrt{-A/2}\,\mbox{coth}(\sqrt{-2A}t^{\se\sm(\sJ\sB\sB)})/E = -\sqrt{1 - 2A\rho^2}/\sqrt{2}E\rho$, 
for model I vi)  $t^{\sg\se\so} =   \sqrt{A/2}\,\mbox{tanh}(\sqrt{-2A}t^{\se\sm(\sJ\sB\sB)})/E  =  \sqrt{-2A\rho^2 - 1}/\sqrt{2}E\rho$ and 
for model I vii),$t^{\sg\se\so} = - \sqrt{A/2}\,\mbox{cot}(  \sqrt{2A}t^{\se\sm(\sJ\sB\sB)})/E = -\sqrt{1 - 2A\rho^2}/\sqrt{2}E\rho $. 
%
%
For the spherical presentation of triangleland model I ii) using $\mI^2$ rather than $\rho^2$,
\beq
t^{\sg\se\so} = \sqrt{\frac{2}{-A}}\left\{\frac{1}{\mI_0} - \frac{1}{\mI}\right\} \mbox{ } . 
\eeq
Thus, inverting, $I = 1/\{1/\mI_0 - \sqrt{-A/2}t^{\sg\se\so}\} = \sqrt{2/-A}/\{{t}^{\sg\se\so}_{\scc} - t^{\sg\se\so}\}$ for 
${t}^{\sg\se\so\sm}_{\scc} := \sqrt{2/-A}/\mI_0$.  
%

\subsubsection{Comments on these RPM solutions}\label{CosSol+}

Note in particular that the cyclic trial models with HO mathematics I vii) of \cite{AF, 08I, 08II, +Tri} do correspond to a known cosmology (Milne in anti de Sitter).  
Also, having some upside-down HO's, rather (also readily tractable), is de Sitter/inflationary in character [I iv), v) and vi)].  
Other models parallel the dynamics of fairly realistic simple models of the early universe involving 
radiation, spatial curvature and cosmological constant type terms.

\begin{subappendices}
\subsection{A range of standard GR isotropic cosmology solutions}\label{GRCos-Solns}

\noindent I) Models with spatial curvature and cosmological constant are as follows (see e.g. \cite{Rindler}).     

\noindent
i)   $k = 0$, $\Lambda = 0$ is $a$ constant: a static universe.  

\noindent
ii)  $k = -1$, $\Lambda = 0$ is $a = t^{\scc\so\sss\sm\si\scc}$.

\noindent
iii) $k = 1$ and $\Lambda \leq 0$ is impossible.  

\noindent
iv)  $k = 0$ and $\Lambda > 0$ is $a = \mbox{exp}(\sqrt{\Lambda/3}t^{\scc\so\sss\sm\si\scc})$: a deSitter/inflationary model with zero curvature.  

\noindent
v)   $k = -1$ and $\Lambda > 0$ is $a = \sqrt{3/\Lambda}\,\mbox{sinh}(\sqrt{\Lambda/3}t^{\scc\so\sss\sm\si\scc})$: a  
de Sitter/inflationary type model with negative curvature.  

\noindent
vi)  $k = 1$ and $\Lambda > 0$ is $a = \sqrt{3/\Lambda}\,\mbox{cosh}(\sqrt{\Lambda/3}t^{\scc\so\sss\sm\si\scc})$: 
a de Sitter/inflationary type model with positive curvature.

\noindent
vii) $k = -1$ and $\Lambda < 0$ is $a = \sqrt{-3/\Lambda}\,\mbox{sin}(\sqrt{-\Lambda/3}t^{\scc\so\sss\sm\si\scc})$ 
-- a `Milne in anti de Sitter' \cite{Rindler} oscillating solution.  

\mbox{ }

\noindent II) Models with spatial curvature and dust are as follows.    

\noindent i)   $k = 1$ is the well-known cycloid solution.  

\noindent ii)  $k = 0$ $a = \{9GM/2\}^{1/3}t^{{\scc\so\sss\sm\si\scc}\,2/3}$.  

\noindent iii) $k = - 1$ is the also well-known hyperbolic analogue of the cycloid. 


\noindent III) Models with radiation and spatial curvature include the following solutions 
\cite{Wald}.    

\noindent i) $k = 1$: $a = \sqrt{2GM}\{1 - \{1 - t^{\scc\so\sss\sm\si\scc}
/\sqrt{2GM}\}^2\}^{1/2}$ (the Tolman model).  

\noindent ii) $k = 0$: $a = \{8GM\}^{1/4}t^{{\scc\so\sss\sm\si\scc}\,1/2}$.

\noindent iii) $k = - 1$: $a = \sqrt{2GM}\{\{1 + t^{\scc\so\sss\sm\si\scc}/\sqrt{2GM}\}^2 - 1\}^{1/2}$.  

\mbox{ } 

\noindent IV) the case of $P = \epsilon/6$ is not exactly integrable except for $k = 0$, in which case the 
solution is $a = \{2GM\}^{13/21}\{4t/3\}^{4/7}$. 

\noindent Finally, I consider further combinations of the well-motivated potential terms. 
E.g., 

\noindent V) the cosmologically standard model comprising dust, spatial curvature and cosmological constant is covered e.g. in \cite{Rindler, Edwards72}.    
Solutions of this include the Lema\^{\i}tre model, a  model in which the Big Bang tends to the Einstein 
static model, the Eddington--Lema\^{\i}tre model, and various oscillatory models including a bounce.  
Solutions for $k$ and $\Lambda$ both nonzero involve in general elliptic integrals.  
Some subcases that are solvable in basic functions are 

\noindent i)   $k = 0$, $\Lambda > 0$, solved by $a = \{\{3GM/\Lambda\}\{\mbox{cosh}(\sqrt{3\Lambda}t^{\scc\so\sss\sm\si\scc}) - 1\}\}^{1/3}$, and 

\noindent ii)  $k = 0$, $\Lambda < 0$, solved by $a = \{\{-3GM/\Lambda\}\{1 - \mbox{cos}(\sqrt{-3\Lambda}t^{\scc\so\sss\sm\si\scc})\}\}^{1/3}$.  

\mbox{ } 

\noindent $|\r^{IJ}|^6$ potential terms for triangleland, can readily be obtained by 
Models with radiation and spatial curvature and cosmological constant include a subcase of what is covered by Harrison \cite{Harrison67} and Vajk \cite{Vajk}.  
Models with all of radiation, dust, spatial curvature and cosmological constant are also a subcase of Harrison's work \cite{Harrison67}.  
They are also treated more explicitly by Coqueraux and Grossmann \cite{CG82} and by Dabrowski and Stelmach \cite{DS86}.
[Note that I am considering models in which dust and radiation contributions do not interact with each other.]
On the other hand, the analogy with ordinary mechanics covers combining a `wrong sign radiation' `central term'  with these other terms (e.g. in the Newton--Hooke problem).

\subsection{Further support from ordinary Mechanics}\label{OM-Supp}

`Wrong sign radiation' in Cosmology clearly corresponds via the Cosmology--Mechanics analogy to just the 
kind of effective potential term that one has for a centrifugal barrier, which is often present and well-studied in ordinary mechanics. 
Thus this case, while cosmologically unusual, does not lead to any unusual calculations either.  
In any case, such sign changes usually do not change exact integrability, but can change the qualitative behaviour in at least some regimes. 
(Consider e.g. trigonometric functions becoming hyperbolic functions under a sign change in some elementary integrals).  
With $E$ and overall angular momentum term (= wrong-sign radiation), III iv),v) are 
impossible cases, while III vi) gives $r = \sqrt{t^2 + L^2}$.  
Some cases of this remain readily tractable if a Kepler--Coulomb potential term is added to this.

\subsection{Can RPM's model the real world at the classical level?}\label{Reality+}

\subsubsection{Could pure-shape RPM describe the real world?}

Pure-shape RPM serves as an intriguing suggestion of how a symmetry principle is capable of locally reproducing 
standard Physics at smaller scales while significantly diverging nonlocally.  
In this model, this can be interpreted as an effect of the `wider matter distribution in the universe' 
(e.g. triangleland's apex particle) affecting the physics of other subsystems (e.g. triangleland's base pair).  
This is interestingly `Machian' (albeit in a distinct sense of the word from that used in the Introduction \cite{Buckets}).  
This may have some capacity to account for deviations from standard physics at larger scales without having to invoke (as many instances of) dark matter (see e.g. the 
discussion in \cite{Westman}).   
E.g. for better explaining (at least at a nonrelativistic level) the rotation curves of galaxies without incurring unacceptably large deviations in Solar System Physics.  
Some problems with modelling the world using SRPM \cite{B03, Piombino} are, however, as follows. 

\noindent 1) Technically it is a lot less straightforward to study in 3-$d$ than in this Article's 1- and 2-$d$ cases.  

\noindent However, scale and rotational relationalism do not interfere with each other, and it is only the latter that is harder 
in 3-$d$, so that the nonrotational RPM can serve as a toy model for the investigation of the specific effects of regarding scale as unphysical in a 3-$d$ world.

\noindent 2) The observed universe's angular momentum does look to balance out to zero (see below).  
However, the observed universe's contents look to be expanding far more than contracting, and we do not 
know of a satisfactory model to fit the details of cosmological redshift without invoking scale.  
Thus zero dilational momentum prediction, unlike zero angular momentum one, would seem to be incompatible with observation, 
at least within the framework of all hitherto-conceived theories that explain the observations.

\subsubsection{Is scaled RPM restrictive at the classical level?}\label{Restrict}\label{06I-CL}

1) Scaled RPM is a Leibnizian/Machian mechanics, and yet is in agreement with a subset of Newtonian Mechanics -- 
the zero total angular momentum universes.\footnote{That   
many of the results and mathematical structures (at least in the simpler cases) previously obtained by standard 
absolutist Physics are recovered from relationalism is an interesting reconciliation as regards the extent of prejudice 
incurred by studying absolutist Physics.   
This Article does however get far enough and detailed enough to list a number of differences (see Secs \ref{QM-Intro}, \ref{RPM-for-QC}).}

\noindent 2) Subsystems need not have zero angular momentum or energy conservation.  
Thus models with such arise in relational setting too, in particular nonzero angular momentum island subuniverses \cite{BB82, LB}.  
Because of this, the restriction to $\underline\scL = 0$ universes is by no means as severe as might be na\"{\i}vely suggested.  

\noindent 3) However, this rests on the $N$-body problem's theoretical framework being such that initially distant subsystems 
do not fall together below any desired finite timescale internal to the subsystem under study.  
This is possible for the equations of mechanics themselves \cite{Pain, Xia}, but is clearly precluded by such as realistic matter modelling, SR and GR.  

\noindent 4) Within the mechanics framework, one can carry out checks on whether $\underline\scL$ is zero in some 
patch of the universe around us.  
Angular momentum can be estimated object-by-object from short-term observations. 
Additionally, $\underline{\scL}$ zero and nonzero (sub)systems are capable of evolving qualitatively differently (see e.g. \cite{HillMoeckel}).

\noindent 5) There is astrophysical and cosmological evidence for zero total angular momentum of the universe. 
Though it does concern a {\sl null outcome}, so it is a relevant further issue as to how accurately is this statement known to hold?  

\noindent Then note that scaled RPM {\sl predicts} this (to the extent that it is a satisfactory cosmological model, see below) 
whereas Newtonian mechanics merely allows for it as a subset of its solutions (which are qualitatively different). 
[To some extent this is a retrodiction since it was known observationally before RPM explained it.    
However, as it relates to a null result, it remains a prediction that this result will {\sl stay null} as our detailed knowledge 
of the universe's contents increases with the scientific progress.]

There is however a loophole in that we do not know whether there is some more complicated way of relationally formulating 
$\underline{\scL} \neq 0$ mechanics.
What is known is that $\underline{\scL} \neq 0$ Newtonian mechanics is a whole lot more complicated at the level of configuration space bundle geometry \cite{LR97}, 
so that, if it were relationally formulable, the complexity of its formulation would likely considerably exceed that of all hitherto-studied RPM's.  

\noindent 6) Another loophole is that, logically, RPM applies to the whole universe, whereas the observations apply to 
the observable universe i.e. within the cosmological horizon, and the latter is very often only a portion of the former 
within the set of cosmological models that are posited to match reality.  

\noindent 7) RPM's lack a practical robustness to modelling that Newtonian mechanics possesses well. 
This is in the sense that particular distinctive features of RPM's can go away, or substantially change, 
if one improves the detail of one's model. 
E.g. if the deeper truth about the 3-particle universe is that those particles are planets, then the 2-$d$ to 3-$d$ indiscernibility 
is absent because the detailed structure of the objects themselves distinguishes these.  
Likewise, the 1 and 2 body problems being meaningless/relationally trivial can cease to make sense if they have internal structure.  
In other words, indiscernibility and relational triviality in fact depend on where one sits in the hierarchy of 
increasingly detailed models of nature. 
This is a case of objects with internal structure allowing for self-relations rather than just mutual relations.

\noindent Overall, the practical advantages of working with an RPM are the modelling of {\sl whole-universe} issues, and the 
provision of an emergent timestandard.

\subsubsection{Is scaled RPM a realistic model of Cosmology?}\label{Realistic}

That scaled RPM itself is a fairly good cosmological model as claimed in Sec \ref{Cos-Mech}  needs some qualification.  
Cosmology is conventionally modelled in terms of the Einstein--Euler system of GR alongside phenomenological fluid matter.  
In this picture, our universe is currently deep into a matter era in which the fluid is dust.  
This extends back to the early universe, at which point, somewhat after the onset of nucleosynthesis, a radiation fluid is additionally required.

On the other hand, Newtonian Mechanics of many particles suffices to model the large-scale matter era of 
Cosmology \cite{Milne,Mc1,Mc2,GP03, BGS03}.  
This makes good physical sense: dust, having no pressure, should be modellable more simply than by a fluid, and not 
require any GR input at least so as to explain many of its basic and long-standingly established features.
That {\sl Newtonian} cosmology itself suffices for this follows from how the scale--shape split of the equations of motion yields 
the `dust case of the Raychaudhuri equation' for the scale of the universe, alongside an equation for the shape part (see e.g. \cite{BGS03}).
The latter is well-known in the Celestial Mechanics literature, albeit in a slightly different context.  
(I.e. for a few particles, whereas cosmology has the large particle number case; these could be galaxy clusters, galaxies, 
stars or even dark matter particles, gravitation being universal and none of these things exerting pressure on each other.)
Also, the above analogy continues to make sense in models with pressure terms \cite{Harrison67, MTW, Pee, Rindler}, allowing for 
radiation models, and with cosmological constant terms. 
The last form of matter model which Sec \ref{Cos-Tri} parallels is sensible as a rough model of a mixture of dust and radiation 
corresponding to the conventional universe model around 60000 years after the Big Bang.  

\mbox{ }

\noindent{\bf A limitation: microlensing} \cite{Ehlers} is a salient example of a fairly large-scale phenomenon that requires additional GR input 
(or perhaps an equivalent but rather probably less insightful input of the universe being an unusual optical medium).  
I see this as a reasonable indication of the extent to which the Newtonian Cosmology paradigm can be pushed.  

\mbox{ } 

\noindent Moreover, the above Newtonian Cosmology paradigm is a zero total angular momentum universe, and therefore can be 
conceived of as scaled RPM.  
Admittedly, viewing it as scaled RPM has no immediate {\sl technical} advantages that I can think of, though it does attribute 
somewhat better explanatory power for features such as the zero total angular momentum and emergent rather than assumed timestandard.  

\noindent On the other hand, this Sec's specific models are obviously related to Cosmology by more tenuous analogies than the above (parallel 
form of the scale dynamics but in a lower dimension, with the benefit of then being reduced/relationalspace formulated).
They have the limitation of not having an underlying concept of energy density. 
[The analogies are at the level of the cosmological equations after making the $\Bigvarepsilon = \Bigvarepsilon$(scale) substitution.  
This is unsurprising given that they are only few-particle models rather than the averaged approximations to GR of the most common cosmological paradigm.  
Newtonian/RPM models would aspire to that, but only as the particle number becomes very large, which lies outside 
of the scope of the present Article's contents and interests.]

%

\noindent
Both these models and the 3-$d$ version itself have a {\bf further analogy with structure formation} in Classical and Quantum Cosmology: 
in each case, one has a simple scale problem coupled to a complicated structure formation shape problem. 
For these papers' models, these shape problems are a lot simpler than GR's, making them further investigable.
On the other hand, the 3-$d$ version's late-time Physics should in many ways coincide with the standard 
cosmology, though in this setting one has to face that the quantum-cosmological origin of the perturbations themselves 
clearly do not belong to the classical matter era setting for which 3-$d$ Newtonian Mechanics/RPM is a good model of GR.  
That is not an issue for this Article 's models since their relation to Cosmology is not meant to exceed a good qualitative analogy. 

\end{subappendices}

\vspace{10in}

\section{Further classical structures}\label{Cl-Str}

This Section is motivated in particular by timeless approaches, for which one's position is `there is no time, so we just 
have instants; what structures remain and are these enough to do some, or all of, Physics?'
How these feature and combine in specific timeless approaches, and various ways in which one might recover 
a semblance of dynamics, I leave to Secs \ref{Cl-POT-Strat}, \ref{Cl-Nihil}, \ref{QM-POT-Strat} and \ref{QM-Nihil}.
Sec \ref{QM-Str} is the quantum counterpart of the present section.  
Histories Theory also has its own parallel of these considerations (see Secs \ref{Cl-POT-Strat} and \ref{QM-POT-Strat}).

\mbox{ }  

\noindent The structures in question include notions of distance, localization, sub-statespace, union of statespaces, 
graining, information, correlation and proposition.
Some of these arise in multiple ways, e.g. notions of localization in space, in statespace and within the treatment of propositions, 
or notions of graining of statespace, of information and of propositions.
Thus there are a lot of levels of structure in instants; moreover there is vast variety of types of structure for many of these levels.
[Though the `candidate time' idea indeed extends to such as candidate distances or candidate measures of information or of the 
closely-related concept of entropy; to some extent this counteracts against the large variety of candidates that can be found in the literature 
and/or by sufficient imagination.]  
Faced with that, I have decided to illustrate the variety in the case of notions of distance between shapes, whilst remaining terse on all other fronts.

\mbox{ }  

\noindent Moreover, I seek {\sl relational versions} of all of the above.  
Being within timeless instants, `relational' boils down to Configurational Relationalism, of which there are two types of implementation as per the next 2 SSSecs.

\subsection{Direct implementation of Configurational Relationalism}

\noindent For certain kinds of objects in simple theories, this is not a hard proposition, by generalizing a means that we encountered before in Sec \ref{Examples}. 
Namely, generalize `potentials that are any functions of the form $V(\underline{r}^{IJ}\cdot\underline{r}^{JK}$ alone) for scaled RPM or of the form 
$V(\mbox{ratios of } \underline{r}^{IJ}\cdot\underline{r}^{JK}$ alone) for pure-shape RPM are relational' to 
`functions taking whichever interpretation that are of the form $V(\underline{r}^{IJ}\cdot\underline{r}^{JK}$ alone) for scaled RPM or of the form 
$V(\mbox{ratios of } \underline{r}^{IJ}\cdot\underline{r}^{JK}$ alone) for pure-shape RPM are relational'.  

\mbox{ } 

\noindent This extends even further to {\sl functionals} with these arguments, albeit subject to the restriction that these are not allowed to involve differentials of these 
variables; we already know from Sec \ref{Intro} that this would substantially complicate one's construction of $\FrG$-invariant entities.

\mbox{ } 

\noindent One can use a different spanning set of $\FrG$-invariant variables if these are available (they are whenever one has a direct r-formulation such as for 1- and 2-$d$ RPM's: 
shape variables, supplemented with scale in the scaled case).
Now, inclusion of differentials of these variables into the structure of one's functionals is permissible.
One can on these terms at least consider a formal counterpart of this in terms of likewise-restricted functionals of 3-geometries for Geometrodynamics or of knots for Nododynamics.

\subsection{$\FrG$-act $\FrG$-all indirect implementation of Configurational Relationalism}

\noindent In generalization of Relationalism 4), one {\bf builds up} compound objects Maps $\circ\,\, \cO$ as bona fide combinations of one's primary objects $\cO$'s 
({\bf tensor/bundle} structure).
This applies to the relational action (Relationalism 4 itself), but also more widely to the production of notions of distance, information, correlation... 
The indirect implementation of this is as follows. 

\mbox{ } 

\noindent {\bf $\FrG$-act $\FrG$-all method} (see e.g. \cite{M09} for a particular case).    
This builds $\FrG$-bundle objects/bibundle objects with one bundling involving $\FrG$.
Schematically, for continuous or discrete isometries (or a mixture), this is a metric background invariantizing (MBI) map
\beq
\mbox{MBI} : \mbox{Maps} \circ \cO \mapsto \cO_{\sFG} = \mbox{\Large S}_{\sFG} \circ \mbox{Maps } \circ \stackrel{\rightarrow}{\FrG} \cO
\eeq
for whichever object $\cO$ upon which $\FrG$ is able to have any kind of group action $\stackrel{\rightarrow}{\FrG}$, 
    whichever Maps render this into a physically-interesting combination, 
and whichever move that runs over all of $\FrG$.	
[Thus it is a {\sl three-fold} generalization of $\cO$.]

\mbox{ } 

\noindent Note 1) This generalizes Best Matching, for which $\FrG$-act is the infinitesimal continuous isometry group action and $\FrG$-all 
is an extremization process by variation of the Principles of Dynamics action with respect to the $\FrG$-generators in the $\FrG$-group-action.\footnote{Extremization 
is in fact one of the less desirable such moves, since an extremum can be a max or a min and 
carries less connotations of existence and, in particular, uniqueness.}  

\noindent Note 2) It also covers what is known in Shape Theory by the picturesque name of {\sl Procrustean procedure} \cite{Kendall}; 
conceptually this is the same as Best Matching as regards bringing pairs of shapes into maximal congruence, but it involves extremizing a distinct object to the RPM action.   

\noindent Note 3) This also generalizes the more basic and well-known notions of group-averaging (sum and divide by order is `all', with `act' being a suitable group action), 
or the similar group-summing/group-integrating:

\noindent
\beq
\mbox{\Large S}_{\sFG} \{ \stackrel{\rightarrow}{\FrG}{\cO} \}    \mbox{ } , \mbox{ } \mbox{ } 
\mbox{\Large S}_{\sFG} = \sum_{\sttg \in \sFG}          \mbox{ } , \mbox{ }
\int_{\sFG}\mathbb{D}\ttg                         \mbox{ } , \mbox{ }
\frac{1}{|\FrG|}\sum_{\sttg\in\sFG}                         \mbox{ } \mbox{ } \mbox{ or } \mbox{ }     \mbox{ } \mbox{ } 
\int_{\sFG}\mathbb{D}\ttg \times\left/\left \{\int_{\sFG}\mathbb{D}\ttg\right\}\right. \mbox{ as appropriate } ,
\eeq
of which Sec \ref{BRGA} has an example with exponentiated adjoint action, and the `democracy by summing over clusters' in Sec \ref{Kensei} 
is an example with the simplest action for the permutation group.
Another case is inf-taking (e.g. in Gromov--Hausdorff type constructions \cite{Gromov}, Sec \ref{Dist9}).  

\noindent Note 4)  See Appendix \ref{Cl-Str}.A for a topological-level counterpart.

\noindent Note 5) Moreover, \K's Principle steers one away from many uses of this due to favouring as early a passage to an r-formulation as is possible, 
thus leaving less space for indirect implementations of a theory's levels of structure.  

\noindent Caveat. This is limited for full GR by not being more than formally implementable for the case of the 3-diffeomorphisms. 

\mbox{ } 

\noindent {\bf Difference 21}: $\FrG$ being (along the lines of) the diffeomorphism group block many an explicit construct from being more than formal.

\subsection{Substatespaces. And Rovelli--Crane versus LMB(-CA)}\label{Cl-Pers}

The configuration spaces of Sec \ref{Q-Geom} are taken to be for whole-universe models.   
However, in Physics, in practise one usually deals with subconfigurations that correspond to subsystems (they are furthermore to be the primary objects in Records Theory). 
One can then define subconfiguration spaces for these in direct parallel to defining configuration spaces for whole systems.  
 
\mbox{ }  

\noindent {\bf Perspectivalism}.  This is the position that taking, say, some particular notion of statespace $\FrS$, to be primary in 
fact amounts to considering some set of subsystems of $\FrS$ rather than just $\FrS$ itself or even {\sl instead of} $\FrS$.
Perspectival carries connotations (especially at the quantum level) of meaning that what is observed depends on the particular specifics of the observer involved, 
e.g. their position, velocity, sensor capacity, and the subsystem--environment split corresponding to the observer.

\mbox{ } 

\noindent Rovelli and Crane have, between them, put forward 4 perspectival postulates.  
Not all have substantial classical forms since some involve observers in (something like) Quantum Theory. 
Thus there is a gap in my classical enumeration of 
these (that is plugged in Sec \ref{QM-Str}).

\mbox{ } 
	
\noindent I use Persp($\FrS$) to mean \{all Sub$\FrS \mbox{ } \underline{\subset} \mbox{ } \FrS$\}.
`Parts'  \cite{LawRose} is a similar categorical concept.\footnote{However, I do not use this name since it sounds too much like `partial', as in `partial observables'...}

\noindent As a simple example, for $\underline{\mbox{\textgoth Sets}}$, Parts is embodied by the notion of {\it power set}.  

\mbox{ }  

\noindent Note 1) Each subset of $\FrS$ would then be elevated to a subspace by possessing its own versions of 
the additional levels of structure that the whole configuration space has.
 
\noindent Note 2) Moreover, in my own presentation, the above `all' has the connotation `all physical'.
What is of interest is, rather, the cases of subconfiguration space that are maintained by the physics 
or which correspond to only partially observing a system as one does from some kind of local perspective. 
E.g. in 2-$d$ the subset comprising of the x-component of one particle and the y-component of another lacks in physical significance as a combination, as compared to, 
say, taking the y component of both.

\noindent Note 3) For triangleland, an example of the more interesting case is keeping track of the length of the base of a triangle $\mI_{\sb\sa\sss\se} = \mI_1$ 
(ignore $\mI_{\sm\se\sd\si\sa\sn} = \mI_2$ and $\Phi$).  
An example of the less interesting case is restricting attention to the isosceles triangles (this is by, as one can see from Sec \ref{Q-Geom}, fixing $\Phi$ to be $\pm \pi/2$; 
however this is prone to not being maintained by the actual dynamics of the system -- perturbative treatment of the equations in Sec \ref{Cl-Soln} for various specific potentials).
\noindent These examples illustrate that the subconfiguration spaces of interest are more likely to be families of subspaces, 
one for each value of irrelevant coordinates rather than the mathematically cleaner, but often less physical, picture of a single subspace of a given space.  

\noindent Note 4) The more interesting example above is, mathematically, a half-line which sits in the radial direction within ${\bigr}(3, 2)$. 

\noindent At least in general, not all values of the ignored quantity correspond to submanifolds of the same type for the observed quantities, 
and for some values, may not even be defined.  
\noindent I note that this example is fairly localized in space by involving the base's cluster, though subconfigurations need not be localized. 

\noindent Note 5) A useful example that illustrates the opposite is the shape--scale split: these are both subconfigurations 
to which each particle in the model contributes, and so they are highly nonlocal in space.
As well as for RPM's, this example clearly applies to inhomogeneous GR and to anisotropic minisuperspace.  

\noindent Note 6) Some subconfigurations have particular meanings pinned on them: studied subsystem 
versus environment, heavy slow h subsystem versus light fast l subsystem...   
Moreover, Nature can hardly be expected to always kindly align h-l, light-fast and studied-environment for us.  
Finally, a useful recollection from Appendix \ref{Cl-Soln}.C is that subconfiguration spaces are not necessarily themselves smaller RPM 
configuration spaces since they can each possess nontrivial properties that the whole RPM universe cannot possess, 
such as nonzero angular momentum (or, in the case of pure-shape RPM, a subconfiguration can possess an overall dilatation/expansion factor).  
Thus e.g. the subconfiguration spaces of a zero total angular momentum universe certainly need not themselves have zero total angular momentum.  

\noindent Note 7) Already at the classical level, Rovelli's partial observables and the `any' time of Relationalism 7-AMR) can be taken to be 
Perspectivalism 3) and 4) respectively.
The Partial Observables Approach is perspectival through pairs of partial observables being required in order to extract entirely physical information.
Partial observables themselves carry `any' connotations:  Perspectivalism 3 and 4) {\sl share} underlying senses of `democracy' and of 
`not needing to detailedly construct', though I argue against these in Sec \ref{Cl-POT}.    

\mbox{ }  

\noindent N.B. that Rovelli and Crane use the word `relational' for what I here term `perspectival'.\footnote{My 
terminology here is a way of maintaining some clarity in a discourse 
that contains large numbers of different ideas that are usually referred to as `relational'.  
I also credit Lee Smolin for having been involved in the conception of these perspectival ideas, and for publicizing some of them in \cite{Smolin08}.
See also e.g. \cite{SaintOurs, ARel2} for Barbour and Rovelli notions compared.}
%
Moreover, Rovelli also uses that in the sense of objects not being located in spacetime (or space) but being located with respect to each other,  
though he does not develop this to the extent of this Article's LMB-CA relational postulates.\footnote{The common ground here is along the lines 
of the sketch that the present Article terms Relationalism 0). 
Rovelli also endorses the Leibnizian timelessness postulate Relationalism 5), albeit, unlike Barbour, he does not make central use the MRI/MPI implementation of it.}
%

Rovelli then speculates that his two uses of `relational' could be linked (p 157 of the online version of \cite{Rovellibook}): 
``{\it Is there a connection... This is of course very vague, and might lead nowhere, but I find the idea intriguing.}" 
The present Article and \cite{ARel} do consider the extent to which the two can be composed (some perspectival postulates can be {\sl added} 
to the LMB(-CA) approach's, but others are revealed to be {\sl alternatives} to some of the LMB(-CA) approach's).
Firstly, Sec \ref{Intro} already explained the `any change'--`all change' distinction in the implementation of Mach's `time is to be abstracted from change' 
(Sec \ref{Cl-POT} analyzes this matter further).   
Secondly, Sec \ref{Cl-POT-Strat} presents Barbour and Rovelli's timelessness as belonging to two separate families of timeless theories 
(Rovelli's being more sui generis whilst Barbour's, for the most part, sits in larger company).
Thirdly, Sec \ref{Cl-POT-Strat} also exposits the partial--Kucha\v{r}--Dirac alternatives to observables, with Barbour's and many others' programmes using the \K or Dirac notion of these.

Some features specific to LMB(-CA) are emphasis on the configuration space, on which Jacobi-type variational principles are defined; 
this then takes Barbour straight to the situation in which no variable is distinguished as time at the kinematic level. 
On the other hand, Rovelli characterizes this as an essential difference between non-relativistic and relativistic mechanics in 
his approach, in Barbour's approach this distinction has dissolved.  
The recovery of Newtonian dynamics from RPM's also has features by which such models are closer in objective 
structure to GR than Rovelli generally holds nonrelativistic mechanics models to be in \cite{Rovellibook}.  
Thus relational particle models provide tractable models outside of the scheme that \cite{Rovellibook} is built around. 
Barbour's relationalism has the virtue of being more in line with Leibniz and Mach's thinking.   
(Thus alongside having attained a concrete mathematical implementation of these ideas it is definitely of interest to the 
Foundations of Physics and Theoretical Physics and so deserves full investigation). 
However, I would not dismiss the possibility that Rovelli's relationalism is {\it also} in line with other (interpretations of)
Leibniz, Mach or other such historical luminaries, and, in any case, has original value and is useful in a major Quantum Gravity 
program (Nododynamics, i.e. LQG \cite{Rovelli, Thiemann}).
Finally, Perspectivalism 1 and 2 are the parts that can be added to LMB(-CA) rather than constituting alternatives to some of its postulates.

\subsection{Imperfect knowledge of a (sub)system's configuration}\label{Imp}

A priori, there are two distinct notions of imperfect knowledge of a (sub)system's configuration.   

\noindent 1) Imperfect knowledge of the system's contents e.g. modelling a labelled triangle \{1, 23\} by \{1, COM(23)\}.  

\noindent 2) Imperfect knowledge of the system's state.  

\noindent However, \{1, COM(23)\} is $\mI_{\sm\se\sd\si\sa\sn}$ with $\mI_{\sb\sa\sss\se}$ and $\Phi$ arbitrary which is clearly a special 
case of imperfect knowledge of state. 
This argument extends to show that 1) is but a subcase of 2), which is in accord with the literature 
usually taking grainings to mean 2) (whether for configuration space, phase space or histories space).  

\mbox{ } 

\noindent Note 1) The above 1) also requires as good a consideration of union of configuration spaces as is possible (hence Sec \ref{Heaps}).  
Though one might restrict oneself to using a space they all sit in and then using a map $M: \FrQ \longrightarrow \mbox{Sub}\FrQ$.  
For this SSSec's example, it sends ${\bigr}(3, 2) = \mathbb{R}^3$ to $\mathbb{R}_+$ which is any radial half-line.  
But some such maps only act on some parts of the configuration space, and may map to multiple types of space 
(mathematical niceness of such a map is not in general guaranteed).

\noindent Note 2) One may well place particular emphasis on localized subsets of $\FrQ$ as regards representing one's imperfect knowledge 
of a (sub)system's state; this requires having a suitable notion of distance on $\FrQ$, and so is not developed until Sec \ref{Loc-SubCon}.

\subsubsection{Examples of physically-and-geometrically significant small regions of RPM configuration spaces}
\label{TessiRegions}

Sec \ref{Tessi} detailed significant points, edges and wedges on the configuration spaces of small RPM's, 
corresponding to such as equilateral, isosceles and regular triangles, collinearities, and various sorts of collisions and mergers.
However, these are zero-measure portions of $\FrQ$.  
Thus one needs approximate notions of these significant properties which cover nonzero-measure portions of $\FrQ$ (see Fig \ref{Fig-2AFc}).  
The work of Kendall \cite{Kendall} is a useful pointer toward how to conceive of such (as well as providing 
substantial statistical machinery for the classical analysis of questions concerning shape, which lie beyond the scope of the present Article).  
As a final piece of motivation for this SSSec, the study of regions of configuration space features in this Article's \NSI (see Secs \ref{QM-POT-Strat}, \ref{QM-Nihil}) 
and Halliwell-type combined scheme (see Secs \ref{Cl-POT-Strat}, \ref{Cl-Hist}, \ref{QM-POT-Strat}, \ref{QM-Combo}) applications.

{            \begin{figure}[ht]
\centering
\includegraphics[width=0.7\textwidth]{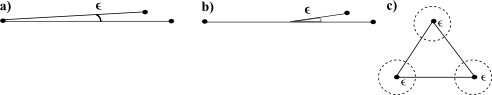}
\caption[Text der im Bilderverzeichnis auftaucht]{        \footnotesize{A meaning in space for a) Kendall's 
\cite{Kendall} $\epsilon$-blunt notion of collinearity (in fact he considers the min for this over all 
choices of construction) and my notions \cite{+Tri} of b) $\epsilon$-collinear 
(adapted to the dynamically useful Jacobi variables) \cite{+Tri}.  
Note: these two $\epsilon$'s are nontrivially related. 
c) $\epsilon$-equilateral \cite{+Tri} (a vertex is to lie within each dashed disc of radius epsilon).      }    }  
\label{Fig-2AFc}\end{figure}            }

\noindent 3-stop metroland has notions of near-collision and near-merger.  
For the pure-shape case, these are arcs centred about the D and M points.
For the scaled case, these are wedges centred about the D and M half-lines, and there is also a disc of near-maximal collision.
4-stop metroland has notions of near-T and near-DD collisions.  
These are spherical caps in the pure-shape case [Fig \ref{Fig-2AFb}a) has an example] or their 
extension in the configuration-space-radial direction to form cones in the scaled case.
There are also now near-D-arc belts [Fig \ref{Fig-2AFb}a)] and (multi)lunes [Fig \ref{Fig-2AFb}b)] in 
the pure-shape case, or their extension in the configuration-space-radial direction to form coins and wedges of solid angle 
respectively in the scaled case.  
One can just as easily construct the same kinds of regions as in the preceding sentence around M-points and M-arcs, 
and there is a sphere of near-maximal collision.  
Triangleland has notions of near-equilaterality, near-D and near-M.
These are spherical caps in the pure-shape case [Fig \ref{Fig-2AFb}c) has an example], or their 
extension in the radial I-direction to form cones in the scaled case.
There is also now a near-collinearity belt [Fig \ref{Fig-2AFb}c)] and near-isoscelesness and near-regularity (multi)lunes [Fig \ref{Fig-2AFb}d)] 
or their extension in the I-direction to form coins and wedges of solid angle respectively in the scaled case.
There is also a sphere of near-maximal collision.
Finally, some questions concern compositions of the above regions under union, intersection and negation. 
(e.g. spherical shells and segments or wedge portions of cones). 

{            \begin{figure}[ht]
\centering
\includegraphics[width=0.9\textwidth]{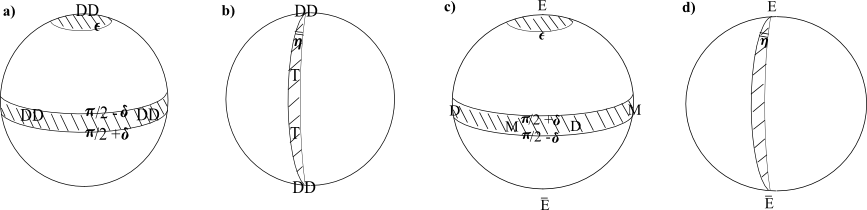}
\caption[Text der im Bilderverzeichnis auftaucht]{        \footnotesize{For 4-stop metroland, a) 
illustrates spherical caps and belts as constant azimuthal angle constructs.  
The spherical cap $\theta \leq \epsilon$ is a notion of closeness to a particular DD collision -- 
that of the \{12\} and \{34\} clusters.  
The belt $\pi/2 - \delta \leq \theta \leq \pi/2 + \delta$ then corresponds to the opposite merger
notion , i.e. to the centres of mass of \{12\} and \{34\} being near each other so that these clusters 
largely overlap (which is, in a certain sense, a more `homogeneous' universe model).

\noindent b)  illustrates a lune, i.e. a constant polar angle construct about a meridian, 
$\phi_0 - \eta \leq \phi \leq \phi_0 + \eta$.
E.g. being in the lune around the Greenwich meridian means that the \{34\} cluster is localized (as 
does the lune about the meridian antipodal to that). 
Being in the bilune perpendicular to that means that the \{12\} cluster is localized.  
Finally, being in the tetralune at $\pi/4$ to all of these signifies that clusters \{12\} and \{34\} are of 
similar size', i.e. $\eta$-close to contents homogeneity.

\noindent For triangleland c) indicates the spherical cap of near-equilaterality $\Theta \leq \epsilon$ 
and the belt of near-collinear configurations correspond to the belt around the equator, 
$\pi/2 - \delta \leq \Theta \leq \pi/2 + \delta$.
d) is an example of lune, $\Phi_{0} - \eta \leq \Phi \leq \Phi_{0} + \eta$;  statements of near-isoscelesness 
and near-regularness correspond to (multi)lunes.                     }        }
\label{Fig-2AFb}\end{figure}            }

\subsubsection{Grainings of universe contents}\label{Cl-Grainings}

Grainings are foremost well-known from SM.
I use $A \preceq B$ for `$A$ is finer-grained than $B$' and $C \succeq D$ for `$C$ is a coarser-grained than $D$'.  
In classical SM, this applies to regions of the classical phase space; in this Article it will apply also to regions of configuration space 
in the case of timeless approaches and, in another sense, within Histories Theory (see Sec \ref{Cl-POT-Strat}).  
In each case, the {\it coarsest} graining of state is the whole state space itself, while the {\it finest} 
graining of state are each of the individual points that make up the space.
Classically, grainings concern limited knowledge in practise (whilst in principle one can have perfect knowledge of a system at the classical level).  

\mbox{ }

\noindent Example 1) In ordinary mechanics or RPM's one can define an operation by which a (localized) cluster is 
approximated by replacing it by the total mass of the cluster at the centre of mass of the cluster.  
This appeared in Sec \ref{Q-Geom} as regards the three coarse-graining triangles constructible from 
a given quadrilateral in a given Jacobi coordinate system.  

\noindent Example 2) Analogously, replacing GR's two anisotropy parameters by a single representative (e.g. the average one).  

\noindent Example 3) Finally, in inhomogeneous GR one can consider defining a local averaging operation that 
approximates complicated/multiple lumps by simple/single ones.

\subsection{Unions of configuration spaces}\label{Heaps}

While each $\FrQ_{\sfA}$ corresponds to a given model with a fixed list of contents, one may not know which model a given (e.g. observed) 
(sub)configuration belongs to, or the theory may admit operations that alter the list of contents of the universe.

\noindent Then one has a collection\footnote{Barbour uses `heap of instants' for these 
\cite{Bararr, B94II, EOT}, though he largely treats this as an unstructured set, whilst I view it as much as 
possible as a more structured space or as structured a union as possible of such \cite{Records}.}
of (sub)configuration spaces of instants, $\FrQ_{\sfA(\sfC)}$, where $\sfC$ parametrizes the collection.
The subconfiguration map idea will at least sometimes be useful here too: if there is a unique biggest true 
space or good enough model, the less detailed ones can all be thought of as subsets of its configuration space.  

\mbox{ }    

\noindent Example 1) use $\bigcup_{N \in \mbox{ \scriptsize $\mathbb{N}_0$ }}\FrQ(N, d)$ for a mechanics 
theory that allows for particle coalescence/splitting or creation/annihilation.
Stratified RPM configuration spaces can at least in some cases be regarded as already being such unions of 
configuration spaces, allowing e.g. for $N$ m-masses and any coalescences among these. 
One would need to do more than that to allow for more general splittings (so that a 3m mass might forget 
it is made up of three 1-m masses and resplit instead into two 1.5m masses.  

\noindent Example 2) use $\bigcup_{\bupSigma \mbox{ \scriptsize compact without boundary}}$ Riem($\bupSigma$) for a 
formulation of GR that allows for spatial topology change (see Appendix \ref{Cl-Str}.A and Sec \ref{TopCha}).  

\noindent One can see that unions of configuration spaces allow for more general physical processes, but can be considerably harder to do mathematics with.  
\noindent Analogy 54) is between the above two examples. 

\mbox{ }

\noindent From the rest of this Sec, I use $\FrQ$ to denote a general (sub)configuration space or union of these.  

\mbox{ }  

\noindent Note 1) As per Secs \ref{Intro} and \ref{Q-Geom}, a configuration space is not just a set, it has many layers of mathematical 
structure that it is not at all clear if and how they might be extended to unions of configuration spaces.

\noindent Note 2) A main modelling point that this SSSec accommodates is that each $\FrQ_{\sfA}$ corresponds to a given model with a fixed list of contents.  
However, one may not know which model a given (e.g. observed) (sub)configuration belongs to, or the theory may admit operations that alter the list of contents of the universe.

\subsection{Notions of distance on configuration spaces} \label{Dist}

In this SSec, I consider distance in the sense of metric spaces; thus the metric space parts of Sec \ref{Q-Geom},  as 
well as the Riemannian metrics part of Sec \ref{Q-Geom} that lead to path metrics are useful again in this SSec, 
though I use a yet wider range of length structures.

\subsubsection{Axioms of distance} 

For $\mbox{Dist}(x, y)$ to be a bona fide {\sl distance function} between points $x$ and $y$ in a set $\FrX$,  
\be
\mbox{Dist}(x, y) \geq 0 \mbox{ } \forall \mbox{ } x, y \in \FrX \mbox{  (non-negativity) } .  
\ee
\be
\mbox{If } \mbox{Dist}(x, y) = 0, \mbox{ then } x = y  \mbox{ (separation) } .
\ee
\be
\mbox{Dist}(x, y) = \mbox{Dist}(y, x) \mbox{ (symmetry) } .
\ee
\be
\mbox{Dist}(x, y) \leq \mbox{Dist}(x, z) + \mbox{Dist}(z, y) \mbox{ (triangle inequality) } .
\ee
$\mbox{Dist}$ is well-known to furnish the following useful things: a notion of limit, closeness, 
openness, continuity and distance-decreasing (Lipschitz) map.
Without separation, it is a so-called {\it pseudometric} [eq (\ref{pseudo}) is an example].

\subsubsection{Some useful infrastructure for notions of distance between shapes}

Infrastructure 1) The Euclidean norm $||\mbox{ }||$ and its generalization to $||\mbox{ }||_{\mbox{\scriptsize\boldmath$M$}}$ norm and 
$(\mbox{ },\mbox{ })_{\mbox{\scriptsize\boldmath$M$}}$ inner product with respect to an array M (in general nonconstant: a 
function of the configuration space objects, which allows for it to be a curved-geometry metric.) 

\noindent Infrastructure 2) Taking an inf or a sup. 

\noindent Infrastructure 3) Integrating over a path (finite models) or a tube (field-theoretic models), or the sums corresponding to each 
in the case of discrete models.   

\noindent Infrastructure 4) Performing intrinsic computations from a single shape, 
\beq
\iota: \FrS \longrightarrow \mathbb{R}^{\sp} \mbox{ } .  
\eeq
\noindent Infrastructure 5) Performing comparison computations involving two shapes.  
\beq
\mbox{Comp} : \FrS \times \FrS \longrightarrow \mathbb{R}^{\sp} \mbox{ } .  
\eeq
Note: some notions of closeness on the collection depend on a fuller notion of comparison {\sl between} 
instants, i.e. their joint consideration rather than a subsequent comparison of real numbers extracted from each individually.   
That may either be a means of judging which instants are similar or of judging which instants can evolve into each other along dynamical trajectories.  
Some criteria to determine which notion should be used are adherence to the axioms of distance, gauge or 3-diffeomorphism invariance as suitable, 
and, for some applications, whether they can be applied to unions of configuration spaces.

\subsubsection{ Examples of distances} 

\noindent Example 1) Using Infrastructure 1), the {\sl Euclidean distance} in Euclidean space is  
\be
\mbox{(Euclidean Dist)}(\ux, \uy) = ||\ux - \uy||^2 \mbox{ } , 
\ee 

\noindent based on the Euclidean norm $|| \mbox{ } ||$, or the generalization of this to mass-weighted and curved-space version $||\mbox{ }||_{\sM}$ 
As per Secs 1--3, these are relevant in simple mechanics configuration spaces configuration space such as particle position 
space $\FrQ(N, d) = \mathbb{R}^{Nd}$ or relative particle position space $\FrR(N, d) = \mathbb{R}^{nd}$.

\mbox{ }

\noindent Moreover, there are problems with generalizing objects based on Infrastructure 1), to be used as notions of distance, 
to the case of semi-Riemannian metrics -- one loses the non-negativity and separation properties.  

\mbox{ } 

\noindent Example 2) Using Infrastructure 2), one has the supremum norm notion: for $f$ a bounded function, 
$\stackrel{\mbox{\scriptsize sup}}{\mbox{\scriptsize $x \in$ \sFrX}}|f(x)|$. 
%

\noindent Example 3) Using Infrastructures 2 and 3 [with Infrastructure 1) in the d$s$]: on a sufficiently smooth Riemannian manifold and
for $\sigma$ a continuous and everywhere positive-definite {\it weight function}, the corresponding {\sl path metric} is 
\be
\mbox{(Path Dist)}[\bg, w](\mbox{point 1, point 2}) = 
\stackrel{\mbox{\scriptsize inf}}{\mbox{\scriptsize $\Gamma$ joining point 1 and point 2}}\left(\int_{\Gamma}w\,\d s\right)  
\label{pathmetric}
\ee
where $\d s$ is the arclength $g_{\mu\nu}\d x\d x^{\nu}$.  

\mbox{ }

\noindent As a generalization, the action for field theories (including on curved spaces) involves integrating over the NOS as well.    
[Insert a $\int_{\sbSigma_p}\d\bupSigma_p$ factor into (\ref{pathmetric}).]
Secs \ref{Dist1} contains numerous further generalizations of this example.  

\mbox{ } 

\noindent Example 4) Plugging Infrastructure 4) [or a finite vector's worth thereof] into Infrastructure 1),  
\be
\mbox{$\iota$-Dist}(x, y) :=  \mbox{(Euclidean Dist)}(\iota(x) - \iota(y)) =  ||\iota(x) - \iota(y)||^2 \mbox{ } ,
\ee 
though N.B. that $\iota$ in general has a nontrivial kernel, by which the candidate $\iota-\mbox{Dist}$ misses out on the separation property.  
If separation fails, one can usually (see e.g. \cite{Gromov}) quotient to leave one with a notion of distance.  
But sometimes this leaves one with a single object, making it a trivial notion of distance, while it is sometimes limited 
or inappropriate to use such a distance as the originally intended space $\FrX$ and not the quotient has significance attached to it.  
Then
\be
\mbox{$\iota$-Dist}(\tilde{x}, \tilde{y}) 
\ee
is a bona fide distance on $\tilde{\FrX} = \FrX$/Ker($\iota$) which may or may not be of use for the originally intended problem on $\FrX$.  
%

\mbox{ } 

\noindent Example 5) Some cases of Infrastructure 5 and 1) hybrids include a prototype of Kendall's two-input finite comparer \cite{Kendall},   
\be
\mbox{(Kendall Proto-Dist)} = (\bfQ, \bfQ)\mbox{}_{\sbfM} \mbox{ } ,  
\label{ProtoKenComp}
\ee
a prototype of Barbour's one-input infinitesimal comparer (c.f. Sec \ref{TR}), 
\be
\mbox{(Barbour Proto-Dist)} = ||\ordial{\bfQ}||_{\sbfM}\mbox{}^2 \mbox{ } , 
\label{ProtoBarComp}
\ee
and a prototype of DeWitt's two-input infinitesimal comparer \cite{DeWitt70},
\be
\mbox{(DeWitt Proto-Dist)} = (\dot{\bfQ}, \dot{\bfQ}^{\prime})_{\sbfM} \mbox{ } .  
\label{ProtoDeWittComp}
\ee
\noindent Note 1) Finite and infinitesimal here are in the same sense as `thick sandwich' and `thin sandwich'.  
    
\noindent Note 2) For the Kendall and DeWitt cases, if $\bfM = \bfM(\fQ)$ then it is ambiguous which of the pair being compared the kinetic term 
is to pertain to.  
In DeWitt's case, he resolved this by `priming half the factors' of each term in the GR $\bM$' \cite{DeWitt70}.   

\noindent Note 3) As presented, these are distance-squared candidates.  
Thus one needs to take their square roots as candidate distances.

\mbox{ } 

\noindent A particular construct is then [combining (\ref{ProtoBarComp}) with Infrastructure 3)]  
\beq
(\mbox{Path Dist})[\bfM, \sqrt{2\fW}](\mbox{Shape 1, Shape 2}) := 
\stackrel{\mbox{\scriptsize inf $\Gamma$ joining}} {\mbox{\scriptsize shape 1 and shape 2}}  \circ 
\left\{ 
\int_{\Gamma} \circ \int_{\sN\sO\sS} \circ \times \, \sqrt{2\fW} \circ \mbox{Sqrt} \circ \mbox{(Barbour Proto-Dist)} 
\right\}   \mbox{ } .  
\eeq
The above bracketed combination was termed $\FS_{\sJ}$  in Sec \ref{Intro}, and the map from $\FrQ$ to it was termed `Jacobi--Synge'.

\subsection{Relational notions of distance}\label{Dist1}

So far, I have not made any configurationally-relational requirement. 
Now also demand one's notion of distance does no depend on physically irrelevant variables, requires 
refining/discarding various of the previously-discussed notions of distance.

\subsubsection{$\FrG$-dependent cores as $\FrG$-act moves to be paired with $\FrG$-all moves}

In cases in which objects are represented redundantly, then the notion of distance used should have the corresponding $\FrG$-invariance. 
Three families of such can be built up from the preceding SSSec's (squared) proto-distances: 
\be
\mbox{(Kendall $\FrG$-Dist)}(\fQ \cdot \fQ^{\prime}) = (\fQ \cdot \stackrel{\longrightarrow}{\FrG_{\sttg}}\fQ^{\prime})_{\sbfM} \mbox{ } ,  
\ee 
\be
\mbox{(Barbour $\FrG$-Dist) } = \stackrel{   \mbox{\scriptsize inf}   }
         {\mbox{\scriptsize $\sordial{\fg} \in$ infinitesimal $\sFG$-transformations}}
         ||\stackrel{\longrightarrow}{\mbox{\FrG}_{\sordial{\sfg}}}\fQ||_{\sbfM}\mbox{}^2 \mbox{ } \mbox{ and }  
\ee
\be
\mbox{(DeWitt $\FrG$-Dist)} = 
        (\stackrel{\longrightarrow}{\mbox{\FrG}_{\sordial{\sfg}}}\fQ, \stackrel{\longrightarrow}{\mbox{\FrG}_{\sordial_{\sfg}}}\fQ^{\prime})_{\sbfM} 
        \mbox{ } .
\ee
\mbox{ } \mbox{ }  One is to regard the above as `$\FrG$-act' half-moves that subsequently need `$\FrG$-all' moves that render the object in question $\FrG$-invariant.  
The most obvious and useful `$\FrG$-all' move in the present context is  
\beq
\stackrel{\mbox{\scriptsize inf}}{\mbox{\scriptsize $\fg \in \sFG$-transformations}}
\eeq
over the finite ones when paired with the Kendall core here and over the infinitesimal ones when paired with the Barbour or DeWitt cores.

\mbox{ } 

\noindent Example 1) The best-matching implementation of Configurational Relationalism is of this form: 
\beq
\mbox{BM}(\FrQ) =       \stackrel{\mbox{\scriptsize extremum}}{\mbox{\scriptsize $\fg \in \sFG$}} \circ 
\left\{ 
\int \circ \int_{\sN\sO\sS} \circ \times\,\sqrt{2\fW} \circ \mbox{Sqrt} 
\right\} 
\circ \mbox{($\FrG$-Barbour Dist)} \mbox{ } .
\eeq
Note 1) The bracketed Maps here is indeed what I previously referred to as JS; also 

\noindent JBB = JS $\circ$ $\FrG$-Bundle($\FrQ$) = Maps $\circ$ ($\FrG$-Barbour Dist).  

\mbox{ } 

\noindent Example 2) The action itself furnishes a candidate path distance,
\beq
(\mbox{Path Dist})[\bfM, \sqrt{2\fW}](\mbox{Shape 1, Shape 2}) := 
\stackrel{\mbox{\scriptsize inf $\Gamma$ joining}}{\mbox{\scriptsize shape 1 and shape 2}} \circ \mbox{ } \mbox{ } 
\stackrel{\mbox{\scriptsize extremum}}{\mbox{\scriptsize $\ttg \in \sFG$}} \circ  \int_{\Gamma} 
\circ \int_{\sN\sO\sS} \circ \times\,\sqrt{2\fW} \circ \mbox{ Sqrt } \circ \mbox{($\FrG$-Barbour Dist)} \mbox{ } .  
\eeq

\subsubsection{The rarely-available r-option}

One could work directly with (or reduce down to) $\FrQ/\FrG$ objects if one has the good fortune of being able to explicitly find and use enough of them.  
Then we have a F($\widetilde{\fQ},\widetilde{\fQ}$) form of object.

\subsubsection{The secondary natural objects option}

\noindent One could base notions of distance on secondary natural objects that do not exhibit the redundancy, 
such as constructions based on eigenvalues of a differential operator.  

\noindent I consider that explicitly group-averaging with respect to Diff would be too tall an order. 
\noindent I more generally make a connection between this and the preceding SSSec's material concerning 
group-averaging notions in QM as relational impositions useful in formulating QM for relational theories.

\subsubsection{Further criteria for a suitable notion of distance}\label{Dist2}

1) Allowing consideration of unions of configuration spaces could reduce the number of useable measures, 
e.g. some could not cope with comparing 3-metrics on 3-spaces with different topologies as per Appendix \ref{Cl-Str}.A.
[This may be too stringent a requirement, however, e.g. conventional Geometrodynamics is of fixed 
topology and much particle mechanics study preclude collisions or even collinearities.]

\mbox{ }

\noindent 2) {\bf Common underlying structures conjecture}  
One should use notions that contain a natural object from the (classical fundamental) laws of Physics.  
One should have preference for dynamical rather than merely kinematical notions, i.e. those that behave well under the action of the natural laws.  
One may consider extending this structural compatibility to involve structures common with such as QM or SM.  

\mbox{ } 

\noindent N.B. one can aim to use the freedom in the preceding SSec's Maps toward obtain compatibilities of this nature.

\subsubsection{Specific examples of relational notions of distance}\label{Dist4}

1) The {\it Kendall comparer} \cite{Kendall84, Kendall} for scaled RPM is
\be
\stackrel{\mbox{\scriptsize min}}{\mbox{\scriptsize $R \in$ Rot($d$)}} \mbox{ } 
(\underline{\rho}^{i} \cdot \underline{\underline{R}} \, \underline{\rho}^{j})  \mbox{ } ,  
\ee
for $R_{\mu}\mbox{}^{\nu}$ the $d$-dimensional rotation matrix (the 2-$d$ form of this plays a part in Sec \ref{Q-Geom}).   

\noindent 2) On the other hand, the infinitesimal core objects are 
\be
||\Circ_{\underline{B}}\brho||\mbox{}^2       \mbox{ } \mbox{ (scaled RPM) }, \mbox{ } 
||\Circ_{\underline{B}, C}\brho||\mbox{}^2    \mbox{ } \mbox{ (pure-shape RPM) }, 
\ee
Some particular objects built out of this are as follows.

\noindent Example 1) The path distance built from the Barbour Rot($d$)-distance based on the scaled RPM action weighting $\sqrt{2W}$ is  
\be
\mbox{(JBB Path Dist)}[||\mbox{ }||, \sqrt{2W}]
(\brho_{\si\sn} \cdot {\brho}_{\sf\si\sn})  =
 \stackrel{\mbox{\scriptsize inf $\Gamma$ joining}}  
         {\mbox{\scriptsize $\brho_{\ti\tn}$ and $\brho_{\tf\ti\tn}$}} \mbox{ } \mbox{ } \left(
\stackrel{\mbox{\scriptsize extremum}}  
         {\mbox{\scriptsize $\underline{B}$ of Rot($d$)}} \left(
\sqrt{2}\int_{\Gamma}||\d_{\underline{B}}\brho||\sqrt{W(\brho)} \right)\right) \mbox{ } .
\ee
In 1-$d$, this is just 
\beq
\mbox{(Jacobi Path Dist)}[||\mbox{ }||, \sqrt{2W}]\mbox{($N$-stop 1, $N$-stop 2)} =
\stackrel{\mbox{\scriptsize inf $\Gamma$ joining}}  
         {\mbox{\scriptsize $N$-stop 1 and $N$-stop 2}}
\left(\sqrt{2}\int_{\Gamma} ||\d\brho||\sqrt{W(\brho)} \right) \mbox{ } .  
\eeq
In indirect form in 2-$d$, it is 
\beq
\mbox{(JBB Path Dist)}[||\mbox{ }||, \sqrt{2W}]\mbox{($N$-a-gon 1, $N$-a-gon 2)} = 
\stackrel{\mbox{\scriptsize inf $\Gamma$ joining}}  
         {\mbox{\scriptsize $N$-a-gon 1 and $N$-a-gon 2}}
\mbox{ } \mbox{ }\left(
\stackrel{\mbox{\scriptsize extremum}}{\mbox{\scriptsize $\underline{B} \in SO(2)$}}
\left(
\sqrt{2}\int_{\Gamma}||\d_{\underline{B}}\brho||\sqrt{W(\brho)}\right)\right) \mbox{ } ,  
\eeq
whilst in r-form it is  
\beq
\mbox{(Jacobi Path Dist)}[\mbox{\boldmath$M$}_{\sC(\sF\sS)}, \sqrt{2W}]\mbox{($N$-a-gon 1, $N$-a-gon 2)} = 
\stackrel{\mbox{\scriptsize inf $\Gamma$ joining}}  
         {\mbox{\scriptsize $N$-a-gon 1 and $N$-a-gon 2}}
\left(\sqrt{2}\int_{\Gamma} 
\sqrt{\d\rho^2 + \rho^2||\d \mbox{\boldmath$Z$}||_{\mbox{\scriptsize\boldmath$M$}_{\tF\tS}}\mbox{}^2  }\sqrt{W(\rho, \mbox{\boldmath$Z$})}\right) \mbox{ } .  
\eeq
For triangleland, after the extremization or straight off in the relationalspace approach, the notion of distance is then
\beq
\mbox{(Jacobi Path Dist)}\big[||\mbox{ }||, \sqrt{2\check{W}}\big](\triangle_1, \triangle_2) = 
\stackrel{\mbox{\scriptsize inf $\Gamma$ joining}}  
         {\mbox{\scriptsize $\triangle_1$  and $\triangle_2$}}\left(
\sqrt{2}\int_{\Gamma}\sqrt{\{ \d{\mI}^2 + \mI^2\{\d{\Theta}^2 + \mbox{sin}^2\Theta\d{\Phi}^2\}\}\check{W}(\mI, \Theta, \Phi)} 
\right) \mbox{ } .
\eeq
\noindent Example 2)  The path distance built from the Barbour Rot$(d)$-distance based on the emergent time weighting $1/\sqrt{2W}$ is
\be
\mbox{(Time Path Dist)}_t(\brho_1, \brho_2) = \lt^{\se\sm}  = 
\stackrel{\mbox{\scriptsize inf $\Gamma$ joining}}  
         {\mbox{\scriptsize $\brho_{\ti\tn}$ and $\brho_{\tf\ti\tn}$}} \mbox{ } \mbox{ }
         \left(
\stackrel{\mbox{\scriptsize extremum}}
{\mbox{\scriptsize $\dot{\underline{B}} \in \mbox{Rot(d)} \mbox{of }$} \tts^{\tE\tR\tP\tM} }
\left(
\int_{\Gamma}||\d_{\underline{B}} \brho||/\sqrt{2W(\brho)} \right)\right) \mbox{ } .
\ee

\subsubsection{The analogous comparers in Geometrodynamics}\label{Dist5}

\noindent These are based on the GR configuration space metric (inverse of DeWitt supermetric) which is natural as regards the physical laws.  

\mbox{ }  

\noindent 1) The Kendall-type comparer for Geometrodynamics is a thick-sandwich comparer.  
It cannot be written more than formally, through not having an explicit formula for the finite action of Diff($\bupSigma$).    

\noindent 2) Barbour-type comparers for Geometrodynamics are built from the infinitesimal group action of Diff($\bupSigma$) generated by $\mF^{\mu}$). 
The infinitesimal core object is then 
\be
||\pa_{\suF}\bh||_{\sbM}\mbox{}^2 \mbox{ } .
\ee

\noindent Example 1) the $\mF^{\mu}$-extremization of the BFO-A action (\ref{BFOA})  
\beq
\mbox{(JBB Path `Dist')}[\bM, \sqrt{2W}](\mbox{geom 1, geom 2}) = 
\stackrel{\mbox{\scriptsize inf $\Gamma$ joining}}  
         {\mbox{\scriptsize geom 1 and geom 2}} \mbox{ } \mbox{ } \left(
\stackrel{    \mbox{\scriptsize extremum}}{\mbox{\scriptsize $\sF^{\mu} \mbox{ }  \in  \mbox{ }  \mbox{\scriptsize Diff}(\bupSigma)$}}
\left(
\int_{\Gamma}
\int_{\sbSigma}\sqrt{2\{\mbox{${\cal R}\mi\mc$}(\ux; \bh] - 2\Lambda\}}||\pa_{F}\bh||_{\sbM}\right)\right) \mbox{ } .  
\eeq
\noindent Note: the non-availability of GR emergent time as an example here due to its not containing a NOS-integral.    

\noindent Example 2) The 
\beq
\int_{\sN\sO\sS} \circ \mbox{ Sqrt}  \hspace{1in} \mbox{ versus } \hspace{1in} \mbox{ } 
\mbox{Sqrt} \circ \int_{\sN\sO\sS} \circ \int_{\sN\sO\sS}
\eeq
ambiguity is a local versus global square root ambiguity in non-prejudgingly attempting to construct relational actions/distances. 
A global square root ordering object gives the DeWitt metric functional which (in velocity of frame presentation) is 
\be
\mbox{(DeWitt Path Dist)}[\bM](\mbox{geom 1}, \mbox{geom 2}) = 
\stackrel{\mbox{\scriptsize inf $\Gamma$ joining}}{\mbox{\scriptsize geom 1 and geom 2}}
\mbox{ } \mbox{ }
\left(
\stackrel{    \mbox{\scriptsize extremum}}{\mbox{\scriptsize $\sF^{\mu} \mbox{ }  \in  \mbox{ }  \mbox{\scriptsize Diff}(\bupSigma)$}}   
\left(
\sqrt{\int_{\sbSigma}\int_{\sbSigma}(\Circ_{\suF}{\bh},\Circ_{\suF}{\bh}^{\prime})_{\sbM}  \sqrt{\mh}\d^3x\sqrt{\mh^{\prime}}\d^3x^{\prime}}
\right)
\right) \mbox{ } .  
\ee 
Note 1) In these examples, it is the local choice that gives good actions (via the Dirac procedure) and the global choice that gives 
good semi-Riemannian geometry (local is not even Finsler by reason of degeneracy). 

\noindent Note 2) in finite theory cases this ambiguity vanishes, e.g. for minisuperspace 
\be
\mbox{(MSS Path Dist)}[M, \sqrt{2\{\mbox{${\cal R}\mi\mc$}(\ux; \bh] - 2\Lambda}\}]( \mbox{MSS geom 1, MSS geom 2}) = 
\stackrel{\mbox{\scriptsize inf $\Gamma$ joining}}{\mbox{\scriptsize MSS geom 1 and MSS geom 2}}
\left( \int_{\Gamma}\sqrt{\{\mbox{${\cal R}\mi\mc$}(\ux; \bh] - 2\Lambda\}}\pa s^{\sG\sR}_{\sM\sS\sS} \right)
\ee  
for $M$ the minisupermetric though it does retain a local versus global sum over particles counterpart in RPM's.  

\noindent Note 3) The emergent time does not furnish a path distance for non-minisuperspace GR since it does not contain a spatial integral.  
This non-generalization is a further reason beyond the extremization leaning on a distinct object 
why to favour action-type path distances over emergent time-type ones.  

\mbox{ }

\noindent There is a serious problem with all of the above, however, namely that the configuration space metric involved is 
indefinite, so these do not furnish bona fide distances.

\mbox{ }

\noindent Difference 22) GR does not have a bona fide notion of distance based on the configuration space metric, as another consequence of this being indefinite for GR.

\subsubsection{Way out 1: pure-shape and Conformogeometrodynamics comparers [Analogy 55]}\label{Dist7}

Consider the object (\ref{Action-ABFO})
One can view this as made from Diff($\bupSigma$) and Conf($\bupSigma$) correcting $||\pa \bh||_{\sbM}$.
However, it also encodes $\pi = 0$, by which we are now free to replace $\mM^{\mu\nu\rho\sigma}$ by 
\be
\mU^{\mu\nu\rho\gamma} := \mh^{\mu\rho}\mh^{\nu\sigma} \mbox{ } ,
\ee
which is indeed a positive-definite supermetric [and on CRiem($\bupSigma$), so it furnishes bona fide distances between the preshapes of GR.  
Thus 
\beq
\FS_{\sA\sB\sF\sO} = \FS_{\sB\sF\sO-\sA}[\upphi^4 \mh_{\mu\nu}]/\mbox{Vol}^{2/3} = 
\int\int_{\sbSigma}\d^3x\sqrt{\mh}\upphi^4
\sqrt{\{\mbox{${\cal R}\mi\mc$}(\ux; \bh] - 8\{\triangle_{\sbh}\upphi\}/  \upphi   \}}||\pa{\bh} + 4\pa{\upphi}\bh/\upphi ||_{\sbU}   /  
 {      \mbox{Vol}^{2/3}     }
\eeq
and the ensuing notion of distance is 

\noindent
$$
\mbox{(CS Barbour Path Dist)}[\sbU, \sqrt{\mbox{${\cal R}\mi\mc$}(\ux; \bh] - 3\{\triangle_{\sbh}\upphi\}/\upphi}](\mbox{conf geom 1, conf geom 2}) = 
$$
\beq
\stackrel{\mbox{\scriptsize inf $\Gamma$ joining}}{\mbox{\scriptsize conf geom 1 and conf geom 2 }} \mbox{ } \mbox{ }
\left(
\stackrel{\mbox{\scriptsize extremum }}{\mbox{\scriptsize F $\in$ Diff($\bupSigma$), $\upphi \in$ Conf($\bupSigma$)}} 
\left(
\int_{\Gamma}  \int_{\sbSigma}\d^3x\sqrt{\mh}\upphi^4
\sqrt{        
            \{\mbox{${\cal R}\mi\mc$}(\ux; \bh] - 8\{\triangle_{\sbh}\upphi\}/  \upphi   \}                         }
            ||\pa{\bh} + 4\pa{\upphi}\bh/\upphi ||_{\sbU}   /   {      \mbox{Vol}^{2/3}     }  
\right)
\right)  \mbox{ } . 
\ee
I also note the `DeWitt' counterpart 

\noindent
$$
\FS_{\sA\sB\sF\sO-\sD} = \FS_{\sB\sF\sO-\sA}[\upphi^4 \mh_{\mu\nu}]/\mV^{2/3} = 
$$
\be
\int\sqrt{   \int_{\sbSigma}\int_{\sbSigma^{\prime}}  \d^3x\d^3x^{\prime}  
\sqrt{\mh}\upphi^4                       \sqrt{ \mbox{${\cal R}\mi\mc$}(\ux; \bh]          - 8\{\triangle_{\sbh}\upphi\}/                     \upphi            }                         
\sqrt{\mh^{\prime}}\upphi^{\prime\,4}    \sqrt{ \mbox{${\cal R}\mi\mc$}(\ux; \bh^{\prime}] - 8\{\triangle_{\sbh^{\prime}}\upphi^{\prime}\}/  \upphi^{\prime}    }                                     
(\pa{\bh} + 4\pa{\upphi}\bh/\upphi, \pa{\bh}^{\prime} + 4\pa{\upphi}^{\prime}\bh^{\prime}/\upphi^{\prime})_{\sbU}                                             }/{\mbox{Vol}^{{2}/{3}}}
\ee

\noindent for $\bU$ here the mixed-index version of the above configuration space metric.

As regards mathematical compatibility, I note that $\bU$ is not in the WDE but it {\sl does} in the 
emergent-time-dependent Tomonaga--Schwinger Einstein--Schr\"{o}dinger equation for the l = shape degrees of freedom of Sec \ref{Semicl}.  
Overall, 

\mbox{ } 

\noindent Analogy 56) Preshape space and CRiem's simple and positive-definite configuration space metrics 
furnish good notions of distance in each case (for GR, being a field theory, these come in distinct Barbour and DeWitt forms), 
and are compatible with physical laws.  

\mbox{ }  

\noindent{\bf Question 27} Extend this study of comparers to Ashtekar variables formulations.  

\mbox{ } 

\noindent The SRPM counterpart is the SRPM action; in indirectly-formulated form in 1-$d$,
\beq
\mbox{(JBB Path Dist)}[||\mbox{ }||, \sqrt{\ttW}]\mbox{($N$-stop 1, $N$-stop 2)} = 
\stackrel{\mbox{\scriptsize inf $\Gamma$ joining}}{\mbox{\scriptsize $N$-stop 1 and $N$-stop 2}}
\mbox{ } \mbox{ }
\left(
\stackrel{\mbox{\scriptsize extremum}}{\mbox{\scriptsize $C \in$ Dil}} 
\left(
\sqrt{2}\int_{\Gamma} ||\d_{{C}}\brho||\sqrt{\ttW(\brho)}
\right)\right)
\eeq
with corresponding r-form  
\beq
\mbox{(Jacobi Path Dist)}[\mbox{\boldmath$M$}_{\sss\sp\sh\se}, \sqrt{2\ttW}]\mbox{($N$-stop 1, $N$-stop 2)} = 
\stackrel{\mbox{\scriptsize inf $\Gamma$ joining}}{\mbox{\scriptsize $N$-stop 1 and $N$-stop 2}}
\left(
\sqrt{2}\int||\d \mbox{\boldmath$\Theta$}||_{\mbox{\scriptsize\boldmath$M$}_{\ts\tp\th\te}}\sqrt{\ttW(\brho)}
\right) \mbox{ } .  
\eeq
In 2-$d$, the indirectly formulated form is 
\beq
\mbox{(JBB Path Dist)}[||\mbox{ }||, \sqrt{2\ttW}]\mbox{($N$-a-gon 1, $N$-a-gon 2)} = 
\stackrel{\mbox{\scriptsize inf $\Gamma$ joining}}{\mbox{\scriptsize $N$-a-gon 1 and $N$-a-gon 2}}
\mbox{ } \mbox{ }
\left(
\stackrel{\mbox{\scriptsize extremum}}{\mbox{\scriptsize $\underline{B} \in SO(2), C \in$ Dil}}
\left(
\sqrt{2}\int_{\Gamma}||\d_{\underline{B},C}\brho||_{\mbox{\scriptsize\boldmath$M$}}\sqrt{\ttW(\brho)}
\right)
\right)
\eeq
with corresponding r-form 
\beq
\mbox{(Jacobi Path Dist)}[\mbox{\boldmath$M$}_{\sF\sS}, \sqrt{2\ttW}]\mbox{($N$-a-gon 1, $N$-a-gon 2)} = 
\left(
\stackrel{\mbox{\scriptsize inf $\Gamma$ joining}}{\mbox{\scriptsize $N$-a-gon 1 and $N$-a-gon 2}}
\left(
\sqrt{2}\int_{\Gamma}||\d\mbox{\boldmath$Z$}||_{\mbox{\scriptsize\boldmath$M$}_{\tF\tS}}\sqrt{\ttW(\mbox{\boldmath$Z$})}
\right)\right) \mbox{ } .  
\eeq
For triangleland, after the extremization or straight off in the relationalspace approach, the notion of distance is then
\beq
\mbox{(Jacobi Path Dist)}[ \mbox{\boldmath$M$}_{\sss\sp\sh\se}, \sqrt{2\ttW}](\triangle_1, \triangle_2) = 
\inf\stackrel{\mbox{\scriptsize $\Gamma$ joining}}{\mbox{\scriptsize $\triangle_1$ and $\triangle_2$}}
\left(
\sqrt{2}\int_{\Gamma}\sqrt{\{\d{\Theta}^2 + \mbox{sin}^2\Theta\d{\Phi}^2\}\ttW(\Theta, \Phi)} 
\right)
\mbox{ } .
\eeq
\noindent {\bf Question 28}. Consider using anisotropy as a notion of distance on mini-CS.

\subsubsection{Way out 2: The Gromov--Hausdorff notion of distance}\label{Dist9}

This approach \cite{Gromov} shuffles both the metrics being compared and is built via an inf rather than an integral. 
This is a {\sl metric space} construction
\beq
\mbox{(Gromov--Hausdorff Dist)} := \stackrel{\mbox{\scriptsize inf}}{\mbox{\scriptsize Met}}\left(\mbox{(Hausdorff Dist)}^{\sZ}(f(X), g(Y))\right) 
\eeq
over all metric spaces Met, with $f$, $g$ being isometric embeddings of $X $and $Y$ into Met, and where  
\beq
\mbox{(Hausdorff Dist)}(A, B) = \stackrel{\mbox{\scriptsize inf}}{\mbox{\scriptsize $\epsilon > 0$}} 
\left\{ B \subset \FrU_{\epsilon}(A), A \subset \FrU_{\epsilon}(B)\right\}
\eeq 
between subsets of a given metric space.  

\mbox{ }

\noindent Note: it is compact metric spaces that (Gromov--Hausdorff Dist) is well-defined for.

\mbox{ } 

\noindent{\bf Question 29}$^*$. Does this kind of construct carry over to Riemannian metrics? 
Certainly a Riemannian metric furnishes a natural metric space notion of metric, via the path metric construction (\ref{pathmetric})
It might however be very impractical to calculate with, needing to consider the set of all possible isometric embeddings...  
One physical price to pay is that it does not relate to the `natural DeWitt structure'.\foo{In the study of spacetimes, there is a counterpart  
of this that serves to compare causal structures \cite{Noldus}.}
%
Are these embeddings of this question isometric in the sense used in Riemannian geometry?
If the preceding parts of this question work well, study GR configuration spaces using such a notion of distance.

\subsubsection{Further constructs for Geometrodynamics}\label{Dist10}

The attitude of this SSSec and the next one is to consider a wide range of structures.  
This is partly to look for analogues in cases which make sense (to see if there are some suitable notions 
that apply to a wide range of (toy) models). 
It is partly due to limitations in some of the structures for some of the theories. 
And it is partly due to the possibility/need for compatibility between notions of 
distance and notions of information and features of the physical laws of each of the models. 
Thus expect reasons of nongenerality or incompatibility will dismiss a number of the structures considered, 
and thus for the moment I provide a substantial list of possibilities.   
I see future work following from the present SSec as involving bigger specific RPM examples and 
midisuperspace examples, as more strenuous tests both of the capacity to compute specific objects and of 
which objects still make clean conceptual sense in these bigger settings.  
And also looking at notions of information (both classical and quantum-mechanical) as the next stage in 
building an explicit records-theoretic framework.    

\mbox{ }

\noindent A) {\bf Poisson sprinkling}\footnote{See e.g. \cite{Sorkin03}, where it is additionally used in the Causal Sets program.} 
is a procedure by which $n$ points can be embedded at random onto a manifold. 
If done for two manifolds, one can then build a (comparer) norm based on just these points, such as a supremum norm or a Euclidean norm.  
An obstacle to this, however, is that different people doing it would get different answers, though one might get 
round this by applying the procedure many times and invoking statistical theory.    
Another problem is that these norms are unrelated to the `natural objects' already present in the physical laws.  
Sprinkling can be on whichever interpretation for a mathematical space: physical space or configuration space.\footnote{By its  
generality, I prefer this scheme to Bookstein's using of special `landmark' points on somewhat symmetric manifold models of biological 
objects \cite{Bookstein}.}   
%
But there may be little motivation to discretely approximate when the manifold is to be interpreted as a {\sl mechanical configuration space}.    

\mbox{ }  

\noindent B) {\bf Curvature invariants}.  
For 3-spaces there are just three built with up to second derivatives of the metric.  
One is the Ricci scalar.  
One could then compute what value this takes on average over the manifold and use that as an $\iota$-map, 
but this will have a big kernel, there being e.g. many vacuum spaces (Ric = 0).  
One could also compare curvature at random points, or consider $\mbox{${\cal R}\mi\mc$} - \langle\mbox{${\cal R}\mi\mc$}\rangle$ or 
$\langle\mbox{${\cal R}\mi\mc$}^2\rangle - \langle\mbox{${\cal R}\mi\mc$}\rangle^2$ rather than 
$\mbox{${\cal R}\mi\mc$}$; these are simple measures of bumpiness/inhomogeneity.    

\mbox{ } 

\noindent C) {\bf Spectral measures} extracted from certain operators.  
These indeed belong at the classical level, since, even if the operators can come from QM equations (most notably the Dirac operator), 
they contain no relative powers of $\hbar$ and thus QM considerations can be dropped in studying their spectra in an $\hbar$-free context.  
Examples (see refs 14-17 of \cite{Seriu1}) demonstrate that there can be isospectrality 
(nonuniqueness of geometries giving the same spectrum, which Kac popularized as `hearing the shape of a drum').  
Thus spectral measures are to be suspected of not obeying the separability axiom.
But is that a problem with all operators or is there some operator (or set of operators) that are entirely spectrally-discerning? 
One benefit in generality of Strategy 3 over Strategy 1 is that comparison between topologies is possible.  

\mbox{ } 

\noindent C.I) {\bf Matzner's spectral measure} \cite{Matzner}. 
Using $\iota = \iota_{\mbox{\scriptsize Matzner}}(\bupSigma, \bh)  :=$ (smallest eigenvalue 
of the Yano--Bochner operator $\mD_{\mu}\{\mD^{\mu}\delta^{\nu}_{\rho} + \mD^{\nu}\delta^{\mu}_{\rho}\}$ on ($\bupSigma, \bh)$) gives 
\be
\mbox{(Matzner Dist)}(\langle\bupSigma, \bh\rangle, \langle\bupSigma^{\prime}, \bh^{\prime}\rangle) = 
|\iota_{\mbox{\scriptsize Matzner}}(\bupSigma, \bh) - \iota_{\mbox{\scriptsize Matzner}}(\bupSigma^{\prime}, 
\bh^{\prime})|^2   
\ee
which obeys all the other distance axioms \cite{Matzner}.  

\mbox{ } 

\noindent C.II) {\bf Seriu's spectral measure} is a comparer, 
\be
\mbox{(Seriu Dist)}(\langle\bupSigma, \bh\rangle, \langle\bupSigma^{\prime}, \bh^{\prime}\rangle| ) = 
\left.
\sum\mbox{}_{\mbox{}_{\mbox{\scriptsize $k$ = 1}}}^{\infty}\mbox{log}\mbox{\LARGE(} 
\sqrt{{\lambda_k}/{\lambda_k^{\prime}}} + \sqrt{{\lambda_k^{\prime}}/{\lambda_k}}
\mbox{\LARGE) } \right/2 \mbox{ } .  
\ee
Here, $\{\lambda_k,\mbox{ $k$ = 0 to } \infty\}$ and $\{\lambda_k^{\prime}, \mbox{ $k$ = 0 to } \infty\}$ 
are the whole spectrum of the Laplace operator on the primed and unprimed manifolds respectively.  
This form is motivated by $\mbox{exp}(-\mbox{(Seriu Dist))}$ being a factor in off-diagonal 
elements of reduced density matrix of the universe, so the larger this comparer's value is, the more 
decoherence there is between corresponding spatial cuts of Einstein--scalar spaces.  
It can be viewed as built out of secondary quantities that are guaranteed to have the suitable invariances.  
This (and spectral measures in general) allow for comparison between geometries with different topology, 
and is in principle of interest in Quantum Cosmology.
However, as a notion of distance, it is not only robbed of separation by the isospectral problem  but also fails to obey the triangle inequality.  

\mbox{ } 

\noindent D) {\bf Inhomogeneity measures}. 
A simple notion of inhomogeneity comes from partitioning space up and computing the energy density 
$\varepsilon$ (GR) or normalized number density $\varepsilon/\int\d\Omega\,\varepsilon$ (RPM) for each.  
However, this has the problem of showing that this is insensitive to how the partitions are carried out.  
That in GR is a major part of the averaging problem, which is particularly acute due to the nonlinearity of the Einstein equations.  

\mbox{ } 

\noindent For the moment, consider the phenomenological matter in terms of which such things have 
already been started to be investigated in the literature.  
Firstly, consider 
\be 
\iota = \iota_{\mbox{\scriptsize density contrast}} = \varepsilon/\langle\varepsilon\rangle
\ee 
or its integrated counterpart, possibly with volume of integration and volume in the average being 
distinct, followed by construction 1 gives\foo{$\varepsilon$ is the energy density. 
2B's notions depend also on $\bupSigma$ and $\mh_{\mu\nu}$ through the integrals involved.}   
\be
\mbox{(Density Contrast Dist)}(\varepsilon_1, \varepsilon_2)  = 
|\iota_{\mbox{\scriptsize density contrast}}(\varepsilon_1) - 
 \iota_{\mbox{\scriptsize density contrast}}(\varepsilon_2)|^2 \mbox{ } .  
\ee
A similar notion which has yet further connotations as a relative information is the Hosoya--Buchert--Morita (HBM) quantity \cite{HBM}
\be
\iota = I_{\sH\sB\sM}(\varepsilon) = \int \d\Omega \varepsilon \mbox{log}(\varepsilon/\langle\varepsilon\rangle)
\ee
The asymmetric input of $\varepsilon$ and $\langle\varepsilon\rangle$ exploits the argument-asymmetric nature of this 
relative information quantity (see Sec \ref{Cl-NOI}), and is a special pairing making triangle inequality violation a non-issue.    
Then use construction 1 to extract
\be
\mbox{(HBM Dist)} = |I_{\sH\sB\sM}(\varepsilon) - I_{\sH\sB\sM}(\varepsilon^{\prime})|^2
\ee
which is guaranteed to obey everything except separation.  
Separation is clearly a problem for inhomogeneities: these take a common value when 
the space is homogeneous (1 for the contrast and 0 for the HBM quantity).  
Using a single spatial topology may help, as may quotienting out the homogeneous spacetimes 
(perturbations about a particular homogeneous universe whose overall shape is an approximate fit for cosmological data). 
This may not remove all of the inhomogeneity kernel: two spaces can be distinct and yet give the same value for each of these 
inhomogeneity indices, via only differing in their `higher moments'. 
Indeed the above notion of distance is poor, but that come from the inhomogeneity quantifier in question 
being very limited in information content; one could well construct distances from sufficiently discerning quantifications of inhomogeneity.  
%

\mbox{ }  

\noindent The corresponding contrast object is 
\be
\varepsilon_{\Pi}/\langle \varepsilon \rangle
\ee
for $\langle A \mbox{ } \rangle = \int A \mathbb{D}\Omega /\int \mathbb{D}\Omega$.
One can then furthermore build the Shannon entropy of $f_{\Pi} \equiv \varepsilon_{\Pi}/\int\varepsilon\mathbb{D}\Omega$, 
\be
I_{\mbox{\scriptsize Shannon}} = \int \mathbb{D}\Omega \frac{\varepsilon_{\Pi}}{\int\varepsilon \mathbb{D}\Omega}
\mbox{log}
\left(
\frac{\varepsilon_{\Pi}}{\int\mathbb{D}\Omega \varepsilon }
\right) \mbox{ } , 
\ee  
and the relative information of $\varepsilon $ versus $\langle \varepsilon  \rangle$
\be
I(f_{\Pi}||\langle f \rangle) = \int f_{\Pi}\mbox{log}
\left(
\frac{f_{\Pi}}{\langle f \rangle}
\right) =
\left\langle
\frac{\varepsilon }{\langle \varepsilon  \rangle}\mbox{log}
\left(
\frac{\varepsilon }{\langle \varepsilon  \rangle}
\right)
\right\rangle \mbox{ } ,
\ee
which by its second form is also an object of contrast form.

Then HBM-type objects are
\be
I_{\sH\sB\sM}(\varepsilon_{\Pi}) = 
\int\mathbb{D}\Omega\varepsilon_{\Pi}\mbox{log}\left(\frac{\varepsilon_{\Pi}}{\langle \varepsilon \rangle}\right)
\mbox{ } \mbox{ (relative information form) } \mbox{ } ,
\label{IHBM}
\ee
\be
I^{\prime}_{\sH\sB\sM} (\varepsilon_{\Pi}) = 
\left\langle \varepsilon_{\Pi} \mbox{log}
\left(
\frac{\varepsilon_{\Pi}}{\langle\varepsilon\rangle}
\right)
\right\rangle
\mbox{ } \mbox{ (expectation form) } \mbox{ } , 
\label{IHBM2}
\ee
which bear the relations 
\be
I_{\sA} =  I_{\sH\sB\sM}\left/\int \mathbb{D}\Omega\varepsilon_{\Pi} \right. = 
           I_{\sH\sB\sM}^{\prime}/\langle \varepsilon \rangle  
\ee
%
(i.e. differ in normalization convention alone).  

\noindent The normalized variance object is  
\be
\{\langle \varepsilon^2\rangle - \langle \varepsilon\rangle^2 \}/{\langle \varepsilon\rangle^2}  \mbox{ } . 
\ee
One can build $\iota$-maps based on each of these or on vectors made out of multiple such. 

\mbox{ }  

\noindent Note 1) Relative information and $I_{\sH\sB\sM}$ are not themselves itself distances -- symmetry and the 
triangle inequality are irrelevant, as it is for comparing a configuration with its average distribution 
(which is a piece of information about shape rather than a distance between two shapes).  
 
\noindent Note 2) One can build $\iota$-maps from the base object, the three simple objects or the interesting 1, 2, 3, 4 objects.

\noindent Note 3) Also note that $I_{\sH\sB\sM}$ is not very discerning: it provides only 1 piece of information, 
but the variety of different possible configurations cannot in general be captured with so little 
information: Ker($I_{\sH\sB\sM}$) is generally nontrivial, continues to be so for the $\iota$ construction. 

\noindent Note 4)
The Gromov--Hausdorff, embedding point, curvature invariant and spectral measure approaches perhaps overly favour geometry over matter.  
The Kendall and Barbour comparers treat the two on an equal footing.  
Finally the HBM object and related such is based just on the matter content.

\subsubsection{Which of the preceding SSec's notions of distance have RPM counterparts? [Analogy 57]}\label{Dist11}

The configuration spaces of RPM's are too simple to furnish much of a classification by curvature or by the 
spectrum of some associated operator.
There are however inhomogeneity (clumping) notions for RPM's partition into pieces of space define number 
density $N_{\Pi}$ for each partition and then build up various significant objects.  

\mbox{ }

\noindent Three subsequent simple compound objects are the {\it normalized number density} $\mn_{\Pi} = 
\mN_{\Pi}/\sum_{\Pi}{\mN_{\Pi}}$, the associated `{\it contrast object}' $\mN_{\Pi}/\langle \mN \rangle$ 
where $\langle A \mbox{ } \rangle = \sum_{\Pi = 1}^{\sn} A_{\Pi}/\mn$ and the {\it normalized variance comparer} 
\be
\frac{\langle \mN_{\Pi}^2\rangle - \langle \mN_{\Pi}\rangle^2 }{\langle \mN_{\Pi}\rangle^2}  \mbox{ } .  
\ee
One can then build further significant composite objects such as 1) the Shannon information of 
$\mn_{\Pi} \equiv \mN_{\Pi}/\sum_{\Lambda} \mN_{\Lambda}$, 
\be
I_{\mbox{\scriptsize Shannon}} = \sum\mbox{}_{\mbox{}_{\mbox{\scriptsize $\Pi$}}}{\mn_{\Pi}} \mbox{log}({\mN_{\Pi}}) \mbox{ } .
\ee  
2) The relative information of $\mN_{\Pi}$ versus $\langle \mN_{\Pi} \rangle$,  
\be
I_{\sr\se\sll} = I(\mn|\langle \mn \rangle) = \sum \mn\,\mbox{log}
\left(
\frac{\mn}{\langle \mn \rangle}
\right) = 
\left\langle
\frac{\mN}{\langle \mN \rangle}\mbox{log}
\left(
\frac{\mN}{\langle \mN \rangle}
\right)
\right\rangle
\ee
by which last form it is thus also an object of contrast form.  

\mbox{ } 

\noindent 3) {\bf HBM-type objects} 
\be
I^{\sm\se\scc\sh}_{\sH\sB\sM}(\mN_{\Pi}(Q_A)) = \sum\mbox{}_{\Pi}\mN_{\Pi}\mbox{log}({\mN_{\Pi}}/{\langle \mN \rangle}) \mbox{ } \mbox{ (relative information form)} 
\ee
\be
\mbox{and } \mbox{ } \mbox{ }
I^{{\sm\se\scc\sh}\prime}_{\sH\sB\sM}(\mN_{\Pi}(Q_A)) =  \langle \mN_{\Pi}\mbox{log}({\mN_{\Pi}}/{\langle \mN \rangle}) \rangle
\mbox{ } \mbox{ (expectation form) } . 
\ee
Note the inter-relations
\be
I^{\sm\se\scc\sh}_{\sA} = {I^{\sm\se\scc\sh}_{\sH\sB\sM}}/{\sum\mbox{}_{\Pi}\mN_{\Pi}} = {I^{\sm\se\scc\sh}_{\sH\sB\sM}\mbox{}^{\prime}}/{\langle \mN \rangle} \mbox{ } .  
\ee

\subsection{Notions of localization in space}

\noindent Example 1) In triangleland, the partial moments of inertia $\mI_{\sb\sa\sss\se} << \mI_{\sm\se\sd\si\sa\sn}$ means that the base can be treated as localized.  

\noindent Example 2) $\langle\mathbb{R}^{d}, ||\mbox{ } ||\rangle$ and $\langle \bupSigma, \bh \rangle$ metrics furnish good distances.

\subsection{Localized subconfiguration spaces}\label{Loc-SubCon}

Now the other hitherto ignored coordinates may in parts become relevant.  
That is through these other coordinates entering the notion of localization itself. 
Following from the above Example 1, consider the base of the triangle as a localized subsystem.  
One notion of localized here is for the apex to have to be outside of the circle of regularity 
(i.e. $\mI_{\sm\se\sd\si\sa\sn} = \mI_2 > \mI_1 = \mI_{\sb\sa\sss\se}$).  
In considering this, one passes from being (in $\FrQ$) on any open half-line emanating from the triple 
collision to being on any such that lies within the D-hemisphere.
I.e., in this example, one passes from the mathematics of the Dirac string to that of the Iwai string 
(c.f. Sec \ref{Guich}).   
Note: I do not mean to imply that these need to be subspaces of configuration space in any 
particular mathematical sense.  
They are physically desirable entities to study but not necessarily mathematically nice.  


\subsection{Classical notions of information}\label{Cl-NOI}

\noindent Information is (more or less: arguments for include those in \cite{Denbigh2, BenNaim}) professed that information is negentropy, so 
that an incipient classical notion is the Boltzmann-like expression\foo{I 
choose units such that Boltzmann's constant is 1.}  
\be
I_{\mbox{\scriptsize Boltzmann}} = - \mbox{log}\,W
\label{Boltz} \mbox{ } . 
\ee
Here, $W$ is the number of microstates, which is evaluated combinatorially in the discrete case or taken to be 
proportional to the phase space volume in the continuous case.  
The Shannon entropy (already encountered in Sec \ref{Dist10}) is then 
\be
I_{\mbox{\scriptsize Shannon}}(p_x) = \sum_x p_x\mbox{log}\,p_x
\label{ShaDisc}
\ee 
for $p_x$ a discrete probability distribution, or  
\be
I_{\mbox{\scriptsize Shannon}}[\sigma] = \int \d\Omega \sigma\,\mbox{log}\,\sigma
\label{ShaCont}
\ee    
for $\sigma$ a continuous probability distribution.  

\mbox{ } 

\noindent Note 1) This is based on the $x$\,log\,$x$ function, which is {\sl the} positive continuous function that is consistent with regraining; 
it also has many further useful properties \cite{Wehrl}). 

\noindent Note 2) If one replaces ln($x$) by ln$_q(x) :=  \frac{x^{1 - q} - 1}{1 - q}$, one has the corresponding more general notion of {\it Tsallis entropy} \cite{Tsallis}.  

\noindent Note 3) More generally also, monotonicity is an often-mentioned property for a bona fide entropy to have. 
[This monotonicity may be connectable to that of time itself via some  `Arrow of Time' demonstration.] 
This has caused problems with gravitational information candidates based on the Weyl tensor \cite{Weyls1, Weyls2, Weyls3, Weyls4, Weyls5}.  
Moreover, information/entropy is characterized by a number of further properties \cite{Wehrl}, and it is 
not clear to me whether the gravitational candidates have been screened for these.  

\noindent Note 4) Examples of cosmologically relevant notions of information proposed to date that are manifestly related to 
conventional notions of information are Rothman--Anninos' \cite{RA} use of the continuous form of 
(\ref{Boltz}) and Brandenberger et al.'s \cite{Brandenberger} continuous version of (\ref{ShaCont}). 
%

\noindent Note 5) In discussing notions of information with a number of other physicists, I have noted that it is important to distinguish between 
observed information as opposed to information content, e.g. the spectrum of the Sun as opposed to the Sun itself.

\subsubsection{Notions of mutual and relative information}

{\bf Mutual information} is one concept of possible use for Records Theory; it is given by  
\be
M_{\mbox{\scriptsize Shannon}}[A, B] = I_{\mbox{\scriptsize Shannon}}[A] + I_{\mbox{\scriptsize Shannon}}[B] - I_{\mbox{\scriptsize Shannon}}[AB]  
\label{Mutua}
\ee
for AB the joint distribution of A and B.  

\mbox{ }  

\noindent {\bf Relative information} is a conceptually-similar quantity given by  
\be
I_{\mbox{\scriptsize relative}}[p, q]         = \sum_x p_x\mbox{log}\left({p_x}/{q_x}\right) 
\mbox{ } \mbox{ (discrete case) ,} \mbox{ } \mbox{ }
I_{\mbox{\scriptsize relative}}[\sigma, \tau] = \int \d\Omega \,\sigma\, \mbox{log}\left({\sigma}/{\tau}\right) 
\mbox{ } \mbox{ (continuous case) ,} 
\label{relen} 
\ee 
This introduced, the mutual information can be cast as a relative information, now between a joint distribution and the product of its marginals \cite{DV12}.
The Hosoya--Buchert--Morita example of Sec \ref{Dist10} can also be viewed as a relative information; it is a subcase of Tsallis relative information 
\cite{Tsallis}, which more general such notion goes further than HBM in incorporating more discerning `higher moment' information.   

\mbox{ } 

\noindent {\bf Question 30}. Does it really make sense for matter to provide a notion of information in GR Physics? 
Does it only make sense subject to some approximation that limits the gravitational sector?

\subsection{Notions of correlation}

Again these can be set up for discrete or continuous cases and occur in an extensive number of branches of Mathematics and Physics.  

\mbox{ } 

\noindent Example 1) I begin with the most basic notions from Statistics.  
For random variables X, Y, 
\beq
\mbox{Pearson's correlation coefficient } \mbox{ }  \rho_{\sP} := \mbox{Cov(X,   Y)}/\sigma_{\sX}\sigma_{\sY} \mbox{ } ,
\eeq
i.e. the normalization of the covariance by the square roots of the also-statistically significant variances. 
In this form it is invariant under exchange of dependent and independent status of the X and Y.  
Moreover, this only captures correlations that approximate a straight line.  
To detect non-linear correlations one can e.g. use Spearman's, or Kendall's-$\tau$, rank correlation coefficients.  
%
%
The already-discussed notion of mutual information indeed also serves as a notion of correlation, albeit it is clearly in sore need of an n-object extension 
in order to address most problems; that straightforwardly gives the {\it total correlation} and {\it dual total correlation} for any n $>$ 2.  
How these fare for {\sl relational} data, i.e. when translations, scalings and rotations of the random variables are specifically not to change the modelling, 
is covered in Sec \ref{Correl1}.  

\noindent Example 2) n-point function correlators are well-known to occur in physical field theories such as Cosmology, QM and QFT.
Some such quantities can come with translational and rotational invariance directly built in via depending only on $\underline{x} - \underline{x}^{\prime}$; 
sometimes they involve taking the dot product with an arbitrary-direction vector and then integrating over all directions, which amounts to a $\FrG$-act, $\FrG$-all manoeuvre.
For now we stick to the classical ones (see Sec \ref{QM-NOI} for the QM counterparts).
In classical cosmology, then, there is a 2-point function for such as mass density or galaxy number density \cite{coscorr}. 
This comes out as a function of inter-particle separation magnitudes. 

\noindent Higher-n-point functions involve higher moments, and that the moments are for many purposes a useful basis for the 
study of correlations.  What of higher moments in statistical correlation theory?  

\noindent Example 3) Addressing some particular questions requires further particular assessors of correlation.  
A fairly well-known example is the search for circles in the microwave background data in Cosmology (under certain circumstances this would be a means of 
determining the large-scale shape of the universe \cite{circles}).  

\noindent Example 4) In the general GR setting, the Zalaletdinov correlation tensor (a comparer) is used in study of inhomogeneity \cite{Zala12}.     
I would like to know if this is diffeomorphism-invariant.

\subsection{Notions of information for RPM's}\label{DrNo}

\noindent Possible {\bf problem of ensembles}  $E$ fixed but $N$ varies if fusion/fission is allowed, and this is a very uncommonly used 
ensemble, rather than microcanonical (both fixed), canonical ($N$ fixed, $E$ free) or grand (both free).  
I get round this for now by not allowing fusion/fission, so that the ensemble is microcanonical in the usual $E$ fixed interpretation, 
and canonical in the varying-$E$ multiverse interpretation.  

\mbox{ }

\noindent {\bf Problem of small particle number effects}.  
One would expect these for the smaller and thus elsewise more tractable RPM's.  
In fact such small-number effects in SM rapidly diminish in importance even for a moderate increase in particle number, but at least nonrelational models of such effects suggest that 
3 or 4 particles are {\sl somewhat} too few (despite not being vastly too few) for such effects not to be very significant.  
By this, it may be a rare case of where using a general $N$ (assumed fairly large) is more straightforward than using some 
concrete small $N$ (and $N$-stop metroland is particularly amenable to such a study, so that beginning to understand 
how to set up a relational SM is a major application of those models).  

\mbox{ } 

\noindent We have hitherto not involved nontrivial $\FrG$ ($\FrG$ trivial models or r-presentation accessible so far). 
However, we do so from here down.  

\mbox{ } 

\noindent Possible {\bf problem of boxes} In setting up a classical SM, one usually proceeds by covering a region in 
imaginary boxes and counting how many particles are in each. 
It may then be problematic to do this in a relational theory due to various combinations of absolutely 
distinct boxes now being indiscernible, so that the construction and the overall counting process may need unusual new relational input.  
Moreover, the primary notion "there is one particle in this imaginary box" is not relational since the box is imaginary and 
the one particle contains no relational information. 
It might then be necessary to have some other kind of primary notion based on multiple particles to avoid the latter, 
and possibly even not on boxes (or only using boxes that are crafted in a relationally invariant manner) to avoid the former.  
This would involve some notion of `these particles are all near neighbours as compared with almost any of the other particles'.
This feeds into how to define suitably relational N on which to subsequently base the classical information formulae.
An indirect alternative would be to consider applying $\FrG$-act, $\FrG$-all moves to a  
non-relational notion of N for particles such as one would use in ordinary mechanics. 
The specifics of this, by either approach, are not yet entirely clear to me.  

\mbox{ } 

\noindent How relational are the mutual and relative notions of information? 
They do carry connotations of the subsystems/perspectival notion of relational.  
But how do they fare as regards the Leibniz--Mach--Barbour type of relationalism underlying the current Article?
Mutual information is a linear combination of already-computed objects and so inherits relational characteristics in cases 
in which these simpler information objects are already-relational. 
Relative information is a linear combination too, though one of the 2 objects is a 2-subsystem index generalization of the 
preceding, necessitating a small amount of extra work.
The outcome is that, if built out of r-formulation objects, relationalism is automatic (they are also timeless objects). 
But, as ever, that is of course a luxury that one cannot afford in the case of GR itself.  
The more general alternative is to consider $\FrG$-act, $\FrG$-all versions of these objects.

\noindent Classical information can be thought of in terms of configuration space or phase space regions. 
Either use the r-approach for this or apply $\FrG$-act $\FrG$-all to one's quantities; 
since $W$ and $p$ are already individually meaningful, $\FrG$-act, $\FrG$-all should be applied at the level of these prior to their 
insertion into the information formulae.

\noindent Gryb and Mercati \cite{GM12} use a Shannon-type entropy in connection with RPM's, although it does not appear to quantify information in a $\FrG$-invariant manner. 
I also question its interpretation there as a Bekenstein--Hawking entropy or of any kind of holography because there is no $\hbar$ in this classical context 
and it is a particle theory and not a field theory.  
I thus put down whatever entropy goes as area result that they get down to coincidence rather 
than any tie between so simple a model of shape and a result that is tied to at least a semiclassical extension of GR gravitational theory (i.e. black hole thermodynamics).

\mbox{ } 

\noindent {\bf Question 31}. What is the entropy of a classical triangle?  
What is the mutual information of the two `base' particles and of the third `apex' particle? 
The relative information for this? 
What are good notions of homogeneity and of correlation for this?  

\mbox{ }

\noindent Analogy 58) with notions of information for GR.  

\mbox{ }  

\noindent Triangleland has some limitations in constructing working models of Records Theories as compared 
to 4-stop metroland, due to the latter possessing splits into two nontrivial (2-particle) subsystems: the H-clusterings.

\subsection{Notions of correlation for RPM's}\label{Correl1}

Suppose we are given a constellation of $N$ points in 2-$d$ -- the physical content of what we term an $N$-a-gon in this Article.
Is it relational to assess this for collinearity using Pearson's correlation coefficient $\rho_{\sP}$? 
Immediately no!\footnote{The everyday rank correlation tests statistics also immediately fail to be rotationally relational, since any two data points can be rotated into a tie.} 
It is very well known that the upward pointing line and the downward pointing one have the opposite extremes of $\rho_{\sP}$.
Thus  $\rho_{\sP}$ cannot be entirely relational.  
In fact, variance and covariance are translation-invariant and Pearson's normalization makes it scale-invariant also, but it indeed fails to be rotationally invariant. 
How can one remedy this for use in relational problems like Kendall's standing stones problem or the study of snapshots from the $N$-a-gonland mechanics? 
One way I found follows from how the {\it covariance matrix}
\beq
\bV = \mbox{\Huge(}\stackrel{\mbox{Var(X)}   \mbox{ } \mbox{ } \mbox{ }  \mbox{ }  \mbox{ } \mbox{ } \mbox{ } \mbox{ }  \mbox{Cov(X, Y)}}
                            {\mbox{Cov(X,Y)}                             \mbox{ }  \mbox{ }                             \mbox{Var(Y)}   }\mbox{\Huge)}
\eeq
is a 2-tensor under the corresponding $SO(2)$ rotations.  
This prompts me to look at the corresponding invariants: det$\bV$ = Var(X)Var(Y) -- Cov(X,Y)$^2$ and tr$\bV$ = Var(X) + Var(Y).
In particular, then 2\,det$\bV$/tr$\bV$ is a scale-invariant ratio of translation-and-rotation invariant quantities and therefore a relationally invariant base-object for 
single-line linear correlations.
It can be rewritten as
\beq
\overline{\rho}_{\sR\se\sll} = 2\frac{\sqrt{1 - \rho^2_{\sP}}}{\omega + \omega^{-1}}
\eeq
for $\omega := \sqrt{\mbox{Var(X)}/\mbox{Var(Y)}}$.  

\mbox{ }

\noindent Note 1) The factor of 2 is present because min($\omega + \omega^{-1}$) = 2, whereas the radicand is bounded by 1, so with the overall factor of 2 inserted, this 
relational notion runs from 0 to 1.  

\noindent Note 2) The bar is there in the function of a `not' since perfect correlation returns the value 0 and not 1, so this quantity is an {\sl uncorrelation} coefficient.

\noindent Note 3) Taking functions of this may allow for more flexibility as regards passing statistics-theoretical hurdles.  

\noindent Note 4) We found a basis of shape quantities for triangleland, so this must be a function of them. 
What does it look like in terms of those? 
If the data is taken to represent the Jacobi vectors, det$\bV \propto$ Area$^2$, as one might anticipate from how Area is attained by noncollinearity (already seen in Sec \ref{Q-Geom}).  
Also, Tr$\bV \propto$ I -- $Aniso$.  
This forewarns us that normalizing by use of I itself is not natural in {\sl statistical} investigations.
All in all, 
\beq
\overline{\rho}_{\sR\se\sll} \propto \frac{area}{1 - aniso} \mbox{ } .  
\eeq
\noindent Note 5) Further study reveals that, just as $area$ = $demo_3$, the square root of the sums of Jacobi subsystems' areas $demo_4$ features as det$\bV$.  
Though for $N > 4$, products of distinct-subsystem areas appear, and I do not yet know whether this continues to coincide with that case's 
democratic invariant, and, if not, which is more interesting as a measure and statistic for noncollinearity.  

\mbox{ } 

\noindent Moreover, seeking a single line of best fit for data is not always appropriate is fairly apparent as regards the study of the distribution of standing stones or quasars.
One consider more sophisticated geometrical propositions about $N$-a-gon constellations such as whether there is a disproportionate number of 
approximately-collinear triples of points within one's $N$-point planar constellation.
Investigating this uses Kendall's shape statistics techniques.
A suitable conceptual notion for triples of points is Kendall's notion of $\epsilon$-flatness (see Fig \ref{Fig-2AFc}); 
he also generalized this for groupings of $N$ particles in 2-$d$ \cite{Kendall84}.
For a yet more comprehensive list of addressable propositions and shape statistics developments, see Kendall's \cite{Kendall84, Kendall89, Kendall} 
(also worked on by his book co-authors Barden, Carne and Le and elsewhere by e.g. Small \cite{Small} and, on rather different premises from the present Article's, 
Bookstein \cite{Bookstein}). 
Moreover, the range of examples investigated intersect strongly with the current Article's predominant emphasis on $N$-a-gonlands.  
Thus the present Article lays down an additional application of the statistical theory of shapes 
-- to the case in which these shapes are whole-universe toy models possessing many features of qualitative interest to Classical and Quantum Cosmology and to the POT 
(see Sec \ref{RPM-for-QC}).   
 
\mbox{ }

\noindent These examples of further kinds of relational correlations should be of qualitative theoretical value as regards being 
sufficiently open-minded in seeking notions of correlation in other background-independent contexts such as LQG.  
At the very least it provides counter-examples to standard n-point function technology being the only approach to characterizing correlations in a physical theory.  

\mbox{ }

\noindent All in all, we have seen how  Information Theory \cite{Preskill}, and notions of complexity and pattern recognition \cite{Small} are relevant to the study of 
subconfigurations of a single instant, which Sec \ref{Cl-POT-Strat} will argue are the basis for Records Theory.

\subsection{Propositions in the classical context}\label{Cl-Prop}

\noindent {\bf Propositioning}.  
If, say, we take the state space $\FrS$ to be primary, Physics is in fact about the propositions about $\FrS$, Prop($\FrS$), rather than about $\FrS$ itself.  

\mbox{ }  

\noindent That is, loosely, that Physics involves asking and answering questions.

\noindent From a more rigorous viewpoint, one concerns oneself with what kind of {\sl propositional logic} these form.  

\mbox{ }

\noindent As far as I am aware, the chain of thought 
that Physics {\sl is} questions which are equivalent to logic was first 
proposed by Mackey \cite{Mackey} (see also \cite{IL2} for suggestions of this for specific POT settings).  
In honour of this, I refer to this viewpoint as `{\bf Mackey's Principle}'.
\noindent [The relationalist will be interested here in those propositions concerning {\sl tangible physics} rather than instances of physically-empty mathematics.]  
Mackey's Principle first entered the POT literature in Isham and Linden's(IL) \cite{IL2} formulation of Histories Theory. 
I have subsequently pointed out its wider applicability \cite{ARel} e.g. to timeless approaches also.

\subsubsection{Atemporal questions} 

\noindent{\bf Questions of Being} generally concern `what is the probability that an (approximate) (sub)configuration has some particular property $P_1$?' 

\noindent Note 1) The totality of these are to form some kind of atemporal logic.  

\noindent Note 2) I take mathematical consistency and adequate physicality as pre-requisites for states, properties and propositions about these to be included 
within one's physical theory.  

%
%

\noindent Note 3) `Approximate' reflects the practicalities of imperfect knowledge of the world, as in Sec \ref{Cl-Grainings} on grainings, and `sub' the modelling of Sec \ref{Cl-Pers}.  

\noindent Note 4) In involving probabilities here, one is to make the distinction between probabilities entering Classical Physics due to imprecise 
knowledge of the system versus probabilities being {\sl inherent} in Quantum Physics (see Sec \ref{QM-Prop} for application of this to the theory of propositions).  
In the latter, a (sub)system has potential properties which may be actualized via (something like) measurement. 

\mbox{ } 

\noindent Among the questions of being, the following are to play a particular role in timeless approaches. 

\mbox{ } 

\noindent {\bf Questions of Conditioned Being}, which involve two properties within a single instant: 
`what is the probability that, given that an (approximate) (sub)configuration $\fQ_1$ has property $P_1$, 
it (or some other $\fQ_2$ within the same instant) has property $P_2$?  

\mbox{ } 

\noindent Note 5) These clearly have significant scientific content as a means of predicting and testing unknowns given knowns.  

\noindent Note 6) Whilst the \NSI and \CPI are inherently quantum, the examples of questions given for these in the Introduction 
serve entirely general as examples of questions of being and of conditioned being respectively.  
For e.g. the relational triangle, one can easily conceive of similar questions based on properties such as the value of the 
moment of inertia, isoscelesness, regularity, contents homogeneity of the subclusters or of notions of uniformity (as  
maximized by equilaterality).
Thus one is already well sorted out for lucid questions to investigate even for so simple 
a whole-universe model, as well as, by Secs 3 and \ref{TessiRegions}, for how these propositions translate to regions of the configuration space.

\subsubsection{Temporal questions}

\noindent Moreover, in the presence of a meaningful notion of time, one can additionally consider the following 
question-types to be primary.

\noindent I) {\bf Questions of Being at a Particular Time} involve Prob($\fQ_1$ has property $P_1$ as timefunction $\ft$ takes a fixed value $\ft_1$). 

\noindent II) {\bf Questions of Becoming}, on the other hand, have the further features of Prob($\fS_1$ dynamically evolves to constitute $\fS_2$).

\mbox{ } 

\noindent Note 1) Some lucid examples of becoming questions are `do highly homogeneous universes become more inhomogeneous?'  
`Do highly homogeneous universes become populated by supermassive black holes?'\footnote{One could further specialize from the 
`at some time in the future' implied here to [composition of I) and II)] `at a given time in the future' or to `becomes permanently'.
`At some time' and `becomes permanently' should probably be shielded by cut-offs due to how it is only physical 
to consider properties that are realized withing finite time.}
%
A specific example of such a question that is integral to the scientific enterprise itself, is as follows.  

\noindent (*) If an experiment is set up in a particular way, what final state does its initial state become due to the active agents of the experiment?  

\noindent Note 2) Without the qualifications of I), questions of being can look very vague.
Compare e.g. `is the universe homogeneous?' to the same but qualified with `today' or `at the time of last scattering'.    

\noindent Note 3) See however Sec \ref{QM-POT-Strat} for how I) and even II) can be rearranged in purely timeless terms, at least in principle.  
If these rearrangements are used in one's conception of the world and of science e.g. (*), one would then expect \cite{Records} 
the resultant form for the physics to strongly reflect the structure of atemporal logic.   
%
 
\noindent Note 4) On the other hand, if this is not attempted, the totality of temporal questions requires a {\it temporal logic} 
that is more complicated than atemporal logic due to the extra constructs present in I) and II).  

\noindent Note 5) there are some practical/physical/mathematical restrictions on questions of becoming. 
E.g. the problem need to be well-posed is well-posed, including $S_1$ being extensive enough 
and well-placed enough to be the only significant input to the process leading to $S_2$. 
E.g. a sufficient chunk $\FrS_1$ of a past Cauchy surface $\bupSigma_1$ is needed to control some future chunk $S_2$ of 
Cauchy surface $\bupSigma_2$, where that sufficiency is dictated by $\FrS_2 \subset D^{+}(\FrS_1) \cup \bupSigma_2$, for $D^+(\FrS)$ 
the future domain of dependence of set $\FrS$.   
[Cauchy surfaces are ensured if the global hyperbolicity of Time-GR7) holds.  

{            \begin{figure}[ht]
\centering
\includegraphics[width=0.4\textwidth]{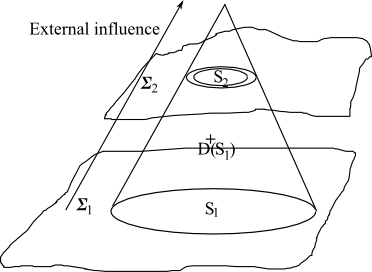}
\caption[Text der im Bilderverzeichnis auftaucht]{        \footnotesize{Propositions concerning becoming of regions. 
Signals from $p_1 \in \bupSigma_1$ outside of $S_1$ cannot influence the physics of  $S_2 \subset \mD^{+}(S_1) \cup \bupSigma_2$.}   }
\label{DOD-for-props} 
\end{figure}  } 

\subsubsection{The logical structure of atemporal questions}\label{log}

I begin with a simple outline of Boolean logic. 
One asks questions the answers to which are YES and NO (or TRUE and FALSE).  
There are then AND, OR and NOT operations, which I denote by $\vee$, $\wedge$ 
and $\neg$ respectively, and a number of relations between these that include the distributivities 
\beq
P \wedge \{ Q \mbox{ } \vee \mbox{ } R \} = 
\{P \wedge Q\} \mbox{ } \vee \mbox{ } \{ P \wedge R \}
\mbox{ } , \mbox{ } \mbox{ }
P \mbox{ } \vee \mbox{ } \{ Q \wedge R \} = 
\{P \mbox{ } \vee \mbox{ } Q\} \wedge \{ P \mbox{ } v \mbox{ } R \}  \mbox{ } .  
\label{Boo-Distrib}
\eeq
and the law of the excluded middle (P AND NOT P are, between them, everything).  
A second layer of structure is the {\it logical implication} operation, which I denote by $\preceq$.  
Mathematically, this is a partial ordering: 
\beq
P \preceq P \mbox{ } \forall \mbox{ } P \mbox{ } \mbox{ (reflexivity) }
\eeq
\beq
\mbox{ If } P \preceq Q \mbox{ and } Q \preceq R \mbox{ then } P \preceq R \mbox{ } \mbox{ (transitivity) }
\eeq
\beq
\mbox{ If } P \preceq Q \mbox{ and } Q \preceq P, \mbox{ then } P = Q \mbox{ } \mbox{ (antisymmetry) } .  
\eeq
Its function in logic amounts to being a comparer of the precision of various propositions/questions (`are 
you six feet tall to the nearest foot' versus `are you six feet tall to the nearest inch').  
Its physical realization is as a graining operation.

\mbox{ } 

\noindent Note: this example shows that a `measure of precision'/notion of distance serves to furnish a partial 
ordering, however not all partial orderings will be underlied by a such.  

\mbox{ }

\noindent Boolean logic suffices for the logical structure of Classical Physics. 
The problem is that the situation is quite different in Quantum Physics (see Sec \ref{QM-Str}).
There one needs a far broader perspective, much like the difference between tacitly modelling the real world as Euclidean 
but then becoming aware of a plethora of differential geometries, which begs the question of investigating what geometry 
the real world has rather than continuing to presume it has the simplest one.
Indeed, a wide variety of logics are available, and it is an open question which of these best suits Quantum Physics (see Sec \ref{QM-Str}).

\begin{subappendices}
\subsection{Topological relationalism?}\label{Top-Rel}

In canonical GR, there is an obvious extra level of background structure (c.f. Sec \ref{Examples}): the fixed spatial 3-topology itself.
The general MBI blueprint extends [Criterion 4)] to this, albeit not the details of many of its implementations: 
\be
\mbox{TBI} = \mbox{\Large S}_{\tFrT} \circ \mbox{Maps } \circ \stackrel{\rightarrow}{\FrT}\cO \mbox{ } .
\ee
This corresponds to Riem =$ \bigcup_{\sbSigma}$Riem($\bupSigma$), which is possibly still a restrictive sum e.g. restricted to the  
1) orientable $\bupSigma$, 2) the closed without boundary $\bupSigma$ or 3) the $\bupSigma$ of a given dimension.\footnote{The 
motivations for these restrictions are 

\noindent 1) an inbuilt time-orientability both for its own sake and for understood and well-defined mathematics of evolution. 

\noindent 2) The putative connection between relationalism/Machianism and closedness, alongside it remaining plausible our universe is closed, and 
alongside the minimality of not having to consider the forms of boundaries and subsequently boundary conditions on them (which, in the lack of 
evidence for boundaries, could be seen as extraneous and un-Machian).  

\noindent 3) 3-$d$ is the smallest dimension that is 
sufficient to describe all experimentally/observationally 
confirmed Physics to date so it is a good minimalistic and non-hypothesis-making choice.  

\noindent However, it could well be that in a deeper theory dimension, orientability and whichever of openness, closedness and boundaries 
are emergent phenomena.  
These could indeed still be viewed as possibly background structure if treated as pre-determined inputs; c.f. the attitude already taken to dimension at the start of Sec \ref{Examples}.  

Observationally the universe could indeed as well be open, and have previously argued that (building on some arguments of Hawking, 
in particular his connecting tubes of negligible action), it is always possible to provide a closed model indiscernible from a given open one 
using current observational sensitivity, and vice versa.  
I now underpin this argument with some relational thinking.  
Sometimes then Occam's razor may favour one of the pair, whilst some relational/Machian thinking might favour the closed one.  
Though the construction used to have that could be as bad as postulating a boundary for no reason, e.g. that an apparently open universe solution 
closes up by curvature or by topological identification on a scale sufficiently in excess of the Hubble radius to not be detectable, or that an 
apparently closed universe has an undetectably thin aperture leading to an open region.  
But to the Leibnizian, these constructs demonstrate that openness or closedness is not true property of the world, since the two are indiscernible.  
It is then just a case that often one of the two will be a simpler description than the other. 
(Both geometrically, which is what the arguments establish, and as regards stability conditions, which notably remains unestablished - do 
undetectably small apertures pinch off, particularly in worlds not constrained to be of fixed spatial topology, and might there be restrictions 
on whole-universe size or on the plausibility of such as topologically-identified FLRW spacetime.)} 


\noindent Note 1) To be clear, the $\FrT$ group is {\sl not} a group of the conceptually simpler kinds that occur in topology (such as topological group - 
both a topological space and a group -, or one of homotopy, homology or cohomology - {\sl characterizers of} topology.  
It is, rather, a `ripping group' of topology-{\sl altering} operations. 
E.g. for 2-$d$ orientable closed without boundary spaces, it is generated by the handle-adding operation that changes the genus.  
In slightly more detail, rippings are the operations of {\it surgery theory}; {\it cobordisms} are a type of ripping: via there being a 
1-$d$ higher manifold linking the two manifolds of distinct topology.  
I do not then know the extent to which cobordisms suffice as regards spanning all types of rippings.

\noindent Note 2) Indirectness means one does not have to face the consequences of this to start off with, e.g. considering each genus in turn. 
One would then add, average or extremize over the genus to render the object topologically background independent.  
A common problem already present in this simple example is that adding and averaging will be infinite contributions (and ratios/other 
compositions of such infinities). 

\noindent Note 3) Inf-taking is definitely accommodating of being over a discrete parameter, since it is free of the specific technical 
limitations of the variational calculus and indeed held to be very widely applicable in mathematics.
But extremization/inf taking looks strange in that it could well pick but one contributor.  
It is far from clear whether it would make any sense to assume the universe has three holes just because the action for a three-holed 
universe is smaller than that for any other number of holes!
This is a discrete parameter extremization of actions selecting dynamical paths involves extremizing continuous parameters.  
At the very least one would have to use an extended version of the calculus of variations to justify this, and how to do so is not clear.  

\noindent Note 4) MBI and TBI can be composed via $\FrT$-act, $\FrG$-act, $\FrG$-all, $\FrT$-all making sense, in that given a particular topology, 
one can establish a $\FrG$-corrected version of $\FrQ$ on it.  

\noindent Note 5) Butterfield pointed out that  $\FrG$-act, $\FrG$-all procedures are interpretable in terms of relations between things, 
and asked whether this carries over for the topological version.  
My answer is that $\FrT$-act is along the lines of cobordisms, which can be interpreted as relations between possibilities for 
the form of physically-relevant space.  
And one can still do something like the Best Matching process between topologically distinct manifolds.  
It is just that one's construct cannot use a unique integral over $\bupSigma$. 
And reducing this to a meaningful and computible quantity is a harder task.
There are possibly some ties between this and Seriu's work.  

\mbox{ } 

\noindent {\bf Question 32} Some theories are {\sl only} topological.  
To what extent has relationalism been considered for these theories?

\mbox{ } 

\noindent {\bf Question 33} Discrete approaches may tie more readily to topological issues, thus furnishing another arena of study.   

\mbox{ } 

\noindent Note 5) All of Matzner, Seriu and some inhomogeneity comparers are useable between manifolds of different spatial topology.  

\mbox{ }

\noindent Basing physical modelling on fixed topology, or even on topology at all is not relational, in the sense that topologically 
distinct spaces with metric structure thereupon and physical entities built on top of that can be physically indiscernible and thus identical.  
One does not know how to calculate without starting by stating the topology, but some results will not be affected.  
The truly physical notion is the large-scale structure of space.  
This is some new kind of mathematics: topology itself coarsened by length concepts, so that, somehow, large handles/tubes count, and small ones do not. 
(`Large' here is with respect to the probing capacity of the observers.)  
Sometimes global effects will serve to discern, but I do not expect these to always be the case.  
I view Seriu's work as an at least qualitatively useful prototype, since observations are tied to observed wavelengths and thus to spectral 
information about the operators on the space that also occur in the physical laws. 
Though it is still indirect: given a metrized topology, compute from the Laplacian, and it still does not maximize the connection to 
observability (or yet establish insensitivity to which of the physical differential operators are used). 
If spectra are everything, the isospectral problem is relevant.  
I posit that a relational view of the isospectral problem is that if spectra are all that is relevant to Physics and there is isospectrality for all operators occurring in Physics, 
then the identity of indiscernibles would have it that we live in an equivalence class of geometries with no physical meaning being ascribable to the mathematically different 
representatives of this.  
This is much like Sec \ref{Examples}'s 3-particle universes having no sense of dimension beyond 1 and `more than 1'. 

\noindent An example to bear in mind in the opposite direction is that it is held that 1 monopole (a defect type) being present in the universe 
would suffice to explain charge quantization (Dirac).  
Here, the conventional argument allows for this monopole to be too weak and far away for direct detection.  
Does this argument hold in the whole-universe GR setting, and could  variants of such render the actually-used mathematically standard notion of topology indirectly detectable.  
E.g. the Dirac argument in curved spacetime, whether the monopole being outside the Hubble radius would affect the result, and, if not, 
is Dirac's argument Machian/Leibnizian (such a monopole acts but isn't at all meaningfully acted upon?)

\mbox{ } 

\noindent As regards RPM's as toy models of this, the most obvious is to use all $N$'s and/or all $d$'s indiscernible to each $N$ for all $N$'s.  
Particle fusion/fission is the RPM analogy of GR's topology change [c.f. Sec \ref{TopCha}].  

\end{subappendices}

\vspace{11in} 

\noindent{\bf \Huge II. CLASSICAL PROBLEM OF TIME}

\section{Facets of the Classical Problem of Time}\label{Cl-POT}

\subsection{A second analysis of clocks including the history of accurate timestandards}

\subsubsection{`Any', `all' and `STLRC' implementations of Mach's Time Principle}

I add the following three independent alternatives to Clock 1) to 5) of Sec \ref{Clocks}. 

\mbox{ }  

\noindent Clock-0A) Any subsystem will do as a clock {\sl or} Clock-0B) some subsystems are better than others as clocks.

\noindent Clock-6A) Clocks are localized subsystems  {\sl or} Clock-6B) (`Clemence's principle' \cite{Clemence}) one is to factor in changes of the actual subsystem under study.  

\noindent Clock-7A) The less coupled one's clock and subsystem under study are, the better \cite{PGP1, PGP2}, or Clock-7B) they fundamentally {\sl have} to be coupled 
(in some formulations there are difficulties as regards a clock keeping time for a subsystem if it is not coupled to it {\sl at all}; Sec \ref{Semicl} is a physical example of this).
With this alternative, a compromise Clock-7C) can be reached: a small but strictly nonzero coupling can be assumed, 
so that the subsystem under study only very gently disturbs the physics of the clock (if nothing else, because nothing can shield gravity). 

\mbox{ }

\noindent Perhaps then time is to be abstracted from {\sl any} change [Clock-0A)].\footnote{This is not to be confused with 
a separate, subsequent idea/program of Rovelli's {\it thermal time} is a more specific, rather than 'any' time, further strand of Rovelli's thinking.  
Here, one uses the $H$, $t$ inter-relation and one supplies the form of the $H$ to be used by inverting Gibbs' thermodynamical formula. 
This could be viewed as another type of emergent time, c.f. the current Sec and \ref{QM-POT}.
However, the {\sl uninverted} Gibbs formula has conceptual and operational primality built into it which itself looks to be far harder to invert than the formula itself.  
In other words, there are deep-seated reasons why thermodynamical entities are held to be secondary to mechanical ones 
(e.g. velocity/momentum being more primary than pressure, heat or temperature).
Moreover, further claims that thermal time is an approach to the Arrow Of Time run into problems with the Gibbs formula only applying to {\sl equilibrium} thermodynamics, 
which is both highly not representative and excludes irreversible processes that are essential for a satisfactory account of the POT.
Finally, thermal time does not reflect how any accurate clocks have hitherto been made (though it remains far from clear what the nature of clocks capable of probing physically-extreme 
regimes right down to conditions requiring Quantum Gravity would be.
I thank Jos Uffink, Julian Barbour and Alexis de Saint Ours for discussions on these point, though the exact words I have chosen to use here are my own.}

\noindent This approach also carries connotations of 6A) [which is far commoner among modern theoretical physicists than 6B)]. 
%

\mbox{ } 

\noindent Perhaps instead time is to be abstracted from {\sl all} change in the universe.  
I term this the {\bf LMB timekeeping} since it is Barbour's \cite{B94I, EOT} whilst leaning heavily on Leibniz 
in amounting to the universe itself being the only perfect clock.\footnote{This is meant at the classical level. 
See e.g. \cite{UW89} for problems with perfect clocks at the quantum level.} 
%
This rests on Clock-0B) and Clock-6B) boosted by 

\mbox{ } 

\noindent Clock-8) `the more change the better', which Barbour takes to its logical extreme:  

\noindent Clock-9) `include all change'.  

\mbox{ }

\noindent Much can be learnt by contrasting these two extremes.  
On the one hand, AMR's Clock-0A) has an underlying sense of `democracy'. 
This is argued to be useful in generic situations (taken to be crucial in GR) that would appear to have no privileged time standard.  
On the other hand, Clock-0B)'s `meritocracy' is in close accord with mankind's history of accurate timekeeping
(both Barbour \cite{Bfqxi} and I invoke this as useful evidence, albeit it leads each of us to different conclusions).  
All of criteria Clock 1--5) entail Clock 0B) as the underlying alternative.
Finally, Clock-9)'s extreme attitude provides, within the context of Clock-6B), an {\sl incontestable} time, in that there is no 
more change elsewhere that can run in concerted ways that cause of one to doubt one's timestandard.  
However, whole-universe and perfect notions for clocks also draw criticisms in Secs \ref{+temJBB} and \ref{QM-Clock}.

\subsubsection{A brief history of timestandards}

\noindent {\bf Sidereal time} is kept by the rotation of the Earth relative to the background of stars, and served 
satisfactorily as a timestandard for two millennia, from Ptolemy \cite{Ptolemy} until the 1890's.  
Barbour pointed out that it reflects Clock-0B), to the extent that by those two millennia's standards, it is reliable for 
making predictions about the other celestial bodies, whereas solar time fails to be. 
[Though this is in fact for {\sl apparent} solar time rather than mean solar time, which itself is both far closer to sidereal 
time and indeed was often also used as a reliable timestandard.]  

\noindent Sidereal time served well because the Earth's rotation is pretty stable (to roughly 1 part in 10$^{8}$),\footnote{This  limitation is 
mostly due to tidal braking as envisaged by Delaunay in 1866 \cite{Delaunay}, though there are quite a number of further 
smaller effects in more detailed considerations of the Earth's interior.}
of three bright stars (Sirius, Arcturus and Aldebaran) with where Hipparcos had catalogued them as being 1800 years previously; 
they were out by over half a degree, i.e. a drift of over 1 second of arc per year.
It also reasonably embodies Clock-6A) in that, as compared to the solar system, the 

\noindent Earth is a fairly localized body.

However, the main point of discussing sidereal time is its demise due to irregularities in the rotation of the Earth.  
This was noted via departures from predicted positions, especially noticeable in the case of the Moon.
These departures were most succinctly accounted for not by modifying lunar theory but by considering the rotation of the Earth to inaccurately 
read off the dynamical time (which to the accuracy of those days' Celestial Mechanics observations, remains conceivable of as the time of Newtonian Mechanics). 
In De Sitter's words from 1927 \cite{DS27}, {\it ``the `astronomical time', given by the Earth's rotation, and used in all practical 
astronomical computations, differs from the `uniform' or `Newtonian' time, which is defined as the independent variable of the equations of celestial mechanics."} 
This was addressed by using the Earth--Moon--Sun system as a superior timestandard provider [hence embodying Clock-0B)], with Clemence's 
eventual proposal in 1952 \cite{Clemence} involving a particular way of iteratively solving for the Earth--Moon--Sun system for an increasingly-accurate 
timestandard that came to be known as the {\bf ephemeris time}.\footnote{Clemence partly credited the English Astronomer Spencer-Jones \cite{SJ39} and the French Astronomer 
Danjon \cite{Dj29}, whilst the name `ephemeris time' itself was coined by another American Astronomer, Brouwer \cite{Brouwer}; significantly, De Sitter and Clemence had 
both been terming it `uniform' or `Newtonian' time, though I would rephrase this as {\sl a relational recovery of Newtonian time}.}
%
Clemence's contributions to the study of time have also been held in high regard from a philosophical perspective by e.g. Whitrow \cite{Whitrow} and Barbour \cite{B94I, Bfqxi}, 
though it is the present Article and \cite{ARel2} that unveils the full extent to which Clemence's conceptualization of time is Machian.
Ephemeris time is a prime example of the highly Machian statements Clock-6B) and Clock-8); moreover, it does not go to the extreme of Clock-9).  
Poincar\'{e}'s work \cite{Poincare98} contained an antecedent of this Machian perspective (again however without any mention of Mach himself).  

\noindent Already in 1948 \cite{Clemence48}, there was a divergence between `civil' and `dynamical' timekeeping, due to the Earth's far more central role in the former.
Everyday life is run according to the Physics of the Earth, despite the small tidal braking on its rotation.
This an important distinction for the reader to be aware of, despite how this Article focuses principally on dynamical time, 
especially as used for the even less Earth-tied space science.  

\mbox{ }

\noindent There have been two significant changes to timekeeping since.  

\mbox{ }

\noindent I) {\bf Relativistic effects} (SR and GR) become significant once timekeeping reaches an accuracy of 1 part in $10^{12}$.  
This accuracy exceeds that which was possible when ephemeris time was introduced in 1952, but such accuracy was attained in the late 1970's.  
Thus one ceased to be able to conceive in terms of Newtonian time, and its relativistic replacement has the novel feature of requiring one to say {\sl where} 
a timestandard holds (by e.g. gravitational time dilation) as well as depending on how the clock in question is moving (SR time dilation).  
These were not issues in the Newtonian model used in practise for timekeeping prior to the 1970's; 
indeed Clemence did not attach his ephemeris definition to a particular point on Earth.

Using the Moon as a clock-hand for reading off Newtonian time only makes sense at some position, since different gravitational 
equipotential surfaces have different notions of time by gravitational time dilation.  
Implicitly, one is defining such a timestandard on the surface of each planet, assuming it is of uniform radius (or using a standardization such as mean sea level).  
One can then see problems with the relative position of the Moon being a primary/clock-hand timestandard, since it does not care about equipotential surfaces near the Earth.  
At best, one should interpret one's observations of the Moon within its own gravitational potential, and then compensate by a gravitational time-dilation  
as regards what timestandards this represents at mean sea level on the Earth itself.

\mbox{ }

\noindent II) {\bf Advent of atomic clocks}. It was these that permitted accuracy to exceed that of relativistic significance 
in the late 1970's; by now we have atomic clocks with a primary timestandard accuracy of 2 parts in $10^{16}$ \cite{LGS11}. 
These accuracies exceed those of astronomical timestandards (whose main limitations come from limited knowledge of accuracy of the contents of the solar system causing various 
fluctuations; it is probably due to this that astronomical timestandards have become rather less well-known among theoretical physicists).
In contrast, atomic clocks, have simple and well-known internal constitution and physics, whilst being selected and further designed for great stability. 
They are small [Clock-2A)], which is all the better for shielding them from disturbances, convenience of obtaining a quick reading, 
and not needing to worry about position-dependent relativistic effects within each atomic clock itself.

However, are atomic clocks a fully independent paradigm or more along the lines of a far more convenient reading-hand to the position of the Moon, 
but which nevertheless need recalibrating?

A fundamental issue is that, independently of how stable the clock is, why should what it it reads out correspond {\sl exactly} to the dynamical time 
(an SR-and-GR-upgraded counterpart of the Newtonian time) of the system under study (see e.g. p4 of \cite{Schegel}, p11 of \cite{Fraser}, \cite{Clemence, Whitrow})? 
(In Engineering terms, one talks of hardware time as distinct from proper time, and terms the difference {\it clock bias}; this is certainly 
known e.g. in the theory of GPS \cite{Bahder}; the contention here then is the possibility that primary time standard atomic clocks may themselves exhibit clock bias.)  
As it happens, from the early days of atomic clocks it was known that they read out ephemeris time to at least 1 part in 10$^9$ \cite{Parry}, 
which substantially eased the transition from ephemeris to atomic timestandards.
However, one might view this as having the status of a {\sl null experiment}, so that one should keep on testing whether this premise continues to 
hold true as accuracy elsewise improves.

After all, taking the Earth {\sl not} to read off the dynamical Newtonian time in the upgrade from sidereal to ephemeris time was of this nature.  
Though it helps here to decompose Clock-6B) into two logically-possible parts: problems with the physics of the clock itself and problems concerning the clock-reading not corresponding 
to the parameter that most simplifies the equations of motion, which are more holistic in depending on the system under study rather than the clock itself.
It should then be noted that the Earth is a `dirty clock' due to its complex and partly unknown internal dynamics, whereas atomic clocks are designed 
from the perspective of cleanliness, with their known, simple and well-shielded dynamics; this is very useful for atomic clocks to attain very great accuracy. 
Unless the holistic effect occurs at some level; have we seen any evidence for this to date?  
For sure, the {\it leap seconds} that one adjusts some years by are not of this nature, since they concern 
adjusting {\sl civil} time to compensate for irregularities in the rotation of the Earth.  
Further adjustments (3 and 5 orders of magnitude smaller) between `barycentric dynamical time' (for space science) 
and `terrestrial time' (for science on Earth) \cite{AlmaSup} are fully accounted for by standard Gravitational and Relativistic Physics.

To that accuracy, at least, modern astronomical timestandards can be readily obtained by shifting atomic timestandards, 
so no holistic realizations of Clemence's principle 2B) are in evidence.
This is worth pointing out since i) elsewise the revelation that Clemence's work on ephemeris time is highly Machian 
might well be taken to carry holistic connotations, for which however to date no evidence is known. 
ii) Nevertheless, one might keep an open mind as to whether holistic connotations would show up at some level much finer than the sidereal to ephemeris time transition of the first 
half of the 20th century, by developing Machian thinking {\sl as well as} the minutiae of yet further improving atomic clocks.  

\mbox{ } 

\noindent Ephemeris time is also a notable exception to Clock-1) via its factoring in irregularities, and also to 
e.g. Einstein's basing of clocks upon periodic motions, though the atomic-clock paradigm is clearly of this nature.

\mbox{ }  

\noindent Taking the above insights on board, with a small amount of very plausible generalization, I arrive at the 
following picture for a practical {\it LMB-CA timekeeping} (where the C stands for `Clemence'), which represents `middle ground' between 
Barbour and Rovelli's positions whilst being maximally in accord with mankind's history of accurate timekeeping.

\subsubsection{Generalized Local Ephemeris Time (GLET) procedure}

{\sl LMB-CA time is best in practise abstracted from a STLRC} [Analogy 59)].  
\noindent This times is a {\it Generalized Local Ephemeris Time (GLET)}, extending [Criterion 4)] astronomers' realizations of ephemeris time.
This concerns not just using a change to abstract a time, but also check whether using this time in the equations of motion for other 
changes suffices to predict these to one's desired accuracy, in accord with Clock-0B), Clock-6B) and Clock-8).   

\mbox{ }  

\noindent Note firstly that, in comparison to AMR, this has the explicit extra element of acknowledging that some times are locally more useful to consider than others.  
For, even in generic situations, one can locally consider a ranking procedure for one's candidate times, alongside a refining procedure until 
the sought-for (or physically maximal) accuracy is attained.  
Whilst these may well give a less accurate timestandard than the highly non-generic Earth--Moon--Sun system, say, these will nevertheless give rise to some kind of extremum.
Thus Rovelli's good idea of genericity is not solely the province of `any time' approaches.
\noindent Secondly, in comparison to LMB, the ephemeris-type method as used in practise for the solar system is entirely adequate without 
having to consider the net physical effect (tidal effect) of distant massive bodies such as Andromeda as per Sec \ref{+temJBB}.  
[This also renders irrelevant our increasing lack of precise knowledge of the masses and positions as one considers more and more distant objects.]   
\noindent By this device, the LMB-CA account of practical timekeeping holds without having to go to LMB's whole-universe extreme Clock-9), 
which is a sensible avoidance to make because I) it is operationally undesirable since we know little of the motions and constitutions of very distant bodies.  
%

The GLET-finding procedure is thus to not just use a change to abstract a time, but also check whether using this time in the equations 
of motion for other changes suffices to predict these to one's desired accuracy.  
If the answer is yes, then we are done.  
If not, consider further locally-significant changes as well/instead in one's operational definition of time (locally-significant is judged by the criterion in Sec \ref{+temJBB}).  
Then if this scheme converges without having to include the entirety of the universe's contents, one has found a LMB-CA time that is locally more robust than just using 
{\sl any} change in order to abstract a time, and it is particularly useful to consider equations of motion in this time and propositions conditional on this time.  
[However, if the scheme requires the entirety of the universe to converge, then LMB time itself is the most robust time, but subject to the above-mentioned impracticalities.]

{\sl Where} in space the timestandard is supposed to hold matters once accuracy exceeds GR and SR corrections.  
Abstracting time from the Earth--Moon--Sun system does {\sl not} mean finding time throughout some box with $L$ of the order of some Earth--Moon--Sun system characteristic length.  
Different objects moving in different ways therein will need specific SR and GR corrections to the timestandard if that is above a certain level of accuracy.

\subsubsection{Other conceptually interesting examples of clocks}

\noindent \bu Einstein's light clocks (reflecting light rays off mirrors) were useful in the conceptual development of SR, and indeed carry over to GR spacetimes.  

\noindent \bu Pulsars may be considered to be clocks, and at least fairly accurate ones.
These furnish an example of a case with {\sl very} negligible gravitational interaction between a clock and the subsystem(s) it `keeps time for'.  
\noindent This then helps envisage a question: is the `locally relevant' in STLRC relevant through physical action or does it also include relevance through visibility alone?
Would one call an earthly timestandard `bad clock' if it failed to be useable to account for pulsar observations?

\subsubsection{Clocks in physically extreme regimes}

One should next consider homing in on GLET procedure in regions under more physically extreme conditions.  

\mbox{ } 

\noindent {\bf Question 34}  Which clocks can survive in various parts of black hole solutions?  
Do black holes themselves provide any further types of notably accurate clocks?

\mbox{ } 

\noindent {\bf Question 35} Which clocks can exist in early universe regimes? 
These need to exist at high temperature and high matter density.  
And which sorts of natural clocks actually did exist there?

\mbox{ } 

\noindent Feats of modern Engineering used in terrestrial timekeeping are rather less relevant here, as opposed to selecting instrumentation or natural physical systems 
which survive, and retain good timekeeping properties, in situations involving large tidal forces due to high curvature (smallness helps in addition to material toughness), 
and large temperatures.

\subsection{The classical Frozen Formalism Problem} \label{FFP}
%
{            \begin{figure}[ht]\centering\includegraphics[width=1.0\textwidth]{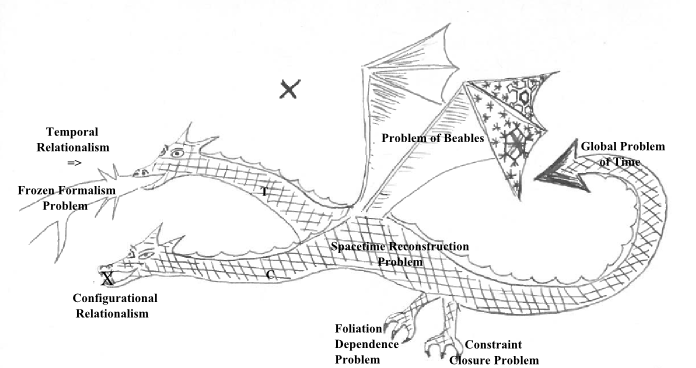}
\caption[Text der im Bilderverzeichnis auftaucht]{        \footnotesize{The Classical POT as a slightly less ferocious Ice Dragon. 
X indicates a missing Facet, so this Ice Dragon is 7/8ths strength.}  } 
\label{Classical-Dragon}\end{figure}          }

\noindent One of this Article's central tenets is that one of the two main relational pillars -- Temporal Relationalism -- 
is the very same as the classical background-independent introduction of what becomes the Frozen Formalism Problem at the quantum level.  
The full argument for this then is as follows. 

\noindent 1) {\bf Temporal Relationalism}.  
Following Leibniz \cite{L} [Relationalism 5)], one postulates that there is to be no time for the universe as a whole.  
{\sl In this sense, there is a classical Frozen Formalism right from the outset, by insisting on modelling closed universes from a background-independent perspective.}
Leibniz's idea can be mathematically implemented by considering reparametrization-invariant actions 
that include no extraneous time-like variables such as absolute Newtonian time or the GR lapse.  
Such actions are then of Jacobi type \cite{Lanczos, BB82, B03} for mechanics or of Baierlein--Sharp--Wheeler type \cite{BSW, RWR, Phan, FEPI} for Geometrodynamics.  
Whilst for now we only consider the Jacobi reformulation of mechanics, 
\beq
S =  2\int\sqrt{T\{E - V\}}\d\lambda \mbox{ } ,  
\eeq
%
this SSec's argument for the Frozen Formalism Problem at the classical level and its formal resolution hold for both RPM's and GR: Analogies 57) and 58).

Then as envisaged by Dirac \cite{Dirac}, reparametrization invariant actions oblige primary constraints to appear 
(i.e. solely due to the structure of the Lagrangian rather than requiring any kind of variational procedure, as per Sec \ref{Dirac-Method}). 
In the present case, the structure of the Lagrangian is such that there arises a single primary constraint that has 
quadratic and no linear dependence in the momenta (as per Sec \ref{Quad-Con}).  
This is the energy constraint $T + V = E$; in the geometrodynamical analogue below, we shall see how the corresponding constraint 
is the Hamiltonian constraint, whose form is well-known to lead to the {\sl quantum} Frozen Formalism Facet of the POT 
(Albeit it is far less well known to arise by the complete chain of reasoning presented here, which is due to Barbour and 
collaborators: \cite{BB82, B94I, RWR, San, FEPI, ARel2} and the present Article.)  

Moreover, one can now obtain at the classical level a Jacobi--Barbour--Bertotti (JBB) emergent time \cite{B94I, SemiclI} (Secs \ref{Intro} and \ref{Examples})
that simplifies the momenta and equations of motion,
\beq
\lt^{\se\sm(\sJ\sB\sB)} = \int\d\lambda\sqrt{T/\{E - V\}}   =  \left. \int\d s \right/\sqrt{2\{E - V\}}
\eeq
(as per Secs \ref{Intro} and \ref{Examples}).  
This is an example of an implementation of Mach's ``time is to be abstracted from change" \cite{M} [Relationalism 6)], which is an emergent-time 
resolution of the classical Frozen Formalism Problem that one is faced with upon adopting Leibniz's Temporal Relationalism postulate.  

\mbox{ }  

\noindent Note 1)  As I showed in Sec \ref{Examples}, emergent JBB time additionally amounts to the relational recovery of a number of well-known notions of time in various 
contexts (Newtonian time, GR proper time, cosmic time), which have the overall property of casting the classical equations in a particularly simple form.

\noindent Note 2) Moreover, it does not unfreeze the frozen formalism, nor does it in any other way directly give quantum equations different from the usual ones.   
\noindent One can view emergent JBB time, rather, as an object that is already present at the classical level that is subsequently to be recovered 
by more bottom-up work at quantum level [since it matches up with the emergent semiclassical (WKB) time) of Sec \ref{Semicl}]. 

\mbox{ }  

\noindent $t^{\se\sm(\sJ\sB\sB)}$ succeeds enough at the classical level and via its semiclassical parallel to strongly motivate its detailed analysis in Sec \ref{+temJBB}.

\subsection{Which other classical POT facets are manifested by Jacobi's form of mechanics}\label{Cl-Jac-POT}
%
{            \begin{figure}[ht]\centering\includegraphics[width=1.0\textwidth]{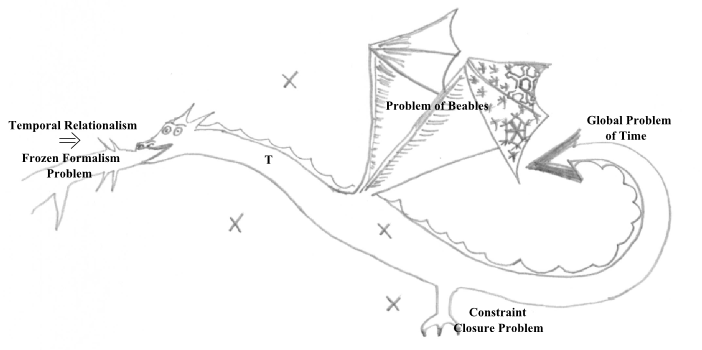}
\caption[Text der im Bilderverzeichnis auftaucht]{        \footnotesize{Classical Jacobi Mechanics POT as a much less ferocious 
Ice Dragon: it is 4/8ths strength.}  } 
\label{Classical-Jac-Dragon}\end{figure}          }

\noindent Looking through the Introduction's outline of the POT facets, I can comment here that this simple model has 

\mbox{ } 

\noindent Useful Difference 23) no Configurational Relationalism to worry about (that generalizing the Best Matching Problem map-ordering wise, with that in turn 
generalizing the Thin Sandwich Problem theory-wise).

\noindent  Useful Difference 24) It has but a trivial notion of \K Observables, since there are no linear constraints as a further consequence of the lack of 
Configurational Relationalism.
As we shall see below, Dirac observables can then be built up for such an approach along the lines given by Halliwell \cite{H03, H09}.

\noindent  Useful Difference 25) There is no Constraint Closure Problem as a consequence of there being just the one constraint.

\noindent  Useful but also limiting Differences 26) and 27) There is no Foliation Dependence Problem or Spacetime Reconstruction Problem, since this model has no GR 
notions of spacetime or foliations.

\noindent This string of differences furthermore carries over to the quantum level.   

\mbox{ } 

\noindent One still has a Global POT as per e.g. Appendix \ref{Examples}.B.5.  

\noindent There is no Multiple Choice Problem yet since that is purely quantum-mechanical; that goes without saying for the rest of Part II.    

\noindent Thus we are assured of being well in control of the classical POT in the relational recovery of Newtonian Mechanics from the Jacobi formulation as envisaged by Barbour.

\subsection{Classical Minisuperspace counterpart}
%
{            \begin{figure}[ht]\centering\includegraphics[width=1.0\textwidth]{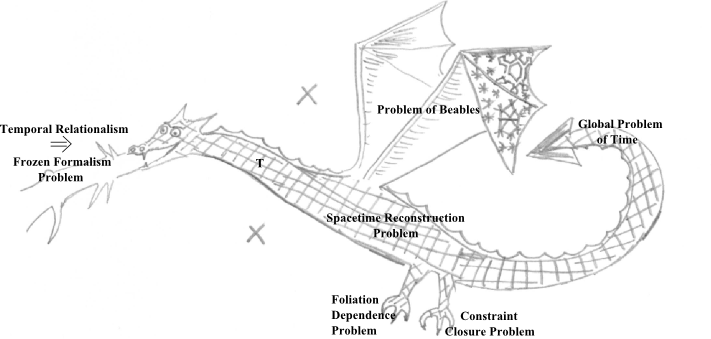}
\caption[Text der im Bilderverzeichnis auftaucht]{        \footnotesize{Classical Minisuperspace POT as an Ice Dragon of 6/8ths strength; 
however, it is around as easily overcome as the Classical Jacobi Mechanics Ice Dragon due to the single constraint and three simple uses of privileged homogeneity.}  } 
\label{Classical-MSS-Dragon}\end{figure}          }

\noindent 1) Here,  
\beq
\lt^{\se\sm(\sJ\sB\sB)} = \int \d s_{\sM\sS\sS}/
\sqrt{2\{Ric(\mbox{\boldmath$h$}) - 2\Lambda\}}
\eeq 
which is aligned with cosmic time in the (approximately) isotropic case.   

\noindent 2) This case also exhibits no nontrivial Configurational Relationalism (the lower X in Fig \ref{Classical-MSS-Dragon}).  

\noindent 3) This case has trivial \K observables once more extendible to form Dirac observables along the lines of Halliwell.

\noindent 4) Again, the Constraint Closure Problem is trivially avoided as a consequence of there being just the one constraint.
Indeed the schematic form $\{\scH, \scH\}$ = (structure function)$^{\mu} \times \scM_{\mu}$ of (\ref{Ham-Ham})
means that no $\scM_{\mu}$ implies closure and as a mere algebra rather than an algebroid.  

\noindent 5), 6) There is homogeneity: all points the same nullifies foliation issues by high symmetry and likely the Spacetime Reconstruction Problem too.  

\noindent 7) One still has a Global POT due to Appendix \ref{Examples}.B.5's POZIN, albeit with a qualitatively different nature to in the preceding SSec.

\subsection{Thin Sandwich Problem to Best Matching Problem to Configurational Relationalism generalizations}

Secs \ref{Examples} and \ref{Q-Geom} gave an outline both of the Thin Sandwich approach to GR and to RPM parallels of this. 
All in all, this is generalized theory-wise by Sec \ref{Best-Match}'s account of Best Matching.

\mbox{ }  

\noindent {\bf Best Matching Problem} \cite{BB82, RWR}.  
If one's theory exhibits Configurational Relationalism, linear constraints $\scL\scI\scN_{\sfZ}$ follow from variation with respect to the auxiliary $\FrG$-variables.   
The Best Matching procedure is then whether one can solve the Lagrangian form of 
$\scL\scI\scN_{\sfZ}\lfloor \fQ^{\sfA}, \dot{\fQ}^{\sfA}\rfloor = 0$ for the $\FrG$-auxiliaries themselves; this is a particular form of reduction.
One is to then substitute the answer back into the action. 
This is clearly a classical-level procedure, an indirect implementation of Configurational Relationalism that is a bringing into  
maximum congruence by keeping $\fQ^{\sfA}_1$ fixed and shuffling $\fQ^{\sfA}_2$ around with $G$-transformations until it is as close to $\fQ^{\sfA}_1$ as possible.
From another perspective, it is establishing a point identification map \cite{Stewart} between configurations.  
It becomes a Problem due to being an often-obstructed calculation.  

\mbox{ }  

\noindent Best Matching is furthermore generalized procedure ordering wise to none other than this Article's second main relational pillar, {\bf Configurational Relationalism}.  

\mbox{ }  

\noindent Note 1) This identification with one of the eight POT facets indeed considerably further motivates the workings in Secs \ref{Intro} and \ref{Examples} 
by which Temporal and Configurational Relationalism are rendered compatible, the workings of Sec \ref{Q-Geom} resolving Best Matching for as many RPM's 
as possible, and the configurationally relational upgrades of all the notions in Sec \ref{Cl-Str}.

\noindent Note 2) It is specifically the resolution of a theory's Best Matching that permits the emergent Jacobi--Barbour--Bertotti time 
\beq
\lft^{\se\sm(\sJ\sB\sB)} = \stackrel{\mbox{\scriptsize extremum $\sfg \mbox{ } \in \mbox{ }$} \sFG}
                                                               {\mbox{\scriptsize of $S^{\tr\te\tl\ta\ttt\ti\to\tn\ta\tl}$}}                                                              
\left(                                                              
                                                               \int||\d_{\sfg}\bQ||_{\sbfM}/\sqrt{W} 
\right)  \mbox{ } 
\eeq
to be explicitly known.  

\mbox{ }  

\noindent Difference 28) is that this extremization is attainable for RPM's in 1- and 2-$d$ by Sec \ref{Q-Geom} but not for Geometrodynamics.  

\mbox{ } 

\noindent Note 3) The significance of Best Matching via the ready consequences of Note 2) and the next SSec strongly motivates asking for 
further details of what is known about this in Geometrodynamics, and for examples of this in further theories and formulations.  
Known results, strategies and open questions in this regard are given in Sec \ref{Q-Geom}.  

\noindent Note 4) Configurational relationalism is part-closed-universe in origin through having no objects external 
to the system with respect to which $\FrG$-transformations could be detected.

\subsection{The Classical Problem of Beables}\label{Cl-POB}

Classical Dirac beables \cite{DiracObs, DeWitt67, +POORef} alias {\bf constants of the motion} alias {\it perennials} \cite{Kuchar93, 
+Perennials1, +Perennials2, +Perennials3, Haj99, HK99, BelEar, Earman02, WuthrichTh, Kouletsis, BF08}  are functionals $\iD = {\cal F}[\ttQ^{\sfA}$, $\ttP^{\sfA}$] 
that Poisson-brackets-commute with all of a theory's constraints,\footnote{One  
may allow for this to be a weak equality \cite{I93}.}
\beq
\{\iD, \scC_{\sfA}\} = 0 \mbox{ } . 
\label{DirObs} 
\eeq  
Classical \K beables \cite{Kuchar93, Kuchar99, BF08} alias {\bf gauge-invariant quantities} (as qualified in Sec \ref{Gau}) are functionals 

\noindent $\iK$ = ${\cal F}$[$\fQ^{\sfA}$, $\fP^{\sfA}$] that Poisson-brackets-commute with all linear constraints, 
\beq
\{\iK, \scL\scI\scN_{\sfZ}\} = 0 \mbox{ } . 
\label{KObs} 
\eeq

\noindent Classical {\bf partial observables} (a point of view that began with \cite{Rov91a, Rov91b, Rov91c, Car9091b} though one might view \cite{DeWitt62, 
PW83, W86} as forerunners in some ways; see also \cite{Rovelli02,Dittrich, Ditt} and the reviews \cite{Rovellibook, Thiemann} do not require commutation with any constraints.
These work via consideration of a pair of them producing a number that can be predicted from the state of the system, even though this cannot be done for individual such observables.
These are in contradistinction to {\bf true observables} \cite{Rov91a, Rov91b, Rov91c} alias {\bf complete observables} \cite{Rovelli02} (which at least Thiemann \cite{Thiemann} also 
calls evolving constant of the motion), that classically involve operations on a system each of which produces a number that can be predicted from the state of the system.  
It is not however clear exactly what objects to compute in this approach in order to investigate whether subsystems exhibit correlation, 
or how a number of other aspects of the POT can be addressed via these \cite{Kuchar92, I93, Kuchar93}.  
Whichever time is used in this approach is argued in e.g. \cite{Thiemann, PS05} to have a canonical generator, i.e. a physical Hamiltonian completely independent from the 
Hamiltonian constraint; here gauge generators and evolution are distinct entities.
%
%
This approach has time-dependent observables, making use of intrinsic coordinate scalars that constitute a gauge fixing.
See \cite{JP09, PSS09, PSS09b, PSS10, AObs} for more.

\mbox{ } 

Thus, for Geometrodynamics  
\beq
\{{\iD}, \scM_{\mu}\} = 0 \mbox{ } ,                     
\label{OHi0}
\eeq
\beq
\{{\iD}, \scH\} =  0 \mbox{ } ,                   
\label{OH0}  
\eeq
whilst only
\beq
\{{\iK}, \scM_{\mu}\} =  0 \mbox{ } 
\label{OHi02}  
\eeq
need hold.  
Justification of the name `constants of the motion' follows e.g. in the GR case from the Hamiltonian taking the form $H[\,\upalpha, \upbeta^{\mu}\,] 
:= \int_{\bupSigma} \d^3x\,(\upalpha\scH + \upbeta^{\mu}\scM_{\mu})$ so that (\ref{OHi0}, \ref{OH0}) imply that
\beq
\frac{\d{\iD}}{\d\mt}[\bh(\mt),\mbox{\boldmath $\uppi$}(\mt)] = \{\iD, H \} 0 \mbox{ } .
\eeq
Thus, observables are automatically constants of the motion with respect to evolution along the foliation associated with any choice of $\upalpha$ and $\upbeta^{\mu}$. 
This straightforwardly generalizes to any other theory (in particular for this Article encompassing RPM's also) via writing the theory's 
total Hamiltonian as $\fH\lfloor \Lambda^{\sfX} \rfloor = \int_{\sS}\d \mS \, \Lambda^{\sfX}\scC_{\sfX}$ for multiplier coordinates $\Lambda^{\sfX}$, in which case
\beq
\frac{\d{\iD}}{\d\ft}[\bfQ(\ft), \bfP(\ft)] = 0 \mbox{ } . 
\eeq
Of course, the properly relational procedure is to use a total OA-Hamiltonian built by OA-Dirac-appending linear constraints with cyclic ordials of auxiliaries, 
$\ordial\fA\lfloor \ordial{\fg}^{\sfZ}, \ordial{\fI} \rfloor = \int_{\sbSigma}\d \bupSigma \{ \ordial \fI \scQ\scU\scA\scD+ \ordial{\fg}^{\sfZ}\scL\scI\scN_{\sfZ}\}$, 
leading to the same conclusion.    

\mbox{ } 

\noindent{\bf The classical Problem of Beables} (more usually called the Problem of Observables \cite{Dirac, Kuchar92, I93, Rovellibook}) 
involves construction of a sufficient set of beables/observables for the physics of one's model, which are then involved in the model's notion of evolution.  

\mbox{ } 

\noindent Clearly, the more stringent one's notion of beables is, the more of a problem this is (Sec \ref{Beables}).  

\mbox{ } 

\noindent Analogy 60) The classical Problem of Beables occurs for RPM's too (see Sec \ref{Beables}).

\subsection{Three classically-overcome POT facets}\label{Overcome}

\noindent Analogies 61, 62, 63) {\bf Classically, the Constraint-Closure, Foliation-Dependence or Spacetime-Reconstruction Problems are} {\boldmath$resolved$} {\bf problems} 
both for Geometrodynamics and for RPM's.      
I note that these all reflect tensions between space-time split approaches and spacetime approaches., but these only become virulent at the quantum level.    

\mbox{ } 

\noindent Caveat: this excludes global sub-facets for the last two (see Sec \ref{Glob-Again}).

\subsubsection{RPM case as regards these Facets}
%
{            \begin{figure}[ht]\centering\includegraphics[width=1.0\textwidth]{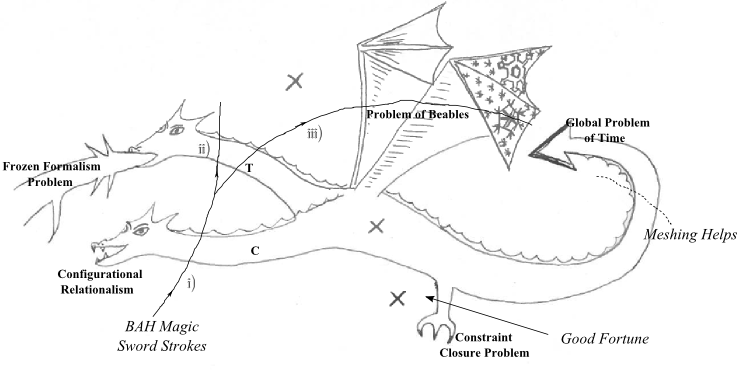}
\caption[Text der im Bilderverzeichnis auftaucht]{\footnotesize{The Classical RPM Ice Dragon is 5/8ths strength with no Inner Product Problem `teeth' behind its 
Frozen Formalism Problem `breath'. 
Weapons are kept visibly separate from facet names by use of slanty script.  
It is to be countered by 

\noindent A) good fortune causing its sole leg to stumble (no Constraint Closure Problem).

\noindent B) A sequence of strokes from the Barbour--A--Halliwell (BAH) `Magic Sword' (see Figs \ref{Magic-Sword}, \ref{Cards} for details).  
Namely, 

\noindent i)   Resolution of the Best Matching Problem form of Configurational Relationalism as posed by Barbour and solved for 1- and 2-$d$ RPM's by me.  

\noindent ii)  Subsequent construction of $t^{\se\sm(\sJ\sB\sB)}$ interpreted in the LMB-CA manner of a `GLET abstracted from a STLRC'. 

\noindent iii) Best Matching Problem resolution also implies the \K observables are known -- functionals of the shapes (and scale) and their conjugate momenta as mathematically 
realized and carefully conceptually interpreted by e.g. Kendall, Dragt and I (see Secs \ref{Gau} and \ref{Beables}).  
One can then obtain a subset of classical Dirac beables from these by Halliwell's procedure (Sec \ref{Cl-Combo-Intro}).
\noindent The immense width this stroke depicts a `flyby slash' from atop one's Winged Unicorn Steed', as per Sec \ref{Beables}.}  } 
\label{Classical-Dragon-RPM}\end{figure}          }

\noindent Classical constraint closure works out fine by  the  Poisson brackets algebra (\ref{first6}) for scaled RPM and (\ref{first6}, \ref{other4}) for pure-shape RPM.  
In the reduced case, this is via {\bf single constraints always closing} (by direct or reduced approach); the Master Constraint Program \cite{Thiemann}
is a distinct attempt at exploiting this universal constraint-closure avoiding strategy.

\mbox{ }  

\noindent Whereas Foliation Dependence and Spacetime Reconstruction are irrelevant to RPM's [Differences 26) and 27)] aspects of GR-as-Geometrodynamics not modelled by RPM's, 
foliation itself does retain some meaning, in that RPM's have a lapse/instant-like notion and a point-identification map \cite{Stewart} 
corresponding to such as the auxiliary rotations in moving along the dynamical curve in the redundant setting.
However, the GR lapse and shift (or instant and frame as per Sec \ref{Examples}) are {\sl more} than just na\"{\i}ve elapsation and point-identification `struts' 
of this nature in that they additionally pack together with the 3-metric configurations to form a unified spacetime 4-metric, which 
then turns out to have a number of additional features such as having the 3-metric embedded into it and being refoliation invariant. 
Thus non-existence of RWR result is cancelled out by not having a spacetime to reconstruct in RPM's.

\subsubsection{Dirac's Trident}\label{Trident}
%
{            \begin{figure}[ht]\centering\includegraphics[width=0.8\textwidth]{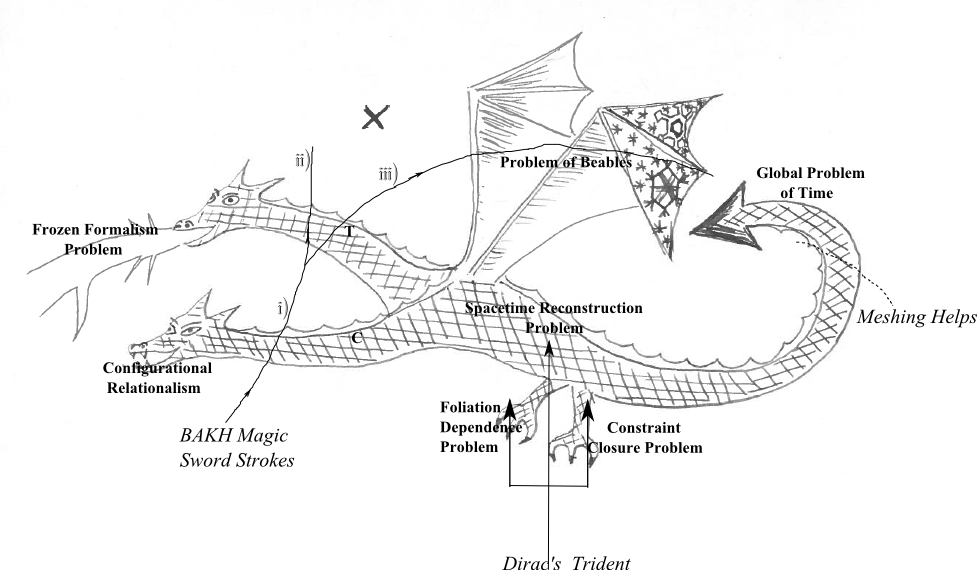}
\caption[Text der im Bilderverzeichnis auftaucht]{        \footnotesize{
The proposed strategy to face this Classical GR POT Ice Dragon is to use the same combination of BAH `Magic Sword' strokes as for the RPM, 
noting that in practise this will require furthering humanity's knowledge of the Thin Sandwich Problem, at least in some formulation and/or alternative theory of Gravity. 
This is picked out as Major Open Question II in the Conclusion.            
Moreover, on this occasion one is facing a 2-legged armoured beast, so one cannot rely upon it benignly stumbling. 
Fortunately, one can strike it with `Dirac's Trident'.}  } 
\label{Classical-Dragon-GR}\end{figure}          }

\noindent For Geometrodynamics itself, the Dirac algebroid indeed locally overcomes all three of these \cite{T73, HKT, RWR, Phan, Lan2, APOT2}. 

\noindent 1) From Dirac's own algebraic structure as formalized by (\ref{Mom-Mom}-\ref{Ham-Ham}), closure is clear.

\noindent  As an aside, that this closure is in terms of structure functions (i.e. as a Lie algebroid) is not a POT-specific feature since  
such algebroids arise in other contexts \cite{Strobl}.  

\noindent 2) In classical GR, one can foliate spacetime in many ways, each corresponding to a different choice of timefunction.  
This is how time in classical GR comes to be `many-fingered', with each finger `pointing orthogonally' to each possible foliation.  
Classical GR then has the remarkable property of being refoliation-invariant \cite{HKT}, so that going between two given spatial geometries 
by means of different foliations in between \cite{T73, HKT} produces the same region of spacetime and thus the same answers to whatever physical questions can be posed therein; 
this is the work of Teitelboim alluded to in Fig \ref{Teit} followed by that of HKT.
This holding does imply validity of the {\sl thick} sandwich (see e.g. p 85 of \cite{Kieferbook}) in the local situation in hand, which does limit the rigour or broadness of the result.  

\noindent 3) The `relativity without relativity' (RWR) approach \cite{RWR} then amounts to, along the relational approach to the POT, having a classical-level spacetime reconstruction.
Here, spacetime structure is not presupposed, but rather only Configurational and Temporal Relationalism, and then the Dirac procedure for constraint 
algebroid consistency  returns GR spacetime as one of very few consistent possibilities from the Dirac procedure, as laid out in Sec \ref{RWR}. 

\mbox{ } 

\noindent Note 1) The Dirac procedure advances via the $\scL\scI\scN$-as-$\scQ\scU\scA\scD$-integrability recovery of GR spacetime.

\noindent Note 2) GR is further picked out among the above-mentioned possibilities by the demand of foliation independence and the demand for a finite nonzero propagation speed.  

\noindent Note 3) RWR includes how, for {\sl simple} \cite{Phan, Lan2} matter, local SR is deduced rather than assumed \cite{RWR}.

\mbox{ } 

\noindent However, at the quantum level one no longer knows any means of guaranteeing these three properties, all of which are conceptually desirable 
(and the last two of which represent aspects of Background Independence).  

\mbox{ } 

\noindent {\bf Question 36$^{*}$}  The Supergravity Constraint Algebra indeed still closes; also see Note 3) of Sec \ref{RPM-Susy}; 
moreover, this algebra can now be constructed from $\scQ$ rather than $\scH$ being the sine qua non starting-point as was the case for GR.  
What becomes of the refoliation-invariance \cite{Teitelboim}, deformation-algebra \cite{HKT} 
and relativity without relativity \cite{RWR, Phan, Lan2} workings under the transition from GR to Supergravity?

\subsubsection{Nododynamical counterpart of the Trident}

\noindent Closure and foliation-independence follow through by the form of the Ashtekar algebroid \cite{Ashtekar, Thiemann} whose function and conceptual meaning 
is similar enough to that of the Dirac algebroid of geometrodynamics.  

\mbox{ } 

\noindent {\bf Question 37} Does the Ashtekar variables form of this manifest any {\sl further} lucid geometrical interpretation 
of the concepts of foliation and refoliation invariance? 

\mbox{ }

\noindent {\bf Question 38} Some aspects of the Spacetime Reconstruction would need to be redone in these new variables. 
The known results are 

\noindent i) that the supermetric coefficient $W$ is fixed to be 1 as one of various options (now not only from constraint propagation but also 
from demanding that the theory be based on a Yang--Mills field strength, giving yet another answer to Wheeler's question about the form of the geometrodynamical $\scH$).  

\noindent ii) in the strong gravity alternative, all values of $W$ work.

\noindent However, the Galilean alternative has not been investigated, nor has there yet been confirmation of a conformal alternative at the level of an RWR working.  
Does the conformal formulation of Ashtekar variables in the recent paper \cite{BST12} also arise as such a conformal alternative in the Ashtekar variables version 
of RWR? 

\noindent Nor has this scheme been investigated with simple matter, which is needed to confirm emergence of SR in the $W = 1$ option.

\subsubsection{Homogeneity resolution of the Trident}\label{Hom-Trident}

This case is very straightforward.
Homogeneity kills the momentum constraint, so there is only one constraint, which self-closes.
Homogeneity does provide a simple privileged foliation, by the spatially homogeneous slices.
Finally, homogeneity kills the Relativity Without Relativity obstruction term; that ensures that the single-constraint criterion for closure is not violated in 
the context of the larger ansatz.

\subsection{Global POT}\label{Glob-Again}

\noindent The {\bf Global POT} alias {\bf Kucha\v{r}'s Embarrassment of Poverty} \cite{Kuchar92} is classically present 
but can be taken to concern how time is but a coordinate in GR and coordinates are but in general locally valid (on charts). 
Thus a single time function is far from often definable over the whole of space, or globally defined in time itself 
(the function in question can `go bad' at some finite value including for no physically plausible reason)
Moreover, we know how to use multiple coordinate patches held together by the meshing conditions of classical differential geometry.
Whilst that is classically straightforward, 

\mbox{ }

\noindent N.B. 1) many of the facets and strategies also have further kinds of globality issues; see Sec \ref{GP5} at the classical level and Sec \ref{QM-Global} at the quantum level.

\noindent N.B. 2) It  will {\sl not} always be a trivial matter to upgrade local results about and solvings of classical {\sl p.d.e.'s} to global ones (see Sec \ref{GP5}).
For the difficulties beyond that from the further elements of essential structure present at the quantum level, see Secs \ref{QM-Global}, \ref{Global}.

\subsection{Difference 29) due to importance of details about diffeomorphisms}

In Isham's words, ``{\it The prime source of the POT in Quantum Gravity is the invariance 
of classical GR under the group $\mbox{Diff}(\FrM)$ of diffeomorphisms of the spacetime manifold $\FrM$}" \cite{I93}.
This is a central and physically sensible property of GR, and it binds together much of the POT `Ice Dragon'.   
It would amount to ignoring most of what has been learnt from GR at a foundational level to simply change to a theory which does 
not have these complications (such as perturbative string theory on a fixed background).  
Both the Nododynamics/LQG program \cite{Thiemann} and the more recent M-theory development of string theory \cite{BerPer} do (aim to) take such complications into account.  
The extent to which RPM's are valuable models is largely constrained by this Difference.
From (\ref{EnCo}), it follows that the Frozen Formalism Facet also occurs for RPM's. 
Nevertheless, RPM's manage to be background-independent within their own theoretical setting.  

\vspace{10in} 

\section{Strategies for the Problem of Time at the classical level}\label{Cl-POT-Strat}

So far, I have outlined a local classical $\ft^{\se\sm(\sJ\sB\sB)}$ resolution of the POT, which is examined in further detail in the next Sec.  
Whilst this resolution has the Thin Sandwich Problem caveat for GR, the main motivation for considering further strategies is that it is 
immediately apparent that $\ft^{\se\sm(\sJ\sB\sB)}$ does not unfreeze at the quantum level. 
This prompts one into looking into further POT strategies, and the present Sec considers the classical roots of these.    
As we shall see in Sec \ref{QM-POT-Strat}, this lack of unfreezing can be bypassed by considering a $\ft^{\se\sm(\sW\sK\sB)}$ that closely parallels 
$\ft^{\se\sm(\sJ\sB\sB)}$ as an implementation of Mach's Time Principle.  
However, this Semiclassical Approach is insufficient in detail; I argue it, Histories Theory and Timeless Records Theory interprotect each other. 
The other strategies considered in this section are present for completeness and to argue why they are not chosen as part of this program.
Thus, all-in-all, we begin by expanding on the classification of the strategies attempted to date.

\subsection{Classify!}

The Introduction outlines one classification of strategies for the Frozen Formalism Problem that is centred about the long-standing philosophical fork (see e.g. \cite{Denbigh})  
between `time is fundamental' and `time should be eliminated from one's conceptualization of the world'.  
Approaches of the second sort are to reduce questions about `being at a time' and `becoming' to, merely, questions of `being'.\footnote{Philosophers 
themselves have considered the further McTaggart series classification \cite{McTaggart}, consisting of the following layers of structure assumed.

\noindent A-series, for which time is dynamic - temporal passage is real: from past to present to future.  

\noindent B series, which involves less: notions of earlier, simultaneous and later (i.e. ordering notions).  

\noindent C-series (less well-known), which is an even more minimalistic timeless set-up.  

\noindent I comment here that McTaggart's B-series is actually a slightly broader criterion than ``being at a time", in that is
in that it explicitly allows for constructs like  ``being at a later time".  
On the other hand, as far as I can tell, dealing merely with questions of being coincides with McTaggart's C-series. 
}
%
A finer classification is Isham and \K's \cite{Kuchar92, I93} into Tempus Ante Quantum, Tempus Post Quantum and Tempus Nihil Est. 

\noindent Their Tempus Nihil Est includes Histories Theory, which I objected to in \cite{APOT} by creating a fourth category: Non Tempus Sed Historia.  

\noindent I then pointed out in \cite{APOT2} 1) that this then naturally splits into Historia Ante Quantum {\sc hq} and Historia Post Quantum {\sc qh}.

\noindent 2) That three more minimalist approaches are possible: {\sc t}, {\sc h} and nothing at all, corresponding to taking the attitude that GR is not after all to be quantized 
(most naturally via the argument that it is only an effective classical theory (much as one would usually not directly quantize the equations of classical fluid mechanics).  
I refer the reader interested in such approaches to Carlip's recent review \cite{Carlip12}.

\noindent I next expand this Nihil--Historia dilemma further, based on the various possible levels of structure of one's theory's primary objects.
One can have i) $\fQ$ alone, ii) $\fQ$ alongside $\ordial\fQ$, iii) finite paths $\upgamma = \fQ(\lambda)$ or iv) histories $\upeta = \fQ(\lambda)$.
%
{            \begin{figure}[ht]\centering\includegraphics[width=0.95\textwidth]{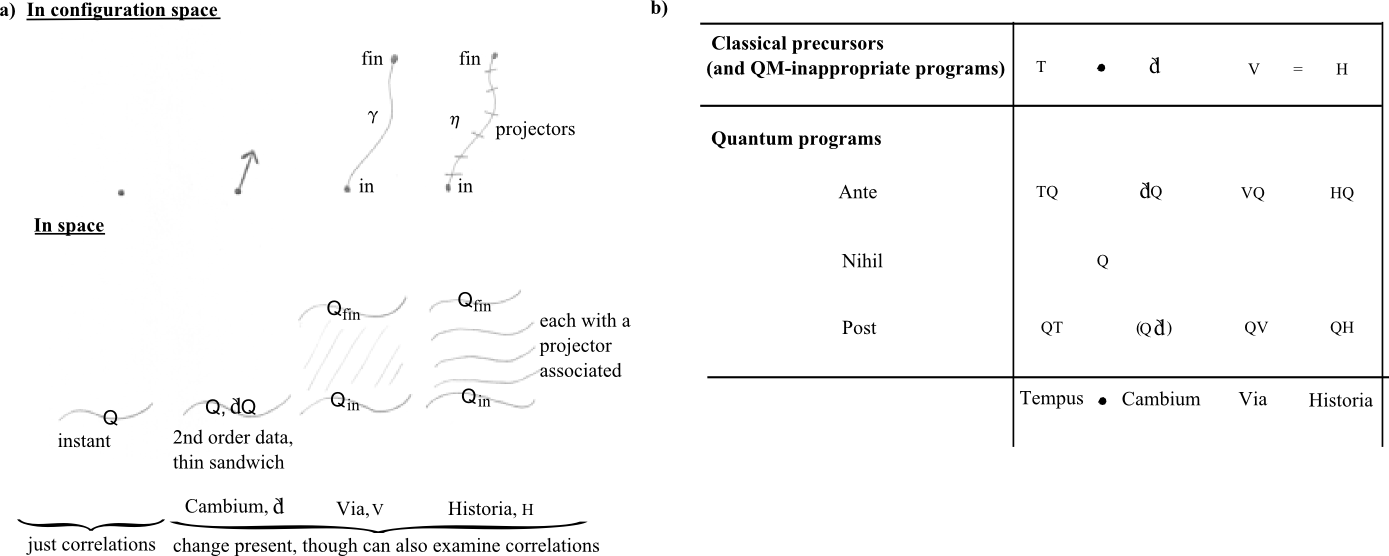}
\caption[Text der im Bilderverzeichnis auftaucht]{        \footnotesize{a) 4 alternative types of primary ontology to `time is primary'. 

\noindent b) 12 ordered combinations of these and quantization/quantum considerations. 
By aiming to involve a notion of time at all, {\sc tq} and {\sc qt} strategies favour ordinary QM, whereas the other approaches are more radical in this respect.
}  } \label{Classi-New}\end{figure}          }

\noindent i) is the most solipsist ontology: a single instant \cite{B94II, EOT} or localized subconfigurations within a single instant \cite{PW83, Page1, Page2}.  
It is `a point in configuration space'.  

\noindent A variant on this is to consider Barbour's ``{\it heap of instants}" \cite{EOT} (this heap being taken to be a set with no further mathematical-space structure).  

\noindent ii) considers ``{\it a point and a direction in configuration space}" (also a well-known saying of Barbour's \cite{EOT}).  
The novel feature from her upwards is that {\sl change} is a valid primary notion.
Thus from here upwards, Mach's Time Principle `Time is to be abstracted from change' can be meaningfully attempted to obtain a time as a secondary notion.
Thus I term these approaches Non Tempus Sed Cambium, and they are followed by Cambium Est Tempus (the former tempus, however, being primary, whilst the latter is secondary).
My shorthand symbol for this in map-orderings is $\sordial$.
For i) however, one can only resort to correlations in order to attempt to obtain a semblance of dynamics/change/path/history.  
My shorthand for this in map-orderings is simply the blank, since it ditches time without replacement at the primary level.  
\noindent i) and ii) are both usually regarded as timeless approaches, though quite clearly they are structurally and conceptually very different; worse, there is 
widespread confusion between the two in both the Physics and Philosophy communities, so in the present Article I make this distinction sharp.  

\noindent iii) The difference between this and ii) is one of infinitesimal versus finite versions. I.e. datum versus evolution or thin versus thick (c.f. the sandwiches).  
The name is then Non Tempus Sed Via.  

\noindent iv) is then Non Tempus sed Historia.  
Whilst some people use path and history interchangeably, I only use History for the histories-theoretic entity -- a path decorated with a string of projectors. 
Thus this is only a distinction at the quantum level.  

\noindent Nobody has to date given a convincing argument for primary change being a purely quantum phenomenon, so `{\sc q}$\sordial$' strategies are omitted by 
one being able to take these to have classical-$\sordial$ precursors.  
Thus, all in all, \K and Isham's trisection has been teased apart into a 12-fold classification as per Fig \ref{Classi-New}.  
%
{            \begin{figure}[ht]\centering\includegraphics[width=1.0\textwidth]{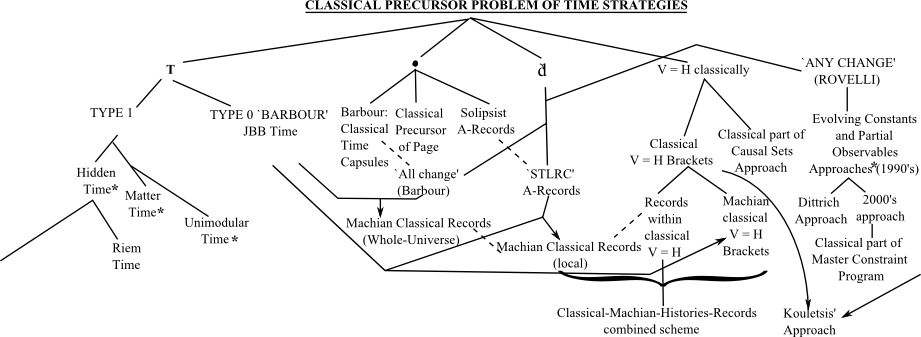}
\caption[Text der im Bilderverzeichnis auftaucht]{        \footnotesize{The various branches of strategy and their relations, in a diagram with `cylindrical topology'. 
* indicates the 4 out of the 10 strategies covered by Kucha\v{r} and Isham's reviews \cite{Kuchar92, I93} that start at the classical level.  
C.f. Fig \ref{Temporary-Web-QM} for the full 10 of these and the QM-level developments since.}  } \label{Temporary-Web-Cl}\end{figure}          }

\noindent Fig \ref{Temporary-Web-Cl} further refines this classification by identifying further varieties of program in various of the 12 categories.  
Whilst indeed over the years, many conceptual strategies have been put forward to resolve the POT,   
none of the ten families proposed to up to 1993 worked upon detailed examination \cite{Kuchar92, I93}. 
As to work since, not all of it has yet been investigated in such detail (\cite{Kuchar99, Kieferbook, Rovellibook, Smolin08, APOT, APOT2} and the present Article cover some more).   
Taking the ten families up to 1993 as precedent, I consider it likely (but not fully demonstrated) that subsequent strategies will also have great 
difficulty in being genuinely satisfactory.
A resolution for the POT being acceptable would entail it to work in the general case for a fully viable theory of gravitation rather than just working for special cases or toy models.  
I in no way claim that this Article is exhaustive as regards what problems various strategies have 
(e.g. \cite{Kuchar92, I93, Kuchar93, Hartle, Kuchar99, Smolin08, APOT, APOT2} between them cover plenty more such issues). 

\noindent Fig \ref{Temporary-Web-Cl} exhibits a trichotomy of Non Tempus Sed Cambium between Rovelli's `any' \cite{Rov91a, Rov91b, Rov91c, Rovelli02, Rov02b, Rovellibook, Dittrich, Rfqxi} 
tied to further developments in LQG,\footnote{Though many of the other types of POT strategy are also considered in this arena, see \cite{APOT} for a selection).}
Barbour's whole-universe `all' \cite{B94I, EOT, Bfqxi} and my local `STLRC' (\cite{Records, ARel2} and Secs \ref{Cl-POT} and \ref{+temJBB}).  
Solipsism itself divides into global-instant (Barbour \cite{B94II, EOT}) and local-instant (Page \cite{Page1, Page2} and some of my \cite{Records}, Secs \ref{Cl-Nihil}, \ref{QM-Nihil}); 
further and even more well-known divisions of this best await the quantum-level treatment of Fig \ref{Temporary-Web-QM}.

\noindent Fig \ref{Temporary-Web-Cl} also exhibits a dichotomy of Tempus Ante Quantum between (Type 0) Barbour -- which is not unfreezing at the quantum level but does re-emerge there instead 
--  and Type 1 due to many other earlier authors, e.g. York, \K, Isham, Brown, Unruh and Wald and subjected to not just technical critiques \cite{Kuchar81, Kuchar92, I93, Torre} 
(and Secs \ref{Cl-POT-Strat}, \ref{Cl-Ante}, \ref{QM-POT-Strat}, \ref{QM-Ante}) but also to relationalist critiques (\cite{APOT} and Secs \ref{Cl-POT-Strat}, \ref{QM-POT-Strat}).  

\noindent Both Tempus Ante Quantum and Tempus Post Quantum include as subcases notions of {\bf emergent time}, as opposed to having an external time to always be present 
(Newtonian or Minkowskian Physics), or for a hidden time to always be present, or for one's sense of time to be appended by inclusion of additional degrees of freedom 
(both of these last two occur in Type 1 Tempus Ante Quantum).

\noindent Diversification of Histories/Paths Approaches since \K and Isham's reviews is mostly quantum-mechanical in nature.  

\noindent The appearance of combinations of some of the types of strategy (see e.g. Secs \ref{R-within-CM}, \ref{Histor} \ref{Cl-Combo}, \ref{QM-Combo} and 
\cite{H03, H09, Kouletsis, Dittrich, GPP, GPPT, APOT}) is also noteworthy; the Figure contains various of these (Kouletsis' aside from the more obviously-named approaches).

\subsubsection{A note on universal strategies}\label{Univ}

Some POT strategies are {\it universal} in the sense that they exist regardless of what the underlying theory is,  at least for a large number of steps.  
By which the RPM and geometrodynamical coverage of the present Article would not substantially change for 
these strategies if one were to pass e.g. to Brans-Dicke, higher curvature theory, Supergravity, M-Theory or the affine-geometrodynamical or nododynamical formulations of quantum GR.  
Thus much of Parts II and IV of the present Article are in fact of much wider interest in classical gravitational theory and QG theory 
rather than just of interest for the arenas used as examples.  
Universality is true of emergent JBB and semiclassical times and of the timeless and histories approaches; 
another case of this is that approaches based on mastering the diffeomorphisms are widely blocked from progress.

\subsubsection{The timelessness-indefiniteness cancellation hypothesis}

Canonical GR has an indefinite kinetic term and time is missing. 
Perhaps then these two defects can cancel each other out.  
We know from the Shape-Definiteness Lemma 13 of Sec \ref{GRAnal} that it is the overall scale which causes indefiniteness. 
So might overall scale, or some close relation of it, be tied to the isolation of, or emergence of, time?  
This is a feature common to a number of the strategies.

\mbox{ }   

\noindent Difference 30) Quite clearly RPM's can not model all aspects of such a cancellation.

\subsubsection{Strategizing about each facet}

Different attitudes to the Problem of Beables have been discussed for quite some time \cite{Kuchar93}, but \cite{APOT2} suggested taking strategization of each facet further.
The basic idea there is in terms of ordering {\sc t}, {\sc q}, {\sc c} and {\sc b} moves (i.e. the order in which one is to face the 3 heads and the wings)
which are moreover very sensitive to ordering (see much of the rest of this Sec and of Sec \ref{QM-POT-Strat}. 
There is a strong tendency to spoil preceding moves as per Kucha\v{r}'s many-gateways intuition or, with the Ice Dragon, that the order in which one confronts the three heads 
and the wings sometimes involves merely temporarily repelling some of these which come back to menace the later stages of one's program.

\subsection{Tempus Ante Quantum ({\sc tq}) for models free of Configurational Relationalism}\label{TAQ}

{\bf Ante Postulate}.  There is a fundamental time to be found at the classical level for the full 
(i.e. untruncated) classical gravitational theory (possibly coupled to suitable matter fields).

\subsubsection{Type 0) Jacobi--Barbour--Bertotti implementation} 

$\ft^{\se\sm(\sJ\sB\sB)}$ is certainly obtained in accordance with the Ante Postulate. 
\noindent I discuss further features of this candidate timefunction in Sec \ref{+temJBB}, covering how to obtain it in detail from within a classical Machian scheme.
However, 

\mbox{ } 

\noindent Main Problem with JBB Time)  This does not in any way unfreeze the quantum GR WDE or its RPM analogue.  
Thus it is only a classical resolution of the POT.

\mbox{ }  

\noindent Moreover, it makes substantial contact with another strategy (the Semiclassical Approach) in which this 
same timestandard (in its approximate form) emerges in an an unfreezing context as the emergent semiclassical (alias WKB) time.  
The JBB approach is also a powerhouse as regards the development of further timeless approaches: such a time that is 
actually made purely out of configurational cambium is sometimes of interest as regards which timeless questions to ask of one's system (as per Sec \ref{Cl-Prop}).

\mbox{ } 

\noindent For the rest of this SSec, I describe, and provide a critique for the distinct Type 1) Tempus Ante strategies.
The conceptual distinction between Type 1 and Type 0 theories is outlined in SSSec \ref{Mach-Antes}.

\subsubsection{Riem time implementation}\label{HSTA}

Riem's DeWitt supermetric has indefinite signature --+++++. 
Perhaps one could then take the indefinite direction to be pick out a timefunction $t^{\sss\su\sp\se\sr}$, 
in parallel to how the indefinite direction in Minkowski spacetime picks out a background timefunction.   
Thus this approach is an obvious attempt at implementing the timelessness-indefiniteness cancellation hypothesis.  
The type of method by which $t^{\sss\su\sp\se\sr}$ would be hoped to produce a quantum Frozen Formalism Problem resolution is via this timefunction being of the more general form 
$\mt^{\sh\sy\sp\se\sr\sb}$, since use of it and its subsequent conjugate momentum $p_{\st^{\th\ty\tp\te\tr\tb}}$ casts $\scH$ into a manifestly naive-hyperbolic form,
\beq
p_{\st^{\th\ty\tp\te\tr\tb}}^2 = H_{\st\sr\su\se} \mbox{ for } \mbox{ }  H_{\st\sr\su\se} \mbox{ quadratic positive-definite in the remaining momenta }.  
\label{Hyperb}
\eeq  
Here, $H^{\st\sr\su\se}$ is then the `true Hamiltonian' for the system.
(\ref{Hyperb}) leads to a Klein--Gordon type TDWE depending on the double derivative with respect to this timefunction, 
\beq
\pa_{\st^{\th\ty\tp\te\tr\tb}}^2 \Psi = \widehat{H}_{\st\sr\su\se} \Psi \mbox{ } .  
\label{KG-type}
\eeq
Probably the most lucid summary of this approach is to note that
\beq
\triangle_{\mbox{\scriptsize\boldmath$M$}} 
\mbox{ }  \mbox{ } \mbox{ is actually a } \mbox{ }   \Box_{\mbox{\scriptsize\boldmath$M$}} \mbox{ } .  
\eeq

This approach was traditionally billed as Tempus Post Quantum, through choosing to make this identification of 
a time after the quantization; however, this approach is essentially unchanged if this identification is made priorly at the classical level, so I present it here.
Moreover, regardless of when this identification is made, however, this approach fails by \cite{Kuchar81, Kuchar91} the GR potential not being as complicit as Klein--Gordon theory's 
simple mass term was (by not respecting the GR configuration space's conformal Killing vector); see Sec \ref{Semicl} for further details.  
The above presentation has been kept free of Configurational Relationalism by being specifically for diagonal minisuperspace, 
which has a 3 $\times$ 3 \mbox{ } --++ metric on it; the 2 $\times$ 2 positive-definite block $M^+$ has its inverse $N_+$ feature 
in $H_{\st\sr\su\se} = N_+^{\sfA\sfB}(t^{\sss\su\sp\se\sr}, Q^{\sfC})P_{\sfA}P_{\sfB} + V(t^{\sss\su\sp\se\sr}, Q^{\sfC})$ .
This approach is more often called `superspace time', though that only makes sense when Riem($\bupSigma$) = Superspace($\bupSigma$) by both being equal to Minisuperspace($\bupSigma$).  
The $Q$'s here are of course this case's shape degrees of freedom: the anisotropy parameters $\beta_{\pm}$

\mbox{ }  

\noindent Difference 31) There is no RPM counterpart of this as a consequence of kinetic term positive-definiteness.

\subsubsection{Implementation by obtaining a parabolic/part-linear form}\label{PLinF}

We now consider starting one's scheme off by finding a way of solving $\scQ\scU\scA\scD$ to obtain a parabolic/part-linear form,  
\beq
\fP_{\sft^{\tp\ta\tr\ta\tb}} + \fH^{\st\sr\su\se}\lfloor \ft^{\sa\sn\st\se}, \fQ_{\so\st\sh\se\sr}^{\sfA}, \fP^{\so\st\sh\se\sr}_{\sfA}\rfloor  = 0 
\mbox{ } .
\label{Para1}
\eeq  
Here, $\mP_{\sft^{\tp\ta\tr\ta\tb}}$ is the momentum conjugate to a candidate classical time variable, $\ft^{\sp\sa\sr\sa\sb}$ 
which is to play a role that closely parallels that of the external classical time of Newtonian mechanics.  
Then, given (\ref{Para1}) such a parabolic form for $\scH$, it becomes possible to apply a conceptually-standard quantization that yields a TDSE (\ref{TEEDEE1}). 
This form occurs in parametrized nonrelativistic particle toy models \cite{Kuchar92}.  
GR examples of part-linear forms are given in various SSSecs below.

\subsubsection{A first Machian comparison between Type 0 and Type 1 Tempus Ante Quantum approaches}\label{Mach-Antes}

Type 1 Tempus Ante can be viewed as the most conservative family of Frozen Formalism Problem-resolving strategies \cite{I93, T97} 
in the sense that it seeks as soon as possible (at the classical level) among the apparent variables of one's theory for 
a substitute for absolute Newtonian time (or likewise-absolute SR time) as per Sec \ref{Examples}.
This is then to play as many as possible of the roles that absolute time played in absolutist classical theory and ordinary quantum 
theory, thus rendering all of these roles straightforwardly conceptually resolved.

This is to be contrasted with how, whilst Tempus Ante Type 0 `Barbour' also obtains its $\ft^{\se\sm(\sJ\sB\sB)}$ as soon as possible, 
it however stands for supplanting absolute time itself for a relational concept which itself {\sl lives on} in an 
unadulterated form in more advanced background-independent theories such as GR.

In seeking among the apparent variables, Type 1 Tempus Ante involves selecting particular variables to be those that provide `the' time.
In contrast, Type 0 `Barbour' Tempus Ante gives all changes the opportunity to contribute to the most dynamically significant notion of time, which is a furtherly Machian feature.
See also Fig \ref{Ante-1-and-2}. 
%
{            \begin{figure}[ht]\centering\includegraphics[width=0.97\textwidth]{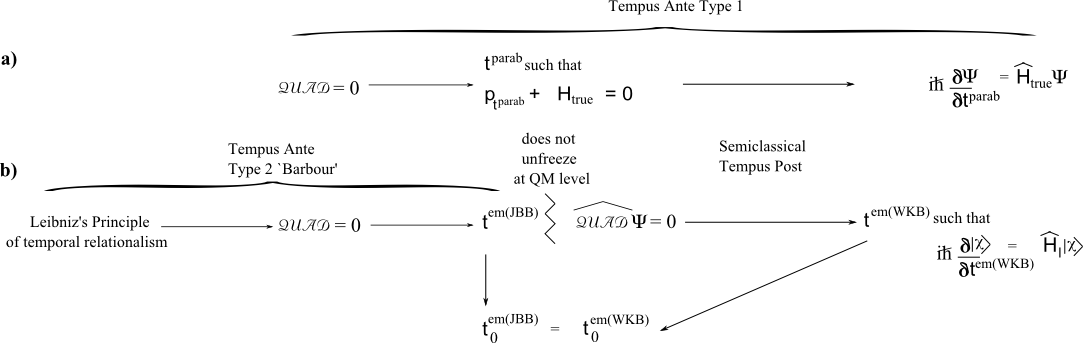}
\caption[Text der im Bilderverzeichnis auftaucht]{        \footnotesize{a)  Schematic form of Type 1 Tempus Ante Quantum approach, in which what is isolated at the 
classical level prevails at the quantum level (I in fact also allow for hyperbolic forms giving Klein--Gordon-type equations rather than a TDSE too, but that does not 
affect the overall shape of the scheme).

b) Barbour's Type 0 Tempus Ante Quantum approach, which rests upon deeper classical foundations, and to which  
all changes have an opportunity to contribute to the emergence of the time (a highly Machian position). 
However, it does not unfreeze the Quantum Physics, for which a second (likewise highly Machian) emergence is required (we need to face off the dragon's 
{\sc t}-head twice, either side of facing the {\sc q}-head.  
Moreover, this semiclassical emergent WKB time is as much of an approximate recovery of the classical emergent JBB time as is possible.
[for, as we shall see in Sec \ref{Semicl}, the unapproximated forms differ because the former's abstraction has additional 
{\sl specifically semiclassical changes} contributing to it.]
Envisaging Type 0 `Barbour' Tempus Ante Quantum and Semiclassical Tempus Post Quantum as a closely-related pair of procedures 
is one of the principal and most fertile insights of the present Article.}  } \label{Ante-1-and-2}\end{figure}          }

\subsubsection{The hidden time Internal Schr\"odinger Approach}\label{ISI}

A first suggested implementation of parabolic Type 1 Tempus Ante Quantum is that there may be a {\it hidden} alias 
{\it internal} time \cite{York72, Kuchar81, Kuchar92, I93, GT11} within one's gravitational theory itself.  
I.e. the apparent frozenness is a formalism-dependent statement that is to be removed via applying some canonical transformation.
This sends GR's (for now, minisuperspace's) spatial 3-geometry configurations to 1 hidden time, so $t^{\sp\sa\sr\sa\sb} = t^{\sh\si\sd\sd\se\sn}$ 
plus 2 `true gravitational degrees of freedom' [which are the form the `other variables' take, and are here `physical' alias `non-gauge'].
The general canonical transformation is here
\beq
(Q^{\sfA}, P_{\sfA}) \longrightarrow 
(t^{\sh\si\sd\sd\se\sn},  p_{t^{\th\ti\td\td\te\tn}}, Q_{\st\sr\su\se}^{{\sfA}}, P^{\st\sr\su\se}_{{\sfA}}) \mbox{ } .
\eeq
Thus one arrives at a hidden-TDSE of form (\ref{TEEDEE1}) for $t^{\sh\si\sd\sd\se\sn} = t^{\sp\sa\sr\sa\sb}$.
Within Geometrodynamics, hidden-time candidates that have been considered are as follows.  


\noindent 1) {\it Intrinsic time candidates}, though these have largely lacked in successful development \cite{Kuchar92}). 

\noindent One family of examples involve using some scale variable from Sec \ref{PS} as a time.
[E.g. Misner time in Quantum Cosmology, or the scalefactor $a$ itself, or the local spatial volume element $\sqrt{h}$ that goes as $a^3$ in the (approximately) isotropic case; 
RPM counterparts are then based on the configuration space radius instead, with Analogies 30, 36, 48 exploitable.] 
These constitute a wider range of obvious attempts at implementing the timelessness-indefiniteness cancellation hypothesis 
[much like in the above-mentioned Riem time, which is indeed also a scale time albeit one that is being used for a hyperbolic rather than parabolic rearrangement].
{\sl The indefinite part of the GR supermetric is intimately related to notions of scale, so that timelessness--indefiniteness cancellation 
is expected to come about from using some kind of time that is closely related to scale, and thus some kind of physical degrees of freedom that are closely related to pure shape}    

\mbox{ }
 
\noindent 2) {\it Extrinsic time candidates}. 
One among these is probably our best candidate for an internal time in GR: {\it York time} \cite{York72, Kuchar81, Kuchar92, I93}, so I present this example in more detail.   
[This is proportional to the constant mean curvature (CMC) of Sec \ref{GR-Gdyn}.]
As per Sec \ref{SS-Split}, perform a size-shape split of GR and then exchange the scale variable $\sqrt{\mh}$ and its 
conjugate relative distance momentum object via a canonical transformation so that the latter is the timestandard,  
\beq
\mt^{\sY\so\sr\sk} := 2\mh_{\mu\nu}\pi^{\mu\nu}/3\sqrt{\mh} 
\mbox{ } \mbox{ } \mbox{ } \mbox{ (proportional to the constant mean curvature)} \mbox{ } .   
\eeq
Then the Hamiltonian constraint is replaced by $\mP_{\sY\so\sr\sk} = H^{\st\sr\su\se} = \sqrt{\mh} = \upphi^6$, where the conformal scale factor $\upphi$ is here being interpreted 
as the solution of the conformally-transformed Hamiltonian constraint, i.e. the Lichnerowicz--York equation (\ref{LYE}).  

\mbox{ } 

\noindent As already covered in the Introduction, scaled RPM's have {\it Euler time} as a close analogue of GR's York Time.  

\mbox{ } 

\noindent Problem with York Time 1).  By Note 1) of Sec \ref{CGdyn} , this is in practise almost never explicitly known, which will hamper many a would-be subsequent working, though 
Note 2) of that SSSec makes clear that minisuperspace does furnish some cases for which this is solvable, allowing for some limited further exploration of such subsequent workings 
\cite{I93}.  

\mbox{ }

\noindent Then quantization gives, formally the York-TDSE (\ref{YorkTDSE}). 
The true degrees of freedom are here again anisotropy shapes, the precise geometry of the space of which depends on the Bianchi type of the homogeneous solution.
The `again' comes from how the canonical transformation has switched the scale and its conjugate around whilst {\sl leaving alone} the pure-shape part of the formulation.  
Thus Dilational times such as York time are also viewable as modified attempts at implementing the timelessness-indefiniteness cancellation hypothesis.  

\mbox{ } 

\noindent {\it Einstein--Rosen time} is another extrinsic time candidate that arises in cylindrical wave midisuperspace models \cite{midi}.  

\mbox{ }

\noindent Problem with Hidden Times 1) Having a {\sl list} of candidates, the Multiple Choice Problem may subsequently strike.  

\mbox{ } 

\noindent As regards RPM toy models of the hidden time approach, I note for now that time candidate status 
can be ascribed to these's scales and dilational objects conjugate to scales (such as $\scD$, which in this context I term the {\it Euler dilational time candidate}). 
I leave the details of this analogy to Sec \ref{Cl-Ante} since they are both quite lengthy and new to the present Article.

\subsubsection{Implementation by unhidden time}\label{BoHo}

There is a higher-derivative theory in which a natural variables set contains an already-explicit internal time \cite{Boulware, Horowitz}.

\subsubsection{Straightforward Matter Time}\label{MattS} 

Typically in minisuperspace examples work that include 1-component scalar matter, one simply isolates the corresponding 
momentum to play the role of the time part of the subsequent TDWE.  
This often taken to be `the' alternative to scale time (e.g. this is often so in the LQC literature, and also in the recent paper \cite{Lawrie}). 
However, the momenta conjugate to each of these represent two {\sl further} possibilities, and then, if if canonical transformations 
are allowed, a whole further host of possibilities become apparent.  

\mbox{ }  

\noindent Whilst using a matter scalar field as a time is often argued to be `relational' \cite{Bojowald05}, 
this is clearly only the AMR `any change' form of the Machian recovery of time, so more scrutiny might be considered. 

\noindent Problem with Straightforward Matter Time 1) Few such models have been checked for their candidate matter time to indeed possess the features expected of a time.   

\noindent Problem with Straightforward Matter Time 2) In the extension to multi-component matter, why should one matter species be given the privilege of constituting the time? 

\noindent Problem  with Straightforward Matter Time 3) Given this awareness of multiplicity (all of which are {\sl equally} relational 
in the above AMR sense) -- will the Multiple Choice Problem appear?

\subsubsection{Implementation by reference-fluid matter} \label{Matt}

One can look to attain the part-linear form (\ref{Para1}) is attained with $\mt^{\sa\sn\st\se} = \mt^{\sm\sa\st\st\se\sr}$ and $\mQ^{\Gamma}_{\so\st\sh\se\sr} =$ $\mh_{\mu\nu}$ 
by extending the geometrodynamical set of variables to include matter variables coupled to these, which then serve so as to label spacetime events \cite{Kuchar92}.
These are sometimes taken to be additional matter fields, including undetectable ones, though sometimes they 
are suggested to be the same as other theoretically-used or practically necessary matter fields.  

\mbox{ }

\noindent The idea of appending matter in order to have a time runs further contrary to relationalism than hidden time due to its externalness which  parallels absolute time. 
This is yet furtherly so in cases in which this matter is undetectable/has no further physical function.  

\mbox{ } 

\noindent As criticisms of Type 1) Tempus Ante Quantum in general, this is not in line with Leibniz timelessness (since the time in use is taken to a priori exist in general at the 
classical level: Tempus Ante Quantum \cite{I93}), nor with the LMB-type take on Mach's time principle, since it involves using one particular change as the time for all the other changes.  
Moreover, matter time schemes themselves are typically additionally suspect along the lines of intangibility, 
via often involving such as `reference fluids', and often endowed with unphysical properties to boot. 
[The intangibility might be taken as cover for the unphysicality, but is itself a conceptually suspect way of handling the POT along the lines exposited in the present Article.]

\subsection{Outline of Classical Solipsist position}\label{TNE}

\noindent Adopting a Tempus Nihil Est approach  saves one from the thorny issue of trying to define time as outlined in Secs \ref{tfns} and \ref{Clocks}.    
However, this is supplanted by three other brier patches.  
One is classical and thus presented here; see Sec \ref{QM-POT-Strat} for the other two.

In practise, this is a case of choosing to face either Saint Augustin's `what is time' or the more recent alternative of `how to cope in Physics without there being a time'.  
Canonical quantum gravity hints at the latter, though there are of course ways around it.
I note that the latter often leads one to a nonstandard interpretation of QM, whereas the former allows for far more conventional  interpretations of QM, at least in principle. 
This is because by possessing a time, which is then to be treated in a sui generis fashion, the former complies with the structural expectations of ordinary QM.

As regards matters of principle, this is an issue of economy (why use a time if it is not necessary) and of primality: 
the Machian `time is an abstraction from change' versus the absolutist `time is a prerequisite for understanding change'.  
I argue that the former is more palatable, particularly as one generalizes one's theory and passes to whole-universe models 
(for which, incidentally, the standard interpretation of Quantum Theory is moot on grounds other than the Frozen Formalism Problem).

\mbox{ }

\noindent {\bf How to explain the semblance of dynamics if the universe is timeless as a whole?}
Dynamics or history are now to be {\sl apparent notions} to be constructed from the instant \cite{GMH, H99, HD, H03} by somehow implementing the following postulate.    

\mbox{ }

\noindent{\bf Solipsism Postulate}. 
One aims to supplant `becoming' with `being' at the primary level (see \cite{Page1, Page2, B94II, EOT, GMH, H99, HT, HD, H03, Records, H09} 
and also e.g. Reichenbach \cite{Reichenbach} for partial antecedents).   
[In this sense, these Timeless Approaches consider the instant/space as primary and change/dynamics/history/spacetime as secondary.] 

\mbox{ } 

\noindent As sample questions, what is Prob(contents homogeneous in scaled 4-stop metroland)? 

\noindent What is Prob(anisotropy is small, $|\beta_{\pm}| < \epsilon$)?  

\mbox{ }

\noindent It is logically possible to think just of questions of being at the classical level, though it is the quantum form that predominates the literature (\NSII, see Sec \ref{NSI}).  


\noindent We next turn to the manoeuvres Sec \ref{Cl-Str} alluded to, by which being at a time and becoming can be supplanted so that timelessness 
ends up being a paradigm for the whole of Physics.

\subsubsection{Supplanting being-at-a-time questions}\label{SupBeAtAT}

`Being at a time' and `becoming' types of questions do not say which time is involved.  
Ordinary Classical Physics has an easy way out: there is an external time belonging to the real numbers, so that each 
configuration space $\FrQ$ is augmented to an extended configuration space $\FrQ \times {\FrT}$.
One key lesson from GR, however, is that there is no such external time.    
Questions along the lines of those above which involve time need here a specification of {\sl which} time.  
Using `just any' time comes with the Multiple Choice and Functional Evolution \cite{Kuchar92, I93} facets of the POT.  
Another way of latching onto some aspects of the above key lesson, which moreover can already be modelled at the level of nonrelativistic 
but temporally-relational mechanical models, is that `being, at a time $\ft_0$' is {\sl by itself} meaningless if one's theory is MRI in the time label.  
Alternatives that render particular times, whether uniquely or in families up to frame embedding variables, 
meaningful are specific internal, emergent or apparent time approaches.
In this scheme, time is but a property that can be read off the (decorated sub)configuration.  
E.g. York's internal time \cite{York72, Kuchar81, Kuchar92, I93} can be thought of in this way, as can the next SSSec's notion of clock time. 
Thus, all question types involving a $t$  are turned into the corresponding question types without one.\foo{It is not clear which as in this setting one can have in principle 
different configurations take the same time value (e.g. through lying on different paths of motion).}  

Perhaps this property concerns a particular subconfiguration lying entirely within the state space in question (`a clock within the subsystem').     
Or perhaps it concerns a subconfiguration lying entirely outside the subsystem under study 
%
%
(`background clock').
Though a clock subsystem could be part-interior and part-exterior to the subsystem under study.     
Indeed, one might consider a universe-time to which all parts of the configuration contribute rather than a clock {\sl sub}system.  
[None of these uses of `clock' necessarily carry any `good clock' connotations. 
In each case that is to be furthermore determined.]  

\mbox{ } 

\noindent Some RPM examples of relevance here are

\noindent 1) contents inhomogeneity.   
For $N = 4$ in H-coordinates, the cross-bar serves as an indicator of relative size of the universe as compared to its subsystems (the posts). 

\noindent 2) For triangleland, relative angles are involved also, e.g. almost-collinear and almost-equilateral notions. 

\noindent As regards Minisuperspace examples `becomes isotropic, flat, large' still make sense. 

\mbox{ }

\noindent One approach here involves {\sl conditioning} on a subconfiguration that plays the role of a clock; 
this involves considering somewhat more complicated propositions than statements of a single subconfiguration being.

\subsubsection{Supplanting becoming questions}\label{SupBec}

\noindent This has been suggested by Page (e.g \cite{PAOT}) and also to some extent by Barbour \cite{EOT}.   
Page's scheme for this is as follows \cite{PAOT, Page1, Page2}.    
It is not the past instant that is involved, but rather this appearing as a memory/subrecord in the present instant, alongside the subsystem itself.   
Thus this is in fact a correlation within the single present instant.  
In this scheme, one does not have a sequence of events.
Rather, there is one present event that contains memories or other evidence of `other events'.
One might view such configurations as e.g. researchers with data sets who remember how they set up the experiment that the data came from (controlled initial conditions, and so on).
See Sec \ref{Cl-Nihil-Intro} for reasons why this scheme might not be adopted or complete.

\subsubsection{Classical Timeless Records Theory}\label{Cl-Nihil-Intro}

Here one one seeks to construct a semblance of dynamics or history from various levels of structure allowed of instants.
The theoreticians have differed somewhat both in how to make the notion of record more precise, and in how they envisage the semblance of dynamics may come about.    
Thus there are in fact a number of Records approaches.

\mbox{ } 

\noindent I. {\bf Page Records} The central idea here is the content of SSSec \ref{SupBec} about supplanting becoming.  

\mbox{ } 

\noindent Page Problem 1) Unfortunately, this is very speculative as regards doing concrete calculations.  
Studying a subsystem S now involves studying a larger subsystem containing multiple imprints of S.
Models involving memories would be particularly difficult to handle. 
It would be expected to be very hard even to toy-model, one would need a working Information Gathering and Utilizing System (IGUS) model \cite{IGUS}.

\noindent Page Problem 2) If one wants a scheme that can additionally explain the Arrow of Time, then Page's scheme looks to be unsatisfactory.  
Single instants could be used to simulate the scientific process as regards `becoming questions'.  
However, N.B. that these single instants correspond to the {\sl latest} stage of the 
investigation (in the `becoming' interpretation), while `earlier instants' will not have this complete information.  
Additionally, important aspects of the scientific enterprise look to be incompletely realized in this approach. 

I.e. as well as the `last instant' playing an important role in the interpretation, initial conditions 
implicit in the `first instant' also look to play a role (see also \cite{Hartle, H99}). 
This could be envisaged as a variation on the theme of Bishop Berkeley's notion of `time as succession of ideas in our minds' 
to `time as an abstraction from the memories in our minds {\sl now}'.
 
\mbox{ } 
 
\noindent Note 1) Page's scheme is minimalist in using subconfigurations of a single instant.  
On the other hand, it need not be minimalist in the distinct sense of being used to investigate multiverse issues.  
I thank Prof. Don Page for this point.  

\noindent Note 2) There may be difficulty with simultaneously accommodating becoming-reduced-to-being and being-at-a-time reduced to conditioned being, 
since the latter would appear to involve multiple instants labelled by distinct clock states being conditioned upon.
 
\mbox{ } 

\noindent II. {\bf Bell--Barbour Records} \cite{Bell,B94II,EOT}.  
Barbour always considers whole-universe configurations.
In one of his approaches \cite{B94II, EOT}, he draws upon the parallel with how $\alpha$-particle tracks form in a bubble chamber to form a 
``{\bf time capsules}" paradigm for Records Theory (this is a quantum-level paradigm, so see Sec \ref{QM-POT-Strat} for more).  
\noindent Perhaps Quantum Cosmology can be studied analogously \cite{B94II, EOT, H03, H09} [though see Sec \cite{Records}) for a strong reason not to bracket Halliwell's papers 
to Barbour's at this point].

\mbox{ } 

\noindent Barbour conjectures a selection principle for records that encode a semblance of dynamics based on the following layers \cite{EOT}.   

\mbox{ } 

\noindent Barbour Records 1) There are some distinctive places in the relational configuration space (possibly linked to its asymmetries or to the most uniform state).  

\noindent Barbour Records 2) The wavefunction of the universe peaks around these places (`mist concentrating'), by which these are probable configurations.   

\noindent Barbour Records 3) These parts of the configuration space contain records (which he terms `time capsules') from which a semblance of dynamics can be extracted. 

\mbox{ } 

\noindent None of these are part of Page's scheme, which also focuses on suppressing becoming rather than upon obtaining a semblance of dynamics from a presumed timeless world.  

\mbox{ }

\noindent XII. One can build a records theory that is localized (like Page's and unlike Barbour's) and which has matching structural levels to those developed in Histories Theory 
(like Gell-Mann--Hartle's but unlike Page's or Barbour's). 
Within the solipsist scheme, this has the following axioms.  
Deficiencies with it are pointed out in the subsequent Non Tempus Sed Cambium Sec.  

\mbox{ }  

\noindent {\bf Records Postulate 1}. Records are information-containing subconfigurations of a single instant that are localized in space (as prepared for in Sec \ref{Imp}).
This is partly so that they are controllable, and partly so that one can have more than one such to compare.  
It also negates signal times within conventional frameworks in which such are relevant. 
However, this is far from necessarily the basis for a criticism; e.g. pp. 225-226 of \cite{Whitrow} points out that relativistic theories make use of a similar notion of locality.
By this `local in space' criterion, Barbour's insistence on whole-universe configurations lies outside of the scheme and thus should be dropped.  

\mbox{ } 

\noindent {\bf Records Postulate 2}. Records are furthermore required to contain useful information. 
I take this to mean information along the lines of single-instant correlations.  
Secondly, however, one would wish for such correlation information to form a basis for a semblance of dynamics or history.  
Information Theory, the study of complexity and of pattern recognition are thus relevant precursors to Records Theory; Secs (\ref{Cl-NOI}--\ref{Cl-Prop}) have left us prepared for this.  

\mbox{ } 

\noindent {\bf Records Postulate 3} Records can be tied to atemporal propositions, which, amount to a suitable logic.
This is classically straightforward by the simpleness of the type of logic involved as per Sec \ref{Cl-Prop}. 
However, at the quantum level, neither the logic (Sec \ref{QM-Prop}) nor the implementation of propositions (Sec \ref{QM-Nihil}).

\mbox{ }

\noindent I define the combined study of the structural levels of Records 1) to 3) to be {\bf Solipsist Pre-Records Theory}.  
In a nutshell, this concerns what questions can be asked about the presence of patterns in subconfigurations of a single instant.

\mbox{ } 

\noindent Note 1) Two similar-size samples of the same kind of sand could be \cite{Reichenbach} a hoof-print and a random pattern due to the wind blowing.
What one requires is a general quantification of a configuration containing a pattern.  
There is at least a partial link between this and information content, in that complicated patterns require a minimum amount of information in order to be realized.   
Records theory is then, intuitively, about drawing conclusions from viewing mutually similar (or elsewise correlated) patterns in different records.    

\noindent Can one tell a game's rules from a series of snapshots of the game's positions? 
Does the theory of complexity help with this?  

\noindent Note 2) The information content versus observable information point of Sec \ref{Cl-NOI} then gives a choice as to which notion is to be used in Records Theory.
Records are in practise often of the latter sort, since one can have under one's nose far more pictures of spectra than 
stars, and likewise many more accounts purporting to be from different times rather than the one instantaneous state of the object itself in that present.
The price to pay is that photos/spectra of stars may well not allow for one to answer all questions about stars themselves.   

\mbox{ }

\noindent Analogy 64) RPM's are useful in the records study of notion of locality in space (leading to notions of inhomogeneity and structure) 

\noindent Analogy 65) This is for Records Theory at the level of notions of 
information/negentropy, including subsystem and mutual notions of such, though these remain very much work in progress  (Sec \ref{Cl-NOI}).  
\noindent By QM solvability allowing one to build up a SM and thus notions of entropy, 
and negentropy is a reasonable characterization of information, RPM's have tractable notions of information, subsystem information, mutual information etc. 

\noindent Analogy 66) RPM's are useful in the records study of notions of information, correlation and patterning . 

\mbox{ }

\noindent In order for this to additionally be a minimalistic {\bf Records Theory}, one has to be able to extract from a semblance of dynamics or history 
from these correlations between same-instant subconfiguration records.  
This is most challenging in the solipsist context. 

\mbox{ }

\noindent {\bf Machian Records Postulate 4} \cite{ARel2} Semblance of dynamics and of time to be abstracted from records are to be of the STLRC and GLET form of Secs \ref{Cl-POT} 
and \ref{+temJBB}. 
This might be by inspection of the emergent notion of change. 
Alternatively, it could be by one's timeless scheme being a computation of timeless correlations within a scheme for which the emergent time.
The latter makes sense for A-Records and Halliwell's scheme [hence XXI) and XXVII) in the Figure] but not for Page's scheme, which provides an alternative semblance, 
nor for Barbour's scheme if one adheres to time capsules semblance or considers it incongruous to study whole-universe instants and then to apply STLRC.  

\mbox{ }

\noindent {\bf Machian Records Postulate 4B} is as above but with `all change' instead.  
This is now more natural for Barbour's scheme (minus time capsules semblance: XIX) as well as still natural for Halliwell's scheme XXIX; it is now less natural for A-Records.  
Moreover, the present article pushes for Postulate 4 and not 4B 
(for all that this analysis has unearthed multiple further programs both as per Fig \ref{Cl-Nihil-Class} and a wider range still at the quantum level as per Fig \ref{Tless-Types}).  
Finally Postulate 4 (and 4B for that matter) are further motivated by their possessing semiclassical counterparts, as covered in Sec \ref{SM-TR}.

\mbox{ } 

\noindent Thus by now we have the following four classification schemes. 

\mbox{ } 

\noindent Timeless Classification 1) By question type: 1-property (NSI) versus multi-property questions (\CPI, Records Theory, Rovelli's Partial Observables Approach).   

\noindent Timeless Classification 2) By any/all/an optimal \cite{APOT2} (rows in Fig \ref{Tless-Types}).  
Thus we see that Rovelli's approach, the CPI, Page's approach and GPP's approach have an AMR attitude to time, 
whilst Barbour Records has LMB time and A-Records has LMB-A time; Halliwell records have LMB-A or LMB time natural to them.
Though one also has to distinguish between `any change' and `any configuration' etc, according to whether one's scheme has a primary notion of change; 
only those versions involving change are a Machian consideration.  

\noindent Timeless Classification 3) by type of semblance (or, in the next SSec, emergence).  

\noindent Timeless Classification 4) by what (if any) additional strategies it is combined with.  

\mbox{ }

\noindent N.B. it remains to be seen whether RPM's are useful as regards demonstrating that the minimalistic Records Theory can by itself produce a semblance of dynamics or history.
E.g. one can investigate the sequence of steps of Barbour's conjecture using RPM's (see Secs \ref{Cl-Nihil} and \ref{QM-Nihil} for more). 
%
{            \begin{figure}[ht]
\centering
\includegraphics[width=0.8\textwidth]{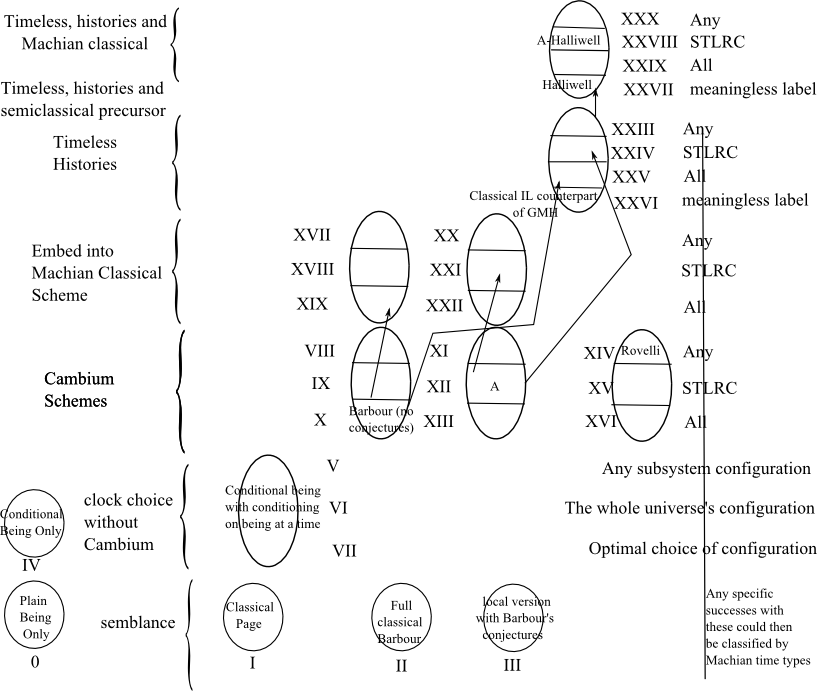}
\caption[Text der im Bilderverzeichnis auftaucht]{        \footnotesize{
Classification of 34 classical timeless strategies, including some new suggestions opened up by the current article's conceptualization and combinations with other schemes.   
Being outside of the box is purely timeless, whilst being inside it involves support by selecting some kind of time.
Dotted edges indicates incompleteness as regards types of question. 
Machian schemes come in three flavours: `any change', `all change' and STLRC as per Sec \ref{Cl-POT} 
Rovelli's scheme XIV and variants are commented on in Secs \ref{Cl-Pers}, \ref{QM-SubS} and \ref{Rov-vs-Other}.
Extendibility to different kinds of Machianity from each original formulation's, and other combineabilities e.g. with Histories Theory.
The labelled boxes have Machianity matching particularly appropriately with their other features.  
The arrows indicate the build-up of combinations of approaches.}        }
\label{Cl-Nihil-Class}\end{figure}            }

\subsection{Non Tempus Sed Cambium}

A missing element stems from not knowing one's configuration beyond a certain level of accuracy. 
An associated notion is coarse graining. 
To some extent one can envisage this within Barbour's heap of instants.

\mbox{ } 

\noindent However, more accurate study involves the space of instants (configuration space or subconfiguration space) having further mathematical structure on it rather than just being a set. 
This kind of setting does well to have the following.

\mbox{ } 

\noindent {\bf Records Postulate 1B}. Records are information-containing subconfigurations of a single instant that are localized in configuration space (as prepared for in 
Sec \ref{Loc-SubCon}).

\mbox{ } 

\noindent A number of such measures then entail a notion of change or even of finite path, taking one outside of the confines of solipsism.  

\noindent Aside 3) Can total randomness and strategic input be discerned between? (E.g. in inferring the rules of Snakes and Ladders versus those of Chess, 
in each case from large amount of data on game positions).  
The key to move forward is significant patterns, as in statistically significant.
For constellations of points in 1- and 2-$d$, corresponding to, among other things, RPM configurations, e.g. 
Roach \cite{Roach} and Kendall \cite{Kendall}'s work is useful in this respect; see Sec \ref{Cl-Nihil} for more.

\mbox{ }

\noindent {\bf A Cambium Pre-Record} obeys postulates 1, 1B, 2 and 3.  

\mbox{ } 

\noindent Once change is primary, abstracting time from it becomes a viable competitor to the semblance of dynamics from pure solipsism.  

\mbox{ }

\noindent One can also {\sl assume} histories theory and then study Pre-Records Theory within that less minimalist scheme now without need for an extraction of a 
semblance of history (IL reformulation of Histories Theory's classical precursor or the classical part of Halliwell's scheme).
This takes one further still away from solipsism, into combined strategies (Sec \ref{TR-PH}).  

\mbox{ }

\noindent Analogy 67) RPM's are useful in the records study of locality in configuration space (Sec \ref{Dist}), and subsequent notions of states that are only approximately known. 
In particular, the kinetic term positive-definiteness lending itself to the construction of

\noindent notions of locality in configuration space.  

\mbox{ } 

\noindent Note: Whilst  attempting to `justify the WKB ansatz' at the semiclassical level motivates involvement of Histories Theory as a third strand, 
the Classical Machian scheme also has the additional simplicity compared to its semiclassical counterpart that there is no counterpart of `justifying the WKB ansatz',  
so that, purely classically, the motivation to combine Machian-classical timeless schemes with histories schemes is rather lacking.

\subsection{Classical Machian--Records Scheme}\label{CM-TR}

Here there is no need for semblance, since an underlying classical Machian scheme is assumed.  
The timeless records theory within this scheme need be no more than a means of handling questions of being.

\subsection{Classical `Paths = Histories' Theory}\label{Histor}

\noindent {\bf Paths Postulate} Perhaps instead it is paths in configuration space that are primary.  
Here $\gamma :=$ $Q^{\sfA}(\ft)$ are taken to be labelled over the entirety of an extensive interval 
(the previous SSec's entities having been taken piecemeal, or infinitesimally so that $\d Q^{\sfA}$ is defined.
The field-theoretic counterpart is straightforward enough, with the emergent-time version carrying a local/multi-fingered label.

\noindent Then Path is the space of paths.

\mbox{ }  

\noindent Note 1) There is the option of the label either being believed to take discrete values or taking a discrete sample of what are believed to be continuous values.

\noindent Note 2) Various additional levels of structure may be postulated, both at the classical level and at the quantum-mechanical level. 
Beyond some point, the primary entity in this construct should be called a history. 
[{\sl Path} integrals are sometimes already called sum over {\sl histories} approaches, but in the present article we adopt a more structured notion of history.] 

\noindent Note 3) Primality of histories in the sense used in this paper is a perspective brought to the GR context by Gell--Mann and, especially, Hartle \cite{GMH, Hartle} and 
subsequently worked on by Isham, Linden, Savvidou and others \cite{IL2, IL, Savvidou02, Savvidou04a, Savvidou04b, AS05, Kouletsis}.  

\noindent Note 4) This article's stance is that path primality is a statement of what the basic objects in one's Principles of Dynamics are to be.  
Thus what are habitually called `histories momentum', `histories bracket' and `histories phase space' should be called {\it path momentum}, {\it paths bracket} and 
{\it path phase space} respectively.  
As regards other structures from `Histories Theory', coarse-graining and finest graining are clearly already defined for paths. 
As is, at the quantum level, the basic notion of decoherence functional.  
After all, Feynman already considered objects of this general nature within his development of {\sl path} integral methods.  
Thus
\beq
\mbox{classical histories} = \mbox{classical paths} \mbox{ } ,
\eeq
and consequently the space of classical histories, Hist, is just the same as Path (and that holds for all other spaces with `Path' as part of their name).  

\noindent Note 5) It is the association to each path of a string of projection operators that is the additional structure distinguishing Histories Theory from path integral approaches. 
This is a quantum-level distinction, and furthermore distinguishes between the Gell-Mann--Hartle work on the one hand and the IL work on the other; 
for, as we shall see in Sec \ref{QM-POT-Strat} each of these uses different kinds of strings of projectors.
For histories, I supplant the symbol $\gamma$ by $\eta$.  
Thus, at the quantum level, one has the additional distinction between the Path Postulate and the following

\mbox{ }  

\noindent {\bf Histories Postulate}. Treat histories rather than paths or configurations as one's primary dynamical entities.

\mbox{ }

\noindent N.B. Not only the Histories Postulate but also the Path Postulate supplant the $\FrQ$-primality of the Primality Postulate [now to be renamed {\bf Primality Alternative i)}, 
with the Path and History Primality Alternatives being {\bf ii)} and {\bf iii)}].
A case can be made that this, like spacetime primality, goes against the tradition of basing physics on dynamics, though on this occasion more similarities with dynamics are preserved, 
as is clear from the paths momenta, paths brackets and paths phase space' names.
One issue is whether paths or histories are as operationally meaningful (i.e. entities that one can directly measure) as primary quantities as configurations are.
Within a paths or histories picture, measurements at one instant of the history constitute records, and the ontology allows for measurements at distinct instants of the history, so 
there is at least a partial sense in which the answer is affirmative: at least {\sl constituent parts}.   

\mbox{ }

\noindent I term $\fQ^{\sfA}(\ft)$ for $\ft = \lambda$ label-time paths.  

\noindent I denote the path momenta conjugate to each of these by $P_{\sfA}(\ft)$ .

\noindent Paths brackets are then the corresponding Poisson brackets
\beq
\{\fQ^{\sfA}(\ft),\fP_{\sfB}(\ft^{\prime})\} = \updelta^{\sfA}\mbox{}_{\sfB}\delta(\ft, \ft^{\prime})\mbox{ } .
\eeq
(such approaches are then often referred to as `histories brackets' approaches) and iii) a histories quadratic constraint,  
\beq
\scQ\scU\scA\scD \lfloor \fQ^{\sfA}(\ft), \fP_{\sfA}(\ft)\rfloor = 0 \mbox{ } .
\eeq
\noindent For the discrete-time case, one has $\ft$, $\ft^{\prime}$ $\longrightarrow$ $\ft_1$, $\ft_2$ and Dirac $\updelta$ $\longrightarrow$ Kronecker $\delta$.

\mbox{ }

\noindent One then has (Path, PathPoint), (PathPhase, PathCan) and (PathRigPhase, PathPoint) defined in the obvious way.  

\mbox{ } 

\noindent One also has {\it path constraints}, for now the path energy constraint, 
\beq
{\scQ\scU\scA\scD}_{\sft}\lfloor\fQ^{\sfA}(\ft), \fP_{\sfA}(\ft)\rfloor = 0
\eeq
(with labels interpreted as continuous intervals or discrete strings).  
Of course, one next has a notion of paths constraint algebra/oid.  

\mbox{ } 

\noindent By dealing with paths at the classical level, we are assuming Via ante Quantum, the classical half of which is of course also Historia ante Quantum ({\sc hq}). 
However, both the quantum half and the post counterpart distinguish between Via and Historia, so all of {\sc hq}, {\sc vq}, {\sc qh} and {\sc qv} are represented as distinct 
options at the quantum level.  

\noindent Moreover, detailed study of QFT makes {\sl joint} use of both canonical and path-integral methods, along the lines of physics-rigourously deriving Feynman rules canonically 
and then manipulating complex calculations using path-integral methods.
This suggests that canonical-or-paths is not a strict alternative, canonical-and-paths being a third alternative. 
By that one might be cautious about ascribing primary ontology exclusively to only one of configurations, paths or histories.   
{\bf A case in which configurations and one of paths or histories have dual primality is also plausible} [{\bf Primality Alternatives iv) and v)}].  
Examples of this include the present Article's program and Halliwell's antecedent of it \cite{H03}, the older idea within more GR-specific work in which concurrently 
canonical--and--generally-covariant classical and quantum schemes are sought, by e.g. Kouletsis, \K and Savvidou \cite{KK02, Savvidou04a, Savvidou04b, Kouletsis08}.  
The latter carry connotations of spacetime-and-space as a third alternative to spacetime or space as the primary ontology of the world.

\mbox{ }
   
\noindent{\bf Question 39}$^{**}$ If one uses an `and' Alternative for one's primary ontology and has a relational mindset, 
one might well wish to use {\sl unsplit spacetime relationalism} as well as split space-time relationalism.
This is an inherently GR/'Diff-type theory' development, since in mechanics one only has space-time rather than spacetime, so it lies outside the scope of the current Article.  

\noindent 1) Develop the unsplit spacetime counterpart of the current paper's relational program. 

\noindent 2) Figure out how to conceptually-coherently have concurrent spacetime--and--space-time relationalism.
Diff({\FrM}) itself is straightforward to handle, but the relation between it and its split with respect to each and every foliation is a more complex issue. 
Nor are all of the current Article's relational developments included in this more obvious and partly already attempted purely group-theoretic issue.
%

\mbox{ } 

\noindent Sub-paths can be pieces of a path or the path traced in a subconfiguration space.
Persp(Path) = $\langle$ the set of physically meaningful SubPath's $\rangle$.   

\noindent Notions of coarse- and fine-graining of histories are then clearly supported on this set of structures.
I use $C_{\bar{\gamma}}$ to denote coarse-graining of paths where $\bar{\gamma}$ is a path consisting of a subsequence of the path $\gamma$'s instants.
Thus, as well as the coarse-graining criteria in Sec \ref{log}, path formulations possess  {\it  coarse graining by probing at less times}.  

\mbox{ } 
 
\noindent Much as time has a list (conflicted) of desirable properties, so do paths/histories and to some lesser extent, instants.
The latter has present, instant, being, being at some time and simultaneity notions, whilst the former additionally has some form of 
time-ordering and causation notions as well, albeit phrased in a paradigm in which the notion of history replaces the notion of time.

\subsubsection{Further versions of Classical `Paths = Histories' Theory}

\noindent {\bf 2 time notions}: Savvidou pointed out \cite{SavThes, AS05} that IL type Histories Theory have a distinct structure 
for each of two conceptually distinct notions of time: 

\noindent I) a kinematical notion of time that labels the paths/histories as sequences of events (the 
`labelling parameter of temporal logic' , taken by \cite{AS05} to mean causal ordering, though see also \cite{IL2}. 

\noindent II) A dynamical notion of time that is generated by the Hamiltonian.  

\noindent Savvidou has argued that having these two distinct notions of time allows for such a Histories Theory to be 
canonical and covariant at once, which is of obvious interest in understanding, and reconciling various viewpoints in, QG.  

\mbox{ } 

\noindent{\bf Kouletsis and Kucha\v{r}} \cite{KK02, Kouletsis08}'s classical work on GR, and on the bosonic string 
as a toy model, involves a {\it space map} as well as a {\it time map} for how the family of geometries along each path/history embed into spacetime. 
This work also makes substantial contact with the Internal Time Approach and with the Problem of Beables.  
Here one goes about quantization  (only carried out to date for simpler models: ordinary mechanics \cite{IL, ILSS}, relativistic mechanics \cite{Savvidou02}, 
minisuperspace \cite{AS05} and quantum field theories \cite{Anastopoulos}) via replacing the canonical group by the histories group.   
Taking time to be continuous in such approaches means that one obtains a 1-$d$ QFT in time even for finite toy model examples.  

\mbox{ }

\noindent {\bf Marolf's alternative} \cite{Marolf} involves a distinct way of obtaining paths brackets: here the Hamiltonian is as an 
extra structure by which the Poisson bracket is extended from being a Lie bracket on Phase to a Lie bracket on PathPhase.  
This is in contrast with the more usual approach in which one puts the equal-time formalism aside at this stage and 
introduces a new phase space in which the Poisson bracket is defined over the space of paths from the beginning.

\subsection{Records within classical `Paths = Histories' Theory}\label{TR-PH}

\noindent XXVI and XXX) {\bf Gell-Mann--Hartle--Halliwell Records} \cite{GMH} found and studied records contained within Histories Theory 
(see the next SSec for details of this combined Histories--Records Approach).
At the classical level, on could consider the classical part of IL's \cite{IL} reformulation of Histories Theory.
Halliwell further studied Records within Histories Theory \cite{H99, HT, H03, H09}, though the last three of these references represent a different approach as regards the timeless 
quantities computed.
[At the classical level, these use Scheme XXIV) in Fig \ref{Cl-Nihil-Class}, i.e. one needs the IL histories scheme's classical precursor to make sense of this at the 
classical level.]

\subsubsection{Classical Machian `Paths = Histories' Combined Approach}\label{CM-H}

Repeat the opening of this subsection with $\ft = \lambda \longrightarrow \ft^{\se\sm(\sJ\sB\sB)}$.  
This has somewhat less purpose than involving  $\ft^{\se\sm(\sJ\sB\sB)}$ in timeless correlation schemes, though it is an interesting choice of labeller of 
histories, and, of course, makes more sense once one considers Machian classical-Records-Histories schemes and Machian-Semiclassical-Records-Histories schemes.  
It is the stance taken henceforth in dealing with histories theories in this Article.

\noindent The point of this is that, whilst classically one can do either label or emergent versions without any supporting relation linking the two, 
semiclassically the label version is to provide the WKB regime that spawns the emergent time with respect to which the histories are then to run.

\mbox{ } 

\noindent Note 1) in the inhomogeneous GR case, this requires labelling one's histories with a {\sl many-fingered} emergent time.  

\noindent Note 2) One can try to use $\mt^{\sY\so\sr\sk}$ (or its analogue $t^{\sE\su\sll\se\sr}$) as label within a histories scheme.  
The argument against that involves how best to implement Mach's Time Principle rather than ease of combineability.

\subsection{Configurational Relationalism generalization of the Best Matching Problem}\label{CR}

Passing from conceiving in terms of Best Matching to conceiving in terms of Configurational Relationalism emphasises that there is no strong 
reason why this facet should be tackled at the level of the classical Lagrangian formalism.  
Nontrivial Configurational Relationalism produces linear constraints, which need to be solved at {\sl some} level; up until that level, 
one may well need to consider other indirectly-formulated $\FrG$-invariant objects (notions of distance, of information... as per Sec \ref{Cl-Str}). 
%
%
Thus one needs to order not just {\sc t} and {\sc q} tuples, but also a {\sc r} = ({\sc r}, nothing) tuple.  
This generalizes `reduced quantization' and `Dirac quantization', which come from a context in which time-choosing was not necessary and simply mean 
{\sc r} ... {\sc q} and {\sc q} ... {\sc r}.
Moreover, this is but one of several further procedural ordering ambiguities.\footnote{See the next Sec for another, whilst {\sc q} itself has many layers  
of internal structure as per Sec \ref{QM-Intro}.}
%
Recall that {\sc t} involves {\sl finding} time whilst $\sordial$ involves there {\sl merely being} change.  
Thus {\sc tr} and {\sc rt} are different procedures, but whether a formulation has change in it or not 
is independent of whether one has reduced out the theory's unphysical degrees of freedom.
Thus the separate programs at the classical level are {\sc tr}, {\sc rt}, {\sc r} in the absence of $\sordial$, 
{\sc r} in the presence of $\sordial$, {\sc hr} = {\sc vr} and {\sc rh} = {\sc vh}.  
For further strategic diversity, one can also consider `$\chi$ or plain' versions of {\sc t} and {\sc h} 
($\chi$ meaning time-and-frame constructions rather than the plain case's single-time constructions). 
However, the {\sc r} ... $\chi$ ordering makes no sense and so is to be discarded. 
On the other hand, {\sc tr} = {\sc rt} since each of these procedures acts on separate parts of one's relational theory.

\subsubsection{Jacobi--Barbour--Bertotti time case}

The emergent JBB time now takes the Configurational Relationalism entwined form 
\beq
t^{\se\sm(\sJ\sB\sB)} = 
\stackrel{\mbox{\scriptsize extremum $\sfg \in \sFrG$}}
         {\mbox{\scriptsize of RPM action}} 
         \int ||\d_{\sfg}\bfQ||_{\sbm}/\sqrt{E - V}  \mbox{ } \mbox{ or } \mbox{ }   
 \mt^{\se\sm(\sJ\sB\sB)} = 
\stackrel{\mbox{\scriptsize extremum $g \in$ Diff($\Sigma$)}}
         {\mbox{\scriptsize of GR relational action}}
         \int_{\bupSigma}\d^3x \sqrt{\mh}\int||\d_{\sg}\bh||_{\sbM}/
		 \sqrt{\mbox{Ric(\ux; \bh]} - 2\Lambda} 
\mbox{  } .
\eeq
for $\FrG$ = Rot($d$) for ERPM or Pl($d$) for SRPM.  

\noindent Thus one has to solve Best Matching first and then one can form the explicit expression for $t^{\se\sm(\sJ\sB\sB)}$. 
This can of course be done explicitly for 1- and 2-$d$ RPM's by Sec \ref{Q-Geom}, whilst giving the Thin Sandwich Problem for Geometrodynamics.

\subsubsection{Thin Sandwich/Best Matching in further arenas}\label{+Sand}

\noindent The strength of this classical POT resolution motivates asking more widely about Thin Sandwich/Best Matching Problems.

\mbox{ } 

\noindent{\bf Major Open Question I}: resolve the Thin Sandwich Problem for at least some observationally viable theories of the background-independent/gravitational gestalt.

\mbox{ } 

\noindent Firstly, Giulini showed that coupling fundamental matter affects the mathematical properties of the Thin Sandwich \cite{TSCG}.

\mbox{ }

\noindent {\bf Question 40} The Einstein--Yang--Mills and Einstein--Dirac thin sandwiches have not been studied to date.  

\mbox{ }

\noindent Secondly, there are two distinct conformal thin sandwiches.
The first is due to York 1999 \cite{CTS}, and the second is based on \cite{ABFKO}. 
The reason they are distinct is because \cite{ABFKO} recovers the York 1972 conformal IVP approach \cite{York72}, whilst York 1999 is a distinct conformal IVP formulation. 
To cast conformal thin sandwiches in useable form for JBB time based approaches, we need them to follow from action principles.
This is ready for \cite{ABFKO}, but to the best of my knowledge no action principle is known from which York 1998 can be derived.

\mbox{ } 

\noindent {\bf Question 41$^*$} Find an action principle for the York 1999 conformal IVP and recast it in relational form.
Account for the differences in relational formulation between this and the ABFKO action.  

\mbox{ } 

\noindent {\bf Question 42$^{*}$} How does each each of the two conformal thin sandwiches fare as a p.d.e. system as compared to the geometrodynamical one?

\mbox{ } 

\noindent [In this case, there is some known work in the case of the York 1999 approach \cite{PY05, W07}.] 

\mbox{ }  

\noindent Thirdly, Komar  \cite{Komar} (see also \cite{FodorTh}) showed that canonical transformations can change the nature of the Thin Sandwich.  

\mbox{ } 

\noindent {\bf Question 43} It is opportune at this point to pose the Nododynamical counterpart of the Thin Sandwich  
in the hope that it turns out to be better-behaved (and allowing for subsequent technical improvements due to the loop and knot variables).
Is it? 

\mbox{ } 

\noindent [The Ashtekar Variables Thin Sandwich Problem has not to my knowledge been developed prior to this v3 version.   
For now, a heuristic outline of the outcome of reduction for Ashtekar variables involves taking $\FrG$ out is passing to loops/holonomies 
and then taking Diff($\bupSigma$) out too involves passing to consideration of knots.]

\subsubsection{Hidden York time cases}\label{Longi}

The general canonical transformation is now 
\beq
(\mh_{\mu\nu}(\ux), \uppi^{\mu\nu}(\ux))  \longrightarrow ({\cal X}^{\mu}(\ux), \Pi_{\mu}(\ux), \mQ_{\mbox{\scriptsize true}}^{\sfA}(\ux), \mP^{\mbox{\scriptsize true}}_{\sfA}(\ux) )
\mbox{ } , 
\label{canex} 
\eeq
for $\mQ_{\mbox{\scriptsize true}}^{\sfA}(\ux)$ the true gravitational degrees of freedom. 

\mbox{ }

\noindent In such an approach, one might well seek to solve $\scL\scI\scN_{\sfZ}$ elsewise than at the Lagrangian level for the auxiliary frame variables;  
e.g. in Conformogeometrodynamics, one's technical preference is to solve $\scM_{\mu}$ for the longitudinal potential $\upzeta^{\mu}$ part of $\mK^{\mu\nu}$ (or of $\uppi^{\mu\nu}$).  
This is defined by the splits $\mK^{\mu\nu} = \mh^{\mu\nu}\mK/3 + \mK_{\sT}^{\mu\nu}$ with $\mK_{\sT} = 0$ (traceless), 
and then $\mK_{\sT}^{\mu\nu} = \mK_{\sT\sT}^{\mu\nu} + \{\mbox{\scriptsize$\mathbb{L}$}\upzeta\}^{\mu\nu}$ such that $\mD_{\mu}\mK_{\sT\sT}^{\mu\nu} = 0$ (transverse) and 
$\mbox{\scriptsize$\mathbb{L}$}$ is the conformal Killing form = trace-removed Killing form $(\{\mbox{\scriptsize$\mathbb{L}$}\upzeta\}^{\mu\nu} = 
\{\mbox{\scriptsize$\mathbb{K}$}\upzeta\}^{\mu\nu} - \mh^{\mu\nu}\{\mbox{\scriptsize$\mathbb{K}$}\upzeta\}/3$.  
This gives e.g. the alternative formal {\sc ctq} scheme in which a single scalar $\mt$ is found by solving the Lichnerowicz--York 
equation with the value of $\upzeta^{\mu}$ found from solving the GR momentum constraint priorly substituted into it.  
Then, formally, 
\beq
\mP_{\mt^{\tY\to\tr\tk}} = \phi = \phi(\widetilde{x}; \mt^{\sY\so\sr\sk}, \mbox{\boldmath${\cal T}$\hspace{-0.04in}rue$^{(3)}$}, 
                                                                          \mbox{\boldmath$\Pi$}^{{\cal T}\hspace{-0.03in}\sr\su\se^{(3)}}] = - \widetilde{\mH_{\st\sr\su\se}}
\mbox{ } , 
\label{Red-York}
\eeq
the $\mbox{\boldmath${\cal T}\hspace{-0.04in}$rue}^{(3)}$ arising from the $\mbox{\boldmath${\cal C}$}^{(3)}$ by the standard interpretation that solving for $\phi$ breaks the conformal symmetry, 
sending one from CS($\bupSigma$) 2/3 of Superspace($\bupSigma$) to the distinct 2/3, True($\bupSigma$).  
$\widetilde{x}$ expresses local dependence in the $\FrG$ reduced configuration space itself.
Then standard quantization formally yields the reduced York TDSE (\ref{Red-York-TDSE}).

\mbox{ } 

\noindent A second possible scheme involves a particular case of the spacetime-vector parabolic form
\beq 
\fP^{{\cal X}}_{\Gamma} + H^{\st\sr\su\se}_{\Gamma}\lfloor {\cal X}^{\Lambda}, \fQ_{\so\st\sh\se\sr}^{\sfA}, \fP^{\so\st\sh\se\sr}_{\sfA}\rfloor = 0 \mbox{ } .
\label{Para4}
\eeq
-- a time-and-frame generalization of the single-time parabolic form (\ref{Para1}).
Here, $\fP^{\cal X}_{\mu}$ are the momenta conjugate to 4 candidate embedding variables ${\cal X}^{\mu}$, 
which form the 4-vector $[\ft^{\st\sr\su\se}, \mX^{\mu}]$ $\mX^{\mu}$ are 3 spatial frame variables. 
$H^{\st\sr\su\se}_{\mu} = [H^{\st\sr\su\se}, \Pi^{\st\sr\su\se}_{\mu}]$, where $\Pi_{\mu}^{\st\sr\su\se}$ is the true momentum flux constraint.
Given such a parabolic form for $\scH$, it is again possible to apply a conceptually-standard quantization, 
this time yielding a time-and-frame dependent Schr\"{o}dinger equation, (\ref{TEEDEE4}).  
This is a $\chi_{\tT}${\sc qr} scheme.  

\noindent Then in the York time case, this involves solving for both $\upphi$ and $\upzeta^{\mu}$ at the classical level, but choosing/being forced to keep the four equations 
rather than substituting the vector solution into the scalar solution as above.
I.e. a system of form
\beq
\mP_{\st^{\tY\to\tr\tk}} = \phi = \phi(\ux; \mt^{\sY\so\sr\sk}, X^{\mu}, \mbox{\boldmath${\cal T}\hspace{-0.04in}$rue$^{(3)}$}, 
                                                                         \mbox{\boldmath$\Pi$}^{{\cal T}\hspace{-0.03in}\sr\su\se^{(3)}}, \upzeta^{\mu}]
                         = - \mH_{\st\sr\su\se} \mbox{ } , 
\eeq
\beq
\mP_{\mu}^{\chi^{\tY\to\tr\tk}} = \upzeta_{\mu} 
                                = \upzeta_{\mu}(\ux; \mt^{\sY\so\sr\sk}, X^{\mu}, \mbox{\boldmath${\cal T}$\hspace{-0.04in}rue$^{(3)}$}, 
								                                                  \mbox{\boldmath$\Pi$}^{{\cal T}\hspace{-0.03in}\sr\su\se^{(3)}}]
                                = - \Pi_{\st\sr\su\se} \mbox{ } .   
\eeq
This passes to the quantum equations (\ref{Dir-York-TDSE}, \ref{Dir-York-FDSE}).

\mbox{ } 

\noindent Note 1) York Time Approaches are in practice hampered by the Lichnerowicz--York equation not being explicitly solvable and by the Torre Impasse (see Sec \ref{GlobStrat}) and by 

\noindent Note 2) This time-and-frame part-linear form occurs for parametrized field theory toy models (see \cite{Kuchar92} and references therein); 
these are the traditional toy models for this approach.

\noindent Note 3) (\ref{Red-York}) has an Mechanics analogue involving the post-reduction version of Euler time (Sec \ref{Cl-Ante}; 
this is simpler than the GR situation because the absolute--relative split is merely algebraic to the longitudinal--transverse split of GR involving differential operators 
[clearly rooted in the principal finite--field-theoretic Difference 4)].
This case does not however model the GR case's symmetry-breaking aspect; see Sec \ref{Cl-Ante} for details.

\noindent Note 4) RPM's also have an analogue of (\ref{Dir-York-TDSE}).  
However, in 1- and 2-$d$, classical reduction is not blocked, so I do not pursue this further.

\noindent Problem with Hidden Time 2) The canonical transformation in question is hard to perform in practise. 
In particular, and as a second example of entwining of linear constraints \cite{Kuchar92}, looking in more detail, the Internal Time approach's evolutionary canonical transformation's 
generating function needs to be a function of the initial and final slices' metrics in the classical configuration representation, implying a need for the {\sl prior} resolution of 
the Thin Sandwich Problem.  
This observation favours the first of the above two approaches over the second one, it not then being clear, at least to me, 
whether one can make do with the more convenient solution of $\scM_{\mu}$ in terms of $\upzeta$ instead of for $\upbeta$.  
If not, one is restricted to the harder system of equations composed of the sandwich equation and the Lichnerowicz--York equation.  

\noindent Note 5) Internal time is not universal, though scale and dilational time are universal within theories possessing scale.
One insight afforded is that a theory possessing scale (and in GR the associated indefiniteness) is not necessarily conceptually poorly.

\noindent Note 6) Work with shape--scale split formulations in the case of Ashtekar variables formulations has mostly only just started \cite{BST12, RW12}.

\subsubsection{Riem time case}

The full GR case would also require the functional- rather than partial-derivative versions throughout.  
E.g. the functional-TDSE based on the functional derivative with respect to the local t($\ux$) 
alongside the functional Laplacian, or the functional Klein--Gordon equation based on the functional wave operator.

\subsubsection{Reference matter time cases}

Here, the part-linear form (\ref{Para1}) is attained with $\mt^{\sa\sn\st\se} = \mt^{\sm\sa\st\st\se\sr}$ and $\mQ^{\Gamma}_{\so\st\sh\se\sr} =$ $\mh_{\mu\nu}$, 
or (rather more commonly and less formally) the 4-component version (\ref{Para4}), 
i.e.  extending the set of variables from the geometrodynamical ones to include also matter variables coupled to these, which then serve so as to label spacetime events.  
Then one passes to the corresponding form of TDSE, (\ref{TEEDEE1}) or (\ref{TEEDEE4}).  
Examples are Gaussian reference fluid \cite{KT91a, Kuchar92} and the reference fluid corresponding to the harmonic gauge \cite{New2}.   

\mbox{ }

\noindent A further type of matter time approach involves additionally forming a quadratic combination of 
constraints for dust \cite{BK94}, more general perfect fluids \cite{BroMa}, and massless scalar fields \cite{KR95}.  
The point of such (see also \cite{Markopoulou}) quadratic combinations is that they result in strongly vanishing Poisson brackets. 

\mbox{ } 

\noindent Problem with Reference Matter 1) As this Article's RPM work for now makes but scant contact with matter clocks/reference matter, 
I refer the reader to \cite{Kuchar92, I93, APOT, ABook} for this approach's problems, except for the comment that some candidates involve 
unphysical matter (physical energy condition violating) and/or intangible matter; clearly then relationalism furnishes an additional objection to the latter.  
Tangible and elsewise physical reference fluid matter candidates are not exhaustively ruled out, however.  

\noindent Problem with Reference Matter 2) This is an objection based on the STLRC implementation of Mach's Time Principle: 
in this approach, gravitational and non-reference matter changes have no opportunity to contribute to the timestandard.

\mbox{ } 

\noindent Note 1) Sec \ref{RPT} represents the first steps in conceiving whether there is an analogue of these for RPM's.

\noindent Note 2) \cite{HP11} exemplifies newer such approaches using both further types/formulations of matter and for GR-as-nododynamics.

\subsubsection{Implementation by unimodular gravity} \label{Unim}

\noindent One might also regard an undetermined cosmological constant $\Lambda$ as a type of reference fluid \cite{UW89}.
Here, one does not consider the lapse to be a variable that is to be varied with respect to.  
Instead, $\scH_{\mu}$ has $\scH_{,\mu}$ as an integrability [c.f. (\ref{Ham-Ham})], leading to $\scH + 2\Lambda = 0$ 
with $\scH$ the vacuum expression and $\Lambda$ now interpreted as a constant of integration (this implementation clearly has no diagonal minisuperspace counterpart).   

\mbox{ }

\noindent Cosmological Problem with Unimodular Approach) There is a further issue of non-correspondence of the unimodular approach with observational cosmology. 
Bertolami resolved this issue in \cite{Bertolami} by generalizing to a nondynamical scalar field model, and after the above critiques.  
However, he admits his resolution is only consistent with a rather special class of metrics (whilst the POT's intended scope is for generic metrics). 

\noindent Machian Problem with Unimodular Time) Almost no changes have any opportunity to contribute to unimodular time.

\noindent Operational Problem with Unimodular Time) Finally, the link between such things as the cosmological constant, the scalefactor, conjugates to these or special, 
unobservable reference fluids, and what actually constitutes the reading-hand and calibration of clocks is largely unclear, 
conceptually or practically, both qualitatively and as regards how none of these things are known to the desired, and used, accuracy in timekeeping.

\subsubsection{Summary of Machian status quo}

Based on {\sl current knowledge} of {\sc t} ... {\sc q} approaches that remain unfreezing 
at the quantum level and how these are all characterized as un-Machian and substantially bereft of accuracy as compared to current timekeeping standards on earth
Thus {\sc q} ... {\sc t} {\sl looks to be presently required for a satisfactory QM-level Machian time} in the STLRC sense.

\subsection{Records Theory with Nontrivial Configuration Relationalism}\label{TNE-2}

\noindent Sec \ref{Cl-Str} also provided us with $\FrG$'d versions of the structures in Records Postulates 1 to 3 (with or without 1B) 
whilst a $\FrG$'d version of STLRC was at least formally laid out in Sec \ref{+temJBB}.  
\noindent Then the Best Matching, or solving the GR momentum constraint for $\upzeta^{\mu}$, {\sc c} moves  
could be followed up by a Tempus Nihil Est approach on the relational configuration space: ({\sc cq}).  

\noindent Consider the idea of using RPM's as toy models for records theory with nontrivial $\FrG$ why they are very much more adequate than minisuperspace: 
Records are about localization and structure.  
This is the content of Sec \ref{Cl-Nihil} (and Sec \ref{QM-Nihil} at the quantum level) in particular for investigating Barbour's conjectures, 
my records approach and combinations with Histories Theory and/or the Classical and Semiclassical Machian Emergent Time Approaches.

\subsection{Classical `Paths = Histories' Formulation with Nontrivial Configuration Relationalism}\label{Histor-2}

\noindent Presenting this in the emergent-time case, there is now 
\beq
\scL\scI\scN_{\sfZ}^{\sft^{\te\tm(\tJ\tB\tB)}} := 
\int \d \ft^{\se\sm(\sJ\sB\sB)} \,\scL\scI\scN_{\sfZ}[\fQ^{\sfA}(\ft^{\se\sm(\sJ\sB\sB)}), \fP_{\sfB}(\ft^{\se\sm(\sJ\sB\sB)})] = 0 \mbox{ } .
\eeq
in addition to the structurally-parallely defined $\scQ\scU\scA\scD^{\sft^{\te\tm(\tJ\tB\tB)}}$ of Sec \ref{Histor}.

\noindent There is also now a ((O)A)-PathPhase distinction, the latter two allowing for MRI and MPI implementations of Temporal Relationalism. 
This is manifest in Appendix \ref{Examples}.A.5's variants on actions.  

\mbox{ }  
  
\noindent One could also follow solving the GR momentum constraint at the Lagrangian Best-Matching level or phase space by finding a single `time-map' Histories Theory ({\sc chq} scheme).  

\noindent One could instead find a classical histories theory with `space-map' and `time-map' \cite{KK02, Kouletsis} , prior to reducing away the 
`space-map' structure to pass to a new `time-map' ($\chi_{\tH}${\sc cq}).
The reduction here could e.g. be a `Best Matching of histories': solve the Lagrangian form of the linear histories constraints 
\beq
\scL\scI\scN^{\sft^{\te\tm(\tJ\tB\tB)}} \lfloor \fQ^{\sfA}, \dot{\fQ}_{\sfA}, \dot{\fG}_{\sfZ}\rfloor = 0 
\eeq
for the histories auxiliary variables so as to remove these from the formalism.
It could however also be a reduction at the Hamiltonian level.

\mbox{ }  

\noindent {\bf Question} 44 Lay out the $\FrG$-nontrivial classical paths = histories theory versions of the layers of structure of physical theory that the current Article exposits  
for configuration-based Physics in Sec \ref{Cl-Str}. 

\mbox{ } 

\noindent Alternatively, one can consider these for the $\ft^{\se\sm(\sJ\sB\sB)} \longrightarrow \ft$ or $\lambda$ of absolutist or label-time theories.

\subsection{Strategizing about the Problem of Beables}\label{Beables}

\noindent Strategy 1) Kucha\v{r} beables are all \cite{Kuchar92, Kuchar93, Haj99, HK99, BelEar, Earman02, WuthrichTh, Kouletsis, BF08}.  
N.B. it is clear that finding these is a timeless pursuit: it involves configuration space or at most phase space but no Hamiltonian and thus no dynamics.
The downside now is that there is still a frozen quadratic energy-type quantum constraint on the wavefunctions, 
so that one has to concoct some kind of Tempus Post Quantum or Tempus Nihil Est manoeuvre to deal with this.  
The tale here is that the Ice Dragon is rendered flightless by disarmament treaty: it concedes not to use its wings in 
exchange for a number of one's strategies ceasing to be useable against the Ice Dragon'.  

\mbox{ }  

\noindent I also use Strategy 1) below as a technical half-way house in the formal and actual construction of Dirac beables.  

\mbox{ } 

\noindent Strategy 2) Dirac beables are necessary as a concept and one needs to know them in order to fully unlock QG.  
This is what Kucha\v{r} considers to be like having the good fortune of possessing a Unicorn \cite{Kuchar93}: 
``{\it Perennials in canonical gravity may have the same ontological status as Unicorns 
-- a priori, these are possible animals, but a posteriori, they are not roaming on the earth}" \cite{Kuchar93}.  
To fit my own mythological mnemonic, this makes most sense if one imagines one's Unicorns (like She Ra's `Swift Wind' \cite{SheRa}) to be winged,  
so as to compensate for the Ice Dragon itself being winged.  

\mbox{ }   

\noindent Strategy 3) In fact it is but partial observables that are necessary.  
Here the Ice Dragon's ever having possessed wings, and the subsequent need for flighted Unicorns to compensate for this, are held to have always been a misunderstanding 
of the true nature of observables, which are in fact commonplace but meaningless other than as regards correlations between more than one such considered at once.  

\noindent Analogy 68) RPM's and GR can both be considered in terms of partial observables.  

\mbox{ } \noindent

\noindent I comment that on the one hand Rovelli's partial observables carry connotations of ``{\it tot tempora quot motus}", but, 
on the other hand, Rovelli showed us a possible means of demonstrating that the Ice Dragon has always been a flightless beast and thus less hard to vanquish. 

\mbox{ } 

\noindent {\bf Question 45}. Relative information was tied to the Partial Observables Approach in \cite{Rov96}. 
Can one more fully exploit this as regards being a specific object to compute/measure in the study of subsystem correlations?  

\mbox{ } 

\noindent Note 1) At the level of counting out options, I consider this to involve introduction of a fourth tuple {\sc b} whose options are, in decreasing 
order of stringency, ({\sc d}, {\sc k}, {\sc p}, --) standing for (Dirac, Kucha\v{r}, partial, none).
The relation-free count of number of possible orderings of maps is by now over 100 at the classical level, already illustrating 
before the even greater proliferation at the quantum level that some major restrictions by suitably general and true/plausible theoretical principles are well in order.  

\noindent Note 2) The {\sc p} (partial) option for observables/beables can in principle be readily carried out at any stage relative to the other moves (that includes {\sc q}).

\noindent Note 3) Some of the known impasses with Strategy 2) for Geometrodynamics are as follows. 

\mbox{ }  

\noindent {\bf \K 1981 no-go Theorem} \cite{Kuchar81b} is for nonlocal objects of the form  
\beq
\int_{\sbSigma} \d^3x {\cal K}_{\mu\nu}(\ux; \bh]\uppi^{\mu\nu}(\ux) \neq \iD \mbox{ } 
\label{K1981}
\eeq
for $\underline{\underline{{\cal K}}}$ some matrix-valued mixed function-functional.  


\noindent {\bf Torre 1993 no-go Theorem} \cite{Torre93} (see also \cite{+Torre1, +Torre2, Carlip01}) is that local functionals 
\beq
{\cal T}(\ux; \bh, \buppi] \neq \iD \mbox{ } .
\eeq

\mbox{ } 

\noindent Note 4) It is clear that finding \K beables is a timeless pursuit: it involves configuration space or 
at most phase space but not the Hamiltonian constraint and thus no dynamics.
The downside now is that there is still a frozen $\scQ\scU\scA\scD$ on the wavefunctions, so that one has to concoct some kind of 
Tempus Post Quantum or Tempus Nihil Est manoeuvre to deal with this.  

\noindent Note 5) LMB(-CA) relationalism favours \K's stance \cite{BF08}, or, at least a partly-Dirac stance \cite{AHall} (i.e. commutativity with $\scH$ still being required alongside 
permission to view $\scH$ as distinct from a gauge constraint, that in itself constituting a deviation from Dirac's writings \cite{Dirac}). 

\noindent Note 6) (A useful entwining with Configurational Relationalism). 
If Configurational Relationalism has by this stage been resolved (e.g. by completing Best Matching), then this alongside a small amount of canonical workings gives a geometrically-lucid 
solution to the problem of finding the Kucha\v{r} beables; I denote a sufficient set of \K beables by $\iK^{\sfA}$.   
Thus for any approaches for which {\sc r} can be attained, {\sc r} ... {\sc k} ordered approaches are then straightforward.

\mbox{ } 

\noindent As a prime example of this, consider the problem of finding \K beables for RPM's. 
I.e., for scaled RPM's, quantities obeying 
\beq
\{\iK, \scL_{\mu}\} = 0
\eeq
and, for pure-shape RPM's, quantities obeying both this and 
\beq
\{\iK, \scD\} = 0 \mbox{ } . 
\eeq
Then Sec \ref{Q-Geom}'s Best Matching Problem resolution for 1- and 2-$d$ RPM's alongside Sec \ref{Dyn1}'s evaluation of the 
momenta conjugate to these, imply as per Sec \ref{Gau} what amounts to a resolution to the classical problem of \K beables for these models.  

\noindent In the pure-shape case, 
\beq
\iK = {\cal F}[\mS^{\sfA}, P^{\sS}_{\sfA}] \mbox{ } , 
\eeq
i.e. they are just the functionals that depend on the shapes and shape momenta alone.

\noindent In the scaled case, 
\beq
\iK = {\cal F}[\mS^{\sfA}, \sigma, P^{\sS}_{\sfA}, 
P^{\sigma}] \mbox{ } , 
\eeq i.e. they are just the functionals that depend on the shapes, 
shape momenta, scale and scale momentum alone.

\noindent This is modulo the global caveat in Sec \ref{GPOB}.  

\noindent On the other hand, for Geometrodynamics, then, the classical \K beables are, formally, coordinate-patch-wise defined functionals of the 
3-geometries and their conjugates alone, 
\beq
\iK = {\cal F}[{\capFrG}^{(3)}, \Pi^{\scapFrG^{(3)}}] \mbox{ } .  
\eeq
For conformogeometrodynamical-type programs, the classical \K beables might on occasion be formally, coordinate-patch-wise defined functionals of the 
conformal 3-geometries and their conjugates alone, 
\beq
\iK = {\cal F}[{\capFrC}^{(3)}, \Pi^{\scapFrC^{(3)}}] \mbox{ } , 
\eeq
or, possibly based on a different 2/3 of Superspace($\bupSigma$): ${\cal T}\hspace{-0.04in}\mbox{rue}(\bupSigma)$.

\noindent Finally, for nododynamics, the classical \K beables are, formally, coordinate-patch-wise defined functionals of the knots and their conjugates alone, 
\beq
\iK = {\cal F}[{\cal K}\mbox{not}, \Pi^{{\cal K}\sn\so\st}] \mbox{ } . 
\eeq
\noindent Note 7) Formal $\FrG$-act, $\FrG$-all expressions for \K beables are also available if Configurational Relationalism has not by this stage been solved.

\mbox{ } 

\noindent Other models for which \K beables are known include 

\noindent 1) models for which the notion is trivial (e.g. minisuperspace).

\noindent 2)  a few midisuperspaces such as the cylindrical wave \cite{Bubble}, spherically symmetry \cite{Kuchar94} and some Gowdy models \cite{TorreGowdy}.  

\noindent 3) Any models for which Dirac beables themselves are known (e.g. this is so for some Gowdy models).  

\mbox{ } 

\noindent Note 8) Halliwell's objects obey \{\scQ, \iB\} = 0 but not necessarily \{$\scL\scI\scN$, \iB\} = 0, or possibly the QM counterpart of this statement 
(in which case these objects are termed {\sl S-matrix quantities}).   
These do not carry background dependence connotations due to corresponding to scattering processes on configuration space rather than on space.
Clearly then at the classical level,
\beq
\mbox{Kucha\v{r} AND Halliwell } \Rightarrow \mbox{Dirac} \mbox{ } .
\label{KSDir}
\eeq

\noindent N.B. this refers to scattering in {\sl configuration space} and not in space, so it is free of relationally questionable assumptions about flat-space spatial infinity; 
Misner was an early pioneer of this perspective, which he used to study anisotropic cosmology \cite{Magic}.   
\noindent Halliwell supplied constructions of these (see \cite{H03, H09} for $\FrG \Leftrightarrow$ $\scL\scI\scN$-free examples, which I generalized \cite{AHall} 
to RPM examples with nontrivial $\FrG$ or $\scL\scI\scN$ included; this is the subject of Secs \ref{Cl-Combo} and \ref{QM-Combo}.  

\noindent Note 9) The Dirac beables problem for scaled RPM's is then to find quantities obeying
\beq
\{\scL_{\mu}, \iD\} = 0 \mbox{ } , \{{\scE}, \iD\}       =  0 \mbox{ } ,
\label{OLE0}
\eeq
and, for pure-shape RPM's, these and
\beq
\{\scD, \iD\} = 0 
\label{OD02}
\eeq
also. 
In the scaled case: the total Hamiltonian takes the form $H[\,\dot{I}, B^{\mu}\,]  = \dot{I}\,\scE + B^{\mu}\scL_{\mu}$ 
Thus, Dirac beables are automatically constants of the motion with respect to the evolution associated with any choice of strut function $\dot{I}$ and point-identifying map $B^{\mu}$. 
In the pure-shape case, the form of the Hamiltonian is $H[\,\dot{I}, B^{\mu}, C\,]  = \dot{I}\,\scE + B^{\mu}\scL_{\mu} + C\,\scD$ so that 
the point-identifying map now involves both $B^{\mu}$ and $C$.  
If the r-version of one's RPM is available [as is the case in 1- or 2-$d$ by Note 6)], the Hamiltonian is $H = \dot{\mI} \, \scE$ and there is no point-identifying map required.                       

\noindent Note 10) A simple partial answer is that in a few cases these include (subsets) of the isometries, i.e. relative angular momenta, relative 
dilational momenta and linear combinations of these with certain shape-valued coefficients.  
E.g. this is a direct analogue of the angular momenta forming a complete set of commuting operators with the Hamiltonian operator, 
provided that the potential is central i.e. itself respecting the isometries of the sphere by being purely radial. 
Thus for the tower of $SO(p)$-symmetric problems (2 $\leq p \leq n$) for $N$-stop metrolands and triangleland, and its more elaborate counterpart for 
quadrilateralland in \cite{QuadII}, we have found in this way some Dirac beables.  
These are not, however, generically present i.e. for arbitrary-potential models. 

\mbox{ } 

\noindent Beables Hidden Problem 1) $\mt^{\st\sr\su\se}$ is required to obey $\{\mt^{\st\sr\su\se}, \scH\} = 0$.

\subsubsection{Machian `Paths = Histories' beables} 

These are, conceptually \cite{Kouletsis, APOT2}, quantities that classical-path-brackets-commute with the path versions of the constraints.  
These are distinct and meaningful concepts [furnishing strategies 4) and 5)] in the Dirac and \K cases, namely 
\beq
\sD_{\tH}^{\sft^{\te\tm(\tJ\tB\tB)}} = {\cal F}[\fQ(\ft), \fP(\ft)]
\eeq 
obeying 
\beq
\{\sD_{\tH}^{\sft^{\te\tm(\tJ\tB\tB)}}, \scC_{\sfA}^{\sft^{\te\tm(\tJ\tB\tB)}}\} = 0 
\label{HDirObs}
\eeq
and 
\beq
\sK{\tH}^{\sft^{\te\tm(\tJ\tB\tB)}} = {\cal F}[\fQ(\ft), \fP(\ft)]
\eeq 
obeying
\beq
\{\sK_{\tH}^{\sft^{\te\tm(\tJ\tB\tB)}}, \scL\scI\scN_{\sfZ}^{\sft^{\te\tm(\tJ\tB\tB)}}\} = 0 \mbox{ } . 
\label{HKObs} 
\eeq
\noindent Alternatively, one can consider these for the $\ft^{\se\sm(\sJ\sB\sB)} \longrightarrow \ft$ or $\lambda$ of absolutist or label-time theories.

\subsection{Machian-Classical--Histories--Records Combined Approaches}\label{Cl-Combo-Intro}

\subsubsection{Motivation}

\noindent Motivation 1) is Motivation 2) of Sec \ref{Cl-Nihil-Intro}.  

\noindent Motivation 2) Halliwell's \cite{H03} (partly with Thorwart \cite{HT}) concerns, in conceptual outline, ``{\it life in an energy eigenstate}".  
Moreover, as well a timeless approach, it is framed within Histories Theory, treated semiclassically, and, 
as I further exposit in the present Article, with connections to some Observables Approaches.

\noindent Motivation 3) The classical precursor to this is already motivated by how Records Theory sits within Histories Theory. 

\noindent Motivation 4) Both histories and timeless approaches lie on the common ground of atemporal logic structures \cite{IL}.   

\noindent Motivation 5) Also, whilst Halliwell did not consider Machian-semiclassical formulations, 
I view these as a significant advance to the conceptual understanding of Quantum Cosmology,
and as such consider the Machian-classical counterpart of Halliwell's work as a temporal-relationalism-resolved formulation.

\subsubsection{Halliwell's classical-level approach itself}\label{Metho}

Here, one considers probability distributions, firstly on classical phase space, $\mw(\bq,\bp)$ and then at the semiclassical level in Sec \ref{QM-POT-Strat}.
At the classical level, for the classical analogue of QM energy eigenstate,
\beq
0 = {\pa \mw}/{\pa t} = \{H, \mw\} \mbox{ } , 
\eeq
so that w is constant along the classical orbits.
Halliwell then considers expressions for  
\beq
P_{\bsFrR} := \mbox{Prob}(\mbox{classical solution will pass through a configuration space region $\bFrR$})  \mbox{ } . 
\eeq
This is a type of implementation of propositions (Sec \ref{Cl-Prop}), which can be considered {\sl via} probabilities such as the above or its 
phase space extension \cite{Omnes2, Ana96, H12}.
%
%
One then makes use of the characteristic function of the region $\bFrR$, denoted $\mbox{Char}_{\bsFrR}(\bq)$. 
Then \cite{H03, AHall} 
\beq
\mbox{Prob(intersection with $\bFrR$)} = \int_{-\infty}^{+\infty}\d t\,\mbox{Char}_{\bsFrR}(\bq^{\scc\sll}(t)) = 
\int \mathbb{D}\bq \, \mbox{Char}_{\bsFrR}(\bq) \, \int_{-\infty}^{+\infty}\d t \, \updelta^{(k)}\big(\bq - \bq^{\scc\sll}(t)\big) = 
\int \mathbb{D}\bq \, \mbox{Char}_{\bsFrR}(\bq) \, A(\bq, \bq_0, \bp_0 )  \mbox{ } ,
\eeq
the `amount of $t$' the trajectory spends in $\FrR$.\footnote{This is, for now, absolute time, though this SSSec can be redone in terms of label time $\lambda$, 
and is subsequently also redone within the classical Machian emergent time scheme, for which we can set $t = t^{\se\sm(\sJ\sB\sB)}$.]  
Recollect also that $k$ = dim($\FrQ$).}
%
It is important to treat the whole path rather than segments of it, since the endpoints of segments contribute right-hand-side terms to 
the attempted commutation with $H$ (Halliwell's specific context being that with no linear constraints, that is $H = \upalpha\scH$).  
\beq
\mbox{Next,} \hspace{2in}
P_{\bsFrR} = \int \int \mathbb{D}\bp_0 \, \mathbb{D}\bq_0 \, \mw(\bq_0,\bp_0) \,\, \theta
\left(
\int_{-\infty}^{+\infty}\d t\,\mbox{Char}_{\bsFrR}(\bq^{\scc\sll}(t)) - \epsilon
\right) \mbox{ } .  \hspace{1.8in}
\eeq
Here, the $\btheta${\bf -function} serves to mathematically implement the restriction the entirety of the phase space 
that is being integrated over to that part in which the corresponding classical trajectory spends time $> \epsilon$ in region $\bFrR$. 
$\epsilon$ is some small positive number that tends to 0, included to avoid ambiguities in the $\theta$-function at zero argument.  

\mbox{ } 

\noindent This has the property that $\{A, H\} = 0$.
[That is indeed a beables, rather than histories beables, condition, since the $t$ has been integrated over.]  
Thus, in the $\FrG$-trivial case, it is a classical Dirac beable.
Hence we can write $A = \iD$.  

\mbox{ }  

\noindent An alternative expression is for the flux through a piece of a \{$k$ -- 1\}-dimensional hypersurface within the configuration space, 
\beq
P_{\Upsilon} = \int\d t\int \mathbb{D}\bp_0 \, 
\mathbb{D}\bq_0 \, \mw(\bq_0, \bp_0) 
\int_{\Upsilon} \mathbb{D}{\Upsilon}(\bq) \,\,\, \bnu\cdot \bfM \cdot \frac{\d \bq^{\scc\sll}(t)}{\d t} \,\,\, \updelta^{(k)}\big(\bq - \bq^{\scc\sll}(t)\big)
= \int \d t \int\mathbb{D}\bp^{\prime} \int_{\Upsilon} \mathbb{D}{\Upsilon}(\bq^{\prime}) \,\,\, 
  \bnu\cdot \bp^{\prime} \,\,\, \mw(\bq^{\prime}, \bp^{\prime}) \mbox{ } , 
\label{33}
\eeq
the latter equality being by passing to $\bq^{\prime} := \bq^{\scc\sll}(t)$ and $\bp^{\prime} := \bp^{\scc\sll}(t)$  coordinates at each $t$.

{            \begin{figure}[ht]
\centering
\includegraphics[width=0.3\textwidth]{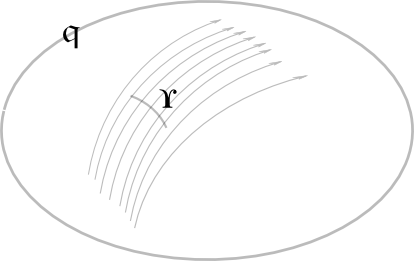}
\caption[Text der im Bilderverzeichnis auftaucht]{        \footnotesize{Halliwell's approach considers propositions corresponding to fluxes through pieces of hypersurfaces 
$\Upsilon$ within configuration space $\FrQ$.}    }
\label{Preliminaria} \end{figure}          }

\subsubsection{Relational versions with classical \K degradeables}

Halliwell \cite{HT, H03} illustrated the above with e.g. a particle in the absolutist $\mathbb{R}^k$ setting.
Sec \ref{Cl-Combo} shall use RPM counterparts of this. 
For now we consider a more general, if formal account. 
$\biK$ is a coordinate vector for configurational \K beables; these are independent and there are the right number $r$ = dim($\FrQ^{\sr}$) of them to span.  
The complete set of \K beables are more general: functionals ${\cal F}[\biK,\biP^{\tK}]$.
Having \K beables explicitly available has close ties with one's model  being reducible; both are in practise exceptional circumstances. 
They do however apply to 1- and 2-$d$ RPM's, which are indeed both reducible and have full sets of \K beables known (c.f. Sec \ref{Dyn1}) and modulo global caveats to be discussed 
later.  
We present an alternative strategy in Sec \ref{Cl-POT-Strat} which, whilst indirect, is more widely applicable in cases without reducibility or knowledge 
of an explicit directly-expressed set of \K beables.  
Full use of \K beables would involve the extension of Halliwell's construct to such as phase space regions since there are 
$\iK = {\cal F}[\mbox{\boldmath$Q$}, \mbox{\boldmath$P$}$] rather than being functional just of the $\mbox{\boldmath$Q$}$. 


\noindent Since such relational examples are additionally whole-universe models, the Machian classical time $t^{\se\sm(\sJ\sB\sB)}$ emerges to fill in the role of $t$ 
and PPSCT) invariance is held to apply as per Appendix \ref{Examples}.B (with further results needed for this SSec in Appendix \ref{Cl-POT}.A). 
By adopting this, this SSSec and its descendants are Machian-Classical-Records-Histories Approaches or their Machian-Semiclassical-Records-Histories successors 
in terms of $t^{\se\sm(\sW\sK\sB)}$.

\mbox{ } 

Then, for the classical analogue of energy eigenstate,
\beq
0 = {\pa \mw}/{\pa t^{\se\sm(\sJ\sB\sB)}} = \{H, \mw\} \mbox{ } ,
\eeq
so $\mw$ is constant along the classical orbits.
I now use a phase space function $A$ that is now not just any $A$ but an $A$ based on Kucha\v{r} observables $\biK$: 
\beq
A(\biK, \biK_0, \biP^{\tiK}_0) = \int_{-\infty}^{+\infty} \d t^{\se\sm(\sJ\sB\sB)} \updelta^{(r)}(\biK - \biK^{\scc\sll}(t^{\se\sm(\sJ\sB\sB)}) ) \mbox{ } . 
\label{candi}
\eeq
\beq
\mbox{Then} \hspace{1.8in} 
\{\scL\scI\scN, A(\biK, \biK_0, \biP^{\tiK}_0)\} = 
\int_{-\infty}^{+\infty} \d t^{\se\sm(\sJ\sB\sB)} \{\scL\scI\scN_{\sfZ}, \updelta^{(r)}(\biK - \biK^{\scc\sll}(t^{\se\sm(\sJ\sB\sB)}) )\} \mbox{ } ,
\label{qirk} \hspace{2.2in} 
\eeq
and $\{ \scL\scI\scN_{\sfZ}, \biK\} = 0$ because the $\biK$ are Kucha\v{r}, so $\{ \scL\scI\scN_{\sfZ}, f(\biK)\} = 0$ by the chain-rule, and so 
\beq
\{\scL\scI\scN_{\sfZ}, A(\biK, \biK_0, \biP^{\tiK}_0)\} = 0  \mbox{ } .
\label{uni2}
\eeq
Moreover, being in terms of a `vector' of Kucha\v{r} observables does not change the argument by which 
\beq
\{ \scQ\scU\scA\scD, A(\biK, \biK_0, \biP^{\tiK}_0)\} = 0 
\label{corn2}
\eeq
(which Halliwell has already demonstrated to be robust to curved configuration space use). 
Thus, by (\ref{uni2},\ref{corn2})  combined, 

\mbox{ } 

\noindent 1) (\ref{candi}) are Dirac, so we can write A = $\iD$.  

\noindent 2) A substantial a set of Dirac beables can be built thus for a theory whose full set of Kucha\v{r} beables are known 
as could be built for Halliwell's simpler non linearly-constrained theories.  
To that extent, one has a formal construction of \K's Unicorn, at least for the range of theories to which Halliwell's construct can be applied.

\noindent N.B. this is a {\sl classical} resolution. 
At the quantum level, this role of A needs to be played out again by a distinct object: the {\sl class operator}, see Sec \ref{QM-POT-Strat}.  

\mbox{ } 

\noindent Note 1) Since the classical \K beables are known for 1- and 2-$d$ RPM's, scaled or unscaled, this ensures that this is an 
{\sl actual} construction for these toy models' toy Unicorn in Sec \ref{Cl-Combo}.

\noindent Note 2) As regards the various no-go theorems, I note that the above avoids Kucha\v{r}'s by not being of form (\ref{K1981});
it avoids the Torre 1993 No-Go Theorem by not being local in space or time.  

\mbox{ }  

Next,
$$ 
\mbox{Prob(intersection with $\bFrR$)} = \int_{-\infty}^{+\infty}\d t^{\se\sm(\sJ\sB\sB)}\,\mbox{Char}_{\bsFrR}(\biK^{\scc\sll}(t^{\se\sm(\sJ\sB\sB)})) = 
\int \mathbb{D}\biK \, \mbox{Char}_{\bsFrR}(\biK) \, \int_{-\infty}^{+\infty}\d t^{\se\sm(\sJ\sB\sB)} \updelta^{(r)}
\big(\biK - \biK^{\scc\sll}(t^{\se\sm(\sJ\sB\sB)})\big) 
$$
\beq
= \int \mathbb{D}\biK \, \mbox{Char}_{\bsFrR}(\biK) \, A(\biK, \biK_0, \biP^{\tiK}_0 ) :
\eeq
the `amount of $t^{\se\sm(\sJ\sB\sB)}$' the trajectory spends in $\FrR$; moreover this physical quantity is constructed to be PPSCT-invariant by (\ref{7iv)}).  
Then,
\beq
P_{\bsFrR} = \int \mathbb{D}\biP^{\tiK}_0 \, \mathbb{D}\biK_0 \, \mw\big(\biK_0,\biP^{\tiK}_0\big) \,\, \theta
\left(
\int_{-\infty}^{+\infty}\d t^{\se\sm(\sJ\sB\sB)} \mbox{Char}_{\bsFrR}\big(\biK^{\scc\sll}(t^{\se\sm(\sJ\sB\sB)})\big) - \epsilon
\right) \mbox{ } , 
\eeq
which is PPSCT-invariant through coming in three factors each of which is PPSCT-invariant by (\ref{Was-viii)}--\ref{Was-ix)}).  

\mbox{ } 

\noindent Note 3) The window function is assumed to fit on a single coordinate system, which in itself limits the procedure to being somewhat local.  
This is entirely fine if one is considering small regions, such as in Sec \ref{TessiRegions}.
The method in question also continues to work approximately for examples with compact relationalspaces (which include pure-shape RPM's as per Sec \ref{Q-Geom}).

\mbox{ }  

\noindent The alternative expression for the flux through a piece of an \{$r$ -- 1\}-$d$ hypersurface within the configuration space is now  
$$
P_{\Upsilon} = \int\d t^{\se\sm(\sJ\sB\sB)}\int\int \mathbb{D}\biP^{\tiK}_0 \, \mathbb{D}\biK_0 \, \mw(\biK_0,\biP^{\tiK}_0) 
\int_{\Upsilon} \d^2{\Upsilon}(\biK^{\prime}) \, \bnu^{\tiK\prime} \cdot \frac{\d \biK^{\scc\sll}(t^{\se\sm(\sJ\sB\sB)})}{\d t^{\se\sm(\sJ\sB\sB)}} 
\, \updelta^{(r)}\big(\biK - \biK^{\scc\sll}(t^{\se\sm(\sJ\sB\sB)})\big)
$$
\beq
= \int \d t^{\se\sm(\sJ\sB\sB)} \int\mathbb{D}\biP^{\sK \prime} \int_{\Upsilon} \mathbb{D}{\Upsilon}(\biK^{\prime})\, 
\bnu^{\tiK\prime}\cdot\biP^{\tiK\prime} \, \mw(\biK^{\prime}, \biP^{\tiK}) \mbox{ } , 
\eeq
the latter equality being by passing to $\biK^{\prime} := \biK^{\scc\sll}(t^{\se\sm(\sJ\sB\sB)})$ and $\biP^{\tiK} 
:= \biP^{\tiK}_{\scc\sll}(t^{\se\sm(\sJ\sB\sB)})$ coordinates at each $t^{\se\sm(\sJ\sB\sB)}$.
This is again PPSCT-invariant, by (\ref{Was-v)}), (\ref{Was-viii)}) and (\ref{Was-ix)}).  

\mbox{ } 

\noindent This and a further implementation at the quantum level in the form of class operators (see below) is conceptually strong: its reparametrization-invariance 
implements Temporal Relationalism \cite{ARel}, the constructs are reasonably universal and have a clear meaning in terms of propositions.  
[Though in this last sense, it would appear to require extension at least to its phase space counterpart.]

\subsubsection{Alternative indirect $\FrG$-act, $\FrG$-all extension}

\beq
\mbox{Now,}  \hspace{1.5in}
{A}^{\sfg-\sf\sr\se\se}(\brho, \mbox{\boldmath$p$}_{0}, \brho_{0}) = \int_{g \in \sFG}\mathbb{D} g \, \stackrel{\rightarrow}{\FrG_g}
\left\{
\int_{-\infty}^{+\infty} \d t^{\se\sm(\sJ\sB\sB)} \, \updelta^{(k)}\big( \brho -\brho^{\scc\sll}(t^{\se\sm(\sJ\sB\sB)})\big)
\right\} 
\mbox{ } .   \hspace{3in}  
\eeq

\subsubsection{Continuation of conceptual analogy between RPM shapes and Nododynamics} 

More ambitiously, might one be able to extend Halliwell's class operator construction to LQG/Nododynamics, so as to be able to, at least formally, write 
down a (perhaps partial) set of complete observables as a subset of the linear constraint complying knots?
Either construct such a set or explain at what stage of relevant generality Halliwell's class operator construction fails.
[I.e. a tentative search for a fully-fledged Winged Unicorn.]
Note that this goes beyond the currently-known geometrodynamical minisuperspace scope of Halliwell's class operators, but is an interesting direction in which to try to extend this scope.

\subsection{The three classically-overcome facets}\label{3-Overcome}

\noindent The Trident counter to the three remaining classical facets mostly means we can stop here at the classical level. 

\mbox{ } 

\noindent If one goes the Type 1 Tempus Ante Quantum route, however, the Trident is not quite perfect as regards classical Spacetime Reconstruction; e.g. one now requires that, 
following up on Beables Hidden Problem 1), internal times be spacetime scalars: Spacetime Reconstruction Problem of Internal Time).
In particular, functions of this form do not have any foliation dependence.
However, the canonical approach to GR uses functionals of the canonical variables, and which there is no a priori reason for such to be scalar fields of this type.  
Thus one is faced with either finding functionals with this property (establishing Foliation Independence by construction 
and standard spacetime interpretation recovery), or with coming up with some new means of arriving at the standard spacetime meaning at the classical level.   

\mbox{ } 

\noindent If one adopts more sparse structure, classical Spacetime Reconstruction can also become an issue, e.g. in the Causal Sets approach 
(see e.g. \cite{RiWa} for recent advances here).

\mbox{ }

\noindent On the other hand, classical knowledge of foliations suffices to cut down the Unimodular approach at the level of mere counting, as follows.  

\mbox{ }

\noindent Unimodular Fatality) The cosmological constant itself plays the role of isolated linear momentum [c.f. eq. (\ref{Para1})] which, at the quantum level gets promoted to the 
derivative with respect to the unimodular internal time function $t^{\sU\sn\si}$ by the presence of which the quantum theory is unfrozen.
However, there is a mismatch between this single time variable and the standard generally-relativistic 
concept of time, which is `many-fingered' with one finger per possible foliation.
[In other words, the derivative with respect to $t^{\sU\sn\si}$ is a partial derivative, as opposed to the 
functional derivative that one expects to be present in a GR POT resolution.]  
The geometrical origin of this mismatch is that a cosmological time measures the 4-volume enclosed between two 
embeddings of the associated internal time functional $t^{\si\sn\st}$.  
However, given one of the embeddings the second is not uniquely determined by the value of $t^{\sU\sn\si}$ 
(since pairs of embeddings that differ by a zero 4-volume are obviously possible due to the Lorentzian signature and cannot be distinguished in this way). 
Bertolami's version of unimodular gravity \cite{Bertolami}  does not attempt to further address this counting mismatch either.

\subsection{Strategizing about the Global POT}\label{GlobStrat}

\noindent Global Attitude 1) Insist on only using globally-valid quantities for timefunctions and beables.

\noindent Whilst this is very naturally covered at the classical level by time being a coordinate and thus only in general locally valid 
and subject to the ordinary differential-geometric meshing condition, the problem now is how to mesh together different unitary evolutions.  
There is also an intermediate problem involving passage from local to global in classical p.d.e. prescriptions (which quite a lot of POT facets and strategies do involve).


\noindent Let us first justify use of RPM's among the examples used to explore the Global POT.  

\mbox{ } 

\noindent Analogy 69)  RPM's have some spatial globality issues due to possessing meaningful notions of localization/clumping.  
Some other types of globality frankly do not care whether one's theory is a field theory or finite (consult the rest of this SSec for evidence!)

\mbox{ } 

\noindent  Global Problem with Scale Time 1) Scale times are not useable globally-in-time in recollapsing universes due to non-monotonicity. 

\mbox{ }

\noindent Passing to dilational times (conjugate to scales) such as York time remedies this problem within Global Attitude 1) and e.g. \cite{Benito} do so by 
patching together multiple timefunctions along the lines of Global Attitude 2).  

\mbox{ } 

\noindent Analogy 70) Equations like GR's CMC LFE (Sec \ref{Rel-CGdyn}) provides guarantees of existence and monotonicity.   
For scaled RPM equations along the lines of the Lagrange--Jacobi relation of Celestial Mechanics provide similar guarantees (see Sec \ref{LJE} for more). 
Both cases work for sizeable classes of examples but not for all examples. 
Sec 10 is required prior to being able to make sharper statements as regards which equations, which GR-RPM analogies and which guaranteed cases.  

\mbox{ } 

\noindent Global Problem with York Time 1)  There is a global POT in the canonical transformation separation into true and embedding (space frame and timefunction) variables 
in the internal time approach.
This is the 

\mbox{ } 

\noindent {\bf Torre Impasse} \cite{Torre}: that the right-hand-side space of (\ref{canex}) is a manifold while the left-hand-side 
space is not (due to the occasional existence of Killing vectors). 
This limits the sense in which such a map could hold (it certainly does not hold globally).

\mbox{ } 

\noindent H\'{a}j\'{\i}\v{c}ek and Kijowski \cite{HK99} have furthermore established that such a map is not unique.  

\mbox{ } 

\noindent Note 1) It is this part of the Global POT that is similar to the Gribov effect in Yang--Mills theory.  

\noindent Note 2) We can skip this if we are on the r-route too: gauge-invariant quantities rather than gauge choices. 
Or by choosing {\sc qt} over {\sc tq}.

\mbox{ } 

\noindent Global Problem with York Time 2) As regards globality in time, for GR, existence of CMC and propagation and monotonicity 
for these \cite{MT80, B88, R96, I95}  does not hold for all examples \cite{B88, R96, I95}.    
Privileged foliation theories can suffer from such shortcomings too.  
This problem is either global in space or global in the candidate time itself.  

\mbox{ }  

\noindent Simpler models' timefunctions can eventually going astray for some other reason that does not involve foliations 
[after all, many simpler models do not have (meaningful) notions of foliation].  
E.g, 

\noindent Machian-Classical Global Problem 1) the allotment of $h$--$l$ status may be only local in configuration space, with the approximate dynamical trajectory
%
%
free to leave that region.
See Sec \ref{+temJBB} for global issues with the further approximations made in this approach.  

\noindent The Global POT is partly avoidable by adopting a timeless approach; however 

\noindent Timeless Global Problem 1) elements of the timeless construct could be only locally defined;
the timeless approach has been argued to concern localized records: these are desirable, but do they cover everything? 
Does this perspective fail to encode any actually-physically-realized topological information, for instance.  
[Barbour Records avoid 6) but at the cost of being impractical to observe with accuracy: we have uncovered that localized versus Barbour Records is a fork, 
in that we now have a global-in-space argument for Barbour Records as well as Sec \ref{Cl-Nihil-Intro}'s arguments for localized records.]

\noindent Timeless Global Problem 2) Is the crucial semblance of dynamics or history itself a merely-local construct?

\noindent Records-within-Histories Theory Global Problem 1) follows from Timeless Global Problem 1) in those cases whose histories are built 
to match such a local conceptualization of the Records.

\subsubsection{Globality issues with five more classical facets}\label{GP5}

\noindent \bu Global Problem with JBB Time 1) POZIN of Sec \ref{POZ} affects relational actions and JBB time; see Sec \ref{+temJBB} for more.  

\noindent  \bu Definition 6 of Sec \ref{Q-Geom}, which allows for geodesics to pass between strata (i.e. extend across a type of obstruction).
 
\noindent \bu Global Problem with Configurational Relationalism resolution).  
A major subcase of this is 

\noindent Global Problem with JBB Time 2): Global Problem with the underlying Best Matching.  
I.e., in the case in which one is seeking a Best Matching Problem resolution of Configurational Relationalism, we have but a localized resolution of Best Matching Problem in general 
(even in cases for which we have {\sl any kind of resolution}, for an important subcase is the geometrodynamical Thin Sandwich Problem.
POZIN feeds into this again, but so does and another technical condition for the Thin Sandwich Conjecture case of geometrodynamics, and one can also not mathematically extend across 
collinearities in $N$-cornerland.

\mbox{ } 

\noindent As regards resolving Configurational Relationalism at the level of the Hamiltonian formulation, the momentum constraint as an equation for the longitudinal potential 
$\upzeta^{\mu}$ is a rather better-behaved elliptical p.d.e.
%

\noindent \bu Global Problems with foliations). 
The preceding SSSec's account of CMC foliations illustrates the wider point that foliation conditions may not be globally well-defined.

\noindent \bu Teitelboim and HKT's demonstration of refoliation-invariance is itself only local.  
This is via its depending on the uniqueness part of the thick sandwich conjecture, and by furtherly assuming locality in time itself.  

\noindent \bu Global problem with Spacetime Reconstructions). 
Cauchy problems can easily come to be locally restricted; the general idea here is that we only can only hold to be building a shallow bucket neighbourhood of a given piece of 
spatial hypersurface unless one can establish that one can do better than that.  
The `Relativity without Relativity' approach is only in general backed by local-in-time evolution guarantees \cite{Wald, CBbook}. 
For global results relevant to GR evolution, see e.g. \cite{RFRev, Rendall, CBbook}.

\subsubsection{Globality issues with the Problem of Beables}\label{GPOB}

\noindent \bu That coordinates are e.g. not globally defined on closed configuration spaces such as the 1- and 2-$d$ shape spaces means that one's set of functionals 
are really only defined coordinate-patch -wise rather than globally. 
Thus one really has a set of \K {\sl degradeables}. 

\mbox{ } 

\noindent \bu {\bf Fashionables} by Bojowald et al \cite{Bojo1, Bojo2, Bojo3, Bojo4} are observables local in time and space.  

\mbox{ } 

\noindent\bu {\bf Degradeables} are beables that are local in time and space, proposed by the Author \cite{AHall}.    

\mbox{ }  

\noindent These are illustrative words for concepts that are local both in time and in space, 
viz `fashionable in Italy', `fashionable in the 1960's', `degradeable within a year' and `degradeable outside of the fridge' all making good sense. 
Also, fashion is in the eye of the beholder -- observer-tied, whereas degradeability is a mere matter of being, rather than of any observing.
Invoking any such approaches exemplifies patching (Global Attitude 2).  
Moreover, the differential geometry of classical Physics offers no problems with meshing these patches together, so this caveat is not severe.  
Patching quite clearly ties well with the Partial Observables Approach [Perspectivalism 3)], though this notion applies also to the \K and Dirac conceptualizations of observables.
Further details of Bojowald et al.'s treatment must await the QM level (see Sec \ref{Global}).  
Specific examples of \K degradeables are the \iK $\longrightarrow$ $\iK\iD$ of Sec \ref{Beables}'s examples.

\mbox{ }  

\noindent Example.  $\Phi$ is a classical \K degradeable for scaled triangleland since it is not well-defined at $\Theta = 0, \pi$.  
On the other hand, the unit polar vectors $dra^{\Gamma}$ are \K beables for this example.    

\mbox{ }

\noindent \bu The above furthermore implies mere locality of Halliwell's Combined Approach's uplift to Dirac observables. 
\noindent Moreover, locality in time enforced in schemes that take this to be the emergent JBB time (which itself has a global-in-time problem) 
prevents the range of integration from being the $-\infty$ to $+\infty$ required for commutation with $\scQ\scU\scA\scD$.

\subsection{Nododynamical counterparts of the POT strategies}\label{Nodo-Strat}

{\bf Question 46}$^{**}$.  Does the Nododynamics (Ashtekar/LQG) formulation substantially alter any of the POT strategies?
Perform a full survey.

\begin{subappendices}
\subsection{Supporting PPSCT results}

We now need the following PPSCT results in addition to those in Appendix \ref{Examples}.B.  

\noindent \bu The characteristic function $\mbox{Char}_{\bsFrR}$ needs to scale as $\Omega^{-2}$ so that the overall combination $\d t \,\mbox{Char}_{\bsFrR}$ is PPSCT-invariant.  

\noindent \bu If one applies a PPSCT to an $r$-manifold with metric $M_{\sfa\sfb}$ containing an \{$r$ -- 1\}-dimensional 
hypersurface with metric $m_{\sfa\sfb}$ and normal $n^{\sfa}_{\tiK}$, then the formula for the induced metric implies that 
\beq
m_{\sfa\sfb} \longrightarrow \overline{m}_{\sfa\sfb} = \Omega^2m_{\sfa\sfb} 
\mbox{ }  \mbox{ and } \mbox{ } 
n_{\sK \sfa} \longrightarrow \overline{n}_{\tiK \sfa} = \Omega \fn_{\tiK \sfa} \label{Was-v)}
\eeq
from which it immediately follows that 
\beq
\sqrt{|m|} \longrightarrow \overline{\sqrt{|m|}} = \Omega^{r - 1}\sqrt{m} 
\mbox{ } \mbox{ and } \mbox{ } 
n_{\tiK}^{\sfa} \longrightarrow \overline{n}_{\tiK}^{\sfa} = \Omega^{-1}n_{\tiK}^{\sfa} \mbox{ } .
\eeq
\beq
\mbox{\bu  } \mbox {The phase space measure does not PPSCT-scale}  \hspace{5in}
\label{Was-viii)}
\eeq
by its momentum space factor scaling oppositely to its configuration space one. 

\noindent \bu Finally, I posit that the classical probability density $\mw$ is PPSCT-invariant, so that
\beq
\int \mathbb{D}\biK\,\mathbb{D}\biP^{\tiK} \mw(\biK, \biP^{\tiK}) \mbox{ is also PPSCT-invariant } .  
\label{Was-ix)}
\eeq
\end{subappendices}

\vspace{10in}

\section{Detailed look at Jacobi--Barbour--Bertotti time}\label{+temJBB}

\subsection{$t^{\se\sm(\sJ\sB\sB)}$ is {\sl NOT the} ephemeris time} 

\noindent Barbour (see e.g. \cite{Bfqxi}) calls emergent JBB time `the ephemeris time'.  
However, this `the' is compromised by how ephemeris time is not unique at the level of detailed procedure.
Additionally (as regards allusions of classical perfection in the LMB approach), only outdated versions of the more general family of 
astronomical timestandards usually come under the name `ephemeris time'.
What makes most sense then, in detail, is that emergent JBB time is qualitatively of ephemeris type; treatment of it has not involved 
any of the specific iterative processes \cite{Clemence, AlmaSup} used to determine ephemeris time (the ones used historically differ as to which 
of three lunar theories were in use for their computations) or any of its `astronomical dynamical time' descendants.
Moreover, as detailed in Sec \ref{temJBB-Approx}, making approximations at the level of the terms in the formula (\ref{mint}) for $t^{\se\sm(\sJ\sB\sB)}$ 
itself will clearly fail to reproduce the particular details of the specific iteration process so as to fit the positions of the 
defining bodies roughly within the error bounds on their positions .  
Any working scheme requires the subtlety of starting with the relational action, varying it, and deciding at the level of the 
forces in the equations of motion which degrees of freedom to keep and which to neglect; this is then reflected in terms of which 
are kept and which are neglected in the approximate formula for $t^{\se\sm(\sJ\sB\sB)}$.
Finally, in this procedure the relational underpinning is doing nothing new for the actual calculation of an ephemeris time type quantity, 
whereas the {\sl practical} knowledge of how to handle the Solar System Physics iteratively to set up the ephemeris, which was developed without 
ever any reference to relationalism, is fully used. 
Thus I do not consider that relationalism contributes anything {\sl practical} to the computation of astronomical timestandards.

\mbox{ }  

\noindent What relationalism does provide, however, is an alternative philosophical underpinning by which such calculations 
can be said to rest on a relational rather than absolute view of the universe.  
I can address this alongside answering a question recently raised by Lawrie \cite{Lawrie} in the QG literature: ``{\it Ordinarily, there 
seems to be a clear sense in which a well-constructed clock reads `10 s' seven seconds after it reads `3 s', and this does not 
appear merely to result from a conspiracy amongst the manufacturers of time-pieces}".

A Newtonian absolute answer to this would be that all good clocks {\sl happen to} display (up to choice of tick-length and of calendar year zero) absolute time to good 
approximation, though this is not a practically-implementable selection procedure.
A stronger answer to this would involve how these manufacturers' professional antecedents separately agreed with {\sl natural timestandards} rather than merely with each other.  
I.e. they had to produce {\sl useful devices} where the measure of that usefulness was agreeing with solar system phenomena such as night and day, where the sun appears in the sky 
from earth, the length of the seasons... 
Clock designs failing to agree with these things were selected against, and there was increasing pressure from people 
wishing to be able to keep successively more accurate appointments, for instance for governing countries or for naval coordination.  
In the Newtonian set-up this might be tied to the preceding by some kind of postulation that the celestial bodies used as natural timestandards 
move in perfect (or very close) accord with absolute time.

The LMB relational answer is that all good clocks march in step with the emergent JBB time, which is then the time according to which the solar system motions are simplest; 
moreover, I argue against Barbour's subsequent explanation of this `marching in step' in Sec \ref{MIS-Demise}.  
I also further qualify that (LMB-CA relational answer) this is not in practise a perfect picture, partly because with the advent of relativity, time becomes position-dependent, 
and partly because $ t^{\se\sm(\sJ\sB\sB)}$ is not useable interchangeably with `the ephemeris' or `the most up-to-date astronomical timestandard'.  
Thus the relational statement is constrained to some accuracy, to something like `all good clocks on/near Earth approximately march in step with the astronomical timestandard that 
can be taken to approximate the emergent JBB time, by which timestandard the solar system's motions are simplest'. 
(Moreover, the scheme's validity will be enhanced by using SR and GR, foremost to model time dilation and gravitational redshift, but also 
further effects to higher precision; this is most naturally taken into account from a relational perspective by anchoring it to the GR $\mt^{\se\sm(\sJ\sB\sB)}$ 
rather than the mechanical one.)
Thus the relational set-up (of LMB or LMB-CA) is then superior through not involving an extra metaphysical entity or a postulate about how the 
celestial spheres then happen to have motions aligned with that metaphysical entity; moreover, the LMB-CA version more truly reflects 
the actual timekeeping procedure used.

\subsection{Heavy--light ($h$-$l$) splits}

\noindent We make an $h$-$l$ split is between heavy degrees of freedom and light degrees of freedom.
This is a classical parallel of the procedure in Molecular Physics by which one solves for the electronic structure under the approximation that the much heavier nuclei stay fixed, 
and of a technically similar approximation procedure from Semiclassical Quantum Cosmology.  
In general, more considerations enter `h--l splits' than just a mass ratio; at the classical level it concerns kinetic term ratio and ratios in the equations of motion.
In doing so, this SSec interpolates (Criterion 3) between the Classical Dynamics of Sec \ref{Cl-Soln} and the Semiclassical Approach to the POT in Sec \ref{Semicl}.

\subsubsection{Subsystem-wise unreduced $h$-$l$ split of RPM's}\label{RPM-H-L}

Assume that e.g. the $\underline{R}^i$  can be split into heavy coordinates $\underline{h}^{\ip}$  with $\ip  = 1$ to $p$ and masses $M_{\ip}$, and 
light coordinates $\underline{l}^{\ipp}$ with $\ipp = p + 1$ to $n$ and cluster masses $m_{\ipp}$ such that
\be
{m_{\ipp}}/{M_{\ip}} =: \epsilon_{\sh\si\se\sr} << 1 \mbox{ }  \mbox{ } \mbox{ ($h$-$l$ mass hierarchy) } .  
\ee
In making such a split\foo{It is to be understood
that the $\epsilon$'s in this Article are small quantities; I use these instead of `$<<$' to keep a more precise 
account of requisite inter-relations and rankings among these small quantities in the approximate approaches under investigation.  
A corresponding `gravitational mass hierarchy' sometimes invoked in motivating such approximations is $M_{\mbox{\scriptsize Planck}} >> M_{\mbox{\scriptsize inflaton}}$; 
another involves the single scale factor dominating over the anisotropic, inhomogeneous modes in GR cosmology.} 
\be
\stackrel{\mbox{\scriptsize max}}{\mbox{\scriptsize $\ip$, $\jp$}} {|M_{\ip} -  M_{\jp}|}/{M_{\ip}}     =: \epsilon_{\Delta \sM} << 1\mbox{ } , \mbox{ }  
\stackrel{\mbox{\scriptsize max}}{\mbox{\scriptsize $\ipp$, $\jpp$}}{|m_{\ipp} -  m_{\jpp}|}/{m_{\ipp}} =: \epsilon_{\Delta \sm} << 1 \mbox{ } .
\ee
\be
\frac{m_{\ipp}  }{  M_{\ip}  } = 
\frac{    \frac{m_{\ipp} - m}{m}  m + m    }{    \frac{M_{\ip} - M}{M}M + M     } = \frac{  \{ \frac{m_{\ipp} - m}{m}   + 1 \}m   }
                                                                                         {   \{\frac{M_{\ip} - M}{M} + 1\}M    } \mbox{ } \mbox{ } \sim \mbox{ } \mbox{ } 
\frac{m}{M}\left\{ 1 +  \frac{m_{\ipp} - m}{m} - \frac{M_{\ip} - M}{M} \right\}
\mbox{ } \mbox{ } \sim \mbox{ } \mbox{ } 
\epsilon_{\sh\si\se\sr} \{ 1 + O( \epsilon_{\Delta \sM}, \epsilon_{\Delta \sm} )\}
\ee
(the third move by the binomial expansion) allow for only one $h$-$l$ mass ratio to feature in subsequent approximations.

Then the classical scaled RPM action is (for its greater cosmological use and natural $h$ = scale split in the next SSec)\foo{
I denote the conjugate momenta by $\mP^{\th}_{i^{\prime}}$ and $\mP^{\tl}_{i^{\prime\prime}}$, and use  
$M_{\th}$, $N_{\th}$, $M_{\tl}$, ${N}_{\tl}$ as configuration space metrics and inverses for each of the $h$ and $l$ parts 
(assumed block-separable as in ordinary mechanics and relative mechanics in Jacobi coordinates).  
I use $E_{\th}$ because an energy-like separation constant $E_{\tl}$ will arise further on in the working.  
Then the fixed universe $E_{\tu\tn\ti} = E_{\th} + E_{\tl}$.}   
\be
\FS^{\sE\sR\sP\sM}_{\sJ\sB\sB} = \sqrt{2}\int 
\sqrt{E_{\sU\sn\si} - V_{h} - V_{l} - J}
\sqrt{||\d_{\underline{B}} \mbox{\boldmath$h$}||_{\mbox{\scriptsize\boldmath$M$}_{h}}\mbox{}^2 + 
||\d_{\underline{B}} \mbox{\boldmath$l$}||_{\mbox{\scriptsize\boldmath$M$}_{l}}\mbox{}^2   }
\mbox{ } ,  
\label{JHL}
\ee
\be
\mbox{ for } \mbox{ } V_{h}  = V_{h}(\underline{h}^{\jp}\cdot\underline{h}^{\kp}  \mbox{ alone} ) \mbox{ } , \mbox{ }
V_{l} = V_{l}(\underline{l}^{\jpp}\cdot\underline{l}^{\kpp} \mbox{ alone} ) \mbox{ } , \mbox{ }
J = J(\underline{h}^{\jp}\cdot\underline{h}^{\kp}, \mbox{ } \underline{l}^{\jpp}\cdot\underline{l}^{\kpp}, 
\mbox{ } \underline{h}^{\lp}\cdot\underline{l}^{\lpp} \mbox{ alone}) , 
\ee
(the `interaction potential' or `forcing term').  
The conjugate momenta are now 
\be
\underline{P}_{\ip}^{h}   = M_{\ip}  \Last_{\underline{B}}  \underline{h}^{\ip} 
\mbox{ } , \mbox{ }
\underline{P}_{\ipp}^{\sll} = m_{\ipp} \Last_{\underline{B}}  \underline{l}^{\ipp}  
\mbox{ } .  
\label{HLmom}
\ee
\mbox{ } \mbox{ } The classical energy constraint now splits into 
\be
\scE := \scE_{h} + \scE_{hl} = E_{h} \mbox{ } , \mbox{ } \mbox{ for } \mbox{ } 
\scE_{hl} := \scE_{l} + J 
\mbox{ } , \mbox{ }
\scE_{h} := ||\nP^{h}||_{\sbN_{h}}\mbox{}^2/2 + V_{h}
\mbox{ } , \mbox{ }
\scE_{l} := ||\nP^{l}||_{\sbN_{l}}\mbox{}^2/2 + V_{l} 
\mbox{ } . 
\ee
The classical zero angular momentum constraint likewise splits into  
\be
\underline\scL = {\underline\scL_{h}} + {\underline\scL_{l}} = 0 \mbox{ } , \mbox{ } \mbox{ for } 
\mbox{ }
{\underline\scL_{h}} = \sum\mbox{}_{\mbox{}_{\mbox{\scriptsize $\ip = 1$}}}^{p}\underline{h}^{\ip} \cr \underline{P}^{h}_{\ip}
\mbox{ } , \mbox{ } 
{\underline\scL_{l}} = \sum\mbox{}_{\mbox{}_{\mbox{\scriptsize $\ipp = p + 1$}}}^{n} \underline{l}^{\ipp} \cr \underline{P}^{l}_{\ipp}  \label{HLAM}
\ee
the $h$- and $l$-subsystems' angular momenta respectively.  
The evolution equations are 
\beq
\Last_{\underline{B}} P_{\ip\mu}^{h}   =  -\pa\{V_{h} + J\}/\pa h^{\ip\mu} 
\mbox{ } , \mbox{ }
\Last_{\underline{B}} P_{\ipp\mu}^{l} =  -\pa\{V_{l} + J\}/\pa  l^{\ipp\mu}  
\mbox{ } .
\eeq
The expression (\ref{JBB}) for emergent JBB time is now
\be
\lt^{\se\sm(\sJ\sB\sB)} = \stackrel{\mbox{\scriptsize extremum $\underline{B}$ of Rot($d$)}}{\mbox{\scriptsize of $\stS_{\mbox{\tiny JBB}}^{\mbox{\tiny ERPM}}$}}
\left(\int\sqrt{        \{   ||\d_{\underline{B}} \mbox{\boldmath$h$}||_{\mbox{\scriptsize\boldmath$M$}_{h}}\mbox{}^2 +  
                             ||\d_{\underline{B}} \mbox{\boldmath$l$}||_{\mbox{\scriptsize\boldmath$M$}_{l}}\mbox{}^2  \}/
                         {    2\{E_{\sU\sn\si} - V_{h} - V_{l} - J\} }            } \right) \mbox{ } . 
\label{mint}
\ee

\subsubsection{Scale--shape $h$-$l$ split of RPM's}\label{RPM-H-L-2}

This SSSec's particular $h$-$l$ split is aligned with shape--scale split of the RPM which has parallels with e.g. the scale--inhomogeneity split in GR 
via Analogies 15) and 36) and the Cosmology--Mechanics analogy of Sec \ref{Cl-Soln}.   

\mbox{ }   
  
\noindent The action is now 
\be
\FS^{\sE\sR\sP\sM}_{\sJ\sB\sB} = \sqrt{2}\int 
\sqrt{E_{\sU\sn\si} - V_{h} - V_{l} - J}
\sqrt{\d h^2 + h^2 ||\d \mbox{\boldmath$l$}||_{\mbox{\boldmath\scriptsize$M$}_{l}}\mbox{}^2   }
\mbox{ } ,  
\label{JHL2}
\ee
(with $\underline{B}$'s hung on the d$l$'s in the uneliminated case) for
\be
V_{h}  = V_{h}( \mh\mbox{ alone} ) = V_{\sigma}(\sigma)
\mbox{ } , \mbox{ }
V_{l} = V_{l}(l^{\sfa} \mbox{ alone} ) = V_{\sS}(\mS^{\sfa} \mbox{ alone} ) \mbox{ } , \mbox{ }
J = J(h, l^{\sfa} \mbox{ alone})= J_{\sigma\sS}(\sigma, \mS^{\sfa} \mbox{ alone}) \mbox{ } . 
\ee
The conjugate momenta are now (with $\Gamma = i\mu$ and a $\underline{B}$ hung on each $\Last l$ in the uneliminated case and $\Gamma = \fa$ in the r-case)
\be
P^{h}   = \Last h^{\ip\mu} 
\mbox{ } , \mbox{ }
P_{\Gamma}^{l} = h^2 M_{\Gamma\Lambda}\Last l^{\Lambda}  
\mbox{ } .  
\label{HLmom2}
\ee
\mbox{ } \mbox{ } The classical energy constraint is now  
\be
\scE := P_{h}^2/2 + ||\mbox{\boldmath$P$}_{l}||_{\mbox{\boldmath\scriptsize$N$}_{l}}\mbox{}^2/2 h^2 + V_{h} + V_{l} + J = E_{\sU\sn\si} 
\label{yet-another-E}
\ee
(alongside, for the uneliminated case 
\be
{\underline\scL_{l}} = \sum\mbox{}_{\mbox{}_{\mbox{\scriptsize $\sfa = 1$}}}^{nd - 1} 
\underline{l}^{\sfa} \cr \underline{P}^{l}_{\sfa}  \label{HLAMSha} \mbox{ } ).
\ee
The evolution equations are [in the same notation as eq. (\ref{HLmom2})] 
\beq
\Last P^{h}   =  h||\Last l||_{\mbox{\boldmath\scriptsize$M$}_l}^2 -\pa\{V_{h} + J\}/\pa h 
\mbox{ } , \mbox{ }
\Last P_{\Gamma}^{l} =  h^2 M_{\Lambda\Sigma,\Gamma} \Last l^{\Lambda}\Last l^{\Sigma} - \pa\{V_{l} + J\}/\pa l^{\Gamma}  
\mbox{ } .
\eeq
\noindent We can treat (\ref{yet-another-E}) in Lagrangian form as an equation for $t^{\se\sm(\sJ\sB\sB)}_{(0)}$ itself.
\noindent In this classical setting, it is coupled to the $l$-equations of motion (and we need the $h$-equation to judge which terms to keep).
For more than one $h$ degree of freedom, there is separate physical content in these.    
\noindent The system is in general composed of the $E$-equation, $k_{h}$ -- 1 $h$-evolution equations and $k_{l}$ $l$-evolution equations system.  
  
\noindent The expression (\ref{JBB}) for emergent JBB time candidate is now (with the $\underline{B}$'s and extremization thereover {\sl absent} in the eliminated case)
\be
\lt^{\se\sm(\sJ\sB\sB)} = \stackrel{\mbox{\scriptsize extremum $\underline{B}$ of Rot($d$)}}
                                                            {\mbox{\scriptsize of $\stS_{\mbox{\tiny JBB}}^{\mbox{\tiny ERPM}}$}}
\left(\int\sqrt{           \{   \d h^2 +  h^2||\d_{\underline{B}} \mbox{\boldmath$l$}||_{\mbox{\boldmath\scriptsize$M$}_l}\mbox{}^2  \}/
                       {    2\{E_{\sU\sn\si} - V_{h} - V_{l} - J\}  }            } \right) 
\mbox{ } .  \label{TorreBruno}
\eeq
(This feature carries over to GR too, via $\mh_{\mu\nu} = \ma^2\muu_{\mu\nu}$ leading to $\{\d - \pounds_{\d{\suF}}\}\{\ma^2\muu_{\mu\nu}\} = \ma^2\{\{{\d \ma}/{\ma}\}\muu_{\mu\nu} + 
\d \muu_{\mu\nu} - \mD^{u}\mbox{}_{(\mu}\d{\mF}_{\nu)} + 0\} = \ma^2\{\d - \pounds_{\d{\suF}}\}\muu_{\mu\nu}$, the 0 arising from the constancy in space of the 
scalefactor-as-conformal-factor killing off the extra conformal connection; here $\mD_{\mu}$ is the covariant derivative associated with $\muu_{\mu\nu}$).       
By this observation, scale--shape split approximate JBB time (and approximate WKB time which coincides with it) avoids the Sandwich/Best Matching Problem.

\subsection{How to use $t^{\se\sm(\sJ\sB\sB)}$ approximately: look at level of the equations of motion} \label{temJBB-Approx}

\noindent It turns out that the JBB time formula is {\sl not} directly implementable.
This does not refer to whether the Best Matching extremization is explicitly solvable. 
Rather, it refers to how approximate use requires approximations to be crude at the level of the equations of motion and not at the level of $\d s$ and $V$.

To fulfil the true content of LMB-CA, all change is given opportunity to contribute to the timestandard. 
However only changes that do so in practise to within the desired accuracy are actually kept. 
This means we need a curious indirect procedure in making such an approximation.  
I.e., one can not simply compare the sizes of the various energy terms, but must rather assess this at the level of the resulting force terms upon variation.

\subsubsection{Subsystem split case}

That procedures at the levels of the equations of motion and at the level of the action are very different 
and that one must do the latter are well brought out by the `Earth--Sun--Andromeda' 3-body problem \cite{SemiclI, ARel2}.  
In $V$, Andromeda being relatively far is rather offset by Andromeda being relatively massive (one is  
comparing $m_{\sA}/|R_{\sE\sA}|$ and $m_{\sS}/|R_{\sE\sS}|$: a ratio of $m/r$ terms).
However, at the level of the equations of motion, the equation for the Earth--Sun separation vector $\rho_{\sE\sS}$ 
gives tidal-type terms for the Earth--Sun--Andromeda system ($m_{\sA}/|R_{\sE\sA}|^3$ and $m_{\sS}/|R_{\sE\sS}|^3$: a ratio of $m/r^3$ terms).  
Now Andromeda being relatively far away very heavily outbalances Andromeda being relatively massive.
I.e., whilst the potential due to Andromeda is {\sl felt} by solar system objects, it is felt {\sl extremely evenly} by them 
all, and thus does not play any appreciable role in the physics of the Solar System itself.

\subsubsection{$h$-approximation version of subsystem split}

We throw away what is sufficiently $l$ and/or away from the subsystem by force-level considerations.

The $h$-approximation to the action (\ref{JHL}) is 
\beq
\FS^{\sE\sR\sP\sM}_{\sJ\sB\sB(\sh)} = \sqrt{2}
\int   ||\d_{\underline{B}}\mbox{\boldmath$h$}||_{\mbox{\scriptsize\boldmath$M$}_{h}}\sqrt{\{E_{h} - V_{h}\}}  \mbox{ } .
\eeq
For this, the conjugate momenta are 
\beq
\underline{P}^{h}_{i^{\prime}} = M_{i^{\prime}}\Last^{h}_{\underline{B}} \underline{h}^{i^{\prime}}  \mbox{ } ,
\eeq
the constraints are 
\beq
\scE_{h} = 0 \mbox{ } , \mbox{ } \mbox{ } \underline\scL_{h} = 0 \mbox{ } 
\eeq
and the evolution equations are 
\beq
\Last^{h}_{\underline{B}}\underline{P}_{\ip}^{h}   =  -\pa V_{h}/\pa\underline{h}^{\ip} \mbox{ } . \mbox{ }
\eeq
This set-up is assuming that  
\beq
|{\pa J}/{\pa h}| << |{\d V_{h}}/{\d h}|
\label{balanci} \mbox{ } , 
\eeq
which enables a closed-$h$ subsystem (i.e. decoupled from l-subsystem to this level of approximation).
$\Last^{h} := \pa/\pa t_{h}^{\se\sm(\sJ\sB\sB)}$ corresponding to [c.f. (\ref{mint})] 
\be
\lt^{\se\sm(\sJ\sB\sB)}_{h} = \stackrel{\mbox{\scriptsize extremum $\underline{B}$ of Rot($d$)}}
{\mbox{\scriptsize of $\stS_{\mbox{\tiny JBB($h$)}}^{\mbox{\tiny ERPM}}$}}  
\left(\int   ||\d_{\underline{B}}\mbox{\boldmath$h$}||_{\mbox{\scriptsize\boldmath$M$}_{h}}/\sqrt{2\{E_{h} - V_{h}\}} \right) \mbox{ } , 
\label{hint} 
\ee
Finally, the first approximation to the l-equations is then 
\beq
P_{\ipp\mu}^{l} = m^{\ipp\jpp\mu\nu} \Last^{h}_{\underline{B}} l_{\jpp\nu} 
\mbox{ } , \mbox{ } \mbox{ }  
\Last^{h}_{\underline{B}} P_{\ipp\mu}^{l} =  -\pa\{V_{l} + J\}/\pa l^{\ipp\mu}  
\mbox{ } . 
\eeq

\subsubsection{$h$ = scale approximation}

The $h$-approximation to the action (\ref{JHL2}) is 
\beq
\FS^{\sE\sR\sP\sM}_{\sJ\sB\sB(\sh)} = \sqrt{2}\int \sqrt{\{E_{h} - V_{h}\}}\d h  \mbox{ } .
\eeq
Then the conjugate momenta are 
\beq
P^{h} = \Last^{h} h  \mbox{ } ,
\label{666}
\eeq
the constraint is  
\beq
\scE_{h}:= P^{h\, 2}/2 + V_{h} = E_{h} \mbox{ } 
\label{667}
\eeq
and the evolution equations are 
\beq
\Last^{h} P^{h}   =  -\pa V_{h}/\pa h^{\ip\mu} 
\mbox{ } . 
\eeq
This assumes (\ref{balanci}) [i.e. $\epsilon_{\sss\sd\sss 2}$ of Sec \ref{Cl-Soln}] and now also 
\beq
||\d S||^2_{\mbox{\boldmath\scriptsize$M$}} << {\sigma}^{-1} {d V_{\sigma}}/{\d \sigma}
\eeq
(or compare it to $\Last^{h}\Last^{h}\sigma$ to make the $\epsilon_{\sss\sd\sss 1}$ of Sec \ref{Cl-Soln}).

(\ref{667}) can be taken as an equation for $t^{\se\sm(\sJ\sB\sB)}$ via momentum--velocity relation (\ref{666}) [and this follows suit in the multi-$h$ case]. 
$\Last^{h} := \pa/\pa t^{\se\sm(\sJ\sB\sB)}_{h}$, now corresponding to [c.f. (\ref{TorreBruno})].
The approximate emergent JBB time candidate is then
\be
\lt^{\se\sm(\sJ\sB\sB)}_{h} = 
\left.
\int   \d h 
\right/
\sqrt{2\{E_{h} - V_{h}\}}  \mbox{ } ,
\label{hint2} 
\ee
which is of the general form 
\beq
\lt^{\se\sm(\sJ\sB\sB)} = {\cal F}[h, \d h] \mbox{ } . 
\label{hdh}
\eeq
N.B. that for this split and to this level of approximation, there is no $\FrG$-correction to be done, 
because the rotations act solely on the shapes and not on the scale; in other words Configurational Relationalism is trivial here.  
The GR counterpart of this constitutes Analogy 71.

Finally, the first approximation to the l-equations is then 
\beq
P^{l}_{\sfa}        =  h^2 \ttM_{\sfa\sfb} \Last^{h}l^{\sfb}
\Last P_{\sfa}^{l}   =  h^2 \ttM_{\sfa\sfb,\sfc} \Last l^{\sfb}\Last l^{\sfc} - \pa\{V_{l} + J\}/\pa l^{\sfa}  
\mbox{ } , 
\eeq
with the same notational interpretation as in Sec \ref{RPM-H-L-2}. 

\mbox{ } 

\noindent Note 1) One can see in this case that the effect of $h$ = size of the universe is not negligible like for $h$ = distance to Andromeda, 
since $h$ = scale enters the force law in a distinct homogeneous way. 

\noindent Note 2) The scale--shape oriented $h$--$l$ split is different in this regard.  
For mechanical models, this scale is the total moment of inertia of the universe, to which all constituent parts of the universe contribute.  
For GR, the situation is more delicate: the scalefactor is indeed now a single gravitational--geometric variable, 
but on the other hand the form it takes is determined by solving field equations that contain averaged matter density 
terms to which all of the contents of the universe contribute.  
In each case, the scale in question then gives rise to the approximate emergent JBB time.

\subsubsection{Extension to the case of multiple $h$ degrees of freedom}\label{Multi-h} 

This applies e.g. to the subsystem-wise split, but also to inhomogeneous GR with homogeneous matter modes and/or anisotropy among the $h$ degrees of freedom.

\mbox{ } 

\noindent Here, we have $t^{\se\sm(\sJ\sB\sB)} = t^{\se\sm(\sJ\sB\sB)}[h_1, h_2, ...]$ as an arc-element integral; 
one can then introduce a physically meaningless label for the practical purposes of performing the integration.
In this case, one should remember that one has also $h$-evolution equations - one less than $h$-degrees-of-freedom.
Assuming that one can decouple these also from the $l$-subsystem, one has at least the right degrees-of-freedom count to subsequently obtain the 
more conventional independent-variable formulation (Sec \ref{NIAG}): $h_1(t^{\se\sm(\sJ\sB\sB)})$, 
$h_2(t^{\se\sm(\sJ\sB\sB)})$, ... by which the feed into the $l$-part of the procedure is of the same qualitative form as before.

\mbox{ }

\noindent N.B. Whenever we get disagreement with experiment, going back to the first, chroniferous formulation should be perceived as a possible option 
(early 20th century `lunar anomalies' being the archetype as per Sec \ref{Cl-POT}).

\subsubsection{Expansion of the isolated emergent-time equation}

\noindent  Pure-$h$ expressions of the general form (\ref{hdh}) are unsatisfactory from a Machian perspective since they do not give $l$-change an opportunity to contribute.  
This deficiency is to be resolved by treating them as zeroth-order approximations in an expansion involving the $l$-physics too.
Expanding (\ref{TorreBruno}), one obtains an expression of the form 
\beq
\lt^{\se\sm(\sJ\sB\sB)}_{(1)} = {\cal F}[h, l, \d h, \d l] \mbox{ } .  
\label{callie}
\eeq
More specifically, 
\beq
\lt^{\se\sm(\sJ\sB\sB)}_{(1)} = \lt^{\se\sm(\sJ\sB\sB)}_{(0)} + \frac{1}{2\sqrt{2}}
\left\{
\int\frac{J_{\rho S}\d \rho}{W_{\rho}^{3/2}} + 
\int\frac{\d\rho}{\sqrt{2W_{\rho}}} 
\left\{
\frac{\d S }{\d \, \mbox{ln} \,\rho}
\right\}^2 
\right\} 
+ O\left(\left\{\frac{J_{\rho S}}{W_{\rho}}\right\}^2\right) + O\left(\left\{\frac{\d S}{\d \,\ml\mn\,\rho}\right\}^4\right) 
\label{Cl-Expansion}
\eeq
so one has an interaction term and an $l$-change term.

\mbox{ }

\noindent The $l$-change term in a simplified GR version of this working reveals a Machian 1 part in $10^{10}$ limit on the accuracy possible for cosmic time as currently standardly 
conceived of (`ephemeris cosmic time effect').
It is certainly conceivable that effects due to other reasons come out at 1 part in $10^5$ (the characteristic number from the inhomogeneity case of scale-dominates-shape as per 
Sec \ref{SSA}) so that the above Machian effect would only show up in the balance of effects at second order.  
It is also relevant to note here that Hubble's constant is presently only claimed to be known to 1 part in $10^2$; for some purposes this bounds accuracy on cosmic time too; 
moreover, this is a bound due to observational limitations, which space-based missions are in the process of significantly improving.  

\mbox{ } 

\noindent This analysis is limited by how the correction terms are themselves determined by solving further equations, so that the emergent-time equation is part of a coupled system.
However the general form (\ref{callie}) itself is unaffected by this coupled nature.
The following perturbative scheme is a simple example of a scheme that takes this further feature into account.

\subsubsection{First Approximation: Machian Classical Scheme}

\noindent Recollecting that one judges what terms one is to keep in practise at the level of the equations of motion rather than at the level of the potential \cite{ARel2}, the idea is 
then to perturbatively expand the energy equation, $l$-evolution equations and $h$-evolution equations (purely as an ancillary judging equation in the case of 1 $h$ degree of freedom).   

\mbox{ } 

\noindent N.B. 1) For the energy equation to serve as a chroniferous equation, the analysis must be carried out at in Lagrangian variables, 
and, for judgement of which contributions are physically negligible, at the level of the equations of motion. 
Contrast with how most classical and essentially all quantum perturbation theory are carried out at the Hamiltonian level.  

\noindent N.B. 2) Moreover, once one has found an accurate enough time for one's purposes, one can of course revert to a Hamiltonian analysis for features within that universe that are 
fine enough to not contribute relevant change to the timestandard; this is very much expected to cover all uses of QM perturbation theory that apply to modelling laboratory experiments.
Thus what is being developed here is a Semiclassical Quantum Cosmology analogue of the astronomers' ephemeris time procedure, in which fairly large-scale 
features of the Universe are expected to contribute a bit in addition to the zeroth-order expansion of the universe and homogeneous-matter-mode contributions.  
 
\noindent N.B. 3) There is a limit on how from ephemeris time schemes themselves, since the iterations in those were at a level of form-fitting rather than a perturbative expansion of 
the equations of motion themselves.  
However, general situations (GLET's, and at the quantum as well as classical level) necessitate a more general analysis.
Though we do not discard the possibility of being able to do form-fitting for the most practically important cosmological situation of an approximately-FLRW universe 
that models our own observed universe.

\subsubsection{On assessing Semiclassical Quantum Cosmology's approximations}

\noindent This SSSec needs to consider the classical counterpart of Semiclassical Quantum Cosmology's status quo before that has in detail been introduced (see Sec \ref{Semicl}). 
Moreover, doing so reveals some pretty glaring discrepancies with conventional practise in the far more carefully studied and experimentally tested arena of classical dynamics.
These do {\sl not} concern the Machianization of Semiclassical Quantum Cosmology itself (which is a constructive import from Dynamics and Astronomy to {\sl whatever form 
Semiclassical Quantum Cosmology should take}), but rather some of the plethora of approximations conventionally made to simplify the Semiclassical Quantum Cosmology equations.  
In this way, this Sec provides one of the slices of `absurdum' that Sec \ref{Semicl} subsequently uses to cast `Reductio ad absurdum' on Semiclassical Quantum Cosmology's status quo.
Thus my temporary entertainment of a Classical Dynamics anathema in this Sec is justified by the same anathema lurking unidentified among the currently accepted Semiclassical 
Quantum Cosmology approximations; this is not highly unexpected because far less work has been done with the latter, especially not in robustness analyses involving not 
making some of the simplifying approximations or in analyzing those approximations within toy model simplifications.  
Essentially, the present Sec and Sec \ref{Semicl} cast Classical Dynamics itself as a suitable toy model for Semiclassical Quantum Cosmology, and this has adverse consequences 
for the justifiability of the latter's highly-approximated, simplified analytically-tractable form.   
The conclusion, for now at the qualitative level, is that Semiclassical Quantum Cosmology looks to need a more highly coupled system of equations that constitute a {\sl numerical},  
rather than analytic, subject. 

\mbox{ } 

\noindent What is the specific classical-level problems have I noticed?  
In addition to the two-body within 3-body issue in Sec \ref{Stab-ClSol}, the present SSec concerns what Semiclassical Quantum Cosmology's neglect of the $T_l$ term implies at 
the classical level.  
Firstly, the presumed motivation for this approximation is as part of decoupling the $h$ and $l$ subsystems, which contributes to rendering them easier to solve analytically. 
But then the classical dynamics version of this (shape--scale split 1- or 2-$d$ RPM version) involves {\bf throwing away the central term}.
And that if course completely changes the character of the solution -- it is best-known as the difference between rectilinear motion's escape to infinity and ellipses' confinedness and 
periodicity in the basic Keplerian analysis of planetary motion. 
And this qualitative difference emblazoned in keeping or ditching the central term indeed carries over to RPM's with relative subsystem angular momentum conservation in place of 
Keplerian angular momentum conservation since these two problems share the same mathematics as established in Secs \ref{Dyn1}, \ref{Cl-Soln}.  
[This analogy is especially direct for 3-, 4-stop metroland and triangleland, though the qualitative significance of central terms does pervade $N$-stop metroland and $N$-a-gonland, 
as more detailedly exemplified in the current Article by quadrilateralland's possessing two independent central terms.]

\subsubsection{Modelling assumptions for the perturbative Classical Machian Scheme}\label{AppA}

\noindent 1) A key feature is that what is conventionally an `independent variable' $t$, is here a $t^{\se\sm(\sJ\sB\sB)}$ in the chronifer interpretation, i.e. 
cast as a highly dependent variable.  
As such it is clear that this itself is to be subjected to perturbations, whereas the conventional $t$ itself is not.  

\noindent 2) Due to the way relevant change is to be judged, we strongly want to keep $\pa_{l}V_{l}$ but not $V_{l}$, and we want to judge $J$ partly via $\pa_{h}J$.

\noindent 3) We need a vector's worth of $\epsilon$'s, $\underline{\epsilon}$ rather than a single small parameter, since we have multiple a-priori independently small quantities.
This is more usual in Theoretical Physics than the previous specific point, e.g. $\lambda \varsigma^3 + \mu \varsigma^4$ interaction potential in toy-model QFT of a scalar field 
$\varsigma$.
The general case will become further complicated if some $\epsilon$'s can be small that they are around the size of nontrivial powers of other $\epsilon$'s.
Three regimes of particular tractability are $\underline{\epsilon} = (0, ..., 0, \epsilon, 0, ... 0)$ (approximation by single-$\epsilon$) and 
$\underline{\epsilon} = (\epsilon, \epsilon , ... \epsilon)$ (`$\epsilon$-democracy', which further models all $\epsilon$'s being {\sl roughly} 
the same size) and the partial vector of same-$\epsilon$ on $> 1$ entries.  

\mbox{ }  

Note: a given $\epsilon$ can be forced to be big by circumstance; then one has a perturbation scheme with one $\epsilon$ less, though the awareness and formalism remain similar.  
Some papers \cite{KS91, BK97, Kiefersugy} investigate Quantum Cosmology by expanding in 1 parameter.  
Padmanabhan \cite{Pad} made a point of there being multiple parameters, though he proceeded by considering which parameter to expand in, whereas I pointed out \cite{SemiclI, SemiclII}, 
rather, that 1-parameter expansions in no matter what parameter will not in general suffice for beyond a corner of the Quantum Cosmology solution space.  
The present Article and \cite{ACos2} then systematize the treatment of this.

\mbox{ } 

\noindent All in all, we take 
\beq
Q^{\sfA} = Q^{\sfA}_{(0)} + \underline{\epsilon}\cdot\underline{Q}^{\sfA}_{(1)} + O(\epsilon^2) \mbox{ } , 
\label{A-1}
\eeq
\beq
t^{\se\sm} = t^{\se\sm}_{(0)} + \underline{\epsilon}\cdot\underline{t}^{\se\sm}_{(1)} + O(\epsilon^2) \mbox{ } ,
\label{A-2}
\eeq
though the $\underline{\epsilon}$ is taken to originate from the primed expressions for the energy equation (\ref{E-prime}) and classical $l$-equations of motion (\ref{l-EOM-prime}).  
Then 

\mbox{ }

\noindent {\bf Lemma 14 (A and Ip)}
\beq
\Last =   \{ 1 - \underline{\epsilon}  \cdot \Last_{(0)}\underline{t}^{\se\sm}_{(1)}  \}  \Last_{(0)} + O(\epsilon^2) \mbox{ } ,
\eeq
\beq
\Last^2 = \{ 1 - \underline{\epsilon}  \cdot \Last_{(0)}\underline{t}^{\se\sm}_{(1)}  \}^2\Last_{(0)}^2 - 
          \underline{\epsilon}  \cdot \Last_{(0)}^2\underline{t}^{\se\sm}_{(1)} \Last_{(0)} + O(\epsilon^2) 
\mbox{ } .   
\eeq
Next, one's classical system is 
\beq
h^{\SStar\SStar} = \epsilon_{-1} h^2 M_{\sfb\sfc} l^{\sfb \SStar} l^{\sfc \SStar} - {\pa V_{h}}/{\pa h} - \epsilon_{0} {\pa J^{\prime}_{hl}}/{\pa h} 
\eeq
\beq
h^{\SStar\, 2} + \epsilon_1 h^2 M^{\prime}_{\sfb\sfc} l^{\sfb \SStar} l^{\sfc \SStar} = 2\{E - V_{h} - \epsilon_2 V_{l}^{\prime} - \epsilon_3 J^{\prime}_{hl}\}  \mbox{ } ,
\label{E-prime}
\eeq
\beq
l^{\sfa \SStar\SStar} + {\Gamma^{\sfa}}_{\sfb\sfc} l^{\sfb \SStar} l^{\sfc \SStar} + {2 h^{\SStar} l^{\sfa \SStar}}/{h} = - h^{-2}N^{\sfa\sfb}
\left\{
{\pa V_{l}}/{\pa l^{\sfb}} + \epsilon_4 {\pa J^{\prime}}/{\pa l^{\sfb}}
\right\} \mbox{ } .  
\label{l-EOM-prime}
\eeq
\noindent Note 1) in absence of $V_{l}$, using an $\epsilon_4$ may be undesirable.
for more than 1 $h$ degree of freedom, $\epsilon_{-1}$ and $\epsilon_0$ are not just judging equations but enter the system, so we have a string of six $\epsilon$'s.

\noindent Note 2) We {\sl decide} to take $h$ as heavy on the basis of the size of $\epsilon_0$, it is $\epsilon_3$ itself that enters the subsequent working.  
Thus classically we have a $\underline{\epsilon}$ of length 4: small $l$-kinetic term, small $l$-potential term, small interaction potential and small interaction force on the $l$-system.

\subsubsection{Interpretation of the six $\epsilon$'s}\label{Cl-adiab}

The cofactor of $\epsilon_1$ is an adiabatic term, related to the order of magnitude estimate    
\be
\epsilon_{\sA\sd} := {\omega_{h}}/{\omega_{l}} = t_{l}/t_{h} 
\ee
for $\omega_{h}$ and $\omega_{l}$ `characteristic frequencies' of the $h$ and $l$ subsystems respectively.   
The cofactor itself the square of the scale-dominates-shape approximation's term \{d shape/ d(ln(scale)\}. 
\noindent $\epsilon_{-1}$ is dimensionally the same, but sometimes whether each can be kept is independent, through $h\, h^{\SStar\SStar}$ not being the same as $h^{\SStar\, 2}$.
These are the two terms that the central term example illustrates have sometimes to be kept to leading order rather than epsilonized.  

\noindent $\epsilon_2$ is $l$ subdominance to $h$ in the potential, i.e. $|V_{h}/V_{l}|$ small.

\noindent $\epsilon_3$, $\epsilon_4$ and $\epsilon_0$ and  are weaknesses of interaction both at level of potential and at the level of the relative forces: 
$|J/V_{h}|$ small, 
$|\pa J/\pa h \, / \, \pa V_{h}/\pa h|$ small and 
$|\pa J/\pa l \, / \, \pa V_{l}/\pa l|$ small.

\subsubsection{Zeroth-order equations}

Applying (\ref{A-1}) for $Q^{\sfA}$ = $h$, $l^{\sfc}$ and (\ref{A-2}) and Taylor-expand $M$, $N$, $\Gamma$, $V_{h}$, $V_{l}$, $J$ and their derivatives
gives back to zeroth order the expected equations [now laced with (0) labels], 
\beq
h_{(0)}^{\SStar_{(0)}\SStar_{(0)}} =  - \pa V_{h}(h_{(0)})/\pa h_{(0)} \mbox{ } , 
\eeq
\beq
h_{(0)}^{\SStar_{(0)} 2} = 2\{E - V_{h}(h_{(0)})\}  \mbox{ } ,
\label{E-zero}
\eeq
\beq
l_{(0)}^{\sfa \SStar_{(0)}\SStar_{(0)}} + {\Gamma^{\sfa}}_{\sfb\sfc}\big(l_{(0)}^{\sfp}\big)l_{(0)}^{\sfb \SStar_{(0)}}l_{(0)}^{\sfc \SStar_{(0)}} + 
{2h_{(0)}^{\SStar_{(0)}}l_{(0)}^{\sfa \SStar_{(0)}}}/{h_{(0)}} = - h_{(0)}^{-2}{N^{\sfa\sfb}\big(l^{\sfp}_{(0)}\big)}
{\pa V_{l}(l^{\sfp}_{(0)})}/{\pa l^{\sfb}_{(0)}} 
\mbox{ } .  
\label{l-EOM-zero}
\eeq

\subsection{The first-order equations in detail}

\noindent These are (A and Ip)
\beq
\underline{\epsilon} \cdot 
\left\{  
h_{(0)}^{\SStar_{(0)}} \{ \underline{h}_{(1)} - 
                            \underline{t}_{(1)}^{\SStar_{(0)}}h_{(0)}^{\SStar_{(0)}} \} + 
\underline{h}_{(1)} {\pa V_{h}(h_{(0)})}/{\pa h_{(0)}}
\right\}
+ {\epsilon_{1}}h_{(0)}^2 M_{\sfc\sfd}^{\prime}\big(l_{(0)}^{\sfp}\big)l_{(0)}^{\sfc \SStar_{(0)}}l_{(0)}^{\sfd \SStar_{(0)}}/2 
+ \epsilon_2V_{l}^{\prime}\big(l_{(0)}^{\sfc}\big) + \epsilon_3 J^{\prime}\big(h, l_{(0)}^{\sfc}\big) \mbox{ } ,   
\label{E-1}
\eeq
$$
\underline{\epsilon}\cdot
\left\{
\underline{l}_{(1)}^{\sfb \SStar_{(0)} \SStar_{(0)}} - \underline{t}_{(1)}^{\SStar_{(0)} \SStar_{(0)}}l_{(0)}^{\sfb \SStar_{(0)}} - 
2 l_{(0)}^{\sfb\SStar_{(0)}\SStar_{(0)}}\underline{t}_{(1)}^{\SStar_{(0)}} + 
\underline{l}_{(1)}^{\sfp} \frac{\pa \Gamma^{\sfb}_{\sfa\sfc}\big(l_{(0)}^{\sfd}\big)}{\pa l^{\sfp}_{(0)}}
l_{(0)}^{\sfa \SStar_{(0)}}l_{(0)}^{\sfc \SStar_{(0)}} + 2\Gamma^{\sfb}_{\sfa\sfc}\big(l_{(0)}^{\sfd}\big) l_{(0)}^{\sfc \SStar_{(0)}}
\{\underline{l}_{(1)}^{\sfa \SStar_{(0)}} - \underline{t}_{(1)}^{\SStar_{(0)}}l_{(0)}^{\sfa \SStar_{(0)}}\} 
\right.
$$
$$
+ \frac{2}{h_{(0)}} 
\left\{
\underline{l}_{(1)}^{\sfb \SStar_{(0)}}h_{(0)}^{\SStar_{(0)}}  + l_{(0)}^{\sfb \SStar_{(0)}}\underline{h}_{(1)}^{\SStar_{(0)}} - l_{(0)}^{\sfb \SStar_{(0)}}h_{(0)}^{\SStar_{(0)}}
\left\{  
\frac{\underline{h}_{(1)}}{h_{(0)}} + 2\underline{t}_{(1)}
\right\}   
\right\}
\left. 
- \frac{N^{\sfa\sfb}\big(l_{(0)}^{\sfq}\big)}{h_{(0)}^2} 
\left\{ 
\frac{2\underline{h}_{(1)}}{h_{(0)}}\frac{\pa V_{l}(l_{(0)}^{\sfq})}{\pa l^{\sfa}_{(0)}}  - 
\underline{l}_{(1)}^{\sfp} \frac{\pa^2V_{l}(l^{\sfq}_{(0)})}{\pa l^{\sfa}_{(0)}\pa l^{\sfp}_{(0)}}
\right\}
\right.
$$
\beq
\left.
+ \frac{\underline{l}_{(1)}^{\sfc}}{h_{(0)}^2}  
\frac{\pa N^{\sfb\sfa}\big(l^{\sfq}_{(0)}\big)}{\pa l^{\sfc}_{(0)}}
\frac{\pa V_{l}\big(l^{\sfq}_{(0)}\big)}{\pa l^{\sfa}_{(0)}}
\right\}  
= - \epsilon_4 \frac{N^{\sfa\sfb}\big(l^{\sfq}_{(0)}\big)}{h_{(0)}} 
\frac{\pa J(h_{(0)}, l^{\sfq}_{(0)})}{\pa l^{\sfa}_{(0)}}
\label{l-EOM-1}  \mbox{ } .  
\eeq
Note 1) one cannot just cancel the epsilons off in general case (unlike for schemes with just the one $\epsilon$).  

\noindent Note 2) This system is indeed is well-determined.
One can take the quantities to be solved for at each step to be as follows. 
\noindent $h(t^{\se\sm(\sJ\sB\sB)})$ = I($t^{\se\sm(\sJ\sB\sB)}$), $l^{\sfa}$($t^{\se\sm(\sJ\sB\sB)}$) = S$^{\sfa}$($t^{\se\sm(\sJ\sB\sB)}$)
\noindent Then $t^{\se\sm(\sJ\sB\sB)}$($h$, $l^{\sfa}$) = $t^{\se\sm(\sJ\sB\sB)}$(I, S$^{\sfa}$).  
\noindent This transcends to geometrodynamics.  
\noindent It is then open to investigation using perturbation theory.
\noindent $\underline{t}^{\se\sm(\sJ\sB\sB)}_{(1)}$ is not a separate entity but rather abstracted from $\underline{h}_{(1)}$ (and $t^{\se\sm(\sJ\sB\sB)}_{(0)}$ or $h_{(0)}$).

\subsection{RPM examples of Machian Classical Scheme}

\subsubsection{3-stop metroland example}

The action is 
\beq
\FS = \sqrt{2}\int \sqrt{\d\rho^2 + \rho^2 \d\varphi^2}\sqrt{E - A\rho^2 - B\rho^2\mbox{}\mbox{cos}\,2\theta}
\eeq
where I am using an HO potential which takes the given form once expressed in the scale--shape coordinates $\rho$, $\varphi$.  
Set I = $h$ and $\varphi$ = $l$. 
Note that this example simplifies by having no $V_{l} = V_{\theta}$ and hence no $\epsilon_2$.

\noindent The $\epsilon$-scheme is now:
\beq
h^{\SStar\SStar} = \epsilon_{-1} h\{\theta^{\SStar\,2}/\epsilon_{-1}\} + 2h\{A + \epsilon_0 B^{\prime}\mbox{cos}\,2l\}
\label{h-prime-3}
\eeq
\beq
h^{\SStar\, 2} + \epsilon_1\{h^2\theta^{\SStar\,2}/\epsilon_1\} = 2\{E - Ah^2 - \epsilon_3 B^{\prime}h^2\mbox{cos}\,2l\}  \mbox{ } ,
\label{E-prime-3}
\eeq
\beq
l^{\SStar\SStar} + 2{h^{\SStar}\theta^{\SStar}}/{h}  = - 2\epsilon_4\,B^{\prime\prime}\,\mbox{sin}\,2l \mbox{ } .  
\label{l-prime-3}
\eeq
\mbox{ } \mbox{ } The zeroth order then gives back the force term judging equation 
\beq
h_{(0)}^{\SStar_{(0)}\SStar_{(0)}} = 2A\,h_{(0)}
\eeq
and the system 
\beq
h_{(0)}^{\SStar_{(0)}\, 2}  = 2\{E - A\,h_{(0)}^2\}  \mbox{ } ,
\label{E-zero-3}
\eeq
\beq
h^{(0)\, 2}\theta^{\SStar} = \sfD  \mbox{ } .  
\label{l-zero-3}
\eeq
Here, a first integral has been performed on the last equation.

\mbox{ } 

\noindent Note 1) This example serves to illustrate the aforementioned  with neglecting the $l$-kinetic term: 
had that been kept, (\ref{l-zero-3}) can then be used to provide an $h$-equation of a qualitatively distinct form, 
\beq
h_{(0)}^{\SStar\, 2} + \sfD^2/h_{(0)}^2 = 2\{E - Ah_{(0)}^2\}  \mbox{ } ,
\eeq
the difference in the outcome at the level of the shapes of the orbits being manifest in Sec \ref{llaut}.  
Also then, 
\beq
l_{(0)}^{\SStar_{(0)}\SStar_{(0)}} + 2{h_{(0)}^{\SStar_{(0)}}l^{\SStar_{(0)}}}/{h_{(0)}}  = -2B\,\mbox{sin}\,2l_{(0)} \mbox{ } .
\eeq
N.B. that the effect of inclusion on this example is not the `1/$r$' potential case's Keplerian ellipses versus straight lines. 
However, the difference between keeping and not keeping the central term is qualitatively significant over the whole set of central force problems rather than just the 
$1/r$ potential case. 
For the present example's HO's, it is the difference between ellipses  centred on the origin and spirals 
\beq
h_{(0)} = \Bigeta/\sqrt{1 + \Biggamma\{\varphi - \varphi_0\}}
\eeq
for $\Bigeta := \sqrt{E/A}, \Biggamma := \{E/\sfD\}^2$.
A forteriori, the spirals are non-periodic, unlike the ellipses.

\noindent Note 2) One might also choose not to use $\epsilon_4$ here so as to support nontrivial physics to first order.  
Then the last equation has a $-2B\,\mbox{sin}\,2l$ right-hand side.
This can be thought of in terms of $V_{l} = 0$ makes lowest order $l$-dynamics trivial.  
Moreover, this is a common feature for scaled RPM's.  
This might well be saying that taking only the lowest iteration in the semiclassical approach for these models is weak, as it may not capture any $l$-dynamics at all.  
\cite{SemiclIII} already went further than that, but under more restrictive assumptions on what is perturbed that themselves lack in Machianity 
(if $t$ is perturbed, so should the $Q$'s from whose d$Q$'s the time is to be abstracted...).  

\noindent Note 3) There can be problems with $J$ and $J_{,l}$ sometimes having the same form here 
(for these trig functions, $l$ near $\pi/4$), so $\epsilon_0$ can cease to be a separate diagnostic.
This teaches us that such schemes really only work out for certain regions (i.e. are local and thus non-global).  
$|B| << A$ helps ensure some $\epsilon$'s are suitably small, but other conditions favour mostly-radial motion 
(i.e. scale dominates shape, so also in accord with cosmological modelling) 
and confinement of these wedges to suitably small values of cos$\,\varphi$ and sin$\,\varphi$.

\subsubsection{Triangleland example}

This is nontrivially configurationally relational.  
For an HO, it has no $\epsilon_3$ or $\epsilon_4$. 
However, the qualitative dynamical analysis of this example is similar to the preceding, so we omit it from this Article.

\subsection{Demise of `marching in step' criterion}\label{MIS-Demise}

Barbour has suggested that in the relational LMB set-up, rather, by good clocks `marching in step' with the emergent JBB time,  
independent observers would be able to `keep appointments' with each other; he argued furthermore that this criterion applies universally, and uniquely to this timefunction.
This is based on \cite{B94I, Bfqxi}, for 1 and 2 indexing two isolated island subuniverses,
\beq
\frac{\delta t^{\se\sm(\sJ\sB\sB)}_1}{\delta t^{\se\sm(\sJ\sB\sB)}_2} = 
\frac{\sqrt{T_1/\{E_1 - V_1\}}\,\d\lambda}{\sqrt{T_2/\{E_2 - V_2\}}\,\d\lambda} = 1 \mbox{ } ,
\eeq
where the cancellation to 1 was argued to follow from use of conservation of energy for each subisland universe.

Unfortunately, `marching is step' is Barbour's language refers to a concept that also carries connotations of the far more often-studied synchronicity \cite{Jammerteneity}.
This matters because the mechanics argument Barbour gives carries straight over to the GR case (i.e. $T_i$  to $\mT_{\sG\sR}$ in each region and $E_i - V_i$ to Ric$(\ux; \bh] - 
2\Lambda$ in each region) for which it cannot hold by how relativistic synchronization requires a {\sl procedure} rather than {\sl just occurring naturally}.   
Alerted thus,\footnote{Chronologically, 
I was first alerted to there being a problem with this claim by $t^{\se\sm(\sJ\sB\sB)}$ not being a PPSCT-scalar as per Appendix \ref{Examples}.B. \label{stiix}}
it is straightforward to spot that the problem with Barbour's argument is a theory-independent circularity: he attains constancy by substituting the energy-type equation into a 
rearrangement of itself (for that is what the formula for $\updelta t^{\se\sm(\sJ\sB\sB)}$ in terms of $T$, $E$, $V$ is, see Sec \ref{JET}).  

\mbox{ } 

\noindent Thus the universal basis for marching in step is lost, so one is left having to consider a procedure for approximately patching disparate observers' GLET's together.

\subsection{Patching together of GLET's}

\noindent Consider modelling two quasi-isolated island subsystems within a universe.
Then Sec \ref{temJBB-Approx}'s approximations applied to each will ensure that the details of the other's contents will contribute negligibly. 
Thus each's timestandard constructed as a GLET would be independent of the other's.
Thus mechanical ephemeris time type constructions do not in practise by themselves appear to provide a common timestandard.

There {\sl is} a JBB timefunction that interpolates between the two: contains inter-particle separations for the joint subsystem.   
However, neither observer would know enough about the other's local solar system so as to be able to at all accurately compute that interpolation.

A more relevant line of enquiry is that observers in different places will have access to (even qualitatively) different changes. 
Whilst I have argued that the GLET's they establish as per Sec \ref{Cl-Str} will not in general march in step with each other, due to being 
in relative motion and in different gravitational potential wells, I now argue that the GLET concept has a physically natural means 
of approximate patching via examination of those changes in the universe that can be observed by both of the observers.
For, by each choosing a time that works well for the sufficient totality of change observed by each observer, both times work well 
to describe the mutually-observed change, so the two timestandards are reasonably accurately synchronized (up to linear transformation 
that gives each observer freedom in choice of time-unit and of `calendar year zero').  
There are two strengths of sharing: of the same subsystem versus of the same {\sl type} of subsystem (see examples below).  

\mbox{ } 

\noindent Example 1) Suppose we use our moon as a clock whilst the Jovians use their moons as a clock.    
Each party could do this without knowing much at all about the specifics of the other's moons. 
Each party would reach the conclusion of the reasonable usefulness of our own moons as clocks by noticing this to furnish 
reasonable timestandards with respect to which other 
physics is simple and reasonably accurately done.
Moreover, in this setting, these moon clocks (allowing for other major solar system bodies, excluding each others' moons) march in step to good accuracy!  
But, unlike in the case of 2 independent civilizations on Earth, this is lacking a why, as $t^{\se\sm(\sJ\sB\sB)}_{\sE\sa\sr\st\sh}$ and 
$t^{\se\sm(\sJ\sB\sB)}_{\sJ\su\sp\si\st\se\sr}$ are a priori not naturally the same 
[and in more ways than just tick-length and calendar year zero conventions that were the sole differences to quite good approximation -- pre-relativistic effects -- 
for the case of two independent civilizations on Earth].    
One could in this case argue instead (to some level of accuracy) for the universality of Keplerian Physics as being useful for a timestandard provider.
One would need to take SR and GR into account for above a certain level of accuracy of synchronicity, due to the Earth and Jupiter's relative motion 
and due to differences in gravitational potential. 

\noindent Example 2) Perhaps what subsystems are locally simplest or locally common may be different in different parts of the universe. 
A chaotic planetary system around a stellar cluster undergoes non-Keplerian motion; can we be sure of what kind of timestandard is locally simplest then?
Nor is there a notion of sidereal time on our Moon. 
Also consider the plight of the Venusians, who would have never observed any Celestial Mechanics due to their planet always being immersed in thick cloud.  
Note that the GLET procedure would still apply as regards what set of most reliable entities were available (e.g. rest-pulses, pendulums or the 
fairly regular variations over time on average in the surface temperature of a cloud-covered planet), in the sense of an iterative procedure to find a 
widely-applicable simplifying time.  
Why should iterated times that are each simplest for most of the motions arising in two separate places march in step with each other?  
That one might come across an abandoned warehouse populated by a batch of {\sl correlatedly}-defective clocks suffices to doubt that general agreement among phenomena in one location 
will produce a timestandard marching in step with that obtained by applying the same kind of iterated `time is abstracted from change' procedure elsewhere.  
This leads to the idea of timestandards being challengeable by considering {\sl further changes}.  

\noindent Example 3) In the case of the {\sl two observers partly sharing observed subsystems}, each's clock will keep roughly in step with the shared subsystems and thus roughly 
in step with each other. 
Pulsars could well often furnish shared observable subsystems.  
This may be the best practical answer to marching in step: to prescribe the time of meeting in terms of pulsar information verifiable from other solar systems/galaxies.

\mbox{ }  

\noindent {\bf Question 47)} Can pulsars serve as `standard clocks`? 
Is there a danger here as regards high accuracy due to some Pulsar Physics analogy of how the physics of the Earth is dirty/unpredictable?
It also makes sense for this procedure to involve the pulsars themselves rather than the pulsars' time re-adjusted by each observer's local gravitational potential.
The `and thus' does require a modicum of knowledge about the other observer, to be able to factor in SR effects. 
In the absence of this, there is an accuracy bound of the order of (peculiar velocity/c)$^2$;  this comes out in the 1 part in $10^{9}$ to $10^{12}$ range. 

\mbox{ }  

\noindent Example 4) One possible way out of not having any suitable sharedly-observable astrophysical objects is that one would however expect observers living within two 
different such quasi-isolated subsystems to be able to make similar large-scale cosmological observations, on which basis they may well develop very similar notions of cosmic time.
[This does partly rely on the Copernican principle holding, at least to good approximation.  
It is also tied to the practical appropriateness of the scale = $h$ identification in the kind of universe that we live in, 
and Sec \ref{RPM-H-L}'s observation of the difference between scale = $h$ and `some subsystem = $h$'.]
And emergent JBB time for GR in the cosmological setting is aligned with cosmic time, so JBB time per se continues to be the right concept, 
it is just that one needs to take care as to what one designates to be $h$ and $l$.  
[This is {not} a shift from Newtonian gravity to GR on scales a few orders of magnitude larger than planetary systems? 
Rather, it is a shift to scale--shape alignment, regardless of whether one's cosmological model is Newtonian, an RPM or generally-relativistic. 
After all, by Appendix \ref{Cl-Soln}.C.3 these give good approximate descriptions of the overall late universe for most purposes; 
in particular, each possesses a satisfactory notion of cosmic time.
Though it should then be said that, with current-Earth technology, cosmic time cannot be set up to be used as an {\sl accurate} timestandard!  
(1 part in 100 may still be generous as regards interpreting cosmological quantities such as the Hubble parameter, and, consequently, the age of the universe.)   
It is also not guaranteed that 2 independent civilizations will single out cosmic time as the most natural time variable for cosmology and inter-stellar appointment keeping.  
For these reasons, I do not in the end view this alternative as being as promising as it might appear at first sight.  

\mbox{ } 

\noindent  Note 1) In extension of the {\sl incontestability} of LMB by including all changes, the LMB-CA practical timestandard 
involves considering a STLRC for one's desired accuracy will be incontestable from the perspective of its own internal physics?  
It may still be contestable by external Physics, e.g. if the Solar System does not keep time with the pulsars, which do we consider to be at fault?  
[This is not an issue in the LMB case itself since everything lies within the universe-system.]    

\mbox{ } 

\noindent{\bf Question 48} Patching together GLET's may well come to have relevance to space programs. 
More generally, what is known quantitatively about the space version of the `longitude at sea' problem?  

\mbox{ } 

\noindent Global Problem with JBB Time 3) Whilst we have a STLRC means of patching, STLRC locality and POZIN locality are logically unrelated.  
However, STLRC locality {\sl is} closely related to $h$--$l$ locality, as is developed in Sec \ref{+temJBB}.

\subsection{Is $t^{\se\sm(\sJ\sB\sB)}$ elsewise globally defined? Monotonic?}\label{JBBGlob}

\subsubsection{POZIN, monotonicity and non-frozenness}\label{globhl}

\noindent \bu If $\ft^{\se\sm(\sJ\sB\sB)}$ the emergent JBB time candidate exists for (a given portion of) a given motion, its 
monotonicity is guaranteed: $\fW > 0$, so $\d{\fI} \geq 0$, so by (\ref{JBB}) $\d{\ft}^{\se\sm(\sJ\sB\sB)}  \geq 0$.

\noindent \bu In terms of the action it emerges from priorly needing to exist, 
it is not in general globally defined, by Sec \ref{POZ}'s problem of `zeros, poles and nonsmoothness': Global Problem with JBB Time 1).  
Then at the level of the emergent JBB time formula itself, sufficiently benign blow-ups in $\d s/\sqrt{\fW}$ (i.e. those retaining integrability so that the emergent JBB time 
candidate does exist) correspond to the $\ft^{\se\sm(\sJ\sB\sB)}$ graph becoming infinite in slope.   
There may also be frozenness: at points for which the graph is horizontal, i.e. $\d \fs = 0$ or $\fW$ infinite.
Both zero and infinite slope may compromise use of $\ft^{\se\sm(\sJ\sB\sB)}$ itself to keep track for some ranges of mechanical motion.  
However, at least in some cases, redefined timestandards may permit motions to be followed through such points.

\noindent Some of Szydlowski's work can be interpreted as a patching approach around the above; this covers mixmaster minisuperspace. 
To counter Burd and Tavakol's argument \cite{TB93}, a patching argument was offered e.g. in \cite{SS94}.
 
\noindent \bu The detailed Sandwich results of Sec \ref{TSC} also count as a global issue with this; conformal, matter and Ashtekar Variables 
extensions as per Sec \ref{+Sand} remain to be considered from this perspective.

\subsubsection{A global problem with $\fh$-$\fl$ approximations themselves}\label{globhl-2}

Global JBB Time Problem 4). 
It should be clear that in passing from the $\fh$--$\fl$ approximation's familiar Celestial and Molecular Physics domain's flat mass metric to a curved configuration space metric 
that the $\fh$ and $\fl$ notions can become merely local in configuration space and thus well capable of breaking down over the course of a given motion.  
Thus use of $\fh$--$\fl$ approximations in general entails a global-in-space problem; this holding equally for RPM's and GR constitutes Analogy 72).  

\begin{subappendices}
\subsection{Consequences of there being a PPSCT-related family of emergent times}\label{EmTime}

\noindent 1) This accounts for Newtonian theory and GR's non-affine transformations of time.   
\noindent Individually, both conformal transformation and non-affine parametrization \cite{Wald, Stewart} complicate the equations of motion. 
\cite{Banal} demonstrated how, nevertheless; the equations of motion are {\sl preserved} when {\sl both} 
of these transformations are applied together.
Then, since PPSCT's are very natural from the relational product-type parageodesic action principle as per Appendix \ref{Examples}.B, 
in the relational context these might be viewed as primary. 
Thus the relationalist can view non-affine reparametrization as a consequence of a very simple property of the form of the relational action.  
 
\noindent 2) Thus, if one's problem requires rescaling {\sl or} non-affinely parametrizing, one's problem may be sufficiently unrestricted so as to 
permit one to `complete' the required transformation to a 3-part conformal transformation. 
By this, the effect of solely rescaling or solely non-affinely parametrizing causing departure from the simple geodesic equation form is circumvented. 
Thus emergent time derivative's being a PPSCT-covector provides a {\sl robustness} result for the property of providing simple equations of motion.  

\noindent 3) Affine transformations send $\ft_{\so\sll\sd}$ to $\ft_{\sn\se\sw}(\ft_{\so\sll\sd})$ subject to 

\noindent I) nonfreezing and monotonicity, so $\d \ft_{\sn\se\sw}/\d \ft_{\so\sll\sd} > 0$ which can be encoded by having it be 
a square of a quantity $\ttf$ with no zeros in the region of use, and 

\noindent II) this derivative and hence $\ttf$ being a physically-reasonable function (to stop the transition damaging the equations of motion).  
But this can be recast as $\ordial/\ordial \ft_{\sn\se\sw} = \ff^{-2}\ordial/\ordial \ft_{\so\sll\sd}$, by which 
(and other properties matching\footnote{Moreover, conventional affine transformations are rather less smooth than is usually assumed of 
conformal transformations (${\cal C}ts^1$ to ${\cal C}ts^{\infty}$).}) 
one is free to identify this $\ff$ with $\Omega$, so any affine transformation is of a form that extends to a PPSCT.  
If one then chooses to `complete' it to a 3-part conformal transformation, the above calculation can be interpreted as the extra non-affine term 
being traded for a $\fT$ term.  
This is by having an accompanying conformal transformation of the kinetic metric, and then this being traded for $\partional^{\sfA}{\fW}$ 
by energy conservation and the compensating transformation of ${\fW}$.
Thus the freedom to affinely-transform the geodesic equation on configuration space can be viewed instead as the freedom to PPSCT a system's equation of motion.  
{\sl Thus the relational approach's simplicity notion for equations of motion has the same mathematical content as prescribing an affine rather than non-affine 
parameter for the geodesic equation on configuration space}. 
Thus the PPSCT-related $\vec{\ft}$ corresponds to `the set of (generally) nonaffine parameters for the geodesic-like equation of motion on configuration space'. 
(However, each is paired with a different, conformally-related ${\bfM}$ and ${\fW}$.  
On, the other hand, $\ft^{\se\sm(\sa\sf\sf-\sg\se\so)}$ indeed remains identified with the much more restricted set (unique up to a 
multiplicative constant time-scale and an additive constant calendar year zero) of affine parameters for the geodesic equation on configuration space.

\noindent 4) It is then a useful probe for any purported arguments for the uniqueness of 
emergent JBB time as to whether PPSCT-related versions of that argument in fact also hold, thus sidestepping that uniqueness (see e.g. footnote \ref{stiix}).
This is of some concern since much of the usual Newtonian picture's freedom of choice of time variable (non-affine parametrizations) was then hidden for many years, 
via the above small (but not so easy to spot) piece of algebra, within the less familiar guise of the PPSCT's/3-part conformal transformations \cite{Banal}.

\subsection{Extension of the Machian Classical Scheme to include fermionic species}

The previously-raised issue -- that only `time is to be abstracted from {\sl bosonic} change' had been demonstrated so far \cite{ARel}, as is clear from only bosons/quadratic 
species and not fermions/linear species entering the na\"{\i}ve expression for $\ft^{\se\sm(\sJ\sB\sB)}$, is countered along the following lines.

One assumes that the linear species are $\fl$. 
Then, whilst they do not feature in $\scQ\scU\scA\scD$, after solving $\scQ\scU\scA\scD$ at zeroth order, one solves the $\fl$-equations of motion and these include the fermionic ones.  
Then the first-order correction of $\scQ\scU\scA\scD$ has these $\fl_{(0)}$'s among its inputs and thus fermionic change does end up having the opportunity to affect 
the corrected timestandard from the perturbative order at which it is usually STLRC upwards.

\subsection{Relationalism and Scale}\label{ScaleDisc}

Whilst many of the more recent papers on relationalism have been positing scale invariance, they have made no suggestions for how to explain Cosmology 
beyond those already in \cite{ABFO, ABFKO}.  
Here I argue contrarily that the following `heterogeneity argument for inclusion of scale can be useful by alignment with choices that would be harder to make without it' 
argument renders scale more presentable from a conceptual perspective.  
It is the scale contribution that gives the indefiniteness in GR kinetic term, a feature not found elsewhere in Physics and which 
causes a number of difficulties. 
(E.g. invalidation of the usual Schr\"{o}dinger interpretation of the inner product, and a whole new theory for the dynamical meaning of zeros 
in the potential factor.)  
I also note that RPM's cannot model this particular feature; minisuperspace does.  
As regards implementations,

\mbox{ }

\noindent 1) Use of the indefiniteness to pick out a `Riem time' fails.  

\noindent 2) Scale as internal time fails (at least globally), and I argue above that use of dilational quantity conjugate to scale as internal time carries bad connotations.

\noindent 3) However, $\fh$--$\fl$ alignment with scale--shape removes ambiguity of how to allot $\fh$ and $\fl$ roles.  
Then the scale part contributes an approximate timestandard with respect to which the shape part runs according to usual 
positive-definite kinetic term physics (`scale-indefiniteness alignment' exploited by aligning h with both).  
This can then be fed into more promising Histories-Records-semiclassical combination strategies.  
 
\end{subappendices}
 
\vspace{10in}

\section{Type-1 Tempus Ante Quantum at the classical level}\label{Cl-Ante}

\subsection{More on candidate scale and dilational times for RPM's}\label{PlayIt}

\noindent One place to seek for an internal time for RPM's is among the theories' natural scalars such as the moment of inertia scale  $\mI$ 
or the dilational quantity $\scD$ \cite{06II}.  
This of course entails the usual belief in canonical transformations.  
Before proceeding with this discussion, I note the following ambiguity.

\mbox{ } 

\noindent {\bf Ambiguity: there is a whole family of scale variables and therefore of dilational conjugates to them}.
This not the obvious fact that positive functions of a scale are also scales. 
It is rather that, once the scales themselves are dismissed as candidate times due to their non-monotonicity in recollapsing or bouncing models 
and one passes on to consideration of their conjugates as time candidates, then the original choice of scale variable turns out to 
nontrivially affect the internal time approach's procedure for solving the Hamiltonian constraint/energy equation. 

\mbox{ } 

\noindent Note: this relies on accepting a nontrivial (i.e. $\fQ^{\sfA}$, $\fP_{\sfA}$ mixing) canonical transformation. 
Thus the approach requires use of (Phase, Can) rather than ($\FrQ$, Point) or (RigPhase, Point).  

\mbox{ } 

\noindent In detail, the passage to using the dilational quantity conjugate to one's scale as a coordinate that is then promoted to a candidate 
timefunction is underlied by a simple canonical transformation by which the dilational quantity becomes a coordinate to 
be identified as a candidate time $\ft^{\sd\si\sll}$, whilst the scale itself becomes $-P_{\sft^{\td\ti\tl}}$.   
Now a multiplicity of possibilities for the scale variable itself reflects itself as a multiplicity of such dilational quantities conjugate to each of these scales.

In RPM's, the in some ways simplest such dilational quantity is $\scD$, which in this context I termed the {\sl Euler time candidate}, $t^{\sE\su\sll\se\sr}$.    
This is conjugate to ln$\,\rho$,   
\beq
\{\mbox{ln}\,\rho, \scD\} = 1 \mbox{ } . 
\eeq
which reveals that the aforementioned analogy between the Euler time candidate and GR's York time candidate (\cite{06II, SemiclI} 
Secs \ref{Intro} and \ref{Examples}) is somewhat loose, by virtue of the above ambiguity. 
Moreover, this observation paves the way to substantially more accurate RPM--GR analogies (Sec \ref{Scream}).

\noindent More generally, consider $f(\rho)$ as a scale variable.  
Then 
\beq
\{f, \scD/L_{\sD}f\} = 1 \mbox{ } ,
\eeq 
for $L_{\sD}$ the linear dilational operator $\rho\,\pa_{\rho}$.  
Likewise (useful in Sec \ref{LJE}) for $F(\mI)$ as a scale variable, 
\beq
\{F, \scD/L_{\sD}f\} = 1 \mbox{ } ,
\eeq 
for $L_{\sD}$ now written in the form $2\,\mI\,\pa_{\sI}$.  

\noindent Some simple examples then are as follows. 

\mbox{ } 

\noindent Example 1) $f^{\prime}(\rho)\rho = 1$ gives back $f$ = ln$\,\rho$ being conjugate to $\scD$ itself.  
By this simpleness and that in the next SSSec, I term ln$\,\rho$ the {\it subsequently simplest scale}.  

\noindent Example 2) $f = \rho$ (configuration space radius scale), the conjugate of which is $\scD/\rho$ 
i.e. indeed just $p_{\rho}$:  radial dilational time candidate $t_{\rho}$.

\noindent Example 3) $f = \mI$ (moment of inertia scale), the conjugate of which is $\scD/2$: MOI dilational time candidate $t_{\sI}$.

\noindent Example 4) $f = 1/\rho := \upsilon$  ({\it reciprocal radius}), the conjugate of which is $-\rho \, \scD$: reciprocal radius dilational time candidate, $t_{\upsilon}$.  

\mbox{ } 

\noindent {\bf Logarithmic impasse with Euler time candidate itself}.
This case's ln$\,\rho$ case gives a part-linear form  with the structure $- P_{t^{\tE\tu\tl\te\tr}} = \mbox{ln}(F(P^{\sS}_{\sfa}, 
\mS^{\sfa}, t^{\sE\su\sll\se\sr}))$ and this logarithm in the `true Hamiltonian' then furthers the operator ordering ambiguities 
and troublesomeness in making rigorous the functional analysis underpinning the form of the candidate true Hamiltonian operator at the quantum level.  

\mbox{ }

\noindent This is apparent in the workings in \cite{SemiclI} and Example 2) of Sec \ref{YaLu} but is very readily bypassable by the above ambiguity.

\subsection{Analysis of monotonicity: Lagrange--Jacobi equation and a generalization}\label{LJE}

For a time candidate to be satisfactory, it is important that this is monotonic.  
For the Euler time candidate $\scD = t^{\sE\su\sll\se\sr}$, this follows \cite{06II} for a number of substantial cases (see below) from the Lagrange--Jacobi identity \cite{Saari2} 
(viewing $\Last$ as $\d/\d t^{\sN\se\sw\st\so\sn}$ in this context) 
\be
{t}^{\sE\su\sll\se\sr\,*} = \mI^{**}/2 = 2T - \mn V =  2 E - \{\mn + 2\} V \mbox{ } , 
\ee
for $V$ homogeneous of degree $\mn$.  
Sums of homogeneous potentials all of which obey a common index inequality also satisfy monotonicity. 
Interpret $\mn$ in that way from now on.

This provides a fairly strong guarantee that $t^{\sE\su\sll\se\sr}$ is monotonic: it is so in a number substantial sectors.

\noindent Sector 1 = $\{ E \geq 0, V \geq 0, \mn \leq -2\}$ and 
\noindent Sector 2 = $\{ E \geq 0, V \leq 0, \mn \geq -2\}$ give  
\beq
t^{\sE\su\sll\se\sr\,*} \geq 0 \mbox{ } .
\eeq
\noindent Sector 3 = $\{ E \leq 0, V \leq 0, \mn \leq -2\}$ and
\noindent Sector 4 = $\{ E \leq 0, V \geq 0, \mn \geq -2\}$ give, using $- {t}^{\sE\su\sll\se\sr}$ as timefunction instead, 
\beq
- t^{\sE\su\sll\se\sr\, *} \geq 0 \mbox{ } .  
\eeq
\mbox{ } \mbox{ } One can immediately check against Sec \ref{Cl-Soln} for which models lie within the above sectors.  
Free models do, whilst HO's do not (not unexpected for oscillators).  


The question then is whether other dilational time candidates are anyway near as good in this respect (particularly due to the 
above `logarithmic impasse' with the Euler time candidate that is absent from dilational candidate times that are conjugate to whichever power of $\rho$.  
I proceed by generalizing the Lagrange--Jacobi equation viewed as the Euler time candidate propagation equation to 
the following general dilational time candidate propagation equation:
\beq
t_{\cal F}^* = \{G\mI^*\}^* = \mI^{**} G(\mI) + G^{\prime}\{\mI^*\}^2 = 2\{2E - \{\mn + 2\}V\}G + 4\scD^2G^{\prime} \mbox{ } , 
\eeq
for $G(\mI) = 1/2L_{\sD}{\cal F}$.  
Thus if $G$ and $G^{\prime}$ are the same sign, there is monotonicity in Sectors 1 and 2.
On the other hand, if $G$ and $G^{\prime}$ are of opposite signs, there is monotonicity in sectors 3 and 4.  
The first of these happens for $F = \mI^{\sn}$ for $\mn < 0$ and the second for $\mn > 0$. 
This makes $\upsilon = 1/\rho$, and minus its conjugate $t^{\upsilon}$ as a candidate time, useful below.

\subsection{GR counterpart}\label{Scream}

Analogy 73) The subsequently simplest scale $2\,\mbox{ln}\,a =$ Mis, a Misner variable.
This has the simplest dilational conjugate,
\beq
\{2\,\mbox{ln}\,a, \pi \} = 1 \mbox{ } .  
\eeq
Then canonically transform such that 
\beq
\mt^{\sss\sss\sss} = \pi \mbox{ } , \mbox{ } \mbox{ }  p_{\st^{\ts\ts\ts}} = - \mbox{Mis} \mbox{ } .  
\eeq
Analogy 74) This has the simplest t-propagation equation, 
\beq
\Circ_{\suF}\mt^{\sss\sss\sss} =  2\sqrt{\mh}\{\dot{\mI}\,\mbox{Ric(\bx; \bh]} - \triangle \dot{\mI}\} \mbox{ } .  
\label{gueri}
\eeq
To maximally align this with Sec \ref{LJE}, rewrite the above as
\beq
\Last_{\suF}\mt^{\sss\sss\sss} =  2\sqrt{\mh}\{\mbox{${\cal R}\mi\mc$}(\ux; \bh]- \{\triangle \d{\mI}\}/\d{\mI}\} \mbox{ }  
\eeq
(noting that a derivative-of-the-instant term has become trapped inside derivatives by an integration by parts move).  
This is then an equation in $a^{**}$ just as the Lagrange--Jacobi equation is in terms of $\mI^{**}$: both are equations for double derivatives of a scale variable.  
Moreover (\ref{gueri}) is the trace of the GR evolution equation, so the cleanest identification of the analogy is between Raychaudhuri and Lagrange--Jacobi equations. 
[These are the dilational time propagation equations corresponding to each theory's subsequently simplest scale.]

\mbox{ }  

\noindent Analogy 75) Each then constitutes a guarantee of monotonicity in certain cases. 
A simple such case for GR is closed minisuperspace (for which scale variables themselves fail to be monotonic): 
\beq
\Circ_{\suF}{\mt}^{\sss\sss\sss} = 2\sqrt{h}\,\dot{\mI}\, Ric(\mbox{\boldmath$h$}) > 0 \mbox{ } 
\eeq
(since $\dot{\mI} > 0$ by the definition of the velocity of the instant, $\sqrt{h} > 0$ by nondegeneracy and $Ric > 0$ for such closed models.

\noindent Analogy 76) Then for other scales $f(h)$, 
\beq
\{f(h), 2G(h)\uppi\} = 1
\eeq
for $G(h) = 1/L_{\sD}f(h)$ and linear dilational operator $L_{\sD} = a\,\pa_a = 6\,h\,\pa_h$.

\noindent Analogy 77) Then upon passing by canonical transformation to $t^{\sd\si\sll} = 2G(\mh)\uppi$, $p_{\st^{\td\ti\tl}} = - f(\mh)$, 
one has the generalized dilational time candidate propagation equation 
\beq
\Circ_{\suF}\mt^{\sd\si\sll} =  2\sqrt{\mh}\{\dot{\mI}\,\mbox{${\cal R}\mi\mc$}(\ux; \bh] - \triangle \dot{\mI}\}G - G^{\prime}\dot{\mI}\,\uppi^2/\sqrt{\mh} \mbox{ } .  
\eeq
This is analogous to the generalized Lagrange--Jacobi equation of mechanics, as is brought out most clearly by casting it in the form 
\beq
\Last_{\suF}\mt^{\sd\si\sll} =  2\sqrt{\mh}\{ \mbox{${\cal R}\mi\mc$}(\ux; \bh] - \{\triangle \d{\mI}\}/\d{\mI}   \}G  -  G^{\prime}\uppi^2/\sqrt{\mh} \mbox{ } .  
\eeq
It retains monotonicity in the above closed minisuperspace context if $G$ and $G^{\prime}$ are of opposite signs (i.e. like in Sectors 3 and 4 of the mechanical counterpart).  
For $f(\mh) = \mh^{k}$, $k > 0$ guarantees this.

A notable example among these is then $f(\mh) = \mh^{1/2}$.  
Then $\mt_{\sd\si\sll} = 2\uppi/3\sqrt{\mh} = \mt_{\sY\so\sr\sk}$, for which 

\noindent 
\beq
\Circ_{\suF}\mt^{\sY\so\sr\sk} =  \frac{4}{3}\left\{ \dot{\mI}\,\mbox{${\cal R}\mi\mc$}(\ux; \bh] - \triangle \dot{\mI} + \frac{\dot{\mI}\,\uppi^2}{4\sqrt{\mh}} \right\}  \mbox{ } , 
\eeq
which is recognizable as the CMC VOTIFE analogue of the CMC LFE (\ref{CMCLFE}).  
The most relational form for this is the CMC `POTIFE', 
\beq
\Last_{\suF}\mt^{\sY\so\sr\sk} =  \frac{4}{3}\left\{\mbox{${\cal R}\mi\mc$}(\ux; \bh] - \frac{\triangle \pa{\mI}}{\pa\mI} + \frac{\uppi^2}{4\sqrt{\mh}} \right\} \mbox{ } .  
\eeq
%
%
Then $\mt^{\sY\so\sr\sk}$ is known to have better monotonicity guarantees than in just the above closed minisuperspace example \cite{York72}.  
Moreover, for GR this particularly good monotonicity case is unhampered by the particular obstruction of the logarithmic impasse 
(which falls, rather, upon the conjugate of the above-mentioned Misner variable.  
Note that this does {\sl not} in general keep time with $\mt^{\se\sm(\sJ\sB\sB)}$.  

\mbox{ }

\noindent {\bf Question 49}. Is the York time uniquely privileged among the family of times in question as regards having particularly good monotonicity properties?  

\mbox{ } 

\noindent Analogy 78) The closest parallel to GR's York time is the conjugate of the pseudo-volume scale.    
This has the misfortune of being particle number-and-dimension dependent (though that is no worse than n-dimensional GR having a different expression for York time for each dimension).  
Here f($\rho$) = $\rho^{\mbox{\scriptsize dim}({\cal R}(N, d))}$, so the conjugate pseudo-York time is $\scD/\mbox{dim}({\cal R}(N, d))\rho^{\mbox{\scriptsize dim}({\cal R}(N, d))} = 
P_{\rho}/\mbox{dim}({\cal R}(N, d))\rho^{\mbox{\scriptsize dim}({\cal R}(N, d)) - 1}$.  

\mbox{ } 

\noindent The idea is then that $\scQ\scU\scA\scD = 0$ is to be interpreted as equation for $P_{\sft}$. 

\mbox{ }

\noindent Difference 32) For RPM's, one has a simpler equation to solve than in GR to obtain a true Hamiltonian -- an algebraic equation of the form 
\beq
\mbox{[rational polynomial]}(\mbox{scale}) = 0 
\eeq
[e.g. eq. \ref{pupsi} and explicitly excluding the subsequently problematic logarithmic case] as opposed to the form 
\beq
\mbox{[rational polynomial]}(\mbox{scale}) = \triangle \mbox{scale}.
\eeq
of the than the quasilinear elliptic Lichnerowicz equation (\ref{LYE}) of GR.  
In each case, scale is interpreted as $-P_{\sft}$, so that solving for it is indeed now linearly isolating one of the momenta as per Sec \ref{PLinF}.
This simpleness permits progress past where the general GR case gets stuck.  
\noindent Which scale variable produces the most palatable powers for algebraic solution.    
However, plenty of other difficulties and absurdities then become apparent, casting further doubt over the sensibleness of internal time programs (see the next SSSec).

\subsection{End-check against hidden time's problems}

\noindent The canonical transformation in question is here simple. 

\noindent Some aspects of the Global POT are, however, present.
E.g. monotonicity can fail to be global in time [though there are a number of significant sectors in which one is protected from that by as per Sec \ref{LJE}]. 
But, being a finite theory, the `global in space' issues are absent.  
Finally,

\mbox{ }
  
\noindent Difference 33) The Torre impasse is absent for this Article's 1- and 2-$d$ RPM's.
This is due to these not having any configuration space stratification.  

\noindent Analogy 79) However, this argument is then clearly bypassed for 3-$d$ RPM's.  

\noindent Analogy 80) [A further Problem with Hidden Time].  
Having looked at a number of toy model examples, it is questionable whether the above conceptually-standard quantization procedure is can be done in practise. 
For, even in the absence of the above constructibility impasse for York time, for toy models such as RPM's, strong gravity  \cite{Rio} and some minisuperspace models, 
for which the Lichnerowicz--York equation is replaced by a solvable algebraic equation, at least some cases would appear to have the following difficulties \cite{Kieferbook, SemiclI}. 
1) well-definedness,
2) negative-probability issues,  
3) both of the preceding are furthermore tangled up with operator-ordering ambiguities too. 
This was already noted by Blyth and Isham \cite{IB75} in the minisuperspace arena.  
Additionally, from my work above, I also note that the internal approach's equations do not look anything like the 
equations encountered in the various conventional approaches to the same problem which are available for comparison for these RPM's.

\mbox{ }

\noindent Since they specifically involve GR spacetime structure, Foliation-Dependence and Spacetime Reconstruction Problems with hidden time are absent from RPM's.  

\noindent There is no a priori time in accord with Relationalism 4) , but a time is found be rearrangement. 
It is not clear if doing so is much of an abstraction from things it is in the sense that it is among the configuration variables and those are material things.
I.e., compliance with Relationalism 6) but it certainly is not from the strongest Leibniz--Barbour perspective since it is but a small amount of the universe degrees 
of freedom that go into it.
And also  Relationalism 7 non-compliance, alongside it does not have the more sensible Relationalism 7M compliance of getting better for 
larger subsystems even if one does not need to take this to the extreme of including the whole universe. 
It is also an interesting counterpoint to Leibniz timelessness for there to exist formulations with an extraneous time that are related to some with no such thing by canonical transformation.  
So rooting out an apparent extraneity may not always be the `right' answer (particularly if canonical transformations are, as standardly, held to be allowed).

\subsection{Reference particle time for RPM's?}\label{RPT}

\noindent Analogy 81) \cite{AKdis10} It is conceivable that there may be notions of reference particles within RPM's, 
associated with gauge-fixing Rot($d$) mathematics much as GR's reference fluids are associated with gauge-fixing the diffeomorphisms.  

\mbox{ }

\noindent {\bf Question 50}.  Can some particles in an RPM be considered as time-defining reference particles for the other particles? 
Does some form of this usefully capture some further elements of the reference fluid approaches to matter time in Geometrodynamics?

\mbox{ }

\noindent A Machian critique of reference fluid matter in GR would be to disbelieve all cases with matter in  insufficient quantities to be tangible.  
One would also suspect use of matter with unphysical equation of state.
Unfortunately these go against much of what has been proposed in this area.

\subsection{No unimodular gravity counterpart for RPM's}\label{UniM}

\noindent Difference 34)  Is it conceivable that an analogue of $\Lambda$ provides a hidden time from the suggested Analogy 51)? 
No!  
This is because this scheme relies on the quadratic constraint being an integrability of the linear constraint (as N or $\dot{\mI}$ is not varied). 
However, firstly, whilst $\scM_{\mu}$ has $\scH_{,\mu}$ as an integrability, which, upon integrating, gives $\scH + 2\Lambda = 0$ with $\scH$ the vacuum expression and $\Lambda$ now 
interpreted as a constant of integration, $\scL_{\mu}$ does not in any way involve $\scD$ [so this Difference follows from the no $\scL\scI\scN$-as-$\scQ\scU\scA\scD$-integrability 
Difference 8)].

Secondly, in relational approaches, variation with respect to the lapse $\ttN$ (or with respect to the instant $\ttI$) 
is replaced by a primary constraint i.e. arising purely from the form of the action with no variation done. 
Thus this way of obtaining a quadratic constraint with a part of it interpreted as a constant of integration 
is not an available option open if one takes a relational approach, even to GR [via a BFO-A type action (\ref{BFOA})].  

\mbox{ }  

\noindent Finally, Gryb \cite{UniGryb} showed that unimodular gravity amounts to the insertion of a background time.  
Conversely, my above work shows that unimodular gravity is not possible within purely relational thinking.

\subsection{Further comparison between Type 0 `Barbour' and Type 1 Tempus Ante Quantum Schemes}

1) Monotonicity and non-frozenness considerations indicate that emergent time can be more widely applicable than hidden dilational Euler time. 
Emergent time also exists for scale-invariant models \cite{B03}, characterized by $\scD = 0$, by which the Euler quantity is frozen and thus unavailable as a timefunction.   
Although some portions of Newtonian Mechanics have guaranteed global monotonicity for $t^{\sE\su\sll\se\sr}$, solutions outside this portion may still 
possess intervals on which $t^{\sE\su\sll\se\sr}$ is monotonic.

\noindent 2) The internal time examples given above work as well for non-interacting hl-systems.  
Though this is not now a conceptual necessity, as h does not now impose a timefunction on l, but rather both contribute to a joint timefunction.
Everything in the universe contributes in this species-by-species way.  
However, note the lack of role within for the potential, which makes it look more artificial or 
imposed, as the details of species from the potential plays no (direct) role in the construction of the timefunction.  

\vspace{10in}

\section{Tempus Nihil Est at the classical level}\label{Cl-Nihil}

\subsection{Records Postulates 1) to 4) for RPM's}

If there is no time, then what physical propositions can one address?  
Can one learn to make do with what one has left. 
Investigating this is one of the main reasons for expositing on the levels of structure of classical theory provided in Sec \ref{Cl-Str}.
Moreover, particular emphasis was placed there on RPM examples, so Sec \ref{Cl-Str} constitutes an answer to what Records 1) (notions of distance), 
Records 2) (notions of information/correlation) and Records 3) (propositions) consist of for RPM's. 
As regards propositions, these are both technically unproblematic for classical RPM's and vividly illustrated with geometrical examples as per Secs \ref{Q-Geom} and \ref{Cl-Str}.
Records 4) (Machian character for the semblance of time) is supplied for classical RPM's by Sec \ref{Cl-POT} subject to the details in Sec \ref{+temJBB}.  

\mbox{ }  

\noindent Two issues remain.  
Firstly, `how can one tell if a given configuration is a record?'
Secondly, we need to discuss approaches to a semblance of dynamics to the extent that is part of one's scheme and classically realizable.

\subsection{How can one tell if a given configuration is a record?}

\noindent Many of Sec \ref{Cl-Str}'s structures applied to RPM's come from Kendall's work on the statistical theory of shape.
I.e. distances between shapes based on shape space geometry, $\epsilon$-collinearity, and notions of information and correlation for shapes. 
It is the last of these that allows for one to culminate Records 2) with diagnostics for whether a given configuration is a record. 
Our point of view is that not all configurations are records.  
Only those with patterns in significant excess of what is to be expected from randomness are held to be records.  

\mbox{ }

\noindent{\bf Clumping--Kendall paradigm for Records Theory}.  
Clumping is already available in 1-$d$; see e.g. Roach \cite{Roach} for a discrete study (interpretable as a coarse-graining of particle model configurations.   
This is a ratios of relative separations issue.  
%

\mbox{ }

\noindent Then the remaining shape information in $d > 1$ concerns relative angles, and Kendall's statistics for the standing stones problem \cite{Kendall80, Kendall84, Kendall} 
(first posed as a problem to be addressed by geometrical statistics by Broadbent \cite{Broadbent}) exemplifies the statistical study of such.  
This concerns {\sl collinearity in threes}, which is a rather more general notion than the single-line regression discussed in Sec \ref{Correl1}.
We need at least the quadrilateral in order to discern between these two notions; 
note also that Kendall's method is based on the minimal fully relational sub-unit for a 2-$d$ model: the constituent triangles.

\mbox{ }  

\noindent Clumping and alignment in threes are {\sl discernible patterns} as compared to what one would expect in a random sample of the positions of the objects in question.  
This is precisely what a Pre-Record is.
%
{            \begin{figure}[ht]
\centering
\includegraphics[width=0.7\textwidth]{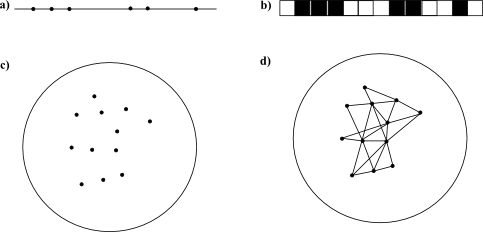}
\caption[Text der im Bilderverzeichnis auftaucht]{        \footnotesize{a) clumping in 1-$d$ and b) a discrete model of it as per \cite{Roach}.  
c) Shape data in 2-$d$.  This could consist of e.g. standing stones or of the particle positions in an RPM.  
d) Is the number of almost-collinear triangles present accountable for by coincidence or is it statistically significant (i.e. a pattern)? }    }
\label{Clumping-Kendall} \end{figure}          }

\mbox{ } 

\noindent That this can be applied to RPM's to cover the postulates Records 1) and 2) (notions of distances and of information/correlation) is a 

\mbox{ }

\noindent {\bf Corollary} [{\bf 10)}] to the Direct = Best-Matched Theorem.
\beq
\mbox{(Statistical Theory of Shape)}   \mbox{ } \mbox{ }        \mbox{ } \mbox{ furnishes items 1) and 2) for} \mbox{ } \mbox{ } \mbox{ } 
\mbox{(Classical Records Program for Barbour 03 RPM)} \mbox{ } ,  
\eeq
\beq
\mbox{(Statistical Theory of Shape and Scale)} \mbox{ } \mbox{ } \mbox{ } \mbox{ furnishes items 1) and 2) for} \mbox{ } \mbox{ } \mbox{ } 
\mbox{(Classical Pre-Records Program for BB82 RPM)} 
\mbox{ } .  
\eeq
\noindent The triangle corresponds to sample size 1 for which no statistics is possible.
The quadrilateral allows for sampling up to 4 triangles;\footnote{The quadrilateral already illustrates that dynamical and geometrical subsystems need not coincide 
-- the H-clustering involves a rhombus subsystem.  
All triangles concern $\big(\stackrel{n}{\mbox{\scriptsize 2}}\big)$ cases; for the quadrilateral these are all covered by the different possible labellings of the K's.  
N large carries more guarantees of reliable shape statistics for structure formation, SM/Information.}  
statistics is now possible, but poorly under control until one reaches a sample size in the region of 30, which occurs first for the heptagon.
C.f. Sec \ref{DrNo}'s account of small particle number SM.  

\mbox{ }  

\noindent Details of such tests are as follows. 
Kendall and collaborators considered this under the assumption that the standing stones lie within a compact convex polygon \cite{Kendall80, Kendall84, Kendall85, KL87, Le87}.  
I comment that for RPM's this is not per se a restriction, since one can always fit mechanics data thusly be rescaling it (which of course does not change the shape).
However, detail of the convex shape in question enters the conclusion.
This is OK if it is a convex approximation to the Cornish coast for the Lands End standing stones problem, but this would clearly be some 
kind of background-dependence in the case of RPM's.
However, this restriction is to cope with uniform independent identically-distributed distributions (intuitively clear to all quantum physicists as a `normalization by boxing'). 
Using distributions that tail off, however, allows one to be free of such a background-independent imprint.   

\mbox{ } 

\noindent Note 1) The collinearity in threes test involves the probability measure on the triangleland shape sphere.

\noindent Note 2) These records peak about the $\mathbb{RP}^k$ of collinearity of the full shape space.
More loosely, one may expect to use $\epsilon$-collinearity notions to find records far away from the collinearity submanifold.
Moreover, thinking about other features (like a gross over-representation of any other angle, one concludes that classical records need not be tied to maximally physically significant 
zones of configuration space.  
It is hard to envisage this happening at the quantum level either, so Barbour's `mists concentrating' conjecture 2) looks to be false in general.  
Contrarily, I add that tight binaries {\sl are} a likely outcome from 3-body dynamics and then these could be paralleled by atom-like constructs at the quantum level; 
these are, however, a clumping notion rather than a relative-angle notion.  

\noindent Note 3) Further technical development of this concerns probability and statistics theory on manifolds, involving a suitable notion of $\sigma$-field, 
geometrical measure, change of variables formula and isometries. 
Relationships with various well-known distributions are considered in this context.  
\cite{Small, Kendall} are useful sources for this material.  

\mbox{ } 

\noindent This SSec's observations are subject to the following caveats.  

\noindent Caveat A) most such work is classical.

\noindent Caveat B) Most such work is 1- or 2-$d$. 
This is not a hindrance within this Article which focuses almost entirely on 1- and 2-$d$ examples... 
but may be a hindrance in yet broader contexts and remains an active field of research [Shape Statistics in 3-$d$].  

\noindent Caveat C) This work's notion of shape is a great deal simpler than GR's conformal 3-geometries.  

\mbox{ } 

\noindent {\bf Question 51$^{**}$}/Analogy 82) to what extent can one pass from statistical analysis of this simple notion of shape to 
statistical analysis of the conformal-geometric notion of shape in GR?  
[I.e., do any of the abovementioned ideas carry over from shape space (and its cone) to CS($\bupSigma$) and \{CS + Vol\}($\bupSigma$)\}?]
 
\mbox{ }  

\noindent {\bf Question} 52) Minisuperspace and 2 + 1 GR may serve as stepping stones in coming to grips with this.  
The shape space geometry of minisuperspace and of the space of anisotropies are known and, like for Kendall's shape spaces, finite.  
What are corresponding shape statistics on this shape space?
Also, the 2 + 1 GR shape space geometry is furtherly known due to reducing to Teichm\"{u}ller space geometry as per Sec \ref{GRCtr}. 
What are corresponding notions of shape statistics on this now-infinite but well-studied space?

\subsection{Some tentative answers toward the semblance of dynamics issue}\label{Semblance}

\subsubsection{The general semblance question}

\noindent{\bf Question 53$^*$}.  How does a record achieve this encodement of a semblance of dynamics?  
Are subconfigurations that encode a semblance of dynamics generic?    
Or are they not generic but, nevertheless, picked out by some selection principle?
[In the absence of either of these, Records Theory by itself could not be all-embracing.]

\mbox{ }

\noindent Comment 1) The answer to the second sub-question is no, in the limited sense that most such will not even have any discernible pattern.
Whether the significantly-patterned subconfigurations are a measure-zero subset of all possible subconfigurations is as far as I know an open question. 
The theory of shape statistics might be able to provide an answer to this in the simple cases for which it yields tractable mathematics.  
In any case, the fourth sub-question looks to be a relevant issue.  

\mbox{ } 

\noindent Comment 2) For the standing stones problem, smaller $\epsilon$ suggests more care, eg astronomical or mystical reason, bigger $\epsilon$ would be things requiring 
less accuracy like marking paths or fences between plots of land \cite{Broadbent}.  
Thus finer detail of such tests at least qualitatively counts as a more detailed reconstruction of history.

\subsubsection{Semblance of dynamics in Barbour's  `time capsules' approach.}\label{Wedge}

\noindent This is the semblance of dynamics conjecture that one can already investigate parts of at the classical level.  

\mbox{ }  

\noindent {\bf Question 54$^*$}. Which characteristics distinguish time capsules from just any instants?  
E.g. are they localized configuration records or a subset thereof, or unrelated? 
E.g. are they characterized by any specific Notion of Information criteria?  
[This question is mostly quantum-mechanical, but does have at least a partial classical counterpart.]  

\mbox{ }

\noindent {\bf Question 55}. how are (sub)configurations with such features characterized geometrically within (sub)configuration space?

\mbox{ } 

\noindent Doubt 1) with Barbour Records)  At least in Barbour's earlier arguments \cite{B94I, EOT}, he suggests 
the wedge-shape of his representation of triangleland could play a role, but (c.f. Sec \ref{Pyramid}) this is a representation-dependent rather than irreducible feature of triangleland.  
I.e. Barbour 1) was originally not well-identified, suggesting there was no correct concrete mechanism underlying the postulation of Barbour 2).  

\mbox{ }

\noindent Perhaps it can then be argued instead to have to do with representation-independent features like the maximal collision itself or, 
more generally, the presence of strata or of a particularly uniform state.  

\mbox{ }

\noindent Modelling 1) 2-$d$ RPM's are likely to benefit from some sophisticated tests for the significance of patterns at the classical level.\foo{In \cite{Kendall} 
and references therein, whether collinearities are statistically significant is studied in the context of 2-$d$ shape spaces.}

\noindent Modelling 2) Specifically investigating whether strata play a role requires potentials that call for excision or 3-$d$ models, 
which lies beyond the scope of the present Article.  

\mbox{ }
  
\noindent Modelling 3) 
%
\noindent{\bf Question 56 (Effect of distinguishability): what is the spectral centre of a triangle?}
%
Ground state eigenfunctions of the Laplacian (and thus of the free TISE) peak about 
`the middle' of the region concerned. 
However, there are multiple notions of centre of a triangle.
Which, if any, out of these does the free QM TISE favour? 
One particular case of interest is the $\pi/2, \pi/3, \pi/3$ isosceles spherical triangle that corresponds to indistinguishable triangleland. 
An even simpler (but non-RPM) toy model case that serves to illustrate this question (without being an RPM model) is the 
$\pi/2, \pi/4, \pi/4$ isosceles triangle in flat space. 
Here, I find the QM/its Laplacian operator  favours a centre of its own that moreover lies rather close to the centroid--orthocentre coincidence that is the most 
geometrically obvious centre for an isosceles triangle, thus producing a `small number' which might have some significance in toy models of small departures from uniformity.
This is also linked to the use of spectral measures in Records Theory, insofar as differential operators encode geometrical information...

\mbox{ }

\noindent If one is willing to assert that records are subconfigurations  and then tries out Barbour's conjecture in 
this setting, the following further objection arises.

\mbox{ } 

\noindent Barbour-Records Doubt 2) In the study of {\bf branching processes}, one learns that Barbour's `how probable a subconfiguration 
is' can depend strongly on the precise extent of its contents.    
As an example (closely paralleling Reichenbach \cite{Reichenbach}), suppose we see two patches of sand exhibiting hoof-shaped cavities. 
Here there are past interactions of these two patches of sand with a third presently unseen subsystem -- a horse  
that has subsequently become quasi-isolated from the two patches.  
By these there is a clear capacity of rendering the individually improbable (low entropy and hence high information) configurations of each 
patch of sand collectively probable (high entropy, low information) for the many sand patches--horse subsystem.  
This still does not explain why useful records appear to be common in nature: a separate argument 
would be needed to account for why branching processes are common.    
The problem Whitrow (p 338) \cite{Whitrow} has with this is that it depends on the entropy of a main system. 
Moreover, then one runs out of being able to resort to such an explanation as one's increases in subsystem size tend to occupying the whole universe.

\subsection{Using Classical Machian and/or Histories approaches instead of a semblance}\label{R-within-CM}

\noindent There is no need for semblance if more structure is assumed, for instance at least one of histories assumed or 
the Classical Machian scheme and its semiclassical Machian counterpart assumed.  
The {\sl three} of these interprotect particularly well (I difer this to Secs \ref{Cl-Combo}, \ref{QM-Halliwell-Intro} and \ref{QM-Combo} 
since most of this interprotection is QM-motivated).

\vspace{10in}

\section{Classical Paths/Histories Theory (including Combined Approaches)}\label{Cl-Hist}
%

\subsection{IL-type formulation of Histories Theory for RPM's (Analogy 83)}

\subsubsection{On the notions of time used to label paths and histories in this Article}

\noindent We use $t^{\se\sm}$ for time, which we assume sufficiently accurately known at least in regions 
in which the Machian classical, and later Machian semiclassical, approaches applies, so we are effectively mounting a path formulation or 
histories theory on top of these Machian approaches.  
N.B. this feature is {\sl not} part of IL's standard package and reflects the relational nature of the examples being considered.

\noindent There are 7 different accuracies to be considered as inputs here.   

\noindent 1) $\ft^{\se\sm(\sJ\sB\sB)}$: the final result of the classical Machian approach.

\noindent 2) $\ft^{\se\sm(\sJ\sB\sB)}_{\{0\}}$: its un-Machian zeroth approximand. 

\noindent 3) $\ft^{\se\sm(\sJ\sB\sB)}_{\{1\}}$: the Machian first approximand.  

\noindent 4) $\ft^{\se\sm(\sW\sK\sB)}$, taken to be formally arbitrarily satisfactory within the premises of the semiclassical approach.

\noindent 5) $\ft^{\se\sm(\sW\sK\sB)}_{(1)}$: the distinct semiclassical quantum Machian first approximand [the two zeroth approximands coincide, 
so I re-name 2) as $\ft^{\se\sm}_{\{0\}}$].

\noindent 6) $\ft^{\se\sm\prime}$, taken to be some formal improvement on  $\ft^{\se\sm(\sW\sK\sB)}$ that is not dependent on any use of semiclassical approximations.

\noindent 7) $\ft^{\se\sm\prime\prime}$, taken to be the arbitrarily satisfactory culmination of the previous.

\mbox{ } 

\noindent Note 1) One may take Halliwell's 1987 \cite{Halliwell87} study of correlations (a timeless notion) {\sl within} the Semiclassical Approach arena of Halliwell--Hawking 1985 
\cite{HallHaw} as a precursor to studying records and histories within Semiclassical Approach arenas. 
Then within this context, labelling histories by semiclassical emergent times becomes a rather natural endeavour.

\subsubsection{Classical $N$-stop metroland Paths Brackets formulation}

For scaled $N$-stop metroland, the classical paths are now $\rho^i(\lambda)$ for $\lambda$ a continuous label time.  
We here take $\lambda = t^{\se\sm(\sJ\sB\sB)}$ as choice of label time \cite{AHall}, by which this Sec's work sits inside the classical Machian scheme.  
I then denote the conjugate paths momenta by $p_i(t^{\se\sm(\sJ\sB\sB)})$.
 
\noindent The nonzero part of the paths brackets algebra is then
\beq
\{\rho^i(t^{\se\sm(\sJ\sB\sB)}_1), p_j(t^{\se\sm(\sJ\sB\sB)}_2)\} = \delta^i_j\updelta(t^{\se\sm(\sJ\sB\sB)}_1 - t^{\se\sm(\sJ\sB\sB)}_2) \mbox{ } . 
\label{N-stop-Cl-Hist-Alg}
\eeq
\noindent The paths energy constraint is in this case  $\scE_{t^{\te\tm(\tJ\tB\tB)}} := \int \d t^{\se\sm(\sJ\sB\sB)} \scE(t^{\se\sm(\sJ\sB\sB)})$ 
for $\scE$ given by the $N$-stop metroland case of (\ref{calE}).
Since there is only one paths constraint, the paths constraint algebra is trivial (the paths-Dirac procedure produces no unexpected secondary paths-constraints).

\subsubsection{Classical scaled triangleland, reduced-style Paths Brackets formulation}

For scaled triangleland, I use that it has the same mathematics in conformally-transformed configuration space as for 3-$d$ mechanics in space, which is covered in \cite{ILSS, SavThes}. 
Physically, the classical paths are now $Dra^{\Gamma}(t^{\se\sm(\sJ\sB\sB)})$, with conjugate paths momenta $P^{Dra}_{\Gamma}(t^{\se\sm(\sJ\sB\sB)})$.

\noindent The nonzero part of the paths brackets algebra is then
\beq
\big\{Dra^{\Gamma}\big(t^{\se\sm(\sJ\sB\sB)}_1\big), P^{Dra}_{\Lambda}\big(t^{\se\sm(\sJ\sB\sB)}_2\big)\big\} = 
\delta^{\Gamma}\mbox{}_{\Lambda}\updelta\big(t^{\se\sm(\sJ\sB\sB)}_1 - t^{\se\sm(\sJ\sB\sB)}_2\big) \mbox{ } , 
\eeq
\beq
\big\{\sfS_{\Gamma}\big(t^{\se\sm(\sJ\sB\sB)}_1\big), \sfS_{\Lambda}\big(t^{\se\sm(\sJ\sB\sB)}_2\big)\big\} = 
\epsilon_{\Gamma\Lambda}\mbox{}^{\Sigma} \sfS_{\Sigma}\big(t^{\se\sm(\sJ\sB\sB)}_1\big)\updelta\big(t^{\se\sm(\sJ\sB\sB)}_1 - t^{\se\sm(\sJ\sB\sB)}_2\big) \mbox{ } , 
\eeq
\beq
\big\{Dra^{\Gamma}\big(t^{\se\sm(\sJ\sB\sB)}_1\big), \sfS_{\Lambda}\big(t^{\se\sm(\sJ\sB\sB)}_2\big)\big\} = 
\epsilon^{\Gamma}\mbox{}_{\Lambda\Sigma}Dra^{\Sigma}\big(t^{\se\sm(\sJ\sB\sB)}_1\big)\updelta\big(t^{\se\sm(\sJ\sB\sB)}_1 - t^{\se\sm(\sJ\sB\sB)}_2\big) \mbox{ } , 
\eeq
\beq
\big\{P^{Dra}_{\Gamma}\big(t^{\se\sm(\sJ\sB\sB)}_1\big), \sfS_{\Lambda}\big(t^{\se\sm(\sJ\sB\sB)}_2\big)\big\} = 
\epsilon_{\Gamma\Lambda}\mbox{}^{\Sigma}P^{Dra}_{\Sigma}\big(t^{\se\sm(\sJ\sB\sB)}_1\big)\updelta\big(t^{\se\sm(\sJ\sB\sB)}_1 - t^{\se\sm(\sJ\sB\sB)}_2\big) \mbox{ } .   
\eeq
The last two of these signify that the paths and their conjugate momenta are $SO(3)$ or $SU(2)$ vectors.  

\noindent The paths energy constraint is in this case $\scE_{t^{\te\tm(\tJ\tB\tB)}} := \int \d t^{\se\sm(\sJ\sB\sB)}\scE(t^{\se\sm(\sJ\sB\sB)})$ 
with $\scE$ given by (\ref{EnDragta}).   
Again, since there is only one paths constraint, the paths constraint algebra is trivial (the paths-Dirac procedure produces no unexpected secondary paths-constraints).

\subsubsection{Classical scaled-triangleland, Dirac style histories}

Here, the classical histories are $\rho^{i\mu}\big(t^{\se\sm(\sJ\sB\sB)}\big)$ with conjugate paths momenta $p_{i\mu}\big(t^{\se\sm(\sJ\sB\sB)}\big)$.
 
\noindent The nonzero part of the paths brackets algebra is then  
\beq
\\big\{\rho^{i\mu}\big(t^{\se\sm(\sJ\sB\sB)}_1\big), p_{j\nu}\big(t^{\se\sm(\sJ\sB\sB)}_2\big)\big\} = \delta^i_j\delta^{\mu}_{\nu}\updelta
\big(t^{\se\sm(\sJ\sB\sB)}_1 - t^{\se\sm(\sJ\sB\sB)}_2\big) \mbox{ } . 
\eeq

\noindent This now comes with the paths total zero angular momentum constraint 
$\scL_{t^{\te\tm(\tJ\tB\tB)}} := \int\d t^{\se\sm(\sJ\sB\sB)} \scL(t^{\se\sm(\sJ\sB\sB)}) = 0$ 
and the paths energy constraint 
$\scE_{t^{\te\tm(\tJ\tB\tB)}} := \int \d t^{\se\sm(\sJ\sB\sB)}\scE(t^{\se\sm(\sJ\sB\sB)})$ with $\scE$ given by the triangleland case of (\ref{calE}). 
Whilst Savvidou and Anastopoulos considered the 3-$d$ non-relativistic particle in this context, 
they proceeded by reduction from this point on, which, for us, would just send us back one SSSec; 
also the most relational approach does not presume 3-$d$-ness of the 3-particle case (c.f. Sec \ref{Examples}).

\noindent The paths constraints algebra is now completely commutative, with with the zeroness of
\beq
\{ \scE_{t^{\te\tm(\tJ\tB\tB)}}, \scL_{t^{\te\tm(\tJ\tB\tB)}} \} = 0
\eeq 
significant in identifying $\scL_{t^{\te\tm(\tJ\tB\tB)}}$ as a paths-conserved quantity (it is not an observable
in this theoretical setting because it then has no operational meaning as an entity fully observable within a given, even specious, instant).  

\noindent Whilst this example is useful as a formal illustration, we have an implicitness problem with it due to $t^{\se\sm(\sJ\sB\sB)}$ not being known unless the 
Best Matching Problem is solved, but, if it is, it is rather more natural to work in the preceding SSSec's reduced approach.

\subsection{Discussion on classical Histories Theory}

\noindent Difference 35) Let us next parallel Kouletsis and Kucha\v{r} \cite{KK02}, or Kouletsis \cite{Kouletsis08}.  
RPM's, like Newtonian Mechanics, have a privileged time (now emergent, but still privileged). 
Thus their equivalent of how to split ${\cal M}^4$ is trivial, due to that privileged slicing. 
For scaled triangleland one has $\mathbb{R} \times \mathbb{R}^6$ in Dirac presentation and $\FrQ = \mathbb{R} \times \mathbb{R}^3$ with nonflat metric in reduced presentation.  
Due to that split, the notion of time map becomes trivial, $\tau: \FrT \times \mathbb{R}^6 \longrightarrow \FrT$, 
as does that of space map $\chi: \FrT \times \mathbb{R}^6 \longrightarrow \mathbb{R}^6$ (space map is even more trivial in the reduced presentation).
There is, moreover no issue of spacetime symmetries and how these are split up.  
Foliation issues and constraint algebra nontrivialities are absent from RPM's.

\subsection{Halliwell-type combined strategies for $N$-stop metroland}\label{Cl-Combo}

This case is limited since its $\FrG$ is trivial.
It is also mathematically the same as Halliwell's own example in Sec \ref{Metho}, but it now has relational and whole-universe connotations; 
the classical part of this now is a Classical {\sl Machian}-Histories-Records scheme.

I now \cite{AHall} use phase space functions  
\beq
A(\brho, \brho_0, \mbox{\boldmath $p$}_0) = \int_{-\infty}^{+\infty} \d t^{\se\sm(\sJ\sB\sB)} \updelta^{(n)}(\brho - \brho^{\scc\sll}(t^{\se\sm(\sJ\sB\sB)}) ) \mbox{ } . 
\label{candi2}
\eeq
These obey 
\beq
\{ \scQ\scU\scA\scD, A(\brho, \brho_0, \mbox{\boldmath $p$}_0)\} = 0 \mbox{ } ,
\label{corn22}
\eeq
so they are classical Dirac observables/beables.  

Next,
$$ 
\mbox{Prob}
\left( 
\stackrel{\mbox{intersection with}}
         {\mbox{some region $\bFrR$}}
\right) = 
\int_{-\infty}^{+\infty}\d t^{\se\sm(\sJ\sB\sB)}\,\mbox{Char}_{\bsFrR}\brho^{\scc\sll}(t^{\se\sm(\sJ\sB\sB)})) = 
\int \mathbb{D}\brho \, \mbox{Char}_{\bsFrR}(\brho) \, \int_{-\infty}^{+\infty}\d t^{\se\sm(\sJ\sB\sB)} \updelta^{(n)}
\big(\brho - \brho^{\scc\sll}(t^{\se\sm(\sJ\sB\sB)})\big) 
$$
\beq
= \int \mathbb{D}\brho \, \mbox{Char}_{\bsFrR}(\brho) \, A(\brho, \brho_0, \mbox{\boldmath $p$}_0 ) :
\eeq
the `amount of $t^{\se\sm(\sJ\sB\sB)}$' the trajectory spends in the relational configuration space region $\bFrR$. 
Then,
\beq
P_{\bsFrR} = \int \mathbb{D}\mbox{\boldmath $p$}_0 \, \mathbb{D}\brho \, \mw \big(\brho_0,\mbox{\boldmath $p$}_0\big) \,\, \theta
\left(
\int_{-\infty}^{+\infty}\d t^{\se\sm(\sJ\sB\sB)} \mbox{Char}_{\bsFrR}\big(\brho^{\scc\sll}(t^{\se\sm(\sJ\sB\sB)})\big) - \epsilon
\right) \mbox{ } .  
\eeq

\noindent The alternative expression for the flux through a piece of an \{$n$ -- 1\}-dimensional hypersurface within the configuration space is now  
$$
P_{\Upsilon} = \int\d t^{\se\sm(\sJ\sB\sB)}\int\int \mathbb{D}\mbox{\boldmath $p$} \, \mathbb{D}\brho_0 \, \mw(\brho_0,\mbox{\boldmath $p$}_0) 
\int_{\Upsilon} \d^2{\Upsilon}(\brho^{\prime}) \, \bnu \cdot \frac{\d \brho^{\scc\sll}(t^{\se\sm(\sJ\sB\sB)})}{\d t^{\se\sm(\sJ\sB\sB)}} 
\, \updelta^{(n)}\big(\brho - \brho^{\scc\sll}(t^{\se\sm(\sJ\sB\sB)})\big)
$$
\beq
= \int \d t^{\se\sm(\sJ\sB\sB)} \int\mathbb{D}\mbox{\boldmath $p$}^{\prime} \int_{\bupSigma} \mathbb{D}{\bupSigma}(\brho^{\prime})\, 
\bnu^{\prime}\cdot\mbox{\boldmath $p$}^{\prime} \, \mw(\brho^{\prime}, \mbox{\boldmath $p$}) \mbox{ } , 
\eeq
where $\bnu$ is the normal to the region in question.
The latter equality is by passing to $\brho^{\prime} := \brho^{\scc\sll}(t^{\se\sm(\sJ\sB\sB)})$ and $\mbox{\boldmath $p$} 
:= \mbox{\boldmath $p$}_{\scc\sll}(t^{\se\sm(\sJ\sB\sB)})$ coordinates at each $t^{\se\sm(\sJ\sB\sB)}$.

\subsection{Halliwell-type combined strategies for scaled triangleland}

I now use functions
\beq
A(\mbox{\scriptsize\boldmath$Dra$}, \mbox{\scriptsize\boldmath$Dra$}_0, \biP^{{\mbox{\tiny$Dra$}}}_0) = \int_{-\infty}^{+\infty}\d t^{\se\sm(\sJ\sB\sB)} \updelta^{(3)}
\big(
\mbox{\scriptsize\boldmath$Dra$} - \mbox{\scriptsize\boldmath$Dra$}^{\scc\sll}(t^{\se\sm(\sJ\sB\sB)})
\big) \mbox{ } .  
\eeq
These commute with $\scL$ and $\scE$ by the argument around equations (\ref{qirk}) and (\ref{corn22}) so they are classical Dirac observables. 
Regions of configuration space for RPM's, includes cases of particularly lucid physical significance as per Sec 3's tessellation interpretation.  
Now,
\beq
P_{\sbFrR} := \mbox{Prob}(\mbox{classical solution will pass through a triangleland configuration space region $\FrR$}) \mbox{ } . 
\eeq 
Next, I evoke $\mbox{Char}_{\sbFrR}(\mbox{\scriptsize\boldmath$Dra$})$ as the characteristic function of the triangleland configuration space region $\FrR$.  
Then  
$$
\mbox{Prob(intersection with $\bFrR$)} = \int_{-\infty}^{+\infty}\d t^{\se\sm(\sJ\sB\sB)}\,\mbox{Char}_{\sbFrR}(\mbox{\scriptsize\boldmath$Dra$}^{\scc\sll}(t^{\se\sm(\sJ\sB\sB)})) = 
$$
\beq
\int \mathbb{D} \mbox{\scriptsize\boldmath$Dra$} \, 
\mbox{Char}_{\sbFrR}(\bq) \, \int_{-\infty}^{+\infty}\d t^{\se\sm(\sJ\sB\sB)} 
\updelta^{(3)}\big(\mbox{\scriptsize\boldmath$Dra$} - \mbox{\scriptsize\boldmath$Dra$}^{\scc\sll}(t^{\se\sm(\sJ\sB\sB)})\big) 
= \int \mathbb{D} \mbox{\scriptsize\boldmath$Dra$} \, \mbox{Char}_{\sbFrR}(\mbox{\scriptsize\boldmath$Dra$}) \, A(\mbox{\scriptsize\boldmath$Dra$}, \mbox{\scriptsize\boldmath$Dra$}_0, \biP^{Dra}_0 ) \mbox{ } :
\eeq
the `amount of $t^{\se\sm(\sJ\sB\sB)}$' the trajectory spends in region $\bFrR$.  
Then 
\beq
P_{\sbFrR} = \int \d^3\biP^{\mbox{\tiny$Dra$}}_0 \, \d^3\mbox{\scriptsize\boldmath$Dra$}_0 \, \mw\big(\mbox{\scriptsize\boldmath$Dra$}_0,\biP^{{\mbox{\tiny$Dra$}}}_0\big) \,\, \theta
\left(
\int_{-\infty}^{+\infty}\d t^{\se\sm(\sJ\sB\sB)} \mbox{Char}_{\sbFrR}\big(\mbox{\scriptsize\boldmath$Dra$}^{\scc\sll}(t^{\se\sm(\sJ\sB\sB)})\big) - \epsilon
\right)   \mbox{ } .  
\eeq
\noindent An alternative expression is for the flux through a piece of a 2-$d$ hypersurface within the configuration space, 
$$
P_{\Upsilon} = \int\d t^{\se\sm(\sJ\sB\sB)}\int \d^3\biP^{{\mbox{\tiny$Dra$}}}_0 \, \d^3\mbox{\scriptsize\boldmath$Dra$}_0 \, \mw(\mbox{\scriptsize\boldmath$Dra$}_0, \biP^{{\mbox{\tiny$Dra$}}}_0) 
\int_{\Upsilon} \d^2{\Upsilon}(\mbox{\scriptsize\boldmath$Dra$}) \, \bn^{{\mbox{\tiny$Dra$}}} \cdot \mbox{\boldmath$M$} \cdot \frac{\d \mbox{\scriptsize\boldmath$Dra$}^{\scc\sll}(t^{\se\sm(\sJ\sB\sB)})}{\d t^{\se\sm(\sJ\sB\sB)}} 
\, \updelta^{(k)}\big(\mbox{\scriptsize\boldmath$Dra$} - \mbox{\scriptsize\boldmath$Dra$}^{\scc\sll}(t^{\se\sm(\sJ\sB\sB)})\big)
$$
\beq
= \int \d t^{\se\sm(\sJ\sB\sB)} \int\mathbb{D}\biP^{{\mbox{\tiny$Dra$}} \prime} \int_{\Upsilon} \mathbb{D}{\bupSigma}(\mbox{\scriptsize\boldmath$Dra$}^{\prime})\, \bn^{{\mbox{\tiny$Dra$}}\prime} 
\cdot \biP^{{\mbox{\tiny$Dra$}}\prime} \, \mw(\mbox{\scriptsize\boldmath$Dra$}^{\prime}, \biP^{{\mbox{\tiny$Dra$}}\prime}) \mbox{ } , 
\eeq
the latter equality being by passing to $\mbox{\scriptsize\boldmath$Dra$}^{\prime} := \mbox{\scriptsize\boldmath$Dra$}^{\scc\sll}(t^{\se\sm(\sJ\sB\sB)})$ and $\biP^{{\mbox{\tiny$Dra$}}\prime} 
:= \biP^{Dra}_{\scc\sll}(t^{\se\sm(\sJ\sB\sB)})$ coordinates at each $t^{\se\sm(\sJ\sB\sB)}$.
Note that the first form,  via the cancelling out of its d$t$'s, possesses MRI and so complies with Temporal Relationalism.

\subsection{Classical Dirac degradeables for RPM's} 

\noindent The preceding 2 SSec's objects comply with \{$\scL$, \iB\} and \{$\scE$, \iB\} = 0 (though the first of these SSecs does so trivially in the case of $\scL\scI\scN_{\sfZ}$).
Explicit calculations for the $\mathbb{CP}^2$ of quadrilateralland are currently under investigation \cite{QuadIV}.  

\mbox{ } 

\noindent Whilst these O's are histories constructs, the final objects themselves are integrals over all times, so they are indeed beables as opposed to histories beables. 

\mbox{ } 

\noindent Global Problem with Halliwell-type Combined Approach) There is an issue of operationally useable for entities that require evaluation over all of history. 
\noindent One needs time to run over all the reals, which clashes with global-in-time obstructions with the $t^{\se\sm(\sJ\sB\sB)}$ in use (Sec \ref{JBBGlob}).  
\noindent  This is to include how the $\int_a^b$ version fails to commute with $H$.  
There are even more causes of global restriction at the quantum level, for which $t^{\se\sm(\sW\sK\sB)}$ is in use (Sec \ref{22-end}).  

\mbox{ }  

\noindent The lack of global compliance from $t^{\se\sm(\sJ\sB\sB)}$ having global-in-configuration space and/or global-in-space issues is 
emblazoned by referring to the entities in question as degradeables. 
Though conceptually "at all times" requiring use of multiple coordinate patches can be anticipated, the calculation is then whether such a patching succeeds in commuting with 
$\scQ\scU\scA\scD$.
Also the Poisson bracket condition is a p.d.e. so need to worry about this at the level of p.d.e. solutions rather than just in terms of classical meshing.  
Though we also need to check the nature of the corresponding entities in Halliwell 2009, rather than just dwelling on the features of Halliwell 2003.  

\vspace{10in}

\noindent{\bf \Huge Part III. Quantum RPM's}\label{QRel}

\section{Quantization, relationalism and RPM's}\label{QM-Intro}

\subsection{Outline of a quantization scheme for $\FrG$-trivial theories}

I consider this along the lines of Isham's scheme \cite{I84}, which, whilst quite general, is by no means all-embracing.  
Being an approach with some ties to geometrical quantization, it is a natural partner for the Jacobi--Synge approach 
to Classical Mechanics as used in Secs \ref{Intro} and \ref{Examples}.  
I take this to consist of the following steps, which compose as per Fig \ref{Mewl}.   
%
{            \begin{figure}[ht]
\centering
\includegraphics[width=0.72\textwidth]{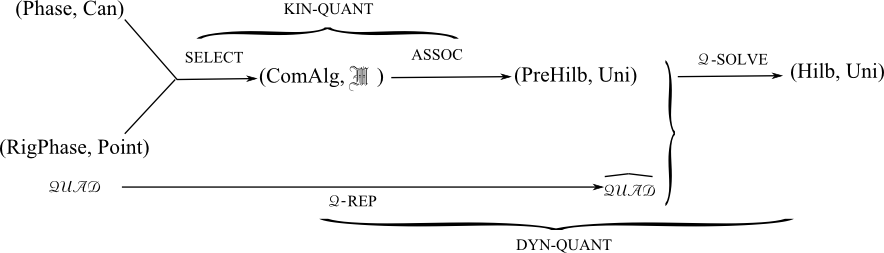}
\caption[Text der im Bilderverzeichnis auftaucht]{        \footnotesize{Breakdown of a simple quantization scheme {\sc quant} ( {\sc q} ) in the 
absence of redundancies.} }
\label{Mewl}\end{figure}            }

\subsubsection{Kinematical quantization ({\sc kin-quant})}

\noindent {\boldmath{\sc select}} is 1) the selection of a set of classical objects $\cF\lfloor\bfQ, \bfP\rfloor$ that are to be promoted to quantum operators 
$\widehat{\cF}\lfloor\bfQ, \bfP\rfloor$. 

\noindent 2) Passing from the classical Poisson bracket algebra: $\langle \bfQ, \bfP, \{ \mbox{ } , \mbox{ } \} \rangle$ 
that is Phase or RigPhase with corresponding morphisms Can and Point respectively, 
to some commutator algebra, ComAl:            $\cF\lfloor \hat{\bfQ}, \hat{\bfP} \rfloor \mbox{ that close under } [ \mbox{ } , \mbox{ } ]$), 
with corresponding commutator-preserving morphisms \textgoth{M}.  
%

\mbox{ } 

\noindent {\boldmath{\sc assoc}} then associates a pre-Hilbert space PreHilb of pre-wavefunctions for these operators to act on; the corresponding morphisms are unitary 
transformations, Uni.  

\mbox{ } 

\noindent {\boldmath{\sc kin-quant}}:= {\sc select} $\circ$ {\sc assoc} is then {\bf kinematical quantization}.   

\mbox{ } 

\noindent {\sc select} is very obvious in the simplest example: $q^{\sfA} \longrightarrow \hat{q}^{\sfA}$, $p_{\sfA} \longrightarrow \hat{p}_{\sfA}$ 
alongside the usual (`correspondence principle') to the fundamental equal-time commutation relations, 
\beq
\mbox{from } \{q^{\sfA}, \mp_{\sfB}\} = \delta^{\sfA}_{\sfB} \mbox{ to } \mbox{ } 
[\hat{q}^{\sfA} \mbox{ } , \mbox{ } \hat{p}_{\sfB}] = i\hbar{\delta_{\sfA}}^{\sfB} \mbox{ } .  
\label{common}
\eeq
Here the two algebras are isomorphic under the obvious `correspondence principle' $\{ \mbox{ } , \mbox{ } \} \longrightarrow \frac{1}{i\hbar}[ \mbox{ } , \mbox{ } ]$.

\noindent However, {\sc select} exhibits a number of subtleties in more general cases (\cite{I84}): 
one has to make a {\it choice} of a preferred subalgebra of Phase objects  \cite{I84} to promote to QM operators.

\noindent Some context for this is that 

\noindent 1) nonlinear systems exhibit the {\bf Groenewold--van Hove phenomenon} (see \cite{Gotay00} and Secs \ref{MCP} and \ref{MCP-Ex}).\footnote{In fact, this result is only  
established for a number of simple phase spaces (see e.g. \cite{Gotay00} and Sec \ref{MCP-Ex}).}
%
By this, classical equivalence can be lost in passage to QM (Fig \ref{GVH}).

{            \begin{figure}[ht]
\centering
\includegraphics[width=0.47\textwidth]{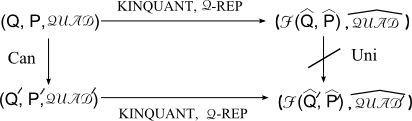}
\caption[Text der im Bilderverzeichnis auftaucht]{        \footnotesize{A major problem in quantization.}   }
\label{GVH} 
\end{figure}  } 

\noindent Example 1) the change of variables $q$, $p$ $\longrightarrow$ $q^3$, $p^3$ is classically harmless (trivially canonical), and yet it yields inequivalent quantum theories.  

\mbox{ }

\noindent There are also global considerations by which the quantum commutator algebra will {\sl not} in general be isomorphic to the classical Poisson bracket algebra. 

\noindent Example 2) Quantization on the half-line already serves to demonstrate this \cite{I84}.  
Here, 
\beq
q > 0
\eeq
impedes $\hat{p}$ as represented by $i\hbar\pa/\pa q$ from being properly self-adjoint.  
On the other hand, $\hat{p}$ represented by $-i\hbar\,{q}\,\pa/\pa q$ does not have this problem, corresponding to the affine commutation relation 
\beq
[\hat{q} \mbox{ } , \mbox{ } \hat{p}] = i\hbar\hat{q} \mbox{ } .
\eeq
\noindent Example 3) $\mathbb{S}^1$ is classically characterized by $\phi_{\sss\sp}, L$ but quantum-mechanically one requires $u_1$ = cos$\,\phi_{\sss\sp}$, 
$u_2$ = sin$\,\phi_{\sss\sp}$ and $L$ i.e. an algebra of dimension {\sl three}, due to noncontinuity of $\phi_{\sss\sp}$ itself.  

\mbox{ } 

\noindent Some further developments in the context of Isham's $\FrQ/\FrG$ example for $\FrG$ a subgroup of $\FrQ$; 
it is important for this Article that all specific RPM's in this Article are nested within this. 
For this, the relevant spaces involved in kinematical quantization can be decomposed as semisimple products $\mathfrakV^*(\FrQ) \mbox{\textcircled{S}} \FrG_{\scc\sa\sn}(\FrQ)$. 
Thus here ComAl = ($\mathfrakV^{*} \mbox{\textcircled{S}} \FrG_{\scc\sa\sn}(\FrQ)$, [ \mbox{ } , \mbox{ }]).
Here, $\FrG_{\scc\sa\sn}(\FrQ)$ is the canonical group and $\mathfrakV^{*}$ is the dual of a linear space $\mathfrakV$ that is natural 
via carrying a linear representation of $\FrQ$ with the property that there is a $\FrQ$-orbit in $\mathfrakV$ that is diffeomorphic to $\FrQ/\FrG$ \cite{I84}.  
Mackey theory is a powerful tool for finding the representations of such semidirect products.  

\mbox{ } 

\noindent Note. One problem here the commutator algebra's closure depends on operator orderings that were trivially equivalent for the brackets' prior closure. 
These can be algebra-altering including in `harsh' ways known as {\bf anomalies} \cite{Dirac, ModernAnomalies1, ModernAnomalies2}, 
by which QM refuses to accept some of what were perfectly good symmetries at the classical level.

\subsubsection{Dynamical Quantization ({\sc dyn-quant})}

\noindent{\boldmath{\scQ}}-{\sc rep} is the promotion $\scQ\scU\scA\scD \longrightarrow \widehat{\scQ\scU\scA\scD}$ as a functional of {\sc kin-quant}'s operators.
This procedure yields a wave equation.

\noindent Note furthermore that there are operator-ordering issues at stake, as well as well-definedness issues (functional derivatives, compositions of operators with no 
functional-analytic reasons to be well-behaved) and regularization issues.  

\mbox{ } 

\noindent {\boldmath{\scQ}}{\sc -solve} is then solving this: passing from PreHilb to Hilb: the Hilbert space within that is annihilated by $\widehat{\scQ\scU\scA\scD}$.  

\noindent Note that Hilb's inner product restricts what are the finally-valid fully physical operators by providing its own notion of self-adjointness. 

\mbox{ }
\beq
\noindent {\mbox{\boldmath{\sc dyn-quant}}} := \scQ\mbox{\sc -solve} \circ \scQ \mbox{\sc -rep} \mbox{ then the {\bf dynamical quantization}: }
(\mbox{PreHilb}, \mbox{\textgoth M}^{\prime}) \longrightarrow \mbox{(Hilb, Uni)} \mbox{ } . 
\eeq
\noindent Finally, in this $\FrG$-free case quantization {\sc q} := {\boldmath{\sc quant}} = {\sc dyn-quant} $\circ$ {\sc kin-quant} is this approach's basic notion of 
{\it quantization}. 
N.B. as formally defined here, this acts on a {\sl triple} such as (Phase, Can, $\scQ\scU\scA\scD$) or (RigPhase, Point, $\scQ\scU\scA\scD$).

\subsubsection{Aside: spaces of mixed states}\label{Mixed-State-Space}

These are much larger than the corresponding spaces of pure states.  
E.g. the discrete up-down system has a whole $\mathbb{S}^2 = \mathbb{CP}^1$ of mixed states; 
more generally the qu$n$it has a $\mathbb{CP}^{n - 1}$ of mixed states, complete with a Fubini--Study metric upon it.

\subsection{Relationalism at the quantum level. I}

\noindent \bu Tangibility, Temporal Relationalism and Configurational Relationalism ideas remain relevant in the quantum arena. 
However, {\sl implementations} are different now, for at the classical level these were implemented at the level of actions, 
whereas in Quantum Theory there usually are no actions.\footnote{There are, however, still actions within which one can implement 
Temporal Relationalism as MPI in path integral approaches to Quantum Theory and, furthermore, in Histories Theory approaches to the POT.}

\noindent \bu Conventionally (and already in good keeping with relationalism), the objects containing tangible physics are (using Dirac notation for the inner product) 
\beq
\langle \Psi_1 |\widehat{O}| \Psi_2 \rangle
\eeq
for $\Psi_1$, $\Psi_2$ wavefunctions and $\widehat{O}$ an operator.
This includes expectations of operators ($\Psi_1 = \Psi_2$), overlap integrals ($\widehat{O}$ = id) and probabilities for regions 
(taking both of these conditions but restricting the integration implicit in the inner product to apply only to a region of configuration 
space; sometimes it is necessary to only consider ratios of these).  
These objects are invariant under unitary transformations $U$ by cancellation of the left action $\Psi^*U$, adjoint action $U\widehat{O}U^{\dagger}$ and right action $U\Psi$.

\bu Temporal Relationalism.
We have argued that quantum timelessness rests on Leibniz's `there is no time for the universe as a whole'.  
The relationalist explanation for this is then Temporal Relationalism $\Rightarrow$ MPI actions $\Rightarrow$ homogeneous quadratic constraint 
${\scQ\scU\scA\scD} \mbox{ } \Rightarrow$ at the quantum level the 


\noindent {\bf Frozen Formalism Problem} facet of the POT, i.e. that the above wave equation $\widehat{\scQ\scU\scA\scD}\Psi = 0$ is timeless, 
in contradistinction from such as a TDSE $i\pa\psi/\pa t = \hat{H}\Psi$, 
Klein--Gordon equation $\Box\psi = \triangle\psi - \pa_{tt}\psi = m^2\psi$ or 
Dirac equation $i\{\gamma^t_{\sfA\sfB}\pa_t\psi^{\sfB} - \gamma^{\mu}_{\sfA\sfB}\}\pa_{\mu}\psi^{\sfB} = m\psi^{\sfA}$.\footnote{Here, I use 
$\psi$ as wavefunction versus $\Psi$ as wavefunction {\sl of the universe}.  
$\psi^{\tfB}$ is a Dirac spinor and $\gamma^{\Gamma}_{\tfA\tfB}$ are Dirac matrices.  
$\Gamma$ takes values $\mu$ (1 to 3) and t.}      
%
each of which is backed by absolutist time notions.
See Sec \ref{QM-POT} for more on the quantum Frozen Formalism Problem.

\subsection{Motivation for studying relational theories at the quantum level}

\subsubsection{Do absolute and relational mechanics lead to different QM?}\label{Min-Motiv}

This motivation follows on from the historical interest in freeing mechanics from absolute structure.  
Would a different sort of QM or generalization thereof have arisen if the conception of mechanics itself 
had been relational \cite{BS, Rovelli, Smolin, 06I, FORD, Cones} rather than absolute?\footnote{This 
issue does not go away in considering relativistic QM or QFT on flat spacetime as these have a privileged Killing vector that effectively 
re-assumes Newtonian time's absolutist role (see also the Preface and Sec \ref{Intro}).  
One can extend this to relativistic particle dynamics in other stationary spacetimes, but, unfortunately, that is as far as it will go \cite{Kuchar91}.  }
%
I.e., has the traditional absolutist approach been misleading us at the quantum level?  
Or does an essentially identical theoretical framework arise from the relational approach too? 
(Perhaps due to some deeper-seated underlying foundation common to both approaches? 
Or perhaps due to curious parallels/technical coincidences?)  

\mbox{ } 

\noindent N.B. what is new in the relational approach is certainly {\sl not} the {\sl use} of relative/relational quantities in calculations -- 
that is already widespread in the standard approach to standard QM.  

\mbox{ }

\noindent In this respect, I find a {\sl lack} of non-standard QM at least in the simpler examples (c.f. the mathematical analogies listed in Sec \ref{Cl-POT-Strat} and \ref{+temJBB}).  
By this means earlier work with absolutist QM gives us plenty of machinery to ease the solving of the simpler examples of relational QM too.  
Note however that the latter but not the former are appropriate as whole-universe models and for detailed analysis of the POT.  
Thus this coincidence is a triumph as regards finding tractable toy models (to the extent that these are useful toy models of generally-relativistic whole-universe models).  

\mbox{ }

\noindent Moreover, there are some differences between absolute and relational QM.  
See Sec \ref{RPM-for-QC} for a summary of those found in this Part.

\subsubsection{Understanding GR and QG LMB(-CA) relationalism {\sl becomes} the quantum POT}

What was classical relationalism becomes, at the quantum level, much of the POT in QG.
This is the subject of Part IV.

\noindent My principal motivation remains the study the quantization of RPM's is that it is likely to be valuable, via the 
RPM--Geometrodynamics analogy for the following Quantum Gravity and Quantum Cosmology investigations. 
Thus what particular features quantum GR has is a bigger thrust here than 'does relationalism paint a different picture of QM?'
We lay things out in general enough terms to be able to (at least formally) discuss GR.

\subsubsection{Tests for techniques for quantization}

Moreover, RPM's also make for interesting examples in which to test out techniques for quantization (\cite{BS, Rovelli, Smolin, 06I, Gryb1, Gryb2, BF08, Banal, GT11} and the 
present Article).

\subsection{Simple $\FrG$-less RPM example of {\sc kin-quant}}\label{ReeRu}

Here, the canonical group $\FrG_{\scc\sa\sn} = \mbox{Isom}({\cal R}(N, 1)) = \mbox{Eucl}(n) = \mbox{Tr}(n) \mbox{\textcircled{S}} \mbox{Rot}(n) =
\mathbb{R}^{n} \mbox{\textcircled{S}} SO(n)$ and an appropriate finite subalgebra is $\mathbb{R}^{n} \mbox{\textcircled{S}} \mbox{Eucl}(n)$.   

\mbox{ }  

\noindent Note 1) In greater generality, from \cite{I84}, $\FrG_{\scc\sa\sn}(\FrQ)$ = Isom($\FrQ$) for the span of all of this Article's specific RPM's (in fact all the 1 and 2 d RPM's). 

\noindent Note 2) Also in greater generality, $\mathfrakV^* = \mathfrakV$ for finite examples.

\mbox{ }

\noindent Then e.g. for scaled 4-stop metroland, the nontrivial commutation relations are then the $U^{\sfA} \longrightarrow \mn^i$, $P_{\sfA} \longrightarrow p_i$ 
and $\sfS_{\sfA} \longrightarrow \sfD_i$ of App \ref{At}. 
These can be recast in a dual form, with $\rho_i = \delta_{ij}\rho^j$ and $\sfD_{ij} = \rho_ip_j - \rho_jp_i$: (\ref{bog-standard}) alongside 
\beq
[\widehat\sfD_{ij},\widehat\sfD_{kl}] = i\hbar\{
\delta_{ik}\widehat\sfD_{jl} + \delta_{jl}\widehat\sfD_{ik} - 
\delta_{il}\widehat\sfD_{jk} - \delta_{jk}\widehat\sfD_{kl}  \} 
\mbox{ } ,
\eeq
\beq
[\widehat{\rho}^{i},\widehat\sfD_{kl}] = i\hbar\{\delta_{lj}\delta^i\mbox{}_k - 
\delta_{lk}\delta^i\mbox{}_j \}\widehat{\rho}^k 
\mbox{ } , \mbox{ } \mbox{ }
[\widehat{p}_{i},\widehat\sfD_{kl}] = i\hbar\{
\delta_{lj}\delta_i\mbox{}^k - \delta_{ij}\delta_l\mbox{}^k \}\widehat{p}_k \mbox{ } ,
\eeq
which has the benefit of extending to all of the higher-$N$ pure-shape $N$-stop metrolands.
3-stop metroland, however, has to be treated differently: the $U^{\sfA} \longrightarrow \mn^i$, $P_{\sfA} \longrightarrow p_i$ and $\sfS \longrightarrow \scD$ of App \ref{Devlan}.

\subsection{Outline of a quantization scheme for $\FrG$-nontrivial, classically-unreduced theories}

Barbour-relationalism's indirect implementation at the classical level gives constraints $\scQ\scU\scA\scD$ and $\scL\scI\scN$.
The reduced/relationalspace r-scheme gives a constraint $\widetilde{\scQ\scU\scA\scD}$. 
Both Dirac and reduced schemes are relational.  
In the preceding literature, Barbour, Rovelli and Smolin \cite{BS,Rovelli,Smolin} had largely not considered the LMB brand of relationalism 
along the above lines beyond the general idea that Dirac Quantization is an indirect scheme that implements Configurational Relationalism. 
Barbour had on the other hand considered a quantum timeless records approach (\cite{B94I, EOT}, Sec \ref{QM-Nihil}) which has some relational features.

The $\FrG$-act, $\FrG$-all method serves for Timeless Records Theory, and \cite{AHall} used it along the lines of Halliwell's combined approach.

Methods such as the group integration approach (essentially a variant/particular implementation of the Dirac quantization approach) are fairly widely used in LQG \cite{Phoenix, DT04}.  
In this sense, LQG does implement Configurational Relationalism at the quantum level.  

{            \begin{figure}[ht]
\centering
\includegraphics[width=0.92\textwidth]{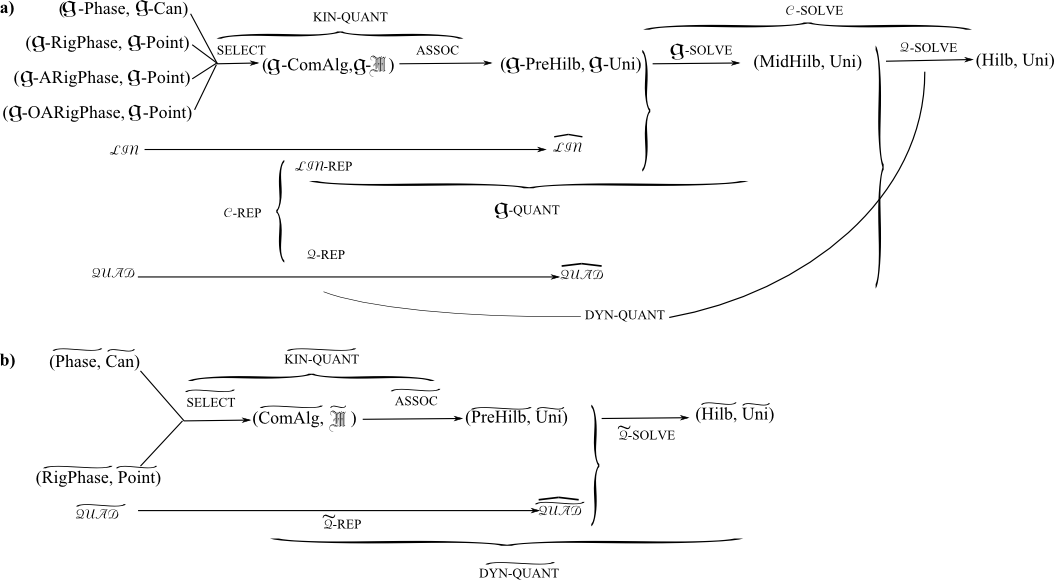}
\caption[Text der im Bilderverzeichnis auftaucht]{        \footnotesize{Breakdown of a simple quantization scheme in the presence 
of a group $\FrG$ of physically-redundant transformations. a) The Dirac-type scheme. b) The reduced-type scheme.
See also Fig \ref{Big-Arrow}.} }
\label{Rawr}\end{figure}            }

\mbox{ }

\noindent In this more general setting [Fig \ref{Rawr}.a)], {\sc q} := {\sc dir-quant} in this case {\sc dyn-quant} $\circ$ $\FrG$-{\sc quant} $\circ$ {\sc kin-quant}.

\noindent {\boldmath{$\FrG$}}-{\sc quant} := $\FrG$-{\sc solve} $\circ$ $\FrG$-{\sc rep}, each of which individual steps is defined in obvious parallel to $\scQ$-{\sc rep} and 
$\scQ$-{\sc solve}.  
The MidHilb, which stands for `middlingly physical Hilbert space', 
i.e. those PreHilb states that are annihilated by $\widehat{\scL\scI\scN}$ but not necessarily by $\widehat{\scQ\scU\scA\scD}$

\noindent Note 1) I also define {\boldmath{$\scC$}}-{\sc rep}   = \scQ-{\sc rep} $\circ$ $\FrG$-{\sc rep} (these maps commute), 
and                     {\boldmath{$\scC$}}-{\sc solve} = \scQ-{\sc solve} $\circ$ $\FrG$-{\sc solve}.  
These maps make particular sense in these approaches that accord equal status to linear and quadratic constraints.  

\mbox{ } 

\noindent Note 2) It thus makes sense to also present $\FrG$-{\sc constrain} as a functor.  
As well as Sec \ref{Q-Geom}'s purely-classical Ante Quantum,
\beq
\mbox{$\FrG$-\mbox{\sc constrain}: ($\FrG$-Phase, Can, $\FrG$-Lin, $\scQ\scU\scA\scD$)}  \longrightarrow 
                       (\widetilde{\mbox{Phase}}, \widetilde{\mbox{Can}}, \widetilde{\scQ\scU\scA\scD}) \mbox{ }    .
\eeq
Post Quantum, 
\beq
\mbox{$\FrG$-{\sc constrain}: (PreHilb, \mbox{\textgoth\Large m}, $\FrG$-{\sc lin}, $\scQ\scU\scA\scD$)}  \longrightarrow (\mbox{MidHilb}, \mbox{Uni}, \scQ\scU\scA\scD^{\sr}) \mbox{ }  
\eeq  
$\scC$ corresponds to the full algebraically-closing set of constraints $\widehat{\scC}_{\sfW} = (\widehat{\scL\scI\scN}_{\sfZ}$, $\widehat{\scQ\scU\scA\scD}$).  

\noindent This entails an operator-ordering choice and that the constraints close under the commutator, 
possibly with a quantum supplement or some strong restriction on what was a free parameter classically -- the dimension in string theory being a good example.  

\noindent $\scC$-{\sc rep} may have more $\widehat{\scC}$'s than it has $\scC$.

\noindent It might also come with anomalies or with strong restrictions that prevent what would otherwise be anomalies.  

\mbox{ }

\noindent Note 1) The a posteriori $\FrQ$--$\FrG$ incompatibility of Relationalism 11) can strike again now, via brackets not closing including due to anomalies.

\subsection{Outline of a quantization scheme for $\FrG$-nontrivial classically-reduced theories}

\noindent Reduced-type formulations also embody Configurational Relationalism.  
These are essentially tilded versions of Fig \ref{Mewl}, available whenever one has the good fortune to, at the classical level, 
beable to directly formulate or reduce away all the redundancies from the indirect formulation.
This is covered in Fig \ref{Rawr}b) and amounts to a tilde-ing of Fig \ref{Mewl} (itself following on from the classical reduction of Sec \ref{Q-Geom} as per Fig \ref{Big-Arrow}).  
\beq
\mbox{\sc red-pre-dyn-quant} = {\scC}\mbox{\sc -constrain} \circ \mbox{\sc kin-quant} \circ \scC\mbox{\sc -rep}   \circ \mbox{\sc assoc}
\mbox{ } , 
\eeq

\subsection{Inequivalence, and GR-RPM comparison}

\noindent One issue then is that $\tilde{\hat{\mbox{ }}}$ and $\hat{\tilde{\mbox{ }}}$ as acting on $\scQ\scU\scA\scD$ are not expected to commute, so the final {\sc dyn-quant} 
step's outcome for each of the above could well be different fir each of these (i.e. the well-known Dirac--reduced quantization inequivalence, {\sc qr} $\neq$ {\sc rq}).  

\noindent Note 1) the above can be written out for both, using r labels for reduced with a smaller {\sc solve} step, 
and using D labels at the Hilbert space level for the versions obtained via the Dirac approach. 

\mbox{ }  

\noindent Analogy 84) In each case, the presence of additional linear constraints permits one to choose to attempt the Dirac quantization approach.  
Geometrodynamics, not being reducible in general, that is one of few schemes that can be attempted in that case. 
There one does not have in general the luxury of reduced/relationalspace quantization.  

\mbox{ }

\noindent Difference 36)  As consequence of Sec \ref{Q-Geom},  one has the good fortune of being able to perform reduce/relationalspace quantization for RPM's; 
such are not open in general for midisuperspace GR and upwards.

\subsection{RPM example of {\sc kin-quant}}

\subsubsection{Dirac approach}\label{Ind-RPM-KinQ}

Take $\FrQ = \Fr(N, 2)$ and $\FrG$ = Rot(2).
One is on $\mathbb{R}^{n d}$, so one has the canonical group $\FrG_{\scc\sa\sn}(\Fr(N, 1)) = \mbox{Isom}(\mathbb{R}^{nd})
\mbox{Eucl}(n d)$ and so $\mathfrakV \mbox{\textcircled{S}}\FrG_{\scc\sa\sn}$ = $\mathbb{R}^{n d}  \mbox{\textcircled{S}} \mbox{Eucl}(n, d)$ (the Heisenberg group).  
Thus the usual canonical commutation relations apply:  
\beq
[\widehat{\rho}^{i\mu}, \widehat{p}_{j\nu}] = i\hbar{\delta^i}_j{\delta^{\mu}}_{\nu} 
\mbox{ }   
\eeq 
alongside the $\rho$'s and $p$'s being $SO(nd)$-vectors and $SO(nd)$ commutation relations among the $\sfS$'s.

\noindent Note: As regards using (O(A))Phase in place of Phase here, in coming from the Poisson bracket, the commutator is not 
affected by the $\FrG$-sector, and so using A or dA does not amount to promoting anything other than configurations and momenta to operators.
The QM concerns only the part-tangible $p$'s and $q$'s.
This is a `Hamiltonian collapse' issue (c.f. Appendix \ref{Examples}.A).

\subsubsection{RPM examples of {\sc kin-quant} in the direct-r approach}\label{QRed}

For $N$-stop metroland's shape space $\FrS(N, 1) = \mathbb{S}^{n - 1}$, the canonical group $\FrG_{\scc\sa\sn}(\FrS(N, 1)) 
= \mbox{Isom}(\mathbb{S}^{n - 1}) =  SO(n)$ and an appropriate finite subalgebra is $SO(n) \mbox{\textcircled{S}} \mathbb{R}^{n}$  (p 1269-70 of \cite{I84}).  
In the present case, this can be taken to be generated by $n\{n - 1\}/2$ $SO(n)$ objects that are 
interpreted as relative distance momenta $\sfD_{\tilde{\Gamma}}$,\footnote{For 4-stop metroland, 
this is the case in H-coordinates, while in K-coordinates these are linear combinations of the $\scD\si\sll_{i}$.  }
alongside $n$ coordinates $u^{\Gamma}$ such that $\sum_{\Gamma}\{u^{\Gamma}\}^2 = 1$, that are 
interpreted as the unit Cartesian directions in the surrounding relational space $\mathbb{R}^{n}$.

For pure-shape $N$-a-gonland's $\FrS(N, 2) = \mathbb{CP}^{n - 1}$ shape space, the canonical group is 
$\FrG_{\scc\sa\sn}(\FrS(N, 2)) = \mbox{Isom}(\mathbb{CP}^{n - 1}) = SU(n)/\mathbb{Z}_n$ 
(perhaps one needs to take care here with including the further quotienting out by $\mathbb{Z}_n$).  
Moreover, since this shape space can also be written as $SU$($n$)/$U$($n$ -- 1), so it is a 
subcase of the general form $\FrQ^{\sr} = \FrQ/\FrG$ for $\FrG$ a subgroup of $\FrQ$ that is considered in \cite{I84}.  
Then a corresponding finite algebra is $\FrG_{\scc\sa\sn} \mbox{\textcircled{S}} \mathfrakV = 
SU(n)/\mathbb{Z}_n \mbox{\textcircled{S}} \mathbb{C}^{n} = SU(n)/\mathbb{Z}_n \mbox{\textcircled{S}} \mathbb{R}^{2n}$.  
However, for the triangle, it is possible to use the more minimalist $\mathbb{R}^3$ in place of $\mathbb{C}^2 = \mathbb{R}^4$.
This choice is as per App \ref{At} with $U^{\sfA} \longrightarrow Dra^{\Gamma}$.  
Moreover, the $\mathbb{C}^2$ includes absolutist information whereas the $\mathbb{R}^3$ is purely relational as is clear from the forms of the Dragt vectors.
Thus especially if one identifies the kinematical quantum algebra and the algebra of quantum \K beables, one wishes to select the former. 

\noindent It is not the immediately clear how to generalize this for the higher $N$-a-gonlands.
This is resolved \cite{QuadIII} by identifying the $\mathbb{R}^3$ as IHP($\mathbb{C}^2$, 2) -- irreducible homogeneous polynomials, complex-valued, of degree 2 --  
via the map relating the spatial 3-vectors and the Pauli matrices.  
Then it is clear that the relational generalization is $\FrG_{\scc\sa\sn} \mbox{\textcircled{S}} \mathfrakV = SU(n)/\mathbb{Z}_n \mbox{\textcircled{S}} \mbox{IHP}(\mathbb{C}^n$, 2).
Moreover, for the quadrilateral and higher, this ceases to be the more minimal-dimensional case, unlike for the triangle (Isham \cite{I84} suggests rather than requires 
minimality, and we also have the relational versus absolute reason to choose the adjoint representation).  

\mbox{ }

\noindent A sometimes useful Lemma for the scaled case is as follows.   

\mbox{ }  

\noindent{\bf Lemma 15}. Given a kinematical quantization algebra $\FC$ for a shape space, then if the corresponding relational 
space has no `extra' symmetries, the kinematical quantization algebra of the corresponding relational space is 
${\FC} \mbox{\textcircled{S}} {\FA}\mathfrak{ff}$ for ${\FA}\mathfrak{ff}$ the `radial'/$\mathbb{R}_+$ problem's affine algebra.  

\mbox{ }

\noindent 
However, this is not applicable to the $N$-stop metroland's relational space ${\cal R}(N, 1) = \mathbb{R}^{n}$ as this has a number of extra symmetries.  
Moreover, this particular case's mathematics is, of course well-known (Sec \ref{ReeRu}).  

It also likewise is not applicable to triangleland's relationalspace $\mathbb{R}^3$, which works like the $n = 3$  
case of the preceding except that the $\rho^i$ and $p_i$ are now, rather $Dra^{\Gamma}$ and $P^{Dra}_{\Gamma}$.  
Note that using e.g. parabolic coordinates instead does not change the underlying canonical group.  
This should be related to coordinate changes merely leading to new bases of wavefunctions that are linear combinations of other coordinatizations'.  

\mbox{ } 

\noindent An appropriate kinematical quantization \cite{I84} for scaled triangleland involves $\mbox{Eucl}(3) \mbox{\textcircled{S}} \mathbb{R}^3$ as per App \ref{At} with 

\noindent $U^{\sfA} \longrightarrow Dra^{\Gamma},  \P_{\sfA} \longrightarrow P^{Dra}_{\Gamma}$.  
A choice of objects is then Dra$^{\Gamma}$ for the first $\mathbb{R}^3$, the translational generators $P^{Dra}_{\Gamma}$ and the $SO(3)$ generators $\sfS_{\Gamma}$.  
Lemma 15 could still be of use since I do not yet know whether C($\mathbb{CP}^2$) and C($\mathbb{CP}^{n}$) have extra symmetries.

For pure-shape 4-stop metroland, the nontrivial commutation relations are then the $U^{\sfA} \longrightarrow \mn^i$ and $\sfS_{\sfA} \longrightarrow \sfD_i$ of App \ref{Duel}.
These can also be written in the following dual form, with $n_i = \delta_{ij}n^j$ and $\sfD_{ij} = n_ip_j - n_jp_i$, 
\beq
[\widehat\sfD_{ij},\widehat\sfD_{kl}] = i\hbar\{\delta_{ik}\widehat\sfD_{jl} + 
\delta_{jl}\widehat\sfD_{ik} - \delta_{il}\widehat\sfD_{jk} - \delta_{jk}\widehat\sfD_{kl}  \} \mbox{ } ,
\eeq
\beq
[\widehat{n}^{i},\widehat\sfD_{kl}] = i\hbar\{\delta_{lj}\delta^i\mbox{}_k - \delta_{lk}\delta^i\mbox{}_j \}\widehat{\mn}^k \mbox{ } , 
\eeq
which has the benefit of extending to all of the higher-$N$ pure-shape $N$-stop metrolands.
For pure-shape triangleland, the spherical presentation's nontrivial commutation relations are the $U^{\sfA} \longrightarrow dra^{\Gamma}$ 
and $\sfS_{\sfA} \longrightarrow \sfS_{\Gamma}$ of App \ref{Duel}.
For scaled triangleland, the spherical presentation's nontrivial commutation relations are then the 
$U^{\sfA} \longrightarrow Dra^{\Gamma}$, $\P_{\sfA} \longrightarrow P^{Dra}_{\Gamma}$ and $\sfS_{\sfA} \longrightarrow \sfS_{\Gamma}$ of Sec \ref{Devlan}.

\subsubsection{{\sc kin-quant} for GR-as-Geometrodynamics}\label{ComRel-KinQ}

Here, one has ${\cal C}\mt\ms^{\infty}(\bupSigma, M(3, \mathbb{R})) \mbox{\textcircled{S}}{\cal C}\mt\ms^{\infty}(\bupSigma, \mbox{GenLin}^+(3, \mathbb{R}))$ where the latter factor is 
closely associated with the mathematical identity of Riem($\bupSigma$) and GenLin stands for `general linear' and $M(3, \mathbb{R})$ are real 3 $\times$ 3 matrices.  

\mbox{ }  

\noindent Analogy 85)  At the level of kinematical quantization, 1 and 2-$d$ RPM's are contained within Isham's work \cite{I84}; so is GR. 

\mbox{ } 

\noindent Difference 37) However, the ensuing representation theory in the GR case is considerably harder and to date largely impassible in any substantial detail.  
(The 1- and 2-$d$ RPM examples involve $SO$($n$) and $SU$($n)/\mathbb{Z}_n$) as canonical groups, and the representation theory of these 
is, of course, elementary and well known from Particle Physics.)

\mbox{ } 

\noindent As regards commutation relations, na\"{\i}vely Geometrodynamics' $h_{\mu\nu}$ and $\uppi^{\mu\nu}$ might follow the simplest (`plain') 
case (\ref{common}): promote to $\hat{\mh}_{\mu\nu}$ and $\hat{\uppi}^{\mu\nu}$ obeying 
\beq
[\widehat{\mh}_{\mu\nu}(x), \widehat{\uppi}^{\rho\sigma}(x^{\prime})] = 2i\,\hbar\,\delta_{(\mu}\mbox{}^{\rho}\delta_{\nu)}\mbox{}^{\sigma}\updelta^{(3)}(\ux, \ux^{\prime}) \mbox{  } .
\eeq
However, classically, there is an inequality on the determinant
\beq
\mbox{det}\, \mh > 0
\label{Taffy}
\eeq
(the nondegeneracy condition). 
This looks more like quantizing $\mathbb{R}_+$ than $\mathbb{R}$, and already that has a distinct kinematical quantization from the na\"{\i}ve one.
The affine Geometrodynamics \cite{IshamKakas1, IshamKakas2, Klauder}  commutation relations that take this into account are
\beq
[\widehat{\mh}_{\mu\nu}(\ux), \widehat{\uppi}_{\rho} \mbox{}^{\sigma}(\ux^{\prime})] = 
2i\,\hbar\,\widehat{\mh}_{\rho(\nu}\delta_{\mu)}\mbox{}^{\sigma}(\ux)\updelta^{(3)}(\ux, \ux^{\prime}) \mbox{ } . 
\eeq
I mention that Otriangleland has a similar inequality 
$Area \geq$ 0, whilst 3-cornerland has an even more similar one: $Area >$ 0. 
I should however caution that det$\,\mh > 0$ is a physical condition from the nature of space whilst there is nothing physically prohibiting the collinear particle configurations.

\subsubsection{Further global issues}\label{+glob}

\noindent \bu There are global issues \cite{I84} that stem from $\pi_1(\FrQ) \neq 0$.  
(However, this is 0 for all complex projective spaces and for spheres other than the circle, and for $\mathbb{R}^k$ though not for this with points deleted as the potential may require.)

\noindent \bu There are also global issues that stem from Chern class nontriviality.  
The first Chern class comes into the classification of the twisted representations.
The second Chern class is related to the instanton number
Here, for pure-shape RPM's second Chern classes are trivial for triangleland's $\mathbb{S}^2 = 
\mathbb{CP}^1$ (and 4-stop metroland) but are nontrivial for quadrilateralland's $\mathbb{CP}^2$ (from Sec \ref{Obgon}).

\noindent \bu For $\FrQ = \FrX/\FrG$ in general, the cocycles of $\FrX$ lead to  a further range of global effects; I have not as yet thought about these for RPM's. 

\noindent See \cite{I84} for the counterparts of the above global issues in GR.   
Note that for a theory like GR, the involvement of $\pi_1(\FrQ)$ is unlikely to bear much connection to one's final quantum theory.   
%
%
\noindent That the configuration spaces in question are homogeneous spaces carries the further 
implication of there being a unique orbit, which allows for group quantization techniques to be straightforward.

\subsection{Relationalism at the quantum level. II.}\label{BRGA}

Barbour-relationalism's indirect implementation at the classical level gives $\scQ\scU\scA\scD$ and $\scL\scI\scN$ \cite{Rovelli, Smolin}.  
The r-approach gives $\scQ\scU\scA\scD^{\sr}$.
Both these schemes are configurationally relational.  

\mbox{ }  

\noindent All in all, Sec \ref{Intro}'s tangibility, Temporal Relationalism and Configurational Relationalism ideas still hold here, but 
{\sl implementations} are different: there are no actions now (unless one goes down the path integral route, as dealt with in Sec \ref{QM-Histories}).   
%

\noindent As what are at least conceptually, variants on Dirac quantization, group-averaging quantization 
looks to be a good {\sl quantum} implementation of Configurational Relationalism. 
See Sec \ref{QM-Str} for more details.

\noindent Group-averaging quantization is essentially a variant/particular implementation of the Dirac quantization approach.  
Methods along these lines (including refined algebraic quantization) are fairly widely used in LQG \cite{Phoenix, DT04}.  
In this sense, LQG is configurationally relational at the quantum level.

\subsection{$\FrQ$-primality at the quantum level}

Recollect that the `minimalist' postulate Relationalism 3) -- that $\FrQ$ is primary -- and its categorized extension that ($\FrQ$, Point) that is primary.  
Then use just ($\FrQ$, Point) or, less limitedly, (RigPhase, Can) with cases with nontrivial $\FrG$ having additional `A' and `OA' options.
Then applying {\sc kin-quant}, we get to (RigComAl, Point), where RigComAl likewise allots distinction to position variables, 
so that the only admissible morphisms are those of the position variables.

\mbox{ } 

\noindent QM $\FrQ$-Primality Motivation 1) On the one hand, one often hears that QM unfolds on configuration space, on the other hand 
most hold that it unfolds equally well on whichever polarization within phase space, of which configuration space is but one.   

\noindent QM $\FrQ$-Primality Motivation 2) I note that Isham's procedure \cite{I84} for kinematic quantization itself favours $\FrQ$: the canonical group comes from just this and then 
the $\mathfrakV$ is controlled by $\FrQ$ and its associated $\FrG$ and $\FrG_{\scc\sa\sn}$ so everything is controlled by configuration space too.    

\noindent QM $\FrQ$-Primality Motivation 3) I expect then also for Point to be relevant to Pre-Hilbert space, and for the operator-orderings of the constraints likewise. 
This enables a connection with a conjecture of DeWitt concerning operator-ordering in Sec \ref{OOP}.  

\mbox{ }  

\noindent Note 1) This SSec's somewhat outlandish suggestion is not, however, the only possible attitude to take to the above Motivations 2 and 3; indeed if Can plays a role, 
Can is an enlargement of Point, and Point is then the `can that act on $\fM_{\sfA\sfB}(\fQ^{\sfC})$' restriction, which suffices to cover DeWitt's issue.

\noindent Note 2) (RigPhase, Point), as classically-motivated and argued to be a sensible implementation of $\FrQ$-primality in Sec \ref{Dyn1}, 
has the same objects and brackets as (Phase, Can) and hence the same {\sc kin-quant} applies to it.  
But in restricting Can to Point as morphisms preserving Poisson brackets and the rigging, then ComAl's morphisms {\textgoth{M}} would likewise be restricted.  
A possible problem with this view is that QM's $\fP$'s and $\fQ$'s may be more alike than their classical counterparts.  

\mbox{ } 

\noindent QM $\FrQ$-Primality Motivation 4) A further reason to question the licitness of all canonical transformations is that classical 
equivalence under canonical transformations is in general broken in the passage to QM as per e.g. the Groenewold--Van Hove phenomenon.  
Moreover, the Groenewold--Van Hove Theorem continues to apply in the RigPhase approach, since one can choose $q^3$ and $p^3$ instead of $q$ and $p$ in that setting too.  
Thus, whilst RigPhase is an illustration of weakening the canonical transformations, it is not the {\sl correct} weakening to take into account the Groenewold--Van Hove 
phenomenon; I leave as an open question what the set of canonical transformations that do become unitary transformations at the quantum level is characterized by.  


\noindent Note 3) Using just Point does {\sl not} resolve out all aspects of unitary inequivalence, though it may help/may be a step in the right direction.
A deeper question is which weakening of the Can's or Point's at the classical level form up into classes that are preserved as unitary equivalence at the quantum level. 
(RigPhase, Point) itself is at the bottom of the hierarchy, whereas the solution of the currently posed problem is likely to involve 
a rather larger proportion of the canonical transformations.   

\mbox{ } 

\noindent Consequences of using RigPhase in place of Phase are as follows. 
Holding canonical transformations in doubt affects Internal Time and Histories Theory approaches to the POT, as well as 
Ashtekar variables and the recent linking theory approach \cite{Kos2, Kos1, GGKM11, Kos3, Gomes, Kos4, GT11, Gom2}. 
The second and fourth of these make {\sl more} than the usual amount of use of canonical transformations;
this Sec suggests, moreover, the possibility of developing Physics in the {\sl opposite} direction.    

\mbox{ } 

\noindent {\bf Paths/Histories primality alternative} Further reasons for these appear at the quantum level, see Sec \ref{QM-POT-Strat}.

\subsection{Arguments against quantization as a functor}

I analyze these by `taking this functor apart' into various simpler constituents as per Figs \ref{Mewl}, \ref{Rawr} and \ref{Big-Arrow}.
A number of these steps already have individual difficulties.  

\noindent Two reasons for the {\sc kin-quant} functor to be `bad' (or, at least, ambiguous as regards the precise procedure to apply to each theory) are  
1) preferred subalgebra choice [and the nontriviality thereof by e.g. Groenewold--Van Hove]  and 2) having to meet global considerations at the quantum level.
 
\noindent 3) A further reason for badness is via the possibility of anomalies arising in the passage from the Poisson brackets algebra to the commutator algebra. 

\noindent 4) In the Geometrodynamics case of $\scQ$-{\sc rep}, there are also well-definedness issues (see Sec \ref{Incl-Well-Def}). 

\noindent 5) On top of these individual difficulties, the order in which some of these procedures are performed affects the outcome 
(e.g. reduce and then conformal-order does not match up with conformal-order, then reduce as per Sec \ref{OOP-3}.  

\noindent This bad functor conclusion, if not all the details of why, matches with Baez's account \cite{Baez12}, which was probably  
arrived at considering other/more quantization approaches than just the Isham 1984 one that the present 
Sec focuses upon.  

\mbox{ } 

\noindent A different line of objection to quantization itself is that it is only desirable in the incipient parts of conceiving a quantum theory 
(so it is tied to an understood classical theory, but QM is more fundamental so one needs to pass to first principles Quantum Theory which one 
then has the sometimes fatal hope of recovery of a recognized  classical limit). 
[For some particular examples, the functor does work out, but the above demonstrates that for suitably general theories there is not at all a standard prescription for quantizing, 
and formatting this as a functor does not in any way help with that.]    

\mbox{ } 

\noindent Moreover, further and distinct use of categories in quantization are as follows.   

\mbox{ } 

\noindent I) quantizing other categories - treating their objects as one usually treats $\FrQ$.

\noindent Simple Example. In the IL approach to Histories Theory, one quantizes the histories and their conjugate momenta themselves \cite{IL2}.

\noindent More general Example.  Isham \cite{Isham03} considered quantization on quite general categories, with the morphisms/arrows now playing the role of momenta.

\noindent II) The Topos approach for modelling the set of quantum propositions as per Sec \ref{QM-Str}.

\subsection{RPM TISE's}\label{Wave-Eqs}

\subsubsection{RPM TISE's in the Dirac formulation}

Dirac quantization involves quantizing and then constraining. 
The general form for the wave equations here is 
\beq
\widehat{\scQ\scU\scA\scD}\Psi = 0 \mbox{ } ,  \mbox{ } \mbox{ } \widehat{\scL\scI\scN}_{\sfZ}\Psi = 0 \mbox{ } . 
\eeq

RPM Dirac-QM study was developed in \cite{Rovelli, Smolin, 06I}.  
For scaled RPM,\footnote{I credit Smolin \cite{Smolin} for an early Article on Dirac Quantization for scaled RPM.  
The first such for pure-shape RPM was in \cite{06II}.}  
\beq
\widehat\scE\Psi = -\frac{\hbar^2}{2}\delta^{\mu\nu}\delta^{ij}
\frac{\pa}{\pa \rho^{i\mu}}\frac{\pa}{\pa \rho^{j\delta}}\Psi + V(\brho)\Psi = E\Psi
\label{zut}
\eeq
(note that this has no operator ordering ambiguity as the configuration space metric is independent of $\rho^{i\mu}$ and the configuration space is flat) 
with the quantum zero total angular momentum constraint 
\beq
\underline{\widehat{\scL}}\Psi = \frac{\hbar}{i}\sum\mbox{}_{\mbox{}_{\mbox{\scriptsize $i = 1$}}}^{n} 
\underline{\rho}^i \cr \frac{\pa}{\pa \underline{\rho}^i}\Psi = 0   \mbox{ } ,
\label{QZAM}
\eeq
which parallels the GR quantum momentum constraint.  
%

For pure-shape RPM in the geometrically natural formulation,  
\beq
\widehat\scE\Psi = -\frac{\hbar^2}{2}\mI\,\delta^{\mu\nu}\delta^{ij}
\frac{\delta}{\delta \rho^{i\mu}}\frac{\delta}{\delta \rho^{j\nu}}\Psi + \fV(\brho)\Psi = \fE\Psi \mbox{ } , 
\label{tuz}
\eeq
alongside the quantum zero total angular momentum constraint and the dilational momentum constraint,  
\beq
\widehat\scD\Psi = \frac{\hbar}{i}\sum\mbox{}_{\mbox{}_{\mbox{\scriptsize $i = 1$}}}^{n}
\underline{\rho}^i \cdot \frac{\pa}{\pa \underline{\rho}^i}\Psi = 0  \mbox{ } .  
\eeq
\noindent Difference 38) The definiteness-indefiniteness Difference 17) causes RPM TISE's to be elliptic-like rather than the hyperbolic-like WDE of GR.  
Thus the skill here is to solve constrained, or curved-space, elliptic-like equations.

\subsubsection{The Dirac approach for Geometrodynamics}\label{Incl-Well-Def}

For GR as Geometrodynamics, the quadratic constraint at the quantum level is the WDE,  
\beq
\hat\scH\Psi := 
- {\hbar^2}
`\left\{\frac{1}{\sqrt{{\mM}}}\frac{\delta}{\delta \mh^{{\mu\nu}}}
\left\{
\sqrt{{\mM}}{\mN}^{\mu\nu\rho\sigma}\frac{\delta\Psi}{\delta \mh^{{\rho\sigma}}}
\right\} 
- \xi \,\mbox{${\cal R}\mi\mc$}_{\sbM}(\ux; \bh]\right\}\mbox{'}\Psi - \sqrt{\mh}\mbox{${\cal R}\mi\mc$}(\ux; \bh]\Psi - {\sqrt{\mh}\Lambda}\Psi  + \hat\scH_{\mbox{\scriptsize matter}}\Psi = 0
\label{WDE2b} \mbox{ } ,   
\eeq
The inverted commas indicate that the Wheeler-DeWitt equation has various technical problems, as follows.   

\mbox{ }  

\noindent WDE Problem 1) There are problems with regularization, which is not at all straightforward for an equation 
for a theory of an infinite number of degrees of freedom in the absence of background structure.  

\noindent WDE Problem 2) The mathematical meaningfulness of functional differential equations is open to question. 

\mbox{ }

\noindent Difference 39) Finiteness dictates that RPM's do not serve to model 1) and 2).

\mbox{ }

\noindent WDE Problem 3) There is an operator-ordering ambiguity [see Sec \ref{OOP} of this, including for an explanation of what $\xi$ is].

\mbox{ }

\noindent The WDE comes alone in basic minisuperspace models, or, more generally, accompanied by the quantum linear momentum constraint 
\beq
\scM_{\mu}\Psi = -2\frac{\hbar}{i}\mh_{\mu\nu}\mD_{\rho}\frac{\delta}{\delta \mh_{\nu\rho}}\Psi + \scM_{\mu}^{\sm\sa\st\st\se\sr}\Psi = 0 \mbox{ } .  
\label{GRmomQ}
\eeq
\noindent Analogy 86) In RPM's, some ordering ambiguities remain: Secs \ref{OOP}, \ref{OOP-2} and \ref{OOP-3}.

\subsubsection{Variants on the WDE}\label{Aff2}

Note 1) The affine approach has a different WDE \cite{IshamKakas1, IshamKakas2, Klauder}.  

\noindent Note 2) Ashtekar variables/LQG/LQC approaches have different-looking WDE's.  
These include discrete versions of the Hamiltonian constraint in LQC \cite{Bojowald}, and come with an additional $SU$(2) constraint equation in the Dirac scheme for Ashtekar variables.  
One may view the loop representation approach \cite{PullinGambini} as providing a more reduced version of this.
However, what is toy-modelled in the present Article are the Geometrodynamics and Conformogeometrodynamics approaches.   
These approaches are fine as regards conceptual consideration of the POT (and are indeed the setting for which this has been the most developed).

\subsubsection{Comments on Nododynamics' Dirac-type quantization scheme}

\noindent There is not much scope in full traditional-variables GR for reduced quantization. 
LQG in principle has some scope in this regard (knot states) \cite{PullinGambini}.  

\mbox{ }

\noindent{\bf Question 58} A useful further project would be to provide a LMB(-CA) relational account of the quantum part of Thiemann's book \cite{Thiemann} 
(and of other schemes for modern Nododynamics/LQG, there being a variety of such schemes).

\subsection{Operator-Ordering Problem I}\label{OOP} 

For now I assume there is no nontrivial $\FrG$.  
The quadratic constraint's (\ref{GEnCo}) classical product combination of configurations and their momenta 
\beq
{\fN}^{\sfA\sfB}({\fQ}^{\sfC}){\fP}_{\sfA}{\fP}_{\sfB} 
\eeq
gives rise to an operator ordering ambiguity upon quantization.  
N.B. how one operator-orders has consequences for the physical predictions of one's theory, and there is 
no established way to prescribe the operator ordering in the case of (toy models of) Quantum Gravity.  
One line of argument which picks out certain operator orderings is as follows. 


\noindent {\bf Einstein's general covariance} of spacetime is a principle for classical Physics: there is to be no dependence of 
classical Physics on how those who study it happen to coordinatize it.  

\mbox{ } 

\noindent It is held to be an uncontroversial extension of this principle for this principle to also apply to space.

\mbox{ } 

\noindent {\bf Wider applicability of general covariance to the associated spaces of the Principles of Dynamics?} 
What about taking this principle to hold for such as configuration space or phase space?  
I furthermore then ask whether there a limit to how far this principle can (or is fruitful to) be taken along the hierarchy of abstract manifolds that one encounters in physical study?  
If Relationalism 3) (primality of $\FrQ$) applies, this could signify a barrier on how far along the sequence of manifolds associated with 
physical systems one should take general covariance.

\mbox{ } 

\noindent {\bf DeWitt's general covariance}. 
DeWitt \cite{DeWitt57} considered elevating the coordinatization-independence of $\FrQ$ to additionally hold at the quantum level.


\noindent [Doubts cast on canonical transformations in Appendix \ref{Dyn1}.C might suggest no further general covariance beyond this point.]  

\mbox{ } 

\noindent {\bf Laplacian operator ordering} 
\beq
\fN^{\sfA\sfB}(\fQ^{\sfC})\fP_{\sfA}\fP_{\sfB}  \longrightarrow \triangle_{\sbfM} = \frac{1}{\sqrt{{\fM}}} \frac{\partional}{\partional {\fQ}_{\sfA}}
\left\{
\sqrt{{\fM}}{\fN}^{\sfA\sfB}(\fQ^{\sfC})\frac{\partional}{\partional{\fQ}_{\sfB}}
\right\} \mbox{ } , 
\eeq
is then an implementation of DeWitt's general covariance.  

\mbox{ } 

\noindent Moreover, it is not a unique implementation: one can include a Ricci scalar curvature term so as to have the

\mbox{ } 

\noindent{\bf $\xi$-operator ordering}
\beq
\triangle_{\sbM}^{\xi} := \triangle_{\sbM} - \xi\,\mbox{${\cal R}\mi\mc$}\lfloor\fM\rfloor
\eeq 
\cite{DeWitt57, HP86,CZ, Halliwell, Moss, RT04} for any real number $\xi$.  
N.B. this gives inequivalent physics for each value of $\xi$.  

\mbox{ }  

\noindent{\bf Barvinsky's approximate equivalence}: the physics for all of these does coincide to $O$($\hbar$) \cite{Barvin, Barvin2, Barvin3}, 
as is relevant to approximate semiclassical Physics; thus the above inequivalence is only relevant if one wishes to investigate the physics to $O$($\hbar^2$) or better. 

\mbox{ }

\noindent{\bf Underlying simplicity}: the above is the extent of the ambiguity only if one excludes more complicated curvature scalars 
e.g. by stipulating no higher-order derivatives nor higher-degree polynomials in the derivatives.  

\mbox{ }

\noindent{\bf Conformal operator ordering}. 
Among the $\xi$-orderings, it is then well-known \cite{Wald} that there is then a unique configuration space dimension $k > 1$-dependent conformally-invariant choice: 
\beq
\triangle_{\sbfM}^{\scc} := \triangle_{\sbfM} - \xi^{\scc} \mbox{${\cal R}\mi\mc$}_{\sbfM} := \triangle_{\sbfM} - \frac{k - 2}{4\{k - 1\}}\mbox{${\cal R}\mi\mc$}_{\sbfM}  \mbox{ } . 
\eeq 
(This furthermore requires that the $\Psi$ itself that it acts upon itself transforms in general tensorially under $\FrQ$-conformal transformations \cite{Wald}).\footnote{$k \leq 1$ 
is a non-issue in relational thinking as per \ref{Quot}.}  
%
What is the underlying conformal invariance in question? 
[E.g. it is {\sl not} of space itself.]

\mbox{ } 

\noindent {\bf Misner's identification} \cite{Magic}: it is that of the Hamiltonian constraint under scaling transformations. 
\beq
\scH = 0 \longrightarrow \widetilde\scH = 0 \mbox{ } .
\eeq
This can be generalized to invariance under 
\beq
\scQ\scU\scA\scD = 0 \longrightarrow {\scQ\scU\scA\scD}^{\sr} = 0 \mbox{ } .
\eeq
\noindent {\bf My identification} \cite{Banal} goes one level deeper to the consideration of actions.  
It then so happens that it is the PPSCT invariance (\ref{Warh}) of the relational product-type parageodesic-type action that most cleanly underlies Misner's identification.
This makes it clear that it is a conformal invariance of the kinetic arc element $\d\cs$ alongside a compensatory conformal invariance in the potential factor $\fW$, 
reflecting that the combination actually present in the action, $\d\widetilde{\cs}$, is not physically meaningfully factorizable, as per Appendix \ref{Examples}.B.\footnote{One 
could also consider the more complicated manifestation (\ref{Whela}) at the level of the more usual difference-type action.  
Misner himself, for all that he used the words ``parageodesic principle" \cite{Magic}, was referring to an action with integrand of the general form Liouville operator -- Hamiltonian.  
My own contribution is to relate this conformal invariance to the {\sl geometrically obvious} parageodesic principle, which is a relational product-type action.}  
As I have previously argued, these are, furthermore, appropriate in the whole-universe context that is the setting for Quantum Cosmology or toy models thereof.  
Thus, demanding conformal operator ordering can be seen as choosing to retain this simple and natural invariance in passing to the quantum level 
(which one is free to do in the absence of unrelated technical caveats to the contrary).  
 
\mbox{ } 

\noindent The TISE following from the above family of orderings is then 
\beq
\scQ\scU\scA\scD \Psi = 0 \Rightarrow \triangle_{\sbfM}^{\xi}\Psi = 2\{V - E\}\Psi/\hbar^2 \mbox{ } .  
\label{DanSci}
\eeq
Surveying the literature, Kucha\v{r} \cite{K73} and Henneaux--Pilati--Teitelboim \cite{HPT} have advocated the Laplacian ordering itself.  
So have Page \cite{Page91}, Louko \cite{Louko} and Barvinsky \cite{Barvin, Barvin2, Barvin3}, however 
their specific examples are 2-dimensional, for which the Laplacian and conformal orderings coincide. 
The conformal ordering, which fixes a particular value of $\xi$, had been previously suggested in the quantum-cosmological context by e.g. 
Misner \cite{Magic}, Halliwell \cite{Halliwell}, Moss \cite{Moss}, Ryan--Turbiner \cite{RT04} and I \cite{Banal}.
Wiltshire advocates both \cite{Wiltshire}. 
Christodoulakis and Zanelli \cite{CZ} consider the arbitrary-$\xi$ case, as do Hawking and 
Page \cite{HP86}, albeit the latter then also pass to a 2-$d$ example for which $\xi$ drops out (see Sec \ref{SimXi} for why).  
\noindent Note that this conformal invariance requires the wavefunction itself to scale as \cite{Wald} 
\beq
\Psi \longrightarrow \overline{\Psi} = \Omega^{\{2 - r\}/2}\Psi \mbox{ } .  
\label{PPSCTwave} \mbox{ } . 
\eeq
\noindent{\bf An additional snag with advocating conformal ordering for Geometrodynamics itself} is that $k$ is infinite so the 
conformally-transformed wavefunction becomes ill-defined.  
However  the $k$'s in working (\ref{FORK}) continue to formally cancel, and it is the outcome of 
this (including its operator expectation counterpart), rather than $\Psi$ itself, that has physical meaning.  
This gives my {\bf conformal-ordered WDE} \cite{Banal},  
\beq
\hbar^2`
\left\{ 
\frac{1}{\sqrt{{\mM}}}\frac{\delta}{\delta \mh_{\mu\nu}}
\left\{
\sqrt{{\mM}}{\mN}^{\mu\nu\rho\sigma}\frac{\delta}{\delta \mh_{\rho\sigma}}
\right\} 
- \frac{1}{4}\mbox{${\cal R}\mi\mc$}_{\sbM}(\ux; \bh] 
\right\}\mbox{}'
\Psi + \sqrt{\mh}\{\mbox{${\cal R}\mi\mc$}(\ux; \bh] - 2\Lambda\}\Psi = 0 \mbox{ } . 
\eeq

\subsection{r-RPM TISE's}\label{RPM-WE}

\subsubsection{Laplacian operators on shape space and relational space}\label{RS-Lap}

For pure-shape RPM's, I denote the shape space Laplacian by 
\beq
\triangle_{\sFS(N, d)} = \frac{1}{\sqrt{M}}\frac{\pa}{\pa \mS^{\sfa}}
\left\{
\sqrt{M} N^{\sfa\sfb}\frac{\pa}{\pa \mS^{\sfb}}
\right\}
\eeq
(or $\triangle_{\sS}$ for a shorthand).  
Then specifically, for $N$-stop metroland, 
\beq
\triangle_{\mathbb{S}^{n - 1}} = \prod_{j = 1}^{n  - 2}\mbox{sin}^{j - n + 1}\theta_j \frac{\pa}{\pa\theta_a}
\left\{
\frac{\prod_{j = 1}^{n - 2}\mbox{sin}^{n - 1 - j}\theta_j}{\prod_{i = 1}^{a - 1}\mbox{sin}^{2}\theta_i}\frac{\pa}{\pa\theta_a}  
\right\} 
= \frac{1}{\mbox{sin}^{n d - 1 - \sfA}\theta_{\sfA}\prod_{i = 1}^{\sfA - 1}\mbox{sin}^2\theta_i}
\frac{\pa}{\pa\theta_{\sfA}}\left\{\mbox{sin}^{n d - 1 - \sfA}\theta_{\sfA}\frac{\pa}{\pa\theta_{\sfA}}\right\} \mbox{ } , 
\eeq
(which result also readily extends to preshape space under $n$ to $nd$), whilst for 
$N$-a-gonland 
\beq
\triangle_{\mathbb{CP}^{n - 1}} = \frac{\{1 + ||{\cal R}||^2\}^{2n - 2}}{\prod_{\barp = 1}^{n - 1}{\cal R}_{\barp}}
\left\{
\frac{\pa}{\pa {\cal R}_{\barp}}
\left\{
\frac{\prod_{\barp = 1}^{n - 1}{\cal R}_{\barp}}{\{1 + ||{\cal R}||^2\}^{2 n - 3}}
\{\delta^{\barp\barq} + {\cal R}^{\barp}{\cal R}^{\barq}\}\frac{\pa}{\pa{\cal R}_{\barq}}
 \right\} + 
\frac{\pa}{\pa {\Theta}_{\tip}}
\left\{
\frac{\prod_{\barp = 1}^{n - 1}{\cal R}_{\barp}}{\{1 + ||{\cal R}||^2\}^{2n - 3}}
\left\{
\frac{\delta^{\tip\tiq}}{{\cal R}_{\barp}^2} + {1|}^{\tip\tiq}
\right\}
\frac{\pa}{\pa{\Theta}_{\tiq}}
\right\}  
\right\} \mbox{ } . 
\label{Lapla} 
\eeq
The triangleland subexample of this is then, in the barred PPSCT representation in spherical coordinates, the 

\noindent $\alpha$, $\chi$ $\longrightarrow$ $\Theta$, $\Phi$ of App \ref{LapSphe}. 
On the other hand, in the tilded PPSCT representation, the triangleland case's Laplacian in the flat coordinates 
(${\cal R}$, $\Phi$) takes the familiar form  
\beq
\triangle_{\mathbb{R}^2} = \frac{1}{{\cal R}} \frac{\pa}{\pa{\cal R}}
\left\{
{\cal R}\frac{\pa }{\pa{\cal R}}
\right\} 
+ \frac{1}{{\cal R}^2}\frac{\pa^2}{\pa\Phi^2} \mbox{ }   
\eeq
(to be used approximately only, so that the global discrepancy is not an issue).

\noindent For scaled RPM's in scale--shape split form,  and using (\ref{qNd})
\beq
\triangle_{{\cal R}(N, d)} = \triangle_{\sC(\sFS(N,d))} = \frac{1}{\sqrt{M}}
\frac{\pa}{\pa Q^{\sfA}}\left\{\sqrt{M}{N}^{\sfA\sfB}\frac{\pa}{\pa Q^{\sfB}}\right\}  = 
\rho^{k(N, d)}\pa_{\rho}\{ \rho^{- k(N, d)}\pa_{\rho}\} + \rho^{-2}\triangle_{\sFS(N,d)} \mbox{ } .
\eeq
We can use this to build all the Laplacians, e.g. in 1-$d$ scaled case  
\beq
- \rho^{1 - n}\{\rho^{n - 1}\Psi_{,\rho}\}_{,\rho} + 
\rho^{-2}\triangle_{\mathbb{S}^{\tn - 1}}    \mbox{ } , 
\label{BigUn}
\eeq
bar the spherical presentation of triangleland one, which is
\beq
\check{\triangle}_{\sC(\mathbb{S}^2)} = \frac{1}{\mI^2}\left\{ \frac{\pa}{\pa\mI}\left\{ \mI^2 \frac{\pa}{\pa\mI} \right\} + 
\triangle_{\mathbb{S}^2} \right\} \mbox{ }  . 
\eeq

\subsubsection{Some simpler cases of $\triangle_{\mbox{\scriptsize\boldmath$M$}}^{\xi}$ operator-ordering}\label{SimXi}

\noindent Simplification 1) For models with 2-$d$ configuration spaces the conformal value of $\xi^{\scc} = \{k - 2\}/4\{k - 1\}$ collapses to 
zero, so this case reduces to  the Laplacian ordering and conformally invariant wavefunctions.  
I.e. $\triangle_{\mbox{\scriptsize\boldmath$M$}}^{\scc} = \triangle_{\mbox{\scriptsize\boldmath$M$}}$ for pure-shape 4-stop metroland and triangleland and for scaled 3-stop metroland.  

\noindent Simplification 2) For models with zero Ricci scalar, all of the $\xi$-orderings reduce to the Laplacian one.  
Thus in particular $\triangle_{\mbox{\scriptsize\boldmath$M$}}^{\scc} = \triangle_{\mbox{\scriptsize\boldmath$M$}}$ for all scaled $N$-stop metrolands.  

\noindent Simplification 3) If a space has constant Ricci scalar, then the effect of a $\xi\, \mbox{${\cal R}\mi\mc$}_{\mbox{\scriptsize\boldmath$M$}}$ term, conformal or otherwise, is 
just something which can be absorbed into redefining the energy in the case of mechanics.
Parallely, were the Ricci scalar constant in a GR model, it could likewise be absorbed into redefining the cosmological constant.
This clearly applies to pure-shape $N$-stop metrolands and $N$-a-gonlands.  

\noindent However, almost all other minisuperspace models and relational particle mechanics models (e.g. \cite{08III}) have configuration space 
dimension $\geq 3$ for which the choice of a value of $\xi$ is required. 
[E.g. C($\mathbb{CP}^2$) is not even conformally flat, see Sec \ref{Obgon2}.]

\subsubsection{The remaining RPM cases of $\triangle_{\mbox{\scriptsize\boldmath$M$}}^{\scc}$}\label{TriC}

For pure-shape $N$-stop metroland, $\triangle^{\scc}_{\mathbb{S}^{n - 1}} = \triangle_{\mathbb{S}^{n - 1}} 
- \{n - 1\}\{n - 3\}/4$ by simplification 3), which discrepancy from the Laplacian is incorporable via the shift 
\beq
\ttE \longrightarrow \ttE^{\scc}_{\mathbb{S}^{n - 1}} = \ttE - \hbar^2\{n - 1\}\{n - 3\}/8\mbox{ } . 
\label{NstopshiftE}
\eeq
For pure-shape $N$-a-gonland, $\triangle^{\scc}_{\mathbb{CP}^{n - 1}} = \triangle_{\mathbb{CP}^{n - 1}} - 2n\{n - 1\}\{n - 2\}/\{2n - 3\}$ 
by simplification 3), which discrepancy from the Laplacian is incorporable via the shift 
\beq
\ttE \longrightarrow \ttE^{\scc}_{\mathbb{CP}^{n - 1}} = \ttE - \hbar^22n\{n - 1\}\{n - 2\}/\{2n - 3\} \mbox{ } . 
\label{NagonlandshiftE}
\eeq
For scaled $N$-a-gonland, 
\beq
\triangle^{\scc}_{\sC(\mathbb{CP}^{n - 1})} = \triangle_{\sC(\mathbb{CP}^{n - 1})}\Psi - \frac{3 n\{2 n - 3\}}{4\{n - 1\}\rho^2}  
\eeq
(see the next SSSec for comments).

\subsubsection{The r-RPM TISE's}\label{Red-RPM-Wave-Eqs}

Direct r-QM study for RPM's was developed in \cite{08II, MGM, ScaleQM, 08III, Banal, SemiclIII, QuadIII}.
A master schematic TISE for pure-shape RPM is
\beq
\triangle_{\sFS(N, d)}^{\scc}\Psi = 2\{\ttV - \ttE\}\Psi/\hbar^2 \mbox{ } .   
\eeq
The $N$-stop metroland case of this is 
\beq
\triangle_{\mathbb{S}^{n - 1}}\Psi - \{n - 1\}\{n - 3\}\Psi/4 = \triangle_{\mathbb{S}^{n - 1}}^{\scc}\Psi = 2\{\ttV - \ttE\}\Psi/\hbar^2 \mbox{ } ,
\eeq
which can be rewritten as
\beq 
\triangle_{\mathbb{S}^{n - 1}}\Psi =  2\{\ttV - \ttE^{\scc}_{\mathbb{S}^{n - 1}}\}\Psi/\hbar^2 \mbox{ } .  
\eeq  
These results also readily extend to the case of wavefunctions on preshape space under $n$ to $nd$.\footnote{The 
above quantum Hamiltonians do involve suitable quantum operators (see also e.g. p 160 of \cite{RS} for an account of the properties 
of the Laplacian operator on $\mathbb{S}^{n - 1}$, and also \cite{Norway, ThisS2A, ThisS2B}).}

The $N$-a-gonland case of this is 
\beq   
\triangle_{\mathbb{CP}^{n - 1}}\Psi - 2n\{n - 1\}\{n - 2\}\Psi/\{2n - 3\}  = \triangle_{\mathbb{CP}^{n - 1}}^{\scc}\Psi 
= 2\{\ttV - \ttE\}\Psi/\hbar^2 \mbox{ } ,
\eeq
which is rewriteable as
\beq
\triangle_{\mathbb{CP}^{n - 1}}\Psi = 2\{\ttV - \ttE^{\scc}_{\mathbb{CP}^{n - 1}}\}\Psi/\hbar^2 \mbox{ } .  
\eeq  
The spherical presentation of triangleland's TISE is  
\beq
\triangle_{\mathbb{S}^{2}}\Psi = 2\{\breve{\fV} - \breve{\fE}\}{\Psi}/\hbar^2 \mbox{ } .  
\label{spheTISE}
\eeq
On the other hand, in plane polar coordinates $\{{\cal R}, \Phi \}$ obtained by passing to stereographic coordinates 
on the sphere and then passing to the `overlined PPSCT representation'  $\overline{\ttT} = \ttT\{1 + {\cal R}^2\}^2$ and 
$\overline{\ttE} - \overline{\ttV} = \{\ttE - \ttV\}/\{1 + {\cal R}^2\}^2$, the TISE is 
\beq
\triangle_{\mathbb{R}^2}^{\scc} = \triangle_{\mathbb{R}^2} = 
\left\{\frac{1}{{\cal R}}\frac{\pa}{\pa{\cal R}}
\left\{
{\cal R}\frac{\pa\Psi}{\pa{\cal R}}
\right\} 
+ \frac{1}{{\cal R}^2}\frac{\pa^2\Psi}{\pa\Phi^2}
\right\} 
= 2\{\overline{\ttV}({\cal R}) - \overline{\ttE}({\cal R})\}\Psi/\hbar^2 \mbox{ }  . 
\label{flatTISE}
\eeq
\mbox{ } \mbox{ } A master schematic TISE for scaled RPM is
\beq
\triangle^{\scc}_{{\cal R}(N, d)}\Psi = 2\{V - E\}\Psi/\hbar^2 \mbox{ } .  
\label{Larq}
\eeq
In particular, for the $N$-stop metroland case, 
\beq
\triangle_{\mathbb{R}^{n}}\Psi = \triangle^{\scc}_{\mathbb{R}^{n}}\Psi = 2\{V - E\}\Psi/\hbar^2 \mbox{ } ,  
\label{TISER}
\eeq
with the special cases (\ref{Inger}) in 2-$d$ polars with $R \longrightarrow \rho$, $\chi \longrightarrow \varphi$ and 
(\ref{Goldf}) in 3-$d$ spherical polars with $R \longrightarrow \rho$, $\alpha \longrightarrow \theta$, $\chi \longrightarrow \phi$.  
For the $N$-a-gonland case, 
\beq
\triangle_{\sC(\mathbb{CP}^{n - 1})}\Psi - \frac{3 n\{2 n - 3\}}{4\{n - 1\}\rho^2}\Psi  = 
\triangle^{\scc}_{\sC(\mathbb{CP}^{n - 1})}\Psi = 2\{V - E\}\Psi/\hbar^2 \mbox{ } .  
\label{TISECCP}
\eeq
Note that the conformal contribution here can no longer simply be used up shifting the energy constant. 
Instead, it adds to the separation constant of the shape part from the scale part in the separable case. 
(This is similar to the difference between monopole harmonics and ordinary spherical harmonics except 
that it applies to the radial equation rather than to the angular equation.)  

\noindent For equation-shortening purposes, define the {\it conformal correction term} by 
\beq
c(N, d) = \mbox{\Huge \{} \stackrel{        \mbox{0} \hspace{0.8in}      }                    
                                   {        {3 n\{2 n - 3\}}/{4\{n - 1\}}} 
\mbox{ } \mbox{ }  
\stackrel{\mbox{for}     \mbox{ }        \mbox{$d$ = 1}         }
         {\mbox{for}     \mbox{ }               d  = 2          }
\eeq
for the range of scaled RPM's covered in detail in the present research program. 
\noindent Then the solvable-scaled-RPM series' quantum energy equation is
\beq
-\hbar^2\{\pa^2_{\rho} + k(N, d)\rho^{-1}\pa_{\rho} - \rho^{-2}c(N, d) + \triangle_{\sFS(N, d)}\}\Psi = 2\{E - V(\rho, S^{\sfa})\}\Psi
\label{Gilthoniel}
\eeq
for particle number $N \geq 3$ for relational nontriviality and $d$ = 1 ($N$-stop metrolands) or 2 ($N$-a-gonlands).

The scaled triangleland TISE is
\beq
\check{\triangle}^{\scc}_{\mathbb{R}^3}\check{\Psi} = \check{\triangle}_{\mathbb{R}^3}\check{\Psi} =
\widehat{\sfT\sfO\sfT}\check{\Psi} = \frac{1}{\mI^2}
\left\{
\frac{\pa}{\pa \mI}
\left\{
\mI^2\frac{\pa\check{\Psi}}{\pa \mI}
\right\} + 
\frac{1}{\mbox{sin}\Theta}\frac{\pa\check{\Psi}}{\pa\Theta}
\left\{
\mbox{sin}\Theta\frac{\pa\check{\Psi}}{\pa\Theta}
\right\} 
+ \frac{1}{\mbox{sin}^2\Theta}\frac{\pa^2\check{\Psi}}{\pa\Phi^2}  
\right\} 
= \frac{2\{\check{\fV}(\Phi, \Theta) - \check{\fE}(\Theta)\}}{\hbar^2}\check{\Psi} \mbox{ } .  
\label{spheTISE2}
\eeq
The non-conformally transformed spherical form for triangleland is, in the conformal-ordered case, using Ric($\mathbb{R}^3$ \mbox{curved}) = $3/2\mI^2$ and 
particular conformal case $\xi_{\scc}Ric_{(\mathbb{R}^3 \, \scc\su\sr\sv\se\sd)} = {3}/{16\mI^2}$, 
\beq
\triangle^{\scc}_{(\mathbb{R}^3(\sn\so\sn-\sf\sll\sa\st))} =  
4\left\{
\frac{1}{\mI^{3/2}}  \frac{\pa}{\pa \mI}  
\left\{
\mI^{3/2}\frac{\pa\Psi}{\pa \mI}
\right\} 
+ \frac{1}{\mI^2}
\left\{
\frac{1}{\mbox{sin}^2\Theta}\frac{\pa\Psi}{\pa\Theta}
\left\{
\mbox{sin}^2\Theta \frac{\pa\Psi}{\pa\Theta}
\right\}
+ \frac{1}{\mbox{sin}^2\Theta} \frac{\pa^2\Psi}{\pa\Phi^2}
\right\}
- \xi_{\scc}Ric_{(\mathbb{R}^3\,{\scc\su\sr\sv\se\sd})}\Psi 
\right\}
= 
2\left\{
V -\frac{E}{\mI}
\right\}
\frac{\Psi}{\hbar^2} \mbox{ } .
\label{SpheTISEn}
\eeq
Also useful below, the parabolic coordinates version of the TISE is 
\beq
{4}
\left\{
\frac{\pa}{\pa\mI_1}
\left\{ 
\mI_1\frac{\pa\overline{\Psi}}{\pa\mI_1}
\right\}
+ \frac{\pa}{\pa\mI_2}
\left\{ 
\mI_2\frac{\pa\overline{\Psi}}{\pa\mI_2}
\right\}
\right\}
+ 
\left\{
\frac{1}{\mI_1} + \frac{1}{\mI_2}
\right\}
\frac{\pa^2}{\pa\Phi^2}\overline{\Psi} = \frac{\{V - E\}}{2\hbar^2}\overline{\Psi}  \mbox{ } .  
\label{JillM}
\eeq

\subsubsection{Closure of RPM quantum equations}\label{RPM-Dirac}

These above schemes are, furthermore, `lucky' in Dirac's sense \cite{Dirac}: the full set of constraints closes at the quantum level. 
Moreover this is in direct parallel with the classical closure [(\ref{first6}) for scaled RPM and (\ref{first6},\ref{other4}) for pure-shape RPM] 
under the correspondence principle $\{\mbox{ } , \mbox{ } \} \longrightarrow \{i\hbar\}^{-1}[ \mbox{ } , \mbox{ } ]$, so I do not present it.

\subsection{Absence of monopole problems in $N$-stop metroland and triangleland}\label{No-Poles}

Following on from Sec \ref{Gau}'s classical monopole considerations [these pertain to Sec \ref{Q-Geom}'s Scheme D)'s split of Newtonian Mechanics itself] 
then the charged particle toy model's position and the canonical momentum are good Hermitian operators.  
However, next in looking to form $SO$(3) objects, $\underline{A}$'s positional dependence complicates the commutation relations.  
This necessitates, beyond what is usual in usual treatments of angular momenta, the introduction of an extra term:  
\beq
\underline{L}^{\se\sx\st\se\sn\sd\se\sd} = 
\underline{x} \cr \{\underline{p} - e\underline{A}\} - q\underline{x}/r^3 \mbox{ } ,   
\eeq
for 
\beq
q = eg = n\hbar c/2 \mbox{ }  
\eeq
(the last equality being the Dirac quantization condition).  
Next, ${L}_{\sT\so\st}^{\se\sx\st\se\sn\sd\se\sd} = \sum_{\mu = 1}^{3}\{{L}_{\mu}^{\se\sx\st\se\sn\sd\se\sd}\}^2 = 
\{\underline{x} \cr \{ \underline{p} - e\underline{A}\}\}^2 + q^2$, and ${L}_3^{\se\sx\st\se\sn\sd\se\sd}\Psi = 
\{-i\pa_{\phi_{\ts\tp}} - q\}\Psi$ in the N-chart and ${L}_3^{\se\sx\st\se\sn\sd\se\sd}\Psi = \{-i\pa_{\phi_{\ts\tp}} + q\}\Psi$ in the S-chart, so $\Psi$ 
has an exp$(i\{\nm \pm q\}\phi_{\sss\sp})$ factor rather than an exp$(i\nm\phi_{\sss\sp})$ factor, with integer m $\pm$ $q$. 
Consequently, the azimuthal part of the TISE then picks up two extra terms as compared to (\ref{stazi}):  
\beq
-  \{\mbox{sin}\,\theta_{\sss\sp}\}^{-1}
\{\mbox{sin}\,\theta_{\sss\sp}\Psi,_{\theta_{\sss\sp}}\},_{\theta_{\sss\sp}}
+  \{\mbox{sin}\,\theta_{\sss\sp}\}^{-2}\{m + q\,\mbox{cos}\,\theta_{\sss\sp}\}\Psi  
= \{\nl\{\nl + 1\} - q^2\}\Psi 
\mbox{ } ,
\eeq 
and so that `monopole harmonics' replace the ordinary spherical harmonics (\cite{YW}; e.g. \cite{ML97} 
has a 2-$d$ counterpart of closer relevance to the present Article).

N.B. however how this working collapses in the case of an uncharged particle.  
The monopole is then not `felt', so one has the mathematically-usual form of the $SO$(3) operator and the mathematically-usual TISE.  
Furthermore, for central potentials (i.e. potentials depending on the radial variable $r$ alone), one has 
the mathematically-usual spherical harmonics.  
This difference from \cite{08II} is one reason why \cite{08III} has had to wait for the below resolution.

Reminder then that total angular momentum plays the role of charge in this analogy, so that the zero angular momentum 
case of relevance to the present Article is not beset by such monopoles issues.

\subsection{Operator-Ordering Problem II: discrepancy between Quantum Cosmology and Molecular Physics} \label{OOP-2} 

There turns out to be a discrepancy between the `relational portion' of Newtonian Mechanics (scheme D) 
and that of the relationalspace-reduced approaches (schemes A-B-C.I).    
In scheme D), there is a chain of transformations as conventionally used in Molecular Physics can be understood.  
This is from particle positions to relative Jacobi coordinates to spherical coordinates plus 
absolute angles to Dragt-type coordinates plus absolute angles.  
[References for this are e.g. \cite{Iwai87, PP87, Mead92} and the much earlier but non-geometrical \cite{Zick}.]   
Consequently Scheme D) has an absolute block in its configuration space metric.  
Then, via the participation of this in the formation of the overall volume element $\sqrt{M_{\sa\sb\sss-\sr\se\sll}}$, this enters the {\sl 
 relational} block's part of the Laplacian, since $\sqrt{M_{\sa\sb\sss-\sr\se\sll}}$ then sits inside the $\pa/\pa Q_{\sr\se\sll}$ derivative.  
Therefore, by this means, if absolute space is assumed, it leaves an imprint on the `relational portion', which is absent if one considers 
a relationally-motivated Lagrangian (as in schemes A-B-C.I).  
%

\mbox{ }

\noindent In any case, it is reasonable from the relational perspective for modelling a molecule in a universe to 
differ from modelling those self-same particles as a whole universe toy model, since the former case possesses an inertial frame concept due to the rest of the universe.
Taking the RPM-Geometrodynamics analogy as primary rather than trying to describe reality as a few (or even very many) 
non-specially-relativistic particles, gives serious reason {\sl not} to use Molecular Physics' quantum equations in the whole-universe model context.  
That mathematical analogy ends with the classical dynamics and the quantum {\sl kinematics}.

In greater detail, firstly note that this difference is a consequence of the nontriviality of the rotations 
(there is no corresponding effect for the translations as the centre of mass motion block does not contribute any further functional dependencies to the volume element.  
The reduced--relationalspace approaches' Laplacian is $\triangle_{\sFS(N, d)}$  in the pure-shape case and  $\triangle_{{\cal R}(N, d)}$ in the scaled case.  
The Laplacian for the only-trivially-reduced convenient starting-point of relative space is $\triangle_{\sFR(N, d)}$.
The reduced--relationalspace versus restriction of relative space distinction is not relevant to scaled $N$-stop metrolands by $\bigr(N, 1) = {\cal R}(N, 1)$.  
However, 
\beq
\triangle^{\scc}_{\sFR(N,1)}|_{\sS-\sr\se\sll \, \sp\sa\sr\st} = 
\triangle_{\sFR(N, 1)}|_{\sS-\sr\se\sll \, \sp\sa\sr\st}     := 
\triangle_{\sFR(N,1)}|_{\pa/\pa\rho = 0} = \rho^2\triangle_{\mathbb{S}^{\tN - 2}}   = \rho^2\triangle_{\sFS(N,1)} 
\neq \rho^2\triangle^{\scc}_{\sFS(N,1) } \mbox{ } , 
\eeq
though the last inequality is but by a constant which can be absorbed into a redefinition of the energy, as per (\ref{NstopshiftE}).  
Also, 
\beq
\check{\triangle}_{{\cal R}(3,2)} = \triangle_{\sFQ(3,2)}|_{\pa/\pa\rho = 0} \mbox{ } , 
\label{Cirith}
\eeq 
which is given by the fourth portion of eq (\ref{spheTISE2}), with $\triangle_{{\cal R}(3,2)}$ is distinct as given by the second 
portion of eq. (\ref{SpheTISEn}), so that these differ by $ 2\pa_{\sI} = \scD/{\mI}$.
Also, Sec \ref{Q-Geom} 's Scheme D) in the usual Molecular Physics context (spatially 3-$d$) gives, for the radial part of the TISE, 
\beq
\pa_{\sI}\mbox{}^2 + 5\,\mI^{-1}\pa_{\sI} \mbox{ } . 
\label{I5}
\eeq
On the other hand, purely relational considerations give the far more usual spherical-type radial part (\ref{Cirith}) 
\beq
\pa_{\sI}\mbox{}^2 + 2\,\mI^{-1}\pa_{\sI} \mbox{ } . 
\eeq 
\noindent Note 1) This difference can be ascribed to an absolutist imprint which occurs at the quantum level 
and even within the drastic $\underline\scL = 0$ simplification within scheme D.   

\noindent Note 2) Also note that the above Molecular Physics literature result is for 3-cornerland due to the molecules being studied living in 3-$d$.  
To be even more relevant to the present Article's examples, one should consider the 2-$d$ case.  
Then I note that the imprint of 2-$d$ absolute space is different from that of 3-$d$ absolute space.  
In that case there is a 3 and not a 5 in (\ref{I5}), corresponding to there being a relational 2 plus now an absolute 1 [for $SO(2)$] rather than an absolute 3 [for $SO(3)$].

\noindent Note 3) In 3-$d$ the shape part also differs between the split of Newtonian Mechanics and the relationalspace approach. 
See e.g. \cite{Iwai87, PP87, Mead92, Zick} for various coordinatizations of the former's Laplacian.  
This is due to this case's absolute block depending on relational angle as well as on scale. 
In 2-$d$ the situation is simpler.  
Thus 3-cornerland's pure-shape workings, as well as workings with scale, differ between scheme A-B-C.I and scheme D.

Now also $\triangle^{\sfc}_{\sFQ(3,2)} = \triangle_{\sFQ(3,2)}$ but $\triangle^{\sfc}_{{\cal R}(3,2)} \neq 
\triangle_{{\cal R}(3, 2)}$ and this is now not just a shift by a constant.
Finally, 
\beq
\triangle_{\sFQ(3, 2)}|_{\sS-\sr\se\sll\,\sp\sa\sr\st} = \triangle_{\sFQ(3,2)}|_{\pa/\pa \zeta = 0, \pa/\pa\rho = 0} = \triangle_{\sFS(3,2)}
\eeq (the above difference is now removed due to containing the new constraint as a factor). 
Though also 
\beq 
\triangle^{\sfc}_{\sFQ(3,2)}|_{\sS-\sr\se\sll\,\sp\sa\sr\st} = 
\triangle^{\sfc}_{\sFQ(3,2)}|_{\pa/\pa \zeta = 0, \pa/\pa\rho = 0} \neq \triangle^{\sfc}_{\sFS(3,2)} \mbox{ } , 
\eeq
albeit the difference is again but an absorbable constant.
(Here, $\zeta$ is an absolute angle). 
 
\mbox{ }

\noindent Finally, I note that the purely relational operator ordering  form that I adopt has the further theoretical advantage of having (well-)known solutions in a number of cases.

\subsection{Operator-Ordering Problem III: Dirac--reduced inequivalence}\label{OOP-3}

Dirac quantization--reduced quantization inequivalence renders all of the above orderings by themselves unsatisfactory (but the Laplacian ordering less so than the conformal ordering).   
The suggested way out is that one can only apply such an ordering prescription to the most reduced configuration space itself 
(this unfortunately then leaves one stuck in the general case, as one has no explicit form for the most reduced configuration space).  

\noindent I will demonstrate below that procedure C.III) [reduction at the level of the QM constraint equations] is capable of 
producing a different answer from procedures A-C.I-II)   [relationalspace approach, indirect approach and configuration space and phase space reductions]. 
If one takes one's relationalism very seriously, one would favour the relationalspace approach A) foremost [and, quite possibly, the conformal ordering]. 
But if one takes relationalism less seriously, it is worth noting that C.III) can often [not always in general but always for this article's models] 
be made to agree with A-C.I-II) by adopting the Laplacian ordering.  

\mbox{ }

\noindent Previously, Barvinsky \cite{Barvin, Barvin2, Barvin3} investigated for what ordering Dirac and reduced approaches coincide.  
On the other hand, e.g. Ashtekar, Horowitz, Romano and Tate \cite{AH,RT} have argued for inequivalence of these two approaches to quantization.  
The Laplace ordering does so to $O(\hbar)$ and to $O(\hbar)$, Laplacian ordering coincides with conformal ordering.

\mbox{ }

\noindent In the Dirac quantization approach for RPM's, the PPSCT rescaling $\scL\scI\scN\,\Psi = 0$ to $\widetilde{\scL\scI\scN}\,
\widetilde{\Psi}$ does cause an alteration since these constraint operators are differential operators and therefore act on the conformal factor of $\widetilde{\Psi}$
The reduced quantization scheme avoids this issue.

This SSec has its own notation: $\fM_{\sfA\sfB}$ and $\ttM_{\sfa\sfb}$ are simply the less and more restricted configuration space metrics with no 
scale--shape connotations intended; they are on a configuration space $\FrQ$ and on a hypersurface $\Upsilon$ therein. 
[For ease of presentation I only give the codimension-1 case; see e.g. \cite{Gergelybranes} for higher-codimension hypersurface split formalism.]    
In more detail, there is an issue of whether $\triangle_{\sbfM}|_{\sFrC}\Psi$ -- the less reduced Laplacian restricted to the constraint surface\footnote{This 
is a less common $\FrQ$, rather than Phase, formulation of constraint surface, based on configuration space reduction.} 
$\FrC$ -- matches up with $\triangle_{\sbfM}\Psi$ -- the more reduced Laplacian. 
For simplicity whilst sufficiently illustrating this argument, I consider the case in which the configuration space dimension goes down by one in the reduction. 
Then there is an (Euclidean-signature for mechanical configuration spaces) ADM-type split of configuration space itself, 
\beq
\fM_{\sfA\sfB} =
\left(
\stackrel{\ttB_{\sfc}\ttB^{\sfc} + \tta^2}{\ttB_{\sfb}} \mbox{ } \mbox{ }   
\stackrel{\ttB_{\sfa}}{\ttM_{\gamma\delta}}
\right) \mbox{ } ,
\mbox{ } 
\eeq
where $\fM_{\sfa\sfb}$ is the kinetic metric on the more reduced configuration space.  
A convenient way of working out the hypersurface geometry for this is to consider Kucha\v{r}'s 
hypersurface formulation \cite{Kuchar76}, which in the present configuration space metric context gives 
${\fM}_{\perp\perp} = 1$, ${\fM}_{\perp\sfa} = 0$, ${\fM}_{\sfa\sfb} = \ttM_{\sfa\sfb}$, and so 
${\fN}^{\perp\perp} = 1$, ${\fN}^{\perp\sfa} = 0$, ${\fN}^{\sfa\sfb} = \ttN^{\sfa\sfb}$ and $\mM = \ttM$.  
Then use the following formulae for the hypersurface split of a covector applied to $\pa_{\sfA}\Psi$, 
\beq
\ttD_{\perp}\pa_{\perp}\Psi = \{\delta_{\sttB}\{\pa_{\perp}\Psi\} + \pa^{\sfa}\Psi\pa_{\sfa}\tta\}/\tta
\eeq
and
\beq
\ttD_{\sfa}\pa_{\sfb}\Psi = \fD_{\sfa}\pa_{\sfb}\Psi - \pa_{\perp}\Psi \ttK_{\sfa\sfb} \mbox{ } .
\eeq
Here, $\fD$ and $\ttD$ are the covariant derivatives corresponding to $\fM$ and $\ttM$ respectively,  
\beq
\ttK_{\sfa\sfb} = - \frac{1}{2\tta}\delta_{\sttB}\ttM_{\sfa\sfb} \mbox{ } 
\eeq
is this situation's extrinsic curvature 
and $\delta_{\sttB} = \mbox{}^{\prime} - \pounds_{\sttB}$ the hypersurface derivative for $\mbox{}^{\prime}$ the derivative with respect to the perpendicular to the hypersurface).

Then 
\beq
\triangle_{\sbfM}\Psi = \fN^{\sfA\sfB}\scD_{\sfA}\pa_{\sfB}\Psi = \triangle_{\sbfM}\Psi - \ttK\pa_{\perp}\Psi + 
\{\delta_{\sttB}\pa_{\perp}\Psi + \pa^{\sfa}\Psi\pa_{\sfa}\tta\}/\tta \mbox{ } .  
\label{Llama}
\eeq
Also the doubly-contracted Gauss Law is (form 1)  
\beq
Ric_{\sbttM} = Ric_{\sbfM} + \ttK^2 - \ttK_{\sfa\sfb}\ttK^{\sfa\sfb} - 2\mbox{Ric}({\ttM})_{\perp\perp}
\eeq
or (form 2) 
\beq
Ric_{\sbttM} = Ric_{\sbfM} + \ttK^2 + \ttK_{\sfa\sfb}K^{\sfa\sfb} + 2\{\delta_{\beta}\ttK - \triangle_{\sbfM}\tta\}/\tta \mbox{ } .     
\eeq
Then form 2 gives that 
\beq
\triangle_{\sbfM}^{\scc}\Psi = \triangle_{\sbfM}^{\scc}\Psi - \frac{\{k - 2\}Ric_{\sbfM}}{4\{k - 1\}}\Psi - 
\frac{k - 2}{4\{k - 1\}}\{ \ttK^2 + \ttK_{\sfa\sfb}\ttK^{\sfa\sfb} \}\Psi + \frac{1}{\tta}
\left\{
\frac{k - 2}{2\{k - 1\}}\big\{\{-\triangle_{\sbfM}\tta+\tta_{\sttB}\ttK\}\Psi+\delta_{\sttB}\pa_{\perp}\Psi+\pa^{\sfc}\Psi\pa_{\sfc}\tta\big\}
\right\} \mbox{ } .
\eeq

\mbox{ }  

Showing that this example's assumptions do permit explicit RPM examples, 1) for the passage from relational space to shape space in 4-stop metroland or triangleland 
($\sigma = \rho$ or $\mI$),  
\beq
2/\sigma^2 = Ric_{\mathbb{S}^2} = \ttK^2 - \ttK_{\sfa\sfb}\ttK^{\sfa\sfb} = \ttK^2/2 \mbox{ } .
\eeq
This is by, respectively, evaluation, Theorema Egregium and isotropy of $K_{\sa\sb}$ due to spherical symmetry alongside the taking of traces.
Thus (\ref{Llama}) reads 
\beq
\triangle_{\mathbb{R}^3}\Psi = \sigma^{-2}\triangle_{\mathbb{S}^2}\Psi + 2\sigma^{-1}\pa_{\sigma}\Psi + \pa_{\sigma}\pa_{\sigma}\Psi \mbox{ } .
\eeq
Thus here 
\beq
\triangle_{\mathbb{R}^3}\Psi|_{\sr\se\sd} = \sigma^{-2}|_{\sigma = \scc\so\sn\sss\st}\triangle_{\mathbb{S}^2}\Psi
\eeq
-- a case which has good agreement (and moreover one for which conformal ordering = Laplacian ordering too).  

\noindent Example 2) The passage from $N$-stop metroland relational space to the shape space \{$n$ -- 1\}-sphere, for which the above working directly generalizes to 
\beq
\{n - 1\}\{n - 2\}/\rho^2 = Ric_{\mathbb{S}^{n - 1}} = \ttK^2 - \ttK_{\sfa\sfb}\ttK^{\sfa\sfb} = \ttK^2\{n - 2\}/\{n - 1\}  
\eeq
(using form 1 of the doubly contracted Gauss equation alongside the higher-$d$ space being flat in place of the Theorema Egregium), so  
\beq
\triangle_{\mathbb{R}^{n}}\Psi = \rho^{-2}\triangle_{\mathbb{S}^{n - 1}}\Psi + \{n - 1\}\,\rho^{-1}\pa_{\rho}\Psi + \pa_{\rho}\pa_{\rho}\Psi \mbox{ } ,
\eeq
\beq
\mbox{ so } \mbox{ } \triangle_{\mathbb{R}^{n}}\Psi|_{\sr\se\sd} = \rho^{-2}|_{\rho = \scc\so\sn\sss\st}\triangle_{\mathbb{S}^{\tn - 1}}\Psi 
\mbox{ } ,  
\eeq
so there is agreement again, but now $\triangle^{\scc}_{\sbttM}\Psi|_{\sr\se\sd}$ is out from $\triangle_{\sbfM}^{\scc}\Psi$ by 
\beq
- \frac{\{n - 3\}}{4\{n - 2\}} Ric_{\mathbb{S}^{\sp}}\Psi \mbox{ } = \mbox{ }  
\left.
\frac{1}{\rho^2}
\right|_{\rho = \scc\so\sn\sss\st} \frac{\{n - 3\}}{4\{n - 2\}}\{n - 1\}\{n - 2\}\Psi \mbox{ } = \mbox{ }  
\left. 
\frac{\{n - 1\}\{n - 3\}}{4}\frac{1}{\rho^2} \right|_{\rho = \scc\so\sn\sss\st}\Psi  \mbox{ } .  
\eeq
[But it is relatively benign in that one can get around it by redefining the energy by an additive  constant.]

\noindent Example 3) The curved, conformally flat $\mathbb{R}^3$ to $\mathbb{S}^2$ and $\mathbb{S}^3$ to $\mathbb{S}^2$ 
cases both send Laplacian ordering to Laplacian ordering as their left hand side spaces are diagonal.  
Thus, via both the relational space route and via the preshape space route, Laplacian ordering is inherited at all levels in the pure-shape triangleland problem.  
This example's steps do {\sl not} preserve conformal ordering which was the case I argued for quantum-cosmologically.   

\noindent Example 4) $\mathbb{R}^{2n} = \mathbb{C}^{n}$ to $\mC(\mathbb{CP}^{n - 1})$, $\mC(\mathbb{CP}^{n - 1})$ to $\mathbb{CP}^{n - 1}$ and $\mathbb{S}^{2n - 1}$ 
to $\mathbb{CP}^{n - 1}$ of $N$-a-gonland \cite{MF03b} all have suitable at-least-block-diagonal form left-hand-side spaces, and therefore preserve Laplacian ordering.  
They do not preserve conformal ordering.  
Thus via both the relational space route and via the preshape space route, Laplacian ordering is inherited at all levels in the pure-shape 
$N$-a-gonland problem, and also in the scaled $N$-a-gonland problem.  


Thus all of this Article's principal concrete examples of RPM's are, nevertheless, Laplacian ordering preserving under passage between Dirac and reduced/relationalspace schemes.   
However, it is the conformal ordering that is relationally motivated, and this is not preserved under this passage even for some of these simple examples.

\mbox{ } 

\noindent {\bf Kucha\v{r}'s argument} applies here.  
Namely, that one would not expect that appending unphysical fields to the reduced description should change any of the physics of the of the true dynamical degrees of freedom.   
Thus, if they differ, one should be more inclined to believe the reduced version.  
See Sec \ref{KP} for further development of this and comments.  

\mbox{ }  

\noindent  Additionally, DeWitt's argument of coordinatization invariance makes best sense in the case which involves just true degrees of freedom 
rather than a mixture of these and gauge degrees of freedom.  

\noindent There is however a modicum of approximate protection.

\mbox{ } 

\noindent {\bf Barvinsky's second approximate equivalence} is that, to 1 loop (first order in $\hbar$ in the Semiclassical Approximation), 
Laplacian ordering coincides for reduced and Dirac schemes.  

\mbox{ } 

\noindent Note 1) This can be composed with Barvinsky's first argument to hold to that accuracy in that regime for any $\xi$-ordering and thus in particular to the 
relationally-motivated conformal operator-ordering.  

\noindent Note 2) This and the preceding SSSec are both cases of how ignoring some of the degrees of freedom in a configuration space can change the QM of the 
remaining subconfigurations by altering the nature of the Laplacian or related differential operators that occur in the QM wave equation.

\mbox{ }

\noindent Analogies 87) and 88) that RPM's suffice to reveal the conformal ordering argument and its midisuperspace-level problem.

\subsection{Summary of inequivalent quantization procedures}

\noindent  
N.B. that in the presence of nontrivial $\FrG$, ${\mbox{Point}}(\FrQ/\FrG)$ invariance trumps Point invariance as a meaningful concept.   
For, proceeding indirectly in the presence of an $\FrG$ really means having $\FrG$-Point invariance, so arguments based na\"{\i}vely 
on DeWitt's use of just Point in this context are indeed open to doubt.  
N.B. these are `midisuperspace' issues and thus invisible in the more usual minisuperspace studies.
Thus, upon encountering discrepancies I for now keep the r-version of the argument as the one I know for now to be correctly formulated and implemented.

In general, once one is dealing with linearly constrained theories, arguments for such as conformal or Laplacian operator ordering 
are insufficient since these differ if applied before or after dealing with the linear constraints.  
Moreover, immediately below is a strong argument for it being the reduced case for which such arguments make sense, and this case is 
less available (in particular it is not available for full Geometrodynamics).
%
{            \begin{figure}[ht]
\centering
\includegraphics[width=0.62\textwidth]{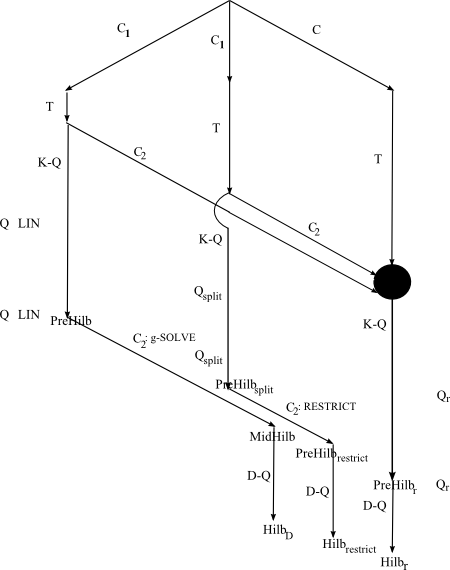}
\caption[Text der im Bilderverzeichnis auftaucht]{        \footnotesize{Sec 3's convergence of classical workings is not in general repeated at the classical level
The {\sc t}-steps are classical; at the quantum level the diagram includes no {\sc t} moves as yet, thus covering Part III's workings.  
Part IV's main focus involves Tempus Post Quantum, Tempus Nihil Est and Non Tempus Sed Historia, 
all of which are well compatible with this diagram or with its obvious histories counterpart.  
Summary of many of the Sec's procedures, as an extension on Sec \ref{Q-Geom}'s opening Figure.} }
\label{Big-Arrow}\end{figure}            }

\vspace{10in}

\subsection{Inner product and adjointness issues} \label{IP-Adj}

Difference 40) Following from the indefiniteness of the GR configuration space metric (Difference 17), GR's inner product is indefinite, rendering 
a Schr\"{o}dinger interpretation inappropriate whilst a Klein--Gordon interpretation fails on further grounds (see Sec \ref{KGI} for details). 
On the other hand, RPM's, like ordinary QM, have a positive-definite configuration space metric, 
giving a positive-definite inner product for which a Schr\"{o}dinger interpretation is appropriate.  

\mbox{ } 

\noindent If one sends $\scQ\scU\scA\scD\Psi = E\Psi$ to $\overline{\scQ\scU\scA\scD}\overline{\Psi} = \overline{E}\overline{\Psi} = \{E/\Omega^2\}\overline{\Psi}$ 
(a move used in the preceding text), one's eigenvalue problem has a weight function $\Omega^{-2}$ which then appears in the inner product: 
\beq
\int_{\FrMgen}\overline{\Psi_1}\mbox{}^*\overline{\Psi_2}\Omega^{-2}\sqrt{\overline{\fM}}d^{k}x \mbox{ } .  
\label{5}
\eeq  
Here, $\langle \FrMgen, \bfM \rangle$ denotes the relationalspace portmanteau at the level of Riemannian geometry.    
This inner product additionally succeeds in being PPSCT invariant, being equal to (c.f. \cite{Magic} for the minisuperspace case)  
\beq
\int_{\FrMgen}\Psi_{1}\mbox{}^*\Omega^{\{2 - k\}/{2}} \Psi_{2}\Omega^{\{2 - k\}/{2}}
             \Omega^{-2}\sqrt{\fM}\Omega^{k}\d^{k}x  = 
\int_{\FrMgen}\Psi_{1}\mbox{}^*\Psi_{2}\sqrt{\fM}\d^{k}x \mbox{ } 
\label{FORK}
\eeq
in the PPSCT representation that is mechanically natural in the sense that $E$ comes with the trivial weight function, 1.

\mbox{ }

\noindent Generally, $\widehat{\overline{\scQ\scU\scA\scD}} = \overline{\widehat{\scQ\scU\scA\scD}}$ is not (in the elementary sense) self-adjoint with respect to $\overline{\langle} \mbox{ } 
| \mbox{ } \overline{\rangle}$, while the mechanically-natural $\widehat{\scQ\scU\scA\scD}$ is, in a simple sense, with respect to $\langle \mbox{ } | \mbox{ } \rangle$.  
More precisely, this is in the sense that 
\beq
\int_{\FrMgen}\sqrt{\fM}\d^kx \Psi^*\triangle\Psi = \int_{\FrMgen}\sqrt{\fM}\d^kx \{\triangle\Psi^*\}\Psi + \mbox{boundary terms },
\eeq 
which amounts to self-adjointness if the boundary terms can be arranged to be zero, whether by the absence of boundaries in the 
configuration spaces for 1- and 2-$d$ RPM's \cite{FORD} or by the usual kind of suitable fall-off conditions on $\Psi$.   
This is not shared by the $\Omega$-inner product as that has an extra factor of $\Omega^{-2}$. 
This in general interferes with the corresponding move by the product rule. 
($\sqrt{\fM}$ does not interfere thus above.  
This is because the Laplacian is built out of derivatives that are covariant with respect to the metric $\fM_{\Gamma\Lambda}$.)
However, on the premise that solving $\widehat{\overline{\scQ\scU\scA\scD}}\overline{\Psi} = \overline{\fE}\overline{\Psi}$ is equivalent to solving 
$\widehat{\scQ\scU\scA\scD}\Psi = \fE\Psi$, the PPSCT might at this level be viewed as a sometimes-useful computational aid.  
The answer would then be placed in the preceding paragraph's PPSCT representation for further physical interpretation.  

\mbox{ }

\noindent For triangleland in the checked representation, $\sqrt{\check{\fM}}$ is the usual spherical Jacobian 
$\mI^2\,\mbox{sin}\,\Theta$, so the checked inner product weight is $\mI\,\mbox{sin}\,\Theta\, /4$. 
Alternatively and equivalently, the plain inner product is $(1/4\mI)^{3/2}\mI^2\mbox{sin}\,\Theta = \mI^{1/2}
\mbox{sin}\,\Theta\, /8$, i.e. differing from the usual spherical one by the obvious conformal factor. 
Also, the checked representation in parabolic coordinates, the inner product weight is just $1/8$.  
x
\mbox{ }

\noindent Conformal transforming prior to solving parallels that which is done by e.g. Iwai, Tachibana and Uwano 
\cite{IwaiUwano1, TachibanaIwai, IwaiUwano2} for a different problem (the 4-$d$ isotropic HO). 
As they explain well, solving QM involves wavefunctions {\sl and} inner product being found, on which grounds the current Article does {\sl not} involve hydrogen.  
For, it has been set up to have the same wavefunctions as hydrogen but the inner products are different.
(So, e.g., normalization is different, as are expectations of operators).  

\vspace{10in}

\subsection{Further quantum cosmological issues modelled by RPM's}\label{IPP} 

\subsubsection{Closed-system/whole universe QM issues} \label{Closed}

Being whole-universe models, RPM's do already exhibit some of these.
For example, RPM's having a fixed `energy of the universe' has the effect of cutting down on the number of valid eigenvalues (a type of 
closed-universe feature that goes back at least to DeWitt \cite{DeWitt67}); see Secs \ref{QM-Intro} and \ref{RPM-for-QC} for more closed-universe features.  

\mbox{ } 
   
\noindent More deeply, {\bf the usual Copenhagen interpretation of quantum mechanics cannot apply to the whole universe}.
In conventional QM, one presupposes that the quantum subsystem under study is immersed in a classical world, crucial parts of which are the observers and/or measuring apparatus.  
In familiar situations, Newtonian Mechanics turns out to give an excellent approximation for this classical world.   
However, there are notable conceptual flaws with extending this `Copenhagen' approach to the whole universe.    
For, observers/measuring apparatus are themselves quantum-mechanical, and are always coupled at some level to the quantum subsystem. 
Treating them as such requires further observers/measuring apparatus so the situation repeats itself.  
But this clearly breaks down once the whole universe is included. 

\mbox{ } 

\noindent An alternative, albeit a somewhat slippery one \cite{KentMW}, is the {\bf many-worlds interpretation}. 
A number of further replacements are tied to various POT strategies such as the \CPII, Histories Theory and Records Theory, as covered in later sections of this Article.

\subsubsection{Structure formation}\label{Str}

\noindent Analogy 89) RPM's are considered to be a qualitative conceptual model of the quantum cosmological seeding of structure formation in a semiclassical regime.  
Particularly for scaled RPM, this parallels the Halliwell--Hawking \cite{HallHaw} approach to GR Quantum Cosmology.  
This is somewhat narrower as a POT strategy (it is an Emergent Semiclassical Time Strategy) but has further conceptual and 
computational applications outside of the POT context too), and of Records Theory \cite{PW83, GMH, B94II, EOT, H99, Records, Records2}.  
I consider this in \cite{MGM, Cones, ScaleQM, 08III, SemiclIII} and Secs \ref{RPM-for-QC} and \ref{Semicl}.

\subsubsection{Timeless Peaking Interpretation and ulterior checks on the Semiclassical Approach}

The present part constitutes of solutions to TISE's for RPM's; these are furtherly interpreted in terms of \NSI and pdf peaking interpretations in Part IV.

RPM's are rare in possessing exact solutions against which semiclassical approximations can be checked.  
This is an extra way in which the present Part supports Part IV.

\subsubsection{Uniform states in (Quantum) Cosmology}\label{Uni}

Analogy 90) Uniformity is of widespread interest in Cosmology.  
It applies to good approximation to the present distribution of galaxies and to the CMB.
There is also the issue of whether there was a considerably more uniform initial state \cite{Penrose}.  
There are further related issues of uniformizing process and how the small perturbations observed today were seeded.
Sec \ref{RPM-for-QC8} outlines how RPM's have qualitative counterparts of these issues, and Sec \ref{QM-Nihil} studies these using the \NSII.

\subsubsection{Robustness of quantization to ignoring some of the degrees of freedom.}\label{QM-Robust}

\noindent {\bf Question 59} [Analogy 91)] RPM's are a toy model for robustness issues, i.e. the consequences of ignoring some of the degrees of freedom.  
This is along the lines in which Kucha\v{r} and Ryan \cite{KR89} question whether Taub microsuperspace  sits stably inside the Mixmaster minisuperspace 
as regards making QM predictions (which is a toy model of whether studying minisuperspace might be fatally flawed due to omitting all of the real universe's inhomogeneous modes).  
This was found to be unstable.  
The RPM (or, for that matter, molecular) counterparts of this have the advantage of possessing a wider range of analytically tractable 
examples with which to carry out such an investigation.  
E.g. is the $N$ = 3 model is stable to the inclusion of 1 further particle, in each of 1-$d$ and 2-$d$? What of triangleland within quadrilateralland?

\subsubsection{Question 60) Do RPM's touch even on some Arrow of Time issues?}\label{RPM-AOT}

E.g. \cite{HH83, EOT, H03, Rovellibook} have some suggestions about associations between this and quantum cosmological issues. 
While I consider this topic to be outside of the usual POT, and make no claim to resolve it or say anything new about it, 
I do make some mention of it in Secs \ref{Semicl} and \ref{Records}.  

\vspace{10in}

\section{The simplest quantum RPM: scaled 3-stop metroland}\label{Q-3-Stop}

Since there are no nontrivial linear constraints in this example, its Dirac and r-quantization coincide.  
%
%
Also, pure-shape RPM is relationally trivial in this case rather than just simpler.  
These features make scaled 3-stop metroland an obvious starting point for the study of quantum RPM's; 
it is used for making a number of basic points about closed-universe QM.

Its study is particularly simple since the mathematics arising is {\sl very} familiar, that of QM in the flat plane.  
However, this mathematics now has an unusual significance that renders it appropriate as a whole universe model, 
and this model already reveals a number of features of RPM models and of closed-universe Physics.
Throughout this Sec, I drop (a) labels.  
I firstly work with subsystem split $\{\rho_1$, $\rho_2\}$ coordinates, which give 2-$d$ Cartesian coordinate mathematics.
I secondly work is scale--shape split $\{\rho$, $\varphi\}$ coordinates, which give plane-polar coordinate mathematics.

\subsection{Scaled 3-stop metroland free quantum problem} \label{QS3Stop1}

\subsubsection{Study in $\{\rho_1$, $\rho_2\}$ coordinates}\label{Q3Stop-Free-1}

In these coordinates, the free problem straightforwardly separates into two copies $i = 1, 2$ of\foo{In this Sec 
$i$ denotes a particular component rather than an index to be summed over.}  
\be
{\hbar^2}\psi_{i,\rho_i^2}/2 +  E_i\psi_i = 0 
\mbox{ } . 
\ee
Thus the wavefunctions are $\psi_i = \mbox{exp}(\pm i \sqrt{2 E_i}\rho_i/\hbar)$, which correspond to a positive continuous spectrum.
The solution of the relational problem may then be reassembled as (up to signs in each exponent)  
\beq
\Psi_{E} = 
\mbox{exp}(\sqrt{2}i\{\sqrt{E_1}\rho_1+\sqrt{E_2}\rho_2\}/\hbar) =
\mbox{exp}
\left(
\frac{\sqrt{2}i}{\hbar}
\left\{
\frac{1}{\sqrt{m_2\hspace{-0.05in}+\hspace{-0.05in}m_3}}
\left\{
\sqrt{E_1m_2m_3}+\sqrt{\frac{E_2m_1}{m_1\hspace{-0.05in}+\hspace{-0.05in}m_2\hspace{-0.05in}+\hspace{-0.05in}m_3}}
\right\}
r_{23}+\sqrt{\frac{E_2m_1\{m_2\hspace{-0.05in}+\hspace{-0.05in}m_3\}}{m_1\hspace{-0.05in}+\hspace{-0.05in}m_2\hspace{-0.05in}+\hspace{-0.05in}m_3}}r_{12}
\right\}
\right) 
\mbox{ } 
\eeq
in terms of straightforwardly relational variables (for use in later discussions).    
There is moreover one unusual relational/closed universe feature: $E_1$ and $E_2$ are not independent: $E_1 + E_2 = E_{\sU\sn\si}$.  
This amounts to the collapse of a larger Hilbert space to a smaller one by the condition that the energy of the whole universe takes a fixed value.

\subsubsection{Study in \{$\rho$, $\varphi$\} coordinates
} \label{Q3Stop-free2}

I separate variables via the ansatz $\Psi = \mG(\rho)\mF(\varphi)$.  
This SSec concerns free models; these require $E > 0$ in order to be physically realized. 
%
%
Then the scale part of the solution is, in terms of Bessel functions, 
\beq
\mG_{\sd}(\rho) \propto \mJ_{\sd}(\sqrt{2E}\rho/\hbar)   \mbox{ } .  
\eeq
The probability density function for this exhibits an infinity of oscillations in the scale direction.

\subsubsection{Interpretation of the shape part}\label{Q3Stop-Int}

For 3-stop metroland in e.g. the \{D, M\} basis corresponding to the (1) cluster, the solutions are 
\beq
\mY_{\sd}(\varphi) \propto \mbox{exp}(i \d \varphi) \mbox{ } , 
\label{16}
\eeq
corresponding to energies $E = \hbar^2 \md^2/2$.   
Here, d $\in \mathbb{Z}$ is a total relative dilational quantum number.    
One often then takes sine and cosine combinations of (\ref{16}).  
Then in terms of shape and size quantities, 
\beq
\mY_{\sd} \propto{\cal T}_{\sd}(\mbox{RelSize}(23)) \mbox{ } ,   
\eeq
for ${\cal T}_{\sd}(X) := \{\mT_{\sd}(X)$ for cosine solutions and $\sqrt{1 - \mT_{\sd}(X)^2}$ 
for sine solutions\}, and $\mT_{\sd}(X)$ the Tchebychev polynomial of the first kind of degree d (see Appendix C).
See Fig \ref{Flowers} for interpretation-by-tessellation of the first few solutions. 
%
{            \begin{figure}[ht]
\centering
\includegraphics[width=0.95\textwidth]{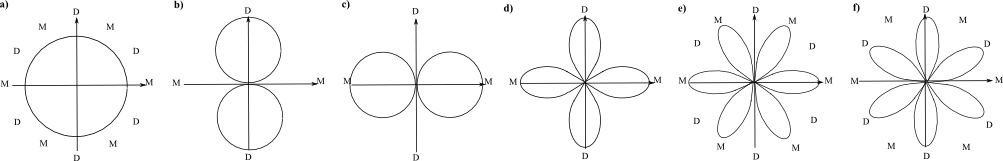}
\caption[Text der im Bilderverzeichnis auftaucht]{        \footnotesize{Pure-shape, or separated-out shape part of, the wavefunction for 3-stop metroland.
The solutions are flowers of 2d petals.
 N.B. the shape-independence of the ground state.}   }
\label{Flowers} 
\end{figure}  } 

\subsection{Scaled 3-stop metroland quantum multi-HO problem} \label{QS3Stop2}

In the isotropic case, there is a choice of separation in subsystem-split \{$\rho_1$, $\rho_2$\} = 
\{Relsize(23), Relsize(1, 23)\} Cartesian coordinates and in scale--shape polars.  
In the non-isotropic case, only the former continues to separate, but after solving in these one can nevertheless 
apply a coordinate change to pass to scale--shape split variables for the purpose of interpretation that parallels Quantum Cosmology.

\subsubsection{Study of multi-HO in $\{\rho_1$, $\rho_2\}$ coordinates} \label{Q3Stop-HO1}

For a single HO, without loss of generality between particles 2 and 3, one has  
\beq
{\hbar^2}\psi_{1,\rho_1\rho_1} - K_1\rho_1^2\psi_1 + 2{E}_1\psi_1 = 0 
\label{Barilla}
\eeq
and the equation in $\rho_2$ from the preceding SSec.
(\ref{Barilla}) is solved by 
\be
\Psi_{\sn_1} \propto \mbox{exp}(\pm i\sqrt{2E_2}\rho_2/\hbar))   
\mbox{H}_{\sn_1}(K_1^{1/4}{\rho_1}^2/\sqrt{\hbar})
\mbox{exp}(-\sqrt{K_1}{\rho_{1}}^2/2\hbar) 
\mbox{ } 
\ee
for H$_{\sn_i}$ the $\mn_i$th Hermite polynomial and with corresponding spectrum $E_i = \sqrt{H_i/\mu_i}\hbar\{{1}/{2} + \mn_i\}$, $\mn_i 
\mbox{ } \in \mbox{ } \mathbb{N}_0$.  
Finally, $\mn := \mn_1$ and $E_2$ are not independent, being related by $\sqrt{K_1}\hbar \{   \mn + {1}/{2}    \}  + E_2 = E_{\sU\sn\si}$.

For 2 or 3 HO potentials, in the special case one obtains two separated-out 1-$d$ HO problems.    
The solution is then  
\be
\Psi_{\sn} \propto  
\mbox{H}_{\sn_1}(K_1^{1/4}{\rho_1}^2/\sqrt{\hbar})
\mbox{exp}(-\sqrt{K_1}{\rho_{1}}^2/2\hbar)
\mbox{H}_{\sn_2}(K_2^{1/4}{\rho_2}^2/\sqrt{\hbar})
\mbox{exp}(-\sqrt{K_2}{\rho_{2}}^2/2\hbar)
\mbox{ } ,   
\ee 
with $\mn_1$ and $\mn_2$ not independent but related, rather, by $\sqrt{K_1}\hbar
\{\mn_1 + {1}/{2}\}  + \sqrt{K_2}\hbar\{\mn_2 + {1}/{2}\} = E_{\sU\sn\si}$.  
The solution here is immediately interpretable using $\rho_1$ = RelSize(23) and $\rho_2$ = RelSize(1,23).   
Early work on quantum RPM's \cite{BS} was hampered by being in terms of $\underline{r}^{IJ}$, which are much less conducive to having separability.

\subsubsection{Isotropic multi-HO ($\Lambda < 0$, $k < 0$ vacuum or wrong-sign radiation) in scale--shape split 
coordinates}\label{Q3Stop-HO2}

For 3-stop metroland with $V = A\rho^2$ $A > 0$, the TISE gives the 2-$d$ isotropic quantum HO problem \cite{Robinett, Schwinger}.   
In scale--shape variables, this is {\sl exactly} solved by \cite{MGM} (\ref{sssplit}, \ref{16}) and a radial factor 
\beq
{\cal I}_{\sN\,\d}(\rho) \propto \rho^{|\sd|}\mL_{\sN}^{\sd}(\omega\rho^2/\hbar)\mbox{exp}(-\omega\rho^2/2\hbar)
\label{Lans}
\eeq
for $\mL^{\sd}_{\tN}(\xi)$ the generalized Laguerre polynomials.  
These solutions correspond to energies $E = \hbar\omega\{2\mN + |\md| + 1\} > 0$ for $\mN \mbox{ } \in 
\mbox{ } \mathbb{N}_0$, $\d \mbox{ } \in \mbox{ } \mathbb{Z}$ and $\omega = \sqrt{K_1} = \sqrt{K_2}$.  
These wavefunctions are finite at 0 and $\infty$ in scale, with N ring nodes in between. 
The ground state, the 2 first excited states and 2 of the second excited states have the plan views of the shapes in Fig \ref{Flowers}, whilst the 2, 0 excited state's 
plan view is a bulge separated from a concentric ring by a ring of absence (`RAF logo shaped').

\subsubsection{Non-isotropic multi-HO transcribed to scale--shape split variables}\label{Q3Stop-HO3}

The solutions are
\beq
\Psi_{\sn_1\sn_2}(\rho, \varphi) \propto 
 \mH_{\sn_1}\big(\sqrt{{K_1}/{\hbar}}\rho\,\mbox{cos}\,\varphi\big) 
 \mH_{\sn_2}\big(\sqrt{{K_1}/{\hbar}}\rho\,\mbox{sin}\,\varphi\big)
\mbox{exp}\big({-\rho^2\{A + B\,\mbox{cos}\,2\varphi\}}/{\hbar}\big) \mbox{ } .  
\eeq
{            \begin{figure}[ht]
\centering
\includegraphics[width=0.7\textwidth]{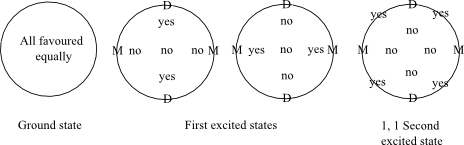}
\caption[Text der im Bilderverzeichnis auftaucht]{        \footnotesize{One can think of these as rectangular/box-shaped arrays of peaks and troughs.
The $B \neq 0$ case (for $|B| < A$) is then a distortion of this (stretched in one direction and squeezed in the perpendicular direction).    }   }
\label{QM-HO-3Stop} 
\end{figure}  } 

\noindent Note 1) There is non-agreement in general with the isotropic case above; this is because solving in Cartesian coordinates amounts to using a different basis of solutions. 
However, the ground state expressions coincide since the ground state is unique and thus independent of which basis one evaluates it in.  

\noindent Note 2) The potential $U\{1 - \mbox{cos}\,2\theta\}$ for $U$ constant occurs in modelling the rotation of a linear molecule in a crystal \cite{P,WS,PW}.  
In the present 2-$d$-like case, this leads to \cite{P} Mathieu's equation \cite{AS} and thus fairly standard Mathematical Physics.

\subsection{Scaled 3-stop metroland quantum problems with further `cosmological' potentials} \label{QS3Stop3}

A number of cases that are quantum cosmologically useful are as per the list in Sec \ref{MechCos} 
(which is only a Lagrangian-level analogy and so is somewhat tenuous).

\subsubsection{Analogues of $\Lambda > 0$, $k < 0$ vacuum models}\label{Q3Stop-+Pots1} 

For $V = A\rho^2$, $A < 0$, the TISE gives instead isotropic upside-down HO mathematics (see e.g. \cite{Robinett}).    
The solutions of these are oscillatory rather than tightly localized.  

\mbox{ } 

\noindent An approximate solution for e.g. the ground state of the isotropic upside-down HO in 3-stop metroland is 
(using separation in Cartesian coordinates \cite{Robinett} and then changing to scale--shape variables)
\beq
\Psi \propto 
\mbox{sin}({\sqrt{-K_1/\hbar}\rho^2\mbox{cos}^2\varphi}/{2})
\mbox{sin}({\sqrt{-K_2/\hbar}\rho^2\mbox{sin}^2\varphi}/{2})
/\rho\sqrt{\mbox{sin}\,2\varphi} \mbox{ } .  
\eeq

\noindent
The probability density function for this is a rectangular grid of humps that decrease in height as one moves out radially.
Getting angular dependence via $B \neq 0$ continues to be straightforward in this case; this readily 
allowing study of models for which the premises of the Semiclassical Approach apply (Sec \ref{Semicl}).  
This is also the case for what is mechanically a mixture of HO's and upside-down HO's.
For other power-law potentials, furthermore approximate or numerical work is required.

\subsubsection{Analogues of $k > 0$ dust model}\label{Q3Stop-+Pots2}

This (restricted to D = 0, i.e. no wrong-sign radiation) is solved by (\ref{16}) and a radial factor
\beq
{\cal I}_{\sN}(\rho) \propto \mL_{\sN - 1 }(2\sqrt{-2E}\rho/\hbar)\mbox{exp}(\sqrt{-2E}\rho/\hbar)
\eeq
corresponding to energies $E = - k^2/2\hbar^2\{\mN - 1/2\}^2$.

\subsubsection{Analogues of right and wrong sign radiation models}\label{Q3Stop-+Pots3}

\noindent In the 2-$d$ case, $\md < \sqrt{2R}/\hbar$ is still a collapse to the maximal collision.  
However, $\md  > \sqrt{2R}/\hbar$ now has a positive root choice, which is finite as $\rho \longrightarrow 0$.
Finally, the critical case $\md = \sqrt{2R}/\hbar$ is now also finite as $\rho \longrightarrow 0$.  
This is somewhat different from the higher-$d$ case covered in Sec \ref{Q-ERPM}.

\subsubsection{Summary of approximate $\rho$ behaviours at the quantum level}\label{Q3Stop++}

Each $O_1$ is bound in an interval including 0 and each $O_2$ is bound in an interval excluding (shielded from) 0.  
$O_3$, $M_1$, $M_2$ are all `tending to free behaviour at large $\rho$, forbidden regions have  
exponential decay, and the $O_2$--forbidden--$O_3$ and $A_1$--$S_1$--$A_2$ slices become tunnelling scenarios.  
This holds also for $N$-stop metroland and for the $\mathbb{CP}^{N - 2}$ presentation of $N$-a-gonland.

\subsection{Interpretation I: closed-universe features}\label{QS3Stop4}

1) There is {\bf energy interlocking} between constituent subsystems.  
E.g. this requires the above na\"{\i}vely free problem to have a segment of linear lattice rather than a quadrant's worth of lattice as its overall eigenspectrum. 
The single HO and free particle system to have a set of points rather than an infinity of lines as its 
overall eigenspectrum, and the coupled HO's give a rather smaller set of points in place of the regular lattice eigenspectrum.  
Moreover, unlike in the usual interpretation of few-particle QM, the energy here is the energy of the universe. 
This is not only fixed, but is also a separate attribute of the universe so that the fixed value it takes need 
bear no relation to the eigenspectra of the universe's contents.  
This leads to the following effects.

\noindent 2) In the multiple HO example, all the energies [$E_1(\nn_1)$ and $E_2(\nn_2)$] are positive.  
Then $E_{\sU\sn\si} < E_1(0) + E_2$ so no wavefunctions exist.  
Universes failing to meet the zero point energy of their content fail to have a wavefunction.  

\noindent 3) Such universes could rather fail to meet the energy required for just some of the states which lie above a given energy that is greater than the zero point energy.  
Then one obtains a {\bf truncation} of the conventional eigenspectrum.  

\noindent 4) In the 2 or 3 HO examples, one is required to solve $k_1\mn_1 + k_2\mn_2 = q$ for $\mn_i \in \mathbb{N}_0$, $k_i = \sqrt{{K_i}}\hbar $ and 
$q = E_{\sU\sn\si} - {\hbar}\{\sqrt{K_1} + \sqrt{K_2}\}/2$.  
Then if $k_i$ and $q$ mismatch through some being rational and some irrational, or even if all are 
integers but the highest common factor of $k_1$ and $k_2$ does not divide $q$, no solutions exist.  
This can also be set up to give {\bf gaps} in what would otherwise look like a truncation of the conventional eigenspectrum.  
Similar effects can be achieved by more complicated matches and mismatches in the other examples' $E_i$ dependences on $\mn_i$.    

\noindent 5) There are also then implications for what is an appropriate ensemble, which I postpone to Sec  \ref{DrNo}.   

\noindent 6) In universes which furthermore contain free particles, because these have continuous spectra, the missing out of some conventional states by 4) cannot occur.  
E.g. this is so for the 1 HO example above, while `tensoring' free particles with the 2 or 3 HO setting alleviates this `missing state' problem 
in the setting of a universe with a slightly larger particle number.    

\noindent 7) If negative energies are possible for some subsystems (which is certainly the case for 
suitable power-law potentials), then subsystems can attain energies higher than $E_{\sU\sn\si}$. 
So if one tensors a subsystem with negative energies with an independent HO pair, one can have less truncation of that HO's spectrum. 
One will still have many, or all, states missing, depending on how the coefficients of the two independent subsystems' potentials are related.  
But if one then tensors in free particles one can have everything up to the truncation.  
Thus truncation can be displaced at least for some universes, by a modest increase in constituent particle number.  
N.B. there is a mismatch between such as the HO which has excited states unbounded 
from above in conventional QM and the atom which has negative energy states bounded from below.  
This apparent difficulty disappears by how very high positive energy states are unlikely to be physically meaningful. 
At the very least the physical validity of the model would break down due to e.g pair production.   
and ultimately the breakdown of spacetime.  
To ensure one's model can attain high enough (but finite) positive energies, wells that are deep and/or numerous enough can be brought in.    

\noindent 8) Unlike 2)--4), energy interlocking {\sl does not} go away with particle number increase or the accommodation of a variety of potentials within one's universe model.
However, the large particle number and high quantum number aspects of semiclassicality are relevant here.  
Subsystems remain well-behaved and all experimental studies in practice involve subsystems.  
But subsystems may be taken to have conventional wavepackets insofar as these {\sl are} products of their constituent separated-out problems' 
wavepackets.\footnote{But why separable cases? 
%
These are, after all, not very general...}
These may still be truncated rather than built out of arbitrarily many eigenfunctions along the lines of 1) to 3) above. 
But, this will be alleviable in practice by ensuring that sufficient additional free particles or particles whose mutual potentials are of an opposing sign to the original subsystem's.  
In such a framework, the correlation of a quantity within a subsystem with another in the rest of the universe is overwhelmingly likely to evade 
detection provided that the rest of the universe contains plenty of other particles.  
And of course the real universe is indeed well-populated with particles.  
This is not a point that the present Article involves itself much with; to us here RPM's are Quantum Cosmology toys rather than likely to be approximate realizations of our universe 
-- it is the `relativity without relativity' form of GR that I consider to be a good candidate for a relational explanation of that.  

\mbox{ } 

\noindent Note: 6) to 8) remove an objection of Barbour and Smolin to RPM's as QM's \cite{BS}. 

\mbox{ }

\noindent 9) If one considers the whole universe, there is a breakdown of the cluster decomposition principle, in which Steven Weinberg places great stock 
-- see e.g. \cite{Weinberg, WeinGHS} in the latter of which he comments that {\it ``I do not see how science is possible without the cluster decomposition principle"}.  
The absence of this principle means that highly separated experiments could in principle yield noticeably correlated results in a small closed-universe setting.
%
%
The chances of this being observed in a large universe like our own, however, is very small.  

\mbox{ } 

\noindent Whether the WKB ansatz may additionally be applied in to whole universes is an issue relevant to the 
Semiclassical Approach to Quantum Cosmology in the general (rather than just RPM) context.  
I mostly leave this issue discussion to Sec \ref{Semicl} \cite{SemiclI, SemiclII, SemiclIII}. 
For the moment I show that increasing the dimension away from 1 reveals further such `small closed universe' effects.  
\noindent After consideration of further examples, I continue this discussion of closed-universe effects in Sec \ref{RPM-for-QC}.

\subsection{Interpretation II: semiclassical issues} \label{QS3StopSemi}

\subsubsection{Characterizations of semiclassicality}   \label{Char-Semi}    

While one common requirement is for the wavelength $\lambda_{\mbox{\scriptsize Q}}$ to be smaller than some characteristic scale 
$l_{\mbox{\scriptsize C}}$ of the problem in question, there is no universal rigorous notion of semiclassical limit for quantum theories.  
Various characterizations are along the following lines (\cite{08I, 08II} and references therein).  

\noindent A) Consider the spread (i.e width) of the wavefunctions (this is essentially what is done in \cite{BS}).    

\noindent B) Furthermore investigate how localized wavepackets are as a whole (as opposed to the spread of each wavefunction in their summand/integrand).

\noindent C) Consider what happens to the system for large quantum numbers (`large occupation number'), for which the wavefunction becomes wiggly on scales much shorter than $l_{\mbox{\scriptsize C}}$.  

\noindent D) Consider a WKB ansatz for the wavefunction and expand in powers of $\lambda_{\sQ}/l_{\sC}$.  

\noindent E) Consider some notion of coherent state.

\subsubsection{Spread of the scaled 3-stop metroland wavefunctions}\label{Spread}

Barbour and Smolin's other objection in \cite{BS}  was to the sensitivity of the spread of the wavefunctions of large-mass particles to the values of the mass of small-mass particles.  
This was in connection with piecewise-constant potential models with two small masses and one large one.  
I clarify why this is not a problem as follows \cite{06I}.

Firstly, in the absolutist quantization of single particles, one thinks primarily of particle 
wavepackets, however in a 1-$d$ relationalist quantization one should think of {\sl relative distance wavepackets}.      
[This is arguably already how one actually thinks in Atomic and Molecular Physics.  
E.g. Lubkin \cite{Lubkin} discusses the approximation by which the proton is considered fixed versus that in which it is not fixed and the spread is in proton--electron separation.
Also consider where in the relatively fixed separations that one encounters in molecules it is that one considers the distribution of `bond electrons' to be. 
[This is an example of shape/relative angle albeit also with the nuclei forming a `fixed background'.]
In the present situation, then, intuitively, once one rephrases one's standard quantum intuition about small masses being more spread out in relational terms. 
Thus it is clear that the position uncertainty of the small mass dominates the relational formulation's relative separation uncertainty between that small mass and a large mass.  

\noindent Barbour and Smolin's specific example has the same content as my Example 1. 
%
%
Then for $m_2, m_3 = m << M = m_1$, the wavefunction goes as 
\be
\Psi \mbox{ } \sim \mbox{ } \mbox{exp}\big(i\big\{\sqrt{E_1} - \sqrt{E_2}\big\}\sqrt{m}r_{23}/\hbar\big)
\mbox{exp}\big(i2\sqrt{E_2}\sqrt{m}r_{12}/\hbar\big) \mbox{ } .  
\ee
Thus, indeed as Barbour and Smolin claim, the small masses dominate all the uncertainties.    
But by my above interpretation, reveals that these uncertainties are in fact in the separations between a big mass 
and a small mass, so this situation conforms to standard quantum intuitions rather than constituting some kind of impasse.  
The truly relevant test to establish whether there is a semiclassical limit problem is rather to check that if  
$m_1, m_3 = M >> m = m_2$, so there is a big mass--big mass relative separation, that this is not influenced much by a small mass somewhere else.  
Now, upon performing the new approximation, and isolating the big mass--big mass separation $r_{13}$ as 
the variable whose spread is of relevance, I find that 
\be
\Psi \mbox{ } \sim \mbox{ } \mbox{exp}\big({ i \sqrt{E_1}\sqrt{M} r_{23}/\hbar}\big)
                            \mbox{exp}\big({-i \sqrt{E_2}\sqrt{m} r_{12}/\hbar}\big) \mbox{ } , 
\ee
so that indeed only the big masses contribute significantly to the spread in the big mass-big mass separation.  
Thus basic conclusion is unaffected by having the two identical masses replaced by merely similar masses.  
Moreover, it holds widely throughout the models presented in this Article when suitable pairs of quantities 
are set to be relatively large and small.\footnote{In pure-shape RPM,  
the corresponding interpretation now additionally involves {\sl spreads in relational ratios} and moreover that now the 
underlying formal mathematics is not standard (at least as far as I know and in the context of the Molecular Physics literature).}

\noindent The Jacobi coordinate presentation renders all of this entirely normal.  

\noindent Also, the application of polar coordinates for $d > 1$ brings out that the standard QM interpretation is close to being relational.  
This is most familiar in the study of the hydrogen atom.  
For this, a simple standard approach is to treat the proton as fixed and then consider the spread of 
the radial separation $r$ between the proton (or more accurately the atom's barycentre) and the electron. 
All that is missing as regards obtaining a fully relational perspective is to consider the position of 
the barycentre not only to be uninteresting but also to be meaningless.  
Then one considers the spread in $\rho_1$.  
The ready availability of this familiar picture is one reason why it is unfortunate that Barbour and Smolin restricted their study to 1-$d$ examples.

\subsubsection{Wavepackets for scaled 3-stop metroland} \label{Packet}

Piecemeal construction of wavepackets for the separated-out 1-$d$ quantum problems does not care whether these arise from separation in 
relational problems or in absolutist ones (see e.g. \cite{Schiff, Robinett}). 
There are, however, limitations building composite wavepackets in the relational case.  
Unlike in the absolutist case, composition of subsystem wavepackets cannot be extended to include the whole system.  
This is due to $\sum_iE_i$ taking a fixed value, $E$.    
By energy interlocking the small universe models built up from individual problems' wavepackets additionally contain a delta function 
\be
\delta
\left(\sum_{\Gamma \in \mbox{\scriptsize subsystems}} E_{\Gamma} - E_{\sU\sn\si}
\right)
\ee
acting inside the sums and integrals required to build it up, which causes it to differ mathematically 
from e.g the direct product of subsystem wavepackets in a fully separable universe.  
This is analogy with conservation of energy--momentum at each vertex in path integral formulation of QFT.

\subsection{Interpretation III: characteristic scales}\label{QS3Stop5}

$K/\rho$ RPM models have a `Bohr moment of inertia' (square of `Bohr configuration space radius') for the model 
universe analogue to (atomic Bohr radius)$^2$, $\mI_0 = \rho_0^2$, which, in the gravitational case, goes like $\hbar^4/G^2m^5$.  
Then $E = - \hbar^2/2\mI_0\{\mN - 1/2\}^2$ in the 3-stop case. 
HO RPM models have characteristic $\mI_{\sH\sO} = \hbar/\omega$.  
Note that the first of these is limited by the breakdown of the approximations used in its derivation, while the second of these has no such problems.

\subsection{Interpretation IV: expectations and spreads of the wavefunctions}\label{QS3Stop6}

\subsubsection{Expectations and spreads of shape operators}\label{4Stop-Spread}

As well as characterization by `modes and nodes' as are evident from figures such as Fig \ref{08III-Fig-4}, 
expectations and spreads of powers of $r$ are used in the study of atoms. 
(See e.g. \cite{Messiah} for elementary use in the study of hydrogen, or \cite{FF} for use in approximate studies of larger atoms).   
These provide further information about the probability distribution function from that in the 
also-studied `modal' quantities (peaks and valleys) that are read off from plots or by the calculus.   
They also represent a wider range of bona fide relational outputs of the quantization procedure 
(like the probability densities but unlike the wavefunctions themselves: only when inner products are used are the outputs physical in this sense).  
E.g. for hydrogen, one obtains from the angular factors of the integrals trivially cancelling and orthogonality and recurrence 
relation properties of Laguerre polynomials in \cite{AS} for the radial factors that (e.g. in the $\ml = 0$ case) 
\beq
\langle\mn\,\ml\,\mm\,|\,r\,|\,\mn\,\ml\,\mm\rangle = \{3\mn^2 - \ml\{\ml + 1\}\}a_0/2 \mbox{ } \mbox{ and } \mbox{ } 
\Delta_{\sn\,\sll\,\sm}r = 
\sqrt{\{\mn^2\{\mn^2 + 2\} - \{\ml\{\ml + 1\}\}^2\}}a_0/2 \mbox{ } ,
\eeq
where $a_0$ is the Bohr radius.
One can then infer from this that a minimal typical size is $3a_0/2$ and that the radius and its spread both become large for large quantum numbers.  
C.f. how the modal estimate of minimal typical size is $a_0$ itself; the slight disagreement between 
these is some indication of the limited accuracy to which either estimate should be trusted. 
Also, the above can be identified as expectations of scale operators, and thus one can next ask whether 
they have pure-shape counterparts in the standard atomic context.

Up to normalization, they are the 3-$\mY$ integrals \cite{LLQM}, the general case of which has been evaluated in terms of Wigner 3j symbols \cite{LLQM} 
(here more properly termed 3d symbols). 
Furthermore, many of the integrals for the present Article's specific cases of interest are provided  case-by case in \cite{Mizushima}. 
%
%
Shape operators for hydrogen are also considered in \cite{AtF} (briefly) and \cite{LSB}.  
Also see \cite{AF} for comments on shape operators elsewhere in Molecular Physics and a start on the corresponding question of shape operators in 'mini- and midi-'superspace.

Moreover, the context in which shape operators occur in Molecular Physics is wider than just the above.  

\mbox{ }  

\noindent Example 1) expectations of cos$\,\upsilon$ for $\upsilon$ a relative angle from inner products between physically meaningful vectors.  
Examples of such are i) between the 2 electron--nucleus relative position vectors in Helium. 
ii) In the characterization of molecules' bonds or in nuclear spin-spin coupling (p 443 of \cite{AF}).  

\noindent Example 2) one also gets expectations of $\mY_{20}(\theta)$ [c.f. form 4 of (\ref{39})] in spin-spin and 
hyperfine interactions (p 437-441) of \cite{AF} (as a shape factor occurring alongside a $1/r^3$ scale factor.  

\noindent Example 3) In the study of the $H_2^+$ molecular ion, one uses fixed nuclear separation as a scale setter.  
Then one has not only 1 relative angle but also 2 ratios forming spheroidal coordinates with 
respect to which this problem separates, and expectations of all these things then make good sense.  

\noindent We contemplate `mini- and midi-'superspace counterparts of such shape operators in the Sec \ref{RPM-for-QC9}.

\mbox{ }  

Overlap integrals $\langle \mD_1\,\md_1\,|\, \widehat{\mbox{Operator}} \,|\,\mD_2\,\md_2\rangle$ are relevant for three applications.

\mbox{ } 

\noindent Application 1) expectation and spread of shape operators (below).  

\noindent Application 2) Time-independent perturbation theory about  very special multi-HO solution. 

\noindent Application 3) Time-dependent perturbation theory on space of shapes with respect to a time provided by the scale 
in the scale--shape split scaled RPM models in semiclassical formulation also makes use of these. 
This parallels Halliwell--Hawking's \cite{HallHaw} approach to Quantum Cosmology and embodies one of the RPM program's eventual goals, 
so I prefer to give details of computing the overlaps to giving details of Application 2).  

\mbox{ } 

\noindent Applications 2) and 3) have the merit of extending to far more general potential terms than the harmonic oscillator-like terms discussed in the present working. 
Also, in the scaled case Application 2) survives as a subproblem in the corresponding time-independent non-semiclassically approximated scale--shape split scaled RPM.

\subsubsection{Multi-HO potential example}\label{3Stop-Spread}

As regards the expectation of the size operator for 3-stop metroland isotropic case in scale--shape 
coordinates, using the normalization result for associated Laguerre polynomials and Gaussian integral 
results, $\langle\,0\,\d\,|\widehat{\mbox{Size}}|\,0\,\d\,\rangle = 
\left(
\stackrel{2|\d| + 1}{\mbox{\scriptsize{$|\d|$}}}
\right)
\frac{\sqrt{\pi}\{|\d| + 1\}}{2^{2|\td| + 1}}\rho_{\sH\sO}$. 
Thus e.g. $\langle\,0\,0\,|\widehat{\mbox{Size}}|\,0\,0\,\rangle = \frac{\sqrt{\pi}}{2}\rho_{\sH\sO}$ 
and the large-$|\d|$ limit for N = 0 is $\sqrt{{\hbar|\d|}/{\omega}}$ which slowly rises to become 
arbitrarily large for configurations possessing more and more relative distance momentum.   
Also, $\langle\, 1\, 0\,| \widehat{\mbox{Size}} | \,1\,0\, \rangle =  
\frac{7\sqrt{\pi}}{8}\rho_{\sH\sO}$. 
In comparison, the mode value for the ground state is $\rho_{\sH\sO}/\sqrt{2}$.

The spreads are an integral made easy by a recurrence relation on the generalized Laguerre polynomials, 
$\langle\, \mN\,\d\,| \widehat{\mbox{Size}}\mbox{}^2 | \,\sN\,\d\, \rangle = \{2\mN + |\d| + 1\}
\rho_{\sH\sO}\mbox{}^2$, minus the square of the expectation.  
Thus $\Delta_{0\,0}(\widehat{\mbox{Size}}) = \frac{4 - \pi}{4}\rho_{\sH\sO}\mbox{}^2 \approx 0.215 
\rho_{\sH\sO}\mbox{}^2$, while the large-$|\d|$ limit with $\mN = 0$ gives $\Delta_{0\,|\sd|}
(\widehat{\mbox{Size}}) \longrightarrow \rho_{\sH\sO}\mbox{}^{2}$.  
Also, $\Delta_{1\,0}(\widehat{\mbox{Size}}) = \frac{192 - 49\pi}{64}\rho_{\sH\sO}\mbox{}^2 
\approx 0.595 \rho_{\sH\sO}\mbox{}^2$.

The expectations of the 3-stop metroland shape operators are all zero by two-angle formulae and the orthogonality of Fourier modes. 
Likewise, the spreads of the shape operators are 1/2 in all states.  
All of the above calculations benefit from factorization into shape and scale parts, with insertion of a pure-shape operator rendering the scale factor trivial and vice versa.

\vspace{10in}

\section{Pure-shape quantum RPM}\label{Q-SRPM}

\subsection{Quantum pure-shape 4-stop metroland}\label{Q4Stop}

In the reduced scheme, the pure-shape case's configuration space geometry is simpler and so this case is treated first.  
For 4-stop metroland, the Laplace ordering and the conformal ordering coincide, both in the Dirac and in the reduced setting.   
Thus there is no trouble in this case with matching up reduced calculations with Dirac ones.

\subsubsection{QM solutions for pure-shape 4-stop metroland}\label{Q4Stop1}

The multi-HO TISE is, for dimensionless constants $\tta := 2\{{\ttA} - {\fE}\}/\hbar^2$, $\ttb := 2{\ttB}/\hbar^2$, $\ttc := 2{\ttC}/\hbar^2$,
\beq
\{\mbox{sin}\,\theta\}^{-1}
\big\{\mbox{sin}\,\theta\,\Psi_{,\theta} 
\big\},_{\theta} 
+  \{\mbox{sin}\,\theta\}^{-2}\Psi_{,\phi\phi} = 
\{\tta + \ttb\,\mbox{cos}\,2\theta + \ttc\,\mbox{sin}^2\theta\,\mbox{cos}\,2\phi\}\Psi \mbox{ } .  
\label{62A}
\eeq

\subsubsection{Solution in very special case}\label{Q4Stop-free1}

The $\ttc = 0$ case of Eq (\ref{62A}) separates to simple harmonic motion and the $\theta$ equation 
\beq
\{\mbox{sin}\,\theta\}^{-1}\{\mbox{sin}\,\theta\,\Psi_{,\theta}\}_{,\theta} - 
\{\mbox{sin}\,\theta\}^{-2}\mD\{\mD + 1\}\Psi  = \tta\Psi + \ttb\,\mbox{cos}\,2\theta\,\Psi \mbox{ } .   
\label{chi}
\eeq
If $\ttb$ = 0 as well (our very special problem), then from Sec \ref{Dyn1}, this has similar mathematics to ordinary QM's central potential problem.  
For this, the quantum Hamiltonian $\hat{H}$, total angular momentum $\widehat{\sfT\sfO\sfT} := \widehat{\scL}_{\sT\so\st} = \sum_{\mu = 1}^3\widehat{L}_{\mu}\mbox{}^2$ and 
magnetic/axial/projected angular momentum $\hat{\scL}_3$ form a complete set of commuting operators.  
Thus, they share eigenvalues and eigenfunctions.  
Recollect that the RPM very special problem is mathematically the same as the rigid rotor, 
for which $\widehat{H}$ {\it is} $\widehat{\sfT\sfO\sfT}$ up to multiplicative and additive constants.  
Thus, effectively one has a complete set of two commuting operators, whose eigenvalues and eigenfunctions are the well-known 
spherical harmonics and, moreover also occur as a separated-out part of the corresponding scaled RPM problem.   
However, the RPM `rigid rotor' is in configuration space rather than in space and with total relative dilational momentum 
$\widehat{\sfT\sfO\sfT} = \sum_{i = 1}^3\widehat{\sfD_i}\mbox{}^2$.
These are in place of total angular momentum and projected relative distance momentum $\widehat{\scD}$ in place of axial angular momentum.  
These then have eigenvalues $\hbar^2\mD\{\mD + 1\}$ and $\hbar\d$ respectively.  
Thus Franzen and I term D and d respectively the {\it total} and {\it projected relative dilational quantum numbers}. 
(These are analogous to the ordinary central force problem/rigid rotor's total and axial/magnetic angular momentum quantum numbers).

As well as the separation involving spherical coordinates, another approach involves 

\noindent 1) using that $\widehat{H}$ and $\rho \, \widehat{\scD}$ form a complete set of commuting operators. 

\noindent 2) Then solve the linear constraint first to get that the wavefunction is independent of $\rho$ and 
then knock out $\rho$ derivatives in the $\widehat{H}$ to get a pure shape problem.  
This approach works for all $N$, showing that $N$-stop metroland indeed has scale--shape split at the quantum level.
Note however that this will not always map to conformal problem, so ($N \geq 5$)-stop metroland does have a constant difference in the energy.  

\mbox{ } 

Our very special problem's TISE separates into simple harmonic motion and the associated Legendre equation (in $X = \mbox{cos}\,\theta$) i.e. the spherical harmonics equations.  
Thus the solutions are 
\beq
\Psi_{\sD\sd}(\theta, \phi) \propto \mY_{\sD\sd}(\theta, \phi) \propto 
\mP_{\sD}^{\sd}(\mbox{cos}\,\theta)\mbox{exp}(\pm i\d\phi) \mbox{ } 
\eeq
for $\mP_{\sD}^{\sd}(X)$ the associated Legendre functions of $X$, D $\in \mathbb{N}_0$ and d such that $|\d| \leq \mD$.  
Also, $\mD\{\mD + 1\} = - \tta$, which, interpreted in terms of the original quantities of the problem, is the condition 
$
\ttE^{\prime} = \ttE - K_3/2 = \hbar^2\mD\{\mD + 1\}/2
$ 
on the model universe's `energy' and inter-cluster effective spring in order to have any quantum solutions. 
($\fE$ is {\sl fixed} as this is a whole-universe model so there is nothing external from which it could gain or lose energy).  
If this is the case, there are then 2D + 1 solutions labelled by d (the preceding sentence cuts down on a given system's solution space, other than in the multiverse sense).

For further interpretation, using a basis with sines and cosines instead of positive and negative exponentials, 
\beq
\Psi_{\sD_{\tN\ta\tm\te}}(\mn^i) \propto \sN\sa\sm\se(\mn^i) \mbox{ } . 
\eeq
Here, the D-label runs over the orbital types ($s$ for D = 0, $p$ for D = 1, $d$ for D = 2 ...) and 
Name is the `naming polynomial' i.e. 1 for $s$, $\mn_x$ for $p_{\sn_x}$, $\mn_x\mn_y$ for $d_{\sn_x\sn_y}$ etc. 
(Note that the name `$z^2$' in $d_{{\sn_z}^2}$ is indeed {\sl shorthand} for $z^2$ -- 1/3; shorthand begins to proliferate if one goes beyond the $d$-orbitals. 
The polynomials arising in this working are also subject to being `nonunique' under $\sum_{i = 1}^3\mn^{i\,2} = 1$.)  
That the wavefunctions are their own naming polynomials is the configuration space analogy of how the orbitals in space historically got their Cartesian names. 
It is also akin to representations (see e.g. \cite{CH}) of the spherical harmonics in terms of homogeneous polynomials.  
Another form for the solution is
\beq
\Psi_{\sD|\sd|}(\mn^i) \propto \mP^{\sd}_{\sD}(\mn_z){\cal T}_{\sd}\left(\mn_x\left/\sqrt{1 - \mn_{z}\mbox{}^2}\right.\right)  
\mbox{ } .  
\eeq
The first factor is purely in Relsize(12,34) whilst the second is a mixed function of this and RelSize(12).  
Also, using the tessellation method, I can interpret the wavefunctions in terms of the 4-stop metroland mechanics 
on the sphere itself, on which they take the particularly familiar `orbital' form.
%
{\begin{figure}[ht]
\centering
\includegraphics[width=1.02\textwidth]{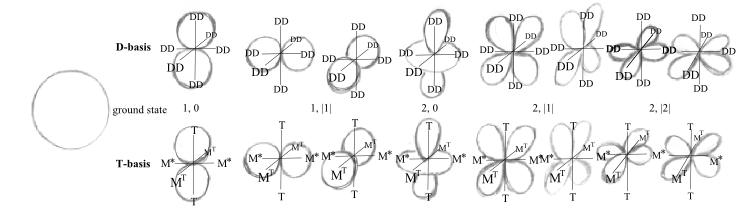}
\caption[Text der im Bilderverzeichnis auftaucht]{\footnotesize{Wavefunctions for triangleland in the DD basis in the text and in the T basis.}} \label{Y-4-Stop}\end{figure} } 
 
\noindent While this case is simple and not general, further motivation for the very special problem is that its exact solution 
serves as something about which one can conduct a more general perturbative treatment (Sec \ref{Q4Stop-Pert}).

\subsubsection{4-stop metroland examples of shape operators}\label{Q4Stop2}

As regards good shape operators for 4-stop metroland, the kinematical quantization carries guarantees that the three $\mn^i$ are promoted to good quantum operators.  
These can be interpreted as RelSize(1,2), RelSize(3,4) and RelSize(12,34).      
Note that $\mn_z$ is not only physically RelSize(12,34) but also mathematically the Legendre variable.

Then $\langle\mD\,\md\,|\,\widehat{\mbox{RelSize(1,2)}}\,|\,\mD\,\md\rangle = 
       \langle\mD\,\md\,|\,\widehat{\mbox{RelSize(3,4)}}\,|\,\mD\,\md\rangle = 0$ and 
$\langle\mD\,\md\,|\,\widehat{\theta}\,|\,\mD\,\md \rangle \approx$
$\langle\mD\,\md\,|\,\widehat{\mbox{RelSize(12,34)}}\,|\,\mD\,\md\rangle = 0$   
as an obvious result of orientational symmetry. 
The useful information starts with the spreads,  
\beq
\Delta_{\sD\,\sd}(\widehat{\mbox{RelSize(1,2)}})  = 
\sqrt{\frac{\mD\{\mD + 1\} + \md^2 - 1}{\{2\mD - 1\}\{2\mD + 3\}}Q_1(\d)}
\mbox{ } \mbox{ } , \mbox{ } \mbox{ }
\Delta_{\sD\,\sd}(\widehat{\mbox{RelSize(3,4)}})  = 
\sqrt{\frac{\mD\{\mD + 1\} + \md^2 - 1}{\{2\mD - 1\}\{2\mD + 3\}}Q_2(\d)} \mbox{ } ,
\eeq
\beq
\Delta_{\sD\,\sd}(\widehat{\theta}) \approx \Delta_{\sD\,\sd}(\widehat{\mbox{RelSize(12,34)}})  = 
\sqrt{\frac{2\{\mD\{\mD + 1\} - \md^2\} - 1}{\{2\mD - 1\}\{2\mD + 3\}}} \mbox{ } .
\eeq
Here, $Q_2(\d) =         1/2 \mbox{ for the } \d \mbox{ cosine solution, }  
                       3/2 \mbox{ for the } \d \mbox{ sine solution, and } 
                       1   \mbox{ otherwise,}$ 
and $Q_1(\d)$ is the sin $\longleftrightarrow$ cos of this. 
One can then readily check that $\langle \widehat{\mn}_x\mbox{}^2 + \widehat{\mn}_y\mbox{}^2 + \widehat{\mn}_z\mbox{}^2\rangle$ = 1, as it should be.

One case of interest is the ground state.
There, the spreads in each are $1/\sqrt{3}$.  
Another case of interest is the large quantum number limit.  
$\Delta_{\sD\,\sd}(\widehat{\theta}) \approx$ $\Delta_{\sD\,\sd}(\widehat{\mbox{RelSize(12,34)}})$ 
which, for the maximal d ($|\md| = \mD$), is equal to $1/\sqrt{2\mD + 3}$ which goes as $1/\sqrt{2\mD} \longrightarrow 0$ for D large.  
The hydrogen counterpart of this result is $\Delta_{\sll\,\sll}\hat{\theta_{\sss\sp}} \approx  1/\sqrt{2\ml} \longrightarrow 0$, 
i.e. restriction to the Kepler--Coulomb plane (e.g. \cite{AF} outlines this, while \cite{LSB} considers it in more detail).   
Back to the RPM problem, this result therefore signifies recovery of the equatorial classical geodesic as the limit of an ever-thinner belt 
in the limit of large maximal projectional relative dilational quantum number $|\md| = \mD$. 
(This is `the rim of the disc' in Sec \ref{Cl-Soln}'s classical representation of this problem, traversed in either direction according to the sign of d.)   
In fact, as for the constant potential one can consider the axis to be wherever one pleases, this leads to recovery of {\sl any} of the classical geodesics.  
Also, for d = 0, $\Delta_{\sD\,0}\widehat{\theta} \approx \Delta_{\sD\,0}(\widehat{\mbox{RelSize(12,34)}}) \longrightarrow 1/\sqrt{2}$ for D large.  
This means that the $s$, $p_{\sn_z}$, $d_{{\sn_z\mbox{}^2}}$ ... sequence of orbitals does not get much narrower as D increases.  
Thus, for these states there is only limited peaking about clusters \{12\} and \{34\} both being small and well apart. 
This is a situation which shall be revisited in the next subsection due to its centrality to the assumptions made in, and applications of, this Article.
The RelSize(1,2) and RelSize(3,4) operators' spreads tend to finite constant values for large D no matter what value d takes.

What of $\hat{\phi}$?  
Now, clearly, by factorization and cancellation of the $\theta$-integrals, the d = 0 states obey the 
uniform distribution over 0 to $2\pi$, with mean $\pi$ and variance $\pi^2/3$ (corresponding to axisymmetry).  
Furthermore, $\langle\mD\,\md\,|\,\hat{\phi}\,|\,\mD\,\md\rangle$ is also $\pi$ and cosine and sine states have 
\beq
\Delta_{\sD\,\sd}(\widehat{\phi}) = \sqrt{{\pi^2}/{3} + {1}/{2\md^2}} 
\mbox{ } \mbox{ and } \mbox{ } 
\Delta_{\sD\,\sd}(\widehat{\phi}) = \sqrt{{\pi^2}/{3} - {1}/{2\md^2}} \mbox{ } ,
\eeq 
which indicate some resemblance to the uniform distribution arising for large d. 
(Mean and variance do not see the multimodality. 
Also, at least, by inspection along the lines of the preceding subsection, it is {\sl regular} multimodality for d maximal.) 
One has, by inspection of the shapes of the standard maximal $s$, $p$, $d$, $f$, $g$ ... orbitals: equatorial flowers of 2D petals.
that get more and more equatorially flat as D gets larger, thus tending to the equatorial great circle classical path (in parallel to \cite{LSB} for hydrogen).

\subsubsection{Solution in special case -- large and small regimes}\label{Q4Stop-As}

Passing to stereographic coordinates and applying a PPSCT to pass to the flat representation and applying the small approximation, the TISE becomes 
\beq
-\{\hbar^2/2\}
\big\{
{\cal R}^{-1}
\{
{\cal R}\Psi_{,{\cal R}}
\}_{,{\cal R}} 
+ {\cal R}^{-2}\Psi_{,\phi\phi}
\big\} = \ttE - {\Omega^2{\cal R}^2}/{2} \mbox{ } ,
\eeq
which is in direct correspondence with the 2-$d$ quantum isotropic harmonic oscillator (see e.g. \cite{Messiah, Schwinger, Robinett} under 
$r \longrightarrow {\cal R}$ (radial coordinate), $1 \longleftrightarrow$ particle mass, and with the above $\omega$ as classical frequency ($\times I$).   
Thus,  
\beq
\ttE = \mn\,\hbar\,\omega     \mbox{ } \mbox{ for } \mbox{ }    \mn := 1 +  2\mN + |\d| 
\label{zin}
\eeq
$\mathbb{N}_0$ and d a `projected' dilational quantum number as in the preceding subsection but now running over $\mathbb{Z}$.
[The `shifted energy' in its usual units, $\ttE^{\prime} = \ttE - \ttA - \ttB$, itself goes as 
$
\ttE^{\prime} = \{\mn^2 \hbar^2/2 \}\{1 + \sqrt{1 - \ttB\{4/\mn\hbar\}^2}\},
$
so for $\mn\,\hbar/\omega << 1$ (small quantum numbers as used below), $\fE^{\prime}\mI \approx \mn\hbar\Omega$ for $\Omega = 2\sqrt{-\ttB}$.] 
%
%
The solutions are then (to suitable approximation) 
\beq
\Psi_{\tN\sd}(\theta, \phi) \propto {\theta}^{|\sd|}\{1 + {|\d|\theta^2}/{12}\}\mbox{exp}
(-{\omega\theta^2}/{8\hbar})\mL_{\tN}^{|\sd|}({\omega\theta^2}/{4\hbar})\mbox{exp}(\pm i\md\phi)
\label{gi2}
\eeq
for $\mL_{a}^b(\xi)$ the associated Laguerre polynomials in $\xi$ (see Appendix C).  
[As regards physical interpretation, the $\phi$-factor of this is rewriteable as before in terms of the $\mn^i$ or RelSize(12,34) and 
RelSize(1,2), while the $\theta$-factor is now a somewhat more complicated function of RelSize(12,34).]

The large regime gives the same eigenvalue condition (\ref{zin}), and (\ref{gi2}) again for wavefunctions except that one now uses the supplementary angle 
$\zeta = \pi - \theta$ in place of $\theta$.   
See Fig \ref{Tulips} for the form and interpretation of the wavefunctions.  

{             \begin{figure}[ht]
\centering
\includegraphics[width=0.8\textwidth]{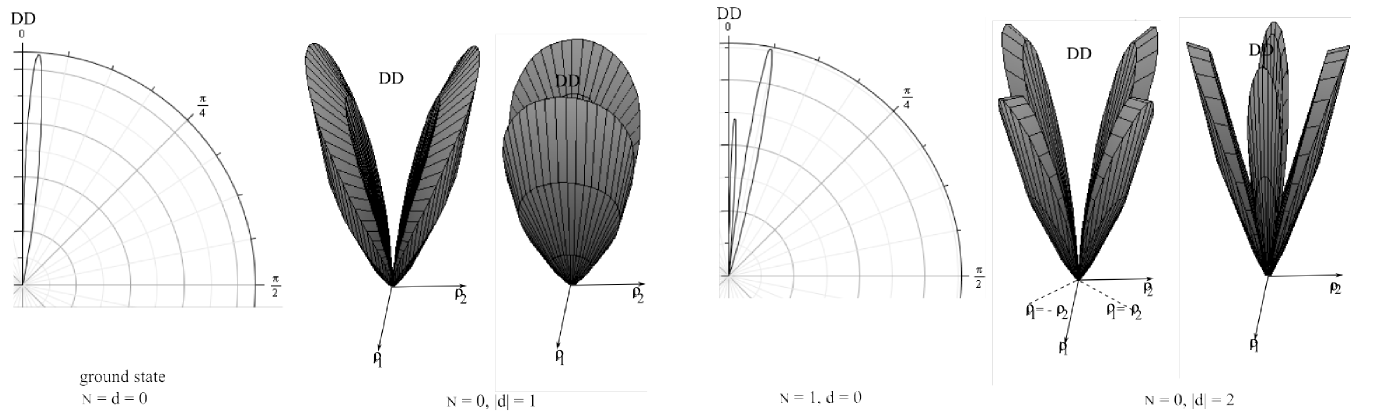}
\caption[Text der im Bilderverzeichnis auftaucht]{        \footnotesize{    Probability density functions for this subsection's problem for $\omega/\hbar$ large, 400, say, 
plotted using Maple \cite{Maple}.  
All d = 0 states are axisymmetric about the \{12,34\} clustering's DD axis, i.e. all relative sizes for cluster \{12\} and for cluster \{34\} are equally favoured.  
$\iota^i$ is my former notation to $\rho^i$.  
The ground state is peaked around the \{12,34\} DD collision.  
It it the surface of revolution of the given curve.  
The $\mN$ = 0, $|\d|$ = 1 solutions are a degenerate pair.  
Each takes the form of a pair of inclined lobes -- `bat ears' -- the cosine solution's is oriented about the $\mn_x =$ 
RelSize(1,2) = 0 D-collision and the sine solution's about the $\mn_y =$ RelSize(3,4) = 0 D-collision.
These next three solutions form a degenerate triplet: the double-band $\mN = 1$, d = 0 solution and the two $\mN$ = 0, $|\d|$ = 2 `tulips of four petals' solutions as indicated.  }   }
\label{Tulips}
\end{figure}            }

In the small regime, the RelSize(1,2) and RelSize(3,4) operators still have zero expectation as each sign for these remains equally probable.  
For D, d substantially smaller than $\omega/\hbar$ so powers of the latter dominate powers of the former. 
[And $\omega/\hbar$ {\sl was} considered to be large, so this works for the kind of quantum numbers in this subsection's specific calculations]. 
Then the following mean and spread results for shape operators are derived using orthogonality of, and 
a recurrence relation for, Laguerre polynomials, as provided in Appendix C.  
\beq
\langle\mN\,\md\,|\,\widehat{\mbox{RelSize(12,34)}}\,|\,\mN\,\md\rangle = 
1 - 2\,\mn\,\hbar/{\omega} \mbox{ } . 
\eeq
$\Delta_{\sN\,\sd}\widehat{\mbox{RelSize(12,34)}}$ is zero to first two orders, beyond which the 
approximations used begin to break down, but it would appear to have leading term proportional to $\hbar/\omega$.  
These results signify that the potential has trapped what was uniform in the preceding example into a narrow area around the \{12,34\} DD collision.  
Furthermore, 
\beq
\Delta_{\sN\,\sd} \widehat{\mn^{\bar{a}}} \approx 
\sqrt{2\,{\mn}\,\hbar Q_{\bar{a}}(\d)/\omega}   
\eeq 
for $\bar{a} = 1, 2$ gives $\Delta_{\sN\,\sd} (\widehat{\mbox{RelSize(1,2)}})$ for $\bar{a} = 1$ and 
$\Delta_{\sN\,\sd} (\widehat{\mbox{RelSize(3,4)}})$ for $\bar{a} = 2$.
So one can obtain strong concentration around the poles by suitable choice of springs, amounting to a tall thick equatorial barrier and polar wells.  
The ground state has the tightest spread in RelSize(1,2) and RelSize(3,4): $\sqrt{2\hbar/\omega}$.  
This has some parallels with how the Bohr radius is an indicator of atomic size, including the hydrogen--isotropic harmonic oscillator correspondence \cite{Schwinger}.   

\mbox{ }  

\noindent N.B. this involved second-order large and small approximations. 
Comparing first and second order results, how can few-peak functions be second approximations when infinite-peak functions are first approximations? 
This is accounted for by the few-peak functions have an extra node-counting quantum number N.  
Thus the first approximation can be seen as a superposition of many different values of N each with its own $Q_2$.   
Thus a multiplicity of peaks builds up.

\subsubsection{Perturbations about the very special solution}\label{Q4Stop-Pert}

Begin by recasting the RPM TISE in Legendre variables $\mn_z = \mbox{cos}\,\theta$, 
\beq
\{
\{1 - \mn_z\mbox{}^2\}\Psi_{,\sn_z} \}_{,\sn_z} 
+ \{1 - \mn_z\mbox{}^2\}^{-1}\Psi_{,\phi\phi} = \{\tta - \ttb + \mn_z\mbox{}^2\{2\ttb - \ttc\mbox{cos}\,2\phi\} + \ttc\mbox{cos}\,2\phi\}\Psi \mbox{ } ,
\label{yt}
\eeq
and then study this using time-independent perturbation theory (see e.g. \cite{LLQM} for derivation of the formulae for this up to second order).  
Applying perturbation theory here means considering 
1) $\ttC$ small, which is high contents homogeneity at the level of each cluster's (Hooke coefficient)/(reduced mass) in the sense that $K_1 - K_2$ is small compared to $\hbar^2$.
2) $\ttB$ small, in the sense that $\hbar^2$ is large compared to $\{K_1 + K_2\}/2 - K_3$, which collapses to $K_1 - K_3$ small in the case of $\ttC = 0$.  
This means that there is little difference between the inter-cluster spring and the intra-cluster springs.

E.g. perturbative study of (\ref{yt}) is amenable to exact calculations though involving various of 
trigonometric and standard/tabulated associated Legendre function integrals, or, alternatively, the aforementioned 3-$\mY$ integrals. 
Furthermore, this continues to be the case if one includes a non-diagonal/non-normal basis' $\ttF$, $\ttG$ and $\ttH$ terms.

For the $\ttb$-perturbation, as both it and the unperturbed Hamiltonian commute with $\sfT\sfO\sfT$
the eigenvalue problem can be solved separately in each subspace $\mathfrakV_{\sd}$ of a given eigenvalue $\d$ of $\sfT\sfO\sfT$. 
As then in each such subspace the spectrum of the unperturbed Hamiltonian is nondegenerate, 
nondegenerate perturbation theory is applicable  (this argument parallels e.g. p 697 of \cite{Messiah}).   
This gives (with the unperturbed problem's $\tta$ playing the role usually ascribed to the energy and $H^{\prime}$ the perturbative term)    
\beq
\tta_{\sD\,\sd}^{\{1\}} = \langle\mD\,\md\,|\, H^{\prime}\,|\,\mD\,\md\rangle  
\label{p1}
\eeq
at first order and 
\beq
\tta_{\sD\,\sd}^{\{2\}} = -\sum\mbox{}_{\mbox{}_{\mbox{\scriptsize $\sD^{\prime},\sd^{\prime} \neq \sD,\sd$}}} 
      |\langle\mD^{\prime}\md^{\prime}|\,{H}^{\prime}\,|\,\mD\,\md\rangle|^2/\{ a_{\tD^{\prime}} - a_{\tD}    \} \mbox{ }   
\label{p2}
\eeq
at second order \cite{LLQM}.
Then e.g. \cite{08II} double use of a standard recurrence relation \cite{AS} gives a $\Delta \md = 0$, 
$\Delta \mD = 0, \pm 2$ `selection rule' paralleling that for the Raman spectrum of a polarized linear rotor.  
Moreover, the terms that survive this take the following forms.    
\beq
\langle\mD\,\d\,|\,\ttb\{2\mn_z\mbox{}^2 - 1\}\,|\,\mD\,\d\rangle = {\ttb\{1 - 4\d^2\}}/{\{2\mD - 1\}\{2\mD + 3\}}
\mbox{ } ,  
\label{*****}
\eeq
which is closely related to the expectation of $\mn_z$ = RelSize(12,34) already computed in Sec \ref{Q4Stop2}.  
Two new overlaps that are more general than expectations are 
\beq
\langle\mD + 2\,\d\,|\, \ttb \{2\mn_z\mbox{}^2 - 1\}\,|\,\mD \,\d\rangle = \frac{2 \ttb}{2\mD + 3}
\sqrt{\frac{\{\{\mD + 2\}^2 - \d^2\}\{\{\mD + 1\}^2 - \d^2\}}{\{2\mD + 5\}\{2\mD + 1\}}}
\mbox{ } , 
\label{***}
\eeq
and then, swapping D for D -- 2, also,  
\beq
\langle\mD - 2\,\d\,|\,\ttb\{2\,\mn_z\mbox{}^2 - 1\}\,|\,\mD\,\d\rangle = \frac{2\ttb}{2\mD - 1}
\sqrt{\frac{\{\{\mD^2 - \d^2\}\{\{\mD - 1\}^2 - \d^2\}}{\{2\mD + 1\}\{2\mD - 3\}}}
\mbox{ } . 
\label{****}
\eeq
Using these then gives the perturbed `pseudoenergies':  
$$
\ttE_{\sD\,\sd}^{\{2\}} = \ttA + {\hbar^2}\mD\{\mD + 1\}/2 + {\ttB\{1 - 4\md^2\}}/{\{2\mD - 1\}\{2\mD + 3\}} + 
$$
\beq
\frac{4\ttB^2\{ \{2\mD + 5\}\{2\mD + 3\}^3\{\mD^2 - \md^2\}\{\{D - 1\}^2 - \md^2\}  - 
             \{2\mD - 1\}^3\{2\mD - 3\}\{\{\mD + 2\}^2 - \md^2\}\{\{\mD + 1\}^2 - \md^2\}\}\}}
     {\hbar^2\{2\mD + 5\}\{2\mD + 3\}^3\{2\mD + 1\}\{2\mD - 1\}^3\{2\mD - 3\}} + O(\ttB^3).
\label{big}
\eeq
Note that d positive and negative remain treated the same, so there is only a partial uplifting of degeneracy.  
Changes to the wavefunction due to the perturbations for the sign of $\ttB$ corresponding to \ref{Q4Stop-As} 
and to second order in $B$ are slight bulges at the poles for the ground state (a bit of $d_{{\sn_z}^2}$ mixed in).

Unlike for triangleland, the $\ttC$-perturbation cannot be turned into a $\ttB$-perturbation with respect to a rotated basis. 
But the ${\cal C}$ perturbation can likewise be studied based on half-way stage overlaps that can be directly transcribed by 
the angular momentum to relative distance momentum analogy from those computed in e.g. \cite{Mizushima}.  
E.g. the surviving terms are found \cite{08II} by a second standard recurrence relation \cite{AS} to obey the selection rule $\Delta\md = \pm 2$, $\Delta \mD = 0$.   
Some noteworthy features of the study of the ${\cal C}$ term are that degenerate perturbation theory is 
now required, there is no first-order contribution as $\Delta \d = \pm 2$ only.  
Now d and --d do get shifted differently corresponding to this perturbation not preserving the axis of symmetry.   
In nondiagonal/nonnormal form, the further $\ttF$ term has the same selection rule to the $\ttC$ term's 
while the $\ttG$ and $\ttH$ terms share the selection rule $\Delta \d = \pm 1$, $\Delta \mD = 0, \pm 2$.  
The above `noteworthy features' apply to these also.

\subsubsection{Molecular Physics analogies for 4-stop metroland }\label{41Mol} 

Analogy A) (\ref{chi}) occurs in Mathematical Physics, e.g. from the separation of the TISE in prolate spherical coordinates \cite{Brief, SMCH, MFII, AS}). 
It also has multiple applications in Molecular Physics studies of which parallel some of the studies in the present Article.    
Examples of this in Molecular Physics are as follows.  


\noindent Analogy A.1) (\ref{chi}) recast in terms of the Legendre variable is 
\beq
\{\{1 - {\mn_z}^2\}\Psi_{,\sn_z} \}_{,\sn_z} - \{1 - {\mn_z}^2\}^{-1}\mD\{\mD + 1\}\Psi = \{\tta - \ttb + 2\,\ttb\,{\mn_z}^2\}\Psi \mbox{ } ,  
\label{ytt}
\eeq
which is the easier of the two spheroidal equations that arise in the study of the $H_2^+$ molecular ion \cite{BLS, Slater, H2+}.    
This and the next two analogies are for $\ttb < 0$, although the aforementioned Mathematical Physics literature covers $\ttb > 0$.
 

\noindent Analogies A.2-3) are Sec \ref{HO-An}'s rotation in a crystal (Pauling) and molecular polarizability (Raman).  

\noindent Analogy A.4) Examples 2) of Sec \ref{Q4Stop2} is another substantially developed area in the Molecular Physics literature. 


\noindent Analogy B) is with the ammonia molecule $NH_3$, in the following rougher but qualitatively valuable sense.   
$NH_3$ has two potential wells separated by a barrier and then is capable of tunnelling between the two at the quantum level (like an umbrella inverting in the wind).  
Our model for $\ttB < 0$ is similar to this, albeit in spherical polar coordinates: there are two polar wells with an equatorial barrier in between.  

\noindent This analogy then gives us some idea about how the separate solutions for the two wells compose. 
For $NH_3$, one can start with separate solutions for each well and additional degeneracies ensue 
(due to the wells being identical and being able to distribute some fixed energies between these in diverse ways). 
However the wavefunctions tend to perturb each other toward breaking these degeneracies, forming symmetric and antisymmetric wavefunctions over the two wells \cite{AF, Slater}.

\subsubsection{Applications of the analogies and developing an overall picture}\label{Q4Stop5}

1) Rotor regime: the $\ttB < 0$ locally stable small or large regimes are the kind of regimes that are termed `rotor-like' in Analogy A.2)'s 
literature; both of the $SO$(3) quantum numbers (for us, dilational quantum numbers) hold good in this regime.  

\noindent 2) Putting together the small and large $\theta$ approximations: for $\ttB < 0$ one can use Analogy B) to form a simple picture of this. 
d remains a good quantum number for the unapproximated problem.  
Thus, one expects to need the North Pole approximation's d and the South Pole approximation's d to 
match and the subsequent perturbations exacted by these two approximations upon each other not to affect d.  
Also, prior to any recombination, one has degeneracies as follows (call the near-North Pole's node-counting quantum number $\mN$ and the near-South Pole's $\mN^{\prime}$).
There is the one ground state    $\mN = \mN^{\prime} = \d = 0$, 
then the degenerate pair         $\mN = \mN^{\prime} = 0, \d = \pm 1$, and 
then the degenerate quadruplet   $\mN = 1, \mN^{\prime} = 0$ 
                            or   $\mN = 0, \mN^{\prime} = 1$ for each of  $\d = \pm 1$.   
Now if $\mN$ and $\mN^{\prime}$ match, $\langle \mbox{RelSize}(12,34) \rangle$ goes to 0 again though the wavefunction's distribution is bimodal 
about both poles. 
If they do not match, $\langle \mbox{RelSize}(12,34)\rangle \neq 0$ due to the peaking near the two poles being different in detail.  
The flip here, as in $NH_3$, is an inversion, i.e. it reverses the orientation, sending 1,2,3,4 to 4,3,2,1. 

\noindent 3) Analogy A.3) is well-known for its Raman-type $\pm 2$ and not $\pm 1$ selection rule, which parallels the results of Sec \ref{Q4Stop-Pert}.  
Analogy A.3) has furthermore been studied perturbatively for what is for us the small $\ttB < 0$ regime.  
This allows us to e.g. check the half-way house results (\ref{*****}, \ref{***}, \ref{****}) against p. 271-273 of \cite{TS}.  

\noindent 4) Analogy A.1)'s references \cite{SMCH, MFII} include i) analysis of this equation's poles in the complex plane. 
ii) How it admits a solution in the form of an infinite series in associated Legendre functions in the vicinity of $\pm 1$ and in Bessel functions in the vicinity of $\infty$.  
iii) How to piece together these different representations. 
It is then appropriate to compare results from the expansion in associated Legendre functions against our perturbative regime 
(this particular working holds regardless of the sign of $\ttB$).
Thus the lowest four cases of (\ref{big}) agree with p 1502-4 of \cite{MFII}.  
This additionally provides the corresponding wavefunctions which are used to first order in $\ttB$ in Sec 
\ref{NSI} to evaluate the \NSI probabilities for these states' model universes being large.  

\noindent 5) Large regime, one cannot really put together the near-polar calculations and the perturbative calculations, because the ``$\ttB$ 
small perturbative condition" goes a long way toward $\omega$ being small and then only a small amount of the wavefunction is near the pole.   
Our near-polar calculations should be compared, rather, with the asymptotics for $\ttB$ large.    
Analogy A.2)'s literature covers this for what is, in the RPM case, the $\ttB$ large negative ($>> \ttE - \ttA - \ttB = \ttE^{\prime}$) regime, giving, via the analogy, 
\beq
\ttE^{\prime} \mbox{ } \widetilde{\mbox{ }} \mbox{ } \mn \hbar\omega_{\sll\sa\sr\sg\se} + O(1/\omega_{\sll\sa\sr\sg\se})
\mbox{ } \mbox{ } \mbox{and}   
\label{HOlike}
\eeq 
\beq
\Psi \propto \mbox{exp}(\omega_{\sll\sa\sr\sg\se}\mbox{cos}\theta/\hbar)
\{\{\mbox{tan$\frac{\theta}{2}$}\}^{2\sN}\{\mbox{sec$\frac{\theta}{2}$}\}^{2\{|\sd| + 1\}} 
+ O(1/\omega_{\sll\sa\sr\sg\se}) \}\mbox{exp}(\pm i\md\phi) \mbox{ } 
\eeq
as the relevant asymptotic solutions, for $\omega_{\sll\sa\sr\sg\se} = 2\sqrt{-\ttB}$.  
Now, from (\ref{oc},\ref{qus}) the small-$\theta$ approximate solution (\ref{gi2})'s $\omega = 8\sqrt{-\ttB} = 4\omega_{\sll\sa\sr\sg\se}$.  
Thus, $\mbox{exp}(\omega_{\sll\sa\sr\sg\se}\mbox{cos}\theta/\hbar)$ in (\ref{HOlike}) $\approx \mbox{const}\times\mbox{exp}(\{\omega/4\hbar\}
\{-\theta^2/2\})$, which is indeed in agreement with the leading and dominant factor of (\ref{gi2}).
For the RPM model, this regime signifies that $K_3 >> K_1, K_2$ i.e. that the inter-cluster spring is much stronger than each of the intra-cluster springs.    
This is a `harmonic oscillator-like regime' -- comparing (\ref{HOlike}) and the standard result for the 2-$d$ isotropic harmonic oscillator makes it clear why.   
d alone is a good quantum number in this regime.  

\noindent 6) the spheroidal equation has led to many hundreds of pages of tabulations \cite{SMCH} and further numerical work e.g. in \cite{AS, 
F57, Meix, Fall, Fall2}, though the most recent of this states that this study is still open in some aspects.

\noindent 7) One can furthermore envisage Analogy A.2) being extended to have further parallels with this model 
via a rotationally-dislocated molecule in a cubic crystal having preferred directions in space of approximately the same form as this model's are in configuration space.  
We do not know if such a study has been done.    

\noindent 8) The polarization analogy A.3) has been extended \cite{WS62,A76} to include the counterparts of the $\ttC$, $\ttF$, $\ttG$ and $\ttH$ terms. 
For, what one has more generally is a symmetric polarization tensor $\underline{\underline{\Bigupalpha}}$ such that $\mu_{\rho} = \Bigupalpha_{\rho\sigma} \lme_{\sigma}$.  
Then for the $CO_2$ model in a diagonal basis $\Bigupalpha_z = \Bigupalpha_{||}$ giving the combination  
-$\Bigupalpha_{\perp}\,\mbox{sin}^2\theta_{\sss\sp} - \Bigupalpha_{||}\,\mbox{cos}^2\theta_{\sss\sp}$ 
[This is a slight improvement of analogy A.3) by inclusion of the smaller $\Bigupalpha_{\perp} = \Bigupalpha_x = \Bigupalpha_y$]. 
Furthermore, this readily rearranges to the special case of the third form of (\ref{39}).  
But for more general groups than just oxygen atoms at each end of the axis (while still remaining in a 
diagonal basis) $\Bigupalpha_x \neq \Bigupalpha_y$, giving $-\Bigupalpha_{x}\,\mbox{sin}^2\theta_{\sss\sp}\,\mbox{cos}^2\phi_{\sss\sp} - 
\Bigupalpha_{y}\,\mbox{sin}^2\theta_{\sss\sp}\,\mbox{cos}^2\phi_{\sss\sp} - \Bigupalpha_{z}\,\mbox{cos}^2\theta_{\sss\sp}$ which is the general case of the second form of (\ref{39}).  
Moreover, in non-diagonal bases, the off-diagonal elements form extra terms directly analogous to those in (\ref{39}).  
Thus there is an extended analogy between this problem and the study of polarization, with 4-stop metroland's Jacobi--Hooke coefficients 
forming a configuration space-indexed analogue of the spatial-indexed polarizability tensor.

\subsection{Extension to QM solutions for pure-shape $N$-stop metroland}\label{Q-ERPM1}

\noindent Here the Laplace/conformal ordered TISE is, with HO-like potentials  
\beq
\frac{1}{\prod_{j = 1}^{A - 1}\mbox{sin}^2\theta_j \mbox{sin}^{n - 1 - A}\theta_{A}}
\frac{\pa}{\pa\theta_{A}}
\left\{
\mbox{sin}^{n - 1 - A}\theta_A \frac{\pa\Psi}{\pa\theta_A}
\right\}
= - \tte\Psi + \sum\mbox{}_{\barp = 1}^{n}\ttK_{\barp}\mn_{\barp}^2\Psi  
\eeq
for $\tte =: 2\fE/\hbar^2$, $\ttK_{\barp} = 2K_{\barp}/\hbar^2$ 
There is also a highly special constant potential case within the multiple harmonic oscillator like potentials.  
This is now a more complicated sequence of associated Gegenbauer equations as explained in Appendix C; 
these are nevertheless also fairly standard and well-documented \cite{AS, GrRy}) and then study perturbations about this.  
Then, if the associated quantum number is not zero, one does not get the Gegenbauer pairings or the right weights straight away.  
Computation of first order perturbations in this case requires the Gegenbauer parameter converting recurrence relation $(\ref{GegRec2})$ 
as well as the polynomial order reducing recurrence relation (\ref{GegRec1}).  
Thus, the calculation is somewhat more complicated in this case.  
%

For constant potential, the TISE is an equation of form $\triangle\Psi = \Lambda\Psi$.  
Then the separation ansatz $\Psi = \prod_{\barp = 1}^{n - 1}\psi_{\barp}(\theta_{\barp})$ yields the simple harmonic motion equation for $\theta_{n - 1}$ 
and $n$ -- 2 equations of form 
\beq
\{1 - X^2_{n - p}\} \frac{\d^2\psi_{n - p}}{\d X_{n - p}^2} - 
\{p - 1\}  X_{n - p}\frac{\d\psi_{n - p}}{\d X_{n - p}} + 
\mj_{p - 1}\{\mj_{p - 1} + p - 2\}\psi_{n - p} - 
\frac{\mj_{p - 2}\{\mj_{p - 2} + p - 3\}}{1 - X^2_{n - p}}\psi_{n - p} = 0
\eeq
under the transformations $X_{\hat{p}} = \mbox{cos}\theta_{\hat{p}}$, $\hat{p} =  1$ to $n$ -- 2.  
These are associated Gegenbauer equations (\ref{AssocGeg}) with parameter $\lambda_{p} = \{p - 2\}/2$, where $j_{p - 1}, j_{p - 2} \in 
\mathbb{N}_0$ are picked out as eigenvalues, so that the \{$n$ -- $p$\}th equation is solved by $C^{\sj_{\p - 2}}_{\sj_{p - 1}}(\mbox{cos}\theta_{n - p}; \{p - 2\}/2)$.  

Then one gets a sequence of quantum numbers taking the ranges $\mj_{p} \in \mathbb{Z}$ such that 
$|\mm_{p}| \leq |\mm_{p - 1}|$ for $\bar{p} =$ 1 to $n$ -- 2 and $\mj_{n - 1} \in \mathbb{N}_0$ 
(a straightforward generalization of the restriction on the two angular momentum quantum numbers in 3-$d$).   

Notation: Top ... $\d$ are a series of $N$ -- 2 relative distance momentum quantum numbers subject to the usual 
kind of restrictions (and which are very well-known for this Article's $N$ = 3 and $N$ = 4 examples).  
I use Top for the top relative distance  momentum quantum number (which is d itself for $N$ = 3 and D for $N$ = 4).

The most special of these in each case has a constant potential and thus gives ultraspherical geodesics 
classically and the ultraspherical rigid rotor quantum-mechanically (solved by ultraspherical harmonics \cite{08II}). 
On the other hand, the next most special of these in each case has \{$N$ -- 1\}-$d$ isotropic harmonic oscillator 
mathematics in its near-polar regime (solved by a power times a Gaussian times an associated Laguerre polynomial).
Establishing a perturbative regime about each most special problem would then appear to be possible e.g. \cite{08II} by recurrence relations of the Gegenbauer polynomials \cite{AS,GrRy}.  
One technical difference is that, if one does use conformal operator ordering, then one can no longer use the configuration space being 2-$d$ to evoke collapse to Laplacian ordering. 
However, hyperspheres are of constant curvature and so of constant Ricci scalar curvature.  
Thus, $\xi\,Ric_{\mathbb{S}^k}$ is just a constant, $\xi\,k\,\{k - 1\}$ (our $\ttE$ has no nonconstant prefactors PPSCT representation having the unit sphere as its configuration space). 
So, even in this case, the sole difference between Laplace and conformal ordering (or any other member 
of the $\triangle^{\xi}$) family of operators) is in what is to be interpreted to be the zero of the energy.  
We also note that for $N$ = 5 the analogy with the Halliwell--Hawking scheme is somewhat tighter, as both involve perturbative expansions in $\mathbb{S}^3$ ultraspherical harmonics.  
Finally, the next most special equation unapproximated can also be mapped to the spheroidal equation, 
so that the fairly standard Mathematical Physics of that equation continues to be of aid in $N$-stop metroland.

\subsection{Quantum pure-shape triangleland}\label{QTri}

As Sec \ref{Q-Geom} makes amply clear this amounts to a different reinterpretation of the mathematics of the sphere to the preceding SSecs
The triangleland TISE (\ref{spheTISE}) is separable under the separation ans\"{a}tze [for (a) or [a] labels, dropped] 
\beq
\Psi({\cal R}, \Phi) = \zeta({\cal R})\eta(\Phi) \mbox{ } \mbox{ or } \mbox{ } 
\Psi(\Theta, \Phi) = \xi(\Theta)\eta(\Phi) \mbox{ } ,
\eeq 
and in each case one obtains simple harmonic motion solved by 
\beq
\eta = \mbox{exp}(\pm i\mj\Phi)
\label{SHM}
\eeq 
for an integer j in an (a) basis, or s in an [a] basis.  
For now we consider the former case, in which j is a relative angular momentum quantum number corresponding to $\sfJ$ being classically conserved.  
This is in analogy with how there is an angular momentum quantum number m in the central-potential 
case of ordinary QM corresponding to the angular momentum $\scL_z$ being classically conserved.   
The accompanying separated-out equation is, in the tilded PPSCT representation in \{${\cal R}$, $\Phi$\} coordinates, the radial equation 
\beq
{\cal R}^2\zeta_{,{\cal RR}} + {\cal R}\zeta_{,{\cal R}} - 
\{2{\cal R}^2\{\widetilde{\ttV}({\cal R}) - \widetilde{\ttE}({\cal R})\}/{\hbar^2} + \mj^2\}
\zeta = 0 \mbox{ } ,  
\label{squib}
\eeq
or, in the barred conformal representation in ($\Theta$, $\Phi$) coordinates, the azimuthal equation 
\beq
\{\mbox{sin}\Theta\}^{-1}\{\mbox{sin}\Theta\xi_{,\Theta}\}_{,\Theta} - 
\{2\overline{\ttV}(\Theta) - \overline{\ttE}\}/\hbar^2 + \mj^2\{\mbox{sin}\Theta\}^{-2}\}\xi = 0  \mbox{ } .  
\eeq
The conformal ordering coincides with the Laplacian one here again because the configuration space dimension $k = 2$.

The TISE for the HO-like potential problem on triangleland is then 
\beq
\widehat{\sfT\sfO\sfT}\Psi = \{a + b\,\widehat{dra}_z\ + c\,\widehat{dra}\mbox{}_x\}\Psi \mbox{ } ,
\eeq
i.e. 
\beq 
\{\mbox{sin}\,\Theta\}^{-1}\{\mbox{sin}\,\Theta\Psi_{,\Theta}\}_{,\Theta} + 
\{\mbox{sin}\,\Theta\}^{-2}\Psi_{,\Phi\Phi} = 
\{a + b\,\mbox{cos}\,\Theta + c\,\mbox{sin}\,\Theta\,\mbox{cos}\,\Phi\}\Psi  \mbox{ }   
\label{spheTISE3}
\eeq
for $a := 2\{A - \fE/4\}/\hbar^2$, $b := 2B/\hbar^2$ and $c := 2C/\hbar^2$.

\subsubsection{Solution in the very special case}\label{VeSiCa}

The very special case $b = c = 0$ then has a potential that balances out to be constant.  
Recollect from Sec \ref{Cl-Soln} that this is mathematically equivalent to the linear rigid rotor, 
for which the Hamiltonian is $\hat{\scL}_{\sT\so\st}$ up to multiplicative and additive constants.  
Thus, effectively, this and $\hat{\scL}_3$ form a complete set of commuting operators whose 
eigenvalues and eigenfunctions are the well-known spherical harmonics, $\mY_{\sll\sm}$.  
Moreover, these also occur as a separated-out part of the corresponding scaled relational particle model problem.

Note that this theoretical possibility for simple mathematics has been known for some time \cite{ACG86}. 
However the idea of using it for toy modelling GR Quantum Cosmology is new (and the simplicity does not match the distinct situation in Molecular Physics itself).   
However, the RPM `rigid rotor' is in configuration space rather than in space.  
Furthermore, it has total relative rational momentum $\widehat{\sfT\sfO\sfT} = \sum_{\Gamma = 1}^3\widehat{\sfS}_{\Gamma}\mbox{}^2$ 
in place of total angular momentum and projected shape momentum $\widehat\sfS_{3}$ in place of magnetic angular momentum.  
These then have eigenvalues $\hbar^2\mS\{\mS + 1\}$ and $\hbar\ms$ respectively; Serna and I therefore 
term $\mS$ and $\ms$ {\it total} and {\it projected `shape momentum quantum numbers'}. 
Our very special problem's TISE thus separates into simple harmonic motion and the associated Legendre equation 
(in $Z = \mbox{cos}\,\Theta$ for any axis system A) that constitute the spherical harmonics equations.   
Thus its solutions are 
\beq
\Psi_{\sS\,\sss}(\Theta, \Phi) \propto \mY_{\sS\sss}(\Theta, \Phi)
\propto \mP_{\sS}^{\sss}(\mbox{cos}\,\Theta)\mbox{exp}(\pm i\ms\,\Phi) \mbox{ } 
\eeq
for $\mP_{\sS}^{\sss}(Z)$ the associated Legendre functions in $Z$, while S $\in \mathbb{N}_0$ and s is such that $|\ms| \leq \mS$.  
Also, $\mS\{\mS + 1\} = - a$, which here signifies that 
\beq
\ttE = 2\hbar^2\mS\{\mS + 1\} + K_2
\label{mabel}
\eeq  
is required of the model universe's energy and inter-cluster effective spring in order for there to be any quantum solutions.   
If this is the case, there are then 2S + 1 solutions labelled by s 
(the preceding sentence cuts down on a given system's solution space, though the more usual larger solution space still exists in the `multiverse' sense).  
From (\ref{mabel}) then, for there to be any chance of solutions one needs $\ttE \geq \{K_1 + K_2\}/4$, so for HO-type models, $\ttE > 0$ is indispensable. 
If there is a S and it is not zero, there are various degenerate solutions corresponding to different values of s.
These correspond to states of different relative angular momentum between the \{2, 3\} and the 1 subsystems.

\noindent I consider various significant bases of orbitals corresponding to the different choices of axis systems in Sec \ref{TessiTri}.  
In particular, the [a] bases with their equilateral principal axis (in the 4 $\times$ $area$ direction) are natural for the very special case.  
E.g. these are well-adapted to questions about equilateral and collinear configurations.   
On the other hand, in the special case with potential $B\,\mbox{cos}\,\Theta^{(1)}$, the (1)-clustering's 
D-M axis [corresponding to $ellip$(1)] is picked out as the sole remaining axis of symmetry for the problem. 
In this case, one would usually prefer to be the principal axis, corresponding to using the (1)-basis (as in \cite{08II}).  
Each (a) basis is associated with the pure relative angular momentum quantity $\sfJ_{(\sa)} = -i\hbar \pa_{\Phi_{(\ta)}}$ for $\Phi_{(\sa)}$ 
a rightness operator, and as such each has a projected pure relative angular angular momentum quantum number $\mj_{(\sa)}$.
In contrast, the [a] bases have `mixed relative distance and relative angular momentum' rational momentum quantity $\hat\sfS_{2} = 
-i\hbar\pa_{\Phi_{[\ta]}}$ for $\Phi_{[\sa]}$ an isoscelesness/regularity quantity about the E$\bar{\mE}$ axis.  
Moreover, this is clustering-independent, so I denote the projected shape quantum number in each case by an unadorned s.      
In this sense these readily afford an extension to $[\gamma]$ bases, for which the quantum number is {\sl still} the same s.
[This permits alignment with the axis picked out by the general case's $B, C \neq 0$ potential via setting $\gamma = \mbox{arctan}(C/B)$ as per Sec \ref{Tri-Norm}.] 
Finally, one can also define a $(\gamma)$ basis by exchanging `y' and `z' axis designations around; in this case one is as best adapted to the potential 
as is possible, but one has a different $\ms_{(\gamma)}$ per $\gamma$ as per Sec \ref{Dyn1}.

In polar coordinate charts, one should not trust the immediate vicinity of the polar axes as the charts go bad there (the angle around the pole becomes undefined).   
Use `cartesianization' near there, e.g. the corresponding stereographic chart `cartesianized'.
E.g. the ground-state probability density function sin$\,\Theta$  gets sent to ${\cal R}/\big\{1 + {\cal R}\mbox{}^2\big\}\mbox{}^2$ and then to 
$1/\big\{1 + X\mbox{}^2 + Y\mbox{}^2\}\big\}\mbox{}^2$

One is then to interpret the familiar orbital basis with respect to the axis system in use as follows.  
I superpose the triangleland tessellation in order to interpret each orbital in terms of triangleland mechanics. 
I compute expectations and spreads of shape operators.

\subsubsection{Wavefunctions in the natural alias equilateral bases [a]}\label{QTri-Ebasis}

The form of the solution is here 
\beq
\Psi_{\sS\sss}\left(\Theta_{[\sa]}, \Phi_{[\sa]}\right) \propto 
\mP_{\sS}^{\sss}(\mbox{4 $\times$ $area$}){\cal T}_{\sss}
\left(
 ellip\mbox{(a)}/\sqrt{1 - \mbox{4 $\times$ $area$}^2}
\right) \mbox{ } .
\eeq
{\begin{figure}[ht]
\centering
\includegraphics[width=1.02\textwidth]{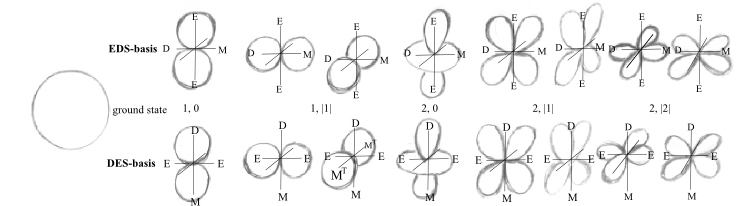}
\caption[Text der im Bilderverzeichnis auftaucht]{\footnotesize{Wavefunctions for triangleland in the EDS alias [ ] and DES alias ( ) bases.}} \label{Y-Tri}\end{figure} } 
%
\subsubsection{Wavefunctions in the clustering-following alias collinear bases (a)}\label{QTri-Dbasis}

The form of the solution is here 
\beq
\Psi_{\sS\sj_{(\ta)}}(\Theta_{(\sa)}, \Phi_{(\sa)}) \propto 
\mP_{\sS}^{\sj_{(\ta)}}(ellip(\ma)){\cal T}_{\sj_{(\ta)}}
\left(
aniso(\ma)/\sqrt{1 - ellip(\ma)^2}
\right) \mbox{ } .   
\eeq

\subsubsection{Wavefunctions in collinear $(\gamma)$ bases}\label{QTri-ColliGamma}

Using the (1)-axis as the axis $\gamma$ is defined about and spherical trigonometry, 
\beq
\Psi_{\sS\sss_{(\gamma)}} \propto 
\mP_{\sS}^{\sss_{(\gamma)}}(ellip(1) \mbox{cos}\,\gamma + aniso(1) \, \mbox{sin}\,\gamma)
{\cal T}_{\sss_{(\gamma)}}
\left( 
\frac{ aniso(1) \mbox{sin}\,\gamma - ellip(1) \mbox{cos}\,\gamma }
{\sqrt{1 - \{aniso(1)\,\mbox{sin}\,\gamma\, + ellip(1) \, \mbox{cos}\,\gamma \}^2}}
\right) \mbox{ } . 
\eeq
Interpretation in terms of mass-weighted space quantities is quite complicated (but can be read off the tessellation).
This basis remains useful for the investigation of the general problem with $B$, $C \neq 0.$

\subsubsection{Wavefunctions in equilateral $[\gamma]$ bases}\label{QTri-EGamma}

By $\Phi_{[\gamma]} = \Phi - \gamma$ and 2-angle formulae, the solutions now take the form
\beq
\Psi_{\sS\sss} \propto \mP_{\sS}^{\sss}(\mbox{4 $\times$ $area$})
\left\{
         {\cal T}_{\sss}
\left(
ellip(1)/\sqrt{1 - \mbox{4 $\times$ $area$}^2}
\right)
\mbox{cos}\,\ms\gamma + \overline{\cal T}_{\sss}
\left(
ellip(1)/\sqrt{1 - \mbox{4 $\times$ $area$}^2}
\right)
\mbox{sin}\,\ms\gamma
\right\} \mbox{ } ,   
\eeq
where $\overline{\cal T}_{\sss}$ of the cosine solution = ${\cal T}_{\sss}$ of the sine solution and 
      $\overline{\cal T}_{\sss}$ of the sine solution  = $-{\cal T}_{\sss}$ of the cosine solution.  
Again, interpretation is complicated, though this basis is again well-adapted to questions concerning near-equilaterality and near-collinearity.
The projected shape quantum number here is as in Sec \ref{VeSiCa}.

\subsubsection{Expectations and spreads of shape operators in the very special case}\label{QTri-Spread}

Here, 
$
\langle \mS \, \mj_{(\sa)}\,|\,\widehat{dra}\mbox{}^{\Gamma}\mbox{}_{(\sa)}\,|\,\mS\,\mj_{(\sa)}\rangle = 0
$, 
i.e. 
\beq
\langle\mS\,\mj_{(\sa)}\,|\,\widehat{aniso(\ma)} \,|\,\mS\,\mj_{(\sa)}\rangle = 
\langle\mS\,\mj_{(\sa)}\,|\,\widehat{\mbox{4 $\times$ $area$}}\,|\,\mS\,\mj_{(\sa)}\rangle = 
\langle\mS\,\mj_{(\sa)}\,|\,\widehat{ellip(\ma)} \,|\,\mS\,\mj_{(\sa)}\rangle = 0 \mbox{ } .  
\label{105}
\eeq
This just means that there is orientation symmetry so for each positive contribution there is a corresponding negative one.  
The useful information starts with the spreads in the relative size shapes 
\beq
\Delta_{\sS\,\sj_{(\ta)}}(\widehat{ellip(\ma)}) = \sqrt{\frac{2\{\mS\{\mS + 1\} - {\mj_{(\sa)}}^2\} - 1}{\{2\mS - 1\}\{2\mS + 3\}}} \mbox{ } , 
\label{110}
\eeq
\beq
\Delta_{\sS\,\sj_{(\ta)}}(\widehat{\mbox{4 $\times$ $area$}}) = 
\sqrt{\frac{\{\mS\{\mS + 1\} + {\mj_{(\sa)}}^2\} - 1}
             {\{2\mS - 1\}\{2\mS + 3\}}                Q_{2}(\mj_{(\sa)})} 
\mbox{ } , \mbox{ }
\Delta_{\sS\,\sj_{(\ta)}}(\widehat{aniso(\ma)}) = 
\sqrt{\frac{\{\mS\{\mS + 1\} + {\mj_{(\sa)}}^2\} - 1}
           {\{2\mS - 1\}\{2\mS + 3\}}                  Q_{1}(\mj_{(\sa)})}
\label{310}
\eeq
for 
$Q_2(\mj_{(\sa)}) =      1/2 \mbox{ for } \mj_{(\sa)} \mbox{ cosine solution, } \mbox{ }  
                         3/2 \mbox{ for } \mj_{(\sa)} \mbox{ sine solution, } \mbox{ }
                         1   \mbox{ otherwise}$, and $Q_1(\mj_{(\sa)})$ the sin $\longleftrightarrow$ cos of this.   
One can then check that indeed $\langle \mS\,\mj_{(\sa)}\,|\,\sum_{k = 1}^3\widehat{dra}\mbox{}^k_{(\sa)}\mbox{}^2\,
|\,\mS\,\mj_{(\sa)}\rangle = 1$, and by (\ref{libellula}) also obtain  $\langle\mS\,\mj_{(\sa)}\,|\,\widehat{sharp(\ma)}\,|\mS\,\mj_{(\sa)}\rangle$ = 
1/2 = $\langle\mS\,\mj_{(\sa)}\,|\,\widehat{flat(\ma)}\,|\mS\,\mj_{(\sa)}\rangle$ 
and $\Delta_{\sS\,\sj_{(\ta)}}(\widehat{sharp(\ma)}) =  \Delta_{\sS\,\sj_{(\ta)}}(\widehat{flat(\ma)}) = \Delta_{\sS\,\sj_{(\ta)}}(\widehat{ellip(\ma)})/2$.

The ground state and very low quantum number states are of interest since I subsequently calculate the nontrivial potential counterparts of the results for these.   
The ground state spreads in $ellip$(a) and $aniso$(a) are $1/\sqrt{3}$ each.

Large quantum number limits are also of interest. 
The spread in $\widehat{ellip(\ma)}$ goes as $1/\sqrt{2\mS}$ for j$_{(\sa)}$ maximal and S large, and as $1/\sqrt{2}$ for j$_{(\sa)}$ = 0 and S large.  
The former amounts to recovery of the equatorial classical geodesic as the limit of an ever-thinner 
belt in the limit of large maximal relative angular momentum quantum number $|\mj_{(\sa)}| = \mS$ 
(This is traversed in either direction according to the sign of j$_{(\sa)}$). 
The latter amounts to the $s$, $p^{(\sa)}_{dra_z}$, $d^{(\sa)}_{dra_z\mbox{}^2}$ ... sequence of orbitals not getting much narrower as 
S increases, so that there is always limited concentration on sharp or flat triangles.

For the isoscelesness/regularness relative angle shape operator $\hat{\Phi}_{(\sa)}$ gives, that, by factorization and cancellation of the $\Theta$-integrals, 
the R0 states obey the uniform distribution over 0 to $2\pi$ i.e. with mean $\pi$ and variance $\pi^2/3$.  
Next, $\langle\mS\,\mj_{(\sa)}\,|\,\hat{\Phi}_{(\sa)}\,|\,\mS\,\mj_{(\sa)}\rangle$ is also $\pi$ and cosine and sine states have, respectively 
\beq 
\Delta_{\sS\,\sj_{(\ta)}}(\hat{\Phi}_{(\sa)}) = \sqrt{\pi^2/3 + 1/2{\mj_{(\sa)}}^2}  \mbox{ }  \mbox{ } , \mbox{ } \mbox{ }
\Delta_{\sS\,\sj_{(\ta)}}(\hat{\Phi}_{(\sa)}) = \sqrt{\pi^2/3 - 1/2{\mj_{(\sa)}}^2}  \mbox{ } , 
\eeq 
which indicate some resemblance to the uniform distribution coming about for large $\mj_{(\sa)}$ (mean and variance do not see the multimodality.  
But at least, by inspection along the lines of the preceding subsection, it is {\sl regular} multimodality for $\mj_{(\sa)}$ maximal -- 
equatorial flowers of 2S petals that gets more and more equatorially flat as S gets larger, thus tending to the equatorial great circle classical path).

For 4 $\times$ $area$ (which is mixed ratio/relative angle information, but nevertheless interesting in its own right).  
$\Delta(\widehat{\mbox{4 $\times$ $area$}})$ goes as $1/\sqrt{2}$ for large quantum number with 
j$_{(\sa)}$ maximal and as 1/2 for large quantum number with j$_{(\sa)}$ = 0.  
That the wavefunctions have increasingly many peaks and valleys does not register unto this overall spread quantifier.  
This also means that there are no limiting collinear states in this basis.  
This $1/\sqrt{2}$ is the largest value it takes, while the smallest, $1/\sqrt{5}$, occurs for the S = 1, j = 0.  
For the ground state, $\Delta_{0\,0} (\widehat{\mbox{4 $\times$ $area$}})$ = $1/\sqrt{3}$.

\subsubsection{Shape operators acting on [a] bases' wavefunctions}\label{QTri-ShapeOps}

In the natural or equilateral basis, one has the 4 $\times$ $area$ $\longleftrightarrow$ $ellip$(a) and 
$\mj_{(\sa)} \rightarrow \ms$ counterpart of eq's (\ref{105}, \ref{110}, \ref{310}).  
Thus in this case one does have a collinear limit for large quantum numbers, e.g. in the case of s 
maximal, $\Delta_{\sS\,\sss}(\widehat{\mbox{4 $\times$ $area$}}) \approx 1/\sqrt{2\mS} \longrightarrow 0$ for s maximal and S large. 
On the other hand, it tends to  $1/\sqrt{2}$ for s = 0 and S large.

\subsubsection{Small asymptotics solutions for the `special multiple HO'}\label{QTri-Small-As}

I work in the tilded PPSCT representation. 
In the special case, note that these equations are self-dual in the sense of \cite{08I}, so that, again, 
direct study of only one of the two asymptotic regimes is necessary and then the other can be read off by mere substitutions.    
This and the next 2 SSecs are interpreted in terms of the (1)-basis.

\subsubsection{Second-order asymptotics}\label{QTri-2ndAs}

The special case has the same mathematical form as the fairly well-known problem of the linear rigid rotor 
in a background homogeneous electric field in the symmetry-adapted basis for that problem.
This then at the quantum level amounts to the present mathematics being analogous to that of the Stark effect for 
the linear rigid rotor, which is well-documented (see e.g. \cite{TS, Hecht, Messiah}.   
Near the \{23\} double collision, to second approximation, this problem gives \cite{08II} 2-$d$ isotropic harmonic oscillator wavefunctions \cite{Messiah, Schwinger, Robinett}. 
(However, it is not the same inner product in detail, due to the curved geometry).
I work in the ${(1)}$-basis that is adapted to this potential, but suppress the (1) labels.    
I require a quantum-mechanically sizeable classical `frequency', $\omega$. 
If the bulk of the wavefunction is to lie where the small-angle approximation holds reasonably.
(This is, dimensionally frequency $\times \mI$ in this Article's formulation.  
$\mI\,\omega/\hbar$ of the order of $10^3$ begins to work well.)
Then the `energies' are
$
\ttE - {K_2}/{2} = \mn\,\hbar\,\omega 
\mbox{ } \mbox{ for } \mbox{ } 
\mn := 1 + 2\mN + |\mj| \mbox{ } .  
$
However, $\omega$ itself depends on the shifted energy, $\ttE^{\prime} := \ttE - A - B$, so 
\beq
\ttE^{\prime} = 2 \big\{\hbar^2\,\mn^2 + \mn\,\hbar\sqrt{\mn^2\hbar^2  - B}\big\}  \mbox{ } , 
\eeq
which, for $\mn\,\hbar/\omega$ small, e.g. $\mn << 10^3$ for the solutions below, goes as $\ttE^{\prime} \approx \mn\,\hbar\,\Omega$ for $\Omega = 2\sqrt{-B}$.)  
The solutions go like 
\beq
\Psi_{\sN\sj}(\Theta, \Phi) \propto {\Theta}\mbox{}^{|\sj|} \left\{    1 + |\mj|\Theta\mbox{}^2/{12}    \right\}    \mbox{exp}\left(- \omega\Theta\mbox{}^2/{8\hbar}\right)
\mL_{\sN}^{|\sj|}\left(    {\omega\Theta\mbox{}^2}/{4\hbar}    \right)                                              \mbox{exp}(\pm i\mj\Phi)
\label{gi}
\eeq
where $\mL_{a}^b(\xi)$ is the associated Laguerre polynomials in $\xi$.  
[As regards interpretation of the $\Phi$ factor, a previous Sec has shown how to rewrite this terms of $aniso$ and $ellip$,    
while the $\Theta$ factor is now a somewhat more complicated function of $ellip$.]  

Again, j = 0 are axisymmetric and as such are totally undiscerning of isoscelesness/collinearity.  
\noindent
For the following results, Fig \ref{Tulips} re-tessellated with the triangleland sphere with DD $\longrightarrow$ D provides the visualization.  
{\mN} = 0, j = 0 favours sharp triangles.  
\noindent
Then {\mN} = 0, j = $\pm 1$ are a degenerate doublet of `bat ears' with the cosine solution favours sharp triangles that are approximately collinear and 
                                                                        the   sine solution favours sharp triangles that are approximately isosceles.
\noindent Next, there is a degenerate triplet.
{\mN} = 1, j = 0 favours two separate bands: the even sharper triangles and the fairly sharp triangles.  
The other two solutions are `tulips of four petals' restrictions on the sharp triangles: the {\mN} = 0, $|\mj| = 2$ cosine solution favours those that are either approximately 
collinear or approximately isosceles, while the {\mN} = 0, $|\mj|$ = 2 sine solution favours those that are neither approximately collinear nor approximately isosceles.

\noindent Note 1) The {\mN} = 0 states have peaks at $\Theta \approx 2\sqrt{\sfJ/\omega}$; c.f. the means in the next SSSec.  

\noindent Note 2) For the opposite sign of $B$, the approximate wavefunctions about the M-pole have the same meaning except that one uses `flat' in place of `sharp'.

\subsubsection{Shape operators acting on $\Theta$ near-polar wavefunctions}\label{QTri-ShapeOps+}

\noindent In the small regime, 4 $\times$ $area$ and 
$aniso$ still have zero expectation, as either sign of these are equally probable.  
For N, $\mj$ substantially smaller than $\omega/\hbar$ so that powers of the latter dominate powers of the former. 
(As $\omega/\hbar$ is considered to be large, this does cover the previous subsection's states.) 
Then the following results hold. 
(This subsection's results come from the orthogonality of, and a recurrence relation for, the Laguerre polynomials provided in an Appendix of \cite{AF}.)
\beq
\langle \mN\,\mj\,|\,\widehat{ellip}\,|\,\mN\,\mj\rangle \approx 
1 - {  \mn\hbar  }/{2\omega}
\eeq
\beq
\Delta_{\sN\,\sj}(\widehat{\mbox{4 $\times$ $area$}}) \approx 
\sqrt{         {\mn\,\hbar\, Q_2(\mj)}/{2\omega}       }
\mbox{ } \mbox{ } , \mbox{ } \mbox{ } 
\Delta_{\sN\,\sj}(\widehat{aniso})   \approx 
\sqrt{        \mn\,\hbar\, Q_1(\mj)/{2\omega}           }
\mbox{ } , 
\eeq
while the spread in ellip goes as $\hbar/\mI\omega$ though the scheme cannot really evaluate its coefficient. 
(The first two orders cancel, and the differential equation that was solved in the first place was only accurate to the first two orders.)

As some idea of what unsigned area is typical, for the equal-masses ground state, this is 
\beq
\Delta_{\sN\,\sj}(physical\mbox{ }area) = \frac{3\mI^2}{4m^2}{\frac{\hbar}{2\omega}} \mbox{ } .
\eeq  
which has an interpretation that bears some parallel to the Bohr radius. 
(This includes the map between the isotropic harmonic oscillator and hydrogen in \cite{Schwinger}.) 
It is a `minimal quantifier' of $area$, albeit of spread of $area$ and in a basis centred on a relative angular momentum interpretable axis.  
Note the change in spread of 4 $\times$ $area$ due to the confining effect of the potential. 
The wide range of areas correspond to a spread of $1/\sqrt{3}$ (where the range is --2 to 2) to 
$\sqrt{\hbar/2\omega}$ for $\omega/\hbar$ large, which is smaller by a factor of $\sqrt{3\hbar/2\omega}$.

Comparison of mean angle (e.g. roughly from the expectation of cos$\,\Theta$) and mode angle (from graphs along the lines of those in \cite{08II}) 
reveals the mean to be larger than the mode, but by not quite as much as occurs radially in hydrogen.  
This reflects that this case's Gaussianity suppresses the mean-shifting tail more than radial hydrogen's mere exponential does.  
Expectations and spreads of $\widehat{\Phi}$ are just like for previous Sec as the $\Theta$-integrals trivially cancel in each case.

\subsubsection{The special triple HO problem treated perturbatively for small $B$}\label{QTri-Pert} 

Let us treat the special triple HO problem as a perturbation about the simple `very special triple HO'.  
For a perturbation $H^{\prime}$ to one's rescaled Hamiltonian [rescaled since the other calligraphic quantities in 
(\ref{p1}, \ref{p2}) are] the first few objects of perturbation theory as used in this Article are as follows \cite{LLQM}. 
For  nondegenerate unperturbed energies
one has counterparts of (\ref{p1}) and (\ref{p2}). 
The key integral underlying time-independent perturbation theory is now 
$
\langle \Psi_{\sS^{\prime}\sj^{\prime}} | \ttV^{\prime} | \Psi_{\sS\sj} \rangle, 
$
%
which has as its nontrivial factor 
$
\int_{-1}^{1}\mP_{\sS^{\prime}}^{\sj^{\prime}}(X)X\mP_{\sS}^{\sj}(X)\d X \mbox{ } , 
$
but there is a recurrence relation (\ref{**}) enabling $X\mP_{\sJ}^{\sj}(X)$ to be turned into a linear combination of $\mP_{\sS^{\prime\prime}}^{\sj^{\prime\prime}}(X)$, 
allowing for evaluation from orthonormality of the associated Legendre functions (\ref{orthog}).  
This calculation parallels that in the derivation of selection rules for electric dipole transitions \cite{AtF, Blum}.  
Then
$
\langle \mS, \mj^{\prime} | b X |\mS, \mj \rangle = 0 
$
since the recurrence relation sends $X\mP_{\sS}^{|\sj|}$ to a sum of $\mP_{\sS^{\prime}}^{|\sj^{\prime}|}$ 
for $\mj^{\prime} \neq \mj$, so each contribution to the integral vanishes by orthogonality.  
Next, for 
$
\langle \mS^{\prime}, \mj^{\prime} | b X |\mS, \mj \rangle
$,   
one similarly needs $\mS^{\prime} =  \mS \pm 1$ and $\mj^{\prime} = \mj$ to avoid it vanishing by orthogonality.  
So two cases survive this `selection rule':   
\beq
\langle \mS + 1, \mj |b X |\mS, \mj \rangle = 
b  \sqrt{\{\{\mS + 1\}^2 - \mj^2\}/{\{2\mS + 1\}\{2\mS + 3\}}}
\label{first} \mbox{ } , 
\eeq
\beq
\langle \mS - 1, \mj |b X |\mS, \mj \rangle = 
\langle \mS, \mj | b X | \mS - 1, \mj\rangle = 
b  \sqrt{\{\mS^2 - \mj^2\}/{\{2\mS - 1\}\{2\mS + 1\}}}  
\eeq
the second just following from the first via using with S -- 1 in place of S (parallelling e.g. \cite{Messiah}).  

The eigenspectrum is thus
\beq
\ttE_{\sS, \sj}^{\{2\}} = 2\hbar^2\mS\{\mS + 1\} + 4A + 
{4B^{2}\{\mS\{\mS + 1\} - 3\mj^2\}}/{\hbar^2\mS\{\mS + 1\}\{2\mS - 1\}\{2\mS + 3\}} + 
O(B^4)  \mbox{ } .  
\label{Diesel}
\eeq
For $\mS = 0$, one needs a separate calculation, which gives 
\beq
\ttE_{0, 0}^{\{2\}} =  4A + {4B^{2}}/{3\hbar^2} + O(B^4)  \mbox{ } .
\label{Diesel2}
\eeq
This calculation can then be checked against its rotor counterpart (originally in \cite{Kronig} and which can also found in e.g. \cite{TS, Hecht, Messiah}).  
The corresponding eigenfunctions can be looked up (e.g. \cite{Proprin}) and reinterpreted in terms of the original problem's mechanical variables.

Bearing this long-term application in mind, note that the following observation in \cite{AF} for 4-stop metroland readily extends to triangleland.
That `halfway-house' overlap integral computation is {\sl common} to building up both the immediate time-independent perturbation theory for 
pure-shape models and to the longer-term time-dependent perturbation theory for scaled models.    
Namely, 

\noindent 1) these overlaps are relevant halfway-house computations for perturbation theory.  

\noindent 2) They are of 3-$\mY$ integral form in terms of what are, mathematically, Wigner 3j symbols' (though for us, physically, they are `3s symbols').  

\noindent 3) That the specific cases relevant to triangleland with harmonic oscillator-like potentials (where the sandwiched $\mY$ is of degree 1) 
are explicitly written out in e.g. \cite{Mizushima}.

\subsubsection{General triple HO pure-shape RPM problem treated perturbatively for small $B$, $C$.}\label{QTri-Pert2}  

A further result (specific to the immediate time-independent perturbation theory of pure-shape triangleland) is the extension of 
\cite{08I}'s special case second order result to the general case by use of rotated bases: 
\beq
\fE_{\sS\,\sss_{(\gamma)}} = 2\hbar^2\mS\{\mS + 1\} + 4A + 
{4\{B^{2} + C^2\}\{\mS\{\mS + 1\} - 3\ms_{(\gamma)}\mbox{}^2\}}/
{\hbar^2\mS\{\mS + 1\}\{2\mS - 1\}\{2\mS + 3\}} + O((B^2 + C^2)^2)  \mbox{ } .  
\label{Diesel3}
\eeq
\mbox{ } \mbox{ } The TISE with general triple HO is harder because the 
angle-dependence of the potential results in nonseparability in these natural coordinates.  
This makes for a useful model of the Semiclassical Approach to the POT \cite{SemiclIII}.

As in \cite{08I}, start again using rotated/normal coordinates at the classical level, or switch to such coordinates at the 
differential equation solving stage, amounting to a choice of basis in which the perturbation is in the (new) axial `$z_{N}$' direction.
While it can be viewed as before in the new rotated/normal coordinates, nevertheless studying the original coordinates' $\Phi$-dependent $\fV$ term remains of interest.  
For, it may well be appropriate for the {\sl original} coordinates to have mechanical attributes or the 
heavy--light subsystem distinction underlying the Semiclassical Approach.\footnote{In 
the laboratory, one might likewise not pick the normal coordinates of the rotor--electric field  if there is e.g. also a magnetic field that picks out a {\sl different} direction.}   
%
Moreover in that case, being able to proceed further in the rotated/normal coordinates can serve as a check on ``standard" procedures in the original coordinates. 
(This is along the lines suggested in \cite{SemiclI}.)

There is now an issue in the projection that there is an additional factor due to the change in area in moving between each patch of sphere and each corresponding patch of plane.  
Namely, the probability density function on the sphere is $\mbox{sin}\,\Theta|\Psi(\Theta, \Phi)|^2$. 
Its  $\mbox{sin}\,\Theta$ part pertains to the sphere itself.  
The probability density function on the stereographic plane is $|\Psi({\cal R}, \Phi)|^2{\cal R}/\{1 + 
{\cal R}^2\}$ of which the ${\cal R}/\{1 + {\cal R}^2\}$ pertains to the stereographic plane itself.  
Unlike doing the very special case in \cite{08II}, I have already looked at that in multiple bases, and so go for the more interesting general multiple HO case.

\subsubsection{Molecular Physics analogies for triangleland} \label{Molecular}

The following are available in the rotor literature (in the spherical presentation) and could thus be straightforwardly transcribed to the 
various presentations for this Article's pure-shape RPM models.   

\noindent 1) For triangleland, one of the main results of \cite{08II} is that the HO-like potential problem on triangleland 
has the same mathematics as the Stark effect for the linear rigid rotor (see e.g. \cite{TS, Hecht, Messiah}).  
In particular, the special case corresponds to the homogeneous electric field pointing in the z-direction. 
The general case to it pointing in the general direction in the zx-plane.  
The very special case is, by this analogy, the undisturbed linear rigid rotor itself.  
This allows for checks of the various regimes ($B$ small, $B$ large, $\Theta_{(1)}$ near-polar 
alongside further knowledge of how to patch together such regimes (see \cite{08II} for a detailed reference list).  
That a 2-$d$ isotropic HO resides within the triangleland RPM multiple HO like problem as a limit 
problem has counterpart in the literature on the linear rotor (\cite{PS57}, see also the figure in \cite{Meyenn}).

\noindent 2) Higher order corrections in $B$ are in the literature for the rotor \cite{Hughes, RPS} and so can be 
transcribed into the relational context [e.g. I was able to write $O(B^4)$ and not $O(B^3)$ in (\ref{Diesel}) due to this].  
Alternative variational methods (using the Hellmann--Feynman and Hypervirial Theorems) appear in \cite{RPS}. 
These cover higher order terms too, and can be used to show generally that only even powers of $B$ occur.  
The calculation for the large $B$ regime has both been done \cite{Proprin} and matched to small $B$ regime calculations. 

\noindent 3) Numerical evaluation of eigenvalues was been done \cite{Hughes, KuschHughes, Shirley, PS57} by the methods of \cite{Lamb, Meyenn, RPS}.  

\noindent 4) The rotor literature also indicates how the $C$ perturbation can be transformed away with a new rotated 
choice of coordinates in which the mathematics is again that of a $B$ type perturbation. 
[This is the quantum application of the normal coordinates trick.]  
While in the laboratory with a rotor one could choose one's axial `$z$' direction to be in the most 
convenient direction, there can in the new context be various POT strategy modelling reasons not just to stay in 
these coordinates in the SRPM case (e.g. allotting h and l statuses to nonseparable coordinates is required in the Semiclassical Approach).  

\vspace{11in}

\section{Scaled quantum RPM's}\label{Q-ERPM}

\noindent This amounts to a re-interpretation of work in $\mathbb{R}^3$ and $\mathbb{R}^{n}$ that is far more often considered from an absolutist perspective.  
This gives a shape part as in the preceding Sec and an extra scale equation.  
I denote the separation ansatz by
\beq
\Psi = \mG(\rho)\mF(\mbox{Shape}) \mbox{ } .
\label{sssplit}
\eeq

\subsection{Quantum scaled 4-stop metroland}\label{QE4Stop}

I consider an (H2) basis, dropping the labels.

\subsubsection{Simple quantum solutions in the small-shape approximation: common separated-out shape part } \label{QSNStop1}

This gives in general the spherical harmonics equation as per the previous Sec.

\subsubsection{Analogues of $k < 0$ vacuum or wrong-sign radiation}\label{QSNStop11}

As mechanics, these are $E > 0$ and may have an approximation to the conformal potential present (to cancel off $\sfT\sfO\sfT$). 
Overall, they are free models with zero and nonzero top relative distance quantum number respectively. 
They are solved by the infinitely-oscillating spherical Bessel functions (see Appendix C) 
 \beq
\mG_{\sD}(\rho) \propto \rho^{-1/2}\mJ_{\sD + 1/2}(\sqrt{2E}\rho/\hbar) \mbox{ } .   
\eeq

\subsubsection{Analogues of very special HO (= $\Lambda < 0$, $k < 0$ vacuum or wrong-sign radiation)}\label{QSNStop2}

For 4-stop metroland with $V = A\rho^2$,  gives the 3-$d$ isotropic quantum HO problem.
This has energies $\fE = \hbar\omega\{2\mN + \mD + 3/2\} > 0$ for $\mN \in \mbox{ } \mathbb{N}_0$, and 
corresponding scale--shape variables N-node wavefunctions  
\beq
\mG_{\sN\,\sD}(\rho) \propto \rho^{\sD}\mL_{\sN}^{\sD + 1/2}(\omega\rho^2/\hbar)
\mbox{exp}(-\omega\rho^2/2\hbar)
\mbox{ } .
\label{Dins}
\eeq

\subsubsection{Analogues of $k > 0$ dust model}\label{QSNStop3}

As mechanics, these are $E < 0$ Newton--Coulomb potential models.
The 4-stop and thus $\mathbb{R}^3$ case of this is well-known (e.g. \cite{LLQM}): it is the analogue of the $l = 0$ hydrogen model, 
\beq
\mG_{\sN}(\rho) \propto \mL_{\sN - 1}^{1}(2\sqrt{-2\ttE}\rho/\hbar)\mbox{exp}(\sqrt{-2\ttE}\rho/\hbar)
\mbox{ } .  
\label{spring1}
\eeq
The 3-stop ($\mathbb{R}^2$) and $N$-stop ($\mathbb{R}^n$) cases then follow by straightforward generalization.  
(The extension to include $\sfT\sfO\sfT > 0$ is simple too, though it corresponds to wrong-sign radiation, so I only present the $\sfT\sfO\sfT = 0$ case in this Article.)

\subsubsection{Analogues of $k \leq 0$ dust models}\label{QSNStop4}

As mechanics, these are $E \geq 0$ Newton--Coulomb potential models.
These are `ionized states' or `scattering problems' and correspond to open cosmologies 
See e.g. \cite{LLQM, CH} for the maths of the ionized atom.

\subsubsection{Analogues of right and wrong sign radiation models}\label{QSNStop5}

Again, for this the 4-stop metroland's 3-$d$ mathematics is well-known \cite{LLQM}. 
It has three cases, 

\noindent i) $\mD + 1/2 < \sqrt{2R}/\hbar$ for which there is collapse to the maximal collision (ground state with $E = - \infty$), which is relevant since the approximate problem's 
scale part can exchange  energy with the shape problem and thus use this energy exchange to run down to the maximal collision. 

\noindent ii) $\mD + 1/2 > \sqrt{2R}/\hbar$ (for which it is not clear that the cutoff in \cite{LLQM} is meaningful in the case of RPM's). 

\noindent iii) The critical case $\mD + 1/2 = \sqrt{2R}/\hbar$ has $\mG$ diverge no worse than $1/\sqrt{\rho}$ as $\rho \longrightarrow 0$.

\subsubsection{The special quadratic potential in 4-stop metroland}\label{QSNStop22}

In $\rho_1, \rho_2, \rho_3$ variables, it is solved in terms of Hermite polynomials and Gaussians, and this extends to the $\ttB \neq 0$ case as well, which is, likewise, a distortion.  
For energies $E = \sum_{i = 3}^{n}\hbar\sqrt{K_1}\{\mn_1 + 1/2\} + \hbar\sqrt{K_2}\{\mn_2 + 1/2\}$ ($k_i = 
\omega_i/\hbar$ and $\mn_1, \mn_2 \mbox{ } \in \mbox{ } \mathbb{N}_0$, is solved standardly in the associated Cartesian coordinates $\rho_i$, i  = 1 to 3 in Hermite polynomials.  
I then express the arguments of this in terms of the scale variable $\rho$ and the shape variable $\theta$, $\phi$ to obtain 
$$
\Psi_{\sn_1\sn_2}(\rho, \theta, \pi) \propto 
\mH_{\sn_1}\big(\sqrt{{K_1}/{\hbar}}\,\rho\,\mbox{sin}\,\theta\,\mbox{cos}\,\phi\big) 
\mH_{\sn_2}\big(\sqrt{{K_1}/{\hbar}}\,\rho\,\mbox{sin}\,\theta\,\mbox{sin}\,\phi\big)
\mH_{\sn_3}\big(\sqrt{{K_1}/{\hbar}}\,\rho\,\mbox{cos}\,\theta\big)
$$
\beq
\times \mbox{exp}({-\rho^2\{\ttA + \ttB\,\mbox{cos}\,2\theta + \ttC\,\mbox{sin}^2\theta \,\mbox{cos}\,2\phi\}}/{\hbar})  \mbox{ } .  
\eeq
\noindent One can again think of these as box-shaped arrays of peaks and troughs, as an obvious extension \cite{ScaleQM} of Sec \ref{QS3Stop2}.

\subsubsection{Interpretation I: characteristic scales}\label{QSNStop-Char}

$K/\rho$ RPM models have a `Bohr moment of inertia' for the model universe analogue to (atomic Bohr radius)$^2$, $\mI_0 = \rho_0^2$, which, in the gravitational case, goes 
like $\hbar^4/G^2m^5$.  
Then $E = - \hbar^2/2\,\mI_0\mN^2$ in the 4-stop case.
Paralleling Sec \ref{QS3Stop5}, HO RPM models have characteristic $\mI_{\sH\sO} = \hbar/\omega$.  
Then the first of these is limited by the breakdown of the approximations used in its derivation,  while the second of these has no such problems.

\subsubsection{Interpretation II: expectations and spreads in analogy with Molecular Physics}\label{QSNStop-Spreads}

Example 1) the most direct counterpart of the atomic working described in Sec \ref{4Stop-Spread}
is the D = 0 approximate Newtonian gravity or attractive Coulomb problem, which is the $a_0 \longrightarrow \rho_0$ of it.   

\mbox{ } 

\noindent Example 2) as regards the expectation of the size operator for 4-stop metroland, isotropic case in scale--shape coordinates, 
again using the normalization result for associated Laguerre polynomials and Gaussian integral results, $\langle \, 0\,\mD\,\d\,| \widehat{\mbox{Size}} | \,0\,\mD\,\d\, \rangle = 
\frac{2^{2\tD + 2}}{\sqrt{\pi}}
\left(
\stackrel{2\sD + 2}{\mbox{\scriptsize $\sD + 1$}}
\right)^{-1} \rho_{\sH\sO}$. 
Thus e.g. $\langle\,0\,0\,0\,|\widehat{\mbox{Size}}|\,0\,0\,0\,\rangle = \frac{1}{2\sqrt{\pi}}\rho_{\sH\sO}$, while the large-D limit (for $\mN =  0$) is 
$\langle\,0\,\mD\,\d\,|\widehat{\mbox{Size}}|\,0\,\mD\,\d\,\rangle$ $\longrightarrow \sqrt{\mD + 1}\rho_{\sH\sO}$. 
Again, the latter gradually grows to infinity along a sequence of configurations with ever-increasing relative distance momentum.  
Also, $\langle\,1\,0\,0\,|\widehat{\mbox{Size}}|\,1\,0\,0\,\rangle = \frac{3}{\sqrt{\pi}}\rho_{\sH\sO} \approx 1.69\,\rho_{\sH\sO}$.   
In comparison, the mode value for the ground state is $\rho_{\sH\sO}$.

The spreads are again an integral made easy by a recurrence relation on the associated Laguerre polynomials, 

\noindent $\langle\, \mN\,\mD\,\d| \widehat{\mbox{Size}}^2 | \,\mN\,\mD\,\d\, \rangle = \{2\sN + \mD + 3/2\}\rho_{\sH\sO}\mbox{}^2$, minus the square of the expectation.  
Thus $\Delta_{0\,0\,0}(\widehat{\mbox{Size}}) =  \frac{6\pi - 1}{4\pi}\rho_{\sH\sO}\mbox{}^2 \approx 1.42\,\rho_{\sH\sO}\mbox{}^2$. 
On the other hand, the large-$\mD$ limit with $\mN$ = 0 gives $\Delta_{0\,\sD\,\sd}(\widehat{\mbox{Size}}) \longrightarrow \rho_{\sH\sO}\mbox{}^2/{2}$, and 
$\Delta_{1\,0\,0}(\widehat{\mbox{Size}}) = \frac{7\pi - 18}{2\pi}\rho_{\sH\sO} \approx 0.635 \, \rho_{\sH\sO}$.

\subsubsection{Further perturbative treatment} \label{QSNStop-Perts}

The Cosmology--RPM analogy in this Article gives (Sec \ref{Cl-Soln}) multiple power-law potential terms with some of them furthermore admitting sensible 
small-shape expansions, which are of interest through being partly tied to the admission of a long-term-stable semiclassical regime.
The negative power cases are not expected to work well already from the classical analysis: as further 
discussed in Sec \ref{QM-Str}, shape just is not secondary in some parts of configuration space in such cases... 
[Also, there are extra layers of control for HO examples: these can be set up to possess semiclassicality everywhere, 
and one can solve these models exactly even in cases for which they are not separable in the $h$--$l$ split variables.]

Next, I note that the above exact solution work can be perturbed about, with a number of cases of this producing standard mathematics as follows.   
The perturbation integrals split into scale parts and shape parts.  
The scale integrals give $\langle\rho^{\sn}\rangle$ for $\mn$ an integer (taken to be positive if one is to avoid classical instability).  
Such integrals are of the same type as those used in the evaluation of expectations and spreads in earlier Secs. 
The shape integrals for 4-stop metroland involve $\theta \mY_{\sD\sd}(\theta,\phi)\mY_{\sD\sd}
(\theta,\phi)$ and $\phi \mY_{\sD\sd}(\theta,\phi)\mY_{\sD\sd}(\theta,\phi)$.   
The latter is trivial and the former is straightforward, at least state-by-state.  
For 3-stop metroland, these are of the form $\varphi\, \mbox{exp}(i \d \varphi)$, and thus trivial again.  
Instead, in \cite{AF} Franzen and I treated the further HO terms not as a small-shape approximation expansion but rather as an exact perturbation.

\subsection{Quantum scaled $N$-stop metroland}\label{QSNStop44} 

The general quantum equation for this is (\ref{BigUn}).
Then for $V$ of the form $V(\rho)$ (or suitably approximated thus), the equation is separable into scale and shape parts via $\Psi(\rho, \theta_{\barr}) = \mG(\rho)\mY(\theta_{\barr})$.

\subsubsection{The common angular part}\label{QSNStop55}  

This is common to all the $V = V(\rho$) problems (as arise e.g. approximately within the scale dominates shape approximation) and is as per Sec \ref{Q-ERPM1}.  
It is now the hyperspherical harmonics equation.

\subsubsection{The general scale equation}  \label{QSNStop-Gen-Sca}

The above working fixes the scale equation's separation constant to be $-\mT\mo\mp\{\mT\mo\mp + N - 3\}$, so that the scale equation is
\beq
-\hbar^2\{\rho^2\mG_{,\rho\rho} + \{N - 2\}\rho\mG_{,\rho} - 
\mT\mo\mp\{\mT\mo\mp + N - 3\}\mG\} + 2\rho^2V(\rho)\mG = 2E\rho^2\mG \mbox{ } .  
\label{scaly}
\eeq
Moreover, this Article considers (subcases of) $V(\rho) = A\rho^2 - K/\rho - R/\rho^2$ arising from the RPM--cosmology analogy.

\subsubsection{Small approximation to the general scale equation} \label{QSNStop6} 

In this regime, (\ref{scaly}) reduces to 
\beq
\rho^2\mG_{,\rho\rho} + \{N - 2\}\rho\mG_{,\rho} + 
\{2R/\hbar^2 - \mT\mo\mp\{\mT\mo\mp + N - 3\}\}\mG = 0 \mbox{ } ,  
\eeq
which is a Cauchy--Euler equation. 
In absence of the R-term, it is solved by
\beq
\mG = \rho^{\sT\so\sp} \mbox{ } , \mbox{ } \mbox{ } \mG = \rho^{-\sT\so\sp - N + 3} \mbox{ } ,
\label{Rless}
\eeq
of which the latter is discarded due to being divergent at the origin (maximal collision).    
[The above excludes $\mT\mo\mp$ = 0 for \mN = 3 (repeated roots) for which the second solution is $\mbox{ln}\,\rho$ which is likewise discarded for being divergent at the origin.]
With the $R$ term present, there are 3 possible behaviours (paralleling \cite{LLQM}): 

\noindent i) $\mD + \{N - 3\}/2 < \sqrt{2R}/\hbar$ for which there is collapse to the maximal collision (ground state with $E = - \infty$).  

This is relevant since the approximate problem's scale part can exchange energy with the shape problem and thus use this energy exchange to run down to the maximal collision. 

\noindent ii) $\mD + \{N - 3\}/2 > \sqrt{2R}/\hbar$ (for which it is not clear that Landau's cutoff \cite{LLQM} is meaningful in RPM's for $N > 3$).

\noindent iii) The critical case $\mD + \{N - 3\}/2 = \sqrt{2R}/\hbar$ has $\mG$ diverge no worse than $1/{\rho}^{\{3 - N\}/2}$ as $\rho \longrightarrow 0$.

\subsubsection{Large approximation to the general scale equation} \label{QSNStop7} 

Let $\mn$ be the highest power in the potential.
For $\mn > 0$, one gets an approximate large solution of the form
\beq
\mG(\rho) \approx \mbox{exp}
\big(- {2\sqrt{2k}} \rho^{1 + \sn/2}/{\hbar\{\mn + 2\}}\big) 
\mbox{ } .  
\eeq
For $\mn \leq 0$, one gets instead an approximate large solution of the form
\beq
\mG(\rho) \approx  \mbox{exp}(-\sqrt{-2E}\rho/\hbar) \mbox{ } .  
\eeq

\subsubsection{Solution in the free case}  \label{QSNStop8}

The scale part's equation is now
\beq
\rho^2\mG_{,\rho\rho} + \{N - 2\}\rho\mG_{,\rho} + 
\{2E\rho^2/\hbar^2 - \mT\mo\mp\{\mT\mo\mp + N - 3\}\}\mG = 0 \mbox{ } .  
\eeq
For $E > 0$, this is solvable by mapping to the Bessel equation (see Appendix C.2) , giving 
\beq
\mG \propto \rho^{\{3 - N\}/2}\mJ_{\tT\to\tp + \{N - 3\}/{2}}(\sqrt{2E}\rho/\hbar) \mbox{ }   
\eeq
(the `ultraspherical' generalization of well-known spherical Bessel functions).
[For $E < 0$, it is instead a modified Bessel function, which has an unphysical blow-up, while for $E = 0$, one gets (\ref{Rless}) again instead.]

\subsubsection{Solution for the $E$ = 0 single arbitrary power law case}\label{QSNStop9}  

Denote this potential by $k\rho^{\sn}$.
The scale part's equation is now
\beq
\rho^2\mG_{,\rho\rho} + \{N - 2\}\rho\mG_{,\rho} + 
\{2k\rho^{2 + \sn}/\hbar^2 - \mT\mo\mp\{\mT\mo\mp + N - 3\}\}\mG = 0 \mbox{ } ,  
\eeq
which for $k$ negative is also solvable by mapping to the Bessel equation, giving 
\beq
\mG \propto \rho^{\{1 - N\}/2}J_{\{2\tT\to\tp + N - 3\}/\{\sn+ 1\}}
\left({2\sqrt{-2k}}\rho^{\{\sn + 2\}/{2}}/{\{\mn + 2\}\hbar}\right) \mbox{ } .  
\eeq
[for $k$ positive this is instead a modified Bessel function (with an unphysical blow-up), while for $k$ = 0, one has (\ref{Rless}) again].

\subsubsection{Solution for the Milne in AdS analogues} \label{QSNStop10} 

As mechanics, these are HO's. 
The scale equation here has the form of a unit-mass \{$N$ -- 1\}-$d$ isotropic HO (with $\rho$ 
as radial variable),
\beq
- \hbar^2 
\left\{
\rho^2\mG_{,\rho\rho} + \{N - 2\}\rho\mG_{,\rho} - 
\mT\mo\mp\{\mT\mo\mp + N - 3\} \mG\right\} 
+ \omega^2\rho^4\mG = E\rho^2\mG 
\mbox{ } .
\label{NdIso}
\eeq
Then using the large and small approximands as factors, $\mG(\rho) =  \rho^{\tT\to\tp}
\mbox{exp}(-\omega\rho^2/2\hbar) f(\rho)$, and a rescaling, this maps to the generalized Laguerre equation, allowing one to read off the solution 
\beq
\mG_{\tN\,\tT\to\tp}(\rho) \propto \rho^{|\tT\to\tp|}\mbox{exp}(-\omega \rho^2/2\hbar)
\mL^{\tT\to\tp + N/2 - 3/2}_{\sN}(\omega \rho^2/\hbar)
\mbox{ } 
\eeq
for the eigenvalues 
\beq
E = \hbar\omega\{2\mN + \mT\mo\mp + \{N - 1\}/2\} \mbox{ } , 
\eeq
where $\mN \mbox{ } \in \mbox{ } \mathbb{N}_0$ is a node-counting quantum number.

\subsubsection{Approximate solution for the dust universe analogues} \label{QSNStop111} 

As mechanics, these are approximations to the Newton--Coulomb problem.
The scale equation here has the form of a unit-mass \{$N$ -- 1\}-dimensional Newtonian Gravity or attractive Coulomb problem with $\rho$ in place of $r$,  
\beq
- \hbar^2\{\rho^2\mG_{,\rho\rho} + \{N - 2\}\,\rho\mG_{,\rho} - 
\mT\mo\mp\{\mT\mo\mp + N - 3\}\mG\} - 2K\rho\,\mG = E\rho^2\mG \mbox{ } . 
\label{NdHydro}
\eeq
For $E > 0$, using the large approximand as a factor (the small one is trivial), $\mG = \mbox{exp}(- \sqrt{-2E}\rho/\hbar)f(\rho)$, 
and a rescaling of the dependent variable, this maps to the generalized Laguerre equation, so this Article's $\mT\mo\mp = 0$ solution is 
\beq
\mG_{\sN}(\rho) \propto \mL^{N - 3}_{\sN - 1}(2\sqrt{-2E}\rho/\hbar)
\mbox{exp}(-\sqrt{-2E}\rho/\hbar)
\mbox{ } 
\eeq
for the eigenvalues 
\beq
E =  - k^2/2\hbar^2\{\mN - \{N - 4\}/2\}^2 = - \hbar^2/2\rho_0^2\{\mN - \{N - 4\}/2\}^2 \mbox{ } , 
\eeq
where $\mN \mbox{ } \in \mbox{ } \mathbb{N}_0$ is a node-counting quantum number and $\rho_0$ is the `Bohr configuration space radius'.

\subsection{Quantum scaled triangleland}\label{QSTri}

While this case is on $\mathbb{R}^3$ again (at lease some times requiring excision of the origin), 
the configuration space geometry is curved (but conformally flat). 
In particular, as per Sec \ref{Q-Geom},  the Ricci scalar is $6/\mI$, and so the conformal and Laplacian (and other $\xi$)  
orderings now differ, and by more than just a constant energy shift.     
In the conformal-ordered case, one is entitled to change PPSCT representation since the QM remains PPSCT-invariant.  
Thus, due to conformal flatness, one can pass to flat 
geometry, in which the kinetic part of the TISE takes a very well-known form.  
For ``central" problems/approximations to more general problems, there is a scale--shape split, with the preceding Sec's working now ocurring as the shape part.

\subsubsection{Very special HO case $B = C = 0$ in spherical coordinates}\label{Pureetee}

I approach the QM via the correspondence with the mathematics for the Kepler--Coulomb problem as laid out by the analogies in Sec \ref{Cl-Soln}.   
It corresponds to positive spatial curvature dust cosmology.  
This gives us the same TISE's as for the atomic problem and thus the usual separation and solvability in spherical and parabolic coordinates.  

\mbox{ } 

\noindent Note: This gives exactly the same wavefunctions as for hydrogen; on the other hand this analogy breaks down at the level of having a different inner product from hydrogen's.  

\mbox{ }

\noindent Hydrogen has principal, angular momentum and magnetic quantum numbers, while the current system has analogues of these, with, again, relative 
shape momentum quantum numbers s and S playing the roles of magnetic and total angular momentum quantum numbers m and l respectively.  
I use $\mN$ for the new principal quantum number that is associated with total moment of inertia of the 
system, which takes values such that (using $K_1 = K_2 = K$ for the very special case) 
\beq
K = E^2/\hbar^2\mN^2 = 4\hbar^2/\mI_0^2\,\mN^2 \mbox{ for } 
\mN \in \mathbb{N} \mbox{ } ,  
\eeq
Here, $4\hbar^2/E = \mI_0 = SIZE_0$ (the meaning of which is explained in Sec \ref{CharSc}).   
Thus one requires
\beq
E/\sqrt{K} = \mN\hbar \mbox{ } , \mbox{ } \mN \in \mathbb{N}
\eeq
as a consistency condition on the universe-model's energy and contents.

The corresponding wavefunctions are of the form 
\beq
\check{\Psi}_{\sN\sS\sj}(\mI, \Theta, \Phi) \propto \mL_{\sN - \sS - 1}^{2\sS + 1}({2\mI}/{\mN\mI_0}) 
\mbox{exp}(-{\mI}/{\mN \mI_0})\{{\mI}/{\mI_0}\}^{\sS}\mP_{\sS}^{\sj}(\mbox{cos}\,\Theta)
\mbox{exp}(i\mj\Phi) \mbox{ } .  
\label{Dale}
\eeq
Or, in terms of scale--shape quantities, 
\beq
\check{\Psi}_{\sN\sS\sj}(SIZE, aniso, ellip) \propto 
\mL_{\sN - \sS - 1}^{2\sS + 1}
\left(
\frac{2\,SIZE}{\mN\,SIZE_0}
\right) 
\mbox{exp}
\left(
-\frac{SIZE}{\mN\,SIZE_0}
\right)
\left\{
\frac{SIZE}{SIZE_0}
\right\}^{\sS}
\mP_{\sS}^{\sj}
(ellip)
{\cal T}_{\sj}
\left(
aniso/\sqrt{1 - ellip^2}
\right) \mbox{ } .  
\eeq
I comment on the differences that the unusual inner product used here makes to the probability densities of various wavefunctions in Fig \ref{08III-Fig-4}.
Thus this Article's analogy is less extensive than \cite{08I, 08II}'s linear rigid rotor analogy.    
In this way the inclusion of scale complicates matters. 
Also, nontrivial PPSCT factors are needed.  
These have more implications for a configuration space of dimension $k = 3$ such as that of the present Sec rather than for 
the $k = 2$ of the previous Sec for which a number of cancellations occur.  

{\begin{figure}[ht]
\centering
\includegraphics[width=0.55\textwidth]{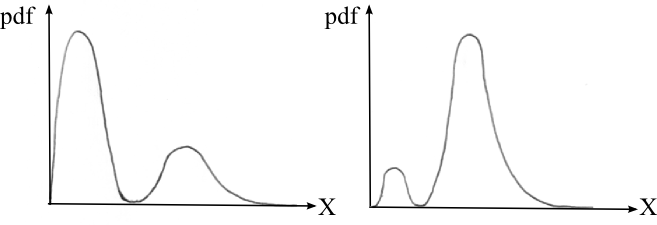}
\caption[Text der im Bilderverzeichnis auftaucht]{\footnotesize{a) For triangleland's `$1s$ orbital' (X = I/I$_0$), the inner rather than the outer pdf lobe is the more significant one.
b) This is in contradistinction to the situation in hydrogen (X = $r/a_0$), where the $2s$ pdf primarily resides outside of the $1s$ pdf.  
Moreover, this feature is new to triangleland, $N$-stop metroland behaving in this respect in qualitative parallel with hydrogen.  
In the current presentation of triangleland QM for which the {\sl wavefunctions} are the same as for hydrogen, this effect is clearly due to the lower power of 
X in triangleland's conformal inner product (1/2) relative to that in the usual hydrogenic inner product (2).
}} \label{08III-Fig-4}\end{figure}} 

While one could proceed to investigate (as in \cite{08II}) the special case in spherical coordinates asymptotically and as a perturbation around the very special case, 
I have however found an alternative {\sl} exact method, which I build up to over the next two SSecs.

\subsubsection{Free problem in parabolic coordinates}

In this case one obtains SHM and two split-out equations in $\mI_1$ and $\mI_2$ that map to the Bessel equation, giving an overall solution 
\beq
\overline{\Psi}(\rho_1, \rho_2, \Phi) \propto 
\mJ_{\sj}(\rho_1/\rho_1^{\scc})\mJ_{\sj}(\rho_2/\rho^{\scc}_{2})\mbox{exp}(i\mj\Phi)
\eeq
for constants $\rho_{\scc}^{\barr} = \sqrt{2}\hbar/\sqrt{ - E_{\barr}}$ and  $E_1 + E_2 = E$.    
The interpretation of the probability density functions has the usual multiplicity of peaks of a free problem. 
These now correspond to a sequence of (cluster) separations that are probable alternating with a sequence that are improbable.
As is usual, have non-normalizability; in any case prefer models with potential terms 
(the latter is probably more primary, and non-normalizability is then a consequence of not doing so).

\subsubsection{Very special and special HO problems in parabolic-type coordinates}

The very special case matches the atomic problem in e.g. \cite{LLQM, Merzbacher, Hecht, Robinett}, under 
the correspondence $\ttE$ to $-K/4$ (which is the right sign to get the bound states) and $e^2/\pi\epsilon_0$ to $E$.

Now, the quantum numbers are given by 
\beq
\mN_{\barp} = -\{|\mj| + 1\}/2 + \mN \mb_{\barp} 
\eeq 
for $\mN$ as before, $\mN_{\barp}$ parabolic quantum numbers and $\mb_{\barp}$ constants such that $\mb_1 + \mb_2 = 1$. 
Here 
\beq
\mN_1 + \mN_2 + |\mj| + 1 = \mN \mbox{ } . \label{sphepara}
\eeq
The TISE separates into simple harmonic motion in $\Phi$, while, 
for the other variables, one gets the same mathematics as for the standard atomic separated-out 1-$d$ parabolic coordinate problem, 
\beq
\frac{\d}{\d\mI_{\barp}}\left\{ \mI_{\barp} \frac{\d}{\d\mI_{\barp}}\Xi_{\barp} \right\} - 
\frac{m^2}{4\mI_{\barp}}\Xi_{\barp} - \frac{K\mI_{\barp}}{8\hbar^2}\Xi_{\barp} = - 
\mb_{\barp}\frac{E}{8\hbar^2}\Xi_{\barp} \mbox{ } .   
\eeq
Thus the wavefunctions for the present problem up to normalization are (see e.g. \cite{LLQM, Hecht}) 
\beq
\overline{\Psi}_{\sN_1 \sN_2 \sj}(\mI_1, \mI_2, \Phi) \propto 
\mL_{\sN_1}^{|\sj|}(\mI_1/\mN \mI_0)
\mL_{\sN_2}^{|\sj|}(\mI_2/\mN \mI_0)
\mbox{exp}
\left(
-\{\mI_1 + \mI_2\}/{2\mN \mI_0}
\right)
\{\mI_1\mI_2\}^{|\sj|/2}\mbox{exp}(i\mj\Phi) \mbox{ } , 
\eeq
for $\mI_0 = 8\hbar^2/E$.

\noindent Or, in terms of shape and size variables,
$$
\overline{\Psi}_{\tN_1\tN_2\sj}(SIZE, aniso, ellip) =
\mL_{\tN_1}^{|\sj|}(\omega_1SIZE\{1 - ellip\}/4\hbar) 
\mL_{\tN_2}^{|\sj|}(\omega_2SIZE\{1 + ellip\}/4\hbar)
\mbox{exp}(    -SIZE\{\{\omega_1 + \omega_2\} + \{\omega_2 - \omega_1\}ellip\}/8\hbar) 
$$
\beq
SIZE^{|\sj|}\{1 - ellip^2\}^{|\sj|/2}
{\cal T}_{\sj}\left({aniso}/{\sqrt{1 - ellip^2}}\right) \mbox{ } ,  \omega_{\barr} = \sqrt{K_{\barr}} \mbox{ } .  
\eeq
\mbox{ } \mbox{ } Next, the special case can done essentially like very special one.  
The separation continues to work out as before except that what is the same energy constant in each 
separated out parabolic problem now takes a different value for each, $-K^{\prime}_{\bari}/4$.\footnote{The 
counterpart of this unusual extension was already remarked upon at the classical level in Sec \ref{Cl-Soln}.}  
%
Now the quantum numbers come out to be as before, though each $\mN_{\barr}$ now has a distinct form of 
$\mN$: 
\beq
\mN_{(\barr)} = E/\sqrt{2K_{\barr} } 2\hbar \mbox{ } , \mbox{ } \mbox{ }  
\eeq
so that there is not a simple relation like (\ref{sphepara}), but, rather, 
\beq
\frac{\mN_1 + \{|\mj| +  1\}/2}{\mN_{(1)}} + \frac{\mN_2 + \{|\mj| +  1\}/2}{\mN_{(2)}} = 1  \mbox{ } . 
\eeq
This corresponds to this less symmetric case not having a principal quantum number analogue. 
The wavefunctions are then    
\beq
\overline{\Psi}_{\sN_1\sN_2\sj}(\mI_1,\mI_2, \Phi) \propto 
\mL_{\sN_1}^{|\sj|}(\mI_1/\mN_{(1)}\mI_0)
\mL_{\sN_2}^{|\sj|}(\mI_2/\mN_{(1)}\mI_0)
\mbox{exp}(- \{\mI_1/\mN_{(1)} + \mI_2/\mN_{(2)}\}/2\mI_0)
\{\mI_1 \mI_2\}^{|\sj|/2}\mbox{exp}(i\mj\Phi) \mbox{ } .  
\label{Lynx}
\eeq
Note 1) compared to the hydrogen ones, the difference in inner product causes the probability density 
functions for these to behave differently for m = 0/j = 0 (out problem's then peak at the origin, while hydrogen's go to zero there).  

\noindent Note 2) the significance of j for parabolic orbitals is as follows. 
j = 0 is the same probability for all relative angles. 

\noindent 
j = 1 has one state favouring the near-collinear configurations and another state favouring the near-right configurations. 

\noindent j = 2 has one state favouring both of these and one state favouring neither of them.

\subsubsection{General HO problem in parabolic-type coordinates}

Take (\ref{Lynx}) with normal coordinate $N$-labels on it and then apply the rotation in Sec \ref{Tri-Norm}.  
Then, in a basis with first axis aligned with the potential and E as second axis [with ($\gamma$) labels suppressed],  
$$
\overline{\Psi}_{\sN_1\sN_2\sss}(SIZE, aniso, ellip) =
\mL_{\sN_1}^{|\sss|}(\omega_1^NSIZE\{f - B\, ellip - C\, aniso\}/2g) 
\mL_{\tN_2}^{|\sss|}(\omega_2^NSIZE\{f + B\, ellip + C\, aniso\}/2g) 
SIZE^{|\sss|}
$$
\beq 
\{f^2 - (B\,ellip + C\,aniso)^2\}^{|\sss|/2}
\mbox{exp}(- SIZE\{f\{\omega_1^N + \omega_2^N\} + 
         \{\omega_2^N - \omega_1^N\}\{B\,ellip + C\,aniso\}\}/2g) 
{\cal T}_{\sr}
\left(
\frac{C\,ellip - B\,aniso}{\sqrt{f^2 - \{B\,ellip + C\,aniso\}^2}}
\right)
\eeq
for $\omega_i^N = \sqrt{K_i^N}$ (normal mode frequencies), $f = \sqrt{B^2 + C^2}$ and $g = 2\hbar\sqrt{B^2 + C^2}$.

\subsubsection{$A < 0$, $E =0$ example later used in Semiclassical Approach}

I.e. the negative-curvature vacuum cosmology analogue.
The exact solution here is
\beq
\Psi_{\sS\,\sj} \propto \mI^{-1/2}\mJ_{2\sS + 1}\left(\frac{\sqrt{-2A}}{\hbar}\mI^{1/2}\right) 
\mP_{\sS}^{\sj}(\mbox{cos}\,\Theta)\mbox{exp}(i\mj\Phi) 
\propto \sqrt{\frac{SIZE_0}{SIZE}}\mJ_{2\sS + 1}\left(\sqrt{\frac{SIZE_0}{SIZE}}\right)
P_{\sS}^{\sj}(ellip){\cal T}_{\sj}\big(aniso/\sqrt{1 - ellip^2}\big) \mbox{ } ,
\eeq
for SIZE$_0$ = $\hbar^2/-2A$.

\subsubsection{Cases with further cosmology-inspired potentials}

A list of potential contributions that directly parallel other work in ScaleQM are 

\noindent 1) the upside-down HO's that map to ionized atoms, and correspond to negative spatial curvature in cosmology. 

\noindent 2) Extra $1/\mI^2$-type potentials, which allow for wrong-sign radiation to approximately become right-sign radiation. 

\noindent 3) $\mI^2$-type potentials, which correspond to cosmological-constant terms of whichever sign.

\noindent  Sec \ref{Silbame} 3) covers 3) in at the level of perturbations about this SSec's main hydrogen-analogue model.

\subsubsection{Interpretation I: Bohr moment of inertia}\label{CharSc}

The mathematical analogy implies the following. 
\beq
 \mbox{ } \mbox{ (Bohr radius) } a_0 = 4\pi\epsilon_0\hbar^2/m_e e^2 
\mbox{ } \longleftrightarrow \mbox{ }  
\mI_0 = {4\hbar^2}/{E} \mbox{ } \mbox{ (a new `Bohr total moment of inertia' scale) } 
\eeq
which has the same kind of interpretation as the typical minimum quantity or effective size -- 
the overall ground state has a moment of inertia distribution with `characteristic width' $\mI_0$.  
N.B. The analogy with hydrogen at the quantum level is part-false, since the inner products do not match up.

\subsubsection{Interpretation II: Expectations and spreads}

Now, in spherical coordinates for the very special case, by the scale--shape split, and the 
orthogonality relation and a recurrence relation for the generalized Laguerre polynomials, 
\beq
\langle \mN\,\mS\,\mj\,|\,\mI\,|\,\mN\,\mS\,\mj \rangle = SIZE_0\mN^2 \mbox{ } , \Delta_{\sN\,\sS\,\sj} = 
\sqrt{ \langle \mN\,\mS\,\mj\,|\,SIZE^2\,|\,\mN\,\mS\,\mj\rangle - \langle \mN\,\mS\,\mj\, |\, SIZE \, 
|\,\mN\,\mS\,\mj \rangle^2 } = SIZE_0\mN\sqrt{\mN^2 + \mS\{\mS - 1\}}/\sqrt{2} \mbox{ } . 
\eeq
Thus, in particular for the ground state,  the expectation is the characteristic Bohr moment of inertia $\mI_0 = SIZE_0$ and the spread is $SIZE_0/\sqrt{2}$.  
Differences from hydrogen's expectation and spread in r are due to the difference in the inner product in each case, noting that the present Article's case is easier to compute. 
(It requires merely one use of the aforementioned recurrence relation. 
Also, the expectation turns out not to depend on the total rational quantum number in this case, unlike for the atomic one.)

On the other hand, in parabolic-type coordinates, for the special case, 
\beq
\langle \mN_1\,\mN_2\,\mj\, |\, SIZE \,|\, \mN_1\,\mN_2\,\mj \rangle = 
SIZE_0\big\{\mN_{(1)}\mN_1 + \mN_{(2)}\mN_2 + \{\mN_{(1)} + \mN_{(2)}\}\{|\mj| + 1\}/2\big\} 
\mbox{ } , 
\eeq
\beq
\Delta_{\sN_1\,\sN_2\,\sj} = SIZE_0\mN\sqrt{  \mN_1^2 + \mN_2^2 + \{\mN - |\mj|\}\{ |\mj| + 1 \} }/\sqrt{2} 
\mbox{ } . 
\eeq
Note that for very special case ground state, these reduce to the above results, as they should.

\subsubsection{Interpretation III: perturbative treatment of additional cosmologically-inspired potential terms}\label{Silbame}

E.g. $|\mr^{IJ}|^6$ terms are cosmology-motivated, by corresponding to cosmological constant terms. 
These can be treated as small perturbations about the preceding Sec's problem.  
The perturbation theory expands in a series of powers of $S$, so then there are also $O$($S^2$) terms 
(and the previously mentioned corrections become $O$($\mbox{S}^{\sfa}, S)$.    
They have a lead scale part $S\,\mI^2$ and then $O$($\mbox{S}^{\sfa}$) terms. 
Then, using the same $\langle \mN\,\mS\,\mj\,|\, \mI^2 |\, \mN\,\mS\,\mj  \rangle$ integral as in the 
preceding SSec one obtains the first-order perturbation correction to the energy analogue to be 
\beq
K = 4\hbar^2/\mN^2\mI_0\mbox{}^2 + S\,\mI_0\mbox{}^2\mN^2\{3\mN^2 + \mS\{\mS - 1\}\}/8 + O(\mbox{S}^{\sfa}, S) + O(S^2) 
\mbox{ } , 
\eeq
so an approximate inversion to look at the effect on $E$ gives that 
\beq
E^{\{1\}} = \mN\,\hbar\,\omega - S\{\hbar\omega\}^3\mN\{3\mN^2 + \mS\{\mS - 1\}\} + O(\mbox{S}^{\sfa}, S) 
+ O(S^2) \mbox{ } .  
\eeq
Thus $S > 0$ lowers $E$, while $S \leq 0$ raises $E$.

\subsection{Extension to quantum quadrilateralland}

For less standard solutions to arise, one would need to check further, more complicated examples, such as the below (which have various further uses as indicated). 

\mbox{ }  

\noindent{\bf Question 61: Free solution}.
Take MacFarlane's \cite{MacFarlane} study of QM on $\mathbb{CP}^2$ and reinterpret this as the plain shape space of quadrilaterals. 
Consider also QM on C($\mathbb{CP}^2$).  
The scale--shape split into the cone now gives MacFarlane's mathematics for the shape part and the standard sort of radial equation already  
familiar from earlier chapters, so I can solve this scaled problem to the same extent that I can solve the preceding.  

\mbox{ }

\noindent{\bf Question 62: HO-like potential solution}\label{QQuad2}.
%
Only the scaled isotropic HO case of this is tractable.

\mbox{ } 

\noindent{\bf Question 63: Perturbations about the free and isotropic HO solutions}\label{QQuad3}.
%
This is be another exercise in the methods of Mathematical Physics, this time concerning recurrence 
relations for, and integrals involving, the Jacobi polynomials \cite{AS, GrRy} and the Wigner D-functions \cite{WigD} that arise from MacFarlane's work \cite{MacFarlane} . 
By this the structure formation counterpart in Quadrilateralland remains around as analytically tractable as it is for triangleland.  

\mbox{ }

\noindent Questions 61--63 are being answered in my collaboration with Sophie Kneller.

\subsection{Otriangleland and 3-cornerland counterparts}

\noindent The Ashtekar Variables approach adjoins degenerate configurations.  
Above experience with kinematical quantization could be viewed as reason to suspect an approach that does that.
On the other hand, it is a new choice of variables with its own kinematical quantization.

\mbox{ }

\noindent Analogy 92)  There are parallels between the O-choice of shapes and the affine approach 
\cite{Klauder, IshamKakas1, IshamKakas2} to Geometrodynamics, as an extension of the half-line toy model for the affine case \cite{I84}.  

\mbox{ }  

\noindent{\bf Question 64} Study Otriangleland as a toy model of affine Geometrodynamics.   

\mbox{ } 

\noindent{\bf Question 65} There is a difference between Otriangleland and 3-cornerland as regards the form and includability of the 
degenerate configurations (the collinear configurations). 
This may be of relevance to the foundations of Nododynamics/LQG.
The situation there is that one extends to include degenerate configurations and then these play an important role in the 
quantization, leading to e.g. the discreteness of quantum area result.  
Are we completely sure then that the nondegenerate part of the metric is being fully taken into account in such an approach?
Are we completely sure that we have an understood and unambiguous procedure as regards quantization of a state space consisting of degenerate and nondegenerate parts?  
[Chris Isham could not point me to an answer to this issue when I discussed it with him in 2009.]

\vspace{11in}

\section{RPM's as toy models for Quantum Cosmology}\label{RPM-for-QC}

\subsection{Further comments on closed-universe effects} \label{RPM-for-QC1}

Closed-universe and finite-universe effects are manifest in RPM's.
Some of these involve collapse in the number of admissible combinations of quantum numbers.

\mbox{ }

\noindent Example 1) The subsystems' energies are interlocked by each model having an overall energy of the universe, 
which leads to gaps, truncations, sometimes even total 
nonexistence of what would elsewise be the eigenspectrum (Sec \ref{QS3Stop4}). 

\noindent Example 2) The subsystems' angular momenta are counterbalanced so as to comply with the zero angular momentum constraint.
This is clearly in line with the solution via second reduction, or third reduction in the Laplace-ordered case.  
For scaled triangleland, Dirac-approach solutions based on base--apex subsystem splits depend on one less quantum number than one might na\"{\i}vely expect \cite{06I}. 
One can get to this via ${L}_i\psi_i = \nm_i\psi_i$ $i = 1, 2$ arising in each separated problem, but then by $0 = \hat{\scL}\Psi = 
\hat{{L}}_1 \otimes \hat{1}_2 \Psi + \hat{1}_1 \otimes \hat{{L}}_2\Psi = \nm_1\Psi + \nm_2\Psi $, so $\nm_1 = - \nm_2 := \nj$, so that 
\be
\Psi_{\sj\sn}(\rho_1,\rho_2, \theta_1, \theta_2) = \mbox{exp}(i\mj\{\theta_1 - \theta_2\})
\mJ_{\sj}(\rho_1/\rho_1^0)\mJ_{-\sj}(\rho_2/\rho_2^0) \mbox{ } . 
\ee
One can alternatively get to this via $p_{\Phi}$ (for $\Phi = \phi_2 - \phi_1$) being cyclic and carrying a relative angular momentum quantum number. 

\noindent Example 3) Is there a dilational counterpart of this in pure-shape RPM's?  
The preceding counterbalance is tied to the very simple form of the phase-angle wavefunctions (exponentials as opposed to exponentials times
polynomials times powers, which produce complicated expressions on being acted upon by quantum constraint operators that are linear in the momenta).   
On the other hand, rewriting in scale--shape split coordinates, the dilational constraint candidly becomes 
\beq
\rho \, \pa_{\rho}\Psi = 0
\eeq 
i.e. that the wavefunction is independent of $\rho$. 
In this case then the only survivor in a separated-out $\rho$-piece is the constant solution.  
Thus this is indeed a collapse in number of admissible quantum numbers but not a counterbalance.  
On this occasion, the argument is independent of all of $N$, $d$ and $V$.

\noindent Example 4) There is still angular momentum counterbalancing in the pure-shape triangleland case. 
This can be demonstrated by classically eliminating the dilational constraint but retaining the zero angular momentum 
constraint which still leaves SHM in the two phase angles, thus continuing to exhibit the characteristics of a counterbalance.  
This is solvable e.g. in $\rho_1$, $\phi_1$ and $\phi_2$, leading to dependence solely on $\Phi = \phi_2 - \phi_1$. 

\mbox{ }

\noindent Note thet Barbour and Smolin suggested \cite{BS, EOT} that (Barbour's) relationalism requires radically different QM theory whilst 
also not possessing a good semiclassical limit, partly on the basis of the above Example 1). 
I argued against this \cite{06I}, by showing how standard QM structure and standard QM solutions made out of the usual Methods of Mathematical Physics 
suffice for the study of the simpler quantum RPM's.
This is achieved via the various key choices of coordinates (Jacobi coordinates, spherical coordinates...) 
and the good fortune of being able to reduce one's way down to the physical configuration space for these examples.
There are still closed-universe effects in the models such as examples 1 to 4, 
but these do not affect the `recovery of everyday Physics' for a large universe with sufficiently varied contents.  
This is via the presence of free particles and of potentials of both signs serving to overcome gaps and truncation. 
And via the practicalities of experimentation involving only small subsystems means that interlocking and counterbalancing would be likely to go unnoticed.
These things said, this Article concerns, rather, RPM's as toy models of Geometrodynamics.

\mbox{ } 

\noindent Example 5) In closed-universe models, the elsewise entirely theoretically solid {\sl cluster decomposition principle} \cite{Weinberg} is globally violated: 
one can not just tensor subsystems together right up to the inclusion of the entire universe if there are closed-universe features such as RPM model universes' 
energy interlocking and overall angular momentum counterbalancing.   

\mbox{ } 

\noindent Analogy 93) There are differences of the subsystem versus the closed whole-system sort between absolute QM and relational QM
Subsystem physics in a relational universe would be expected to be linked by global restrictions due to its  constraint equations.
Such differences were already found to be present in the GR case by DeWitt \cite{DeWitt67}.

\subsection{Quantum Cosmology versus Atomic/Molecular Physics}\label{RPM-for-QC4}

This Article analogue-models some fairly conventional Quantum Cosmology models.  
Despite this Article forming many bridges between Molecular Physics and its underlying classical kinematics and Quantum Cosmology [exemplifying RPM's value by Criterion 3)], 
it is unsurprising that this is different in at least some of the details from modelling commonly encountered molecules/ions such as $H_2^+$ and $NH_3$ \cite{AtF}.  
Some underlying similarities and differences are as follows. 

\noindent Difference I) Molecules are but tiny and unisolated pieces of a much larger universe. 
(In particular, in \cite{+Tri}, I consider Molecular Physics as studied in laboratories on Earth, where there is good control over initial 
conditions but certainly not isolation from the rest of the universe, in particular in the aspect of determining what the inertial frames are). 
Thus in Molecular Physics, some coordinates can effectively refer to the larger set-up.  
E.g. this is true of `electric field parallel to the z-axis' in the Stark effect for the linear rigid rotor.  
It is also true if one uses the spatial angles ($\theta_{\sp\sh}$, $\phi_{\sp\sh}$) for the same mathematics as in 
triangleland or 4-stop metroland but now laced with absolute rather than purely relational meaning.  
These features are clearly less desirable for GR-like quantum cosmological modelling than RPM's relational context. 

\noindent Similarity I) Useful coordinate choices/kinematics are largely shared between quantum cosmological RPM's and molecular models, 
e.g. the use of Jacobi coordinates and of Dragt-type coordinates in the present Article. 

\noindent Similarity II) Then, at the level of the solutions, scaled models with multi-HO potentials have the obvious standard mathematics.  
Pure-shape 4-stop metroland with multi-HO-type potentials has mathematical analogies with Pauling's 
study of rotations of molecules in crystals, the polarization theory behind the Raman effect and the $H_2^+$ problem.
Scalefree triangleland with multi-HO-type potentials has a mathematical analogy with the Stark effect for the linear rigid rotor.  
Scaled $N$-stop metroland models with a range of cosmologically-motivated potentials also give 
rise to a range of further mathematical models that are (fairly) common in QM textbooks.
Examples of such are upside-down HO's, and \{$N$ -- 1\}--$d$ isotropic HO's, the ordinary HO, and hydrogen-type problems.  
Scaled triangleland with the very special multi-HO-type potential's wavefunctions can be set up to coincide mathematically with hydrogen's.
Here, the special and general multi-HO-type potentials continuing to be accessible by extension of the hydrogen problem's well-known solvability in parabolic coordinates.  
Finally, interpreting many of the above solutions involves work similar to/identical with the 
evaluation of radial expectations and spreads for hydrogen, and of the corresponding shape integrals (usually known as 3-$\mY$ integrals).  

\noindent Difference II) However, molecules have rather tightly-restricted Physics. 
E.g. they are invariably held together by Coulombic forces. 
In contrast, as regards cosmological models, different matter types contribute to different mechanical-analogue power-law potential contributions \cite{Cones}.  
Nor by any means is there a concept of fixed bond length in analogue models of Quantum Cosmology.  
Furthermore, molecules also have a precise mass hierarchy following from the large ratio between the electron and nucleon masses -- of the order of 1 part in 2000.  
This is by no means appropriate for quantum cosmological modelling (though there is also a somewhat less precise, somewhat time-dependent and rather smaller small ratio, 
namely the 1 part in $10^5$ concerning the inhomogeneities).
Moreover, one would like this to possess {\sl some sort of} mass hierarchy if one is attempting to obtain time as an emergent semiclassical concept, c.f. Sec \ref{Semicl}). 
Also, molecules' constituent elementary particles have a precise character as regards statistics and distinguishability, 
while such properties would appear to be more optional than requisite for (analogue) Quantum Cosmology models.  

\noindent Difference III) In various cases, the analogies break down if pushed far enough.  For instance, 

\noindent a) the 2-body approximation only sits unstably inside the 3-body problem for negative power-law potential $N$-stop metroland.  

\noindent b) The very special HO for triangleland has the same wavefunctions as for hydrogen but not the same inner product. 

\noindent c) There are differences in the ways in which the next levels of structure produce complications.
E.g. various nuclear and SR effects occur in molecules \cite{LLQM, Weinberg}, 
while one may wish to build in certain analogue-GR/quantum comological details into RPM's \cite{Cones} (see also SSec \ref{RPM-for-QC6}).   

\noindent Differences IV) to VI) consist of the Operator Ordering, absolutist imprints and multiverse interpretation SSecs below.  

\mbox{ }  

\noindent Difference 41) I do not know of any meaningful analogue of the Hartle--Hawking type condition on $\Psi$ in RPM's.  

\subsection{`Multiverse' differences between RPM's and molecular models}\label{RPM-for-QC6}

One difference between this Article's specific models and their Molecular Physics analogues is at the level of which `multiverses' correspond to each. 
(For Molecular Physics models the corresponding multiverse can be thought of as a collection of laboratory setups with different parameter values. 
Multiverses in the RPM context are at least aesthetically more appealing in distancing themselves from Copenhagen interpretation connotations.) 

\mbox{ } 

\noindent Example 1) the set-up for the Stark effect for the linear rigid rotor can have an electric field in any 3-$d$ direction, 
while the corresponding direction in the triangleland harmonic oscillator problem cannot be rotated out of the collinearity plane.  
Moreover, in triangleland there are 3 (DM) axes of particular physical significance within this plane.   
Thus each corresponds to a different multiverse setting.  

\noindent Example 2) Both the 4-stop metroland harmonic oscillator and the crystal problem have privileged directions; these can be set up in 
particularly close correspondence if the crystal had cubic symmetry.  

\noindent Example 3) On the other hand, the `Raman' multiverse does not have such a tight similarity with the 4-stop metroland multiverse.
What the theory of Raman spectroscopy does have \cite{TS, WS62, A76} is further analogies which  
extend to the general problem [i.e. the $\ttC$, $\ttF$, $\ttG$ and $\ttH$ terms of Sec \ref{Cl-Soln}] \cite{AF}.   

\mbox{ } 

\noindent As regards the issue \cite{DeWitt67, EOT} of whether stationary quantum universes have single or multiple states, I comment that 
this Article's models do exhibit some degenerate states (both among simple exact solutions and perturbatively to second order).  
However one needs a much more extensive study of RPM model universes before one can begin to say whether these are, however, non-generic.

\noindent A related issue concerns placing a closed-universe interpretation on the perturbed problem.
Sec \ref{Cl-POT}, \ref{Cl-POT-Strat} and \ref{+temJBB}'s solutions are universes of fixed energy $E_{\sU\sn\si}$ and not modes within a particular universe.  
%
%
Sometimes \cite{DeWitt67} this corresponds to no allowed S, j, sometimes to 1 and at least sometimes to more than 1.  
E.g. $E_{0, 0} = E_{1, 0}$ for $B = \sqrt{15/2}\hbar^2$, which is perturbatively acceptable provided that $A >> B$.  
I use this further in Sec \ref{QM-Nihil} toward SM (SM)/Entropy/Information construction.

\subsection{Relational nontriviality at the quantum level}\label{Leela}

As regards a classical system with only one degree of freedom: in passing to the quantum level, there may be further latent quantum degrees of 
freedom such as spin, and via 1 degree of freedom sufficing to have a nontrivial wavefunction and subsequently multiple observables such as the 
nth moments, which could furnish enough of a sense of correlation to have meaningful Physics.  
Since the point particle has gained structure such as spread and, in mixed states, the possibility of multimodality, a lesser number 
of quantum particle-waves can suffice to furnish nontrivially relational Physics [I thank Don Page for first making me aware of this point]. 
This parallels Appendix \ref{Cl-Soln}.C.2's point 6)'s that looking closer at point particles revealing that they are planets alters which type of relational model one requires.

\subsection{Discussion of absolutist imprints}\label{Abs-Imprint}

1) There is no physical distinction between the relational sector of  $\underline\scL = 0$ absolute mechanics and RPM's.
Nor as regards QM monopole effects (unlike in $\underline\scL \neq 0$ Newtonian Mechanics, for which such show up). 

\noindent 2) There is however a distinction between absolute and relational associated linear spaces at the level of kinematical quantization.

\noindent 3) There is also quantum-level distinction between the relational sector of $\underline\scL = 0$ absolute mechanics 
(taken to be a subsystem study) and RPM's (taken to be a whole-universe setting) at the level of which operator orderings are to be used in forming the TISE.  
(In being based on an overall determinant having a factor coming in from a different sector, there is a loose parallel between this 
and the Fadde'ev--Popov determinants being imprints of the ghost species that `live' along the unphysical orbits \cite{Singer}.) 
Most of the Molecular Physics literature does not, as far as I have seen, make a connection between the ordering used in e.g. the 3-body problem and the absolute versus relative motion. 
Aharonov and Kaufherr \cite{Aharonov} do however show awareness of this issue; comparing that work and the present Article would be interesting.
This difference accounts for differences at the quantum level between RPM whole-universe models and conventional molecular models, 
which bear an absolute imprint by which they are implicitly subsystems within a much larger universe.  

\noindent 4) Point suffices as regards various key quantum steps such as Isham's kinematic quantization and DeWitt's family of operator orderings.    

\mbox{ }  

\noindent{\bf Question 66} In the light of 3-cornerland being less relational than OTriangleland, interpret the differences between the two models 
at the quantum level as another form of absolutist imprint.

\subsection{Uniform states}\label{RPM-for-QC8}

\noindent RPM's are also a useful toy model for notions of uniformity that are of widespread interest in Cosmology.
This applies to good approximation to the present distribution of galaxies and to the CMB. 
There is also the issue of whether there was a considerably more uniform quantum-cosmological initial state \cite{Penrose}.  
Finally, there are related issues of uniformizing process and how the small perturbations observed today were seeded. 

\noindent There is no distinction between scaled and pure-shape theories' notions of uniformity as these are pure-shape notions 
and scaled theories admit a scale--shape split (excluding the maximal collision from the definition of uniform in the scaled case).   

\noindent Study of the 3-stop metroland case makes for a good parent for bigger models' larger complex of notions of uniformity.  
In the equal-mass case of this, there is one notion of merger M and this coincides with equally-separated-out particles 
(which is the most natural-looking notion of uniform for an $N$-stop metroland).  
The corresponding notion of least uniformity are the double collisions D.  
4-stop metroland then has numerous notions of merger as per Sec \ref{Q-Geom},  one point among which additionally involves the four equal masses being equally spaced out.
   
\noindent For triangleland with equal masses there are the equilateral triangle states at the poles that are the most uniform states.  
This is a 2-$d$ version of equally-spaced-out particles, and it is now, more satisfactory, a cluster-independent i.e. democracy invariant notion.
That is characterized by it being the maximum of 4 $\times$ area; note that this is a democratic invariant, $|$demo(3)$|$.  
It is a unique maximum (in plain shape case, signed area is involved, and there is a maximum and a minimum value of demo(3) corresponding to the two 
orientations of the equilateral triangle.
Then the collinear configurations are a well-defined opposite notion, corresponding to the minimum value of $|$demo(3)$|$.
One could then use the 3-stop metroland version of least uniform to further discern the least uniform of these.    
Uniformity is further investigated using the \NSI in Sec \ref{NSI}. 
\noindent There is also a notion of correlation characterizer along these lines as per Sec \ref{DrNo}.  
\noindent Finally see \cite{LostaglioMA} for use of a distinct potential-dependent notion of uniformity.  

\noindent In the quadrilateralland case, there are 3 squares in each hemi-$\mathbb{CP}^2$ as opposed to the single equilateral triangle in each 
hemisphere of triangleland; these are the obvious particularly uniform configurations.  
That there are three per hemisphere reflects the presence of a further 3-fold symmetry in quadrilateralland; 
choosing to use indistinguishable particles then quotients this out.
These are also characterized as the extremum value of the corresponding democracy invariant, demo(4),however in this case they 
are not the only extrema: there is a whole extremal curve per hemi-$\mathbb{CP}^2$.   
In $\{\mN^e$-$aniso(e)$\} coordinates, this is given by $aniso$(1) = 0 = $aniso$(2) and $\left| aniso(3) \right|$ = 1, i.e. (1)-isosceles, (2)-isosceles and maximally 
(3)-right or (3)-left i.e. (3)-collinear, with $\mN_3$ = 1/2 and $\mN_1$, $\mN_2$ varying (but such that the on-$\mathbb{S}^5$ condition $\mN_1 + \mN_2 + \mN_3 = 1$ holds).  
In Gibbons--Pope type coordinates, the uniformity condition is $\phi = 0$, $\psi = 4\pi$, $\chi = \pi/4$ and $\beta$ free, 
i.e., in the H-coordinates case, freedom in the contents inhomogeneity i.e. size of subsystem 1 relative to the size of subsystem 2.
On the other hand, in the K-coordinates case, $\beta$-free signifies freedom in how tall one makes the selected \{12, 3\} cluster's triangle.

\subsection{Structure formation}\label{Str-Form}

Does structure formation in the universe have a quantum-mechanical origin?  
In GR, studying this requires midisuperspace or at least inhomogeneous perturbations about minisuperspace, which are of great difficulty.     
There are also various difficulties associated with closed system physics and observables, speculations on initial conditions, the 
meaning and form of $\Psi$ (e.g. the discussion of uniformity below) and the origin of the arrow of time. 
RPM's Semiclassical Approach scheme are useful toy models of midisuperspace Quantum Cosmology models 
that investigate the origin of structure formation in the universe. 
(E.g. the Halliwell--Hawking model toward Quantum Cosmology seeding galaxy formation and CMB inhomogeneities).
Thus RPM's are valuable conceptually and to test whether one should be {\sl qualitatively} confident in the assumptions and approximations 
made in such schemes.
This involves the shape wavefunctions provided as the l-physics, alongside $t^{\se\sm(\sJ\sB\sB)}$-dependent perturbations of these, a small start on which is made in Sec \ref{Semicl}.  

%

To address some questions, one {\sl does} want to consider specific clusters (e.g. the model has 2 heavy particles and a light one and the 
observer considers the light one to be their galaxy).  
The advantages of 4-stop metroland are in its cleaner concept of contents inhomogeneity, which arises 
through being able to partition the universe into 2 clusters of 2 particles each.  
In triangleland, one can only make such considerations by comparing clusters of 2 particles that overlap as regards which clusters they include.  
This point is further developed in the \NSI Sec \ref{NSI}.
To combine this with nontrivial linear constraints, use quadrilateralland \cite{QuadI, QuadIII}.  

\mbox{ } 

\noindent Note 1) Upgrading to $N$-stop metroland and $N$-a-gonland models is likely to improve one's capacity to use particle clumps to model inhomogeneities.  

\noindent Note 2) The spherical presentation of triangleland is of limited use due to not extending to higher $N$-a-gonlands. 
However, one use for it is in allowing the shape part to be studied in $\mathbb{S}^2$ terms which more closely parallel the Halliwell--Hawking \cite{HallHaw} 
analysis of GR inhomogeneities over $\mathbb{S}^3$.  
This application will be more fully investigated in \cite{SemiclIII}.

\noindent Note 3) Moreover, Kucha\v{r} has also argued \cite{Kuchar99} that Halliwell and Hawking's perturbative treatment of inhomogeneous 
GR is likewise only a model of Quantum Cosmology, because quantum fluctuations are far smaller than the classical universe's inhomogeneities.  
In models with inflation, this criticism may at least in part be circumvented.
See Sec 22's RPM's (and \cite{ACos2, QuadIV} as a study in that direction.

\subsection{Parallel mini- and midi-superspace investigations}\label{RPM-for-QC9}

Another longer-term goal would be to export insights acquired by the RPM program to `mini and midi'superspace; are there then useful analogues 
of shape operators for these (anisotropy operators, inhomogeneity operators?)  

\noindent Anisotropy operators for GR are considered in e.g. \cite{KR89, Amsterdamski,AB}.  

\mbox{ } 

\noindent The form and usefulness of inhomogeneous operators for GR I leave as an open question.  

\begin{subappendices}
\subsection{Various attitudes to environments in Quantum Cosmological models}\label{App-Env}  

\noindent For the RPM's considered, scale dominates shape so as to model the small inhomogeneous fluctuations of the cosmological arena.
There is then a fork as to whether to model this with a notion of environment. 

\mbox{ }

\noindent Attitude 1) On the one hand, the 3-particle RPM has hitherto been taken as a whole-universe model.  
This most ideal interpretation is a lot less robust to assuming existence of additional particles whose contributions are then 
to be traced over (as compared with  minisuperspace modelling having little problem with inclusion of an environment since one 
assumes there that this model sits in some kind of neglected environment of small inhomogeneous fluctuations).  
The issue then is justifying the latter parts of \cite{H03, HT} (or, even more so, the elsewise more correct \cite{H09, H11}) 
in the absence of an environment; this costs us e.g. lines (\ref{pipupi}-\ref{whiffle}) in the triangleland example. 
\noindent I note however that Halliwell's 2009 parallel \cite{H09} of that working, configuration space regions that are large enough need no environment.\footnote{Here,  
large means compared to wavelength, which is solution-dependent. 
I also note that spherical geometry cases of this as per regions of the shapes of Fig \ref{Fig-Noob}c) are not really any harder to 
treat than the usual flat space cases.}

\noindent Attitude 2) Alternatively, study scale alone and use shape as an environment.   
This suffers from over-simpleness of the original system, but this is alleviated once one considers the shape perturbations about this.  

\noindent Attitude 3) Study a small set of particles (say a triangle of particles) that are taken to dominate over a cloud of lesser particles, 
which contribute in a small and averaged-out manner.  
These other particles are taken to be negligible in terms of most of their physical effects, but are still available as an environment for decoherence and accompanying approximate 
information storage; even 1 particle's worth of environment can serve as a nontrivial environment \cite{H99}.    
This offers a second resolution to `losing the environment' of \cite{HT}, i.e. arguing that it was hitherto negligible in the study but is 
nevertheless available at this stage of the study as an environment.  

\end{subappendices}

\vspace{10in}

\section{Levels of mathematical structure at the quantum level}\label{QM-Str}

If one is considering an r-formulation, one can use the standard notions for each of these. 
If not, one can at least formally consider $\FrG$-act $\FrG$-all versions.

\subsection{A further implementation of Configurational Relationalism at the quantum level}\label{QM-CR}

\noindent Given a non-configurationally-relational QM operator $\widehat{O}$, 
\beq
\widehat{O}_{\sFG} :=
\int_{\sfg \, \in \, \sFG} \mathbb{D}\ttg\,\mbox{exp}
\left(
i\sum\mbox{}_{\mbox{}_{\sfZ}} \stackrel{\rightarrow}{\FrG_{\sttg^{\tfZ}}}
\right) 
\, \widehat{O} \, 
\mbox{exp}
\left(
- i\sum\mbox{}_{\mbox{}_{\sfZ}} \stackrel{\rightarrow}{\FrG_{\sttg^{\tfZ}}}
\right) 
\label{G-op}
\eeq
(for $\stackrel{\rightarrow}{\FrG_{\sttg^{\tfZ}}}$ the action of infinitesimal generators) is a configurationally-relational counterpart. 
This is of course another example of $\FrG$-act, $\FrG$-all procedure, now based on the exponentiated adjoint group action followed by integration over the group (see also \cite{M09}).  
One can do similarly to pass from non-configurationally relational kets $|\Psi\rangle$ to configurationally-relational ones 
\beq
|\Psi_{\sFG}\rangle := 
\int_{\sfg \, \in \, \sFG}
\mathbb{D}\ttg\,
\mbox{exp}
\left(
i\sum\mbox{}_{\mbox{}_{\sfZ}} \stackrel{\rightarrow}{\FrG_{\sttg^{\tfZ}}}
\right) |\Psi\rangle \mbox{ } .  
\label{G-ket}
\eeq
The logic here is that, although the averaging might sometimes be carried out at a subsequent level of structure such as the expectation (if $\FrG$ is redundant expectations 
will be $\FrG$-invariant), other applications of operators and wave functions such as operator eigenvalue finding or $\FrG$-invariant wavefunction basis construction have 
enough partial meaning to on occasion be considered for individual $\FrG$-act $\FrG$-all moves.  

\noindent Admittedly, there are well-definedness issues in general as regards the measure $\mathbb{D}\ttg$ over the group $\FrG$. 
In particular, it is not clear how the diffeomorphisms could be treated explicitly in this way.  

\mbox{ } 

\noindent As regards the indirect definition of rotation-averaged operators,  triangleland's $SO(2) = U(1)$ is a particularly simple case, for which  
\beq
\int_{g \in \sFG}\mathbb{D} g = \int_{\zeta \in \mathbb{S}^1} \mathbb{D} \mathbb{\zeta} = \int_{\zeta = 0}^{2\pi}\d\zeta
\eeq
and $\stackrel{\rightarrow}{\FrG_g}$ is the infinitesimal 2-$d$ rotation action via the matrix $\underline{\underline{R}}\mbox{}_{\zeta}$ acting on the vectors of the model.
$\zeta$ is here the absolute rotation.

\subsection{Substatespaces Sub$\FrS$ at the quantum level}\label{QM-SubS}

Localized Sub$\FrS$ plays a further underlying role in Crane's thinking \cite{Crane} at the quantum level.  
This involves the postulate that I term 

\mbox{ } 

\noindent Perspectivalism 1) {\bf QM ONLY makes sense for subsystems} (my caps).
In the quantum GR arena, he deals with this by considering subsystem--environment splits of the universe with the observer residing on the surface of 
that split.\footnote{In fact, Crane \cite{Crane93} defines the observer to be such a boundary of a localized region, though I caution that not all 
boundaries will have actual observers upon them and that sizeable boundaries would need to be populated by many observers, forming `shells of 
observers'/a shell array of detectors.
I also note that, at least as far as I can tell, Crane's view (and Rovelli's less specific one \cite{Rov96}) of observers both strip them of any 
connotations of animateness or manifestly capacitated for processing information.
I term this the {\it introspective implementation}, as it involves a shell of inward-looking observers.}
%
Each of these splits then has its own Hilbert space, SubHilb.  
\noindent Then, at least formally, SubQuant: (SubPhase, SubCan) $\longrightarrow$ (SubHilb, SubUni), or the (SubRigPhase, SubPoint) counterpart of this.
Crane maintains that the whole universe has no Hilbert space (so his version of what I term Persp($\FrQ$) has $\subset$ in place of my $\underline{\subset}$), 
though a Hilbert space is to be recovered in a semiclassical limit.  

\mbox{ } 

\noindent I decompose Perspectivalism 1) as follows due to not agreeing will all parts of it.  

\noindent Crane 1) QM makes sense for subsystems (true, and obvious); each has its own Hilbert space, within which the standard interpretation of QM applies.  

\noindent Crane 2) Almost all QM concerns subsystems (true at an elementary level, but sometimes forgotten in Quantum Cosmology).  

\noindent Crane 3) According to Crane, what would elsewise be QM for the whole universe does not possess a Hilbert space or the standard interpretation of QM. 

\noindent A--Crane Difference 1) Here I disagree about the lack of Hilbert space (shrunken as it may be due to closed-universe effects as per Secs \ref{Q-3-Stop} and 
\ref{RPM-for-QC}), 
but agree about the subsequent non-applicability of the standard interpretation (though this is well known and plenty of different groups of authors have been 
willing to work under such conditions). 
That is why I use $\underline{\subset}$ and not $\subset$; this difference also accounts for my replacing `all' by `almost'.  
I envisage the scalefactor of the universe as a possible whole-universe property that can plausibly enter one's physical propositions; at the very least, scale is 
not a {\sl localized} subsystem.  

\noindent A--Crane Difference 2) Finally, Crane did not stipulate his  Sub$\FrQ$ to carry the same local physical connotations as mine (as per Sec \ref{Cl-Pers}.  

\noindent Crane 4) QM makes sense for the universe as a whole in some kind of semiclassical limit. 
Here I agree with Crane, though in detail we may well not both mean exactly the same thing by a `semiclassical limit'.
Conceptually, his semiclassical means lack of correlations; technically it concerns the observers associated with a triangulation of the model in question.  
Moreover Crane 4) is less necessary from my point of view (it applies just to interpretation, rather than to the possession of a Hilbert space as well).

\mbox{ } 

\noindent Perspectivalism 1W) (W for `weakened') shall be used to refer to Crane 1, 2) and my side of A-Crane Differences 1, 2).  

\mbox{ } 

\noindent Note 1) Perspectivalism 1W) is what I posit as the perspectival expansion of Relationalism 3) 
[though one {\sl could} instead elect to use Perspectivalism 1) in such a role]. 

\mbox{ } 

\noindent {\bf Question 67} On the one hand there is multiplicity by inequivalent quantum theories 
and on the other hand multiplicity from there being multiple observer perspectives.  
Whilst I do not present any evidence for this, I ask about the extent to which there are whether there are connections between these two multiplicities.  

\mbox{ }

\noindent Perspectivalism 2) Crane also allows for the QM of {\bf observers observing other observers observing subsystems} (see also Rovelli \cite{Rov96}).
This provides a further level of structure between the plethora of Hilbert spaces associated with all the contexts of Perspectivalism 1).  
It has no nontrivial classical counterpart.  

\mbox{ }  

\noindent Note 2) Considering subsystems, whether by its particular practicality or as a matter of no choice, takes one away from where RPM 
is an important part of the modelling as opposed to the highly unrestricted emergent Newtonian Mechanics of localized subsystems as per Appendix \ref{Cl-Soln}.C.
Using specifically subsystems of small  RPM's allows one to investigate what happens as the subsystems one considers get to have enough 
contents to be the whole-universe model, whilst keeping the mathematics (if not necessarily the standardness of the QM interpretation) simple.  

\noindent Note 3) In RPM's, the perspectival approach can be toy-modelled using the notion of a localized fictitious test-observers who 
can measure nearby inter-particle distances and perhaps relative angles about their particular position (c.f. Appendices 3.E and 4.B). 
Archetypes of this are the base subsystem of a triangleland or the base alongside the relative angle 
(the `fictitious test-observer more or less at the centre of mass of a localized binary'). 
This can accommodate Perspectivalism 1) to some extent, but cannot meaningfully incorporate Perspectivalism 2) which explicitly requires 
non-negligibility of observer material properties.
%

\noindent Note 4) As regards relational examples, not much happens if one uses an r-scheme, except that subsystem energies will usually have freedom even 
if $E_{\sU\sn\si}$ is taken to be fixed.
As regards $\FrG$-nontrivial theories treated in the Dirac way, e.g. for scaled RPM, each subsystem's AM is free except that they have to sum to zero.
Thus, it would appear that even if one neglects the complement subsystem in all other respects one still has to do the $\FrG$-variation for the whole universe. 
E.g. for a 2-island universe model with each island conserving its own angular momentum, the study of one island needs 
to involve $(\underline{\rho} \cr \underline{p})_{\mbox{\scriptsize subsystem}} = 
          - (\underline{\rho} \cr \underline{p})_{\mbox{\scriptsize environment}}$ which is then to take a free (or observed) value.  
This reflects that the best-matching argument has an inherently whole-universe character: rotation has to be of the whole universe 
rather than just of the subsystem under study, since rotation that subsystem and not the complement involves an in-principle discernible.
This amounts to there being a tension  between relationalism and Perspectivalism 1.  
It should also be clear from the above that this SSec's use of Sub$\FrQ$ is to include variable $E$ and $\sfJ$ 
parameters as part of its specification (with SubPhase, SubHilb etc. following suit).

\noindent Note 5) The Crane set-up then involves {\sl the entire set of SubQuant's'}, with Perspectivalism 2 necessitating yet further structure 
beyond that, as well as nontrivial modelling of observers as quantum systems (an idea nobody can really carry out for now, which also features 
in the thinking of Page \cite{Page1, Page2} and of Hartle \cite{IGUS}; the state of the art here is Hartle's IGUS model for observers (see also \cite{GMH11}). 

\mbox{ } 

\noindent The concepts embodied by Perspectivalism 3) (partial observables) and 4) of Sec\ref{Cl-Str}(any time) carry over straightforwardly to the quantum level.  
A form of Perspectivalism 4) purely at the QM level had previously been proposed by Page and Wootters \cite{PW83}, and this position 
(if not some of the accompanying technical details) was also taken up by Gambini--Porto--Pullin \cite{GPP03, GPP04a}.  
\noindent Also, Rovelli's partial observables and Bojowald et al's \cite{Bojo1, Bojo2, Bojo3, Bojo4} type of notion of 
localized clocks are part of a perspectival picture and have ties to Perspectivalism 1) or 1W).

\mbox{ } 

\noindent Considering subspaces of Hilbert space (rather than Hilbert space for a subsystem) is less of an issue in relational/perspectival theory.
Some possible uses of this are the subspace of pure states or the subspace of the naive Hilbert space that has been adjusted by such as $E_{\sU\sn\si}$ being fixed.  
On the whole, this means recycling Part I's objects buildup for (Sub)$\FrQ$ rather than building up structures on (Sub)Hilb.

\subsection{Grainings}\label{QM-Grainings}

\noindent Even more generally, a distinct quantum SM is now based on QM.  
Now imperfect knowledge is required in principle due to the nature of QM (see e.g. \cite{GMH06}).

\noindent Some notions of coarse graining here carry over from the classical level because the same spaces (Sub)$\FrS$ are in use.  
However, other notions will be new, via Sec \ref{Cl-Grainings}'s 2) now referring to quantum state spaces.
\noindent Coarse-graining calculations here consist of tracing out modes in (Sub)Hilb.

\subsection{State space unions and topology change}\label{TopCha}

Again, some such notions will carry over from the classical level because the same spaces (Sub)$\FrS$ are in use.  

\mbox{ } 

\noindent Analogy 94) Then one expects topological interaction terms much as one has particle-nonconserving interaction terms.   
These are unclear at the classical level (except possibly in a hydrodynamical limit with blobs), but at the quantum level, QFT is obviously of 
that nature.  
So, can we have a particle creation-annihilation incorporating theory of finite number of particles via some partly-QFT-like version of RPM's?  
Is there any way of classically modelling variable-number RPM?
This gives the idea that sum over topologies is a free theory, and the interacting theory would have topology-changing `ripping' operations. 
Sum over topologies is common in path integral approaches, as is transition amplitude for topology change.
Thus these are a starting point. 
Can lessons be learnt from those which are applicable to configuration space? 
The path integral form is however already ready-tailored for Histories Theory.  
This is one advantage of such which I point out as notable from perspective that fixed topology is itself a deeper absolutist intrusion into physics.  
But there is a problem insofar as the interpretation of second-quantizing as model universe creation rather than as particle creation? 
Or are both somehow possible interpretations of second-quantizing the RPM?
Second-quantizing nonrelativistic mechanics is well known.  
It is a QFT in the standard sense: infinite, variable-particle-number. 
One would postulate that the second $\psi$ has the wavefunction of the universe interpretation now.  
This is the opposite of second-quantizing as a toy model of third-quantizing as a POT scheme. 
It is nonstandard in being constrained (Dirac), or in being a curved-space theory (r-formulation).

\mbox{ } 

\noindent Note: the above allows for consideration of topology change.  
This is problematic to study from a canonical perspective, but standard to study using Feynman path integral approaches to Quantum Theory and to String Theory.

\mbox{ } 

\noindent See \cite{Duston} for another take on topological relationalism within (an extension of) Nododynamics.   
This approach considers not spin networks but `topspin networks'; assuming at least some of the structure of Nododynamics, 
one can then consider not just holonomies along loops but monodromies as well, relationalizing at least some topological features.

\subsection{State space distances for use in QM}\label{QM-Dist}

\noindent It is already clear from Sec \ref{QM-SubS} that localization in space also ties in well with basic aspects of Crane's program.
\noindent There is also the issue of notions of distance between QM states themselves, i.e. on such as Hilb, $\FrG$-MidHilb or SubHilb. 
On the one hand, new space, new study of suitable notions of distance!  
On the other hand, at least crude possibilities may be inherited from the classical $\FrQ$ 
(particularly if primality is giveable to this) as in e.g. the \NSII.  
It is a fair point that this is beyond where one should really have to face fermions, as such I place an extra problem on the \NSII: how does its 
configuration space region specification face up to generalizing to models including fermions?  
In particular, whether and how does its usual canonical GR specification carry over to Einstein--Dirac theory.
All in all, I leave this as

\mbox{ }

\noindent{\bf Question 68$^*$} What are the quantum state spaces for RPM's and for GR (at least formally in the latter case)? 
What are suitable notions of distance on these? 
Can such be chosen to comply with the common structures conjecture as regards the next SSec's quantum notions of information?
What about the larger spaces of mixed quantum states (c.f. Sec \ref{Mixed-State-Space}) for RPM's and for GR?

\subsection{Quantum Information}\label{QM-NOI}

Everitt \cite{EverittTh} considered using Shannon information itself to characterize quantum measurements.  
However, this misses out quantum effects.
A more solid quantum analogue of Shannon information (\ref{ShaCont} or \ref{ShaDisc}) is {\it von Neumann information}, 
\be
I_{\mbox{\scriptsize von Neumann}}[\BigupRho] = \mbox{Tr}(\BigupRho\,\mbox{log}\,\BigupRho)
\ee 
for $\BigupRho$ the density matrix of the quantum system.  
For, both are based on the $x$\,log\,$x$ function, and there is a decent classical--QM correspondence between them as well (see e.g. \cite{Wehrl, Mackey}).  

\mbox{ } 

\noindent Note 1) Furthermore the von Neumann notion survives the transition to relativistic QM, and that to QFT modulo a short-distance cutoff \cite{CC04, Wald}.  
As regards GR, it has been used in the context of black holes (see e.g. \cite{Wald}), though it should be stated that classical (never mind quantum) gravitational entropy 
is not a well-understood concept \cite{Weyls1, Weyls2, Weyls3, Weyls4, Weyls5, Smolin85, Brandenberger, RA, HBM, Wald}. 
Thus, merely formally,
\be
I[\BigupRho_{\mbox{\scriptsize QGrav}}] = 
\mbox{Tr}\BigupRho_{\mbox{\scriptsize QGrav}}\mbox{log}\BigupRho_{\mbox{\scriptsize QGrav}} 
\mbox{ } . 
\label{Winged-Pig}
\ee
However, either the quantum-gravitational density matrix $\BigupRho_{\sQ\sG\sr\sa\sv}$ is an unknown object since the underlying microstates are unknown.

\noindent Note 2) Rather than a notion of gravitational information that is completely general, a notion of entropy 
suitable for approximate classical and quantum cosmologies may suffice for the present study. 
While many candidate objects of this kind have been proposed at the classical level, it is unclear how many of these would arise from QG.

\noindent Note 3) At the quantum level, having the underlying QM generally allows one to obtain a corresponding SM. 
Moreover, perturbative solution of the quantum theory suffices \cite{LLSM} in order to build an approximate SM.
This provides further uses for the perturbative treatments of Secs 15-16, giving RPM's advantageous constructs as compared to inhomogeneous GR.  

\mbox{ }

\noindent The perturbative model context from which one would build a perturbative SM from Secs \ref{Cl-POT-Strat} and \ref{+temJBB}'s perturbative QM requires the canonical ensemble.  

\mbox{ }

\noindent {\bf Quantum non-impasse} Given a QM, one can in principle construct the corresponding SM.  
Furthermore, by Part II, a wide range of quantum RPM models have been solved, and are thus available to construct ensuing SM's 
which should inherit their relationalism from coming from a relationalist QM.  
The levels of structure beyond the wavefunctions and inner products solved for are firstly density matrices and then 
combinations of these such as the von Neumann entropy and the relative information.  
I do note that for now not much of the solving was localized subsystem-wise (e.g. finding separate wavefunctions for 
the base pair and the apex of the triangle) as required as inputs for relative-type notions of information (see however Appendix \ref{QM-Str}.B). 
(But, at least for metrolands and triangleland, the coordinates in question for doing so are clear enough -- relative Jacobi coordinates 
for the former and parabolic-type coordinates for the latter -- and are at least sometimes analytically tractable.  
Thus, while Part II's study is not specifically geared toward SM/quantum information, it is reasonably straightforward to 
proceed in that way for somesuch models).  
Moreover, by Note 2) of Sec \ref{QM-NOI}, the perturbative results of Secs \ref{Q4Stop-Pert} and \ref{QTri-Pert} continue 
to be relevant here as regards furnishing nontrivial but at least to start off with computable examples of approximate SM construction; 
due to the approximate nature of $E$ in these perturbative examples, this should be treated within a canonical ensemble.  
Moreover, these have not been explored beyond this point.  

\mbox{ }

\noindent{\bf Question 69} Is the counterpart of Question 31 at the quantum level.

\mbox{ } 

\noindent{\bf Question 70}$^{**}$ Also, Kendall's work directly relates only to classical RPM.  
How much of it can be used at the quantum level/has constructible quantum analogues?  
 
\mbox{ }

\noindent For now we simply conceptualize about further notions of information at the QM level rather than building concrete RPM examples for each.  

\mbox{ } 

\noindent {\bf Mutual information} (\ref{Mutua}) also applies to van Neumann information \cite{Preskill, DV12}, 
though now it is of the form
\beq
M_{\mbox{\scriptsize von Neumann}}[\BigupRho_A, \BigupRho_B, \BigupRho_{AB}] = 
I_{\mbox{\scriptsize von Neumann}}[\BigupRho_A] + 
I_{\mbox{\scriptsize von Neumann}}[\BigupRho_B] - 
I_{\mbox{\scriptsize von Neumann}}[\BigupRho_{AB}] \mbox{ } . 
\eeq
This does include quantum effects.

\mbox{ }  

\noindent {\bf Relative information} at the quantum level is \cite{NC, DV12}
\be
I_{\mbox{\scriptsize relative}}[\BigupRho_1, \BigupRho_2] = \mbox{Tr}(\BigupRho_1\{ \mbox{log}\BigupRho_1 - \mbox{log}\BigupRho_2\})
\ee
This can be interpreted as a state space distance.
Mutual information can then be seen as a distance between $\BigupRho_{AB}$ and the uncorrelated state $\BigupRho_{A} \bigotimes \BigupRho_B$. 
In ordinary QM, one can view this as a quantifier of entanglement.  

\mbox{ } 

\noindent Note 2) There is also an obvious quantum analogue of Tsallis information \cite{Tsallis}.      

\noindent Note 3) The above notions may well be of use in setting up the quantum cosmological counterpart of Hosoya--Buchert--Morita type inhomogeneity measures.  

\mbox{ }  

\noindent Finally, from (\ref{G-ket}) it is clear that configurationally-relational density matrices can be defined by 
\beq
\BigupRho_{\sFG} := \int_{\sfg \, \in \, \sFG} \mathbb{D}\ttg\mbox{exp}
\left(
i\sum\mbox{}_{\mbox{}_{\sfZ}} \stackrel{\rightarrow}{\FrG_{\sttg^{\tfZ}}}
\right) 
\, \BigupRho \, 
\mbox{exp}
\left(
- i\sum\mbox{}_{\mbox{}_{\sfZ}} \stackrel{\rightarrow}{\FrG_{\sttg^{\tfZ}}}
\right) \mbox{ } ;  
\eeq
clearly then projectors come out built the same way whether assembled out of states or treated as a subcase of operators.  
Then e.g. a configurationally relational version of von Neumann information is 
\beq
I^{\sFG}_{\mbox{\scriptsize von Neumann}} := 
\int_{\sfg \, \in \, \sFG} \mathbb{D}\ttg\,\mbox{exp}
\left(
i\sum\mbox{}_{\mbox{}_{\sfZ}} \stackrel{\rightarrow}{\FrG_{\sttg^{\tfZ}}}
\right) 
\, \BigupRho \, 
\mbox{exp}
\left(- 
i\sum\mbox{}_{\mbox{}_{\sfZ}} \stackrel{\rightarrow}{\FrG_{\sttg^{\tfZ}}}
\right) \mbox{ ln }
\left(
\int_{\sfg \, \in \, \sFG} \mathbb{D}\ttg\,\mbox{exp}
\left(
i\sum\mbox{}_{\mbox{}_{\sfZ}} \stackrel{\rightarrow}{\FrG_{\sttg^{\tfZ}}}
\right) 
\, \BigupRho \, 
\mbox{exp}
\left(- 
i\sum\mbox{}_{\mbox{}_{\sfZ}} \stackrel{\rightarrow}{\FrG_{\sttg^{\tfZ}}}
\right)
\right) \mbox{ } , 
\eeq
with configurationally relational versions of quantum mutual, relative and Tsallis information following suit in an obvious way.

\subsection{Correlations at the quantum level}\label{QM-Corr}

\noindent Example 1) Ordinary (Minkowski spacetime) QFT has n-point functions \cite{qftcorr} of the same well-known kind as in Example 1) of Sec \ref{Correl1}, 
where $\langle \mbox{ } \rangle$ now includes inserting the ground-state wavefunction at each end.      
This notion carries over to n-point functions for a simple `Mukhanov variables QFT' picture of Quantum Cosmology \cite{qftcoscorr}.    
Giddings--Marolf--Hartle furthermore present a useful treatment of correlators for Quantum Cosmology in  \cite{Giddings}; some aspects of this 
(and the relational underpinning of Example 1) go back to DeWitt \cite{DeWitt62, DeWitt67}).  
Need to be careful because at least some forms of n-point function are not manifestly already-relational.  
In such a case one could at least formally apply $\FrG$-act $\FrG$-all with the $\FrG$-all move being integration over the $\FrG$ in question,
\beq
\langle \varsigma(\ux_1) ... \varsigma(\ux_1) \rangle_{\sg} :=
\int_{\sg \in \sFG} \mathbb{D}\mg \int_{\FrQ} \stackrel{\rightarrow}{\FrG}_{\sg}\left\{ \mathbb{D} \varsigma \, \mbox{exp}(- \FS[\varsigma]) \varsigma(\ux_1) ... \varsigma(\ux_n)\right\} 
\left/
\int_{\sg \in \sFG} \mathbb{D}\mg \int_{\FrQ} \stackrel{\rightarrow}{\FrG}_{\sg}\left\{ \mathbb{D} \varsigma \, \mbox{exp}(- \FS[\varsigma])                        \right\} 
\right. 
\mbox{ } .  
\eeq
%

\noindent Example 2) Due to entanglement, QM has an extra kind of correlations that classical theory does not have.  
This leads to the concept of {\sl discord} = QM correlations - classical correlations, for which expressions using von Neumann and Shannon 
informations can be used \cite{DV12}; this clearly quantifies how Shannon entropy does not suffice at the quantum level. 

\noindent Example 3) Nododynamics/LQG has difficulties with this topic through standard machinery for building n-point functions being unavailable on account of background-independence.
The recent thesis \cite{AlesciTh} lists some works in that direction.

\subsection{Propositions at the quantum level}\label{QM-Prop}

Propositions now have an inherently probabilistic nature, i.e. they have to concern `Prob($X$) is' rather than `$X$ is'. 

\noindent Representing propositions at the quantum level by projectors is a key move, c.f. Sec \ref{PPA}. 
In ordinary Quantum Theory, for state $\BigupRho$ and proposition $P$ implemented by projector $\widehat{\mP}$, Prob($P$; $\BigupRho$) = tr($\widehat{\BigupRho}\widehat{\mP}$) 
with Gleason's Theorem providing strong uniqueness criteria for this choice of object from the perspective of satisfying the basic axioms of probabilities (see e.g. \cite{Ibook}).
Furthermore, in some approaches to the POT, one goes beyond (see \cite{PW83, IL2} or Secs \ref{CPI}, \ref{IL-Hist} 
the usual context and interpretation that are ascribed to projectors in ordinary Quantum Theory. 
E.g. desiring a projector implementation of propositions also led to IL's reformulation of Histories Theory (Sec \ref{IL-Hist}). 
The \CPI and Records within Histories Theory approach is also more satisfactory from this perspective.

On the other hand, the \NSI and Halliwell approaches are still using integration over classical regions as their implementation.   
Prima facie, this is suspect (see Sec \ref{QM-POT-Strat} and the Conclusion for more).  

\mbox{ }  

\noindent A more general structure brought in at the quantum level is a {\it lattice of propositions} \cite{IL2}.  
In fact, the minimal such structure is an {\it ortholagebra}, in which $P \preceq R$ iff $\exists S \in L$ such that $R = P \oplus 
S := P \vee S$ in cases in which $P$ and $S$ are disjoint, $\vee$  not being defined in other cases.
Passing to an orthoalgebra matters as regards supporting a satisfactory tensor product operation to support the Proposition--Projector association of Sec \ref{QM-POT-Strat}. 

\noindent In one approach, there has been consideration of quantum logics \cite{I94, Ibook, IL2} [which, in contradistinction to (\ref{Boo-Distrib}) are not distributive]. 
%

\noindent Another approach makes nontrivial use of Topos Theory \cite{ToposI, ToposTalk, ToposRev} (where answers can be multi-valued rather than YES/NO, as well 
as this {\it valuation} `differing from place to place', for these are geometrical logics whose valuations are now somewhat 
like locally-valid charts in differential geometry to Boolean logic's globally valid valuations being somewhat like flat space).  

\mbox{ } 

\noindent One of the many ways of envisaging a {\bf topos} is as a category with three extra structures: 
1) finite limits and colimits 2) power objects and 3) a subobject classifier \cite{LawRose, ToposRev}.
See ultimately \cite{Johnstone} for further ways.

From this perspective, however, classical Physics is easily included as a trivial case: the standard means of handling propositions in classical Physics turns out to be the category 
of sets alongside the subobject classifier YES/NO 
of the Boolean logic of classical Physics, which forms a trivial example of a topos (also here the power objects are just the power set -- set of all subsets).  
[Indeed it is the idealization that topoi are endowed with structure so as to resemble as closely as is possible for a wide range of 
branches of mathematics.]  
More concretely, classical propositions can be considered in terms of (Borel) subsets of $\FrQ$, Sub$\FrQ$, Sub$\FrS$, with 
intersection, union and complement playing the roles of AND, OR and NOT.  

\noindent One possible mathematical implementation of propositioning along the lines of 1) and 3) is {\bf Isham--Doering (ID)-type approaches} \cite{ToposI, ToposRev} 
to using Topos Theory  in Physics provides a large amount of candidate structure as regards realizing Prop($\FrS$) at the quantum level. 
The logic here is internal logic, which ID term `intuitionistic logic' \cite{bigcite2,ToposI,ToposTalk,ToposRev}.

\mbox{ }

\noindent Note 1) open (or clopen: simultaneously closed and open) subsets are used in Topos Theory, so that 
the most natural notion of localization in the simple Topos Theory applied to physics is topological rather than metric.
This is at some odds with Sec \ref{Cl-Str}'s conceptualization of localization.

\noindent Note 2) This approach once again gets by with a distributive logic. 

\noindent Note 3) Given a state, truth values are assigned to all propositions about the quantum system, in which sense this is a `neo-realist approach'.  

\noindent Note 4) {\it ``Constructing a theory of physics is then equivalent to finding a representation in a topos 
of a certain formal language that is attached to the system"} 
\cite{ToposI} (see also \cite{JLBell}).  
One considers maps from state objects in $\bupSigma$ to quantity-value objects in $R$, for $\bupSigma$ is the `linguistic precursor' of the state space $\FrS$.  
Moreover, one can make use to some extent of a formal language which is independent of theory type (classical/quantum) though $R$ clearly 
does depend on this: the propositional language PL($\FrS$).
In fact, ID study's mathematics has 2 languages at the quantum level: as well as the propositional language PL($\FrS$), it has a higher-order typed language, HOTL($\FrS$).  
The propositional language is simpler and more directly related to standard quantum logic. 

\noindent Note 5) In this approach, there is a map that maps projectors to subobjects of the spectral presheaf, which is the QM analogue of a classical state space.\footnote{A presheaf 
is a mathematical implementation of the idea of attaching local data to a structure; as such it ought to be of considerable interest 
in Records Theory and in Histories Theory, for all that hitherto these have been modelled with more (metric) structure.} 
%
Moreover, the topos perspective itself favours the HOTL($\FrS$) as the natural kind of language that can be represented in a topos; it is `internal' to it. 

\mbox{ } 

\noindent Note 6) This approach does require radical revision of QM itself \cite{ToposI, ToposRev}, amounting to avoiding the obstruction in standard QM due to 
the {\bf Kochen--Specker Theorem}, which is, schematically (see e.g. \cite{Ibook}),
\beq
\mbox{Function(Value($\widehat{A}$)) \mbox{ } need not be the same as \mbox{ } Value(Function($\widehat{A}$)) for Hilbert spaces of dimension $> 2$ } .   
\eeq
\noindent Note 7) The role of logic played in classical physics by the Boolean algebra is played more generally in QM-relevant topoi by 
a Heyting algebra \cite{ToposRev} (this and the Boolean algebra are both distributive lattices, the difference being that the general Heyting 
algebra, unlike the Boolean Algebra has no law of the excluded middle).  

\noindent Note 8) QM ends up involving the topos $\underline{\mbox{\textgoth Sets}}\mbox{}^{\mbox{\scriptsize\textgoth V}(\sH\si\sll\sb)^{\to\tp}}$, i.e. the topos of contravariant (`opposite') 
valued functors on the poset category \textgoth{V}(Hilb) of commutative subalgebras of the algebra of bounded operators on the Hilbert space of the system, Hilb.  

\noindent Note 9) The $\FrG$'d counterpart of this is for now only {\sl posed} in the Conclusion.

\vspace{10in}  

\noindent{\huge\bf IV QUANTUM PROBLEM OF TIME}

\section{Facets of the Quantum Problem of Time}\label{QM-POT}

The greater part of the POT in QG \cite{Battelle, Kuchar81, Kuchar91, Kuchar92, I93, Kuchar99, Kieferbook, Rovellibook, 
Smolin08, APOT, APOT2} occurs because the `time' of GR and the `time' of Quantum Theory are mutually incompatible notions.
This causes difficulty in trying to replace these two branches of physics with a single framework in regimes in which neither Quantum Theory 
nor GR can be neglected, such as is needed in parts of the study of black holes or of the very early universe.
The POT is pervasive throughout sufficiently GR-like attempts at doing Quantum Gravity, both at the quantum and at the classical level of the foundations in question.    
As outlined in Sec \ref{Intro} and studied in detail at the classical level in Sec \ref{Cl-POT}, it is multi-faceted.
Resolving this incompatibility is clearly of importance as regards Theoretical Physics forming a coherent whole. 
Study of the POT is also important toward acquiring more solid foundations for the gradually-developing discipline of Quantum Cosmology (see e.g. \cite{HT, H03, Wiltshire, Kiefer99}).

\subsection{Quantum clocks}\label{QM-Clock}

There are reasons other than non-whole-universe considerations not to believe in perfect clocks at the quantum level.   

\noindent See e.g. Unruh and Wald's account \cite{UW89} in which it is demonstrated that all quantum clocks occasionally run backwards.   

\mbox{ }  

\noindent Also, for a quantum clock\footnote{This is valid for single analogue, as opposed to digital, quantum clocks.} of mass $M$ 
to run for a maximum time interval ${\cal T}$ with an accuracy (minimal discernible times) $\tau$, the Salecker--Wigner inequalities hold
\beq
\mbox{linear spread } \mbox{ }  \lambda    \geq    2\sqrt{\frac{\hbar {\cal T}}{M}} \mbox{ } , 
\eeq

\noindent
\beq
M                               \geq    \frac{4\hbar}{c^2\tau}\,\frac{{\cal T}}{\tau} \mbox{ } .  
\eeq
These apply to atomic clocks and also for Einsteinian mirror clocks; the latter are useable in the GR context, e.g. by combining the Schwarzschild radius with the inequalities. 
This gives (e.g. \cite{Ng}) an alternative derivation of the Hawking lifetime 
\beq
{\cal T} \leq  \tau_{\sP\sll\sa\sn\scc\sk} \left\{\frac{M}{M_{\sP\sll\sa\sn\scc\sk}}\right\}^3 =: {\cal T}_{\sH\sa\sw\sk\si\sn\sg}
\eeq
for a black hole as an upper bound on the longevity of a black hole cast in the role of a clock.

\subsection{The Frozen Formalism Facet of the POT} 

The canonical approach at the classical level gives a constraint that is quadratic in the momenta whilst containing no linear dependence on the momenta.  
For GR, this is the Hamiltonian constraint, which, including matter, takes the form\footnote{ 
$\tcH^{\mbox{\tiny matter}}$ is proportional to the energy density of the universe model's matter, 
and $\tcM_{\mu}^{\mbox{\tiny matter}}$ is proportional to the momentum flux of the universe model's matter.}
\beq
\scH := {\mN}_{\mu\nu\rho\sigma}\uppi^{\mu\nu}\uppi^{\rho\sigma}/\sqrt{\mh} - \sqrt{\mh}\{\mbox{${\cal R}\mi\mc$}(\ux; \bh] - 2\Lambda\} + 
\scH^{\sm\sa\st\st\se\sr} = 0  \mbox{ } .  
\eeq
Then, recapping on Sec \ref{POTiQG}, promoting an equation with a momentum dependence of this kind to the quantum level does not give 
a TDWE such as (for some notion of time $t$ and some quantum Hamiltonian $\widehat{\fH}$), 
\beq
i\hbar\partional\Psi/\ordial t = \widehat{\fH}\Psi
\eeq
as one might expect, but rather a stationary, i.e. frozen, i.e. timeless equation 
\beq
\widehat\scH\Psi = 0 \mbox{ } .  
\label{calH}
\eeq
In the case of GR, this is a WDE,  
\beq
\hat\scH\Psi := -\hbar^2 \mbox{`}\left\{\frac{1}{\sqrt{{\mM}}}\frac{\delta}{\delta \mh^{{\mu\nu}}}
\left\{
\sqrt{{\mM}}{\mN}^{\mu\nu\rho\sigma}\frac{\delta\Psi}{\delta \mh^{{\rho\sigma}}}
\right\} 
- \xi \,\mbox{${\cal R}\mi\mc$}_{\sbM}(\ux; \bh]\right\}\mbox{'}\,\Psi -  \sqrt{\mh}\mbox{${\cal R}\mi\mc$}(\ux; \bh]\Psi + {\sqrt{\mh}2\Lambda   }\Psi  
+ \hat\scH^{\mbox{\scriptsize matter}}\Psi = 0  \label{WDE3} \mbox{ }      
\eeq
modulo the discussion in Sec \ref{Wave-Eqs}.
This suggests, in apparent contradiction with everyday experience, that noting at all {\sl happens} in the universe! 
Thus one is faced with having to explain the origin of the notions of time in the laws of Physics that 
appear to apply in the universe; this paper reviews a number of strategies for such explanations.   
[Moreover timeless equations such as the WDE apply {\sl to the universe as a whole}, whereas the more ordinary laws of 
Physics apply to small subsystems {\sl within} the universe, which does suggest that this is an apparent, rather than actual, paradox. 
See also the comments on Perspectivalism in Secs \ref{Cl-Str} and \ref{QM-Str} as regards the pervasive role of subsystems in Physics.]


\noindent Note 1) This can all be taken to rest on the classical Frozen Formalism Problem, which comes about from Leibniz's principle of 
Temporal Relationalism for the universe as a whole.

\noindent Note 2) Whilst this is classically resolved by $\ft^{\se\sm(\sJ\sB\sB)}$, this however fails to unfreeze the quantum equation, so one needs to start afresh.  

\mbox{ }

\noindent Analogy 95) For RPM's, the energy constraint (\ref{EnCo}) likewise manifests the Frozen Formalism Facet of the POT.   

\noindent Analogy 96) $\ft^{\se\sm(\sJ\sB\sB)}$ does not unfreeze this for RPM's or GR.  

\mbox{ }

\noindent As per Sec \ref{POTiQG}, one follow-on Facet from the Frozen Formalism Facet is the {\bf Hilbert Space/Inner Product Problem}, 
i.e. how to turn the space of solutions of the frozen equation in question into a Hilbert space.
See Sec \ref{TPQ} for more about this.

\subsubsection{Scope of Jacobi mechanics and minisuperspace toys at the quantum level}\label{Q-Jac-POT}

At the QM level, these are models for the Frozen Formalism Problem, Dirac beables the Global POT and the Multiple Choice Problem.  

{            \begin{figure}[ht]\centering\includegraphics[width=0.97\textwidth]{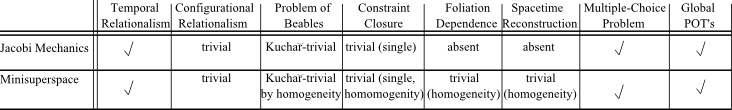}
\caption[Text der im Bilderverzeichnis auftaucht]{        \footnotesize{A start on a more compact presentation (`bestiary') of which models exhibit which facets. 
}  } \label{Temporary-Tab}\end{figure}          }

{            \begin{figure}[ht]\centering\includegraphics[width=0.8\textwidth]{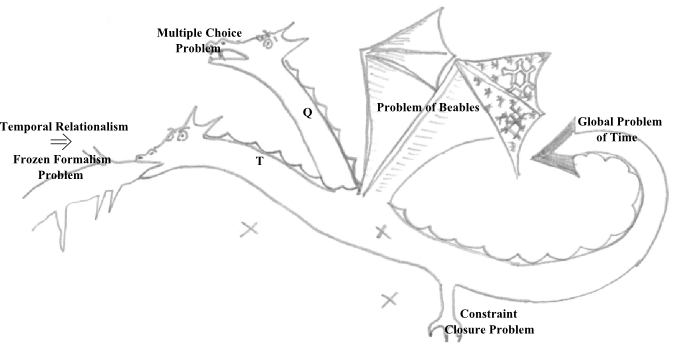}
\caption[Text der im Bilderverzeichnis auftaucht]{        \footnotesize{The Quantum Jacobi Mechanics Toy Ice Dragon is 5/8ths strength}  } \label{Temporary-Tab2}\end{figure}          }

{            \begin{figure}[ht]\centering\includegraphics[width=0.72\textwidth]{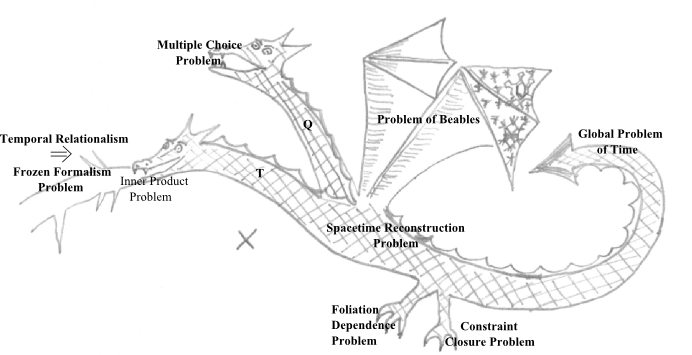}
\caption[Text der im Bilderverzeichnis auftaucht]{        \footnotesize{The Quantum Minisuperspace Mechanics Toy Ice Dragon is 7/8ths strength \cite{AMini1}, 
though homogeneity continues to easily overcome the bottom 3 facets.}  } \label{Temporary-Tab3}\end{figure}          }

\subsection{The further facets in detail}\label{FacetsDet}

\subsubsection{Configurational relationalism}\label{QM-CR-bis}

\noindent This is already taken care of at the classical level in most of the schemes favoured by \K's principle in this Article.   
\noindent If not, one has a Dirac quantization scheme, whether or not with classical time and classical observables available at its outset.    
\noindent There are also $\FrG$'d versions of quantum structure, as per Sec \ref{QM-Str}.

\subsubsection{Problem of Beables}\label{QM-Beables}

The QM versions of the definitions are now in terms not of Poisson brackets but of commutators, 
\beq
\mbox{Quantum Dirac beables: $\widehat{\iD}$ such that}  [\widehat{\iD}, \widehat\scC_{\sfA}] = 0 \mbox{ } .
\eeq
\beq
\mbox{Quantum \K beables and degradeables $\widehat{\iK}$ such that}  [\widehat{\iK}, \widehat{\scL\scI\scN}_{\sfZ}] = 0 \mbox{ } .
\eeq
Quantum partial observables are defined exactly as before too, except that `produce a number' now carries inherent probabilistic connotations.

Thus, for Geometrodynamics, 
\beq
[\widehat{\iD}, \widehat\scM_{\mu}] = 0 \mbox{ } , 
[\widehat{\iD}, \widehat\scH] = 0       \mbox{ } , 
\eeq
and
\beq
[\widehat{\iK}, \widehat{\scM}_{\mu}] = 0 \mbox{ } .
\eeq
The {\bf Problem of Quantum Beables} is then that it is hard to come up with a sufficient set of these for QG Physics.  

\mbox{ } 

\noindent Three reasons why a classical resolution of the Problem of Beables may not pass over to the quantum level are as follows.

\mbox{ }  

\noindent Subfacet 1) We know from Sec \ref{QM-Intro} that the commutator algebroids are not necessarily the same as Poisson algebroids for a given system. 

\noindent Subfacet 2) We also know from Sec \ref{QM-Intro} that we need to select a subalgebra/oid of classical objects to promote to quantum operators; 
this applies to whatever collections of \K or Dirac observables that one may have at the classical level.  

\noindent Subfacet 3) Whether a given object's brackets with the constraints contribute to equate to zero will clearly depend on 
how one operator-orders the candidate quantum observable corresponding to that object as well as on how one operator-orders the quantum constraints.  

\noindent Subfacet 4) is Note 3) of \ref{MCP}.

\subsubsection{Foliation Dependence Problem}\label{QM-Fol-Dep}

\noindent For all that a quantization of GR retaining the nice classical property of refoliation invariance would be widely seen as conceptually sound and appealing, 
 at the quantum level there ceases to be an established way of guaranteeing this.    
If not, as \K states \cite{Kuchar92}, {\it ``When one starts with the same initial state} $\Psi_{\si\sn}$ 
{\it on the initial hypersurface and develops it to the final hypersurface along two different routes} 
\beq
\Psi_{\sf\si\sn-1} \neq \Psi_{\sf\si\sn-2}  
\label{fol-dep}
\eeq
{\it on the final hypersurface.  
Such a situation certainly violates what one would expect of a relativistic theory."}  
I refer to eq. (\ref{fol-dep}) as the {\it quantum foliation-dependence criterion}.

\mbox{ } 

\noindent That this is obviously a time problem follows from how each foliation by spacelike hypersurfaces is orthogonal to a GR timefunction: each slice can be interpreted as an instant of time 
for a cloud of observers distributed over the slice, and each foliation corresponds to the these moving in a particular way.  

\mbox{ } 

\noindent Difference 42) RPM's do not have anything like the Foliation Dependence Problem, since the foliation of spacetime/embedding into spacetime meaning of GR's Dirac 
algebroid of constraints is lost through toy-modelling it with rotations (and/or dilations).  

\mbox{ } 

\noindent Note: whilst the two preceding sentences are already true at the classical level, however there the Foliation Dependence is a {\sl resolved} Problem...

\subsubsection{Constraint Closure Problem now reduces back to the Partional-Evolution Problem}\label{FEP-Intro}

This is the quantum-level problem that the commutator of constraints is capable of manifesting, involving breakdown of the 
immediate closure of the constraint algebroid that occurs at the classical level:
\beq
\{\scC_{\Gamma}, \scC_{\Lambda}\} = 0 \not{\vspace{-0.04in}\Rightarrow}  \mbox{ } [\widehat\scC_{\Gamma}, \widehat\scC_{\Lambda}] = 0
\eeq
(the equals are `perhaps weakly').

\mbox{ } 

\noindent One factor affecting this is that, for a given system, the commutator algebroid is not necessarily the same algebroid as the Poisson-brackets algebroid.   

\mbox{ } 

\noindent Moreover, how one operator-orders the constraints in promoting these to the quantum level affects whether and how these close under commutators.

\noindent Breakdown of constraint algebroid closure at the quantum level is termed the {\bf Functional Evolution Problem} in \cite{Kuchar92, I93}. 
However, this name carries field-theoretic connotations (from the type of derivative that features in the formulation of the brackets); this is, rather, 
a partial derivative for finite theories, so the overall portmanteau is the partional derivative, and so {\bf Partional Evolution Problem} is a more theory-independent name.
Finally, to include the possibility of classical counterparts of this problem, it is best to generalize further to {\bf Constraint Closure Problem}.  

\mbox{ }

\noindent Avoiding this problem via anomalies arising at the quantum level, from Dirac's perspective  \cite{Dirac}, requires `luck'.   

\noindent Moreover, only some \cite{Kuchar89, Torre07} of the anomalies that one finds in physics \cite{ModernAnomalies1, ModernAnomalies2} are time-related.
However, in the present case of the quantum counterpart of the Dirac algebroid of GR constraints, this is a time issue due to what these constraints signify.
In particular, non-closure here is a way in which the Foliation Dependence Problem can manifest itself, through the non-closure becoming entwined with details of the foliation. 

\mbox{ } 

\noindent N.B. following on from (c.f. Sec \ref{Q-G-Comp}), this is a second round of a posteriori rejection of a $\FrG$ at the level of the quantum constraint algebroid.  

\mbox{ } 

\noindent Difference 43) 1- and 2-$d$ RPM's {\sl are} `lucky in Dirac's sense': one can formulate these in a form that is free of anomalies and all other aspects of the 
$\partional$-Evolution Problem.   
This is the reduced from.
Lostaglio \cite{LostaglioMA} has shown that if one uses a Dirac approach (*) and Sec 2.3.1's conceptualization (**) for pure-shape RPM, the classically strongly cancellable 
obstruction term spawns an anomaly. 
Moreover, this leads not only to emergent scale but to another type of emergent time as well.  
However, (*) and (**) are substantial caveats; using sec 2.3.2's conceptualization and/or an r-approach, one sees no such anomaly as per Part III. 
In order to extend this sort of emergent scale-and-time to GR, one needs to work with a formulation containing strong obstructions, i.e. the conformogeometrodynamical 
version of RWR (not yet done as per Question 2).

\subsubsection{Spacetime Reconstruction (or Replacement) Problem}\label{QM-SRP}

There are further issues involving properties of spacetime being problematical at the quantum level.  

\noindent 1) At the at the quantum level, fluctuations of the dynamical entities are unavoidable, 
i.e. here fluctuations of 3-geometry, and these are then too numerous to be embedded within a single spacetime (see e.g. \cite{Battelle}).  
This is by the breakdown of the beautiful geometrical way that classical GR manages to be reslicing invariant.   
Thus (something like) the superspace picture (considering the set of possible 3-geometries) might be expected to take over from the spacetime picture at the quantum level.  
It is then not clear what becomes of causality (or of locality, if one believes that the quantum replacement for spacetime is `foamy' \cite{Battelle}).  
in particular, microcausality is violated in some such approaches \cite{Savvidou04a, Savvidou04b}.

\noindent 2) There is also an issue of recovering continuity in suitable limits in approaches that treat space or spacetime as discrete at the most fundamental level .  

\noindent 3) {\bf Recovering continuity and}, a forteriori, {\bf something that looks like spacetime} (e.g. as regards dimensionality) is 
\noindent an issue in discrete or bottom-up approaches to Quantum Gravity (see e.g. \cite{Regge, Perez03, Sorkin03, disc1, disc2, disc3, disc4}).
This is not a given, since some approaches give unclassical entities or too low a continuum dimension.  
x
\noindent The choice of the `scales' for this part of the POT `Ice Dragon' is lucid insofar as 

\noindent A) classical spacetime a particularly solid and intermeshed structure as per Sec \ref{Examples}.  

\noindent B) The last task faced is often to see the extent to which one's approach successfully deals with this reconstruction, 
so it makes sense to view this as the innermost layer of defense for those who manage to dance past the frozen breath, 
teeth, wings, tail and claws; unfortunately none of that virtuosity counts for much if one cannot plant one's lance through this final barrier...  

\mbox{ } 

\noindent Difference 44) There is no Spacetime Reconstruction Problem for RPM's since they have no nontrivial GR-like spacetime notion.    

\mbox{ } 

\noindent Issues in Nododynamics/LQG concerning the recovery of semiclassical limit, or of the flat spacetime limit for the purpose of the recovery of standard Particle Physics results, 
could be viewed along such lines.

\subsubsection{Multiple Choice Problem}\label{MCP}

\noindent The {\bf Multiple Choice Problem} alias {\bf Kucha\v{r}'s Embarrassment of Riches} \cite{Kuchar92} is the purely 
quantum-mechanical problem that different choices of time variable may give inequivalent quantum theories.  
[The riches are then the multiplicity of such inequivalent quantum theories.]  
It is a subcase of how making different choices of sets of variables to quantize may give inequivalent quantum theories, which follows from e.g. the Groenewold--van Hove Theorem.  

\mbox{ }

\noindent Analogy 97) The Groenewold--Van Hove phenomenon (see \cite{GVH1, GVH2} for the originals, \cite{GvH+RecentDiscuss1, GvH+RecentDiscuss3} for more recent considerations and 
\cite{Gotay95, Gotay96, Gotay96b, Gotay97, Gotay99, Gotay00} for more recent technical papers on particular examples, the last of which reviews all of the others) indeed already occurs 
for finite theories and thus can be expected to indeed occur for RPM's just as it occurs in minisuperspace \cite{I93}.    
Both RPM's and Geometrodynamics manifest the Multiple Choice Problem facet of the POT \cite{Kuchar92, I93}. 
[Despite the fairly widespread `belief' otherwise this phenomenon is not QFT-specific but already occurs in finite systems.]

\mbox{ }

\noindent There is a notion of time among the variables; the Multiple Choice Problem also strikes at the beables in {\sc b ... q} approaches.

\mbox{ }  

\noindent Note 1) Foliation Dependence is one of the ways in which the Multiple Choice Problem can manifest itself.

\noindent Note 2) Moreover, the Multiple Choice Problem is known to occur even in some finite toy models \cite{Kuchar92}. 

\noindent Note 3) Observables/beables themselves suffer from a Multiple Choice Problem if they are to be chosen as a subalgebra of the classical observables/beables.

\mbox{ } 

\noindent This is the `teeth' of the {\sc q}-head in the full-blown POT Ice Dragon, making dealing with the {\sc q}-head 
(and whatever one does prior to dealing with the {\sc q}-head altogether more difficult).

\subsubsection{Global POT}\label{QM-Global}

The equivalent of `meshing charts' in using multiple time coordinates is even less clear at the quantum level, where one is now `meshing together' {\sl unitary evolutions}. 

\mbox{ }  

\noindent Many of the facets and strategies also have further kinds of globality issues; at the QM level, some of these are as follows (see \cite{AGlob} for a more detailed account).

\noindent \bu Globality in time read off a given clock is limited in sufficiently long-lived universes 
and/or sufficiently accurate clocks by the Salecker--Wigner/Hawking lifetime bounds.

\noindent \bu One has to be careful to have a globally-valid choice of subalgebra of quantum objects, which in some approaches will contain some objects with time and/or 
frame connotations; furthermore this is the case for each choice made in investigating the Multiple Choice Problem.

\noindent \bu Quantum observables/beables may be localized notions (see Secs \ref{Global} and \ref{QM-KO}), in which case one has to think how to pass from one set of these to another.  

\noindent \bu Contrast kinematical quantization's operators being global with how some conceptualizations of observables/beables can have local connotations. 
Thus identifying these two algebras two up may be difficult. 

\noindent \bu Anomalies have some topological connotations; by that, the quantum Constraint Closure Problem is at least in part global in nature.  

\noindent \bu Quantum-level Spacetime Reconstruction may well continue to be only locally demonstrable/protected-by theorems, that is locally 
both in space and in time, given that this already was the case at the classical level.

\subsection{Aside: the Arrow of Time}\label{AOT}

\noindent One further issue that is usually considered to be not part of the POT mismatch but rather a further time problem is the {\bf Arrow of Time}.  
This is a further issue since it concerns not `what happens to the notion of time upon jointly considering Quantum Theory and GR' but rather `why is there a 
(theory-independent and in practise observed) consistent direction to time corresponding to a very tangible distinction between past and future?' 
\noindent The various Arrows of Time are thermodynamical, cosmological, radiative (electromagnetic radiation not being a mixture of advanced and retarded), causal, 
quantum-mechanical (due to collapse of the wavefunction), weak-force and psychological \cite{AOT92, Savitt}.  
The issue is how and why these are aligned, and whether there is a coherent primitive explanation of this.  
Quantum Gravity or Quantum Cosmology are then sometimes evoked in attempts to better explain this than one has managed to do with more everyday physical theories \cite{AOT92}.  
As regards RPM's being useful for investigating the Arrow of Time, Barbour conjectures about this in \cite{EOT}, however I am not aware of any concrete evidence for this.  
See Sec \ref{Records} for a few more comments.

\subsection{Diffeomorphism caveats are a a major frontier}\label{QM-Diffeos} 

\noindent Difference 45) Following on from Difference 29) about the significance of Diffeomorphisms, 
Isham and Kucha\v{r} \cite{IK85} and much of Isham's review \cite{I84} involve mostly quantum issues that are not captured by RPM's.  
2 + 1 GR and the bosonic string embody more of the character of the diffeomorphisms, with midisuperspace bearing an even closer parallel.  

\vspace{10in}

\section{QM parts of POT Strategies}\label{QM-POT-Strat}
%
{            \begin{figure}[ht]\centering\includegraphics[width=1.0\textwidth]{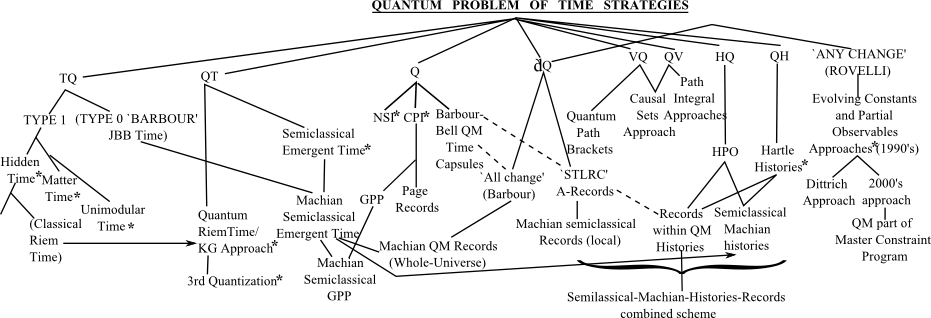}
\caption[Text der im Bilderverzeichnis auftaucht]{        \footnotesize{The quantum-level extension of Fig \ref{Temporary-Web-Cl}'s POT strategies. 
* indicates the 10 strategies covered by Kucha\v{r} and Isham's classic reviews \cite{Kuchar92, I93}.}  } \label{Temporary-Web-QM}\end{figure}          }

\subsection{QM Part of Tempus Ante Quantum ({\sc tq})} 

One has
\beq
\hat\scH\Psi = 0
\label{WDE-AGAIN} \mbox{ } ,   
\eeq
(Wheeler--DeWitt equation: WDE) in place of ordinary QM's TDSE, 
\beq
i\pa\Psi/\pa t = \hat{H}\Psi \mbox{ }  .
\eeq  
Here, $H$ is a Hamiltonian, $\Psi$ is the wavefunction of the universe and $t$ is absolute Newtonian time [or a local GR-type generalization $\mt(\ux)$, in which case one also requires 
a functional derivative $\delta/\delta \mt(\ux)$ in place of the ordinary derivative in the given equation].   
This clearly manifests the quantum Frozen Formalism Problem.

\subsubsection{Jacobi--Barbour--Bertotti time (Type 0 tempus ante)}

The above can be taken to rest on Leibnizian Temporal Relationalism present ab initio at the classical level.  
One can resolve the subsequent classical Frozen Formalism Problem by classical Machian emergent time (Sec 7), but that does not alter the frozen form of the WDE. 
Thus (as per Fig \ref{Ante-1-and-2}) one needs to start afresh at the quantum level (Sec \ref{Semicl-Intro}).

\subsubsection{Quantum unhidden time}

What about using alternative theories in place of GR? 
E.g. \cite{Brout, EOT, H00} argue that successors to the WDE through GR being supplanted would likely continue to have a Frozen Formalism Problem. 
I largely agree with this, for all that the rare counterexample in Sec \ref{BoHo} produces an unhidden-TDSE.

\subsubsection{Quantum part of Type 1 Tempus Ante Quantum}

\noindent The TDSE's for this have in principle Schr\"{o}dinger inner products and interpretation of QM leaning on the classical time in parallel to the Newtonian case.  
Here, the functional dependence of $H^{\st\sr\su\se}$ on the other variables is not known (or at best known implicitly via the solution of such a partial differential equation). 
And thus the quantum `true Hamiltonian' $\widehat{H^{\st\sr\su\se}}$ cannot be explicitly defined as an operator 
(or, at best, is very prone to having the horrendous operator-ordering and well-definedness issues that only knowing its form implicitly entails.)
It is also the case that simpler models than full Geometrodynamics give severe difficulties just beyond where Geometrodynamics has this impasse.    

\mbox{ } 

\noindent The general classical parabolic form (\ref{Para1}) for the formerly purely-quadratic constraint in for now the $\FrG$-trivial context (minisuperspace) yields the TDSE 
\beq
i\hbar \pa\Psi/\pa t^{\sa\sn\st\se} = \hat{H}_{\st\sr\su\se}\lfloor t^{\sa\sn\st\se}, 
Q_{\so\st\sh\se\sr}^{\sfA}, P^{\so\st\sh\se\sr}_{\sfA}\rfloor\Psi \mbox{ } . 
\label{TEEDEE1}
\eeq 
%
%
This  unfreezing is accompanied, at least formally, by the obvious associated Schr\"{o}dinger inner product, which frontal assault renders the Ice Dragon's {\sc t}-head toothless.  

\noindent A particular example of this is then the minisuperspace York-TDSE
\beq
i\hbar \pa\Psi/\pa t^{\sY\so\sr\sk}  = \widehat{H^{\st\sr\su\se}}(Q_{\st\sr\su\se}, P^{\st\sr\su\se}, t^{\sY\so\sr\sk} ) \Psi 
\mbox{ }   .
\label{YorkTDSE}
\eeq

\subsubsection{Quantum part of straightforward and reference matter time approaches}

The minisuperspace versions of these are likewise TDSE's with respect to their candidate times.

\subsection{Tempus Post Quantum ({\sc qt})}\label{TPQ}

\noindent {\bf Post Postulate} In strategies in which time is not always present at the fundamental level, time is nevertheless capable of emerging in the quantum regime.  
Because this is an emergence, it means that the Hilbert space structure of the final quantum theory is capable of being (largely) unrelated to that of the WDE-type quantum 
theory that one starts with.   
Such emergent strategies are of the following types.

\subsubsection{Attempting a Schr\"{o}dinger Interpretation} \label{KGI}

\noindent Schemes based on a Schr\"{o}dinger-type inner product fail due to the indefiniteness of the supermetric underlying the WDE.  

\mbox{ } 

\noindent Difference 46) There is (mostly) No Inner Product Problem for RPM's, since these come with positive-definite kinetic metrics, 
so the corresponding natural Schr\"{o}dinger inner product suffices.

\subsubsection{Attempting a Klein-Gordon Interpretation based on Riem time}\label{KGI2}

\noindent The preceding SSSec suggests thinking about the WDE not as a TISE but as an analogue of the Klein--Gordon TDWE (\ref{KG-type}) 
with a corresponding Klein--Gordon type inner product.  
In this way, this approach has QFT, and thus spacetime, undertones.
However, there are the following issues with this strategy. 

\mbox{ } 

\noindent Problem with Riem time 1)
Superspace null cones are not respected by superspace trajectories, limiting the analogy. 

\noindent Problem with Riem time 2) 
{\bf Attempting a Klein-Gordon Interpretation based on Riem time} fails regardless of whether the scheme is {\sc QT} or {\sc TQ}.  
This is because the scheme has an Inner Product Problem with its candidate Klein--Gordon inner product\footnote{The capital 
Latin indices arise from DeWitt's 2-index to 1-index map.  
$\stackrel{\longleftrightarrow}{\delta_B}$ denotes functional derivative with both backwards as well as forwards action with respect to $\mh^{B} = \mh_{\mu\nu}$.} 
\beq
\langle\Psi_1[\bh]|\Psi_2[\bh]\rangle  = \frac{1}{2i}\prod\mbox{}_{\mbox{}_{\mbox{\scriptsize $x \in \Sigma$}}}
\int_{\mbox{\scriptsize Riem}(\bupSigma)}
\mathbb{D}\Upsilon_{A}{\mM}^{AB}(\bh)
\left\{
\Psi_1[\bh]\stackrel{\longleftrightarrow}{\delta_{B} }\Psi_2[\bh]
\right\} 
\mbox{ } , 
\eeq
where $\mathbb{D}\Upsilon_A$ is a suitable directed `hypersurface volume element' in Riem($\bupSigma$) at the point $x \in \bupSigma$.\footnote{I believe 
I have improved the clarity of presentation by using a distinct letter $\Upsilon$ for `hypersurface in configuration space', for all that $\Upsilon$ bears 
the stated {\sl labelling} relation to $\Sigma$.  
Isham \cite{I93} further asserts that what I call $\Upsilon$ need furthermore be spacelike with respect to $\bM$, 
and that making this inner product rigorous is difficult. 
[It is certainly only intended as a formal expression which has yet to take into account the momentum 
constraint, for instance, which he would formally do by `projecting the inner product down to' Superspace($\bupSigma$).   
This is of course brushed under the carpet in this approach's ubiquitous minisuperspace examples.]  
The conceptual core, however, is clear: the expression is {\it ``invariant under deformations of the `spatial' hypersurface in Riem($\bupSigma$)"} \cite{I93}, 
which is (paraphrasing) the QG analogue of the normal Klein--Gordon inner product's time-independence property.}   
%
The problem is then due to a breakdown of the analogy between Geometrodynamics and stationary-spacetime Klein--Gordon Theory.  
\noindent While there is a conformal Killing vector on superspace \cite{Kuchar81}, the GR potential does not in general respect this, and so this scheme fails.  

\mbox{ } 

\noindent Difference 47)  Due to Difference 31) in definiteness of the kinetic terms, the Schr\"{o}dinger inner product scheme is available for RPM's but not for GR.  
Also, the Klein--Gordon-type scheme (is not available for RPM's.  

\noindent Problem with Riem time 3) The positive--negative modes split of states in the usual Klein--Gordon theory arises from the presence of a privileged time; 
thus, without such a privileged time in the general GR case, one's quantization scheme will not have this familiar and useful feature.    

\mbox{ }  

\noindent Difference 48) Finally, each of definiteness and indefiniteness furnishes qualitatively different mathematics: 
elliptic operators for RPM's versus hyperbolic ones furnishing Riem time for GR (e.g. in the minisuperspace case, or some infinite-dimensional analogue for full GR).  

\mbox{ }

\noindent Noting however the parallel with Klein--Gordon failing as a first-quantization leading to its reinterpretation as a second-quantized QFT, one might then try the following.

\subsubsection{Third Quantization}\label{3rd} 

This suggestion then is that the solutions $\Psi[\bh]$ of the WDE might be turned into operators, so that one now has an equation
\beq
\widehat\scH\widehat{\Psi}\psi = 0 \mbox{ } .  
\eeq

\noindent
[{\sl Second and third} quantizations are maps between (Hilb, Uni) type spaces.  
It is only  {\sl first} quantization and overall quantization are formally maps between (Phase, Can) and (Hilb, Uni).] 
While Third Quantization is of interest as regards various technical issues, it was not held to provide a satisfactory approach to the POT  
up to the early 90's \cite{Kuchar92, I93}, and I am not aware of any subsequent advances in this respect. 

\mbox{ }

\noindent Difference 49) Also, the third quantization scheme makes no sense in RPM's, due to these being finite rather than field-theoretic.   
Second quantization is the RPM analogue (the key is that wavefunctions of the universe are themselves quantized, not how many quantizations are needed to get to that stage. 
This is technically like QFT, but interpretationally the wavefunction of the finite model universe has been elevated to a quantum operator 
(which parallels the status of the wavefunction of the universe for infinite theories in third quantization).  
Moreover, even second quantization is unnecessary and absurd given that the Schr\"{o}dinger inner product works just fine here. 

\mbox{ }

\noindent Also, Third Quantization unfortunately turns out not to shed much light on the POT \cite{Kuchar92}, and, whilst Third Quantization recurs \cite{GFT} in the Group Field Theory 
approach to Spin Foams (\cite{Perez}, Sec \ref{Histor}), its use there is not known to extend to furnish a POT strategy either.

\mbox{ } 

\noindent All strategies discussed from now on in this Sec are universal as per Sec \ref{Univ}.

\subsubsection{Semiclassical Approach: emergent semiclassical time}\label{Semicl-Intro}

\noindent Paralleling Sec \ref{+temJBB}'s classical treatment, perhaps one has slow, heavy `$\fh$'  variables that provide an approximate timestandard with respect to which the other 
fast, light `$\fl$' degrees of freedom evolve (\cite{DeWitt67, LR79, Banks, HallHaw, BV89, KieS91, Kuchar92, I93, SCB3, Parentani, Kiefer93, Kiefer94, Kiefer99, Kieferbook, SemiclI, 
SemiclIII} and Sec \ref{Intro}).  

\noindent In addition to being an emergent time strategy toward resolving the POT, it is specifically the Semiclassical Approach to Quantum Cosmology (along the lines of e.g. 
Halliwell and Hawking \cite{HallHaw} that goes toward acquiring more solid foundations for other aspects of Quantum Cosmology as per e.g. Sec \ref{Q-Cos}.  
In this context, h is scale (and homogeneous matter modes) and the l-part are small inhomogeneities, and part of the point of the present Article is that one can toy-model 
the above by considering the regime in which scaled RPM's have a scale variable, such as the configuration space radius playing the role of $h$ and the pure-shape degrees of freedom 
playing the role of $l$.

\mbox{ } 

\noindent Analogy 98) RPM's generalize previously-studied absolute particle models of the Semiclassical Approach \cite{Banks, BriggsRost, Pad, PadSingh} 
by inclusion of auxiliary terms and subsequently of linear constraints.    
(Reduced RPM is a subcase of Datta's generalized curved configuration space mechanics work \cite{Datta97}.)

\noindent Analogy 99) is use of $\ft^{\se\sm(\sW\sK\sB)}$.

\mbox{ } 

\noindent As a counterpoint, evoking semiclassicality only makes sense if one's goals are somewhat modest as compared to some more general goals in Quantum Gravity programs.  

\mbox{ } 

\noindent Problem with Semiclassical Approach 1)  Having invoked a WDE results in inheriting some of its problems \cite{Kuchar92, I93}.  

\noindent Note 1) The first approximation is rather un-Machian (in the STLRC sense) via deriving its change just from scale (plus possibly homogeneous isotropic matter modes, 
or, more widely, from the usually-small subset of $\fh$ degrees of freedom). 
However, as we shall see in Sec \ref{22}, the second approximation remedies this by allowing anisotropic and inhomogeneous changes (or, more widely, 
whatever else the $\fl$-degrees of freedom may be) to contribute to the corrected emergent time.  
This is to be contrasted with the criticism at the end of Sec \ref{TAQ}.  

\noindent Note 2) Moreover, that the detailed emergent time here is {\sl not} exactly the same as the detailed classical $\ft^{\se\sm(\sJ\sB\sB)}$, 
as is to be expected on Machian grounds due to how {\sl quantum} change corrections influence this new quantum-level version.  

\noindent Problems 2 and 3 with Semiclassical Approach). Whilst the WKB Problem with the Semiclassical Approach \cite{Zeh88, BV89, BS, Kuchar92, I93, B93, Kiefer94, SemiclI} 
has been prominently mentioned so far in this Article, it is now time to say that the number of approximations concurrently made in the Semiclassical Approach are legion 
(Multiple Approximations Problem).   
Specifically, there are non-adiabaticities, other (including higher) emergent time derivatives, and averaged terms \cite{SemiclI, MP98, Arce}.  
%
%
These have the effect of obscuring tests of the validity of the WKB approximation -- the truth involves vast numbers of different possible regimes, 
so tests of validity are likely contingent on a whole list of approximations made.  

\noindent Problem with Semiclassical Approach 4) The imprecision due to omitted terms means deviation from exact unitarity.
This gives problems with the probability interpretation \cite{I93, Kuchar92}. of these approximations (and how to make sense of the sequence of these corresponding to increasingly 
accurate modelling, up to and including their relation with the probability interpretation for the unapproximated WDE itself; 
e.g. some of these involve other than Schr\"{o}dinger equations and thus require new inner products and new probability interpretations based thereupon).  
Thus the Hilbert space structure of the final theory may be related only very indirectly (if at all) to that of the quantum theory with which the construction starts.
[One still has a Hilbert space, but one does not know a priori {\sl which} Hilbert space one will end up working with...]
%

\noindent Note 5) I consider the Semiclassical Approach to be the most promising branch among those that amount to using the timelessness-indefiniteness cancellation hypothesis.   
One now does not take scale {\sl as} a time, but rather considers scale to be part of some set of heavy, slow, global variables which go into providing an approximate timestandard 
rather than into the fast, light, local variables that deal with actually-observed subsystem physics.
This then possesses a far more familiar, and probably conceptually more satisfactory positive-definite kinetic term.  
While it will certainly not always happen that the scale physics will be heavy and slow, this is fortunately the case in early-universe cosmology, 
so that one can look into the Quantum Cosmology arena from this Semiclassical Approach perspective.  

\mbox{ }

\noindent Overall in the Semiclassical Approach, there is still an Ice Dragon but it may mostly only menace outlandish places.  
I.e. there is some chance that the realm of Quantum Cosmology is free from it, at least as regards the 
{\sl practical} task of explaining the origin of the CMB inhomogeneities and the seeding of the galaxies.  
%
{            \begin{figure}[ht]
\centering
\includegraphics[width=0.65\textwidth]{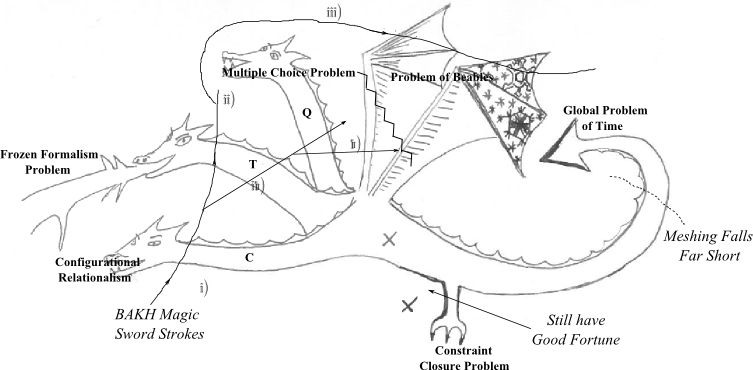}
\caption[Text der im Bilderverzeichnis auftaucht]{        \footnotesize{Now, upon passing to the quantum level, the first head ({\sc c}) indeed remains hewn.
But the second head ({\sc t}) re-grows.

\noindent iv) It is now hewn off by the Machian Semiclassical Approach's emergent WKB time.

\noindent v) One can then attempt a second Unicorn flyby slash involving the \K beables concept and Halliwell's Semiclassical--Histories--Records construct for selection 
of a subset of Dirac Beables as consistently embedded within my Machian-Semiclassical scheme.   
However, now the third ({\sc q}) head (quantizing does not imply conquering the quantum domain!) blocks this slash before it gets too far, 
catching the blade in its vicious Multiple Choice Problem teeth; this is indicated with the jagged line of impasse.}        }
\label{Full-Dragon-RPM-QM}\end{figure}            }
%
{            \begin{figure}[ht]
\centering
\includegraphics[width=0.65\textwidth]{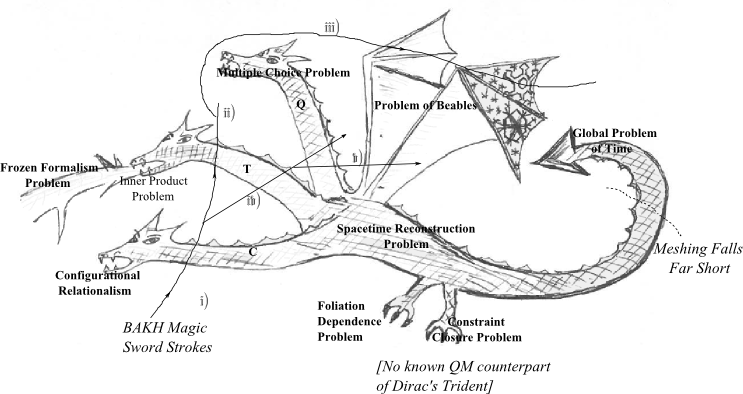}
\caption[Text der im Bilderverzeichnis auftaucht]{        \footnotesize{Fighting the actual Quantum GR POT Ice Dragon.  
The pattern of `magic sword' strokes is as before (and subject to the aforementioned classical caveats). 
Now the resurgent {\sc T} head additionally sports Inner Product Problem teeth, and stroke v) faces untested geometrical implementation issues as well as the Multiple Choice Problem.  

\noindent The most notable new problem as compared to fighting the Classical GR POT Ice Dragon is the absence of a known quantum analogue of Dirac's Trident, 
even at the semiclassical level. 
This is picked out as Major Open Question II in the Conclusion.}        }
\label{Full-Dragon-GR-QM}\end{figure}            }
%
\vspace{10in}
%
\subsection{Quantum Solipsism}\label{QM-Nihil-Intro}

Here, I continue the discussion of Sec \ref{Cl-Nihil-Intro} with two more purely QM brier patches.

\mbox{ }
  
\noindent Brier patch 2) {\bf Nonstandard Interpretation of Quantum Theory}.  
All of Solipsism, Cambium, Via and Historia both soon become entwined in general questions about the interpretation of Quantum Theory, 
in particular as regards whole-universe replacements for standard Quantum Theory's Copenhagen Interpretation.
I note that some criticisms of Timeless Approaches \cite{Kuchar92, Kuchar99} are subject to wishing to preserve aspects of the Copenhagen Interpretation, 
which may not be appropriate for Quantum Cosmology and Quantum Gravity.

\noindent Brier patch 3) {\bf WDE Dilemma}. Such approaches either 

\noindent (horn 1) invoke the WDE and so inherit some of its problems, 

\noindent or do not, thus risking the alternative problem 

\noindent (horn 2) of being incompatible with the WDE, so that the action of the Wheeler--DeWitt operator kicks purported solutions out of the physical solution space \cite{Kuchar92}. 

\noindent {\bf Nonstandard Interpretation of Quantum Theory}.  
Type 1 Tempus Nihil Est and Histories Theory both come to involve in general questions about the interpretation of Quantum Theory, 
in particular as regards whole-universe replacements for standard Quantum Theory's Copenhagen Interpretation.

\subsubsection{As regards Records Postulate 3: propositions}

\noindent The form of logical structure for Quantum Theory remains open to debate (Sec \ref{QM-Prop}).  
The notion of localization in configuration space may well furnish the graining/partial order/logical implication operation.    

\noindent The Projector--Proposition Association is preferred at the quantum level.  

\noindent Also at the QM level, Isham and Doering's work \cite{ToposI} can be viewed as basis for implementation of a Records theory, 
though this is a bridge in the process of being built \cite{AF13}.

\subsubsection{Regions Implementation of quantum propositions}

This involves integration over some classical configuration space region, or perhaps a phase space region (alternatives to phase space are not relevant in this application).  
Examples are the \NSI and the combined Semiclassical-Machian--Histories--Records Approach.

\subsubsection{Na\"{\i}ve Schr\"{o}dinger Interpretation} \label{Sec: NSI}

\noindent 0. The \NSI is an Interpretation of Quantum Theory for the whole universe originally due to Hawking \cite{Hawking84} (see also \cite{HP86, HP88, UW89}.)
It concerns probabilities of `being'.  
N.B. in the \NSII, one makes no attempt to collapse all questions to questions to being, but is temporally trivial as 
result of discarding the other questions so as to concentrate on demonstrating that some interesting questions can be answered.  
E.g. what is the probability that the universe is large? 
The subsequent minimalist's question is then is whether using (RigPhase, Point) can cover all physical propositions 
(the simple rationale here being `what about propositions involving momenta'?) 

\mbox{ }

\noindent One obtains answers to such questions via consideration of the probability that 
the universe belongs to region $\bFrR$ of the configuration space that corresponds to a quantification of a particular such property, 
\beq
\mbox{Prob}(\bFrR) \propto \int_{\bsFrR}|\Psi|^2  \mathbb{D}\fQ \mbox{ } , 
\eeq 
for $\mathbb{D}\fQ$ the configuration space volume element.

\mbox{ }

\noindent In the case of GR-as-geometrodynamics,  
\beq
\mbox{Prob($\bh$ has property P)} = \int_{\mbox{\scriptsize P-affirmative subset of Riem}(\sbSigma)}|\Psi[\bh]|^2\mathbb{D}\bh
\mbox{ } .  
\label{NSIIP}
\eeq
[This should be taken with a pinch of salt as regards the measure on Riem($\bupSigma$) and the P-affirmative subsets  
happening to form a suitable region of integration; one is also dealing with {\sl relative} probabilities here.]  
For GR-as-nododynamics, it is the connections (or, better, holonomies) that play the role of configuration variables
\beq
\mbox{Prob($\bA$ has property P)} = \int_{\mbox{\scriptsize P-affirmative subset of Hol}(\sbSigma)}|\Psi[\bA]|^2\mathbb{D}\bA
\mbox{ } 
\eeq
(with a similar pinch of salt, and noting that $\FrQ$-primality here assumes the nododynamical picture holds from first principles rather than being arrived at from 
geometrodynamics via a canonical transformation).  

\mbox{ }

\noindent Analogy 100) RPM's admit toy models of the \NSI \cite{HP86,UW89}. 
Here, simple (and ``geometrically nice") questions of being are addressed, are listed in Sec \ref{Cl-Prop} and examples of which are computed in Sec \ref{NSI}.
The evaluation is now e.g. for the r-presentation of scaled RPM
\beq
\mbox{Prob($\brho$ has property P)} = \int_{\mbox{\scriptsize P-affirmative subset of relationalspace}}|\Psi[\bS, \rho]|^2\mathbb{D}[\bS, \rho]\mbox{ } .  
\eeq
\noindent Set theory's complement $\mbox{}^{\scc}$, union $\bigcup$, intersection $\bigcap$ and inclusion $\underline{\subset}$ 
are a realization of conventional propositional logic's $\neg$, $\vee$, $\wedge$ and $\preceq$.
Continuous regions of a manifold are then one example of such a set, and so these implement conventional propositional logic.  

\mbox{ }

\noindent The \NSI can be taken to be a ($\FrQ$, Point) scheme: $\psi(\FrQ)$, and probabilities built solely out of those and a characterization of regions of $\FrQ$.  
Thus it is very much under the domain of $\FrQ$-primality.  
Solipsist Pre-Records Theory can be taken to be likewise.

\mbox{ }

\noindent The \NSI has various fairly obvious problems.

\mbox{ }  

\noindent \NSI Problem 1)  
It is indeed of limited use via not accommodating questions of being at a particular time, or of becoming \cite{Kuchar92}; 
nor are subsequent SSecs' routes around this available within the \NSI picture.  

\noindent \NSI Problem 2) It is termed `na\"{\i}ve' due to it not using any further features of the constraint equations. 
[This is less severe when r-formulations are available, however it fails to deal with $\scQ\scU\scA\scD$ also, and $\scQ\scU\scA\scD$ is very symptomatic of the Frozen Formalism Problem, 
so doubt should be cast on a strategy that purports to handle Frozen Formalism Problem whilst leaving $\scQ\scU\scA\scD$ not bypassed but simply {\sl unaddressed}.]   

\noindent \NSI Problem 3) It is subsequently menaced by horn 2 of the WDE Dilemma via its thus-named `na\"{\i}ve' inner product postulation.  

\noindent \NSI Problem 4) It also involves generally non-normalizable probabilities, which, however, support finite {\sl ratios} of probabilities.  

\noindent \NSI Problem 5) As already delineated in Sec \ref{QM-Prop}, this logical structure for the propositions 
is a questionable one to use in a quantum-mechanical context due to its {\sl classical} form.

\noindent \NSI Problem 6) In the geometrodynamical case, time enters the scheme as an internal coordinate function of $h_{\mu\nu}$. 
Therefore it is represented by an operator.  
However, as pointed out in e.g. \cite{I93}, there are problems with representing time as an operator.

\subsubsection{Proposition--Projector Association}\label{PPA}

The aim here is to implement propositions at the quantum level by projectors, including beyond \cite{IL2} 
the usual context and interpretation that these are ascribed in ordinary Quantum Theory. 

\mbox{ }

\noindent This is in contradistinction to the above kind of attempts at representation via classical regions of integration 
that have the defect that classical regions combine Booleanly but quantum propositions do not in general combine Booleanly.

\subsubsection{Conditional probabilities in ordinary Quantum Theory}

\noindent In ordinary quantum theory, for state $\BigupRho$ and proposition $P$ implemented by projector $\widehat{\mP}$, Prob($P$; $\BigupRho$) = tr($\widehat{\BigupRho}\widehat{\mP}$) 
with Gleason's Theorem providing strong uniqueness criteria for this choice of object from the perspective of satisfying the basic axioms of probabilities (see e.g. \cite{Ibook}).

\noindent The formula for conditional probability in ordinary Quantum Theory is then \cite{Ibook} 
\beq
\mbox{Prob}(B \in b \mbox{ at } \ft = \ft_2 | A\in a \mbox{ at } \ft = \ft_1;   \BigupRho) = 
\frac{     \mbox{Tr}\big(\mP^B_b(\ft_2)\,\mP^A_a(\ft_1)\,\BigupRho\,\mP^A_a(\ft_1)\big)         }
     {     \mbox{Tr}\big(\mP^A_a(\ft_1)\,\BigupRho\big)                                     } \mbox{ } .      
\label{Prob-Std}
\eeq
Here, I denote the projection operator for an observable $A$ an observable and $a$ a subset of the values that this can take is denoted by $\mP^A_a$.
N.B. that this is in the 2-time context, i.e. to be interpreted as {\sl subsequent} measurements.
It also follows that 
\beq
\mbox{Prob}(B \in b \mbox{ at } \ft_2 \mbox{ and } A \in a \mbox{ at } \ft_1) = 
\mbox{Tr}\big(\mP^B_b(\ft_2)\,\mP^A_a(\ft_1)\,\BigupRho\,\mP^A_a(\ft_1)\big) \mbox{ } , 
\eeq
and this extends in the obvious way to $p$ propositions at times $\ft_1$ to $\ft_p$.

\mbox{ } 

\noindent {\bf Supplant at-a-time by value of a particular `clock' subconfiguration}.  

\mbox{ } 

\noindent Note that not all uses of this necessarily carry `good clock' connotations: 
in the literature, this is done for whichever of the `any', `all' and `sufficient local' connotations.

\subsubsection{Conditional Probabilities Interpretation of Timeless Quantum Theory} \label{CPI}

IV. The \CPI was proposed by Page and Wootters \cite{PW83} (see also the comments, criticisms and variants in \cite{Kuchar92, Kuchar99, Giddings, Dolby, PGP1, PGP2, Arce}).   

\mbox{ } 

\noindent Conceptually, it is a refinement of the \NSI that extends the range of questions it can answer [thus it is an improvement as regards \NSI Problem 1)]. 
In particular, it addresses questions concerning conditioned being: conditional probabilities for the results of a pair of observables $A$ and $B$, 
in particular concern correlations between $A$ and $B$ at a single instant in time.  
E.g. ``what is the probability that the universe is almost-flat {\sl given} that it is almost-isotropic?"

\mbox{ } 

\noindent Analogy 101) RPM's admit toy models of the Conditional Probabilities Interpretation \cite{PW83}, in which questions concerning pairs of propositions of being are addressed. 
E.g., `what is the probability that the triangle model universe has a large area per unit moment of inertia given that it is approximately isosceles?'  

\mbox{ } 

\noindent Moreover, technically and as interpretations of QM, the two schemes are highly distinct.      

\noindent In particular,

\noindent 1) it specifically implements propositions at the quantum level by use of projectors.

\noindent 2) It addresses questions concerning conditioned being via {\sl postulating} the relevance of conditional probabilities,  
\beq
\mbox{Prob}(B\in b|A\in a;\BigupRho) = \frac{\mbox{Tr}\big(\mP^B_b\,\mP^A_a\,\BigupRho\,\mP^A_a\big)}{\mbox{Tr}\big(\mP^A_a\,\BigupRho\big)}                    \mbox{ } ,      
\label{Prob-BArho}
\eeq
for finding $B$ in the subset $b$, given that $A$ lies in the subset $a$ for a (sub)system in state $\BigupRho$. 
N.B. these occurring within the one instant rather than ordered in time (one measurement and {\sl then} another measurement) places this postulation outside the conventional formalism 
of Quantum Theory, for all that(\ref{Prob-BArho}) superficially resembles (\ref{Prob-Std}).

\mbox{ }

\noindent Next, the \CPI can be used in principle to replace questions of `being at a time' by simple questions of conditioned being. 
This is via one subsystem $A$ being useable as a timefunction, so that the above question about $A$ and $B$ can be rephrased to concern what value $B$ takes 
when the timefunction-giving $A$ indicates a particular time \cite{PW83, I93}.  

\mbox{ } 

\noindent Note 1) The \CPI is based on an AMR type notion of clock, albeit configuration-based rather than change based: `the clock reads three o'clock.' 

\mbox{ }

\noindent New Structure 1) the \CPII's traditional development did not set up a scheme of logical propositions 
(it came historically before awareness of that began to enter the POT community via \cite{IL2, IL}).    
However, adding that layer of structure to the \CPI for reason of first principles (Mackey's Principle) can be seen as a cleaner successor to my reasoning \cite{Records, Records2} in 
suggesting Records Theory outside of Histories Theory have such a structure via how Records Theory that sits inside of Histories Theory inherits such a structure from Histories Theory.  
In the \CPI scheme, then, the projectors used are a solid anchor for embodying propositions, and the structure of the atemporal part of quantum logic is then a reasonable first working 
guess for this newly-proposed extra level of logical structure, though one would need to think carefully about whether \CPII-Copenhagen differences affect this in any way.  

\mbox{ }

\noindent \CPI Problem 1) The means of replacement of `being at a time' has the practical limitation 
that $A$ may not happen to have characteristics that make it suitable as a sufficiently good timefunction.  

\noindent Problem with \CPI 2) supplanting being-at-a-time by being is as far as the original Conditional Probabilities Interpretation goes.  
It is not a full resolution of Problem with \NSI 1) since it does not per se address questions of becoming; 
for a further extension to address those too, see the `Records A)' scheme by Page in the next SSec.  

\noindent Problem with \CPI 3) \K pointed out that horn 1 of the WDE Dilemma also applies here \cite{Kuchar92}.  

\noindent Non? Problem with \CPI 4) I also comment that some of Kucha\v{r}'s critiques \cite{Kuchar92, Kuchar99} of the \NSI and the \CPI can be interpreted as not accepting 
ab initio a separate `being' position rather than constituting conceptual or technical problems once one has adapted such a position.    
\noindent E.g. \K criticized the \CPI in \cite{Kuchar92, Kuchar99} for its leading to incorrect forms for propagators.
Page answered that this is a timeless conceptualization of the world, so it does not need 2-time entities such as propagators; see also the next SSSec for a distinct resolution of this.

\subsubsection{Gambini--Porto--Pullin approach}\label{GPP}

Vb. This \cite{PGP1, PGP2} is built upon conditional probabilities, which are now of the form 
\beq
\mbox{Prob}
\left(
\stackrel{\mbox{\normalsize beable ${\iB}$ lies in interval $\Delta{\iB}$}} 
         {\mbox{provided that clock variable $\tau$ lies in interval $\Delta\tau$}} 
\right) = 
\stackrel{\mbox{lim}}{T \longrightarrow 0} \frac{    \int_0^T\d \ft \mbox{Tr}(    \mP_{\Delta {B}}(\ft)  \mP_{\Delta \sft}(\ft) \BigupRho_0 \mP_{\Delta \sft}(\ft)  )}
                                                {    \int_0^T\d \ft \mbox{Tr}(\BigupRho_0 \mP_{\Delta \sft}(\ft))}  \mbox{ } .
\label{NewP}
\eeq
Here, the P($\ft$)'s here are Heisenberg time evolutions of projectors P: $\mP(\ft) = \mbox{exp}(iH\ft)\,\mP\,\mbox{exp}(-iH\ft)$.  
This approach is also for now based on the `any change' notion of relational clocks (that can be changed \cite{ARel2} to the STLRC notion).
Moreover, these clocks are taken to be non-ideal at the QM level, giving rise to 

\noindent 1) a decoherence mechanism (distinct from that in Histories Theory; there is also a non-ideal rod source of decoherence postulated; 
the general form of the argument is in terms of imprecise knowledge due to imprecise clocks and rods). 
%
%
Another feature of this approach is that employing a non-ideal clock in itself gives rise to decoherence (though that needs to be checked case by case rather than assumed \cite{AH08}).  

\noindent 2) A modified version of the Heisenberg equations of motion of the Lindblad type, taking the semiclassical form (here this carries coherent-state connotations)
\beq
i\hbar\frac{\partional \BigupRho}{\partional \ft} = [H, \BigupRho] + R[\BigupRho] \mbox{ } .
\label{Lindblad}
\eeq
The only case of this explicitly explored so far is the finite $\pa$-derivative subcase.  
The form of the $R$-term here is $\sigma(\ft)[H, [H, \BigupRho]]$  where $\sigma(\ft)$ is dominated by the rate of change of width of the probability distribution.
By the presence of this `emergent becoming' equation (\ref{Lindblad}), this approach looks to be more promising in practise than Page's extended form of \CPII. 
This approach also has the advantage over the original \CPI of producing consistent propagators.  
By the presence of the $R$-term, the evolution is not unitary, which is all right insofar as it represents a system with imprecise knowledge (c.f. Sec \ref{Cl-Pers}).    
Dirac observables for this scheme are furthermore considered in their paper with Toterolo \cite{GPPT}.  

\mbox{ } 

\noindent Considering sets of propositions and using $\ft^{\se\sm(\sW\sK\sB)}$ for $\ft$ above, this scheme can be cast to comply with Mach's Time Principle and Mackey's Principle.  

\mbox{ }

\noindent Analogy 102) This Gambini--Porto--Pullin scheme and the just-mentioned reconceptualization of it exist also for RPM's.  

\mbox{ }

\noindent From this point onward, the present Article uses the \NSI as a simple example and then looks at Records Theory, 
thus for now missing out on giving detailed examples of the \CPII.  
However, the \CPI does give tractable examples comparable to the \NSI ones , albeit these do require  a bit more work/space to present.  

\mbox{ }

\noindent New Structure) One could in this case condition on JBB time, and, more accurately on WKB time, 
rendering this scheme time-Machian at the expense of coupling it more tightly to the semiclassical approach [combined schemes Vc)--VIIc)].
I view this option as possible for Gambini--Porto--Pullin schemes (unlike standard \CPII) due to their less solipsist attitude (limiting case of a non-solipsist construct, 
a place in the theoretical scheme for those propagators).

\subsubsection{Solipsist Records Theory schemes}\label{Records} 

\noindent {\bf Supplanting becoming questions} in the quantum context has been considered e.g. in \cite{Page1, Page2, EOT, Records}.    

\mbox{ }

\noindent I. {\bf Page Records} The Projector--Proposition Association obviously carries over from the \CPI to the Page Records scheme, bringing it into accord with Mackey's Principle.
And then whether to postulate the relevance of, and evaluate, the Conditional Probabilities Interpretation/Page Records 
scheme's concrete conditional probability object is to be computed and interpreted, or to concentrate on some other kind of object. 

\noindent Resolution of Problem with \CPI 2) In Page's particular version, one does not have a sequence of a set of events; rather the scheme is a 
single instant that contains memories or other evidence of `past events'. 

\mbox{ }

\noindent Note: One can have a {\bf Mackey--Page Program}.  
However, a Machian-Semiclassical Mackey-\CPI would have the Semiclassical part being used to handle becoming questions rather than Page's own way of doing this in principle, 
so I omit such a mixed option from my classification.  

\mbox{ } 

\noindent II. {\bf Bell--Barbour Records}.  \cite{Bell,B94II,EOT} (see \cite{ButterBar} for further differences between these authors) reinterpret Mott's calculation \cite{Mott} 
of how $\alpha$-particle tracks form in a bubble chamber as a ``{\bf time capsules}" paradigm for Records Theory.  
N.B. that this calculation is based on a TISE and is therefore timeless.  
Barbour has then argued for Quantum Cosmology to be studied analogously, with somewhat similar arguments 
being made by Halliwell, various of his students and Castagnino--Laura \cite{H99, H00, H03, HT, HD, CastAsym, CaLa, HW}. 
However, Halliwell considers tracks in configuration space rather than in space, along the lines of 
Misner's vision \cite{Magic} of Cosmology as a scattering problem in configuration space.
Tracks in configuration space do represent more than individual instants, so these approaches have more structure than the most minimalist timeless approaches, 
albeit less than in Histories Theory.  

\noindent Upgrading the Bell--Barbour scheme likewise would first require deciding whether to adopt the Projector--Proposition Association within that context.   
(Without such a concrete specification, the Bell--Barbour scheme remains vague from the perspective of actually doing calculations.)  
This includes a higher level of vagueness than in the Page Records scheme, in which {\sl what} to compute is at least clear, 
for all that we have no idea how to evaluate such a thing for a nontrivially-functional model of becoming-supplantation).

\noindent Barbour's timeless approach is based on $\FrQ$ primary and LMB time.  
If this is given an LMB--CA upgrade XII, as well as a local rather than Leibniz-whole-universe interpretation, one transits to my conception of Records Theory.  

\noindent To use (\ref{Winged-Pig}) for one's Records Theory, one would need to import a procedure for obtaining this, such as how to solve and interpret the WDE, 
which would be fraught with numerous further technical and conceptual problems.   

\mbox{ } 

\noindent XII. {\bf My records approach, and its tie to LMB-A time}  The most minimalistic scheme is Mackey-A-Records. 
There is also an Machian-Semiclassical-Mackey-A-Records alternative XXIV.  
Each of these makes use of Notions of Distance, Notions of Information, Proposition--Projector association.  

\mbox{ }  

\noindent Additionally, much of Isham and Doering's work \cite{ToposI} can be also taken to be a purveyor of structure to solipsist approaches.  

\mbox{ }

\noindent Barbour favours his `being' perspective so as to be open to the possibility of explaining the Arrow of Time as emergent from that.
On the other hand, Castagnino's scheme \cite{CastAsym} (which builds in a time asymmetry in the choice of admitted solutions), and 
Page's scheme (which privileges a `final' instant as the place within which to find a sufficiency of correlations), are not open to such a possibility.    
Nor is the Gell-Mann--Hartle--Halliwell's scheme (XXVI in Fig \ref{Tless-Types}) of Records within Histories Theory, 
since its presupposition of histories comes with a direction of time.

\subsubsection{Classifications of Quantum Solipsism and Quantum Cambium Approaches and Composites}\label{QM-Tless-Classi}

To the Timeless Classifications in Sec \ref{Cl-Nihil-Intro}, I now add the following.

\mbox{ } 

\noindent Timeless Classification 0) \cite{APOT} Freestyle versus insisting on the Heisenberg picture and particular stances observables 
(Rovelli-type: column 1 in Fig \ref{Tless-Types}).
The cause of this distinction is Rovelli having markedly different ideas from his timeless-approach predecessors; both schools of thought remain active. 

{            \begin{figure}[ht]
\centering
\includegraphics[width=0.93\textwidth]{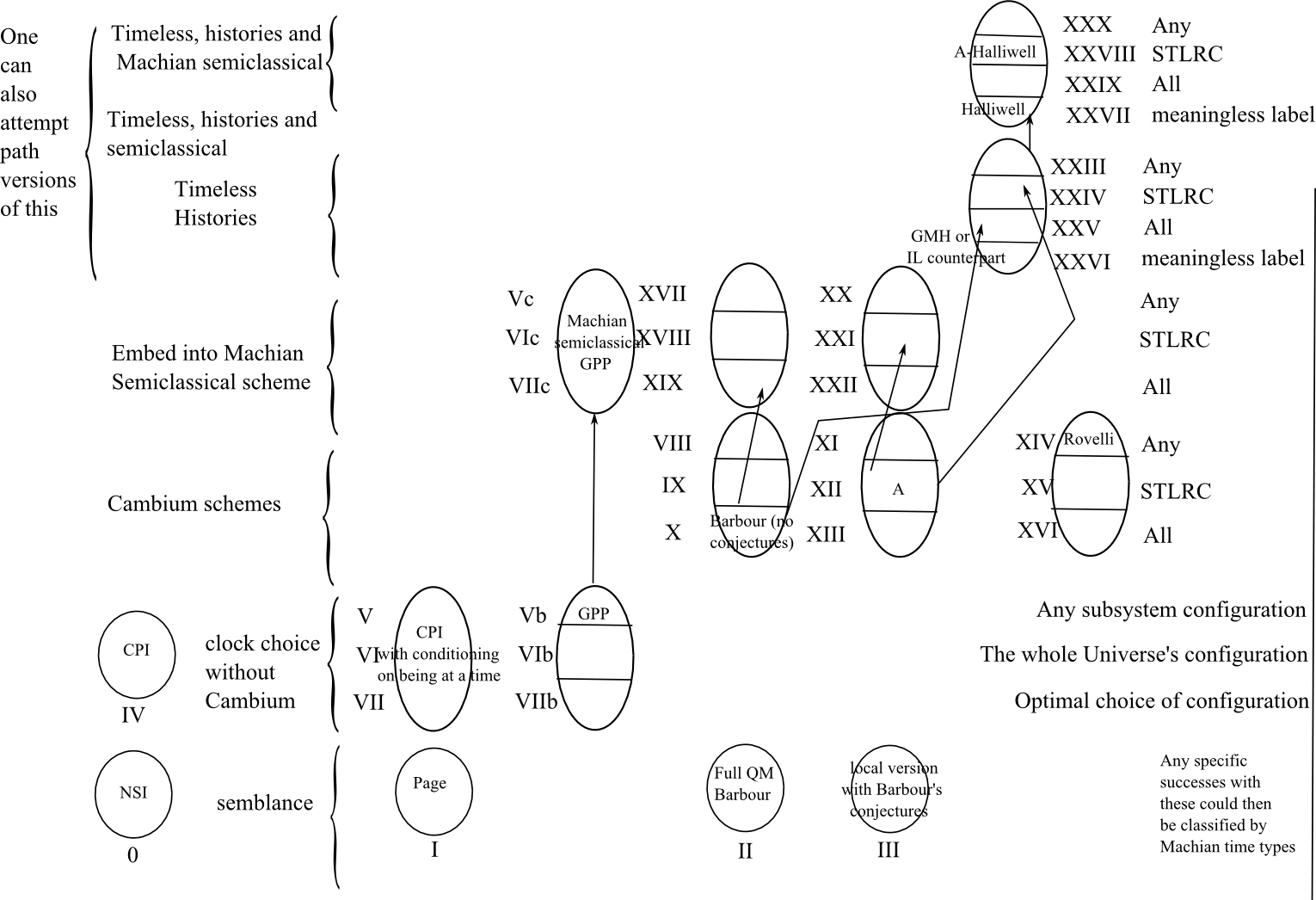}
\caption[Text der im Bilderverzeichnis auftaucht]{        \footnotesize{Classification of timeless strategies; 
the main new feature as compared to the classical version is the \CPI and variations thereupon, including Page, Gambini--Porto--Pullin and Machianized schemes.
Quite a few of these 39 schemes are new to this Article. }        }
\label{Tless-Types}\end{figure}            }
 
\noindent Timeless Classification 5) by interpretation of QM, due to whole-universe and quantum-gravitational nonstandardnesses.

\noindent Timeless Classification 6) by compatibility with the Semiclassical Approach (see Fig \ref{Tless-Types}).  

\mbox{ } 

\noindent N.B. the semblance of dynamics is the principal point at which the differences between different records schemes enter. 
The Page approach proceeds via valid-in-principle reduction of becoming to being, Gambini--Porto--Pullin's approach via their Lindblad equation, Barbour via attempting to demonstrate 
his conjecture, and Gell-Mann--Hartle--Halliwell by already sitting within a Histories Theory (though one might still ask if that can be attained without presupposition). 
Machian-classical considerations may also be invoked (or indeed a distinct three-way unification of some records scheme with some 
histories scheme and some semiclassical scheme, so that these cancel out each others' deficiencies, see Secs \ref{QM-Combo} and \ref{Conclusion}).  
These become Machian-semiclassical considerations at the quantum level.

\subsection{Machian-Semiclassical Correlations Schemes}\label{SM-TR}

The basic idea is to follow up having a semiclassical scheme with consider a scheme for timeless correlations within it.  
That I am aware of, this started with Halliwell's 1987 follow-up \cite{Halliwell87} to Halliwell--Hawking's 1985 Semiclassical Quantum Cosmology.  
There is also early work by Kiefer \cite{Kiefer87} (see also \cite{Kieferbook}). 
It has the benefit of not requiring a purely timeless semblance of dynamics, since the overarching semiclassical scheme provides an emergent time.
Moreover, this idea can then be carried over to the specifically Machian formulation of Semiclassical Quantum Cosmology of Sec \ref{Semicl}.  
In the present Article, I show how to repackage quite a number of other programs into this 
form, and moreover do so in a manner compliant with both Mach's Time Principle and with Mackey's Principle.  
These are new `MSM-' variants of existing programs, and each represents a new POT strategy in its own right, so the present Article is a {\sl substantial} advancement in 
timeless, timeless-semiclassical and timeless-semiclassical-histories programs.
MSM stands here for Machian-Semiclassical-Mackey- indicating the incorporation of Mach's ideas about time and of Mackey's ideas about propositions.  

\mbox{ } 

\noindent There are various options as regards handling the timeless questions themselves.  
This could be considered from a \NSI perspective \cite{HP86}, though that has the stated limitations, or from a \CPI perspective.  
It is however an alternative to Page's extension of the \CPII, since that concerns establishing a semblance of dynamics from purely timeless terms.
The Gambini--Porto--Pullin scheme is also amenable to being `embedded' in this way within a semiclassical approach, at the cost of expanding 
somewhat on GPP's own meaning of `semiclassical'.  
There is also Halliwell's own scheme \cite{Halliwell87}, though its descendants transcend into including also Histories Theory as per below.    

\mbox{ } 

\noindent The problem with most of these schemes is that the timeless approaches involved do not provide a reason to support the use of the crucial WKB ansatz.  
The exception to that is Gambini--Porto--Pullin's scheme, insofar as it has a decoherence mechanism.  
The usual answer for other schemes is to unify with histories theory so as to obtain a distinct means of decoherence.  

\vspace{10in}

\subsection{QM path integral approaches}\label{Path-Int}

\noindent 1) The usual (Feynman) Path Integral Formulation of QM involves   
The path integral is, schematically,  
\beq
\int_{\gamma} \mathbb{D} \bfP \, \mathbb{D} \bfQ \, \mathbb{D}\d{\fI} \,  \mbox{det}\,{\cal F}[\bfQ, \d{\fI}] 
              \updelta[{\cal F}[\bfQ, \d{\fI}]] \mbox{exp}(i\FS[\bfQ, \bfP, \d{\fI}]) \mbox{ } .
\label{PI-3-prelim}
\eeq
and is a Via Post Quantum approach.  
Here, the $\updelta$ is a Dirac delta function imposing the basis set of gauge-fixing conditions ${\cal F}_{\sfB}$ and that the $\mathbb{D}$'s collectively form the 
Liouville measure on Phase.
Also, det$\,{\cal F}$ is the {\it Fadde'ev--Popov determinant} \cite{Fadde'ev, +Fadde'ev}.  
This can be seen to arise as the Jacobian in the transformation of  $\fP$ and $\fQ$ variables by which the positions, 
momenta and cyclic ordials are formally split into the gauge-fixing-conditions-as-momenta, their conjugate coordinates and a further independent set of 
%
%
coordinates and momenta.

\mbox{ }

\noindent {\bf Question 71)} What are the implications as to how to proceed here in each of the cases of considering $\scQ\scU\scA\scD$ to be or not to be a gauge constraint? 
If $\scQ\scU\scA\scD$ is not gauge, $\FrG$-triviality would imply trivial Fadde'ev--Popov factor, but does one then have a {\sl satisfactory} path integral approach (per se and 
in terms of equivalences with other quantum approaches such as the canonical one).  

\mbox{ }  

\noindent 2) The Paths Brackets Formulation is the classical half of a Via Ante Quantum (or Historia Ante Quantum, but for now that is not counted) approach.  

\noindent  N.B. 1) and 2) involve two different senses of quantum path theory: the plain Path Integral Approach and the new idea of additionally assuming quantum paths bracket 
structure without assuming projectors, i.e. the specifically Via Ante Quantum variant of IL's scheme.  
This difference is included in the name `Brackets', which is {\sl not} taken to supplant the word `Integral' -- a quantum Paths Brackets Formulation is shorthand for a 
quantum Paths Brackets AND Path Integral formulation.

\noindent 3) One can also have a Via Post Quantum approach in which quantum paths bracket structure is postulated ab initio, though this does look somewhat less natural than each 
of the preceding -- if paths merit their own bracket structure, why would one not already consider paths bracket structure at the classical level too?

\mbox{ } 

\noindent Gravitational theory faces some additional challenges as compared to flat-spacetime QFT. 

\mbox{ }

\noindent Problems with Path Integral Formulation 1 to 3) There are now various problems with using complex methods.  
These no longer greatly improve the behaviour of the action in the exponent, due to the complex rotation nullifying the Lorentzian indefiniteness but not the DeWittian indefiniteness.
Furthermore, the guarantee of a complex 4-$d$ curved spacetime containing both suitable Lorentzian and Euclidean real 4-manifolds as submanifolds is a nontrivial issue in place of 
a flat-spacetime triviality.
Next, even for minisuperspace, the contours between (c.f. Wick rotation) the real-Euclidean and real-Lorentzian 4-manifolds are ambiguous and with a substantial multiplicity \cite{HL90}.    

\noindent Problem with Path Integral Formulation 4) There is also a {\bf Measure Problem}.  
The basic effect of passing from a canonical formulation to a path-integral one is to 

\noindent i) posit a non tempus sed via resolution of the Frozen Formalism Problem (which begs further ontological questions as well as the above questions with the habitual technique 
used to handle path integrals, and does not address any of the other 7 facets of the POT -- search the rest of this Sec for `Problem with Path Integral Formulation'). 

\noindent ii) To trade having an Inner Product Problem with one's TDWE for having a new Measure Problem.

\mbox{ }

\noindent Comment 1) Much of the conceptual value attributed in Quantum Gravity to Path Integral Approaches over canonical approaches unfortunately follows from the misapprehension 
that the terrible Problem of Time {\sl is} the Frozen Formalism Problem, which then the path integral approach `resolves' rather than trades for other in-general insurmountable 
difficulties.

\noindent Comment 2) The bad behaviour of the continuum GR path integral formulation {\sl is} ameliorated by passing to discrete approaches.  
However, wishing Nature to be calculable does not make Her discrete, and arguments for discreteness of time, space and spacetime do not hold as much water 
as some people suppose.  
That being too far off-topic, I refer to \cite{Disc-Rev}.  
What I say for now is that Discrete\footnote{Or Almost-Discrete, e.g. involving an ancillary continuum sampling space in the Causal Sets Approach or eventually taking a continuum limit 
in the Causal Dynamical Triangulation Approach.} 
approaches tend to have a poor track record as regards the usually-last and especially unsurmountable POT facet that is the Spacetime Reconstruction Problem.

\noindent Comment 3) Concerning canonical-and-path-integral approaches, (e.g. QFT's in the usual Minkowski spacetime context for which canonical and path-integral approaches 
{\sl support} each other). 
%
%
As well as then having {\sl both} the canonical and the path-integral sets of difficulties, 
it is not clear whether the large amount of path-integral--canonical compatibility in e.g. flat-spacetime QFT continues to hold in QG.
And, if not, whether {\sl either} approach can be self-sufficient (or gain sufficiency from combination with some {\sl third} program such as algebraic QFT \cite{AQFT}.  
If one tries to combine diverse such approaches for QG, then one would expect most of the problems of {\sl each} to be present, including a wide range of time-related problems.

\mbox{ } 

\noindent N.B. the Path Integral Approach already possesses the notions of coarse graining and finest possible graining that are more usually used in, and attributed to, 
Histories Theory.

\subsection{QM part of Histories Theory}\label{QM-Histories}

\noindent Let us now conceive of histories at the quantum level, whether because one is doing `Historia Post Quantum' ({\sc qh}) 
or one is promoting the classical part of `Historia ante Quantum' to the quantum level.  

\mbox{ } 

\noindent One thinks at the quantum level level along the lines of Feynman path integrals, moreover building up much further structure.

\subsubsection{Hartle-type Histories Theory}  

1) Individual histories are built out of strings of projectors $\mP^{A_i}_{a_i}(\ft_i)$, i = 1 to $N$ at times $\ft_i$, 
\beq
c_{\eta} := \mP^{A_N}_{a_N}(\ft_N) . . . \mP^{A_2}_{a_2}(\ft_2) \mP^{A_1}_{a_1}(\ft_1) \mbox{ } .  
\label{ceta}
\eeq 
N.B. these do not imply measurement; Histories Theory is intended to have other than  the standard interpretation of QM. 

\mbox{ } 

\noindent Assigning probabilities to histories does not work in Quantum Theory.  
For, if $a(t)$ has amplitude $A[a] = \mbox{exp}(i\FS[a])$ and $b(t)$ has amplitude $B[b] = \mbox{exp}(i\FS[b])$, 
then these are nonadditive since in general $|A[a] + B[b]|^2 \neq |A[a]|^2 + |B[b]|^2$ in general.
In fact, Histories Theory involves \cite{GMH, Hartle} a far-reaching \cite{I93} extension of normal Quantum Theory 
to a form outside its conventional Copenhagen interpretation: it is a many-worlds or beables type scheme. 

\noindent One then considers the notion of fine/coarse graining corresponding to different levels of imperfection of knowledge, 
by which families of histories are partitioned exhaustively and exclusively into subfamilies.  
I use $C_{\bar{\eta}}$ for coarse-graining, where $\bar{\eta}$ is a subsequence of $\eta$'s times and each projector in the new string may concern a less precise proposition.
Thus, as well as the coarse-graining criteria in Sec \ref{log}, Histories Theory possesses {\it coarse graining by probing at less times}.  

\mbox{ } 

\noindent 2) Notions of coarse-graining and finest graining are held to apply

\mbox{ } 

\noindent 3) The final traditional ingredient of the Histories Theory approach is the {\bf decoherence functional} between a pair of histories $\eta$, $\eta^{\prime}$,
\beq
{\cal D}\me\mc[c_{\eta^{\prime}}, c_{\eta}] := \mbox{tr}(c_{\eta^{\prime}}\BigupRho c_{\eta}) \mbox{ } .
\eeq 
This is useful as a `measure' of interference between $c_{\eta^{\prime}}$ and $c_{\eta}$.  
It is zero for perfectly consistent theories.  
It has the following properties.
\noindent      
\beq 
{\cal D}\me\mc[c_{\eta^{\prime}}, c_{\eta}] = {\cal D}\me\mc[c_{\eta}, c_{\eta^{\prime}}]         \mbox{ } \mbox{ (Hermeticity) ,}
\label{dec1}
\eeq
\beq
{\cal D}\me\mc[c_{\eta^{\prime}}, c_{\eta}] \geq 0                     \mbox{ } \mbox{ (Positivity) ,} 
\label{dec2}
\eeq
\beq
\sum\mbox{}_{\mbox{}_{\mbox{\scriptsize $c_{\eta^{\prime}}, c_{\eta}$}  }} {\cal D}[c_{\eta^{\prime}}, c_{\eta}] = 1   \mbox{ } \mbox{ (Normalization) ,}
\label{dec3}
\eeq
\beq
{\cal D}\me\mc[c_{\eta^{\prime}}, c_{\eta}] = \sum\mbox{}_ {\mbox{}_{\mbox{\scriptsize $\eta{\prime} \in \bar{\eta}^{\prime}, 
\eta \in \bar{\eta} $}}}  {\cal D}[c_{\eta^{\prime}}, c_{\eta}]     \mbox{ } \mbox{ (Superposition property) .}
\label{dec4}
\eeq
A new probability postulate for this scheme is that 
\beq
{\cal D}\me\mc[c_{\eta^{\prime}}, c_{\eta}] = \delta_{\eta^{\prime}, \eta}\mbox{Prob}(a_N \ft_N, a_{N - 1} \ft_{N - 1}, ... a_{1}\ft_1; \BigupRho_0) \mbox{ } . 
\eeq
Approximate consistency is held to be sufficient.

\mbox{ } 

\noindent The issue with whether just paths can support a notion of decoherence functional boils down to whether an entity with enough properties to serve in this role 
can be constructed from some purely path-theoretic replacement $c_{\gamma}$ of the histories-theoretic $c_{\eta}$ of eq (\ref{ceta}).

\mbox{ }

\noindent Problem with Histories Theory 1) It is not obvious whether this scheme is in correspondence with an algebra of propositions.  

\mbox{ } 

\noindent At least insofar as Gell-Mann--Hartle histories are the product of Heisenberg picture projection operators, 
and such products are usually not themselves a projection operators, so this fails to implement proposition-as-projector.
They do however have a disjoint sum of histories OR and a NOT operation.   
\noindent What about looking to implement propositions along the lines of Hartle's consideration of how each history intersects with a given region in configuration space?  
Hartle uses proper time spent in that volume for that purpose, a notion which is independent of canonical slicing.  
However (in parallel with my questioning the \NSI's use of classical regions to pose its quantum-mechanical questions), I do not 
expect this to cover all physically-relevant propositions and the inter-relations between them at the quantum level.   
Additionally, Kucha\v{r} has argued that it is possibly problematic that this approach takes a variable that is not dynamical and not quantized 
and effectively gives it meaning, thus breaking a feature of ordinary quantization without as yet providing enough justification/interpretation for this.  
However the next SSec bypasses Problem with Histories Theory 1) by finding a different way in which to anchor the Projector--Proposition Association to Histories Theory.

\mbox{ } 

\noindent{\bf Question 72} Can the notion of a decoherence functional (or a weakening thereof but still sufficient to support the establishment of WKB regimes)  
be furnished in the absence of associating chains of projectors to one's paths by which it is rendered a history in the sense of `Histories Theory'? 
%

\subsubsection{Isham--Linden approach to Histories Theory}\label{IL-Hist}

I motivate this as demonstration that a somewhat different match-up between histories and questions/propositions can in fact be made.

\mbox{ }  

\noindent{\bf Isham and Linden} (IL) \cite{IL2, IL} (see also e.g. \cite{Savvidou1, Savvidou02, Savvidou04a, Savvidou04b, AS05, Kessari} 
established this by considering the role of $c_{\eta}$ to now be played by the following {\sl tensor products} of the projectors 
\beq
c_{\eta} := \mP^{A_N}_{a_N}(\ft_N) \otimes . . . \otimes \mP^{A_2}_{a_2}(\ft_2) \otimes \mP^{A_1}_{a_1}(\ft_1) \mbox{ } ;
\eeq 
these trivially inherit the projector axioms from the individual projectors.
Thus an alias for this program is the 

\noindent{\bf histories projection operator (HPO)} approach.    
The above match-up can then readily be completed with notions of negation and disjoint sum to form an orthoalgebra/lattice of propositions, $\FrU\FrP$.

This has enlarged the Hilbert space Hilb to a tensor product Hilb $\otimes$ Hilb $\otimes$ ... $\otimes$ Hilb, 
and then this approach encounters the problem of there being no natural time translation operator to move between the copies.
IL found, however, that they could extend their tensor product to an infinite-dimensional continuum limit, 
for which this issue of the lack of a natural time translation operator is overcome. 
This more technical, albeit still time-tied, issue pushed them away from Hartle's discrete time steps to histories with continuous time.  

\mbox{ } 

\noindent Questions about histories are then another simplified form of logical structure as compared to temporal logic \cite{IL2, Flori10}.  

\mbox{ } 

\noindent In the HPO scheme, the decoherence functional is regarded as a functional ${\FrU\FrP} \times {\FrU\FrP} \longrightarrow \mathbb{C}$.    
Is this additional structure a significant further mathematical construct? 
Some partial answers to this are as follows. 

\noindent 1) The structure of this map parallels that of maps representable by matrices, so that it then makes sense to talk in terms of decoherence involving negligibility of 
off-diagonal elements. 

\noindent 2) Decoherence functionals are the analogues of quantum states in the parallel given by IL's between Histories Theory and ordinary QM.

\noindent See e.g. \cite{AS05} for the specific form of the decoherence functional in IL's formalism for Histories Theory.   

\noindent The HPO scheme is a QFT {\sl in the (label or emergent) time}, even for `conventionally finite' models.  

\mbox{ } 
 
\noindent Analogy 103) HPO schemes can be set up for RPM's with reasonable parallel to in the case of GR (see Sec \ref{QM-Hist}). 

\mbox{ }

\noindent In the HPO scheme, which is usually taken to be {\sc hq}, there is a kinematical commutator algebroid of histories, and a quantum histories quadratic constraint.  
\noindent This is regardless of whether it is HPO or of which {\sc h}, {\sc q} ordering is used, the following additional layers of structure are considered.

\noindent There continue to be histories-theoretic notions of coarse- and fine-graining at the quantum level.  

\mbox{ } 

\noindent Note 1) `Historia ante Quantum' ({\sc hq}) approaches can be seen as providing a second opportunity to a number of {\sc tq} 
approaches and group/geometrical quantization methods.  
E.g. new kinematical quantization, commutator and constraint-bracket algebroids. 


\noindent Problems with Histories Theory 2 to 5) Most of the Problems with Path Integral Formulations continue to be relevant here by Histories Theory involving Path Integrals plus extra 
layers of structure that, whilst they serve various other purposes (decoherence, proposition implementing), do not ameliorate the problems with the path integrals involved themselves.  
 
\noindent Problem with Histories 6)  Which degrees of freedom decohere which is unclear in the GR context \cite{H03}.  

\noindent Possible Problem with Histories 7)  Is the generalization of QM involved in Histories Theory self-consistent and meaningful?  
Does it reduce to ordinary QM in cases testable by experiment?
Kucha\v{r} has argued \cite{Kuchar99} that it does not, in the case of a relativistic particle.  
Elsewise, it has been commented on by Kent and Dowker \cite{DK94, Ketc} that future and past can easily fall apart in this scheme, 
possibly compromising the capacity to do science in such a universe!  


\noindent Note 2) The records scheme sitting within a Histories Theory \cite{GMH, GMH11, H99, HT, H03, HW, H09, H11} is independent of the Gell-Mann--Hartle versus IL 
distinction because these involve the single-time histories, i.e. a single projector, and then the ordinary and tensor products of a single projector obviously coincide and indeed 
trivially constitute a projector.  
Thus one can apply the Projector--Proposition Association and nicely found a propositional logic structure on this.

\noindent Note 3) like `$\FrQ$ is primary', `Hist is primary' can also be expanded categorically, propositionally {\sl and} perspectivally.  
Considering Prop(Hist) was an important motivation for IL. 

\mbox{ } 

\noindent {\bf Question 73} Are the momenta conjugate to histories themselves implemented by projectors? (One may well want to make propositions concerning these too).  

\mbox{ } 

\noindent The perspectival version of histories, Persp(Hist) = $\langle$ set of physical SubHist's $\rangle$ is somewhat different from that of configurations.  
In the Causal Sets approach, this was covered by Markopoulou \cite{MrkP}.
In jest, and yet with some lucidity, this `concerns histories that depend on who the historian is  
(rather than history just being written from the victor's perspective) and historians studying histories that are partly protagonized by other historians 
(perhaps via some of the mighty choosing to heed history's lessons rather than being doomed to repeat them)'.
What is less clear is whether historians are to be held to be some kind of privileged entities like observers are held to be in conventional QM. 
If historians are those who observe correlations in records (and thus deduce likely histories), then the two notions conflate.  

\mbox{ }  

\noindent Note 4) Considering Prop(Persp(Hist)) combines IL's program part of with Crane's.  
More generally, 

\noindent Prop(Persp( )) combines Mackey's Principle with part of Crane's program.

\mbox{ } 

\noindent {\bf Question 74} Does the Partial Observables Approach offer simpler ways of doing this than Isham-type approaches?    

\mbox{ }  

\noindent {\bf Question 75} What is a suitable notion of distance between quantum histories, i.e. on the space Hist?

\mbox{ }

\noindent Note 5) Notions of information for quantum histories have been considered by Hartle \cite{Hartle95}, Kent \cite{Kent97} and IL \cite{IL96}.

\noindent Note 6) As regards a notion of correlation between histories, note that the decoherence functional is already a key structural element for this.

\subsubsection{Further types of Histories Theory at the quantum level}

\noindent Problem with Histories Theory 8)/{\bf Question 76$^*$} incompleteness of development: the Kouletsis--Kucha\v{r} approach remains unfinished at the quantum level. 
Does the histories group approach really succeed where the canonical group approach got stuck?  

\mbox{ }  

\noindent {\bf Causal Sets Approach} \cite{Sorkin03, Surya} is like a Histories Theory in being path-based and spacetime-presumed. 

\noindent However, it does not assume strings of projectors, so it does not strictly meet the current Article's specifications for a Histories Theory.

\noindent It certainly is not timeless due to the emphasis on spacetime structure and on the primality of causality in particular,   
which is taken in this approach to be the fundamental notion to be kept in QG. 

\noindent Spacetime Reconstruction difficulties do however follow from the Causal Sets Approach's insistence on very sparse structure; already problematic 
for this approach at the classical level (Sec \ref{3-Overcome}), this issue remains very largely unexplored at the quantum level.  

\noindent Since it avoids using projectors, it is worth commenting that this approach uses a {\sl third} approach to the implementation of propositions, at the level of regions in Path.

\subsubsection{Computation of decoherence functionals in the $\FrG$-free case}

\beq
\langle \fQ^{\sf\si\sn} ||C_{\eta}|| \fQ^{\si\sn} \rangle = 
\sum\mbox{}_{\eta \, \in \, \sfQ^{\si\sn} c_{\eta} \sfQ^{\sf\si\sn} } \mbox{exp}(i\FS[\eta])
\label{Har0}
\eeq
for a given coarse-graining $C_{\eta}$ consisting of classes $\{c_{\eta}\}$, and summing over all paths beginning at configuration $\fQ^{\si\sn}$ 
and ending at configuration $\fQ^{\sf\si\sn}$.  
{\sl This} formulation for the class function can be interpreted in terms of paths without association of projectors. 
Then ($i$, $j$ are clearly not particle labels here)
\beq
{\cal D}\me\mc[\eta, \eta^{\prime}] = {\cal N}\sum\mbox{}_{\mbox{}_{\mbox{\scriptsize $ij$}}}\mbox{Prob}_i^{\sf\si\sn} 
                             \langle \psi_i^{\sf\si\sn} | C_{\eta^{\prime}} | \psi_j^{\si\sn} \rangle 
                             \langle \psi_i^{\sf\si\sn} | C_{\eta}          | \psi_j^{\si\sn} \rangle^* \mbox{Prob}^{\si\sn}_j \mbox{ } ,  
\label{Har1}
\eeq
where the probabilities are initial and final inputs alongside the initial and final states.  
\beq
{\cal N} :=  1\big/\sum\mbox{}_{\mbox{}_{\mbox{\scriptsize $ij$}}}
\mbox{Prob}^{\sf\si\sn}_i|\langle \psi_i^{\sf\si\sn}| C_S |\psi^{\si\sn}_j \rangle|^2\mbox{Prob}^{\si\sn}_j  
\label{Norm1}
\eeq
where $C_S$ is the sum of all the paths in (\ref{Har0}).
This is built in terms of a `kernel' path integral, 
\beq
\langle \psi_i^{\sf\si\sn} | C_{\eta} | \psi^{\si\sn}_j \rangle := 
        \psi_i^{\sf\si\sn}(\fQ^{\sf\si\sn}) \circ \langle 
        \fQ^{\sf\si\sn} || C_{\eta} || \fQ^{\si\sn} \rangle \circ \psi^{\si\sn}_j(\fQ^{\si\sn})
\label{Nin} \mbox{ } , 
\eeq
where $\circ$ is some Hermitian but not necessarily positive inner product, $\psi^{\si\sn}    = \psi^{\si\sn}   (\fQ^{\si\sn})   $ and 
$\psi^{\sf\si\sn} = \psi^{\sf\si\sn}(\fQ^{\sf\si\sn})$ are initial- and final-condition wavefunctions.    
Finally, the above `kernel' path integral takes e.g. the MRI form
$
\langle \fQ^{\sf\si\sn} || C_{\eta} || \fQ^{\si\sn} \rangle =
$
(\ref{PI-3-prelim}).
I differ from Hartle's review \cite{Hartle} in using $\d\fI$ in place of the Lagrange multiplier counterpart $\fN$.  
This distinction renders Histories Theory compatible with this Article's general ethos of Temporal Relationalism implemented by Temporal Relationalism via MRI/MPI; it also means that 
the precise form that the action is taken to have here is, as per App \ref{Examples}.A.5, $(\int_{\sbSigma}\d\bupSigma)\int\{\fP_{\sfB}\ordial{\fQ}^{\sfB} - \ordial\fA\}$.

\subsubsection{Quantum Histories Theory on a Machian temporally-relational footing}\label{HarRPM2}

As the input action to be used in Histories theory is that related to the Hamiltonian by a Legendre transformation, for 

\noindent theories spanning both RPM's and GR, by the last form in eq (\ref{Aac-1}), this can be built from the $\ft^{\se\sm(\sW\sK\sB)}$
However, the temporally relational actions in question are linear in $\d/\d\ft^{\se\sm(\sJ\sB\sB)}$ 
and in $\d\ft^{\se\sm(\sJ\sB\sB)}$ and so in fact they are MRI/MPI, and so actually one is free to use any other label time $\lambda$ if one so wishes. 
So there is in fact no hard content to using the emergent time in the construction of the decoherence functional.  
This is clear e.g. from the second form of eq (\ref{Aac-1}).
This is also the nature of the time used in the next SSec.  

\noindent In the $\FrG$-free case, conventionally, starting from (HistPhase, HistCan), one applies {\sc kin-quant}, {\sc assoc} and $\scQ$-{\sc solve} maps to arrive at 
[at least formally] (HistHilb, HistUni) [meant in the enlarged sense of IL].  
This is held to be useful because 1) given the subsequent histories group, its unitary representations specifically allow one access to the 
orthoalgebra of projection operators as propositions about histories of the theory.
2) The histories scheme involves a fresh set of algebraic entities and thus provides a second opportunity for the corresponding quantization scheme to work out in practise.  
[Though the price to pay may be significant: it is less conceptually well-established that one can base a canonical quantization scheme on histories rather than on configurations.]

\noindent In terms of the answers to physical questions, there would appear to be no difference between IL's scheme and Gell-Mann--Hartle's.  
IL is a technical refinement useful for establishing theorems, and also a conceptual improvement due to its manifestly fitting the Projector--Proposition Association.

 \subsubsection{Further layers of structure for Histories Theory}

Paralleling Secs \ref{Cl-Str} and \ref{QM-Str}, one can define notions of 

\noindent 1) subhistories,

\noindent 2) distances between histories and corresponding sense of localization, 

\noindent 3) histories correlators: quantum mechanically the decoherence functional is a type of such, but what about classically?

\noindent Next, histories propositions/logic are still an atemporal construct. 
Once again, this is an area that mostly only becomes interesting at the quantum level (see e.g. \cite{I97, Flori10}).

\noindent Finally, since in Records Theory, the histories brackets reduce to the usual Poisson brackets, records approaches give rise to the {\sl usual} 
canonical group with which so much exasperation was had as regards quantization \cite{I84, IK85}.

\subsection{$\FrG$'d versions of POT strategies at the quantum level}\label{KP}

First I rephrase the argument from Sec \ref{OOP-3} and provide more detailed commentary on it.  

\mbox{ }  

\noindent{\bf Kucha\v{r}'s Argument}  The reduced ordering {\sc r} ... {\sc q} is the physical ordering.  
This is because, firstly, {\sc r} ... {\sc q} and {\sc q} ... {\sc r} do not always agree. 
Secondly, one would not expect that appending unphysical fields to the reduced description should change any of the physics of the of the true dynamical degrees of freedom.   
Thus, if they do differ, one should go with the reduced version.  

\mbox{ } 

\noindent Note 1)  \K told me he believes a particular operator ordering of the Dirac approach exists for which its outcome and that of the reduced quantization coincide.  
For this ordering, therefore, the Dirac approach is just as physical. 
The trouble is, I note, how one would determine this operator ordering in cases for which the reduced quantization is technically inaccessible.  
Another question then is whether this lies within, say, DeWitt's family of operator orderings say, for which there is separate motivation.

\noindent Note 2) This is a set-up in which classical \K degradeables are known and quantum \K degradeables are determined up to the Multiple Choice Problem 
and are useable up to the extent that we know how to patch degradeables at the quantum level.  

\noindent Note 3) This argument logically extends \cite{APOT2} to its histories counterpart:  {\sc h} ... {\sc q} is the physical ordering.  

\mbox{ } 

\noindent {\bf Thiemann's pro-Dirac argument} \cite{Thie06}.  
This is in opposition to \K's pro-reduced or Dirac-only-if-aligned-with-reduced argument.  
I present this, and my counterargument to it in the next SSec, since it also involves issues of observables/beables.  

\mbox{ }

\noindent I next consider various specific strategies for nontrivial-$\FrG$ models.

\subsubsection{Type 1 Tempus Ante Quantum schemes}

In {\sc rtq} schemes, the general classical parabolic form (\ref{Para1}) yields the TDSE for the reduced configuration space evolution, 
\beq
i\hbar \partional\Psi/\partional \ft^{\sa\sn\st\se} = \hat{\fH}_{\st\sr\su\se}\lfloor \ft^{\sa\sn\st\se}, 
\widetilde{\bfQ}_{\so\st\sh\se\sr}^{\sfA}, \widetilde{\bfP}^{\so\st\sh\se\sr}_{\sfA}\rfloor\Psi \mbox{ } . 
\label{TEEDEE12}
\eeq 
A particular example of such is then the reduced--York-TDSE
\beq
i\hbar \delta\Psi/\delta \mt^{\sY\so\sr\sk}  = \widehat{\mH^{\st\sr\su\se}}( \widetilde{x}; \mt^{\sY\so\sr\sk}, \widetilde{\bQ}_{\st\sr\su\se}, \widetilde{\bP}^{\st\sr\su\se}] \Psi 
\mbox{ }   , 
\label{Red-York-TDSE}
\eeq
which is contingent on succeeding in classically having solved the accompanying $\scM_{\mu}$ so that true are the true gravitational degrees of freedom.

\noindent On the other hand, in classical internal or matter time-and-frame finding {\sc tqr} schemes, the generic spacetime-vector time-and-frame-dependent Schr\"{o}dinger equation is
\beq
i\delta\Psi/\delta {\cal X}^{\mu} = \widehat{\fH^{\st\sr\su\se}}_{\mu}\lfloor {\cal X}^{\mu}, \bfQ_{\so\st\sh\se\sr}, \bfP^{\so\st\sh\se\sr}_{\sfA}\rfloor\Psi \mbox{ } .    
\label{TEEDEE4}
\eeq 
The Dirac--York case of this is 
\beq
i\hbar\delta \Psi/\delta \mt^{\sY\so\sr\sk} = 
\widehat{\mH^{\st\sr\su\se}}(\ux; \mt^{\sY\so\sr\sk}, X^{\mu}, \mbox{\boldmath${\cal T}$\hspace{-0.04in}rue$^{(3)}$}, 
\mbox{\boldmath$\Pi$}^{{\cal T}\hspace{-0.03in}\sr\su\se^{(3)}}, \upzeta^{\mu}]      \mbox{ } , 
\label{Dir-York-TDSE}
\eeq
\beq
i\hbar\delta \Psi_{\mu}/\delta^{\chi^{\tY\to\tr\tk}} = 
\widehat{\Pi_{\st\sr\su\se}} (\ux; \mt^{\sY\so\sr\sk}, X^{\mu}, \mbox{\boldmath${\cal T}\hspace{-0.04in}$rue$^{(3)}$}, 
\mbox{\boldmath$\Pi$}^{{\cal T}\hspace{-0.03in}\sr\su\se^{(3)}}] \mbox{ } .   
\label{Dir-York-FDSE}
\eeq
\noindent Note 1) It is worth emphasizing that Problems with York Time 1 and 2) these are only formal because $\mH^{\st\sr\su\se}$ is only implicitly known as per Sec \ref{CGdyn}, 
and the explicit elimination of $\scM_{\mu}$ is also formal (if not so technically hard) as per Sec \ref{Longi}.

\noindent Note 2) Reference matter approaches usually also end up in with a quantum equation of the form (\ref{TEEDEE4}).

\subsubsection{Semiclassical schemes}

\noindent One could likewise face the linear constraints in an emergent semiclassical time approach 
that has been enlarged to include $\fh$- and $\fl$-linear constraints ({\sc qrt} scheme \cite{Kieferbook, SemiclI}.  

Here, the linear constraints $\scL\scI\scN_{\sfZ}$ contribute their own $\fh$ and $\fl$ equations.  
These do not enter the specification of the timestandard (though this can acquire the presence of the corresponding auxiliary cyclic velocity $\dot{\mc}^{\sfZ}$, 
in which case one needs to append a variational prescription that frees one of this, as per Sec \ref{Aux}).
Via the properly auxiliary-corrected version of the momentum--velocity relation, $\scL\scI\scN_{\sfZ}$ also enters the $\fl$-TDSE: 
\beq
i\hbar\{\partional/\partional\ft^{\se\sm} -  \scL\scI\scN_{\sfZ}\partional{\mc}^{\sfZ}/\partional\ft^{\se\sm} \}|\chi\rangle = \widehat{\fH}_{\sfl}|\chi\rangle \mbox{ } .
\eeq
This parallels the GR version of the Tomonaga--Schwinger equation (emergent-time-dependent Tomonaga--Schwinger--Einstein--Schr\"{o}dinger equation) with respect to bubble time 
\cite{Bubble}, 
\beq
i\hbar\{\delta/\delta\mt^{\se\sm} -  \scM_{\mu}\delta{\mF}^{\mu}/\delta\mt^{\se\sm} \}|\chi\rangle = \widehat{\mH}_{\sll}^{\sG\sR}|\chi\rangle \mbox{ } .  
\label{Tom-Schwi}  
\eeq
See Sec \ref{Dir-Semi} for further development of this.  

\mbox{ }

\noindent Note: the sufficiently-Machian emergent time is then infested with a $\FrG$-act that needs cancelling off with a $\FrG$-all, as detailed also in Sec \ref{Dir-Semi}.  

\mbox{ } 

\noindent This is bypassed by working with a reduced version, for which the TDSE is 
\beq
i\hbar\{\partional/\partional\ft^{\se\sm}  = \widehat{\widetilde{\fH}}_{\fl}|\chi\rangle \mbox{ } .
\eeq

\subsubsection{Paths and Histories schemes} 

\noindent 1) Configurational Relationalism Problem with Path Integral Formulation).  
Configurational Relationalism remains (in fact Wheeler \cite{WheelerGRT} originally envisaged the Sandwich Problems as a classical precursor to the Path Integral Formulation.  
One might be wary that path integrals approaches are a quantum counterpart of the already poorly-behaved thick sandwich in classical Geometrodynamics.

\noindent 2) For a $\FrG$-constrained formulation of some physical theory, the path integral takes the most very relational MPI form 
\beq
\langle \bfQ^{\sf\si\sn} || C_{\gamma} || \bfQ^{\si\sn} \rangle = 
\int_{\eta} \mathbb{D} \bfP \, \mathbb{D} \bfQ \, \mathbb{D}\d{\fI} \, \mathbb{D} \d{\fc} \, \mbox{det}{\cal F}[\bfQ, \d{\fI}, \d{\fc}^{\sfZ}] 
              \updelta[{\cal F}[\bfQ, \d{\fI}, \d{\fc}^{\sfZ}]] \mbox{exp}(i\FS[\bfQ, \bfP, \d{\fI}, \d{\fc}^{\sfZ}]) \mbox{ } .
\label{PI-3}
\eeq
\noindent Now, the $\updelta$ is a Dirac delta function imposing the basis set of gauge-fixing conditions ${\cal F}_{\sfB}$ and that the $\mathbb{D}$'s collectively form the 
OA-Liouville measure on OAPhase.

\noindent Also, in (\ref{PI-3}) $\mbox{det}{\cal F}$ is now the OA-version of the Fadde'ev--Popov determinant \cite{Fadde'ev, +Fadde'ev}.  
This determinant can be seen to arise as the Jacobian now in the OAPhase transformation by which the positions, momenta and cyclic ordials are formally split into the 
gauge-fixing-conditions-as-momenta, their conjugate coordinates and a further independent set of coordinates and momenta.

From its mathematical origin it is clear that in constraint-less cases (whether never-constrained or successfully reduced), the Fadde'ev--Popov factor is multiplicatively trivial.  
For use in the surviving cases, this determinant is, computationally det($\fM_{\sfA\sfB}$) for 
\beq
\fM_{\sfA\sfB}\delta(\ft_1 - \ft_2) = \frac{\delta}{\delta u(\ft_1)} \frac{\delta}{\delta v(\ft_2)} 
\int \{ \scC_{\sfA}[v], {\cal F}_{\sfB}[u]\} \d \ft \mbox{ } , 
\eeq
for whichever notion of $\ft$ the paths come with (most usually $\lambda$ or $\ft^{\se\sm(\sW\sK\sB)}$).  
Also here, $\scC_{\sfA}$ are secondary first-class constraints and ${\cal F}_{\sfB}$ are gauge-fixing conditions, which are assumed to obey the closure 
\beq
\{{\cal F}_{\sfB}, {\cal F}_{\sfC}\} = 0 \mbox{ } .  
\eeq
For RPM's the finiteness (i.e. lack of spatial extent) simplifies this working somewhat to
\beq
\mbox{det}(M_{\sfA\sfB}) = \int\{\scC_{\sfA}, {\cal F}_{\sfB}\}\d t^{\se\sm(\sW\sK\sB)}   \mbox{ } .  
\eeq

\mbox{ }  

\noindent Note: Histories Theory may still have linear constraints at this stage ({\sc qr} orderings), in which case there is a nontrivial histories commutator constraint algebra.  
If this is an {\sc hqr} ordering, the kinematical commutator algebra is obtained by selecting a subalgebra of the classical histories quantities, 
whose commutator suitably reflects global considerations, and the quantum histories constraints are some operator-ordering of their classical counterparts.
[This will not always form the same constraint algebroid that the histories Poisson brackets of the classical constraints did.]

\subsection{Problem of Beables at the quantum level}

\noindent The {\sc qbrt} classification,\footnote{It is pronounced `Cubert'.} 
i.e. how to order {\sc q}, {\sc b}, {\sc r}, {\sc t} tuples in dealing with the quantum version of a theory with Background Independence.  
It concerns in which order to hew at the necks and wings of the Ice Dragon.  
The relation-free count of number of possible orderings of maps is by now several hundred, further illustrating the value of restricting 
these via firstly some of the maps commuting and secondly by further selection principles.

\noindent As an example of applying this understanding about multiple orderings of maps, consider Thiemann's argument for `Dirac' rather than `reduced' \cite{Thie06}. 
The argument is based on [Fig \ref{ThieDir} a)] extra freedom in clock choices in the Dirac picture has its fluctuations suppressed in the reduced case, 
rendering the reduced case less physical.\footnote{\cite{Carlip01} 
also talks of clock fluctuations in this way, motivated e.g. by the status quo approach to path integrals for constrained theories.} 

\noindent My counterargument to this is that it specifically involves favouring {\sc tqr} over {\sc trq} rather than for {\sc q} ... {\sc r} over {\sc r} ... {\sc q} in general 
that standard useage of `reduced quantization' actually implies [Fig \ref{ThieDir} b)].  
Moreover, whilst the {\sc tqr} scheme having more variables does mean that it has a larger variety of such clock variables, these extra degrees of freedom 
are clearly unphysical and thus fluctuations in them are physically irrelevant  by applying \K's Argument to go in the opposite direction.
I.e. If one has the reduced system and adjoins auxiliary variables to it, it is not desirable for this unphysical adjunction to 
influence the physics, and this unphysical adjunction should not increase the variety of physical choices of clock variables in a way that is then to have physical consequences. 
%
{            \begin{figure}[ht]
\centering
\includegraphics[width=1.02\textwidth]{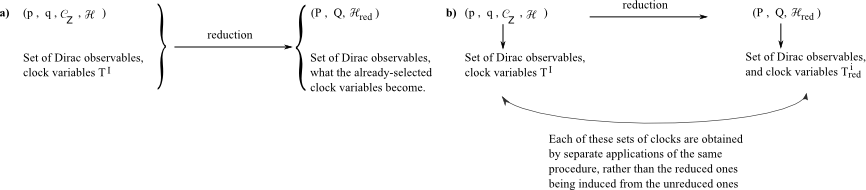}
\caption[Text der im Bilderverzeichnis auftaucht]{        \footnotesize{
\noindent a) Thiemann's notion of reduction versus b) the simpler and at least formerly standard notion, which logically precedes 
and mention of considering what to use for clock variables.}        }
\label{ThieDir}\end{figure}            }

\noindent Note 1) This alongside \K's argument makes for a strong case for {\sc r} ... {\sc q} approaches, contrary to much current literature.

\noindent Note 2) The general suggestion is that gauge theory is unfortunately no longer physically viable in the presence of a $\scQ\scU\scA\scD$ that corresponds to the temporal 
relationalism aspect of background independence.  
Thus a leading challenge for quantum gravity would appear to be how to redo classical gauge theory in gauge-invariant terms and follow through on quantizing that.
RPM's in 1- and 2-$d$ have the good fortune of being tractable in this way, but, for now, the extension of that to field theoretical models remains a major obstacle.  

\mbox{ } 

\noindent Furthermore, out of the remaining {\sc qbrt}'s, I argue in favour of {\sc rqbt}, {\sc rqtb}, {\sc rqbh}, {\sc rqhb} and {\sc rqb} using {\sc k} or {\sc d} for the {\sc b}'s 
and noting that all the {\sc t}'s and {\sc h}'s are single times and not time-frames, and basing my choice on Kucha\v{r}'s argument and the Machian Status Quo.  
Thus, provided that {\sc t}-{\sc h}-Nihil combinations are not being sought (for these it is equitable to set up {\sc t} and {\sc h} and the 
same relative position to {\sc q}, {\sc b}, {\sc r}), {\sc rhqb}, {\sc hrqb} and $\chi_{\tH}${\sc rqb} Histories Brackets Approaches also fit these criteria.

\mbox{ }  

\noindent I next focus on beables aspects of various particular strategies.  

\noindent \bu For RPM's the present Article resolves Configurational Relationalism at the classical level, and that stays solved, so this SSec is not elaborated upon strategy by strategy.  
\noindent This means that classical \K beables are known.  
Does this suffice to have quantum \K beables? Unfortunately, passage from classical \K beables to quantum \K beables requires 
choice of a subalgebra of them that are to be promoted to QM operators, which is in general a nontrivial procedure.
Moreover, as we shall see in the next SSec, my and Halliwell's classical construct for classical Dirac beables from classical \K beables does not render objects known to be promoteable 
to the quantum level as quantum Dirac beables.  
And yet there is a new semiclassical construct for objects obeying the quantum Dirac beables conditions, so what is not promoteable is in any case supplanted.

\noindent \bu Beables Problem with Unimodular Approach) The 3-metric operator does not commute with the approach's constraints, by which this scheme's interpretation 
of its wavefunction of the universe as a probability distribution for the 3-metric is not tenable.
I further contribute to this in Sec \ref{UniM} from a relational perspective.  

\noindent \bu As regards Rovelli-type approaches,   

\noindent 1) I recommend that the lattice of propositions structure also be incorporated into these.  
Indeed the partial observables notion would seem to fit in well with this as notions of correlation/mutual information 
corresponding to the physical questions that one can ask. 

\noindent 2) Rovelli states that his approach is {\sl required} to be in the Heisenberg picture \cite{RovPMPATPS}.    
Is this absolutely obligatory?  If so, what does this imply anything about changes in the foundations/interpretation of QM? 
(Ordinary QM has equivalence between the Heisenberg and Schr\"{o}dinger pictures; how is the implication that this is lost in the passage to QG to be understood?)  

\noindent \bu Beables Problem with Path Integral Formulation and with Histories Theory). 
One now has Path/Histories beables to find, as well as the issue of whether these are operationally meaningful.

\subsection{Halliwell's Combined Histories-Records-Semiclassical Approach}\label{QM-Halliwell-Intro}

\subsubsection{Quantum-level motivations for this unification}\label{Motiv}

Some motivation for this scheme (XXX and then XXXII in Fig \ref{Tless-Types}) is as follows.

\noindent Pro 1) has already been mentioned above: that both histories and timeless approaches lie on the common ground of atemporal logic structures \cite{IL}.   

\noindent Pro 2) As already encountered at the classical level, there is a Records Theory within Histories Theory, as pointed out by Gell-Mann and Hartle \cite{GMH} and 
further worked on by them \cite{GMH11}, and by Halliwell and collaborators \cite{H99, HT, H03, HT, HD, HW, HY, Yearsley, H09, H11}.     
Thus Histories Theory supports Records Theory by providing guidance as to the form a working Records Theory would take.
This also allows for these two to be jointly cast as a mathematically-coherent package (as already illustrated in preceding subsections).
Moreover, at the quantum level, as Gell-Mann and Hartle say \cite{GMH}, 
\beq
\mbox{Records are ``{\it somewhere in the universe where information is stored when histories decohere}"}. 
\label{situ2}
\eeq
Anti 1) If one alternatively considers records from first principles from which histories are to be derived, there is a problem that the 
decoherence functional collapses out of Records Theory via the path integrals involved ceasing to be defined as the history collapses to a single instant.
I.e. there is some chance that being more minimalistic than Gell-Mann--Hartle and Halliwell could itself succeed.  

\noindent Pro 3) Histories decohereing is a leading (but as yet {\sl not fully established}) way by which a semiclassical regime's WKB 
approximation could be legitimately obtained in the first place.  
Thus Histories Theory could support the Semiclassical Approach by freeing it of a major weakness.

\noindent Pro 4) The Semiclassical Approach and/or Histories Theory could plausibly support Records Theory by providing a mechanism for the semblance of dynamics \cite{H03, Kieferbook} 
(though the possibility of a practically useable such occurring within pure Records Theory has not been overruled).  
Such would go a long way towards Records Theory being complete.  
I note that emergent semiclassical time amounts to an approximate semiclassical recovery \cite{SemiclI} of Barbour's classical emergent time \cite{B94I}, 
which is an encouraging result as regards making such a Semiclassical--Timeless Records combination. 

\noindent Pro 5) The elusive question of which degrees of freedom decohere which should be answerable through where in the universe the information is actually stored, 
i.e. where the records thus formed are \cite{GMH, H03}.  
In this way, Records Theory could in turn support Histories Theory. 

\noindent Pro 6) The Semiclassical approach aids in the computation of timeless probabilities of histories entering given configuration space regions.
This is by the WKB assumption giving a semiclassical flux into each region \cite{H03} in terms of $S(h)$ and the Wigner functional (defined in Sec \ref{Wigi}).  
Such schemes go beyond the standard Semiclassical Approach, and as such there may be some chance that further objections to the Semiclassical Approach 
(problems inherited from the WDE and Spacetime Reconstruction Problems) would be absent from the new unified  strategy.  

\noindent Pro 7) Halliwell's approach has further foundational value for practical Quantum Cosmology.  

\noindent Semiclassical provision of a timefunction frees one from much of the Multiple Choice Problem with Histories or any further mishaps arising from having 
to import internal time concepts, with decoherence covering why semiclassical and records covering the issue of what decoheres what.  

\noindent Caveat. For now I know of no way in which this combined scheme can ameliorate the Histories Problems 2) to 5) that it inherits.

\noindent The question to be addressed now is: for an eigenstate of the Hamiltonian, what is the probability of finding the system in a region (or series of regions) 
of configuration space without reference to time \cite{H03}? 

\noindent How does making the Semiclassical Approach more Machian impinge on the combined scheme?  

\mbox{ } 

\noindent Note 1) Halliwell's approach's attitude to observables/beables is a Dirac one (see Sec \ref{Examples}), though he does not for now consider the effect of linear constraints;  
the present Article's study is the first to consider examples possessing linear constraints.  

\noindent Note 2) The Gambini--Porto--Pullin approach \cite{GPP, GPP04a, GPP04b, PGP1} is a competitor insofar as it is a timeless/semiclassical combination 
(their work with Torterolo \cite{GPPT} furthermore combines this with ideas about beables).   
Thus it is similarly motivated to Halliwell's approach and similarly applicable to POT and other issues in the foundations of Quantum Cosmology.  
It can also be investigated in the RPM arena. 

\noindent Note 3) {\bf Deteriorated/doctored records}.  
Information can be lost from a record `after its formative event' -- the word ``stored" in (\ref{situ2}).      
Photos yellow with age and can be defaced or doctored. 
Some characteristics of the microwave `background' radiation that we observe have in part been 
formed since last scattering by such as the integrated Sachs--Wolfe effect or foreground effects \cite{coscorr}.

\subsubsection{Outline of Halliwell's investigations so far}

\noindent \bu Halliwell works within Histories Theory, doing so in the Gell-Mann--Hartle formalism.  
This picks up the POT criticisms from the end of the last section albeit these are very largely absent from the present article's RPM's.

\noindent \bu Halliwell alludes to the Mott--Bell--Barbour bubble chamber paradigm, and to Dirac observables (though he has not to date considered the effect of linear constraints).    
He implements propositions via regions of configuration space.
In fact, Hartle and Halliwell both separately consider such schemes, via various versions of {\bf class operators} that relate to Prob(enters a 
region $\bFrR$ of configuration space); $C_{\bsFrR}$ refers to an $\eta$ comprising of those histories that involves crossing over into region $\bFrR$.
For RPM's, this includes cases of particularly lucid physical significance via Sec \ref{TessiRegions}'s tessellation interpretation.\footnote{In fact, 
Prob(does not enter a region $\bFrR$) plays the main role in Halliwell's program.}
%
The extent to which the regions criticism of the \NSI carries over to this is covered in the Conclusion.  
 
\noindent \bu In 1999, Halliwell investigated Pro 2) and 5) in \cite{H99} for a very simple toy model: a heavy particle moving through a medium consisting of a few light particles. 
The heavy particle disturbs these into motion. 
Subsequent instants consist of the particles' positions and momenta. 
It is these instants which are the records, and the dynamics or history of the large particle can then be reconstructed (perhaps to some approximation) from them. 
A particular feature of this calculation is that indeed a very small environment of l-particles suffices in order to have a nontrivial notion of imperfect record.
This points to the possibility of using finite toy models with additional desirable theoretical parallels, which is central to the present Article.  
Halliwell \cite{H99} also makes (and reasonably establishes, in the context of that paper's very simple models) the information-theoretic and Histories--Records abridging conjecture 
that the number of bits required to describe a set of decoherent histories is approximately equal to number of bits thrown away to environment 
(this ties in well with the aforementioned use of notions of information).
See Sec 25 for more details, with RPM-example comments.

\noindent \bu Halliwell investigated Pro 6) in 2003 \cite{H03} for a free particle, addressing, for an energy eigenstate, 
what is the probability of finding the system in a series of regions of configuration space without reference to time.

\noindent \bu Halliwell \cite{H09, H11} has recently (2009, 2011) investigated decohereing histories as a possible means of constructing the probability distribution for the WDE. 
(This also uses Semiclassical Approach techniques and could be useful for avoiding problems involving how to interpret the Semiclassical Approach's WDE).
Some intermediate/supporting steps in this program were co-authored with his students Dodd, Thorwart, Wallden and Yearsley \cite{HD, HT, HW, HY, Yearsley}.


\subsubsection{Semiclassical--Histories--Records Combined Scheme}\label{Wigi}

\noindent The last alternative has further parallel \cite{Halliwell87} at the semiclassical level with the {\bf Wigner function} \cite{Wigner32}
(see also \cite{Hillery, Carruthers, Tatarski, BJ84, BB97, B-Bbook}).  
This is a QM phase space distribution function. 
It is but a quasiprobability distribution because it can take negative values; the preceding three references deal in detail with its physical interpretation.  
To give the reader a conceptual outline of the Wigner function, a form for it in the flat-configuration-space $K$-dimensional form in the Cartesian case is
\beq
\mbox{Wig}[\bq, \bp] = \frac{1}{p} \int\int \d^K{\by} \,\langle \psi(\bq + {\by} | \mbox{exp}(2i\,{\by} \cdot \bq ) |\psi(\bq - {\by})\rangle  \mbox{ } . 
\eeq
This set-up is such that its integral over $\bp$ gives $|\psi(\bq)|^2$ and its integral over $\bq$ gives $|\psi(\bp)|^2$.  
A distinguishing characteristic of the Wigner functional is that its equation of motion is very similar to the classical one; in Liouville form, in the Cartesian coordinate case, 
one has just a correction proportional to (making $\hbar$ explicit) $\hbar^2V^{\prime\prime\prime}\pa^3 S/\pa p^3$ for $V$ the potential \cite{B-Bbook}. 

\noindent Semiclassicality helps at this particular point with explicit construction of class functional in what seems to be an unambiguous manner 
(this includes \cite{H09, H11}'s modifications).  
Next, starting from the WKB ansatz (\ref{WKB}), Halliwell \cite{Halliwell87} established a further result, 
\beq
\mbox{Wig}[\bq, \bp] \approx |\chi(\bq)|^2\updelta^{(K)}(\bp - {\mbox{\boldmath{$\pa$}}}S) \mbox{ } , 
\label{Halliwell87}
\eeq
with $\bp$ being equal to ${\mbox{\boldmath{$\pa$}}}S$ at the purely classical level (as per Hamilton--Jacobi Theory).   
Then (based on \cite{HP86, Halliwell87, 92})
\beq
P_{\Upsilon}^{\sss\se\sm\si\scc\sll} \approx  
\int \d t \int_{\Upsilon}\mathbb{D}{\Upsilon}(\bq) \,\,\, \bnu \cdot {\mbox{\boldmath{$\pa$}}} S \, |\chi(\bq)|^2 \mbox{ } .
\label{36}
\eeq
Halliwell's treatment continues within the standard framework of decoherent histories.
The key step for this continuation is the construction of class operators, which uplifts a number of features of the preceding structures.

\subsubsection{Class operators}

\noindent Class operators, which I denote by $\mC_{\bsFrR}$, concern Prob(enters a region $\bFrR$ of configuration space); $\mC_{\bsFrR}$ refers to an $\eta$ comprising 
of those histories that involves crossing over into region $\bFrR$.\footnote{This is based on the idea of scattering of classical trajectories in the region.  
There are various versions of class operators as per the distinction between Hartle and Halliwell versions, below and \cite{H09, H11}/Paper II's version.
Issues with this are then as follows.  
i)    A slight spreading occurs. 
ii)   The $P_{\tFrRR}$ notion is open to difficulties in general due to chaos.
iii)  Harshness of the $\theta$ function's edges causes a quantum Zeno problem; this is resolved by the construction in \cite{H09} and \cite{AHall2}.
In fact, via this resolution Prob(does not enter a region $\bFrR$) ends up playing the main role in Halliwell's program.\label{Zeno}}
%
Halliwell's versions \cite{H03, H09, H11} use integrals over time to resolve the issue of compatibility with the Hamiltonian constraint; 
that is why I choose to use Halliwell's over Hartle's.  

\mbox{ } 

\noindent Note 1) (Conceptually-plausible Weakening).  At least this conceptualization of class operators suggests being able to have a path-only (no projectors) form of class operator 
(now referring to a $\gamma$), this itself being the key entity in this approach and, more widely, a means of constructing decoherence functionals.

\mbox{ }

\noindent {\bf Mathematical implementation of class operators}.  
One cannot use the most obvious 
\beq
\mC_{\bsFrR}(\bq_{\sf}, \bq_0) = \int_{-\infty}^{+\infty} \d \lambda\, \mbox{exp}(-i E\lambda) \, 
\int \mathbb{D}\bq(t)\mbox{exp}(iS[\bq(t)]) \, \theta 
\left(   
\int_0^{\lambda} \d t \, \mbox{Char}_{\bsFrR}(\bq(t))  - \epsilon 
\right) 
\eeq
(whose $E$-factor' here comes from \cite{HT} assuming the {\it Rieffel inner product} \cite{HM97, H03}) because of non-commutation with $H$.  

\mbox{ }

\noindent However, the `sharpened' (and `sharp-edged', which turns out to be inconvenient later on) version,  
\beq
{\widehat{\iC}}^{{\sharp}}_{\bsFrR} = \theta
\left(   
\int_{-\infty}^{\infty} \d t \, \mbox{Char}_{\bsFrR}(\bq(t))  - \epsilon 
\right) \fP(\bq_{\sf}, \bq_0) \,   \mbox{exp}(i\,\fA(\bq_{\sf}, \bq_0)) \mbox{ } , \label{gyr}
\eeq
is satisfactory, both conceptually and as regards commutation with $H$.  
The cofactor of the $\theta$-function above is the standard semiclassical approximation to the unrestricted path integral. 
The nature of the prefactor $\fP$ is described in \cite{HT} and references therein.  
This is not the end of the story since (\ref{gyr}) is technically unsatisfactory for the reason given in footnote \ref{Zeno}, 
as resolved in \cite{H09} (and covered in  Paper II), but the above form does suffice as a conceptual-and-technical 
start for RPM version of Halliwell-type approaches and amounting to an  extension of them to cases including also linear constraints.  

\mbox{ } 

\noindent Note 2) Semiclassicality helps at this particular point with explicit construction of class functional in what seems to be an unambiguous manner.

\subsubsection{Decoherence functionals} 

\noindent The decoherence functional is of the form 
\beq
{\cal D}\me\mc[\eta, \eta^{\prime}] = \int_{\eta}\mathbb{D} \bq \int_{\eta^{\prime}} \mathbb{D} \bq^{\prime} 
\mbox{exp}(i\{S[\bq(t)] - S[\bq^{\prime}(t)]\}\BigupRho(\bq_{0}, \bq_{0}^{\prime}) \mbox{ } .
\eeq
\noindent Class operators are then fed into the expression for the decoherence functional:  
\beq
{\cal D}\me\mc[\eta, \eta^{\prime}] = \int\int\int \mathbb{D}\bq_{\sf}\mathbb{D}\bq_{0} \mathbb{D}\bq^{\prime}_{\sf} \, 
{\widehat{\iC}}^{{\sharp}}_{\eta}[\bq_{\sf}, \bq_{0}] \, 
{\widehat{\iC}}^{{\sharp}}_{\eta^{\prime}}[\bq^{\prime}_{\sf}, \bq^{\prime}_{0}] 
\Psi(\bq_{0})  
\Psi(\bq^{\prime}_{0}) \mbox{ } .  
\label{dunno}
\eeq
The decoherence functional can be recast in terms of the {\it influence functional} ${\cal I}$ \cite{FH} as 
\beq
{\cal D}\me\mc[\eta, \eta^{\prime}] = \int\int\int \mathbb{D} \bq_{\sf} \mathbb{D} \bq_{0} \mathbb{D} \bq_{0}^{\prime}
{\widehat{\iC}}^{{\sharp}}_{\eta}[\bq_{\sf}, \bq_0]  
{\widehat{\iC}}^{{\sharp}}_{\eta}[\bq_{\sf}, \bq_0^{\prime}]
{\cal I}[\bq_{\sf}, \bq_0, \bq_0^{\prime}]
\Psi(\bq_0)
\Psi^*(\bq_0^{\prime}) \mbox{ } .
\eeq 
Then, if \cite{HT}'s conditions hold (involving environment-system interactions), the influence functional takes the form (in the Cartesian case) 
\beq
{\cal I}[\bq_{\sf}, \bq_0, \bq_0^{\prime}] = \mbox{exp}(i \bq \cdot \bGamma + \bq \cdot \bsigma \cdot \bq)
\eeq
for          ${\bq}^{-} := {\bq} - {\bq}^{\prime}$ and $\Gamma_{\Gamma}$, $\sigma_{\Gamma\Lambda}$ are real coefficients 
depending on ${\bq} + {\bq}^{\prime}$ alone and with $\bsigma$ a non-negative matrix.
Using $\bq^+ := \{\bq_0 + \bq_0^{\prime}\}/2$ as well, the Wigner functional is 
\beq
\mbox{Wig}[\bq, \bp] = \frac{1}{\{2\pi\}^K}\int \mathbb{D}\bq \, \mbox{exp}(-i\bp\cdot\bq)\rho(\bq^+ + \bq^-/2, \bq^+ - \bq^-/2)  \mbox{ } . 
\eeq
\beq
\mbox{Then} \hspace{1.7in} P_{\bsFrR} = \int\int \mathbb{D} \bp_0 \mathbb{D}\bq \, \, \theta
\left(
\int_{-\infty}^{+\infty} \d  t \,\mbox{Char}_{\bsFrR}\big( \bq^{+\scc\sll}\big) - \epsilon
\right)
\widetilde{\mbox{Wig}}[\bq_0^+, \bp_0]  \hspace{4in}  
\eeq 
for $\bq^{+\scc\sll}(t)$ the classical path with initial data $\bq^+_0, \bp_0$ and Gaussian-smeared Wigner function
\beq 
\widetilde{\mbox{Wig}}[\bq^+_0, \bp_0] = \int \mathbb{D} \bp \, \mbox{exp}(-\frac{1}{2}
\{\bp_0 - \bp - \Gamma\}\cdot\bsigma\cdot \{\bp_0 - \bp - \bGamma\}) \mbox{Wig}[\bq_0^+, \bp_0]  \mbox{ } .  
\eeq 
\noindent Note 1) Moreover, by involving a $t$-integral, Halliwell's object is not local in time, which, as for the classical counterpart, 
would however be a desirable property in a beable that can be used in practise.  

\noindent Note 2) There may be a region implementation of the propositions problem.  
The ${\widehat{\iC}}^{\sharp}$ expression's dependence on regions is Boolean: $\mR_1$ OR $\mR_2$ is covered by $f_{R_1 \bigcup R_2}$.  
However, it is not then clear that this is desirable as regards considering the entirety of the quantum propositions and how these combine, as explained in Sec \ref{QM-Str}.

\subsubsection{Machian Time version sitting within Semiclassical Approach}

This involves the $t \longrightarrow t^{\se\sm(\sW\sK\sB)} \approx t^{\se\sm(\sW\sK\sB)}_{\{1\}}$ version of the preceding 3 SSSecs.

\subsection{Extension of Halliwell's work to $\FrG$-nontrivial theories}

\subsubsection{Semiclassical quantum working}

\noindent Extension to consideration of the Wigner functional in curved space is required.  
As well as previous considerations of volume elements, this has the further subtlety that the sums inside the bra and ket are no longer trivially defined.  
This was resolved by Winter, Calzetta, Habib, Hu and Kandrup \cite{WKCHH1, WKCHH2, WKCHH3, WKCHH4}, by Fonarev \cite{Fon} and by Liu and Qian \cite{LQ} in the case of 
Riemannian configuration space geometry via local geodesic constructions.

\noindent Underhill's earlier study \cite{Underhill} works with just affine structure assumed.
(In searching for this topic in the literature, it is useful to note that the Wigner functional is closely related to the Weyl transformation; 
see also Sec 2.3 of review \cite{Landsman}.)     
Liu and Qian also extended their work \cite{LQ} to principal bundles over Riemannian manifolds, thus covering what is required to extend Halliwell's 2003 approach's 
treatment in terms of Wigner functionals.
Because of this, I specifically take `Wigner functionals in curved space' in the sense of Liu and Qian when in need of sufficiently detailed considerations.\footnote{Many  
issues in geometrical quantization and similar (e.g. deformation quantization) are lurking just behind this door...}  
%
Finally, I emphasize again that Wigner functionals are only temporary passengers in the present program due to their being used in Halliwell 2003 
types implementations of class operators but no longer in Halliwell 2009 ones, by which I keep the account of this SSec's subtleties brief.   

\mbox{ }  

\noindent The preceding alternative expression has further parallel with the Wigner functional at the semiclassical level.  
Next, \cite{Halliwell87}'s straightforward approximations in deriving (\ref{Halliwell87}) locally carry over \cite{AHall}, so 
\beq
\mbox{Wig}[\biK, \biP^{\tiK}] \approx |\chi(\biK)|^2\updelta^{(r)}(\biP^{\tiK} - {\mbox{\boldmath{$\pa$}}}^{\tiK}S)
\label{Wig0}
\eeq
($\biP^{\tiK}$ being ${\mbox{\boldmath{$\pa$}}}^{\tiK} S$ for classical trajectories). 
Then Halliwell'-type heuristic move is then to replace $\mw$ by Wig in (\ref{clvers3}), giving
\beq
P_{\Upsilon}^{\sss\se\sm\si\scc\sll} \approx \int\d t^{\se\sm(\sW\sK\sB)}\int_{\Upsilon}\d\Upsilon(\biK) \bnu^{\tiK} \cdot \frac{\pa S}{\pa\biK} |\chi(\biK)|^2 \mbox{ } .  
\eeq
This remains PPSCT-invariant as the quantum inner product and the classical $\int \d\biP^{\tiK} \mw(\biK, \biP^{\tiK})$ both scale  equally as $\Omega^{-r}$.

\subsubsection{Class operators as quantum Dirac degradeables}

The Halliwell-type treatment continues \cite{AHall} within the framework of decoherent histories, which I take as formally standard for this setting too.  
The key step for this continuation is the construction of class operators, which uplifts a number of features of the preceding structures.
One now uses  
\beq
{\widehat{\iC}}^{\sharp}_{\bsFrR} = \theta
\left(   
\int_{-\infty}^{+\infty} \d t^{\se\sm(\sW\sK\sB)}_{\{1\}} \mbox{Char}_{\bsFrR}(\biK(t^{\se\sm(\sW\sK\sB)}))  - \epsilon 
\right) 
P(\biK_{\sf}, \biK_0) \,   \mbox{exp}(iA(\biK_{\sf}, \biK_0)) \mbox{ } , \label{spoo}
\eeq
which obey 
\beq
[ \widehat{\scQ\scU\scA\scD}, \widehat{\iC}^{\sharp}_{\bsFrR} ] = 0 
\eeq
by construction and 
\beq
[ \widehat{\scL\scI\scN}, \widehat{\iC}^{\sharp}_{\bsFrR} ] = 0 
\eeq
by being a function of the \K beables.
The cofactor of $\theta$ being some approximand to the quantum wavefunction, it PPSCT-scales as $\Omega^{\{2 - r\}/2}$.

Again, this class operator is not the end of the story since it is technically unsatisfactory, as resolved in \cite{H09, H11} (and to be covered in Paper II). 
Nevertheless, the above form serves as a conceptual-and-technical start for RPM version of the work and extension to cases with 
linear constraints, and for the present conceptual, whole-universe and linear-constraint extending paper, this is as far as we shall go.  

\mbox{ }  

\noindent In general these are just degradeables since they concern localized regions; nor have they for now in my account been built with global or even widely non-local 
geometrical considerations in mind.

\subsubsection{Decoherence functional}

The decoherence functional is of the form 
\beq
{\cal D}\me\mc[\eta, \eta^{\prime}] = \int_{\eta}\mathbb{D} \biK \int_{\eta^{\prime}} \mathbb{D} \biK^{\prime} 
\mbox{exp}(i\{S[\biK(t^{\se\sm(\sW\sK\sB)})] - S[\biK^{\prime}(t^{\se\sm(\sW\sK\sB)})]\}\BigupRho(\biK_{0}, \biK_{0}^{\prime})\omega_{\sd} \mbox{ } .
\eeq
For this to be PPSCT-invariant as befits a physical quantity, it needs to have its own weight $\omega_{\sd}$, PPSCT-scaling as 
$\omega_{\sd} \longrightarrow \overline{\omega}_{\sd} = \Omega^{-r}\omega_{\sd}$.

Class operators are then fed into the expression for the decoherence functional, giving 
\beq
{\cal D}\me\mc[\eta, \eta^{\prime}] = \int \d^3\biK_{\sf}\d^3\biK_{0} \d^3\biK^{\prime}_{\sf}
\widehat{\iC}^{{\sharp}}_{\eta}[\biK_{\sf}, \biK_{0}] 
\widehat{\iC}^{{\sharp}}_{\eta^{\prime}}[\biK^{\prime}_{\sf}, \biK^{\prime}_{0}]
\Psi(\biK_{0})  
\Psi(\biK^{\prime}_{0})
\omega^2\omega_{\sd} \mbox{ } .  
\label{dunno2}
\eeq 
\noindent The $\omega^2$ factor has one $\omega$ arise from the density matrix and the other from the 2-wavefunction approximand expressions from the two $\mC^{\prime}$'s.
If the universe contains a classically-negligible but quantum-non-negligible environment as per Appendix \ref{App-Env}, the influence functional ${\cal I}$ 
makes conceptual sense and one can rearrange (\ref{dunno2}) in terms of this into the form  
\beq
{\cal D}\me\mc[\eta, \eta^{\prime}] = \int\int\int \mathbb{D}\biK_{\sf} \mathbb{D}\biK_{0} \mathbb{D}\biK_{0}^{'}
\widehat{\iC}_{\eta}^{\sharp}[\biK_{\sf}, \biK_0]  
\widehat{\iC}_{\eta}^{\sharp}[\biK_{\sf}, \biK_0^{'}] 
{\cal I}[\biK_{\sf}, \biK_0, \biK_0^{'}]
\Psi(\biK_0)
\Psi^*(\biK_0^{'})
\omega^2\omega_{\sd} \mbox{ } .  
\label{83}
\eeq

\subsubsection{Alternative indirect $\FrG$-act, $\FrG$-all implementation}

Here,
\beq
{\widehat{\iC}}^{{\sharp}\sfg-\sf\sr\se\se}[\brho_{\sf}, \brho_{0}] = \int_{\sfg \in \sFG}\mathbb{D} \fg \, \stackrel{\rightarrow}{\FrG_{\sfg}}
\left(
\theta
\left(
\int_{-\infty}^{+\infty} \d t^{\se\sm(\sW\sK\sB)} \, \mbox{Char}_{\bsFrR}\big(\brho_0^{\sf}(t^{\se\sm})) - \epsilon
\right)
P(\brho_{\sf}, \brho_0) \mbox{exp}(i A(\brho_f, \brho_0)
\right) \mbox{ } . \hspace{3.3in}
\eeq
It is indeed physically desirable for these to already be individually $\FrG$-invariant.
Then making the decoherence functional out of ${\widehat{\iC}}^{\sharp\sfg-\sf\sr\se\se}_{\bsFrR}$ (and noting there is an issue of then needing to average multiple times, 
though at least $\FrG$-averaging a $\FrG$-average has no further effect, making this procedure somewhat less ambiguous than it would have been otherwise),
$$
\mbox{${\cal D}$ec}[\eta, \eta^{\prime}] = \int_{\sfg \in \sFG}\mathbb{D} \fg \, \stackrel{\rightarrow}{\FrG_{\sfg}} \left\{
\int\int\int \mathbb{D}\bq_{\sf}\,\mathbb{D}\bq_0\,\mathbb{D}\bq_0^{\prime}\, 
{\widehat{\iC}}^{{\sharp}\sfg-\sf\sr\se\se}_{\eta}[\bq_{\sf}, \bq_0] 
{\widehat{\iC}}^{{\sharp}\sfg-\sf\sr\se\se}_{\eta^{\prime}}[\bq_{\sf}, \bq_0^{\prime}]
\Psi(\bq_0)\Psi(\bq_0^{\prime})\right\} 
$$
\beq
= \int_{\sfg \in \sFG}\mathbb{D} \fg \, \stackrel{\rightarrow}{\FrG_{\sfg}} \left\{
\int\int\int \mathbb{D}\bq_{\sf}\,\mathbb{D}\bq_0\,\mathbb{D}\bq_0^{\prime}\, 
{{\widehat{\iC}}}^{{\sharp}\sfg-\sf\sr\se\se}_{\eta}[\bq_{\sf}, \bq_0] 
{{\widehat{\iC}}}^{{\sharp}\sfg-\sf\sr\se\se}_{\eta^{\prime}}[\bq_{\sf}, \bq_0^{\prime}]
{\cal I}[\bq_{\sf}, \bq_0, \bq_0^{\prime}]\Psi(\bq_0)\Psi(\bq_0^{\prime})\right\}  \mbox{ } .  
\eeq
\noindent A problem with this alternative approach is that it becomes blocked early on as regards more-than-formality for the case of the 3-diffeomorphisms. 

\mbox{ }  

\noindent Analogy. The {\sl arrival time problem} has similar mathematics \cite{HY, Yearsley, AT1, AT2, AT3, AT4}.

\subsubsection{Comparison with a few more minimalist combined schemes} 

\noindent Note 1) One advantage of Gambini--Porto--Pullin's scheme over the Halliwell-type one is that it is based on the 
Projector-Proposition Association rather than on a classical regions implementation. 

\noindent Note 2) Arce \cite{Arce} has also provided a combined \CPII--semiclassical scheme [for yet another notion of semiclassicality, 
that keeps some non-adiabatic terms and constituting an example of self-consistent approach, c.f. Sec \ref{Semi-Avs}].  

\vspace{10in}

\subsection{Strategizing about the Quantum Constraint Closure/Partional Evolution Problem}\label{FEP}

Closure becomes complicated at the quantum level.
Firstly, one has to bear in mind that the classical and quantum algebras/oids are not in fact necessarily related due to global and Multiple Choice Problem considerations \cite{I84}.  

\noindent Suppose then that we experience non-closure.  

\noindent 1) One could try to blame this on operator-ordering ambiguities, and continue to try to obtain closure by ordering differently.  

\noindent 2) Another countermeasure sometimes available is to set some numerical factor of the obstructing term to zero, like how 
String Theory acquires its particular dimensionalities (26 for the bosonic string or 10 for the superstring).  
This is termed applying a {\sl strong} restriction.

\noindent 3) Another possibility is to accept the QM-level loss of what had been a symmetry at the classical level. 

\noindent 4) Yet another possibility is that including additional matter fields could cancel off the anomaly hitherto found; some supersymmetric field theories and 
Supergravity theories exhibit this feature. 

\noindent 5) A final possibility is to consider a new constraint algebroid, in particular one involving an algebra on a smaller set of objects than the 
Dirac algebroid's set of GR constraints, along the lines of e.g. the Master Constraint Program.  
This cuts down the number of constraints so far that it would free the theory of all anomalies.  
However, that is in itself suspect since the same conceptual packaging of constraints into a Master Constraint would then appear to be applicable to the flat-spacetime gauge theory of 
standard Particle Physics, and yet these have not been declared to no longer have anomaly concerns due to the advent of the Master Constraint Program...

\mbox{ } 

\noindent On the other hand, other variants of constraint algebroid are enlarged, e.g. more traditional forms of LQG algebroid 
(anomaly analysis for which is in e.g. \cite{LQGAnomal1, LQGAnomal2}), histories algebroid \cite{Kouletsis} or a linking theory algebroid \cite{Kos4}.  

\noindent Non-closure can be due to 

\noindent a) foliation-dependent terms.

\noindent b) If scale is included among the physically-irrelevant variables, then there can also be conformal anomalies; these may plague such as the Linking Theory of Shape Dynamics.
This amount to `frame' now involving a conformal factor as well as a shift.

\mbox{ } 

\noindent Some problems with particular strategies are as follows.

\noindent Constraint Closure Problem with Path Integral Formulation 1).  
Path integrals do not directly involve constraint closure; however, the Measure Problem's Fadde'ev--Popov factors would be affected if constraint closure were not to hold.  

\noindent Diffeomorphism-based complications with Measure Problem with Path Integral Formulation.  
I take this opportunity to point out that, much like the Inner Product Problem 
canonical sequel of the Frozen Formalism Problem itself has connotations of time, when one instead skirts the Frozen Formalism Problem by passing to a path integral approach and the 
sequel problem is traded for the Measure Problem, then the diffeomorphism-based complications to the Measure Problem ensure that this replacement sequel remains time-related.  

\noindent Constraint Closure Problem with Path Integral Formulation 2)
Moreover, QFT is certainly often formulated in terms of path integral formulated and yet this is not claimed to free one from the scourge of anomalies.

\noindent I partly credit \cite{Kuchar92} for this block.  

\mbox{ }

\noindent Next, I point out the concept of `{\it beables anomalies}', i.e. finding that promoting classical beables to QM operators produces objects that no longer 
commutator-commute with the quantum constraints. 
Schematically,
\beq
\{\iD, \scC_{\Gamma}\} = 0 \mbox{ } \mbox{ } \not{\hspace{-0.05in}\Rightarrow} \mbox{ } \mbox{ } [\hat{\iD}, \hat\scC_{\Gamma}] = 0
\eeq
so classical Dirac beables are far from necessarily promoted to quantum Dirac beables, and 
\beq
\{\iK, \scL\scI\scN_{\sfZ}\} = 0 \mbox{ } \mbox{ } \not{\hspace{-0.05in}\Rightarrow} \mbox{ } \mbox{ } [\hat{\iK}, \widehat{\scL\scI\scN}_{\sfZ}] = 0
\eeq
so likewise for \K beables.
These statements are moreover dependent {\sl twice over} on operator-ordering ambiguities (in the beable operator and in the constraint operators).
This will tend to add to the futility of {\sc b} ... {\sc q} schemes, since classical Problem of Beables solutions 
need not be straightforwardly promoteable to a quantum Problem of Beables solution.  
Perfectly good classical beables fail to be quantum beables, as some kind of parallel of perfectly good classical symmetries failing to be quantum symmetries 
in the study of anomalies proper.  
\K beables may well be largely exempt due to the nice properties of Lie groups under quantization schemes, but, in cases going beyond that 
(Dirac observables, the classical Dirac algebroid...) one may find one has to solve the QM Problem of Beables again. 
See the Halliwell-type combined scheme of Sec \ref{Cl-Combo} for a good example, there are separate classical and QM implementations for Dirac-type beables, 
whilst the \K beables translate over straightforwardly for the triangle RPM model \cite{AHall}.  

\noindent Such a complication might be avoidable by making strong restrictions at the level of getting the algebraic structure of the beables to close. 

\noindent Finally, if the Constraint Closure Problem leads to $\scC${\sc -rep} may have more $\widehat{\scC}$'s than there were $\scC$'s, 
the definition of quantum beables is furtherly stringent due to requiring commutation with the further $\widehat{\scC}'s$.  

\vspace{10in}

\subsection{Strategizing about the Quantum Foliation Dependence Problem}\label{Fol}

\noindent In the continuum GR case, the normal to the spatial hypersurface, $\mn^{\Gamma}$ of Sec \ref{GR-Gdyn} can furthermore be interpreted as the foliation 4-vector 
(with a label running along the foliation, $\mn^{\Gamma}(\lambda)$.

\noindent One should also point out at this stage for analogues of spatial slices and then of foliations for discrete-type approaches to Quantum Gravity.
E.g. in the Causal Sets program, the instants are {\it antichains}; this makes good sense, since these are sets of causally {\sl unrelated} points.   
Moreover, since points and causal relations are {\sl all} of the structure in this approach, these antichains are just unstructured sets of points.
However, they can be slightly thickened so as to have enough relations to be structured.                                        
One can then imagine in each of these discrete approaches for layered structures of each's means of modelling an instant, and then pose 
questions of `foliation dependence' about this (or some suitable limit of this).  

\mbox{ } 

\noindent Attitude 1) Demand classical GR's foliation independence and refoliation invariance continue to hold at the quantum level.  
One then has to face that the commutator algebroid at the quantum level is almost certainly distinct from classical GR's Dirac algebroid, 
(whilst not getting any simpler as regards having structure functions) and may contain foliation-dependent anomalies too \cite{Kuchar89, Torre07}.  
  
\mbox{ }  
  
\noindent Attitude 2) Background-dependent/privileged slicing alternatives are approaches with times with sufficient significance 
imposed on them that they cannot be traded for other times (these are more like the ordinary Quantum Theory notion of time than the conventional lore of time in GR.  
As the below examples show, the amount of theorizing involving Attitude 2) has been on the increase.

\mbox{ }

\noindent However, I first present a few cautions. 
The obvious example of a background-dependent theory is perturbative string theory.  
This manifests no POT.  
However, technical issues then drive one to seek nonperturbative background-independent strategies and then POT issues resurface. 
As regards privileged slicing theories, these are affected by how the classically significant notion of diffeomorphisms is to have a meaningful counterpart.  
Thus background structures and privileged slicings are to be viewed at least with quite some suspicion.  
LQG has a greater degree of background independence than perturbative String Theory. 
Namely, it is independent of background {\sl metric} structure. 
Ridding ourselves of further background structure, e.g. to the point of having no background spatial topology, is probably too ambitious for present-day Physics. 
(See however Appendix \ref{Cl-Str}.A.)   
However, it may be necessary in order to genuinely free oneself of unwanted background structures. 
(There are few reasons why the wish to rid oneself of background metric structure should not also 
extend to background topological structure, other than considerable mathematical inconvenience).
M-theory is also expected to be, in various senses, background-independent.   

\mbox{ }

\noindent Note 1) One should disentangle Attitude 2) from how special highly-symmetric solutions {\sl can} have geometrically-preferred  foliations.
However, GR is about generic solutions, and even perturbations about highly-symmetric solutions cease to have geometrically-privileged foliations 
to the perturbative order of precision \cite{Kuchar99}.  
Additionally, even highly-symmetric solutions admitting a privileged foliation in GR are {\sl refoliable}, so the below problem with losing refoliation invariance does not apply.  

\noindent Note 2) \K also states that \cite{Kuchar92} {\it ``The foliation fixing prevents one from asking what would happen if 
one attempted to measure the gravitational degrees of freedom  on an arbitrary hypersurface. 
Such a solution amounts to conceding that one can quantize gravity only by giving up GR: to say that a quantum theory makes sense only when one fixes the 
foliation is essentially the same thing as saying that quantum gravity makes sense only in one coordinate system."}    
%

\noindent Note 3) On the other hand, one cannot press too hard with envisaging different foliations as corresponding to various motions of a cloud of observers distributed throughout 
space, since that can be `dually' modelled by multiple congruences of curves without any reference to multiple foliations (the threading formulation \cite{HE}).  

\noindent Note 4) Fixed foliations are a type of background-dependence, which, from the relational perspective, is undesirable {\sl from the outset},
 for all that investigating the effects of making {\sl just} this concession is  indeed also of theoretical interest.

\mbox{ }    

\noindent Example 1) The linking theory \cite{GGKM11, Kos4, GrybTh} has a fixed CMC foliation. 
Here, one ``{\it trades refoliation invariance for a conformal symmetry}".
[It is a fair point that not everybody working in this area is holding out for a fixed-foliation interpretation.  
The situation is an enlarged phase space `linking theory', for which one gauge-fixing produces a GR sector and another gauge-fixing produces a CMC-fixed sector.
Then from the perspective of the linking theory, GR and CMC-fixed sector are gauge-related.]  

\noindent Example 2) Foliation Problem with Hidden Times). The bubble time version describes a hypersurface in spacetime only after the classical equations have been solved. 
Thus its significance at the quantum level is not in fact understood due to the unclear status of foliations at the quantum level \cite{Kieferbook}.   

\noindent Example 3) Foliation Dependence Problem with Path Integral Formulation).
The path integral's definition furnishes a direct counterpart of Foliation Independence (the construction of intermediate surfaces in Fig \ref{Teit}b); 
the Path Integral Formulation is an interesting place to look for whether this probably physically desirable result of classical GR extends to the quantum level, 
rather than a formulation in which this issue has somehow become irrelevant or resolved.  

\noindent Example 4) Foliation Dependence Problem with Histories Theory).  GR time's many-fingeredness brings in foliation-dependence issues to Histories Theory.  

\mbox{ }

\noindent Some constructive examples toward Attitude 1) are as follows. 

 
\noindent Example 5) Kouletsis and \K \cite{KK02, Kouletsis} provided a means of including the set of foliations into an extension of the ADM phase space that is generally covariant.
This amounts to extending phase space to include embeddings so as to take into account the discrepancy between the Diff({\cal M}) algebra and 
the Dirac algebroid (itself an older idea of Isham and Kucha\v{r} \cite{IK85}), here effectuated by construction of a time map and a space map.
Thus this is a $\chi_{\tH}$ {\sc k}... approach; it remains unstudied for GR past the classical level.
 
\noindent Example 6) In Savvidou's approach to HPO, she pointed out that the space of histories has implicit dependence on the 
foliation vector (the unit vector $\mn^{\Gamma}$ mentioned in Sec \ref{Examples} as being orthogonal to a given hypersurface).
With this view of HPO admitting 2 types of time transformation, the histories algebroid turns out to now be foliation-dependent. 
However, the probabilities that are the actual physical quantities, are {\sl not} foliation-dependent, so this approach avoids having a Foliation Dependence Problem at its end.

\noindent Example 7) Isham and Savvidou also considered the possibility of quantizing the classical foliation vector itself \cite{Savvidou02, IS01101}.  
Here,  
\beq
\widehat{\mn}_{\Gamma}\Psi        = \mn_{\Gamma}\Psi \mbox{ } , \mbox{ } 
\widehat{\Pi}^{\Gamma\Delta} \Psi = i \left\{ \mn_{\Gamma}\frac{\pa}{\pa \mn_{\Delta}} - \mn_{\Gamma}\frac{\pa}{\pa \mn_{\Delta}}  \right\} \Psi \mbox{ } ,  
\eeq 
where the antisymmetric $\Pi^{\Gamma\Delta}$ is the conjugate of $\mn^{\Gamma}$ and satisfies the Lorentz algebra.
They then apply a group-theoretic quantization to the configuration space of all foliation vectors for the Minkowski spacetime toy model of the HPO approach.

\subsection{Strategizing about the Quantum Spacetime Reconstruction Problem}\label{SRP}

\noindent Programs are indeed often designed so that the recovery of a continuum with spacetime properties is the last facet to face.  

\mbox{ } 

\noindent Example 1) The status of the Spacetime Reconstruction Problem remains unclear for the Semiclassical Approach  in detail \cite{Kuchar92, I93}.  
This is tied to the bizarre sequence of TDWE's, inner products and probability interpretations that one is to face in passing from less toward more accurate 
versions along the many chains of approximations involved.  

\mbox{ }  

\noindent {\bf Question 77} Is there a semiclassical relativity without relativity? E.g. a first-order counterpart for which the classical RWR was a zeroth order?

\mbox{ }

\noindent The next examples involve bottom-up approaches, i.e. less structure assumed.  

\mbox{ } 

\noindent Example 2) Getting back a semblance of dynamics or a notion of history from Timeless Approaches counts as a type of spacetime reconstruction. 
This is in contradistinction to histories-with-time-assumed approaches.

\mbox{ } 

\noindent Example 3) In Nododynamics/LQG, semiclassical {\sl weave states} have been considered by e.g. Ashtekar, Rovelli, Smolin, Arnsdorf and Bombelli; 
see \cite{Thiemann} for a brief review and critique of these works. 
Subsequent semiclassical reconstruction work has mostly involved instead (a proposed counterpart of the notion of) 
{\it coherent states} constructed by the complexifier method \cite{Thiemann}; see \cite{FL11} for a different take on such.  
In Lorentzian spin foams, semiclassical limits remain a largely open problem; e.g. \cite{Baretal} is a treatment of Lorentzian spin foams that also covers how the Regge action emerges 
as a semiclassical limit in the Euclidean case, whilst almost all the semiclassical treatment in the most review \cite{Perez} remains Euclidean.  
The further LQC truncation \cite{Bojowald05} does possess solutions that look classical at later times (amidst larger numbers of solutions that do not, which are, for now, 
discarded due to not looking classical at later times, which is somewhat unsatisfactory in replacing predictivity by what ammounts to a {\sl future} boundary condition). 
It also possesses further features of a semiclassical limit \cite{Bojowald06} (meaning a WKB regime with powers of both $\hbar$ and 
Immirzi's $\gamma$ neglected, whilst still subject to open questions about correct expectation values of operators in semiclassical states).

\mbox{ } 

\noindent Example 5) Spacetime Reconstruction Problems with Path Integral Formulation).
Whilst (almost) discrete approaches render the path integral itself tamer, Spacetime Reconstruction Problems poses significant difficulties for such approaches.
E.g. Spacetime Reconstruction difficulties are to be expected from the Causal Sets Approach's insistence on very sparse structure;  for recent advances here, see e.g. \cite{MRS} 
for a recovery of a spacetime-like notion of topology, or \cite{RiWa} for a recovery of a metric notion (at the classical level, so rebalance).   

\mbox{ }  

\noindent Recovery of microcausality at the quantum level is possible in the Savvidou \cite{Savvidou04a, Savvidou04b} or Kouletsis \cite{Kouletsis} canonical-covariant formalisms
(though this work has as yet very largely not been extended to the quantum level).  

\vspace{10in}

\subsection{Strategizing about the Multiple Choice Problem}\label{MC}

Whilst patchings involve multiple times, and \cite{Bojo1, Bojo2, Bojo3, Bojo4} claim that as a Multiple Choice Problem solver too, this does not address whole of this problem.  
Generally, the Multiple Choice Problem is multi-faceted -- due to a heterogeneous collection of mathematical causes.    
Multiplicity of times is also inherent in Rovelli's `any change' and `partial observables' based relationalism, and in my STLRC 
(where one tests one's way among the many to find the locally-best).

\mbox{ }

\noindent 1) Some cases of Multiple Choice Problem reflect foliation dependence.  
It has been suggested that one way out of the Multiple Choice Problem is to specify the lapse $\upalpha$ and shift $\upbeta^{\mu}$ a priori (e.g. in \cite{unim-SUGRA}). 
However, such amounts to yet another case of foliation fixing (so return to Sec \ref{Fol}).   
Also, foliation issues are not the only source of the Multiple Choice Problem, as the below testifies.    
%

\noindent 2) Perhaps some cases of Multiple Choice Problem could be due to differences in perspective: 
different observers observing different subsystems that have different notions of time; this fits in well with the partial observables and patching paradigms.  

\noindent 3) Some cases of Multiple Choice Problem come from how choosing different times at the classical level for subsequent use 
(`promotion to operators') at the quantum level can lead to unitarily-inequivalent QM's, even if at the classical level they are canonically-related, 
by the Groenewold--Van Hove phenomenon. 
This is a time problem if one's search for time takes it to lie among the objects in the classical algebra (i.e. {\sc b ... t ... q} approaches). 

\noindent 4) There is a further Multiple Choice Problem concerning frames at the QM level in absence of reduction.

\mbox{ }  

\noindent 5) Bojowald et al's \cite{Bojo1} Patching Approach purports to resolve the Multiple Choice Problem as well as Inner Product Problem and working toward resolving the Global POT.  
It addresses the Inner Product Problem via use of a moments expansion.
There are considerable doubts about whether it addresses the Multiple Choice Problem due to it not making contact with such as the Gronewold-Van Hove phenomenon (Patching Problem 1); 
each patch having its own time {\sl manifests} some aspects of Multiple-Choice but is not a scheme that per se resolves all aspects of the Multiple Choice Problem.  
A couple of further possible problems relevant at this stage are as follows (see also the next SSec for Patching and the Global POT).

\mbox{ } 

\noindent Possible Patching Problem 2) the moments expansion approach looks to rely on classical and quantum having the same bracket algebroid, but this is generally untrue \cite{I84}.  

\noindent Possible Patching Problem 3) The moments approach's use of all polynomials (up to a given degree in the approximate case detailed in \cite{Bojo1, Bojo2, Bojo3, Bojo4}) 
could well place it at further odds with the Groenewold--Van Hove phenomenon (rather than this approach somehow turning out to be able to deal with this as-yet unaddressed 
aspect of the Multiple Choice Problem).  

\mbox{ }  

\noindent 6) There is a Multiple Choice Problem on selection of a subalgebra of classical beables to promote to quantum beables.  

\noindent 7) There is  a Multiple Choice Problem with the Semiclassical Approach when studied in detail \cite{Kuchar92}, though the 
Groenewold--Van Hove phenomenon will sometimes be avoided here due to some subalgebra/oid choices and unitary inequivalences being of 
negligible order in semiclassical expansions [e.g. O($\hbar$) smaller than the smallest terms kept).  

\mbox{ } 

\noindent 8) Multiple Choice Problem with Path Integral Formulation 1).  There is no longer a time whose selection causes this problem, but there often does remain a choice of frame, 
and always a choice of beables, and each of these is rather likely to be afflicted with this problem.

\noindent Multiple Choice Problem with Path Integral Formulation 2) The measures involved in gravitational path integrals also look prone to Multiple Choice issues \cite{Kuchar92}. 

\mbox{ }  

\noindent 9) The previous two carry over to Histories Theory, which also suffers the following extra problem.  

\noindent Multiple Choice Problem with Histories Theory 3) Histories Theory is usually done in gauge-dependent form, with each gauge being equipped with an  
internal time and so many internal time problems come back, a particular instance being the Multiple Choice Problem. 
In the present Article, we break this internal time use by using emergent Machian time, but that still carries no guarantee of freedom from Multiple Choice 
(multiple WKB regimes each with their own time, or different approximations for the emergent WKB time within the one regime).  

\vspace{10in}

\subsection{Strategizing about the Global Problems of Time}\label{Global}

\noindent Hidden Time Quantum Global Problem) The domain of defineability of the `true quantum Hamiltonian' itself may well be localized (see e.g. Sec \ref{203} for an example).  

\noindent Semiclassical Global Problems) is that Machian-Classical Global Problem 1)'s forcedly local nature of $h$--$l$ split designations carries 
over to the quantum level, as do Sec \ref{+temJBB}'s other locality restrictions on underlying classical approximations. 
Moreover there are now a host of further quantum and semiclassical approximations as per Sec \ref{Semicl}, none of which come with any guarantees of global applicability 
(Sec \ref{22-end}).  

\noindent Timeless Records Theory Global Problems 1 and 2) are in practise mostly to be considered at the quantum level, due to the importance of quantum conceptualization 
of records and of the role of (some variant on) quantum theory in whatever actually-proposed mechanisms for obtaining a semblance of dynamics or of history.

\noindent Global Problems with Path Integral Formulations).  
It suffices to note that the Gribov ambiguity in Yang--Mills theory remains present in Path Integral Formulations to establish that these are not some magic cure to global 
issues.  
Most of these carry over to Quantum Histories Theory too [Problem 1) for that].

\noindent Global Problem  with Quantum Histories Theory 2) Histories Theory is usually done in gauge-dependent form, with each gauge being equipped with an  
internal time and so many internal time problems come back, a particular instance being global aspects. 
We break this internal time use by using emergent Machian time, but that still carries no global guarantees, since the emergent Machian time itself 
lacks in a number of guarantees of global definedness. 
%

\noindent Global Problem with rendering Halliwell's Combined Approach Machian).  
Semiclassical considerations mean that $t^{\se\sm(\sW\sK\sB)}$ is only locally defined, but the S-matrix quantity step in the Dirac beables construction of this scheme is 
utterly dependent on the time variable in use running from $-\infty$ to $+\infty$.

\mbox{ }

\noindent {\bf Quantum part of Bojowald et al's patching approach}. Here, this approach uses a moments expansion to bypass the Inner Product Problem.
It is also a Semiclassical Approach, in their sense that they neglect O($\hbar^2$) and moment-polynomials above some degree.  
Within this regime, Bojowald et al's specific fashionables implementation is then based on time going complex around turning points within their notion of semiclassical regime. 
%
{            \begin{figure}[ht]
\centering
\includegraphics[width=0.3\textwidth]{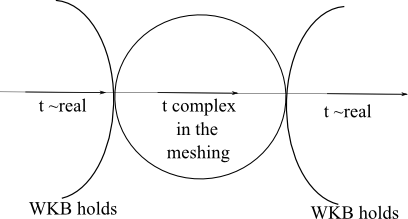}
\caption[Text der im Bilderverzeichnis auftaucht]{        \footnotesize{Time in the Patching Approach} }
\label{Bojowald}\end{figure}            }

\noindent However (Patching Problem 4) the geometrical interpretation of the transition between these fashionables is, for now, at least to me, unclear.

\noindent Also (Patching Problem 5) making time complex as in this program (and others) may interfere with such a notion of time having enough of the standard property list pass muster as a time  
(or at least require extra work to demonstrate that such properties can be established in this less standard context).  

\noindent Finally (Patching Problem 6) use of Complex Methods for detailed path or canonical type calculations has a poor track record as per e.g. 
Sec \ref{Path-Int} and \cite{Kuchar93}. 

\noindent Note that at least 5) and 6) are problems {\sl shared} with the Semiclassical Approach.

\begin{subappendices}
%
\subsection{Strategies for the Problem of Time in Affine Geometrodynamics}\label{Affine3}

In affine geometrodynamics \cite{IshamKakas1, IshamKakas2, Klauder}, one has a distinct form for the unreduced Hilbert space and for the detailed structure of the WDE. 
Tempus Ante Quantum approaches have no plain--affine distinction at the classical level at which the timefunctions are found.   
There may however still be some scope for different commutation relations and operator orderings subsequently arising at the quantum level. 
The affine approach still has an analogue of Riem time (as the signature of the wave equation is unaltered by passing to the affine approach).  
The subsequent inner product issue involving the potential not respecting the conformal Killing vector is at the 
classical level, so that the non-useability of a Klein--Gordon inner product carries through to the affine case.  
I am not aware of Third Quantization having been tried in the affine case, but a number of the reasons for it 
not appearing to be very promising as a POT resolution do look to carry over to the affine case.

The \NSI should not care about such a change, since this approach does not make use of commutation relations or of the WDE.  
As the \CPI does work in a Wheeler--DeWitt framework \cite{Kuchar92}, this will change in detail.  
classical Records Theory is unaffected by this change, though subsequent changes would then be expected at the quantum level.  
A Histories Theory approach to Affine Geometrodynamics was considered by Kessari \cite{Kessari}.
\cite{I93} indicates that affine version of evolving constants of the motion exists.  

\vspace{10in}

All in all, one seldom gets far enough \cite{I09-10} in non-formal detail with POT approaches to Geometrodynamics 
for the ordinary versus affine geometrodynamics distinction to give notable changes in behaviour.  

\mbox{ } 

\noindent {\bf Question 78}$^*$ use the model of Question 64 to further investigate affine-type differences in the above significantly different POT approaches.

\subsection{Further PPSCT support for generalization of Halliwell's combined approach}\label{Affine}

\noindent If (\ref{PPSCTwave}) applies to a wavefunction obeying the BO-WKB ans\"{a}tze form (\ref{BO}, \ref{WKB}), 
then preservation of the physically-significant $\fh$--$\fl$ split under PPSCT transformations requires it to be entirely the $\chi($\fh$, $\fl$)$ factor that PPSCT-scales, 
\beq
\chi \longrightarrow \overline{\chi} = \Omega^{\{2 - r\}/2}\chi
\eeq
since $\Omega$ itself is in general a function of $\fh$ and $\fl$, and so would map $S$(\fh) out of the functions of $\fh$ alone.
Likewise it is the $\fl$-inner product that carries a $\Omega^{-2}$ factor.

\noindent Then the outer rather than inner product of two wavefunctions necessitates the same weight function; this will of course be used to build density matrices.

\subsection{Removing a sometimes-given motivation for Records Theory}

\noindent One of the arguments used toward Records Theory is that this considers the entities that are actually operationally meaningful.  
For the study of records is how one does science (and history) in practise, whether or not one ascribes reality to whatever 
secondary frameworks one reconstructs from this (such as histories, spacetimes or the local semblance of dynamics)?   
However, I argue that Histories Theory being primarily a reconstruction from records is {\sl not a motivation for} Records Theory but, rather, {\sl an assertion to be demonstrated}.  
For, the records present in nature may not in general be of sufficient quality (localization, accessibility, 
sufficiently high retrieveable information content of the right kind) to be able to reconstruct history.
Thus one should seek to 

\noindent A) pin down where the ``somewhere" in (\ref{situ2}) is (which is the central motivation in some of Halliwell's papers \cite{H99, HD}).    

\noindent B) Determine whether the record at this location is {\sl useful}.  
E.g. Gell-Mann and Hartle assert that what they call records ``{\it may not represent records in the usual sense of being constructed from quasiclassical variables accessible to us}" 
(p 3353 of \cite{GMH}), which, in my parlance, amounts to insightfully allowing for the possibility that some of nature's records may not be very useful to us.  

\noindent  Another issue is that the $\alpha$-particle track in the Mott--Bell--Barbour bubble chamber example may well be atypical in its neatness and localization.  
For, bubble chambers are carefully selected environments for revealing tracks -- much human trial and error has gone into finding a piece of apparatus that does just that.  
$\alpha$-tracks being useful records could then hinge on this careful pre-selection, records in general then being expected to be poorer, 
perhaps far poorer, as suggested e.g. by the Joos--Zeh paradigm of a dust particle decohering due to the microwave background photons \cite{JZ}.  
In this situation, records are exceedingly diffuse as the information is spread around by the CMB photons and ends up  ``{\it in the vastness of cosmological space}".   
It is then a fair point that bubble chambers are atypical in being specifically constructed as specialized 
pieces of equipment for seeing clear-cut tracks, so that the dust-grain paradigm should be more typical of what is found in nature.    
It is a further fair point that what happens depends furthermore on the magnitude of the object. 
Dust grains are bigger than $\alpha$-particles and, likewise, macroscopic bodies are bigger than dust grains; 
These typically self-decohere, and on an extremely short timescale, which is a further paradigm and presumably 
dominant over both the preceding as regards cosmological structure formation.  
There is then a worry that records could be too problematic to access and/or of unsuitable information content in the case of QG 
(e.g. via high nonlocality, any useful cites for programs in which QG is highly nonlocal).  
Though my own goals are for now the more modest origin of quantum-cosmological structure and 
assessing which of the recoveries of late-universe subsystem physics stand up to careful scrutiny.  

\noindent C) I note that, most well-known old arguments about the physics being in the correlations in fact specifically do not refer to timeless approaches. 
E.g. what Wigner actually said on this subject was (\cite{Wigner} p 145, see also \cite{JWY}) 
``{\it quantum mechanics only furnishes us with correlations between SUBSEQUENT observations}" (my caps), and Wheeler wholly agreed with this (p 295 of \cite{Battelle}).  
Were one to wish to draw motivation from these statements, they would need to be rephrased and the original context would be lost.    

\vspace{10in}

\subsection{Rovelli-type versus other timeless approaches}\label{Rov-vs-Other} 

\noindent 1) Rovelli-type timeless approaches use the AMR notion of time, Barbour's timeless approach use the LMB notion of time, other timeless approaches 
like the \CPI and Page's also use the AMR notion of time.
This Article's main Combined Scheme is formulated with the LMB-A notion of time, as argued to be the most conceptually and practically satisfactory time out of the three.  

\noindent 2) Cautions about requiring the observables/beables cast in the role of clocks/timefunctions to actually be (often hard to find) {\sl good} clocks/timefunctions 
apply to both families of approaches (taken in this Appendix to mean Rovelli and non-Rovelli families).

\noindent 3) Rovelli's approach and Conditional Probabilities Interpretation-type approaches have particular resemblance due to both involve notions of 
observables/beables, and of subsequent pairings of such, with the one then serving as a clock for the other.
However, Rovelli's scheme is less specific in not postulating a rigid formula like the \CPI one for the extraction of the information about the correlation between the two observables.  

\noindent 4) Rovelli also insisted that only the Heisenberg picture remains meaningful for Quantum Gravity, by which Type 1 Approaches such as the 
Conditional Probabilities Interpretation/Records Theory could not be just the same physics as Rovelli-type but in a different picture.
[Though it is not clear, at least to me, whether this insistence is absolutely necessary; attempting to lift it might reveal further approaches 
to the POT, or further connections between existing approaches, such as between Rovelli-type and Type 1 Timeless Approaches.]  

\noindent 5) The other timeless approaches involve nonstandard interpretation of QM; is this also the case for the Rovelli-type ones? 
One would suspect that the Heisenberg picture alone remaining meaningful indeed itself entail some interpretational changes.
Also, Thiemann's combination of Rovelli-type with Histories Theory certainly involves bringing in nonstandardness of interpretation of Quantum Theory.  

\noindent 6) Criterion 3) applies to both Rovelli's approach and Barbour's approach, in that each allows for mechanics and GR to be formulated more similarly than is usual, 
albeit in each case in a distinct sense.  

\noindent 7) $t^{\se\sm(\sJ\sB\sB)}$ (whether emergent or as the form that the semblance of dynamics is to take) is distinguished by its substantially simplifying both the 
momentum-velocity relations and the evolution equations, and amounts to an emergent recovery of other well-known notions of time: Newtonian, proper, cosmic and emergent WKB, 
in various contexts.   
This gives Barbour-type approaches an advantage over Rovelli-type ones by Criterion 2), though it should also be noted that Barbour-type approaches do no better as regards 
providing a precise prescription of how to find an `ephemeris-type' time within a given region of the universe, by which this advantage is only modest.  

\noindent (A useful clarification as regards using Celestial Mechanics examples for illustrative purposes as in \cite{Rovelli}, is that the 

\noindent mathematically-simple choice of time 
variable used there is specifically to illustrate partial observables in a context that is 
mathematically-familiar even to the newest students of Physics, but {\sl not} however how Celestial Mechanics, or time-keeping as a maximally-accurate pursuit, is actually done.     
More accurate timekeeping involves, rather, a GLET and the possibility of having to recalibrate to include further relevant changes as one's need for accuracy increases.)

\noindent 8) Barbour-type relational approaches additionally, they satisfy Criterion 4), by suggesting further alternative theories 
[for all that some such like the theory in \cite{ABFO} are then questionable on logically-independent grounds].  
The partial observables formulation, however, can be carried out for absolutely any theory that has multiple degrees of freedom, and therefore does not exhibit Criterion 4).  

\noindent 9)  RPM's are not covered in Rovelli's book \cite{Rovellibook}, in part because he does not there consider 
where in his `order of things'/`space and categorification of theories' such toy models would sit.
The present Article has amply demonstrated that, whilst RPM's are not special-relativistic, RPM models share a large number of other features with GR, 
thus laying out a particular arena whose study helps patch up these gaps.  
By this, the current Article can be viewed as substantially complementary to Rovelli's book; it is suggested that those genuinely interested in Foundations of Physics, 
particularly in foundations for QG, read both and only then decide whether the one, the other, or parts of both, best suit their needs.  

\end{subappendices}

\vspace{11in}

\section{RPM cases of Quantum part of {\sc tq}}\label{QM-Ante}

\subsection{3-stop metroland examples of difficult and likely inequivalent TDSE's}\label{YaLu}

These suffice to establish the desired features.

\noindent Example 1) I use the conjugate to reciprocal radius $\upsilon$ as a time since this suits the combination of physical relevance 
and monotonicity. 
[The free case at the classical level requires $E > 0$ and thus a same-sign monotonicity sector, which, in the case of the scale 
being a power of the configuration space radius, requires that power to be negative.  
Then we have 
\beq
t_{\upsilon}^2p_{t_{\upsilon}}\mbox{}^4 + p_{\varphi}^2p_{t_{\upsilon}}\mbox{}^2 - 2E = 0 \mbox{ } , 
\eeq
which is solved by 
\beq
p_{t_{\upsilon}} = 
\pm\sqrt{        \frac{    - p_{\varphi}\mbox{}^2 \pm\sqrt{   p_{\varphi}\mbox{}^2 + 8 E t_{\upsilon}\mbox{}^2   }    }
                      {    2t_{\upsilon}\mbox{}^2    }        } \mbox{ } .
\label{pupsi}
\eeq      
The inner sign needs to be `+' for classical consistency, and I argue in Sec \ref{nEye} for the outer sign to be `--' as regards approximate recovery of a close-to-conventional QM.
(\ref{pupsi}) is then promoted to the $\upsilon$-TDSE
\beq
i\hbar\frac{\pa\Psi}{\pa t_{\upsilon}} = 
-\sqrt{        \frac{    \hbar^2 {\pa^2}/{\pa\varphi^2} + \sqrt{  \hbar^4{\pa^4}/{\pa\varphi^4}  + 8Et_{\upsilon}\mbox{}^2  }    }    
                    {    2t_{\upsilon}\mbox{}^2    }        }\Psi
\eeq
unambiguously orderwise since the sole constituents of their radicand $p_{\varphi}$ and $t_{\upsilon}$ commute.  

\mbox{ }

\noindent Example 2) On the other hand, working with Euler time, 
\beq
t^{\sE\su\sll\se\sr\,2} + p_{\varphi}\mbox{}^2 = 2E\,\mbox{exp}(-2p_{t^{\tE\tu\tl\te\tr}})
\eeq
which is solved by 
\beq
p_{t^{\tE\tu\tl\te\tr}} = - \frac{1}{2}\mbox{ln}
\left(
\frac{t^{\sE\su\sll\se\sr}\mbox{}^2 + p_{\varphi}\mbox{}^2}{2E}
\right)
\eeq
This is then likewise unambiguously promoted to the Euler-TDSE  
\beq      
i\hbar \frac{\pa\Psi}{\pa t^{\sE\su\sll\se\sr}} = \frac{1}{2}\mbox{ln}
\left(
\frac{t^{\sE\su\sll\se\sr\,2} - \hbar^2{\pa^2}/{\pa\varphi^2}}{2E}
\right)\Psi   \mbox{ } . 
\eeq      
Having this pair of approaches leading to such unusual and disparate equations illustrates that it does appear to make a big difference what one chooses as scale 
(and neither looks anything like the exactly soluble form (or the semiclassical form, c.f. the next SSSec).  
Thus, as claimed in Sec \ref{PlayIt} the ambiguity in defining scales is nontrivial.

\subsection{An approximate approach to internal time}\label{nEye} 

\noindent One can apply the method of approximating in a series at the classical level and only then promoting 
the outcome of that to quantum operators (and Sec \ref{TAQ}), which are then rather better defined and less ambiguous. 
This has parallels to the treatment of relativistic wave equations (done before in the Semiclassical Approach but not as far as I know for the Hidden Time Approach).  
Suppose then that we expand in powers of $p_{\varphi}$ prior to promotion to operators (which is close to expanding in powers of $\hbar$ 
and thus is semiclassical-like and likely to benefit from lessons from approximations made in the study of relativistic QM).     
Then, for the given choice of sign,
\beq
i\hbar\frac{\pa\Psi}{\pa t_{\upsilon}} = -\frac{\hbar^2}{2}\frac{\pa^2\Psi}{\pa^2\varphi^2} - \frac{\{2E\}^{1/4}}{\sqrt{t_{\upsilon}}}\Psi
\eeq   
[+ $O(\hbar^4) \times$ (a 4th order derivative)]
which is an ordinary TDSE with a particular $t_{\upsilon}$-dependent potential and, moreover, soluble to give 
\beq
\Psi \propto \mbox{exp}(i\,\md \, \varphi) \mbox{exp}
\left(
i\hbar t_{\upsilon}
\left\{\frac{\md^2}{2}  + \frac{2}{\hbar^2\sqrt{t}}\{2E\}^{1/4}
\right\}
\right)
\eeq
which for $t_{\upsilon}$ big takes the form of a free wave, with some distortion for $t_{\upsilon}$ small.

The Euler time one gives, with respect to a  {\it rectified time} (i.e. a redefinition by which the TDSE
takes on a simplified form by the time absorbing prefactors of the configuration variables' derivatives) 
$t^{\sR\se\scc} = \int \d t/2Et_{\sE\su\sll\se\sr}\mbox{}^2 = - 1/2Et_{\sE\su\sll\se\sr}$
\beq
i\hbar\frac{\pa\Psi}{\pa t^{\sR\se\scc}} = - \frac{\hbar^2}{2}\frac{\pa\Psi^2}{\pa \varphi^2} - \frac{1}{4E\,t^{\sR\se\scc\,2}}\mbox{ln}(8E^3t^{\sR\se\scc\,2}) 
\mbox{ } , 
\eeq
which can also be solved to give
\beq
\Psi \propto \mbox{exp}(i\,\md\,\varphi)
\mbox{exp}
\left(
i
\left\{ 
\frac{\hbar\,\md^2}{4Et_{\sE\su\sll\se\sr}} + \frac{t_{\sE\su\sll\se\sr}}{2\hbar}
\left\{
\mbox{ln}(8E^3) + 2 - 2\,\mbox{ln}(-2\,E\,t_{\sE\su\sll\se\sr})
\right\}
\right\}
\right) \mbox{ } .  
\eeq
By examining the approximate approach, somewhat more standard mathematics appears.

\noindent Time-dependent potentials are hard to interpret here; 
such are usually held to correspond to non-conservative Physics, but, unlike in that setting, the model is a whole-universe so there is nothing for energy to dissipate into.

\subsection{RPM examples of operator-ordering and well definedness problems with internal times}\label{203}

\noindent These examples need the presence of shape potentials to confer operator-ordering ambiguity. 
An upside-down HO example will do (so $A < 0$). $E > 0$, $k = -2 < 2$, $V < 0$. 
$\upsilon$-time then does nicely.  
Then
\beq
p_{t_{\upsilon}} = \pm \sqrt{
\frac{p_{\varphi}^2 + 2\{A + B\,\mbox{cos}\,2\varphi\} \pm \sqrt{\{p_{\varphi}\mbox{}^2  + 2\{A + B\,\mbox{cos}\,2\varphi\}\}^2 + 8 E t_{\upsilon}\mbox{}^2}}
     {2E}                   }  \mbox{ } ,  
\eeq
which clearly has $p_{\varphi}f(\varphi)$ products [specifically $p_{\varphi}\,\mbox{cos}\,2\varphi$ ones] causing nontrivial operator-ordering ambiguities.  
Also in this example not everything under square roots is positive, so one is away from nice guarantees of being able to promote such to an operator.  

\mbox{ }  

\noindent Note: there is a Spectral Theorem result for well-definedness of positive combinations under square root signs, but in the case of Example 2) above, 
I do not know of a guarantee that this Hamiltonian is well-defined.  

\mbox{ } 

\noindent Thus, well-definedness and operator-ordering issues are present in the dilational time approach.

\subsection{End-check against hidden time's problems}

\noindent Multiple Choice Problems with Hidden Time 1) the Multiple Choice Problem is bolstered by the significant ambiguity arising from the choice of 
scale variable, which then corresponds to selecting different entities to be among one's subalgebra of quantum observables. 

\noindent 2) The ambiguity in Sec \ref{PlayIt} can be envisaged as sourcing a further aspect of the Multiple Choice Problem once one passes to the quantum level.  

\noindent N.B. that this Sec's mathematics is highly unlike Part III's, reflecting insertion of a {\sc t}-move in before the {\sc Q}-move, qualitatively changing the nature of the 
TDWE one is to face, and completely changing the technical content of the approach.

\subsection{An RPM analogue of matter time (Analogy 104)}\label{RAT}

Consider for instance scaled triangleland in the subsystem-centric parabolic coordinates for a fixed relative angular momentum situation/$\Phi$-independent potential.  
Then one can attempt to consider $\Phi$ as a time variable, leading to an elliptic (Klein--Gordon analogue) equation 
\beq
- \hbar^2\frac{\pa^2\Psi}{\pa\Phi^2} = \frac{4\mI_1\mI_2}{\mI_1 + \mI_2}
\left\{
\hbar^2
\left\{
\frac{\pa}{\pa \mI_1}\left\{\mI_1\frac{\pa\Psi}{\pa\mI_1}\right\}
\frac{\pa}{\pa \mI_2}\left\{\mI_2\frac{\pa\Psi}{\pa\mI_2}\right\}
\right\} 
+ 2\{E - V\}\Psi
\right\}
\eeq
\noindent or the TISE
\beq
- i\hbar\frac{\pa\Psi}{\pa\Phi} = \sqrt{\frac{4\mI_1\mI_2}{\mI_1 + \mI_2}
\left\{
\hbar^2
\left\{
\frac{\pa}{\pa \mI_1}\left\{\mI_1\frac{\pa }{\pa\mI_1}\right\}
\frac{\pa}{\pa \mI_2}\left\{\mI_2\frac{\pa }{\pa\mI_2}\right\}
\right\} + 2\{E - V\}
\right\}   } \Psi
\eeq
with the extra benefit that the $\Phi$-independence of the potential and kinetic terms supports equivalence between the two.
Monotonicity is assured in this case by (\ref{posiassur}) with $\sfJ$ being fixed and partial moments of inertia being positive.  
One problem is why this angle and not another one, which ambiguity grows as the model increases in size.
Choosing $\mI_1$ or $\mI_2$ as a time turns out to lack a number of these benefits.  
This can be thought of to some extent as a `reference matter particle' example.  

\vspace{11in}

\section{Semiclassical Approach to RPM's}\label{Semicl}

\noindent This scheme starts from the $h$--$l$ split of Sec \ref{+temJBB}) and is that Sec's bona fide successor, 
including at the level of implementing Mach's Time Principle in the GLET is to be abstracted from STLRC way.  
The current Sec interpolates (Criterion 3) between classical and quantum forms of perturbation theory.  
I choose to use reduced scale = $h$, shape = $l$ models for most detailed work, so I usually start from the reduced scale--shape split TISE (\ref{Larq}).

\subsection{Born--Oppenheimer (BO) scheme and its quantum-cosmological analogue} 

I take this first step of the Semiclassical Approach to mean ansatz (\ref{BO}) alongside the following approximations.

\subsubsection{BO Approximation} 

\noindent Let $\widehat{C} := \widehat{{T_{h}^{c}}} := \hat{H}  -  \hat{T}_{h}$ denote the complement of the heavy kinetic term.
%
%
%
%
The $|\chi\rangle$-wavefunction expectation value (integrated over the l degrees of freedom, i.e. `$l$-averaged') is then 
\beq
c := \langle\chi|\widehat{C}|\chi\rangle = \int_{\sFS(N, d)}\chi^{*}(h, l^{\sfa}) \, \widehat{C}(h, l^{\sfa}, p^{l}_{\sfa}) \, 
\chi(h, l^{\sfa})\, \mathbb{D} l \mbox{ }  
\eeq 
with the associated integration being over the $l$ degrees of freedom and thus in the present context over shape space, 
and for $\mathbb{D}\ml$ the measure over the shape space $\FrS(N, d)$].  
The latter may also be regarded as the `$h$-parameter-dependent eigenvalue' via  
\be
\hat{C}(h, l^{\sfa}, p^{l}_{\sfa}) |\chi(h, l^{\sfa}) \rangle = c(h) |\chi(h, l^{\sfa})\rangle \mbox{ } .  
\ee
The $|\chi\rangle$ sometimes requires suffixing by its quantum numbers, which I take to be multi-indexed by a single straight letter, j.  
Thus the above $c$ is, strictly, $c_{\sj\sj}$ and there is an obvious off-diagonal equivalent 
\be
c_{\sj\sk} := \langle\chi_{\sj}|\hat{C}|\chi_{\sk}\rangle \mbox{ } .  
\ee 
{\bf The BO approximation} alias `diagonal dominance condition' is then that 
\be
\mbox{for } \mj \neq \mk \mbox{ } , \left|c_{{\sj}{\sk}}/c_{{\sj}{\sj}}  \right| =: \epsilon_{\sB\sO} << 1 \mbox{ } .  
\ee
Assuming that this holds, one then considers $\langle\chi| \times$ the TISE with the Born--Oppenheimer (BO) ansatz substituted in.

\subsubsection{Adiabatic assumptions}

\noindent Because the ansatz is a product in which both factors depend on $h$, the $h$-derivatives acting upon it produce multiple terms by the product rule.  
In particular, both BO's own atomic-and-molecular-physics case and the quantum-cosmological case contain second derivatives in $h$, for which the product rule produces three terms, 
schematically 
\beq
|\chi\rangle \pa_{h}^2\psi \mbox{ } , \mbox{ } \mbox{ }  \pa_{h}\psi \pa_{h}  |\chi\rangle \mbox{ } , \mbox{ } \mbox{ } \psi\pa^2_{h}  |\chi\rangle \mbox{ } .  
\eeq
The first is always kept.  
BO themselves discarded the next two as being far smaller than the first (this is due to a first kind of {\bf adiabatic assumption}, 
by which $h$-changes in $\chi$ are considered to be much smaller than $h$-changes in $\psi$).   
However, as explained in Sec \ref{VaNiel}, the emergent semiclassical time approach to the Problem of Time requires at least one such cross-term to be kept. 
(This is a case in which qualitative importance for reasons given in Sec \ref{VaNiel} overrides smallness.) 
Likewise, if there is a linear derivative term in the $\scH\Psi = 0$ (curvilinear coordinates or curved space are conducive toward this), 
the product rule produces two terms, schematically 
\beq
|\chi\rangle  \pa_{h}\psi \mbox{ } , \mbox{ }  \mbox{ } \psi\pa_{h}  |\chi\rangle \mbox{ } ,
\eeq  
and the second of these is discarded likewise due to the above kind of adiabatic assumption.

\subsubsection{Analogy 105) Commentary on adiabatic-type terms}

Beyond those types of adiabatic terms already present at the classical level as covered in Sec \ref{+temJBB},   
there are two different `pure forms' that adiabaticity can take at the quantum level.  
Firstly, there are quantities that are small through $|\chi\rangle$ being far less sensitive to changes 
in $h$-subsystem physics than to changes in $l$-subsystem physics, which I label `a($l$)' with the $l$ standing for `internal to the $l$-subsystem'.  
Secondly, there are quantities that are small through $|\chi\rangle$ being far less sensitive to changes in $l$-subsystem physics than $\psi$ is sensitive to changes in $h$-subsystem 
physics, which I label by `a($m$)' with the $m$ standing for `mutual between the $h$ and $l$ subsystems'.  

\mbox{ }

\vspace{10in}

\noindent Note 1) None of the above in general follow from the smallness of the classical adiabatic dimensional parameter $\epsilon_{\sA\sd}$.
For, some wavefunctions can be very steep or wiggly even for slow processes, e.g. the 1000th Hermite function for the slower oscillator.  
However, high wiggliness is related to high occupation number via quantum states increasing in number of nodes 
as one increases the corresponding quantum numbers, and high occupation number itself is a characterization of semiclassicality (Sec \ref{Char-Semi}). 

\noindent Note 2) Inspection of the $h$ and $l$ equations furthermore reveals that both a($l$) and a($m$) occur in terms also containing an $\epsilon_{hl} = |J/V_{h}|$ type factor.  
Thus, overall, these terms in the equations are particularly small.  

\noindent Note 3) Massar and Parentani's work on inclusion of non-adiabatic effects \cite{Parentani} in the minisuperspace arena is similar in spirit to the present Sec 
as regards considering the qualitative effects of retaining terms usually neglected in the Quantum Cosmology literature.  
In this case, the effects found are expanding universe--contracting universe matter state couplings and a quantum-cosmological 
case of the Klein paradox (backward-travelling waves being generated from an initially forward-travelling wave).

\subsubsection{Consequences of using a single $h$ degree of freedom}
%

{\bf Berry phase} \cite{Berry84, Berryrev} (as explained in Appendix \ref{Semicl}.A) is a further consideration that is in general necessary in setting 
up the preceding SSSec on geometrical grounds and with physical consequences.

\mbox{ } 

\noindent {\bf Brout--Venturi Preclusion} \cite{BV89} (Incentive 1) for 1 $h$ degree of freedom.)   
If the $h$-part is 1-$d$ as is the main case in this Article  -- the scale part of the scale--shape split -- 
however, there is no Berry phase issue due to the reparametrizability of 1-$d$ geometry.  

\mbox{ }

\noindent Note 1) Datta's position (Appendix \ref{Semicl}.A) may circumnavigate Brout--Venturi Preclusion.  

\noindent Note 2) Berry phase is in any case generally a relevant consideration in the case of $N_h$ heavy particles and $N_l$ 
light particles since Brout-Venturi preclusion does not apply for $dN_h \geq 2$.

\subsection{The WKB scheme}

I take this to consist of the subsequent ansatz \label{WKB-An} for the $h$-wavefunction alongside the following approximations.
The contentiousness of this ansatz in the quantum-cosmological context is addressed in Sec \ref{Cruxi}.  
For physical interpretation, I rewrite the principal function $S$ by isolating a heavy mass $M$, $S(h) = M F(h)$. 
[For 1 $h$ degree of freedom, this is trivial; for more than 1 $h$ degree of freedom it still makes sense if the sharply-peaked hierarchy condition $\epsilon_{\Delta M}$ holds.]
The associated {\bf WKB approximation} is the negligibility of second derivatives, 
\be
\left|\frac{\hbar}{M}\frac{\pa_{h}^2F}{|\pa_{h}F|^2}\right| << 1 \mbox{ } . 
\ee
The associated dimensional expression is
\beq
\hbar/MF =: \epsilon_{\sW\sK\sB^{\prime}} << 1 \mbox{ } . 
\eeq
This is to be interpreted as (quantum of action) $<<$ (classical action) via the reinterpretability of $S$ as classical action (see e.g. \cite{Lanczos}), 
which has clear semiclassical connotations.  

\mbox{ }  

\noindent Incentive 2) for using 1 $h$ degree of freedom is that this trivially gets round having to explicitly solve nonseparable Hamilton--Jacobi equations, 
which practical problem generally plagues the case of $> 1$ $h$ degrees of freedom \cite{Goldstein, BenFra}).

\subsection{The BO--WKB scheme's scale--shape split RPM $h$- and $l$-equations}

\noindent Then the $N$-stop metroland or $N$-a-gonland r-presentation {\boldmath$h$}{\bf-equation} $\langle\chi| \times$ [TISE (\ref{Gilthoniel})], 
with the BO and WKB ans\"{a}tze substituted in, is \cite{QuadIII, ACos2}
$$ 
\{\pa_{h}S\}^2 - i\hbar \, \pa_{h}\mbox{}^2S - 2i\hbar \, \pa_{h}S\langle\chi|\pa_{h}|\chi\rangle - \hbar^2\big\{  \langle\chi|  \pa_{h}\mbox{}^2  |\chi\rangle + 
k(N, d)h^{-1}  \langle\chi|\pa_{h}|\chi\rangle   \big\} - i\hbar h^{-1}k(N, d)\pa_{h}S 
$$
\beq
+ \hbar^2 h^{-2}\{c(N, d) - \langle\chi|\triangle_{l}|\chi\rangle\} + 2 V_{h}(h) + 2\langle\chi| J(h, l^{\sfa})|\chi\rangle = 2 E \mbox{ } \mbox{ } ;  
\eeq
\noindent Also, the $N$-stop metroland or $N$-a-gonland r-presentation {\boldmath$l$}{\bf-equation} is given by [TISE (\ref{Gilthoniel})) -- $|\chi\rangle \times$ 
($h$ equation), and takes for now the fluctuation equation form \cite{QuadIII, ACos2} 
\beq
\{1 - \mP_{\chi}\}    \big\{ - 2i\hbar \, \pa_{h}  |\chi\rangle  \pa_{h} S - \hbar^2\big\{  \pa_{h}\mbox{}^2  |\chi\rangle + k(N, d)h^{-1}  \pa_{h}  |\chi\rangle  + 
h^{-2}\triangle_{l}\}  |\chi\rangle + 2\{V_{l}(l^{\sfa}) + J(h, l^{\sfa})\}  |\chi\rangle  \big\} = 0  \mbox{ } .  
\label{l-TDSE-prime}
\eeq
These equations result from the line of work of Banks \cite{Banks} and Halliwell--Hawking \cite{HallHaw} via various specializations in e.g. \cite{Datta97, Kieferbook, SemiclIII}.

\subsection{Emergent WKB time}

It is then standard in the semiclassical approach to use that $\pa_{h}\mbox{}^2 S$ is negligible by the WKB approximation to remove the second term from the $h$-equation, and apply 
\beq
\pa_{h} S = p_{h} = \mbox{\Last}h
\label{lance2}
\eeq
by identifying $S$ as Hamilton's function, and using the expression for momentum in the Hamilton--Jacobi formulation,  
the momentum-velocity relation, and the chain-rule to recast $\pa_{h}$ as $\pa_{h}t\, \Last$. 

\mbox{ }  

\noindent{\bf WKB--JBB Time-Alignment Lemma 16)} (Analogy 106) \footnote{That these coincide is an expansion \cite{SemiclI} 
on comments by Barbour \cite{Barbourcom}, Kiefer \cite{Kiefer93} and Datta \cite{Datta97}.
Also, Sec \ref{+temJBB}'s use of ( ) for orders of perturbation theory now clashes with use of ( ) for bases, so I use \{ \} instead for the former.}
%
$t^{\se\sm(\sW\sK\sB)}_{\{0\}} = t^{\se\sm(\sJ\sB\sB)}_{\{0\}}$, so the notation can be simplified to $t^{\se\sm}_{\{0\}}$.
This much was already clear in \cite{Barbourcom, Kiefer93, Datta97}.  
It then follows from this identification and Sec \ref{Examples} that the approximate emergent WKB time is aligned with Newtonian time, proper time and cosmic time in the various 
contexts but can additionally be regarded on a relational footing, and that Sec \ref{+temJBB}'s properties and critiques extend to approximate emergent WKB time.  

\mbox{ } 

\noindent The full (except for $\pa_{h}\mbox{}^2 S$ neglected) Machian $h$-equation is then \cite{ACos2}
$$ 
\{\Last h\}^2 - 2i\hbar    \langle\chi|  \Diamond \Last |\chi\rangle - 
\hbar^2 \{ \langle \chi | \Heart^2  | \chi \rangle + k(N, d) h^{-1} \langle \chi | \Heart | \chi \rangle \} 
$$
\beq
- i\hbar k(N, d) h^{-1} \Last h  +  \hbar^2 h^{-2} \{ k(\xi) - \langle \chi |  \triangle_{l}  | \chi \rangle \} = 
2\{  E - V_{h}(h)  -  \langle\chi|  V_{l}      |\chi\rangle  -  \langle\chi|  J_{h, l}   |\chi\rangle        \} 
\mbox{ } \mbox{ } ;  
\label{TimeSet}
\eeq
here $\Diamond := \Last  - \Last l \, \pa_{l}$ and $\Heart := \Diamond/{\Last h}$.  
\cite{SemiclI} had an antecedent, it recognized the derivative combination but only gave an example of it rather than writing the $l$-equation with all occurrences of this.

Then neglecting the second, third, fourth, fifth, sixth and eighth terms and the $\Last l\,\pa_{l}$ contributions 
(see Sec \ref{Approx-types} for various possible justifications), this $h$-equation collapses to the standard semiclassical approach's HJE,    
\beq
\{\pa_{h}S\}^2 = 2\{E - V_{h}\} \mbox{ } , \mbox{ or } \mbox{ } 
\label{Sicut}
{\Last} h^2 = 2\{E - V_{h}\} \mbox{ } . 
\eeq
The second form is by (\ref{lance2}), and is especially justified because $S$ is a standard HJ function.

\mbox{ } 

\noindent Note 1) In the more general case, one is assuming that the QM-corrected HJE is solved by something which still behaves like a standard HJ function.  

\noindent Note 2) A reformulation of the latter is of use in further discussions in this article is the analogue Friedmann equation, 
\beq
\left\{
\frac{\Last h}{h}\right\}^2 = \frac{2E}{h^2} - \frac{2V_{h}}{h^{2}} \hspace{0.3in} \mbox{ (general: $h$ = } \rho)  \hspace{0.3in} \mbox{  or }  \hspace{0.3in}
\left\{ 
\frac{    \check{\Last} {h} }{h}
\right\}^2 
=  \frac{E}{2h^3} - \frac{V_{h}}{2 h^3}  \hspace{0.3in} \mbox{ (triangleland: $h$ = I) } .
\label{Este}
\eeq                            
The energy equation form is then solved by 
\beq
{\lt}^{\se\sm} = \frac{1}{\sqrt{2}}           \int   \frac{\d h}        {\sqrt{E - V_{h}}}  \hspace{0.3in} (h = \rho)  \hspace{0.3in} \mbox{ or }  \hspace{0.3in}
\check{\lt}^{\se\sm}             = \sqrt{2}\int   \frac{\sqrt{h}\d h}{\sqrt{E - V_{h}}}  \hspace{0.3in} (h = \mI)   \mbox{ } \mbox{ } .  
\label{Jesu}
\eeq

\subsubsection{Types of contribution to the Machian semiclassical time}

\noindent Expanding out and keeping up to 1 power of $\hbar$,
\beq
\lt^{\se\sm(\sW\sK\sB)} = \lt^{\se\sm(\sW\sK\sB)}_{\{0\}} + \frac{1}{2\sqrt{2}}\int\frac{\langle\chi| J |\chi\rangle}{W_{h}^{3/2}}\d h +  
i\hbar  \int   \frac{\d h}{\sqrt{2W_{h}}}  \left\{  \frac{k(N, d)}{h} + 2\langle\chi|\pa_{h}|\chi\rangle \right\} + O(\hbar^2) \mbox{ } .   
\label{QM-expansion}
\eeq
I.e., with comparison with the classical counterpart (\ref{Cl-Expansion}) an `expectation of interaction' term in place of an interaction term, 
and an operator-ordering term and an expectation term in place of a classical $l$-change term.  

\mbox{ }  

\noindent Regime 0) Thus even if expectation terms are small, there is a novel operator-ordering term. 
Incorporating this does {\sl not} require coupling the chronifer procedure to the quantum $l$-equation; 
it is a quantum correction to the nature of the scale physics itself rather than a Machian $l$-subsystem change contribution.  

\noindent This working suffices to show that the Machian emergent time finding procedure is capable of returning {\sl complex} answers in the 
semiclassical, and, more generally, fully quantum, regimes.  
This opens up questions of interpretation.
Complex entities are common enough in quantum theory (e.g. slightly deformed contour integrals in expressions for propagators), however what complex methods are 
well-established to work in flat geometries encounter further difficulties in passing to curved-geometry cases required by GR (Sec \ref{Path-Int}).

\noindent This correction term is readily evaluable for some simple examples. 
It is $\frac{i\hbar k(N, d)}{\sqrt{2E}}\mbox{ln}\, h$ for free RPM problems, though it continues to make sense to not neglect central terms for nonzero total shape 
momentum, by which this becomes, e.g. for the $\sfD \neq 0$ 3-stop metroland free problem 
$\frac{i\hbar}{|\sfD|\sqrt{E}} \mbox{ln}\left(   \sqrt{E} h + \sqrt{E h^2 - \sfD^2}  \right)$. 
$N$-stop metroland and $N$-a-gonland extensions of this working cannot include central terms in a separated-out fashion since these now involve functions of $l$.
The isotropic HO problem (or approximation) has,  for zero shape momentum or neglecting central terms, 
$\frac{-i\hbar k(N, d)}{\sqrt{2A}}  \,  \mbox{ln}\left(\frac{\sqrt{A} h}{\sqrt{E  - A h^2}} \right)$.   
With nonzero shape momentum central term included for 3-stop metroland, this is still analytically evaluable albeit in terms of an elliptic function.  
%

\mbox{ }

\noindent To explicitly evaluate the other 2 terms here, we need coupling to the $\fl$-equation to have $|\chi\rangle$ (see Sec \ref{Det-Back}); 
the above expansion suffices, however, to demonstrate the Machian character of the emergent WKB time now indeed give the QM $l$-subsystem an opportunity to contribute:
\beq
\lt^{\se\sm(\sW\sK\sB)} = {\cal F}\lfloor h, l, \d h, |\chi(h, l) \rangle \rfloor \mbox{ } .  
\eeq
\noindent In greater generality than the above $\hbar$ expansion,
\beq
\lt^{\se\sm(\sW\sK\sB)} = \int\frac{2\,\d h}{-B \pm \sqrt{B^2 - 4C}}
\eeq
for 
\beq
B = - i\hbar\{2 \langle\chi|\pa_{h}|\chi\rangle +  h^{-1}k(N, d)\} \mbox{ } ,
\eeq
\beq
C = -2\{W_{h} - \langle\chi| J |\chi\rangle\} + \hbar^2\{   h^{-1}k(N, d)\langle\chi|\pa_{h}|\chi\rangle - \langle\chi|\pa_{h}^2|\chi\rangle + h^{-2}c(N, d)\} \mbox{ } .  
\eeq
So what were $\pm$ pairs of solutions classically are turned into more distinct complex pairs, which splitting is mediated by operator-ordering and expectation 
contributions to first order in $\hbar$. 
One also sees that the second-order contributions are another expectation, another ordering term and one that has one factor's worth of each.

\subsection{Extension to the case of multiple $h$ degrees of freedom}\label{Multi-h2} 

Here, we are initially solving for a wavefunction which straightforwardly accommodates multiple $h$-variables into its dependency.  
Upon stripping this down to a classical equation, one gets the HJE in $S$. 

\noindent One moreover has to argue that one is allowed to interpret this in the kind of way that it is in classical theory, by which it encodes all information about the $h$-system (c.f. Sec \ref{Multi-h}). 
Then suppose that the classical evolution equations are recovered.  
One can then take all bar one of these and the HJE itself cast into Lagrangian energy equation form as an equation for 
$t^{\se\sm(\sW\sK\sB)})$ to obtain an at least well-determined set of equations to be able to obtain, inversions permitting, $h_1(t^{\se\sm(\sW\sK\sB)})), h_2(t^{\se\sm(\sW\sK\sB)})), ...$ 
Then one can interpret the approximate $l$-equation as an emergent time-dependent perturbation of a TISE just as before; this can be indicated 
by writing an $h$-subconfiguration-space-valued $h(t^{\se\sm(\sW\sK\sB)})$].  

\mbox{ }  

\noindent {\bf Question 79)} A suite of open problems is then as follows. 

\noindent 1) Can, and if so how, the quantum-corrected $h$-equation be taken as effectuating a semiclassical quantum equivalent of `the HJE encodes all information about the $h$-system'.  
I.e. by what procedure is one to obtain semiclassical-quantum-corrected counterparts of equations of motion from a semiclassical-quantum-corrected HJE?    
I anticipate that this is it is surmountable, though how uniquely is unclear and may well depend on the detailed sense in which `semiclassical' is to be mathematically implemented.

\noindent 2) What happens at the quantum level to the role of the forces determining which approximations to make in the GLET is to be abstracted from STLRC procedure as laid 
out in Sec 9?    
[Force equations are far less common at the quantum level; which formalism is one to use for these?]

\subsection{Recasting of the $l$-fluctuation equation as a TDSE}\label{VaNiel}

\noindent One passes for a fluctuation equation to a semiclassical emergent TDWE via the crucial chroniferous move 
\beq
N^{hh}  i\hbar  \frac{\pa S}{\pa h}               \frac{\pa \left| \chi\right \rangle}{\pa h} = 
i\hbar \, N^{hh}  p_{h}                                    \frac{\pa \left| \chi\right \rangle}{\pa h} =
i\hbar \, N^{hh}  M_{hh}  \Last h      \frac{\pa \left| \chi\right \rangle}{\pa h} = 
i\hbar \frac{  \pa h       }{  \pa t^{\se\sm(\sW\sK\sB)}  }\frac{\pa \left| \chi\right \rangle}{\pa h} = 
i\hbar                                                     \frac{\pa \left| \chi\right\rangle }{  \pa t^{\se\sm(\sW\sK\sB)}  }    \mbox{ } , 
\eeq
which proceeds via (\ref{lance2}) and the chain-rule in reverse.

\mbox{ }  

\noindent Note 1) In this paper's case, $N_{hh} = 1 = M^{hh}$; I include these, however, to show the greater generality of the working; in particular this is needed in GR examples. 

\noindent Note 2) This move as displayed is only approximate in assuming the zeroth order $l$-independence of $t^{\se\sm(\sW\sK\sB)}$.
If one wishes to work more accurately than that in this respect, one needs to use $\Diamond$ in place of $\Last$ \cite{SemiclI} above.   

\mbox{ } 

\noindent The full emergent semiclassical TDWE is then
\beq
i\hbar\{1 - \mP_{\chi}\} \Diamond |\chi\rangle = 
\{1 - \mP_{\chi}\}  \{ - \{\hbar^2/2\} \{ \Heart^2 + k(N, d) h^{-1} \Heart + h^{-2} \triangle_{l} \} + V_{l} +  J\} |\chi\rangle 
\label{Em-TDWE} \mbox{ } . 
\eeq
\noindent Note 3) One is then using one of eq's (\ref{Sicut}--\ref{Jesu}) to express $h$ as an explicit function of $t^{\se\sm}$ 
This does require invertibility (at least on some intervals of the mechanical motion) in order to set up the $t^{\se\sm}$-dependent perturbation equation and more generally 
have a time provider equation followed by an explicit time-dependent rather than heavy degree of freedom dependent equation,  which I now denote
\be
h = h(t^{\se\sm(\sW\sK\sB)}) \mbox{ } . \label{tazenda}
\ee
Even for 1 $h$ variable, this is not in general guaranteed, but the examples in question do possess it.  
The inversion can also be used to convert $h$-derivatives to $t^{\se\sm(\sW\sK\sB)}$-derivatives, so one has a bona fide $l$-equation. 

\mbox{ } 

\noindent (\ref{Em-TDWE}) is usually approximated by a semiclassical emergent TDSE, 
\beq
i\hbar\Last{\pa\left|\chi\right\rangle} = H_{l}|\chi\rangle =
- \{\hbar^2/2\}{\triangle_{\sFS(N, d)}    }\left|\chi\right\rangle/{    h^2(t^{\se\sm})    }  + V_{l}\left|\chi\right\rangle \mbox{ } . \label{SETDSE}
\eeq
(see Secs \ref{Semi-Avs}, \ref{Semi-ID} for various possible justifications of the approximations made).   

\noindent Note 4) (\ref{SETDSE} is, modulo the $h$--$l$ coupling term, `ordinary relational $l$-physics'.  
In turn, this is `ordinary $l$-physics' modulo the effect of the angular momentum correction term. 
[This is itself absent in 1-$d$ or if one repeats the above working in a spatially nonrelational setting].  
Thus the purported simple situation has `the scene set' by the $h$-subsystem for the $l$-subsystem to have dynamics. 
This dynamics is furthermore slightly perturbed by the $h$-subsystem, while neglecting the back-reaction of the $l$-subsystem on the $h$-subsystem.  
One might even argue for the interaction term to be quantitatively negligible as regards the observed $l$-physics. 

\mbox{ }  

\noindent Thus, overall,  the fluctuation $l$-equation (\ref{l-TDSE-prime}) can be rearranged to obtain a TDSE with respect to an emergent time that is `provided by the h-subsystem'. 
The full (except for $\pa_{h}\mbox{}^2 S$ neglected) Machian $l$-equation is then \cite{ACos2} 

\mbox{ } 

\noindent The main idea is then to consider (\ref{TimeSet}) and (\ref{l-TDSE-prime}) as a pair of equations to solve for the unknowns $t^{\se\sm}$ and $|\chi\rangle$. 

\mbox{ } 

\noindent Regime 1)  One might argue for the interaction term to be quantitatively negligible as regards the observed $l$-physics.   

\noindent Regime 2)  Keeping the interaction term, (\ref{Rec-TDSE}) has not only a time provided by the $h$-subsystem but also a time-dependent imprint on the $l$-subsystem's 
physics due to the $h$-subsystem's physics.  
This amounts to neglecting the averaged terms and the unaveraged first and second terms (see Sec \ref{Approx-types} for various possible justifications).

\subsection{Use of rectified time, and that this amounts to working on shape space}

The TDWE or its TDSE core (\ref{SETDSE}) simplifies if one furthermore chooses the rectified time given by \cite{Cones} 
\beq
h^2\Last := h^2 \pa/\pa t^{\se\sm} = \pa/\pa t^{\sr\se\scc} =: \mbox{\textcircled{$\star$}} 
\label{Tu}
\eeq
We define this to arbitrary order, though firstly we consider the zeroth order version, i.e. 
\beq
t^{\sr\se\scc}_{\{0\}} - t^{\sr\se\scc}_{\{0\}}(0) =  
\int\d t^{\se\sm(\sW\sK\sB)}_{\{0\}}/h^2(t^{\se\sm(\sW\sK\sB)}_{\{0\}}) = \frac{1}{\sqrt{2}}\int \frac{\d h}{\rho^2\sqrt{{E - V(h)}}} 
\hspace{1in} \mbox{ ($h$ = $\rho$) } , 
\eeq
\beq
\mbox{ or } \hspace{1in}   
                                   = \int\d \check{t}^{\se\sm(\sW\sK\sB)}_{\{0\}}/ h^2(\check{t}^{\se\sm(\sW\sK\sB)}_{\{0\}})^2  = 
								     \sqrt{2}\int  \frac{\d h}{h^{2}}\sqrt{\frac{h}{E - V(h)}} 
\mbox{ } \hspace{1in} (h = \mI) \mbox{ } .  
\label{Rek}
\eeq
As regards interpreting the rectified timefunction, in each case using $t^{\sr\se\scc}$ amounts to 
working on the shape space itself, which amounts to using the geometrically natural presentation.     

\noindent Note 1) I use $\lt^{\sr\se\scc}$ for $t^{\sr\se\scc}$ up to a constant [always including $t^{\sr\se\scc}(0)$ 
and sometimes including the constant of integration from the other side of (\ref{Rek}ii)].  

\noindent Note 2) $t^{\sr\se\scc}$ is very similar to the $t_{\sg\se\so}$ of Sec (\ref{+PPSCT}) referring to the shape geometry, which is the natural geometry for the $l$-physics, 
since this Sec's study has $l$ aligned with the shapes.
The difference lies in that $\5Star = \mI \sqrt{W/T_{\sss\sh\sa\sp\se}} \Circ$, whilst \textcircled{$\star$} = $\mI\sqrt{{W}/{T}} \Circ$ 
i.e. the same conformal factor but a difference in totality of the kinetic term involved.  

\mbox{ } 

\noindent {\bf Lemma 17}: suppose $\lt^{\se\sm}$ is monotonic.  
Then the rectified time $\lt^{\sr\se\scc}$ is also monotonic.

\beq
\underline{\mbox{Proof}} \hspace{1in} \frac{\d \lt^{\sr\se\scc}}{\d \rho} = \frac{\d \lt^{\sr\se\scc}}{\d t^{\se\sm(\sW\sK\sB)}}\frac{\d t^{\se\sm(\sW\sK\sB)}}{\d \rho} = 
\frac{1}{\rho(t^{\se\sm(\sW\sK\sB)})^2}\frac{\d t^{\se\sm}(\sW\sK\sB)}{\d\rho} \geq 0 \hspace{1in} 
\eeq
by the chain-rule in step 1, (\ref{Tu}) in step 2 and positivity of squares and the assumed monotonicity of $t^{\se\sm(\sW\sK\sB)}$ in step 3 $\Box$.
%

\mbox{ } 

\noindent The full rectified $l$-TDWE is then
\beq
i\hbar\{1 - \mP_{\chi}\} \Club |\chi\rangle = 
\{1 - \mP_{\chi}\} \{ - \{\hbar^2/2\} \{\Spade^2 + k(N, d) \Spade  + \triangle_{l} \} +  V^{\sr\se\scc}_{l} + J^{\sr\se\scc}  \}   |\chi\rangle . 
\label{Rec-TDSE}
\eeq
for $\Club    := \Rec  - \Rec  l\pa_{l}$  and $\Spade := \Club/\Rec  \, \mbox{ln} \, h(t^{\sr\se\scc})$.  

\mbox{ }  

\noindent This is most commonly viewed as (perturbations about) a TDSE, 
\beq
i\hbar\mbox{\textcircled{$\star$}}|\chi\rangle = -\{\hbar^2/2\}\triangle_{l}\left|\chi\right\rangle + 
V^{\sr\se\scc}_{l} \left|\chi\right\rangle + {J}^{\sr\se\scc}\left|\chi\right\rangle \mbox{ \{+ further perturbation terms\} } . 
\eeq
\noindent Note 3) The rectified time's simplification of the emergent-TDSE equation can be envisages as passing from the emergent time that is natural to the whole relational space to a 
time that is natural on the shape space of the $l$-degrees of freedom themselves, i.e. to working on the shape space of the $l$-physics itself.    

\noindent Note 4) The $l$-subsystem's simplest time is {\sl not} immediately the one provided by the $h$-subsystem.  
%

\noindent Caveat)  The $V$--$J$ split is not preserved by the rectifying operation; thus in subsequent working it is not necessarily clear whether some, 
all or none of this should carry an $\epsilon$.  
$J^{\sr\se\scc}$ denotes $h^2\big(\lt^{\sr\se\scc}_{\{0\}}\big)\{V_{l} + J\}$ 
(we do not cover how a $1/\rho^2$ or $1/\mI^2$ term will give a new separated-out $V_{l}$ under rec-ing as this does not occur in our examples).

\noindent Note 5) This article's specific examples of rectified emergent TDSE's -- for 3-stop, 4-stop, triangleland and the $N$-stop extension -- are mathematically familiar equations. 
They are TDSE counterparts of Sec 16's TISE's, 
with (essentially) the same, separated-out shape parts as in Secs 14 and 15.  
\cite{QuadIII} carries the quadrilateralland counterpart of this.

\mbox{ } 

\noindent Thus Regime 1) amounts to solving an HJE and then a non-interacting TDSE, whereas Regime 2 amounts to solving an HJE and then an interacting TDSE 
(e.g. as a time-dependent perturbation about the non-interacting TDSE).  

\mbox{ } 

\noindent Regimes 1b) and 2b) extend these two systems by allowing for back-reaction of the $l$-subsystem on the $h$-subsystem, via the $h$-equation including the term 
$\langle\chi|J|\chi\rangle$.  
Only the case with interaction {\sl and} back-reaction makes detailed and/or long-term sense from the perspective of energy-balance `book-keeping'.  
Namely that, energy transitions in the one system have to be compensated by opposing energy transitions in the other subsystem 
(the long-term part of that then being due to secular build-up).  
[Backreactions mediated by other terms are not precluded.]

\subsection{Discussion of the other conventionally small terms, by qualitative types}\label{Approx-types}

\subsubsection{Back-reactions (Analogy 107)}\label{Semi-Back}

\noindent One interesting feature is that the $l$-subsystem can back-react on the $h$-subsystem rather than just merely receive a time-standard from it.  
Some early literature on this point is by Brout and collaborators \cite{BV89} (see also \cite{Kieferbook} for a review).
The $l$-equation is now coupled to a less approximate chroniferous $h$-equation containing operator ordering and expectation quantum terms.
Here, the perturbations of expectation type having input from the $l$-subsystem (they are expectation values in the $l$-subsystem's wavefunction).  
Clearly then the previously-suggested simple procedure of solving the $h$-HJE first is insufficient by itself to capture this level of detail.  
The current paper's scheme does allow for such terms and, moreover, points out the significance of the expectation term corrections to the $h$-equation as implementing 
Mach's Time Principle in a STLRC way - giving the $l$-subsystem the opportunity to contribute to the final more accurate estimate of the emergent timefunction (Motivation 1).  

\noindent The presence of these corrections makes further physical sense, as follows.

\noindent Motivation 2) the HJE approximation here depicts a conservative system, 
but if the $h$-subsystem interacts with the $l$-subsystem one is to expect it to have a more general form than the conservative one 
Then indeed, expectation terms can be seen as functionals of $|\chi(l^{\sfa}, t^{\se\sm}_{\{0\}})\rangle$ with the integration involved indeed 
not removing the $t^{\se\sm}_{\{0\}}$ dependence, so the $h$-equation containing these corrections is indeed dissipative rather than conservative.

\noindent Motivation 3) The preceding SSec's book-keeping argument.
 
\noindent Motivation 4) {\bf Back-reaction is conceptually central to GR}.  
Matter back-reacting on the geometry is a conceptually important part of GR  (both at the level of what the 
Einstein field equations mean and in GR's aspect as supplanter of absolute structure). 
Thus toy models that include back-reaction are conceptually desirable in schemes that concentrate on better understanding GR.

\subsubsection{Averaged terms (Analogy 108)}\label{Semi-Avs}

Expectation/averaged terms are often dropped in the Quantum Cosmology literature.\footnote{Incidentally, the idea of neglecting averaged terms 
is a brutal illustration of how dimensional analysis is not all, since averaged and unaveraged versions of a quantity clearly have the same 
`dimensionless groups' (meant in the sense of fluid mechanics).}  
%
The usual line given for this in that literature is that these are argued to be negligible by the 

\mbox{ } 

\noindent{\bf Riemann--Lebesgue Theorem}, which is the mathematics corresponding to the physical idea of {\bf destructive interference}.  

\mbox{ } 

\noindent I add that Quantum Cosmology practitioners probably do not want such terms to be around due to non-amenability to exact 
treatment that they confer upon the equations if included.  
However, here are some reasons to keep it.   

\mbox{ }  

\noindent 1) Some RPM counterexamples to these terms being small are as follows. 
For 3-stop metroland's analogue of the central problem, $\langle\pa_{\varphi}^2\rangle|\chi\rangle$ and $\pa_{\varphi}^2|\chi\rangle$ are of the same size since 
the wavefunctions in question are eigenfunctions of this operator.
This type of example generalizes to $N$-stop metroland and triangleland too, in general concerning eigenfunctions of the Laplacian.  

\noindent 2) Moreover, then $H_l|\chi\rangle = \{\triangle_l - \langle\triangle_l\rangle\}|\chi\rangle$ gives zero rather than $\triangle_l|\chi\rangle$.
Still, the solution to the unaveraged equation solves the averaged equation too, and constitutes a proper eigenfunction (unlike 0).  
This approach suggests keeping all average terms in the $l$-equation together.

\noindent 3) I have pointed out \cite{SemiclIII} an analogy with Atomic/Molecular Physics, where the counterparts of such terms require a 

\noindent {\bf self-consistent} variational--numerical approach.  

\mbox{ } 

\noindent An example of this is the iterative technique of the {\bf Hartree--Fock approach}. 
In Atomic/Molecular Physics, it is is then conceded that this ensuing non-exactly tractable mathematics is necessary so as to get passably correct answers 
(comparison with experiments confirms this).  
I investigate the quantum-cosmological counterpart of this in more detail in \cite{SemiclIV}.

While there are a number of differences between Molecular Physics and Quantum Cosmology, Hartree--Fock theory in fact is known to span those differences.
E.g. it is available for $\ft$-dependent physics, and involving a plain rather than antisymmetrized wavefunction, 
and for field theory (c.f. the Condensed Matter Physics literature \cite{Cond1, Cond2, Cond3, Cond4, Cond5, Cond6, Cond7}).  

\mbox{ }

\noindent Regime 3: $t^{\se\sm}_{(0)}$ is satisfactory, then apply an HF procedure on the $l$-equation with average terms kept.  

\mbox{ } 

\noindent{\bf Question 80}. Variationally justify this quantum-cosmological analogue of Hartree--Fock self-consistent procedure, and then study its outcome.   
(This is being covered in \cite{QuadIV}.)

\mbox{ } 

\noindent However, this is insufficient if one's scheme is to comply with Mach's Time Principle.

\mbox{ } 

\noindent{\bf Question 81$^{*}$}.  It is then not as yet clear how to extend the self-consistent treatment to this non-negligible back-reaction case (Regime 3b).
Schematically, this is of the form 
\beq
\mbox{\Huge\{}  \stackrel{\mbox{(chroniferous expectation-corrected variant of the Hamilton--Jacobi equation)}}
                         {\mbox{(emergent-time-dependent Hartree--Fock scheme)}}
                          \mbox{ } , 
\label{cyril}  
\eeq
which is probably this time a new type of system from a Mathematical Physics perspective.  
Can this be anchored on a variational principle?  
As such, this investigation is not just of qualitative confidence in the Halliwell--Hawking scheme but important also as regards 
the detailed robustness of the Semiclassical Approach's time-emergence itself. 

\mbox{ } 

\noindent Note 1) If a Hartree--Fock scheme is deemed to be appropriate for Quantum Cosmology, whether it iteratively converges will become an issue; 
moreover, the nonstandardness of the system (\ref{cyril}) will likely make this hard to check for.

\noindent Note 2) Is there any promise to additionally incorporating the approximate atom-by-atom product form of 
Hartree--Fock wavefunctions to make a clump-by-clump analysis of inhomogeneous cosmology? 
The nonlinearity of GR ultimately will cause problems here, though this clump-by-clump analysis might just be possible within the RPM toy models themselves.

\subsubsection{Higher derivative terms (Analogy 109)}\label{Semi-Deri}

One often neglects the extra $t^{\se\sm(\sW\sK\sB)}$-derivative terms whether by discarding them prior to noticing they are also convertible 
into $t^{\se\sm(\sW\sK\sB)}_{\{0\}}$-derivatives or by arguing that $\hbar^2$ is small or $\rho$ variation is slow. 
Moreover there is a potential danger in ignoring higher derivative terms even if they are small (c.f. Navier--Stokes equation versus Euler equation in fluid dynamics).

\mbox{ } 

\noindent One would expect there to be some regions 
of configuration space where the emergent TDWE behaves more like a 
Klein--Gordon equation than a TDSE, albeit in full it is more general.    
Thus the guarantee of appropriate interpretability that accompanies TDSE's is replaced by a difficult study of a more general TDWE.  
It is then worth noting that Klein--Gordon-like but more complicated equations are prone to substantial extra impasses (see, e.g. \cite{Kuchar81}).

\mbox{ } 

\noindent Kiefer and Singh's expansion \cite{KS91} treats higher derivative terms along the lines of the next-order correction to the TDSE from the Klein--Gordon equation.

\subsubsection{Identifying the nature of each term present}\label{Semi-ID}

\noindent There are 14 terms that are often neglected in the reduced system \cite{SemiclI}.
$h$ and $l$ in the enumeration denote which equation these terms feature in. 

\vspace{10in}

\noindent $h$1) $\hbar \pa_{h}S  \langle \pa_{h} \rangle$ is a back-reaction, O($\hbar$), a time-derivative, an average and an $l$-diabatic term.

\noindent $h$2) $\hbar^2\langle \pa_{h}^2\rangle$ is a back-reaction, O($\hbar^2$), a {\sl higher} time-derivative, an average and an $l$-diabatic term. 

\noindent $h$3) $\hbar^2 h^{-1}\langle \pa_{h}\rangle$ is a back-reaction, O($\hbar^2$), a time-derivative, an average and an $l$-diabatic term.  

\noindent $h$4) $\langle J\rangle$ is a back-reaction, an average and small for $J$ perturbatively small (compared to $V_{h}$)).  

\noindent $h$5) $\hbar^2 h^{-2}k(\xi)$ is O($\hbar^2$).  

\noindent $h$6) $\hbar\pa_{h}^2S$ is O($\hbar$) and the term usually neglected by the WKB approximation itself.

\noindent $l$1)  $\hbar \pa_{h}S$ is O($\hbar$), a time-derivative, an average and an $m$-diabatic term.  

\noindent $l$2) $\hbar^2\langle \pa_{h}^2\rangle$ is O($\hbar^2$), a {\sl higher} time-derivative, an average and an $m$-diabatic term.  

\noindent $l$3) $\hbar^2h^{-1}\langle \pa_{h}\rangle$ is O($\hbar^2$), a time-derivative, an average and an $m$-diabatic term.  

\noindent $l$4) $\langle J\rangle$ is an average and small for $J$ perturbatively small (compared to $V_{h}$).  

\noindent $l$7) $J|\chi\rangle$ is small for $J$ perturbatively small.

\noindent $l$8) $\hbar^2\langle \triangle_{l}\rangle$ is O($\hbar^2$) and an average.  

\noindent $l$9) $\hbar^2\pa_{h}^2|\chi\rangle$ is O($\hbar^2$), a {\sl higher} time-derivative (often the most significant of these through being unaveraged) and an $m$-diabatic term.  

\noindent $l$10) $\hbar^2h^{-1}\pa_{h}|\chi\rangle$  is O($\hbar^2$), a time-derivative and an $m$-diabatic term.

\noindent There are a number of further terms that carry 1 or 2 factors of 
\beq
\frac{\pa l}{\pa t}\frac{\frac{\pa |\chi\rangle}{\pa l}}{\frac{\pa |\chi\rangle}{\pa h}} = \frac{\ma(l)}{\ma} := \epsilon_i\mbox{ } ,
\eeq 
i.e. a ratio of quantum $l$-subsystem adiabaticity to classical mutual adiabaticity.
This is another type of term with which dimensional analysis is of no use.

\subsection{Detail of the negligible-back-reaction regime}

\noindent Example 1) \cite{SemiclI} considered this for the subsystem-aligned $h$--$l$ split for the 3-stop metroland HO, 
whose Cartesian coordinates are of course very highly tractable as per Sec \ref{Q-3-Stop}.  

\noindent Example 2) \cite{SemiclIII} considered this for the more quantum-cosmologically adept scale--shape-aligned $h$--$l$ split for the 3-stop metroland HO.  
Here, 
the emergent-TDSE is 
\beq
i\hbar\frac{\pa|\chi_{\{1\}}\rangle}{\pa \lt^{\sr\se\scc}_{\{0\}}} = - \frac{\hbar^2}{2}\frac{\pa^2|\chi_{\{1\}}\rangle}{\pa\varphi^2} + 
\frac{  B\,E^2\,\mbox{cos}\,2\varphi  }{A^2\left\{1 + 2E^2\lt^{\sr\se\scc\, 2}_{\{0\}}/ A\right\}^2  }  |\chi_{\{1\}}\rangle \mbox{ } .
\label{main}
\eeq
For $B$ small in relation to e.g. the $A$-term or the $\pa_{\varphi}\mbox{}^2$ term and corresponding to $\epsilon$ small in this example, 
i.e. $K_1 \approx K_2$, poses, about a very simple quantum equation, a (fairly analytically tractable) $\lt^{\sr\se\scc}$-dependent perturbation problem.  
See \cite{SemiclIII} for the solution to this, and for a simple triangleland upside-down HO example also.  

\mbox{ }  

\noindent Note 1) Here, one has a $t^{\se\sm}_{\{0\}}$-dependent perturbation theory problem. 
\noindent The $J$ renders the $l$-equation nonseparable as is required for the Semiclassical Approach to work. 
\noindent It also constitutes an {\sl an imprint from} the heavy chroniferous subsystem unto the light subsystem.  

\noindent Note 2)  There are no averaged terms in the $h$-equation, i.e. no back-reaction and the system is decoupled: 
one can solve the $h$-equation first as no $|\chi\rangle$'s are present in it.  

\noindent Note 3) Of the omitted terms, 
$
\hbar^2 \pa^2 S \approx O(\hbar)\pa^2 S, \{\hbar^2/\rho\}\langle \pa_{\rho} \rangle \approx O(\hbar^3)\{B/A\}^2, \mbox{ and }   
\{\hbar^2\}\langle \pa_{\rho}^2 \rangle \approx O(\hbar^4)\{B/A\}^2. 
$
\noindent Note 4) However, this scheme is not satisfactory from the perspective of Mach's Time Principle, 
since it does not give the $l$-subsystem an opportunity to contribute to the timestandard.  
This is amended by passing to small but non-negligible back-reaction schemes such as in the next SSec.

\subsection{Detail of the small but non-negligible back-reaction}\label{Det-Back}  

In the case that $J \approx 0$ suffices in the $l$-equation, the other two low-order terms in (\ref{QM-expansion}) are 1) 
\beq
2i\hbar\int\frac{\d h}{\sqrt{2W_{h}}}\langle\pa_{h}\rangle =  2i\hbar \int\frac{\d h}{\sqrt{2W_{h}}}R^{*}\frac{\d R}{d t^{\sr\se\scc}}\frac{\d t^{\sr\se\scc}}{\d h} = 
\int\frac{\d h}{h^2}\,\frac{\hbar^2\md^2/2 + Ah^4}{\{E - Ah^2\}}
= - \left\{h + \frac{\hbar^2\md^2}{2Eh}\right\} + \left\{ \sqrt{\frac{E}{A}} + \sqrt{\frac{A}{E^3}} \frac{\hbar^2\md^2}{2}  \right\}\mbox{artanh}\left(\sqrt{\frac{A}{E}}h\right)
\label{correction-1}
\eeq
for $R$ the separated-out scale/time part of $|\chi\rangle$.
\beq
2) \hspace{1in} \frac{1}{2\sqrt{2}}\int\langle J \rangle \frac{\d h}{W_{h}^{3/2}} = 
\frac{B\langle d| \mbox{cos}^2\varphi |d \rangle}{2\sqrt{2}}  \int \frac{h^2\d h}{W_{h}^{3/2}} = 
\frac{1}{2\sqrt{2A}}\frac{B}{A}  \left\{ \frac{\sqrt{A}h}{\sqrt{E - Ah^2}} - \mbox{arcsin}\left(\sqrt{\frac{A}{E}}h\right)\right\} \mbox{ } .  
\hspace{1in} 
\eeq 
These are for 3-stop metroland, but it is not hard to generalize them to $N$-stop and the $N$-a-gon. 
This is via the useful integral (\ref{*****}) for 4-stop metroland; for triangleland the integral is zero by symmetry/selection rule, so this term is unavailable 
as a contributor to back-reaction; see \cite{QuadIII} for the quadrilateralland counterpart of this.  
All these further examples that are nonzero pick up dependence on the quantum numbers of the system.   

\noindent Finally, in the above 3-stop metroland example, there is no problem adding a classical central correction to the potential; 
\ref{correction-1}'s integral remains in terms of basic functions by virtue of $v = \rho^2$ substitution and completing the square. 
 
\mbox{ }

\noindent Next, consider a more in-depth treatment in terms of expansions. 
Apply $t = t_{\{0\}} + \epsilon t_{\{1\}}$ and $|\chi\rangle = |\chi_{\{0\}}\rangle + \epsilon|\chi_{\{1\}}\rangle$ as per \cite{SemiclIII}.  
Here the $\epsilon$ originally comes from being a split-out factor in front of the interaction term $J$.  

Usually the first $J$ should be kept, since elsewise the $l$-subsystem's energy changes without the $h$-system responding, violating conservation of energy.    
But if this is just looked at for a ``short time" (few transitions, the drift may not be great, and lie within the 
uncertainty to which an internal observer would be expected to know their universe's energy.  
Here, I do not explicitly perturbatively expand the last equation as it is a decoupled problem of a standard form:
a $\lt^{\sr\se\scc}_{(0)}$-dependent perturbation of a simple and well-known $\lt^{\sr\se\scc}_{(0)}$-dependent perturbation equation.

My particular RPM proposal that is a simple modelling of back-reaction is as follows (the first of the above).  
Look to solve 
\beq
\frac{1}{2}
\left\{
\frac{\pa h}{\pa t^{\se\sm(\sW\sK\sB)}}
\right\}^2 + 
\epsilon\langle\chi|\, J\,|\chi\rangle = E - V \mbox{ } , 
\eeq
\beq
i\hbar\frac{\pa|\chi\rangle}{\pa t^{\se\sm(\sW\sK\sB)}} = - \frac{\hbar^2}{2}\frac{\triangle_{\sFS(N, d)}|\chi\rangle}{h^2(t^{\se\sm(\sW\sK\sB)})} 
+ \epsilon J|\chi\rangle \mbox{ } . 
\eeq
\noindent Note: if there were a separate $V_{l}$, rectification leads to this becoming part of ${J}^{\sr\se\scc}$, and, in any case, I am considering $V$ 's that are homogeneous in 
scale variable, thus I do not write down a separate $V_{l}$: the isotropic $V_{h}$ and the direction-dependent interaction term $J$ contain everything that ends up to be of relevance.    

\noindent 
Then in the small but non-negligible-back-reaction regime, there is a $t_{\{1\}}$ equation to solve, and the $|\chi_{\{1\}}\rangle$ stage then gives the following set of equations 
\beq
\d h^2 = 2\{E - V(h)\}\d t^{\se\sm(\sW\sK\sB)\,2}_{\{0\}} \mbox{ } ,
\eeq
\beq
i\hbar\frac{\pa|\chi_{\{0\}}\rangle}{  \pa \lt^{\sr\se\scc}_{\{0\}}  } = -\frac{\hbar^2}{2}\triangle_{l}|\chi_{\{0\}}\rangle \mbox{ } ,
\eeq
\beq
\d t_{\{1\}}^{\se\sm(\sW\sK\sB)}\{E - V(h)\} = \langle\chi_{\{0\}}|\,J\,|\chi_{\{0\}}\rangle \d t^{\se\sm(\sW\sK\sB)}_{\{0\}} \mbox{ } ,
\label{three}
\eeq
\beq
i\hbar\frac{\pa|\chi_{\{1\}}\rangle}{  \pa \lt^{\sr\se\scc}_{\{0\}}  } = -\frac{\hbar^2}{2}\triangle_{\FrS(N, d)}|\chi_{\{1\}}\rangle + 
\left\{ 
J^{\sr\se\scc} - \frac{\hbar^2}{2}\frac{\d \lt^{\sr\se\scc}_{\{1\}}}{\d \lt^{\sr\se\scc}_{\{0\}}}\triangle_{\FrS(N, d)}
\right\}
|\chi_{\{0\}}\rangle
\mbox{ } .
\eeq 
One can then solve (\ref{three}) for $t_{\{1\}}$ using knowledge of the solutions of the first two decoupled equations, 
\beq
\lt_{\{1\}}^{\se\sm(\sW\sK\sB)} = \frac{1}{2} \int_{t^{\prime}_{\{0\}} = t_{\{0\}}(0)}^{t_{\{0\}}}
\left\{
\frac{1}{E - V\big(h_{\{0\}}\big(t^{\prime}_{\{0\}}\big)\big)}\int \chi_{\{0\}}^*(t^{\prime}_{\{0\}}, \mS^u)\, J(t^{\prime}_{\{0\}}, l^u) \,\chi_{\{0\}}(t_{\{0\}}, l^u) 
\mathbb{D} l \, \d t_{\{0\}}^{\prime}
\right\}
\eeq
(This use of small star denotes complex conjugate, not emergent-time derivative). 
Then also using 
\beq
\frac{\d \lt^{\sr\se\scc}_{\{1\}}}{\d \lt^{\sr\se\scc}_{\{0\}}} = \frac{\pa t^{\se\sm(\sW\sK\sB)}_{\{1\}}}{\pa t^{\se\sm(\sW\sK\sB)}_{\{0\}}} = 
\frac{1}
                    {4\{E - V\}}\int \chi_{\{0\}}^*\, J \,\chi_{\{0\}} \mathbb{D} l
\eeq
to render the fourth equation into the form 
\beq
i\hbar\frac{\pa|\chi_{\{1\}}\rangle}{  \pa \lt^{\sr\se\scc}_{\{0\}}  } = -\frac{\hbar^2}{2}\triangle_{\FrS(N, d)}|\chi_{\{1\}}\rangle 
+ \left\{
J^{\sr\se\scc}  - \frac{    \hbar^2\langle\chi_{\{0\}}    |\,J \,|\chi_{\{0\}}  \rangle \triangle_{\FrS(N,d)}   } 
                       {    4 \big\{   E - V\big(h\big(\lt^{\sr\se\scc}_{\{0\}}\big)\big)\big\}                 } 
\right\}
|\chi_{\{0\}}\rangle \mbox{ } . 
\label{Flaot}
\eeq
\noindent The last term with the big bracket is by this stage a known, 
so this is just an inhomogeneous version of the second equation and therefore amenable to the method of Green's functions. 
\beq
|\chi_{\{1\}}\rangle =  \int_{\mbox{\normalsize $t$}^{\tr\te\tc\,\prime} = 0}^{\mbox{\normalsize $t^{\tr\te\tc}$}}   
\int_{\FrS(N, d)}
G\left(\lt^{\sr\se\scc}_{\{0\}}, l^u; \lt^{\sr\se\scc}_{\{0\}}\mbox{}^{\prime}, l^{u\,\prime}\right)
f\left(\lt^{\sr\se\scc\,\prime}, \ml^{u\prime}\right)
\mathbb{D} l^{\prime} \d\lt^{\sr\se\scc\,\prime}
\label{Munch}
\eeq
modulo additional boundary terms/complementary function terms. 
Now, this is a very standard linear operator for this article's simple RPM examples (time-dependent 1-$d$ and 2-$d$ rotors).
However, 

\noindent i) the region in question is less standard (an annulus or spherical shell with the time variable playing the role of  radial thickness).  

\noindent ii) Nor is it clear what prescription to apply at the boundaries. 
On these grounds, I do not for now provide explicit expressions for these Green's functions, though this should be straightforward enough once ii) is accounted for.

\subsection{Semiclassical emergent Machian time: perturbative scheme}

Regime 3) 
We are led to solve for $t = t(h, l^{\sfc})$ -- no independent notion of time and for $|\chi(l, t)\rangle$ (standard QM for the $l$-subsystem with respect to the emergent time).
As these are the functions to solve for, they must be perturbed.  
In classical theory, the $Q$'s are perturbed and this is required here since $t$ is in terms of them.
This differs then from standard QM perturbation theory in which the $Q$'s are not perturbed.  

\noindent All in all, we now take 
\beq
Q^{\sfA} = Q^{\sfA}_{\{0\}} + \underline{\epsilon}\cdot\underline{Q}^{\sfA}_{\{1\}} + O(\epsilon^2) \mbox{ } , \mbox{ for both $h$ and $l$, }
\label{B-1}
\eeq
\beq
t^{\se\sm(\sW\sK\sB)} = t^{\se\sm(\sW\sK\sB)}_{\{0\}} + \underline{\epsilon}\cdot\underline{t}^{\se\sm}_{\{1\}} + O(\epsilon^2) \mbox{   and    }
\label{B-2}
\eeq
\beq
|\chi\rangle = |\chi_{\{0\}}\rangle +  \underline{\epsilon}\cdot |\underline{\chi}_{\{1\}}\rangle + O(\epsilon^2) \mbox{ } .
\label{B-3}
\eeq
I.e. {\sl simultaneous} consideration of Sec \ref{+temJBB}'s perturbations and the current Sec's perturbations in a Mach's Time Principle context 
(both being given the opportunity to contribute to $t^{\se\sm}$ perturbations).  

\noindent Note: for some purposes (a set of relevant $\epsilon_u$) some of the corresponding responses (e.g. $t_{\{1\}u}$, $l_{\{1\}u}$, $q_{\{1\}u}$) would be expected to be negligible.
For instance, we can turn on a small electric field in our laboratory in order to study the Stark effect in atoms without expecting this to in any way 
significantly affect the timestandard.  
Once we are sure this is the case for a particular set-up, it can be modelled by a rather less all-encompassing set of perturbed quantities than is considered above.
The full system would only be expected to be used for quantum-cosmological applications in which an accurate emergent time is required.  
We next need some lemmas about derivatives.

\mbox{ }  

\noindent Lemma 18) $\pa_{h} = \Heart$.  

\noindent This builds on the realization in \cite{SemiclI} that the chronifer chain-rule is not exact; they were not used in \cite{SemiclIII}, by which some equations written down there 
were implicitly only valid to zeroth and first order, though that only matters for future papers since both \cite{SemiclIII} and the present paper at most work to first order.

\mbox{ }

\noindent Lemma 19) 
\beq
\frac{\pa}{\pa l^{\sfc}} = 
\left\{
\delta^{\sfe}_{\sfc} - \underline{\epsilon} \cdot \frac{\pa\underline{l}^{\sfe}_{\{1\}}}{\pa l_{\{0\}}^{\sfc}}
\right\}
\frac{\pa}{\pa l^{\sfe}_{\{0\}}} + O(\epsilon^2) \mbox{ } , 
\eeq
$$
\triangle_{l} = \triangle_{l_{\{0\}}}
+ \underline{\epsilon} \cdot 
\left\{
- \frac{\underline{l}^{\sfp}_{\{1\}}}{2{M}\big(l^{\sfq}_{\{0\}}\big)}
\frac{\pa{M}\big(l^{\sfq}_{\{0\}}\big)}{\pa l^{\sfp}_{\{0\}}}
\triangle_{l_{\{0\}}}
- \frac{\pa\underline{l}^{\sfe}_{\{1\}}}{\pa l^{\sfc}_{\{0\}}}\frac{1}{\sqrt{{M}(l^{\sfq}_{\{0\}})}}
\frac{\pa}{\pa l^{\sfe}_{\{0\}}}
\left\{
\sqrt{{M}\big(l^{\sfq}_{\{0\}}\big)}  {N}^{\sfc\sfd}\big(l^{\sfq}_{\{0\}}\big)
\frac{\pa}{\pa l^{\sfd}_{\{0\}}}
\right\}
+ \frac{1}{\sqrt{{M}\big(l^{\sfq}_{\{0\}}\big)}}\times
\right.
$$
\beq
\left.
\frac{\pa}{\pa l_{\{0\}}^{\sfc}}
\left\{
\sqrt{{M}\big(l^{\sfq}_{\{0\}}\big)}
\left\{
{N}^{\sfc\sfd}\big(l^{\sfq}_{\{0\}}\big)
\left\{
\frac{\underline{l}^{\sfs}_{\{1\}}}{2{M}\big(l^{\sfq}_{\{0\}}\big)}
\frac{\pa{M}\big(l^{\sfq}_{\{0\}}\big)}{\pa l^{\sfs}_{\{0\}}}
\frac{\pa}{\pa l^{\sfd}_{\{0\}}} 
- \frac{\pa\underline{l}^{\sff}_{\{1\}}}{\pa l^{\sfd}_{\{0\}}}
\frac{\pa}{\pa l^{\sff}_{\{0\}}}
\right\}
+ \frac{\pa{N}^{\sfc\sfd}\big(l^{\sfq}_{\{0\}}\big)}{\pa l^{\sfs}_{\{0\}}}
\underline{l}_{\{1\}}^{\sfs}
\frac{\pa}{\pa l^{\sfd}_{\{0\}}}
\right\}
\right\}
\right\}
+ O(\epsilon^2) \mbox{ } .  
\eeq

\subsubsection{epsilonized $h$ and $l$ equations}

Use each term's label to also label the corresponding $\epsilon$, giving 
$$ 
\{\Last\mh\}^2 - 
2i\epsilon_a  \left\{\frac{\hbar}{\epsilon_a} \langle\chi|   \{    \Last  - \epsilon_i\{\Last l \pa_{l}/\epsilon_i\}  \} |\chi\rangle \right\} - 
\left.
\epsilon_b \mbox{\LARGE $\langle$} \chi \mbox{\LARGE $|$}\frac{\hbar^2}{\epsilon_b} 
\mbox{\LARGE $\{$}\frac{\Last  - \epsilon_i\{\Last l\pa_{l}/\epsilon_i\}}{\Last h}\mbox{\LARGE $\}$}^2   \mbox{\LARGE $|$}\chi\mbox{\LARGE $\rangle$} -
\frac{k(N, d)}{h} 
\right\{
\epsilon_c  
\mbox{\LARGE $\{$}\frac{\hbar^2}{\epsilon_c} 
\mbox{\LARGE $\langle$}\chi\mbox{\LARGE $|$}
\frac{\Last  - \epsilon_i\{\Last l \pa_{l}/\epsilon_i\}}{\Last h}\mbox{\LARGE $|$}\chi\mbox{\LARGE $\rangle$}  \mbox{\LARGE $\}$}
$$
\beq
\left.
+ \epsilon_d \mbox{\LARGE $\{$} \frac{i\hbar}{\epsilon_d} {\Last h} \mbox{\LARGE $\}$}
\right\}
+ \epsilon_e
\left\{
\frac{\hbar^2}{\epsilon_e}  \frac{c(N, d)}{h^2}
\right\} 
- \epsilon_f  
\left\{ 
\frac{\hbar^2}{\epsilon_f}  \frac{\langle\chi|\triangle_{l}|\chi\rangle}{h^{2}} 
\right\}
+ 2V_{h}(h) 
+ 2\epsilon_g  \langle\chi|  V^{\prime}              |\chi\rangle 
+ 2\epsilon_h  \langle\chi|  J^{\prime}  |\chi \rangle 
= 2E 
\mbox{ } \mbox{ } ,   
\label{epsi-TimeSet}
\eeq
$$
2i\hbar\{\Last - \epsilon_i\{\Last l \pa_{l}/\epsilon_i\}\}|\chi\rangle 
- \epsilon_q 
\left\{
\frac{2i\hbar}{\epsilon_q} \mP_{\chi} \{\Last - \epsilon_i\{\Last l \pa_{l}/\epsilon_i\}\}|\chi\rangle 
\right\} = 
- \epsilon_r 
\left\{
\frac{\hbar^2}{\epsilon_r}\frac{\Last - \epsilon_i\{\Last l \pa_{l}/\epsilon_i\}}{\Last h}
\right\}^2     |\chi\rangle 
$$
$$
- \epsilon_s 
\left\{
\frac{\hbar^2}{\epsilon_s}   \frac{k(N, d)}{h}  \frac{\Last - \epsilon_i\{\Last l \pa_{l}/\epsilon_i\}}{\Last h}  |\chi\rangle 
\right\} 
- \hbar^2  \frac{\triangle_{l}}{h^2} |\chi\rangle + 2V_{l}|\chi\rangle + 2\epsilon_u J^{\prime}|\chi\rangle + \epsilon_v 
\left\{
\frac{\hbar^2}{\epsilon_v} \mP_{\chi} 
\left\{
\frac{\Last - \epsilon_i\{\Last l \pa_{l}/\epsilon_i\}}{\Last h}
\right\}^2     
|\chi\rangle 
\right\}
$$
\beq
\left. 
+\epsilon_w 
\left\{
\frac{\hbar^2}{\epsilon_w}  \frac{k(N, d)}{h}  \mP_{\chi}  \frac{\Last - \epsilon_i\{\Last l \pa_{l}/\epsilon_i\}}{\Last h}  |\chi\rangle 
\right\}
- \epsilon_x 
\left\{
\frac{\hbar^2}{\epsilon_x}  \mP_{\chi} \frac{\triangle_{l}}{h^2}  |\chi\rangle
\right\} 
- 2\epsilon_y  \mP_{\chi}  V^{\prime}_{l}     |\chi\rangle 
- 2\epsilon_z  \mP_{\chi}  J^{\prime}_{hl}  |\chi\rangle 
\right\}
\mbox{ } . 
\label{epsi-Em-TDSE}
\eeq
\noindent \mbox{ } \mbox{ } Then zeroth order simply returns
\beq
\{\Last_{\{0\}}h_{\{0\}}\}^2 = 2\{E - V_{h}(h_{\{0\}})\} \mbox{ } ,
\eeq
\beq
i\hbar\Last_{\{0\}}|\chi_{\{0\}}\rangle = - \{\hbar^2/2\}\triangle_{l_{\{0\}}} |\chi_{\{0\}}\rangle + V_{l}\big(l^{\sfc}_{\{0\}}\big) |\chi_{\{0\}}\rangle \mbox{ } .  
\eeq

\vspace{10in}

\noindent The first-order equations are then
$$
\underline{\epsilon}\cdot
\mbox{\Huge\{}
i\hbar
\left\{
\Last_{\{0\}}|\underline{\chi}_{\{1\}}\rangle - \Last_{\{0\}}\underline{t}_{\{1\}}| \chi_{\{0\}} \rangle 
\right\}
-\frac{\hbar^2}{2h_{\{0\}}^2}
\left\{
\frac{\underline{h}_{\{1\}}}{h_{\{0\}}}\triangle_{l_{\{0\}}}|\chi_{\{0\}}\rangle - \triangle_{l_{\{0\}}}|\underline{\chi}_{\{1\}}\rangle 
\right\}
$$
$$
- \frac{\underline{l}_{\{1\}}^p}{2M(l_{\{0\}}^q)}  \frac{\pa M(l_{\{0\}}^q)}{\pa l_{\{0\}}^p} \triangle_{l_{\{0\}}}|\chi_{\{0\}}\rangle - 
\frac{\pa\underline{l}_{\{1\}}^e}{\pa l^c_{\{0\}}}  \frac{1}{\sqrt{M\big(l_{\{0\}}^q\big)}}  \frac{\pa}{\pa l^{e}_{\{0\}}}
\left\{ 
\sqrt{M\big(l_{\{0\}}^q\big)}  N^{cd}\big(l_{\{0\}}^q\big)  \frac{\pa | \chi_{\{0\}} \rangle}{\pa l^{d}_{\{0\}}}
\right\}
$$
$$
+ \frac{1}{\sqrt{M\big(l_{\{0\}}^q\big)}}\frac{\pa}{\pa l_{\{0\}}}
\left\{
\sqrt{M\big(l_{\{0\}}^q\big)}
\left\{
N^{cd}\big(l_{\{0\}}^q\big)
\left\{
\frac{\underline{l}_{\{1\}}^s}{2M\big(l_{\{0\}}^q\big)} \frac{\pa M\big(l_{\{0\}}^q\big)}{\pa l_{\{0\}}^s} \frac{\pa |\chi_{\{0\}}\rangle}{\pa l^d_{\{0\}}} - 
\frac{\pa \underline{l}_{\{1\}}^f}{\pa l_{\{0\}}^d} \frac{\pa |\chi_{\{0\}}\rangle}{\pa l_{\{0\}}^f}
\right\}
+ \frac{\pa  N^{cd}\big(l_{\{0\}}^q\big)}{\pa l_{\{0\}}^s}  \underline{l}_{\{1\}}^s \frac{\pa | \chi_{\{0\}} \rangle}{\pa l^d_{\{0\}}   }
\right\}
\right\}
$$
$$
- V_{l}\big(l^{q}_{\{0\}}\big) |\underline{\chi}_{\{1\}}\rangle  - \underline{l}_{\{1\}}^c \frac{\pa V \big(l^q_{\{0\}}\big)}{\pa l^c_{\{0\}}}  |\chi_{\{0\}}\rangle
\mbox{\Huge\}} = 
$$
$$
i\hbar
\left\{
\epsilon_i
\left\{
\frac{\Last_{\{0\}}  l_{\{0\}}^{c} \pa_{l_{\{0\}}^c} |\chi_{\{0\}}\rangle}{\epsilon_i}
\right\}
+ 
\epsilon_q
\left\{
\frac{\mP_{\chi_{\{0\}}}\Last_{\{0\}} |\chi_{\{0\}}\rangle}{\epsilon_q}
\right\}
\right\}
- \epsilon_r
\left\{
\frac{\hbar^2}{2\epsilon_{r}}\big\{  \pa_{h_{\{0\}}}t_{\{0\}}\Last_{\{0\}}  \big\}^2 |\chi_{\{0\}}\rangle
\right\}
- \epsilon_s
\left\{
\frac{\hbar^2}{2\epsilon_{s}}\frac{k(N, d)}{h_{\{0\}}}   \pa_{h_{\{0\}}}t_{\{0\}}\Last_{\{0\}} |\chi_{\{0\}}\rangle
\right\}
$$
$$
+ \epsilon_u J^{\prime}(h_{\{0\}}, l_{\{0\}}^q) |\chi_{\{0\}}\rangle + 
\mP_{\chi_{\{0\}}}  \epsilon_v
\left\{
\frac{\hbar^2}{2\epsilon_{v}}\big\{  \pa_{h_{\{0\}}}t_{\{0\}}\Last_{\{0\}}  \big\}^2 |\chi_{\{0\}}\rangle
\right\}
+ \mP_{\chi_{\{0\}}}\epsilon_w
\left\{
\frac{\hbar^2}{2\epsilon_{w}}\frac{k(N, d)}{h_{\{0\}}}   \pa_{h_{\{0\}}}t_{\{0\}}\Last_{\{0\}} |\chi_{\{0\}}\rangle
\right\}
$$
\beq
- \mP_{\chi_{\{0\}}}\epsilon_{x}
\left\{
\frac{\hbar^2}{2\epsilon_x}\frac{\triangle_{l_{\{0\}}}}{h_{\{0\}}^2}|\chi_{\{0\}}\rangle 
\right\}
- \epsilon_y \mP_{\chi_{\{0\}}} V_l^{\prime}(l_{\{0\}}^q) |\chi_{\{0\}}\rangle   - \epsilon_z \mP_{\chi_{\{0\}}} J^{\prime}(h_{\{0\}}, l^q_{\{0\}}) |\chi_{\{0\}}\rangle \mbox{ } .
\eeq
\noindent Note 1) If we need $k_{h} - 1$ equations for classical $h$, we suggest we use $k_{h} - 1$ of the $\dot{p}_{\sfa\prime} = -\pa H/\pa h^{\sfa\prime}$ Hamilton's 
equations that follow from the $h$-equation as a QM-corrected Hamiltonian.
However, classical equation of motion analogues arising from such a scheme is open to some other prescription as regards inclusion of QM terms, so this step might be revised 
in a more accurate version once we know how to be more careful.  
In any case, this paper restricts attention to a single $h$-degree of freedom, for which this problem does not occur.  
This observation has value as regards robustness of the scheme proposed in the present paper.  
This is also how to do the analogue of force-judging in consideration of a semiclassical system.

\noindent Note 2) Including the $\hbar$ correction from operator-ordering at least conceptually composes with all the other considerations in SSecs 22.9--10
  
\noindent Note 3) It is straightforward to specialize this system of equations to the 3-stop metroland and triangleland cases.

\subsection{Pairing a $t^{\sr\se\scc}$-dependent $h$ equation with the $t^{\sr\se\scc}$-dependent $l$ equation}

Maximally simplifying the $l$-equation changes the timestandard from a shape-and-scale geometrically natural one to a pure-shape geometrically natural one. 
It may then be advantageous to use this single timefunction rather than a mixture of it and emergent WKB time.
The full equation for this is 
$$
\{\Rec \, \mbox{ln} \, h\}^2 - 2i\hbar\langle \chi |  \Club \Rec  | \chi \rangle - \hbar^2\{ \langle \chi | \Spade^2 | \chi \rangle + k(N, d) \langle \chi | \Spade | \chi \rangle \} 
$$
\beq
- i\hbar k(N, d) \Rec \, \mbox{ln} \, h + \hbar^2\{ k(\xi) - \langle \triangle_l \rangle = 
2\{E^{\sr\se\scc} -  V_h^{\sr\se\scc} - \langle \chi | V_l^{\sr\se\scc} | \chi \rangle - \langle \chi | J^{\sr\se\scc} | \chi \rangle \}\mbox{ } .  
\eeq
The simplest truncation of this is 
\beq
\{\Rec \, \mbox{ln} \, h\}^2   = 2\{E^{\sr\se\scc} -  V_h^{\sr\se\scc} \} \mbox{ } , 
\eeq
which rearranges to 
\beq
\lt^{\sr\se\scc} = \int\frac{\d h}{h^2\sqrt{2W_h}} \mbox{ } .  
\eeq
The principal corrections of this are as computible as before.

\noindent E.g., the operator-ordering correction is $- i\hbar k(N, d)/\sqrt{2}h^2$ in the free $\sfS = 0$ case, 
\beq
- \frac{i\hbar k(N, d)}{2\sqrt{2}E}  \left\{ 
\frac{\sqrt{E - Ah^2}}{h^2} + \frac{A}{\sqrt{E}}\mbox{ln}
\left( 
\frac{\sqrt{E}}{h}\{\sqrt{E} + \sqrt{E - Ah^2}\}
\right)
\right\}
\eeq
in the HO $\sfS = 0$ case, and $- i\hbar \sqrt{E^2 - \sfD^2}/\sqrt{2}\sfD^2h$ in the 3-stop metroland free case with nontrivial central term.

\noindent Also,
\beq
2i\hbar  \int  \frac{\d h}{h^2\sqrt{2W_{h}}} \langle \pa_{h} \rangle =  
\int  \frac{\d h}{h^2} \, \frac{\hbar^2\md^2/2 + Ah^4}{E - Ah^2} = - 
\left\{
\frac{\hbar^2\md^2}{2Eh}
\left\{ 
\frac{1}{3h^2} + \frac{A}{E}
\right\}
+ \sqrt{A}{E}
\left\{
\frac{\hbar^2\md^2 A}{2E^2}  + 1
\right\}
\mbox{artanh} \left(\sqrt{\frac{A}{E}}h
\right)
\right\}
 \mbox{ } , 
\label{correction-1b}
\eeq

\beq
\frac{1}{2\sqrt{2}}\int\langle J \rangle \frac{\d h}{h^2W_{h}^{3/2}} = 
\frac{B\langle d| \mbox{cos}^2\varphi |d \rangle}{2\sqrt{2}}  \int \frac{\d h}{W_{h}^{3/2}} = \frac{Bh}{2\sqrt{2}E\sqrt{E - Ah^2}}  \mbox{ } .  
\hspace{1in} 
\eeq 
{\bf Question 82} Does working purely in rectified time improve the clarity of the no-backreaction, small-backreaction and fully perturbative schemes? 


\noindent N.B. once correction terms to which the $l$-physics contributes are included, rectified time is clearly as Machian as emergent WKB time; 
in fact the two are PPSCT-related and so lie within the same theoretical scheme from the Machian perspective.  
This suggests that, whilst emergent WKB time follows on as a quantum-corrected form of emergent JBB time, the mathematics of the 
quantum system dictates passage to the rectified time instead as regards semiclassical quantum-level calculations.
This amounts to studying the $l$-physics on its most natural configuration space -- shape space -- 
rather than the restriction to the shape part of the relationalspace cone over shape space.

\subsection{Dirac scheme for current practical use in GR-as-geometrodynamics}\label{Dir-Semi}

The general case here is, for a shape-nonshape $\fh$--$\fl$ split 
\beq
i\hbar
\left\{ 
\frac{\partional}{\partional \mt} - \frac{\partional\fc^{\sfZ} }{\partional \ft} \widehat{\scL\scI\scN}_{\mu}           
             \right\}|\chi\rangle \propto - \frac{\hbar^2}{2}\triangle^{\scc}_{\mbox{\scriptsize Preshape}}|\chi\rangle + ... \label{Roth}
\eeq
for $\ft$ an emergent, or, possibly, rectified, time. 
This is accompanied by
\beq
\mbox{\Large t}^{\se\sm(\sW\sK\sB)} = 2\stackrel{\mbox{\scriptsize extremum $\fg \in \FrG$}}
                                                {\mbox{\scriptsize of $\stS_{\ts\te\tm\ti}$}}
\left(
\int ||\d_{\sfg}h||/\{-B \pm \sqrt{B^2 - 4C}\} 
\right)
\eeq
where the extremization is unnecessary in shape--scale split RPM's and minisuperspace but involving an as-yet undetailed object $\FS_{\sss\se\sm\si}$. 
This unknown object reduces to the relational action $\FS_{\sr\se\sll}$ in the classical limit but presumably contains quantum corrections.  
Identifying this object is a second objective for the upcoming work on variational methods for Quantum Cosmology. 
$B$ and $C$ are generalizations of the previous specific example of forms for these.
Finally, these equations are further accompanied by $h$- and $l$-${\scL\scI\scN}$ equations,
\beq
\langle \widehat{\scL\scI\scN}_{\sfZ} \rangle = 0 \mbox{ } , \mbox{ } \mbox{ } \{1 - \mP_{}\chi\}\widehat{\scL\scI\scN}_{\sfZ}|\chi\rangle = 0 \mbox{ } .  
\eeq
The form of (\ref{Roth}) specifically for ERPM is then
\beq
i\hbar
\left\{
\frac{\pa}{\pa\lt^{\sr\se\scc}} - \frac{\pa\underline{B}}{\pa\lt^{\sr\se\scc}}\cdot\hat{\underline\scL}
\right\}|\chi\rangle = - \frac{\hbar^2}{2}\triangle^{\scc}_{\sFrP(N, d)}|\chi\rangle + ... = - \frac{\hbar^2}{2}\triangle^{\scc}_{\mathbb{S}^{nd - 1}}|\chi\rangle + ... \mbox{ } , 
\eeq 
and for Geometrodynamics, it is,  following on from (\ref{Tom-Schwi}),
\beq
i\hbar
\left\{
\frac{\delta}{\delta{\cal \mt^{\sr\se\scc}}} - \frac{\delta\underline{\mF}^{\mu}}{\delta \mt^{\sr\se\scc}}\hat{\scM_{\mu}}
\right\}|\chi\rangle = - {\hbar^2}\triangle^{\scc}_{\sC\sR\si\se\sm(\sbSigma)}                                      |\chi\rangle         + ...   =   
                       - {\hbar^2}\triangle^{\scc}_{\sbU}                          |\chi\rangle                                          + ...       \mbox{ } . 
\eeq
%
Note that this features $\bU$ as promised compatibility-wise in Sec \ref{Dist7}.     
Here $\mt^{\sr\se\scc}$ bears a slightly different relation to $\mt^{\se\sm(\sW\sK\sB)}$ due to power freedom in scale variables as per Sec 10.

\subsection{Discussion}\label{22-end}

\subsubsection{The WKB procedure is crucial but unjustified}\label{Cruxi}

It is easy to argue out the applicability of the WKB scheme to Quantum Cosmology to general Physics 
audiences in a way which sounds plausible and familiar and thus not requiring careful scrutiny.
Unfortunately, the familiarity stems from the commonplace occurrence of the WKB approximation in basic ordinary QM, where it rests safely on 
the Copenhagen interpretation and on controlled lab conditions, neither of which are meaningful in Quantum Cosmology \cite{Battelle}.

\noindent 1) In ordinary QM, one often assumes the WKB ansatz as a consequence of the pre-existence of a surrounding 
classical large system \cite{LLQM}, which is no longer applicable for the whole universe.

\noindent 2) In ordinary QM, the WKB ansatz can be justified by the lab set-up being a ``pure incoming wave".
But if one assumes a pure wave in the quantum cosmological context, its wavefronts orthogonally pick 
out a direction which serves as timefunction, so this amounts to `supposing time' rather than a `bona fide emergence of time' as required to resolve the POT. 

\noindent 3) Arguing for constructive interference which underlies classicality \cite{WheelerGRT, Battelle}; however this amounts to imposing, rather than {\sl deducing} 
(semi)classicality, as does using (semi)classicality as a `final condition' restriction on quantum-cosmological solutions.\footnote{LQC  
can also be argued to not address this point either, there aggravated by the number of rejected solutions becoming larger than usual.
This is also akin to how Griffiths and Omn\`{e}s remove by hand the superposition states that they 
term ``grotesque universes" \cite{Gri} due to their behaviour being much unlike that we experience today.}    

\noindent On the other hand, as argued in Quantum Cosmology,  firstly, the WKB ansatz is not general or a priori natural.  
The $S$-function will solve an $h$-equation that is mathematically a HJE (or at least approximated by a such).  
However, HJE's have 2 solutions $S^{\pm}$.
(For the case in which the velocities feature solely homogeneous-quadratically in the kinetic term,  
these are $\pm$ the same thing, but more generally the 2 solutions are $\pm$ in the sense of being a complex conjugate pair, c.f. \ref{QM-expansion})
Thus in general one would expect not exp(i$S/\hbar$) but \cite{Zeh88, BS, B93, I93, Kuchar92, Giu, Zehbook} a superposition, and not just of exp$(\pm S/\hbar)$: 
%
%
\be
\phi(h) = A_{+}\mbox{exp}(i\, S_+(h)/\hbar) + A_{-}\mbox{exp}({i\, S_-}(h)/\hbar)  \mbox{ } .
\label{Bicycle-Repair-Man} 
\ee
Secondly, the reasons for use of the WKB approximation in ordinary QM do not carry over. 

\mbox{ }  

\noindent Not being able to justify the WKB ansatz is a particular problem \cite{Zeh86, Zeh88, BS, Kuchar92, B93, I93, B94II, EOT}.for the Semiclassical Approach since its trick by 
which the chroniferous cross-term becomes the time-derivative part of a TDSE is exclusive to wavefunctions obeying the WKB ansatz.

If the many-worlds interpretation of QM is adopted, each piece of the typical superposition is realized in a different branch.   
However, one may be able to object that this is not in accord with actually experiencing such a superposition, or undermine the many-worlds 
interpretation on grounds of unnecessary richness or, possibly, of conceptual difficulties. 
This is another case of trading off time problems for interpretation of QM problems.

Though some reservations about this have also been expressed (e.g. \cite{Kuchar92, I93, Giu, Kiefer93, HT, H03} are between far from and not entirely optimistic about this). 
This includes getting the decoherence from the machinery of Histories Theory.  

\noindent Decoherence in Quantum Cosmology (see e.g. \cite{Kiefer99, Kieferbook}) has somewhat distinct features from its more well-known QM counterpart (App \ref{2-Decs}).     
Such decoherence has been suggested to justify the WKB ansatz by collapse of the superposition (\ref{Bicycle-Repair-Man}) to a single exp(i$S/\hbar$).

\subsubsection{The many approximations problem.} 

Some of this was already present at the classical (Sec \ref{+temJBB}) but it gets worse upon passing to the quantum level.  
Moreover, at the semiclassical quantum level, this problem means that it is hard to meaningfully isolate 
the testing of whether the WKB condition applies due to the plethora of other approximations made.  
It would seem that at best one can make a number of such and then test whether the WKB approximation 
holds in the small region of the configuration space where all those approximations are applicable.  
Thus it would in general be very drawn-out to carry out the above-suggested programs.    

\mbox{ } 
 
\noindent Some papers \cite{KieS91, SCB3, Kiefersugy} investigate Quantum Cosmology by expanding in 1 parameter.  
That however there are multiple parameters was pointed out by Padmanabhan \cite{Pad}, and is investigated explicitly in the present Article.
While \cite{Pad} proceeded by considering which parameter to expand in, in the present Article I point out rather that 
1-parameter expansions in no matter what parameter will not in general suffice for beyond a corner of the Quantum Cosmology solution space.   
In general one would have to expand in many independent parameters.
Careful theoretical arguments may however then match certain frameworks with less parameters to certain relevant situations to various degrees of accuracy. 
For some consideration as to what regimes are required in GR Cosmology, see \cite{SemiclII}.  

\mbox{ }

\noindent With increasing accuracy, higher-order WKB techniques \cite{KS91} become relevant when the associated small quantity 
$\epsilon_{\sW\sK\sB}$ is insufficiently small for ${\epsilon_{\sW\sK\sB}}^2$ to be entirely negligible. 
Does the chroniferous role and the difficulty in justifying the semiclassical regime pervade all of this?

\subsubsection{Testing approximations by ulterior exact solvability}

A different perspective to postulating an unestablished ansatz is putting that ansatz to the test, by investigation for classes of ulteriorly exactly-soluble models 
(i.e. ones which are soluble by techniques outside the Semiclassical Approach.) 
This application combines Part III with the present Sec's material in a further way.    
An extension of this involves use e.g. the \NSI to provide the relative probabilities of experiencing a WKB regime within a given range of model universes.     
%

\mbox{ } 
 
\noindent However, I find that even addressing the question of where this regime holds for simple toy models is not in practice clear-cut, 
due to there being of the order of  additional, distinct quantum-cosmological approximations required alongside it in order to be able to complete practical calculations.  
In the existing literature, these are mostly tacit or not touched upon by leaving vague the full extent of 
what the words `adiabatic' and `WKB' need to mean to fully cover all the simplifications required.  
Various of the below considerations also act to worsen this proliferation of approximations.  
Thus even explicit investigations in ulteriorly exactly soluble models only concern small pieces of configuration space characterized by all the other approximations being made.  
[Thus RPM's are valuable conceptually and to test whether one should be {\sl qualitatively} confident in the assumptions and approximations made in such schemes.] 


\noindent The basic scheme is that one can do further perturbation terms with standard mathematics to begin to investigate the effects of not neglecting 
each of the commonly neglected terms and Part II has exact comparers.  
Thus the present Article contains a  mathematically-simple laboratory for qualitative investigation of how reliable our understanding of the quantum-cosmological origin of the 
structure in the universe is.  
%

\subsubsection{End-check against Emergent Semiclassical Time Approach's problems}

\noindent WDE-inherited Problems with Semiclassical Approach). Invoking a WDE at the base of the semiclassical approach results in inheriting some of its problems \cite{Kuchar92, I93}.  
Of course, for RPM's this is less severe: there is no inner product problem, no functional derivatives or need of regularization.  
It is furthermore unclear how to relate the probability interpretation of the approximation with that for the underlying WDE itself \cite{I93, Kuchar92}. 
Considering various less approximated equations will help with this; here the qualitative changes due to inclusion of higher-derivative terms are significant.     
I have not looked into this yet, but RPM's are a suitable arena in which to do so.  

\mbox{ } 

\noindent Multiple Choice Problem with Semiclassical Approach).   This remains present \cite{Kuchar92}.  
Moreover, ditching $O(\hbar^2)$ may occasionally enhance equivalence, paralleling how it does so as regards operator-ordering ambiguities in Sec \ref{QM-Intro}.

\mbox{ }  

\noindent Global Problem with Semiclassical Approach 1) The WKB approximation only holds in general in certain regions.
Thus the Semiclassical Approach is not a complete resolution of the Problem of Time, nor even of the `paradox' of the timeless appearance of the quantum theory.  
(Though there may not necessarily be any need to resolve this in regions outside `quotidian experience' 
-- one cannot testify that there is a semblance of dynamics in regions in which semiclassical Quantum Cosmology does not apply).  

\mbox{ }  

\noindent JBB-WKB alignment extends the relevance of the issues raised with JBB time in Sec \ref{+temJBB}.   
WKB time is globally limited by $S$ having zeros (a Problem of Zeros). 
Often one will have oscillatory behaviour on the one side of these and decaying behaviour on the other.  
The WKB procedure is then invalid at each zero and a very poor approximation near each zero.
A distinct approximation is required around each zero (as per the theory of connection functions in the ordinary differential equation case).
Thus a time arising from a WKB procedure cannot be claimed to be generically applicable over configuration space.   
Rather, one should expect a number of patches in configuration space where a different regime applies, for which emergent WKB time is not a valid answer to the POT.  
Additionally, if the zeros are sufficiently near to each other, there is no room for a WKB regime in the region between them, 
so applicability of a WKB procedure is scarce in such a region of configuration space. 
From a physical perspective, (semi)classicality in the sense of WKB does need not occur everywhere or `everywhen' in a mechanical motion.  
I note that the tessellation by the physical interpretation method helps with identifying such regions for RPM's.   

\mbox{ } 

\noindent{\bf Question 83}. What type of equation arises in the connecting regions, and is there any resolution of the POT within these?  
If so, is it patchable to the emergent WKB time resolution in the other regions? 

\mbox{ }  

\noindent{\bf Question 84$^*$}. Does some version of the connection formula procedure for patching together regions remain suitable at the level of prolonging QM evolution?   

\mbox{ }

\noindent The 3-stop metroland example provided in (\ref{3stoppot}) is stable to small angular disturbances about 
$\varphi$ for some cases of HO/

\noindent cosmological constant, but it is unstable to small angular disturbances for the gravity/dust sign of inverse-power potential. 

\noindent Positive-power potentials are finite-minimum wells, but cease to be exactly soluble, other than the HO considered here, making this statement weak...   
For instance we saw that in Newtonian Gravity/dust models (or negative power potentials more generally), 
near the corresponding lines of double collision, the potential has abysses or infinite peaks. 
%
%
Thus the scale-dominates-shape approximation, which represents a standard part of the approximations 
made in setting up the Semiclassical Approach to Quantum Cosmology, fails in the region around these lines.  
It is also then possible that dynamics that is set up to originally run in such a region leaves it, so a 
more detailed stability analysis is required to ascertain whether semiclassicality is stable in such models.   
This issue can be interpreted as a conflict between the procedure used in the Semiclassical Approach and the 
example of trying to approximate a 3-body problem by a 2-body one (see around Fig \ref{Fig--5} for more on this).  
The difference in the analogy between Cosmology and the spherical presentation of triangleland causes the potentials for that which are studied in 
this article (and the wider range of these considered in \cite{08III}) to be purely positive powers for which the preceding problem does not occur.  

\mbox{ } 

\noindent Spacetime Reconstruction Problem with the Semiclassical Approach). 
The status of this is unclear.    
Moreover, RPM's cannot address this issue since these have no spacetime to reconstruct.

\vspace{10in}

\begin{subappendices}
%
\subsection{Geometric phase issues}  

\subsubsection{Berry phase}\label{Dat}

%
\be
{D_{\sB\se\sr\sr\sy(h)}}^{\mu\ip} = {\pa_{h}}^{\mu \ip} + i{\mA_{h}}^{\mu\ip} \mbox{ } , \mbox{ } \mbox{ } 
{D_{\sB\se\sr\sr\sy(h)}^*}^{\mu\ip} = {\pa_{h}}^{\mu \ip} - i{\mA_{h}}^{\mu\ip} 
\label{Bercov}
\ee
denote the Berry covariant derivative and its conjugate. 
Here, ${\mA_{h}}^{\mu\ip}$ is the Berry connection \cite{Berry84, Simon83}, i.e. the vector gauge 
potential induced by $l$-physics on $h$-physics of a nondegenerate quantum state that corresponds to its $U$(1) freedom in phase, 
\be
{\mA_{h}}^{\mu\ip} = -i\left< \chij  \right| {\pa_{h}}^{\mu\ip} \left|\chij \right>  
\mbox{ } .
\ee
$h$-equations in place of the Born-Oppenheimer equation are e.g. Mead--Truhlar's \cite{MT79} or Berry--Simon's \cite{Berry84, Simon83, Berry89} geometrical form)\cite{Berry89}.
One merit of Berry's scheme as opposed to BO's is that it is one means of including back-reaction.  
Moreover, Berry's scheme includes this in a geometrically insightful way which is more precise and indeed more correct in laboratory situations. 
(Effects have been observed in experiments \cite{LH}, the explanation of which has been found to rest on geometric phase \cite{MT79, Berry84, Simon83}.)  

\mbox{ }

\noindent {\bf Question 85}.  Provide a clear geometrical characterization of what happens to the geometrization of phase after the WKB ansatz is applied to it.

\subsubsection{Absolute or relative geometrical phase?}\label{AORGP}

Using $l$-equations with no averaging corrections amounts to removing phase, and this is in general 
inadmissible \cite{Datta97} due to this phase possessing a physical meaning.

This issue is moreover quite subtle, requiring quite some conceptualization and nomenclature to discuss. 
The point of including this Appendix is that this issue is tied both to time issues and to relational 
considerations, and moreover may override the preceding SSSec's no-go as concerns the present Article's concrete modelling.
Let $\gamma_{\mbox{\scriptsize TNS}}$ be the {\sl total phase} in the case in which the presence of an external time causes this to naturally split into
\be
\gamma_{\mbox{\scriptsize TNS}} = \gamma_{\mbox{\scriptsize geometric}} + \gamma_{\mbox{\scriptsize evolutionary}} \mbox{ } .
\ee
In the case in which there is no such external time, however, one has, rather, a total phase 
$\gamma_{\sT}$ for which there is no natural such split; moreover this case can be thought of as entirely due to quantum phase geometry,  
\be
\gamma_{\sT} = \gamma_{\sg\se\so\sm\se\st\sr\si\scc}^{\prime} \mbox{ } . 
\ee
Datta's argument is that the primed and unprimed notions of quantum phase geometry above are geometrically distinct.   
In particular, $\gamma_{\sT}$ is both gauge invariant (as befits its nature as a geometric quantum 
phase) and MRI (as befits a total dynamical phase of a theory with no external time, of which Quantum Cosmology is surely an example.) 
This means overall that the primed, {\it relative quantum phase geometry} has more unphysical 
quantities ('overall gauge freedom') than the unprimed {\it absolute quantum phase geometry}.  
This means that more terms `are gauge rather than physical' if relative quantum phase is in use.  
Thus there are differences in each case as regards what is purportedly physical, so one should take due care to use whichever is appropriate for the situation in hand.  
Non-external time requires relative rather than absolute quantum phase geometry.  
Assuming this argument is correct, this Sec's treatment would need to be upgraded.  

\mbox{ } 

\noindent One consequence of Datta's relative phase is that gauge choice in the original sense also causes the zero-point energy to be shifted so that $J$ is renormalized to 
$J - \langle J \rangle$.  
By this, $\langle J \rangle$ drops out of the $h$-equation.
The physics is then only in the fluctuating quantities, invalidating the discarding of averages in the $l$-equations also.  
If this argument holds, it invalidates Regimes 1b) and 2b).
Note that this would invalidate the particular back-reaction mechanism currently in use (Sec \ref{Det-Back}). 
However we argued that a second term that can serve as a backreaction: $\langle \chi |  \pa_{\rho}  | \chi \rangle$,  
and computed its effect in a basic scheme.  

\mbox{ } 

\noindent{\bf Question 86} Use RPM's to further investigate how this alternative type of back-reaction works, at the level of a small perturbation.  

\mbox{ } 

\noindent A second consequence is that the enlargement of what is gauge circumvents the argument that a 1-$d$ 
configuration space for the $h$-subsystem entails automatic neglect of connection terms. 
Thus the very simplest toy models can be investigated to see if relative phase effects cause major or minor alterations to `conventional wisdom'.  
That includes RPM's due to their implementation of Temporal Relationalism, even in the easiest cases in which nontrivial spatial relationalism is absent.  

\mbox{ } 

\noindent{\bf Question 87}.  Carry out this investigation for simple RPM's.  
What is the relative phase geometry for these?
How are this Sec's equations to be modified in this context, and how does this affect the nature of the emergent Semiclassical Approach's equations and interpretation?

\subsection{Extension to include fermions}

Proceed as per the classical counterpart. 
Now the zeroth-order $l$-TDSE contains only RPM-fermion potential, 
so $\Psi$ can be separated into $\Psi$(bose, $t_{\{0\}}$) and $\Psi$(fermi) factors. 
But then the separated-out fermi part reads (for $C$ the constant of separation) 
\beq
\mbox{\{$V$(fermi)$ - C\}\Psi$(fermi) = 0} \mbox{ } , 
\label{Triv-Obstr}
\eeq
which at most has algebraic-root solutions.  
Thus no fermion-$l$-nontriviality can feed into the first-order system, so there is a breakdown of `giving an opportunity to all species'.  
However, the interest in Nature of linear/fermionic species is {\sl field-theoretic}, and there (\ref{Triv-Obstr}) is a p.d.e. via the potential 
containing spatial derivatives, and thus $\Psi$ does pick up nontrivial fermi-l dependence at zeroth order, and so one continues to reside 
within the GLET is to be abstracted from STLRC interpretation that the present Article exposits.

\subsection{Semiclassical time versus internal time}

This is the sequel of Type 0 `Barbour' versus Type 1 Tempus Ante Quantum as per Fig \ref{Ante-1-and-2}.

\noindent In the internal time approach, the QM interpretation is formally standard; unlike the Semiclassical Approach, there are no towers of approximations.  

\noindent But there are well-definedness and operator-ordering issues in the Internal Time Approach. 

\noindent Also, out of the semiclassical and internal-time TDSE's, only the semiclassical one is interpretable in terms of dissipation into the $\fh$-system. 
This renders the status of $\mt^{\sa\sn\st\se}$-dependent $\mH_{\st\sr\su\se}$ mysterious, 
leaving open the possibility that hidden time is an ill-conceived scheme even at the conceptual level [against Criterion 1)].

\noindent As regards Global POT's, some patching issues have semiclassical regime counterparts in the standard patching together of WKB regions using connection formulae.
However the two situations as they currently exist in the literature are chrono-geometrically distinct. 
(This is partly because the scope of Bojowald et al's examples is for now minisuperspace, with patching in time rather 
than in space, so some parts of looking to tie together these two approaches will have to await the treatment of inhomogeneous-type models.  
On the other hand, the issue of how to define observables at the quantum level for models requiring patches could well still 
have some useful counterpart in the semiclassical WKB patching approach.)

\subsection{Decoherence in QM and in Quantum Cosmology, and differences between them}\label{2-Decs}

Why do we not observe superpositions of macroscopic objects, such as of dead and live cats? 
An explanation is that interaction with the environment swiftly measures such a system, reducing it to an entirely live or dead state. 
On the other hand, QM phenomena indicate that very small things are not swiftly reduced in such a way. 
For example, the ammonia molecule behaves as a superposition of umbrellas and wind-blown umbrellas.  
There is then the question of for what sort of size do such superpositions quickly become reduced.  
The outcome of this is that sugar molecules are observed to stay in one chirality, unlike ammonia. 
Therefore, the boundary for such behaviour lies somewhere between the characteristic size of ammonia molecules and that of sugar molecules (which are much smaller than cats). 
For standard QM decoherence, see e.e. \cite{Giu} and references within.

{            \begin{figure}[ht]
\centering
\includegraphics[width=0.45\textwidth]{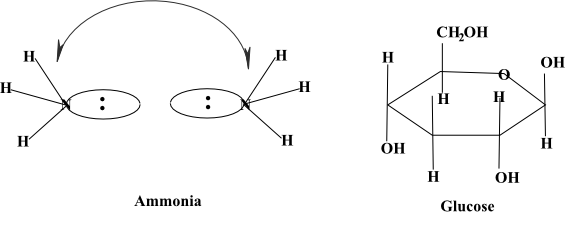}
\caption[Text der im Bilderverzeichnis auftaucht]{        \footnotesize{There is a QM behaviour 
size interface somewhere between the sizes of the depicted molecules.} }
\label{AmmoGluc}\end{figure}            }

\noindent Decoherence in Quantum Cosmology (see e.g. \cite{Kiefer99, Kieferbook}) has somewhat distinct features. 
For instance, there is no pre-existing time in which {\sl to} decohere...

\mbox{ } 

\noindent{\bf Question 88} and Analogy 110) Decoherence time is the (suitable notion of) time taken for off-diagonal components of the density matrix to effectively vanish. 
This is typically extremely short for everyday, macroscale process within the usual QM framework.  
Some physical examples are useful in understanding the nature of this quantity.  
$t_{\sD\se\scc} = \hbar/2\pi|J|^2$ for $J$ an interaction function \cite{CaLo05}. 
$t_{\sD\se\scc}$ for a macroscopic body is within a few orders of magnitude of the Planck time.  
In cases involving a thermal bath, $t_{\sD\se\scc} = \hbar/4\pi K_{\sB}T$.  
Generally, $ML^2/\tau$ is classical action and $\hbar$ is what it is compared to.   
Then $t_{\sD\se\scc}$ goes as $\tau\hbar^2/ML^2KT$, i.e. as $\tau(\lambda_{\sD\sB}/L_{\scc\sh\sa\sr})^2$ where `DB' stands for De Broglie.
These examples are, moreover general enough to apply e.g. to the triangle with various potentials.  
This will go into questions concerning e.g. the base decohereing the position of the apex. 
These are, in turn, as good a detector model as is possible within the triangle.
Investigate. 

\end{subappendices}

\vspace{10in}

\section{Quantum \K beables for RPM's}\label{QM-KO} 

\noindent One can contemplate taking Sec \ref{QM-Intro}'s choice of kinematical quantization as a selection of relational beables.
I.e. a subset of \K beables that has the following five properties.

\noindent 1) They cover all the relational information.  

\noindent 2) They obey suitable global continuity conditions. 

\noindent 3) They close under the QM's commutation relations.  

\noindent 4) They are allowed some classical redundancy (meaning more relational functions than there are independent pieces of relational information -- 
not to be confused with including unphysical/gauge/non-relational information).

\mbox{ } 

\noindent Note: the semiclassical case is more straightforward through just requiring ordering and closure of the chosen subalgebra up to O($\hbar^2$).  

\mbox{ }

\noindent Example 1) Properties 2) to 4) are already evident in using sin$\,\varphi$ and cos$\,\varphi$ rather than $\varphi$ in the quantization of scaled 3-stop metroland. 
$\varphi$ itself was already established to be a degradeable even at the classical level through not being defined for $\theta = 0, \pi$; 
moreover, $\varphi$ fails the globally continuity condition 2), unlike the suitably-periodic sin$\,\varphi$ and cos$\,\varphi$ 
(a degradeables versus beables reinterpretation of the na\"{\i}ve angular variable versus the kinematical quantization objects).  

\mbox{ } 

\noindent 5) Excessive polynomiality is to be avoided if at all possible due to the threat of the Gronewold-van Hove phenomenon.
Cubic combinations are afflicted by this whilst the homogeneous linear polynomials in $\underline{\mn}^i$ are too simple to encode relational angles; the homogeneous quadratic polynomials, 
however, comply with 1) and 5) at once.  

\mbox{ } 

\noindent Example 2)  N-stop metroland is straightforward as regards 5).  
Everything that is picked is sub-cubic: $\hat{\mn}$, $\hat\sfD$ for the pure-shape case and $\hat{\rho}$, $\hat{p}$ and $\hat\scD$ for the scaled one.     
Take these as a particular and non-conflicting set of quantum \K beables spanning each problem's relational quantities. 
(They are degradeables, in fact, since they are not valid everywhere on the shape sphere - no coordinate-based quantity is).  

\noindent Example 3) For Triangleland viewed from the $\underline{\mn}^i$ perspective, the $\sfS^{\Gamma}$ are themselves also homogeneous quadratic as regards 5).  
\noindent Thus Triangleland is also a sub-cubic pick, less obvious a prior but clear enough given the 
machinery of Sec \ref{Q-Geom}: $\hat{\mbox{\scriptsize dra}}$, $\sfS$ for the pure-shape case and $\hat{\mbox{\scriptsize Dra}}$, $\hat{\mbox{\scriptsize P}}\mbox{}^{\tD\tr\ta}$ 
and $\hat\sfS$ for the scaled case.  
One again, there are a particular and non-conflicting set of quantum \K degradeables spanning each problem

\noindent Example 4) For quadrilateralland, as regards 5) the ${\cal U}$'s and ${\cal V}$'s do pose a problem as regards quadraticity, 
but do note that these enter via the {\sl canonical group} which is rigidly prescribed for problems of this nature.   
The Gronewold--van Hove phenomenon has had relatively little concrete exemplification as per Sec \ref{MCP-Ex}, so determining whether it occurs for  
theories based on the $SU(3)$ canonical group is very likely to be new and difficult material.
At the end of the day, we do not here do {\sl a lot} of work with our selected beables, nor are we trying to 'reproduce reality'.  

\vspace{10in}

\section{Type 1 Tempus Nihil Est for RPM's}\label{QM-Nihil} 

We focus on `peaking interpretations'. 
Simple such include 

\noindent a) modes-and-nodes as already used to explain Part III's TISE's, 

\noindent b) The \NSII.  

\noindent There are further peaking interpretations (see Sec \ref{+Peak}), and other interpretations that share the \NSII's use of classical regions to implement propositions 
(Sec \ref{QM-Combo}).

\subsection{Na\"{\i}ve Schr\"{o}dinger Interpretation for RPM's}\label{NSI}

\subsubsection{Pure-shape 4-stop metroland examples of \NSI} 

\noindent Example 1) Consider quantifying Prob(universe is large), in the sense that the two clusters under study are but specks in the firmament, by 
\beq
\mbox{Prob(model universe is $\epsilon$-close to the \{12,34\} DD collision)} \mbox{ } .  
\eeq
This means, at the level of the configurations themselves, that the magnitude of $\sqrt{\mn_1\mbox{}^2 + \mn_2\mbox{}^2}/\mn_3$ lies between 1 and 1 -- $\epsilon^2/2$. 
[I drop (H2) cluster labels in this SSec.]
In configuration space terms, this means that one is in the $\epsilon$-caps about each pole [Fig \ref{Fig-2AFb}.a)].  
Then from the latter and by the \NSI, this quantity is 
\beq
\propto\int_{\epsilon\mbox{\scriptsize --caps in $\FrS(4, 1) = \mathbb{S}^2$}}|\Psi|^2\mathbb{D}\mS = \int_{\phi = 0}^{2\pi}
\left\{\int_{\theta = 0}^{\epsilon} + \int_{\theta = \pi - \epsilon}^{\pi}\right\}|\Psi(\theta, \phi)|^2
\mbox{sin}\,\theta\,\d\theta\,\d\phi \mbox{ } .
\eeq  
So, e.g. for the free/very special HO ground state and first excited state, this is  
$\propto \epsilon^2 + O(\epsilon^4)$  with the D-contribution adding to the negativeness of the correction term.
Additionally, the D = 1, $|$d$|$ = 1 states, this is $\propto\epsilon^4$ + $O(\epsilon^6)$.  
A general interpretational note of use throughout this SSSec and the next is that the minimum amount of powers of $\epsilon$, $\delta$ and $\eta$ simply represent the 
size of the region (2: area, 1: width, 1: width respectively). 
Thus more attention should be place on suppression factors above that and to corrections to that; hence the other details given above.  
Suppression by 2 extra powers is the norm when suppression occurs throughout these examples; it clearly corresponds to the region in question being centred about a node.  

\noindent Example 2) Consider quantifying Prob(the two clusters nominally under study are in fact merged), which is connected to localization and isolated systems issues,
by 
\beq
\mbox{Prob(model universe is $\delta$-close to \{12,34\} merger)} \mbox{ } .
\eeq
This means, at the level of the configurations themselves, that the size of $\mbox{n}_3$  does not 
exceed the small number $\delta$, and, in configuration space terms, that one is in the $\delta$-belt around the equator [Fig \ref{Fig-2AFb}.a)].  
The \NSI then gives this quantity to be
\beq
\propto \int_{\delta\mbox{\scriptsize --belt in } \sFS(4, 1) = \mathbb{S}^2}|\Psi|^2\mathbb{D}\mS 
\mbox{ } ,
\eeq 
which, in the free/very special HO case, works out to be $\propto \delta^3 + O(\delta^5)$ for the D = 1 d = 0 state and to $\delta + O(\delta)^3$ for the other three lowest-lying states.

\noindent Example 3) Consider quantifying Prob(universe is contents-homogeneous) in the sense that the two clusters 
under study in Jacobi H-coordinates are similar to each other, by the magnitude of $\mn_1/\mn_2$ departing from 1 by no more than $2\eta$.  
This is qualitatively relevant to structure formation.   
The region of configuration space that this corresponds to consists of four lunes of width $\eta$ [Fig \ref{Fig-2AFb}b)], so the \NSI gives
\beq
\mbox{Prob(universe is $\eta$-contents-homogeneous)} \propto \int_{\mbox{\scriptsize 4 lunes of width } 
\eta \mbox{ } \mbox{\scriptsize in $\sFS(4, 1) = \mathbb{S}^2$}}
|\Psi|^2\mathbb{D}\mS \mbox{ } , 
\eeq
which, in the very special case, comes out as proportional to $\eta$ for all four of the lowest-lying states.

\noindent Example 4) Example 1's question also makes sense for the small-regime special HO solution. 
One now obtains proportionality to $\epsilon^2 \sqrt{\omega/\hbar}$ to leading order. 
I.e., one has the same `(small)$^2$' factor as in the very special problem but now with an opposing `$\sqrt{\mbox{large}}$' factor. 
This amounts to the small regime's potential well (Fig \ref{AF-Fig3}) concentrating the wavefunction near the 
poles, i.e. in the region of the configuration space that corresponds to large universes in the sense described above.

\noindent Example 5) repeating Example 1 for the wavefunctions with first order perturbative corrections in $\ttB$ included, 
there is now proportionality to $\epsilon^2\{1 - 8\,\ttB\,\mI^2/9\hbar^2  \} + O(\ttB^2) + O(\epsilon^4)$ for the ground state, to 
$\epsilon^2\{1 - 8\,\ttB\,\mI^2/25\hbar^2 \} + O(\ttB^2) + O(\epsilon^4)$ for D = 1, d = 0, and 
$\epsilon^4\{1 - 36\,\mI^2\ttB/25\hbar^2\} + O(\ttB^2) + O(\epsilon^6)$ for the D = 1, d = 1 states.  
The signs of these corrections conform with intuition.  
For, (Fig \ref{AF-Fig3}) $\ttB > 0$ corresponds to placing a potential barrier at the poles and a well around the equator.  
This should indeed decrease the amount of wavefunction there, i.e. making large universes less probable, and vice versa for $\ttB < 0$.

\subsubsection{Pure-shape triangleland examples of the \NSI}\label{Kensei}

\noindent Example 1) Consider quantifying Prob(model universe is uniform), for which triangleland has particularly clean equilateral and almost-equilateral criteria.  
A suitable notion of uniformity (Sec \ref{RPM-for-QC8} is maximization of the democratic invariant $demo(3) = 
dra_2 = 4 \times$ (area per unit moment of inertia), which peaks about the equilateral triangle.  
[I drop (1)-cluster and [1]-basis subscripts in this SSSec.] 
Fig \ref{Fig-2AFb}c) then presents a notion of $\epsilon$-equilaterality which quantifies approximate uniformity. 
This corresponds to the polar cap region indicated in Fig \ref{Fig-2AFb}c).   
Thus, the \NSI gives
\beq
\mbox{Prob(triangular model universe is $\epsilon$-equilateral)} \propto  
\int_{\epsilon\mbox{\scriptsize -caps in $\sFS(3, 2) = \mathbb{S}^2$}}|\Psi|^2 \mathbb{D} \mS = \int_{\Phi = 0}^{2\pi}
\left\{
\int_{\Theta = 0}^{\epsilon} + \int_{\pi - \epsilon}^{\pi}
\right\}
|\Psi(\Theta, \Phi)|^2 \mbox{sin}\,\Theta\,\d \Theta \, \d\Phi \mbox{ } , 
\eeq
which, for the free/very special HO case, is e.g. $\propto$ $\epsilon^2 + O(\epsilon^4)$ for the ground state and the S = 1, s = 0 state, and 
$\propto$ $\epsilon^4$ + $O(\epsilon^6)$ for the S = 1 = $|$s$|$ state.  
This conforms with S = 1, $|$s$|$ = 1 pointing along axes in the plane of collinearity.  

\noindent Example 2) Collinear configurations have zero area per unit moment of inertia, and so represent a nonunique opposite of the preceding notion of uniformity.  
Fig \ref{Fig-2AFb}d) then presents a notion of $\delta$-collinearity.    
%
%
This corresponds to the equatorial belt region indicated in Fig \ref{Fig-2AFb}d).
Thus, the \NSI gives
\beq
\mbox{Prob(triangular model universe is $\delta$-collinear)} 
\propto \int_{\delta\mbox{\scriptsize -belt in $\sFS(3, 2) = \mathbb{S}^2$}}|\Psi|^2 \mathbb{D}\mathbb{S}^2 \mbox{ } , 
\eeq
which, for the free/the very special HO case is e.g. $\propto$ $\delta + O(\delta^3)$ for the ground 
state and the S = 1 = $|$s$|$ state, and $\propto \delta^3$ for the S = 1, s = 0 state.  
This conforms with S = 1, s = 0 pointing along the E$\bar{\mE}$ axis with a node in the plane of collinearity.

\noindent Example 2)   continues to make sense for the small regime of the special problem.  
The result here for the lowest 4 states is, using the (1)-basis and a $\eta$-collinear bilune, 
\beq
\mbox{Prob(triangular model is $\eta$-collinear)} \propto \eta\sqrt{\omega/\hbar} \mbox{ } .  
\eeq
Thus this case has a sizeable concentrating factor $\sqrt{\omega/\hbar}$ as compared to the very special 
case, i.e. the potential is trapping more of the wavefunction near the collinearity plane. 
[The lack of a third-order correction is down to the small difference in area between the bilune used here and the belt used above.]

\noindent Example 3) Consider Prob(model universe is isosceles).  
It is useful to first consider Prob(model universe is sharp isosceles).    
A zone quantifying this with respect to the (1) clustering is the $\eta$-lune centred on the Greenwich meridian, so   
\beq
\mbox{Prob(triangular model universe is $\eta$-(sharp isosceles) with regard to the (1)-clustering)} 
\propto \int_{\eta\mbox{\scriptsize -lune in $\sFS(3, 2) = \mathbb{S}^2$}}|\Psi|^2 \mathbb{D} \mathbb{S}^2 \mbox{ } . 
\eeq 
For the free/very special HO solution, this gives $\propto$ $\eta$ for the ground state and the S = 1, s = 0 state, 
$\propto$ $\eta + O(\eta^3)$ for the S = 1, $|$s$|$ = 1 cosine solution and $\propto \eta^3 + O(\eta^5)$ for the S = 1, $|$s$|$ = 1 sine solution. 
(This last solution is particularly lacking in intersection with that lune).  
For the (2)- and (3)-clusterings, the relevant lunes are centred around meridians at $\pm \pi/3$ to the principal one.  
Then all four of the above states give $\propto \eta + O(\eta^3)$.
As regards Prob(triangular model universe is $\eta$-(sharp isosceles)) i.e. Prob(triangular model universe is $\eta$-(sharp isosceles) 
for {\sl any} clustering), one should sum over the three clusterings (or equivalently integrate over the obvious trilune), 
which, for these examples, retains the form $\propto \eta + O(\eta^3)$.
This exemplifies the less neat but more widely available notion of `democracy by summing over all clusterings', which is another use 
of group invariance by summing over the actions  of all of the relevant group's elements.
%
%
Prob(universe is contents homogeneous) in the preceding SSSec also attains democracy in this way.

Next, for each of the above $\eta$-sharp results there is a corresponding $\eta$-flat result of the same form, 
obtained by using the antipodal counterparts of each meridian. 
Then sum over both of these to obtain Prob(triangular model universe is $\eta$-isosceles) can now be obtained by summing 
the $\eta$-sharp and $\eta$-flat results.  
[Here, the meaning in space of $\eta$-isosceles is that the magnitude of $aniso$ weighted by $1/\sqrt{1 - \mbox{4 $\times$ $area$}^2}$ 
does not exceed the small number $\eta$.]

Finally, note that investigating Prob(model universe is regular) involves a hexalune at $\pi/6$ to the above.  
This is a more interesting case since it quantifies `size' relative to contents and also whether there are isolated model island universes.

\subsubsection{Scaled RPM examples of the \NSI}

The preceding calculations hold again in the scaled case by the scale--shape split. 
One can now also investigate questions about the scale of the model universe. 
E.g. what is Prob(moment of inertia of the model universe lies between 0 and $\mI$): a direct quantifier of size.  
I consider this question for the ground states of two scale-shape separable RPM models.

\noindent Example 1) for scaled 3-stop metroland model, this is 
\beq
\propto \int_{\rho^{\prime} \leq \rho = \sqrt{\sI}}|\Psi(\rho, \varphi)|^2\rho\,\d\rho\,\d\varphi 
\propto \int_{\rho^{\prime} = 0}^{\rho}\mG^2(\rho^{\prime}) \rho{\prime} \d \rho^{\prime}
\eeq
which is, for the special multi-HO, by (\ref{Lans}), $\propto 1 - \mbox{exp}(-2\mI/\mI_{\sH\sO})\{1 + 2\mI/\mI_{\sH\sO}\}$. 
Thus it tends to 0 for $\mI << \mI_{\sH\sO}$ (the characteristic value of the moment of inertia for 
this particular problem) and to 1 for $\mI >> \mI_{\sH\sO}$.

\noindent 
Example 2) for scaled triangleland, this is 
\beq 
\propto \int_{I^{\prime} \leq I}
|\Psi(\mI^{\prime},\Theta, \Phi)|^2 \mI^{\prime}\mbox{sin}\,\Theta\, \d \mI^{\prime} \,\d\Theta\,\d\Phi \propto 
\int_{\sI^{\prime} = 0}^{\sI}\mG^2(\mI^{\prime}) \mI^{\prime} \d \mI^{\prime}
\eeq 
which is, for the special multi-HO, by (\ref{Dale}) $\propto 1 - \mbox{exp}(-2\mI/\mI_0)\{1 - \mI/\mI_0\}$.
The small and large $\mI$ analysis for this parallels that of Example 1.

\subsubsection{Extensions to bigger RPM models}

While \cite{QSub} covers quadrilateralland at the level of characterizing regions, it awaits \cite{QuadIII} 
for wavefunction inputs so as to be able to finish these \NSI calculations off.

\subsection{Quantum Records Theory}

\subsubsection{Overview}

Secs \ref{Cl-Str} and \ref{QM-Str} already have material of value for this: notions of distance on the classical configuration space over which QM unfolds, 
quantum notions of information and of correlation.  

\noindent An at least crude QM pdf peaking analysis can be built upon Part III's work.  
Part III's wavefunctions can also be used to assemble density matrices which can then be substituted into Sec \ref{QM-Str}'s notions of information.  
\noindent Perturbative wavefunctions are good enough for this application, and can be fed into quantum SM too \cite{LLSM}, adding to the value of Part III's perturbative workings.

\subsubsection{More on QM Peaking Interpretation}\label{+Peak}

\noindent Barbour's statement about mist concentration is to be identified as a particular conjecture about pdf peaking. 

\noindent Moreover, no basis for this conjecture is found among the current Article's simple solved concrete models.  
Relationalspace geometry does not drastically affect the distribution of $\Psi$ to substantially peak about configurations of the required sort.

\noindent Quantitative detail of how to assess mist concentration from a statistical perspective remains unclear, 
but the clumping--Kendall use of shape statistics at the classical level may be of some value as guidance.  

\mbox{ } 

\noindent Addressing Barbour's Conjectures in full requires the configuration space to be conformally non-flat. 
[In conformally-flat spaces, configuration space geometry can be entirely sent to a redefined potential factor by a PPSCT.  
The potential factor, at least locally away from its zeros, can be sent to a PPSCT-redefined configuration space geometry.  
But in non-conformally-flat spaces, the configuration space geometry has an irreducible part that is {\sl not} re-encodeable as potential.
Moreover, the present Article makes it clear that the simplest such relational dynamics is quadrilateralland, which itself lies beyond the scope of the specific examples presented here.] 
As regards RPM's producing elliptic TISE's, these are a class of problem known to be {\sl capable} of producing some sort of pattern which reflects the underlying shape 
\cite{Spots, Than}. 

\mbox{ } 

\noindent Barbour-Records Modelling 4) Specifically investigating whether maximal collisions play a role requires scaled models.  
The atom-like model of triangleland has this effectively excised by the potential being singular there.  
The wavefunction is zero there, though a number of wavefunctions are {\sl centred about} there.

\noindent Barbour-Records Modelling 5) Specifically investigating whether notions of uniform state play a role is looked into in this Article. 
For the moment, I note that my simple pure-shape RPM model would not seem to exhibit very heavy peaking around its equilateral configuration.

\subsubsection{Arguments against further parts of Barbour's conjectures}

\noindent Barbour Records 3) (Bubble chamber paradigm) has doubt cast upon it if If Joos--Zeh's dust grain is a more generic paradigm; 
can the purported selection principle sufficiently compensate for such an offset?  

\noindent Consider also the situation in which information in a curve or in a wave pulse that is detectable by/storeable in a detector in terms of approximands or modes.  
Detectors could happen to be tuned to pick up the harmonics that are principal contributors in the signal.  
Much like a bubble chamber is attuned to seeing tracks, a detector will often only detect certain (expected) frequencies.  
In this way one can obtain a good approximation to a curve from relatively little information.  
E.g. compare the square wave with the almost-square wave that is comprised of the first 10 harmonics of the square wave.  
That is clearly specific information as opposed to information storage capacity in general.  
This overall middles between the Joos--Zeh and Mott--Bell--Barbour paradigms, suggesting that 
Joos--Zeh is not too dominant for some natural Physics to fall under the Mott--Bell--Barbour paradigm.  

\noindent Concerning Barbour-Records Doubt 3), Conjecture Barbour 2) needs qualifying, in the sense that obviously the form of the potential and total energy 
also contribute (and are in fact far more well-known for having this effect at the elementary level).  
Barbour agrees with this \cite{Barbourcom}.
The global problems with the $\fh$-$\fl$ split \ref{globhl} in Secs \ref{+temJBB} and \ref{Semicl} \cite{SemiclI} sharpens one part of Barbour's 
assertion of the importance of the configuration space geometry \cite{EOT}.

\noindent Concerning Barbour-Records Doubt 4), in modelling small atoms and cosmic strings \cite{KS91}, the effect on the dynamics of 
representation-invariant features such as stratifications, while non-negligible, only impart a small distortion on the wavefunction.
[In the case of small atoms, these approximations are of reasonable success in comparison with observation.]

\noindent Barbour-Records Doubt 5). Nobody has been able to supply any evidence for Barbour 3).  
Butterfield already commented on the lack of evidence for time capsules being probable in \cite{ButterBar}; Healey has 
commented more widely on the lack of clarity and consistency of Barbour's writings \cite{Healey}.

\mbox{ }

\noindent Finally, the current Article's RPM's do not suffice to investigate a number of features of Barbour's conjectures.  
Mirror image-identified RPM's and 3-$d$ RPM's suffice to cover some of the latter.

\subsubsection{Semiclassical Approach or Histories Theory in place of semblance}

\noindent In considering semblance of dynamics in purely timeless and histories approaches, one is mainly down to catchy adages, c.f. the bubble chamber versus the 
dust grain `life in an energy eigenstate' and `somewhere in the universe where information is stored when histories decohere'.
I finally suggest one more: 
\beq
\mbox{\it correlation does not imply causation}.  
\eeq
This comes from basic statistics/general science; the suggestion then is that this  may continue to apply when correlation 
and causation pick up specific Theoretical Physics meanings.  
This gives a hunch that the more minimalist approaches documented in this Article may not be enough.

\subsubsection{Quantum timeless correlations for RPM's}

Investigating the \CPI with RPM's is also possible.  
Given explicit wavefunctions such as this Article's, one can build up projectors and mixed states as required for the \CPI (including with environment portions traced out). 
Then one can construct conditional probabilities \cite{PW83} for pairs of universe properties. 
In the standard Conditional Probabilities Interpretation, one of the correlated subsystems is now a clock for the other.
Thus one could here investigate the goodness of various possible clock variables for such as $N$-stop metroland or triangleland.  
Some suggestions for clocks and rulers are as follows.  
Scaled triangleland could be viewed as one ruler, one clock variable and one piece of Physics.  
E.g. $\Phi$ could be a clock-hand angle, the base could play the role of ruler, leaving the physical subsystem under study to be the ellipticity of the triangle.
However, none of the relative angle, the base or the overall scale come with any widespread guarantees of monotonicity. 
See the next SSec for a number of further developments in timeless strategy classification and new conceptual combinations.   

\mbox{ } 

\noindent {\bf Question 89}/Analogy 111) The Gambini--Porto--Pullin version of \CPI can also be investigated using RPM's, 
with the usual value of being a closed-universe model with parallels to structure formation.  
As things stand, I view this as the more promising branch of the Conditional Probabilities Interpretation, 
making this one of the more likely questions I will work on in the future.  
%

\mbox{ } 

\noindent End-Note 1) Barbour and Page are more general and minimalistic.
The semiclassical way has less generality, and Gell-Mann--Hartle--Halliwell entirely rests on Histories theory (and the Halliwell part of it on semiclassicality as well).  

\noindent End-Note 2) the current Article's RPM's do not suffice to investigate Page's approach.

\begin{subappendices}
%
\subsection{Simple RPM example of subsystem wavefunctions}\label{App-A}

For the scaled triangleland model with $\Phi$-independent, and concentrating on observation around the base, the energy equation and $\sfJ$ conservation equation are all, 
giving a quadrature for the orbit shape and another for the approximate emergent timefunction in parallel to whichever 2 degree of freedom working in Sec \ref{Cl-Soln}.  
The free quantum problem is then solved by 
\beq
\Psi^{\sb\sa\sss\se}_{\sm} \propto \mbox{sin}(\nm\Phi)\mJ(\rho_{\sb\sa\sss\se}/\rho^0_{\sb\sa\sss\se})
\eeq
for m the usual kind of angular momentum quantum number.  Thus these form a SubHilb, and the ones concentrating on observation about the median do also.  
The HO problem with an HO along the base is a more convenient example by boundedness; its subsystem-wavefunctions are given by
\beq
\Psi^{\sb\sa\sss\se}_{\sm\,\sN_{\tb\ta\ts\te}} \propto 
\mbox{sin}(\nm\Phi)\mL^{|\sm|}_{\sN_{\tb\ta\ts\te}}(\mI_{\sb\sa\sss\se}/\mN_{\sb\sa\sss\se}\mI^0_{\sb\sa\sss\se})\mbox{exp}(-\mI_{\sb\sa\sss\se}/2\mN_{\sb\sa\sss\se}\mI^0_{\sb\sa\sss\se})
\mI_{\sb\sa\sss\se}^{|\sm|/2} \mbox{ } . 
\eeq
In each of these cases, these lie within a whole-universe Hilbert space as per Sec \ref{QM-Str} (which does exhibit less of a spectrum by 
closed-universe effects, and does necessitate some nonstandard interpretations, but would still nevertheless appear to be characterizable as a Hilbert space).
\end{subappendices}

\vspace{11in}

\section{Quantum Path Approaches and Histories Theory for RPM's} \label{QM-Hist}

I work in the scaled case, to aid the unification with semiclassical ideas and to better parallel GR Quantum Cosmology.  

\mbox{ } 

\noindent Analogy 112) Decoherence functional computation can be attempted in both GR and RPM's.

\subsection{Path Integral Approaches}\label{PI-App}

\subsubsection{$\FrG$-trivial case} 

Brown and York's work \cite{BY1, BY2} amounts to this for MRI mechanics, investigating whether the sort of transition amplitude 
approach that Wheeler envisaged as arising from resolving thin sandwich schemes is fruitful.  

\mbox{ } 

\noindent 
The 1-$d$ $N$-particle case of this can immediately be reinterpreted in $N$-stop metroland terms, within which one can don $t^{\se\sm(\sW\sK\sB)}$ as an emergent privileged label.
Classically, the configurations are $N$ particles on a line, as described by $n$ relative separations 
$\rho^i$ and the histories are then sequences of these at times $t_1^{\se\sm(\sW\sK\sB)}$, $t_2^{\se\sm(\sW\sK\sB)}$, ... $t_p^{\se\sm(\sW\sK\sB)}$ or a continuum equivalent.  
Quantum-mechanically, dynamical paths become pdf's such as wavepackets.  
The path integral in this case is (\ref{PI-3-prelim}).
We can then use I = $t^{\se\sm(\sJ\sB\sB)}$ in the Machianized version of this scheme.  

\noindent See \cite{Gryb1} for treatment of such path integrals in more general reparametrization-invariant gauges.  

The decoherence functional (\ref{Har1}, \ref{Har2b}) is then as follows.
$\psi^{\si\sn} = \psi^{\si\sn}(\brho^{\si\sn})$ and $\psi^{\sf\si\sn} = \psi^{\sf\si\sn}(\brho^{\sf\si\sn})$.  
The $\mathbb{D}\brho$ that plays the role of $\mathbb{D} \bfQ$ is 
trivial as the corresponding configuration space is flat with the $\rho^i$ playing the role of Cartesian coordinates (Sec \ref{SQNS}).
The associated $\mathbb{D} \mp$ that plays the role of $\mathbb{D} \fP$ is likewise trivial. 
Use action (\ref{Aac-3}), in particular in its second, MRI form.
The Fadde'ev--Popov and gauge-condition-imposing factors are trivial for never-constrained theories.
Thus, 
\beq
\langle \rho^{\sf\si\sn} || C_{\gamma}|| \rho^{\si\sn}\rangle = \int_{\eta} \mathbb{D} \brho\, \mathbb{D} \mbox{\boldmath$p$} \, \mathbb{D} \d{I}\,
\mbox{exp}(i \FS[\brho, \mbox{\boldmath$p$}, \d{I}]) 
\label{Melia} \mbox{ } ,  
\eeq
and then ${\cal D}$ec[$\gamma$, $\gamma^{\prime}$] is built out of this from (\ref{Har1}, \ref{Norm1}, \ref{Nin}) under $\fQ^{\sfA} \longrightarrow \brho$.

\subsubsection{r-presentation of $\FrG$-nontrivial case} 

These are some specific curved-geometry generalizations of the previous.

\noindent I use the triangleland example, though admittedly this example is atypically castable in terms of flat geometry. 
Quadrilateralland is in that sense a better example, but must await \cite{QuadIV}.  
Classically, the configurations are triangles, as described by e.g. three Dragt-type coordinates $\mbox{\boldmath$Dra$}$ 
and the histories are then sequences of these at (e.g. emergent WKB) times $t_1$, $t_2$, ... $t_p$ or a continuum equivalent.  
The path integral here is
\beq
\langle \mbox{\boldmath$Dra$}^{\sf\si\sn} || C_{\gamma} || \mbox{\boldmath$Dra$}^{\si\sn} \rangle = 
\int_{\eta} \mathbb{D} \mbox{\boldmath$P$}^{Dra} \, \mathbb{D} \mbox{\boldmath$Dra$} \, \mathbb{D}\d{I}  \mbox{exp}(i\FS[\mbox{\boldmath$Dra$}, \mbox{\boldmath$P$}^{Dra}, \d{I}]) \mbox{ } .
\label{PI-2}
\eeq
We can then use I = $t^{\se\sm(\sJ\sB\sB)}$ in the Machianized version of this scheme.  

\noindent The decoherence functional (\ref{Har1}, \ref{Har2b}) is then as follows.
$\psi^{\si\sn}    = \psi^{\si\sn}   (\mbox{\boldmath $Dra$}_{\si\sn})$ and 
$\psi^{\sf\si\sn} = \psi^{\sf\si\sn}(\mbox{\boldmath $Dra$}^{\Gamma}_{\sf\si\sn})$.  
The $\mathbb{D} \mbox{\boldmath$Dra$}$ that plays the role of $\mathbb{D} \fQ$ in these equations is trivial as the corresponding 
configuration space is flat and the $\mbox{\boldmath$Dra$}$ play the role of Cartesian coordinates for it (Sec \ref{CPST}).
The associated $\mathbb{D} \mbox{\boldmath$P$}^{\mbox{\scriptsize$Dra$}}$ that plays the role of $\mathbb{D} \fP$ is likewise trivial. 
Use action (\ref{Aac-6}), in particular in its second, MRI form.
The Fadde'ev--Popov and gauge-condition-imposing factors are trivial for the r-formulation of this model.
Thus, 
\beq
\langle \mbox{\boldmath$Dra$}_{\sf\si\sn} || C_{\eta}|| \mbox{\boldmath$Dra$}_{\si\sn} \rangle = 
\int_{\eta} \mathbb{D}\mbox{\boldmath$Dra$}\,\mathbb{D}\mbox{\boldmath$P$}^{\mbox{$Dra$}}\,\mathbb{D}\d{I} \,    
              \mbox{exp}(i \FS[\mbox{\boldmath$Dra$}, \mbox{\boldmath$P$}^{Dra}, \d{I}]) \mbox{ } ,
\label{Melian}
\eeq
and then ${\cal D}$ec[$\eta$, $\eta^{\prime}$] is built out of this from  (\ref{Har1}, \ref{Norm1}, \ref{Nin}) 
under $\bfQ \longrightarrow \mbox{\boldmath$Dra$}$.

\subsubsection{Dirac presentation of $\FrG$-nontrivial case} 

\noindent As an example, consider the Dirac case of the path formulation for triangleland.  
This more clearly parallels full GR/

\noindent midisuperspace (in involving rotations and associated linear constraints 
in parallel to GR involving 3-diffeomorphisms and associated linear constraints), for all that then making progress with it is more formal; see the subsequent example for that).  
%

The configurations are now triangles redundantly described by the four components of the pair of relative 
Jacobi vectors, and then the classical histories are built up from sequences of these.
The $\scL = 0$ constraint remains to be imposed.  
The path integral in this case is 
\beq
\langle \brho^{\sf\si\sn} || C_{\gamma} || \brho^{\si\sn} \rangle = 
\int_{\eta} \mathbb{D} \mbox{\boldmath$p$} \, \mathbb{D} \brho \, \mathbb{D}\d{I} \, \mathbb{D} \d{B} \, \fD_{\cal F}[\brho, \d{I}, \d{B}] 
              \updelta[{\cal F}[\brho, \d{I}, \d{B}^{\sfZ}]] \mbox{exp}(i\FS[\brho, \mbox{\boldmath$p$}, \d I, \d{\B}]) \mbox{ } .
\label{Har2b}
\eeq
The decoherence functional (\ref{Har1}) can then be computed as follows.
$\psi^{\si\sn}    = \psi^{\si\sn}({\brho}\mbox{}_{\si\sn})$ and 
$\psi^{\sf\si\sn} = \psi^{\sf\si\sn}({\brho}\mbox{}_{\sf\si\sn})$.  
The $\mathbb{D} \brho$ that plays the role of $\mathbb{D} \fQ$ in equations (\ref{Har1}) 
is trivial as the corresponding configuration space is flat and the ${\brho}$ play the role of Cartesian coordinates for it (Sec \ref{Rel-Jac}).
The associated $\mathbb{D} \mbox{\boldmath$p$}$ that plays the role of $\mathbb{D} \fP$ is likewise trivial. 
Use action (\ref{Aac-5}), in particular in its second, MRI form.
The DA-Fadde'ev--Popov and gauge-condition-imposing factors are then in general nontrivial. 
Explicit evaluation of these requires making a gauge-fixing choice for scaled triangleland.
I consider the base = x gauge's ${\cal F}_{\sM}:= \theta_1 = 0$ (see App \ref{Guich}). 
Then $\{{\cal F}_{\sM}, \scL\} = 1$, so the Fadde'ev-Popov determinant does not contribute any nontriviality to the integration in this gauge.
App \ref{Guich}  also makes it clear that this is not a globally valid gauge.  
Thus one should also consider the median = x gauge's ${\cal F}_{\sD}:= \theta_2 = 0$, which gives the same paths bracket 
as above, and split the region of integration into two charts within each of which one of these two gauges is entirely valid.  
As regards the delta function, ${\cal F}$ can be written in Cartesian coordinates as arctan$(\rho_y^1/\rho_x^1) = 0$, i.e. $\rho_y^1 = 0$ 
provided that $\rho^1_x \neq 0$, which is guaranteed by the choice of region of integration in which this gauge-fixing is applied.  
Thus, using $\theta_{\sM} := \theta_{1}$ and $\theta_{\sD} := \theta_{2}$ and similarly for the other 1- and 2-coordinates, 
and denoting M or D collectively by the index F, for $\eta$ lying entirely within one F, 
\beq
\langle {\brho}_1 || C_{\gamma}|| {\brho}_2 \rangle = 
\int_{\gamma \mbox{ } \mbox{\scriptsize represented in F-chart}} \mathbb{D}{\brho} \, \mathbb{D} {\bp} \,
 \mathbb{D} \d{I}\, \mathbb{D}\d{B}\,\updelta[\theta_{\suF}({\brho}^{\suF})] \mbox{exp}(i\FS[{\brho}, {\bp}, \d{I}, \d{B}]) \label{Meelia} \mbox{ } .  
\eeq
Then ${\cal D}$ec[$\gamma$, $\gamma^{\prime}$] is built out of this via  (\ref{Har1}, \ref{Norm1}, \ref{Nin}) 
using $\fQ^{\sfA} \longrightarrow \underline{\rho}^i$ and $\mc^{\sfZ} \longrightarrow B$.
I certainly in general see exclusion of histories that require multiple charts for their description as a problem here, 
unless one can somehow patch multiple charts (`Gribov regions') together.  
I leave that as an open question for now.    

\mbox{ }  

\noindent In parallel, for full GR, the configurations are now 3-geometries redundantly described by the six components 
of the 3-metric $h_{\mu\nu}$, and then the classical histories are built up from sequences of these.
The $\scM_{\mu} = 0$ constraint remains to be imposed.  

The decoherence functional (\ref{Har1}) is then further computed as follows.
$\psi^{\si\sn}    = \psi^{\si\sn}(\bh^{\si\sn})$ and 
$\psi^{\sf\si\sn} = \psi^{\sf\si\sn}(\bh^{\sf\si\sn})$.
Use action (\ref{Aac-2}), in particular in its second form.
The $\mathbb{D} \mh$ that plays the role of $\mathbb{D} \fQ$ in equation (\ref{Har1}) and the associated 
$\mathbb{D} \uupi$ that plays the role of $\mathbb{D} \fP$ need to be left formal here, and I leave the PA-Fadde'ev--Popov and gauge-condition-imposing factors general too.  
Thus, 
\beq
\langle \bh^{\sf\si\sn}, \varsigma^{\sf\si\sn} || C_{\gamma} || \bh^{\si\sn}, \varsigma^{\si\sn} \rangle = 
\int_{\gamma} \mathbb{D}\bh \, \mathbb{D}\varsigma \, \mathbb{D}\buppi \, \mathbb{D}\uppi_{\varsigma} \, \mathbb{D}\d{\mI} \, \mathbb{D}\d{\mF} \, 
\fD_{{\cal F}}[\bh, \buppi, \varsigma, \uppi_{\varsigma}, \d{\mI}, \d{\uF}] 
\updelta[{\cal F}^{\gamma}[\bh, \varsigma, \d{\mF}\mbox{}^{\Gamma}]]\mbox{exp}\{i\FS[\uupi, \buppi_{\varsigma}, \bh, \varsigma, \d{\mF}^{\Gamma}, \d{\mI}]\}  \mbox{ } .
\eeq
Then ${\cal D}$ec[$\eta$, $\eta^{\prime}$] is built out of this via  (\ref{Har1}, \ref{Norm1}, \ref{Nin}) using $\fQ^{\sfA} \longrightarrow \bh$. 
One would in general expect a Gribov-type problem here too.

\subsection{Quantum Paths Brackets Theory for RPM's}\label{PaB-RPM}

This has the following quantum brackets, followed by the same material as in the previous SSec; they are viewed as the kinematical quantization
I present the continuous-time Machian-time version due to the direction of further development in this Article.

\subsubsection{Scaled $4$-stop metroland}

\noindent $\FrG_{\scc\sa\sn}$ is here an Eucl(3) isometry group applying at each time-value, of histories of physically relative dilational momentum $SO(3)$ objects 
$\scD(t^{\se\sm(\sW\sK\sB)})$ and $\mathbb{R}^3$ objects $p_i(t^{\se\sm(\sW\sK\sB)})$, and the corresponding linear space $\mathfrakV^*$ is an $\mathbb{R}^3$ 
at each time-value, of histories of non-angular ratio objects $\rho^i(t^{\se\sm(\sW\sK\sB)})$.  
I assume that these continue to be combineable as a semi-direct product with no problems in considering it to be the standard 
$\FrG_{\scc\sa\sn} \mbox{\textcircled{S}} \mathfrakV$ at each time-value, forming an associated 1-$d$ QFT.

\mbox{ } 

\noindent There is scope for a global POT subfacet appearing here via the semiclassical regime breakdown causing the 
realised $t^{\se\sm(\sW\sK\sB)}$'s to belong to a half-line or an interval; this could well alter the nature of the QFT, particularly as such a region is approached.  
It is not clear to me yet what effect it would have, or whether it could be countered by patching.   

\mbox{ } 
 
\noindent The scaled $4$-stop metroland classical paths brackets (\ref{N-stop-Cl-Hist-Alg}) are then promoteable to (ignoring the above possible global problem)
the quantum commutator algebra whose nontrivial part is 
\beq
\big[\widehat{\rho}^i\big(t^{\se\sm(\sW\sK\sB)}_1\big), \widehat{p}_j\big(t^{\se\sm(\sW\sK\sB)}_2\big)\big] = 
i\hbar\,\delta^i_j\updelta\big(t^{\se\sm(\sW\sK\sB)}_1 - t^{\se\sm(\sW\sK\sB)}_2\big) \mbox{ } . 
\label{4-stop-Cl-Hist-Alg}
\eeq
\beq
\big[\widehat{\sfD}_{i}\big(t^{\se\sm(\sW\sK\sB)}_1\big), \widehat{\sfD}_{j}\big(t^{\se\sm(\sW\sK\sB)}_2\big)\big] = 
i\hbar\epsilon_{ij}\mbox{}^{k} \widehat{\sfD}_{k}\big(t^{\se\sm(\sW\sK\sB)}_1\big)\updelta\big(t^{\se\sm(\sW\sK\sB)}_1 - t^{\se\sm(\sW\sK\sB)}_2\big) \mbox{ } , 
\eeq
\beq
\big[\widehat{\rho}^{i}\big(t^{\se\sm(\sW\sK\sB)}_1), \widehat{\sfD}_{j}\big(t^{\se\sm(\sW\sK\sB\big)}_2\big] = 
i\hbar\epsilon^{i}\mbox{}_{jk}\widehat{\rho}^{k}\big(t^{\se\sm(\sW\sK\sB)}_1\big)\updelta\big(t^{\se\sm(\sW\sK\sB)}_1 - t^{\se\sm(\sW\sK\sB)}_2\big) \mbox{ } , 
\eeq
\beq
\big[\widehat{p}_{i}\big(t^{\se\sm(\sW\sK\sB)}_1\big), \widehat{\sfD}_{j}\big(t^{\se\sm(\sW\sK\sB)}_2\big)\big] = 
i\hbar\epsilon_{ij}\mbox{}^{k}\widehat{p}_{k}\big(t^{\se\sm(\sW\sK\sB)}_1)\updelta\big(t^{\se\sm(\sW\sK\sB)}_1 - t^{\se\sm(\sW\sK\sB)}_2\big) \mbox{ } .   
\eeq
The promotion requires discounting the possibility of a central term \cite{ILSS}; this working out entails a benign absence of constraint algebra complications.

\noindent This is a known structure, matching the mathematics of the 1-$d$ absolutist model \cite{IL} that has one particle more.  

\noindent The quantum histories energy constraint is then $\widehat\scE_{t^{\te\tm}} := \int \d t^{\se\sm(\sW\sK\sB)} \widehat\scE(t^{\se\sm(\sW\sK\sB)})$.
Since there is only one histories constraint, the quantum histories constraint algebra is trivial: i.e, there is no constraint closure problem facet of the POT here.

\subsubsection{Reduced formulation of scaled triangleland}

\noindent $\FrG_{\scc\sa\sn}$ is also here an Eucl(3) isometry group applying at each time-value, though the constituent objects have a distinct physical 
interpretation: the objects are now the histories of mixed relative dilational momenta and relative angular momenta: the $SO(3)$ objects $\sfS(t^{\se\sm(\sW\sK\sB)})$ and 
the $\mathbb{R}^3$ objects $\mbox{\boldmath $p$}^{Dra}(t^{\se\sm(\sW\sK\sB)})$.
The corresponding linear space $\fV^*$ is an $\mathbb{R}^3$ at each time-value, of objects $\mbox{\boldmath $Dra$}(t^{\se\sm(\sW\sK\sB)})$ that are the histories of $ellip$,  
$area$ and $aniso$.
Again, I assume that this continues to be combineable as a semi-direct product with no problems in considering it to be the standard 
$\FrG_{\scc\sa\sn} \mbox{\textcircled{S}} \mathfrakV$ at each time-value, which again forms an associated 1-$d$ QFT.

The scaled $N$-stop metroland classical histories brackets (\ref{N-stop-Cl-Hist-Alg}) are less straightforwardly promoteable to the quantum commutator algebra whose nontrivial part is 
\beq
\big[\widehat{\mbox{\scriptsize $Dra$}}^{\Gamma}\big(t^{\se\sm(\sW\sK\sB)}_1\big), \widehat{\mbox{\scriptsize$P$}}^{Dra}_{\Lambda}\big(t^{\se\sm(\sW\sK\sB)}_2\big)\big] = 
i\hbar\updelta^{\Gamma}\mbox{}_{\Lambda}\updelta\big(t^{\se\sm(\sW\sK\sB)}_1 - t^{\se\sm(\sW\sK\sB)}_2\big) \mbox{ } , 
\eeq
\beq
\big[\widehat\sfS_{\Gamma}\big(t^{\se\sm(\sW\sK\sB)}_1\big), \widehat\sfS_{\Lambda}\big(t^{\se\sm(\sW\sK\sB)}_2\big)\big] = 
i\hbar\epsilon_{\Gamma\Lambda}\mbox{}^{\Sigma} \widehat\sfS_{\Sigma}\big(t^{\se\sm(\sW\sK\sB)}_1\big)\updelta\big(t^{\se\sm(\sW\sK\sB)}_1 - t^{\se\sm(\sW\sK\sB)}_2\big) \mbox{ } , 
\eeq
\beq
\big[\widehat{\mbox{\scriptsize $Dra$}}^{\Gamma}\big(t^{\se\sm(\sW\sK\sB)}_1\big), \widehat\sfS_{\Lambda}\big(t^{\se\sm(\sW\sK\sB)}_2\big)\big] = 
i\hbar\epsilon^{\Gamma}\mbox{}_{\Lambda\Sigma}\widehat{\mbox{\scriptsize$Dra$}}^{\Sigma}\big(t^{\se\sm(\sW\sK\sB)}_1\big)
\updelta\big(t^{\se\sm(\sW\sK\sB)}_1 - t^{\se\sm(\sW\sK\sB)}_2\big) \mbox{ } , 
\eeq
\beq
\big[\widehat{\mbox{\scriptsize$P$}}^{Dra}_{\Gamma}\big(t^{\se\sm(\sW\sK\sB)}_1\big), \widehat\sfS_{\Lambda}\big(t^{\se\sm(\sW\sK\sB)}_2\big)\big] = 
i\hbar\epsilon_{\Gamma\Lambda}\mbox{}^{\Sigma}\widehat{\mbox{\scriptsize$P$}}^{Dra}_{\Sigma}\big(t^{\se\sm(\sW\sK\sB)}_1\big)
\updelta\big(t^{\se\sm(\sW\sK\sB)}_1 - t^{\se\sm(\sW\sK\sB)}_2\big) \mbox{ } .   
\eeq
The last two of these signify that the histories and the momenta are $SO(3)$ or $SU(2)$ vectors.  

\noindent The histories energy constraint is in this case $\widehat\scE_{t^{\te\tm}} := \int \d t^{\se\sm(\sW\sK\sB)}\widehat\scE(t^{\se\sm(\sW\sK\sB)})$ 
with $\scE$ given by (\ref{EnDragta}).   
Since there is only one histories constraint, the quantum histories constraint algebra is once again trivial and so there is no constraint closure problem facet of the POT here

\mbox{ } 

\noindent{\bf Question 90}.  The quadrilateralland counterpart of this central check is mathematically new.

\subsection{Hartle-type formulation Histories Theory for RPM's}\label{HarRPM}

Hartle had workings with physical outcomes paralleling those of Sec \ref{PI-App}'s formalism but 

\noindent 1) cast in terms of the commonplace multipliers rather than the most highly relationalism-implementing cyclic ordials. 

\noindent 2) Allowing for time to be treated discretely (and thus the subsequently-attached projectors at each time forming finite strings, 
moreover held together by plain multiplication).  

\noindent 3) Projectors being thus evoked, the paths $\gamma$ so decorated are to be re-denoted as histories $\eta$.  

\noindent 4) Histories are here plain products of projectors and thus not themselves projectors, meaning a lack of Projectors Implementation of Propositions for whole histories.

\subsection{IL type formulation of Histories Theory for RPM's}

This one has the brackets of Sec \ref{PaB-RPM} plus tying a continuous limit of tensor-product strings of projector operators 
(which object is itself a projector and therefore implements a histories proposition) to each of the preceding SSec's paths, by which these are regarded as quantum histories.  

\noindent This additionally allows for further computational techniques for the decoherence functional object.

\subsection{Further types of Histories Theory}

\noindent{\bf Question 91}.  Consider RPM's as a theoretical probe of Savvidou's 2-times and canonical-and-covariant-at-once scheme: 
how much of that exists in a background-independent but non-spacetime scheme? 

\mbox{ } 

\noindent Note 1) This Article's simple RPM models are free of the following Histories Theory versions of Problems: Measure Problem, 
Functional Evolution Problem, Thin Sandwich Problem, Diffeomorphism-specific Problems and Foliation Dependence Problem. 
Moreover the reduced approach to the scaled triangle is also free of global issues (unlike the Dirac approach to the scaled triangle). 
Thus, at least for this model, one looks to have a rather good resolution of the POT, upon which the next Sec
builds an even better resolution.

\subsection{Records-within-Histories scheme extended to RPM's}

While Halliwell's particular HO example's mathematics can be directly adapted to e.g. a $N = 3$, $d = 2$ scaled RPM setting, 
in doing so the $h$--$l$ split is not scale--shape aligned, so I prefer to start with a new example. 
To date only a handful of models \cite{H99, HD} have been cast in records formulation, and all have external time.  
With GR in mind, this is an important deficiency to resolve, and the RPM's look to be a useful setting for doing this. 
How time enters this parallels how time enters Histories Theory.

\subsubsection{Sketch of the steps involved}\label{QR-in-QH}

\noindent Given exact wavefunctions (e.g. free and HO ones from Sec 9.3) one can construct decoherence functionals from them.
\noindent It is also not clear to me what kind of notion of temperature might exist within the context of whole-universe RPM's. 
Halliwell makes use of the Von Neumann information of the environment (c.f. Sec \ref{Cl-NOI}).  

\noindent Analogy 113) Imperfect records are realized within both RPM's and GR.   

\noindent This approach considers Records within Histories Theory.  
In particular, imperfect records are still possible in settings with very small environments (even 1 degree of freedom).  
This makes this work compatible with small RPM's in their aspect as small models. 
Conditional probability of records $\underline{\kappa}$ [I use the notation $\underline{\kappa} = (\kappa_1, \kappa_2, ... , \kappa_n)$ given the past alternatives $\underline{\eta}$ is
\beq
\mbox{Prob}(\underline{\kappa}| \underline{\kappa}) =  
\mbox{Prob}(\underline{\kappa}, \underline{\kappa})/\mbox{Prob}(\underline{\eta}) = 
\mbox{Tr}(R_{\underline{\kappa}} \BigupRho_{\se\sf\sf}(\underline{\eta})) \mbox{ } ,
\eeq
for
\beq
\BigupRho_{\se\sf\sf}(\underline{\eta})) = \frac{\mC_{\underline{\eta}}\BigupRho \mC_{\underline{\eta}}^{\dagger}}
                                             {\mbox{Tr}(\mC_{\underline{\eta}}\BigupRho \mC_{\underline{\eta}}^{\dagger})} \mbox{ } .
\eeq
Here, $R_{\underline{\kappa}}$ denotes a {\bf records projector}, which can be envisaged as the obvious subcase of the general histories  projector. 
[Note also the relation
\beq
\mbox{Prob}(\underline{\eta}) = {\mbox{Tr}(\mC_{\underline{\eta}}\BigupRho \mC_{\underline{\eta}}^{\dagger})} = 
\mbox{Tr}(R_{\underline{\eta}}\BigupRho(t^{\se\sm(\sW\sK\sB)}_n)) \mbox{ } ] \mbox{ } .
\eeq
Perfect correlation between records and past alternatives is only guaranteed if $\mbox{Prob}(\underline{\kappa} \mbox{ } | \mbox{ } \underline{\eta}) = 1$, 
which only occurs for $\BigupRho_{\se\sf\sf}(\underline{\eta})$ pure. 
For the more general case of $\BigupRho_{\se\sf\sf}(\underline{\eta})$ mixed, $\mbox{Prob}(\underline{\kappa}| \underline{\eta}) < 1$ and this correlation is imperfect.  
{\sl Thus in general one can expect the presence of imperfect records.}
\noindent This approach would amount to considering detectors within the RPM setting: a base pair serving as detector for a passing apex.  

\mbox{ }  

\noindent Analogy 114) Records within histories theory occur in both RPM's and in GR.

\subsubsection{Open questions}

\noindent {\bf Question 92}. As already alluded to in Sec 20, Halliwell's 1999 approach contains his `information storage conjecture': 
information stored is related to extent of coarse graining.
which of the many notions of information in \ref{Cl-NOI}, \ref{QM-NOI} is most appropriate for this?  

\mbox{ }   

\noindent {\bf Question 93} Consider RPM's in the large-$N$ limit so as to consider models with sizeable environments and sizeable amounts 
of records/records information within them. 

\noindent The $N$-stop case of this has been mostly developed in this Article, though it has not for now been uplifted to Histories Theory schemes. 
On the other hand, $N$-a-gonland is a long haul and of starred difficulty besides, die to increasing unfamiliarity with the Methodsof Mathematical Physics involved 
and with such as consideration of central terms in the histories-theoretical kinematical quantization.  

\noindent The SM of the large-$N$ $N$-a-gon is another application (with some inter-relations between the two expected).  

\vspace{10in}  

\section{Halliwell's Combined Approach for RPM's} \label{QM-Combo}

\noindent Analogy 115) scaled RPM is well-suited for investigating the combination of Histories, Records and Semiclassical Approaches.  
This is because it is a simple enough toy model to get far in its study, whilst nevertheless possessing 
sufficient features to do a reasonable job of toy-modelling midisuperspace.  
I want to consider these together because there are indications that these are able to prop each other 
up toward finally providing a resolution of the POT and associated issues in Quantum Cosmology \cite{H03}.

\subsection{Scaled $4$-stop metroland example}

Halliwell's example (Sec \ref{QM-Halliwell-Intro}) is a particle in the $n$-dimensional absolutist context. 
Whilst this shares mathematics with $N$-stop metroland, I wish for the study to be able to rest upon the IL formulation of Histories Theory, 
and the currently worked-out state of that restricts me to $N \leq 4$.
Thus the below $n$ is for now to be taken to be 2 (a case for which explicit semiclassical approach calculations have been done) or 3 (a case whose configuration space 
mathematics has the `sphere' parallel the next SSec's triangleland example, but for which explicit semiclassical approach calculations have not yet been done).   

\noindent The preceding alternative expression has further parallel with the Wigner functional at the semiclassical level.  
Here,
\beq
\mbox{Wig}[\brho, \mbox{\boldmath $p$}] \approx |\chi(\brho)|^2\updelta^{(n)}(\mbox{\boldmath $p$} - {\mbox{\boldmath{$\pa$}}}S)
\label{Wig0B}
\eeq
($\mbox{\boldmath $p$}$ being ${\mbox{\boldmath{$\pa$}}} S$ for classical trajectories). 
Then Halliwell'-type heuristic move is then to replace $\mw$ by Wig in (\ref{clvers3}), giving 
\beq
P_{\Upsilon}^{\sss\se\sm\si\scc\sll} \approx \int\d t^{\se\sm(\sW\sK\sB)}_{\{1\}}\int_{\Upsilon}\d\Upsilon(\biK) \bnu \cdot \frac{\pa S}{\pa\biK} 
|\chi(\biK)|^2 \mbox{ } .  
\eeq
This is written to the semiclassical accuracy that the current paper establishes.  
$\bnu$ is the normal to the hypersurface $\Upsilon$.

\subsubsection{Class operators as quantum Dirac degradeables}

The Halliwell-type treatment continues \cite{AHall} within the framework of decoherent histories, which I take as formally standard for this setting too.  
The key step for this continuation is the construction of class operators, which uplifts a number of features of the preceding structures.
One now uses  
\beq
\widehat{\iC}^{\sharp}_{\bsFrR} = \theta
\left(   
\int_{-\infty}^{+\infty} \d t^{\se\sm(\sW\sK\sB)}_{\{1\}} \mbox{Char}_{\bsFrR}(\brho(t^{\se\sm(\sW\sK\sB)}_{\{1\}}))  - \epsilon 
\right) P(\brho_{\sf}, \brho_0) \,   \mbox{exp}(i A(\brho_{\sf}, \brho_0)) \mbox{ } , \label{spoo2}
\eeq
which obeys
\beq
[ \widehat{\scQ\scU\scA\scD}, \widehat{\iC}^{\sharp}_{\bsFrR} ] = 0 
\eeq
by construction and there are no linear constraints in this example.

\noindent As regards why in general these are just degradeables, 1) we firstly need to know if this scheme really does not $h$-$l$ split its degrees of freedom in the 
$\widehat{\iC}_{\bsFrR}^{\sharp}$. 
If it does, sphere coords are limited.  
2) Also, there is a global-in-time aspect that degradeables may well not cover: the finite-interval case ceases to commute with $\widehat{\scQ\scU\scA\scD}$.  

\mbox{ } 

\noindent Analogy 116) Halliwell-type class operators can be constructed for RPM's and for, at least, minisuperspace GR examples.

\subsubsection{Decoherence functional}

The decoherence functional is of the form 
\beq
{\cal D}\me\mc[\eta, \eta^{\prime}] = \int_{\eta}\mathbb{D} \brho \int_{\eta^{\prime}} \mathbb{D} \brho^{\prime} 
\mbox{exp}\big(i\big\{S\big[\brho(t^{\se\sm(\sW\sK\sB)}_{\{1\}}\big)\big] - S\big[\brho^{\prime}(t^{\se\sm(\sW\sK\sB)}_{\{1\}}\big)\big]\big\}
\BigupRho(\brho_{0}, \brho_{0}^{\prime})\omega_{\sd} \mbox{ } .
\eeq

Class operators are then fed into the expression for the decoherence functional, giving 
\beq
{\cal D}\me\mc[\eta, \eta^{\prime}] = \int \d^3\brho_{\sf}\d^3\brho_{0} \d^3\brho^{\prime}_{\sf}
\widehat{\iC}^{{\sharp}}_{\eta}[\brho_{\sf}, \brho_{0}] 
\widehat{\iC}^{{\sharp}}_{\eta^{\prime}}[\brho^{\prime}_{\sf}, \brho^{\prime}_{0}]
\Psi(\brho_{0})  
\Psi(\brho^{\prime}_{0})
\omega^2\omega_{\sd} \mbox{ } .  
\label{dunno22}
\eeq 
If the universe contains a classically-negligible but QM-non-negligible environment as per Appendix \ref{App-Env}, 
the influence functional ${\cal I}$ makes conceptual sense and one can rearrange (\ref{dunno22}) in terms of this into the form  
\beq
{\cal D}\me\mc[\eta, \eta^{\prime}] = \int\int\int \mathbb{D}\brho_{\sf} \mathbb{D}\brho_{0} \mathbb{D}\brho_{0}^{'}
\widehat{\iC_{\eta}}^{\sharp}[\brho_{\sf}, \brho_0]  
\widehat{\iC}_{\eta}^{\sharp}[\brho_{\sf}, \brho_0)^{'}] 
{\cal I}[\brho_{\sf}, \brho_0, \brho_0^{'}]
\Psi(\brho_0)
\Psi^*(\brho_0^{'})
\omega^2\omega_{\sd} \mbox{ } .  
\label{83b}
\eeq

\subsection{Semiclassical quantum working for r-presentation of triangleland}\label{Wigi2}

The last alternative above further parallel at the semiclassical level with the Wigner function. 
Now, including a power of the PPSCT conformal factor,
\beq
\mbox{Wig}[\mbox{\scriptsize\boldmath$p$}^{\mbox{\tiny $Dra$}}, \mbox{\scriptsize\boldmath$Dra$}] \approx |\chi(\mbox{\scriptsize\boldmath $Dra$})|^2
\delta^{(3)}(\mbox{\scriptsize\boldmath$p$}^{\mbox{\tiny $Dra$}} - {\mbox{\boldmath{$\pa$}}}^{Dra}S)/\{4\mI\}^{3/2} \mbox{ } 
\label{Wig2}
\eeq
($\mbox{\scriptsize\boldmath$p$}^{\mbox{\tiny $Dra$}} = {\mbox{\boldmath{$\pa$}}}^{Dra}S$ for classical trajectories).
then for triangleland, $|\chi\rangle = |\chi(\Theta, \Phi, \mI)\rangle$ and $S = S(\mI)$. 
Halliwell's heuristic move is then to replace $\mw$ by Wig in (\ref{clvers3})
\beq
P_{\Upsilon}^{\sss\se\sm\si\scc\sll} \approx \int \d t^{\se\sm(\sW\sK\sB)}_{\{1\}}\int_{\Upsilon}\mathbb{D}\Upsilon(\mbox{\scriptsize$Dra$}) \bnu^{\mbox{\tiny $Dra$}} 
\frac{\pa S}{\pa\mbox{\scriptsize\boldmath$Dra$}} |\chi(\mbox{\scriptsize\boldmath $Dra$})|^2 \mbox{ } . \label{lallo}  
\eeq
\noindent The RPM case of most interest is that with the radial `scale of the universe' direction having particular $h$-significance, by which the configurational 2-surface element is 
a piece of sphere with a number of these carrying lucid significance by Sec \ref{Dyn1} and the 3-momentum 3-surface element is the spherical polars one (modulo conformal factors).
Moreover, the $\FrS$ makes the evaluation of this in spherical polars natural, even if $\Upsilon$ itself is unaligned with those 
(though it simplifies the calculation if there is such an alignment).

So rewriting (\ref{lallo}) in conformal--spherical polar coordinates, e.g. for 
\beq
\mbox{\scriptsize Prob(universe attains size $\mI_0 \pm \delta\mI$ whilst being $\epsilon$-equilateral)} \approx  
\int_{t^{\se\sm}_{\{0\}} = t_0 - \delta t}^{t_0 + \delta t} \d t^{\se\sm}_{\{0\}} 
\int_{\Phi = 0}^{2\pi}\int_{\Theta = 0}^{\epsilon} 
\mI^2 \mbox{sin}\,\Theta\,\d\,\Theta\d\,\Phi 
\frac{\d S(\mI)}{\d \mI}
|\chi(\mI, \Theta, \Phi)|^2 \mbox{ } .  
\eeq
This made use of this question being addressed by the $\epsilon$-cap about the E-pole, which is very simply parametrized by the coordinates in use 
[see Fig \ref{Fig-2AFb}c) for this cap and the below belt].  
It also makes use of the approximation $t^{\se\sm(\sW\sK\sB)} \approx t^{\se\sm(\sJ\sB\sB)}$.   
Also, $S = S(\mI)$ alone, so $\bnu^{Dra\prime} {\pa S}/{\pa\mbox{\scriptsize\boldmath $Dra$}^{\prime}}$ becomes a radial $\bnu^{\sI}\,\pa S/\pa\mI$ factor and two zero components. 
To proceed, $\d t^{\se\sm(\sJ\sB\sB)} {\d S}/{\d \sI}$ = $\d t^{\se\sm(\sJ\sB\sB)} \frac{\d \sI}{\d t^{\se\sm(\sJ\sB\sB)}} = \d \mI$ by the Hamilton--Jacobi expression for momentum, 
the momentum-velocity relation and the chain-rule, so we do not need to explicitly evaluate $t_0 \pm \delta t$ in terms of $\mI \pm \delta I$.
Then e.g. for the approximate semiclassical wavefunction from the explicit triangleland example in \cite{SemiclIII} (the upside-down harmonic oscillator for the universe at zero energy), 
\beq
\mbox{\scriptsize{Prob(universe attains size I $\pm \delta$I whilst being $\epsilon$-equilateral)}} \approx  
2 \, \mI_0^2 \, \delta\mI   
\int_{\Theta = 0}^{\epsilon}\int_{\Phi = 0}^{2\pi}   |\mY_{\sS\,\ms}(\Theta, \Phi)|^2\mbox{sin}\,\Theta\,\d\Theta\, \d\Phi + O(\delta \mI^2)
\label{clvers3} \mbox{ } .
\eeq
Here, the $\mY_{\sS\sss}$ are the [\mbox{ }]-basis spherical harmonics of Sec \ref{QTri-Ebasis}.
The answer then comes out with leading term proportional to e.g. $\mI_0^2\delta\mI \, \epsilon^2$ for all the axially-symmetric wavefuntions $\chi_{\sS\,0} \propto \mY_{\sS\,0}$ and to 
$\mI_0^2\delta\mI\,\epsilon^4$ for the first non-axial wavefunctions (the sine and cosine combinations corresponding to the quantum numbers S and $|\ms|$ = 1).  
These answers make good sense as regards the axisymmetric wavefuntions being peaked around the equilateral triangle whilst the equilateral 
triangle is nodal for the first non-axisymmetric wavefunctions.  

\mbox{ }  

\noindent Note 1) This is an \NSII-type construct, though it is for the semiclassical $l$-part, so there is some kind of semiclassical imprint left on it.

\noindent Note 2) {Prob}(universe attains size I$_0$ $\pm \delta$I whilst being $\epsilon$-D) is given likewise but for a particular D being given 
by the same in the corresponding ( ) basis (Sec \ref{QTri-Dbasis}), and the words ``one orientation" or ``a particular D" being suppressible by summing over various such integrals.  
 
\noindent Note 3) As a final example,  
\beq
\mbox{\scriptsize Prob(universe attains size $\mI_{0} \pm \delta$I whilst being $\epsilon$-collinear)}  \mbox{ } 
\mbox{ has $\epsilon$-cap replaced by $\epsilon$-belt about the equator in the [ ] basis} \mbox{ } ,
\eeq
and the answer goes as as $\mI_0^2\delta\mI\,\epsilon^3$ for the odd-$\mS$ axisymmetric wavefunctions and as $\mI_0^2\delta\mI\,\epsilon$ for the even-$\mS$ axisymmetric 
wavefunctions and the first non-axisymmetric ones.   
The extra $\epsilon^2$ factor can again be explained in terms of peaks and nodes: a nodal plane of collinearity as compared to peaks on all or part of it.

\subsubsection{Class operators}

Again, one uses the modified version, which here takes the form  
\beq
{\iC}_{\bsFrR}^{\sharp} = \theta
\left(   
\int_{-\infty}^{+\infty} \d t \, \mbox{Char}_{\bsFrR}(\mbox{\scriptsize\boldmath $Dra$}(t^{\se\sm(\sW\sK\sB)}_{\{1\}}))  - \epsilon 
\right) B(\mbox{\scriptsize\boldmath $Dra$}_{\sf}, \mbox{\scriptsize\boldmath $Dra$}_0) \,   \mbox{exp}(iA(\mbox{\scriptsize\boldmath $Dra$}_{\sf}, \mbox{\scriptsize\boldmath $Dra$}_0)) \mbox{ }  .  
\eeq
This obeys 
\beq
[\widehat{\scE}, \widehat{\iC}^{\sharp}] = 0 \mbox{ } , \mbox{ } \mbox{ } [\widehat{\scL}, \widehat{\iC}^{\sharp}] = 0 \mbox{ } .
\eeq

\subsubsection{Decoherence functionals}

Next,   
\beq
{\cal D}\me\mc[\eta, \eta^{\prime}] = \int_{\eta} \d^3\mbox{\scriptsize\boldmath $Dra$}  \int_{\eta^{\prime}} \d^3\mbox{\scriptsize\boldmath $Dra$}^{\prime} 
\mbox{exp}\big(i\big\{S\big[\mbox{\scriptsize\boldmath $Dra$}\big(t^{\se\sm(\sW\sK\sB)}_{\{1\}}\big)\big] - 
                      S\big[\mbox{\scriptsize\boldmath $Dra$}\big(t^{\se\sm(\sW\sK\sB)}_{\{1\}}\big)\big]\big)\big\}
\BigupRho(\mbox{\scriptsize\boldmath $Dra$}_{\si\sn}, \mbox{\scriptsize\boldmath $Dra$}_{\si\sn}^{\prime}) \mbox{ } ,
\eeq
\beq
\mbox{giving} \hspace{0.7in}
{\cal D}\me\mc[\eta, \eta^{\prime}] = 
\int\int\int \d^3\mbox{\scriptsize\boldmath $Dra$}_{\sf}\d^3\mbox{\scriptsize\boldmath $Dra$}_{0} \d^3\mbox{\scriptsize\boldmath $Dra$}^{\prime}_{\sf} \, 
{\iC}^{\sharp}_{\eta}(\mbox{\scriptsize\boldmath $Dra$}_{\sf}, \mbox{\scriptsize\boldmath $Dra$}_{0}) \, 
{\iC}^{\sharp}_{\eta^{\prime}}(\mbox{\scriptsize\boldmath $Dra$}^{\prime}_{\sf}, 
\mbox{\scriptsize\boldmath $Dra$}^{\prime}_{0})
\Psi(\mbox{\scriptsize\boldmath $Dra$}_{\si\sn})  
\Psi(\mbox{\scriptsize\boldmath $Dra$}^{\prime}_{\si\sn}) 
\mbox{ } . 
\label{dunno3b} \hspace{3in}
\eeq
Under classically insignificant, QM significant environment assumption under which the influence functional is justified,\footnote{Note 
however that the eventual target of paralleling \cite{H09} target differs in not requiring environments, 
at least for `larger regions', so not having an alternative at this stage is not a long-term hindrance to the present program.  
It is more a case of \cite{H03} coming with 
1) environment-based reservations (not optimal for a fully closed system study) and 
2) the  quantum Zeno problem addressed in \cite{H09} and the next SSec;  
both of these issues go away upon passing to the more advanced \cite{H09} construction in \cite{AHall2}.
If there is no environment, we lose (\ref{pipupi}), and if \cite{HT}'s justification fails we lose eq's (\ref{snarl}-\ref{whiffle}).}  
\beq
{\cal D}\me\mc[\eta, \eta^{\prime}] = \int\int\int \d^3 \mbox{\scriptsize\boldmath $Dra$}_{\sf} \d^3 \mbox{\scriptsize\boldmath $Dra$}_{0} \d^3 \mbox{\scriptsize\boldmath $Dra$}_{0}^{\prime} \, 
{\iC}^{\sharp}_{\eta}(\mbox{\scriptsize\boldmath $Dra$}_{\sf}, \mbox{\scriptsize\boldmath $Dra$}_0) \, {\iC}^{\sharp}_{\eta}(\mbox{\scriptsize\boldmath $Dra$}_{\sf}, \mbox{\scriptsize\boldmath $Dra$}_0^{\prime}) 
{\cal I}[\mbox{\scriptsize\boldmath $Dra$}_{\sf}, \mbox{\scriptsize\boldmath $Dra$}_0, \mbox{\scriptsize\boldmath $Dra$}_0^{\prime}]\Psi(\mbox{\scriptsize\boldmath $Dra$}_0)\Psi^*(\mbox{\scriptsize\boldmath $Dra$}_0^{\prime}) \mbox{ } .  
\label{pipupi}
\eeq 
\noindent Then if \cite{HT}'s conditions apply [which they do according to Attitude 3) of Sec \ref{App-Env}],   
\beq
{\cal I}[\mbox{\scriptsize\boldmath $Dra$}_{\sf}, \mbox{\scriptsize\boldmath $Dra$}_0, \mbox{\scriptsize\boldmath $Dra$}_0^{\prime}] = \mbox{exp}(i\,\mbox{\scriptsize\boldmath$Dra$} \cdot \bGamma + 
\mbox{\scriptsize\boldmath $Dra$} \cdot \bsigma \cdot \mbox{\scriptsize\boldmath $Dra$}/2  )  \mbox{ } .  
\label{snarl}
\eeq
If the above step holds, then the below makes sense too. 
Here,          $\mbox{\scriptsize\boldmath$Dra$}^{-} := \mbox{\scriptsize\boldmath$Dra$} - \mbox{\scriptsize\boldmath$Dra$}^{\prime}$ and $\Gamma_{\Lambda}$, $\sigma_{\Gamma\Lambda}$ 
real coefficients depending on $\mbox{\scriptsize\boldmath$Dra$} + \mbox{\scriptsize\boldmath$Dra$}^{\prime}$ alone and with $\bsigma$ a non-negative matrix.
Using $\mbox{\scriptsize\boldmath$Dra$}^+ := \{\mbox{\scriptsize\boldmath$Dra$}_0 + \mbox{\scriptsize\boldmath$Dra$}_0^{\prime}\}/2$ as well, the Wigner functional is 
\beq
\mbox{Wig}[\mbox{\scriptsize\boldmath$Dra$}, \mbox{\scriptsize\boldmath$p$}^{\mbox{\tiny $Dra$}}] = 
\frac{1}{\{2\pi\}^3}\int \d^3\mbox{\scriptsize\boldmath$Dra$} \, \mbox{exp}(-i \mbox{\scriptsize\boldmath$p$}^{\mbox{\tiny $Dra$}} 
\cdot \mbox{\scriptsize\boldmath$Dra$})\rho(\mbox{\scriptsize\boldmath$Dra$}^+ + \mbox{\scriptsize\boldmath$Dra$}^-/2, \mbox{\scriptsize\boldmath$Dra$}^+ - 
\mbox{\scriptsize\boldmath$Dra$}^-/2) \mbox{ } . 
\eeq
\beq
\mbox{Then} \hspace{0.7in}
P_{\bsFrR} = \int\int \d^3 \mbox{\scriptsize\boldmath$p$}^{\mbox{\tiny $Dra$}}_0\d^3\mbox{\scriptsize\boldmath$Dra$} \, \, \theta
\left(
\int_{-\infty}^{+\infty} \d  t^{\se\sm(\sW\sK\sB)}  \mbox{Char}_{\bsFrR}\big( \mbox{\scriptsize\boldmath$Dra$}^{+\scc\sll}\big) - \epsilon
\right)
\widetilde{\mbox{Wig}}[\mbox{\scriptsize\boldmath$Dra$}_0^+, \mbox{\scriptsize\boldmath$Dra$}_0]  \hspace{3in}
\eeq 
for $\mbox{\scriptsize\boldmath$Dra$}^{+\scc\sll}(t^{\se\sm(\sW\sK\sB)})$ the classical path with initial data $\mbox{\scriptsize\boldmath$Dra$}^+_0, \biiP^{Dra}_0$ and Gaussian-smeared 
Wigner function
\beq 
\widetilde{\mbox{Wig}}[\mbox{\scriptsize\boldmath$Dra$}^+_0, \biiP^{Dra}_0] = \int \d^3 \mbox{\scriptsize\boldmath$p$}^{\mbox{\tiny $Dra$}} \, \mbox{exp}(-\frac{1}{2}
\{\mbox{\scriptsize\boldmath$p$}^{\mbox{\tiny $Dra$}}_0 - \mbox{\scriptsize\boldmath$p$}^{\mbox{\tiny $Dra$}} - \bGamma\} \cdot \bsigma \cdot 
\{\mbox{\scriptsize\boldmath$p$}^{\mbox{\tiny $Dra$}}_0 - \mbox{\scriptsize\boldmath$p$}^{\mbox{\tiny $Dra$}} - \bGamma\})  
\mbox{Wig}[\mbox{\scriptsize\boldmath$Dra$}_0^+, \mbox{\scriptsize\boldmath$p$}^{\mbox{\tiny $Dra$}}_0] \mbox{ } .  
\label{whiffle}
\eeq 
This final step is the one in which Halliwell's setting gives a good classical recovery with a smeared Wigner functional in place of a classical probability distribution.  

\mbox{ } 

\noindent {\bf Question 94}.  Extend these examples to contain harmonic oscillator potentials.

\subsection{$\FrG$-act, $\FrG$-all version}

For the triangleland example, $\FrG = SO(2) = U(1)$, so 
\beq
\int_{g \in \sFG}\mathbb{D} g = \int_{\zeta \in \mathbb{S}^1} \mathbb{D} \mathbb{\zeta} = \int_{\zeta = 0}^{2\pi}\d\zeta
\eeq
with $\stackrel{\rightarrow}{\FrG_g}$ being the infinitesimal 2-$d$ rotation action via the matrix $\underline{\underline{R}}\mbox{}_{\zeta}$ acting on the vectors of the model.
$\zeta$ is here the absolute rotation.

\subsection{Some concern with passage from \K to Dirac beables}

\noindent One issue here is whether the cat gets out of the bag if one subsequently forms functionals out of the \K variables (specifically, class functionals).  
That provides commutation with $\scQ\scU\scA\scD$, but now a functional of observable is not necessarily an observable (or an observable compatible with those 
one already has), so the Multiple Choice Problem is looming as a reason for breakdown of the Halliwell procedure being applied to produce nontrivially-\K Dirac observables. 
And yet, $\scL\scI\scN$ is gone here, so it is not [ $\widehat{\scL\scI\scN}, \widehat{\sfK}$] = 0 failing to imply [$\widehat{\scL\scI\scN}$, $\widehat{{\cal F}(\ux; \sfK]}$] = 0 
also, but rather an issue of whether class functionals themselves furnish a compatible set {\sl among themselves} if they are to now be regarded as observables.

\subsection{Halliwell 2009 type extension}  

\noindent Extending Halliwell's 2009 work to the RPM arena mostly concerns defining class operators somewhat differently so as to get these to 
be better-behaved as regards the quantum Zeno effect.  
In detail, \cite{H03} used a sharp-edged window function, whilst the subsequent \cite{H09, H11} uses a smoothed-out one. 
This amounts to the region in question being taken to contain a potential, with the class operator being the corresponding S-matrix and the 
smoothed-out case representing a softening in the usual sense of scattering theory (albeit in configuration space rather than in space).  
[The smoothed-out case manages to avoid the quantum Zeno effect in addition to managing to still be compatible with the (quadratic, not for now linear) constraint.]

Moreover, whilst conceptually illustrative, the above SSecs' class functional itself has technical problems, in particular it suffers from the Quantum Zeno problem. 
This can be dealt with firstly by considering the following reformulation \cite{HW}:
\beq
\mP(N\epsilon) ... \mP(2\epsilon)\mP(\epsilon) = \mbox{exp}(-iHt) \mbox{ } ,
\eeq 
%
%
as well as now conceptualizing in terms of probabilities of {\sl never} entering the region.  
Next, one applies Halliwell's `softening' (in the sense of scattering theory), passing to (in for now his nonspecific-$t$ context)
\beq
\mP(N\epsilon) ... \mP(2\epsilon)\mP(\epsilon) = \mbox{exp}(-i(H - iV_0 P)t)
\eeq
for $\epsilon V_0 \approx 1$.  
The class operator for {\sl not} passing through region $\bFrR$ is then
\beq
\iC_{\overline{\bsFrR}} = \stackrel{\mbox{lim}}{\mbox{\scriptsize{$t_1 \longrightarrow -\infty, t_2 \longrightarrow \infty$}}} 
\mbox{exp}(i H t_2) \mbox{exp}(-i\{H - iV\}\{t_2 - t_1\}) \mbox{exp}(- i H t_1)
\mbox{ } .  
\eeq
This now does not suffer from the Quantum Zeno Problem whilst remaining an S-matrix quantity and thus commuting with $\widehat{\scQ\scU\scA\scD}$.

\mbox{ }

\noindent One would then identify the above $t$'s as $t^{\se\sm(\sW\sK\sB)}$'s in the Machianized version of the updated Combined Approach.  

\noindent Finally, the class functional (\ref{spoo2}) is indeed a locally interpretable concept: select a region, so using such a 
construct to solve the Problem of (Dirac) Observables/Beables at the quantum level is indeed compatible with the basic ethos of fashionables/degradeables.

\subsection{Nododynamics counterpart posed}

\noindent {\bf Question 95$^*$}/Analogy 117) More ambitiously [1.iii)], might one be able to extend Halliwell's class operator construction to Nododynamics, 
so as to be able to, at least formally, write down a (perhaps partial) set of complete observables as a subset of the linear constraint complying knots?
Either construct such a set or explain at what stage of relevant generality Halliwell's class operator construction fails.
[I.e. a tentative search for a fully-fledged Unicorn.]
Note that this goes beyond the currently-known geometrodynamical minisuperspace scope of Halliwell's class operators, but is an interesting direction in which to try to extend this scope.   

\mbox{ } 

\noindent For now, Schroeren \cite{Schroeren} has considered {\sl Hartle}-type class functions for Nododynamics.
Discussions between me and him since have begun to consider whether these calculations could be extended to Halliwell-type class functions, e.g. in the mathematically simpler case of LQC.  

\vspace{11in}

\section{Conclusion: Relationalism and the POT}\label{Conclusion}

\subsection{Development of Relationalism}

This Article further develops, composes and contrasts various notions of relationalism. 
It completes the Barbour/LMB approach and provides extensions of it. 
The first such direction is the more localized LMB-CA approach (the C stands for Clemence \cite{Clemence}).  
The second such direction shows how this can to some extent be composed with Rovelli and Crane's ideas. 
Moreover, in some places these form alternatives (i.e. one has to choose a Barbour--A postulate or a Rovelli--Crane one).  

\noindent Example 1) Different attitudes to implementing Mach's Time Principle: the `any', `all' and `STLRC' to GLET attitudes to Mach's Time Principle.

\noindent Example 2) Whether to use partial instead of \K or Dirac observables/beables.

\subsection{I argued for intimate ties between this paper's notion of relationalism and the POT}

The POT mostly concerns Background Independence at the quantum level.  
As argued in the preface, GR can be viewed as a gestalt of reconciliation of relativity and gravitation {\sl and} a freeing of Physics from background structure.
Barbour's work and my extension of it are background-independent at the classical level, unearthing classical counterparts of 7 of the 8 POT facets. 
The argument then is that Background Independence is philosophically and physically desirable, classical relationalism is the first manifestation of this, which 
then sheds some further light on the meaning of, and strategization for, the quantum-level POT.
Moreover the background-independent formulation of Mechanics already has 5/8ths of the POT facets at the classical level and 6/8ths of them at the quantum level, 
rendering it a useful toy model for quite a few POT investigations.  
This study is to be complemented \cite{I93, ABook} with models that nontrivially involve diffeomorphisms and GR spacetime-like notions, since the missing 2/8ths are of that nature, 
and making the Configurational Relationalism involve diffeomorphisms technically complicates most of the other facets too.
Classical GR has all 7/8ths of the classically-available facets, the missing one being the Multiple Choice Problem.

\subsection{Concrete RPM models can be solved}

\noindent Ultimately, the Barbour 2003 pure-shape mechanics is equivalent to the Jacobi--Synge construction applied to Kendall's geometry of the space of shapes.
\noindent It is then straightforward to also show that BB82 is equivalent to the Jacobi--Synge construction applied to the cone over Kendall's shape geometry (the space of scaled shapes).

\noindent I showed how the Barbour--Bertotti 1982 and Barbour 2003 theories transcend formulation in a way in which the Barbour--Bertotti 1977 theory fails to, 
which provides a further strong reason to favour the first two (and readily leads to formulation of various other theories with similar limitations to Barbour--Bertotti 1977). 

\noindent The simplest configuration space geometries for 1- and 2-$d$ RPM's are $\mathbb{S}^{n - 1}$ and $\mathbb{CP}^{n - 1}$ (pure shapes), 
and $\mC(\mathbb{S}^{n - 1}) = \mathbb{R}^{n}$ and $\mC(\mathbb{CP}^{n - 1})$, the first three of which are very well known as geometries 
and as regards subsequent classical and quantum mechanics thereupon and supporting linear methods of Mathematical Physics.  
These render many QM and POT strategy calculations tractable and available for comparison with each other, which is a rarity in the latter field.  
Triangleland is further aided in this way by $\mathbb{CP}^{1} = \mathbb{S}^{2}$ and $C(\mathbb{CP}^1) = \mathbb{R}^3$ albeit the latter is not flat; it is, however, conformally flat.
The simpleness of the ensuing mathematics, even well into the usually complex rearrangements necessary for the investigation of POT strategies, is a major asset in this RPM toy model 
arena, securing many computational successes beyond the usual points at which these break down for full GR/many other toy models.  
We demonstrated solvability, by mixture of basic maths and interdisciplinarity with statistical theory of shape, 
Molecular Physics and a few other areas for quadrilateralland and higher-$N$ $N$-a-gons (Particle Physics, instantons).  

\noindent This Article has considered lucid coordinatizations of shape space, the momenta conjugate to shape coordinates.
\noindent Kinematics can be uplifted from Celestial Mechanics and Molecular Physics, many of whose names and meanings have been enhanced by the present Article. 
And from further places such as geometry and gravitational instantons the quadrilateral and the $N$-a-gon more generally.  
The shapes split irreducibly into relative angle and relative distance (non-angular ratio) quantities, and this is paralleled by their  
conjugate momenta splitting into relative angular momenta and relative distance momenta.

\noindent RPM isometry groups have atomic/molecular physics analogies ($SO(3) = SU(2)/\mathbb{Z}_2$ for triangleland) and particle physics analogies (the $SU(3)$/$\mathbb{Z}_3$ 
for quadrilateralland is identical to colour group and shares the same Lie algebra with approximate flavour physics too).  

\noindent I presented a Cosmology-Mechanics analogy in which (the square root of) the moment of inertia plays the role of the scalefactor and 
the shape degrees of freedom play the role of the small inhomogeneities.  
\noindent I also presented exact QM solutions that parallel Atomic and Molecular Physics results in their mathematical form, but now with a new relational whole-universe interpretation.

\noindent Finally I considered the levels of structure needed for a full exposition of all practical aspects of physics: notions of distance, of information, of correlation, 
and of proposition, both at the classical and at the quantum levels.
I exemplified many of these using RPM's.

\subsubsection{Application 1: Absolute versus relational QM}\label{AORM-Concl}

\noindent Much of the mathematics of these, especially for simpler cases, coincides (albeit of course carrying entirely different physical meanings).
The differences are as follows.  

\noindent 1) 3 particles treated as relationally as possible can only distinguish between $d = 1$ and $d > 1$; thus the additional distinctive features of 
3 particles in 3-$d$ are absolutist vestiges from attributing physical meaning to the dimension of space itself.  

\noindent 2) There are differences between absolutist and relational quantizations at the level of the kinematical quantization; the first cases for which 
this occurs are triangleland and quadrilateralland.

\noindent 3) There are differences between absolutist and relational QM at the level of the operator ordering of the wave equation.

\noindent 4) There are subsequently some differences at level of solutions for given types of potentials, e.g. how for triangleland more of the `2$s$ orbital' 
probability density is inside, rather than outside, the `1$s$ orbital' one.

\noindent 5) There are plenty of eigenspectrum restrictions, though these follow from the closed-universe status of the RPM's rather than from any further RPM features.

\subsubsection{Application 2: further directions in RPM modelling} 

This Article also laid out the beginnings of the study of 

\noindent 1) alternative mirror-image-identified and indistinguishable-particle RPM's up to the level of configuration space geometry.  
The configuration spaces for these are quotients of the preceding by $\mathbb{Z}_2$ and by permutation groups; as configuration space geometries, these include 
$\mathbb{RP}^{n - 1}$, so-called weighted projective spaces, spherical primes and generalizations thereof, and the 3-$d$ half-space with and without edge.    
The mirror-image-identified case has applications as regards quantizing manifolds with edge and being a toy model of affine Geometrodynamics.  
The indistinguishable particle case is tied to arrow structure and arrow statistics (I use the word `arrow' instead of `spin' so as to not 
carry any angular momentum connotations in models that do not possess any notion of angular momentum in physical space).  

\noindent 2) Fermionic RPM's and whether supersymmetric RPM's exist.

\subsubsection{Application 3: Quantum Cosmology/POT Modelling}

\noindent This is by far the largest and most fruitful application.  
\noindent This RPM study has rendered a fair amount of POT and other foundational issues in Quantum Cosmology approachable using (little more than) 
the amount of maths needed for basic study of the atom (i.e. as a TISE rather than as a Dirac equation or even more detailed modelling).  

\noindent In RPM's, linear constraints and inhomogeneities are logically-independent features, in contrast with how in minisuperspace both are 
simultaneously trivialized by the derivative operator becoming meaningless.
Thus, overall, RPM's permit study of each of the midisuperspace-like features of linear constraints and inhomogeneities 
without these complicating each other or being complicated by the infiniteness and mathematical complexity of midisuperspace. 
By including such features, study of RPM's brings into sharp focus a number of issues outside of those found and studied by use of minisuperspace.  

\noindent Uses of RPM's in these sorts of contexts include unveiling operator ordering problems, toy modelling structure formation, further investigation of effect of linear 
constraints on various POT strategies and the exhibition of a number of closed-universe effects. 
%
{            \begin{figure}[ht]
\centering
\includegraphics[width=0.9\textwidth]{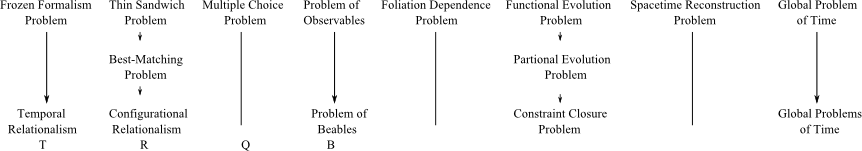}
\caption[Text der im Bilderverzeichnis auftaucht]{\footnotesize{Evolution of conceptualization and nomenclature of POT Facets over the course of this Article.
The Multiple Choice Problem is entirely quantum, the other 7/8ths are already present in classical GR.
The 5/8ths of the POT already present in Classical Mechanics suitably reformulated are Temporal Relationalism, Configurtional Relationalism, the Problem of Beables, 
the Constraint Closure Problem and (quite a lot of) the Global POT's.} }
\label{Evol-Fac}\end{figure}            }

\vspace{10in}

{            \begin{figure}[ht]
\centering
\includegraphics[width=0.89\textwidth]{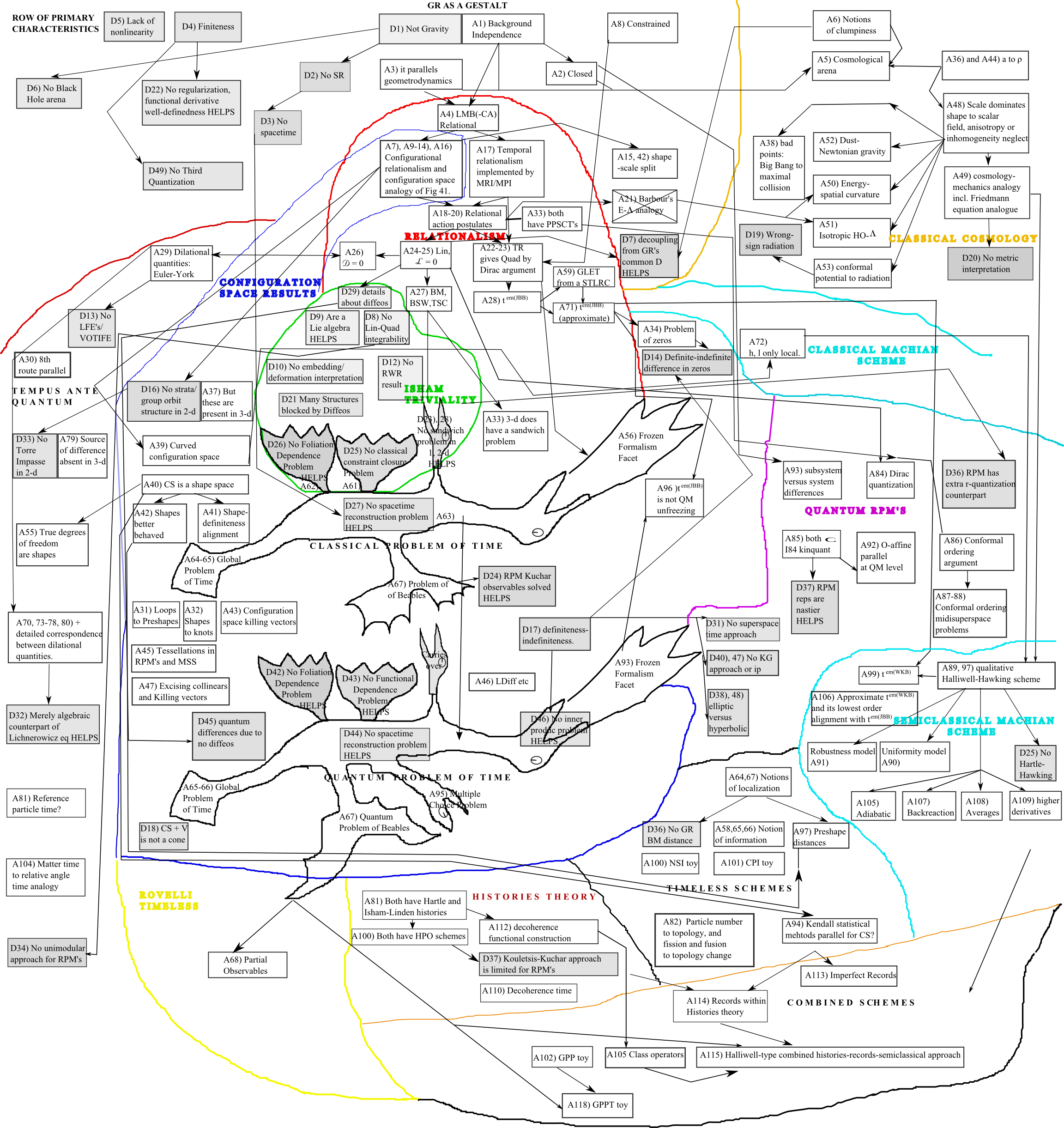}
\caption[Text der im Bilderverzeichnis auftaucht]{        \footnotesize{
\noindent The inter-relations between the 118 analogies and 49 differences between RPM's and GR.  
A superseded analogy is crossed out, differences are in shaded boxes and labels can then be pdf-searched for in the text for full statement of each.  
To explore this figure, blow up its pdf version by around 400 percent.
\noindent Note what is left of the Ice Dragon here: frozen breath but no teeth, one leg (Multiple Choice), a tail, only small wings and no scales.  
The lower parts of the figure reflect what strategies RPM models can be specifically geared toward.  
Arrows with no box at one end are from the entire region at that end.    }        }
\label{LastFig}\end{figure}            }
%
\noindent {\bf Question 96$^*$} (Long) Analyze the other toy models in \cite{Kuchar92} and the Preface to this extent, represent them by similar figures and compare them.  
In particular, conduct this analysis for Minisuperspace and inhomogeneous perturbations thereabout along the lines of Halliwell--Hawking \cite{HallHaw}.  
%
\subsection{POT Facets}\label{Concl-Facets}

\noindent [Secs \ref{Concl-Facets} to \ref{Concl-Strat} are the basis of the forthcoming article \cite{A13}: 
my Invited Talk at the December 2012 Cape Town Conference on Time `Do we need a Physics of Passage?']  


\noindent In this Article, I laid out the role of time in the various accepted theories of Physics.
An improved understanding of the POT facets based partly on relationalism and partly on various further strands of conceptual thinking and suitable generalizations was obtained, 
as summarized in Fig \ref{Evol-Fac}.  

\noindent The Frozen Formalism Problem follows directly from the quadratic constraint which arises from the MRI/MPI implementation of Temporal Relationalism. 
The Inner Product Problem is linked to relationalism at a very simple level (only inner products of wavefunctions are measurable and thus tangible); 
it is a time problem via its connection with conserving probability during evolution. 
I view the Inner Product Problem as a subfacet of the quantum-level Frozen Formalism problem, due to getting physics out of QM involving the inner product as well as the $\Psi$ 
that solves the (here frozen) wave equation; thus the inner product is the next level of structure after the wave equation.

\noindent Accompanying linear constraints are a direct consequence of Configurational Relationalism; in the GR case. 
The Thin Sandwich Problem itself is tied to 3-diffeomorphisms; in this Article I argued that this is a subcase of the Best Matching Problem 
(and showed that for the RPM's studied in detail here, it is a resolved problem), and that this itself is a subcase of Configurational Relationalism.    

\noindent The Problem of Beables consists of finding objects which brackets-commute with all the constraints (Dirac beables) or perhaps just with the linear constraints.
Especially the former is renownly difficult to approach in a satisfactory manner in the case of GR, even at the classical level. 

\noindent The GR Hamiltonian and momentum constraints form the split-diffeomorphism (alias Dirac alias hypersurface-deformation) algebroid.
\noindent The Functional Evolution Problem is viewed as part of the a posteriori compatibility for relational models, and is fortunately absent in this Article's RPM's. 
I generalize this to simply the Constraint Closure Problem so as to include the classical case which the Dirac algebroid's closure indeed resolves for classical GR.
Foliation-independence also classically guaranteed by the Dirac algebroid
As regards the Spacetime Reconstruction Problem, space/configurations/dynamics are primary, and spacetime may not exist as a meaningful concept at the level of Quantum Gravity.
Whether reconstruction procedures themselves are particularly relational needs somewhat more thought.  
This problem's classical incarnation has some ties to the split-diffeomorphism algebroid. 

\noindent One can see on the basis of the above summary that most of these facets are well-connected to relational ideas, 
so that relationalism may give a deeper account of whence some of these facets. 

\noindent The Multiple Choice Problem is purely quantum-mechanical.
\noindent Placing emphasis on only a subset of the canonical transformations may well help with this,  
but it is not concretely clear which subset would achieve subsequent entirely faithful representation by unitary transformations.  
Alternatively, there might just be some hope that in the perspectival approach, different unitarily-inequivalent quantum theories correspond to different observers' perspectives.  
This is likely much easier to refute by counterexample than it would be to establish it for a class of theories. 
\noindent Finally, as regards the Global POT's, the perspectival approach places less importance on this, e.g. via Bojowald's fashionables: 
observables perspectivally tied to a clock being useable in that region and epoch (see Sec \ref{ObsApp} for more).
The main view of relationalism in the present article for now has little to say about these last two POT facets (though these do lie largely beyond the present article's scope).

\subsection{RPM's as a POT arena: which facets they manifest}\label{Concl-RPM-POT}

\noindent RPM's have no Foliation Dependence Problem or Spacetime Reconstruction Problem due to foliations and spacetime not being meaningful concepts in this arena.

\noindent Also, an inner product is needed but there is no Inner Product Problem since RPM's, as mechanics theories, 
have positive-definite kinetic term and so the Schr\"{o}dinger approach will suffice.

\noindent All the other facets -- Temporal Relationalism, Configurational Relationalism, Problem of Beables, Constraint Closure Problem, Multiple Choice Problem, and Global POT's 
are represented.  
This is a case of background structure and quantization having these features for a wider range of theories than GR/Diff based theories.  

\noindent Moreover, upon inspection, RPM's are found to have trivial resolution of the Constraint Closure Problem, and, in 1 or 2-$d$ (which suffices for RPM's to toy-model 
Geometrodynamics whilst ensuring the mathematics remains tractable , the Best Matching Problem generalization of the Sandwich Problem has a fairly straightforward resolution 
(\cite{TriCl, FORD, Cones}, Sec \ref{Q-Geom}), and this permits one to have an explicit $t^{\se\sm(\sJ\sB\sB)}$ resolution of Frozen Formalism Problem as well as a full 
set of \K beables.

\noindent We shall see that these absences and fortunate resolutions fit in well with Halliwell's approach fit well together as regards eliminating  most of the  facets of the POT.

\subsection{POT Strategies}\label{Concl-Strat}

This article's `{\sc qbrt}' classification in terms of order of maps. 
{\sc q} itself has multiple steps though and is by itself in general a bad functor.  
I do not expect that to be ameliorated by {\sc qbrt} combinations...

This Article's particular suggested strategy for a local Problem of Time resolution is as follows.

\subsubsection{Level 1 Machian Approach}\label{Level-1}

\noindent This involves synthesizing Barbour's first work of 1994 \cite{B94I} concerning $\ft^{\se\sm(\sJ\sB\sB)}$ into the earlier \K and Isham POT scheme.
This gives {\it emergent relational} versions of Newtonian, proper and cosmic time in various contexts. 
The plain formula for this in the absence of Configurational Relationalism is 
\beq
\lft^{\se\sm(\sJ\sB\sB)}             =            \int||\d \bfQ||_{\sbfM}/\sqrt{\fW(\fQ)} \mbox{ } .
\eeq
It is a whole-universe LMB notion of time in principle, but not as regards practical and sufficiently accurate computations;   
taking these into account it is a LMB-CA notion of time, which then plays a substantial role in this Article's Tempus Post Quantum and Tempus Nihil Est treatments too.   
In the latter form, it is a `GLET is to be abstracted from STLRC' realization of ``Mach's time is to be abstracted from change", where STLRC stands for 
sufficient totality of locally relevant change and GLET for generalized local ephemeris time. 
I provided a perturbative Machian classical scheme for this in Sec \ref{+temJBB}.  
Here the configuration variables are split into slow heavy `$h$' ones and light fast `$l$' ones, of interest in all of island universes, Classical Cosmology and Quantum Cosmology.
This scheme is only fully Machian once one passes from the zeroth-order emergent times whose form is $F[h, \d h]$ to at least-first order emergent times of from $F[h, l, \d h, \d l]$, 
i.e. giving the $l$ degrees of freedom the opportunity to contribute.
The main application at present being cosmological, this Article mostly considers $h = I$, leaving $l = S^{\sfa}$, the shape degrees of freedom.
The zeroth-order emergent times remain relational recoveries of elsewise well-known timestandards, with the small first-order corrections to these then being 
Machian predictions for limitations on the accuracies on these (and what form the corrected timestandards are to take).  

\mbox{ }

\noindent To be more than formal for Configurational-Relationalism-nontrivial theories, $\ft^{\se\sm(\sJ\sB\sB)}$ requires the Best Matching Problem to have been explicitly resolved.  
I.e.  
\beq
\lft^{\se\sm(\sJ\sB\sB)} = \stackrel{\mbox{\scriptsize extremum $\sfg \mbox{ } \in \mbox{ }$} \sFG}
                                                               {\mbox{\scriptsize of $S^{\tr\te\tl\ta\ttt\ti\to\tn\ta\tl}$}}                                                              
\left(                                                              
                                                               \int||\d_{\sfg}\bfQ||_{\sbfM}/\sqrt{\fW(\bfQ)} 
\right)  \mbox{ } .
\eeq
\noindent The Best Matching Problem {\sl is} resolved for 1- and 2-$d$ RPM's (Sec \ref{Q-Geom}).  

\mbox{ } 

\noindent {\bf Major Open Question I}. On the other hand, Best-Matching Problem for GR is the Thin Sandwich Problem, which mostly remains unresolved.   
I suggest this Problem be very seriously attacked, including in broad-minded ways since some of its features are formalism and theory dependent.  
I am indeed aware that resolving Thin Sandwich Problem for GR is a tall order -- it is after all more specific than `merely' providing a full set of classical \K beables for 
GR, i.e. a specific construct of the Diff(3)-invariant 3-geometries that are (in the conventional sense) the gauge-invariant dynamical entities for GR. 
Somewhat shorter-term goals along these lines include consideration of the Halliwell--Hawking arena and of various non-perturbative midisuperspaces like 
the Einstein--Rosen cylindrical wave, spherical model and Gowdy models; these have a {\sl fair} track record for Sandwich-solving and beables finding \cite{Bubble, Kuchar94}.  

\mbox{ }

\noindent (\cite{ABook} shall report progress in the current Article's seven Major Open Questions; be welcome to contact me if you wish to work on any of these over the 
next few years...)

\mbox{ } 

\noindent Best Matching Problem Resolution Consequence 1) classical \K beables are then automatically available, particularly at the classical level.

\noindent Best Matching Problem Resolution Consequence 2) The Constraint Closure Problem is then resolved by there being only one constraint (per space point in field-theoretic case) 
-- the reduced $\scQ\scU\scA\scD$ -- which then straightforwardly closes with itself.  

\noindent Does the reduced version therefore require only a Bident? 
Not quite, since the Spacetime Reconstruction part's Dirac procedure is capable of bringing about further secondary constraints as integrabilities of $\widetilde{\scH}$, 
so away from the classical GR case one in general returns to needing to use a Trident.

\mbox{ } 

\noindent Aside 1) This Article's principal approach is a rival to the hidden/York time approach; this rivalry goes back to Wheeler's Thin Sandwich approach of the early 1960's 
and how York chose for his work in the early 1970's to diverge from that on technical grounds.  
Wheeler championed {\sl both} of these as Machian \cite{WheelerGRT, W88}, and indeed the various program of Barbour et al. since the early 2000's continue to afford 
both of these options as providing distinct Machian foundations for GR (\cite{RWR, ABFKO} and the present Article's Sec \ref{Examples}).  

\noindent Aside 2) Moreover, hidden time itself has been argued in the present Article to be less Machian; \cite{ABFKO}'s Machian conformogeometrodynamics {\sl need not} 
place any fundamental significance on the York time, but rather on the conformogeometrodynamical case of the emergent JBB time).
Matter time and unimodular time have also been argued to be less Machian, all these arguments are different conceptual arguments as opposed to the technical difficulties already 
known to plague most of these strategies.  
This Machian counter-argument comes about because some species of change are {\sl not} given opportunity to contribute to the purported timestandard.

\noindent Aside 3) Rovelli-type Timeless Approach-based combinations are not taken up in the present Article also on these more detailed Machian grounds and due to taking an 
entirely \K or Dirac stance toward beables.\footnote{I restrict
the principal combinations involving these alternatives to the present footnote.  

\noindent i) Kouletsis' \cite{Kouletsis08} tie between Histories Theory and the Internal Time Approach. 
This approach has for now the setback of mostly only having been investigated at the classical level so far. 
I encourage further work at the quantum level for this strategy!

\noindent ii)  There are a number of ties between Rovelli-type Timeless Approaches and the Internal Time Approach.  
Internal time can provide the time/clock in schemes requiring a such if one is able or willing to pay for the usual inconveniences of a such (e.g. Dittrich \cite{Ditt, Dittrich}).
Note that here these do constitute at least an emergent Schr\"{o}dinger picture. 
How does this square with Rovelli's previous insistence on solely the Heisenberg picture being of use in QG?

\noindent iii) The above triangle of ties can be completed by combining Rovelli-type Timeless Approaches and 
Histories Theory as done e.g. by Thiemann \cite{Thiemann}; \cite{APOT} carries some discussion of this.  }

\subsubsection{Level 2 Machian Approach}\label{Level-2}

Emergent JBB time fails to carry over as a Frozen Formalism Problem resolver at the quantum level.
However, emergent semiclassical time is its sequel. 
I recast this in Machian form in the present Article and in \cite{ACos2}.
Sec \ref{Semicl} carries a perturbative Semiclassical Machian Scheme for the RPM analogue of semiclassical Hartle--Hawking-style Quantum Cosmology.
In this scheme, $\ft^{\se\sm(\sW\sK\sB)}$ is $\ft^{\se\sm(\sJ\sB\sB)}$ to zeroth order but not so to higher (perturbative) order, as is clear from 
{\sl quantum} change being given the opportunity to enter the former Machian timestandard.  
One passes from an emergent Machian time of the form ${\cal F}[h,l, \d h, \d l]$ to one of the form ${\cal F}[h, l, \d h, |\chi(h, l)\rangle]$ which takes into account 
that the light subsystem has passed from a classical to a quantum description, so that {\sl quantum} change is now being given an opportunity to contribute.
%

\noindent Also, Configurational Relationalism {\sl remains} resolved: having reduced at the classical level, quantization does not unreduce the system.  
The classical restriction of the \K beables to a set of Dirac beables has to be abandoned. 
However, the quantum \K beables are obtained by promoting some subalgebra of the classical \K beables to quantum operators.  
And then Halliwell also provided a semiclassical construct for objects commutator-commuting with the quadratic constraint, which is now to be used to construct 
a set of quantum Dirac beables as functionals of the quantum \K beables.
Next, Constraint Closure remains a non-issue at the quantum level for RPM's.  
Foliation and spacetime reconstruction issues being absent from RPM's, that means that we are done as regards providing {\sl a local} resolution of the POT for this RPM's?

\mbox{ } 
  
\noindent Problems with Level 2 are as follows; the first is a major problem, whilst the first and the second suggest conceptual and technical cautiousness be 
exercised before `going to town' with solving any particular perturbative schemes (whilst the input maths is `as easy as hydrogen' term by term, it is not yet clear 
which terms it is truly consistent to discard in each physically relevant regime...)

\mbox{ }

\noindent 1) Developing this strand by itself runs into `why WKB ansatz' problems.  
The applicability of the WKB ansatz is truly important in the sense that if this does not apply, one no longer gets anything like a TDSE for the $l$-physics.     
Moreover, I showed that this ansatz does not hold in all regions of configuration space.  

\noindent In my opinion and that of a number of other authors \cite{Kieferbook, Zehbook, GPP}, the `why WKB ansatz' debacle requires investing in histories and/or timeless approaches.
I build these up to Level 1 (classical) and Level 2 (quantum, suitable for subsequent combination with semiclassical) as separate strands and then consider combinations among them.

\mbox{ } 

\noindent 2) The failure to incorporate central terms in the classical counterpart scheme, and 2-body problem within 3-body problem, Euler within Navier--Stokes and 
emergent-TDSE within an emergent-time-dependent Hartree--Fock scheme qualitative examples that I discussed in this Article substantiate 
qualitative suspicions of some of the other approximations being made in the Semiclassical Approach.  
As such it may be worth questioning for the moment whether one actually believes in semiclassical quantum cosmological calculations, and this leads to my envisaging these 
will have to be replaced with a more complicated coupled `self-consistent' scheme much as was required to correctly start to predict atomic and molecular spectra in the 1930's. 

\mbox{ } 

\noindent 3) The order of quantization and declaration of beables stated above is also cause for at least some concern.
Quantization complicates status of beables via the subalgebra selection criterion (and its own associated Multiple Choice Problem).  
Are the quantum Dirac beables {\sl well-defined} as functionals of the quantum \K beables? 
%
%
What algebra do these quantum Dirac beables obey among themselves?  
It does not look at all likely that this is a subalgebra of the algebra the quantum \K beables obey among themselves; is this a problem? 
Finally, this also a case in which the kinematical quantization algebra and the two quantum observables algebras are very likely to be separate 
mathematical entities.

\subsubsection{Combination of 3 mutually-supporting approaches}

I prop up the principal `why WKB' deficiency using Histories Theory.  
This is best done balancing the three of Timeless Records, Histories Theory and Machian schemes.\footnote{It may be possible to supplant histories with just paths.}
This requires starting from Level 1 again in each of the Timeless Records and Histories Theory strands.  
There is no obstruction to doing this with RPM's.  
The IL version of Histories Theory starts classically and is readily combineable with my classical POT resolution along lines of Barbour 
-- I view these as extra constructs in the usual manner except that they are now based on emergent JBB time.  
Gell-Mann--Hartle's version of Histories Theory starts at the quantum level, though there is sizeable conceptual and technical convergence between the two approaches 
-- I view IL's work as an improved formalism for the main Gell-Mann--Hartle tenets that just also happens to come with a classical precursor.
\noindent Finally, if histories are assumed, the classical part of Halliwell's construction promotes a particular subset of the Kucha\v{r} beables to Dirac beables.  
%

{            \begin{figure}[ht]
\centering
\includegraphics[width=0.82\textwidth]{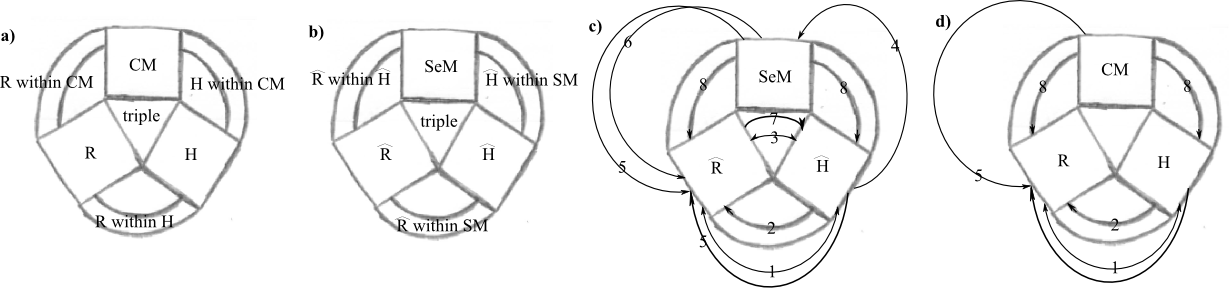}
\caption[Text der im Bilderverzeichnis auftaucht]{\footnotesize{a) I represent these with three squares back to back for the pure programs, 
annular sectors between these for the three pairwise combinations and the filling between the three squares for the triple combination.  
This is at the classical level. 
CM = Classical Machian Scheme (\cite{ACos2}, Secs \ref{Cl-POT} and \ref{+temJBB}), 
R = Timeless Records (Secs \ref{TNE}, \ref{TNE-2} and \ref{Cl-Nihil}), 
H = classical Isham--Linden (IL) type Histories Theory (\cite{IL}, Secs \ref{Histor}, \ref{Histor-2} and \ref{Cl-Hist}).
R within CM is as per Fig \ref{Cl-Nihil-Class},  
H within CM is the particular specialization of Secs \ref{CM-H} and \ref{Cl-Hist}, and 
R within H is the classical Records-within-IL-Histories 
analogue of Gell-Mann--Hartle's better-known quantum inclusion \cite{GMH}, Sec \ref{TNE}.  
The triple combination is my Machianization (\cite{AHall}, Secs \ref{Cl-Combo-Intro}, \ref{Cl-Combo}) of Halliwell's classical prequel \cite{H03}.  

\noindent b) The quantum counterpart.
SeM = Semiclassical Machian scheme (\cite{ACos2}, Secs \ref{QM-POT} and \ref{Semicl}), 
$\widehat{\mR}$ = Quantum Records (Secs \ref{QM-Nihil-Intro} and \ref{QM-Nihil}) and 
$\widehat{\mH}$ = Quantum IL type Histories Theory (\cite{IL}, Secs \ref{QM-Histories} and \ref{QM-Hist}).  
$\widehat{\mR}$ within SeM is as per Secs \ref{QM-Tless-Classi} and \ref{QM-Nihil} , 
$\widehat{\mH}$ within SeM is the particular specialization of Secs \ref{HarRPM2} and \ref{QM-Hist}, and 
$\widehat{\mR}$ within $\widehat{\mH}$ is the IL version of Gell-Mann--Hartle's better-known quantum inclusion \cite{GMH}, Sec \ref{QR-in-QH}. 
The triple combination is my Machianization (Secs \ref{QM-Halliwell-Intro}, \ref{QM-Combo}) of Halliwell's semiclassical working in \cite{H03}.  

\noindent a) and b) are the main structural basis of the largest `Magic Swords', c.f. Figs \ref{Magic-Sword} and \ref{GR-Magic-Sword}.

\noindent c) Detail of the actual bindings and interprotections between the three pure programs at the quantum level. 
1 = both rest on atemporal logic (`Mackey's Principle'). 
2 = $\widehat{\mR}$ within $\widehat{\mH}$ (one of the annular segments)
3 = Gell-Mann Hartle's `records are somewhere in the universe where information is stored when histories decohere'.
4 = `decoherence $\Rightarrow$ WKB' -- the crucial support needed to further justify the original SM scheme past Level 2.  
5 = `provide a semblance of dynamics or history' -- overcoming present-day pure records theory's principal weakness.  
6 = semiclassical regime aiding in the computation of timeless probabilities (i.e. Halliwell's scheme).
7 = answering the elusive question of `what decoheres what' from what the records are.
8 = providing a Machian scheme for QR and QH to sit inside (the other two annular segments).  

\noindent d) Those of the bindings that still hold at the classical level, though, as is evidenced by comparison with b), the triple combination is 
primarily motivated by quantum-level interrelations.  
One is lost due to its involving quantum probabilities, and three more are lost due to concerning decoherence.   
Moreover, item 1 is far less significant classically, since logical nontriviality is primarily a quantum issue.  
 
\noindent Those parts of c) and d) that are not annular segments represent the structure of the triple combination's quantum `blade' and the simpler 
triple combination's `handgrip', c.f. Fig \ref{Magic-Sword}.
The basic construct in these figures is a `card house': cards do not stand alone but can prop each other up to forms a standing structure.} }
\label{Cards}\end{figure}            }

\subsubsection{Level 1 and 2 Timeless Approaches}

\noindent Quantum-level Records are also a very natural successor to classical relationalism.

\mbox{ } 

\noindent Example 1) LMB relationalism to Barbour-records (with or without additional conjectures). 

\noindent Example 2) LMB-CA relationalism to localized A-Records, though one can also think of the \CPII, Gambini--Porto--Pullin and Halliwell approaches in LMB-CA terms.  
LMB-CA permits Classical Machian--A-Records and Semiclassical Machian---A-Records combinations.  

\mbox{ } 

\noindent This Article included a wider Classification of timeless approaches, leading to a large expansion on number of known types.
The ones I use lie within Type 1 timeless as opposed to Rovelli-type timeless.  


\noindent The specific features of timeless records approaches were handled as follows. 

\noindent \bu I consider records to be localized subconfigurations of a single instant that are useful as regards reconstruction of a semblance of dynamics.  
Notions of localization in space and in configuration space are useful here.   
I discuss what criteria a notion of distance should satisfy if it is to apply to a relational theory: axioms of distance, $\FrG$-invariance, sufficient generality, naturality (dynamical 
rather than just kinematical, compatible with the physical laws of that system), perhaps the possession of a QM counterpart, or a structural analogue in QM or SM).  
Notions of distance include configuration space metric based ones such as the RPM action itself.   
However, these are indefinite in the case of GR, therefore not at all serving as a notion of distance in that case [against 1.0)].  
One possible resolution of that is that the pure-shape part of an object of that nature serves for both RPM's and GR however -- 
a new interpretation of the conformal gravity action \cite{ABFO}. 

\noindent \bu A suitable notion of localization in space and in configuration space may be hard to come by and/or to use for Quantum Gravity in general.  
I.e., `where' particular records are can be problematic to quantify, and the records can be problematic to access and use too, 
since the relevant information may be `all over the place'.  
 
\noindent \bu Information may additionally in practise be problematic firstly as it may be of too poor a quality to reconstruct the history. 
Secondly, because a suitably general notion of information is missing from our current understanding of classical gravity 
(ref Sec \ref{Cl-NOI} for references), never mind Quantum Gravity with its unknown microstates. 
(Mechanical toy models are useful in not having this last obstruction).  
Finally, the further Records Theory notions of significant correlation patterns and how one is to deduce dynamics/history from them looks to be a difficult and unexplored area 
even in simpler contexts than gravitation.  
 
\noindent \bu In addition to clumping patterns being capable of constituting useful records, 
Kendall's work is a tour de force in finding statistically-significant relative angle based patterns in 2-$d$ data in what can certainly be described as a relational setting.  

\noindent \bu As regards semblance of dynamics, the two purely atemporal alternatives known are 

\noindent 1) Barbour's conjecture based on configuration space shape somehow bringing about 
bubble chamber track-like phenomena with sufficient frequency to account for the richness of records around us. 

\noindent 2) Page's in-principle but not in-practise approach involving a single configuration containing `memories of `previous configurations'.  
I provided in particular a step-by-step critique of the former 
(the main problem with the latter is that one knows very little as to how to calculate, or even toy-model, with such a scheme).  

\noindent 3) See also the combined strategy section for non-purely atemporal ways round the problem of whither the semblance of dynamics.

\subsubsection{Timeless--Machian--(semi)classical combos}

\noindent In this article I also considered the embedding of timeless approaches within Machian classical and Machian semiclassical schemes.  

\noindent The prototype of this is how Halliwell--Hawking 1985 \cite{HallHaw} semiclassical scheme was followed up by Halliwell 1987 \cite{Halliwell87} 
on timeless correlations within such a scheme.

\noindent The current Article takes this sort of scheme further by considering the classical precursor, Machianizing both the classical and semiclassical versions of the 
scheme and using A-records (and/or \CPI and variants in the quantum half of the scheme).

\subsubsection{The other two combinations of two strategies}

\noindent \bu The current article used $\ft^{\se\sm(\sW\sK\sB)}$ and $\ft^{\se\sm(\sJ\sB\sB)}$ as emergent times with respect to which classical and semiclassical Histories Theory 
is to run.  

\noindent \bu To complete the pairwise approaches, Gell-Mann--Hartle \cite{GMH} and Halliwell \cite{H99} considered timeless Records within Histories Theory.  
One needs the IL version of Histories Theory to do this at the classical level, but there is no difference between Records within Gell-Mann--Hartle's scheme 
and within IL's scheme, so the former's notion of included Records Theory directly carries over to the latter.  

\noindent \bu At the quantum level, the IL formulation of Histories Theory is chosen for use in my scheme since it succeeds in implementing the Projectors Implementation 
of Propositions.

\subsubsection{Level 3 Machian Approach: Machian variant of Halliwell's work for the Triangle}
%
\noindent The final triple combination is a Machianized version of Halliwell's work also extended to nontrivial Configurational Relationalism.  
This is the `blade of the Magic Sword', the further interprotective `magic' of which can be read off Figs \ref{Cards}c),d) and \ref{Composorda}.  
%
{            \begin{figure}[ht]
\centering
\includegraphics[width=0.7\textwidth]{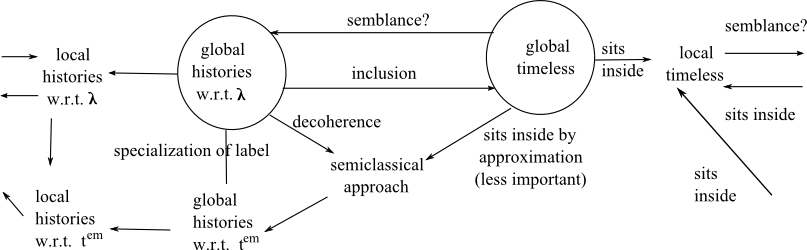}
\caption[Text der im Bilderverzeichnis auftaucht]{        \footnotesize{I now want to logically order the interprotections in an `unfolded' version of the blade.  
This cuts the Gordian knot of which of the 3 strategies in the combined approach comes first.
The result is: {\sl meaningless label} histories comes first, this gives the Semiclassical Approach and then this gives the emergent-time version of the histories approach.  
Then localized timeless approaches sit inside the last two of these.  
On the other hand, the Semiclassical Approach sits inside the global timeless approach as is manifest from comparing 
Parts III and IV of the present article, but global timeless approach can be taken to sit within global meaningless label time histories approach, so down both strands this is primary.  
}    }
\label{Composorda} \end{figure}          }
%
{            \begin{figure}[ht]
\centering
\includegraphics[width=0.65\textwidth]{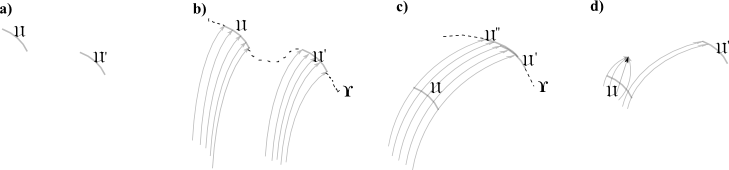}
\caption[Text der im Bilderverzeichnis auftaucht]{        \footnotesize{a) Consider now two regions $\FrU$ and $\FrU^{\prime}$ -- pieces of hypersurfaces, 
generally distinct, $\Upsilon$ and $\Upsilon^{\prime}$, within configuration space $\FrQ$.
b) Sometimes there will exist a single section (global section-hypersurface) $\Upsilon$ orthogonal to the classical flow that includes both 
$\FrU$ and $\FrU^{\prime}$.  
Then via (\ref{33}) or (\ref{36}), P$_{\FrU\,\,\,\sO\sR\,\,\, \FrU^{\prime}}$ = P$_{\FrU\,\bigcup\, \FrU^{\prime}}$ just as for the \NSII.   
c) However, if $\FrU$ and $\FrU^{\prime}$ do not lie in the same section, it is not directly like the situation for the \NSII. 
Now one can consider the flow-evolution $\FrU^{\prime\prime}$ of region $\FrU$, say, so that it lies in a section $\Upsilon$ that extends 
$\FrU^{\prime}$; one can now compose $\FrU^{\prime}$ and $\FrU^{\prime\prime}$ just like one composed in case a). 
d) Moreover there need not always exist a section that extends $\FrU^{\prime}$ whilst also containing a flow evolution image $\FrU^{\prime\prime}$ of $\FrU$. 
The construction of counterexamples, however, can avoid this case. 
What this case represents is that the composition of Halliwell's implementation is not always reducible to a parallel of the \NSII's composition.
Sometimes local sections and meshing conditions between them suffice for composition, but this does not always hold either. 
These non-existences reflect that some flows can be pathological/exhibit breakdowns in well-definedness or smoothness.}    }
\label{FigLast} \end{figure}          }
 
\vspace{10in} 
 
\noindent Caveat 1 with the Combined Approach) Whilst Halliwell's scheme does not a priori use a Na\"{\i}ve Schr\"{o}dinger inner product, it still does use a Na\"{\i}ve
Schr\"{o}dinger-type implementation of propositions by configuration space regions (or maybe a generalization to phase space regions).  
This suffers from probabilities corresponding to regions composing too simply (Booleanly) for one to be able to represent all QM propositions in such a form.  
E.g. quantum propositions may be taken to obey quantum logic \cite{QLog, I94, IL2}, which is nondistributive, unlike the composition of 
configuration space regions, which is distributive [and elsewise Boolean].  
This is immediately obvious for the \NSII, but requires a bit more thought to establish in the case of a Halliwell-type program.  
This is can be seen from how having probability 1 for a slice of a flux tube implies probability 1 in other slices of the tube.  
I proceed here by taking sections of configuration space; on these, the Booleanness of region composition is as clear as for 
the \NSII.  
As Fig \ref{FigLast} explains, such sections may not always be available, even for composing regions with flow-evolutions of other 
regions along tubes, but they are available enough to be able to have counter-examples (cases in which QM propositions do not obey a Boolean algebra but a 
section {\sl is} available so as to carry out composition of probabilities within Halliwell's scheme).  
See Sec \ref{VII} for possible way out (Problem VII).  Problems I) to VI) can also be seen as caveats to the Machianized Halliwell Triple Combination [I) to IV) just for 
GR, but V) to VII) even for Mechanics].  
%
{            \begin{figure}[ht]
\centering
\includegraphics[width=0.95\textwidth]{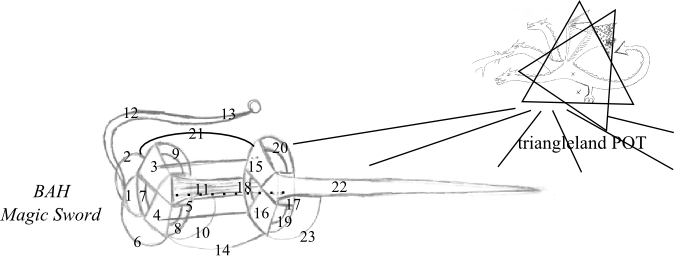}
\caption[Text der im Bilderverzeichnis auftaucht]{        \footnotesize{This is the `Magic Sword' for slaying the local 1- or 2-$d$ RPM Toy `Ice Dragon', 
i.e. the functional rather than developmental or discursive core of the present Article.
The `pommel' 1 is the resolution of Configurational Relationalism by explicit completion of Best-Matching (\cite{FORD, Cones}, Sec \ref{Q-Geom}).
2 is $\ft^{\se\sm(\sJ\sB\sB)}$ as the subsequent explicit resolution of Temporal Relationalism's classical Frozen Formalism Problem (\cite{B94I} and Sec \ref{Cl-POT})).  
3, 4, 5, 7, 8, 9 are covered by Figure \ref{Cards}a).
6 is the trivial possession of a set of classical \K beables as a second consequence of 1 (\cite{AHall}, Secs \ref{Gau} and \ref{Beables}).
10 is finding a subset of Dirac beables among these using Halliwell's classical construct (\cite{H03, AHall} and Secs \ref{Cl-Combo-Intro}, \ref{Cl-Combo}).
The handgrip 11 is covered by Figure \ref{Cards}c).  
The three thick black lines (one dotted) are the promotion of each of CM, R and H to their (semiclassical) quantum counterparts as per Secs \ref{QM-Intro}, \ref{QM-POT}, 
\ref{QM-POT-Strat}, \ref{Semicl}.  
12 and 13 form the `handguard of good fortune' by which the classical and quantum constraint algebras close (Secs \ref{Examples} and \ref{QM-Intro}).  
21 is $\ft^{\se\sm(\sW\sK\sB)}$ as the subsequent explicit resolution of Temporal Relationalism's quantum Frozen Formalism Problem \cite{DeWitt67, HallHaw, SemiclIII, ACos2}.    
15, 16, 17, 18, 19, 20 are covered by Fig \ref{Cards}b). 
14 is the promotion of a subalgebra of classical \K beables to quantum \K beables (Sec \ref{QM-KO}).
23 is finding a subset of Dirac beables among these using Halliwell's semiclassical construct (\cite{H03, AHall} and Secs \ref{QM-Halliwell-Intro} and \ref{QM-Combo}).
The `blade' 22 is my Machianized version of Halliwell's quantum triple combination as covered by Fig \ref{Cards}d).   
The logical ordering of the explanations of the parts of the sword are 1, 2, 3, 21, 15 and 1, 6, 14 (Levels 1 and 2 of Secs \ref{Level-1} and \ref{Level-2}), which then 
need reinforcing by the Combination of 3 mutually-supporting approaches' tripling that turns 3 into the complex 3, 4, 5, 7, 8 , 9, 11 and 15 into the complex 15, 16, 17, 18, 19, 20, 22.    
Then, provided that at least 4, 15 and 16 are in play, the `tassel' 6,14 can be expanded upon to form the `rein' 6, 10, 14, 23 by adding the `Dirac links' 10 and 23 that 
are crucial for the Toy Unicorn (Toy RPM, Unicorn in the sense of \K) to accept to be ridden. 
The `crossguard' protects one from this Toy Ice Dragon's Constraint Closure Problem claw, and, locally, one is done since this Toy Ice Dragon has no 
Foliation-Dependence claw or Spacetime Reconstruction scales.
By counting the occurrences of the authors involved in its pieces, I term this the BAH Magic Sword, where the B stands for `Barbour' and the H for `Halliwell'. } }
\label{Magic-Sword}\end{figure}            }

\subsubsection{GR counterparts of the handgrip and the blade}

\noindent{\bf Major Open Question II}. The Dirac beables procedure from Halliwell's scheme has not been carried out for midisuperspace or beyond.  

\mbox{ } 

\noindent{\bf Major Open Question III}. The triple combination of Halliwell and its Machianized conceptual upgrade have not been done for midisuperspace or beyond.

\subsubsection{Comment on more minimalist Magic Swords}\label{Mini-Swords}
%
%
%
{            \begin{figure}[ht]
\centering
\includegraphics[width=0.75\textwidth]{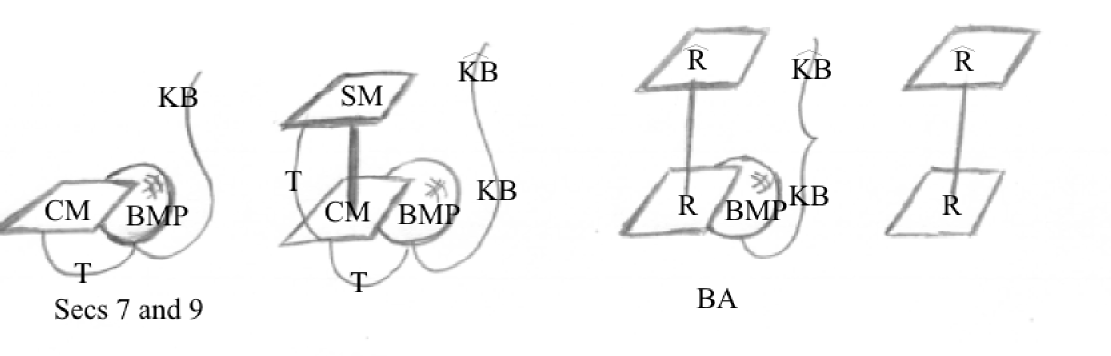}
\caption[Text der im Bilderverzeichnis auftaucht]{        \footnotesize{a) is Secs 7 and 9's level-1 Machian scheme.

\noindent b) is the preceding boosted by Sec \ref{QM-POT} and \ref{Semicl}'s level-2 semiclassical Machian scheme.

\noindent a) and b) were commented upon as the logical core of the full `Magic Sword' in Fig \ref{Cards}. 

\noindent c) and d) represent using instead just timeless elements. c) is Barbour's Records scheme boosted by my 
localization and observations about \K beables. Or, alternatively, my localized rather than whole-universe version of the same (A-Records).  
d) is Page's scheme.
} }  \label{+Minimalist-Magic-Swords}\end{figure}            }

\noindent One can take Page-Records or Barbour-Records or my local version of the latter as a timeless-only approach.  
This is the most minimal way of attempting Level 3, without using the above attempt at Level 2 via the semiclassical.  
Barbour's set-up \cite{B94I, B94II, EOT} is of this form. 
Page's set-up, on the other hand, never considers a classical precursor.  
My STLRC and then localized records approach can be assembled as a more practical version of the \cite{B94I, B94II, EOT} set-up.  
This Article has ample amounts of examples that could be used to make (a localized version of) \cite{B94I, B94II} that is made out of 
mathematical concreteness rather than conjecture and speculation.  
Minimalism of whichever of these kinds, however, runs into a wall. 
Page's approach is not practically implementable for becoming replaced by being.  
What concrete mathematics the present Article has evaluated does not reproduce Barbour's conjectured mechanism.  

\mbox{ } 

\noindent The Gambini--Porto--Pullin approach can also, at least at a conceptual level, be cast as similar, by (not like its own authors) 
considering it to be how to handle correlations in something like the Halliwell--Hawking arena. 
Details of compatibilities of this remain open.  
The reason the Gambini--Porto--Pullin approach per se isn't as minimalist and general as Page is that it requires a sense of semiclassicality, 
so it is a combined scheme, if not usually cast as a Machian such.
One can view Arce's work along these lines too -- another computation of conditional proposition objects within a semiclassical regime, but using somewhat different concept 
and method -- a different selection of ansatz/semiclassical conditions.  
I note that, as things stand, among these more minimalist Level 3 approaches only the Gambini--Porto--Pullin one provides a decoherence mechanism justification of WKB.
Arce's has extra lacks as a very new candidate -- it is not clear yet if it is compatible as a correlation-finder {\sl within a Halliwell--Hawking semiclassical scheme}. 
It is also plausible that elements of these four 2-strand Level 3 attempts may be combineable -- e.g. other schemes within might inherit Gambini--Porto--Pullin's 
means of attaining decoherence if that is held to be a desirable feature.

\mbox{ }

\noindent Note 1) (Blade fault) The lack of Histories Theory (or perhaps Path Integrals) takes away the most plausible WKB explanation.  
There is some chance to be able to do so based on the Gambini--Porto--Pullin approach instead.  
Here one would take a specific \CPII-Gambini--Porto--Pullin stance on the procedures for a quantum A-records scheme embedded within semiclassical Machian scheme.

\noindent Note 2) This approach is not known to have a quantum Dirac-beables either.  
Gambini--Porto--Pullin--Torterolo \cite{GPPT} do at least consider notions of observables for this scheme.
As this lies within the scope of what can be done for RPM's, it is Analogy 118).

\noindent Note 3) Everything in the blade can indeed be defined just for paths, at least at the conceptual level. 
At the technical level, the jury is still out as to whether this more minimalist view is possible.

\subsubsection{The next three facets for GR}

\noindent At the classical level for GR, one can down all three of these -- Constraint Closure Problem, Foliation Dependence Problem and Spacetime Reconstruction Problem 
-- using Dirac's Trident, as summarized in the below figure.  
That these are absent from QBRT is already an indication of the rough factorization of `a local' into `{\sc qbrt}' and `{\sc cfs}' components..
%
{            \begin{figure}[ht]
\centering
\includegraphics[width=0.7\textwidth]{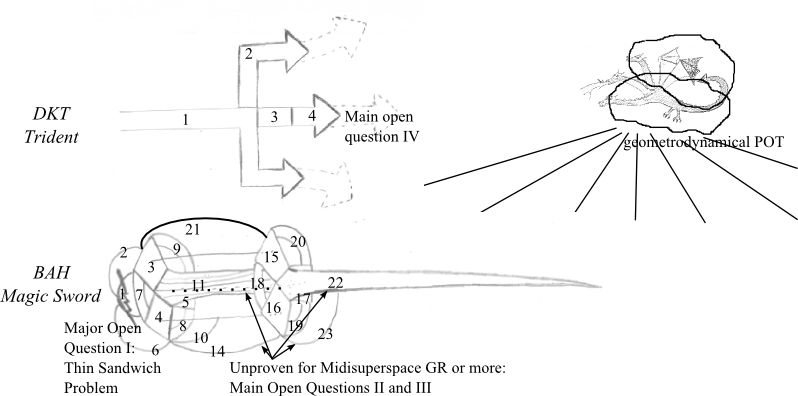}
\caption[Text der im Bilderverzeichnis auftaucht]{        \footnotesize{This is the `Magic Sword' and `Trident' armament for slaying a local GR `Ice Dragon'.  
Its tentativeness picks out the major open questions posed in this Article (last paragraph of caption). 

\noindent The enumeration of the sword is as before, though references and comments change.  
1 is the Thin Sandwich Problem, posed in \cite{BSW, WheelerGRT}, worked on in \cite{TSC1, TSC2} and commented on in Secs 3.16 and 8.8.
2 is the classical Machian resolution of the Frozen Formalism problem built up from \cite{B94I, SemiclI} in Secs \ref{Intro} and \ref{Examples}). 
3 is the classical Machian scheme of \cite{ACos2} and Secs \ref{Cl-POT} and \ref{+temJBB}.  
4 is A-Records (\cite{Records}, Secs \ref{TNE}, \ref{TNE-2} and \ref{Cl-Nihil}) which owes a fair amount to the precursor references listed in Sec \ref{TNE}.
5 is the IL form of classical histories theory for GR of (\cite{IL}, Sec 12).
6 is for now only formal (Sec 8.11's identification of the GR observables, which goes back to Wheeler: \cite{Battelle} and references therein).
7 is the classical IL counterpart of inclusion within Gell-Mann-Hartle's scheme of timeless records, Sec 8.7.  
8 is Sec 12's classical Machian histories for GR.
9 is records within Machian classical scheme for GR as posed in Sec \ref{Cl-Nihil-Class}.  
10 is Halliwell's construction \cite{H03} extended to the case in terms of \K beables (\cite{AHall}, Sec  \ref{Cl-Combo-Intro}) so as to construct Dirac beables, 
now extended to the GR case.  
11 involves the overlaps of Fig \ref{Cards} d).  
14 Quantum \K beables are now but doubly formal (Sec 23).
15 The Machian semiclassical scheme for GR as per Secs 20.2.4 and \ref{Semicl}.
16 Quantum A-records for GR as posed in \ref{QM-Nihil-Intro} and \ref{QM-Nihil}).
17 IL quantum histories for GR (\cite{IL}, Secs \ref{QM-Histories} and \ref{QM-Hist})
18 Gell-Mann--Hartle records within histories uplifted to the quantum IL scheme of Sec \ref{QM-Hist}.
19 is SSec \ref{HarRPM2} and \ref{QM-Hist}'s Machian semiclassical histories for GR.
20 is Sec \ref{QM-Tless-Classi} and \ref{QM-Nihil}'s A-records within Machian classical scheme for GR.
21 involves using the Semiclassical Approach as Frozen Formalism Problem resolver.
22 is my Machianization of Halliwell's triple combination for GR (Secs \ref{QM-Halliwell-Intro}, \ref{QM-Combo}) of Halliwell's semiclassical working in \cite{H03}).  
23 is Halliwell's semiclassical construction of objects commuting with the quadratic constraint likewise extended to the GR case.  

\noindent The enumeration on the trident is 
1) Dirac's algebroid of the GR constraints and its classical closure \cite{Dirac}; 
casting it in ADM \cite{ADM} and \K \cite{Kuchar76} forms helps with some of the following developments.  
2) Teitelboim \cite{T73} and Hojman--\K--Teitelboim \cite{HKT}'s foliation-independence.
3) Hojman-\K-Teitelboim \cite{HKT}'s obtaining $\scH$ from less assumptions via the nature of the deformation algebroid manifestation of the Dirac algebroid.  
4) Barbour--Foster--O Murchadha and I (\cite{RWR, Phan, Lan2} and Sec 2.6.7) obtaining $\scH$ from even less assumptions; by now classical spacetime is being 
deduced rather than supposed.  
This has earlier roots as per Secs 2.6.2 and 2.6.6. 
%

\noindent The Major Open Problems depicted here are as follows.  

\noindent I) The `pommel' 1 being the resolution of Configurational Relationalism by explicit completion of Best-Matching, for GR this is presently `cracked' 
since this is now the resolution of the Thin Sandwich Problem, and only limited progress has been made on this by humanity since it was posed in 
the early 1960's by John Archibald Wheeler.  
This renders t$^{\se\sm(\sJ\sB\sB)}$ and the \K beables only formal for now.   

\noindent II) The `rein' is unproven for midisuperspace or better as a Problem of Beables Resolution by Unicorn in the sense of \K.  

\noindent III) The `handgrip and blade' of Halliwell's classical and semiclassical schemes are likewise unproven at this level of modelling complexity.  

\noindent IV) The Trident is only established classically; one requires it to be semiclassical (and in a sense compatible with the semiclassical Machian approach).

\noindent By counting the occurrences of the authors involved in its pieces, I term this the BAH-DKT armament, where the D stands for `Dirac', K for `\K' and T for `Teitelboim'.    
} }
\label{GR-Magic-Sword}\end{figure}            }
%
%
\noindent {\bf Major Open Question IV}.  
The Sword--Trident factorization of one's armament/strategization is clear-cut, at least to good approximation at the conceptual level.  
It is based on just the Trident involving the whole of the split-diffeomorphism algebra (one can treat Dirac beables conditions piecemeal or try to stick with \K beables conditions).
The question then is whether a semiclassical version of Dirac's Trident can be built.  

\mbox{ } 

\noindent The Homogeneity Trident for classical and quantum Minisuperspace in Sec \ref{Cl-POT} is a case that is straightforwardly resolved both classically and quantum-mechanically.

\mbox{ }

\noindent {\bf Question 97$^{**}$} Look to make first progress on Major Open Question IV by looking at Halliwell--Hawking scheme \cite{HallHaw}. 
Here have nontrivial diffeomorphism information only to first order: the zeroth order needs none and the second order is discarded; 
as well as this simplification, life is also made simpler by the scheme being merely semiclassical.  
If still stuck, one might switch to the analogue of this scheme for such as 2 + 1 GR or the bosonic string, though these schemes are of less value to quantum, and 
-- via inflationary expansion of the QM fluctuations to CMB inhomogeneities, galaxies -- classical, Cosmology.

\mbox{ }  

\noindent For some other directions with constructing what I term Dirac's Trident at the quantum level, see \cite{Bojo12}.  

\mbox{ }  

\noindent N.B. that so far we are asking for {\sl a} local resolution of POT, which looks to be a coherent if incomplete subquestion to investigate.  

\vspace{10in}

\subsubsection{Major Open Question V. Multiple Choice Problem}\label{MCP-Ex}

From Groenewold \cite{GVH1} to Gotay's review \cite{Gotay00}, the specific examples worked out are (they are phase spaces): 
$\mathbb{R}^{2n}$ \cite{GVH1, GVH2, Gotay99}, 
$\mathbb{S}^2$ \cite{Gotay96}, 
$\mT^*\mathbb{S}^1$ \cite{Gotay97} which do exhibit the Multiple Choice Problem,
and $\mathbb{T}^2$ \cite{Gotay95} which does not; 
$\mT^*\mathbb{R}_+$ admits a polynomial quantization \cite{GotayR+}.

\mbox{ } 

\noindent This settles that MCP does happen for some RPM's, which, conversely, pose a number of interesting extensions to the Multiple Choice Problem (MCP) 
analysis so far.  

\mbox{ }

\noindent{\bf Scaled RPM}

         \underline{\hspace{1.5in}     $N = 3$                      \hspace{1.6in}      $N \geq 4$ \hspace{1in}   } 
		 
$d$ = 1  \hspace{1.1in}     Phase = $\mathbb{R}^4$                  \hspace{1.2in}      Phase = $\mathbb{R}^{2n}$

         \hspace{2.2in}                  All of these exhibit MCP

$d$ = 2  \hspace{1.1in}     Phase = $\mathbb{R}^6$                  \hspace{1.2in}      Phase = T$^*$C($\mathbb{CP}^{n - 1}$)

         \hspace{1.47in}     Exhibits MCP                            \hspace{1.05in}      MCP status unknown: {\bf Question 98}

\mbox{ } 
					
\noindent{\bf Pure-shape RPM}

\underline{         \hspace{1.4in}     $N = 3$                      \hspace{1.5in}      $N \geq 4$ \hspace{1in}   }

$d$ = 1  \hspace{1.0in}     Phase = $\mT^*\mathbb{S}^1$             \hspace{1.2in}      Phase = $\mT^*\mathbb{S}^{n - 1}$

         \hspace{1.4in}     Exhibits MCP                            \hspace{1.17in}      MCP status unknown: {\bf Question 99} 
                  
         \hspace{1.38in}     but is relationally trivial

$d$ = 2  \hspace{1.0in}     Phase = $\mT^*\mathbb{S}^2$             \hspace{1.2in}      Phase = T$^*$($\mathbb{CP}^{n - 1}$)

\hspace{1.4in}              MCP status unknown                      \hspace{0.73in}      MCP status unknown: {\bf Question 100}

\hspace{1.36in}   (subcase of Question 99)
					  
\mbox{ }

\noindent{\bf Question 101}  What is known about MCP for systems that retain linear constraints at the quantum level?

\subsubsection{Major Open Question VI: Global Problems of Time}

\noindent Classical meshings of charts are easy, meshings of classical p.d.e. solutions are not, meshings of QM unitary evolutions are probably not even conceptually understood yet.  

\mbox{ }  

\noindent Global POT's are common facet by facet and strategy by strategy, consult the lists in Secs \ref{GP5}, \ref{GlobStrat}, \ref{QM-Global} and \ref{Global}, 
noting that the Machian-Classical, Semiclassical, Histories, Paths and Records approaches central to this Conclusion.  
Moreover, combinations of strategies look to be more-than-additively vulnerable.
For instance, the Magic Sword has a lot of global problems: the combination of 3 mutually-supporting approaches is a local stabilization, 
but the global problems with each strand add up.
And in fact the combination gets extra global problems beyond those with each strand: 
by involving a $t$-integral, Halliwell's object is not local in time, which would however be a desirable property in a practically useable beable.
Moreover, it has to be from $t$ = $-\infty$ to $+\infty$ in order to commute with $\scH$, without which one is left bereft of both classical and quantum Dirac beables.

\mbox{ }

\noindent RPM's and other comparably-simple models may serve to get to grips with the multiple subfacets of the Global POT \cite{AGlob}.

\subsubsection{Major Open Question VII.  Topos foundations for Timeless as well as Histories Approaches}\label{VII}

\noindent One way out of the Combined Approach's Caveat 1) is that QM may require modifying such as to obey an intuitionistic logic 
\cite{BI, ToposI, ToposTalk, ToposRev}, which {\sl is} distributive. 
I also note that regions are conceptually different in this kind of approach (they are now primary themselves rather than `made up of' 
some notion of points that are held to be primary), so that this possibility lies outside the scope of the present Article. 
It is also worth mentioning that [in e.g. (\ref{CPI}), \cite{IL}] proposition--projector association makes better sense than evoking regions, at the quantum level.
This covers e.g. the case of the IL approach to Histories Theory (\cite{IL2, IL} see also \cite{FloriTh, Flori11}), the \CPI 
[c.f. (\ref{CPI})], or Records within Histories theory \cite{GMH, H99}. 
The main atemporal part of Isham and Doering \cite{ToposI, ToposTalk, ToposRev} could also be viewed an underlying structure for Records Theory. 
I am working with Cecilia Flori on this \cite{AF13}.  

\mbox{ }

\noindent Level 4 Mach--Mackey scheme for RPM POT's would then be a topos-rooted combined scheme with globalities for the structures at all levels sorted out and the 
Multiple Choice Problem addressed.  

\mbox{ } 

\noindent {\bf Question 102}  In this case, do not presently even have $\mathbb{R}$, $\mathbb{R}^n$ in the standard/absolutist interpretation 
(the latter has the added complication of the angular momentum operators in the kinematical quantization scheme). 

\mbox{ } 

\noindent {\bf Question 103} Make $\FrQ$ curved; for this, $\mathbb{S}^2$ is a convenient example.  

\mbox{ } 

\noindent {\bf Question 104} Interpret $\mathbb{R}^n$, $\mathbb{R}^3$ cases from a whole-universe relational perspective.

\mbox{ } 

\noindent {\bf Question 105} Consider cases with linear constraints, e.g. $\mathbb{R}^4$ configuration space with $\scL$ and/or $\scD$.

\mbox{ }

\noindent {\bf Question 106} For RPM completeness, deal with $\mathbb{CP}^{n - 1}$ and C($\mathbb{CP}^{n - 1}$) cases from this perspective also.  

\mbox{ }

\noindent {\bf Question 107$^{**}$}  (is a long shot.) Does this Topos Approach shed any light on the Gronewold-van Hove phenomenon that underlies most of the Multiple Choice Problem?   

\mbox{ }

\noindent {\bf Question 108} Give a theory of the history of a quadrilateral.  

\mbox{ } 

\noindent {\bf Question 109$^*$}  Give the quadrilateral as an example of combined Halliwell-type scheme.

\subsection{This Article and Nododynamics/LQG} 

\noindent Involving Ashtekar Variables offer various {\sl technical advantages} in setting up QG \cite{Ashtekar, Thiemann}.

\noindent From a relational perspective, Nododynamics is a better name than Loop Quantum Gravity (LQG).

\noindent Also from a foundational perspective, most uses of the word `relational' in the current Nododynamics/LQG are what this Article terms `perspectival'.
I argued in this Article that the Ashtekar Variables formulation of GR is relational in Barbour's sense too 
(at the usual metric level; whether background-independence and relationalism should be taken to apply at the topological level as well remains a question largely for the 
future which may well take the line of thinking beyond Nododynamics into some new arena).  
However, so far the Nododynamics/LQG literature does not systematically apply LMB(-CA) relationalism 
(though there is some evidence for individual pieces of such thinking in various formalisms and techniques used).
Doing so could give the subject more conceptual awareness.  

\noindent I also argued that Geometrodynamics has a particularly solid foundation resting on primality of spatial geometry, which itself already applies to mechanics, 
whilst Ashtekar variables has a distinct one resting on electromagnetism and Yang--Mills theory. 
Moreover deriving Ashtekar variables rests upon extending the set of configurations and upon performing a canonical transformation.    
I provided moderate arguments for alternatives to each of these practises.
  
\noindent Finally, do Ashtekar Variables programs' advantages offer any better ways round the actual salient problems with geometrodynamics exposed by this paper? 
I.e. the Major Open Questions I to VII.  
The Thin Sandwich Problem has known flexibility under canonical transformations.
Schroeren's work \cite{Schroeren} for now ends one step away from beginning to check II, III for LQC at least, whilst Bojowald's work \cite{Bojo12} begins to consider IV.   

\subsection{This Article and M-Theory} 

Background-independence brings a POT, so it it is exceedingly likely that there will be one of those in M-Theory too.  
Geometrodynamics (very easily generalizable to arbitrary-$d$) could be seen as a first toy model of spatially 10-$d$ M-Theory.  
There may be better RPM's and Geometrodynamics models for this particular setting, e.g. supersymmetric RPM's and the geometrodynamics corresponding to 
spatially 10-$d$ Supergravity (which is a low-energy/`semiclassical' limit of M-theory). 
Moreover, many features of the POT are universal, so these will occur here too unless features such as supersymmetry and/or extended objects specifically conspire to eliminate facets.  
They were not however designed to do so, so this is a more stringent theoretical test for these notions/structures.   
Good physical theories work no matter how one approaches them [Criterion 1)], rather than just have the aspects and properties they were constructed for. 
It would thus be a pleasant surprise if, in addition to all the different usefulnesses of these structures, they also happened to cure the POT. 
But it is also possible that this will not happen, or that the POT may require further or somewhat different structures.

\mbox{ } 

\noindent Passing from studying particle mechanics to RPM's is, in a sense, conceptually similar to 
passing from studying point particles to strings; however in the RPM case, one proceeds here by making less rather than more hypotheses. 
Technically they are mostly very different, though, as this Article has pointed out, orbifolds and weighted projective spaces occur in both. 
\noindent In this Article, considering String Theory led to a refinement of how Barbour-type-relationalism should be defined.  
Note also how dimension ceases to be a primary input in RPM's, for all that this only so for particle number.
Finally, supersymmetry \cite{{Arelsusy}} and the situation with extended objects that occupy/furnish a multiplicity of notions of space are interesting frontiers in which to 
further explore the relational postulates.  

\mbox{ } 

\noindent \cite{APOT} listed a few approaches to time in M-Theory, though such works are as yet few and far between.  
As regards POT (and assuming Supersymmetry, Supergravity and M-Theory are all suitably relational for the current article's program to make sense for these theories), particular issues are as follows.
Sandwich features are not robust to theory change, so we might just `get lucky'.  
Are not near having a Histories Theory formulation of M-Theory.
Supergravity is a semiclassical limit of M-Theory; one would then need further semiclassical approximations on that (e.g. WKB).
There is some chance of getting a better-behaved (semiclassical) quantum trident e.g. by anomaly-cancelling properties of the supersymmetry.

\subsection{Epilogue: POT as only a part of the GR-QM inconsistency}\label{ObsApp}

Some observer issues have already been touched upon this Article through study of perspectivalism.    
Some further issues are as follows (I do not claim completeness, and these may in some ways reinforce the form of perspectivalism).  

\noindent Observers are modelled in GR as negligible energy-momentum entities, whilst, on the other hand, in QM, these are usually held to be much larger than the system in question.
Thus the theory of the large and of the small is not just a straightforward comparison but also a comparison of each with the sizes and sensitivities of the observers.  
GR observers are moreover idealized as regards their internal constitution is not posited; 
this is also the case for any clocks involved (this is also true in the below-mentioned case of rods in GR). 
As such one should believe little in these idealizations.  
Is the QM way of handling these things extendible to subsystems within the GR setting? 

\noindent More generally, QG may be a {\it perspectival theory} with both observers and their times being required in the posing and answering of questions.  

\mbox{ }

\noindent {\bf Question 110$^*$}. Aside from time, clocks and observers, are there any other kinds of conceptual clashes between QM and GR/due to Background Independence 
that are not readily attributable just to technical differences?  

\mbox{ }

\noindent Some answers to this are space, length, rods and frames.  
Though some of these do enter some POT aspects, they can also be considered for their own sake, which is not the same perspective.  
Kochen's \cite{Kochen} or Bene's \cite{Bene} brands of perspectival QM involve quantum reference systems.

\mbox{ }

\noindent {\bf Space} \cite{JammerSpace} is in a number of ways rather less controversial than time.  
It is time and not space or length that has a peculiar role in QM.  
But then GR looks to put time and space on an equal footing.  
Choices then are to challenge the peculiarities of time in QM or challenge the equal footing of GR. 
Tempus ante quantum and Semiclassical Approaches go for the latter, whilst abolishing time goes for the former.

\mbox{ } 

\noindent There is a Newtonian Mechanics to SR gap in the theory of {\bf rods} (rigid rods are meaningless in relativity) and the actual nature of rods is ignored in setting up GR.
These are arguably a less deep notion/apparatus clocks e.g. via Bondi's argument \cite{Bondi59} that they are made out of quantum matter which is ultimately underlaid by frequencies.  
Also, by their nature and function, rods are necessarily macroscopic \cite{SW} and so interact with their 
environment in uncontrollable ways, whilst microscopic clocks are indeed possible.  

\mbox{ }

\noindent Linking Theory \cite{Kos1, GrybTh, GK12} and Doubly General Relativity \cite{GomKos12} are relevant here as examples of attempting to treat time and scale 
on a more equal footing.

\mbox{ }

\noindent Finally, paralleling the question of what is a good clock, what is a good rod?  
It should be at rest in an unambiguous frame.  
At a fixed temperature, with high control of the temperature/a low thermal expansion. 
To measure elsewhere, we copy the master standard and take it with us (with less control).
It is worth mentioning also that in the Gambini--Porto--Pullin scheme \cite{PGP1, PGP2}, decoherence effects due to use of real clocks 
are accompanied likewise by decoherence effects due to use of real rods.  

\vspace{10in}

\appendix\section{Fonts}

\noindent {\bf The 5 basic fonts for different functional dependencies}

\mbox{ } 

\noindent $ABCDEFGHIJKLMNOPQRSTUVWXYZ$ $abcdefghijklmnopqrstuvwxyz$ (slanty Roman font) for finite-model primary quantities and functions thereof. 

\noindent ABCDEFGHIJKLMNOPQRSTUVWXYZ abcdefghijklmnopqrstuvwxyz (Straight Roman font) field theory primary quantities.

\noindent {\sffamily ABCDEFGHIJKLMNOPQRSTUVWXYZ abcdefghijklmnopqrstuvwxyz} (sans serif font) for portmanteaux of last two.

\noindent ${\cal ABCDEFGHIJKLMNOPQRSTUVWXYZ}$ (slightly curly calligraphic font) for functionals. 

\noindent {\calligra ABCDEFGHIJKLMNOPQRSTUVWXYZ abcdefghijklmnopqrstuvwxyz} (very curly calligraphic font)  for function-functional portmanteaux.

\mbox{ }

\noindent Underline denotes space vector, with double-underline denoting space matrix.  
Bold jointly denotes configuration space vector/matrix 

\mbox{ }

\noindent {\bf The further theoretical structure's 5 undersized fonts}

\mbox{ } 

\noindent $\mbox{\scriptsize${\cal ABCDEFGHIJKLMNOPQRSTUVWXYZ}$}$ (undersized slightly curly calligraphics) for constraints.

\noindent $\mbox{\scriptsize{\sffamily ABCDEFGHIJKLMNOPQRSTUVWXYZ abcdefghijklmnopqrstuvwxyz}}$ (undersized sans-serif font) for conserved quantities.

\noindent {\scriptsize $ABCDEFGHIJKLMNOPQRSTUVWXYZ$ $abcdefghijklmnopqrstuvwxyz$}  (undersized slanty Romans) for observables/beables 

\noindent /fashionables/degradeables.

\noindent {\scriptsize\tt $ABCDEFGHIJKLMNOPQRSTUVWXYZ$ $abcdefghijklmnopqrstuvwxyz$} (undersized typeface) for kinematical quantization algebra objects 
that are not conserved quantities and are not necessarily observables/beables/fashionables/degradeables.

\noindent {\scriptsize ABCDEFGHIJKLMNOPQRSTUVWXYZ abcdefghijklmnopqrstuvwxyz} (undersized straight Romans) for POT facets and quantization steps, 
used in discussing how to order these procedures in map language.

\mbox{ } 

\noindent The above are united as objects that obey algebras above and beyond those of one's theory's $\fQ$'s and $\fP$'s.  
Which of the {\sl previous} 5 types these are is emphasized upon introducing each such object. 

\mbox{ } 

\noindent{\bf Fonts for mathematical spaces} 

\mbox{ }

\noindent $\mathfrak{ABCDEFGHIJKLMNOPQRSTUVWXYZ abcdefghijklmnopqrstuvwxyz}$ (Mathfrak font) for spaces.  

\noindent $\mathbb{ABCDEFGHIJKLMNOPQRSTUVWXYZ}$ (boldmath font) for individual examples of topological/Riemannian manifolds.  

\noindent \textgoth{ABCDEFGHIJKLMNOPQRSTUVWXYZ} (capital Gothic letters) for categories.

\noindent Oversized capital Gothic letters for functors.   

\mbox{ }

\noindent {\bf The reserve font (typeface):} {\tt $ABCDEFGHIJKLMNOPQRSTUVWXYZ$ $abcdefghijklmnopqrstuvwxyz$}

\mbox{ } 

\noindent This is used to mop up the plethora of slanty symbols needed for this Article's exposition of RPM's without quite so much use of barring, priming, subscripting etc.  
In particular, my geometrically-natural (as opposed to Barbour's mechanically-conventional) picture for the pure-shape theory/part uses the obvious typeface counterparts.

\mbox{ } 

\noindent {\bf The few exceptions to the font rules} 

\mbox{ } 

\noindent ${\cal R}$ for stereographic radius and multipolar radial variables ($\bar{\ttp}$-indexed).  
${\cal U} = 1/{\cal R}$ extension of the former.  

\noindent I, $\mI^i$, $\mN^i$ and $\mn^i$ (MOI: moment of inertia, partial MOI, normalized partial MOI, square root of normalized partial MOI) 
to not confuse with finite instant $I$, particle number $N$ and $n = N - 1$.  

\noindent S for reduced shape variables, since $S$ is already in use for scalefree variables.

\vspace{10in}

\section{Acronyms}

\noindent A     as a prefix on Principles of Dynamics entities stands for `almost'.

\noindent ABFO   A--Barbour--Foster--O Murchadha (alternative relational theory of conformal 3-geometry)

\noindent ABFKO  A--Barbour--Foster--Kelleher--O Murchadha (relational variational formulation of York's Conformogeometrodynamics initial value problem)

\noindent ADM    Arnowitt--Deser--Misner (formulation of geometrodynamics)

\noindent BFO-A  Barbour--Foster--O Murchadha--A (relational formulation of geometrodynamics)

\noindent BSW    Baierlein--Sharp--Wheeler 

\noindent DOTI   Differential of the Instant

\noindent EFE's   Einstein Field Equations

\noindent FEP     Free end point

\noindent FESH    Free end spatial hypersurface

\noindent GLET    Generalized local ephemeris time

\noindent HJE     Hamilton--Jacobi equation

\noindent HKT    Hojman--Kucha\v{r}--Teitelboim

\noindent HPO     Histories Projection Operator (a particular type of quantum Histories Theory)

\noindent IL      Isham--Linden approach to classical and quantum Histories Theory.

\noindent IVP     Initial-value problem

\noindent JBB     Jacobi--Barbour--Bertotti (action, time)

\noindent LFE     Lapse-fixing equation

\noindent LMB     Leibniz--Mach--Barbour (relational program)

\noindent MRI     Manifest reparametrization invariance

\noindent MPI     Manifest parametrization irrelevance

\noindent NOS     Notion of space

\noindent OA       ordial almost. (O(A)) means ` any of ordial almost, almost or plain'

\noindent OOTI    Ordial of the Instant

\noindent POTI    Partial-differential of the Instant

\noindent PPSCT   Parageodesic Principle Split Conformal Transformation

\noindent POT     Problem of Time

\noindent POZIN   Problem of Zeros, Infinities and Non-smoothnesses

\noindent QG      Quantum Gestalt (of Relativistic Gravitation and Background Independence)

\noindent RPM     Relational particle mechanics

\noindent SM      Statistical Mechanics

\noindent STLRC   Sufficient totality of locally-significant change

\noindent TDSE    Time-dependent Schr\"{o}dinger equation

\noindent TDWE    Time-dependent wave equation

\noindent TISE    Time-independent Schr\"{o}dinger equation

\noindent VOTI    Velocity of the Instant (relational analogue of Lapse)

\mbox{ }

\noindent{\bf Extensions of previous Acronyms}

\mbox{ } 

\noindent DOTIFE   DOTI-fixing equation, similarly for OOTIFE, POTIFE, VOTIFE.

\noindent ERPM     Euclidean RPM

\noindent FENOS    Free end NOS

\noindent LMB-CA   My extension of LMB, including paying closer attention to Clemence's work.  

\noindent SRPM     Similarity RPM

\vspace{10in}
    
\section{Some Classical Dynamics results} \label{B}

\subsection{The general mechanics problem in arbitrary dimension in Cartesian coordinates}\label{Land}

We assume a Jacobi-type action (though that is not necessary so as to get all of these results).  
The kinetic term is
\beq
T = \sum\mbox{}_{\mbox{}_{\mbox{\scriptsize \sfA = 1}}}^k \dot{X}^{\sfA\, 2}/2 \mbox{ } . 
\eeq
Then the conjugate momenta are  
\beq 
P_{\sfA} = X_{\sfA}^{*} \mbox{ } .
\eeq
These momenta obey as a primary constraint the quadratic `energy equation'
\beq
\scE := \sum\mbox{}_{\mbox{}_{\mbox{\scriptsize \sfA = 1}}}^k P_{\sfA}^2/2 + V = E
\mbox{ } ,   
\label{QConst}
\eeq
the middle expression of which also serves as the classical Hamiltonian $H$ for the system. 
The evolution equations are
\beq
X_{\sfA}^{**} = - \pa V/\pa X^{\sfA} \mbox{ } , 
\eeq
one of which may be replaced by the Lagrangian form of (\ref{QConst}), 
\beq
\sum\mbox{}_{\mbox{}_{\mbox{\scriptsize \sfA = 1}}}^k {X}^{\sfA *\,2}/2 + V = E \mbox{ } . 
\eeq

\subsection{The general 2-$d$ mechanics problem in polar coordinates}\label{Hope}

The kinetic term is
\beq
T = \{\dot{R}^2 + R^2\dot{\chi}^2\}/2 \mbox{ } . 
\eeq
Then the conjugate momenta are  
\beq 
P_{R} = R^*   \mbox{ } , \mbox{ } \mbox{ } 
P_{\chi} = R^2\chi^*  = \sfS \mbox{ } , \mbox{ } \mbox{ constant if $V$ is independent of $\chi$ } . 
\eeq
These momenta obey as a primary constraint the quadratic `energy equation'
\beq
\scE := \{P_R^2 + P_{\chi}^2/R^2\}/2 + V  = E
\mbox{ } ,   
\label{QConst2}
\eeq
the middle expression of which also serves as the classical Hamiltonian $H$ for the system. 
The evolution equations are
\beq
R^{**}  - R \,\chi^{*\,2} = - \pa V/\pa R \mbox{ } , \mbox{ } \mbox{ } {\cal S} := \{R^2\chi^*\}^* = - \pa V/\pa \chi \mbox{ } ,
\eeq  
one of which may be replaced by the Lagrangian form of (\ref{QConst2}), 
\beq
\{R^{*\,2} + R^2\chi^{*\,2} \}/2 + V = E \mbox{ } . 
\eeq
If ${\cal S}$ is a conserved quantity, one can rewrite the last equation as 
\beq
\{R^{*\,2} + \sfS^2/R^2 \}/2 + V = E \mbox{ } , 
\eeq
which can be integrated to give the quadrature
\beq
t^{\se\sm} - t^{\se\sm}(0) = \int\d R/\sqrt{2\{E - V(R)\} - \sfS^2/R^2} \mbox{ } .
\eeq
One can also readily obtain a quadrature for the shape of the orbit (for the nontrivial case of $\sfS \neq 0$): 
\beq
\phi - \phi(0) = \sfS\int\d R/R^2\sqrt{2\{E - V(R)\} - \sfS^2/R^2} \mbox{ } .
\eeq

\subsection{The general 3-$d$ mechanics problem in spherical polar coordinates}\label{Glory}

The kinetic term is
\beq
T = \{  \dot{R}^2 + R^2  \{\dot{\alpha}^2 + \mbox{sin}^2\alpha  \,  \dot{\chi}^2  \} \}/2 \mbox{ } . 
\eeq
Then the conjugate momenta are  
\beq 
P_{R} = R^*                                \mbox{ } , \mbox{ } \mbox{ } 
P_{\alpha} = R^2\alpha^*                    \mbox{ } , \mbox{ } \mbox{ }
P_{\chi}   = R^2\mbox{sin}^2\alpha\chi^*  = \sfS_3 \mbox{ } , \mbox{ } \mbox{ constant, if $V$ is independent of $\chi$ } . 
\eeq
These momenta obey as a primary constraint the quadratic `energy equation'
\beq
\scE := \{P_R^2 + \{P_{\alpha}^2 + P_{\chi}^2/\mbox{sin}^2\alpha\}/R^2\}/2 + V  = E
\mbox{ } ,   
\label{QConst3}
\eeq
the middle expression of which also serves as the classical Hamiltonian $H$ for the system. 
The evolution equations are
\beq
\Last\Last R - R \{\{\Last \alpha\}^2 + \mbox{sin}^2\alpha \{\Last\chi\}^2\} = - \pa V/\pa R \mbox{ } , 
\eeq
\beq
\Last\{R^2\mbox{sin}^2\alpha \Last \chi\} = - \pa V/\pa \chi \mbox{ } , 
\eeq
\beq
\Last\{R^2\Last \alpha\} - R^2\mbox{sin}\,\alpha\,\mbox{cos}\,\alpha \{\Last\chi\}^2 = -\pa V/\pa\alpha 
\mbox{ } ,
\eeq
one of which may be replaced by the Lagrangian form of (\ref{QConst3}), 
\beq
\{\{\Last R\}^2 + R^2\chi^{*\,2} + \mbox{sin}^2\alpha \alpha^{*\,2}\}/2 + V = E \mbox{ } . 
\eeq
If all three $\sfS_{\Gamma}$ are conserved quantities, one can rewrite this last equation as 
\beq
\{  R^{*\,2} + \sfT\sfO\sfT//R^2 \}^{2}/2 + V = E \mbox{ } ,
\eeq
which can be integrated to give the quadrature
\beq
t^{\se\sm} - t^{\se\sm}(0) = \int\d R/\sqrt{2\{E - V(R)\} - \sfT\sfO\sfT/R^2} \mbox{ } .
\eeq
Finding the form of the orbits is somewhat less straightforward in this case.

\subsection{The general $\mathbb{S}^2$ mechanics in spherical coordinates}\label{DynSphe}

The kinetic term is 
\beq
\ttT = \{\dot{\alpha}^3 + \mbox{sin}^2\alpha\,\dot{\chi}^2\}/2 \mbox{ } .  
\eeq
Then the conjugate momenta are  
\beq
p_{\alpha} = \alpha^{\star} \mbox{ } \mbox{ and } \mbox{ } p_{\chi} = \mbox{sin}^2\alpha\,\chi^{\star} \mbox{ } .
\eeq
The $p_{\Gamma} = u^{\Gamma\,{\star}}$ conjugate to the $u^{\Gamma}$ are also used below.  
These momenta obey as a primary constraint the quadratic `energy equation'
\beq
\scE := {p_{\alpha}}^2/2 + {p_{\chi}}^2/2\,\mbox{sin}^2\alpha + \ttV(\alpha, \chi) = \ttE \mbox{ } ,
\label{LAF}
\eeq
the middle expression of which also serves as the classical Hamiltonian $H$ for the system.

The equations of motion for the general sphere are
\beq
\alpha^{{{\star}{\star}}} - \mbox{sin}\,\alpha\,\mbox{cos}\,\alpha\, \chi^{{{\star}}2} = 
- {\pa{\ttV}}/{\pa\alpha} \mbox{ } \mbox{ } , \mbox{ } \mbox{ } 
\{\mbox{sin}^2\alpha\, \chi^{{{\star}}}\}^{{{\star}}} = - {\pa{\ttV}}/{\pa\chi}
\label{Su} \mbox{ } ,
\eeq
one of which can be replaced by the Lagrangian form of (\ref{LAF}), 
\beq
{\alpha^{{{\star}}2}}/{2} + {\mbox{sin}^2\alpha\,\chi^{{{\star}}2}}/{2} + \ttV = \ttE \mbox{ } .
\label{MPW}
\eeq
The general conserved quantity is (for $\fV$ $\chi$-independent) 
\beq
p_{\chi} = \mbox{sin}^2\alpha\,\chi^{\star} = \sfS_3 \mbox{ } .  
\label{CF}
\eeq
If V is $\alpha$-independent also, $\sfT\sfO\sfT = \sum_{\Gamma = 1}^3\sfS_{\Gamma}\mbox{}^2$ is conserved as well.  
Combining (\ref{MPW}) and (\ref{CF}) gives 
\beq
\alpha^{\star\,2}/2 + \sfT\sfO\sfT/2\,\mbox{sin}^2\alpha + \ttV = \ttE \mbox{ } ,
\eeq
and, so, integrating, the quadrature for the shapes of the orbits,  
\beq
\chi - \chi_0 = \int 
{\sfS}\,\d\alpha
\left/
\mbox{sin}\,\alpha\sqrt{2\{\ttE - \ttV(\alpha)\}\mbox{sin}^2\alpha - \sfT\sfO\sfT}  
\right.  
%
%
\mbox{ } ,
\label{quad}
\eeq
while integrating (\ref{MPW}) itself gives the quadrature for the time-traversal, 
\beq
t - t_0 = \int
\mbox{sin}\,\alpha\,\d\alpha
\left/
\sqrt{2\{\ttE - \ttV(\alpha)\} \mbox{sin}^2\alpha - {\sfT\sfO\sfT}  }   \right.  
\mbox{ } . 
\eeq

\vspace{10in}

\section{Some QM results}\label{D}

These are presented generally, though in each case a very common physical interpretation is attached. 
This Article makes use of physical interpretations other than the usual ones; these are in the main part of the Article.

\subsection{Commutation relations/kinematical quantization for $\mathbb{S}^2$}\label{Duel}

We have three $\mbox{\scriptsize $U$}^{\sfA}$ such that $\sum_{\sfA = 1}^3\mbox{\scriptsize $U$}^{\sfA \,2} = 1$ forming a $\mathfrakV = \mathbb{R}^3$ and 
$\FrG_{\scc\sa\sn} = \mbox{Isom}(\mathbb{S}^2) = SO(3)$ whose objects I denote by $\sfS_{\sfC}$. 
Thus, overall $\FrG_{\scc\sa\sn} \mbox{\textcircled{S}} \mathfrakV = SO(3) \mbox{\textcircled{S}} \mathbb{R}^3$.  
The commutation relations are 
\beq
[\widehat\sfS_{\sfA}, \widehat\sfS_{\sfB}] = i\hbar\epsilon_{\sfA\sfB}\mbox{}^{\sfC}\widehat\sfS_{\sfC} 
\mbox{ } , \mbox{ } \mbox{ } 
[\widehat{\mbox{\scriptsize $U$}}^{\sfA}, \widehat\sfS_{\sfB}] = i\hbar\epsilon^{\sfA}\mbox{}_{\sfB\sfC}\widehat{\mbox{\scriptsize $U$}}^{\sfC} \mbox{ } .  
\eeq

\subsection{Commutation relations/kinematical quantization for $\mathbb{R}^3$}\label{At}

We have three $\mbox{\scriptsize $U$}^{\sfA}$ forming $\mathfrakV = \mathbb{R}^3$, three $\mbox{\scriptsize $P$}_{\sfA}$ conjugate to these and three $SO(3)$ objects $\sfS_{\sfC}$, 
these last two triples forming $\FrG_{\scc\sa\sn} = \mbox{Isom}(\mathbb{R}^3) = \mbox{Eucl}(3)$.  
Thus overall $\FrG_{\scc\sa\sn} \mbox{\textcircled{S}} \mathfrakV = \mbox{Eucl}(3) \mbox{\textcircled{S}} \mathbb{R}^3$, the Heisenberg group.  
The commutation relations are
\beq
[\widehat{\mbox{\scriptsize $U$}}^{\sfA}, \widehat{\mbox{\scriptsize $P$}}_{\sfB}] = i\hbar\delta^{\sfA}\mbox{}_{\sfB} \mbox{ } , \mbox{ } \mbox{ }
\label{bog-standard}
\eeq
\beq
[\widehat\sfS_{\sfA}, \widehat\sfS_{\sfB}] = i\hbar\epsilon_{\sfA\sfB}\mbox{}^{\sfC}\widehat\sfS_{\sfC} 
\mbox{ } , \mbox{ } \mbox{ } 
[\widehat{\mbox{\scriptsize $U$}}^{\sfA}, \widehat\sfS_{\sfB} ] = i\hbar\epsilon^{\sfA}\mbox{}_{\sfB\sfC}\widehat{\mbox{\scriptsize $U$}}_{\sfC} 
\mbox{ } , \mbox{ } \mbox{ } 
[\widehat{\mbox{\scriptsize $P$}}_{\sfA}, \widehat\sfS_{\sfB} ] = i\hbar\epsilon^{\sfA}\mbox{}^{\sfC}\widehat{\mbox{\scriptsize $P$}}_{\sfC}  \mbox{ }  .  
\label{Squire}
\eeq

\subsection{Commutation relations/kinematical quantization for $\mathbb{R}^2$}\label{Devlan}

We now have two $\mbox{\scriptsize $U$}^{\sfA}$ forming $\mathfrakV = \mathbb{R}^2$, two $\mbox{\scriptsize $P$}_{\sfA}$ conjugate to these and one $SO(2)$ object $\sfS$, 
these last two items forming $\FrG_{\scc\sa\sn} = \mbox{Isom}(\mathbb{R}^2) = \mbox{Eucl}(2)$.  
Thus overall $\FrG_{\scc\sa\sn} \mbox{\textcircled{S}} \mathfrakV = \mbox{Eucl}(3) \mbox{\textcircled{S}} \mathbb{R}^2$
The commutation relations are (\ref{Squire}) with $\sfS_3 = \sfS$, (\ref{bog-standard}) and 
\beq
[\widehat{\mbox{\scriptsize $U$}}^1, \widehat\sfS] = - i\hbar\widehat{\mbox{\scriptsize $U$}}^2  \mbox{ } , \mbox{ } \mbox{ }
[\widehat{\mbox{\scriptsize $U$}}^2, \widehat\sfS] =   i\hbar\widehat{\mbox{\scriptsize $U$}}^1  \mbox{ } , \mbox{ } \mbox{ }
[\widehat{\mbox{\scriptsize $P$}}_1, \widehat\sfS] = - i\hbar\widehat{\mbox{\scriptsize $P$}}_2      \mbox{ } , \mbox{ } \mbox{ }
[\widehat{\mbox{\scriptsize $P$}}_2, \widehat\sfS] =   i\hbar\widehat{\mbox{\scriptsize $P$}}_1      \mbox{ } . 
\eeq

\subsection{TISE and Laplacian for $\mathbb{S}^2$} \label{LapSphe}

\beq
 \triangle_{\mathbb{S}^2}\Psi = 
2\{\fV - \fE\}\Psi/\hbar^2 
\eeq
for
\beq
\triangle_{\mathbb{S}^2} = \mbox{sin}^{-1}\alpha \, \pa_{\alpha}\{ \mbox{sin}\,\alpha {\pa}_{\alpha}\} + 
{\mbox{sin}^{-2}\alpha} \pa_{\chi^2} \mbox{ } .  
\eeq

\subsection{TISE and Laplacian for $\mathbb{R}^3$ in spherical polar coordinates} \label{R3Lap}

\beq
 \triangle_{\mathbb{R}^3}\Psi = 2\{V - E\}\Psi/\hbar^2 
\eeq
for
\beq
\triangle_{\mathbb{R}^3} = R^{-2} \{\pa_R \{ R^2 \pa_R \} + \triangle_{\mathbb{S}^2} \} \mbox{ } .  
\label{Goldf}
\eeq

\subsection{TISE and Laplacian for $\mathbb{R}^2$ in plane polar coordinates} \label{R2Lap}

\beq
\triangle_{\mathbb{R}^2}\Psi = 2\{V - E\}\Psi/\hbar^2
\eeq
for
\beq
\triangle_{\mathbb{R}^2} = R^{-1} \pa_R \{ R  \, \pa_R \} + R^{-2} \pa_{\chi^2} \mbox{ } .  
\label{Inger}
\eeq

\vspace{10in}

\section{Linear Methods used}\label{E}

\subsection{The Bessel equation}

\noindent The Bessel equation of order $p$,
\beq
v^2w_{,vv} + vw_{,v} +\{v^2 - p^2\}w = 0 \mbox{ } ,  
\label{Bess}
\eeq
is solved by the Bessel functions.
I denote Bessel functions of the first kind by $J_{\sp}(v)$.  
The family of equations 
\beq
x^2y_{,xx} + \{ 1 - 2\Bigalpha \} xy_{,x} + \{ \Bigalpha^2 + \Bigbeta^2\{ k^2 x^{2\normalbeta} - p^2\}\}y = 0
\eeq
map to the Bessel equation under the transformations $w = x^{-\normalalpha}y$ and $v = kx^{\normalbeta}$.  
The subcase of this with $\Bigalpha = 1/2$, $\Bigbeta = 1$ and $p = \ml + 1/2$ for $\ml \in \mathbb{N}$ are the well-known spherical Bessel functions \cite{AS}.

\subsection{Spherical harmonics and the (associated) Legendre equation}

The (3-$d$) spherical harmonics equation arising as angular part of e.g. hydrogen-type problem and of 3-$d$ isotropic HO.  
Its equation is
\beq
\mbox{sin}^{-1}\theta      \{    \mbox{sin}\,\theta \,  \mU_{,\theta}     \}_{,\theta} +  \mbox{sin}^{-2}\theta  \mU_{,\phi\phi} + \lambda \mU = 0
\eeq
It is straightforwardly separable by the ansatz
\beq
\mU = U(\theta)P(\phi)
\eeq
into simple harmonic motion for $\phi$ and 
\beq
\mbox{sin}^{-1}\theta\{\mbox{sin}\theta \mU_{,\theta}\}_{,\theta} + \{\nJ\{\nJ + 1\} - \nj^2\mbox{sin}^{-2}\theta\}\mU = 0 \mbox{ } ,
\label{stazi}
\eeq
which is equivalent, under the transformation $X = \mbox{cos}\theta$), to the associated Legendre equation
\beq
\{1 - X^2\}\mY_{,XX} - 2X\mY_{,X} + \{\mJ\{\mJ + 1\} - \mj^2\{1 - X^2\}^{-1}\}\mY = 0 \mbox{ } .
\label{Leg}
\eeq

This is solved by the associated Legendre functions $\mP_{\sJ}^{|\sj|}(X)$ (and unbounded second solutions), for J $\in \mathbb{N}_0$, j $\in \mathbb{Z}$, $|\mj| \leq$ J.
We use the standard convention that
\beq 
\mP^{\sj}_{\sJ}(X) = \{-1\}^{\sj}\{1 - X^2\}^{\frac{\sj}{2}}\frac{\d^{\sj}}{\d X^{\sj}}
\left\{
\frac{1}{2^{\sJ}\mJ !}\frac{\d^{\sJ}}{\d X^{\sJ}}\{X^2 - 1\}^{\sJ}
\right\} \mbox{ } , 
\eeq
by which 
\beq
\left\{ 
\sqrt{    \frac{2\mJ + 1}{2}  \frac{    \{\mJ - |\mj|\}!    }{    \{\mJ + |\mj|\}!    }    }
\mP^{|\sj|}_{\sJ}(X)
\right\}
\label{orthog}
\eeq
is a complete set of orthonormal functions for $X \in$ [--1, 1].  
We also require the recurrence relations \cite{GrRy, AS} 
\beq
X\mP^{|\sj|}_{\sJ}(X) = 
\frac{\{\mJ - |\mj| + 1\}\mP^{|\sj|}_{\sJ + 1}(X) + \{\mJ + |\mj|\}\mP^{|\sj|}_{\sJ - 1}(X)}{2\mJ + 1} 
\mbox{ } , 
\label{*1}
\eeq
\beq
\sqrt{1 - X^2}\mP_{\sJ}^{\sj - 1} = \frac{\mP^{\sj}_{\sJ - 1} - \mP^{\sj}_{\sJ + 1}}{2\mJ + 1} \mbox{ } .  
\label{**}
\eeq

\subsection{Ultraspherical harmonics and the (associated) Gegenbauer (ultraspherical) equation}

The \{k $>$ 3\}-$d$ ultraspherical harmonics equation arising as angular part of higher-$d$ problems (see Sec \ref{Q-ERPM1}) is 
straightforwardly separable by the ansatz-cum-change-of-variables into simple harmonic motion and a sequence of Gegenbauer problems.
The Gegenbauer equation 
\beq
\{1 - X^2\} \mY_{,XX} - \{2\lambda + 1\}X\mY_{,X} + \mJ\{\mJ + 2\lambda\}\mY = 0 
\label{Geg}
\eeq
is solved boundedly by the Gegenbauer Polynomials $\mC_{\sJ}(X;\lambda)$.    
Normalization for these is provided in e.g. \cite{AS, GrRy}; the weight function is 
$\{1 - X^2\}^{\lambda - {1}/{2}}$ between equal-$\lambda$ Gegenbauer polynomials.  
These furthermore obey the recurrence relations \cite{AS, GrRy}
\beq
X\mC_{\sJ}(X;\lambda) = \frac{  \{\mJ + 1\}\mC_{\sJ + 1}(X; \lambda) + 
\{2\lambda + \mJ - 1\}\mC_{\sJ - 1}(X; \lambda)}{2\{\mJ + \lambda\}} \mbox{ } , 
\label{GegRec1}
\eeq
\beq
\mC_{\sJ + 1}(X; \lambda) = \frac{\lambda\{\mC_{\sJ + 1}(X; \lambda + 1) - \mC_{\sJ - 1}(X; \lambda + 1)\}}
                            {\mJ + \lambda + 1}
\label{GegRec2} \mbox{ } .  
\eeq
The associated Gegenbauer equation 
\beq
\{1 - X^2\}\mY_{,XX} - \{2\lambda + 1\}X\mY_{,X} + \mJ\{\mJ + 2\lambda\}\mY - \mj\{\mj + 2\lambda - 1\}\{1 - X^2\}^{-2}\mY = 0 
\label{AssocGeg}
\eeq
is solved boundedly by the associated Gegenbauer functions $\mC_{\sJ}^{\sj}(X; \lambda)$.  
These are re-expressible in terms of Gegenbauer polynomials via \cite{Norway} 
\beq
\mC_{\sJ}^{\sj}(X; \lambda) \propto \{1 - X^2\}^{{\sj}/{2}}\mC_{\sJ - \sj}(X; \lambda/2 - 1 + \mj) \mbox{ } . 
\label{conv}
\eeq
With this conversion, the recurrence relations between Gegenbauer polynomials (\ref{GegRec1}, \ref{GegRec2}) turn out to suffice for this Article.  
For $\lambda = 1/2$, (\ref{Geg}) is the Legendre equation solved by $P_{\sJ}(X) = \mC_{\sJ}(X;1/2)$, 
(\ref{GegRec1}) becomes (\ref{*1}), and (\ref{AssocGeg}) becomes the associated Legendre equation 
(\ref{Leg}) solved by $\mP^{-\sj}_{\sJ} \propto \mC^{\sj}_{\sJ}(X; 1/2)$, by (\ref{conv}) and \cite{AS} 
\beq
\mC_{\sJ}(X;\lambda) \propto \{X^2 - 1\}^{{1}/{4} - {\lambda}/{2}}\mP^{{1}/{2} - \lambda}_{\sJ + \lambda - {1}/{2}}(X) \mbox{ } .  
\eeq

\subsection{(Associated) Laguerre equation and isotropic HO wavefunctions}

The associated Laguerre equation, 
\beq
x\my_{,xx} + \{\alpha + 1 - x\}\my_{,x} + \mn \, \my = 0 \mbox{ } , 
\label{Laguerre}
\eeq  
is solved by the associated Laguerre polynomials $\mL_{\sn}^{\alpha}(x)$ (and unbounded second solutions).  
They obey \cite{AS} the orthogonality relation
\beq
\int_0^{\infty}x^{\alpha}\mbox{exp}(- x)\mL^{\alpha}_{\sn}(x)\mL^{\alpha}_{\sn^{\prime}}(x)
\d x = 0 \mbox{ unless } \mn = \mn^{\prime} 
\eeq
and the recurrence relation
\beq
x\mL^{\alpha}_{\mn}(x) = \{2\mn + \alpha + 1\}\mL_{\sn}^{\alpha}(x) - 
                               \{\mn + 1\}          \mL_{\sn + 1}^{\alpha}(x) - 
                               \{\mn + \alpha\}     \mL^{\alpha}_{\sn - 1}(x) \mbox{ } .  
\eeq
The 2-$d$ quantum isotropic harmonic oscillator's radial equation for a particle of mass $\mu$ and classical oscillator frequency $\omega$, 
\beq
-\{{\hbar^2}/{2\mu}\}\{R_{,rr} + {R_{,r}}/{r} - {\nm^2R}/{r^2}\} + {\mu\omega^2r^2R}/{2} = E R \mbox{ } , 
\label{isoho}
\eeq
maps to the associated Laguerre equation under the asymptotically-motivated transformations 
\beq
R = 
\left\{
{\hbar x}/{\mu\omega}
\right\}^{{|\sm|}/{2}}
\mbox{e}^{- x/2}y(x) \mbox{ } , \mbox{ }  
x = {\mu\omega r^2}/{\hbar}
\label{isohotrans}
\eeq
and so is solved by 
\beq
R \propto r^{|\sm|}\mbox{e}^{\mu\omega r^2/2\hbar}L_{\sr}^{|\sm|}
\left(
\mu\omega r^2/{\hbar}
\right)
\label{isohosoln}
\eeq
corresponding to the discrete energies $E = \{|\nm| + 2\nr + 1\}\hbar\omega$ for radial quantum number $\nr \in \mathbb{N}_0$ \cite{Schwinger, Robinett}.

\subsection{The Tchebychev equation}

The Tchebychev equation 
\beq
\{1 - x^2\}\my_{xx} - x \my_x + \mn^2\my = 0  
\eeq
is solved by the Tchebychev polynomials T$_{\sn}(x) = \mbox{cos(\mn}\,\mbox{arccos}(x))$.

\vspace{10in}


\end{document}